\documentclass[12pt]{article}

\usepackage{amssymb,amsmath,amsfonts,eurosym,geometry,ulem,graphicx,caption,color,setspace,sectsty,comment,footmisc,caption,natbib,pdflscape,subcaption,array,hyperref, threeparttable,booktabs}

\usepackage{amsthm}
\usepackage{tcolorbox}
\tcbuselibrary{skins,breakable}
\usepackage{tikz}
\usepackage[T1]{fontenc}
\usetikzlibrary{arrows.meta}
\usetikzlibrary{fit, decorations.pathreplacing, positioning,arrows}

\newcolumntype{L}[1]{>{\raggedright\let\newline\\arraybackslash\hspace{0pt}}m{#1}}
\newcolumntype{C}[1]{>{\centering\let\newline\\arraybackslash\hspace{0pt}}m{#1}}
\newcolumntype{R}[1]{>{\raggedleft\let\newline\\arraybackslash\hspace{0pt}}m{#1}}
\theoremstyle{definition}
\newtheorem{definition}{Definition}
\usepackage{subcaption}
 
\newcommand{\kprob}{\kappa^{prob}}
\newcommand{\kdet}{\kappa^{fit}}
\newcommand{\rff}{r^{f\!f}}
\newcommand{\rprior}{r^{prior}}
\newcommand{\rprob}{r^{prob}}
\newcommand{\fbar}{\bar f}
\newcommand{\RM}{\mathcal{R}_{M,d}}
\newcommand{\SM}{\mathcal{S}_{M,d}}
\newcommand{\LM}{\Lambda_{d}} 
\newcommand{\PiM}{\Pi_{M,d}}

\newcommand{\E}{\mathbb{E}}

\newtheorem{proposition}{Proposition}

\theoremstyle{remark}

\tikzset{
  unitsq/.style={x=3.0cm,y=3.0cm,line cap=round,line join=round},
  frame/.style={black,thin},
  curveM/.style={black,thick},
  refpt/.style={draw=black,fill=white,thin},
  bead/.style={fill=black},
  distseg/.style={black,densely dashed,thin},
}
 
\begin{document}

\begin{titlepage}
\title{Comparing the Empirical Content of Economic Theories\thanks{I  would like to thank Andreas Bjerre-Nielsen and Rosa Lavelle-Hill for their insightful comments and valuable feedback on early drafts of this manuscript} \\ \Large Completeness, Restrictiveness, and Scope}
\author{Johanna Einsiedler\thanks{University of Basel, University of Copenhagen}}

\date{\today}
\maketitle
\begin{abstract}
A fitted theory that predicts well may do so because of the structure it commits to or because it was fitted to the data it is evaluated on. We propose measures that separate the two by operationalising a theory as a functional form, a support restriction on behavior and a distribution over the parameters, with no parameter fitted to the evaluation sample. Probabilistic completeness is the share of the predictable variation that a rule drawn from the theory captures out of sample. Probabilistic restrictiveness is how much conceivable behavior such a rule excludes; it separates into what the functional form can reach and how the parameter distribution concentrates on it. The parameter distribution can be interpreted as uncertainty about a representative agent or as the distribution of parameters across heterogeneous agents, and the measures adapt to both interpretations through the choice of scoring rule, baseline and ceiling. A scope layer prices the domains a theory claims and compares theories across domains. We apply the measures to expected utility, cumulative prospect theory and disappointment aversion on certainty equivalents for binary lotteries. The functional forms exclude little conceivable behavior, while the literature's parameter distributions exclude most of what the forms leave open. A rule drawn from a literature-informed prior keeps two thirds of fitted prospect theory's completeness, most of expected utility's and none of disappointment aversion's. The flexibility-fit trade-off found for functional forms disappears once the theories are held with their literature priors. Read as populations, disappointment aversion and expected utility cannot produce the dispersion the subjects show, and the literature priors are less well placed for losses than for gains.
\end{abstract}

\setcounter{page}{0}
\thispagestyle{empty}
\end{titlepage}
\pagebreak \newpage
\providecommand{\setres}[4]{\expandafter\gdef\csname res@\detokenize{#1@#2@#3}\endcsname{#4}}
\providecommand{\res}[3]{\ifcsname res@\detokenize{#1@#2@#3}\endcsname\csname res@\detokenize{#1@#2@#3}\endcsname\else\textbf{??}\fi}
\setres{gains}{COH_deterministic}{kappa}{1.000}
\setres{gains}{COH_deterministic}{kappa_se}{0.000}
\setres{gains}{COH_deterministic}{mse}{61.70}
\setres{gains}{COH_deterministic}{nrmse_pct}{16.9}
\setres{gains}{COH_deterministic}{r_ff}{0.046}
\setres{gains}{COH_deterministic}{r_prior}{1.0000}
\setres{gains}{COH_deterministic}{r_prob}{1.000}
\setres{gains}{COH_deterministic}{r_prob_se}{0.000}
\setres{gains}{COH_uniform}{kappa}{-1.434}
\setres{gains}{COH_uniform}{kappa_prior_mean}{0.068}
\setres{gains}{COH_uniform}{kappa_prior_mean_se}{0.037}
\setres{gains}{COH_uniform}{kappa_se}{0.109}
\setres{gains}{COH_uniform}{mse}{236.96}
\setres{gains}{COH_uniform}{mse_prior_mean}{128.81}
\setres{gains}{COH_uniform}{nrmse_pct}{17.9}
\setres{gains}{COH_uniform}{pred_variance}{108.15}
\setres{gains}{COH_uniform}{r_ff}{0.046}
\setres{gains}{COH_uniform}{r_ff_ev_check}{0.484}
\setres{gains}{COH_uniform}{r_ff_se}{0.001}
\setres{gains}{COH_uniform}{r_prior}{0.0000}
\setres{gains}{COH_uniform}{r_prior_se}{0.000}
\setres{gains}{COH_uniform}{r_prob}{0.046}
\setres{gains}{COH_uniform}{r_prob_se}{0.001}
\setres{gains}{COH_uniform}{share_refs_coherent}{0}
\setres{gains}{COH_uniform}{spread_penalty}{1.502}
\setres{gains}{COH_uniform}{spread_penalty_se}{0.111}
\setres{gains}{CPT}{det_pct}{97}
\setres{gains}{CPT}{gap_lit}{0.331}
\setres{gains}{CPT}{gap_uniform}{1.467}
\setres{gains}{CPT}{one_minus_rff}{0.860}
\setres{gains}{CPT}{passthrough_lit}{0.051}
\setres{gains}{CPT}{passthrough_uniform}{0.485}
\setres{gains}{CPT}{prior_contrib_lit}{0.816}
\setres{gains}{CPT}{prior_contrib_uniform}{0.443}
\setres{gains}{CPT}{share_lit}{66}
\setres{gains}{CPT}{share_uniform}{-51}
\setres{gains}{CPT_deterministic}{kappa}{0.972}
\setres{gains}{CPT_deterministic}{kappa_se}{0.008}
\setres{gains}{CPT_deterministic}{mse}{63.70}
\setres{gains}{CPT_deterministic}{nrmse_pct}{18.2}
\setres{gains}{CPT_deterministic}{r_ff}{0.140}
\setres{gains}{CPT_deterministic}{r_ff_se}{0.001}
\setres{gains}{CPT_deterministic}{r_prior}{1.0000}
\setres{gains}{CPT_deterministic}{r_prior_se}{---}
\setres{gains}{CPT_deterministic}{r_prob}{1.000}
\setres{gains}{CPT_deterministic}{r_prob_se}{0.000}
\setres{gains}{CPT_deterministic}{volume_ess}{---}
\setres{gains}{CPT_lit}{kappa}{0.641}
\setres{gains}{CPT_lit}{kappa_prior_mean}{0.853}
\setres{gains}{CPT_lit}{kappa_prior_mean_se}{0.035}
\setres{gains}{CPT_lit}{kappa_se}{0.050}
\setres{gains}{CPT_lit}{mse}{87.56}
\setres{gains}{CPT_lit}{mse_prior_mean}{72.28}
\setres{gains}{CPT_lit}{nrmse_pct}{19.6}
\setres{gains}{CPT_lit}{pred_variance}{15.28}
\setres{gains}{CPT_lit}{r_ff}{0.140}
\setres{gains}{CPT_lit}{r_ff_se}{0.001}
\setres{gains}{CPT_lit}{r_prior}{0.9486}
\setres{gains}{CPT_lit}{r_prior_se}{0.000}
\setres{gains}{CPT_lit}{r_prob}{0.956}
\setres{gains}{CPT_lit}{r_prob_se}{0.000}
\setres{gains}{CPT_lit}{spread_penalty}{0.212}
\setres{gains}{CPT_lit}{spread_penalty_se}{0.016}
\setres{gains}{CPT_lit}{volume_ess}{55{,}652}
\setres{gains}{CPT_uniform}{kappa}{-0.495}
\setres{gains}{CPT_uniform}{kappa_prior_mean}{0.811}
\setres{gains}{CPT_uniform}{kappa_prior_mean_se}{0.040}
\setres{gains}{CPT_uniform}{kappa_se}{0.129}
\setres{gains}{CPT_uniform}{mse}{169.35}
\setres{gains}{CPT_uniform}{mse_prior_mean}{75.31}
\setres{gains}{CPT_uniform}{nrmse_pct}{19.4}
\setres{gains}{CPT_uniform}{pred_variance}{94.04}
\setres{gains}{CPT_uniform}{r_ff}{0.140}
\setres{gains}{CPT_uniform}{r_ff_se}{0.001}
\setres{gains}{CPT_uniform}{r_prior}{0.5147}
\setres{gains}{CPT_uniform}{r_prior_se}{0.003}
\setres{gains}{CPT_uniform}{r_prob}{0.583}
\setres{gains}{CPT_uniform}{r_prob_se}{0.003}
\setres{gains}{CPT_uniform}{spread_penalty}{1.306}
\setres{gains}{CPT_uniform}{spread_penalty_se}{0.096}
\setres{gains}{CPT_uniform}{volume_ess}{55{,}652}
\setres{gains}{DA}{det_pct}{63}
\setres{gains}{DA}{gap_cpt_lit}{0.621}
\setres{gains}{DA}{gap_lit}{0.606}
\setres{gains}{DA}{gap_r_cpt_lit}{0.000}
\setres{gains}{DA}{gap_uniform}{1.453}
\setres{gains}{DA}{one_minus_rff}{0.719}
\setres{gains}{DA}{passthrough_lit}{0.062}
\setres{gains}{DA}{passthrough_uniform}{0.133}
\setres{gains}{DA}{prior_contrib_lit}{0.674}
\setres{gains}{DA}{prior_contrib_uniform}{0.623}
\setres{gains}{DA}{share_lit}{3}
\setres{gains}{DA}{share_uniform}{-132}
\setres{gains}{DA_deterministic}{kappa}{0.626}
\setres{gains}{DA_deterministic}{kappa_se}{0.062}
\setres{gains}{DA_deterministic}{mse}{88.64}
\setres{gains}{DA_deterministic}{nrmse_pct}{22.6}
\setres{gains}{DA_deterministic}{r_ff}{0.281}
\setres{gains}{DA_deterministic}{r_ff_se}{0.004}
\setres{gains}{DA_deterministic}{r_prior}{1.0000}
\setres{gains}{DA_deterministic}{r_prior_se}{---}
\setres{gains}{DA_deterministic}{r_prob}{1.000}
\setres{gains}{DA_deterministic}{r_prob_se}{0.000}
\setres{gains}{DA_deterministic}{volume_ess}{---}
\setres{gains}{DA_lit}{kappa}{0.020}
\setres{gains}{DA_lit}{kappa_prior_mean}{0.340}
\setres{gains}{DA_lit}{kappa_prior_mean_se}{0.113}
\setres{gains}{DA_lit}{kappa_se}{0.136}
\setres{gains}{DA_lit}{mse}{132.27}
\setres{gains}{DA_lit}{mse_prior_mean}{109.23}
\setres{gains}{DA_lit}{nrmse_pct}{23.9}
\setres{gains}{DA_lit}{pred_variance}{23.04}
\setres{gains}{DA_lit}{r_ff}{0.281}
\setres{gains}{DA_lit}{r_ff_se}{0.004}
\setres{gains}{DA_lit}{r_prior}{0.9383}
\setres{gains}{DA_lit}{r_prior_se}{0.000}
\setres{gains}{DA_lit}{r_prob}{0.956}
\setres{gains}{DA_lit}{r_prob_se}{0.000}
\setres{gains}{DA_lit}{spread_penalty}{0.320}
\setres{gains}{DA_lit}{spread_penalty_se}{0.024}
\setres{gains}{DA_lit}{volume_ess}{436{,}452}
\setres{gains}{DA_uniform}{kappa}{-0.827}
\setres{gains}{DA_uniform}{kappa_prior_mean}{0.060}
\setres{gains}{DA_uniform}{kappa_prior_mean_se}{0.149}
\setres{gains}{DA_uniform}{kappa_se}{0.212}
\setres{gains}{DA_uniform}{mse}{193.20}
\setres{gains}{DA_uniform}{mse_prior_mean}{129.35}
\setres{gains}{DA_uniform}{nrmse_pct}{24.1}
\setres{gains}{DA_uniform}{pred_variance}{63.85}
\setres{gains}{DA_uniform}{r_ff}{0.281}
\setres{gains}{DA_uniform}{r_ff_se}{0.004}
\setres{gains}{DA_uniform}{r_prior}{0.8666}
\setres{gains}{DA_uniform}{r_prior_se}{0.001}
\setres{gains}{DA_uniform}{r_prob}{0.904}
\setres{gains}{DA_uniform}{r_prob_se}{0.001}
\setres{gains}{DA_uniform}{spread_penalty}{0.887}
\setres{gains}{DA_uniform}{spread_penalty_se}{0.065}
\setres{gains}{DA_uniform}{volume_ess}{436{,}452}
\setres{gains}{EU}{det_pct}{62}
\setres{gains}{EU}{gap_lit}{0.095}
\setres{gains}{EU}{gap_uniform}{0.826}
\setres{gains}{EU}{one_minus_rff}{0.664}
\setres{gains}{EU}{passthrough_lit}{0.186}
\setres{gains}{EU}{passthrough_uniform}{0.422}
\setres{gains}{EU}{prior_contrib_lit}{0.541}
\setres{gains}{EU}{prior_contrib_uniform}{0.384}
\setres{gains}{EU}{share_lit}{85}
\setres{gains}{EU}{share_uniform}{-34}
\setres{gains}{EU_deterministic}{kappa}{0.616}
\setres{gains}{EU_deterministic}{kappa_se}{0.061}
\setres{gains}{EU_deterministic}{mse}{89.31}
\setres{gains}{EU_deterministic}{nrmse_pct}{23.0}
\setres{gains}{EU_deterministic}{r_ff}{0.336}
\setres{gains}{EU_deterministic}{r_ff_se}{0.005}
\setres{gains}{EU_deterministic}{r_prior}{1.0000}
\setres{gains}{EU_deterministic}{r_prior_se}{---}
\setres{gains}{EU_deterministic}{r_prob}{1.000}
\setres{gains}{EU_deterministic}{r_prob_se}{0.000}
\setres{gains}{EU_deterministic}{volume_ess}{---}
\setres{gains}{EU_lit}{kappa}{0.521}
\setres{gains}{EU_lit}{kappa_prior_mean}{0.611}
\setres{gains}{EU_lit}{kappa_prior_mean_se}{0.068}
\setres{gains}{EU_lit}{kappa_se}{0.074}
\setres{gains}{EU_lit}{mse}{96.17}
\setres{gains}{EU_lit}{mse_prior_mean}{89.72}
\setres{gains}{EU_lit}{nrmse_pct}{23.1}
\setres{gains}{EU_lit}{pred_variance}{6.45}
\setres{gains}{EU_lit}{r_ff}{0.336}
\setres{gains}{EU_lit}{r_ff_se}{0.005}
\setres{gains}{EU_lit}{r_prior}{0.8143}
\setres{gains}{EU_lit}{r_prior_se}{0.001}
\setres{gains}{EU_lit}{r_prob}{0.877}
\setres{gains}{EU_lit}{r_prob_se}{0.001}
\setres{gains}{EU_lit}{spread_penalty}{0.090}
\setres{gains}{EU_lit}{spread_penalty_se}{0.007}
\setres{gains}{EU_lit}{volume_ess}{172{,}086}
\setres{gains}{EU_uniform}{kappa}{-0.209}
\setres{gains}{EU_uniform}{kappa_prior_mean}{0.298}
\setres{gains}{EU_uniform}{kappa_prior_mean_se}{0.118}
\setres{gains}{EU_uniform}{kappa_se}{0.154}
\setres{gains}{EU_uniform}{mse}{148.75}
\setres{gains}{EU_uniform}{mse_prior_mean}{112.21}
\setres{gains}{EU_uniform}{nrmse_pct}{23.2}
\setres{gains}{EU_uniform}{pred_variance}{36.53}
\setres{gains}{EU_uniform}{r_ff}{0.336}
\setres{gains}{EU_uniform}{r_ff_se}{0.005}
\setres{gains}{EU_uniform}{r_prior}{0.5782}
\setres{gains}{EU_uniform}{r_prior_se}{0.001}
\setres{gains}{EU_uniform}{r_prob}{0.720}
\setres{gains}{EU_uniform}{r_prob_se}{0.002}
\setres{gains}{EU_uniform}{spread_penalty}{0.507}
\setres{gains}{EU_uniform}{spread_penalty_se}{0.037}
\setres{gains}{EU_uniform}{volume_ess}{172{,}086}
\setres{gains}{EV_deterministic}{kappa}{0.493}
\setres{gains}{EV_deterministic}{kappa_se}{0.064}
\setres{gains}{EV_deterministic}{mse}{98.19}
\setres{gains}{EV_deterministic}{nrmse_pct}{23.4}
\setres{gains}{EV_deterministic}{r_ff}{0.484}
\setres{gains}{EV_deterministic}{r_ff_se}{---}
\setres{gains}{EV_deterministic}{r_prior}{1.0000}
\setres{gains}{EV_deterministic}{r_prior_se}{---}
\setres{gains}{EV_deterministic}{r_prob}{1.000}
\setres{gains}{EV_deterministic}{r_prob_se}{0.000}
\setres{gains}{EV_deterministic}{volume_ess}{---}
\setres{gains}{reference}{lookup_nrmse_pct}{16.9}
\setres{gains}{reference}{midpoint_mse}{133.69}
\setres{gains}{reference}{midpoint_nrmse_pct}{27.2}
\setres{gains}{reference}{n_lotteries}{25}
\setres{gains}{reference}{n_obs}{4{,}459}
\setres{gains}{reference}{n_subjects}{179}
\setres{gains}{reference}{oracle_mse}{61.70}
\setres{signed}{CPT}{det_pct}{83}
\setres{signed}{CPT}{gap_lit}{0.810}
\setres{signed}{CPT}{gap_uniform}{2.065}
\setres{signed}{CPT}{one_minus_rff}{0.805}
\setres{signed}{CPT}{passthrough_lit}{0.095}
\setres{signed}{CPT}{passthrough_uniform}{0.467}
\setres{signed}{CPT}{prior_contrib_lit}{0.729}
\setres{signed}{CPT}{prior_contrib_uniform}{0.429}
\setres{signed}{CPT}{share_lit}{2}
\setres{signed}{CPT}{share_uniform}{-150}
\setres{signed}{CPT_deterministic}{kappa}{0.828}
\setres{signed}{CPT_deterministic}{kappa_se}{0.027}
\setres{signed}{CPT_deterministic}{mse}{77.11}
\setres{signed}{CPT_deterministic}{nrmse_pct}{18.3}
\setres{signed}{CPT_deterministic}{r_ff}{0.195}
\setres{signed}{CPT_deterministic}{r_ff_se}{0.002}
\setres{signed}{CPT_deterministic}{r_prior}{1.0000}
\setres{signed}{CPT_deterministic}{r_prior_se}{---}
\setres{signed}{CPT_deterministic}{r_prob}{1.000}
\setres{signed}{CPT_deterministic}{r_prob_se}{0.000}
\setres{signed}{CPT_deterministic}{volume_ess}{---}
\setres{signed}{CPT_lit}{kappa}{0.018}
\setres{signed}{CPT_lit}{kappa_prior_mean}{0.470}
\setres{signed}{CPT_lit}{kappa_prior_mean_se}{0.054}
\setres{signed}{CPT_lit}{kappa_se}{0.079}
\setres{signed}{CPT_lit}{mse}{132.69}
\setres{signed}{CPT_lit}{mse_prior_mean}{101.64}
\setres{signed}{CPT_lit}{nrmse_pct}{20.1}
\setres{signed}{CPT_lit}{pred_variance}{31.05}
\setres{signed}{CPT_lit}{r_ff}{0.195}
\setres{signed}{CPT_lit}{r_ff_se}{0.002}
\setres{signed}{CPT_lit}{r_prior}{0.9054}
\setres{signed}{CPT_lit}{r_prior_se}{0.001}
\setres{signed}{CPT_lit}{r_prob}{0.924}
\setres{signed}{CPT_lit}{r_prob_se}{0.001}
\setres{signed}{CPT_lit}{spread_penalty}{0.452}
\setres{signed}{CPT_lit}{spread_penalty_se}{0.033}
\setres{signed}{CPT_lit}{volume_ess}{52{,}676}
\setres{signed}{CPT_uniform}{kappa}{-1.238}
\setres{signed}{CPT_uniform}{kappa_prior_mean}{0.152}
\setres{signed}{CPT_uniform}{kappa_prior_mean_se}{0.056}
\setres{signed}{CPT_uniform}{kappa_se}{0.140}
\setres{signed}{CPT_uniform}{mse}{218.86}
\setres{signed}{CPT_uniform}{mse_prior_mean}{123.51}
\setres{signed}{CPT_uniform}{nrmse_pct}{20.3}
\setres{signed}{CPT_uniform}{pred_variance}{95.35}
\setres{signed}{CPT_uniform}{r_ff}{0.195}
\setres{signed}{CPT_uniform}{r_ff_se}{0.002}
\setres{signed}{CPT_uniform}{r_prior}{0.5327}
\setres{signed}{CPT_uniform}{r_prior_se}{0.003}
\setres{signed}{CPT_uniform}{r_prob}{0.624}
\setres{signed}{CPT_uniform}{r_prob_se}{0.002}
\setres{signed}{CPT_uniform}{spread_penalty}{1.389}
\setres{signed}{CPT_uniform}{spread_penalty_se}{0.100}
\setres{signed}{CPT_uniform}{volume_ess}{52{,}676}
\setres{signed}{DA}{det_pct}{58}
\setres{signed}{DA}{gap_cpt_lit}{1.256}
\setres{signed}{DA}{gap_lit}{1.820}
\setres{signed}{DA}{gap_r_cpt_lit}{-0.004}
\setres{signed}{DA}{gap_uniform}{1.799}
\setres{signed}{DA}{one_minus_rff}{0.665}
\setres{signed}{DA}{passthrough_lit}{0.121}
\setres{signed}{DA}{passthrough_uniform}{0.224}
\setres{signed}{DA}{prior_contrib_lit}{0.585}
\setres{signed}{DA}{prior_contrib_uniform}{0.516}
\setres{signed}{DA}{share_lit}{-212}
\setres{signed}{DA}{share_uniform}{-209}
\setres{signed}{DA_deterministic}{kappa}{0.583}
\setres{signed}{DA_deterministic}{kappa_se}{0.067}
\setres{signed}{DA_deterministic}{mse}{93.94}
\setres{signed}{DA_deterministic}{nrmse_pct}{22.0}
\setres{signed}{DA_deterministic}{r_ff}{0.335}
\setres{signed}{DA_deterministic}{r_ff_se}{0.003}
\setres{signed}{DA_deterministic}{r_prior}{1.0000}
\setres{signed}{DA_deterministic}{r_prior_se}{---}
\setres{signed}{DA_deterministic}{r_prob}{1.000}
\setres{signed}{DA_deterministic}{r_prob_se}{0.000}
\setres{signed}{DA_deterministic}{volume_ess}{---}
\setres{signed}{DA_lit}{kappa}{-1.238}
\setres{signed}{DA_lit}{kappa_prior_mean}{-0.834}
\setres{signed}{DA_lit}{kappa_prior_mean_se}{0.160}
\setres{signed}{DA_lit}{kappa_se}{0.187}
\setres{signed}{DA_lit}{mse}{218.87}
\setres{signed}{DA_lit}{mse_prior_mean}{191.15}
\setres{signed}{DA_lit}{nrmse_pct}{25.9}
\setres{signed}{DA_lit}{pred_variance}{27.72}
\setres{signed}{DA_lit}{r_ff}{0.335}
\setres{signed}{DA_lit}{r_ff_se}{0.003}
\setres{signed}{DA_lit}{r_prior}{0.8792}
\setres{signed}{DA_lit}{r_prior_se}{0.001}
\setres{signed}{DA_lit}{r_prob}{0.920}
\setres{signed}{DA_lit}{r_prob_se}{0.001}
\setres{signed}{DA_lit}{spread_penalty}{0.404}
\setres{signed}{DA_lit}{spread_penalty_se}{0.029}
\setres{signed}{DA_lit}{volume_ess}{277{,}116}
\setres{signed}{DA_uniform}{kappa}{-1.216}
\setres{signed}{DA_uniform}{kappa_prior_mean}{-0.128}
\setres{signed}{DA_uniform}{kappa_prior_mean_se}{0.114}
\setres{signed}{DA_uniform}{kappa_se}{0.188}
\setres{signed}{DA_uniform}{mse}{217.41}
\setres{signed}{DA_uniform}{mse_prior_mean}{142.69}
\setres{signed}{DA_uniform}{nrmse_pct}{23.8}
\setres{signed}{DA_uniform}{pred_variance}{74.72}
\setres{signed}{DA_uniform}{r_ff}{0.335}
\setres{signed}{DA_uniform}{r_ff_se}{0.003}
\setres{signed}{DA_uniform}{r_prior}{0.7757}
\setres{signed}{DA_uniform}{r_prior_se}{0.001}
\setres{signed}{DA_uniform}{r_prob}{0.851}
\setres{signed}{DA_uniform}{r_prob_se}{0.001}
\setres{signed}{DA_uniform}{spread_penalty}{1.089}
\setres{signed}{DA_uniform}{spread_penalty_se}{0.079}
\setres{signed}{DA_uniform}{volume_ess}{277{,}116}
\setres{signed}{EU}{det_pct}{44}
\setres{signed}{EU}{gap_lit}{0.585}
\setres{signed}{EU}{gap_uniform}{1.567}
\setres{signed}{EU}{one_minus_rff}{0.573}
\setres{signed}{EU}{passthrough_lit}{0.385}
\setres{signed}{EU}{passthrough_uniform}{0.447}
\setres{signed}{EU}{prior_contrib_lit}{0.353}
\setres{signed}{EU}{prior_contrib_uniform}{0.317}
\setres{signed}{EU}{share_lit}{-32}
\setres{signed}{EU}{share_uniform}{-255}
\setres{signed}{EU_deterministic}{kappa}{0.442}
\setres{signed}{EU_deterministic}{kappa_se}{0.066}
\setres{signed}{EU_deterministic}{mse}{103.59}
\setres{signed}{EU_deterministic}{nrmse_pct}{23.4}
\setres{signed}{EU_deterministic}{r_ff}{0.427}
\setres{signed}{EU_deterministic}{r_ff_se}{0.004}
\setres{signed}{EU_deterministic}{r_prior}{1.0000}
\setres{signed}{EU_deterministic}{r_prior_se}{---}
\setres{signed}{EU_deterministic}{r_prob}{1.000}
\setres{signed}{EU_deterministic}{r_prob_se}{0.000}
\setres{signed}{EU_deterministic}{volume_ess}{---}
\setres{signed}{EU_lit}{kappa}{-0.143}
\setres{signed}{EU_lit}{kappa_prior_mean}{0.172}
\setres{signed}{EU_lit}{kappa_prior_mean_se}{0.085}
\setres{signed}{EU_lit}{kappa_se}{0.104}
\setres{signed}{EU_lit}{mse}{143.73}
\setres{signed}{EU_lit}{mse_prior_mean}{122.11}
\setres{signed}{EU_lit}{nrmse_pct}{23.4}
\setres{signed}{EU_lit}{pred_variance}{21.63}
\setres{signed}{EU_lit}{r_ff}{0.427}
\setres{signed}{EU_lit}{r_ff_se}{0.004}
\setres{signed}{EU_lit}{r_prior}{0.6148}
\setres{signed}{EU_lit}{r_prior_se}{0.001}
\setres{signed}{EU_lit}{r_prob}{0.779}
\setres{signed}{EU_lit}{r_prob_se}{0.002}
\setres{signed}{EU_lit}{spread_penalty}{0.315}
\setres{signed}{EU_lit}{spread_penalty_se}{0.023}
\setres{signed}{EU_lit}{volume_ess}{154{,}983}
\setres{signed}{EU_uniform}{kappa}{-1.126}
\setres{signed}{EU_uniform}{kappa_prior_mean}{-0.592}
\setres{signed}{EU_uniform}{kappa_prior_mean_se}{0.136}
\setres{signed}{EU_uniform}{kappa_se}{0.170}
\setres{signed}{EU_uniform}{mse}{211.18}
\setres{signed}{EU_uniform}{mse_prior_mean}{174.58}
\setres{signed}{EU_uniform}{nrmse_pct}{24.4}
\setres{signed}{EU_uniform}{pred_variance}{36.60}
\setres{signed}{EU_uniform}{r_ff}{0.427}
\setres{signed}{EU_uniform}{r_ff_se}{0.004}
\setres{signed}{EU_uniform}{r_prior}{0.5531}
\setres{signed}{EU_uniform}{r_prior_se}{0.001}
\setres{signed}{EU_uniform}{r_prob}{0.744}
\setres{signed}{EU_uniform}{r_prob_se}{0.002}
\setres{signed}{EU_uniform}{spread_penalty}{0.533}
\setres{signed}{EU_uniform}{spread_penalty_se}{0.039}
\setres{signed}{EU_uniform}{volume_ess}{154{,}983}
\setres{signed}{EV_deterministic}{kappa}{0.441}
\setres{signed}{EV_deterministic}{kappa_se}{0.066}
\setres{signed}{EV_deterministic}{mse}{103.63}
\setres{signed}{EV_deterministic}{nrmse_pct}{23.3}
\setres{signed}{EV_deterministic}{r_ff}{0.488}
\setres{signed}{EV_deterministic}{r_ff_se}{---}
\setres{signed}{EV_deterministic}{r_prior}{1.0000}
\setres{signed}{EV_deterministic}{r_prior_se}{---}
\setres{signed}{EV_deterministic}{r_prob}{1.000}
\setres{signed}{EV_deterministic}{r_prob_se}{0.000}
\setres{signed}{EV_deterministic}{volume_ess}{---}
\setres{signed}{reference}{lookup_nrmse_pct}{17.5}
\setres{signed}{reference}{midpoint_mse}{133.93}
\setres{signed}{reference}{midpoint_nrmse_pct}{27.1}
\setres{signed}{reference}{n_lotteries}{50}
\setres{signed}{reference}{n_obs}{8{,}906}
\setres{signed}{reference}{n_subjects}{179}
\setres{signed}{reference}{oracle_mse}{65.29}
\setres{pdG}{CPT}{det_pct}{97}
\setres{pdG}{CPT}{gap_lit}{0.331}
\setres{pdG}{CPT}{gap_uniform}{1.467}
\setres{pdG}{CPT}{one_minus_rff}{0.860}
\setres{pdG}{CPT}{passthrough_lit}{0.051}
\setres{pdG}{CPT}{passthrough_uniform}{0.485}
\setres{pdG}{CPT}{prior_contrib_lit}{0.816}
\setres{pdG}{CPT}{prior_contrib_uniform}{0.443}
\setres{pdG}{CPT}{share_lit}{66}
\setres{pdG}{CPT}{share_uniform}{-51}
\setres{pdG}{CPT_deterministic}{kappa}{0.972}
\setres{pdG}{CPT_deterministic}{kappa_se}{0.008}
\setres{pdG}{CPT_deterministic}{mse}{63.70}
\setres{pdG}{CPT_deterministic}{r_ff}{0.140}
\setres{pdG}{CPT_deterministic}{r_ff_se}{0.001}
\setres{pdG}{CPT_deterministic}{r_prior}{1.0000}
\setres{pdG}{CPT_deterministic}{r_prior_se}{---}
\setres{pdG}{CPT_deterministic}{r_prob}{1.000}
\setres{pdG}{CPT_deterministic}{r_prob_se}{0.000}
\setres{pdG}{CPT_deterministic}{volume_ess}{---}
\setres{pdG}{CPT_lit}{kappa}{0.641}
\setres{pdG}{CPT_lit}{kappa_prior_mean}{0.853}
\setres{pdG}{CPT_lit}{kappa_prior_mean_se}{0.035}
\setres{pdG}{CPT_lit}{kappa_se}{0.050}
\setres{pdG}{CPT_lit}{mse}{87.56}
\setres{pdG}{CPT_lit}{mse_prior_mean}{72.28}
\setres{pdG}{CPT_lit}{pred_variance}{15.28}
\setres{pdG}{CPT_lit}{r_ff}{0.140}
\setres{pdG}{CPT_lit}{r_ff_se}{0.001}
\setres{pdG}{CPT_lit}{r_prior}{0.9486}
\setres{pdG}{CPT_lit}{r_prior_se}{0.000}
\setres{pdG}{CPT_lit}{r_prob}{0.956}
\setres{pdG}{CPT_lit}{r_prob_se}{0.000}
\setres{pdG}{CPT_lit}{spread_penalty}{0.212}
\setres{pdG}{CPT_lit}{spread_penalty_se}{0.016}
\setres{pdG}{CPT_lit}{volume_ess}{55{,}652}
\setres{pdG}{CPT_uniform}{kappa}{-0.495}
\setres{pdG}{CPT_uniform}{kappa_prior_mean}{0.811}
\setres{pdG}{CPT_uniform}{kappa_prior_mean_se}{0.040}
\setres{pdG}{CPT_uniform}{kappa_se}{0.129}
\setres{pdG}{CPT_uniform}{mse}{169.35}
\setres{pdG}{CPT_uniform}{mse_prior_mean}{75.31}
\setres{pdG}{CPT_uniform}{pred_variance}{94.04}
\setres{pdG}{CPT_uniform}{r_ff}{0.140}
\setres{pdG}{CPT_uniform}{r_ff_se}{0.001}
\setres{pdG}{CPT_uniform}{r_prior}{0.5147}
\setres{pdG}{CPT_uniform}{r_prior_se}{0.003}
\setres{pdG}{CPT_uniform}{r_prob}{0.583}
\setres{pdG}{CPT_uniform}{r_prob_se}{0.003}
\setres{pdG}{CPT_uniform}{spread_penalty}{1.306}
\setres{pdG}{CPT_uniform}{spread_penalty_se}{0.096}
\setres{pdG}{CPT_uniform}{volume_ess}{55{,}652}
\setres{pdG}{DA}{det_pct}{63}
\setres{pdG}{DA}{gap_cpt_lit}{0.621}
\setres{pdG}{DA}{gap_lit}{0.606}
\setres{pdG}{DA}{gap_r_cpt_lit}{0.000}
\setres{pdG}{DA}{gap_uniform}{1.453}
\setres{pdG}{DA}{one_minus_rff}{0.719}
\setres{pdG}{DA}{passthrough_lit}{0.062}
\setres{pdG}{DA}{passthrough_uniform}{0.133}
\setres{pdG}{DA}{prior_contrib_lit}{0.674}
\setres{pdG}{DA}{prior_contrib_uniform}{0.623}
\setres{pdG}{DA}{share_lit}{3}
\setres{pdG}{DA}{share_uniform}{-132}
\setres{pdG}{DA_deterministic}{kappa}{0.626}
\setres{pdG}{DA_deterministic}{kappa_se}{0.062}
\setres{pdG}{DA_deterministic}{mse}{88.64}
\setres{pdG}{DA_deterministic}{r_ff}{0.281}
\setres{pdG}{DA_deterministic}{r_ff_se}{0.004}
\setres{pdG}{DA_deterministic}{r_prior}{1.0000}
\setres{pdG}{DA_deterministic}{r_prior_se}{---}
\setres{pdG}{DA_deterministic}{r_prob}{1.000}
\setres{pdG}{DA_deterministic}{r_prob_se}{0.000}
\setres{pdG}{DA_deterministic}{volume_ess}{---}
\setres{pdG}{DA_lit}{kappa}{0.020}
\setres{pdG}{DA_lit}{kappa_prior_mean}{0.340}
\setres{pdG}{DA_lit}{kappa_prior_mean_se}{0.113}
\setres{pdG}{DA_lit}{kappa_se}{0.136}
\setres{pdG}{DA_lit}{mse}{132.27}
\setres{pdG}{DA_lit}{mse_prior_mean}{109.23}
\setres{pdG}{DA_lit}{pred_variance}{23.04}
\setres{pdG}{DA_lit}{r_ff}{0.281}
\setres{pdG}{DA_lit}{r_ff_se}{0.004}
\setres{pdG}{DA_lit}{r_prior}{0.9383}
\setres{pdG}{DA_lit}{r_prior_se}{0.000}
\setres{pdG}{DA_lit}{r_prob}{0.956}
\setres{pdG}{DA_lit}{r_prob_se}{0.000}
\setres{pdG}{DA_lit}{spread_penalty}{0.320}
\setres{pdG}{DA_lit}{spread_penalty_se}{0.024}
\setres{pdG}{DA_lit}{volume_ess}{436{,}452}
\setres{pdG}{DA_uniform}{kappa}{-0.827}
\setres{pdG}{DA_uniform}{kappa_prior_mean}{0.060}
\setres{pdG}{DA_uniform}{kappa_prior_mean_se}{0.149}
\setres{pdG}{DA_uniform}{kappa_se}{0.212}
\setres{pdG}{DA_uniform}{mse}{193.20}
\setres{pdG}{DA_uniform}{mse_prior_mean}{129.35}
\setres{pdG}{DA_uniform}{pred_variance}{63.85}
\setres{pdG}{DA_uniform}{r_ff}{0.281}
\setres{pdG}{DA_uniform}{r_ff_se}{0.004}
\setres{pdG}{DA_uniform}{r_prior}{0.8666}
\setres{pdG}{DA_uniform}{r_prior_se}{0.001}
\setres{pdG}{DA_uniform}{r_prob}{0.904}
\setres{pdG}{DA_uniform}{r_prob_se}{0.001}
\setres{pdG}{DA_uniform}{spread_penalty}{0.887}
\setres{pdG}{DA_uniform}{spread_penalty_se}{0.065}
\setres{pdG}{DA_uniform}{volume_ess}{436{,}452}
\setres{pdG}{EU}{det_pct}{62}
\setres{pdG}{EU}{gap_lit}{0.095}
\setres{pdG}{EU}{gap_uniform}{0.826}
\setres{pdG}{EU}{one_minus_rff}{0.664}
\setres{pdG}{EU}{passthrough_lit}{0.186}
\setres{pdG}{EU}{passthrough_uniform}{0.422}
\setres{pdG}{EU}{prior_contrib_lit}{0.541}
\setres{pdG}{EU}{prior_contrib_uniform}{0.384}
\setres{pdG}{EU}{share_lit}{85}
\setres{pdG}{EU}{share_uniform}{-34}
\setres{pdG}{EU_deterministic}{kappa}{0.616}
\setres{pdG}{EU_deterministic}{kappa_se}{0.061}
\setres{pdG}{EU_deterministic}{mse}{89.31}
\setres{pdG}{EU_deterministic}{r_ff}{0.336}
\setres{pdG}{EU_deterministic}{r_ff_se}{0.005}
\setres{pdG}{EU_deterministic}{r_prior}{1.0000}
\setres{pdG}{EU_deterministic}{r_prior_se}{---}
\setres{pdG}{EU_deterministic}{r_prob}{1.000}
\setres{pdG}{EU_deterministic}{r_prob_se}{0.000}
\setres{pdG}{EU_deterministic}{volume_ess}{---}
\setres{pdG}{EU_lit}{kappa}{0.521}
\setres{pdG}{EU_lit}{kappa_prior_mean}{0.611}
\setres{pdG}{EU_lit}{kappa_prior_mean_se}{0.068}
\setres{pdG}{EU_lit}{kappa_se}{0.074}
\setres{pdG}{EU_lit}{mse}{96.17}
\setres{pdG}{EU_lit}{mse_prior_mean}{89.72}
\setres{pdG}{EU_lit}{pred_variance}{6.45}
\setres{pdG}{EU_lit}{r_ff}{0.336}
\setres{pdG}{EU_lit}{r_ff_se}{0.005}
\setres{pdG}{EU_lit}{r_prior}{0.8143}
\setres{pdG}{EU_lit}{r_prior_se}{0.001}
\setres{pdG}{EU_lit}{r_prob}{0.877}
\setres{pdG}{EU_lit}{r_prob_se}{0.001}
\setres{pdG}{EU_lit}{spread_penalty}{0.090}
\setres{pdG}{EU_lit}{spread_penalty_se}{0.007}
\setres{pdG}{EU_lit}{volume_ess}{172{,}086}
\setres{pdG}{EU_uniform}{kappa}{-0.209}
\setres{pdG}{EU_uniform}{kappa_prior_mean}{0.298}
\setres{pdG}{EU_uniform}{kappa_prior_mean_se}{0.118}
\setres{pdG}{EU_uniform}{kappa_se}{0.154}
\setres{pdG}{EU_uniform}{mse}{148.75}
\setres{pdG}{EU_uniform}{mse_prior_mean}{112.21}
\setres{pdG}{EU_uniform}{pred_variance}{36.53}
\setres{pdG}{EU_uniform}{r_ff}{0.336}
\setres{pdG}{EU_uniform}{r_ff_se}{0.005}
\setres{pdG}{EU_uniform}{r_prior}{0.5782}
\setres{pdG}{EU_uniform}{r_prior_se}{0.001}
\setres{pdG}{EU_uniform}{r_prob}{0.720}
\setres{pdG}{EU_uniform}{r_prob_se}{0.002}
\setres{pdG}{EU_uniform}{spread_penalty}{0.507}
\setres{pdG}{EU_uniform}{spread_penalty_se}{0.037}
\setres{pdG}{EU_uniform}{volume_ess}{172{,}086}
\setres{pdG}{EV_deterministic}{kappa}{0.493}
\setres{pdG}{EV_deterministic}{kappa_se}{0.064}
\setres{pdG}{EV_deterministic}{mse}{98.19}
\setres{pdG}{EV_deterministic}{r_ff}{0.484}
\setres{pdG}{EV_deterministic}{r_ff_se}{---}
\setres{pdG}{EV_deterministic}{r_prior}{1.0000}
\setres{pdG}{EV_deterministic}{r_prior_se}{---}
\setres{pdG}{EV_deterministic}{r_prob}{1.000}
\setres{pdG}{EV_deterministic}{r_prob_se}{0.000}
\setres{pdG}{EV_deterministic}{volume_ess}{---}
\setres{pdLasis}{CPT}{det_pct}{98}
\setres{pdLasis}{CPT}{gap_lit}{1.238}
\setres{pdLasis}{CPT}{gap_uniform}{3.038}
\setres{pdLasis}{CPT}{one_minus_rff}{0.856}
\setres{pdLasis}{CPT}{passthrough_lit}{0.050}
\setres{pdLasis}{CPT}{passthrough_uniform}{0.459}
\setres{pdLasis}{CPT}{prior_contrib_lit}{0.813}
\setres{pdLasis}{CPT}{prior_contrib_uniform}{0.463}
\setres{pdLasis}{CPT}{share_lit}{-26}
\setres{pdLasis}{CPT}{share_uniform}{-210}
\setres{pdLasis}{CPT_deterministic}{kappa}{0.980}
\setres{pdLasis}{CPT_deterministic}{kappa_se}{0.010}
\setres{pdLasis}{CPT_deterministic}{mse}{70.22}
\setres{pdLasis}{CPT_deterministic}{r_ff}{0.144}
\setres{pdLasis}{CPT_deterministic}{r_ff_se}{0.002}
\setres{pdLasis}{CPT_deterministic}{r_prior}{1.0000}
\setres{pdLasis}{CPT_deterministic}{r_prior_se}{---}
\setres{pdLasis}{CPT_deterministic}{r_prob}{1.000}
\setres{pdLasis}{CPT_deterministic}{r_prob_se}{0.000}
\setres{pdLasis}{CPT_deterministic}{volume_ess}{---}
\setres{pdLasis}{CPT_lit}{kappa}{-0.258}
\setres{pdLasis}{CPT_lit}{kappa_prior_mean}{-0.019}
\setres{pdLasis}{CPT_lit}{kappa_prior_mean_se}{0.098}
\setres{pdLasis}{CPT_lit}{kappa_se}{0.110}
\setres{pdLasis}{CPT_lit}{mse}{151.01}
\setres{pdLasis}{CPT_lit}{mse_prior_mean}{135.36}
\setres{pdLasis}{CPT_lit}{pred_variance}{15.64}
\setres{pdLasis}{CPT_lit}{r_ff}{0.144}
\setres{pdLasis}{CPT_lit}{r_ff_se}{0.002}
\setres{pdLasis}{CPT_lit}{r_prior}{0.9500}
\setres{pdLasis}{CPT_lit}{r_prior_se}{0.000}
\setres{pdLasis}{CPT_lit}{r_prob}{0.957}
\setres{pdLasis}{CPT_lit}{r_prob_se}{0.000}
\setres{pdLasis}{CPT_lit}{spread_penalty}{0.240}
\setres{pdLasis}{CPT_lit}{spread_penalty_se}{0.021}
\setres{pdLasis}{CPT_lit}{volume_ess}{52{,}317}
\setres{pdLasis}{CPT_uniform}{kappa}{-2.058}
\setres{pdLasis}{CPT_uniform}{kappa_prior_mean}{-0.577}
\setres{pdLasis}{CPT_uniform}{kappa_prior_mean_se}{0.107}
\setres{pdLasis}{CPT_uniform}{kappa_se}{0.203}
\setres{pdLasis}{CPT_uniform}{mse}{268.51}
\setres{pdLasis}{CPT_uniform}{mse_prior_mean}{171.84}
\setres{pdLasis}{CPT_uniform}{pred_variance}{96.66}
\setres{pdLasis}{CPT_uniform}{r_ff}{0.144}
\setres{pdLasis}{CPT_uniform}{r_ff_se}{0.002}
\setres{pdLasis}{CPT_uniform}{r_prior}{0.5409}
\setres{pdLasis}{CPT_uniform}{r_prior_se}{0.003}
\setres{pdLasis}{CPT_uniform}{r_prob}{0.607}
\setres{pdLasis}{CPT_uniform}{r_prob_se}{0.003}
\setres{pdLasis}{CPT_uniform}{spread_penalty}{1.481}
\setres{pdLasis}{CPT_uniform}{spread_penalty_se}{0.127}
\setres{pdLasis}{CPT_uniform}{volume_ess}{52{,}317}
\setres{pdLasis}{DA}{det_pct}{87}
\setres{pdLasis}{DA}{gap_cpt_lit}{0.638}
\setres{pdLasis}{DA}{gap_lit}{1.768}
\setres{pdLasis}{DA}{gap_r_cpt_lit}{-0.018}
\setres{pdLasis}{DA}{gap_uniform}{2.520}
\setres{pdLasis}{DA}{one_minus_rff}{0.808}
\setres{pdLasis}{DA}{passthrough_lit}{0.075}
\setres{pdLasis}{DA}{passthrough_uniform}{0.189}
\setres{pdLasis}{DA}{prior_contrib_lit}{0.748}
\setres{pdLasis}{DA}{prior_contrib_uniform}{0.656}
\setres{pdLasis}{DA}{share_lit}{-103}
\setres{pdLasis}{DA}{share_uniform}{-189}
\setres{pdLasis}{DA_deterministic}{kappa}{0.872}
\setres{pdLasis}{DA_deterministic}{kappa_se}{0.017}
\setres{pdLasis}{DA_deterministic}{mse}{77.23}
\setres{pdLasis}{DA_deterministic}{r_ff}{0.192}
\setres{pdLasis}{DA_deterministic}{r_ff_se}{0.002}
\setres{pdLasis}{DA_deterministic}{r_prior}{1.0000}
\setres{pdLasis}{DA_deterministic}{r_prior_se}{---}
\setres{pdLasis}{DA_deterministic}{r_prob}{1.000}
\setres{pdLasis}{DA_deterministic}{r_prob_se}{0.000}
\setres{pdLasis}{DA_deterministic}{volume_ess}{---}
\setres{pdLasis}{DA_lit}{kappa}{-0.896}
\setres{pdLasis}{DA_lit}{kappa_prior_mean}{-0.472}
\setres{pdLasis}{DA_lit}{kappa_prior_mean_se}{0.141}
\setres{pdLasis}{DA_lit}{kappa_se}{0.169}
\setres{pdLasis}{DA_lit}{mse}{192.63}
\setres{pdLasis}{DA_lit}{mse_prior_mean}{164.95}
\setres{pdLasis}{DA_lit}{pred_variance}{27.68}
\setres{pdLasis}{DA_lit}{r_ff}{0.192}
\setres{pdLasis}{DA_lit}{r_ff_se}{0.002}
\setres{pdLasis}{DA_lit}{r_prior}{0.9251}
\setres{pdLasis}{DA_lit}{r_prior_se}{0.001}
\setres{pdLasis}{DA_lit}{r_prob}{0.940}
\setres{pdLasis}{DA_lit}{r_prob_se}{0.000}
\setres{pdLasis}{DA_lit}{spread_penalty}{0.424}
\setres{pdLasis}{DA_lit}{spread_penalty_se}{0.037}
\setres{pdLasis}{DA_lit}{volume_ess}{30{,}044}
\setres{pdLasis}{DA_uniform}{kappa}{-1.647}
\setres{pdLasis}{DA_uniform}{kappa_prior_mean}{-0.336}
\setres{pdLasis}{DA_uniform}{kappa_prior_mean_se}{0.129}
\setres{pdLasis}{DA_uniform}{kappa_se}{0.221}
\setres{pdLasis}{DA_uniform}{mse}{241.68}
\setres{pdLasis}{DA_uniform}{mse_prior_mean}{156.06}
\setres{pdLasis}{DA_uniform}{pred_variance}{85.62}
\setres{pdLasis}{DA_uniform}{r_ff}{0.192}
\setres{pdLasis}{DA_uniform}{r_ff_se}{0.002}
\setres{pdLasis}{DA_uniform}{r_prior}{0.8114}
\setres{pdLasis}{DA_uniform}{r_prior_se}{0.001}
\setres{pdLasis}{DA_uniform}{r_prob}{0.848}
\setres{pdLasis}{DA_uniform}{r_prob_se}{0.001}
\setres{pdLasis}{DA_uniform}{spread_penalty}{1.312}
\setres{pdLasis}{DA_uniform}{spread_penalty_se}{0.113}
\setres{pdLasis}{DA_uniform}{volume_ess}{30{,}044}
\setres{pdLasis}{EU}{det_pct}{59}
\setres{pdLasis}{EU}{gap_lit}{1.076}
\setres{pdLasis}{EU}{gap_uniform}{2.729}
\setres{pdLasis}{EU}{one_minus_rff}{0.622}
\setres{pdLasis}{EU}{passthrough_lit}{0.213}
\setres{pdLasis}{EU}{passthrough_uniform}{0.483}
\setres{pdLasis}{EU}{prior_contrib_lit}{0.490}
\setres{pdLasis}{EU}{prior_contrib_uniform}{0.321}
\setres{pdLasis}{EU}{share_lit}{-82}
\setres{pdLasis}{EU}{share_uniform}{-363}
\setres{pdLasis}{EU_deterministic}{kappa}{0.590}
\setres{pdLasis}{EU_deterministic}{kappa_se}{0.081}
\setres{pdLasis}{EU_deterministic}{mse}{95.66}
\setres{pdLasis}{EU_deterministic}{r_ff}{0.378}
\setres{pdLasis}{EU_deterministic}{r_ff_se}{0.006}
\setres{pdLasis}{EU_deterministic}{r_prior}{1.0000}
\setres{pdLasis}{EU_deterministic}{r_prior_se}{---}
\setres{pdLasis}{EU_deterministic}{r_prob}{1.000}
\setres{pdLasis}{EU_deterministic}{r_prob_se}{0.000}
\setres{pdLasis}{EU_deterministic}{volume_ess}{---}
\setres{pdLasis}{EU_lit}{kappa}{-0.486}
\setres{pdLasis}{EU_lit}{kappa_prior_mean}{-0.387}
\setres{pdLasis}{EU_lit}{kappa_prior_mean_se}{0.136}
\setres{pdLasis}{EU_lit}{kappa_se}{0.142}
\setres{pdLasis}{EU_lit}{mse}{165.89}
\setres{pdLasis}{EU_lit}{mse_prior_mean}{159.42}
\setres{pdLasis}{EU_lit}{pred_variance}{6.47}
\setres{pdLasis}{EU_lit}{r_ff}{0.378}
\setres{pdLasis}{EU_lit}{r_ff_se}{0.006}
\setres{pdLasis}{EU_lit}{r_prior}{0.7872}
\setres{pdLasis}{EU_lit}{r_prior_se}{0.001}
\setres{pdLasis}{EU_lit}{r_prob}{0.868}
\setres{pdLasis}{EU_lit}{r_prob_se}{0.001}
\setres{pdLasis}{EU_lit}{spread_penalty}{0.099}
\setres{pdLasis}{EU_lit}{spread_penalty_se}{0.009}
\setres{pdLasis}{EU_lit}{volume_ess}{133{,}991}
\setres{pdLasis}{EU_uniform}{kappa}{-2.139}
\setres{pdLasis}{EU_uniform}{kappa_prior_mean}{-1.577}
\setres{pdLasis}{EU_uniform}{kappa_prior_mean_se}{0.209}
\setres{pdLasis}{EU_uniform}{kappa_se}{0.248}
\setres{pdLasis}{EU_uniform}{mse}{273.77}
\setres{pdLasis}{EU_uniform}{mse_prior_mean}{237.11}
\setres{pdLasis}{EU_uniform}{pred_variance}{36.66}
\setres{pdLasis}{EU_uniform}{r_ff}{0.378}
\setres{pdLasis}{EU_uniform}{r_ff_se}{0.006}
\setres{pdLasis}{EU_uniform}{r_prior}{0.5166}
\setres{pdLasis}{EU_uniform}{r_prior_se}{0.001}
\setres{pdLasis}{EU_uniform}{r_prob}{0.699}
\setres{pdLasis}{EU_uniform}{r_prob_se}{0.003}
\setres{pdLasis}{EU_uniform}{spread_penalty}{0.562}
\setres{pdLasis}{EU_uniform}{spread_penalty_se}{0.048}
\setres{pdLasis}{EU_uniform}{volume_ess}{133{,}991}
\setres{pdLasis}{EV_deterministic}{kappa}{0.384}
\setres{pdLasis}{EV_deterministic}{kappa_se}{0.085}
\setres{pdLasis}{EV_deterministic}{mse}{109.08}
\setres{pdLasis}{EV_deterministic}{r_ff}{0.480}
\setres{pdLasis}{EV_deterministic}{r_ff_se}{---}
\setres{pdLasis}{EV_deterministic}{r_prior}{1.0000}
\setres{pdLasis}{EV_deterministic}{r_prior_se}{---}
\setres{pdLasis}{EV_deterministic}{r_prob}{1.000}
\setres{pdLasis}{EV_deterministic}{r_prob_se}{0.000}
\setres{pdLasis}{EV_deterministic}{volume_ess}{---}
\setres{pdLadapt}{CPT}{det_pct}{98}
\setres{pdLadapt}{CPT}{gap_lit}{1.272}
\setres{pdLadapt}{CPT}{gap_uniform}{2.016}
\setres{pdLadapt}{CPT}{one_minus_rff}{0.856}
\setres{pdLadapt}{CPT}{passthrough_lit}{0.096}
\setres{pdLadapt}{CPT}{passthrough_uniform}{0.397}
\setres{pdLadapt}{CPT}{prior_contrib_lit}{0.774}
\setres{pdLadapt}{CPT}{prior_contrib_uniform}{0.516}
\setres{pdLadapt}{CPT}{share_lit}{-30}
\setres{pdLadapt}{CPT}{share_uniform}{-106}
\setres{pdLadapt}{CPT_deterministic}{kappa}{0.980}
\setres{pdLadapt}{CPT_deterministic}{kappa_se}{0.010}
\setres{pdLadapt}{CPT_deterministic}{mse}{70.22}
\setres{pdLadapt}{CPT_deterministic}{r_ff}{0.144}
\setres{pdLadapt}{CPT_deterministic}{r_ff_se}{0.002}
\setres{pdLadapt}{CPT_deterministic}{r_prior}{1.0000}
\setres{pdLadapt}{CPT_deterministic}{r_prior_se}{---}
\setres{pdLadapt}{CPT_deterministic}{r_prob}{1.000}
\setres{pdLadapt}{CPT_deterministic}{r_prob_se}{0.000}
\setres{pdLadapt}{CPT_deterministic}{volume_ess}{---}
\setres{pdLadapt}{CPT_lit}{kappa}{-0.293}
\setres{pdLadapt}{CPT_lit}{kappa_prior_mean}{0.176}
\setres{pdLadapt}{CPT_lit}{kappa_prior_mean_se}{0.086}
\setres{pdLadapt}{CPT_lit}{kappa_se}{0.111}
\setres{pdLadapt}{CPT_lit}{mse}{153.25}
\setres{pdLadapt}{CPT_lit}{mse_prior_mean}{122.66}
\setres{pdLadapt}{CPT_lit}{pred_variance}{30.60}
\setres{pdLadapt}{CPT_lit}{r_ff}{0.144}
\setres{pdLadapt}{CPT_lit}{r_ff_se}{0.002}
\setres{pdLadapt}{CPT_lit}{r_prior}{0.9043}
\setres{pdLadapt}{CPT_lit}{r_prior_se}{0.001}
\setres{pdLadapt}{CPT_lit}{r_prob}{0.918}
\setres{pdLadapt}{CPT_lit}{r_prob_se}{0.001}
\setres{pdLadapt}{CPT_lit}{spread_penalty}{0.469}
\setres{pdLadapt}{CPT_lit}{spread_penalty_se}{0.040}
\setres{pdLadapt}{CPT_lit}{volume_ess}{52{,}317}
\setres{pdLadapt}{CPT_uniform}{kappa}{-1.037}
\setres{pdLadapt}{CPT_uniform}{kappa_prior_mean}{0.736}
\setres{pdLadapt}{CPT_uniform}{kappa_prior_mean_se}{0.039}
\setres{pdLadapt}{CPT_uniform}{kappa_se}{0.152}
\setres{pdLadapt}{CPT_uniform}{mse}{201.83}
\setres{pdLadapt}{CPT_uniform}{mse_prior_mean}{86.08}
\setres{pdLadapt}{CPT_uniform}{pred_variance}{115.74}
\setres{pdLadapt}{CPT_uniform}{r_ff}{0.144}
\setres{pdLadapt}{CPT_uniform}{r_ff_se}{0.002}
\setres{pdLadapt}{CPT_uniform}{r_prior}{0.6030}
\setres{pdLadapt}{CPT_uniform}{r_prior_se}{0.003}
\setres{pdLadapt}{CPT_uniform}{r_prob}{0.660}
\setres{pdLadapt}{CPT_uniform}{r_prob_se}{0.003}
\setres{pdLadapt}{CPT_uniform}{spread_penalty}{1.773}
\setres{pdLadapt}{CPT_uniform}{spread_penalty_se}{0.152}
\setres{pdLadapt}{CPT_uniform}{volume_ess}{52{,}317}
\setres{pdLadapt}{DA}{det_pct}{87}
\setres{pdLadapt}{DA}{gap_cpt_lit}{1.751}
\setres{pdLadapt}{DA}{gap_lit}{2.916}
\setres{pdLadapt}{DA}{gap_r_cpt_lit}{0.000}
\setres{pdLadapt}{DA}{gap_uniform}{1.802}
\setres{pdLadapt}{DA}{one_minus_rff}{0.808}
\setres{pdLadapt}{DA}{passthrough_lit}{0.102}
\setres{pdLadapt}{DA}{passthrough_uniform}{0.166}
\setres{pdLadapt}{DA}{prior_contrib_lit}{0.726}
\setres{pdLadapt}{DA}{prior_contrib_uniform}{0.674}
\setres{pdLadapt}{DA}{share_lit}{-234}
\setres{pdLadapt}{DA}{share_uniform}{-107}
\setres{pdLadapt}{DA_deterministic}{kappa}{0.872}
\setres{pdLadapt}{DA_deterministic}{kappa_se}{0.017}
\setres{pdLadapt}{DA_deterministic}{mse}{77.23}
\setres{pdLadapt}{DA_deterministic}{r_ff}{0.192}
\setres{pdLadapt}{DA_deterministic}{r_ff_se}{0.002}
\setres{pdLadapt}{DA_deterministic}{r_prior}{1.0000}
\setres{pdLadapt}{DA_deterministic}{r_prior_se}{---}
\setres{pdLadapt}{DA_deterministic}{r_prob}{1.000}
\setres{pdLadapt}{DA_deterministic}{r_prob_se}{0.000}
\setres{pdLadapt}{DA_deterministic}{volume_ess}{---}
\setres{pdLadapt}{DA_lit}{kappa}{-2.043}
\setres{pdLadapt}{DA_lit}{kappa_prior_mean}{-1.560}
\setres{pdLadapt}{DA_lit}{kappa_prior_mean_se}{0.210}
\setres{pdLadapt}{DA_lit}{kappa_se}{0.243}
\setres{pdLadapt}{DA_lit}{mse}{267.53}
\setres{pdLadapt}{DA_lit}{mse_prior_mean}{235.96}
\setres{pdLadapt}{DA_lit}{pred_variance}{31.57}
\setres{pdLadapt}{DA_lit}{r_ff}{0.192}
\setres{pdLadapt}{DA_lit}{r_ff_se}{0.002}
\setres{pdLadapt}{DA_lit}{r_prior}{0.8981}
\setres{pdLadapt}{DA_lit}{r_prior_se}{0.001}
\setres{pdLadapt}{DA_lit}{r_prob}{0.918}
\setres{pdLadapt}{DA_lit}{r_prob_se}{0.001}
\setres{pdLadapt}{DA_lit}{spread_penalty}{0.484}
\setres{pdLadapt}{DA_lit}{spread_penalty_se}{0.042}
\setres{pdLadapt}{DA_lit}{volume_ess}{30{,}044}
\setres{pdLadapt}{DA_uniform}{kappa}{-0.930}
\setres{pdLadapt}{DA_uniform}{kappa_prior_mean}{0.584}
\setres{pdLadapt}{DA_uniform}{kappa_prior_mean_se}{0.070}
\setres{pdLadapt}{DA_uniform}{kappa_se}{0.190}
\setres{pdLadapt}{DA_uniform}{mse}{194.87}
\setres{pdLadapt}{DA_uniform}{mse_prior_mean}{96.02}
\setres{pdLadapt}{DA_uniform}{pred_variance}{98.84}
\setres{pdLadapt}{DA_uniform}{r_ff}{0.192}
\setres{pdLadapt}{DA_uniform}{r_ff_se}{0.002}
\setres{pdLadapt}{DA_uniform}{r_prior}{0.8339}
\setres{pdLadapt}{DA_uniform}{r_prior_se}{0.001}
\setres{pdLadapt}{DA_uniform}{r_prob}{0.866}
\setres{pdLadapt}{DA_uniform}{r_prob_se}{0.001}
\setres{pdLadapt}{DA_uniform}{spread_penalty}{1.514}
\setres{pdLadapt}{DA_uniform}{spread_penalty_se}{0.130}
\setres{pdLadapt}{DA_uniform}{volume_ess}{30{,}044}
\setres{pdLadapt}{EU}{det_pct}{59}
\setres{pdLadapt}{EU}{gap_lit}{1.098}
\setres{pdLadapt}{EU}{gap_uniform}{1.421}
\setres{pdLadapt}{EU}{one_minus_rff}{0.622}
\setres{pdLadapt}{EU}{passthrough_lit}{0.408}
\setres{pdLadapt}{EU}{passthrough_uniform}{0.554}
\setres{pdLadapt}{EU}{prior_contrib_lit}{0.368}
\setres{pdLadapt}{EU}{prior_contrib_uniform}{0.278}
\setres{pdLadapt}{EU}{share_lit}{-86}
\setres{pdLadapt}{EU}{share_uniform}{-141}
\setres{pdLadapt}{EU_deterministic}{kappa}{0.590}
\setres{pdLadapt}{EU_deterministic}{kappa_se}{0.081}
\setres{pdLadapt}{EU_deterministic}{mse}{95.66}
\setres{pdLadapt}{EU_deterministic}{r_ff}{0.378}
\setres{pdLadapt}{EU_deterministic}{r_ff_se}{0.006}
\setres{pdLadapt}{EU_deterministic}{r_prior}{1.0000}
\setres{pdLadapt}{EU_deterministic}{r_prior_se}{---}
\setres{pdLadapt}{EU_deterministic}{r_prob}{1.000}
\setres{pdLadapt}{EU_deterministic}{r_prob_se}{0.000}
\setres{pdLadapt}{EU_deterministic}{volume_ess}{---}
\setres{pdLadapt}{EU_lit}{kappa}{-0.508}
\setres{pdLadapt}{EU_lit}{kappa_prior_mean}{-0.184}
\setres{pdLadapt}{EU_lit}{kappa_prior_mean_se}{0.122}
\setres{pdLadapt}{EU_lit}{kappa_se}{0.143}
\setres{pdLadapt}{EU_lit}{mse}{167.34}
\setres{pdLadapt}{EU_lit}{mse_prior_mean}{146.16}
\setres{pdLadapt}{EU_lit}{pred_variance}{21.18}
\setres{pdLadapt}{EU_lit}{r_ff}{0.378}
\setres{pdLadapt}{EU_lit}{r_ff_se}{0.006}
\setres{pdLadapt}{EU_lit}{r_prior}{0.5916}
\setres{pdLadapt}{EU_lit}{r_prior_se}{0.002}
\setres{pdLadapt}{EU_lit}{r_prob}{0.746}
\setres{pdLadapt}{EU_lit}{r_prob_se}{0.002}
\setres{pdLadapt}{EU_lit}{spread_penalty}{0.325}
\setres{pdLadapt}{EU_lit}{spread_penalty_se}{0.028}
\setres{pdLadapt}{EU_lit}{volume_ess}{133{,}991}
\setres{pdLadapt}{EU_uniform}{kappa}{-0.831}
\setres{pdLadapt}{EU_uniform}{kappa_prior_mean}{0.122}
\setres{pdLadapt}{EU_uniform}{kappa_prior_mean_se}{0.102}
\setres{pdLadapt}{EU_uniform}{kappa_se}{0.168}
\setres{pdLadapt}{EU_uniform}{mse}{188.41}
\setres{pdLadapt}{EU_uniform}{mse_prior_mean}{126.22}
\setres{pdLadapt}{EU_uniform}{pred_variance}{62.18}
\setres{pdLadapt}{EU_uniform}{r_ff}{0.378}
\setres{pdLadapt}{EU_uniform}{r_ff_se}{0.006}
\setres{pdLadapt}{EU_uniform}{r_prior}{0.4460}
\setres{pdLadapt}{EU_uniform}{r_prior_se}{0.002}
\setres{pdLadapt}{EU_uniform}{r_prob}{0.655}
\setres{pdLadapt}{EU_uniform}{r_prob_se}{0.003}
\setres{pdLadapt}{EU_uniform}{spread_penalty}{0.953}
\setres{pdLadapt}{EU_uniform}{spread_penalty_se}{0.082}
\setres{pdLadapt}{EU_uniform}{volume_ess}{133{,}991}
\setres{pdLadapt}{EV_deterministic}{kappa}{0.384}
\setres{pdLadapt}{EV_deterministic}{kappa_se}{0.085}
\setres{pdLadapt}{EV_deterministic}{mse}{109.08}
\setres{pdLadapt}{EV_deterministic}{r_ff}{0.480}
\setres{pdLadapt}{EV_deterministic}{r_ff_se}{---}
\setres{pdLadapt}{EV_deterministic}{r_prior}{1.0000}
\setres{pdLadapt}{EV_deterministic}{r_prior_se}{---}
\setres{pdLadapt}{EV_deterministic}{r_prob}{1.000}
\setres{pdLadapt}{EV_deterministic}{r_prob_se}{0.000}
\setres{pdLadapt}{EV_deterministic}{volume_ess}{---}
\setres{pdL}{CPT}{det_pct}{98}
\setres{pdL}{CPT}{gap_lit}{1.272}
\setres{pdL}{CPT}{gap_uniform}{3.038}
\setres{pdL}{CPT}{one_minus_rff}{0.856}
\setres{pdL}{CPT}{passthrough_lit}{0.096}
\setres{pdL}{CPT}{passthrough_uniform}{0.459}
\setres{pdL}{CPT}{prior_contrib_lit}{0.774}
\setres{pdL}{CPT}{prior_contrib_uniform}{0.463}
\setres{pdL}{CPT}{share_lit}{-30}
\setres{pdL}{CPT}{share_uniform}{-210}
\setres{pdL}{CPT_deterministic}{kappa}{0.980}
\setres{pdL}{CPT_deterministic}{kappa_se}{0.010}
\setres{pdL}{CPT_deterministic}{mse}{70.22}
\setres{pdL}{CPT_deterministic}{r_ff}{0.144}
\setres{pdL}{CPT_deterministic}{r_ff_se}{0.002}
\setres{pdL}{CPT_deterministic}{r_prior}{1.0000}
\setres{pdL}{CPT_deterministic}{r_prior_se}{---}
\setres{pdL}{CPT_deterministic}{r_prob}{1.000}
\setres{pdL}{CPT_deterministic}{r_prob_se}{0.000}
\setres{pdL}{CPT_deterministic}{volume_ess}{---}
\setres{pdL}{CPT_lit}{kappa}{-0.293}
\setres{pdL}{CPT_lit}{kappa_prior_mean}{0.176}
\setres{pdL}{CPT_lit}{kappa_prior_mean_se}{0.086}
\setres{pdL}{CPT_lit}{kappa_se}{0.111}
\setres{pdL}{CPT_lit}{mse}{153.25}
\setres{pdL}{CPT_lit}{mse_prior_mean}{122.66}
\setres{pdL}{CPT_lit}{pred_variance}{30.60}
\setres{pdL}{CPT_lit}{r_ff}{0.144}
\setres{pdL}{CPT_lit}{r_ff_se}{0.002}
\setres{pdL}{CPT_lit}{r_prior}{0.9043}
\setres{pdL}{CPT_lit}{r_prior_se}{0.001}
\setres{pdL}{CPT_lit}{r_prob}{0.918}
\setres{pdL}{CPT_lit}{r_prob_se}{0.001}
\setres{pdL}{CPT_lit}{spread_penalty}{0.469}
\setres{pdL}{CPT_lit}{spread_penalty_se}{0.040}
\setres{pdL}{CPT_lit}{volume_ess}{52{,}317}
\setres{pdL}{CPT_uniform}{kappa}{-2.058}
\setres{pdL}{CPT_uniform}{kappa_prior_mean}{-0.577}
\setres{pdL}{CPT_uniform}{kappa_prior_mean_se}{0.107}
\setres{pdL}{CPT_uniform}{kappa_se}{0.203}
\setres{pdL}{CPT_uniform}{mse}{268.51}
\setres{pdL}{CPT_uniform}{mse_prior_mean}{171.84}
\setres{pdL}{CPT_uniform}{pred_variance}{96.66}
\setres{pdL}{CPT_uniform}{r_ff}{0.144}
\setres{pdL}{CPT_uniform}{r_ff_se}{0.002}
\setres{pdL}{CPT_uniform}{r_prior}{0.5409}
\setres{pdL}{CPT_uniform}{r_prior_se}{0.003}
\setres{pdL}{CPT_uniform}{r_prob}{0.607}
\setres{pdL}{CPT_uniform}{r_prob_se}{0.003}
\setres{pdL}{CPT_uniform}{spread_penalty}{1.481}
\setres{pdL}{CPT_uniform}{spread_penalty_se}{0.127}
\setres{pdL}{CPT_uniform}{volume_ess}{52{,}317}
\setres{pdL}{DA}{det_pct}{87}
\setres{pdL}{DA}{gap_cpt_lit}{1.751}
\setres{pdL}{DA}{gap_lit}{2.916}
\setres{pdL}{DA}{gap_r_cpt_lit}{0.000}
\setres{pdL}{DA}{gap_uniform}{2.520}
\setres{pdL}{DA}{one_minus_rff}{0.808}
\setres{pdL}{DA}{passthrough_lit}{0.102}
\setres{pdL}{DA}{passthrough_uniform}{0.189}
\setres{pdL}{DA}{prior_contrib_lit}{0.726}
\setres{pdL}{DA}{prior_contrib_uniform}{0.656}
\setres{pdL}{DA}{share_lit}{-234}
\setres{pdL}{DA}{share_uniform}{-189}
\setres{pdL}{DA_deterministic}{kappa}{0.872}
\setres{pdL}{DA_deterministic}{kappa_se}{0.017}
\setres{pdL}{DA_deterministic}{mse}{77.23}
\setres{pdL}{DA_deterministic}{r_ff}{0.192}
\setres{pdL}{DA_deterministic}{r_ff_se}{0.002}
\setres{pdL}{DA_deterministic}{r_prior}{1.0000}
\setres{pdL}{DA_deterministic}{r_prior_se}{---}
\setres{pdL}{DA_deterministic}{r_prob}{1.000}
\setres{pdL}{DA_deterministic}{r_prob_se}{0.000}
\setres{pdL}{DA_deterministic}{volume_ess}{---}
\setres{pdL}{DA_lit}{kappa}{-2.043}
\setres{pdL}{DA_lit}{kappa_prior_mean}{-1.560}
\setres{pdL}{DA_lit}{kappa_prior_mean_se}{0.210}
\setres{pdL}{DA_lit}{kappa_se}{0.243}
\setres{pdL}{DA_lit}{mse}{267.53}
\setres{pdL}{DA_lit}{mse_prior_mean}{235.96}
\setres{pdL}{DA_lit}{pred_variance}{31.57}
\setres{pdL}{DA_lit}{r_ff}{0.192}
\setres{pdL}{DA_lit}{r_ff_se}{0.002}
\setres{pdL}{DA_lit}{r_prior}{0.8981}
\setres{pdL}{DA_lit}{r_prior_se}{0.001}
\setres{pdL}{DA_lit}{r_prob}{0.918}
\setres{pdL}{DA_lit}{r_prob_se}{0.001}
\setres{pdL}{DA_lit}{spread_penalty}{0.484}
\setres{pdL}{DA_lit}{spread_penalty_se}{0.042}
\setres{pdL}{DA_lit}{volume_ess}{30{,}044}
\setres{pdL}{DA_uniform}{kappa}{-1.647}
\setres{pdL}{DA_uniform}{kappa_prior_mean}{-0.336}
\setres{pdL}{DA_uniform}{kappa_prior_mean_se}{0.129}
\setres{pdL}{DA_uniform}{kappa_se}{0.221}
\setres{pdL}{DA_uniform}{mse}{241.68}
\setres{pdL}{DA_uniform}{mse_prior_mean}{156.06}
\setres{pdL}{DA_uniform}{pred_variance}{85.62}
\setres{pdL}{DA_uniform}{r_ff}{0.192}
\setres{pdL}{DA_uniform}{r_ff_se}{0.002}
\setres{pdL}{DA_uniform}{r_prior}{0.8114}
\setres{pdL}{DA_uniform}{r_prior_se}{0.001}
\setres{pdL}{DA_uniform}{r_prob}{0.848}
\setres{pdL}{DA_uniform}{r_prob_se}{0.001}
\setres{pdL}{DA_uniform}{spread_penalty}{1.312}
\setres{pdL}{DA_uniform}{spread_penalty_se}{0.113}
\setres{pdL}{DA_uniform}{volume_ess}{30{,}044}
\setres{pdL}{EU}{det_pct}{59}
\setres{pdL}{EU}{gap_lit}{1.098}
\setres{pdL}{EU}{gap_uniform}{2.729}
\setres{pdL}{EU}{one_minus_rff}{0.622}
\setres{pdL}{EU}{passthrough_lit}{0.408}
\setres{pdL}{EU}{passthrough_uniform}{0.483}
\setres{pdL}{EU}{prior_contrib_lit}{0.368}
\setres{pdL}{EU}{prior_contrib_uniform}{0.321}
\setres{pdL}{EU}{share_lit}{-86}
\setres{pdL}{EU}{share_uniform}{-363}
\setres{pdL}{EU_deterministic}{kappa}{0.590}
\setres{pdL}{EU_deterministic}{kappa_se}{0.081}
\setres{pdL}{EU_deterministic}{mse}{95.66}
\setres{pdL}{EU_deterministic}{r_ff}{0.378}
\setres{pdL}{EU_deterministic}{r_ff_se}{0.006}
\setres{pdL}{EU_deterministic}{r_prior}{1.0000}
\setres{pdL}{EU_deterministic}{r_prior_se}{---}
\setres{pdL}{EU_deterministic}{r_prob}{1.000}
\setres{pdL}{EU_deterministic}{r_prob_se}{0.000}
\setres{pdL}{EU_deterministic}{volume_ess}{---}
\setres{pdL}{EU_lit}{kappa}{-0.508}
\setres{pdL}{EU_lit}{kappa_prior_mean}{-0.184}
\setres{pdL}{EU_lit}{kappa_prior_mean_se}{0.122}
\setres{pdL}{EU_lit}{kappa_se}{0.143}
\setres{pdL}{EU_lit}{mse}{167.34}
\setres{pdL}{EU_lit}{mse_prior_mean}{146.16}
\setres{pdL}{EU_lit}{pred_variance}{21.18}
\setres{pdL}{EU_lit}{r_ff}{0.378}
\setres{pdL}{EU_lit}{r_ff_se}{0.006}
\setres{pdL}{EU_lit}{r_prior}{0.5916}
\setres{pdL}{EU_lit}{r_prior_se}{0.002}
\setres{pdL}{EU_lit}{r_prob}{0.746}
\setres{pdL}{EU_lit}{r_prob_se}{0.002}
\setres{pdL}{EU_lit}{spread_penalty}{0.325}
\setres{pdL}{EU_lit}{spread_penalty_se}{0.028}
\setres{pdL}{EU_lit}{volume_ess}{133{,}991}
\setres{pdL}{EU_uniform}{kappa}{-2.139}
\setres{pdL}{EU_uniform}{kappa_prior_mean}{-1.577}
\setres{pdL}{EU_uniform}{kappa_prior_mean_se}{0.209}
\setres{pdL}{EU_uniform}{kappa_se}{0.248}
\setres{pdL}{EU_uniform}{mse}{273.77}
\setres{pdL}{EU_uniform}{mse_prior_mean}{237.11}
\setres{pdL}{EU_uniform}{pred_variance}{36.66}
\setres{pdL}{EU_uniform}{r_ff}{0.378}
\setres{pdL}{EU_uniform}{r_ff_se}{0.006}
\setres{pdL}{EU_uniform}{r_prior}{0.5166}
\setres{pdL}{EU_uniform}{r_prior_se}{0.001}
\setres{pdL}{EU_uniform}{r_prob}{0.699}
\setres{pdL}{EU_uniform}{r_prob_se}{0.003}
\setres{pdL}{EU_uniform}{spread_penalty}{0.562}
\setres{pdL}{EU_uniform}{spread_penalty_se}{0.048}
\setres{pdL}{EU_uniform}{volume_ess}{133{,}991}
\setres{pdL}{EV_deterministic}{kappa}{0.384}
\setres{pdL}{EV_deterministic}{kappa_se}{0.085}
\setres{pdL}{EV_deterministic}{mse}{109.08}
\setres{pdL}{EV_deterministic}{r_ff}{0.480}
\setres{pdL}{EV_deterministic}{r_ff_se}{---}
\setres{pdL}{EV_deterministic}{r_prior}{1.0000}
\setres{pdL}{EV_deterministic}{r_prior_se}{---}
\setres{pdL}{EV_deterministic}{r_prob}{1.000}
\setres{pdL}{EV_deterministic}{r_prob_se}{0.000}
\setres{pdL}{EV_deterministic}{volume_ess}{---}
\setres{pdSG}{CPT}{det_pct}{97}
\setres{pdSG}{CPT}{gap_lit}{0.528}
\setres{pdSG}{CPT}{gap_uniform}{1.764}
\setres{pdSG}{CPT}{one_minus_rff}{0.860}
\setres{pdSG}{CPT}{passthrough_lit}{0.103}
\setres{pdSG}{CPT}{passthrough_uniform}{0.464}
\setres{pdSG}{CPT}{prior_contrib_lit}{0.771}
\setres{pdSG}{CPT}{prior_contrib_uniform}{0.461}
\setres{pdSG}{CPT}{share_lit}{46}
\setres{pdSG}{CPT}{share_uniform}{-81}
\setres{pdSG}{CPT_deterministic}{kappa}{0.972}
\setres{pdSG}{CPT_deterministic}{kappa_se}{0.008}
\setres{pdSG}{CPT_deterministic}{mse}{63.70}
\setres{pdSG}{CPT_deterministic}{r_ff}{0.140}
\setres{pdSG}{CPT_deterministic}{r_ff_se}{0.001}
\setres{pdSG}{CPT_deterministic}{r_prior}{1.0000}
\setres{pdSG}{CPT_deterministic}{r_prior_se}{---}
\setres{pdSG}{CPT_deterministic}{r_prob}{1.000}
\setres{pdSG}{CPT_deterministic}{r_prob_se}{0.000}
\setres{pdSG}{CPT_deterministic}{volume_ess}{---}
\setres{pdSG}{CPT_lit}{kappa}{0.444}
\setres{pdSG}{CPT_lit}{kappa_prior_mean}{0.873}
\setres{pdSG}{CPT_lit}{kappa_prior_mean_se}{0.031}
\setres{pdSG}{CPT_lit}{kappa_se}{0.063}
\setres{pdSG}{CPT_lit}{mse}{101.72}
\setres{pdSG}{CPT_lit}{mse_prior_mean}{70.81}
\setres{pdSG}{CPT_lit}{pred_variance}{30.91}
\setres{pdSG}{CPT_lit}{r_ff}{0.140}
\setres{pdSG}{CPT_lit}{r_ff_se}{0.001}
\setres{pdSG}{CPT_lit}{r_prior}{0.8966}
\setres{pdSG}{CPT_lit}{r_prior_se}{0.001}
\setres{pdSG}{CPT_lit}{r_prob}{0.911}
\setres{pdSG}{CPT_lit}{r_prob_se}{0.001}
\setres{pdSG}{CPT_lit}{spread_penalty}{0.429}
\setres{pdSG}{CPT_lit}{spread_penalty_se}{0.032}
\setres{pdSG}{CPT_lit}{volume_ess}{55{,}652}
\setres{pdSG}{CPT_uniform}{kappa}{-0.792}
\setres{pdSG}{CPT_uniform}{kappa_prior_mean}{0.846}
\setres{pdSG}{CPT_uniform}{kappa_prior_mean_se}{0.027}
\setres{pdSG}{CPT_uniform}{kappa_se}{0.105}
\setres{pdSG}{CPT_uniform}{mse}{190.67}
\setres{pdSG}{CPT_uniform}{mse_prior_mean}{72.76}
\setres{pdSG}{CPT_uniform}{pred_variance}{117.91}
\setres{pdSG}{CPT_uniform}{r_ff}{0.140}
\setres{pdSG}{CPT_uniform}{r_ff_se}{0.001}
\setres{pdSG}{CPT_uniform}{r_prior}{0.5364}
\setres{pdSG}{CPT_uniform}{r_prior_se}{0.003}
\setres{pdSG}{CPT_uniform}{r_prob}{0.601}
\setres{pdSG}{CPT_uniform}{r_prob_se}{0.003}
\setres{pdSG}{CPT_uniform}{spread_penalty}{1.638}
\setres{pdSG}{CPT_uniform}{spread_penalty_se}{0.121}
\setres{pdSG}{CPT_uniform}{volume_ess}{55{,}652}
\setres{pdSG}{DA}{det_pct}{63}
\setres{pdSG}{DA}{gap_cpt_lit}{0.953}
\setres{pdSG}{DA}{gap_lit}{1.135}
\setres{pdSG}{DA}{gap_r_cpt_lit}{0.034}
\setres{pdSG}{DA}{gap_uniform}{1.069}
\setres{pdSG}{DA}{one_minus_rff}{0.719}
\setres{pdSG}{DA}{passthrough_lit}{0.077}
\setres{pdSG}{DA}{passthrough_uniform}{0.151}
\setres{pdSG}{DA}{prior_contrib_lit}{0.664}
\setres{pdSG}{DA}{prior_contrib_uniform}{0.610}
\setres{pdSG}{DA}{share_lit}{-81}
\setres{pdSG}{DA}{share_uniform}{-71}
\setres{pdSG}{DA_deterministic}{kappa}{0.626}
\setres{pdSG}{DA_deterministic}{kappa_se}{0.062}
\setres{pdSG}{DA_deterministic}{mse}{88.64}
\setres{pdSG}{DA_deterministic}{r_ff}{0.281}
\setres{pdSG}{DA_deterministic}{r_ff_se}{0.004}
\setres{pdSG}{DA_deterministic}{r_prior}{1.0000}
\setres{pdSG}{DA_deterministic}{r_prior_se}{---}
\setres{pdSG}{DA_deterministic}{r_prob}{1.000}
\setres{pdSG}{DA_deterministic}{r_prob_se}{0.000}
\setres{pdSG}{DA_deterministic}{volume_ess}{---}
\setres{pdSG}{DA_lit}{kappa}{-0.509}
\setres{pdSG}{DA_lit}{kappa_prior_mean}{-0.177}
\setres{pdSG}{DA_lit}{kappa_prior_mean_se}{0.175}
\setres{pdSG}{DA_lit}{kappa_se}{0.199}
\setres{pdSG}{DA_lit}{mse}{170.33}
\setres{pdSG}{DA_lit}{mse_prior_mean}{146.46}
\setres{pdSG}{DA_lit}{pred_variance}{23.88}
\setres{pdSG}{DA_lit}{r_ff}{0.281}
\setres{pdSG}{DA_lit}{r_ff_se}{0.004}
\setres{pdSG}{DA_lit}{r_prior}{0.9234}
\setres{pdSG}{DA_lit}{r_prior_se}{0.000}
\setres{pdSG}{DA_lit}{r_prob}{0.945}
\setres{pdSG}{DA_lit}{r_prob_se}{0.000}
\setres{pdSG}{DA_lit}{spread_penalty}{0.332}
\setres{pdSG}{DA_lit}{spread_penalty_se}{0.024}
\setres{pdSG}{DA_lit}{volume_ess}{436{,}452}
\setres{pdSG}{DA_uniform}{kappa}{-0.443}
\setres{pdSG}{DA_uniform}{kappa_prior_mean}{0.735}
\setres{pdSG}{DA_uniform}{kappa_prior_mean_se}{0.057}
\setres{pdSG}{DA_uniform}{kappa_se}{0.143}
\setres{pdSG}{DA_uniform}{mse}{165.57}
\setres{pdSG}{DA_uniform}{mse_prior_mean}{80.80}
\setres{pdSG}{DA_uniform}{pred_variance}{84.76}
\setres{pdSG}{DA_uniform}{r_ff}{0.281}
\setres{pdSG}{DA_uniform}{r_ff_se}{0.004}
\setres{pdSG}{DA_uniform}{r_prior}{0.8493}
\setres{pdSG}{DA_uniform}{r_prior_se}{0.001}
\setres{pdSG}{DA_uniform}{r_prob}{0.892}
\setres{pdSG}{DA_uniform}{r_prob_se}{0.001}
\setres{pdSG}{DA_uniform}{spread_penalty}{1.177}
\setres{pdSG}{DA_uniform}{spread_penalty_se}{0.087}
\setres{pdSG}{DA_uniform}{volume_ess}{436{,}452}
\setres{pdSG}{EU}{det_pct}{62}
\setres{pdSG}{EU}{gap_lit}{0.284}
\setres{pdSG}{EU}{gap_uniform}{0.784}
\setres{pdSG}{EU}{one_minus_rff}{0.664}
\setres{pdSG}{EU}{passthrough_lit}{0.364}
\setres{pdSG}{EU}{passthrough_uniform}{0.485}
\setres{pdSG}{EU}{prior_contrib_lit}{0.423}
\setres{pdSG}{EU}{prior_contrib_uniform}{0.342}
\setres{pdSG}{EU}{share_lit}{54}
\setres{pdSG}{EU}{share_uniform}{-27}
\setres{pdSG}{EU_deterministic}{kappa}{0.616}
\setres{pdSG}{EU_deterministic}{kappa_se}{0.061}
\setres{pdSG}{EU_deterministic}{mse}{89.31}
\setres{pdSG}{EU_deterministic}{r_ff}{0.336}
\setres{pdSG}{EU_deterministic}{r_ff_se}{0.005}
\setres{pdSG}{EU_deterministic}{r_prior}{1.0000}
\setres{pdSG}{EU_deterministic}{r_prior_se}{---}
\setres{pdSG}{EU_deterministic}{r_prob}{1.000}
\setres{pdSG}{EU_deterministic}{r_prob_se}{0.000}
\setres{pdSG}{EU_deterministic}{volume_ess}{---}
\setres{pdSG}{EU_lit}{kappa}{0.333}
\setres{pdSG}{EU_lit}{kappa_prior_mean}{0.633}
\setres{pdSG}{EU_lit}{kappa_prior_mean_se}{0.065}
\setres{pdSG}{EU_lit}{kappa_se}{0.087}
\setres{pdSG}{EU_lit}{mse}{109.74}
\setres{pdSG}{EU_lit}{mse_prior_mean}{88.15}
\setres{pdSG}{EU_lit}{pred_variance}{21.60}
\setres{pdSG}{EU_lit}{r_ff}{0.336}
\setres{pdSG}{EU_lit}{r_ff_se}{0.005}
\setres{pdSG}{EU_lit}{r_prior}{0.6364}
\setres{pdSG}{EU_lit}{r_prior_se}{0.001}
\setres{pdSG}{EU_lit}{r_prob}{0.758}
\setres{pdSG}{EU_lit}{r_prob_se}{0.002}
\setres{pdSG}{EU_lit}{spread_penalty}{0.300}
\setres{pdSG}{EU_lit}{spread_penalty_se}{0.022}
\setres{pdSG}{EU_lit}{volume_ess}{172{,}086}
\setres{pdSG}{EU_uniform}{kappa}{-0.167}
\setres{pdSG}{EU_uniform}{kappa_prior_mean}{0.695}
\setres{pdSG}{EU_uniform}{kappa_prior_mean_se}{0.051}
\setres{pdSG}{EU_uniform}{kappa_se}{0.110}
\setres{pdSG}{EU_uniform}{mse}{145.73}
\setres{pdSG}{EU_uniform}{mse_prior_mean}{83.65}
\setres{pdSG}{EU_uniform}{pred_variance}{62.08}
\setres{pdSG}{EU_uniform}{r_ff}{0.336}
\setres{pdSG}{EU_uniform}{r_ff_se}{0.005}
\setres{pdSG}{EU_uniform}{r_prior}{0.5146}
\setres{pdSG}{EU_uniform}{r_prior_se}{0.002}
\setres{pdSG}{EU_uniform}{r_prob}{0.678}
\setres{pdSG}{EU_uniform}{r_prob_se}{0.003}
\setres{pdSG}{EU_uniform}{spread_penalty}{0.862}
\setres{pdSG}{EU_uniform}{spread_penalty_se}{0.064}
\setres{pdSG}{EU_uniform}{volume_ess}{172{,}086}
\setres{pdSG}{EV_deterministic}{kappa}{0.493}
\setres{pdSG}{EV_deterministic}{kappa_se}{0.064}
\setres{pdSG}{EV_deterministic}{mse}{98.19}
\setres{pdSG}{EV_deterministic}{r_ff}{0.484}
\setres{pdSG}{EV_deterministic}{r_ff_se}{---}
\setres{pdSG}{EV_deterministic}{r_prior}{1.0000}
\setres{pdSG}{EV_deterministic}{r_prior_se}{---}
\setres{pdSG}{EV_deterministic}{r_prob}{1.000}
\setres{pdSG}{EV_deterministic}{r_prob_se}{0.000}
\setres{pdSG}{EV_deterministic}{volume_ess}{---}
\setres{pdSL}{CPT}{det_pct}{98}
\setres{pdSL}{CPT}{gap_lit}{1.433}
\setres{pdSL}{CPT}{gap_uniform}{2.016}
\setres{pdSL}{CPT}{one_minus_rff}{0.856}
\setres{pdSL}{CPT}{passthrough_lit}{0.099}
\setres{pdSL}{CPT}{passthrough_uniform}{0.397}
\setres{pdSL}{CPT}{prior_contrib_lit}{0.771}
\setres{pdSL}{CPT}{prior_contrib_uniform}{0.516}
\setres{pdSL}{CPT}{share_lit}{-46}
\setres{pdSL}{CPT}{share_uniform}{-106}
\setres{pdSL}{CPT_deterministic}{kappa}{0.980}
\setres{pdSL}{CPT_deterministic}{kappa_se}{0.010}
\setres{pdSL}{CPT_deterministic}{mse}{70.22}
\setres{pdSL}{CPT_deterministic}{r_ff}{0.144}
\setres{pdSL}{CPT_deterministic}{r_ff_se}{0.002}
\setres{pdSL}{CPT_deterministic}{r_prior}{1.0000}
\setres{pdSL}{CPT_deterministic}{r_prior_se}{---}
\setres{pdSL}{CPT_deterministic}{r_prob}{1.000}
\setres{pdSL}{CPT_deterministic}{r_prob_se}{0.000}
\setres{pdSL}{CPT_deterministic}{volume_ess}{---}
\setres{pdSL}{CPT_lit}{kappa}{-0.453}
\setres{pdSL}{CPT_lit}{kappa_prior_mean}{0.024}
\setres{pdSL}{CPT_lit}{kappa_prior_mean_se}{0.095}
\setres{pdSL}{CPT_lit}{kappa_se}{0.121}
\setres{pdSL}{CPT_lit}{mse}{163.74}
\setres{pdSL}{CPT_lit}{mse_prior_mean}{132.56}
\setres{pdSL}{CPT_lit}{pred_variance}{31.18}
\setres{pdSL}{CPT_lit}{r_ff}{0.144}
\setres{pdSL}{CPT_lit}{r_ff_se}{0.002}
\setres{pdSL}{CPT_lit}{r_prior}{0.9012}
\setres{pdSL}{CPT_lit}{r_prior_se}{0.001}
\setres{pdSL}{CPT_lit}{r_prob}{0.915}
\setres{pdSL}{CPT_lit}{r_prob_se}{0.001}
\setres{pdSL}{CPT_lit}{spread_penalty}{0.478}
\setres{pdSL}{CPT_lit}{spread_penalty_se}{0.041}
\setres{pdSL}{CPT_lit}{volume_ess}{52{,}317}
\setres{pdSL}{CPT_uniform}{kappa}{-1.037}
\setres{pdSL}{CPT_uniform}{kappa_prior_mean}{0.736}
\setres{pdSL}{CPT_uniform}{kappa_prior_mean_se}{0.039}
\setres{pdSL}{CPT_uniform}{kappa_se}{0.152}
\setres{pdSL}{CPT_uniform}{mse}{201.83}
\setres{pdSL}{CPT_uniform}{mse_prior_mean}{86.08}
\setres{pdSL}{CPT_uniform}{pred_variance}{115.74}
\setres{pdSL}{CPT_uniform}{r_ff}{0.144}
\setres{pdSL}{CPT_uniform}{r_ff_se}{0.002}
\setres{pdSL}{CPT_uniform}{r_prior}{0.6030}
\setres{pdSL}{CPT_uniform}{r_prior_se}{0.003}
\setres{pdSL}{CPT_uniform}{r_prob}{0.660}
\setres{pdSL}{CPT_uniform}{r_prob_se}{0.003}
\setres{pdSL}{CPT_uniform}{spread_penalty}{1.773}
\setres{pdSL}{CPT_uniform}{spread_penalty_se}{0.152}
\setres{pdSL}{CPT_uniform}{volume_ess}{52{,}317}
\setres{pdSL}{DA}{det_pct}{87}
\setres{pdSL}{DA}{gap_cpt_lit}{1.590}
\setres{pdSL}{DA}{gap_lit}{2.916}
\setres{pdSL}{DA}{gap_r_cpt_lit}{0.002}
\setres{pdSL}{DA}{gap_uniform}{1.802}
\setres{pdSL}{DA}{one_minus_rff}{0.808}
\setres{pdSL}{DA}{passthrough_lit}{0.102}
\setres{pdSL}{DA}{passthrough_uniform}{0.166}
\setres{pdSL}{DA}{prior_contrib_lit}{0.726}
\setres{pdSL}{DA}{prior_contrib_uniform}{0.674}
\setres{pdSL}{DA}{share_lit}{-234}
\setres{pdSL}{DA}{share_uniform}{-107}
\setres{pdSL}{DA_deterministic}{kappa}{0.872}
\setres{pdSL}{DA_deterministic}{kappa_se}{0.017}
\setres{pdSL}{DA_deterministic}{mse}{77.23}
\setres{pdSL}{DA_deterministic}{r_ff}{0.192}
\setres{pdSL}{DA_deterministic}{r_ff_se}{0.002}
\setres{pdSL}{DA_deterministic}{r_prior}{1.0000}
\setres{pdSL}{DA_deterministic}{r_prior_se}{---}
\setres{pdSL}{DA_deterministic}{r_prob}{1.000}
\setres{pdSL}{DA_deterministic}{r_prob_se}{0.000}
\setres{pdSL}{DA_deterministic}{volume_ess}{---}
\setres{pdSL}{DA_lit}{kappa}{-2.043}
\setres{pdSL}{DA_lit}{kappa_prior_mean}{-1.560}
\setres{pdSL}{DA_lit}{kappa_prior_mean_se}{0.210}
\setres{pdSL}{DA_lit}{kappa_se}{0.243}
\setres{pdSL}{DA_lit}{mse}{267.53}
\setres{pdSL}{DA_lit}{mse_prior_mean}{235.96}
\setres{pdSL}{DA_lit}{pred_variance}{31.57}
\setres{pdSL}{DA_lit}{r_ff}{0.192}
\setres{pdSL}{DA_lit}{r_ff_se}{0.002}
\setres{pdSL}{DA_lit}{r_prior}{0.8981}
\setres{pdSL}{DA_lit}{r_prior_se}{0.001}
\setres{pdSL}{DA_lit}{r_prob}{0.918}
\setres{pdSL}{DA_lit}{r_prob_se}{0.001}
\setres{pdSL}{DA_lit}{spread_penalty}{0.484}
\setres{pdSL}{DA_lit}{spread_penalty_se}{0.042}
\setres{pdSL}{DA_lit}{volume_ess}{30{,}044}
\setres{pdSL}{DA_uniform}{kappa}{-0.930}
\setres{pdSL}{DA_uniform}{kappa_prior_mean}{0.584}
\setres{pdSL}{DA_uniform}{kappa_prior_mean_se}{0.070}
\setres{pdSL}{DA_uniform}{kappa_se}{0.190}
\setres{pdSL}{DA_uniform}{mse}{194.87}
\setres{pdSL}{DA_uniform}{mse_prior_mean}{96.02}
\setres{pdSL}{DA_uniform}{pred_variance}{98.84}
\setres{pdSL}{DA_uniform}{r_ff}{0.192}
\setres{pdSL}{DA_uniform}{r_ff_se}{0.002}
\setres{pdSL}{DA_uniform}{r_prior}{0.8339}
\setres{pdSL}{DA_uniform}{r_prior_se}{0.001}
\setres{pdSL}{DA_uniform}{r_prob}{0.866}
\setres{pdSL}{DA_uniform}{r_prob_se}{0.001}
\setres{pdSL}{DA_uniform}{spread_penalty}{1.514}
\setres{pdSL}{DA_uniform}{spread_penalty_se}{0.130}
\setres{pdSL}{DA_uniform}{volume_ess}{30{,}044}
\setres{pdSL}{EU}{det_pct}{59}
\setres{pdSL}{EU}{gap_lit}{1.259}
\setres{pdSL}{EU}{gap_uniform}{1.421}
\setres{pdSL}{EU}{one_minus_rff}{0.622}
\setres{pdSL}{EU}{passthrough_lit}{0.417}
\setres{pdSL}{EU}{passthrough_uniform}{0.554}
\setres{pdSL}{EU}{prior_contrib_lit}{0.363}
\setres{pdSL}{EU}{prior_contrib_uniform}{0.278}
\setres{pdSL}{EU}{share_lit}{-113}
\setres{pdSL}{EU}{share_uniform}{-141}
\setres{pdSL}{EU_deterministic}{kappa}{0.590}
\setres{pdSL}{EU_deterministic}{kappa_se}{0.081}
\setres{pdSL}{EU_deterministic}{mse}{95.66}
\setres{pdSL}{EU_deterministic}{r_ff}{0.378}
\setres{pdSL}{EU_deterministic}{r_ff_se}{0.006}
\setres{pdSL}{EU_deterministic}{r_prior}{1.0000}
\setres{pdSL}{EU_deterministic}{r_prior_se}{---}
\setres{pdSL}{EU_deterministic}{r_prob}{1.000}
\setres{pdSL}{EU_deterministic}{r_prob_se}{0.000}
\setres{pdSL}{EU_deterministic}{volume_ess}{---}
\setres{pdSL}{EU_lit}{kappa}{-0.669}
\setres{pdSL}{EU_lit}{kappa_prior_mean}{-0.337}
\setres{pdSL}{EU_lit}{kappa_prior_mean_se}{0.132}
\setres{pdSL}{EU_lit}{kappa_se}{0.153}
\setres{pdSL}{EU_lit}{mse}{177.81}
\setres{pdSL}{EU_lit}{mse_prior_mean}{156.16}
\setres{pdSL}{EU_lit}{pred_variance}{21.65}
\setres{pdSL}{EU_lit}{r_ff}{0.378}
\setres{pdSL}{EU_lit}{r_ff_se}{0.006}
\setres{pdSL}{EU_lit}{r_prior}{0.5835}
\setres{pdSL}{EU_lit}{r_prior_se}{0.002}
\setres{pdSL}{EU_lit}{r_prob}{0.741}
\setres{pdSL}{EU_lit}{r_prob_se}{0.002}
\setres{pdSL}{EU_lit}{spread_penalty}{0.332}
\setres{pdSL}{EU_lit}{spread_penalty_se}{0.029}
\setres{pdSL}{EU_lit}{volume_ess}{133{,}991}
\setres{pdSL}{EU_uniform}{kappa}{-0.831}
\setres{pdSL}{EU_uniform}{kappa_prior_mean}{0.122}
\setres{pdSL}{EU_uniform}{kappa_prior_mean_se}{0.102}
\setres{pdSL}{EU_uniform}{kappa_se}{0.168}
\setres{pdSL}{EU_uniform}{mse}{188.41}
\setres{pdSL}{EU_uniform}{mse_prior_mean}{126.22}
\setres{pdSL}{EU_uniform}{pred_variance}{62.18}
\setres{pdSL}{EU_uniform}{r_ff}{0.378}
\setres{pdSL}{EU_uniform}{r_ff_se}{0.006}
\setres{pdSL}{EU_uniform}{r_prior}{0.4460}
\setres{pdSL}{EU_uniform}{r_prior_se}{0.002}
\setres{pdSL}{EU_uniform}{r_prob}{0.655}
\setres{pdSL}{EU_uniform}{r_prob_se}{0.003}
\setres{pdSL}{EU_uniform}{spread_penalty}{0.953}
\setres{pdSL}{EU_uniform}{spread_penalty_se}{0.082}
\setres{pdSL}{EU_uniform}{volume_ess}{133{,}991}
\setres{pdSL}{EV_deterministic}{kappa}{0.384}
\setres{pdSL}{EV_deterministic}{kappa_se}{0.085}
\setres{pdSL}{EV_deterministic}{mse}{109.08}
\setres{pdSL}{EV_deterministic}{r_ff}{0.480}
\setres{pdSL}{EV_deterministic}{r_ff_se}{---}
\setres{pdSL}{EV_deterministic}{r_prior}{1.0000}
\setres{pdSL}{EV_deterministic}{r_prior_se}{---}
\setres{pdSL}{EV_deterministic}{r_prob}{1.000}
\setres{pdSL}{EV_deterministic}{r_prob_se}{0.000}
\setres{pdSL}{EV_deterministic}{volume_ess}{---}
\setres{share}{CPT}{delta_separate_gains}{96.1}
\setres{share}{CPT}{delta_separate_losses}{98.9}
\setres{share}{CPT}{delta_shared_gains}{132.9}
\setres{share}{CPT}{delta_shared_losses}{133.8}
\setres{share}{CPT}{kappa_avg}{0.827}
\setres{share}{CPT}{kappa_avg_se}{0.027}
\setres{share}{CPT}{kappa_corr}{0.646}
\setres{share}{CPT}{kappa_gains}{0.836}
\setres{share}{CPT}{kappa_gains_se}{0.029}
\setres{share}{CPT}{kappa_gains_separate}{0.972}
\setres{share}{CPT}{kappa_losses}{0.818}
\setres{share}{CPT}{kappa_losses_se}{0.031}
\setres{share}{CPT}{kappa_losses_separate}{0.980}
\setres{share}{CPT}{mse_gains}{73.496}
\setres{share}{CPT}{mse_losses}{80.735}
\setres{share}{CPT}{n_infeasible}{0}
\setres{share}{CPT}{n_infeasible_separate_gains}{0}
\setres{share}{CPT}{n_infeasible_separate_losses}{0}
\setres{share}{CPT}{n_sim}{1{,}000}
\setres{share}{CPT}{normaliser_gains}{684.1}
\setres{share}{CPT}{normaliser_losses}{685.5}
\setres{share}{CPT}{r_ff_avg}{0.195}
\setres{share}{CPT}{r_ff_gains}{0.194}
\setres{share}{CPT}{r_ff_gains_se}{0.002}
\setres{share}{CPT}{r_ff_gains_separate}{0.141}
\setres{share}{CPT}{r_ff_joint}{0.195}
\setres{share}{CPT}{r_ff_joint_se}{0.002}
\setres{share}{CPT}{r_ff_losses}{0.195}
\setres{share}{CPT}{r_ff_losses_se}{0.002}
\setres{share}{CPT}{r_ff_losses_separate}{0.144}
\setres{share}{CPT}{theta_star}{1.00, 0.40, 1.00}
\setres{share}{CPT}{theta_star_str}{1.00, 0.40, 1.00}
\setres{share}{DA}{delta_separate_gains}{191.8}
\setres{share}{DA}{delta_separate_losses}{132.2}
\setres{share}{DA}{delta_shared_gains}{232.8}
\setres{share}{DA}{delta_shared_losses}{226.3}
\setres{share}{DA}{kappa_avg}{0.580}
\setres{share}{DA}{kappa_avg_se}{0.068}
\setres{share}{DA}{kappa_corr}{0.702}
\setres{share}{DA}{kappa_gains}{0.624}
\setres{share}{DA}{kappa_gains_se}{0.062}
\setres{share}{DA}{kappa_gains_separate}{0.626}
\setres{share}{DA}{kappa_losses}{0.536}
\setres{share}{DA}{kappa_losses_se}{0.085}
\setres{share}{DA}{kappa_losses_separate}{0.872}
\setres{share}{DA}{mse_gains}{88.756}
\setres{share}{DA}{mse_losses}{99.140}
\setres{share}{DA}{n_infeasible}{0}
\setres{share}{DA}{n_infeasible_separate_gains}{0}
\setres{share}{DA}{n_infeasible_separate_losses}{0}
\setres{share}{DA}{n_sim}{1{,}000}
\setres{share}{DA}{normaliser_gains}{684.1}
\setres{share}{DA}{normaliser_losses}{685.5}
\setres{share}{DA}{r_ff_avg}{0.335}
\setres{share}{DA}{r_ff_gains}{0.340}
\setres{share}{DA}{r_ff_gains_se}{0.005}
\setres{share}{DA}{r_ff_gains_separate}{0.280}
\setres{share}{DA}{r_ff_joint}{0.335}
\setres{share}{DA}{r_ff_joint_se}{0.003}
\setres{share}{DA}{r_ff_losses}{0.330}
\setres{share}{DA}{r_ff_losses_se}{0.005}
\setres{share}{DA}{r_ff_losses_separate}{0.193}
\setres{share}{DA}{theta_star}{1.03, 0.48}
\setres{share}{DA}{theta_star_str}{1.03, 0.48}
\setres{share}{EU}{delta_separate_gains}{229.8}
\setres{share}{EU}{delta_separate_losses}{259.0}
\setres{share}{EU}{delta_shared_gains}{283.2}
\setres{share}{EU}{delta_shared_losses}{301.0}
\setres{share}{EU}{kappa_avg}{0.441}
\setres{share}{EU}{kappa_avg_se}{0.067}
\setres{share}{EU}{kappa_corr}{0.638}
\setres{share}{EU}{kappa_gains}{0.460}
\setres{share}{EU}{kappa_gains_se}{0.064}
\setres{share}{EU}{kappa_gains_separate}{0.616}
\setres{share}{EU}{kappa_losses}{0.422}
\setres{share}{EU}{kappa_losses_se}{0.084}
\setres{share}{EU}{kappa_losses_separate}{0.590}
\setres{share}{EU}{mse_gains}{100.561}
\setres{share}{EU}{mse_losses}{106.626}
\setres{share}{EU}{n_infeasible}{0}
\setres{share}{EU}{n_infeasible_separate_gains}{0}
\setres{share}{EU}{n_infeasible_separate_losses}{0}
\setres{share}{EU}{n_sim}{1{,}000}
\setres{share}{EU}{normaliser_gains}{684.1}
\setres{share}{EU}{normaliser_losses}{685.5}
\setres{share}{EU}{r_ff_avg}{0.427}
\setres{share}{EU}{r_ff_gains}{0.414}
\setres{share}{EU}{r_ff_gains_se}{0.006}
\setres{share}{EU}{r_ff_gains_separate}{0.336}
\setres{share}{EU}{r_ff_joint}{0.427}
\setres{share}{EU}{r_ff_joint_se}{0.004}
\setres{share}{EU}{r_ff_losses}{0.439}
\setres{share}{EU}{r_ff_losses_se}{0.006}
\setres{share}{EU}{r_ff_losses_separate}{0.378}
\setres{share}{EU}{theta_star}{1.03}
\setres{share}{EU}{theta_star_str}{1.03}
\setres{share}{reference}{n_lotteries}{50}
\setres{share}{reference}{n_lotteries_gains}{25}
\setres{share}{reference}{n_lotteries_losses}{25}
\setres{share}{reference}{n_subjects}{179}
\setres{repl}{CPT}{kappa_ev}{0.945}
\setres{repl}{CPT}{kappa_ev_se}{0.017}
\setres{repl}{CPT}{kappa_mid}{0.972}
\setres{repl}{CPT}{kappa_mid_se}{0.008}
\setres{repl}{CPT}{published_kappa}{0.950}
\setres{repl}{CPT}{published_kappa_se}{0.020}
\setres{repl}{CPT}{published_r}{0.280}
\setres{repl}{CPT}{published_r_se}{0.003}
\setres{repl}{CPT}{r_b}{0.274}
\setres{repl}{CPT}{r_b_se}{0.003}
\setres{repl}{CPT}{r_c}{0.258}
\setres{repl}{CPT}{r_c_se}{0.003}
\setres{repl}{CPT}{r_d}{0.262}
\setres{repl}{CPT}{r_d_se}{0.003}
\setres{repl}{CPT}{r_e}{0.290}
\setres{repl}{CPT}{r_e_se}{0.003}
\setres{repl}{CPT}{r_f}{0.561}
\setres{repl}{CPT}{r_f_se}{0.005}
\setres{repl}{CPT}{r_g}{0.160}
\setres{repl}{CPT}{r_g_se}{0.002}
\setres{repl}{CPT}{r_h}{0.140}
\setres{repl}{CPT}{r_h_se}{0.001}
\setres{repl}{DA}{kappa_ev}{0.262}
\setres{repl}{DA}{kappa_ev_se}{0.058}
\setres{repl}{DA}{kappa_mid}{0.626}
\setres{repl}{DA}{kappa_mid_se}{0.062}
\setres{repl}{DA}{published_kappa}{0.270}
\setres{repl}{DA}{published_kappa_se}{0.060}
\setres{repl}{DA}{published_r}{0.470}
\setres{repl}{DA}{published_r_se}{0.006}
\setres{repl}{DA}{r_b}{0.494}
\setres{repl}{DA}{r_b_se}{0.005}
\setres{repl}{DA}{r_c}{0.382}
\setres{repl}{DA}{r_c_se}{0.005}
\setres{repl}{DA}{r_d}{0.446}
\setres{repl}{DA}{r_d_se}{0.007}
\setres{repl}{DA}{r_e}{0.581}
\setres{repl}{DA}{r_e_se}{0.008}
\setres{repl}{DA}{r_f}{1.125}
\setres{repl}{DA}{r_f_se}{0.023}
\setres{repl}{DA}{r_g}{0.320}
\setres{repl}{DA}{r_g_se}{0.006}
\setres{repl}{DA}{r_h}{0.281}
\setres{repl}{DA}{r_h_se}{0.004}
\setres{repl}{derived}{midpoint_rule_rff}{0.250}
\setres{repl}{derived}{midpoint_rule_rff_const}{0.285}
\setres{box}{CPT_alpha_lo05__lit}{kappa}{0.641}
\setres{box}{CPT_alpha_lo05__lit}{kappa_prior_mean}{0.853}
\setres{box}{CPT_alpha_lo05__lit}{kappa_se}{0.050}
\setres{box}{CPT_alpha_lo05__lit}{r_ff}{0.140}
\setres{box}{CPT_alpha_lo05__lit}{r_ff_se}{0.001}
\setres{box}{CPT_alpha_lo05__lit}{r_prior}{0.9483}
\setres{box}{CPT_alpha_lo05__lit}{r_prior_se}{0.001}
\setres{box}{CPT_alpha_lo05__lit}{r_prob}{0.956}
\setres{box}{CPT_alpha_lo05__lit}{r_prob_se}{0.001}
\setres{box}{CPT_alpha_lo05__lit}{volume_ess}{11{,}135}
\setres{box}{CPT_alpha_lo05__uniform}{kappa}{-0.412}
\setres{box}{CPT_alpha_lo05__uniform}{kappa_prior_mean}{0.854}
\setres{box}{CPT_alpha_lo05__uniform}{kappa_se}{0.119}
\setres{box}{CPT_alpha_lo05__uniform}{r_ff}{0.140}
\setres{box}{CPT_alpha_lo05__uniform}{r_ff_se}{0.001}
\setres{box}{CPT_alpha_lo05__uniform}{r_prior}{0.5009}
\setres{box}{CPT_alpha_lo05__uniform}{r_prior_se}{0.005}
\setres{box}{CPT_alpha_lo05__uniform}{r_prob}{0.571}
\setres{box}{CPT_alpha_lo05__uniform}{r_prob_se}{0.004}
\setres{box}{CPT_alpha_lo05__uniform}{volume_ess}{11{,}135}
\setres{box}{CPT_alpha_lo20__lit}{kappa}{0.641}
\setres{box}{CPT_alpha_lo20__lit}{kappa_prior_mean}{0.853}
\setres{box}{CPT_alpha_lo20__lit}{kappa_se}{0.050}
\setres{box}{CPT_alpha_lo20__lit}{r_ff}{0.140}
\setres{box}{CPT_alpha_lo20__lit}{r_ff_se}{0.001}
\setres{box}{CPT_alpha_lo20__lit}{r_prior}{0.9484}
\setres{box}{CPT_alpha_lo20__lit}{r_prior_se}{0.001}
\setres{box}{CPT_alpha_lo20__lit}{r_prob}{0.956}
\setres{box}{CPT_alpha_lo20__lit}{r_prob_se}{0.000}
\setres{box}{CPT_alpha_lo20__lit}{volume_ess}{11{,}135}
\setres{box}{CPT_alpha_lo20__uniform}{kappa}{-0.149}
\setres{box}{CPT_alpha_lo20__uniform}{kappa_prior_mean}{0.962}
\setres{box}{CPT_alpha_lo20__uniform}{kappa_se}{0.086}
\setres{box}{CPT_alpha_lo20__uniform}{r_ff}{0.140}
\setres{box}{CPT_alpha_lo20__uniform}{r_ff_se}{0.001}
\setres{box}{CPT_alpha_lo20__uniform}{r_prior}{0.5851}
\setres{box}{CPT_alpha_lo20__uniform}{r_prior_se}{0.004}
\setres{box}{CPT_alpha_lo20__uniform}{r_prob}{0.643}
\setres{box}{CPT_alpha_lo20__uniform}{r_prob_se}{0.004}
\setres{box}{CPT_alpha_lo20__uniform}{volume_ess}{11{,}135}
\setres{box}{CPT_delta_hi2__lit}{kappa}{0.643}
\setres{box}{CPT_delta_hi2__lit}{kappa_prior_mean}{0.852}
\setres{box}{CPT_delta_hi2__lit}{kappa_se}{0.050}
\setres{box}{CPT_delta_hi2__lit}{r_ff}{0.140}
\setres{box}{CPT_delta_hi2__lit}{r_ff_se}{0.001}
\setres{box}{CPT_delta_hi2__lit}{r_prior}{0.9495}
\setres{box}{CPT_delta_hi2__lit}{r_prior_se}{0.001}
\setres{box}{CPT_delta_hi2__lit}{r_prob}{0.957}
\setres{box}{CPT_delta_hi2__lit}{r_prob_se}{0.000}
\setres{box}{CPT_delta_hi2__lit}{volume_ess}{11{,}135}
\setres{box}{CPT_delta_hi2__uniform}{kappa}{-0.518}
\setres{box}{CPT_delta_hi2__uniform}{kappa_prior_mean}{0.480}
\setres{box}{CPT_delta_hi2__uniform}{kappa_se}{0.159}
\setres{box}{CPT_delta_hi2__uniform}{r_ff}{0.140}
\setres{box}{CPT_delta_hi2__uniform}{r_ff_se}{0.001}
\setres{box}{CPT_delta_hi2__uniform}{r_prior}{0.6394}
\setres{box}{CPT_delta_hi2__uniform}{r_prior_se}{0.004}
\setres{box}{CPT_delta_hi2__uniform}{r_prob}{0.690}
\setres{box}{CPT_delta_hi2__uniform}{r_prob_se}{0.003}
\setres{box}{CPT_delta_hi2__uniform}{volume_ess}{11{,}135}
\setres{box}{CPT_delta_hi5__lit}{kappa}{0.641}
\setres{box}{CPT_delta_hi5__lit}{kappa_prior_mean}{0.853}
\setres{box}{CPT_delta_hi5__lit}{kappa_se}{0.050}
\setres{box}{CPT_delta_hi5__lit}{r_ff}{0.140}
\setres{box}{CPT_delta_hi5__lit}{r_ff_se}{0.001}
\setres{box}{CPT_delta_hi5__lit}{r_prior}{0.9596}
\setres{box}{CPT_delta_hi5__lit}{r_prior_se}{0.000}
\setres{box}{CPT_delta_hi5__lit}{r_prob}{0.965}
\setres{box}{CPT_delta_hi5__lit}{r_prob_se}{0.000}
\setres{box}{CPT_delta_hi5__lit}{volume_ess}{8{,}658}
\setres{box}{CPT_delta_hi5__uniform}{kappa}{-0.793}
\setres{box}{CPT_delta_hi5__uniform}{kappa_prior_mean}{0.866}
\setres{box}{CPT_delta_hi5__uniform}{kappa_se}{0.113}
\setres{box}{CPT_delta_hi5__uniform}{r_ff}{0.140}
\setres{box}{CPT_delta_hi5__uniform}{r_ff_se}{0.001}
\setres{box}{CPT_delta_hi5__uniform}{r_prior}{0.5326}
\setres{box}{CPT_delta_hi5__uniform}{r_prior_se}{0.005}
\setres{box}{CPT_delta_hi5__uniform}{r_prob}{0.598}
\setres{box}{CPT_delta_hi5__uniform}{r_prob_se}{0.005}
\setres{box}{CPT_delta_hi5__uniform}{volume_ess}{8{,}658}
\setres{box}{CPT_gamma_lo20__lit}{kappa}{0.641}
\setres{box}{CPT_gamma_lo20__lit}{kappa_prior_mean}{0.851}
\setres{box}{CPT_gamma_lo20__lit}{kappa_se}{0.050}
\setres{box}{CPT_gamma_lo20__lit}{r_ff}{0.140}
\setres{box}{CPT_gamma_lo20__lit}{r_ff_se}{0.001}
\setres{box}{CPT_gamma_lo20__lit}{r_prior}{0.9494}
\setres{box}{CPT_gamma_lo20__lit}{r_prior_se}{0.001}
\setres{box}{CPT_gamma_lo20__lit}{r_prob}{0.957}
\setres{box}{CPT_gamma_lo20__lit}{r_prob_se}{0.000}
\setres{box}{CPT_gamma_lo20__lit}{volume_ess}{11{,}135}
\setres{box}{CPT_gamma_lo20__uniform}{kappa}{-0.374}
\setres{box}{CPT_gamma_lo20__uniform}{kappa_prior_mean}{0.812}
\setres{box}{CPT_gamma_lo20__uniform}{kappa_se}{0.127}
\setres{box}{CPT_gamma_lo20__uniform}{r_ff}{0.140}
\setres{box}{CPT_gamma_lo20__uniform}{r_ff_se}{0.001}
\setres{box}{CPT_gamma_lo20__uniform}{r_prior}{0.6394}
\setres{box}{CPT_gamma_lo20__uniform}{r_prior_se}{0.004}
\setres{box}{CPT_gamma_lo20__uniform}{r_prob}{0.690}
\setres{box}{CPT_gamma_lo20__uniform}{r_prob_se}{0.003}
\setres{box}{CPT_gamma_lo20__uniform}{volume_ess}{11{,}135}
\setres{box}{CPT_main__lit}{kappa}{0.641}
\setres{box}{CPT_main__lit}{kappa_prior_mean}{0.853}
\setres{box}{CPT_main__lit}{kappa_se}{0.050}
\setres{box}{CPT_main__lit}{r_ff}{0.140}
\setres{box}{CPT_main__lit}{r_ff_se}{0.001}
\setres{box}{CPT_main__lit}{r_prior}{0.9483}
\setres{box}{CPT_main__lit}{r_prior_se}{0.001}
\setres{box}{CPT_main__lit}{r_prob}{0.956}
\setres{box}{CPT_main__lit}{r_prob_se}{0.001}
\setres{box}{CPT_main__lit}{volume_ess}{11{,}135}
\setres{box}{CPT_main__uniform}{kappa}{-0.495}
\setres{box}{CPT_main__uniform}{kappa_prior_mean}{0.811}
\setres{box}{CPT_main__uniform}{kappa_se}{0.129}
\setres{box}{CPT_main__uniform}{r_ff}{0.140}
\setres{box}{CPT_main__uniform}{r_ff_se}{0.001}
\setres{box}{CPT_main__uniform}{r_prior}{0.5122}
\setres{box}{CPT_main__uniform}{r_prior_se}{0.005}
\setres{box}{CPT_main__uniform}{r_prob}{0.581}
\setres{box}{CPT_main__uniform}{r_prob_se}{0.004}
\setres{box}{CPT_main__uniform}{volume_ess}{11{,}135}
\setres{box}{DA_alpha_lo05__lit}{kappa}{0.020}
\setres{box}{DA_alpha_lo05__lit}{kappa_prior_mean}{0.340}
\setres{box}{DA_alpha_lo05__lit}{kappa_se}{0.136}
\setres{box}{DA_alpha_lo05__lit}{r_ff}{0.281}
\setres{box}{DA_alpha_lo05__lit}{r_ff_se}{0.004}
\setres{box}{DA_alpha_lo05__lit}{r_prior}{0.9384}
\setres{box}{DA_alpha_lo05__lit}{r_prior_se}{0.000}
\setres{box}{DA_alpha_lo05__lit}{r_prob}{0.956}
\setres{box}{DA_alpha_lo05__lit}{r_prob_se}{0.000}
\setres{box}{DA_alpha_lo05__lit}{volume_ess}{87{,}113}
\setres{box}{DA_alpha_lo05__uniform}{kappa}{-0.754}
\setres{box}{DA_alpha_lo05__uniform}{kappa_prior_mean}{0.115}
\setres{box}{DA_alpha_lo05__uniform}{kappa_se}{0.205}
\setres{box}{DA_alpha_lo05__uniform}{r_ff}{0.281}
\setres{box}{DA_alpha_lo05__uniform}{r_ff_se}{0.004}
\setres{box}{DA_alpha_lo05__uniform}{r_prior}{0.8712}
\setres{box}{DA_alpha_lo05__uniform}{r_prior_se}{0.001}
\setres{box}{DA_alpha_lo05__uniform}{r_prob}{0.907}
\setres{box}{DA_alpha_lo05__uniform}{r_prob_se}{0.001}
\setres{box}{DA_alpha_lo05__uniform}{volume_ess}{87{,}113}
\setres{box}{DA_alpha_lo20__lit}{kappa}{0.060}
\setres{box}{DA_alpha_lo20__lit}{kappa_prior_mean}{0.361}
\setres{box}{DA_alpha_lo20__lit}{kappa_se}{0.132}
\setres{box}{DA_alpha_lo20__lit}{r_ff}{0.281}
\setres{box}{DA_alpha_lo20__lit}{r_ff_se}{0.004}
\setres{box}{DA_alpha_lo20__lit}{r_prior}{0.9430}
\setres{box}{DA_alpha_lo20__lit}{r_prior_se}{0.000}
\setres{box}{DA_alpha_lo20__lit}{r_prob}{0.959}
\setres{box}{DA_alpha_lo20__lit}{r_prob_se}{0.000}
\setres{box}{DA_alpha_lo20__lit}{volume_ess}{87{,}113}
\setres{box}{DA_alpha_lo20__uniform}{kappa}{-0.523}
\setres{box}{DA_alpha_lo20__uniform}{kappa_prior_mean}{0.301}
\setres{box}{DA_alpha_lo20__uniform}{kappa_se}{0.180}
\setres{box}{DA_alpha_lo20__uniform}{r_ff}{0.281}
\setres{box}{DA_alpha_lo20__uniform}{r_ff_se}{0.004}
\setres{box}{DA_alpha_lo20__uniform}{r_prior}{0.9036}
\setres{box}{DA_alpha_lo20__uniform}{r_prior_se}{0.001}
\setres{box}{DA_alpha_lo20__uniform}{r_prob}{0.931}
\setres{box}{DA_alpha_lo20__uniform}{r_prob_se}{0.001}
\setres{box}{DA_alpha_lo20__uniform}{volume_ess}{87{,}113}
\setres{box}{DA_eta_hi5__lit}{kappa}{0.020}
\setres{box}{DA_eta_hi5__lit}{kappa_prior_mean}{0.340}
\setres{box}{DA_eta_hi5__lit}{kappa_se}{0.136}
\setres{box}{DA_eta_hi5__lit}{r_ff}{0.280}
\setres{box}{DA_eta_hi5__lit}{r_ff_se}{0.004}
\setres{box}{DA_eta_hi5__lit}{r_prior}{0.9405}
\setres{box}{DA_eta_hi5__lit}{r_prior_se}{0.000}
\setres{box}{DA_eta_hi5__lit}{r_prob}{0.957}
\setres{box}{DA_eta_hi5__lit}{r_prob_se}{0.000}
\setres{box}{DA_eta_hi5__lit}{volume_ess}{87{,}474}
\setres{box}{DA_eta_hi5__uniform}{kappa}{-1.101}
\setres{box}{DA_eta_hi5__uniform}{kappa_prior_mean}{-0.371}
\setres{box}{DA_eta_hi5__uniform}{kappa_se}{0.248}
\setres{box}{DA_eta_hi5__uniform}{r_ff}{0.280}
\setres{box}{DA_eta_hi5__uniform}{r_ff_se}{0.004}
\setres{box}{DA_eta_hi5__uniform}{r_prior}{0.8946}
\setres{box}{DA_eta_hi5__uniform}{r_prior_se}{0.001}
\setres{box}{DA_eta_hi5__uniform}{r_prob}{0.924}
\setres{box}{DA_eta_hi5__uniform}{r_prob_se}{0.001}
\setres{box}{DA_eta_hi5__uniform}{volume_ess}{87{,}474}
\setres{box}{DA_eta_lo00__lit}{kappa}{-0.066}
\setres{box}{DA_eta_lo00__lit}{kappa_prior_mean}{0.126}
\setres{box}{DA_eta_lo00__lit}{kappa_se}{0.155}
\setres{box}{DA_eta_lo00__lit}{r_ff}{0.281}
\setres{box}{DA_eta_lo00__lit}{r_ff_se}{0.004}
\setres{box}{DA_eta_lo00__lit}{r_prior}{0.9674}
\setres{box}{DA_eta_lo00__lit}{r_prior_se}{0.000}
\setres{box}{DA_eta_lo00__lit}{r_prob}{0.977}
\setres{box}{DA_eta_lo00__lit}{r_prob_se}{0.000}
\setres{box}{DA_eta_lo00__lit}{volume_ess}{87{,}113}
\setres{box}{DA_eta_lo00__uniform}{kappa}{-0.928}
\setres{box}{DA_eta_lo00__uniform}{kappa_prior_mean}{-0.611}
\setres{box}{DA_eta_lo00__uniform}{kappa_se}{0.244}
\setres{box}{DA_eta_lo00__uniform}{r_ff}{0.281}
\setres{box}{DA_eta_lo00__uniform}{r_ff_se}{0.004}
\setres{box}{DA_eta_lo00__uniform}{r_prior}{0.9435}
\setres{box}{DA_eta_lo00__uniform}{r_prior_se}{0.000}
\setres{box}{DA_eta_lo00__uniform}{r_prob}{0.959}
\setres{box}{DA_eta_lo00__uniform}{r_prob_se}{0.000}
\setres{box}{DA_eta_lo00__uniform}{volume_ess}{87{,}113}
\setres{box}{DA_eta_lo50__lit}{kappa}{0.026}
\setres{box}{DA_eta_lo50__lit}{kappa_prior_mean}{0.284}
\setres{box}{DA_eta_lo50__lit}{kappa_se}{0.139}
\setres{box}{DA_eta_lo50__lit}{r_ff}{0.281}
\setres{box}{DA_eta_lo50__lit}{r_ff_se}{0.004}
\setres{box}{DA_eta_lo50__lit}{r_prior}{0.9471}
\setres{box}{DA_eta_lo50__lit}{r_prior_se}{0.000}
\setres{box}{DA_eta_lo50__lit}{r_prob}{0.962}
\setres{box}{DA_eta_lo50__lit}{r_prob_se}{0.000}
\setres{box}{DA_eta_lo50__lit}{volume_ess}{87{,}113}
\setres{box}{DA_eta_lo50__uniform}{kappa}{-0.814}
\setres{box}{DA_eta_lo50__uniform}{kappa_prior_mean}{-0.372}
\setres{box}{DA_eta_lo50__uniform}{kappa_se}{0.228}
\setres{box}{DA_eta_lo50__uniform}{r_ff}{0.281}
\setres{box}{DA_eta_lo50__uniform}{r_ff_se}{0.004}
\setres{box}{DA_eta_lo50__uniform}{r_prior}{0.9169}
\setres{box}{DA_eta_lo50__uniform}{r_prior_se}{0.000}
\setres{box}{DA_eta_lo50__uniform}{r_prob}{0.940}
\setres{box}{DA_eta_lo50__uniform}{r_prob_se}{0.000}
\setres{box}{DA_eta_lo50__uniform}{volume_ess}{87{,}113}
\setres{box}{DA_eta_lo75__lit}{kappa}{0.031}
\setres{box}{DA_eta_lo75__lit}{kappa_prior_mean}{0.321}
\setres{box}{DA_eta_lo75__lit}{kappa_se}{0.137}
\setres{box}{DA_eta_lo75__lit}{r_ff}{0.281}
\setres{box}{DA_eta_lo75__lit}{r_ff_se}{0.004}
\setres{box}{DA_eta_lo75__lit}{r_prior}{0.9411}
\setres{box}{DA_eta_lo75__lit}{r_prior_se}{0.000}
\setres{box}{DA_eta_lo75__lit}{r_prob}{0.958}
\setres{box}{DA_eta_lo75__lit}{r_prob_se}{0.000}
\setres{box}{DA_eta_lo75__lit}{volume_ess}{87{,}113}
\setres{box}{DA_eta_lo75__uniform}{kappa}{-0.773}
\setres{box}{DA_eta_lo75__uniform}{kappa_prior_mean}{-0.203}
\setres{box}{DA_eta_lo75__uniform}{kappa_se}{0.219}
\setres{box}{DA_eta_lo75__uniform}{r_ff}{0.281}
\setres{box}{DA_eta_lo75__uniform}{r_ff_se}{0.004}
\setres{box}{DA_eta_lo75__uniform}{r_prior}{0.8972}
\setres{box}{DA_eta_lo75__uniform}{r_prior_se}{0.001}
\setres{box}{DA_eta_lo75__uniform}{r_prob}{0.926}
\setres{box}{DA_eta_lo75__uniform}{r_prob_se}{0.001}
\setres{box}{DA_eta_lo75__uniform}{volume_ess}{87{,}113}
\setres{box}{DA_main__lit}{kappa}{0.020}
\setres{box}{DA_main__lit}{kappa_prior_mean}{0.340}
\setres{box}{DA_main__lit}{kappa_se}{0.136}
\setres{box}{DA_main__lit}{r_ff}{0.281}
\setres{box}{DA_main__lit}{r_ff_se}{0.004}
\setres{box}{DA_main__lit}{r_prior}{0.9383}
\setres{box}{DA_main__lit}{r_prior_se}{0.000}
\setres{box}{DA_main__lit}{r_prob}{0.956}
\setres{box}{DA_main__lit}{r_prob_se}{0.000}
\setres{box}{DA_main__lit}{volume_ess}{87{,}113}
\setres{box}{DA_main__uniform}{kappa}{-0.827}
\setres{box}{DA_main__uniform}{kappa_prior_mean}{0.060}
\setres{box}{DA_main__uniform}{kappa_se}{0.212}
\setres{box}{DA_main__uniform}{r_ff}{0.281}
\setres{box}{DA_main__uniform}{r_ff_se}{0.004}
\setres{box}{DA_main__uniform}{r_prior}{0.8666}
\setres{box}{DA_main__uniform}{r_prior_se}{0.001}
\setres{box}{DA_main__uniform}{r_prob}{0.904}
\setres{box}{DA_main__uniform}{r_prob_se}{0.001}
\setres{box}{DA_main__uniform}{volume_ess}{87{,}113}
\setres{box}{EU_alpha_lo05__lit}{kappa}{0.521}
\setres{box}{EU_alpha_lo05__lit}{kappa_prior_mean}{0.611}
\setres{box}{EU_alpha_lo05__lit}{kappa_se}{0.074}
\setres{box}{EU_alpha_lo05__lit}{r_ff}{0.336}
\setres{box}{EU_alpha_lo05__lit}{r_ff_se}{0.005}
\setres{box}{EU_alpha_lo05__lit}{r_prior}{0.8140}
\setres{box}{EU_alpha_lo05__lit}{r_prior_se}{0.001}
\setres{box}{EU_alpha_lo05__lit}{r_prob}{0.876}
\setres{box}{EU_alpha_lo05__lit}{r_prob_se}{0.001}
\setres{box}{EU_alpha_lo05__lit}{volume_ess}{34{,}727}
\setres{box}{EU_alpha_lo05__uniform}{kappa}{-0.118}
\setres{box}{EU_alpha_lo05__uniform}{kappa_prior_mean}{0.339}
\setres{box}{EU_alpha_lo05__uniform}{kappa_se}{0.145}
\setres{box}{EU_alpha_lo05__uniform}{r_ff}{0.336}
\setres{box}{EU_alpha_lo05__uniform}{r_ff_se}{0.005}
\setres{box}{EU_alpha_lo05__uniform}{r_prior}{0.6076}
\setres{box}{EU_alpha_lo05__uniform}{r_prior_se}{0.002}
\setres{box}{EU_alpha_lo05__uniform}{r_prob}{0.739}
\setres{box}{EU_alpha_lo05__uniform}{r_prob_se}{0.002}
\setres{box}{EU_alpha_lo05__uniform}{volume_ess}{34{,}727}
\setres{box}{EU_alpha_lo20__lit}{kappa}{0.522}
\setres{box}{EU_alpha_lo20__lit}{kappa_prior_mean}{0.611}
\setres{box}{EU_alpha_lo20__lit}{kappa_se}{0.074}
\setres{box}{EU_alpha_lo20__lit}{r_ff}{0.336}
\setres{box}{EU_alpha_lo20__lit}{r_ff_se}{0.005}
\setres{box}{EU_alpha_lo20__lit}{r_prior}{0.8145}
\setres{box}{EU_alpha_lo20__lit}{r_prior_se}{0.001}
\setres{box}{EU_alpha_lo20__lit}{r_prob}{0.877}
\setres{box}{EU_alpha_lo20__lit}{r_prob_se}{0.001}
\setres{box}{EU_alpha_lo20__lit}{volume_ess}{34{,}727}
\setres{box}{EU_alpha_lo20__uniform}{kappa}{0.189}
\setres{box}{EU_alpha_lo20__uniform}{kappa_prior_mean}{0.488}
\setres{box}{EU_alpha_lo20__uniform}{kappa_se}{0.113}
\setres{box}{EU_alpha_lo20__uniform}{r_ff}{0.336}
\setres{box}{EU_alpha_lo20__uniform}{r_ff_se}{0.005}
\setres{box}{EU_alpha_lo20__uniform}{r_prior}{0.6994}
\setres{box}{EU_alpha_lo20__uniform}{r_prior_se}{0.002}
\setres{box}{EU_alpha_lo20__uniform}{r_prob}{0.800}
\setres{box}{EU_alpha_lo20__uniform}{r_prob_se}{0.002}
\setres{box}{EU_alpha_lo20__uniform}{volume_ess}{34{,}727}
\setres{box}{EU_main__lit}{kappa}{0.521}
\setres{box}{EU_main__lit}{kappa_prior_mean}{0.611}
\setres{box}{EU_main__lit}{kappa_se}{0.074}
\setres{box}{EU_main__lit}{r_ff}{0.336}
\setres{box}{EU_main__lit}{r_ff_se}{0.005}
\setres{box}{EU_main__lit}{r_prior}{0.8140}
\setres{box}{EU_main__lit}{r_prior_se}{0.001}
\setres{box}{EU_main__lit}{r_prob}{0.876}
\setres{box}{EU_main__lit}{r_prob_se}{0.001}
\setres{box}{EU_main__lit}{volume_ess}{34{,}727}
\setres{box}{EU_main__uniform}{kappa}{-0.209}
\setres{box}{EU_main__uniform}{kappa_prior_mean}{0.298}
\setres{box}{EU_main__uniform}{kappa_se}{0.154}
\setres{box}{EU_main__uniform}{r_ff}{0.336}
\setres{box}{EU_main__uniform}{r_ff_se}{0.005}
\setres{box}{EU_main__uniform}{r_prior}{0.5777}
\setres{box}{EU_main__uniform}{r_prior_se}{0.002}
\setres{box}{EU_main__uniform}{r_prob}{0.719}
\setres{box}{EU_main__uniform}{r_prob_se}{0.003}
\setres{box}{EU_main__uniform}{volume_ess}{34{,}727}
\setres{kern}{diagonal__CPT_lit}{r_ff}{0.125}
\setres{kern}{diagonal__CPT_lit}{r_ff_se}{0.001}
\setres{kern}{diagonal__CPT_uniform}{r_ff}{0.125}
\setres{kern}{diagonal__CPT_uniform}{r_ff_se}{0.001}
\setres{kern}{diagonal__DA_lit}{r_ff}{0.243}
\setres{kern}{diagonal__DA_lit}{r_ff_se}{0.003}
\setres{kern}{diagonal__DA_uniform}{r_ff}{0.243}
\setres{kern}{diagonal__DA_uniform}{r_ff_se}{0.003}
\setres{kern}{diagonal__EU_lit}{r_ff}{0.289}
\setres{kern}{diagonal__EU_lit}{r_ff_se}{0.004}
\setres{kern}{diagonal__EU_uniform}{r_ff}{0.289}
\setres{kern}{diagonal__EU_uniform}{r_ff_se}{0.004}
\setres{kern}{linear__CPT_lit}{r_ff}{0.032}
\setres{kern}{linear__CPT_lit}{r_ff_se}{0.001}
\setres{kern}{linear__CPT_uniform}{r_ff}{0.032}
\setres{kern}{linear__CPT_uniform}{r_ff_se}{0.001}
\setres{kern}{linear__DA_lit}{r_ff}{0.123}
\setres{kern}{linear__DA_lit}{r_ff_se}{0.004}
\setres{kern}{linear__DA_uniform}{r_ff}{0.123}
\setres{kern}{linear__DA_uniform}{r_ff_se}{0.004}
\setres{kern}{linear__EU_lit}{r_ff}{0.152}
\setres{kern}{linear__EU_lit}{r_ff_se}{0.004}
\setres{kern}{linear__EU_uniform}{r_ff}{0.152}
\setres{kern}{linear__EU_uniform}{r_ff_se}{0.004}
\setres{kern}{matern52__CPT_lit}{r_ff}{0.047}
\setres{kern}{matern52__CPT_lit}{r_ff_se}{0.001}
\setres{kern}{matern52__CPT_uniform}{r_ff}{0.047}
\setres{kern}{matern52__CPT_uniform}{r_ff_se}{0.001}
\setres{kern}{matern52__DA_lit}{r_ff}{0.169}
\setres{kern}{matern52__DA_lit}{r_ff_se}{0.004}
\setres{kern}{matern52__DA_uniform}{r_ff}{0.169}
\setres{kern}{matern52__DA_uniform}{r_ff_se}{0.004}
\setres{kern}{matern52__EU_lit}{r_ff}{0.195}
\setres{kern}{matern52__EU_lit}{r_ff_se}{0.004}
\setres{kern}{matern52__EU_uniform}{r_ff}{0.195}
\setres{kern}{matern52__EU_uniform}{r_ff_se}{0.004}
\setres{kern}{rbf__CPT_lit}{r_ff}{0.038}
\setres{kern}{rbf__CPT_lit}{r_ff_se}{0.001}
\setres{kern}{rbf__CPT_uniform}{r_ff}{0.038}
\setres{kern}{rbf__CPT_uniform}{r_ff_se}{0.001}
\setres{kern}{rbf__DA_lit}{r_ff}{0.157}
\setres{kern}{rbf__DA_lit}{r_ff_se}{0.004}
\setres{kern}{rbf__DA_uniform}{r_ff}{0.157}
\setres{kern}{rbf__DA_uniform}{r_ff_se}{0.004}
\setres{kern}{rbf__EU_lit}{r_ff}{0.183}
\setres{kern}{rbf__EU_lit}{r_ff_se}{0.004}
\setres{kern}{rbf__EU_uniform}{r_ff}{0.183}
\setres{kern}{rbf__EU_uniform}{r_ff_se}{0.004}
\setres{mid}{CPT}{one_minus_rff}{0.405}
\setres{mid}{CPT}{passthrough_lit}{0.074}
\setres{mid}{CPT}{passthrough_uniform}{0.696}
\setres{mid}{CPT}{prior_contrib_lit}{0.376}
\setres{mid}{CPT}{prior_contrib_uniform}{0.123}
\setres{mid}{CPT_deterministic}{r_ff}{---}
\setres{mid}{CPT_deterministic}{r_ff_se}{---}
\setres{mid}{CPT_deterministic}{r_prior}{1.0000}
\setres{mid}{CPT_deterministic}{r_prob}{1.000}
\setres{mid}{CPT_deterministic}{r_prob_se}{---}
\setres{mid}{CPT_lit}{r_ff}{0.595}
\setres{mid}{CPT_lit}{r_ff_se}{0.006}
\setres{mid}{CPT_lit}{r_prior}{0.9262}
\setres{mid}{CPT_lit}{r_prob}{0.970}
\setres{mid}{CPT_lit}{r_prob_se}{0.000}
\setres{mid}{CPT_uniform}{r_ff}{0.595}
\setres{mid}{CPT_uniform}{r_ff_se}{0.006}
\setres{mid}{CPT_uniform}{r_prior}{0.3036}
\setres{mid}{CPT_uniform}{r_prob}{0.718}
\setres{mid}{CPT_uniform}{r_prob_se}{0.004}
\setres{mid}{DA}{gap_r_cpt_lit}{0.032}
\setres{mid}{DA}{one_minus_rff}{-0.033}
\setres{mid}{DA}{passthrough_lit}{0.072}
\setres{mid}{DA}{passthrough_uniform}{0.155}
\setres{mid}{DA}{prior_contrib_lit}{-0.031}
\setres{mid}{DA}{prior_contrib_uniform}{-0.028}
\setres{mid}{DA_deterministic}{r_ff}{---}
\setres{mid}{DA_deterministic}{r_ff_se}{---}
\setres{mid}{DA_deterministic}{r_prior}{1.0000}
\setres{mid}{DA_deterministic}{r_prob}{1.000}
\setres{mid}{DA_deterministic}{r_prob_se}{---}
\setres{mid}{DA_lit}{r_ff}{1.033}
\setres{mid}{DA_lit}{r_ff_se}{0.017}
\setres{mid}{DA_lit}{r_prior}{0.9283}
\setres{mid}{DA_lit}{r_prob}{1.002}
\setres{mid}{DA_lit}{r_prob_se}{0.001}
\setres{mid}{DA_uniform}{r_ff}{1.033}
\setres{mid}{DA_uniform}{r_ff_se}{0.017}
\setres{mid}{DA_uniform}{r_prior}{0.8450}
\setres{mid}{DA_uniform}{r_prob}{1.005}
\setres{mid}{DA_uniform}{r_prob_se}{0.003}
\setres{mid}{EU}{one_minus_rff}{-0.748}
\setres{mid}{EU}{passthrough_lit}{0.406}
\setres{mid}{EU}{passthrough_uniform}{0.923}
\setres{mid}{EU}{prior_contrib_lit}{-0.444}
\setres{mid}{EU}{prior_contrib_uniform}{-0.058}
\setres{mid}{EU_deterministic}{r_ff}{---}
\setres{mid}{EU_deterministic}{r_ff_se}{---}
\setres{mid}{EU_deterministic}{r_prior}{1.0000}
\setres{mid}{EU_deterministic}{r_prob}{1.000}
\setres{mid}{EU_deterministic}{r_prob_se}{---}
\setres{mid}{EU_lit}{r_ff}{1.748}
\setres{mid}{EU_lit}{r_ff_se}{0.027}
\setres{mid}{EU_lit}{r_prior}{0.5935}
\setres{mid}{EU_lit}{r_prob}{1.304}
\setres{mid}{EU_lit}{r_prob_se}{0.011}
\setres{mid}{EU_uniform}{r_ff}{1.748}
\setres{mid}{EU_uniform}{r_ff_se}{0.027}
\setres{mid}{EU_uniform}{r_prior}{0.0769}
\setres{mid}{EU_uniform}{r_prob}{1.691}
\setres{mid}{EU_uniform}{r_prob_se}{0.025}
\setres{mid}{EV_deterministic}{r_ff}{1.936}
\setres{mid}{EV_deterministic}{r_ff_se}{---}
\setres{mid}{EV_deterministic}{r_prior}{1.0000}
\setres{mid}{EV_deterministic}{r_prob}{1.000}
\setres{mid}{EV_deterministic}{r_prob_se}{---}
\setres{mid}{const__CPT_deterministic}{r_ff}{0.160}
\setres{mid}{const__CPT_deterministic}{r_ff_se}{0.002}
\setres{mid}{const__CPT_deterministic}{r_prior}{1.0000}
\setres{mid}{const__CPT_deterministic}{r_prob}{1.000}
\setres{mid}{const__CPT_lit}{r_ff}{0.160}
\setres{mid}{const__CPT_lit}{r_ff_se}{0.002}
\setres{mid}{const__CPT_lit}{r_prior}{0.9486}
\setres{mid}{const__CPT_lit}{r_prob}{0.957}
\setres{mid}{const__CPT_uniform}{r_ff}{0.160}
\setres{mid}{const__CPT_uniform}{r_ff_se}{0.002}
\setres{mid}{const__CPT_uniform}{r_prior}{0.5147}
\setres{mid}{const__CPT_uniform}{r_prob}{0.592}
\setres{mid}{const__DA_deterministic}{r_ff}{0.320}
\setres{mid}{const__DA_deterministic}{r_ff_se}{0.006}
\setres{mid}{const__DA_deterministic}{r_prior}{1.0000}
\setres{mid}{const__DA_deterministic}{r_prob}{1.000}
\setres{mid}{const__DA_lit}{r_ff}{0.320}
\setres{mid}{const__DA_lit}{r_ff_se}{0.006}
\setres{mid}{const__DA_lit}{r_prior}{0.9383}
\setres{mid}{const__DA_lit}{r_prob}{0.958}
\setres{mid}{const__DA_uniform}{r_ff}{0.320}
\setres{mid}{const__DA_uniform}{r_ff_se}{0.006}
\setres{mid}{const__DA_uniform}{r_prior}{0.8666}
\setres{mid}{const__DA_uniform}{r_prob}{0.909}
\setres{mid}{const__EU_deterministic}{r_ff}{0.382}
\setres{mid}{const__EU_deterministic}{r_ff_se}{0.007}
\setres{mid}{const__EU_deterministic}{r_prior}{1.0000}
\setres{mid}{const__EU_deterministic}{r_prob}{1.000}
\setres{mid}{const__EU_lit}{r_ff}{0.382}
\setres{mid}{const__EU_lit}{r_ff_se}{0.007}
\setres{mid}{const__EU_lit}{r_prior}{0.8143}
\setres{mid}{const__EU_lit}{r_prob}{0.885}
\setres{mid}{const__EU_uniform}{r_ff}{0.382}
\setres{mid}{const__EU_uniform}{r_ff_se}{0.007}
\setres{mid}{const__EU_uniform}{r_prior}{0.5782}
\setres{mid}{const__EU_uniform}{r_prob}{0.739}
\setres{mid}{const__EV_deterministic}{r_ff}{0.551}
\setres{mid}{const__EV_deterministic}{r_ff_se}{---}
\setres{mid}{const__EV_deterministic}{r_prior}{1.0000}
\setres{mid}{const__EV_deterministic}{r_prob}{1.000}
\setres{mid}{midpoint__CPT_deterministic}{r_ff}{0.561}
\setres{mid}{midpoint__CPT_deterministic}{r_ff_se}{0.005}
\setres{mid}{midpoint__CPT_deterministic}{r_prior}{1.0000}
\setres{mid}{midpoint__CPT_deterministic}{r_prob}{1.000}
\setres{mid}{midpoint__CPT_lit}{r_ff}{0.561}
\setres{mid}{midpoint__CPT_lit}{r_ff_se}{0.005}
\setres{mid}{midpoint__CPT_lit}{r_prior}{0.9486}
\setres{mid}{midpoint__CPT_lit}{r_prob}{0.977}
\setres{mid}{midpoint__CPT_uniform}{r_ff}{0.561}
\setres{mid}{midpoint__CPT_uniform}{r_ff_se}{0.005}
\setres{mid}{midpoint__CPT_uniform}{r_prior}{0.5147}
\setres{mid}{midpoint__CPT_uniform}{r_prob}{0.787}
\setres{mid}{midpoint__DA_deterministic}{r_ff}{1.125}
\setres{mid}{midpoint__DA_deterministic}{r_ff_se}{0.023}
\setres{mid}{midpoint__DA_deterministic}{r_prior}{1.0000}
\setres{mid}{midpoint__DA_deterministic}{r_prob}{1.000}
\setres{mid}{midpoint__DA_lit}{r_ff}{1.125}
\setres{mid}{midpoint__DA_lit}{r_ff_se}{0.023}
\setres{mid}{midpoint__DA_lit}{r_prior}{0.9383}
\setres{mid}{midpoint__DA_lit}{r_prob}{1.008}
\setres{mid}{midpoint__DA_uniform}{r_ff}{1.125}
\setres{mid}{midpoint__DA_uniform}{r_ff_se}{0.023}
\setres{mid}{midpoint__DA_uniform}{r_prior}{0.8666}
\setres{mid}{midpoint__DA_uniform}{r_prob}{1.017}
\setres{mid}{midpoint__EU_deterministic}{r_ff}{1.343}
\setres{mid}{midpoint__EU_deterministic}{r_ff_se}{0.027}
\setres{mid}{midpoint__EU_deterministic}{r_prior}{1.0000}
\setres{mid}{midpoint__EU_deterministic}{r_prob}{1.000}
\setres{mid}{midpoint__EU_lit}{r_ff}{1.343}
\setres{mid}{midpoint__EU_lit}{r_ff_se}{0.027}
\setres{mid}{midpoint__EU_lit}{r_prior}{0.8143}
\setres{mid}{midpoint__EU_lit}{r_prob}{1.064}
\setres{mid}{midpoint__EU_uniform}{r_ff}{1.343}
\setres{mid}{midpoint__EU_uniform}{r_ff_se}{0.027}
\setres{mid}{midpoint__EU_uniform}{r_prior}{0.5782}
\setres{mid}{midpoint__EU_uniform}{r_prob}{1.145}
\setres{mid}{midpoint__EV_deterministic}{r_ff}{1.936}
\setres{mid}{midpoint__EV_deterministic}{r_ff_se}{---}
\setres{mid}{midpoint__EV_deterministic}{r_prior}{1.0000}
\setres{mid}{midpoint__EV_deterministic}{r_prob}{1.000}
\setres{anat}{CPT}{d_c2}{0.0}
\setres{anat}{CPT}{dc2_over_V}{0.00}
\setres{anat}{CPT}{location}{1.00}
\setres{anat}{CPT}{r_ff}{0.140}
\setres{anat}{CPT}{rff_central_rule}{0.250}
\setres{anat}{CPT}{size}{0.56}
\setres{anat}{CPT}{size_mc}{0.561}
\setres{anat}{CPT}{size_reduction_pct}{44}
\setres{anat}{CPT}{theta_centre}{0.983 0.010 1.025}
\setres{anat}{CPT_lit__COH_uniform}{channelwise}{yes}
\setres{anat}{CPT_lit__COH_uniform}{more_restrictive}{CPT\_lit}
\setres{anat}{CPT_lit__CPT_uniform}{channelwise}{yes}
\setres{anat}{CPT_lit__CPT_uniform}{more_restrictive}{CPT\_lit}
\setres{anat}{CPT_lit__DA_lit}{ahead_on}{r\_prior}
\setres{anat}{CPT_lit__DA_lit}{channelwise}{no}
\setres{anat}{CPT_lit__DA_lit}{more_restrictive}{CPT\_lit}
\setres{anat}{CPT_lit__DA_lit}{weight_factor}{1.0}
\setres{anat}{CPT_lit__DA_lit}{weight_ratio}{1.0}
\setres{anat}{CPT_lit__DA_uniform}{ahead_on}{r\_prior}
\setres{anat}{CPT_lit__DA_uniform}{channelwise}{no}
\setres{anat}{CPT_lit__DA_uniform}{more_restrictive}{CPT\_lit}
\setres{anat}{CPT_lit__DA_uniform}{weight_factor}{5.3}
\setres{anat}{CPT_lit__DA_uniform}{weight_ratio}{5.3}
\setres{anat}{CPT_lit__EU_lit}{ahead_on}{r\_prior}
\setres{anat}{CPT_lit__EU_lit}{channelwise}{no}
\setres{anat}{CPT_lit__EU_lit}{more_restrictive}{CPT\_lit}
\setres{anat}{CPT_lit__EU_lit}{weight_factor}{5.0}
\setres{anat}{CPT_lit__EU_lit}{weight_ratio}{5.0}
\setres{anat}{CPT_lit__EU_uniform}{ahead_on}{r\_prior}
\setres{anat}{CPT_lit__EU_uniform}{channelwise}{no}
\setres{anat}{CPT_lit__EU_uniform}{more_restrictive}{CPT\_lit}
\setres{anat}{CPT_lit__EU_uniform}{weight_factor}{8.2}
\setres{anat}{CPT_lit__EU_uniform}{weight_ratio}{8.2}
\setres{anat}{CPT_uniform__COH_uniform}{channelwise}{yes}
\setres{anat}{CPT_uniform__COH_uniform}{more_restrictive}{CPT\_uniform}
\setres{anat}{DA}{d_c2}{147.9}
\setres{anat}{DA}{dc2_over_V}{0.87}
\setres{anat}{DA}{location}{1.87}
\setres{anat}{DA}{r_ff}{0.281}
\setres{anat}{DA}{rff_central_rule}{0.467}
\setres{anat}{DA}{size}{0.60}
\setres{anat}{DA}{size_mc}{0.603}
\setres{anat}{DA}{size_reduction_pct}{40}
\setres{anat}{DA}{theta_centre}{3.158 3.000}
\setres{anat}{DA_lit__COH_uniform}{channelwise}{yes}
\setres{anat}{DA_lit__COH_uniform}{more_restrictive}{DA\_lit}
\setres{anat}{DA_lit__CPT_lit}{channelwise}{yes}
\setres{anat}{DA_lit__CPT_lit}{more_restrictive}{DA\_lit}
\setres{anat}{DA_lit__CPT_uniform}{channelwise}{yes}
\setres{anat}{DA_lit__CPT_uniform}{more_restrictive}{DA\_lit}
\setres{anat}{DA_lit__DA_uniform}{channelwise}{yes}
\setres{anat}{DA_lit__DA_uniform}{more_restrictive}{DA\_lit}
\setres{anat}{DA_lit__EU_lit}{ahead_on}{r\_prior}
\setres{anat}{DA_lit__EU_lit}{channelwise}{no}
\setres{anat}{DA_lit__EU_lit}{more_restrictive}{DA\_lit}
\setres{anat}{DA_lit__EU_lit}{weight_factor}{14.0}
\setres{anat}{DA_lit__EU_lit}{weight_ratio}{14.0}
\setres{anat}{DA_lit__EU_uniform}{ahead_on}{r\_prior}
\setres{anat}{DA_lit__EU_uniform}{channelwise}{no}
\setres{anat}{DA_lit__EU_uniform}{more_restrictive}{DA\_lit}
\setres{anat}{DA_lit__EU_uniform}{weight_factor}{24.4}
\setres{anat}{DA_lit__EU_uniform}{weight_ratio}{24.4}
\setres{anat}{DA_uniform__COH_uniform}{channelwise}{yes}
\setres{anat}{DA_uniform__COH_uniform}{more_restrictive}{DA\_uniform}
\setres{anat}{DA_uniform__CPT_uniform}{channelwise}{yes}
\setres{anat}{DA_uniform__CPT_uniform}{more_restrictive}{DA\_uniform}
\setres{anat}{DA_uniform__EU_lit}{ahead_on}{r\_prior}
\setres{anat}{DA_uniform__EU_lit}{channelwise}{no}
\setres{anat}{DA_uniform__EU_lit}{more_restrictive}{DA\_uniform}
\setres{anat}{DA_uniform__EU_lit}{weight_factor}{4.2}
\setres{anat}{DA_uniform__EU_lit}{weight_ratio}{4.2}
\setres{anat}{DA_uniform__EU_uniform}{ahead_on}{r\_prior}
\setres{anat}{DA_uniform__EU_uniform}{channelwise}{no}
\setres{anat}{DA_uniform__EU_uniform}{more_restrictive}{DA\_uniform}
\setres{anat}{DA_uniform__EU_uniform}{weight_factor}{14.6}
\setres{anat}{DA_uniform__EU_uniform}{weight_ratio}{14.6}
\setres{anat}{EU}{d_c2}{160.5}
\setres{anat}{EU}{dc2_over_V}{0.94}
\setres{anat}{EU}{location}{1.94}
\setres{anat}{EU}{r_ff}{0.336}
\setres{anat}{EU}{rff_central_rule}{0.485}
\setres{anat}{EU}{size}{0.69}
\setres{anat}{EU}{size_mc}{0.692}
\setres{anat}{EU}{size_reduction_pct}{31}
\setres{anat}{EU}{theta_centre}{1.146}
\setres{anat}{EU_lit__COH_uniform}{channelwise}{yes}
\setres{anat}{EU_lit__COH_uniform}{more_restrictive}{EU\_lit}
\setres{anat}{EU_lit__CPT_uniform}{channelwise}{yes}
\setres{anat}{EU_lit__CPT_uniform}{more_restrictive}{EU\_lit}
\setres{anat}{EU_lit__DA_uniform}{ahead_on}{r\_ff}
\setres{anat}{EU_lit__DA_uniform}{channelwise}{no}
\setres{anat}{EU_lit__DA_uniform}{more_restrictive}{EU\_lit}
\setres{anat}{EU_lit__DA_uniform}{weight_factor}{1.0}
\setres{anat}{EU_lit__DA_uniform}{weight_ratio}{1.0}
\setres{anat}{EU_lit__EU_uniform}{channelwise}{yes}
\setres{anat}{EU_lit__EU_uniform}{more_restrictive}{EU\_lit}
\setres{anat}{EU_uniform__COH_uniform}{channelwise}{yes}
\setres{anat}{EU_uniform__COH_uniform}{more_restrictive}{EU\_uniform}
\setres{anat}{EU_uniform__CPT_uniform}{ahead_on}{r\_ff}
\setres{anat}{EU_uniform__CPT_uniform}{channelwise}{yes}
\setres{anat}{EU_uniform__CPT_uniform}{more_restrictive}{EU\_uniform}
\setres{anat}{EU_uniform__CPT_uniform}{weight_factor}{18.9}
\setres{anat}{EU_uniform__CPT_uniform}{weight_ratio}{0.1}
\setres{anat}{EV}{d_c2}{162.2}
\setres{anat}{EV}{dc2_over_V}{0.95}
\setres{anat}{EV}{location}{1.95}
\setres{anat}{EV}{r_ff}{0.484}
\setres{anat}{EV}{rff_central_rule}{0.488}
\setres{anat}{EV}{size}{1.00}
\setres{anat}{EV}{size_mc}{0.993}
\setres{anat}{EV}{size_reduction_pct}{0}
\setres{anat}{EV}{theta_centre}{}
\setres{anat}{linkage}{CPT_one_domain}{0.191}
\setres{anat}{linkage}{CPT_one_domain_se}{0.002}
\setres{anat}{linkage}{CPT_one_minus_two_se}{0.002}
\setres{anat}{linkage}{CPT_two_domains}{0.141}
\setres{anat}{linkage}{CPT_two_domains_se}{0.001}
\setres{anat}{linkage}{DA_one_domain}{0.332}
\setres{anat}{linkage}{DA_one_domain_se}{0.003}
\setres{anat}{linkage}{DA_one_minus_two_se}{0.002}
\setres{anat}{linkage}{DA_two_domains}{0.236}
\setres{anat}{linkage}{DA_two_domains_se}{0.002}
\setres{anat}{linkage}{EU_one_domain}{0.418}
\setres{anat}{linkage}{EU_one_domain_se}{0.004}
\setres{anat}{linkage}{EU_one_minus_two_se}{0.003}
\setres{anat}{linkage}{EU_two_domains}{0.354}
\setres{anat}{linkage}{EU_two_domains_se}{0.004}
\setres{anat}{linkage}{n_sim}{1{,}000}
\setres{anat}{mirror}{CPT_losses}{0.144}
\setres{anat}{mirror}{CPT_losses_mirrored}{0.142}
\setres{anat}{mirror}{CPT_losses_mirrored_se}{0.001}
\setres{anat}{mirror}{CPT_losses_se}{0.002}
\setres{anat}{mirror}{EU_losses}{0.378}
\setres{anat}{mirror}{EU_losses_mirrored}{0.347}
\setres{anat}{mirror}{EU_losses_mirrored_se}{0.005}
\setres{anat}{mirror}{EU_losses_se}{0.006}
\setres{anat}{mirror}{n_differ}{1}
\setres{anat}{mirror}{n_sim}{1{,}000}
\setres{anat}{mirror}{share_V_pct}{20}
\setres{anat}{reference}{V}{170.7}
\setres{anat}{reference}{centre_rule_rff}{0.250}
\setres{anat}{reference}{rho2}{512.0}
\setres{anat}{steps}{DA_over_CPT}{2.0}
\setres{anat}{steps}{DA_over_CPT_location}{1.87}
\setres{anat}{steps}{DA_over_CPT_size}{1.08}
\setres{anat}{steps}{EU_over_CPT}{2.4}
\setres{anat}{steps}{EU_over_CPT_location}{1.94}
\setres{anat}{steps}{EU_over_CPT_size}{1.23}
\setres{anat}{steps}{EU_over_DA}{1.2}
\setres{anat}{steps}{EU_over_DA_location}{1.04}
\setres{anat}{steps}{EU_over_DA_size}{1.15}
\setres{anat}{steps}{EV_over_EU}{1.4}
\setres{anat}{steps}{EV_over_EU_location}{1.01}
\setres{anat}{steps}{EV_over_EU_size}{1.43}
\setres{anat}{steps}{EV_over_midpoint}{1.9}
\setres{anat}{summary}{frontier_channelwise}{CPT\_lit, DA\_lit, EU\_lit}
\setres{anat}{summary}{frontier_product}{CPT\_lit}
\setres{anat}{summary}{n_channelwise}{13}
\setres{anat}{summary}{n_frontier_channelwise}{3}
\setres{anat}{summary}{n_frontier_product}{1}
\setres{anat}{summary}{n_pairs}{21}
\setres{boxfit}{CPT_gains__alpha_hi2}{alpha_at_edge}{0}
\setres{boxfit}{CPT_gains__alpha_hi2}{alpha_star}{0.80}
\setres{boxfit}{CPT_gains__alpha_hi2}{kappa}{0.972}
\setres{boxfit}{CPT_gains__alpha_hi2}{kappa_se}{0.008}
\setres{boxfit}{CPT_gains__alpha_hi2}{log_vol}{12.75}
\setres{boxfit}{CPT_gains__alpha_hi2}{log_vol_se}{0.001}
\setres{boxfit}{CPT_gains__alpha_hi2}{r_ff}{0.142}
\setres{boxfit}{CPT_gains__alpha_hi2}{r_ff_se}{0.001}
\setres{boxfit}{CPT_gains__alpha_hi2}{theta_star}{0.80, 0.40, 1.00}
\setres{boxfit}{CPT_gains__alpha_hi2}{volume_ess}{464{,}350}
\setres{boxfit}{CPT_gains__box}{alpha_at_edge}{0}
\setres{boxfit}{CPT_gains__box}{alpha_star}{0.80}
\setres{boxfit}{CPT_gains__box}{kappa}{0.972}
\setres{boxfit}{CPT_gains__box}{kappa_se}{0.008}
\setres{boxfit}{CPT_gains__box}{log_vol}{12.49}
\setres{boxfit}{CPT_gains__box}{log_vol_se}{0.001}
\setres{boxfit}{CPT_gains__box}{r_ff}{0.149}
\setres{boxfit}{CPT_gains__box}{r_ff_se}{0.001}
\setres{boxfit}{CPT_gains__box}{theta_star}{0.80, 0.40, 1.00}
\setres{boxfit}{CPT_gains__box}{volume_ess}{632{,}669}
\setres{boxfit}{CPT_gains__domain}{alpha_at_edge}{0}
\setres{boxfit}{CPT_gains__domain}{alpha_star}{0.80}
\setres{boxfit}{CPT_gains__domain}{kappa}{0.972}
\setres{boxfit}{CPT_gains__domain}{kappa_se}{0.008}
\setres{boxfit}{CPT_gains__domain}{log_vol}{12.85}
\setres{boxfit}{CPT_gains__domain}{log_vol_se}{0.004}
\setres{boxfit}{CPT_gains__domain}{r_ff}{0.140}
\setres{boxfit}{CPT_gains__domain}{r_ff_se}{0.001}
\setres{boxfit}{CPT_gains__domain}{theta_star}{0.80, 0.40, 1.00}
\setres{boxfit}{CPT_gains__domain}{volume_ess}{55{,}652}
\setres{boxfit}{CPT_gains__gain}{box_volume_share_pct}{70}
\setres{boxfit}{CPT_gains__gain}{kappa_gain}{0.000}
\setres{boxfit}{CPT_gains__gain}{lifted_volume_share_pct}{90}
\setres{boxfit}{CPT_gains__gain}{r_ff_gain}{0.008}
\setres{boxfit}{CPT_gains__gain}{range_bits}{0.52}
\setres{boxfit}{CPT_gains__gain}{range_bits_form}{0.01}
\setres{boxfit}{CPT_gains__gain}{range_bits_lifted}{0.15}
\setres{boxfit}{CPT_losses__alpha_hi2}{alpha_at_edge}{0}
\setres{boxfit}{CPT_losses__alpha_hi2}{alpha_star}{1.40}
\setres{boxfit}{CPT_losses__alpha_hi2}{kappa}{0.980}
\setres{boxfit}{CPT_losses__alpha_hi2}{kappa_se}{0.010}
\setres{boxfit}{CPT_losses__alpha_hi2}{log_vol}{12.76}
\setres{boxfit}{CPT_losses__alpha_hi2}{log_vol_se}{0.001}
\setres{boxfit}{CPT_losses__alpha_hi2}{r_ff}{0.149}
\setres{boxfit}{CPT_losses__alpha_hi2}{r_ff_se}{0.002}
\setres{boxfit}{CPT_losses__alpha_hi2}{theta_star}{1.40, 0.40, 0.90}
\setres{boxfit}{CPT_losses__alpha_hi2}{volume_ess}{443{,}020}
\setres{boxfit}{CPT_losses__box}{alpha_at_edge}{1}
\setres{boxfit}{CPT_losses__box}{alpha_star}{1.00}
\setres{boxfit}{CPT_losses__box}{kappa}{0.953}
\setres{boxfit}{CPT_losses__box}{kappa_se}{0.018}
\setres{boxfit}{CPT_losses__box}{log_vol}{12.50}
\setres{boxfit}{CPT_losses__box}{log_vol_se}{0.001}
\setres{boxfit}{CPT_losses__box}{r_ff}{0.161}
\setres{boxfit}{CPT_losses__box}{r_ff_se}{0.002}
\setres{boxfit}{CPT_losses__box}{theta_star}{1.00, 0.40, 1.40}
\setres{boxfit}{CPT_losses__box}{volume_ess}{617{,}748}
\setres{boxfit}{CPT_losses__domain}{alpha_at_edge}{0}
\setres{boxfit}{CPT_losses__domain}{alpha_star}{1.40}
\setres{boxfit}{CPT_losses__domain}{kappa}{0.980}
\setres{boxfit}{CPT_losses__domain}{kappa_se}{0.010}
\setres{boxfit}{CPT_losses__domain}{log_vol}{12.86}
\setres{boxfit}{CPT_losses__domain}{log_vol_se}{0.004}
\setres{boxfit}{CPT_losses__domain}{r_ff}{0.144}
\setres{boxfit}{CPT_losses__domain}{r_ff_se}{0.002}
\setres{boxfit}{CPT_losses__domain}{theta_star}{1.40, 0.40, 0.90}
\setres{boxfit}{CPT_losses__domain}{volume_ess}{52{,}317}
\setres{boxfit}{CPT_losses__gain}{box_volume_share_pct}{70}
\setres{boxfit}{CPT_losses__gain}{kappa_gain}{0.026}
\setres{boxfit}{CPT_losses__gain}{lifted_volume_share_pct}{91}
\setres{boxfit}{CPT_losses__gain}{r_ff_gain}{0.017}
\setres{boxfit}{CPT_losses__gain}{range_bits}{0.51}
\setres{boxfit}{CPT_losses__gain}{range_bits_form}{0.03}
\setres{boxfit}{CPT_losses__gain}{range_bits_lifted}{0.14}
\setres{boxfit}{DA_beijing__eta3}{eta_star}{3.000}
\setres{boxfit}{DA_beijing__eta3}{kappa}{0.337}
\setres{boxfit}{DA_beijing__eta3}{theta_star}{6.07, 3.00}
\setres{boxfit}{DA_beijing__eta30}{eta_star}{30.000}
\setres{boxfit}{DA_beijing__eta30}{kappa}{0.795}
\setres{boxfit}{DA_beijing__eta30}{theta_star}{9.66, 30.00}
\setres{boxfit}{DA_gains__alpha_hi2}{alpha_at_edge}{0}
\setres{boxfit}{DA_gains__alpha_hi2}{alpha_star}{1.36}
\setres{boxfit}{DA_gains__alpha_hi2}{kappa}{0.626}
\setres{boxfit}{DA_gains__alpha_hi2}{kappa_se}{0.062}
\setres{boxfit}{DA_gains__alpha_hi2}{log_vol}{9.89}
\setres{boxfit}{DA_gains__alpha_hi2}{log_vol_se}{0.001}
\setres{boxfit}{DA_gains__alpha_hi2}{r_ff}{0.264}
\setres{boxfit}{DA_gains__alpha_hi2}{r_ff_se}{0.003}
\setres{boxfit}{DA_gains__alpha_hi2}{theta_star}{1.36, 1.35}
\setres{boxfit}{DA_gains__alpha_hi2}{volume_ess}{539{,}629}
\setres{boxfit}{DA_gains__box}{alpha_at_edge}{1}
\setres{boxfit}{DA_gains__box}{alpha_star}{1.00}
\setres{boxfit}{DA_gains__box}{kappa}{0.625}
\setres{boxfit}{DA_gains__box}{kappa_se}{0.061}
\setres{boxfit}{DA_gains__box}{log_vol}{9.82}
\setres{boxfit}{DA_gains__box}{log_vol_se}{0.001}
\setres{boxfit}{DA_gains__box}{r_ff}{0.258}
\setres{boxfit}{DA_gains__box}{r_ff_se}{0.003}
\setres{boxfit}{DA_gains__box}{theta_star}{1.00, 0.47}
\setres{boxfit}{DA_gains__box}{volume_ess}{560{,}294}
\setres{boxfit}{DA_gains__domain}{alpha_at_edge}{0}
\setres{boxfit}{DA_gains__domain}{alpha_star}{1.36}
\setres{boxfit}{DA_gains__domain}{kappa}{0.626}
\setres{boxfit}{DA_gains__domain}{kappa_se}{0.062}
\setres{boxfit}{DA_gains__domain}{log_vol}{9.98}
\setres{boxfit}{DA_gains__domain}{log_vol_se}{0.001}
\setres{boxfit}{DA_gains__domain}{r_ff}{0.281}
\setres{boxfit}{DA_gains__domain}{r_ff_se}{0.004}
\setres{boxfit}{DA_gains__domain}{theta_star}{1.36, 1.35}
\setres{boxfit}{DA_gains__domain}{volume_ess}{436{,}452}
\setres{boxfit}{DA_gains__eta3}{eta_star}{1.350}
\setres{boxfit}{DA_gains__eta3}{kappa}{0.626}
\setres{boxfit}{DA_gains__eta3}{theta_star}{1.36, 1.35}
\setres{boxfit}{DA_gains__eta30}{eta_star}{1.350}
\setres{boxfit}{DA_gains__eta30}{kappa}{0.626}
\setres{boxfit}{DA_gains__eta30}{theta_star}{1.36, 1.35}
\setres{boxfit}{DA_gains__gain}{box_volume_share_pct}{85}
\setres{boxfit}{DA_gains__gain}{kappa_gain}{0.001}
\setres{boxfit}{DA_gains__gain}{lifted_volume_share_pct}{91}
\setres{boxfit}{DA_gains__gain}{r_ff_gain}{-0.023}
\setres{boxfit}{DA_gains__gain}{range_bits}{0.23}
\setres{boxfit}{DA_gains__gain}{range_bits_form}{-0.05}
\setres{boxfit}{DA_gains__gain}{range_bits_lifted}{0.13}
\setres{boxfit}{DA_losses__alpha_hi2}{alpha_at_edge}{1}
\setres{boxfit}{DA_losses__alpha_hi2}{alpha_star}{2.00}
\setres{boxfit}{DA_losses__alpha_hi2}{kappa}{0.652}
\setres{boxfit}{DA_losses__alpha_hi2}{kappa_se}{0.067}
\setres{boxfit}{DA_losses__alpha_hi2}{log_vol}{9.15}
\setres{boxfit}{DA_losses__alpha_hi2}{log_vol_se}{0.001}
\setres{boxfit}{DA_losses__alpha_hi2}{r_ff}{0.309}
\setres{boxfit}{DA_losses__alpha_hi2}{r_ff_se}{0.005}
\setres{boxfit}{DA_losses__alpha_hi2}{theta_star}{2.00, -0.45}
\setres{boxfit}{DA_losses__alpha_hi2}{volume_ess}{371{,}191}
\setres{boxfit}{DA_losses__box}{alpha_at_edge}{1}
\setres{boxfit}{DA_losses__box}{alpha_star}{1.00}
\setres{boxfit}{DA_losses__box}{kappa}{0.533}
\setres{boxfit}{DA_losses__box}{kappa_se}{0.085}
\setres{boxfit}{DA_losses__box}{log_vol}{8.77}
\setres{boxfit}{DA_losses__box}{log_vol_se}{0.001}
\setres{boxfit}{DA_losses__box}{r_ff}{0.345}
\setres{boxfit}{DA_losses__box}{r_ff_se}{0.005}
\setres{boxfit}{DA_losses__box}{theta_star}{1.00, 0.47}
\setres{boxfit}{DA_losses__box}{volume_ess}{328{,}679}
\setres{boxfit}{DA_losses__domain}{alpha_at_edge}{0}
\setres{boxfit}{DA_losses__domain}{alpha_star}{7.96}
\setres{boxfit}{DA_losses__domain}{kappa}{0.872}
\setres{boxfit}{DA_losses__domain}{kappa_se}{0.017}
\setres{boxfit}{DA_losses__domain}{log_vol}{9.87}
\setres{boxfit}{DA_losses__domain}{log_vol_se}{0.006}
\setres{boxfit}{DA_losses__domain}{r_ff}{0.192}
\setres{boxfit}{DA_losses__domain}{r_ff_se}{0.002}
\setres{boxfit}{DA_losses__domain}{theta_star}{7.96, -0.98}
\setres{boxfit}{DA_losses__domain}{volume_ess}{30{,}044}
\setres{boxfit}{DA_losses__eta3}{eta_star}{-0.978}
\setres{boxfit}{DA_losses__eta3}{kappa}{0.872}
\setres{boxfit}{DA_losses__eta3}{theta_star}{7.96, -0.98}
\setres{boxfit}{DA_losses__eta30}{eta_star}{-0.978}
\setres{boxfit}{DA_losses__eta30}{kappa}{0.872}
\setres{boxfit}{DA_losses__eta30}{theta_star}{7.96, -0.98}
\setres{boxfit}{DA_losses__gain}{box_volume_share_pct}{33}
\setres{boxfit}{DA_losses__gain}{kappa_gain}{0.340}
\setres{boxfit}{DA_losses__gain}{lifted_volume_share_pct}{49}
\setres{boxfit}{DA_losses__gain}{r_ff_gain}{0.153}
\setres{boxfit}{DA_losses__gain}{range_bits}{1.59}
\setres{boxfit}{DA_losses__gain}{range_bits_form}{0.30}
\setres{boxfit}{DA_losses__gain}{range_bits_lifted}{1.04}
\setres{boxfit}{DA_shared06__eta3}{eta_star}{3.000}
\setres{boxfit}{DA_shared06__eta3}{kappa}{0.899}
\setres{boxfit}{DA_shared06__eta3}{theta_star}{4.01, 3.00}
\setres{boxfit}{DA_shared06__eta30}{eta_star}{30.000}
\setres{boxfit}{DA_shared06__eta30}{kappa}{0.971}
\setres{boxfit}{DA_shared06__eta30}{theta_star}{7.40, 30.00}
\setres{boxfit}{EU_gains__alpha_hi2}{alpha_at_edge}{0}
\setres{boxfit}{EU_gains__alpha_hi2}{alpha_star}{0.76}
\setres{boxfit}{EU_gains__alpha_hi2}{kappa}{0.616}
\setres{boxfit}{EU_gains__alpha_hi2}{kappa_se}{0.061}
\setres{boxfit}{EU_gains__alpha_hi2}{log_vol}{5.03}
\setres{boxfit}{EU_gains__alpha_hi2}{log_vol_se}{0.001}
\setres{boxfit}{EU_gains__alpha_hi2}{r_ff}{0.379}
\setres{boxfit}{EU_gains__alpha_hi2}{r_ff_se}{0.005}
\setres{boxfit}{EU_gains__alpha_hi2}{theta_star}{0.76}
\setres{boxfit}{EU_gains__alpha_hi2}{volume_ess}{660{,}856}
\setres{boxfit}{EU_gains__box}{alpha_at_edge}{0}
\setres{boxfit}{EU_gains__box}{alpha_star}{0.76}
\setres{boxfit}{EU_gains__box}{kappa}{0.616}
\setres{boxfit}{EU_gains__box}{kappa_se}{0.061}
\setres{boxfit}{EU_gains__box}{log_vol}{4.75}
\setres{boxfit}{EU_gains__box}{log_vol_se}{0.000}
\setres{boxfit}{EU_gains__box}{r_ff}{0.437}
\setres{boxfit}{EU_gains__box}{r_ff_se}{0.006}
\setres{boxfit}{EU_gains__box}{theta_star}{0.76}
\setres{boxfit}{EU_gains__box}{volume_ess}{833{,}022}
\setres{boxfit}{EU_gains__domain}{alpha_at_edge}{0}
\setres{boxfit}{EU_gains__domain}{alpha_star}{0.76}
\setres{boxfit}{EU_gains__domain}{kappa}{0.616}
\setres{boxfit}{EU_gains__domain}{kappa_se}{0.061}
\setres{boxfit}{EU_gains__domain}{log_vol}{5.53}
\setres{boxfit}{EU_gains__domain}{log_vol_se}{0.002}
\setres{boxfit}{EU_gains__domain}{r_ff}{0.336}
\setres{boxfit}{EU_gains__domain}{r_ff_se}{0.005}
\setres{boxfit}{EU_gains__domain}{theta_star}{0.76}
\setres{boxfit}{EU_gains__domain}{volume_ess}{172{,}086}
\setres{boxfit}{EU_gains__gain}{box_volume_share_pct}{46}
\setres{boxfit}{EU_gains__gain}{kappa_gain}{0.000}
\setres{boxfit}{EU_gains__gain}{lifted_volume_share_pct}{60}
\setres{boxfit}{EU_gains__gain}{r_ff_gain}{0.101}
\setres{boxfit}{EU_gains__gain}{range_bits}{1.13}
\setres{boxfit}{EU_gains__gain}{range_bits_form}{0.24}
\setres{boxfit}{EU_gains__gain}{range_bits_lifted}{0.73}
\setres{boxfit}{EU_losses__alpha_hi2}{alpha_at_edge}{0}
\setres{boxfit}{EU_losses__alpha_hi2}{alpha_star}{1.41}
\setres{boxfit}{EU_losses__alpha_hi2}{kappa}{0.590}
\setres{boxfit}{EU_losses__alpha_hi2}{kappa_se}{0.081}
\setres{boxfit}{EU_losses__alpha_hi2}{log_vol}{5.02}
\setres{boxfit}{EU_losses__alpha_hi2}{log_vol_se}{0.001}
\setres{boxfit}{EU_losses__alpha_hi2}{r_ff}{0.381}
\setres{boxfit}{EU_losses__alpha_hi2}{r_ff_se}{0.005}
\setres{boxfit}{EU_losses__alpha_hi2}{theta_star}{1.41}
\setres{boxfit}{EU_losses__alpha_hi2}{volume_ess}{659{,}246}
\setres{boxfit}{EU_losses__box}{alpha_at_edge}{0}
\setres{boxfit}{EU_losses__box}{alpha_star}{1.00}
\setres{boxfit}{EU_losses__box}{kappa}{0.384}
\setres{boxfit}{EU_losses__box}{kappa_se}{0.085}
\setres{boxfit}{EU_losses__box}{log_vol}{4.75}
\setres{boxfit}{EU_losses__box}{log_vol_se}{0.000}
\setres{boxfit}{EU_losses__box}{r_ff}{0.427}
\setres{boxfit}{EU_losses__box}{r_ff_se}{0.006}
\setres{boxfit}{EU_losses__box}{theta_star}{1.00}
\setres{boxfit}{EU_losses__box}{volume_ess}{832{,}987}
\setres{boxfit}{EU_losses__domain}{alpha_at_edge}{0}
\setres{boxfit}{EU_losses__domain}{alpha_star}{1.41}
\setres{boxfit}{EU_losses__domain}{kappa}{0.590}
\setres{boxfit}{EU_losses__domain}{kappa_se}{0.081}
\setres{boxfit}{EU_losses__domain}{log_vol}{5.40}
\setres{boxfit}{EU_losses__domain}{log_vol_se}{0.003}
\setres{boxfit}{EU_losses__domain}{r_ff}{0.378}
\setres{boxfit}{EU_losses__domain}{r_ff_se}{0.006}
\setres{boxfit}{EU_losses__domain}{theta_star}{1.41}
\setres{boxfit}{EU_losses__domain}{volume_ess}{133{,}991}
\setres{boxfit}{EU_losses__gain}{box_volume_share_pct}{52}
\setres{boxfit}{EU_losses__gain}{kappa_gain}{0.206}
\setres{boxfit}{EU_losses__gain}{lifted_volume_share_pct}{69}
\setres{boxfit}{EU_losses__gain}{r_ff_gain}{0.049}
\setres{boxfit}{EU_losses__gain}{range_bits}{0.93}
\setres{boxfit}{EU_losses__gain}{range_bits_form}{0.12}
\setres{boxfit}{EU_losses__gain}{range_bits_lifted}{0.54}
\setres{boxfit}{evenodds_gains}{ce_minus_ev}{-16.4}
\setres{boxfit}{evenodds_gains}{ev}{75}
\setres{boxfit}{evenodds_gains}{mean_ce}{58.6}
\setres{boxfit}{evenodds_gains}{p}{0.50}
\setres{boxfit}{evenodds_gains}{z1}{150}
\setres{boxfit}{evenodds_gains}{z2}{0}
\setres{boxfit}{evenodds_losses}{ce_minus_ev}{-16.5}
\setres{boxfit}{evenodds_losses}{ev}{-75}
\setres{boxfit}{evenodds_losses}{mean_ce}{-91.5}
\setres{boxfit}{evenodds_losses}{p}{0.50}
\setres{boxfit}{evenodds_losses}{z1}{-150}
\setres{boxfit}{evenodds_losses}{z2}{0}
\setres{mirror}{gains}{arc_length}{5.000}
\setres{mirror}{gains}{kl}{0.1518}
\setres{mirror}{gains}{r_ff}{0.416}
\setres{mirror}{gains}{r_prior}{0.1409}
\setres{mirror}{losses}{arc_length}{15.000}
\setres{mirror}{losses}{kl}{0.9251}
\setres{mirror}{losses}{r_ff}{0.016}
\setres{mirror}{losses}{r_prior}{0.6035}
\setres{mirror}{lottery1}{ce_edge}{5}
\setres{mirror}{lottery1}{ce_one}{1.32}
\setres{mirror}{lottery1}{ce_own}{4.45}
\setres{mirror}{lottery1}{f_ref}{4.45}
\setres{mirror}{lottery1}{hi}{20}
\setres{mirror}{lottery1}{lo}{0}
\setres{mirror}{lottery1}{p}{0.25}
\setres{mirror}{lottery1}{p_better}{0.25}
\setres{mirror}{lottery1}{z_high}{20}
\setres{mirror}{lottery1}{z_low}{0}
\setres{mirror}{lottery2}{ce_edge}{-15}
\setres{mirror}{lottery2}{ce_one}{-11.38}
\setres{mirror}{lottery2}{ce_own}{-9.63}
\setres{mirror}{lottery2}{f_ref}{-9.63}
\setres{mirror}{lottery2}{hi}{0}
\setres{mirror}{lottery2}{lo}{-20}
\setres{mirror}{lottery2}{p}{0.75}
\setres{mirror}{lottery2}{p_better}{0.25}
\setres{mirror}{lottery2}{z_high}{-20}
\setres{mirror}{lottery2}{z_low}{0}
\setres{mirror}{mirrored}{arc_length}{7.071}
\setres{mirror}{mirrored}{kl}{0.2177}
\setres{mirror}{mirrored}{r_prior}{0.1956}
\setres{mirror}{one}{arc_length}{16.796}
\setres{mirror}{one}{arc_length_ratio}{3.3592}
\setres{mirror}{one}{kl}{0.7964}
\setres{mirror}{one}{max_mirror_error}{0.000}
\setres{mirror}{one}{r_ff}{0.331}
\setres{mirror}{one}{r_prior}{0.5491}
\setres{mirror}{reference}{alpha_star_gains}{0.922}
\setres{mirror}{reference}{alpha_star_losses}{0.394}
\setres{mirror}{reference}{alpha_star_one}{0.510}
\setres{mirror}{reference}{dist_gains}{0.00}
\setres{mirror}{reference}{dist_losses}{0.00}
\setres{mirror}{reference}{dist_one}{6.43}
\setres{mirror}{reference}{dist_two_domains}{0.00}
\setres{mirror}{reference}{n_grid}{2{,}001}
\setres{mirror}{reference}{n_ref}{10{,}000}
\setres{mirror}{reference}{normaliser}{133.3}
\setres{mirror}{reference}{seed}{54}
\setres{mirror}{reference2}{alpha_star_gains}{1.000}
\setres{mirror}{reference2}{alpha_star_losses}{1.000}
\setres{mirror}{reference2}{alpha_star_one}{1.000}
\setres{mirror}{reference2}{dist_gains}{9.00}
\setres{mirror}{reference2}{dist_losses}{9.00}
\setres{mirror}{reference2}{dist_one}{9.00}
\setres{mirror}{reference2}{dist_two_domains}{9.00}
\setres{mirror}{reference2}{f_ref_gain}{8}
\setres{mirror}{reference2}{f_ref_loss}{-18}
\setres{evid}{CPT}{gap_fit_minus_lit_mean}{1.018}
\setres{evid}{CPT}{gap_fit_minus_lit_median}{0.475}
\setres{evid}{CPT}{gap_fit_minus_transfer_max}{4.771}
\setres{evid}{CPT}{gap_fit_minus_transfer_mean}{0.212}
\setres{evid}{CPT}{gap_fit_minus_transfer_median}{0.032}
\setres{evid}{CPT}{gap_fit_minus_uniform_mean}{3.803}
\setres{evid}{CPT}{gap_transfer_minus_lit_mean}{0.807}
\setres{evid}{CPT}{n_transfer_above_lit}{30}
\setres{evid}{CPT}{n_transfer_at_or_above_fit}{5}
\setres{evid}{CPT}{n_transfer_below_fit}{25}
\setres{evid}{CPT}{one_minus_rff}{0.801}
\setres{evid}{CPT}{passthrough_lit}{0.080}
\setres{evid}{CPT}{passthrough_uniform}{0.485}
\setres{evid}{CPT}{prior_contrib_lit}{0.737}
\setres{evid}{CPT}{prior_contrib_uniform}{0.412}
\setres{evid}{CPT}{r_ff}{0.199}
\setres{evid}{CPT_deterministic}{country_max}{Nicaragua}
\setres{evid}{CPT_deterministic}{country_min}{Nigeria}
\setres{evid}{CPT_deterministic}{kappa_max}{0.997}
\setres{evid}{CPT_deterministic}{kappa_mean}{0.963}
\setres{evid}{CPT_deterministic}{kappa_mean_se}{0.007}
\setres{evid}{CPT_deterministic}{kappa_median}{0.971}
\setres{evid}{CPT_deterministic}{kappa_min}{0.790}
\setres{evid}{CPT_deterministic}{kappa_sd}{0.036}
\setres{evid}{CPT_deterministic}{n_countries}{30}
\setres{evid}{CPT_deterministic}{n_pos}{30}
\setres{evid}{CPT_deterministic}{r_ff}{0.199}
\setres{evid}{CPT_deterministic}{r_prior}{1.0000}
\setres{evid}{CPT_deterministic}{r_prob}{1.000}
\setres{evid}{CPT_lit}{country_max}{USA}
\setres{evid}{CPT_lit}{country_min}{Nigeria}
\setres{evid}{CPT_lit}{kappa_max}{0.732}
\setres{evid}{CPT_lit}{kappa_mean}{-0.055}
\setres{evid}{CPT_lit}{kappa_mean_se}{0.207}
\setres{evid}{CPT_lit}{kappa_median}{0.485}
\setres{evid}{CPT_lit}{kappa_min}{-11.792}
\setres{evid}{CPT_lit}{kappa_sd}{2.251}
\setres{evid}{CPT_lit}{n_countries}{30}
\setres{evid}{CPT_lit}{n_pos}{24}
\setres{evid}{CPT_lit}{r_ff}{0.199}
\setres{evid}{CPT_lit}{r_prior}{0.9203}
\setres{evid}{CPT_lit}{r_prob}{0.936}
\setres{evid}{CPT_lit}{r_prob_se}{0.000}
\setres{evid}{CPT_transfer_others}{country_max}{Kyrgyzstan}
\setres{evid}{CPT_transfer_others}{country_min}{Nigeria}
\setres{evid}{CPT_transfer_others}{kappa_max}{0.987}
\setres{evid}{CPT_transfer_others}{kappa_mean}{0.752}
\setres{evid}{CPT_transfer_others}{kappa_mean_se}{0.100}
\setres{evid}{CPT_transfer_others}{kappa_median}{0.944}
\setres{evid}{CPT_transfer_others}{kappa_min}{-3.982}
\setres{evid}{CPT_transfer_others}{kappa_sd}{0.897}
\setres{evid}{CPT_transfer_others}{n_countries}{30}
\setres{evid}{CPT_transfer_others}{n_pos}{29}
\setres{evid}{CPT_transfer_others}{r_ff}{0.199}
\setres{evid}{CPT_transfer_others}{r_prior}{1.0000}
\setres{evid}{CPT_transfer_others}{r_prob}{1.000}
\setres{evid}{CPT_uniform}{country_max}{USA}
\setres{evid}{CPT_uniform}{country_min}{Nigeria}
\setres{evid}{CPT_uniform}{kappa_max}{-0.815}
\setres{evid}{CPT_uniform}{kappa_mean}{-2.840}
\setres{evid}{CPT_uniform}{kappa_mean_se}{0.440}
\setres{evid}{CPT_uniform}{kappa_median}{-1.582}
\setres{evid}{CPT_uniform}{kappa_min}{-28.796}
\setres{evid}{CPT_uniform}{kappa_sd}{5.005}
\setres{evid}{CPT_uniform}{n_countries}{30}
\setres{evid}{CPT_uniform}{n_pos}{0}
\setres{evid}{CPT_uniform}{r_ff}{0.199}
\setres{evid}{CPT_uniform}{r_prior}{0.5146}
\setres{evid}{CPT_uniform}{r_prob}{0.611}
\setres{evid}{CPT_uniform}{r_prob_se}{0.003}
\setres{evid}{DA}{gap_fit_minus_lit_mean}{2.347}
\setres{evid}{DA}{gap_fit_minus_lit_median}{1.338}
\setres{evid}{DA}{gap_fit_minus_transfer_max}{2.446}
\setres{evid}{DA}{gap_fit_minus_transfer_mean}{0.135}
\setres{evid}{DA}{gap_fit_minus_transfer_median}{0.026}
\setres{evid}{DA}{gap_fit_minus_uniform_mean}{4.656}
\setres{evid}{DA}{gap_r_cpt_lit}{-0.015}
\setres{evid}{DA}{gap_transfer_minus_lit_mean}{2.212}
\setres{evid}{DA}{n_transfer_above_lit}{30}
\setres{evid}{DA}{n_transfer_at_or_above_fit}{7}
\setres{evid}{DA}{n_transfer_below_fit}{23}
\setres{evid}{DA}{one_minus_rff}{0.686}
\setres{evid}{DA}{passthrough_lit}{0.115}
\setres{evid}{DA}{passthrough_uniform}{0.216}
\setres{evid}{DA}{prior_contrib_lit}{0.607}
\setres{evid}{DA}{prior_contrib_uniform}{0.538}
\setres{evid}{DA}{r_ff}{0.314}
\setres{evid}{DA_deterministic}{country_max}{Guatemala}
\setres{evid}{DA_deterministic}{country_min}{Nigeria}
\setres{evid}{DA_deterministic}{kappa_max}{0.966}
\setres{evid}{DA_deterministic}{kappa_mean}{0.682}
\setres{evid}{DA_deterministic}{kappa_mean_se}{0.120}
\setres{evid}{DA_deterministic}{kappa_median}{0.917}
\setres{evid}{DA_deterministic}{kappa_min}{-4.797}
\setres{evid}{DA_deterministic}{kappa_sd}{1.040}
\setres{evid}{DA_deterministic}{n_countries}{30}
\setres{evid}{DA_deterministic}{n_pos}{29}
\setres{evid}{DA_deterministic}{r_ff}{0.314}
\setres{evid}{DA_deterministic}{r_prior}{1.0000}
\setres{evid}{DA_deterministic}{r_prob}{1.000}
\setres{evid}{DA_lit}{country_max}{Germany}
\setres{evid}{DA_lit}{country_min}{Nigeria}
\setres{evid}{DA_lit}{kappa_max}{0.198}
\setres{evid}{DA_lit}{kappa_mean}{-1.665}
\setres{evid}{DA_lit}{kappa_mean_se}{0.424}
\setres{evid}{DA_lit}{kappa_median}{-0.444}
\setres{evid}{DA_lit}{kappa_min}{-26.492}
\setres{evid}{DA_lit}{kappa_sd}{4.782}
\setres{evid}{DA_lit}{n_countries}{30}
\setres{evid}{DA_lit}{n_pos}{3}
\setres{evid}{DA_lit}{r_ff}{0.314}
\setres{evid}{DA_lit}{r_prior}{0.8846}
\setres{evid}{DA_lit}{r_prob}{0.921}
\setres{evid}{DA_lit}{r_prob_se}{0.001}
\setres{evid}{DA_transfer_others}{country_max}{Guatemala}
\setres{evid}{DA_transfer_others}{country_min}{Nigeria}
\setres{evid}{DA_transfer_others}{kappa_max}{0.973}
\setres{evid}{DA_transfer_others}{kappa_mean}{0.547}
\setres{evid}{DA_transfer_others}{kappa_mean_se}{0.158}
\setres{evid}{DA_transfer_others}{kappa_median}{0.891}
\setres{evid}{DA_transfer_others}{kappa_min}{-7.243}
\setres{evid}{DA_transfer_others}{kappa_sd}{1.481}
\setres{evid}{DA_transfer_others}{n_countries}{30}
\setres{evid}{DA_transfer_others}{n_pos}{29}
\setres{evid}{DA_transfer_others}{r_ff}{0.314}
\setres{evid}{DA_transfer_others}{r_prior}{1.0000}
\setres{evid}{DA_transfer_others}{r_prob}{1.000}
\setres{evid}{DA_uniform}{country_max}{Germany}
\setres{evid}{DA_uniform}{country_min}{Nigeria}
\setres{evid}{DA_uniform}{kappa_max}{-0.993}
\setres{evid}{DA_uniform}{kappa_mean}{-3.975}
\setres{evid}{DA_uniform}{kappa_mean_se}{0.633}
\setres{evid}{DA_uniform}{kappa_median}{-2.153}
\setres{evid}{DA_uniform}{kappa_min}{-41.616}
\setres{evid}{DA_uniform}{kappa_sd}{7.265}
\setres{evid}{DA_uniform}{n_countries}{30}
\setres{evid}{DA_uniform}{n_pos}{0}
\setres{evid}{DA_uniform}{r_ff}{0.314}
\setres{evid}{DA_uniform}{r_prior}{0.7841}
\setres{evid}{DA_uniform}{r_prob}{0.852}
\setres{evid}{DA_uniform}{r_prob_se}{0.001}
\setres{evid}{EU}{gap_fit_minus_lit_mean}{0.658}
\setres{evid}{EU}{gap_fit_minus_lit_median}{0.279}
\setres{evid}{EU}{gap_fit_minus_transfer_max}{1.254}
\setres{evid}{EU}{gap_fit_minus_transfer_mean}{0.084}
\setres{evid}{EU}{gap_fit_minus_transfer_median}{0.016}
\setres{evid}{EU}{gap_fit_minus_uniform_mean}{2.817}
\setres{evid}{EU}{gap_transfer_minus_lit_mean}{0.575}
\setres{evid}{EU}{n_transfer_above_lit}{30}
\setres{evid}{EU}{n_transfer_at_or_above_fit}{7}
\setres{evid}{EU}{n_transfer_below_fit}{23}
\setres{evid}{EU}{one_minus_rff}{0.633}
\setres{evid}{EU}{passthrough_lit}{0.205}
\setres{evid}{EU}{passthrough_uniform}{0.475}
\setres{evid}{EU}{prior_contrib_lit}{0.503}
\setres{evid}{EU}{prior_contrib_uniform}{0.332}
\setres{evid}{EU}{r_ff}{0.367}
\setres{evid}{EU_deterministic}{country_max}{Thailand}
\setres{evid}{EU_deterministic}{country_min}{Nigeria}
\setres{evid}{EU_deterministic}{kappa_max}{0.942}
\setres{evid}{EU_deterministic}{kappa_mean}{0.482}
\setres{evid}{EU_deterministic}{kappa_mean_se}{0.182}
\setres{evid}{EU_deterministic}{kappa_median}{0.840}
\setres{evid}{EU_deterministic}{kappa_min}{-8.104}
\setres{evid}{EU_deterministic}{kappa_sd}{1.633}
\setres{evid}{EU_deterministic}{n_countries}{30}
\setres{evid}{EU_deterministic}{n_pos}{29}
\setres{evid}{EU_deterministic}{r_ff}{0.367}
\setres{evid}{EU_deterministic}{r_prior}{1.0000}
\setres{evid}{EU_deterministic}{r_prob}{1.000}
\setres{evid}{EU_lit}{country_max}{USA}
\setres{evid}{EU_lit}{country_min}{Nigeria}
\setres{evid}{EU_lit}{kappa_max}{0.806}
\setres{evid}{EU_lit}{kappa_mean}{-0.177}
\setres{evid}{EU_lit}{kappa_mean_se}{0.273}
\setres{evid}{EU_lit}{kappa_median}{0.514}
\setres{evid}{EU_lit}{kappa_min}{-15.268}
\setres{evid}{EU_lit}{kappa_sd}{2.889}
\setres{evid}{EU_lit}{n_countries}{30}
\setres{evid}{EU_lit}{n_pos}{24}
\setres{evid}{EU_lit}{r_ff}{0.367}
\setres{evid}{EU_lit}{r_prior}{0.7947}
\setres{evid}{EU_lit}{r_prob}{0.870}
\setres{evid}{EU_lit}{r_prob_se}{0.001}
\setres{evid}{EU_transfer_others}{country_max}{Thailand}
\setres{evid}{EU_transfer_others}{country_min}{Nigeria}
\setres{evid}{EU_transfer_others}{kappa_max}{0.941}
\setres{evid}{EU_transfer_others}{kappa_mean}{0.398}
\setres{evid}{EU_transfer_others}{kappa_mean_se}{0.199}
\setres{evid}{EU_transfer_others}{kappa_median}{0.811}
\setres{evid}{EU_transfer_others}{kappa_min}{-9.358}
\setres{evid}{EU_transfer_others}{kappa_sd}{1.856}
\setres{evid}{EU_transfer_others}{n_countries}{30}
\setres{evid}{EU_transfer_others}{n_pos}{29}
\setres{evid}{EU_transfer_others}{r_ff}{0.367}
\setres{evid}{EU_transfer_others}{r_prior}{1.0000}
\setres{evid}{EU_transfer_others}{r_prob}{1.000}
\setres{evid}{EU_uniform}{country_max}{USA}
\setres{evid}{EU_uniform}{country_min}{Nigeria}
\setres{evid}{EU_uniform}{kappa_max}{-0.170}
\setres{evid}{EU_uniform}{kappa_mean}{-2.335}
\setres{evid}{EU_uniform}{kappa_mean_se}{0.484}
\setres{evid}{EU_uniform}{kappa_median}{-0.977}
\setres{evid}{EU_uniform}{kappa_min}{-30.723}
\setres{evid}{EU_uniform}{kappa_sd}{5.470}
\setres{evid}{EU_uniform}{n_countries}{30}
\setres{evid}{EU_uniform}{n_pos}{0}
\setres{evid}{EU_uniform}{r_ff}{0.367}
\setres{evid}{EU_uniform}{r_prior}{0.5250}
\setres{evid}{EU_uniform}{r_prob}{0.699}
\setres{evid}{EU_uniform}{r_prob_se}{0.002}
\setres{evid}{EV_deterministic}{country_max}{UK}
\setres{evid}{EV_deterministic}{country_min}{Nigeria}
\setres{evid}{EV_deterministic}{kappa_max}{0.933}
\setres{evid}{EV_deterministic}{kappa_mean}{0.416}
\setres{evid}{EV_deterministic}{kappa_mean_se}{0.185}
\setres{evid}{EV_deterministic}{kappa_median}{0.778}
\setres{evid}{EV_deterministic}{kappa_min}{-8.694}
\setres{evid}{EV_deterministic}{kappa_sd}{1.733}
\setres{evid}{EV_deterministic}{n_countries}{30}
\setres{evid}{EV_deterministic}{n_pos}{29}
\setres{evid}{EV_deterministic}{r_ff}{0.400}
\setres{evid}{EV_deterministic}{r_prior}{1.0000}
\setres{evid}{EV_deterministic}{r_prob}{1.000}
\setres{evid}{pairs}{n_cpt_da_fit}{29}
\setres{evid}{pairs}{n_cpt_da_lit}{30}
\setres{evid}{pairs}{n_cpt_da_transfer}{27}
\setres{evid}{pairs}{n_cpt_da_uniform}{30}
\setres{evid}{pairs}{n_cpt_eu_fit}{30}
\setres{evid}{pairs}{n_cpt_eu_lit}{9}
\setres{evid}{pairs}{n_cpt_eu_transfer}{29}
\setres{evid}{pairs}{n_cpt_eu_uniform}{1}
\setres{evid}{pairs}{n_da_eu_fit}{28}
\setres{evid}{pairs}{n_da_eu_lit}{0}
\setres{evid}{pairs}{n_da_eu_transfer}{25}
\setres{evid}{pairs}{n_da_eu_uniform}{0}
\setres{evid}{pairs}{n_ev_above_eu_fit}{7}
\setres{evid}{pairs}{share_cpt_da_fit}{97}
\setres{evid}{pairs}{share_cpt_da_lit}{100}
\setres{evid}{pairs}{share_cpt_da_transfer}{90}
\setres{evid}{pairs}{share_cpt_da_uniform}{100}
\setres{evid}{pairs}{share_cpt_eu_fit}{100}
\setres{evid}{pairs}{share_cpt_eu_lit}{30}
\setres{evid}{pairs}{share_cpt_eu_transfer}{97}
\setres{evid}{pairs}{share_cpt_eu_uniform}{3}
\setres{evid}{pairs}{share_da_eu_fit}{93}
\setres{evid}{pairs}{share_da_eu_lit}{0}
\setres{evid}{pairs}{share_da_eu_transfer}{83}
\setres{evid}{pairs}{share_da_eu_uniform}{0}
\setres{evid}{reference}{countries}{Australia, Belgium, Brazil, Cambodia, Chile, China, Colombia, Costa\_Rica, Czech\_Republic, Ethiopia, France, Germany, Guatemala, India, Japan, Kyrgyzstan, Malaysia, Nicaragua, Nigeria, Peru, Poland, Russia, Saudi\_Arabia, South\_Africa, Spain, Thailand, Tunisia, UK, USA, Vietnam}
\setres{evid}{reference}{n_countries}{30}
\setres{evid}{reference}{n_countries_full_design}{29}
\setres{evid}{reference}{n_lotteries}{14}
\setres{evid}{reference}{n_obs}{40{,}886}
\setres{evid}{reference}{n_subjects}{2{,}939}
\setres{evid}{reference}{subjects_max}{204}
\setres{evid}{reference}{subjects_min}{61}
\setres{evidZ06}{CPT}{det_pct}{100}
\setres{evidZ06}{CPT}{gap_lit}{0.095}
\setres{evidZ06}{CPT}{gap_uniform}{0.415}
\setres{evidZ06}{CPT}{share_lit}{91}
\setres{evidZ06}{CPT}{share_uniform}{58}
\setres{evidZ06}{CPT_deterministic}{kappa}{1.000}
\setres{evidZ06}{CPT_deterministic}{kappa_se}{0.003}
\setres{evidZ06}{CPT_deterministic}{mse}{21.31}
\setres{evidZ06}{CPT_deterministic}{nrmse_pct}{13.7}
\setres{evidZ06}{CPT_deterministic}{r_ff}{---}
\setres{evidZ06}{CPT_deterministic}{r_ff_se}{---}
\setres{evidZ06}{CPT_deterministic}{r_prior}{---}
\setres{evidZ06}{CPT_deterministic}{r_prior_se}{---}
\setres{evidZ06}{CPT_deterministic}{r_prob}{---}
\setres{evidZ06}{CPT_deterministic}{r_prob_se}{---}
\setres{evidZ06}{CPT_deterministic}{theta_star}{1.20, 0.50, 0.70}
\setres{evidZ06}{CPT_deterministic}{volume_ess}{---}
\setres{evidZ06}{CPT_lit}{kappa}{0.905}
\setres{evidZ06}{CPT_lit}{kappa_prior_mean}{0.954}
\setres{evidZ06}{CPT_lit}{kappa_prior_mean_se}{0.013}
\setres{evidZ06}{CPT_lit}{kappa_se}{0.016}
\setres{evidZ06}{CPT_lit}{mse}{31.52}
\setres{evidZ06}{CPT_lit}{mse_prior_mean}{26.26}
\setres{evidZ06}{CPT_lit}{nrmse_pct}{14.7}
\setres{evidZ06}{CPT_lit}{pred_variance}{5.26}
\setres{evidZ06}{CPT_lit}{r_ff}{---}
\setres{evidZ06}{CPT_lit}{r_ff_se}{---}
\setres{evidZ06}{CPT_lit}{r_prior}{---}
\setres{evidZ06}{CPT_lit}{r_prior_se}{---}
\setres{evidZ06}{CPT_lit}{r_prob}{---}
\setres{evidZ06}{CPT_lit}{r_prob_se}{---}
\setres{evidZ06}{CPT_lit}{spread_penalty}{0.049}
\setres{evidZ06}{CPT_lit}{spread_penalty_se}{0.003}
\setres{evidZ06}{CPT_lit}{volume_ess}{---}
\setres{evidZ06}{CPT_transfer_zurich03}{kappa}{0.977}
\setres{evidZ06}{CPT_transfer_zurich03}{kappa_prior_mean}{0.977}
\setres{evidZ06}{CPT_transfer_zurich03}{kappa_prior_mean_se}{0.007}
\setres{evidZ06}{CPT_transfer_zurich03}{kappa_se}{0.007}
\setres{evidZ06}{CPT_transfer_zurich03}{mse}{23.78}
\setres{evidZ06}{CPT_transfer_zurich03}{mse_prior_mean}{23.78}
\setres{evidZ06}{CPT_transfer_zurich03}{nrmse_pct}{13.0}
\setres{evidZ06}{CPT_transfer_zurich03}{pred_variance}{0.00}
\setres{evidZ06}{CPT_transfer_zurich03}{r_ff}{---}
\setres{evidZ06}{CPT_transfer_zurich03}{r_ff_se}{---}
\setres{evidZ06}{CPT_transfer_zurich03}{r_prior}{---}
\setres{evidZ06}{CPT_transfer_zurich03}{r_prior_se}{---}
\setres{evidZ06}{CPT_transfer_zurich03}{r_prob}{---}
\setres{evidZ06}{CPT_transfer_zurich03}{r_prob_se}{---}
\setres{evidZ06}{CPT_transfer_zurich03}{spread_penalty}{0.000}
\setres{evidZ06}{CPT_transfer_zurich03}{spread_penalty_se}{0.000}
\setres{evidZ06}{CPT_transfer_zurich03}{theta_star}{1.10, 0.40, 0.80}
\setres{evidZ06}{CPT_transfer_zurich03}{transfer}{zurich03}
\setres{evidZ06}{CPT_transfer_zurich03}{volume_ess}{---}
\setres{evidZ06}{CPT_uniform}{kappa}{0.585}
\setres{evidZ06}{CPT_uniform}{kappa_prior_mean}{0.951}
\setres{evidZ06}{CPT_uniform}{kappa_prior_mean_se}{0.009}
\setres{evidZ06}{CPT_uniform}{kappa_se}{0.013}
\setres{evidZ06}{CPT_uniform}{mse}{65.97}
\setres{evidZ06}{CPT_uniform}{mse_prior_mean}{26.54}
\setres{evidZ06}{CPT_uniform}{nrmse_pct}{14.0}
\setres{evidZ06}{CPT_uniform}{pred_variance}{39.43}
\setres{evidZ06}{CPT_uniform}{r_ff}{---}
\setres{evidZ06}{CPT_uniform}{r_ff_se}{---}
\setres{evidZ06}{CPT_uniform}{r_prior}{---}
\setres{evidZ06}{CPT_uniform}{r_prior_se}{---}
\setres{evidZ06}{CPT_uniform}{r_prob}{---}
\setres{evidZ06}{CPT_uniform}{r_prob_se}{---}
\setres{evidZ06}{CPT_uniform}{spread_penalty}{0.366}
\setres{evidZ06}{CPT_uniform}{spread_penalty_se}{0.021}
\setres{evidZ06}{CPT_uniform}{volume_ess}{---}
\setres{evidZ06}{DA}{det_pct}{90}
\setres{evidZ06}{DA}{gap_cpt_lit}{0.106}
\setres{evidZ06}{DA}{gap_lit}{0.099}
\setres{evidZ06}{DA}{gap_uniform}{0.205}
\setres{evidZ06}{DA}{share_lit}{89}
\setres{evidZ06}{DA}{share_uniform}{77}
\setres{evidZ06}{DA_deterministic}{kappa}{0.899}
\setres{evidZ06}{DA_deterministic}{kappa_se}{0.019}
\setres{evidZ06}{DA_deterministic}{mse}{32.20}
\setres{evidZ06}{DA_deterministic}{nrmse_pct}{16.9}
\setres{evidZ06}{DA_deterministic}{r_ff}{---}
\setres{evidZ06}{DA_deterministic}{r_ff_se}{---}
\setres{evidZ06}{DA_deterministic}{r_prior}{---}
\setres{evidZ06}{DA_deterministic}{r_prior_se}{---}
\setres{evidZ06}{DA_deterministic}{r_prob}{---}
\setres{evidZ06}{DA_deterministic}{r_prob_se}{---}
\setres{evidZ06}{DA_deterministic}{theta_star}{4.01, 3.00}
\setres{evidZ06}{DA_deterministic}{volume_ess}{---}
\setres{evidZ06}{DA_lit}{kappa}{0.799}
\setres{evidZ06}{DA_lit}{kappa_prior_mean}{0.830}
\setres{evidZ06}{DA_lit}{kappa_prior_mean_se}{0.032}
\setres{evidZ06}{DA_lit}{kappa_se}{0.033}
\setres{evidZ06}{DA_lit}{mse}{42.88}
\setres{evidZ06}{DA_lit}{mse_prior_mean}{39.58}
\setres{evidZ06}{DA_lit}{nrmse_pct}{17.4}
\setres{evidZ06}{DA_lit}{pred_variance}{3.30}
\setres{evidZ06}{DA_lit}{r_ff}{---}
\setres{evidZ06}{DA_lit}{r_ff_se}{---}
\setres{evidZ06}{DA_lit}{r_prior}{---}
\setres{evidZ06}{DA_lit}{r_prior_se}{---}
\setres{evidZ06}{DA_lit}{r_prob}{---}
\setres{evidZ06}{DA_lit}{r_prob_se}{---}
\setres{evidZ06}{DA_lit}{spread_penalty}{0.031}
\setres{evidZ06}{DA_lit}{spread_penalty_se}{0.002}
\setres{evidZ06}{DA_lit}{volume_ess}{---}
\setres{evidZ06}{DA_transfer_zurich03}{kappa}{0.875}
\setres{evidZ06}{DA_transfer_zurich03}{kappa_prior_mean}{0.875}
\setres{evidZ06}{DA_transfer_zurich03}{kappa_prior_mean_se}{0.017}
\setres{evidZ06}{DA_transfer_zurich03}{kappa_se}{0.017}
\setres{evidZ06}{DA_transfer_zurich03}{mse}{34.78}
\setres{evidZ06}{DA_transfer_zurich03}{mse_prior_mean}{34.78}
\setres{evidZ06}{DA_transfer_zurich03}{nrmse_pct}{21.1}
\setres{evidZ06}{DA_transfer_zurich03}{pred_variance}{0.00}
\setres{evidZ06}{DA_transfer_zurich03}{r_ff}{---}
\setres{evidZ06}{DA_transfer_zurich03}{r_ff_se}{---}
\setres{evidZ06}{DA_transfer_zurich03}{r_prior}{---}
\setres{evidZ06}{DA_transfer_zurich03}{r_prior_se}{---}
\setres{evidZ06}{DA_transfer_zurich03}{r_prob}{---}
\setres{evidZ06}{DA_transfer_zurich03}{r_prob_se}{---}
\setres{evidZ06}{DA_transfer_zurich03}{spread_penalty}{0.000}
\setres{evidZ06}{DA_transfer_zurich03}{spread_penalty_se}{0.000}
\setres{evidZ06}{DA_transfer_zurich03}{theta_star}{4.83, 3.00}
\setres{evidZ06}{DA_transfer_zurich03}{transfer}{zurich03}
\setres{evidZ06}{DA_transfer_zurich03}{volume_ess}{---}
\setres{evidZ06}{DA_uniform}{kappa}{0.694}
\setres{evidZ06}{DA_uniform}{kappa_prior_mean}{0.845}
\setres{evidZ06}{DA_uniform}{kappa_prior_mean_se}{0.029}
\setres{evidZ06}{DA_uniform}{kappa_se}{0.038}
\setres{evidZ06}{DA_uniform}{mse}{54.23}
\setres{evidZ06}{DA_uniform}{mse_prior_mean}{38.02}
\setres{evidZ06}{DA_uniform}{nrmse_pct}{17.9}
\setres{evidZ06}{DA_uniform}{pred_variance}{16.21}
\setres{evidZ06}{DA_uniform}{r_ff}{---}
\setres{evidZ06}{DA_uniform}{r_ff_se}{---}
\setres{evidZ06}{DA_uniform}{r_prior}{---}
\setres{evidZ06}{DA_uniform}{r_prior_se}{---}
\setres{evidZ06}{DA_uniform}{r_prob}{---}
\setres{evidZ06}{DA_uniform}{r_prob_se}{---}
\setres{evidZ06}{DA_uniform}{spread_penalty}{0.151}
\setres{evidZ06}{DA_uniform}{spread_penalty_se}{0.009}
\setres{evidZ06}{DA_uniform}{volume_ess}{---}
\setres{evidZ06}{EU}{det_pct}{83}
\setres{evidZ06}{EU}{gap_lit}{0.002}
\setres{evidZ06}{EU}{gap_uniform}{0.012}
\setres{evidZ06}{EU}{share_lit}{100}
\setres{evidZ06}{EU}{share_uniform}{99}
\setres{evidZ06}{EU_deterministic}{kappa}{0.832}
\setres{evidZ06}{EU_deterministic}{kappa_se}{0.030}
\setres{evidZ06}{EU_deterministic}{mse}{39.32}
\setres{evidZ06}{EU_deterministic}{nrmse_pct}{18.0}
\setres{evidZ06}{EU_deterministic}{r_ff}{---}
\setres{evidZ06}{EU_deterministic}{r_ff_se}{---}
\setres{evidZ06}{EU_deterministic}{r_prior}{---}
\setres{evidZ06}{EU_deterministic}{r_prior_se}{---}
\setres{evidZ06}{EU_deterministic}{r_prob}{---}
\setres{evidZ06}{EU_deterministic}{r_prob_se}{---}
\setres{evidZ06}{EU_deterministic}{theta_star}{0.95}
\setres{evidZ06}{EU_deterministic}{volume_ess}{---}
\setres{evidZ06}{EU_lit}{kappa}{0.830}
\setres{evidZ06}{EU_lit}{kappa_prior_mean}{0.831}
\setres{evidZ06}{EU_lit}{kappa_prior_mean_se}{0.031}
\setres{evidZ06}{EU_lit}{kappa_se}{0.031}
\setres{evidZ06}{EU_lit}{mse}{39.59}
\setres{evidZ06}{EU_lit}{mse_prior_mean}{39.43}
\setres{evidZ06}{EU_lit}{nrmse_pct}{18.8}
\setres{evidZ06}{EU_lit}{pred_variance}{0.15}
\setres{evidZ06}{EU_lit}{r_ff}{---}
\setres{evidZ06}{EU_lit}{r_ff_se}{---}
\setres{evidZ06}{EU_lit}{r_prior}{---}
\setres{evidZ06}{EU_lit}{r_prior_se}{---}
\setres{evidZ06}{EU_lit}{r_prob}{---}
\setres{evidZ06}{EU_lit}{r_prob_se}{---}
\setres{evidZ06}{EU_lit}{spread_penalty}{0.001}
\setres{evidZ06}{EU_lit}{spread_penalty_se}{0.000}
\setres{evidZ06}{EU_lit}{volume_ess}{---}
\setres{evidZ06}{EU_transfer_zurich03}{kappa}{0.801}
\setres{evidZ06}{EU_transfer_zurich03}{kappa_prior_mean}{0.801}
\setres{evidZ06}{EU_transfer_zurich03}{kappa_prior_mean_se}{0.028}
\setres{evidZ06}{EU_transfer_zurich03}{kappa_se}{0.028}
\setres{evidZ06}{EU_transfer_zurich03}{mse}{42.65}
\setres{evidZ06}{EU_transfer_zurich03}{mse_prior_mean}{42.65}
\setres{evidZ06}{EU_transfer_zurich03}{nrmse_pct}{21.7}
\setres{evidZ06}{EU_transfer_zurich03}{pred_variance}{0.00}
\setres{evidZ06}{EU_transfer_zurich03}{r_ff}{---}
\setres{evidZ06}{EU_transfer_zurich03}{r_ff_se}{---}
\setres{evidZ06}{EU_transfer_zurich03}{r_prior}{---}
\setres{evidZ06}{EU_transfer_zurich03}{r_prior_se}{---}
\setres{evidZ06}{EU_transfer_zurich03}{r_prob}{---}
\setres{evidZ06}{EU_transfer_zurich03}{r_prob_se}{---}
\setres{evidZ06}{EU_transfer_zurich03}{spread_penalty}{0.000}
\setres{evidZ06}{EU_transfer_zurich03}{spread_penalty_se}{0.000}
\setres{evidZ06}{EU_transfer_zurich03}{theta_star}{2.05}
\setres{evidZ06}{EU_transfer_zurich03}{transfer}{zurich03}
\setres{evidZ06}{EU_transfer_zurich03}{volume_ess}{---}
\setres{evidZ06}{EU_uniform}{kappa}{0.820}
\setres{evidZ06}{EU_uniform}{kappa_prior_mean}{0.827}
\setres{evidZ06}{EU_uniform}{kappa_prior_mean_se}{0.032}
\setres{evidZ06}{EU_uniform}{kappa_se}{0.032}
\setres{evidZ06}{EU_uniform}{mse}{40.61}
\setres{evidZ06}{EU_uniform}{mse_prior_mean}{39.93}
\setres{evidZ06}{EU_uniform}{nrmse_pct}{19.0}
\setres{evidZ06}{EU_uniform}{pred_variance}{0.68}
\setres{evidZ06}{EU_uniform}{r_ff}{---}
\setres{evidZ06}{EU_uniform}{r_ff_se}{---}
\setres{evidZ06}{EU_uniform}{r_prior}{---}
\setres{evidZ06}{EU_uniform}{r_prior_se}{---}
\setres{evidZ06}{EU_uniform}{r_prob}{---}
\setres{evidZ06}{EU_uniform}{r_prob_se}{---}
\setres{evidZ06}{EU_uniform}{spread_penalty}{0.006}
\setres{evidZ06}{EU_uniform}{spread_penalty_se}{0.000}
\setres{evidZ06}{EU_uniform}{volume_ess}{---}
\setres{evidZ06}{EV_deterministic}{kappa}{0.833}
\setres{evidZ06}{EV_deterministic}{kappa_se}{0.030}
\setres{evidZ06}{EV_deterministic}{mse}{39.25}
\setres{evidZ06}{EV_deterministic}{nrmse_pct}{17.9}
\setres{evidZ06}{EV_deterministic}{r_ff}{---}
\setres{evidZ06}{EV_deterministic}{r_ff_se}{---}
\setres{evidZ06}{EV_deterministic}{r_prior}{---}
\setres{evidZ06}{EV_deterministic}{r_prior_se}{---}
\setres{evidZ06}{EV_deterministic}{r_prob}{---}
\setres{evidZ06}{EV_deterministic}{r_prob_se}{---}
\setres{evidZ06}{EV_deterministic}{volume_ess}{---}
\setres{evidZ06}{reference}{lookup_nrmse_pct}{12.7}
\setres{evidZ06}{reference}{midpoint_mse}{128.88}
\setres{evidZ06}{reference}{midpoint_nrmse_pct}{23.4}
\setres{evidZ06}{reference}{n_lotteries}{15}
\setres{evidZ06}{reference}{n_obs}{1{,}756}
\setres{evidZ06}{reference}{n_subjects}{118}
\setres{evidZ06}{reference}{oracle_mse}{21.29}
\setres{evidZ06}{reference}{sample}{Zurich2006\_shared}
\setres{evidZ03}{CPT}{det_pct}{100}
\setres{evidZ03}{CPT}{gap_lit}{0.215}
\setres{evidZ03}{CPT}{gap_uniform}{0.512}
\setres{evidZ03}{CPT}{share_lit}{78}
\setres{evidZ03}{CPT}{share_uniform}{49}
\setres{evidZ03}{CPT_deterministic}{kappa}{0.999}
\setres{evidZ03}{CPT_deterministic}{kappa_se}{0.003}
\setres{evidZ03}{CPT_deterministic}{mse}{43.86}
\setres{evidZ03}{CPT_deterministic}{nrmse_pct}{16.9}
\setres{evidZ03}{CPT_deterministic}{r_ff}{---}
\setres{evidZ03}{CPT_deterministic}{r_ff_se}{---}
\setres{evidZ03}{CPT_deterministic}{r_prior}{---}
\setres{evidZ03}{CPT_deterministic}{r_prior_se}{---}
\setres{evidZ03}{CPT_deterministic}{r_prob}{---}
\setres{evidZ03}{CPT_deterministic}{r_prob_se}{---}
\setres{evidZ03}{CPT_deterministic}{theta_star}{1.10, 0.40, 0.80}
\setres{evidZ03}{CPT_deterministic}{volume_ess}{---}
\setres{evidZ03}{CPT_lit}{kappa}{0.785}
\setres{evidZ03}{CPT_lit}{kappa_prior_mean}{0.851}
\setres{evidZ03}{CPT_lit}{kappa_prior_mean_se}{0.040}
\setres{evidZ03}{CPT_lit}{kappa_se}{0.045}
\setres{evidZ03}{CPT_lit}{mse}{60.91}
\setres{evidZ03}{CPT_lit}{mse_prior_mean}{55.63}
\setres{evidZ03}{CPT_lit}{nrmse_pct}{19.6}
\setres{evidZ03}{CPT_lit}{pred_variance}{5.27}
\setres{evidZ03}{CPT_lit}{r_ff}{---}
\setres{evidZ03}{CPT_lit}{r_ff_se}{---}
\setres{evidZ03}{CPT_lit}{r_prior}{---}
\setres{evidZ03}{CPT_lit}{r_prior_se}{---}
\setres{evidZ03}{CPT_lit}{r_prob}{---}
\setres{evidZ03}{CPT_lit}{r_prob_se}{---}
\setres{evidZ03}{CPT_lit}{spread_penalty}{0.067}
\setres{evidZ03}{CPT_lit}{spread_penalty_se}{0.005}
\setres{evidZ03}{CPT_lit}{volume_ess}{---}
\setres{evidZ03}{CPT_transfer_zurich06}{kappa}{0.968}
\setres{evidZ03}{CPT_transfer_zurich06}{kappa_prior_mean}{0.968}
\setres{evidZ03}{CPT_transfer_zurich06}{kappa_prior_mean_se}{0.018}
\setres{evidZ03}{CPT_transfer_zurich06}{kappa_se}{0.018}
\setres{evidZ03}{CPT_transfer_zurich06}{mse}{46.33}
\setres{evidZ03}{CPT_transfer_zurich06}{mse_prior_mean}{46.33}
\setres{evidZ03}{CPT_transfer_zurich06}{nrmse_pct}{18.1}
\setres{evidZ03}{CPT_transfer_zurich06}{pred_variance}{0.00}
\setres{evidZ03}{CPT_transfer_zurich06}{r_ff}{---}
\setres{evidZ03}{CPT_transfer_zurich06}{r_ff_se}{---}
\setres{evidZ03}{CPT_transfer_zurich06}{r_prior}{---}
\setres{evidZ03}{CPT_transfer_zurich06}{r_prior_se}{---}
\setres{evidZ03}{CPT_transfer_zurich06}{r_prob}{---}
\setres{evidZ03}{CPT_transfer_zurich06}{r_prob_se}{---}
\setres{evidZ03}{CPT_transfer_zurich06}{spread_penalty}{0.000}
\setres{evidZ03}{CPT_transfer_zurich06}{spread_penalty_se}{0.000}
\setres{evidZ03}{CPT_transfer_zurich06}{theta_star}{1.20, 0.50, 0.70}
\setres{evidZ03}{CPT_transfer_zurich06}{transfer}{zurich06}
\setres{evidZ03}{CPT_transfer_zurich06}{volume_ess}{---}
\setres{evidZ03}{CPT_uniform}{kappa}{0.488}
\setres{evidZ03}{CPT_uniform}{kappa_prior_mean}{0.986}
\setres{evidZ03}{CPT_uniform}{kappa_prior_mean_se}{0.006}
\setres{evidZ03}{CPT_uniform}{kappa_se}{0.036}
\setres{evidZ03}{CPT_uniform}{mse}{84.46}
\setres{evidZ03}{CPT_uniform}{mse_prior_mean}{44.93}
\setres{evidZ03}{CPT_uniform}{nrmse_pct}{18.4}
\setres{evidZ03}{CPT_uniform}{pred_variance}{39.53}
\setres{evidZ03}{CPT_uniform}{r_ff}{---}
\setres{evidZ03}{CPT_uniform}{r_ff_se}{---}
\setres{evidZ03}{CPT_uniform}{r_prior}{---}
\setres{evidZ03}{CPT_uniform}{r_prior_se}{---}
\setres{evidZ03}{CPT_uniform}{r_prob}{---}
\setres{evidZ03}{CPT_uniform}{r_prob_se}{---}
\setres{evidZ03}{CPT_uniform}{spread_penalty}{0.498}
\setres{evidZ03}{CPT_uniform}{spread_penalty_se}{0.040}
\setres{evidZ03}{CPT_uniform}{volume_ess}{---}
\setres{evidZ03}{DA}{det_pct}{81}
\setres{evidZ03}{DA}{gap_cpt_lit}{0.231}
\setres{evidZ03}{DA}{gap_lit}{0.252}
\setres{evidZ03}{DA}{gap_uniform}{0.391}
\setres{evidZ03}{DA}{share_lit}{69}
\setres{evidZ03}{DA}{share_uniform}{51}
\setres{evidZ03}{DA_deterministic}{kappa}{0.806}
\setres{evidZ03}{DA_deterministic}{kappa_se}{0.032}
\setres{evidZ03}{DA_deterministic}{mse}{59.19}
\setres{evidZ03}{DA_deterministic}{nrmse_pct}{22.1}
\setres{evidZ03}{DA_deterministic}{r_ff}{---}
\setres{evidZ03}{DA_deterministic}{r_ff_se}{---}
\setres{evidZ03}{DA_deterministic}{r_prior}{---}
\setres{evidZ03}{DA_deterministic}{r_prior_se}{---}
\setres{evidZ03}{DA_deterministic}{r_prob}{---}
\setres{evidZ03}{DA_deterministic}{r_prob_se}{---}
\setres{evidZ03}{DA_deterministic}{theta_star}{4.83, 3.00}
\setres{evidZ03}{DA_deterministic}{volume_ess}{---}
\setres{evidZ03}{DA_lit}{kappa}{0.554}
\setres{evidZ03}{DA_lit}{kappa_prior_mean}{0.596}
\setres{evidZ03}{DA_lit}{kappa_prior_mean_se}{0.081}
\setres{evidZ03}{DA_lit}{kappa_se}{0.084}
\setres{evidZ03}{DA_lit}{mse}{79.19}
\setres{evidZ03}{DA_lit}{mse_prior_mean}{75.89}
\setres{evidZ03}{DA_lit}{nrmse_pct}{23.4}
\setres{evidZ03}{DA_lit}{pred_variance}{3.30}
\setres{evidZ03}{DA_lit}{r_ff}{---}
\setres{evidZ03}{DA_lit}{r_ff_se}{---}
\setres{evidZ03}{DA_lit}{r_prior}{---}
\setres{evidZ03}{DA_lit}{r_prior_se}{---}
\setres{evidZ03}{DA_lit}{r_prob}{---}
\setres{evidZ03}{DA_lit}{r_prob_se}{---}
\setres{evidZ03}{DA_lit}{spread_penalty}{0.042}
\setres{evidZ03}{DA_lit}{spread_penalty_se}{0.003}
\setres{evidZ03}{DA_lit}{volume_ess}{---}
\setres{evidZ03}{DA_transfer_zurich06}{kappa}{0.775}
\setres{evidZ03}{DA_transfer_zurich06}{kappa_prior_mean}{0.775}
\setres{evidZ03}{DA_transfer_zurich06}{kappa_prior_mean_se}{0.048}
\setres{evidZ03}{DA_transfer_zurich06}{kappa_se}{0.048}
\setres{evidZ03}{DA_transfer_zurich06}{mse}{61.68}
\setres{evidZ03}{DA_transfer_zurich06}{mse_prior_mean}{61.68}
\setres{evidZ03}{DA_transfer_zurich06}{nrmse_pct}{21.9}
\setres{evidZ03}{DA_transfer_zurich06}{pred_variance}{0.00}
\setres{evidZ03}{DA_transfer_zurich06}{r_ff}{---}
\setres{evidZ03}{DA_transfer_zurich06}{r_ff_se}{---}
\setres{evidZ03}{DA_transfer_zurich06}{r_prior}{---}
\setres{evidZ03}{DA_transfer_zurich06}{r_prior_se}{---}
\setres{evidZ03}{DA_transfer_zurich06}{r_prob}{---}
\setres{evidZ03}{DA_transfer_zurich06}{r_prob_se}{---}
\setres{evidZ03}{DA_transfer_zurich06}{spread_penalty}{0.000}
\setres{evidZ03}{DA_transfer_zurich06}{spread_penalty_se}{0.000}
\setres{evidZ03}{DA_transfer_zurich06}{theta_star}{4.01, 3.00}
\setres{evidZ03}{DA_transfer_zurich06}{transfer}{zurich06}
\setres{evidZ03}{DA_transfer_zurich06}{volume_ess}{---}
\setres{evidZ03}{DA_uniform}{kappa}{0.415}
\setres{evidZ03}{DA_uniform}{kappa_prior_mean}{0.620}
\setres{evidZ03}{DA_uniform}{kappa_prior_mean_se}{0.078}
\setres{evidZ03}{DA_uniform}{kappa_se}{0.095}
\setres{evidZ03}{DA_uniform}{mse}{90.22}
\setres{evidZ03}{DA_uniform}{mse_prior_mean}{74.00}
\setres{evidZ03}{DA_uniform}{nrmse_pct}{23.0}
\setres{evidZ03}{DA_uniform}{pred_variance}{16.22}
\setres{evidZ03}{DA_uniform}{r_ff}{---}
\setres{evidZ03}{DA_uniform}{r_ff_se}{---}
\setres{evidZ03}{DA_uniform}{r_prior}{---}
\setres{evidZ03}{DA_uniform}{r_prior_se}{---}
\setres{evidZ03}{DA_uniform}{r_prob}{---}
\setres{evidZ03}{DA_uniform}{r_prob_se}{---}
\setres{evidZ03}{DA_uniform}{spread_penalty}{0.205}
\setres{evidZ03}{DA_uniform}{spread_penalty_se}{0.016}
\setres{evidZ03}{DA_uniform}{volume_ess}{---}
\setres{evidZ03}{EU}{det_pct}{64}
\setres{evidZ03}{EU}{gap_lit}{0.033}
\setres{evidZ03}{EU}{gap_uniform}{0.053}
\setres{evidZ03}{EU}{share_lit}{95}
\setres{evidZ03}{EU}{share_uniform}{92}
\setres{evidZ03}{EU_deterministic}{kappa}{0.641}
\setres{evidZ03}{EU_deterministic}{kappa_se}{0.059}
\setres{evidZ03}{EU_deterministic}{mse}{72.32}
\setres{evidZ03}{EU_deterministic}{nrmse_pct}{23.7}
\setres{evidZ03}{EU_deterministic}{r_ff}{---}
\setres{evidZ03}{EU_deterministic}{r_ff_se}{---}
\setres{evidZ03}{EU_deterministic}{r_prior}{---}
\setres{evidZ03}{EU_deterministic}{r_prior_se}{---}
\setres{evidZ03}{EU_deterministic}{r_prob}{---}
\setres{evidZ03}{EU_deterministic}{r_prob_se}{---}
\setres{evidZ03}{EU_deterministic}{theta_star}{2.05}
\setres{evidZ03}{EU_deterministic}{volume_ess}{---}
\setres{evidZ03}{EU_lit}{kappa}{0.607}
\setres{evidZ03}{EU_lit}{kappa_prior_mean}{0.609}
\setres{evidZ03}{EU_lit}{kappa_prior_mean_se}{0.077}
\setres{evidZ03}{EU_lit}{kappa_se}{0.077}
\setres{evidZ03}{EU_lit}{mse}{74.96}
\setres{evidZ03}{EU_lit}{mse_prior_mean}{74.80}
\setres{evidZ03}{EU_lit}{nrmse_pct}{22.9}
\setres{evidZ03}{EU_lit}{pred_variance}{0.15}
\setres{evidZ03}{EU_lit}{r_ff}{---}
\setres{evidZ03}{EU_lit}{r_ff_se}{---}
\setres{evidZ03}{EU_lit}{r_prior}{---}
\setres{evidZ03}{EU_lit}{r_prior_se}{---}
\setres{evidZ03}{EU_lit}{r_prob}{---}
\setres{evidZ03}{EU_lit}{r_prob_se}{---}
\setres{evidZ03}{EU_lit}{spread_penalty}{0.002}
\setres{evidZ03}{EU_lit}{spread_penalty_se}{0.000}
\setres{evidZ03}{EU_lit}{volume_ess}{---}
\setres{evidZ03}{EU_transfer_zurich06}{kappa}{0.622}
\setres{evidZ03}{EU_transfer_zurich06}{kappa_prior_mean}{0.622}
\setres{evidZ03}{EU_transfer_zurich06}{kappa_prior_mean_se}{0.074}
\setres{evidZ03}{EU_transfer_zurich06}{kappa_se}{0.074}
\setres{evidZ03}{EU_transfer_zurich06}{mse}{73.78}
\setres{evidZ03}{EU_transfer_zurich06}{mse_prior_mean}{73.78}
\setres{evidZ03}{EU_transfer_zurich06}{nrmse_pct}{23.5}
\setres{evidZ03}{EU_transfer_zurich06}{pred_variance}{0.00}
\setres{evidZ03}{EU_transfer_zurich06}{r_ff}{---}
\setres{evidZ03}{EU_transfer_zurich06}{r_ff_se}{---}
\setres{evidZ03}{EU_transfer_zurich06}{r_prior}{---}
\setres{evidZ03}{EU_transfer_zurich06}{r_prior_se}{---}
\setres{evidZ03}{EU_transfer_zurich06}{r_prob}{---}
\setres{evidZ03}{EU_transfer_zurich06}{r_prob_se}{---}
\setres{evidZ03}{EU_transfer_zurich06}{spread_penalty}{0.000}
\setres{evidZ03}{EU_transfer_zurich06}{spread_penalty_se}{0.000}
\setres{evidZ03}{EU_transfer_zurich06}{theta_star}{0.95}
\setres{evidZ03}{EU_transfer_zurich06}{transfer}{zurich06}
\setres{evidZ03}{EU_transfer_zurich06}{volume_ess}{---}
\setres{evidZ03}{EU_uniform}{kappa}{0.588}
\setres{evidZ03}{EU_uniform}{kappa_prior_mean}{0.596}
\setres{evidZ03}{EU_uniform}{kappa_prior_mean_se}{0.079}
\setres{evidZ03}{EU_uniform}{kappa_se}{0.080}
\setres{evidZ03}{EU_uniform}{mse}{76.53}
\setres{evidZ03}{EU_uniform}{mse_prior_mean}{75.85}
\setres{evidZ03}{EU_uniform}{nrmse_pct}{23.2}
\setres{evidZ03}{EU_uniform}{pred_variance}{0.68}
\setres{evidZ03}{EU_uniform}{r_ff}{---}
\setres{evidZ03}{EU_uniform}{r_ff_se}{---}
\setres{evidZ03}{EU_uniform}{r_prior}{---}
\setres{evidZ03}{EU_uniform}{r_prior_se}{---}
\setres{evidZ03}{EU_uniform}{r_prob}{---}
\setres{evidZ03}{EU_uniform}{r_prob_se}{---}
\setres{evidZ03}{EU_uniform}{spread_penalty}{0.009}
\setres{evidZ03}{EU_uniform}{spread_penalty_se}{0.001}
\setres{evidZ03}{EU_uniform}{volume_ess}{---}
\setres{evidZ03}{EV_deterministic}{kappa}{0.624}
\setres{evidZ03}{EV_deterministic}{kappa_se}{0.071}
\setres{evidZ03}{EV_deterministic}{mse}{73.61}
\setres{evidZ03}{EV_deterministic}{nrmse_pct}{23.4}
\setres{evidZ03}{EV_deterministic}{r_ff}{---}
\setres{evidZ03}{EV_deterministic}{r_ff_se}{---}
\setres{evidZ03}{EV_deterministic}{r_prior}{---}
\setres{evidZ03}{EV_deterministic}{r_prior_se}{---}
\setres{evidZ03}{EV_deterministic}{r_prob}{---}
\setres{evidZ03}{EV_deterministic}{r_prob_se}{---}
\setres{evidZ03}{EV_deterministic}{volume_ess}{---}
\setres{evidZ03}{reference}{lookup_nrmse_pct}{16.7}
\setres{evidZ03}{reference}{midpoint_mse}{123.14}
\setres{evidZ03}{reference}{midpoint_nrmse_pct}{22.8}
\setres{evidZ03}{reference}{n_lotteries}{15}
\setres{evidZ03}{reference}{n_obs}{2{,}673}
\setres{evidZ03}{reference}{n_subjects}{179}
\setres{evidZ03}{reference}{oracle_mse}{43.82}
\setres{evidZ03}{reference}{sample}{Zurich2003\_shared}
\setres{domB}{CPT}{det_pct}{99}
\setres{domB}{CPT}{gap_lit}{0.898}
\setres{domB}{CPT}{gap_uniform}{1.062}
\setres{domB}{CPT}{share_lit}{9}
\setres{domB}{CPT}{share_uniform}{-7}
\setres{domB}{CPT_deterministic}{kappa}{0.992}
\setres{domB}{CPT_deterministic}{kappa_se}{0.009}
\setres{domB}{CPT_deterministic}{mse}{5.60}
\setres{domB}{CPT_deterministic}{nrmse_pct}{17.4}
\setres{domB}{CPT_deterministic}{r_ff}{---}
\setres{domB}{CPT_deterministic}{r_ff_se}{---}
\setres{domB}{CPT_deterministic}{r_prior}{---}
\setres{domB}{CPT_deterministic}{r_prior_se}{---}
\setres{domB}{CPT_deterministic}{r_prob}{---}
\setres{domB}{CPT_deterministic}{r_prob_se}{---}
\setres{domB}{CPT_deterministic}{theta_star}{1.10, 0.30, 0.90}
\setres{domB}{CPT_deterministic}{volume_ess}{---}
\setres{domB}{CPT_lit}{kappa}{0.094}
\setres{domB}{CPT_lit}{kappa_prior_mean}{0.225}
\setres{domB}{CPT_lit}{kappa_prior_mean_se}{0.192}
\setres{domB}{CPT_lit}{kappa_se}{0.208}
\setres{domB}{CPT_lit}{mse}{11.03}
\setres{domB}{CPT_lit}{mse_prior_mean}{10.23}
\setres{domB}{CPT_lit}{nrmse_pct}{20.5}
\setres{domB}{CPT_lit}{pred_variance}{0.79}
\setres{domB}{CPT_lit}{r_ff}{---}
\setres{domB}{CPT_lit}{r_ff_se}{---}
\setres{domB}{CPT_lit}{r_prior}{---}
\setres{domB}{CPT_lit}{r_prior_se}{---}
\setres{domB}{CPT_lit}{r_prob}{---}
\setres{domB}{CPT_lit}{r_prob_se}{---}
\setres{domB}{CPT_lit}{spread_penalty}{0.132}
\setres{domB}{CPT_lit}{spread_penalty_se}{0.015}
\setres{domB}{CPT_lit}{volume_ess}{---}
\setres{domB}{CPT_transfer_zurich03}{kappa}{0.894}
\setres{domB}{CPT_transfer_zurich03}{kappa_prior_mean}{0.894}
\setres{domB}{CPT_transfer_zurich03}{kappa_prior_mean_se}{0.052}
\setres{domB}{CPT_transfer_zurich03}{kappa_se}{0.052}
\setres{domB}{CPT_transfer_zurich03}{mse}{6.19}
\setres{domB}{CPT_transfer_zurich03}{mse_prior_mean}{6.19}
\setres{domB}{CPT_transfer_zurich03}{nrmse_pct}{17.8}
\setres{domB}{CPT_transfer_zurich03}{pred_variance}{0.00}
\setres{domB}{CPT_transfer_zurich03}{r_ff}{---}
\setres{domB}{CPT_transfer_zurich03}{r_ff_se}{---}
\setres{domB}{CPT_transfer_zurich03}{r_prior}{---}
\setres{domB}{CPT_transfer_zurich03}{r_prior_se}{---}
\setres{domB}{CPT_transfer_zurich03}{r_prob}{---}
\setres{domB}{CPT_transfer_zurich03}{r_prob_se}{---}
\setres{domB}{CPT_transfer_zurich03}{spread_penalty}{0.000}
\setres{domB}{CPT_transfer_zurich03}{spread_penalty_se}{0.000}
\setres{domB}{CPT_transfer_zurich03}{theta_star}{0.80, 0.40, 1.00}
\setres{domB}{CPT_transfer_zurich03}{transfer}{zurich03}
\setres{domB}{CPT_transfer_zurich03}{volume_ess}{---}
\setres{domB}{CPT_uniform}{kappa}{-0.070}
\setres{domB}{CPT_uniform}{kappa_prior_mean}{0.910}
\setres{domB}{CPT_uniform}{kappa_prior_mean_se}{0.046}
\setres{domB}{CPT_uniform}{kappa_se}{0.159}
\setres{domB}{CPT_uniform}{mse}{12.02}
\setres{domB}{CPT_uniform}{mse_prior_mean}{6.10}
\setres{domB}{CPT_uniform}{nrmse_pct}{17.8}
\setres{domB}{CPT_uniform}{pred_variance}{5.92}
\setres{domB}{CPT_uniform}{r_ff}{---}
\setres{domB}{CPT_uniform}{r_ff_se}{---}
\setres{domB}{CPT_uniform}{r_prior}{---}
\setres{domB}{CPT_uniform}{r_prior_se}{---}
\setres{domB}{CPT_uniform}{r_prob}{---}
\setres{domB}{CPT_uniform}{r_prob_se}{---}
\setres{domB}{CPT_uniform}{spread_penalty}{0.980}
\setres{domB}{CPT_uniform}{spread_penalty_se}{0.113}
\setres{domB}{CPT_uniform}{volume_ess}{---}
\setres{domB}{DA}{det_pct}{34}
\setres{domB}{DA}{gap_cpt_lit}{0.702}
\setres{domB}{DA}{gap_lit}{0.946}
\setres{domB}{DA}{gap_uniform}{1.211}
\setres{domB}{DA}{share_lit}{-180}
\setres{domB}{DA}{share_uniform}{-259}
\setres{domB}{DA_deterministic}{kappa}{0.337}
\setres{domB}{DA_deterministic}{kappa_se}{0.125}
\setres{domB}{DA_deterministic}{mse}{9.56}
\setres{domB}{DA_deterministic}{nrmse_pct}{22.5}
\setres{domB}{DA_deterministic}{r_ff}{---}
\setres{domB}{DA_deterministic}{r_ff_se}{---}
\setres{domB}{DA_deterministic}{r_prior}{---}
\setres{domB}{DA_deterministic}{r_prior_se}{---}
\setres{domB}{DA_deterministic}{r_prob}{---}
\setres{domB}{DA_deterministic}{r_prob_se}{---}
\setres{domB}{DA_deterministic}{theta_star}{6.07, 3.00}
\setres{domB}{DA_deterministic}{volume_ess}{---}
\setres{domB}{DA_lit}{kappa}{-0.608}
\setres{domB}{DA_lit}{kappa_prior_mean}{-0.521}
\setres{domB}{DA_lit}{kappa_prior_mean_se}{0.318}
\setres{domB}{DA_lit}{kappa_se}{0.328}
\setres{domB}{DA_lit}{mse}{15.27}
\setres{domB}{DA_lit}{mse_prior_mean}{14.75}
\setres{domB}{DA_lit}{nrmse_pct}{26.1}
\setres{domB}{DA_lit}{pred_variance}{0.52}
\setres{domB}{DA_lit}{r_ff}{---}
\setres{domB}{DA_lit}{r_ff_se}{---}
\setres{domB}{DA_lit}{r_prior}{---}
\setres{domB}{DA_lit}{r_prior_se}{---}
\setres{domB}{DA_lit}{r_prob}{---}
\setres{domB}{DA_lit}{r_prob_se}{---}
\setres{domB}{DA_lit}{spread_penalty}{0.087}
\setres{domB}{DA_lit}{spread_penalty_se}{0.010}
\setres{domB}{DA_lit}{volume_ess}{---}
\setres{domB}{DA_transfer_zurich03}{kappa}{-0.636}
\setres{domB}{DA_transfer_zurich03}{kappa_prior_mean}{-0.636}
\setres{domB}{DA_transfer_zurich03}{kappa_prior_mean_se}{0.335}
\setres{domB}{DA_transfer_zurich03}{kappa_se}{0.335}
\setres{domB}{DA_transfer_zurich03}{mse}{15.44}
\setres{domB}{DA_transfer_zurich03}{mse_prior_mean}{15.44}
\setres{domB}{DA_transfer_zurich03}{nrmse_pct}{26.8}
\setres{domB}{DA_transfer_zurich03}{pred_variance}{0.00}
\setres{domB}{DA_transfer_zurich03}{r_ff}{---}
\setres{domB}{DA_transfer_zurich03}{r_ff_se}{---}
\setres{domB}{DA_transfer_zurich03}{r_prior}{---}
\setres{domB}{DA_transfer_zurich03}{r_prior_se}{---}
\setres{domB}{DA_transfer_zurich03}{r_prob}{---}
\setres{domB}{DA_transfer_zurich03}{r_prob_se}{---}
\setres{domB}{DA_transfer_zurich03}{spread_penalty}{0.000}
\setres{domB}{DA_transfer_zurich03}{spread_penalty_se}{0.000}
\setres{domB}{DA_transfer_zurich03}{theta_star}{1.36, 1.35}
\setres{domB}{DA_transfer_zurich03}{transfer}{zurich03}
\setres{domB}{DA_transfer_zurich03}{volume_ess}{---}
\setres{domB}{DA_uniform}{kappa}{-0.874}
\setres{domB}{DA_uniform}{kappa_prior_mean}{-0.448}
\setres{domB}{DA_uniform}{kappa_prior_mean_se}{0.305}
\setres{domB}{DA_uniform}{kappa_se}{0.354}
\setres{domB}{DA_uniform}{mse}{16.87}
\setres{domB}{DA_uniform}{mse_prior_mean}{14.30}
\setres{domB}{DA_uniform}{nrmse_pct}{24.3}
\setres{domB}{DA_uniform}{pred_variance}{2.57}
\setres{domB}{DA_uniform}{r_ff}{---}
\setres{domB}{DA_uniform}{r_ff_se}{---}
\setres{domB}{DA_uniform}{r_prior}{---}
\setres{domB}{DA_uniform}{r_prior_se}{---}
\setres{domB}{DA_uniform}{r_prob}{---}
\setres{domB}{DA_uniform}{r_prob_se}{---}
\setres{domB}{DA_uniform}{spread_penalty}{0.425}
\setres{domB}{DA_uniform}{spread_penalty_se}{0.049}
\setres{domB}{DA_uniform}{volume_ess}{---}
\setres{domB}{EU}{det_pct}{-12}
\setres{domB}{EU}{gap_lit}{0.358}
\setres{domB}{EU}{gap_uniform}{0.399}
\setres{domB}{EU}{share_lit}{391}
\setres{domB}{EU}{share_uniform}{424}
\setres{domB}{EU_deterministic}{kappa}{-0.123}
\setres{domB}{EU_deterministic}{kappa_se}{0.189}
\setres{domB}{EU_deterministic}{mse}{12.34}
\setres{domB}{EU_deterministic}{nrmse_pct}{28.4}
\setres{domB}{EU_deterministic}{r_ff}{---}
\setres{domB}{EU_deterministic}{r_ff_se}{---}
\setres{domB}{EU_deterministic}{r_prior}{---}
\setres{domB}{EU_deterministic}{r_prior_se}{---}
\setres{domB}{EU_deterministic}{r_prob}{---}
\setres{domB}{EU_deterministic}{r_prob_se}{---}
\setres{domB}{EU_deterministic}{theta_star}{3.59}
\setres{domB}{EU_deterministic}{volume_ess}{---}
\setres{domB}{EU_lit}{kappa}{-0.481}
\setres{domB}{EU_lit}{kappa_prior_mean}{-0.478}
\setres{domB}{EU_lit}{kappa_prior_mean_se}{0.311}
\setres{domB}{EU_lit}{kappa_se}{0.311}
\setres{domB}{EU_lit}{mse}{14.50}
\setres{domB}{EU_lit}{mse_prior_mean}{14.48}
\setres{domB}{EU_lit}{nrmse_pct}{25.6}
\setres{domB}{EU_lit}{pred_variance}{0.02}
\setres{domB}{EU_lit}{r_ff}{---}
\setres{domB}{EU_lit}{r_ff_se}{---}
\setres{domB}{EU_lit}{r_prior}{---}
\setres{domB}{EU_lit}{r_prior_se}{---}
\setres{domB}{EU_lit}{r_prob}{---}
\setres{domB}{EU_lit}{r_prob_se}{---}
\setres{domB}{EU_lit}{spread_penalty}{0.003}
\setres{domB}{EU_lit}{spread_penalty_se}{0.000}
\setres{domB}{EU_lit}{volume_ess}{---}
\setres{domB}{EU_transfer_zurich03}{kappa}{-0.469}
\setres{domB}{EU_transfer_zurich03}{kappa_prior_mean}{-0.469}
\setres{domB}{EU_transfer_zurich03}{kappa_prior_mean_se}{0.309}
\setres{domB}{EU_transfer_zurich03}{kappa_se}{0.309}
\setres{domB}{EU_transfer_zurich03}{mse}{14.43}
\setres{domB}{EU_transfer_zurich03}{mse_prior_mean}{14.43}
\setres{domB}{EU_transfer_zurich03}{nrmse_pct}{25.4}
\setres{domB}{EU_transfer_zurich03}{pred_variance}{0.00}
\setres{domB}{EU_transfer_zurich03}{r_ff}{---}
\setres{domB}{EU_transfer_zurich03}{r_ff_se}{---}
\setres{domB}{EU_transfer_zurich03}{r_prior}{---}
\setres{domB}{EU_transfer_zurich03}{r_prior_se}{---}
\setres{domB}{EU_transfer_zurich03}{r_prob}{---}
\setres{domB}{EU_transfer_zurich03}{r_prob_se}{---}
\setres{domB}{EU_transfer_zurich03}{spread_penalty}{0.000}
\setres{domB}{EU_transfer_zurich03}{spread_penalty_se}{0.000}
\setres{domB}{EU_transfer_zurich03}{theta_star}{0.76}
\setres{domB}{EU_transfer_zurich03}{transfer}{zurich03}
\setres{domB}{EU_transfer_zurich03}{volume_ess}{---}
\setres{domB}{EU_uniform}{kappa}{-0.522}
\setres{domB}{EU_uniform}{kappa_prior_mean}{-0.508}
\setres{domB}{EU_uniform}{kappa_prior_mean_se}{0.316}
\setres{domB}{EU_uniform}{kappa_se}{0.318}
\setres{domB}{EU_uniform}{mse}{14.75}
\setres{domB}{EU_uniform}{mse_prior_mean}{14.67}
\setres{domB}{EU_uniform}{nrmse_pct}{26.3}
\setres{domB}{EU_uniform}{pred_variance}{0.08}
\setres{domB}{EU_uniform}{r_ff}{---}
\setres{domB}{EU_uniform}{r_ff_se}{---}
\setres{domB}{EU_uniform}{r_prior}{---}
\setres{domB}{EU_uniform}{r_prior_se}{---}
\setres{domB}{EU_uniform}{r_prob}{---}
\setres{domB}{EU_uniform}{r_prob_se}{---}
\setres{domB}{EU_uniform}{spread_penalty}{0.014}
\setres{domB}{EU_uniform}{spread_penalty_se}{0.002}
\setres{domB}{EU_uniform}{volume_ess}{---}
\setres{domB}{EV_deterministic}{kappa}{-0.426}
\setres{domB}{EV_deterministic}{kappa_se}{0.291}
\setres{domB}{EV_deterministic}{mse}{14.17}
\setres{domB}{EV_deterministic}{nrmse_pct}{25.9}
\setres{domB}{EV_deterministic}{r_ff}{---}
\setres{domB}{EV_deterministic}{r_ff_se}{---}
\setres{domB}{EV_deterministic}{r_prior}{---}
\setres{domB}{EV_deterministic}{r_prior_se}{---}
\setres{domB}{EV_deterministic}{r_prob}{---}
\setres{domB}{EV_deterministic}{r_prob_se}{---}
\setres{domB}{EV_deterministic}{volume_ess}{---}
\setres{domB}{reference}{lookup_nrmse_pct}{17.0}
\setres{domB}{reference}{midpoint_mse}{11.60}
\setres{domB}{reference}{midpoint_nrmse_pct}{22.0}
\setres{domB}{reference}{n_lotteries}{14}
\setres{domB}{reference}{n_obs}{2{,}112}
\setres{domB}{reference}{n_subjects}{151}
\setres{domB}{reference}{oracle_mse}{5.55}
\setres{domB}{reference}{sample}{Beijing2005\_gains}
\setres{domLV}{CPT}{det_pct}{97}
\setres{domLV}{CPT}{gap_lit}{0.621}
\setres{domLV}{CPT}{gap_uniform}{3.098}
\setres{domLV}{CPT}{share_lit}{36}
\setres{domLV}{CPT}{share_uniform}{-220}
\setres{domLV}{CPT_deterministic}{kappa}{0.970}
\setres{domLV}{CPT_deterministic}{kappa_se}{0.002}
\setres{domLV}{CPT_deterministic}{mse}{16.30}
\setres{domLV}{CPT_deterministic}{nrmse_pct}{22.7}
\setres{domLV}{CPT_deterministic}{r_ff}{---}
\setres{domLV}{CPT_deterministic}{r_ff_se}{---}
\setres{domLV}{CPT_deterministic}{r_prior}{---}
\setres{domLV}{CPT_deterministic}{r_prior_se}{---}
\setres{domLV}{CPT_deterministic}{r_prob}{---}
\setres{domLV}{CPT_deterministic}{r_prob_se}{---}
\setres{domLV}{CPT_deterministic}{theta_star}{1.00, 0.60, 0.90}
\setres{domLV}{CPT_deterministic}{volume_ess}{---}
\setres{domLV}{CPT_lit}{kappa}{0.349}
\setres{domLV}{CPT_lit}{kappa_prior_mean}{0.697}
\setres{domLV}{CPT_lit}{kappa_prior_mean_se}{0.017}
\setres{domLV}{CPT_lit}{kappa_se}{0.022}
\setres{domLV}{CPT_lit}{mse}{21.49}
\setres{domLV}{CPT_lit}{mse_prior_mean}{18.58}
\setres{domLV}{CPT_lit}{nrmse_pct}{23.9}
\setres{domLV}{CPT_lit}{pred_variance}{2.91}
\setres{domLV}{CPT_lit}{r_ff}{---}
\setres{domLV}{CPT_lit}{r_ff_se}{---}
\setres{domLV}{CPT_lit}{r_prior}{---}
\setres{domLV}{CPT_lit}{r_prior_se}{---}
\setres{domLV}{CPT_lit}{r_prob}{---}
\setres{domLV}{CPT_lit}{r_prob_se}{---}
\setres{domLV}{CPT_lit}{spread_penalty}{0.348}
\setres{domLV}{CPT_lit}{spread_penalty_se}{0.007}
\setres{domLV}{CPT_lit}{volume_ess}{---}
\setres{domLV}{CPT_transfer_zurich03}{kappa}{0.793}
\setres{domLV}{CPT_transfer_zurich03}{kappa_prior_mean}{0.793}
\setres{domLV}{CPT_transfer_zurich03}{kappa_prior_mean_se}{0.010}
\setres{domLV}{CPT_transfer_zurich03}{kappa_se}{0.010}
\setres{domLV}{CPT_transfer_zurich03}{mse}{17.78}
\setres{domLV}{CPT_transfer_zurich03}{mse_prior_mean}{17.78}
\setres{domLV}{CPT_transfer_zurich03}{nrmse_pct}{24.0}
\setres{domLV}{CPT_transfer_zurich03}{pred_variance}{0.00}
\setres{domLV}{CPT_transfer_zurich03}{r_ff}{---}
\setres{domLV}{CPT_transfer_zurich03}{r_ff_se}{---}
\setres{domLV}{CPT_transfer_zurich03}{r_prior}{---}
\setres{domLV}{CPT_transfer_zurich03}{r_prior_se}{---}
\setres{domLV}{CPT_transfer_zurich03}{r_prob}{---}
\setres{domLV}{CPT_transfer_zurich03}{r_prob_se}{---}
\setres{domLV}{CPT_transfer_zurich03}{spread_penalty}{0.000}
\setres{domLV}{CPT_transfer_zurich03}{spread_penalty_se}{0.000}
\setres{domLV}{CPT_transfer_zurich03}{theta_star}{0.80, 0.40, 1.00}
\setres{domLV}{CPT_transfer_zurich03}{transfer}{zurich03}
\setres{domLV}{CPT_transfer_zurich03}{volume_ess}{---}
\setres{domLV}{CPT_uniform}{kappa}{-2.129}
\setres{domLV}{CPT_uniform}{kappa_prior_mean}{-0.046}
\setres{domLV}{CPT_uniform}{kappa_prior_mean_se}{0.033}
\setres{domLV}{CPT_uniform}{kappa_se}{0.066}
\setres{domLV}{CPT_uniform}{mse}{42.22}
\setres{domLV}{CPT_uniform}{mse_prior_mean}{24.79}
\setres{domLV}{CPT_uniform}{nrmse_pct}{26.6}
\setres{domLV}{CPT_uniform}{pred_variance}{17.43}
\setres{domLV}{CPT_uniform}{r_ff}{---}
\setres{domLV}{CPT_uniform}{r_ff_se}{---}
\setres{domLV}{CPT_uniform}{r_prior}{---}
\setres{domLV}{CPT_uniform}{r_prior_se}{---}
\setres{domLV}{CPT_uniform}{r_prob}{---}
\setres{domLV}{CPT_uniform}{r_prob_se}{---}
\setres{domLV}{CPT_uniform}{spread_penalty}{2.083}
\setres{domLV}{CPT_uniform}{spread_penalty_se}{0.042}
\setres{domLV}{CPT_uniform}{volume_ess}{---}
\setres{domLV}{DA}{det_pct}{85}
\setres{domLV}{DA}{gap_cpt_lit}{1.262}
\setres{domLV}{DA}{gap_lit}{1.765}
\setres{domLV}{DA}{gap_uniform}{3.786}
\setres{domLV}{DA}{share_lit}{-107}
\setres{domLV}{DA}{share_uniform}{-344}
\setres{domLV}{DA_deterministic}{kappa}{0.852}
\setres{domLV}{DA_deterministic}{kappa_se}{0.009}
\setres{domLV}{DA_deterministic}{mse}{17.28}
\setres{domLV}{DA_deterministic}{nrmse_pct}{23.1}
\setres{domLV}{DA_deterministic}{r_ff}{---}
\setres{domLV}{DA_deterministic}{r_ff_se}{---}
\setres{domLV}{DA_deterministic}{r_prior}{---}
\setres{domLV}{DA_deterministic}{r_prior_se}{---}
\setres{domLV}{DA_deterministic}{r_prob}{---}
\setres{domLV}{DA_deterministic}{r_prob_se}{---}
\setres{domLV}{DA_deterministic}{theta_star}{1.69, 1.45}
\setres{domLV}{DA_deterministic}{volume_ess}{---}
\setres{domLV}{DA_lit}{kappa}{-0.913}
\setres{domLV}{DA_lit}{kappa_prior_mean}{-0.342}
\setres{domLV}{DA_lit}{kappa_prior_mean_se}{0.048}
\setres{domLV}{DA_lit}{kappa_se}{0.058}
\setres{domLV}{DA_lit}{mse}{32.05}
\setres{domLV}{DA_lit}{mse_prior_mean}{27.27}
\setres{domLV}{DA_lit}{nrmse_pct}{29.5}
\setres{domLV}{DA_lit}{pred_variance}{4.77}
\setres{domLV}{DA_lit}{r_ff}{---}
\setres{domLV}{DA_lit}{r_ff_se}{---}
\setres{domLV}{DA_lit}{r_prior}{---}
\setres{domLV}{DA_lit}{r_prior_se}{---}
\setres{domLV}{DA_lit}{r_prob}{---}
\setres{domLV}{DA_lit}{r_prob_se}{---}
\setres{domLV}{DA_lit}{spread_penalty}{0.571}
\setres{domLV}{DA_lit}{spread_penalty_se}{0.012}
\setres{domLV}{DA_lit}{volume_ess}{---}
\setres{domLV}{DA_transfer_zurich03}{kappa}{0.728}
\setres{domLV}{DA_transfer_zurich03}{kappa_prior_mean}{0.728}
\setres{domLV}{DA_transfer_zurich03}{kappa_prior_mean_se}{0.016}
\setres{domLV}{DA_transfer_zurich03}{kappa_se}{0.016}
\setres{domLV}{DA_transfer_zurich03}{mse}{18.32}
\setres{domLV}{DA_transfer_zurich03}{mse_prior_mean}{18.32}
\setres{domLV}{DA_transfer_zurich03}{nrmse_pct}{23.4}
\setres{domLV}{DA_transfer_zurich03}{pred_variance}{0.00}
\setres{domLV}{DA_transfer_zurich03}{r_ff}{---}
\setres{domLV}{DA_transfer_zurich03}{r_ff_se}{---}
\setres{domLV}{DA_transfer_zurich03}{r_prior}{---}
\setres{domLV}{DA_transfer_zurich03}{r_prior_se}{---}
\setres{domLV}{DA_transfer_zurich03}{r_prob}{---}
\setres{domLV}{DA_transfer_zurich03}{r_prob_se}{---}
\setres{domLV}{DA_transfer_zurich03}{spread_penalty}{0.000}
\setres{domLV}{DA_transfer_zurich03}{spread_penalty_se}{0.000}
\setres{domLV}{DA_transfer_zurich03}{theta_star}{1.36, 1.35}
\setres{domLV}{DA_transfer_zurich03}{transfer}{zurich03}
\setres{domLV}{DA_transfer_zurich03}{volume_ess}{---}
\setres{domLV}{DA_uniform}{kappa}{-2.934}
\setres{domLV}{DA_uniform}{kappa_prior_mean}{-1.295}
\setres{domLV}{DA_uniform}{kappa_prior_mean_se}{0.068}
\setres{domLV}{DA_uniform}{kappa_se}{0.096}
\setres{domLV}{DA_uniform}{mse}{48.96}
\setres{domLV}{DA_uniform}{mse_prior_mean}{35.24}
\setres{domLV}{DA_uniform}{nrmse_pct}{31.5}
\setres{domLV}{DA_uniform}{pred_variance}{13.71}
\setres{domLV}{DA_uniform}{r_ff}{---}
\setres{domLV}{DA_uniform}{r_ff_se}{---}
\setres{domLV}{DA_uniform}{r_prior}{---}
\setres{domLV}{DA_uniform}{r_prior_se}{---}
\setres{domLV}{DA_uniform}{r_prob}{---}
\setres{domLV}{DA_uniform}{r_prob_se}{---}
\setres{domLV}{DA_uniform}{spread_penalty}{1.639}
\setres{domLV}{DA_uniform}{spread_penalty_se}{0.033}
\setres{domLV}{DA_uniform}{volume_ess}{---}
\setres{domLV}{EU}{det_pct}{76}
\setres{domLV}{EU}{gap_lit}{0.426}
\setres{domLV}{EU}{gap_uniform}{2.264}
\setres{domLV}{EU}{share_lit}{44}
\setres{domLV}{EU}{share_uniform}{-198}
\setres{domLV}{EU_deterministic}{kappa}{0.760}
\setres{domLV}{EU_deterministic}{kappa_se}{0.014}
\setres{domLV}{EU_deterministic}{mse}{18.05}
\setres{domLV}{EU_deterministic}{nrmse_pct}{23.8}
\setres{domLV}{EU_deterministic}{r_ff}{---}
\setres{domLV}{EU_deterministic}{r_ff_se}{---}
\setres{domLV}{EU_deterministic}{r_prior}{---}
\setres{domLV}{EU_deterministic}{r_prior_se}{---}
\setres{domLV}{EU_deterministic}{r_prob}{---}
\setres{domLV}{EU_deterministic}{r_prob_se}{---}
\setres{domLV}{EU_deterministic}{theta_star}{0.95}
\setres{domLV}{EU_deterministic}{volume_ess}{---}
\setres{domLV}{EU_lit}{kappa}{0.334}
\setres{domLV}{EU_lit}{kappa_prior_mean}{0.478}
\setres{domLV}{EU_lit}{kappa_prior_mean_se}{0.025}
\setres{domLV}{EU_lit}{kappa_se}{0.028}
\setres{domLV}{EU_lit}{mse}{21.62}
\setres{domLV}{EU_lit}{mse_prior_mean}{20.41}
\setres{domLV}{EU_lit}{nrmse_pct}{25.2}
\setres{domLV}{EU_lit}{pred_variance}{1.21}
\setres{domLV}{EU_lit}{r_ff}{---}
\setres{domLV}{EU_lit}{r_ff_se}{---}
\setres{domLV}{EU_lit}{r_prior}{---}
\setres{domLV}{EU_lit}{r_prior_se}{---}
\setres{domLV}{EU_lit}{r_prob}{---}
\setres{domLV}{EU_lit}{r_prob_se}{---}
\setres{domLV}{EU_lit}{spread_penalty}{0.145}
\setres{domLV}{EU_lit}{spread_penalty_se}{0.003}
\setres{domLV}{EU_lit}{volume_ess}{---}
\setres{domLV}{EU_transfer_zurich03}{kappa}{0.624}
\setres{domLV}{EU_transfer_zurich03}{kappa_prior_mean}{0.624}
\setres{domLV}{EU_transfer_zurich03}{kappa_prior_mean_se}{0.020}
\setres{domLV}{EU_transfer_zurich03}{kappa_se}{0.020}
\setres{domLV}{EU_transfer_zurich03}{mse}{19.19}
\setres{domLV}{EU_transfer_zurich03}{mse_prior_mean}{19.19}
\setres{domLV}{EU_transfer_zurich03}{nrmse_pct}{24.6}
\setres{domLV}{EU_transfer_zurich03}{pred_variance}{0.00}
\setres{domLV}{EU_transfer_zurich03}{r_ff}{---}
\setres{domLV}{EU_transfer_zurich03}{r_ff_se}{---}
\setres{domLV}{EU_transfer_zurich03}{r_prior}{---}
\setres{domLV}{EU_transfer_zurich03}{r_prior_se}{---}
\setres{domLV}{EU_transfer_zurich03}{r_prob}{---}
\setres{domLV}{EU_transfer_zurich03}{r_prob_se}{---}
\setres{domLV}{EU_transfer_zurich03}{spread_penalty}{0.000}
\setres{domLV}{EU_transfer_zurich03}{spread_penalty_se}{0.000}
\setres{domLV}{EU_transfer_zurich03}{theta_star}{0.76}
\setres{domLV}{EU_transfer_zurich03}{transfer}{zurich03}
\setres{domLV}{EU_transfer_zurich03}{volume_ess}{---}
\setres{domLV}{EU_uniform}{kappa}{-1.504}
\setres{domLV}{EU_uniform}{kappa_prior_mean}{-0.518}
\setres{domLV}{EU_uniform}{kappa_prior_mean_se}{0.051}
\setres{domLV}{EU_uniform}{kappa_se}{0.068}
\setres{domLV}{EU_uniform}{mse}{36.99}
\setres{domLV}{EU_uniform}{mse_prior_mean}{28.74}
\setres{domLV}{EU_uniform}{nrmse_pct}{29.7}
\setres{domLV}{EU_uniform}{pred_variance}{8.25}
\setres{domLV}{EU_uniform}{r_ff}{---}
\setres{domLV}{EU_uniform}{r_ff_se}{---}
\setres{domLV}{EU_uniform}{r_prior}{---}
\setres{domLV}{EU_uniform}{r_prior_se}{---}
\setres{domLV}{EU_uniform}{r_prob}{---}
\setres{domLV}{EU_uniform}{r_prob_se}{---}
\setres{domLV}{EU_uniform}{spread_penalty}{0.986}
\setres{domLV}{EU_uniform}{spread_penalty_se}{0.020}
\setres{domLV}{EU_uniform}{volume_ess}{---}
\setres{domLV}{EV_deterministic}{kappa}{0.752}
\setres{domLV}{EV_deterministic}{kappa_se}{0.015}
\setres{domLV}{EV_deterministic}{mse}{18.12}
\setres{domLV}{EV_deterministic}{nrmse_pct}{24.2}
\setres{domLV}{EV_deterministic}{r_ff}{---}
\setres{domLV}{EV_deterministic}{r_ff_se}{---}
\setres{domLV}{EV_deterministic}{r_prior}{---}
\setres{domLV}{EV_deterministic}{r_prior_se}{---}
\setres{domLV}{EV_deterministic}{r_prob}{---}
\setres{domLV}{EV_deterministic}{r_prob_se}{---}
\setres{domLV}{EV_deterministic}{volume_ess}{---}
\setres{domLV}{reference}{lookup_nrmse_pct}{22.6}
\setres{domLV}{reference}{midpoint_mse}{24.41}
\setres{domLV}{reference}{midpoint_nrmse_pct}{27.2}
\setres{domLV}{reference}{n_lotteries}{14}
\setres{domLV}{reference}{n_obs}{40{,}886}
\setres{domLV}{reference}{n_subjects}{2{,}939}
\setres{domLV}{reference}{oracle_mse}{16.04}
\setres{domLV}{reference}{sample}{LV2019\_pooled}
\setres{domAvg}{CPT_deterministic}{kappa_max}{0.992}
\setres{domAvg}{CPT_deterministic}{kappa_mean}{0.978}
\setres{domAvg}{CPT_deterministic}{kappa_mean_se}{0.004}
\setres{domAvg}{CPT_deterministic}{kappa_min}{0.970}
\setres{domAvg}{CPT_deterministic}{kappa_zurich_only}{0.324}
\setres{domAvg}{CPT_deterministic}{n_domains}{3}
\setres{domAvg}{CPT_deterministic}{r_prob_mean}{1.000}
\setres{domAvg}{CPT_deterministic}{r_prob_zurich_only}{0.333}
\setres{domAvg}{CPT_lit}{kappa_max}{0.641}
\setres{domAvg}{CPT_lit}{kappa_mean}{0.361}
\setres{domAvg}{CPT_lit}{kappa_mean_se}{0.072}
\setres{domAvg}{CPT_lit}{kappa_min}{0.094}
\setres{domAvg}{CPT_lit}{kappa_zurich_only}{0.214}
\setres{domAvg}{CPT_lit}{n_domains}{3}
\setres{domAvg}{CPT_lit}{r_prob_mean}{0.958}
\setres{domAvg}{CPT_lit}{r_prob_zurich_only}{0.319}
\setres{domAvg}{CPT_transfer_zurich03}{kappa_max}{0.972}
\setres{domAvg}{CPT_transfer_zurich03}{kappa_mean}{0.886}
\setres{domAvg}{CPT_transfer_zurich03}{kappa_mean_se}{0.018}
\setres{domAvg}{CPT_transfer_zurich03}{kappa_min}{0.793}
\setres{domAvg}{CPT_transfer_zurich03}{kappa_zurich_only}{0.324}
\setres{domAvg}{CPT_transfer_zurich03}{n_domains}{3}
\setres{domAvg}{CPT_transfer_zurich03}{r_prob_mean}{1.000}
\setres{domAvg}{CPT_transfer_zurich03}{r_prob_zurich_only}{0.333}
\setres{domAvg}{CPT_uniform}{kappa_max}{-0.070}
\setres{domAvg}{CPT_uniform}{kappa_mean}{-0.898}
\setres{domAvg}{CPT_uniform}{kappa_mean_se}{0.072}
\setres{domAvg}{CPT_uniform}{kappa_min}{-2.129}
\setres{domAvg}{CPT_uniform}{kappa_zurich_only}{-0.165}
\setres{domAvg}{CPT_uniform}{n_domains}{3}
\setres{domAvg}{CPT_uniform}{r_prob_mean}{0.673}
\setres{domAvg}{CPT_uniform}{r_prob_zurich_only}{0.194}
\setres{domAvg}{DA_deterministic}{kappa_max}{0.852}
\setres{domAvg}{DA_deterministic}{kappa_mean}{0.605}
\setres{domAvg}{DA_deterministic}{kappa_mean_se}{0.047}
\setres{domAvg}{DA_deterministic}{kappa_min}{0.337}
\setres{domAvg}{DA_deterministic}{kappa_zurich_only}{0.209}
\setres{domAvg}{DA_deterministic}{n_domains}{3}
\setres{domAvg}{DA_deterministic}{r_prob_mean}{1.000}
\setres{domAvg}{DA_deterministic}{r_prob_zurich_only}{0.333}
\setres{domAvg}{DA_lit}{kappa_max}{0.020}
\setres{domAvg}{DA_lit}{kappa_mean}{-0.500}
\setres{domAvg}{DA_lit}{kappa_mean_se}{0.120}
\setres{domAvg}{DA_lit}{kappa_min}{-0.913}
\setres{domAvg}{DA_lit}{kappa_zurich_only}{0.007}
\setres{domAvg}{DA_lit}{n_domains}{3}
\setres{domAvg}{DA_lit}{r_prob_mean}{0.954}
\setres{domAvg}{DA_lit}{r_prob_zurich_only}{0.319}
\setres{domAvg}{DA_transfer_zurich03}{kappa_max}{0.728}
\setres{domAvg}{DA_transfer_zurich03}{kappa_mean}{0.239}
\setres{domAvg}{DA_transfer_zurich03}{kappa_mean_se}{0.114}
\setres{domAvg}{DA_transfer_zurich03}{kappa_min}{-0.636}
\setres{domAvg}{DA_transfer_zurich03}{kappa_zurich_only}{0.209}
\setres{domAvg}{DA_transfer_zurich03}{n_domains}{3}
\setres{domAvg}{DA_transfer_zurich03}{r_prob_mean}{1.000}
\setres{domAvg}{DA_transfer_zurich03}{r_prob_zurich_only}{0.333}
\setres{domAvg}{DA_uniform}{kappa_max}{-0.827}
\setres{domAvg}{DA_uniform}{kappa_mean}{-1.545}
\setres{domAvg}{DA_uniform}{kappa_mean_se}{0.141}
\setres{domAvg}{DA_uniform}{kappa_min}{-2.934}
\setres{domAvg}{DA_uniform}{kappa_zurich_only}{-0.276}
\setres{domAvg}{DA_uniform}{n_domains}{3}
\setres{domAvg}{DA_uniform}{r_prob_mean}{0.907}
\setres{domAvg}{DA_uniform}{r_prob_zurich_only}{0.301}
\setres{domAvg}{EU_deterministic}{kappa_max}{0.760}
\setres{domAvg}{EU_deterministic}{kappa_mean}{0.418}
\setres{domAvg}{EU_deterministic}{kappa_mean_se}{0.067}
\setres{domAvg}{EU_deterministic}{kappa_min}{-0.123}
\setres{domAvg}{EU_deterministic}{kappa_zurich_only}{0.205}
\setres{domAvg}{EU_deterministic}{n_domains}{3}
\setres{domAvg}{EU_deterministic}{r_prob_mean}{1.000}
\setres{domAvg}{EU_deterministic}{r_prob_zurich_only}{0.333}
\setres{domAvg}{EU_lit}{kappa_max}{0.521}
\setres{domAvg}{EU_lit}{kappa_mean}{0.125}
\setres{domAvg}{EU_lit}{kappa_mean_se}{0.107}
\setres{domAvg}{EU_lit}{kappa_min}{-0.481}
\setres{domAvg}{EU_lit}{kappa_zurich_only}{0.174}
\setres{domAvg}{EU_lit}{n_domains}{3}
\setres{domAvg}{EU_lit}{r_prob_mean}{0.904}
\setres{domAvg}{EU_lit}{r_prob_zurich_only}{0.292}
\setres{domAvg}{EU_transfer_zurich03}{kappa_max}{0.624}
\setres{domAvg}{EU_transfer_zurich03}{kappa_mean}{0.257}
\setres{domAvg}{EU_transfer_zurich03}{kappa_mean_se}{0.105}
\setres{domAvg}{EU_transfer_zurich03}{kappa_min}{-0.469}
\setres{domAvg}{EU_transfer_zurich03}{kappa_zurich_only}{0.205}
\setres{domAvg}{EU_transfer_zurich03}{n_domains}{3}
\setres{domAvg}{EU_transfer_zurich03}{r_prob_mean}{1.000}
\setres{domAvg}{EU_transfer_zurich03}{r_prob_zurich_only}{0.333}
\setres{domAvg}{EU_uniform}{kappa_max}{-0.209}
\setres{domAvg}{EU_uniform}{kappa_mean}{-0.745}
\setres{domAvg}{EU_uniform}{kappa_mean_se}{0.120}
\setres{domAvg}{EU_uniform}{kappa_min}{-1.504}
\setres{domAvg}{EU_uniform}{kappa_zurich_only}{-0.070}
\setres{domAvg}{EU_uniform}{n_domains}{3}
\setres{domAvg}{EU_uniform}{r_prob_mean}{0.785}
\setres{domAvg}{EU_uniform}{r_prob_zurich_only}{0.240}
\setres{domAvg}{EV_deterministic}{kappa_max}{0.752}
\setres{domAvg}{EV_deterministic}{kappa_mean}{0.273}
\setres{domAvg}{EV_deterministic}{kappa_mean_se}{0.099}
\setres{domAvg}{EV_deterministic}{kappa_min}{-0.426}
\setres{domAvg}{EV_deterministic}{kappa_zurich_only}{0.164}
\setres{domAvg}{EV_deterministic}{n_domains}{3}
\setres{domAvg}{EV_deterministic}{r_prob_mean}{1.000}
\setres{domAvg}{EV_deterministic}{r_prob_zurich_only}{0.333}
\setres{domAvg}{pairs}{n_cpt_da_fit}{3}
\setres{domAvg}{pairs}{n_cpt_da_lit}{3}
\setres{domAvg}{pairs}{n_cpt_da_transfer}{3}
\setres{domAvg}{pairs}{n_cpt_da_uniform}{3}
\setres{domAvg}{pairs}{n_cpt_eu_fit}{3}
\setres{domAvg}{pairs}{n_cpt_eu_lit}{3}
\setres{domAvg}{pairs}{n_cpt_eu_transfer}{3}
\setres{domAvg}{pairs}{n_cpt_eu_uniform}{1}
\setres{domAvg}{pairs}{n_da_eu_fit}{3}
\setres{domAvg}{pairs}{n_da_eu_lit}{0}
\setres{domAvg}{pairs}{n_da_eu_transfer}{2}
\setres{domAvg}{pairs}{n_da_eu_uniform}{0}
\setres{domAvg}{reference}{domains}{zurich03, beijing, lv\_pooled}
\setres{domAvg}{reference}{n_domains}{3}
\setres{domZ06}{CPT}{det_pct}{100}
\setres{domZ06}{CPT}{gap_lit}{0.094}
\setres{domZ06}{CPT}{gap_uniform}{0.377}
\setres{domZ06}{CPT}{share_lit}{91}
\setres{domZ06}{CPT}{share_uniform}{62}
\setres{domZ06}{CPT_deterministic}{kappa}{0.997}
\setres{domZ06}{CPT_deterministic}{kappa_se}{0.002}
\setres{domZ06}{CPT_deterministic}{mse}{36.81}
\setres{domZ06}{CPT_deterministic}{nrmse_pct}{13.1}
\setres{domZ06}{CPT_deterministic}{r_ff}{---}
\setres{domZ06}{CPT_deterministic}{r_ff_se}{---}
\setres{domZ06}{CPT_deterministic}{r_prior}{---}
\setres{domZ06}{CPT_deterministic}{r_prior_se}{---}
\setres{domZ06}{CPT_deterministic}{r_prob}{---}
\setres{domZ06}{CPT_deterministic}{r_prob_se}{---}
\setres{domZ06}{CPT_deterministic}{theta_star}{0.80, 0.50, 0.90}
\setres{domZ06}{CPT_deterministic}{volume_ess}{---}
\setres{domZ06}{CPT_lit}{kappa}{0.904}
\setres{domZ06}{CPT_lit}{kappa_prior_mean}{0.952}
\setres{domZ06}{CPT_lit}{kappa_prior_mean_se}{0.012}
\setres{domZ06}{CPT_lit}{kappa_se}{0.014}
\setres{domZ06}{CPT_lit}{mse}{58.07}
\setres{domZ06}{CPT_lit}{mse_prior_mean}{47.20}
\setres{domZ06}{CPT_lit}{nrmse_pct}{14.7}
\setres{domZ06}{CPT_lit}{pred_variance}{10.88}
\setres{domZ06}{CPT_lit}{r_ff}{---}
\setres{domZ06}{CPT_lit}{r_ff_se}{---}
\setres{domZ06}{CPT_lit}{r_prior}{---}
\setres{domZ06}{CPT_lit}{r_prior_se}{---}
\setres{domZ06}{CPT_lit}{r_prob}{---}
\setres{domZ06}{CPT_lit}{r_prob_se}{---}
\setres{domZ06}{CPT_lit}{spread_penalty}{0.048}
\setres{domZ06}{CPT_lit}{spread_penalty_se}{0.002}
\setres{domZ06}{CPT_lit}{volume_ess}{---}
\setres{domZ06}{CPT_transfer_zurich03}{kappa}{0.974}
\setres{domZ06}{CPT_transfer_zurich03}{kappa_prior_mean}{0.974}
\setres{domZ06}{CPT_transfer_zurich03}{kappa_prior_mean_se}{0.007}
\setres{domZ06}{CPT_transfer_zurich03}{kappa_se}{0.007}
\setres{domZ06}{CPT_transfer_zurich03}{mse}{42.17}
\setres{domZ06}{CPT_transfer_zurich03}{mse_prior_mean}{42.17}
\setres{domZ06}{CPT_transfer_zurich03}{nrmse_pct}{13.2}
\setres{domZ06}{CPT_transfer_zurich03}{pred_variance}{0.00}
\setres{domZ06}{CPT_transfer_zurich03}{r_ff}{---}
\setres{domZ06}{CPT_transfer_zurich03}{r_ff_se}{---}
\setres{domZ06}{CPT_transfer_zurich03}{r_prior}{---}
\setres{domZ06}{CPT_transfer_zurich03}{r_prior_se}{---}
\setres{domZ06}{CPT_transfer_zurich03}{r_prob}{---}
\setres{domZ06}{CPT_transfer_zurich03}{r_prob_se}{---}
\setres{domZ06}{CPT_transfer_zurich03}{spread_penalty}{0.000}
\setres{domZ06}{CPT_transfer_zurich03}{spread_penalty_se}{0.000}
\setres{domZ06}{CPT_transfer_zurich03}{theta_star}{0.80, 0.40, 1.00}
\setres{domZ06}{CPT_transfer_zurich03}{transfer}{zurich03}
\setres{domZ06}{CPT_transfer_zurich03}{volume_ess}{---}
\setres{domZ06}{CPT_uniform}{kappa}{0.620}
\setres{domZ06}{CPT_uniform}{kappa_prior_mean}{0.967}
\setres{domZ06}{CPT_uniform}{kappa_prior_mean_se}{0.005}
\setres{domZ06}{CPT_uniform}{kappa_se}{0.017}
\setres{domZ06}{CPT_uniform}{mse}{122.52}
\setres{domZ06}{CPT_uniform}{mse_prior_mean}{43.62}
\setres{domZ06}{CPT_uniform}{nrmse_pct}{14.4}
\setres{domZ06}{CPT_uniform}{pred_variance}{78.90}
\setres{domZ06}{CPT_uniform}{r_ff}{---}
\setres{domZ06}{CPT_uniform}{r_ff_se}{---}
\setres{domZ06}{CPT_uniform}{r_prior}{---}
\setres{domZ06}{CPT_uniform}{r_prior_se}{---}
\setres{domZ06}{CPT_uniform}{r_prob}{---}
\setres{domZ06}{CPT_uniform}{r_prob_se}{---}
\setres{domZ06}{CPT_uniform}{spread_penalty}{0.347}
\setres{domZ06}{CPT_uniform}{spread_penalty_se}{0.016}
\setres{domZ06}{CPT_uniform}{volume_ess}{---}
\setres{domZ06}{DA}{det_pct}{92}
\setres{domZ06}{DA}{gap_cpt_lit}{0.111}
\setres{domZ06}{DA}{gap_lit}{0.126}
\setres{domZ06}{DA}{gap_uniform}{0.209}
\setres{domZ06}{DA}{share_lit}{86}
\setres{domZ06}{DA}{share_uniform}{77}
\setres{domZ06}{DA_deterministic}{kappa}{0.918}
\setres{domZ06}{DA_deterministic}{kappa_se}{0.013}
\setres{domZ06}{DA_deterministic}{mse}{54.75}
\setres{domZ06}{DA_deterministic}{nrmse_pct}{16.4}
\setres{domZ06}{DA_deterministic}{r_ff}{---}
\setres{domZ06}{DA_deterministic}{r_ff_se}{---}
\setres{domZ06}{DA_deterministic}{r_prior}{---}
\setres{domZ06}{DA_deterministic}{r_prior_se}{---}
\setres{domZ06}{DA_deterministic}{r_prob}{---}
\setres{domZ06}{DA_deterministic}{r_prob_se}{---}
\setres{domZ06}{DA_deterministic}{theta_star}{1.68, 1.20}
\setres{domZ06}{DA_deterministic}{volume_ess}{---}
\setres{domZ06}{DA_lit}{kappa}{0.792}
\setres{domZ06}{DA_lit}{kappa_prior_mean}{0.815}
\setres{domZ06}{DA_lit}{kappa_prior_mean_se}{0.028}
\setres{domZ06}{DA_lit}{kappa_se}{0.029}
\setres{domZ06}{DA_lit}{mse}{83.36}
\setres{domZ06}{DA_lit}{mse_prior_mean}{78.16}
\setres{domZ06}{DA_lit}{nrmse_pct}{19.2}
\setres{domZ06}{DA_lit}{pred_variance}{5.20}
\setres{domZ06}{DA_lit}{r_ff}{---}
\setres{domZ06}{DA_lit}{r_ff_se}{---}
\setres{domZ06}{DA_lit}{r_prior}{---}
\setres{domZ06}{DA_lit}{r_prior_se}{---}
\setres{domZ06}{DA_lit}{r_prob}{---}
\setres{domZ06}{DA_lit}{r_prob_se}{---}
\setres{domZ06}{DA_lit}{spread_penalty}{0.023}
\setres{domZ06}{DA_lit}{spread_penalty_se}{0.001}
\setres{domZ06}{DA_lit}{volume_ess}{---}
\setres{domZ06}{DA_transfer_zurich03}{kappa}{0.896}
\setres{domZ06}{DA_transfer_zurich03}{kappa_prior_mean}{0.896}
\setres{domZ06}{DA_transfer_zurich03}{kappa_prior_mean_se}{0.018}
\setres{domZ06}{DA_transfer_zurich03}{kappa_se}{0.018}
\setres{domZ06}{DA_transfer_zurich03}{mse}{59.88}
\setres{domZ06}{DA_transfer_zurich03}{mse_prior_mean}{59.88}
\setres{domZ06}{DA_transfer_zurich03}{nrmse_pct}{17.5}
\setres{domZ06}{DA_transfer_zurich03}{pred_variance}{0.00}
\setres{domZ06}{DA_transfer_zurich03}{r_ff}{---}
\setres{domZ06}{DA_transfer_zurich03}{r_ff_se}{---}
\setres{domZ06}{DA_transfer_zurich03}{r_prior}{---}
\setres{domZ06}{DA_transfer_zurich03}{r_prior_se}{---}
\setres{domZ06}{DA_transfer_zurich03}{r_prob}{---}
\setres{domZ06}{DA_transfer_zurich03}{r_prob_se}{---}
\setres{domZ06}{DA_transfer_zurich03}{spread_penalty}{0.000}
\setres{domZ06}{DA_transfer_zurich03}{spread_penalty_se}{0.000}
\setres{domZ06}{DA_transfer_zurich03}{theta_star}{1.36, 1.35}
\setres{domZ06}{DA_transfer_zurich03}{transfer}{zurich03}
\setres{domZ06}{DA_transfer_zurich03}{volume_ess}{---}
\setres{domZ06}{DA_uniform}{kappa}{0.710}
\setres{domZ06}{DA_uniform}{kappa_prior_mean}{0.838}
\setres{domZ06}{DA_uniform}{kappa_prior_mean_se}{0.026}
\setres{domZ06}{DA_uniform}{kappa_se}{0.032}
\setres{domZ06}{DA_uniform}{mse}{102.18}
\setres{domZ06}{DA_uniform}{mse_prior_mean}{73.10}
\setres{domZ06}{DA_uniform}{nrmse_pct}{18.8}
\setres{domZ06}{DA_uniform}{pred_variance}{29.08}
\setres{domZ06}{DA_uniform}{r_ff}{---}
\setres{domZ06}{DA_uniform}{r_ff_se}{---}
\setres{domZ06}{DA_uniform}{r_prior}{---}
\setres{domZ06}{DA_uniform}{r_prior_se}{---}
\setres{domZ06}{DA_uniform}{r_prob}{---}
\setres{domZ06}{DA_uniform}{r_prob_se}{---}
\setres{domZ06}{DA_uniform}{spread_penalty}{0.128}
\setres{domZ06}{DA_uniform}{spread_penalty_se}{0.006}
\setres{domZ06}{DA_uniform}{volume_ess}{---}
\setres{domZ06}{EU}{det_pct}{91}
\setres{domZ06}{EU}{gap_lit}{0.076}
\setres{domZ06}{EU}{gap_uniform}{0.115}
\setres{domZ06}{EU}{share_lit}{92}
\setres{domZ06}{EU}{share_uniform}{87}
\setres{domZ06}{EU_deterministic}{kappa}{0.908}
\setres{domZ06}{EU_deterministic}{kappa_se}{0.014}
\setres{domZ06}{EU_deterministic}{mse}{57.11}
\setres{domZ06}{EU_deterministic}{nrmse_pct}{16.8}
\setres{domZ06}{EU_deterministic}{r_ff}{---}
\setres{domZ06}{EU_deterministic}{r_ff_se}{---}
\setres{domZ06}{EU_deterministic}{r_prior}{---}
\setres{domZ06}{EU_deterministic}{r_prior_se}{---}
\setres{domZ06}{EU_deterministic}{r_prob}{---}
\setres{domZ06}{EU_deterministic}{r_prob_se}{---}
\setres{domZ06}{EU_deterministic}{theta_star}{1.22}
\setres{domZ06}{EU_deterministic}{volume_ess}{---}
\setres{domZ06}{EU_lit}{kappa}{0.832}
\setres{domZ06}{EU_lit}{kappa_prior_mean}{0.837}
\setres{domZ06}{EU_lit}{kappa_prior_mean_se}{0.026}
\setres{domZ06}{EU_lit}{kappa_se}{0.026}
\setres{domZ06}{EU_lit}{mse}{74.27}
\setres{domZ06}{EU_lit}{mse_prior_mean}{73.22}
\setres{domZ06}{EU_lit}{nrmse_pct}{18.7}
\setres{domZ06}{EU_lit}{pred_variance}{1.04}
\setres{domZ06}{EU_lit}{r_ff}{---}
\setres{domZ06}{EU_lit}{r_ff_se}{---}
\setres{domZ06}{EU_lit}{r_prior}{---}
\setres{domZ06}{EU_lit}{r_prior_se}{---}
\setres{domZ06}{EU_lit}{r_prob}{---}
\setres{domZ06}{EU_lit}{r_prob_se}{---}
\setres{domZ06}{EU_lit}{spread_penalty}{0.005}
\setres{domZ06}{EU_lit}{spread_penalty_se}{0.000}
\setres{domZ06}{EU_lit}{volume_ess}{---}
\setres{domZ06}{EU_transfer_zurich03}{kappa}{0.847}
\setres{domZ06}{EU_transfer_zurich03}{kappa_prior_mean}{0.847}
\setres{domZ06}{EU_transfer_zurich03}{kappa_prior_mean_se}{0.024}
\setres{domZ06}{EU_transfer_zurich03}{kappa_se}{0.024}
\setres{domZ06}{EU_transfer_zurich03}{mse}{70.87}
\setres{domZ06}{EU_transfer_zurich03}{mse_prior_mean}{70.87}
\setres{domZ06}{EU_transfer_zurich03}{nrmse_pct}{17.8}
\setres{domZ06}{EU_transfer_zurich03}{pred_variance}{0.00}
\setres{domZ06}{EU_transfer_zurich03}{r_ff}{---}
\setres{domZ06}{EU_transfer_zurich03}{r_ff_se}{---}
\setres{domZ06}{EU_transfer_zurich03}{r_prior}{---}
\setres{domZ06}{EU_transfer_zurich03}{r_prior_se}{---}
\setres{domZ06}{EU_transfer_zurich03}{r_prob}{---}
\setres{domZ06}{EU_transfer_zurich03}{r_prob_se}{---}
\setres{domZ06}{EU_transfer_zurich03}{spread_penalty}{0.000}
\setres{domZ06}{EU_transfer_zurich03}{spread_penalty_se}{0.000}
\setres{domZ06}{EU_transfer_zurich03}{theta_star}{0.76}
\setres{domZ06}{EU_transfer_zurich03}{transfer}{zurich03}
\setres{domZ06}{EU_transfer_zurich03}{volume_ess}{---}
\setres{domZ06}{EU_uniform}{kappa}{0.793}
\setres{domZ06}{EU_uniform}{kappa_prior_mean}{0.823}
\setres{domZ06}{EU_uniform}{kappa_prior_mean_se}{0.027}
\setres{domZ06}{EU_uniform}{kappa_se}{0.029}
\setres{domZ06}{EU_uniform}{mse}{83.15}
\setres{domZ06}{EU_uniform}{mse_prior_mean}{76.48}
\setres{domZ06}{EU_uniform}{nrmse_pct}{19.5}
\setres{domZ06}{EU_uniform}{pred_variance}{6.67}
\setres{domZ06}{EU_uniform}{r_ff}{---}
\setres{domZ06}{EU_uniform}{r_ff_se}{---}
\setres{domZ06}{EU_uniform}{r_prior}{---}
\setres{domZ06}{EU_uniform}{r_prior_se}{---}
\setres{domZ06}{EU_uniform}{r_prob}{---}
\setres{domZ06}{EU_uniform}{r_prob_se}{---}
\setres{domZ06}{EU_uniform}{spread_penalty}{0.029}
\setres{domZ06}{EU_uniform}{spread_penalty_se}{0.001}
\setres{domZ06}{EU_uniform}{volume_ess}{---}
\setres{domZ06}{EV_deterministic}{kappa}{0.894}
\setres{domZ06}{EV_deterministic}{kappa_se}{0.018}
\setres{domZ06}{EV_deterministic}{mse}{60.32}
\setres{domZ06}{EV_deterministic}{nrmse_pct}{16.6}
\setres{domZ06}{EV_deterministic}{r_ff}{---}
\setres{domZ06}{EV_deterministic}{r_ff_se}{---}
\setres{domZ06}{EV_deterministic}{r_prior}{---}
\setres{domZ06}{EV_deterministic}{r_prior_se}{---}
\setres{domZ06}{EV_deterministic}{r_prob}{---}
\setres{domZ06}{EV_deterministic}{r_prob_se}{---}
\setres{domZ06}{EV_deterministic}{volume_ess}{---}
\setres{domZ06}{reference}{lookup_nrmse_pct}{12.7}
\setres{domZ06}{reference}{midpoint_mse}{263.32}
\setres{domZ06}{reference}{midpoint_nrmse_pct}{25.0}
\setres{domZ06}{reference}{n_lotteries}{20}
\setres{domZ06}{reference}{n_obs}{2{,}337}
\setres{domZ06}{reference}{n_subjects}{118}
\setres{domZ06}{reference}{oracle_mse}{36.22}
\setres{domZ06}{reference}{sample}{Bruhin2010\_gains}
\setres{rLV}{CPT}{one_minus_rff}{0.801}
\setres{rLV}{CPT}{passthrough_lit}{0.080}
\setres{rLV}{CPT}{passthrough_uniform}{0.485}
\setres{rLV}{CPT}{prior_contrib_lit}{0.737}
\setres{rLV}{CPT}{prior_contrib_uniform}{0.412}
\setres{rLV}{CPT_lit}{r_ff}{0.199}
\setres{rLV}{CPT_lit}{r_ff_se}{0.002}
\setres{rLV}{CPT_lit}{r_prior}{0.9203}
\setres{rLV}{CPT_lit}{r_prior_se}{0.001}
\setres{rLV}{CPT_lit}{r_prob}{0.936}
\setres{rLV}{CPT_lit}{r_prob_se}{0.000}
\setres{rLV}{CPT_lit}{volume_ess}{62{,}726}
\setres{rLV}{CPT_uniform}{r_ff}{0.199}
\setres{rLV}{CPT_uniform}{r_ff_se}{0.002}
\setres{rLV}{CPT_uniform}{r_prior}{0.5146}
\setres{rLV}{CPT_uniform}{r_prior_se}{0.003}
\setres{rLV}{CPT_uniform}{r_prob}{0.611}
\setres{rLV}{CPT_uniform}{r_prob_se}{0.003}
\setres{rLV}{CPT_uniform}{volume_ess}{62{,}726}
\setres{rLV}{DA}{gap_r_cpt_lit}{-0.015}
\setres{rLV}{DA}{one_minus_rff}{0.686}
\setres{rLV}{DA}{passthrough_lit}{0.115}
\setres{rLV}{DA}{passthrough_uniform}{0.216}
\setres{rLV}{DA}{prior_contrib_lit}{0.607}
\setres{rLV}{DA}{prior_contrib_uniform}{0.538}
\setres{rLV}{DA_lit}{r_ff}{0.314}
\setres{rLV}{DA_lit}{r_ff_se}{0.003}
\setres{rLV}{DA_lit}{r_prior}{0.8846}
\setres{rLV}{DA_lit}{r_prior_se}{0.001}
\setres{rLV}{DA_lit}{r_prob}{0.921}
\setres{rLV}{DA_lit}{r_prob_se}{0.001}
\setres{rLV}{DA_lit}{volume_ess}{410{,}881}
\setres{rLV}{DA_uniform}{r_ff}{0.314}
\setres{rLV}{DA_uniform}{r_ff_se}{0.003}
\setres{rLV}{DA_uniform}{r_prior}{0.7841}
\setres{rLV}{DA_uniform}{r_prior_se}{0.001}
\setres{rLV}{DA_uniform}{r_prob}{0.852}
\setres{rLV}{DA_uniform}{r_prob_se}{0.001}
\setres{rLV}{DA_uniform}{volume_ess}{410{,}881}
\setres{rLV}{EU}{one_minus_rff}{0.633}
\setres{rLV}{EU}{passthrough_lit}{0.205}
\setres{rLV}{EU}{passthrough_uniform}{0.475}
\setres{rLV}{EU}{prior_contrib_lit}{0.503}
\setres{rLV}{EU}{prior_contrib_uniform}{0.332}
\setres{rLV}{EU_lit}{r_ff}{0.367}
\setres{rLV}{EU_lit}{r_ff_se}{0.004}
\setres{rLV}{EU_lit}{r_prior}{0.7947}
\setres{rLV}{EU_lit}{r_prior_se}{0.001}
\setres{rLV}{EU_lit}{r_prob}{0.870}
\setres{rLV}{EU_lit}{r_prob_se}{0.001}
\setres{rLV}{EU_lit}{volume_ess}{132{,}427}
\setres{rLV}{EU_uniform}{r_ff}{0.367}
\setres{rLV}{EU_uniform}{r_ff_se}{0.004}
\setres{rLV}{EU_uniform}{r_prior}{0.5250}
\setres{rLV}{EU_uniform}{r_prior_se}{0.002}
\setres{rLV}{EU_uniform}{r_prob}{0.699}
\setres{rLV}{EU_uniform}{r_prob_se}{0.002}
\setres{rLV}{EU_uniform}{volume_ess}{132{,}427}
\setres{rLV}{EV_deterministic}{r_ff}{0.400}
\setres{rLV}{EV_deterministic}{r_ff_se}{---}
\setres{rLV}{EV_deterministic}{r_prior}{1.0000}
\setres{rLV}{EV_deterministic}{r_prior_se}{---}
\setres{rLV}{EV_deterministic}{r_prob}{1.000}
\setres{rLV}{EV_deterministic}{r_prob_se}{0.000}
\setres{rLV}{EV_deterministic}{volume_ess}{---}
\setres{rLV}{reference}{design}{LV2019\_pooled}
\setres{rLV}{reference}{n_lotteries}{14}
\setres{rLV}{reference}{source}{l'Haridon and Vieider (2019), 30 countries}
\setres{rS15}{CPT}{one_minus_rff}{0.847}
\setres{rS15}{CPT}{passthrough_lit}{0.034}
\setres{rS15}{CPT}{passthrough_uniform}{0.332}
\setres{rS15}{CPT}{prior_contrib_lit}{0.818}
\setres{rS15}{CPT}{prior_contrib_uniform}{0.566}
\setres{rS15}{CPT_lit}{r_ff}{0.153}
\setres{rS15}{CPT_lit}{r_ff_se}{0.002}
\setres{rS15}{CPT_lit}{r_prior}{0.9657}
\setres{rS15}{CPT_lit}{r_prior_se}{0.000}
\setres{rS15}{CPT_lit}{r_prob}{0.971}
\setres{rS15}{CPT_lit}{r_prob_se}{0.000}
\setres{rS15}{CPT_lit}{volume_ess}{87{,}437}
\setres{rS15}{CPT_uniform}{r_ff}{0.153}
\setres{rS15}{CPT_uniform}{r_ff_se}{0.002}
\setres{rS15}{CPT_uniform}{r_prior}{0.6677}
\setres{rS15}{CPT_uniform}{r_prior_se}{0.002}
\setres{rS15}{CPT_uniform}{r_prob}{0.719}
\setres{rS15}{CPT_uniform}{r_prob_se}{0.002}
\setres{rS15}{CPT_uniform}{volume_ess}{87{,}437}
\setres{rS15}{DA}{gap_r_cpt_lit}{0.000}
\setres{rS15}{DA}{one_minus_rff}{0.718}
\setres{rS15}{DA}{passthrough_lit}{0.040}
\setres{rS15}{DA}{passthrough_uniform}{0.068}
\setres{rS15}{DA}{prior_contrib_lit}{0.690}
\setres{rS15}{DA}{prior_contrib_uniform}{0.669}
\setres{rS15}{DA_lit}{r_ff}{0.282}
\setres{rS15}{DA_lit}{r_ff_se}{0.003}
\setres{rS15}{DA_lit}{r_prior}{0.9602}
\setres{rS15}{DA_lit}{r_prior_se}{0.000}
\setres{rS15}{DA_lit}{r_prob}{0.971}
\setres{rS15}{DA_lit}{r_prob_se}{0.000}
\setres{rS15}{DA_lit}{volume_ess}{351{,}503}
\setres{rS15}{DA_uniform}{r_ff}{0.282}
\setres{rS15}{DA_uniform}{r_ff_se}{0.003}
\setres{rS15}{DA_uniform}{r_prior}{0.9321}
\setres{rS15}{DA_uniform}{r_prior_se}{0.000}
\setres{rS15}{DA_uniform}{r_prob}{0.951}
\setres{rS15}{DA_uniform}{r_prob_se}{0.000}
\setres{rS15}{DA_uniform}{volume_ess}{351{,}503}
\setres{rS15}{EU}{one_minus_rff}{0.687}
\setres{rS15}{EU}{passthrough_lit}{0.058}
\setres{rS15}{EU}{passthrough_uniform}{0.104}
\setres{rS15}{EU}{prior_contrib_lit}{0.647}
\setres{rS15}{EU}{prior_contrib_uniform}{0.615}
\setres{rS15}{EU_lit}{r_ff}{0.313}
\setres{rS15}{EU_lit}{r_ff_se}{0.003}
\setres{rS15}{EU_lit}{r_prior}{0.9422}
\setres{rS15}{EU_lit}{r_prior_se}{0.000}
\setres{rS15}{EU_lit}{r_prob}{0.960}
\setres{rS15}{EU_lit}{r_prob_se}{0.000}
\setres{rS15}{EU_lit}{volume_ess}{671{,}385}
\setres{rS15}{EU_uniform}{r_ff}{0.313}
\setres{rS15}{EU_uniform}{r_ff_se}{0.003}
\setres{rS15}{EU_uniform}{r_prior}{0.8957}
\setres{rS15}{EU_uniform}{r_prior_se}{0.000}
\setres{rS15}{EU_uniform}{r_prob}{0.928}
\setres{rS15}{EU_uniform}{r_prob_se}{0.000}
\setres{rS15}{EU_uniform}{volume_ess}{671{,}385}
\setres{rS15}{EV_deterministic}{r_ff}{0.712}
\setres{rS15}{EV_deterministic}{r_ff_se}{---}
\setres{rS15}{EV_deterministic}{r_prior}{1.0000}
\setres{rS15}{EV_deterministic}{r_prior_se}{---}
\setres{rS15}{EV_deterministic}{r_prob}{1.000}
\setres{rS15}{EV_deterministic}{r_prob_se}{0.000}
\setres{rS15}{EV_deterministic}{volume_ess}{---}
\setres{rS15}{reference}{design}{Zurich2003\_shared}
\setres{rS15}{reference}{n_lotteries}{15}
\setres{rS15}{reference}{source}{Bruhin et al. (2010), Zurich 2003 and 2006, gain lotteries run in both}
\setres{rBJ}{CPT}{one_minus_rff}{0.837}
\setres{rBJ}{CPT}{passthrough_lit}{0.021}
\setres{rBJ}{CPT}{passthrough_uniform}{0.207}
\setres{rBJ}{CPT}{prior_contrib_lit}{0.819}
\setres{rBJ}{CPT}{prior_contrib_uniform}{0.664}
\setres{rBJ}{CPT_lit}{r_ff}{0.163}
\setres{rBJ}{CPT_lit}{r_ff_se}{0.002}
\setres{rBJ}{CPT_lit}{r_prior}{0.9789}
\setres{rBJ}{CPT_lit}{r_prior_se}{0.000}
\setres{rBJ}{CPT_lit}{r_prob}{0.982}
\setres{rBJ}{CPT_lit}{r_prob_se}{0.000}
\setres{rBJ}{CPT_lit}{volume_ess}{102{,}466}
\setres{rBJ}{CPT_uniform}{r_ff}{0.163}
\setres{rBJ}{CPT_uniform}{r_ff_se}{0.002}
\setres{rBJ}{CPT_uniform}{r_prior}{0.7927}
\setres{rBJ}{CPT_uniform}{r_prior_se}{0.001}
\setres{rBJ}{CPT_uniform}{r_prob}{0.827}
\setres{rBJ}{CPT_uniform}{r_prob_se}{0.001}
\setres{rBJ}{CPT_uniform}{volume_ess}{102{,}466}
\setres{rBJ}{DA}{gap_r_cpt_lit}{0.002}
\setres{rBJ}{DA}{one_minus_rff}{0.685}
\setres{rBJ}{DA}{passthrough_lit}{0.022}
\setres{rBJ}{DA}{passthrough_uniform}{0.049}
\setres{rBJ}{DA}{prior_contrib_lit}{0.670}
\setres{rBJ}{DA}{prior_contrib_uniform}{0.651}
\setres{rBJ}{DA_lit}{r_ff}{0.315}
\setres{rBJ}{DA_lit}{r_ff_se}{0.004}
\setres{rBJ}{DA_lit}{r_prior}{0.9778}
\setres{rBJ}{DA_lit}{r_prior_se}{0.000}
\setres{rBJ}{DA_lit}{r_prob}{0.985}
\setres{rBJ}{DA_lit}{r_prob_se}{0.000}
\setres{rBJ}{DA_lit}{volume_ess}{171{,}764}
\setres{rBJ}{DA_uniform}{r_ff}{0.315}
\setres{rBJ}{DA_uniform}{r_ff_se}{0.004}
\setres{rBJ}{DA_uniform}{r_prior}{0.9508}
\setres{rBJ}{DA_uniform}{r_prior_se}{0.000}
\setres{rBJ}{DA_uniform}{r_prob}{0.966}
\setres{rBJ}{DA_uniform}{r_prob_se}{0.000}
\setres{rBJ}{DA_uniform}{volume_ess}{171{,}764}
\setres{rBJ}{EU}{one_minus_rff}{0.648}
\setres{rBJ}{EU}{passthrough_lit}{0.055}
\setres{rBJ}{EU}{passthrough_uniform}{0.099}
\setres{rBJ}{EU}{prior_contrib_lit}{0.613}
\setres{rBJ}{EU}{prior_contrib_uniform}{0.584}
\setres{rBJ}{EU_lit}{r_ff}{0.352}
\setres{rBJ}{EU_lit}{r_ff_se}{0.004}
\setres{rBJ}{EU_lit}{r_prior}{0.9453}
\setres{rBJ}{EU_lit}{r_prior_se}{0.000}
\setres{rBJ}{EU_lit}{r_prob}{0.965}
\setres{rBJ}{EU_lit}{r_prob_se}{0.000}
\setres{rBJ}{EU_lit}{volume_ess}{683{,}357}
\setres{rBJ}{EU_uniform}{r_ff}{0.352}
\setres{rBJ}{EU_uniform}{r_ff_se}{0.004}
\setres{rBJ}{EU_uniform}{r_prior}{0.9011}
\setres{rBJ}{EU_uniform}{r_prior_se}{0.000}
\setres{rBJ}{EU_uniform}{r_prob}{0.936}
\setres{rBJ}{EU_uniform}{r_prob_se}{0.000}
\setres{rBJ}{EU_uniform}{volume_ess}{683{,}357}
\setres{rBJ}{EV_deterministic}{r_ff}{0.688}
\setres{rBJ}{EV_deterministic}{r_ff_se}{---}
\setres{rBJ}{EV_deterministic}{r_prior}{1.0000}
\setres{rBJ}{EV_deterministic}{r_prior_se}{---}
\setres{rBJ}{EV_deterministic}{r_prob}{1.000}
\setres{rBJ}{EV_deterministic}{r_prob_se}{0.000}
\setres{rBJ}{EV_deterministic}{volume_ess}{---}
\setres{rBJ}{reference}{design}{Beijing2005\_gains}
\setres{rBJ}{reference}{n_lotteries}{14}
\setres{rBJ}{reference}{source}{Bruhin et al. (2010), Beijing 2005}
\setres{learn}{CPT_post_m0}{ess}{732}
\setres{learn}{CPT_post_m0}{kappa_mean}{-0.091}
\setres{learn}{CPT_post_m0}{kappa_mean_se}{0.211}
\setres{learn}{CPT_post_m0}{kappa_median}{0.465}
\setres{learn}{CPT_post_m0}{kappa_prior_mean_median}{0.768}
\setres{learn}{CPT_post_m0}{m}{0}
\setres{learn}{CPT_post_m0}{n_countries}{30}
\setres{learn}{CPT_post_m0}{n_pos}{23}
\setres{learn}{CPT_post_m0}{n_subjects_seen}{0}
\setres{learn}{CPT_post_m0}{r_ff}{0.199}
\setres{learn}{CPT_post_m0}{r_prior}{0.9191}
\setres{learn}{CPT_post_m0}{r_prob}{0.935}
\setres{learn}{CPT_post_m0}{r_prob_se}{0.000}
\setres{learn}{CPT_post_m0}{theta_mean}{0.70, 0.73, 1.08}
\setres{learn}{CPT_post_m0}{theta_sd}{0.14, 0.18, 0.28}
\setres{learn}{CPT_post_m10}{ess}{55}
\setres{learn}{CPT_post_m10}{kappa_mean}{0.512}
\setres{learn}{CPT_post_m10}{kappa_mean_se}{0.120}
\setres{learn}{CPT_post_m10}{kappa_median}{0.806}
\setres{learn}{CPT_post_m10}{kappa_prior_mean_median}{0.837}
\setres{learn}{CPT_post_m10}{m}{10}
\setres{learn}{CPT_post_m10}{n_countries}{30}
\setres{learn}{CPT_post_m10}{n_pos}{29}
\setres{learn}{CPT_post_m10}{n_subjects_seen}{10}
\setres{learn}{CPT_post_m10}{r_ff}{0.199}
\setres{learn}{CPT_post_m10}{r_prior}{0.9941}
\setres{learn}{CPT_post_m10}{r_prob}{0.995}
\setres{learn}{CPT_post_m10}{r_prob_se}{0.000}
\setres{learn}{CPT_post_m10}{theta_mean}{0.84, 0.68, 0.99}
\setres{learn}{CPT_post_m10}{theta_sd}{0.07, 0.09, 0.12}
\setres{learn}{CPT_post_m100}{ess}{3}
\setres{learn}{CPT_post_m100}{kappa_mean}{0.706}
\setres{learn}{CPT_post_m100}{kappa_mean_se}{0.106}
\setres{learn}{CPT_post_m100}{kappa_median}{0.924}
\setres{learn}{CPT_post_m100}{kappa_prior_mean_median}{0.932}
\setres{learn}{CPT_post_m100}{m}{100}
\setres{learn}{CPT_post_m100}{n_countries}{30}
\setres{learn}{CPT_post_m100}{n_pos}{29}
\setres{learn}{CPT_post_m100}{n_subjects_seen}{100}
\setres{learn}{CPT_post_m100}{r_ff}{0.199}
\setres{learn}{CPT_post_m100}{r_prior}{0.9997}
\setres{learn}{CPT_post_m100}{r_prob}{1.000}
\setres{learn}{CPT_post_m100}{r_prob_se}{0.000}
\setres{learn}{CPT_post_m100}{theta_mean}{0.95, 0.66, 0.93}
\setres{learn}{CPT_post_m100}{theta_sd}{0.04, 0.03, 0.06}
\setres{learn}{CPT_post_m1000}{ess}{1}
\setres{learn}{CPT_post_m1000}{kappa_mean}{0.753}
\setres{learn}{CPT_post_m1000}{kappa_mean_se}{0.090}
\setres{learn}{CPT_post_m1000}{kappa_median}{0.945}
\setres{learn}{CPT_post_m1000}{kappa_prior_mean_median}{0.951}
\setres{learn}{CPT_post_m1000}{m}{1000}
\setres{learn}{CPT_post_m1000}{n_countries}{30}
\setres{learn}{CPT_post_m1000}{n_pos}{29}
\setres{learn}{CPT_post_m1000}{n_subjects_seen}{1{,}000}
\setres{learn}{CPT_post_m1000}{r_ff}{0.199}
\setres{learn}{CPT_post_m1000}{r_prior}{0.9999}
\setres{learn}{CPT_post_m1000}{r_prob}{1.000}
\setres{learn}{CPT_post_m1000}{r_prob_se}{0.000}
\setres{learn}{CPT_post_m1000}{theta_mean}{0.98, 0.58, 0.90}
\setres{learn}{CPT_post_m1000}{theta_sd}{0.00, 0.00, 0.00}
\setres{learn}{CPT_post_m3}{ess}{191}
\setres{learn}{CPT_post_m3}{kappa_mean}{0.104}
\setres{learn}{CPT_post_m3}{kappa_mean_se}{0.213}
\setres{learn}{CPT_post_m3}{kappa_median}{0.659}
\setres{learn}{CPT_post_m3}{kappa_prior_mean_median}{0.734}
\setres{learn}{CPT_post_m3}{m}{3}
\setres{learn}{CPT_post_m3}{n_countries}{30}
\setres{learn}{CPT_post_m3}{n_pos}{27}
\setres{learn}{CPT_post_m3}{n_subjects_seen}{3}
\setres{learn}{CPT_post_m3}{r_ff}{0.199}
\setres{learn}{CPT_post_m3}{r_prior}{0.9766}
\setres{learn}{CPT_post_m3}{r_prob}{0.981}
\setres{learn}{CPT_post_m3}{r_prob_se}{0.000}
\setres{learn}{CPT_post_m3}{theta_mean}{0.79, 0.56, 1.07}
\setres{learn}{CPT_post_m3}{theta_sd}{0.09, 0.14, 0.18}
\setres{learn}{CPT_post_m30}{ess}{12}
\setres{learn}{CPT_post_m30}{kappa_mean}{0.627}
\setres{learn}{CPT_post_m30}{kappa_mean_se}{0.113}
\setres{learn}{CPT_post_m30}{kappa_median}{0.868}
\setres{learn}{CPT_post_m30}{kappa_prior_mean_median}{0.884}
\setres{learn}{CPT_post_m30}{m}{30}
\setres{learn}{CPT_post_m30}{n_countries}{30}
\setres{learn}{CPT_post_m30}{n_pos}{29}
\setres{learn}{CPT_post_m30}{n_subjects_seen}{30}
\setres{learn}{CPT_post_m30}{r_ff}{0.199}
\setres{learn}{CPT_post_m30}{r_prior}{0.9987}
\setres{learn}{CPT_post_m30}{r_prob}{0.999}
\setres{learn}{CPT_post_m30}{r_prob_se}{0.000}
\setres{learn}{CPT_post_m30}{theta_mean}{0.89, 0.66, 0.91}
\setres{learn}{CPT_post_m30}{theta_sd}{0.04, 0.05, 0.07}
\setres{learn}{CPT_post_m300}{ess}{2}
\setres{learn}{CPT_post_m300}{kappa_mean}{0.719}
\setres{learn}{CPT_post_m300}{kappa_mean_se}{0.102}
\setres{learn}{CPT_post_m300}{kappa_median}{0.935}
\setres{learn}{CPT_post_m300}{kappa_prior_mean_median}{0.942}
\setres{learn}{CPT_post_m300}{m}{300}
\setres{learn}{CPT_post_m300}{n_countries}{30}
\setres{learn}{CPT_post_m300}{n_pos}{29}
\setres{learn}{CPT_post_m300}{n_subjects_seen}{300}
\setres{learn}{CPT_post_m300}{r_ff}{0.199}
\setres{learn}{CPT_post_m300}{r_prior}{0.9999}
\setres{learn}{CPT_post_m300}{r_prob}{1.000}
\setres{learn}{CPT_post_m300}{r_prob_se}{0.000}
\setres{learn}{CPT_post_m300}{theta_mean}{0.97, 0.59, 0.91}
\setres{learn}{CPT_post_m300}{theta_sd}{0.03, 0.02, 0.04}
\setres{learn}{CPT_post_mall}{ess}{1}
\setres{learn}{CPT_post_mall}{kappa_mean}{0.758}
\setres{learn}{CPT_post_mall}{kappa_mean_se}{0.090}
\setres{learn}{CPT_post_mall}{kappa_median}{0.944}
\setres{learn}{CPT_post_mall}{kappa_prior_mean_median}{0.950}
\setres{learn}{CPT_post_mall}{m}{all}
\setres{learn}{CPT_post_mall}{n_countries}{30}
\setres{learn}{CPT_post_mall}{n_pos}{29}
\setres{learn}{CPT_post_mall}{n_subjects_seen}{2{,}939}
\setres{learn}{CPT_post_mall}{r_ff}{0.199}
\setres{learn}{CPT_post_mall}{r_prior}{0.9999}
\setres{learn}{CPT_post_mall}{r_prob}{1.000}
\setres{learn}{CPT_post_mall}{r_prob_se}{0.000}
\setres{learn}{CPT_post_mall}{theta_mean}{0.98, 0.63, 0.90}
\setres{learn}{CPT_post_mall}{theta_sd}{0.00, 0.00, 0.00}
\setres{learn}{DA_post_m0}{ess}{439}
\setres{learn}{DA_post_m0}{kappa_mean}{-1.698}
\setres{learn}{DA_post_m0}{kappa_mean_se}{0.427}
\setres{learn}{DA_post_m0}{kappa_median}{-0.470}
\setres{learn}{DA_post_m0}{kappa_prior_mean_median}{0.036}
\setres{learn}{DA_post_m0}{m}{0}
\setres{learn}{DA_post_m0}{n_countries}{30}
\setres{learn}{DA_post_m0}{n_pos}{3}
\setres{learn}{DA_post_m0}{n_subjects_seen}{0}
\setres{learn}{DA_post_m0}{r_ff}{0.314}
\setres{learn}{DA_post_m0}{r_prior}{0.8842}
\setres{learn}{DA_post_m0}{r_prob}{0.920}
\setres{learn}{DA_post_m0}{r_prob_se}{0.001}
\setres{learn}{DA_post_m0}{theta_mean}{0.58, 0.30}
\setres{learn}{DA_post_m0}{theta_sd}{0.19, 0.45}
\setres{learn}{DA_post_m10}{ess}{31}
\setres{learn}{DA_post_m10}{kappa_mean}{0.136}
\setres{learn}{DA_post_m10}{kappa_mean_se}{0.253}
\setres{learn}{DA_post_m10}{kappa_median}{0.697}
\setres{learn}{DA_post_m10}{kappa_prior_mean_median}{0.712}
\setres{learn}{DA_post_m10}{m}{10}
\setres{learn}{DA_post_m10}{n_countries}{30}
\setres{learn}{DA_post_m10}{n_pos}{28}
\setres{learn}{DA_post_m10}{n_subjects_seen}{10}
\setres{learn}{DA_post_m10}{r_ff}{0.314}
\setres{learn}{DA_post_m10}{r_prior}{0.9940}
\setres{learn}{DA_post_m10}{r_prob}{0.996}
\setres{learn}{DA_post_m10}{r_prob_se}{0.000}
\setres{learn}{DA_post_m10}{theta_mean}{0.89, 0.17}
\setres{learn}{DA_post_m10}{theta_sd}{0.06, 0.14}
\setres{learn}{DA_post_m100}{ess}{2}
\setres{learn}{DA_post_m100}{kappa_mean}{0.396}
\setres{learn}{DA_post_m100}{kappa_mean_se}{0.200}
\setres{learn}{DA_post_m100}{kappa_median}{0.859}
\setres{learn}{DA_post_m100}{kappa_prior_mean_median}{0.861}
\setres{learn}{DA_post_m100}{m}{100}
\setres{learn}{DA_post_m100}{n_countries}{30}
\setres{learn}{DA_post_m100}{n_pos}{29}
\setres{learn}{DA_post_m100}{n_subjects_seen}{100}
\setres{learn}{DA_post_m100}{r_ff}{0.314}
\setres{learn}{DA_post_m100}{r_prior}{0.9997}
\setres{learn}{DA_post_m100}{r_prob}{1.000}
\setres{learn}{DA_post_m100}{r_prob_se}{0.000}
\setres{learn}{DA_post_m100}{theta_mean}{0.98, 0.14}
\setres{learn}{DA_post_m100}{theta_sd}{0.01, 0.03}
\setres{learn}{DA_post_m1000}{ess}{1}
\setres{learn}{DA_post_m1000}{kappa_mean}{0.395}
\setres{learn}{DA_post_m1000}{kappa_mean_se}{0.201}
\setres{learn}{DA_post_m1000}{kappa_median}{0.860}
\setres{learn}{DA_post_m1000}{kappa_prior_mean_median}{0.862}
\setres{learn}{DA_post_m1000}{m}{1000}
\setres{learn}{DA_post_m1000}{n_countries}{30}
\setres{learn}{DA_post_m1000}{n_pos}{29}
\setres{learn}{DA_post_m1000}{n_subjects_seen}{1{,}000}
\setres{learn}{DA_post_m1000}{r_ff}{0.314}
\setres{learn}{DA_post_m1000}{r_prior}{0.9999}
\setres{learn}{DA_post_m1000}{r_prob}{1.000}
\setres{learn}{DA_post_m1000}{r_prob_se}{0.000}
\setres{learn}{DA_post_m1000}{theta_mean}{0.99, 0.15}
\setres{learn}{DA_post_m1000}{theta_sd}{0.00, 0.00}
\setres{learn}{DA_post_m3}{ess}{80}
\setres{learn}{DA_post_m3}{kappa_mean}{-0.251}
\setres{learn}{DA_post_m3}{kappa_mean_se}{0.318}
\setres{learn}{DA_post_m3}{kappa_median}{0.570}
\setres{learn}{DA_post_m3}{kappa_prior_mean_median}{0.620}
\setres{learn}{DA_post_m3}{m}{3}
\setres{learn}{DA_post_m3}{n_countries}{30}
\setres{learn}{DA_post_m3}{n_pos}{24}
\setres{learn}{DA_post_m3}{n_subjects_seen}{3}
\setres{learn}{DA_post_m3}{r_ff}{0.314}
\setres{learn}{DA_post_m3}{r_prior}{0.9822}
\setres{learn}{DA_post_m3}{r_prob}{0.988}
\setres{learn}{DA_post_m3}{r_prob_se}{0.000}
\setres{learn}{DA_post_m3}{theta_mean}{0.82, 0.13}
\setres{learn}{DA_post_m3}{theta_sd}{0.09, 0.23}
\setres{learn}{DA_post_m30}{ess}{10}
\setres{learn}{DA_post_m30}{kappa_mean}{0.305}
\setres{learn}{DA_post_m30}{kappa_mean_se}{0.220}
\setres{learn}{DA_post_m30}{kappa_median}{0.815}
\setres{learn}{DA_post_m30}{kappa_prior_mean_median}{0.822}
\setres{learn}{DA_post_m30}{m}{30}
\setres{learn}{DA_post_m30}{n_countries}{30}
\setres{learn}{DA_post_m30}{n_pos}{29}
\setres{learn}{DA_post_m30}{n_subjects_seen}{30}
\setres{learn}{DA_post_m30}{r_ff}{0.314}
\setres{learn}{DA_post_m30}{r_prior}{0.9982}
\setres{learn}{DA_post_m30}{r_prob}{0.999}
\setres{learn}{DA_post_m30}{r_prob_se}{0.000}
\setres{learn}{DA_post_m30}{theta_mean}{0.94, 0.25}
\setres{learn}{DA_post_m30}{theta_sd}{0.03, 0.09}
\setres{learn}{DA_post_m300}{ess}{1}
\setres{learn}{DA_post_m300}{kappa_mean}{0.395}
\setres{learn}{DA_post_m300}{kappa_mean_se}{0.201}
\setres{learn}{DA_post_m300}{kappa_median}{0.858}
\setres{learn}{DA_post_m300}{kappa_prior_mean_median}{0.860}
\setres{learn}{DA_post_m300}{m}{300}
\setres{learn}{DA_post_m300}{n_countries}{30}
\setres{learn}{DA_post_m300}{n_pos}{29}
\setres{learn}{DA_post_m300}{n_subjects_seen}{300}
\setres{learn}{DA_post_m300}{r_ff}{0.314}
\setres{learn}{DA_post_m300}{r_prior}{0.9999}
\setres{learn}{DA_post_m300}{r_prob}{1.000}
\setres{learn}{DA_post_m300}{r_prob_se}{0.000}
\setres{learn}{DA_post_m300}{theta_mean}{0.99, 0.15}
\setres{learn}{DA_post_m300}{theta_sd}{0.00, 0.00}
\setres{learn}{DA_post_mall}{ess}{1}
\setres{learn}{DA_post_mall}{kappa_mean}{0.417}
\setres{learn}{DA_post_mall}{kappa_mean_se}{0.196}
\setres{learn}{DA_post_mall}{kappa_median}{0.861}
\setres{learn}{DA_post_mall}{kappa_prior_mean_median}{0.862}
\setres{learn}{DA_post_mall}{m}{all}
\setres{learn}{DA_post_mall}{n_countries}{30}
\setres{learn}{DA_post_mall}{n_pos}{29}
\setres{learn}{DA_post_mall}{n_subjects_seen}{2{,}939}
\setres{learn}{DA_post_mall}{r_ff}{0.314}
\setres{learn}{DA_post_mall}{r_prior}{0.9999}
\setres{learn}{DA_post_mall}{r_prob}{1.000}
\setres{learn}{DA_post_mall}{r_prob_se}{0.000}
\setres{learn}{DA_post_mall}{theta_mean}{0.99, 0.15}
\setres{learn}{DA_post_mall}{theta_sd}{0.00, 0.00}
\setres{learn}{EU_post_m0}{ess}{98}
\setres{learn}{EU_post_m0}{kappa_mean}{-0.178}
\setres{learn}{EU_post_m0}{kappa_mean_se}{0.273}
\setres{learn}{EU_post_m0}{kappa_median}{0.513}
\setres{learn}{EU_post_m0}{kappa_prior_mean_median}{0.632}
\setres{learn}{EU_post_m0}{m}{0}
\setres{learn}{EU_post_m0}{n_countries}{30}
\setres{learn}{EU_post_m0}{n_pos}{24}
\setres{learn}{EU_post_m0}{n_subjects_seen}{0}
\setres{learn}{EU_post_m0}{r_ff}{0.367}
\setres{learn}{EU_post_m0}{r_prior}{0.7947}
\setres{learn}{EU_post_m0}{r_prob}{0.870}
\setres{learn}{EU_post_m0}{r_prob_se}{0.001}
\setres{learn}{EU_post_m0}{theta_mean}{0.70}
\setres{learn}{EU_post_m0}{theta_sd}{0.14}
\setres{learn}{EU_post_m10}{ess}{35}
\setres{learn}{EU_post_m10}{kappa_mean}{0.225}
\setres{learn}{EU_post_m10}{kappa_mean_se}{0.250}
\setres{learn}{EU_post_m10}{kappa_median}{0.814}
\setres{learn}{EU_post_m10}{kappa_prior_mean_median}{0.818}
\setres{learn}{EU_post_m10}{m}{10}
\setres{learn}{EU_post_m10}{n_countries}{30}
\setres{learn}{EU_post_m10}{n_pos}{28}
\setres{learn}{EU_post_m10}{n_subjects_seen}{10}
\setres{learn}{EU_post_m10}{r_ff}{0.367}
\setres{learn}{EU_post_m10}{r_prior}{0.9380}
\setres{learn}{EU_post_m10}{r_prob}{0.961}
\setres{learn}{EU_post_m10}{r_prob_se}{0.000}
\setres{learn}{EU_post_m10}{theta_mean}{0.83}
\setres{learn}{EU_post_m10}{theta_sd}{0.05}
\setres{learn}{EU_post_m100}{ess}{13}
\setres{learn}{EU_post_m100}{kappa_mean}{0.381}
\setres{learn}{EU_post_m100}{kappa_mean_se}{0.204}
\setres{learn}{EU_post_m100}{kappa_median}{0.833}
\setres{learn}{EU_post_m100}{kappa_prior_mean_median}{0.834}
\setres{learn}{EU_post_m100}{m}{100}
\setres{learn}{EU_post_m100}{n_countries}{30}
\setres{learn}{EU_post_m100}{n_pos}{29}
\setres{learn}{EU_post_m100}{n_subjects_seen}{100}
\setres{learn}{EU_post_m100}{r_ff}{0.367}
\setres{learn}{EU_post_m100}{r_prior}{0.9786}
\setres{learn}{EU_post_m100}{r_prob}{0.986}
\setres{learn}{EU_post_m100}{r_prob_se}{0.000}
\setres{learn}{EU_post_m100}{theta_mean}{0.92}
\setres{learn}{EU_post_m100}{theta_sd}{0.02}
\setres{learn}{EU_post_m1000}{ess}{5}
\setres{learn}{EU_post_m1000}{kappa_mean}{0.380}
\setres{learn}{EU_post_m1000}{kappa_mean_se}{0.204}
\setres{learn}{EU_post_m1000}{kappa_median}{0.823}
\setres{learn}{EU_post_m1000}{kappa_prior_mean_median}{0.823}
\setres{learn}{EU_post_m1000}{m}{1000}
\setres{learn}{EU_post_m1000}{n_countries}{30}
\setres{learn}{EU_post_m1000}{n_pos}{29}
\setres{learn}{EU_post_m1000}{n_subjects_seen}{1{,}000}
\setres{learn}{EU_post_m1000}{r_ff}{0.367}
\setres{learn}{EU_post_m1000}{r_prior}{0.9926}
\setres{learn}{EU_post_m1000}{r_prob}{0.995}
\setres{learn}{EU_post_m1000}{r_prob_se}{0.000}
\setres{learn}{EU_post_m1000}{theta_mean}{0.92}
\setres{learn}{EU_post_m1000}{theta_sd}{0.01}
\setres{learn}{EU_post_m3}{ess}{51}
\setres{learn}{EU_post_m3}{kappa_mean}{-0.050}
\setres{learn}{EU_post_m3}{kappa_mean_se}{0.303}
\setres{learn}{EU_post_m3}{kappa_median}{0.741}
\setres{learn}{EU_post_m3}{kappa_prior_mean_median}{0.762}
\setres{learn}{EU_post_m3}{m}{3}
\setres{learn}{EU_post_m3}{n_countries}{30}
\setres{learn}{EU_post_m3}{n_pos}{27}
\setres{learn}{EU_post_m3}{n_subjects_seen}{3}
\setres{learn}{EU_post_m3}{r_ff}{0.367}
\setres{learn}{EU_post_m3}{r_prior}{0.9066}
\setres{learn}{EU_post_m3}{r_prob}{0.941}
\setres{learn}{EU_post_m3}{r_prob_se}{0.000}
\setres{learn}{EU_post_m3}{theta_mean}{0.81}
\setres{learn}{EU_post_m3}{theta_sd}{0.07}
\setres{learn}{EU_post_m30}{ess}{23}
\setres{learn}{EU_post_m30}{kappa_mean}{0.315}
\setres{learn}{EU_post_m30}{kappa_mean_se}{0.225}
\setres{learn}{EU_post_m30}{kappa_median}{0.823}
\setres{learn}{EU_post_m30}{kappa_prior_mean_median}{0.825}
\setres{learn}{EU_post_m30}{m}{30}
\setres{learn}{EU_post_m30}{n_countries}{30}
\setres{learn}{EU_post_m30}{n_pos}{29}
\setres{learn}{EU_post_m30}{n_subjects_seen}{30}
\setres{learn}{EU_post_m30}{r_ff}{0.367}
\setres{learn}{EU_post_m30}{r_prior}{0.9599}
\setres{learn}{EU_post_m30}{r_prob}{0.975}
\setres{learn}{EU_post_m30}{r_prob_se}{0.000}
\setres{learn}{EU_post_m30}{theta_mean}{0.84}
\setres{learn}{EU_post_m30}{theta_sd}{0.03}
\setres{learn}{EU_post_m300}{ess}{8}
\setres{learn}{EU_post_m300}{kappa_mean}{0.377}
\setres{learn}{EU_post_m300}{kappa_mean_se}{0.206}
\setres{learn}{EU_post_m300}{kappa_median}{0.829}
\setres{learn}{EU_post_m300}{kappa_prior_mean_median}{0.829}
\setres{learn}{EU_post_m300}{m}{300}
\setres{learn}{EU_post_m300}{n_countries}{30}
\setres{learn}{EU_post_m300}{n_pos}{29}
\setres{learn}{EU_post_m300}{n_subjects_seen}{300}
\setres{learn}{EU_post_m300}{r_ff}{0.367}
\setres{learn}{EU_post_m300}{r_prior}{0.9869}
\setres{learn}{EU_post_m300}{r_prob}{0.992}
\setres{learn}{EU_post_m300}{r_prob_se}{0.000}
\setres{learn}{EU_post_m300}{theta_mean}{0.93}
\setres{learn}{EU_post_m300}{theta_sd}{0.01}
\setres{learn}{EU_post_mall}{ess}{3}
\setres{learn}{EU_post_mall}{kappa_mean}{0.398}
\setres{learn}{EU_post_mall}{kappa_mean_se}{0.199}
\setres{learn}{EU_post_mall}{kappa_median}{0.811}
\setres{learn}{EU_post_mall}{kappa_prior_mean_median}{0.811}
\setres{learn}{EU_post_mall}{m}{all}
\setres{learn}{EU_post_mall}{n_countries}{30}
\setres{learn}{EU_post_mall}{n_pos}{29}
\setres{learn}{EU_post_mall}{n_subjects_seen}{2{,}939}
\setres{learn}{EU_post_mall}{r_ff}{0.367}
\setres{learn}{EU_post_mall}{r_prior}{0.9957}
\setres{learn}{EU_post_mall}{r_prob}{0.997}
\setres{learn}{EU_post_mall}{r_prob_se}{0.000}
\setres{learn}{EU_post_mall}{theta_mean}{0.94}
\setres{learn}{EU_post_mall}{theta_sd}{0.00}
\setres{learn}{reference}{kind}{common}
\setres{learn}{reference}{m_grid}{0, 3, 10, 30, 100, 300, 1000, all}
\setres{learn}{reference}{n_countries}{30}
\setres{learn}{reference}{start}{lit}
\setres{learnZ}{CPT_post_m0}{ess}{732}
\setres{learnZ}{CPT_post_m0}{kappa}{0.902}
\setres{learnZ}{CPT_post_m0}{kappa_prior_mean}{0.954}
\setres{learnZ}{CPT_post_m0}{kappa_se}{0.016}
\setres{learnZ}{CPT_post_m0}{mse}{31.78}
\setres{learnZ}{CPT_post_m0}{n_subjects_seen}{0}
\setres{learnZ}{CPT_post_m0}{spread_penalty}{0.052}
\setres{learnZ}{CPT_post_m0}{theta_mean}{0.70, 0.73, 1.08}
\setres{learnZ}{CPT_post_m0}{theta_sd}{0.14, 0.18, 0.28}
\setres{learnZ}{CPT_post_m10}{ess}{66}
\setres{learnZ}{CPT_post_m10}{kappa}{0.968}
\setres{learnZ}{CPT_post_m10}{kappa_prior_mean}{0.974}
\setres{learnZ}{CPT_post_m10}{kappa_se}{0.007}
\setres{learnZ}{CPT_post_m10}{mse}{24.73}
\setres{learnZ}{CPT_post_m10}{n_subjects_seen}{10}
\setres{learnZ}{CPT_post_m10}{spread_penalty}{0.006}
\setres{learnZ}{CPT_post_m10}{theta_mean}{0.74, 0.56, 1.19}
\setres{learnZ}{CPT_post_m10}{theta_sd}{0.12, 0.05, 0.13}
\setres{learnZ}{CPT_post_m100}{ess}{4}
\setres{learnZ}{CPT_post_m100}{kappa}{0.983}
\setres{learnZ}{CPT_post_m100}{kappa_prior_mean}{0.985}
\setres{learnZ}{CPT_post_m100}{kappa_se}{0.006}
\setres{learnZ}{CPT_post_m100}{mse}{23.15}
\setres{learnZ}{CPT_post_m100}{n_subjects_seen}{100}
\setres{learnZ}{CPT_post_m100}{spread_penalty}{0.002}
\setres{learnZ}{CPT_post_m100}{theta_mean}{0.87, 0.43, 0.93}
\setres{learnZ}{CPT_post_m100}{theta_sd}{0.09, 0.01, 0.06}
\setres{learnZ}{CPT_post_m3}{ess}{194}
\setres{learnZ}{CPT_post_m3}{kappa}{0.955}
\setres{learnZ}{CPT_post_m3}{kappa_prior_mean}{0.971}
\setres{learnZ}{CPT_post_m3}{kappa_se}{0.006}
\setres{learnZ}{CPT_post_m3}{mse}{26.11}
\setres{learnZ}{CPT_post_m3}{n_subjects_seen}{3}
\setres{learnZ}{CPT_post_m3}{spread_penalty}{0.016}
\setres{learnZ}{CPT_post_m3}{theta_mean}{0.73, 0.54, 1.23}
\setres{learnZ}{CPT_post_m3}{theta_sd}{0.12, 0.08, 0.19}
\setres{learnZ}{CPT_post_m30}{ess}{13}
\setres{learnZ}{CPT_post_m30}{kappa}{0.944}
\setres{learnZ}{CPT_post_m30}{kappa_prior_mean}{0.947}
\setres{learnZ}{CPT_post_m30}{kappa_se}{0.009}
\setres{learnZ}{CPT_post_m30}{mse}{27.34}
\setres{learnZ}{CPT_post_m30}{n_subjects_seen}{30}
\setres{learnZ}{CPT_post_m30}{spread_penalty}{0.003}
\setres{learnZ}{CPT_post_m30}{theta_mean}{0.78, 0.43, 1.17}
\setres{learnZ}{CPT_post_m30}{theta_sd}{0.10, 0.02, 0.09}
\setres{learnZ}{CPT_post_mall}{ess}{4}
\setres{learnZ}{CPT_post_mall}{kappa}{0.977}
\setres{learnZ}{CPT_post_mall}{kappa_prior_mean}{0.979}
\setres{learnZ}{CPT_post_mall}{kappa_se}{0.007}
\setres{learnZ}{CPT_post_mall}{mse}{23.80}
\setres{learnZ}{CPT_post_mall}{n_subjects_seen}{179}
\setres{learnZ}{CPT_post_mall}{spread_penalty}{0.003}
\setres{learnZ}{CPT_post_mall}{theta_mean}{0.90, 0.41, 0.90}
\setres{learnZ}{CPT_post_mall}{theta_sd}{0.05, 0.02, 0.03}
\setres{learnZ}{DA_post_m0}{ess}{439}
\setres{learnZ}{DA_post_m0}{kappa}{0.798}
\setres{learnZ}{DA_post_m0}{kappa_prior_mean}{0.830}
\setres{learnZ}{DA_post_m0}{kappa_se}{0.033}
\setres{learnZ}{DA_post_m0}{mse}{42.97}
\setres{learnZ}{DA_post_m0}{n_subjects_seen}{0}
\setres{learnZ}{DA_post_m0}{spread_penalty}{0.032}
\setres{learnZ}{DA_post_m0}{theta_mean}{0.58, 0.30}
\setres{learnZ}{DA_post_m0}{theta_sd}{0.19, 0.45}
\setres{learnZ}{DA_post_m10}{ess}{75}
\setres{learnZ}{DA_post_m10}{kappa}{0.805}
\setres{learnZ}{DA_post_m10}{kappa_prior_mean}{0.811}
\setres{learnZ}{DA_post_m10}{kappa_se}{0.030}
\setres{learnZ}{DA_post_m10}{mse}{42.26}
\setres{learnZ}{DA_post_m10}{n_subjects_seen}{10}
\setres{learnZ}{DA_post_m10}{spread_penalty}{0.005}
\setres{learnZ}{DA_post_m10}{theta_mean}{0.70, -0.37}
\setres{learnZ}{DA_post_m10}{theta_sd}{0.16, 0.11}
\setres{learnZ}{DA_post_m100}{ess}{9}
\setres{learnZ}{DA_post_m100}{kappa}{0.827}
\setres{learnZ}{DA_post_m100}{kappa_prior_mean}{0.827}
\setres{learnZ}{DA_post_m100}{kappa_se}{0.030}
\setres{learnZ}{DA_post_m100}{mse}{39.95}
\setres{learnZ}{DA_post_m100}{n_subjects_seen}{100}
\setres{learnZ}{DA_post_m100}{spread_penalty}{0.001}
\setres{learnZ}{DA_post_m100}{theta_mean}{0.93, -0.14}
\setres{learnZ}{DA_post_m100}{theta_sd}{0.04, 0.05}
\setres{learnZ}{DA_post_m3}{ess}{153}
\setres{learnZ}{DA_post_m3}{kappa}{0.800}
\setres{learnZ}{DA_post_m3}{kappa_prior_mean}{0.816}
\setres{learnZ}{DA_post_m3}{kappa_se}{0.031}
\setres{learnZ}{DA_post_m3}{mse}{42.86}
\setres{learnZ}{DA_post_m3}{n_subjects_seen}{3}
\setres{learnZ}{DA_post_m3}{spread_penalty}{0.017}
\setres{learnZ}{DA_post_m3}{theta_mean}{0.66, -0.36}
\setres{learnZ}{DA_post_m3}{theta_sd}{0.17, 0.20}
\setres{learnZ}{DA_post_m30}{ess}{27}
\setres{learnZ}{DA_post_m30}{kappa}{0.765}
\setres{learnZ}{DA_post_m30}{kappa_prior_mean}{0.768}
\setres{learnZ}{DA_post_m30}{kappa_se}{0.031}
\setres{learnZ}{DA_post_m30}{mse}{46.58}
\setres{learnZ}{DA_post_m30}{n_subjects_seen}{30}
\setres{learnZ}{DA_post_m30}{spread_penalty}{0.003}
\setres{learnZ}{DA_post_m30}{theta_mean}{0.81, -0.51}
\setres{learnZ}{DA_post_m30}{theta_sd}{0.11, 0.06}
\setres{learnZ}{DA_post_mall}{ess}{4}
\setres{learnZ}{DA_post_mall}{kappa}{0.826}
\setres{learnZ}{DA_post_mall}{kappa_prior_mean}{0.826}
\setres{learnZ}{DA_post_mall}{kappa_se}{0.030}
\setres{learnZ}{DA_post_mall}{mse}{40.01}
\setres{learnZ}{DA_post_mall}{n_subjects_seen}{179}
\setres{learnZ}{DA_post_mall}{spread_penalty}{0.000}
\setres{learnZ}{DA_post_mall}{theta_mean}{0.96, -0.14}
\setres{learnZ}{DA_post_mall}{theta_sd}{0.03, 0.03}
\setres{learnZ}{EU_post_m0}{ess}{98}
\setres{learnZ}{EU_post_m0}{kappa}{0.830}
\setres{learnZ}{EU_post_m0}{kappa_prior_mean}{0.831}
\setres{learnZ}{EU_post_m0}{kappa_se}{0.031}
\setres{learnZ}{EU_post_m0}{mse}{39.59}
\setres{learnZ}{EU_post_m0}{n_subjects_seen}{0}
\setres{learnZ}{EU_post_m0}{spread_penalty}{0.001}
\setres{learnZ}{EU_post_m0}{theta_mean}{0.70}
\setres{learnZ}{EU_post_m0}{theta_sd}{0.14}
\setres{learnZ}{EU_post_m10}{ess}{50}
\setres{learnZ}{EU_post_m10}{kappa}{0.833}
\setres{learnZ}{EU_post_m10}{kappa_prior_mean}{0.833}
\setres{learnZ}{EU_post_m10}{kappa_se}{0.031}
\setres{learnZ}{EU_post_m10}{mse}{39.29}
\setres{learnZ}{EU_post_m10}{n_subjects_seen}{10}
\setres{learnZ}{EU_post_m10}{spread_penalty}{0.000}
\setres{learnZ}{EU_post_m10}{theta_mean}{0.85}
\setres{learnZ}{EU_post_m10}{theta_sd}{0.08}
\setres{learnZ}{EU_post_m100}{ess}{16}
\setres{learnZ}{EU_post_m100}{kappa}{0.833}
\setres{learnZ}{EU_post_m100}{kappa_prior_mean}{0.833}
\setres{learnZ}{EU_post_m100}{kappa_se}{0.030}
\setres{learnZ}{EU_post_m100}{mse}{39.24}
\setres{learnZ}{EU_post_m100}{n_subjects_seen}{100}
\setres{learnZ}{EU_post_m100}{spread_penalty}{0.000}
\setres{learnZ}{EU_post_m100}{theta_mean}{0.96}
\setres{learnZ}{EU_post_m100}{theta_sd}{0.02}
\setres{learnZ}{EU_post_m3}{ess}{75}
\setres{learnZ}{EU_post_m3}{kappa}{0.832}
\setres{learnZ}{EU_post_m3}{kappa_prior_mean}{0.833}
\setres{learnZ}{EU_post_m3}{kappa_se}{0.031}
\setres{learnZ}{EU_post_m3}{mse}{39.40}
\setres{learnZ}{EU_post_m3}{n_subjects_seen}{3}
\setres{learnZ}{EU_post_m3}{spread_penalty}{0.001}
\setres{learnZ}{EU_post_m3}{theta_mean}{0.78}
\setres{learnZ}{EU_post_m3}{theta_sd}{0.11}
\setres{learnZ}{EU_post_m30}{ess}{20}
\setres{learnZ}{EU_post_m30}{kappa}{0.833}
\setres{learnZ}{EU_post_m30}{kappa_prior_mean}{0.833}
\setres{learnZ}{EU_post_m30}{kappa_se}{0.030}
\setres{learnZ}{EU_post_m30}{mse}{39.24}
\setres{learnZ}{EU_post_m30}{n_subjects_seen}{30}
\setres{learnZ}{EU_post_m30}{spread_penalty}{0.000}
\setres{learnZ}{EU_post_m30}{theta_mean}{0.94}
\setres{learnZ}{EU_post_m30}{theta_sd}{0.03}
\setres{learnZ}{EU_post_mall}{ess}{10}
\setres{learnZ}{EU_post_mall}{kappa}{0.833}
\setres{learnZ}{EU_post_mall}{kappa_prior_mean}{0.833}
\setres{learnZ}{EU_post_mall}{kappa_se}{0.030}
\setres{learnZ}{EU_post_mall}{mse}{39.24}
\setres{learnZ}{EU_post_mall}{n_subjects_seen}{179}
\setres{learnZ}{EU_post_mall}{spread_penalty}{0.000}
\setres{learnZ}{EU_post_mall}{theta_mean}{0.97}
\setres{learnZ}{EU_post_mall}{theta_sd}{0.02}
\setres{learnZ}{reference}{n_source_subjects}{179}
\setres{learnZ}{reference}{sigma}{6.6}
\setres{learnRZ}{CPT_post_m0}{ess}{732}
\setres{learnRZ}{CPT_post_m0}{n_subjects_seen}{0}
\setres{learnRZ}{CPT_post_m0}{r_ff}{0.153}
\setres{learnRZ}{CPT_post_m0}{r_ff_se}{0.002}
\setres{learnRZ}{CPT_post_m0}{r_prior}{0.9652}
\setres{learnRZ}{CPT_post_m0}{r_prior_se}{0.000}
\setres{learnRZ}{CPT_post_m0}{r_prob}{0.971}
\setres{learnRZ}{CPT_post_m0}{r_prob_se}{0.000}
\setres{learnRZ}{CPT_post_m0}{theta_mean}{0.70, 0.73, 1.08}
\setres{learnRZ}{CPT_post_m0}{theta_sd}{0.14, 0.18, 0.28}
\setres{learnRZ}{CPT_post_m0}{volume_ess}{87{,}437}
\setres{learnRZ}{CPT_post_m10}{ess}{66}
\setres{learnRZ}{CPT_post_m10}{n_subjects_seen}{10}
\setres{learnRZ}{CPT_post_m10}{r_ff}{0.153}
\setres{learnRZ}{CPT_post_m10}{r_ff_se}{0.002}
\setres{learnRZ}{CPT_post_m10}{r_prior}{0.9954}
\setres{learnRZ}{CPT_post_m10}{r_prior_se}{0.000}
\setres{learnRZ}{CPT_post_m10}{r_prob}{0.996}
\setres{learnRZ}{CPT_post_m10}{r_prob_se}{0.000}
\setres{learnRZ}{CPT_post_m10}{theta_mean}{0.74, 0.56, 1.19}
\setres{learnRZ}{CPT_post_m10}{theta_sd}{0.12, 0.05, 0.13}
\setres{learnRZ}{CPT_post_m10}{volume_ess}{87{,}437}
\setres{learnRZ}{CPT_post_m100}{ess}{4}
\setres{learnRZ}{CPT_post_m100}{n_subjects_seen}{100}
\setres{learnRZ}{CPT_post_m100}{r_ff}{0.153}
\setres{learnRZ}{CPT_post_m100}{r_ff_se}{0.002}
\setres{learnRZ}{CPT_post_m100}{r_prior}{0.9996}
\setres{learnRZ}{CPT_post_m100}{r_prior_se}{0.000}
\setres{learnRZ}{CPT_post_m100}{r_prob}{1.000}
\setres{learnRZ}{CPT_post_m100}{r_prob_se}{0.000}
\setres{learnRZ}{CPT_post_m100}{theta_mean}{0.87, 0.43, 0.93}
\setres{learnRZ}{CPT_post_m100}{theta_sd}{0.09, 0.01, 0.06}
\setres{learnRZ}{CPT_post_m100}{volume_ess}{87{,}437}
\setres{learnRZ}{CPT_post_m3}{ess}{194}
\setres{learnRZ}{CPT_post_m3}{n_subjects_seen}{3}
\setres{learnRZ}{CPT_post_m3}{r_ff}{0.153}
\setres{learnRZ}{CPT_post_m3}{r_ff_se}{0.002}
\setres{learnRZ}{CPT_post_m3}{r_prior}{0.9857}
\setres{learnRZ}{CPT_post_m3}{r_prior_se}{0.000}
\setres{learnRZ}{CPT_post_m3}{r_prob}{0.988}
\setres{learnRZ}{CPT_post_m3}{r_prob_se}{0.000}
\setres{learnRZ}{CPT_post_m3}{theta_mean}{0.73, 0.54, 1.23}
\setres{learnRZ}{CPT_post_m3}{theta_sd}{0.12, 0.08, 0.19}
\setres{learnRZ}{CPT_post_m3}{volume_ess}{87{,}437}
\setres{learnRZ}{CPT_post_m30}{ess}{13}
\setres{learnRZ}{CPT_post_m30}{n_subjects_seen}{30}
\setres{learnRZ}{CPT_post_m30}{r_ff}{0.153}
\setres{learnRZ}{CPT_post_m30}{r_ff_se}{0.002}
\setres{learnRZ}{CPT_post_m30}{r_prior}{0.9988}
\setres{learnRZ}{CPT_post_m30}{r_prior_se}{0.000}
\setres{learnRZ}{CPT_post_m30}{r_prob}{0.999}
\setres{learnRZ}{CPT_post_m30}{r_prob_se}{0.000}
\setres{learnRZ}{CPT_post_m30}{theta_mean}{0.78, 0.43, 1.17}
\setres{learnRZ}{CPT_post_m30}{theta_sd}{0.10, 0.02, 0.09}
\setres{learnRZ}{CPT_post_m30}{volume_ess}{87{,}437}
\setres{learnRZ}{CPT_post_mall}{ess}{4}
\setres{learnRZ}{CPT_post_mall}{n_subjects_seen}{179}
\setres{learnRZ}{CPT_post_mall}{r_ff}{0.153}
\setres{learnRZ}{CPT_post_mall}{r_ff_se}{0.002}
\setres{learnRZ}{CPT_post_mall}{r_prior}{0.9996}
\setres{learnRZ}{CPT_post_mall}{r_prior_se}{0.000}
\setres{learnRZ}{CPT_post_mall}{r_prob}{1.000}
\setres{learnRZ}{CPT_post_mall}{r_prob_se}{0.000}
\setres{learnRZ}{CPT_post_mall}{theta_mean}{0.90, 0.41, 0.90}
\setres{learnRZ}{CPT_post_mall}{theta_sd}{0.05, 0.02, 0.03}
\setres{learnRZ}{CPT_post_mall}{volume_ess}{87{,}437}
\setres{learnRZ}{DA_post_m0}{ess}{439}
\setres{learnRZ}{DA_post_m0}{n_subjects_seen}{0}
\setres{learnRZ}{DA_post_m0}{r_ff}{0.282}
\setres{learnRZ}{DA_post_m0}{r_ff_se}{0.003}
\setres{learnRZ}{DA_post_m0}{r_prior}{0.9600}
\setres{learnRZ}{DA_post_m0}{r_prior_se}{0.000}
\setres{learnRZ}{DA_post_m0}{r_prob}{0.971}
\setres{learnRZ}{DA_post_m0}{r_prob_se}{0.000}
\setres{learnRZ}{DA_post_m0}{theta_mean}{0.58, 0.30}
\setres{learnRZ}{DA_post_m0}{theta_sd}{0.19, 0.45}
\setres{learnRZ}{DA_post_m0}{volume_ess}{351{,}503}
\setres{learnRZ}{DA_post_m10}{ess}{75}
\setres{learnRZ}{DA_post_m10}{n_subjects_seen}{10}
\setres{learnRZ}{DA_post_m10}{r_ff}{0.282}
\setres{learnRZ}{DA_post_m10}{r_ff_se}{0.003}
\setres{learnRZ}{DA_post_m10}{r_prior}{0.9871}
\setres{learnRZ}{DA_post_m10}{r_prior_se}{0.000}
\setres{learnRZ}{DA_post_m10}{r_prob}{0.991}
\setres{learnRZ}{DA_post_m10}{r_prob_se}{0.000}
\setres{learnRZ}{DA_post_m10}{theta_mean}{0.70, -0.37}
\setres{learnRZ}{DA_post_m10}{theta_sd}{0.16, 0.11}
\setres{learnRZ}{DA_post_m10}{volume_ess}{351{,}503}
\setres{learnRZ}{DA_post_m100}{ess}{9}
\setres{learnRZ}{DA_post_m100}{n_subjects_seen}{100}
\setres{learnRZ}{DA_post_m100}{r_ff}{0.282}
\setres{learnRZ}{DA_post_m100}{r_ff_se}{0.003}
\setres{learnRZ}{DA_post_m100}{r_prior}{0.9992}
\setres{learnRZ}{DA_post_m100}{r_prior_se}{0.000}
\setres{learnRZ}{DA_post_m100}{r_prob}{0.999}
\setres{learnRZ}{DA_post_m100}{r_prob_se}{0.000}
\setres{learnRZ}{DA_post_m100}{theta_mean}{0.93, -0.14}
\setres{learnRZ}{DA_post_m100}{theta_sd}{0.04, 0.05}
\setres{learnRZ}{DA_post_m100}{volume_ess}{351{,}503}
\setres{learnRZ}{DA_post_m3}{ess}{153}
\setres{learnRZ}{DA_post_m3}{n_subjects_seen}{3}
\setres{learnRZ}{DA_post_m3}{r_ff}{0.282}
\setres{learnRZ}{DA_post_m3}{r_ff_se}{0.003}
\setres{learnRZ}{DA_post_m3}{r_prior}{0.9716}
\setres{learnRZ}{DA_post_m3}{r_prior_se}{0.000}
\setres{learnRZ}{DA_post_m3}{r_prob}{0.980}
\setres{learnRZ}{DA_post_m3}{r_prob_se}{0.000}
\setres{learnRZ}{DA_post_m3}{theta_mean}{0.66, -0.36}
\setres{learnRZ}{DA_post_m3}{theta_sd}{0.17, 0.20}
\setres{learnRZ}{DA_post_m3}{volume_ess}{351{,}503}
\setres{learnRZ}{DA_post_m30}{ess}{27}
\setres{learnRZ}{DA_post_m30}{n_subjects_seen}{30}
\setres{learnRZ}{DA_post_m30}{r_ff}{0.282}
\setres{learnRZ}{DA_post_m30}{r_ff_se}{0.003}
\setres{learnRZ}{DA_post_m30}{r_prior}{0.9954}
\setres{learnRZ}{DA_post_m30}{r_prior_se}{0.000}
\setres{learnRZ}{DA_post_m30}{r_prob}{0.997}
\setres{learnRZ}{DA_post_m30}{r_prob_se}{0.000}
\setres{learnRZ}{DA_post_m30}{theta_mean}{0.81, -0.51}
\setres{learnRZ}{DA_post_m30}{theta_sd}{0.11, 0.06}
\setres{learnRZ}{DA_post_m30}{volume_ess}{351{,}503}
\setres{learnRZ}{DA_post_mall}{ess}{4}
\setres{learnRZ}{DA_post_mall}{n_subjects_seen}{179}
\setres{learnRZ}{DA_post_mall}{r_ff}{0.282}
\setres{learnRZ}{DA_post_mall}{r_ff_se}{0.003}
\setres{learnRZ}{DA_post_mall}{r_prior}{0.9996}
\setres{learnRZ}{DA_post_mall}{r_prior_se}{0.000}
\setres{learnRZ}{DA_post_mall}{r_prob}{1.000}
\setres{learnRZ}{DA_post_mall}{r_prob_se}{0.000}
\setres{learnRZ}{DA_post_mall}{theta_mean}{0.96, -0.14}
\setres{learnRZ}{DA_post_mall}{theta_sd}{0.03, 0.03}
\setres{learnRZ}{DA_post_mall}{volume_ess}{351{,}503}
\setres{learnRZ}{EU_post_m0}{ess}{98}
\setres{learnRZ}{EU_post_m0}{n_subjects_seen}{0}
\setres{learnRZ}{EU_post_m0}{r_ff}{0.313}
\setres{learnRZ}{EU_post_m0}{r_ff_se}{0.003}
\setres{learnRZ}{EU_post_m0}{r_prior}{0.9422}
\setres{learnRZ}{EU_post_m0}{r_prior_se}{0.000}
\setres{learnRZ}{EU_post_m0}{r_prob}{0.960}
\setres{learnRZ}{EU_post_m0}{r_prob_se}{0.000}
\setres{learnRZ}{EU_post_m0}{theta_mean}{0.70}
\setres{learnRZ}{EU_post_m0}{theta_sd}{0.14}
\setres{learnRZ}{EU_post_m0}{volume_ess}{671{,}385}
\setres{learnRZ}{EU_post_m10}{ess}{50}
\setres{learnRZ}{EU_post_m10}{n_subjects_seen}{10}
\setres{learnRZ}{EU_post_m10}{r_ff}{0.313}
\setres{learnRZ}{EU_post_m10}{r_ff_se}{0.003}
\setres{learnRZ}{EU_post_m10}{r_prior}{0.9710}
\setres{learnRZ}{EU_post_m10}{r_prior_se}{0.000}
\setres{learnRZ}{EU_post_m10}{r_prob}{0.980}
\setres{learnRZ}{EU_post_m10}{r_prob_se}{0.000}
\setres{learnRZ}{EU_post_m10}{theta_mean}{0.85}
\setres{learnRZ}{EU_post_m10}{theta_sd}{0.08}
\setres{learnRZ}{EU_post_m10}{volume_ess}{671{,}385}
\setres{learnRZ}{EU_post_m100}{ess}{16}
\setres{learnRZ}{EU_post_m100}{n_subjects_seen}{100}
\setres{learnRZ}{EU_post_m100}{r_ff}{0.313}
\setres{learnRZ}{EU_post_m100}{r_ff_se}{0.003}
\setres{learnRZ}{EU_post_m100}{r_prior}{0.9909}
\setres{learnRZ}{EU_post_m100}{r_prior_se}{0.000}
\setres{learnRZ}{EU_post_m100}{r_prob}{0.994}
\setres{learnRZ}{EU_post_m100}{r_prob_se}{0.000}
\setres{learnRZ}{EU_post_m100}{theta_mean}{0.96}
\setres{learnRZ}{EU_post_m100}{theta_sd}{0.02}
\setres{learnRZ}{EU_post_m100}{volume_ess}{671{,}385}
\setres{learnRZ}{EU_post_m3}{ess}{75}
\setres{learnRZ}{EU_post_m3}{n_subjects_seen}{3}
\setres{learnRZ}{EU_post_m3}{r_ff}{0.313}
\setres{learnRZ}{EU_post_m3}{r_ff_se}{0.003}
\setres{learnRZ}{EU_post_m3}{r_prior}{0.9566}
\setres{learnRZ}{EU_post_m3}{r_prior_se}{0.000}
\setres{learnRZ}{EU_post_m3}{r_prob}{0.970}
\setres{learnRZ}{EU_post_m3}{r_prob_se}{0.000}
\setres{learnRZ}{EU_post_m3}{theta_mean}{0.78}
\setres{learnRZ}{EU_post_m3}{theta_sd}{0.11}
\setres{learnRZ}{EU_post_m3}{volume_ess}{671{,}385}
\setres{learnRZ}{EU_post_m30}{ess}{20}
\setres{learnRZ}{EU_post_m30}{n_subjects_seen}{30}
\setres{learnRZ}{EU_post_m30}{r_ff}{0.313}
\setres{learnRZ}{EU_post_m30}{r_ff_se}{0.003}
\setres{learnRZ}{EU_post_m30}{r_prior}{0.9884}
\setres{learnRZ}{EU_post_m30}{r_prior_se}{0.000}
\setres{learnRZ}{EU_post_m30}{r_prob}{0.992}
\setres{learnRZ}{EU_post_m30}{r_prob_se}{0.000}
\setres{learnRZ}{EU_post_m30}{theta_mean}{0.94}
\setres{learnRZ}{EU_post_m30}{theta_sd}{0.03}
\setres{learnRZ}{EU_post_m30}{volume_ess}{671{,}385}
\setres{learnRZ}{EU_post_mall}{ess}{10}
\setres{learnRZ}{EU_post_mall}{n_subjects_seen}{179}
\setres{learnRZ}{EU_post_mall}{r_ff}{0.313}
\setres{learnRZ}{EU_post_mall}{r_ff_se}{0.003}
\setres{learnRZ}{EU_post_mall}{r_prior}{0.9945}
\setres{learnRZ}{EU_post_mall}{r_prior_se}{0.000}
\setres{learnRZ}{EU_post_mall}{r_prob}{0.996}
\setres{learnRZ}{EU_post_mall}{r_prob_se}{0.000}
\setres{learnRZ}{EU_post_mall}{theta_mean}{0.97}
\setres{learnRZ}{EU_post_mall}{theta_sd}{0.02}
\setres{learnRZ}{EU_post_mall}{volume_ess}{671{,}385}
\setres{learnRZ}{reference}{sigma}{6.6}
\setres{learnU}{CPT_post_m0}{ess}{8{,}000}
\setres{learnU}{CPT_post_m0}{kappa_mean}{-2.803}
\setres{learnU}{CPT_post_m0}{kappa_mean_se}{0.434}
\setres{learnU}{CPT_post_m0}{kappa_median}{-1.563}
\setres{learnU}{CPT_post_m0}{kappa_prior_mean_median}{0.177}
\setres{learnU}{CPT_post_m0}{m}{0}
\setres{learnU}{CPT_post_m0}{n_countries}{30}
\setres{learnU}{CPT_post_m0}{n_pos}{0}
\setres{learnU}{CPT_post_m0}{n_subjects_seen}{0}
\setres{learnU}{CPT_post_m0}{r_ff}{0.199}
\setres{learnU}{CPT_post_m0}{r_prior}{0.5146}
\setres{learnU}{CPT_post_m0}{r_prob}{0.611}
\setres{learnU}{CPT_post_m0}{r_prob_se}{0.003}
\setres{learnU}{CPT_post_m0}{theta_mean}{0.50, 0.50, 1.55}
\setres{learnU}{CPT_post_m0}{theta_sd}{0.29, 0.29, 0.84}
\setres{learnU}{CPT_post_m10}{ess}{58}
\setres{learnU}{CPT_post_m10}{kappa_mean}{0.479}
\setres{learnU}{CPT_post_m10}{kappa_mean_se}{0.121}
\setres{learnU}{CPT_post_m10}{kappa_median}{0.791}
\setres{learnU}{CPT_post_m10}{kappa_prior_mean_median}{0.819}
\setres{learnU}{CPT_post_m10}{m}{10}
\setres{learnU}{CPT_post_m10}{n_countries}{30}
\setres{learnU}{CPT_post_m10}{n_pos}{29}
\setres{learnU}{CPT_post_m10}{n_subjects_seen}{10}
\setres{learnU}{CPT_post_m10}{r_ff}{0.199}
\setres{learnU}{CPT_post_m10}{r_prior}{0.9937}
\setres{learnU}{CPT_post_m10}{r_prob}{0.995}
\setres{learnU}{CPT_post_m10}{r_prob_se}{0.000}
\setres{learnU}{CPT_post_m10}{theta_mean}{0.89, 0.66, 0.92}
\setres{learnU}{CPT_post_m10}{theta_sd}{0.08, 0.09, 0.14}
\setres{learnU}{CPT_post_m100}{ess}{2}
\setres{learnU}{CPT_post_m100}{kappa_mean}{0.719}
\setres{learnU}{CPT_post_m100}{kappa_mean_se}{0.103}
\setres{learnU}{CPT_post_m100}{kappa_median}{0.926}
\setres{learnU}{CPT_post_m100}{kappa_prior_mean_median}{0.932}
\setres{learnU}{CPT_post_m100}{m}{100}
\setres{learnU}{CPT_post_m100}{n_countries}{30}
\setres{learnU}{CPT_post_m100}{n_pos}{29}
\setres{learnU}{CPT_post_m100}{n_subjects_seen}{100}
\setres{learnU}{CPT_post_m100}{r_ff}{0.199}
\setres{learnU}{CPT_post_m100}{r_prior}{0.9998}
\setres{learnU}{CPT_post_m100}{r_prob}{1.000}
\setres{learnU}{CPT_post_m100}{r_prob_se}{0.000}
\setres{learnU}{CPT_post_m100}{theta_mean}{0.97, 0.66, 0.90}
\setres{learnU}{CPT_post_m100}{theta_sd}{0.02, 0.03, 0.02}
\setres{learnU}{CPT_post_m1000}{ess}{1}
\setres{learnU}{CPT_post_m1000}{kappa_mean}{0.753}
\setres{learnU}{CPT_post_m1000}{kappa_mean_se}{0.090}
\setres{learnU}{CPT_post_m1000}{kappa_median}{0.945}
\setres{learnU}{CPT_post_m1000}{kappa_prior_mean_median}{0.951}
\setres{learnU}{CPT_post_m1000}{m}{1000}
\setres{learnU}{CPT_post_m1000}{n_countries}{30}
\setres{learnU}{CPT_post_m1000}{n_pos}{29}
\setres{learnU}{CPT_post_m1000}{n_subjects_seen}{1{,}000}
\setres{learnU}{CPT_post_m1000}{r_ff}{0.199}
\setres{learnU}{CPT_post_m1000}{r_prior}{0.9999}
\setres{learnU}{CPT_post_m1000}{r_prob}{1.000}
\setres{learnU}{CPT_post_m1000}{r_prob_se}{0.000}
\setres{learnU}{CPT_post_m1000}{theta_mean}{0.98, 0.58, 0.90}
\setres{learnU}{CPT_post_m1000}{theta_sd}{0.00, 0.00, 0.00}
\setres{learnU}{CPT_post_m3}{ess}{348}
\setres{learnU}{CPT_post_m3}{kappa_mean}{-0.049}
\setres{learnU}{CPT_post_m3}{kappa_mean_se}{0.214}
\setres{learnU}{CPT_post_m3}{kappa_median}{0.523}
\setres{learnU}{CPT_post_m3}{kappa_prior_mean_median}{0.616}
\setres{learnU}{CPT_post_m3}{m}{3}
\setres{learnU}{CPT_post_m3}{n_countries}{30}
\setres{learnU}{CPT_post_m3}{n_pos}{26}
\setres{learnU}{CPT_post_m3}{n_subjects_seen}{3}
\setres{learnU}{CPT_post_m3}{r_ff}{0.199}
\setres{learnU}{CPT_post_m3}{r_prior}{0.9529}
\setres{learnU}{CPT_post_m3}{r_prob}{0.962}
\setres{learnU}{CPT_post_m3}{r_prob_se}{0.000}
\setres{learnU}{CPT_post_m3}{theta_mean}{0.78, 0.47, 1.16}
\setres{learnU}{CPT_post_m3}{theta_sd}{0.15, 0.15, 0.36}
\setres{learnU}{CPT_post_m30}{ess}{13}
\setres{learnU}{CPT_post_m30}{kappa_mean}{0.633}
\setres{learnU}{CPT_post_m30}{kappa_mean_se}{0.110}
\setres{learnU}{CPT_post_m30}{kappa_median}{0.866}
\setres{learnU}{CPT_post_m30}{kappa_prior_mean_median}{0.885}
\setres{learnU}{CPT_post_m30}{m}{30}
\setres{learnU}{CPT_post_m30}{n_countries}{30}
\setres{learnU}{CPT_post_m30}{n_pos}{29}
\setres{learnU}{CPT_post_m30}{n_subjects_seen}{30}
\setres{learnU}{CPT_post_m30}{r_ff}{0.199}
\setres{learnU}{CPT_post_m30}{r_prior}{0.9987}
\setres{learnU}{CPT_post_m30}{r_prob}{0.999}
\setres{learnU}{CPT_post_m30}{r_prob_se}{0.000}
\setres{learnU}{CPT_post_m30}{theta_mean}{0.94, 0.66, 0.85}
\setres{learnU}{CPT_post_m30}{theta_sd}{0.04, 0.05, 0.08}
\setres{learnU}{CPT_post_m300}{ess}{1}
\setres{learnU}{CPT_post_m300}{kappa_mean}{0.727}
\setres{learnU}{CPT_post_m300}{kappa_mean_se}{0.100}
\setres{learnU}{CPT_post_m300}{kappa_median}{0.935}
\setres{learnU}{CPT_post_m300}{kappa_prior_mean_median}{0.944}
\setres{learnU}{CPT_post_m300}{m}{300}
\setres{learnU}{CPT_post_m300}{n_countries}{30}
\setres{learnU}{CPT_post_m300}{n_pos}{29}
\setres{learnU}{CPT_post_m300}{n_subjects_seen}{300}
\setres{learnU}{CPT_post_m300}{r_ff}{0.199}
\setres{learnU}{CPT_post_m300}{r_prior}{0.9999}
\setres{learnU}{CPT_post_m300}{r_prob}{1.000}
\setres{learnU}{CPT_post_m300}{r_prob_se}{0.000}
\setres{learnU}{CPT_post_m300}{theta_mean}{0.97, 0.59, 0.90}
\setres{learnU}{CPT_post_m300}{theta_sd}{0.01, 0.02, 0.01}
\setres{learnU}{CPT_post_mall}{ess}{1}
\setres{learnU}{CPT_post_mall}{kappa_mean}{0.758}
\setres{learnU}{CPT_post_mall}{kappa_mean_se}{0.090}
\setres{learnU}{CPT_post_mall}{kappa_median}{0.944}
\setres{learnU}{CPT_post_mall}{kappa_prior_mean_median}{0.950}
\setres{learnU}{CPT_post_mall}{m}{all}
\setres{learnU}{CPT_post_mall}{n_countries}{30}
\setres{learnU}{CPT_post_mall}{n_pos}{29}
\setres{learnU}{CPT_post_mall}{n_subjects_seen}{2{,}939}
\setres{learnU}{CPT_post_mall}{r_ff}{0.199}
\setres{learnU}{CPT_post_mall}{r_prior}{0.9999}
\setres{learnU}{CPT_post_mall}{r_prob}{1.000}
\setres{learnU}{CPT_post_mall}{r_prob_se}{0.000}
\setres{learnU}{CPT_post_mall}{theta_mean}{0.98, 0.63, 0.90}
\setres{learnU}{CPT_post_mall}{theta_sd}{0.00, 0.00, 0.00}
\setres{learnU}{DA_post_m0}{ess}{1{,}600}
\setres{learnU}{DA_post_m0}{kappa_mean}{-3.931}
\setres{learnU}{DA_post_m0}{kappa_mean_se}{0.629}
\setres{learnU}{DA_post_m0}{kappa_median}{-2.153}
\setres{learnU}{DA_post_m0}{kappa_prior_mean_median}{-0.488}
\setres{learnU}{DA_post_m0}{m}{0}
\setres{learnU}{DA_post_m0}{n_countries}{30}
\setres{learnU}{DA_post_m0}{n_pos}{0}
\setres{learnU}{DA_post_m0}{n_subjects_seen}{0}
\setres{learnU}{DA_post_m0}{r_ff}{0.314}
\setres{learnU}{DA_post_m0}{r_prior}{0.7815}
\setres{learnU}{DA_post_m0}{r_prob}{0.850}
\setres{learnU}{DA_post_m0}{r_prob_se}{0.001}
\setres{learnU}{DA_post_m0}{theta_mean}{0.50, 1.00}
\setres{learnU}{DA_post_m0}{theta_sd}{0.29, 1.15}
\setres{learnU}{DA_post_m10}{ess}{23}
\setres{learnU}{DA_post_m10}{kappa_mean}{0.147}
\setres{learnU}{DA_post_m10}{kappa_mean_se}{0.251}
\setres{learnU}{DA_post_m10}{kappa_median}{0.690}
\setres{learnU}{DA_post_m10}{kappa_prior_mean_median}{0.705}
\setres{learnU}{DA_post_m10}{m}{10}
\setres{learnU}{DA_post_m10}{n_countries}{30}
\setres{learnU}{DA_post_m10}{n_pos}{29}
\setres{learnU}{DA_post_m10}{n_subjects_seen}{10}
\setres{learnU}{DA_post_m10}{r_ff}{0.314}
\setres{learnU}{DA_post_m10}{r_prior}{0.9958}
\setres{learnU}{DA_post_m10}{r_prob}{0.997}
\setres{learnU}{DA_post_m10}{r_prob_se}{0.000}
\setres{learnU}{DA_post_m10}{theta_mean}{0.94, 0.25}
\setres{learnU}{DA_post_m10}{theta_sd}{0.06, 0.15}
\setres{learnU}{DA_post_m100}{ess}{1}
\setres{learnU}{DA_post_m100}{kappa_mean}{0.397}
\setres{learnU}{DA_post_m100}{kappa_mean_se}{0.200}
\setres{learnU}{DA_post_m100}{kappa_median}{0.858}
\setres{learnU}{DA_post_m100}{kappa_prior_mean_median}{0.860}
\setres{learnU}{DA_post_m100}{m}{100}
\setres{learnU}{DA_post_m100}{n_countries}{30}
\setres{learnU}{DA_post_m100}{n_pos}{29}
\setres{learnU}{DA_post_m100}{n_subjects_seen}{100}
\setres{learnU}{DA_post_m100}{r_ff}{0.314}
\setres{learnU}{DA_post_m100}{r_prior}{0.9998}
\setres{learnU}{DA_post_m100}{r_prob}{1.000}
\setres{learnU}{DA_post_m100}{r_prob_se}{0.000}
\setres{learnU}{DA_post_m100}{theta_mean}{0.99, 0.15}
\setres{learnU}{DA_post_m100}{theta_sd}{0.01, 0.02}
\setres{learnU}{DA_post_m1000}{ess}{1}
\setres{learnU}{DA_post_m1000}{kappa_mean}{0.395}
\setres{learnU}{DA_post_m1000}{kappa_mean_se}{0.201}
\setres{learnU}{DA_post_m1000}{kappa_median}{0.860}
\setres{learnU}{DA_post_m1000}{kappa_prior_mean_median}{0.862}
\setres{learnU}{DA_post_m1000}{m}{1000}
\setres{learnU}{DA_post_m1000}{n_countries}{30}
\setres{learnU}{DA_post_m1000}{n_pos}{29}
\setres{learnU}{DA_post_m1000}{n_subjects_seen}{1{,}000}
\setres{learnU}{DA_post_m1000}{r_ff}{0.314}
\setres{learnU}{DA_post_m1000}{r_prior}{0.9999}
\setres{learnU}{DA_post_m1000}{r_prob}{1.000}
\setres{learnU}{DA_post_m1000}{r_prob_se}{0.000}
\setres{learnU}{DA_post_m1000}{theta_mean}{0.99, 0.15}
\setres{learnU}{DA_post_m1000}{theta_sd}{0.00, 0.00}
\setres{learnU}{DA_post_m3}{ess}{79}
\setres{learnU}{DA_post_m3}{kappa_mean}{-0.298}
\setres{learnU}{DA_post_m3}{kappa_mean_se}{0.324}
\setres{learnU}{DA_post_m3}{kappa_median}{0.527}
\setres{learnU}{DA_post_m3}{kappa_prior_mean_median}{0.584}
\setres{learnU}{DA_post_m3}{m}{3}
\setres{learnU}{DA_post_m3}{n_countries}{30}
\setres{learnU}{DA_post_m3}{n_pos}{23}
\setres{learnU}{DA_post_m3}{n_subjects_seen}{3}
\setres{learnU}{DA_post_m3}{r_ff}{0.314}
\setres{learnU}{DA_post_m3}{r_prior}{0.9846}
\setres{learnU}{DA_post_m3}{r_prob}{0.989}
\setres{learnU}{DA_post_m3}{r_prob_se}{0.000}
\setres{learnU}{DA_post_m3}{theta_mean}{0.88, 0.23}
\setres{learnU}{DA_post_m3}{theta_sd}{0.10, 0.28}
\setres{learnU}{DA_post_m30}{ess}{6}
\setres{learnU}{DA_post_m30}{kappa_mean}{0.317}
\setres{learnU}{DA_post_m30}{kappa_mean_se}{0.217}
\setres{learnU}{DA_post_m30}{kappa_median}{0.824}
\setres{learnU}{DA_post_m30}{kappa_prior_mean_median}{0.830}
\setres{learnU}{DA_post_m30}{m}{30}
\setres{learnU}{DA_post_m30}{n_countries}{30}
\setres{learnU}{DA_post_m30}{n_pos}{29}
\setres{learnU}{DA_post_m30}{n_subjects_seen}{30}
\setres{learnU}{DA_post_m30}{r_ff}{0.314}
\setres{learnU}{DA_post_m30}{r_prior}{0.9990}
\setres{learnU}{DA_post_m30}{r_prob}{0.999}
\setres{learnU}{DA_post_m30}{r_prob_se}{0.000}
\setres{learnU}{DA_post_m30}{theta_mean}{0.97, 0.30}
\setres{learnU}{DA_post_m30}{theta_sd}{0.02, 0.08}
\setres{learnU}{DA_post_m300}{ess}{1}
\setres{learnU}{DA_post_m300}{kappa_mean}{0.395}
\setres{learnU}{DA_post_m300}{kappa_mean_se}{0.201}
\setres{learnU}{DA_post_m300}{kappa_median}{0.858}
\setres{learnU}{DA_post_m300}{kappa_prior_mean_median}{0.860}
\setres{learnU}{DA_post_m300}{m}{300}
\setres{learnU}{DA_post_m300}{n_countries}{30}
\setres{learnU}{DA_post_m300}{n_pos}{29}
\setres{learnU}{DA_post_m300}{n_subjects_seen}{300}
\setres{learnU}{DA_post_m300}{r_ff}{0.314}
\setres{learnU}{DA_post_m300}{r_prior}{0.9999}
\setres{learnU}{DA_post_m300}{r_prob}{1.000}
\setres{learnU}{DA_post_m300}{r_prob_se}{0.000}
\setres{learnU}{DA_post_m300}{theta_mean}{0.99, 0.15}
\setres{learnU}{DA_post_m300}{theta_sd}{0.00, 0.00}
\setres{learnU}{DA_post_mall}{ess}{1}
\setres{learnU}{DA_post_mall}{kappa_mean}{0.417}
\setres{learnU}{DA_post_mall}{kappa_mean_se}{0.196}
\setres{learnU}{DA_post_mall}{kappa_median}{0.861}
\setres{learnU}{DA_post_mall}{kappa_prior_mean_median}{0.862}
\setres{learnU}{DA_post_mall}{m}{all}
\setres{learnU}{DA_post_mall}{n_countries}{30}
\setres{learnU}{DA_post_mall}{n_pos}{29}
\setres{learnU}{DA_post_mall}{n_subjects_seen}{2{,}939}
\setres{learnU}{DA_post_mall}{r_ff}{0.314}
\setres{learnU}{DA_post_mall}{r_prior}{0.9999}
\setres{learnU}{DA_post_mall}{r_prob}{1.000}
\setres{learnU}{DA_post_mall}{r_prob_se}{0.000}
\setres{learnU}{DA_post_mall}{theta_mean}{0.99, 0.15}
\setres{learnU}{DA_post_mall}{theta_sd}{0.00, 0.00}
\setres{learnU}{EU_post_m0}{ess}{200}
\setres{learnU}{EU_post_m0}{kappa_mean}{-2.334}
\setres{learnU}{EU_post_m0}{kappa_mean_se}{0.484}
\setres{learnU}{EU_post_m0}{kappa_median}{-0.979}
\setres{learnU}{EU_post_m0}{kappa_prior_mean_median}{-0.116}
\setres{learnU}{EU_post_m0}{m}{0}
\setres{learnU}{EU_post_m0}{n_countries}{30}
\setres{learnU}{EU_post_m0}{n_pos}{0}
\setres{learnU}{EU_post_m0}{n_subjects_seen}{0}
\setres{learnU}{EU_post_m0}{r_ff}{0.367}
\setres{learnU}{EU_post_m0}{r_prior}{0.5250}
\setres{learnU}{EU_post_m0}{r_prob}{0.699}
\setres{learnU}{EU_post_m0}{r_prob_se}{0.002}
\setres{learnU}{EU_post_m0}{theta_mean}{0.51}
\setres{learnU}{EU_post_m0}{theta_sd}{0.29}
\setres{learnU}{EU_post_m10}{ess}{41}
\setres{learnU}{EU_post_m10}{kappa_mean}{0.229}
\setres{learnU}{EU_post_m10}{kappa_mean_se}{0.251}
\setres{learnU}{EU_post_m10}{kappa_median}{0.805}
\setres{learnU}{EU_post_m10}{kappa_prior_mean_median}{0.813}
\setres{learnU}{EU_post_m10}{m}{10}
\setres{learnU}{EU_post_m10}{n_countries}{30}
\setres{learnU}{EU_post_m10}{n_pos}{29}
\setres{learnU}{EU_post_m10}{n_subjects_seen}{10}
\setres{learnU}{EU_post_m10}{r_ff}{0.367}
\setres{learnU}{EU_post_m10}{r_prior}{0.9299}
\setres{learnU}{EU_post_m10}{r_prob}{0.956}
\setres{learnU}{EU_post_m10}{r_prob_se}{0.000}
\setres{learnU}{EU_post_m10}{theta_mean}{0.86}
\setres{learnU}{EU_post_m10}{theta_sd}{0.06}
\setres{learnU}{EU_post_m100}{ess}{14}
\setres{learnU}{EU_post_m100}{kappa_mean}{0.383}
\setres{learnU}{EU_post_m100}{kappa_mean_se}{0.203}
\setres{learnU}{EU_post_m100}{kappa_median}{0.829}
\setres{learnU}{EU_post_m100}{kappa_prior_mean_median}{0.830}
\setres{learnU}{EU_post_m100}{m}{100}
\setres{learnU}{EU_post_m100}{n_countries}{30}
\setres{learnU}{EU_post_m100}{n_pos}{29}
\setres{learnU}{EU_post_m100}{n_subjects_seen}{100}
\setres{learnU}{EU_post_m100}{r_ff}{0.367}
\setres{learnU}{EU_post_m100}{r_prior}{0.9769}
\setres{learnU}{EU_post_m100}{r_prob}{0.985}
\setres{learnU}{EU_post_m100}{r_prob_se}{0.000}
\setres{learnU}{EU_post_m100}{theta_mean}{0.93}
\setres{learnU}{EU_post_m100}{theta_sd}{0.02}
\setres{learnU}{EU_post_m1000}{ess}{5}
\setres{learnU}{EU_post_m1000}{kappa_mean}{0.380}
\setres{learnU}{EU_post_m1000}{kappa_mean_se}{0.204}
\setres{learnU}{EU_post_m1000}{kappa_median}{0.822}
\setres{learnU}{EU_post_m1000}{kappa_prior_mean_median}{0.822}
\setres{learnU}{EU_post_m1000}{m}{1000}
\setres{learnU}{EU_post_m1000}{n_countries}{30}
\setres{learnU}{EU_post_m1000}{n_pos}{29}
\setres{learnU}{EU_post_m1000}{n_subjects_seen}{1{,}000}
\setres{learnU}{EU_post_m1000}{r_ff}{0.367}
\setres{learnU}{EU_post_m1000}{r_prior}{0.9926}
\setres{learnU}{EU_post_m1000}{r_prob}{0.995}
\setres{learnU}{EU_post_m1000}{r_prob_se}{0.000}
\setres{learnU}{EU_post_m1000}{theta_mean}{0.92}
\setres{learnU}{EU_post_m1000}{theta_sd}{0.01}
\setres{learnU}{EU_post_m3}{ess}{62}
\setres{learnU}{EU_post_m3}{kappa_mean}{-0.100}
\setres{learnU}{EU_post_m3}{kappa_mean_se}{0.318}
\setres{learnU}{EU_post_m3}{kappa_median}{0.746}
\setres{learnU}{EU_post_m3}{kappa_prior_mean_median}{0.767}
\setres{learnU}{EU_post_m3}{m}{3}
\setres{learnU}{EU_post_m3}{n_countries}{30}
\setres{learnU}{EU_post_m3}{n_pos}{27}
\setres{learnU}{EU_post_m3}{n_subjects_seen}{3}
\setres{learnU}{EU_post_m3}{r_ff}{0.367}
\setres{learnU}{EU_post_m3}{r_prior}{0.8970}
\setres{learnU}{EU_post_m3}{r_prob}{0.935}
\setres{learnU}{EU_post_m3}{r_prob_se}{0.000}
\setres{learnU}{EU_post_m3}{theta_mean}{0.85}
\setres{learnU}{EU_post_m3}{theta_sd}{0.09}
\setres{learnU}{EU_post_m30}{ess}{24}
\setres{learnU}{EU_post_m30}{kappa_mean}{0.318}
\setres{learnU}{EU_post_m30}{kappa_mean_se}{0.224}
\setres{learnU}{EU_post_m30}{kappa_median}{0.813}
\setres{learnU}{EU_post_m30}{kappa_prior_mean_median}{0.816}
\setres{learnU}{EU_post_m30}{m}{30}
\setres{learnU}{EU_post_m30}{n_countries}{30}
\setres{learnU}{EU_post_m30}{n_pos}{29}
\setres{learnU}{EU_post_m30}{n_subjects_seen}{30}
\setres{learnU}{EU_post_m30}{r_ff}{0.367}
\setres{learnU}{EU_post_m30}{r_prior}{0.9575}
\setres{learnU}{EU_post_m30}{r_prob}{0.973}
\setres{learnU}{EU_post_m30}{r_prob_se}{0.000}
\setres{learnU}{EU_post_m30}{theta_mean}{0.85}
\setres{learnU}{EU_post_m30}{theta_sd}{0.03}
\setres{learnU}{EU_post_m300}{ess}{8}
\setres{learnU}{EU_post_m300}{kappa_mean}{0.377}
\setres{learnU}{EU_post_m300}{kappa_mean_se}{0.206}
\setres{learnU}{EU_post_m300}{kappa_median}{0.827}
\setres{learnU}{EU_post_m300}{kappa_prior_mean_median}{0.828}
\setres{learnU}{EU_post_m300}{m}{300}
\setres{learnU}{EU_post_m300}{n_countries}{30}
\setres{learnU}{EU_post_m300}{n_pos}{29}
\setres{learnU}{EU_post_m300}{n_subjects_seen}{300}
\setres{learnU}{EU_post_m300}{r_ff}{0.367}
\setres{learnU}{EU_post_m300}{r_prior}{0.9865}
\setres{learnU}{EU_post_m300}{r_prob}{0.991}
\setres{learnU}{EU_post_m300}{r_prob_se}{0.000}
\setres{learnU}{EU_post_m300}{theta_mean}{0.93}
\setres{learnU}{EU_post_m300}{theta_sd}{0.01}
\setres{learnU}{EU_post_mall}{ess}{3}
\setres{learnU}{EU_post_mall}{kappa_mean}{0.398}
\setres{learnU}{EU_post_mall}{kappa_mean_se}{0.199}
\setres{learnU}{EU_post_mall}{kappa_median}{0.810}
\setres{learnU}{EU_post_mall}{kappa_prior_mean_median}{0.811}
\setres{learnU}{EU_post_mall}{m}{all}
\setres{learnU}{EU_post_mall}{n_countries}{30}
\setres{learnU}{EU_post_mall}{n_pos}{29}
\setres{learnU}{EU_post_mall}{n_subjects_seen}{2{,}939}
\setres{learnU}{EU_post_mall}{r_ff}{0.367}
\setres{learnU}{EU_post_mall}{r_prior}{0.9957}
\setres{learnU}{EU_post_mall}{r_prob}{0.997}
\setres{learnU}{EU_post_mall}{r_prob_se}{0.000}
\setres{learnU}{EU_post_mall}{theta_mean}{0.95}
\setres{learnU}{EU_post_mall}{theta_sd}{0.00}
\setres{learnU}{reference}{kind}{common}
\setres{learnU}{reference}{m_grid}{0, 3, 10, 30, 100, 300, 1000, all}
\setres{learnU}{reference}{n_countries}{30}
\setres{learnU}{reference}{start}{uniform}
\setres{learnM}{CPT}{kappa_fit}{0.971}
\setres{learnM}{CPT}{kappa_lit}{0.489}
\setres{learnM}{CPT}{kappa_m0}{-1.545}
\setres{learnM}{CPT}{kappa_m10}{0.876}
\setres{learnM}{CPT}{kappa_m100}{0.962}
\setres{learnM}{CPT}{kappa_m3}{0.672}
\setres{learnM}{CPT}{kappa_m30}{0.944}
\setres{learnM}{CPT}{kappa_mall}{0.962}
\setres{learnM}{CPT}{m_within_fit}{100}
\setres{learnM}{CPT}{m_within_lit}{3}
\setres{learnM}{CPT}{n_countries}{30}
\setres{learnM}{CPT}{r_prob_lit}{0.935}
\setres{learnM}{CPT}{r_prob_m0}{0.611}
\setres{learnM}{CPT}{r_prob_m10}{0.996}
\setres{learnM}{CPT}{r_prob_m100}{1.000}
\setres{learnM}{CPT}{r_prob_m3}{0.984}
\setres{learnM}{CPT}{r_prob_m30}{0.999}
\setres{learnM}{CPT}{r_prob_mall}{1.000}
\setres{learnM}{CPT}{share_all_fit}{77}
\setres{learnM}{CPT}{share_all_lit}{100}
\setres{learnM}{CPT}{share_m3_fit}{0}
\setres{learnM}{CPT}{share_m3_lit}{83}
\setres{learnM}{CPT}{z_kappa_fit}{0.972}
\setres{learnM}{CPT}{z_kappa_lit}{0.639}
\setres{learnM}{CPT}{z_kappa_m0}{-0.459}
\setres{learnM}{CPT}{z_kappa_m10}{0.928}
\setres{learnM}{CPT}{z_kappa_m100}{0.978}
\setres{learnM}{CPT}{z_kappa_m3}{0.752}
\setres{learnM}{CPT}{z_kappa_m30}{0.954}
\setres{learnM}{CPT}{z_kappa_mall}{0.977}
\setres{learnM}{CPT}{z_m_within_fit}{100}
\setres{learnM}{CPT}{z_m_within_lit}{3}
\setres{learnM}{CPT}{z_r_prob_lit}{0.955}
\setres{learnM}{CPT}{z_r_prob_m0}{0.582}
\setres{learnM}{CPT}{z_r_prob_m10}{0.998}
\setres{learnM}{CPT}{z_r_prob_m100}{1.000}
\setres{learnM}{CPT}{z_r_prob_m3}{0.991}
\setres{learnM}{CPT}{z_r_prob_m30}{0.999}
\setres{learnM}{CPT}{z_r_prob_mall}{1.000}
\setres{learnM}{DA}{kappa_fit}{0.917}
\setres{learnM}{DA}{kappa_lit}{-0.434}
\setres{learnM}{DA}{kappa_m0}{-2.182}
\setres{learnM}{DA}{kappa_m10}{0.801}
\setres{learnM}{DA}{kappa_m100}{0.869}
\setres{learnM}{DA}{kappa_m3}{0.644}
\setres{learnM}{DA}{kappa_m30}{0.849}
\setres{learnM}{DA}{kappa_mall}{0.869}
\setres{learnM}{DA}{m_within_fit}{never}
\setres{learnM}{DA}{m_within_lit}{3}
\setres{learnM}{DA}{n_countries}{30}
\setres{learnM}{DA}{r_prob_lit}{0.920}
\setres{learnM}{DA}{r_prob_m0}{0.850}
\setres{learnM}{DA}{r_prob_m10}{0.997}
\setres{learnM}{DA}{r_prob_m100}{1.000}
\setres{learnM}{DA}{r_prob_m3}{0.989}
\setres{learnM}{DA}{r_prob_m30}{0.999}
\setres{learnM}{DA}{r_prob_mall}{1.000}
\setres{learnM}{DA}{share_all_fit}{37}
\setres{learnM}{DA}{share_all_lit}{100}
\setres{learnM}{DA}{share_m3_fit}{0}
\setres{learnM}{DA}{share_m3_lit}{100}
\setres{learnM}{DA}{z_kappa_fit}{0.626}
\setres{learnM}{DA}{z_kappa_lit}{0.018}
\setres{learnM}{DA}{z_kappa_m0}{-0.867}
\setres{learnM}{DA}{z_kappa_m10}{0.591}
\setres{learnM}{DA}{z_kappa_m100}{0.624}
\setres{learnM}{DA}{z_kappa_m3}{0.432}
\setres{learnM}{DA}{z_kappa_m30}{0.587}
\setres{learnM}{DA}{z_kappa_mall}{0.624}
\setres{learnM}{DA}{z_m_within_fit}{10}
\setres{learnM}{DA}{z_m_within_lit}{3}
\setres{learnM}{DA}{z_r_prob_lit}{0.955}
\setres{learnM}{DA}{z_r_prob_m0}{0.903}
\setres{learnM}{DA}{z_r_prob_m10}{0.996}
\setres{learnM}{DA}{z_r_prob_m100}{1.000}
\setres{learnM}{DA}{z_r_prob_m3}{0.993}
\setres{learnM}{DA}{z_r_prob_m30}{0.998}
\setres{learnM}{DA}{z_r_prob_mall}{1.000}
\setres{learnM}{EU}{kappa_fit}{0.840}
\setres{learnM}{EU}{kappa_lit}{0.518}
\setres{learnM}{EU}{kappa_m0}{-0.948}
\setres{learnM}{EU}{kappa_m10}{0.816}
\setres{learnM}{EU}{kappa_m100}{0.838}
\setres{learnM}{EU}{kappa_m3}{0.767}
\setres{learnM}{EU}{kappa_m30}{0.828}
\setres{learnM}{EU}{kappa_mall}{0.837}
\setres{learnM}{EU}{m_within_fit}{10}
\setres{learnM}{EU}{m_within_lit}{3}
\setres{learnM}{EU}{n_countries}{30}
\setres{learnM}{EU}{r_prob_lit}{0.870}
\setres{learnM}{EU}{r_prob_m0}{0.699}
\setres{learnM}{EU}{r_prob_m10}{0.975}
\setres{learnM}{EU}{r_prob_m100}{0.989}
\setres{learnM}{EU}{r_prob_m3}{0.966}
\setres{learnM}{EU}{r_prob_m30}{0.982}
\setres{learnM}{EU}{r_prob_mall}{0.989}
\setres{learnM}{EU}{share_all_fit}{90}
\setres{learnM}{EU}{share_all_lit}{100}
\setres{learnM}{EU}{share_m3_fit}{43}
\setres{learnM}{EU}{share_m3_lit}{100}
\setres{learnM}{EU}{z_kappa_fit}{0.616}
\setres{learnM}{EU}{z_kappa_lit}{0.522}
\setres{learnM}{EU}{z_kappa_m0}{-0.195}
\setres{learnM}{EU}{z_kappa_m10}{0.600}
\setres{learnM}{EU}{z_kappa_m100}{0.616}
\setres{learnM}{EU}{z_kappa_m3}{0.533}
\setres{learnM}{EU}{z_kappa_m30}{0.591}
\setres{learnM}{EU}{z_kappa_mall}{0.616}
\setres{learnM}{EU}{z_m_within_fit}{10}
\setres{learnM}{EU}{z_m_within_lit}{3}
\setres{learnM}{EU}{z_r_prob_lit}{0.877}
\setres{learnM}{EU}{z_r_prob_m0}{0.720}
\setres{learnM}{EU}{z_r_prob_m10}{0.972}
\setres{learnM}{EU}{z_r_prob_m100}{0.991}
\setres{learnM}{EU}{z_r_prob_m3}{0.981}
\setres{learnM}{EU}{z_r_prob_m30}{0.983}
\setres{learnM}{EU}{z_r_prob_mall}{0.993}
\setres{learnM}{reference}{m_grid}{0, 3, 10, 30, 100, all}
\setres{learnM}{reference}{n_countries}{30}
\setres{learnO}{CPT_post_m0}{kappa_mean}{-0.050}
\setres{learnO}{CPT_post_m0}{kappa_mean_se}{1.525}
\setres{learnO}{CPT_post_m0}{kappa_median}{0.489}
\setres{learnO}{CPT_post_m0}{kappa_prior_mean_median}{0.783}
\setres{learnO}{CPT_post_m0}{m}{0}
\setres{learnO}{CPT_post_m0}{n_countries}{30}
\setres{learnO}{CPT_post_m0}{n_pos}{23}
\setres{learnO}{CPT_post_m0}{n_subjects_seen}{0}
\setres{learnO}{CPT_post_m0}{r_ff}{0.199}
\setres{learnO}{CPT_post_m0}{r_prior}{0.9191}
\setres{learnO}{CPT_post_m0}{r_prob}{0.935}
\setres{learnO}{CPT_post_m10}{kappa_mean}{0.828}
\setres{learnO}{CPT_post_m10}{kappa_mean_se}{0.171}
\setres{learnO}{CPT_post_m10}{kappa_median}{0.876}
\setres{learnO}{CPT_post_m10}{kappa_prior_mean_median}{0.903}
\setres{learnO}{CPT_post_m10}{m}{10}
\setres{learnO}{CPT_post_m10}{n_countries}{30}
\setres{learnO}{CPT_post_m10}{n_pos}{29}
\setres{learnO}{CPT_post_m10}{n_subjects_seen}{10}
\setres{learnO}{CPT_post_m10}{r_ff}{0.199}
\setres{learnO}{CPT_post_m10}{r_prior}{0.9958}
\setres{learnO}{CPT_post_m10}{r_prob}{0.997}
\setres{learnO}{CPT_post_m100}{kappa_mean}{0.949}
\setres{learnO}{CPT_post_m100}{kappa_mean_se}{0.055}
\setres{learnO}{CPT_post_m100}{kappa_median}{0.963}
\setres{learnO}{CPT_post_m100}{kappa_prior_mean_median}{0.970}
\setres{learnO}{CPT_post_m100}{m}{100}
\setres{learnO}{CPT_post_m100}{n_countries}{30}
\setres{learnO}{CPT_post_m100}{n_pos}{30}
\setres{learnO}{CPT_post_m100}{n_subjects_seen}{87}
\setres{learnO}{CPT_post_m100}{r_ff}{0.199}
\setres{learnO}{CPT_post_m100}{r_prior}{0.9997}
\setres{learnO}{CPT_post_m100}{r_prob}{1.000}
\setres{learnO}{CPT_post_m1000}{kappa_mean}{0.952}
\setres{learnO}{CPT_post_m1000}{kappa_mean_se}{0.019}
\setres{learnO}{CPT_post_m1000}{kappa_median}{0.963}
\setres{learnO}{CPT_post_m1000}{kappa_prior_mean_median}{0.970}
\setres{learnO}{CPT_post_m1000}{m}{1000}
\setres{learnO}{CPT_post_m1000}{n_countries}{30}
\setres{learnO}{CPT_post_m1000}{n_pos}{30}
\setres{learnO}{CPT_post_m1000}{n_subjects_seen}{98}
\setres{learnO}{CPT_post_m1000}{r_ff}{0.199}
\setres{learnO}{CPT_post_m1000}{r_prior}{0.9997}
\setres{learnO}{CPT_post_m1000}{r_prob}{1.000}
\setres{learnO}{CPT_post_m3}{kappa_mean}{0.340}
\setres{learnO}{CPT_post_m3}{kappa_mean_se}{1.318}
\setres{learnO}{CPT_post_m3}{kappa_median}{0.769}
\setres{learnO}{CPT_post_m3}{kappa_prior_mean_median}{0.818}
\setres{learnO}{CPT_post_m3}{m}{3}
\setres{learnO}{CPT_post_m3}{n_countries}{30}
\setres{learnO}{CPT_post_m3}{n_pos}{29}
\setres{learnO}{CPT_post_m3}{n_subjects_seen}{3}
\setres{learnO}{CPT_post_m3}{r_ff}{0.199}
\setres{learnO}{CPT_post_m3}{r_prior}{0.9890}
\setres{learnO}{CPT_post_m3}{r_prob}{0.991}
\setres{learnO}{CPT_post_m30}{kappa_mean}{0.852}
\setres{learnO}{CPT_post_m30}{kappa_mean_se}{0.486}
\setres{learnO}{CPT_post_m30}{kappa_median}{0.943}
\setres{learnO}{CPT_post_m30}{kappa_prior_mean_median}{0.952}
\setres{learnO}{CPT_post_m30}{m}{30}
\setres{learnO}{CPT_post_m30}{n_countries}{30}
\setres{learnO}{CPT_post_m30}{n_pos}{29}
\setres{learnO}{CPT_post_m30}{n_subjects_seen}{30}
\setres{learnO}{CPT_post_m30}{r_ff}{0.199}
\setres{learnO}{CPT_post_m30}{r_prior}{0.9990}
\setres{learnO}{CPT_post_m30}{r_prob}{0.999}
\setres{learnO}{CPT_post_m300}{kappa_mean}{0.952}
\setres{learnO}{CPT_post_m300}{kappa_mean_se}{0.017}
\setres{learnO}{CPT_post_m300}{kappa_median}{0.963}
\setres{learnO}{CPT_post_m300}{kappa_prior_mean_median}{0.970}
\setres{learnO}{CPT_post_m300}{m}{300}
\setres{learnO}{CPT_post_m300}{n_countries}{30}
\setres{learnO}{CPT_post_m300}{n_pos}{30}
\setres{learnO}{CPT_post_m300}{n_subjects_seen}{98}
\setres{learnO}{CPT_post_m300}{r_ff}{0.199}
\setres{learnO}{CPT_post_m300}{r_prior}{0.9997}
\setres{learnO}{CPT_post_m300}{r_prob}{1.000}
\setres{learnO}{CPT_post_mall}{kappa_mean}{0.952}
\setres{learnO}{CPT_post_mall}{kappa_mean_se}{0.019}
\setres{learnO}{CPT_post_mall}{kappa_median}{0.963}
\setres{learnO}{CPT_post_mall}{kappa_prior_mean_median}{0.970}
\setres{learnO}{CPT_post_mall}{m}{all}
\setres{learnO}{CPT_post_mall}{n_countries}{30}
\setres{learnO}{CPT_post_mall}{n_pos}{30}
\setres{learnO}{CPT_post_mall}{n_subjects_seen}{98}
\setres{learnO}{CPT_post_mall}{r_ff}{0.199}
\setres{learnO}{CPT_post_mall}{r_prior}{0.9997}
\setres{learnO}{CPT_post_mall}{r_prob}{1.000}
\setres{learnO}{DA_post_m0}{kappa_mean}{-1.622}
\setres{learnO}{DA_post_m0}{kappa_mean_se}{3.221}
\setres{learnO}{DA_post_m0}{kappa_median}{-0.434}
\setres{learnO}{DA_post_m0}{kappa_prior_mean_median}{0.099}
\setres{learnO}{DA_post_m0}{m}{0}
\setres{learnO}{DA_post_m0}{n_countries}{30}
\setres{learnO}{DA_post_m0}{n_pos}{3}
\setres{learnO}{DA_post_m0}{n_subjects_seen}{0}
\setres{learnO}{DA_post_m0}{r_ff}{0.314}
\setres{learnO}{DA_post_m0}{r_prior}{0.8842}
\setres{learnO}{DA_post_m0}{r_prob}{0.920}
\setres{learnO}{DA_post_m10}{kappa_mean}{0.333}
\setres{learnO}{DA_post_m10}{kappa_mean_se}{1.668}
\setres{learnO}{DA_post_m10}{kappa_median}{0.800}
\setres{learnO}{DA_post_m10}{kappa_prior_mean_median}{0.813}
\setres{learnO}{DA_post_m10}{m}{10}
\setres{learnO}{DA_post_m10}{n_countries}{30}
\setres{learnO}{DA_post_m10}{n_pos}{29}
\setres{learnO}{DA_post_m10}{n_subjects_seen}{10}
\setres{learnO}{DA_post_m10}{r_ff}{0.314}
\setres{learnO}{DA_post_m10}{r_prior}{0.9948}
\setres{learnO}{DA_post_m10}{r_prob}{0.996}
\setres{learnO}{DA_post_m100}{kappa_mean}{0.470}
\setres{learnO}{DA_post_m100}{kappa_mean_se}{1.485}
\setres{learnO}{DA_post_m100}{kappa_median}{0.868}
\setres{learnO}{DA_post_m100}{kappa_prior_mean_median}{0.871}
\setres{learnO}{DA_post_m100}{m}{100}
\setres{learnO}{DA_post_m100}{n_countries}{30}
\setres{learnO}{DA_post_m100}{n_pos}{29}
\setres{learnO}{DA_post_m100}{n_subjects_seen}{87}
\setres{learnO}{DA_post_m100}{r_ff}{0.314}
\setres{learnO}{DA_post_m100}{r_prior}{0.9996}
\setres{learnO}{DA_post_m100}{r_prob}{1.000}
\setres{learnO}{DA_post_m1000}{kappa_mean}{0.471}
\setres{learnO}{DA_post_m1000}{kappa_mean_se}{1.443}
\setres{learnO}{DA_post_m1000}{kappa_median}{0.869}
\setres{learnO}{DA_post_m1000}{kappa_prior_mean_median}{0.871}
\setres{learnO}{DA_post_m1000}{m}{1000}
\setres{learnO}{DA_post_m1000}{n_countries}{30}
\setres{learnO}{DA_post_m1000}{n_pos}{29}
\setres{learnO}{DA_post_m1000}{n_subjects_seen}{98}
\setres{learnO}{DA_post_m1000}{r_ff}{0.314}
\setres{learnO}{DA_post_m1000}{r_prior}{0.9996}
\setres{learnO}{DA_post_m1000}{r_prob}{1.000}
\setres{learnO}{DA_post_m3}{kappa_mean}{-0.217}
\setres{learnO}{DA_post_m3}{kappa_mean_se}{2.639}
\setres{learnO}{DA_post_m3}{kappa_median}{0.652}
\setres{learnO}{DA_post_m3}{kappa_prior_mean_median}{0.698}
\setres{learnO}{DA_post_m3}{m}{3}
\setres{learnO}{DA_post_m3}{n_countries}{30}
\setres{learnO}{DA_post_m3}{n_pos}{27}
\setres{learnO}{DA_post_m3}{n_subjects_seen}{3}
\setres{learnO}{DA_post_m3}{r_ff}{0.314}
\setres{learnO}{DA_post_m3}{r_prior}{0.9827}
\setres{learnO}{DA_post_m3}{r_prob}{0.988}
\setres{learnO}{DA_post_m30}{kappa_mean}{0.346}
\setres{learnO}{DA_post_m30}{kappa_mean_se}{2.056}
\setres{learnO}{DA_post_m30}{kappa_median}{0.843}
\setres{learnO}{DA_post_m30}{kappa_prior_mean_median}{0.849}
\setres{learnO}{DA_post_m30}{m}{30}
\setres{learnO}{DA_post_m30}{n_countries}{30}
\setres{learnO}{DA_post_m30}{n_pos}{29}
\setres{learnO}{DA_post_m30}{n_subjects_seen}{30}
\setres{learnO}{DA_post_m30}{r_ff}{0.314}
\setres{learnO}{DA_post_m30}{r_prior}{0.9984}
\setres{learnO}{DA_post_m30}{r_prob}{0.999}
\setres{learnO}{DA_post_m300}{kappa_mean}{0.471}
\setres{learnO}{DA_post_m300}{kappa_mean_se}{1.443}
\setres{learnO}{DA_post_m300}{kappa_median}{0.869}
\setres{learnO}{DA_post_m300}{kappa_prior_mean_median}{0.871}
\setres{learnO}{DA_post_m300}{m}{300}
\setres{learnO}{DA_post_m300}{n_countries}{30}
\setres{learnO}{DA_post_m300}{n_pos}{29}
\setres{learnO}{DA_post_m300}{n_subjects_seen}{98}
\setres{learnO}{DA_post_m300}{r_ff}{0.314}
\setres{learnO}{DA_post_m300}{r_prior}{0.9996}
\setres{learnO}{DA_post_m300}{r_prob}{1.000}
\setres{learnO}{DA_post_mall}{kappa_mean}{0.471}
\setres{learnO}{DA_post_mall}{kappa_mean_se}{1.443}
\setres{learnO}{DA_post_mall}{kappa_median}{0.869}
\setres{learnO}{DA_post_mall}{kappa_prior_mean_median}{0.871}
\setres{learnO}{DA_post_mall}{m}{all}
\setres{learnO}{DA_post_mall}{n_countries}{30}
\setres{learnO}{DA_post_mall}{n_pos}{29}
\setres{learnO}{DA_post_mall}{n_subjects_seen}{98}
\setres{learnO}{DA_post_mall}{r_ff}{0.314}
\setres{learnO}{DA_post_mall}{r_prior}{0.9996}
\setres{learnO}{DA_post_mall}{r_prob}{1.000}
\setres{learnO}{EU_post_m0}{kappa_mean}{-0.168}
\setres{learnO}{EU_post_m0}{kappa_mean_se}{2.016}
\setres{learnO}{EU_post_m0}{kappa_median}{0.518}
\setres{learnO}{EU_post_m0}{kappa_prior_mean_median}{0.635}
\setres{learnO}{EU_post_m0}{m}{0}
\setres{learnO}{EU_post_m0}{n_countries}{30}
\setres{learnO}{EU_post_m0}{n_pos}{24}
\setres{learnO}{EU_post_m0}{n_subjects_seen}{0}
\setres{learnO}{EU_post_m0}{r_ff}{0.367}
\setres{learnO}{EU_post_m0}{r_prior}{0.7947}
\setres{learnO}{EU_post_m0}{r_prob}{0.870}
\setres{learnO}{EU_post_m10}{kappa_mean}{0.381}
\setres{learnO}{EU_post_m10}{kappa_mean_se}{1.460}
\setres{learnO}{EU_post_m10}{kappa_median}{0.820}
\setres{learnO}{EU_post_m10}{kappa_prior_mean_median}{0.829}
\setres{learnO}{EU_post_m10}{m}{10}
\setres{learnO}{EU_post_m10}{n_countries}{30}
\setres{learnO}{EU_post_m10}{n_pos}{29}
\setres{learnO}{EU_post_m10}{n_subjects_seen}{10}
\setres{learnO}{EU_post_m10}{r_ff}{0.367}
\setres{learnO}{EU_post_m10}{r_prior}{0.9578}
\setres{learnO}{EU_post_m10}{r_prob}{0.973}
\setres{learnO}{EU_post_m100}{kappa_mean}{0.445}
\setres{learnO}{EU_post_m100}{kappa_mean_se}{1.389}
\setres{learnO}{EU_post_m100}{kappa_median}{0.838}
\setres{learnO}{EU_post_m100}{kappa_prior_mean_median}{0.839}
\setres{learnO}{EU_post_m100}{m}{100}
\setres{learnO}{EU_post_m100}{n_countries}{30}
\setres{learnO}{EU_post_m100}{n_pos}{29}
\setres{learnO}{EU_post_m100}{n_subjects_seen}{87}
\setres{learnO}{EU_post_m100}{r_ff}{0.367}
\setres{learnO}{EU_post_m100}{r_prior}{0.9819}
\setres{learnO}{EU_post_m100}{r_prob}{0.989}
\setres{learnO}{EU_post_m1000}{kappa_mean}{0.447}
\setres{learnO}{EU_post_m1000}{kappa_mean_se}{1.387}
\setres{learnO}{EU_post_m1000}{kappa_median}{0.838}
\setres{learnO}{EU_post_m1000}{kappa_prior_mean_median}{0.839}
\setres{learnO}{EU_post_m1000}{m}{1000}
\setres{learnO}{EU_post_m1000}{n_countries}{30}
\setres{learnO}{EU_post_m1000}{n_pos}{29}
\setres{learnO}{EU_post_m1000}{n_subjects_seen}{98}
\setres{learnO}{EU_post_m1000}{r_ff}{0.367}
\setres{learnO}{EU_post_m1000}{r_prior}{0.9823}
\setres{learnO}{EU_post_m1000}{r_prob}{0.989}
\setres{learnO}{EU_post_m3}{kappa_mean}{0.092}
\setres{learnO}{EU_post_m3}{kappa_mean_se}{2.221}
\setres{learnO}{EU_post_m3}{kappa_median}{0.758}
\setres{learnO}{EU_post_m3}{kappa_prior_mean_median}{0.780}
\setres{learnO}{EU_post_m3}{m}{3}
\setres{learnO}{EU_post_m3}{n_countries}{30}
\setres{learnO}{EU_post_m3}{n_pos}{29}
\setres{learnO}{EU_post_m3}{n_subjects_seen}{3}
\setres{learnO}{EU_post_m3}{r_ff}{0.367}
\setres{learnO}{EU_post_m3}{r_prior}{0.9367}
\setres{learnO}{EU_post_m3}{r_prob}{0.960}
\setres{learnO}{EU_post_m30}{kappa_mean}{0.431}
\setres{learnO}{EU_post_m30}{kappa_mean_se}{1.394}
\setres{learnO}{EU_post_m30}{kappa_median}{0.832}
\setres{learnO}{EU_post_m30}{kappa_prior_mean_median}{0.835}
\setres{learnO}{EU_post_m30}{m}{30}
\setres{learnO}{EU_post_m30}{n_countries}{30}
\setres{learnO}{EU_post_m30}{n_pos}{29}
\setres{learnO}{EU_post_m30}{n_subjects_seen}{30}
\setres{learnO}{EU_post_m30}{r_ff}{0.367}
\setres{learnO}{EU_post_m30}{r_prior}{0.9724}
\setres{learnO}{EU_post_m30}{r_prob}{0.982}
\setres{learnO}{EU_post_m300}{kappa_mean}{0.447}
\setres{learnO}{EU_post_m300}{kappa_mean_se}{1.387}
\setres{learnO}{EU_post_m300}{kappa_median}{0.838}
\setres{learnO}{EU_post_m300}{kappa_prior_mean_median}{0.839}
\setres{learnO}{EU_post_m300}{m}{300}
\setres{learnO}{EU_post_m300}{n_countries}{30}
\setres{learnO}{EU_post_m300}{n_pos}{29}
\setres{learnO}{EU_post_m300}{n_subjects_seen}{98}
\setres{learnO}{EU_post_m300}{r_ff}{0.367}
\setres{learnO}{EU_post_m300}{r_prior}{0.9823}
\setres{learnO}{EU_post_m300}{r_prob}{0.989}
\setres{learnO}{EU_post_mall}{kappa_mean}{0.447}
\setres{learnO}{EU_post_mall}{kappa_mean_se}{1.387}
\setres{learnO}{EU_post_mall}{kappa_median}{0.838}
\setres{learnO}{EU_post_mall}{kappa_prior_mean_median}{0.839}
\setres{learnO}{EU_post_mall}{m}{all}
\setres{learnO}{EU_post_mall}{n_countries}{30}
\setres{learnO}{EU_post_mall}{n_pos}{29}
\setres{learnO}{EU_post_mall}{n_subjects_seen}{98}
\setres{learnO}{EU_post_mall}{r_ff}{0.367}
\setres{learnO}{EU_post_mall}{r_prior}{0.9823}
\setres{learnO}{EU_post_mall}{r_prob}{0.989}
\setres{learnO}{reference}{m_grid}{0, 3, 10, 30, 100, 300, 1000, all}
\setres{learnO}{reference}{n_countries}{30}
\setres{learnO}{reference}{source}{own}
\setres{learnO}{reference}{start}{lit}
\setres{learnUO}{CPT_post_m0}{kappa_mean}{-2.761}
\setres{learnUO}{CPT_post_m0}{kappa_mean_se}{3.412}
\setres{learnUO}{CPT_post_m0}{kappa_median}{-1.545}
\setres{learnUO}{CPT_post_m0}{kappa_prior_mean_median}{0.174}
\setres{learnUO}{CPT_post_m0}{m}{0}
\setres{learnUO}{CPT_post_m0}{n_countries}{30}
\setres{learnUO}{CPT_post_m0}{n_pos}{0}
\setres{learnUO}{CPT_post_m0}{n_subjects_seen}{0}
\setres{learnUO}{CPT_post_m0}{r_ff}{0.199}
\setres{learnUO}{CPT_post_m0}{r_prior}{0.5146}
\setres{learnUO}{CPT_post_m0}{r_prob}{0.611}
\setres{learnUO}{CPT_post_m10}{kappa_mean}{0.809}
\setres{learnUO}{CPT_post_m10}{kappa_mean_se}{0.198}
\setres{learnUO}{CPT_post_m10}{kappa_median}{0.876}
\setres{learnUO}{CPT_post_m10}{kappa_prior_mean_median}{0.912}
\setres{learnUO}{CPT_post_m10}{m}{10}
\setres{learnUO}{CPT_post_m10}{n_countries}{30}
\setres{learnUO}{CPT_post_m10}{n_pos}{29}
\setres{learnUO}{CPT_post_m10}{n_subjects_seen}{10}
\setres{learnUO}{CPT_post_m10}{r_ff}{0.199}
\setres{learnUO}{CPT_post_m10}{r_prior}{0.9948}
\setres{learnUO}{CPT_post_m10}{r_prob}{0.996}
\setres{learnUO}{CPT_post_m100}{kappa_mean}{0.951}
\setres{learnUO}{CPT_post_m100}{kappa_mean_se}{0.047}
\setres{learnUO}{CPT_post_m100}{kappa_median}{0.962}
\setres{learnUO}{CPT_post_m100}{kappa_prior_mean_median}{0.967}
\setres{learnUO}{CPT_post_m100}{m}{100}
\setres{learnUO}{CPT_post_m100}{n_countries}{30}
\setres{learnUO}{CPT_post_m100}{n_pos}{30}
\setres{learnUO}{CPT_post_m100}{n_subjects_seen}{87}
\setres{learnUO}{CPT_post_m100}{r_ff}{0.199}
\setres{learnUO}{CPT_post_m100}{r_prior}{0.9998}
\setres{learnUO}{CPT_post_m100}{r_prob}{1.000}
\setres{learnUO}{CPT_post_m1000}{kappa_mean}{0.949}
\setres{learnUO}{CPT_post_m1000}{kappa_mean_se}{0.029}
\setres{learnUO}{CPT_post_m1000}{kappa_median}{0.962}
\setres{learnUO}{CPT_post_m1000}{kappa_prior_mean_median}{0.967}
\setres{learnUO}{CPT_post_m1000}{m}{1000}
\setres{learnUO}{CPT_post_m1000}{n_countries}{30}
\setres{learnUO}{CPT_post_m1000}{n_pos}{30}
\setres{learnUO}{CPT_post_m1000}{n_subjects_seen}{98}
\setres{learnUO}{CPT_post_m1000}{r_ff}{0.199}
\setres{learnUO}{CPT_post_m1000}{r_prior}{0.9998}
\setres{learnUO}{CPT_post_m1000}{r_prob}{1.000}
\setres{learnUO}{CPT_post_m3}{kappa_mean}{0.192}
\setres{learnUO}{CPT_post_m3}{kappa_mean_se}{1.380}
\setres{learnUO}{CPT_post_m3}{kappa_median}{0.672}
\setres{learnUO}{CPT_post_m3}{kappa_prior_mean_median}{0.739}
\setres{learnUO}{CPT_post_m3}{m}{3}
\setres{learnUO}{CPT_post_m3}{n_countries}{30}
\setres{learnUO}{CPT_post_m3}{n_pos}{28}
\setres{learnUO}{CPT_post_m3}{n_subjects_seen}{3}
\setres{learnUO}{CPT_post_m3}{r_ff}{0.199}
\setres{learnUO}{CPT_post_m3}{r_prior}{0.9802}
\setres{learnUO}{CPT_post_m3}{r_prob}{0.984}
\setres{learnUO}{CPT_post_m30}{kappa_mean}{0.834}
\setres{learnUO}{CPT_post_m30}{kappa_mean_se}{0.562}
\setres{learnUO}{CPT_post_m30}{kappa_median}{0.944}
\setres{learnUO}{CPT_post_m30}{kappa_prior_mean_median}{0.955}
\setres{learnUO}{CPT_post_m30}{m}{30}
\setres{learnUO}{CPT_post_m30}{n_countries}{30}
\setres{learnUO}{CPT_post_m30}{n_pos}{29}
\setres{learnUO}{CPT_post_m30}{n_subjects_seen}{30}
\setres{learnUO}{CPT_post_m30}{r_ff}{0.199}
\setres{learnUO}{CPT_post_m30}{r_prior}{0.9989}
\setres{learnUO}{CPT_post_m30}{r_prob}{0.999}
\setres{learnUO}{CPT_post_m300}{kappa_mean}{0.948}
\setres{learnUO}{CPT_post_m300}{kappa_mean_se}{0.028}
\setres{learnUO}{CPT_post_m300}{kappa_median}{0.962}
\setres{learnUO}{CPT_post_m300}{kappa_prior_mean_median}{0.968}
\setres{learnUO}{CPT_post_m300}{m}{300}
\setres{learnUO}{CPT_post_m300}{n_countries}{30}
\setres{learnUO}{CPT_post_m300}{n_pos}{30}
\setres{learnUO}{CPT_post_m300}{n_subjects_seen}{98}
\setres{learnUO}{CPT_post_m300}{r_ff}{0.199}
\setres{learnUO}{CPT_post_m300}{r_prior}{0.9998}
\setres{learnUO}{CPT_post_m300}{r_prob}{1.000}
\setres{learnUO}{CPT_post_mall}{kappa_mean}{0.948}
\setres{learnUO}{CPT_post_mall}{kappa_mean_se}{0.029}
\setres{learnUO}{CPT_post_mall}{kappa_median}{0.962}
\setres{learnUO}{CPT_post_mall}{kappa_prior_mean_median}{0.968}
\setres{learnUO}{CPT_post_mall}{m}{all}
\setres{learnUO}{CPT_post_mall}{n_countries}{30}
\setres{learnUO}{CPT_post_mall}{n_pos}{30}
\setres{learnUO}{CPT_post_mall}{n_subjects_seen}{98}
\setres{learnUO}{CPT_post_mall}{r_ff}{0.199}
\setres{learnUO}{CPT_post_mall}{r_prior}{0.9998}
\setres{learnUO}{CPT_post_mall}{r_prob}{1.000}
\setres{learnUO}{DA_post_m0}{kappa_mean}{-3.957}
\setres{learnUO}{DA_post_m0}{kappa_mean_se}{4.985}
\setres{learnUO}{DA_post_m0}{kappa_median}{-2.182}
\setres{learnUO}{DA_post_m0}{kappa_prior_mean_median}{-0.521}
\setres{learnUO}{DA_post_m0}{m}{0}
\setres{learnUO}{DA_post_m0}{n_countries}{30}
\setres{learnUO}{DA_post_m0}{n_pos}{0}
\setres{learnUO}{DA_post_m0}{n_subjects_seen}{0}
\setres{learnUO}{DA_post_m0}{r_ff}{0.314}
\setres{learnUO}{DA_post_m0}{r_prior}{0.7815}
\setres{learnUO}{DA_post_m0}{r_prob}{0.850}
\setres{learnUO}{DA_post_m10}{kappa_mean}{0.351}
\setres{learnUO}{DA_post_m10}{kappa_mean_se}{1.652}
\setres{learnUO}{DA_post_m10}{kappa_median}{0.801}
\setres{learnUO}{DA_post_m10}{kappa_prior_mean_median}{0.818}
\setres{learnUO}{DA_post_m10}{m}{10}
\setres{learnUO}{DA_post_m10}{n_countries}{30}
\setres{learnUO}{DA_post_m10}{n_pos}{29}
\setres{learnUO}{DA_post_m10}{n_subjects_seen}{10}
\setres{learnUO}{DA_post_m10}{r_ff}{0.314}
\setres{learnUO}{DA_post_m10}{r_prior}{0.9961}
\setres{learnUO}{DA_post_m10}{r_prob}{0.997}
\setres{learnUO}{DA_post_m100}{kappa_mean}{0.473}
\setres{learnUO}{DA_post_m100}{kappa_mean_se}{1.486}
\setres{learnUO}{DA_post_m100}{kappa_median}{0.869}
\setres{learnUO}{DA_post_m100}{kappa_prior_mean_median}{0.871}
\setres{learnUO}{DA_post_m100}{m}{100}
\setres{learnUO}{DA_post_m100}{n_countries}{30}
\setres{learnUO}{DA_post_m100}{n_pos}{29}
\setres{learnUO}{DA_post_m100}{n_subjects_seen}{87}
\setres{learnUO}{DA_post_m100}{r_ff}{0.314}
\setres{learnUO}{DA_post_m100}{r_prior}{0.9997}
\setres{learnUO}{DA_post_m100}{r_prob}{1.000}
\setres{learnUO}{DA_post_m1000}{kappa_mean}{0.473}
\setres{learnUO}{DA_post_m1000}{kappa_mean_se}{1.443}
\setres{learnUO}{DA_post_m1000}{kappa_median}{0.869}
\setres{learnUO}{DA_post_m1000}{kappa_prior_mean_median}{0.871}
\setres{learnUO}{DA_post_m1000}{m}{1000}
\setres{learnUO}{DA_post_m1000}{n_countries}{30}
\setres{learnUO}{DA_post_m1000}{n_pos}{29}
\setres{learnUO}{DA_post_m1000}{n_subjects_seen}{98}
\setres{learnUO}{DA_post_m1000}{r_ff}{0.314}
\setres{learnUO}{DA_post_m1000}{r_prior}{0.9997}
\setres{learnUO}{DA_post_m1000}{r_prob}{1.000}
\setres{learnUO}{DA_post_m3}{kappa_mean}{-0.304}
\setres{learnUO}{DA_post_m3}{kappa_mean_se}{2.835}
\setres{learnUO}{DA_post_m3}{kappa_median}{0.644}
\setres{learnUO}{DA_post_m3}{kappa_prior_mean_median}{0.692}
\setres{learnUO}{DA_post_m3}{m}{3}
\setres{learnUO}{DA_post_m3}{n_countries}{30}
\setres{learnUO}{DA_post_m3}{n_pos}{26}
\setres{learnUO}{DA_post_m3}{n_subjects_seen}{3}
\setres{learnUO}{DA_post_m3}{r_ff}{0.314}
\setres{learnUO}{DA_post_m3}{r_prior}{0.9833}
\setres{learnUO}{DA_post_m3}{r_prob}{0.989}
\setres{learnUO}{DA_post_m30}{kappa_mean}{0.353}
\setres{learnUO}{DA_post_m30}{kappa_mean_se}{2.058}
\setres{learnUO}{DA_post_m30}{kappa_median}{0.849}
\setres{learnUO}{DA_post_m30}{kappa_prior_mean_median}{0.855}
\setres{learnUO}{DA_post_m30}{m}{30}
\setres{learnUO}{DA_post_m30}{n_countries}{30}
\setres{learnUO}{DA_post_m30}{n_pos}{29}
\setres{learnUO}{DA_post_m30}{n_subjects_seen}{30}
\setres{learnUO}{DA_post_m30}{r_ff}{0.314}
\setres{learnUO}{DA_post_m30}{r_prior}{0.9990}
\setres{learnUO}{DA_post_m30}{r_prob}{0.999}
\setres{learnUO}{DA_post_m300}{kappa_mean}{0.473}
\setres{learnUO}{DA_post_m300}{kappa_mean_se}{1.443}
\setres{learnUO}{DA_post_m300}{kappa_median}{0.869}
\setres{learnUO}{DA_post_m300}{kappa_prior_mean_median}{0.871}
\setres{learnUO}{DA_post_m300}{m}{300}
\setres{learnUO}{DA_post_m300}{n_countries}{30}
\setres{learnUO}{DA_post_m300}{n_pos}{29}
\setres{learnUO}{DA_post_m300}{n_subjects_seen}{98}
\setres{learnUO}{DA_post_m300}{r_ff}{0.314}
\setres{learnUO}{DA_post_m300}{r_prior}{0.9997}
\setres{learnUO}{DA_post_m300}{r_prob}{1.000}
\setres{learnUO}{DA_post_mall}{kappa_mean}{0.474}
\setres{learnUO}{DA_post_mall}{kappa_mean_se}{1.444}
\setres{learnUO}{DA_post_mall}{kappa_median}{0.869}
\setres{learnUO}{DA_post_mall}{kappa_prior_mean_median}{0.871}
\setres{learnUO}{DA_post_mall}{m}{all}
\setres{learnUO}{DA_post_mall}{n_countries}{30}
\setres{learnUO}{DA_post_mall}{n_pos}{29}
\setres{learnUO}{DA_post_mall}{n_subjects_seen}{98}
\setres{learnUO}{DA_post_mall}{r_ff}{0.314}
\setres{learnUO}{DA_post_mall}{r_prior}{0.9997}
\setres{learnUO}{DA_post_mall}{r_prob}{1.000}
\setres{learnUO}{EU_post_m0}{kappa_mean}{-2.284}
\setres{learnUO}{EU_post_m0}{kappa_mean_se}{3.722}
\setres{learnUO}{EU_post_m0}{kappa_median}{-0.948}
\setres{learnUO}{EU_post_m0}{kappa_prior_mean_median}{-0.102}
\setres{learnUO}{EU_post_m0}{m}{0}
\setres{learnUO}{EU_post_m0}{n_countries}{30}
\setres{learnUO}{EU_post_m0}{n_pos}{0}
\setres{learnUO}{EU_post_m0}{n_subjects_seen}{0}
\setres{learnUO}{EU_post_m0}{r_ff}{0.367}
\setres{learnUO}{EU_post_m0}{r_prior}{0.5250}
\setres{learnUO}{EU_post_m0}{r_prob}{0.699}
\setres{learnUO}{EU_post_m10}{kappa_mean}{0.403}
\setres{learnUO}{EU_post_m10}{kappa_mean_se}{1.417}
\setres{learnUO}{EU_post_m10}{kappa_median}{0.816}
\setres{learnUO}{EU_post_m10}{kappa_prior_mean_median}{0.827}
\setres{learnUO}{EU_post_m10}{m}{10}
\setres{learnUO}{EU_post_m10}{n_countries}{30}
\setres{learnUO}{EU_post_m10}{n_pos}{29}
\setres{learnUO}{EU_post_m10}{n_subjects_seen}{10}
\setres{learnUO}{EU_post_m10}{r_ff}{0.367}
\setres{learnUO}{EU_post_m10}{r_prior}{0.9608}
\setres{learnUO}{EU_post_m10}{r_prob}{0.975}
\setres{learnUO}{EU_post_m100}{kappa_mean}{0.449}
\setres{learnUO}{EU_post_m100}{kappa_mean_se}{1.384}
\setres{learnUO}{EU_post_m100}{kappa_median}{0.838}
\setres{learnUO}{EU_post_m100}{kappa_prior_mean_median}{0.839}
\setres{learnUO}{EU_post_m100}{m}{100}
\setres{learnUO}{EU_post_m100}{n_countries}{30}
\setres{learnUO}{EU_post_m100}{n_pos}{29}
\setres{learnUO}{EU_post_m100}{n_subjects_seen}{87}
\setres{learnUO}{EU_post_m100}{r_ff}{0.367}
\setres{learnUO}{EU_post_m100}{r_prior}{0.9821}
\setres{learnUO}{EU_post_m100}{r_prob}{0.989}
\setres{learnUO}{EU_post_m1000}{kappa_mean}{0.450}
\setres{learnUO}{EU_post_m1000}{kappa_mean_se}{1.384}
\setres{learnUO}{EU_post_m1000}{kappa_median}{0.838}
\setres{learnUO}{EU_post_m1000}{kappa_prior_mean_median}{0.839}
\setres{learnUO}{EU_post_m1000}{m}{1000}
\setres{learnUO}{EU_post_m1000}{n_countries}{30}
\setres{learnUO}{EU_post_m1000}{n_pos}{29}
\setres{learnUO}{EU_post_m1000}{n_subjects_seen}{98}
\setres{learnUO}{EU_post_m1000}{r_ff}{0.367}
\setres{learnUO}{EU_post_m1000}{r_prior}{0.9824}
\setres{learnUO}{EU_post_m1000}{r_prob}{0.989}
\setres{learnUO}{EU_post_m3}{kappa_mean}{0.010}
\setres{learnUO}{EU_post_m3}{kappa_mean_se}{2.520}
\setres{learnUO}{EU_post_m3}{kappa_median}{0.767}
\setres{learnUO}{EU_post_m3}{kappa_prior_mean_median}{0.784}
\setres{learnUO}{EU_post_m3}{m}{3}
\setres{learnUO}{EU_post_m3}{n_countries}{30}
\setres{learnUO}{EU_post_m3}{n_pos}{29}
\setres{learnUO}{EU_post_m3}{n_subjects_seen}{3}
\setres{learnUO}{EU_post_m3}{r_ff}{0.367}
\setres{learnUO}{EU_post_m3}{r_prior}{0.9468}
\setres{learnUO}{EU_post_m3}{r_prob}{0.966}
\setres{learnUO}{EU_post_m30}{kappa_mean}{0.439}
\setres{learnUO}{EU_post_m30}{kappa_mean_se}{1.386}
\setres{learnUO}{EU_post_m30}{kappa_median}{0.828}
\setres{learnUO}{EU_post_m30}{kappa_prior_mean_median}{0.830}
\setres{learnUO}{EU_post_m30}{m}{30}
\setres{learnUO}{EU_post_m30}{n_countries}{30}
\setres{learnUO}{EU_post_m30}{n_pos}{29}
\setres{learnUO}{EU_post_m30}{n_subjects_seen}{30}
\setres{learnUO}{EU_post_m30}{r_ff}{0.367}
\setres{learnUO}{EU_post_m30}{r_prior}{0.9712}
\setres{learnUO}{EU_post_m30}{r_prob}{0.982}
\setres{learnUO}{EU_post_m300}{kappa_mean}{0.450}
\setres{learnUO}{EU_post_m300}{kappa_mean_se}{1.384}
\setres{learnUO}{EU_post_m300}{kappa_median}{0.837}
\setres{learnUO}{EU_post_m300}{kappa_prior_mean_median}{0.838}
\setres{learnUO}{EU_post_m300}{m}{300}
\setres{learnUO}{EU_post_m300}{n_countries}{30}
\setres{learnUO}{EU_post_m300}{n_pos}{29}
\setres{learnUO}{EU_post_m300}{n_subjects_seen}{98}
\setres{learnUO}{EU_post_m300}{r_ff}{0.367}
\setres{learnUO}{EU_post_m300}{r_prior}{0.9824}
\setres{learnUO}{EU_post_m300}{r_prob}{0.989}
\setres{learnUO}{EU_post_mall}{kappa_mean}{0.450}
\setres{learnUO}{EU_post_mall}{kappa_mean_se}{1.384}
\setres{learnUO}{EU_post_mall}{kappa_median}{0.837}
\setres{learnUO}{EU_post_mall}{kappa_prior_mean_median}{0.838}
\setres{learnUO}{EU_post_mall}{m}{all}
\setres{learnUO}{EU_post_mall}{n_countries}{30}
\setres{learnUO}{EU_post_mall}{n_pos}{29}
\setres{learnUO}{EU_post_mall}{n_subjects_seen}{98}
\setres{learnUO}{EU_post_mall}{r_ff}{0.367}
\setres{learnUO}{EU_post_mall}{r_prior}{0.9824}
\setres{learnUO}{EU_post_mall}{r_prob}{0.989}
\setres{learnUO}{reference}{m_grid}{0, 3, 10, 30, 100, 300, 1000, all}
\setres{learnUO}{reference}{n_countries}{30}
\setres{learnUO}{reference}{source}{own}
\setres{learnUO}{reference}{start}{uniform}
\setres{learnOZ}{CPT_post_m0}{ess}{732}
\setres{learnOZ}{CPT_post_m0}{kappa}{0.639}
\setres{learnOZ}{CPT_post_m0}{kappa_crps}{0.605}
\setres{learnOZ}{CPT_post_m0}{kappa_crps_se}{0.030}
\setres{learnOZ}{CPT_post_m0}{kappa_prior_mean}{0.851}
\setres{learnOZ}{CPT_post_m0}{kappa_se}{0.050}
\setres{learnOZ}{CPT_post_m0}{mse}{87.82}
\setres{learnOZ}{CPT_post_m0}{n_subjects_seen}{0}
\setres{learnOZ}{CPT_post_m0}{r_ff}{0.140}
\setres{learnOZ}{CPT_post_m0}{r_ff_se}{0.001}
\setres{learnOZ}{CPT_post_m0}{r_prior}{0.9477}
\setres{learnOZ}{CPT_post_m0}{r_prior_se}{0.000}
\setres{learnOZ}{CPT_post_m0}{r_prob}{0.955}
\setres{learnOZ}{CPT_post_m0}{r_prob_se}{0.000}
\setres{learnOZ}{CPT_post_m0}{spread_penalty}{0.213}
\setres{learnOZ}{CPT_post_m0}{theta_mean}{0.70, 0.73, 1.08}
\setres{learnOZ}{CPT_post_m0}{theta_sd}{0.14, 0.18, 0.28}
\setres{learnOZ}{CPT_post_m0}{volume_ess}{55{,}652}
\setres{learnOZ}{CPT_post_m10}{ess}{20}
\setres{learnOZ}{CPT_post_m10}{kappa}{0.929}
\setres{learnOZ}{CPT_post_m10}{kappa_crps}{0.663}
\setres{learnOZ}{CPT_post_m10}{kappa_crps_se}{0.019}
\setres{learnOZ}{CPT_post_m10}{kappa_prior_mean}{0.942}
\setres{learnOZ}{CPT_post_m10}{kappa_se}{0.016}
\setres{learnOZ}{CPT_post_m10}{mse}{66.92}
\setres{learnOZ}{CPT_post_m10}{n_subjects_seen}{10}
\setres{learnOZ}{CPT_post_m10}{r_ff}{0.140}
\setres{learnOZ}{CPT_post_m10}{r_ff_se}{0.001}
\setres{learnOZ}{CPT_post_m10}{r_prior}{0.9980}
\setres{learnOZ}{CPT_post_m10}{r_prior_se}{0.000}
\setres{learnOZ}{CPT_post_m10}{r_prob}{0.998}
\setres{learnOZ}{CPT_post_m10}{r_prob_se}{0.000}
\setres{learnOZ}{CPT_post_m10}{spread_penalty}{0.012}
\setres{learnOZ}{CPT_post_m10}{theta_mean}{0.76, 0.56, 1.15}
\setres{learnOZ}{CPT_post_m10}{theta_sd}{0.07, 0.04, 0.13}
\setres{learnOZ}{CPT_post_m10}{volume_ess}{55{,}652}
\setres{learnOZ}{CPT_post_m100}{ess}{2}
\setres{learnOZ}{CPT_post_m100}{kappa}{0.978}
\setres{learnOZ}{CPT_post_m100}{kappa_crps}{0.687}
\setres{learnOZ}{CPT_post_m100}{kappa_crps_se}{0.014}
\setres{learnOZ}{CPT_post_m100}{kappa_prior_mean}{0.983}
\setres{learnOZ}{CPT_post_m100}{kappa_se}{0.004}
\setres{learnOZ}{CPT_post_m100}{mse}{63.44}
\setres{learnOZ}{CPT_post_m100}{n_subjects_seen}{100}
\setres{learnOZ}{CPT_post_m100}{r_ff}{0.140}
\setres{learnOZ}{CPT_post_m100}{r_ff_se}{0.001}
\setres{learnOZ}{CPT_post_m100}{r_prior}{0.9998}
\setres{learnOZ}{CPT_post_m100}{r_prior_se}{0.000}
\setres{learnOZ}{CPT_post_m100}{r_prob}{1.000}
\setres{learnOZ}{CPT_post_m100}{r_prob_se}{0.000}
\setres{learnOZ}{CPT_post_m100}{spread_penalty}{0.006}
\setres{learnOZ}{CPT_post_m100}{theta_mean}{0.80, 0.42, 0.94}
\setres{learnOZ}{CPT_post_m100}{theta_sd}{0.04, 0.02, 0.06}
\setres{learnOZ}{CPT_post_m100}{volume_ess}{55{,}652}
\setres{learnOZ}{CPT_post_m3}{ess}{80}
\setres{learnOZ}{CPT_post_m3}{kappa}{0.777}
\setres{learnOZ}{CPT_post_m3}{kappa_crps}{0.563}
\setres{learnOZ}{CPT_post_m3}{kappa_crps_se}{0.029}
\setres{learnOZ}{CPT_post_m3}{kappa_prior_mean}{0.803}
\setres{learnOZ}{CPT_post_m3}{kappa_se}{0.033}
\setres{learnOZ}{CPT_post_m3}{mse}{77.89}
\setres{learnOZ}{CPT_post_m3}{n_subjects_seen}{3}
\setres{learnOZ}{CPT_post_m3}{r_ff}{0.140}
\setres{learnOZ}{CPT_post_m3}{r_ff_se}{0.001}
\setres{learnOZ}{CPT_post_m3}{r_prior}{0.9941}
\setres{learnOZ}{CPT_post_m3}{r_prior_se}{0.000}
\setres{learnOZ}{CPT_post_m3}{r_prob}{0.995}
\setres{learnOZ}{CPT_post_m3}{r_prob_se}{0.000}
\setres{learnOZ}{CPT_post_m3}{spread_penalty}{0.026}
\setres{learnOZ}{CPT_post_m3}{theta_mean}{0.82, 0.58, 1.31}
\setres{learnOZ}{CPT_post_m3}{theta_sd}{0.08, 0.08, 0.17}
\setres{learnOZ}{CPT_post_m3}{volume_ess}{55{,}652}
\setres{learnOZ}{CPT_post_m30}{ess}{5}
\setres{learnOZ}{CPT_post_m30}{kappa}{0.955}
\setres{learnOZ}{CPT_post_m30}{kappa_crps}{0.665}
\setres{learnOZ}{CPT_post_m30}{kappa_crps_se}{0.013}
\setres{learnOZ}{CPT_post_m30}{kappa_prior_mean}{0.962}
\setres{learnOZ}{CPT_post_m30}{kappa_se}{0.015}
\setres{learnOZ}{CPT_post_m30}{mse}{65.06}
\setres{learnOZ}{CPT_post_m30}{n_subjects_seen}{30}
\setres{learnOZ}{CPT_post_m30}{r_ff}{0.140}
\setres{learnOZ}{CPT_post_m30}{r_ff_se}{0.001}
\setres{learnOZ}{CPT_post_m30}{r_prior}{0.9994}
\setres{learnOZ}{CPT_post_m30}{r_prior_se}{0.000}
\setres{learnOZ}{CPT_post_m30}{r_prob}{1.000}
\setres{learnOZ}{CPT_post_m30}{r_prob_se}{0.000}
\setres{learnOZ}{CPT_post_m30}{spread_penalty}{0.007}
\setres{learnOZ}{CPT_post_m30}{theta_mean}{0.77, 0.43, 1.15}
\setres{learnOZ}{CPT_post_m30}{theta_sd}{0.05, 0.02, 0.09}
\setres{learnOZ}{CPT_post_m30}{volume_ess}{55{,}652}
\setres{learnOZ}{CPT_post_mall}{ess}{2}
\setres{learnOZ}{CPT_post_mall}{kappa}{0.977}
\setres{learnOZ}{CPT_post_mall}{kappa_crps}{0.685}
\setres{learnOZ}{CPT_post_mall}{kappa_crps_se}{0.012}
\setres{learnOZ}{CPT_post_mall}{kappa_prior_mean}{0.983}
\setres{learnOZ}{CPT_post_mall}{kappa_se}{0.005}
\setres{learnOZ}{CPT_post_mall}{mse}{63.48}
\setres{learnOZ}{CPT_post_mall}{n_subjects_seen}{179}
\setres{learnOZ}{CPT_post_mall}{r_ff}{0.140}
\setres{learnOZ}{CPT_post_mall}{r_ff_se}{0.001}
\setres{learnOZ}{CPT_post_mall}{r_prior}{0.9998}
\setres{learnOZ}{CPT_post_mall}{r_prior_se}{0.000}
\setres{learnOZ}{CPT_post_mall}{r_prob}{1.000}
\setres{learnOZ}{CPT_post_mall}{r_prob_se}{0.000}
\setres{learnOZ}{CPT_post_mall}{spread_penalty}{0.006}
\setres{learnOZ}{CPT_post_mall}{theta_mean}{0.80, 0.43, 0.93}
\setres{learnOZ}{CPT_post_mall}{theta_sd}{0.04, 0.01, 0.06}
\setres{learnOZ}{CPT_post_mall}{volume_ess}{55{,}652}
\setres{learnOZ}{DA_post_m0}{ess}{439}
\setres{learnOZ}{DA_post_m0}{kappa}{0.018}
\setres{learnOZ}{DA_post_m0}{kappa_crps}{0.254}
\setres{learnOZ}{DA_post_m0}{kappa_crps_se}{0.059}
\setres{learnOZ}{DA_post_m0}{kappa_prior_mean}{0.370}
\setres{learnOZ}{DA_post_m0}{kappa_se}{0.136}
\setres{learnOZ}{DA_post_m0}{mse}{132.47}
\setres{learnOZ}{DA_post_m0}{n_subjects_seen}{0}
\setres{learnOZ}{DA_post_m0}{r_ff}{0.281}
\setres{learnOZ}{DA_post_m0}{r_ff_se}{0.004}
\setres{learnOZ}{DA_post_m0}{r_prior}{0.9380}
\setres{learnOZ}{DA_post_m0}{r_prior_se}{0.000}
\setres{learnOZ}{DA_post_m0}{r_prob}{0.955}
\setres{learnOZ}{DA_post_m0}{r_prob_se}{0.000}
\setres{learnOZ}{DA_post_m0}{spread_penalty}{0.352}
\setres{learnOZ}{DA_post_m0}{theta_mean}{0.58, 0.30}
\setres{learnOZ}{DA_post_m0}{theta_sd}{0.19, 0.45}
\setres{learnOZ}{DA_post_m0}{volume_ess}{436{,}452}
\setres{learnOZ}{DA_post_m10}{ess}{27}
\setres{learnOZ}{DA_post_m10}{kappa}{0.594}
\setres{learnOZ}{DA_post_m10}{kappa_crps}{0.329}
\setres{learnOZ}{DA_post_m10}{kappa_crps_se}{0.052}
\setres{learnOZ}{DA_post_m10}{kappa_prior_mean}{0.604}
\setres{learnOZ}{DA_post_m10}{kappa_se}{0.060}
\setres{learnOZ}{DA_post_m10}{mse}{91.05}
\setres{learnOZ}{DA_post_m10}{n_subjects_seen}{10}
\setres{learnOZ}{DA_post_m10}{r_ff}{0.281}
\setres{learnOZ}{DA_post_m10}{r_ff_se}{0.004}
\setres{learnOZ}{DA_post_m10}{r_prior}{0.9954}
\setres{learnOZ}{DA_post_m10}{r_prior_se}{0.000}
\setres{learnOZ}{DA_post_m10}{r_prob}{0.997}
\setres{learnOZ}{DA_post_m10}{r_prob_se}{0.000}
\setres{learnOZ}{DA_post_m10}{spread_penalty}{0.010}
\setres{learnOZ}{DA_post_m10}{theta_mean}{0.74, -0.13}
\setres{learnOZ}{DA_post_m10}{theta_sd}{0.11, 0.16}
\setres{learnOZ}{DA_post_m10}{volume_ess}{436{,}452}
\setres{learnOZ}{DA_post_m100}{ess}{4}
\setres{learnOZ}{DA_post_m100}{kappa}{0.622}
\setres{learnOZ}{DA_post_m100}{kappa_crps}{0.360}
\setres{learnOZ}{DA_post_m100}{kappa_crps_se}{0.050}
\setres{learnOZ}{DA_post_m100}{kappa_prior_mean}{0.623}
\setres{learnOZ}{DA_post_m100}{kappa_se}{0.063}
\setres{learnOZ}{DA_post_m100}{mse}{89.02}
\setres{learnOZ}{DA_post_m100}{n_subjects_seen}{100}
\setres{learnOZ}{DA_post_m100}{r_ff}{0.281}
\setres{learnOZ}{DA_post_m100}{r_ff_se}{0.004}
\setres{learnOZ}{DA_post_m100}{r_prior}{0.9997}
\setres{learnOZ}{DA_post_m100}{r_prior_se}{0.000}
\setres{learnOZ}{DA_post_m100}{r_prob}{1.000}
\setres{learnOZ}{DA_post_m100}{r_prob_se}{0.000}
\setres{learnOZ}{DA_post_m100}{spread_penalty}{0.001}
\setres{learnOZ}{DA_post_m100}{theta_mean}{0.95, 0.38}
\setres{learnOZ}{DA_post_m100}{theta_sd}{0.03, 0.07}
\setres{learnOZ}{DA_post_m100}{volume_ess}{436{,}452}
\setres{learnOZ}{DA_post_m3}{ess}{41}
\setres{learnOZ}{DA_post_m3}{kappa}{0.451}
\setres{learnOZ}{DA_post_m3}{kappa_crps}{0.242}
\setres{learnOZ}{DA_post_m3}{kappa_crps_se}{0.058}
\setres{learnOZ}{DA_post_m3}{kappa_prior_mean}{0.470}
\setres{learnOZ}{DA_post_m3}{kappa_se}{0.067}
\setres{learnOZ}{DA_post_m3}{mse}{101.32}
\setres{learnOZ}{DA_post_m3}{n_subjects_seen}{3}
\setres{learnOZ}{DA_post_m3}{r_ff}{0.281}
\setres{learnOZ}{DA_post_m3}{r_ff_se}{0.004}
\setres{learnOZ}{DA_post_m3}{r_prior}{0.9918}
\setres{learnOZ}{DA_post_m3}{r_prior_se}{0.000}
\setres{learnOZ}{DA_post_m3}{r_prob}{0.994}
\setres{learnOZ}{DA_post_m3}{r_prob_se}{0.000}
\setres{learnOZ}{DA_post_m3}{spread_penalty}{0.019}
\setres{learnOZ}{DA_post_m3}{theta_mean}{0.77, -0.31}
\setres{learnOZ}{DA_post_m3}{theta_sd}{0.12, 0.15}
\setres{learnOZ}{DA_post_m3}{volume_ess}{436{,}452}
\setres{learnOZ}{DA_post_m30}{ess}{14}
\setres{learnOZ}{DA_post_m30}{kappa}{0.587}
\setres{learnOZ}{DA_post_m30}{kappa_crps}{0.336}
\setres{learnOZ}{DA_post_m30}{kappa_crps_se}{0.052}
\setres{learnOZ}{DA_post_m30}{kappa_prior_mean}{0.592}
\setres{learnOZ}{DA_post_m30}{kappa_se}{0.060}
\setres{learnOZ}{DA_post_m30}{mse}{91.59}
\setres{learnOZ}{DA_post_m30}{n_subjects_seen}{30}
\setres{learnOZ}{DA_post_m30}{r_ff}{0.281}
\setres{learnOZ}{DA_post_m30}{r_ff_se}{0.004}
\setres{learnOZ}{DA_post_m30}{r_prior}{0.9975}
\setres{learnOZ}{DA_post_m30}{r_prior_se}{0.000}
\setres{learnOZ}{DA_post_m30}{r_prob}{0.998}
\setres{learnOZ}{DA_post_m30}{r_prob_se}{0.000}
\setres{learnOZ}{DA_post_m30}{spread_penalty}{0.005}
\setres{learnOZ}{DA_post_m30}{theta_mean}{0.75, -0.16}
\setres{learnOZ}{DA_post_m30}{theta_sd}{0.10, 0.14}
\setres{learnOZ}{DA_post_m30}{volume_ess}{436{,}452}
\setres{learnOZ}{DA_post_mall}{ess}{2}
\setres{learnOZ}{DA_post_mall}{kappa}{0.623}
\setres{learnOZ}{DA_post_mall}{kappa_crps}{0.351}
\setres{learnOZ}{DA_post_mall}{kappa_crps_se}{0.051}
\setres{learnOZ}{DA_post_mall}{kappa_prior_mean}{0.624}
\setres{learnOZ}{DA_post_mall}{kappa_se}{0.063}
\setres{learnOZ}{DA_post_mall}{mse}{88.97}
\setres{learnOZ}{DA_post_mall}{n_subjects_seen}{179}
\setres{learnOZ}{DA_post_mall}{r_ff}{0.281}
\setres{learnOZ}{DA_post_mall}{r_ff_se}{0.004}
\setres{learnOZ}{DA_post_mall}{r_prior}{0.9999}
\setres{learnOZ}{DA_post_mall}{r_prior_se}{0.000}
\setres{learnOZ}{DA_post_mall}{r_prob}{1.000}
\setres{learnOZ}{DA_post_mall}{r_prob_se}{0.000}
\setres{learnOZ}{DA_post_mall}{spread_penalty}{0.001}
\setres{learnOZ}{DA_post_mall}{theta_mean}{0.97, 0.41}
\setres{learnOZ}{DA_post_mall}{theta_sd}{0.02, 0.05}
\setres{learnOZ}{DA_post_mall}{volume_ess}{436{,}452}
\setres{learnOZ}{EU_post_m0}{ess}{98}
\setres{learnOZ}{EU_post_m0}{kappa}{0.522}
\setres{learnOZ}{EU_post_m0}{kappa_crps}{0.354}
\setres{learnOZ}{EU_post_m0}{kappa_crps_se}{0.051}
\setres{learnOZ}{EU_post_m0}{kappa_prior_mean}{0.611}
\setres{learnOZ}{EU_post_m0}{kappa_se}{0.074}
\setres{learnOZ}{EU_post_m0}{mse}{96.25}
\setres{learnOZ}{EU_post_m0}{n_subjects_seen}{0}
\setres{learnOZ}{EU_post_m0}{r_ff}{0.336}
\setres{learnOZ}{EU_post_m0}{r_ff_se}{0.005}
\setres{learnOZ}{EU_post_m0}{r_prior}{0.8143}
\setres{learnOZ}{EU_post_m0}{r_prior_se}{0.001}
\setres{learnOZ}{EU_post_m0}{r_prob}{0.877}
\setres{learnOZ}{EU_post_m0}{r_prob_se}{0.001}
\setres{learnOZ}{EU_post_m0}{spread_penalty}{0.089}
\setres{learnOZ}{EU_post_m0}{theta_mean}{0.70}
\setres{learnOZ}{EU_post_m0}{theta_sd}{0.14}
\setres{learnOZ}{EU_post_m0}{volume_ess}{172{,}086}
\setres{learnOZ}{EU_post_m10}{ess}{24}
\setres{learnOZ}{EU_post_m10}{kappa}{0.603}
\setres{learnOZ}{EU_post_m10}{kappa_crps}{0.339}
\setres{learnOZ}{EU_post_m10}{kappa_crps_se}{0.052}
\setres{learnOZ}{EU_post_m10}{kappa_prior_mean}{0.606}
\setres{learnOZ}{EU_post_m10}{kappa_se}{0.061}
\setres{learnOZ}{EU_post_m10}{mse}{90.41}
\setres{learnOZ}{EU_post_m10}{n_subjects_seen}{10}
\setres{learnOZ}{EU_post_m10}{r_ff}{0.336}
\setres{learnOZ}{EU_post_m10}{r_ff_se}{0.005}
\setres{learnOZ}{EU_post_m10}{r_prior}{0.9602}
\setres{learnOZ}{EU_post_m10}{r_prior_se}{0.000}
\setres{learnOZ}{EU_post_m10}{r_prob}{0.974}
\setres{learnOZ}{EU_post_m10}{r_prob_se}{0.000}
\setres{learnOZ}{EU_post_m10}{spread_penalty}{0.003}
\setres{learnOZ}{EU_post_m10}{theta_mean}{0.83}
\setres{learnOZ}{EU_post_m10}{theta_sd}{0.03}
\setres{learnOZ}{EU_post_m10}{volume_ess}{172{,}086}
\setres{learnOZ}{EU_post_m100}{ess}{7}
\setres{learnOZ}{EU_post_m100}{kappa}{0.616}
\setres{learnOZ}{EU_post_m100}{kappa_crps}{0.347}
\setres{learnOZ}{EU_post_m100}{kappa_crps_se}{0.051}
\setres{learnOZ}{EU_post_m100}{kappa_prior_mean}{0.616}
\setres{learnOZ}{EU_post_m100}{kappa_se}{0.062}
\setres{learnOZ}{EU_post_m100}{mse}{89.47}
\setres{learnOZ}{EU_post_m100}{n_subjects_seen}{100}
\setres{learnOZ}{EU_post_m100}{r_ff}{0.336}
\setres{learnOZ}{EU_post_m100}{r_ff_se}{0.005}
\setres{learnOZ}{EU_post_m100}{r_prior}{0.9866}
\setres{learnOZ}{EU_post_m100}{r_prior_se}{0.000}
\setres{learnOZ}{EU_post_m100}{r_prob}{0.991}
\setres{learnOZ}{EU_post_m100}{r_prob_se}{0.000}
\setres{learnOZ}{EU_post_m100}{spread_penalty}{0.000}
\setres{learnOZ}{EU_post_m100}{theta_mean}{0.76}
\setres{learnOZ}{EU_post_m100}{theta_sd}{0.01}
\setres{learnOZ}{EU_post_m100}{volume_ess}{172{,}086}
\setres{learnOZ}{EU_post_m3}{ess}{27}
\setres{learnOZ}{EU_post_m3}{kappa}{0.562}
\setres{learnOZ}{EU_post_m3}{kappa_crps}{0.322}
\setres{learnOZ}{EU_post_m3}{kappa_crps_se}{0.054}
\setres{learnOZ}{EU_post_m3}{kappa_prior_mean}{0.566}
\setres{learnOZ}{EU_post_m3}{kappa_se}{0.062}
\setres{learnOZ}{EU_post_m3}{mse}{93.36}
\setres{learnOZ}{EU_post_m3}{n_subjects_seen}{3}
\setres{learnOZ}{EU_post_m3}{r_ff}{0.336}
\setres{learnOZ}{EU_post_m3}{r_ff_se}{0.005}
\setres{learnOZ}{EU_post_m3}{r_prior}{0.9617}
\setres{learnOZ}{EU_post_m3}{r_prior_se}{0.000}
\setres{learnOZ}{EU_post_m3}{r_prob}{0.975}
\setres{learnOZ}{EU_post_m3}{r_prob_se}{0.000}
\setres{learnOZ}{EU_post_m3}{spread_penalty}{0.004}
\setres{learnOZ}{EU_post_m3}{theta_mean}{0.92}
\setres{learnOZ}{EU_post_m3}{theta_sd}{0.04}
\setres{learnOZ}{EU_post_m3}{volume_ess}{172{,}086}
\setres{learnOZ}{EU_post_m30}{ess}{15}
\setres{learnOZ}{EU_post_m30}{kappa}{0.594}
\setres{learnOZ}{EU_post_m30}{kappa_crps}{0.332}
\setres{learnOZ}{EU_post_m30}{kappa_crps_se}{0.052}
\setres{learnOZ}{EU_post_m30}{kappa_prior_mean}{0.595}
\setres{learnOZ}{EU_post_m30}{kappa_se}{0.060}
\setres{learnOZ}{EU_post_m30}{mse}{91.08}
\setres{learnOZ}{EU_post_m30}{n_subjects_seen}{30}
\setres{learnOZ}{EU_post_m30}{r_ff}{0.336}
\setres{learnOZ}{EU_post_m30}{r_ff_se}{0.005}
\setres{learnOZ}{EU_post_m30}{r_prior}{0.9758}
\setres{learnOZ}{EU_post_m30}{r_prior_se}{0.000}
\setres{learnOZ}{EU_post_m30}{r_prob}{0.984}
\setres{learnOZ}{EU_post_m30}{r_prob_se}{0.000}
\setres{learnOZ}{EU_post_m30}{spread_penalty}{0.001}
\setres{learnOZ}{EU_post_m30}{theta_mean}{0.86}
\setres{learnOZ}{EU_post_m30}{theta_sd}{0.02}
\setres{learnOZ}{EU_post_m30}{volume_ess}{172{,}086}
\setres{learnOZ}{EU_post_mall}{ess}{6}
\setres{learnOZ}{EU_post_mall}{kappa}{0.616}
\setres{learnOZ}{EU_post_mall}{kappa_crps}{0.339}
\setres{learnOZ}{EU_post_mall}{kappa_crps_se}{0.052}
\setres{learnOZ}{EU_post_mall}{kappa_prior_mean}{0.616}
\setres{learnOZ}{EU_post_mall}{kappa_se}{0.062}
\setres{learnOZ}{EU_post_mall}{mse}{89.49}
\setres{learnOZ}{EU_post_mall}{n_subjects_seen}{179}
\setres{learnOZ}{EU_post_mall}{r_ff}{0.336}
\setres{learnOZ}{EU_post_mall}{r_ff_se}{0.005}
\setres{learnOZ}{EU_post_mall}{r_prior}{0.9900}
\setres{learnOZ}{EU_post_mall}{r_prior_se}{0.000}
\setres{learnOZ}{EU_post_mall}{r_prob}{0.993}
\setres{learnOZ}{EU_post_mall}{r_prob_se}{0.000}
\setres{learnOZ}{EU_post_mall}{spread_penalty}{0.000}
\setres{learnOZ}{EU_post_mall}{theta_mean}{0.76}
\setres{learnOZ}{EU_post_mall}{theta_sd}{0.01}
\setres{learnOZ}{EU_post_mall}{volume_ess}{172{,}086}
\setres{learnOZ}{reference}{n_subjects}{179}
\setres{learnOZ}{reference}{sigma}{7.8}
\setres{learnUOZ}{CPT_post_m0}{ess}{8{,}000}
\setres{learnUOZ}{CPT_post_m0}{kappa}{-0.459}
\setres{learnUOZ}{CPT_post_m0}{kappa_crps}{0.500}
\setres{learnUOZ}{CPT_post_m0}{kappa_crps_se}{0.024}
\setres{learnUOZ}{CPT_post_m0}{kappa_prior_mean}{0.807}
\setres{learnUOZ}{CPT_post_m0}{kappa_se}{0.128}
\setres{learnUOZ}{CPT_post_m0}{mse}{166.83}
\setres{learnUOZ}{CPT_post_m0}{n_subjects_seen}{0}
\setres{learnUOZ}{CPT_post_m0}{r_ff}{0.140}
\setres{learnUOZ}{CPT_post_m0}{r_ff_se}{0.001}
\setres{learnUOZ}{CPT_post_m0}{r_prior}{0.5144}
\setres{learnUOZ}{CPT_post_m0}{r_prior_se}{0.003}
\setres{learnUOZ}{CPT_post_m0}{r_prob}{0.582}
\setres{learnUOZ}{CPT_post_m0}{r_prob_se}{0.003}
\setres{learnUOZ}{CPT_post_m0}{spread_penalty}{1.267}
\setres{learnUOZ}{CPT_post_m0}{theta_mean}{0.50, 0.50, 1.55}
\setres{learnUOZ}{CPT_post_m0}{theta_sd}{0.29, 0.29, 0.84}
\setres{learnUOZ}{CPT_post_m0}{volume_ess}{55{,}652}
\setres{learnUOZ}{CPT_post_m10}{ess}{27}
\setres{learnUOZ}{CPT_post_m10}{kappa}{0.928}
\setres{learnUOZ}{CPT_post_m10}{kappa_crps}{0.659}
\setres{learnUOZ}{CPT_post_m10}{kappa_crps_se}{0.019}
\setres{learnUOZ}{CPT_post_m10}{kappa_prior_mean}{0.941}
\setres{learnUOZ}{CPT_post_m10}{kappa_se}{0.016}
\setres{learnUOZ}{CPT_post_m10}{mse}{67.04}
\setres{learnUOZ}{CPT_post_m10}{n_subjects_seen}{10}
\setres{learnUOZ}{CPT_post_m10}{r_ff}{0.140}
\setres{learnUOZ}{CPT_post_m10}{r_ff_se}{0.001}
\setres{learnUOZ}{CPT_post_m10}{r_prior}{0.9972}
\setres{learnUOZ}{CPT_post_m10}{r_prior_se}{0.000}
\setres{learnUOZ}{CPT_post_m10}{r_prob}{0.998}
\setres{learnUOZ}{CPT_post_m10}{r_prob_se}{0.000}
\setres{learnUOZ}{CPT_post_m10}{spread_penalty}{0.014}
\setres{learnUOZ}{CPT_post_m10}{theta_mean}{0.75, 0.55, 1.19}
\setres{learnUOZ}{CPT_post_m10}{theta_sd}{0.09, 0.04, 0.18}
\setres{learnUOZ}{CPT_post_m10}{volume_ess}{55{,}652}
\setres{learnUOZ}{CPT_post_m100}{ess}{2}
\setres{learnUOZ}{CPT_post_m100}{kappa}{0.978}
\setres{learnUOZ}{CPT_post_m100}{kappa_crps}{0.686}
\setres{learnUOZ}{CPT_post_m100}{kappa_crps_se}{0.014}
\setres{learnUOZ}{CPT_post_m100}{kappa_prior_mean}{0.983}
\setres{learnUOZ}{CPT_post_m100}{kappa_se}{0.004}
\setres{learnUOZ}{CPT_post_m100}{mse}{63.45}
\setres{learnUOZ}{CPT_post_m100}{n_subjects_seen}{100}
\setres{learnUOZ}{CPT_post_m100}{r_ff}{0.140}
\setres{learnUOZ}{CPT_post_m100}{r_ff_se}{0.001}
\setres{learnUOZ}{CPT_post_m100}{r_prior}{0.9997}
\setres{learnUOZ}{CPT_post_m100}{r_prior_se}{0.000}
\setres{learnUOZ}{CPT_post_m100}{r_prob}{1.000}
\setres{learnUOZ}{CPT_post_m100}{r_prob_se}{0.000}
\setres{learnUOZ}{CPT_post_m100}{spread_penalty}{0.006}
\setres{learnUOZ}{CPT_post_m100}{theta_mean}{0.81, 0.42, 0.93}
\setres{learnUOZ}{CPT_post_m100}{theta_sd}{0.04, 0.02, 0.06}
\setres{learnUOZ}{CPT_post_m100}{volume_ess}{55{,}652}
\setres{learnUOZ}{CPT_post_m3}{ess}{139}
\setres{learnUOZ}{CPT_post_m3}{kappa}{0.752}
\setres{learnUOZ}{CPT_post_m3}{kappa_crps}{0.542}
\setres{learnUOZ}{CPT_post_m3}{kappa_crps_se}{0.030}
\setres{learnUOZ}{CPT_post_m3}{kappa_prior_mean}{0.782}
\setres{learnUOZ}{CPT_post_m3}{kappa_se}{0.034}
\setres{learnUOZ}{CPT_post_m3}{mse}{79.71}
\setres{learnUOZ}{CPT_post_m3}{n_subjects_seen}{3}
\setres{learnUOZ}{CPT_post_m3}{r_ff}{0.140}
\setres{learnUOZ}{CPT_post_m3}{r_ff_se}{0.001}
\setres{learnUOZ}{CPT_post_m3}{r_prior}{0.9896}
\setres{learnUOZ}{CPT_post_m3}{r_prior_se}{0.000}
\setres{learnUOZ}{CPT_post_m3}{r_prob}{0.991}
\setres{learnUOZ}{CPT_post_m3}{r_prob_se}{0.000}
\setres{learnUOZ}{CPT_post_m3}{spread_penalty}{0.031}
\setres{learnUOZ}{CPT_post_m3}{theta_mean}{0.80, 0.57, 1.43}
\setres{learnUOZ}{CPT_post_m3}{theta_sd}{0.12, 0.08, 0.32}
\setres{learnUOZ}{CPT_post_m3}{volume_ess}{55{,}652}
\setres{learnUOZ}{CPT_post_m30}{ess}{6}
\setres{learnUOZ}{CPT_post_m30}{kappa}{0.954}
\setres{learnUOZ}{CPT_post_m30}{kappa_crps}{0.664}
\setres{learnUOZ}{CPT_post_m30}{kappa_crps_se}{0.013}
\setres{learnUOZ}{CPT_post_m30}{kappa_prior_mean}{0.961}
\setres{learnUOZ}{CPT_post_m30}{kappa_se}{0.015}
\setres{learnUOZ}{CPT_post_m30}{mse}{65.14}
\setres{learnUOZ}{CPT_post_m30}{n_subjects_seen}{30}
\setres{learnUOZ}{CPT_post_m30}{r_ff}{0.140}
\setres{learnUOZ}{CPT_post_m30}{r_ff_se}{0.001}
\setres{learnUOZ}{CPT_post_m30}{r_prior}{0.9994}
\setres{learnUOZ}{CPT_post_m30}{r_prior_se}{0.000}
\setres{learnUOZ}{CPT_post_m30}{r_prob}{0.999}
\setres{learnUOZ}{CPT_post_m30}{r_prob_se}{0.000}
\setres{learnUOZ}{CPT_post_m30}{spread_penalty}{0.007}
\setres{learnUOZ}{CPT_post_m30}{theta_mean}{0.77, 0.42, 1.16}
\setres{learnUOZ}{CPT_post_m30}{theta_sd}{0.06, 0.02, 0.10}
\setres{learnUOZ}{CPT_post_m30}{volume_ess}{55{,}652}
\setres{learnUOZ}{CPT_post_mall}{ess}{2}
\setres{learnUOZ}{CPT_post_mall}{kappa}{0.977}
\setres{learnUOZ}{CPT_post_mall}{kappa_crps}{0.685}
\setres{learnUOZ}{CPT_post_mall}{kappa_crps_se}{0.012}
\setres{learnUOZ}{CPT_post_mall}{kappa_prior_mean}{0.983}
\setres{learnUOZ}{CPT_post_mall}{kappa_se}{0.005}
\setres{learnUOZ}{CPT_post_mall}{mse}{63.49}
\setres{learnUOZ}{CPT_post_mall}{n_subjects_seen}{179}
\setres{learnUOZ}{CPT_post_mall}{r_ff}{0.140}
\setres{learnUOZ}{CPT_post_mall}{r_ff_se}{0.001}
\setres{learnUOZ}{CPT_post_mall}{r_prior}{0.9998}
\setres{learnUOZ}{CPT_post_mall}{r_prior_se}{0.000}
\setres{learnUOZ}{CPT_post_mall}{r_prob}{1.000}
\setres{learnUOZ}{CPT_post_mall}{r_prob_se}{0.000}
\setres{learnUOZ}{CPT_post_mall}{spread_penalty}{0.006}
\setres{learnUOZ}{CPT_post_mall}{theta_mean}{0.81, 0.43, 0.92}
\setres{learnUOZ}{CPT_post_mall}{theta_sd}{0.04, 0.01, 0.06}
\setres{learnUOZ}{CPT_post_mall}{volume_ess}{55{,}652}
\setres{learnUOZ}{DA_post_m0}{ess}{1{,}600}
\setres{learnUOZ}{DA_post_m0}{kappa}{-0.867}
\setres{learnUOZ}{DA_post_m0}{kappa_crps}{0.064}
\setres{learnUOZ}{DA_post_m0}{kappa_crps_se}{0.067}
\setres{learnUOZ}{DA_post_m0}{kappa_prior_mean}{0.150}
\setres{learnUOZ}{DA_post_m0}{kappa_se}{0.214}
\setres{learnUOZ}{DA_post_m0}{mse}{196.17}
\setres{learnUOZ}{DA_post_m0}{n_subjects_seen}{0}
\setres{learnUOZ}{DA_post_m0}{r_ff}{0.281}
\setres{learnUOZ}{DA_post_m0}{r_ff_se}{0.004}
\setres{learnUOZ}{DA_post_m0}{r_prior}{0.8649}
\setres{learnUOZ}{DA_post_m0}{r_prior_se}{0.001}
\setres{learnUOZ}{DA_post_m0}{r_prob}{0.903}
\setres{learnUOZ}{DA_post_m0}{r_prob_se}{0.001}
\setres{learnUOZ}{DA_post_m0}{spread_penalty}{1.017}
\setres{learnUOZ}{DA_post_m0}{theta_mean}{0.50, 1.00}
\setres{learnUOZ}{DA_post_m0}{theta_sd}{0.29, 1.15}
\setres{learnUOZ}{DA_post_m0}{volume_ess}{436{,}452}
\setres{learnUOZ}{DA_post_m10}{ess}{34}
\setres{learnUOZ}{DA_post_m10}{kappa}{0.591}
\setres{learnUOZ}{DA_post_m10}{kappa_crps}{0.326}
\setres{learnUOZ}{DA_post_m10}{kappa_crps_se}{0.052}
\setres{learnUOZ}{DA_post_m10}{kappa_prior_mean}{0.604}
\setres{learnUOZ}{DA_post_m10}{kappa_se}{0.060}
\setres{learnUOZ}{DA_post_m10}{mse}{91.26}
\setres{learnUOZ}{DA_post_m10}{n_subjects_seen}{10}
\setres{learnUOZ}{DA_post_m10}{r_ff}{0.281}
\setres{learnUOZ}{DA_post_m10}{r_ff_se}{0.004}
\setres{learnUOZ}{DA_post_m10}{r_prior}{0.9943}
\setres{learnUOZ}{DA_post_m10}{r_prior_se}{0.000}
\setres{learnUOZ}{DA_post_m10}{r_prob}{0.996}
\setres{learnUOZ}{DA_post_m10}{r_prob_se}{0.000}
\setres{learnUOZ}{DA_post_m10}{spread_penalty}{0.013}
\setres{learnUOZ}{DA_post_m10}{theta_mean}{0.74, -0.12}
\setres{learnUOZ}{DA_post_m10}{theta_sd}{0.14, 0.21}
\setres{learnUOZ}{DA_post_m10}{volume_ess}{436{,}452}
\setres{learnUOZ}{DA_post_m100}{ess}{2}
\setres{learnUOZ}{DA_post_m100}{kappa}{0.624}
\setres{learnUOZ}{DA_post_m100}{kappa_crps}{0.363}
\setres{learnUOZ}{DA_post_m100}{kappa_crps_se}{0.050}
\setres{learnUOZ}{DA_post_m100}{kappa_prior_mean}{0.625}
\setres{learnUOZ}{DA_post_m100}{kappa_se}{0.063}
\setres{learnUOZ}{DA_post_m100}{mse}{88.90}
\setres{learnUOZ}{DA_post_m100}{n_subjects_seen}{100}
\setres{learnUOZ}{DA_post_m100}{r_ff}{0.281}
\setres{learnUOZ}{DA_post_m100}{r_ff_se}{0.004}
\setres{learnUOZ}{DA_post_m100}{r_prior}{0.9999}
\setres{learnUOZ}{DA_post_m100}{r_prior_se}{0.000}
\setres{learnUOZ}{DA_post_m100}{r_prob}{1.000}
\setres{learnUOZ}{DA_post_m100}{r_prob_se}{0.000}
\setres{learnUOZ}{DA_post_m100}{spread_penalty}{0.001}
\setres{learnUOZ}{DA_post_m100}{theta_mean}{0.98, 0.43}
\setres{learnUOZ}{DA_post_m100}{theta_sd}{0.02, 0.04}
\setres{learnUOZ}{DA_post_m100}{volume_ess}{436{,}452}
\setres{learnUOZ}{DA_post_m3}{ess}{52}
\setres{learnUOZ}{DA_post_m3}{kappa}{0.432}
\setres{learnUOZ}{DA_post_m3}{kappa_crps}{0.233}
\setres{learnUOZ}{DA_post_m3}{kappa_crps_se}{0.059}
\setres{learnUOZ}{DA_post_m3}{kappa_prior_mean}{0.455}
\setres{learnUOZ}{DA_post_m3}{kappa_se}{0.068}
\setres{learnUOZ}{DA_post_m3}{mse}{102.68}
\setres{learnUOZ}{DA_post_m3}{n_subjects_seen}{3}
\setres{learnUOZ}{DA_post_m3}{r_ff}{0.281}
\setres{learnUOZ}{DA_post_m3}{r_ff_se}{0.004}
\setres{learnUOZ}{DA_post_m3}{r_prior}{0.9905}
\setres{learnUOZ}{DA_post_m3}{r_prior_se}{0.000}
\setres{learnUOZ}{DA_post_m3}{r_prob}{0.993}
\setres{learnUOZ}{DA_post_m3}{r_prob_se}{0.000}
\setres{learnUOZ}{DA_post_m3}{spread_penalty}{0.022}
\setres{learnUOZ}{DA_post_m3}{theta_mean}{0.78, -0.30}
\setres{learnUOZ}{DA_post_m3}{theta_sd}{0.15, 0.18}
\setres{learnUOZ}{DA_post_m3}{volume_ess}{436{,}452}
\setres{learnUOZ}{DA_post_m30}{ess}{18}
\setres{learnUOZ}{DA_post_m30}{kappa}{0.587}
\setres{learnUOZ}{DA_post_m30}{kappa_crps}{0.339}
\setres{learnUOZ}{DA_post_m30}{kappa_crps_se}{0.052}
\setres{learnUOZ}{DA_post_m30}{kappa_prior_mean}{0.593}
\setres{learnUOZ}{DA_post_m30}{kappa_se}{0.060}
\setres{learnUOZ}{DA_post_m30}{mse}{91.57}
\setres{learnUOZ}{DA_post_m30}{n_subjects_seen}{30}
\setres{learnUOZ}{DA_post_m30}{r_ff}{0.281}
\setres{learnUOZ}{DA_post_m30}{r_ff_se}{0.004}
\setres{learnUOZ}{DA_post_m30}{r_prior}{0.9971}
\setres{learnUOZ}{DA_post_m30}{r_prior_se}{0.000}
\setres{learnUOZ}{DA_post_m30}{r_prob}{0.998}
\setres{learnUOZ}{DA_post_m30}{r_prob_se}{0.000}
\setres{learnUOZ}{DA_post_m30}{spread_penalty}{0.006}
\setres{learnUOZ}{DA_post_m30}{theta_mean}{0.77, -0.14}
\setres{learnUOZ}{DA_post_m30}{theta_sd}{0.13, 0.18}
\setres{learnUOZ}{DA_post_m30}{volume_ess}{436{,}452}
\setres{learnUOZ}{DA_post_mall}{ess}{1}
\setres{learnUOZ}{DA_post_mall}{kappa}{0.624}
\setres{learnUOZ}{DA_post_mall}{kappa_crps}{0.353}
\setres{learnUOZ}{DA_post_mall}{kappa_crps_se}{0.051}
\setres{learnUOZ}{DA_post_mall}{kappa_prior_mean}{0.625}
\setres{learnUOZ}{DA_post_mall}{kappa_se}{0.063}
\setres{learnUOZ}{DA_post_mall}{mse}{88.90}
\setres{learnUOZ}{DA_post_mall}{n_subjects_seen}{179}
\setres{learnUOZ}{DA_post_mall}{r_ff}{0.281}
\setres{learnUOZ}{DA_post_mall}{r_ff_se}{0.004}
\setres{learnUOZ}{DA_post_mall}{r_prior}{0.9999}
\setres{learnUOZ}{DA_post_mall}{r_prior_se}{0.000}
\setres{learnUOZ}{DA_post_mall}{r_prob}{1.000}
\setres{learnUOZ}{DA_post_mall}{r_prob_se}{0.000}
\setres{learnUOZ}{DA_post_mall}{spread_penalty}{0.001}
\setres{learnUOZ}{DA_post_mall}{theta_mean}{0.99, 0.44}
\setres{learnUOZ}{DA_post_mall}{theta_sd}{0.01, 0.02}
\setres{learnUOZ}{DA_post_mall}{volume_ess}{436{,}452}
\setres{learnUOZ}{EU_post_m0}{ess}{200}
\setres{learnUOZ}{EU_post_m0}{kappa}{-0.195}
\setres{learnUOZ}{EU_post_m0}{kappa_crps}{0.206}
\setres{learnUOZ}{EU_post_m0}{kappa_crps_se}{0.060}
\setres{learnUOZ}{EU_post_m0}{kappa_prior_mean}{0.306}
\setres{learnUOZ}{EU_post_m0}{kappa_se}{0.155}
\setres{learnUOZ}{EU_post_m0}{mse}{147.78}
\setres{learnUOZ}{EU_post_m0}{n_subjects_seen}{0}
\setres{learnUOZ}{EU_post_m0}{r_ff}{0.336}
\setres{learnUOZ}{EU_post_m0}{r_ff_se}{0.005}
\setres{learnUOZ}{EU_post_m0}{r_prior}{0.5781}
\setres{learnUOZ}{EU_post_m0}{r_prior_se}{0.001}
\setres{learnUOZ}{EU_post_m0}{r_prob}{0.720}
\setres{learnUOZ}{EU_post_m0}{r_prob_se}{0.002}
\setres{learnUOZ}{EU_post_m0}{spread_penalty}{0.500}
\setres{learnUOZ}{EU_post_m0}{theta_mean}{0.51}
\setres{learnUOZ}{EU_post_m0}{theta_sd}{0.29}
\setres{learnUOZ}{EU_post_m0}{volume_ess}{172{,}086}
\setres{learnUOZ}{EU_post_m10}{ess}{26}
\setres{learnUOZ}{EU_post_m10}{kappa}{0.600}
\setres{learnUOZ}{EU_post_m10}{kappa_crps}{0.337}
\setres{learnUOZ}{EU_post_m10}{kappa_crps_se}{0.052}
\setres{learnUOZ}{EU_post_m10}{kappa_prior_mean}{0.604}
\setres{learnUOZ}{EU_post_m10}{kappa_se}{0.061}
\setres{learnUOZ}{EU_post_m10}{mse}{90.60}
\setres{learnUOZ}{EU_post_m10}{n_subjects_seen}{10}
\setres{learnUOZ}{EU_post_m10}{r_ff}{0.336}
\setres{learnUOZ}{EU_post_m10}{r_ff_se}{0.005}
\setres{learnUOZ}{EU_post_m10}{r_prior}{0.9578}
\setres{learnUOZ}{EU_post_m10}{r_prior_se}{0.000}
\setres{learnUOZ}{EU_post_m10}{r_prob}{0.972}
\setres{learnUOZ}{EU_post_m10}{r_prob_se}{0.000}
\setres{learnUOZ}{EU_post_m10}{spread_penalty}{0.004}
\setres{learnUOZ}{EU_post_m10}{theta_mean}{0.83}
\setres{learnUOZ}{EU_post_m10}{theta_sd}{0.04}
\setres{learnUOZ}{EU_post_m10}{volume_ess}{172{,}086}
\setres{learnUOZ}{EU_post_m100}{ess}{7}
\setres{learnUOZ}{EU_post_m100}{kappa}{0.616}
\setres{learnUOZ}{EU_post_m100}{kappa_crps}{0.347}
\setres{learnUOZ}{EU_post_m100}{kappa_crps_se}{0.051}
\setres{learnUOZ}{EU_post_m100}{kappa_prior_mean}{0.616}
\setres{learnUOZ}{EU_post_m100}{kappa_se}{0.062}
\setres{learnUOZ}{EU_post_m100}{mse}{89.47}
\setres{learnUOZ}{EU_post_m100}{n_subjects_seen}{100}
\setres{learnUOZ}{EU_post_m100}{r_ff}{0.336}
\setres{learnUOZ}{EU_post_m100}{r_ff_se}{0.005}
\setres{learnUOZ}{EU_post_m100}{r_prior}{0.9866}
\setres{learnUOZ}{EU_post_m100}{r_prior_se}{0.000}
\setres{learnUOZ}{EU_post_m100}{r_prob}{0.991}
\setres{learnUOZ}{EU_post_m100}{r_prob_se}{0.000}
\setres{learnUOZ}{EU_post_m100}{spread_penalty}{0.000}
\setres{learnUOZ}{EU_post_m100}{theta_mean}{0.76}
\setres{learnUOZ}{EU_post_m100}{theta_sd}{0.01}
\setres{learnUOZ}{EU_post_m100}{volume_ess}{172{,}086}
\setres{learnUOZ}{EU_post_m3}{ess}{19}
\setres{learnUOZ}{EU_post_m3}{kappa}{0.533}
\setres{learnUOZ}{EU_post_m3}{kappa_crps}{0.304}
\setres{learnUOZ}{EU_post_m3}{kappa_crps_se}{0.055}
\setres{learnUOZ}{EU_post_m3}{kappa_prior_mean}{0.537}
\setres{learnUOZ}{EU_post_m3}{kappa_se}{0.063}
\setres{learnUOZ}{EU_post_m3}{mse}{95.44}
\setres{learnUOZ}{EU_post_m3}{n_subjects_seen}{3}
\setres{learnUOZ}{EU_post_m3}{r_ff}{0.336}
\setres{learnUOZ}{EU_post_m3}{r_ff_se}{0.005}
\setres{learnUOZ}{EU_post_m3}{r_prior}{0.9721}
\setres{learnUOZ}{EU_post_m3}{r_prior_se}{0.000}
\setres{learnUOZ}{EU_post_m3}{r_prob}{0.981}
\setres{learnUOZ}{EU_post_m3}{r_prob_se}{0.000}
\setres{learnUOZ}{EU_post_m3}{spread_penalty}{0.004}
\setres{learnUOZ}{EU_post_m3}{theta_mean}{0.96}
\setres{learnUOZ}{EU_post_m3}{theta_sd}{0.04}
\setres{learnUOZ}{EU_post_m3}{volume_ess}{172{,}086}
\setres{learnUOZ}{EU_post_m30}{ess}{16}
\setres{learnUOZ}{EU_post_m30}{kappa}{0.591}
\setres{learnUOZ}{EU_post_m30}{kappa_crps}{0.331}
\setres{learnUOZ}{EU_post_m30}{kappa_crps_se}{0.052}
\setres{learnUOZ}{EU_post_m30}{kappa_prior_mean}{0.592}
\setres{learnUOZ}{EU_post_m30}{kappa_se}{0.061}
\setres{learnUOZ}{EU_post_m30}{mse}{91.26}
\setres{learnUOZ}{EU_post_m30}{n_subjects_seen}{30}
\setres{learnUOZ}{EU_post_m30}{r_ff}{0.336}
\setres{learnUOZ}{EU_post_m30}{r_ff_se}{0.005}
\setres{learnUOZ}{EU_post_m30}{r_prior}{0.9750}
\setres{learnUOZ}{EU_post_m30}{r_prior_se}{0.000}
\setres{learnUOZ}{EU_post_m30}{r_prob}{0.983}
\setres{learnUOZ}{EU_post_m30}{r_prob_se}{0.000}
\setres{learnUOZ}{EU_post_m30}{spread_penalty}{0.001}
\setres{learnUOZ}{EU_post_m30}{theta_mean}{0.87}
\setres{learnUOZ}{EU_post_m30}{theta_sd}{0.02}
\setres{learnUOZ}{EU_post_m30}{volume_ess}{172{,}086}
\setres{learnUOZ}{EU_post_mall}{ess}{6}
\setres{learnUOZ}{EU_post_mall}{kappa}{0.616}
\setres{learnUOZ}{EU_post_mall}{kappa_crps}{0.339}
\setres{learnUOZ}{EU_post_mall}{kappa_crps_se}{0.052}
\setres{learnUOZ}{EU_post_mall}{kappa_prior_mean}{0.616}
\setres{learnUOZ}{EU_post_mall}{kappa_se}{0.062}
\setres{learnUOZ}{EU_post_mall}{mse}{89.49}
\setres{learnUOZ}{EU_post_mall}{n_subjects_seen}{179}
\setres{learnUOZ}{EU_post_mall}{r_ff}{0.336}
\setres{learnUOZ}{EU_post_mall}{r_ff_se}{0.005}
\setres{learnUOZ}{EU_post_mall}{r_prior}{0.9900}
\setres{learnUOZ}{EU_post_mall}{r_prior_se}{0.000}
\setres{learnUOZ}{EU_post_mall}{r_prob}{0.993}
\setres{learnUOZ}{EU_post_mall}{r_prob_se}{0.000}
\setres{learnUOZ}{EU_post_mall}{spread_penalty}{0.000}
\setres{learnUOZ}{EU_post_mall}{theta_mean}{0.76}
\setres{learnUOZ}{EU_post_mall}{theta_sd}{0.01}
\setres{learnUOZ}{EU_post_mall}{volume_ess}{172{,}086}
\setres{learnUOZ}{reference}{n_subjects}{179}
\setres{learnUOZ}{reference}{sigma}{7.8}
\setres{learnOJ}{CPT_post_m0}{ess}{732}
\setres{learnOJ}{CPT_post_m0}{kappa}{0.797}
\setres{learnOJ}{CPT_post_m0}{kappa_crps}{0.642}
\setres{learnOJ}{CPT_post_m0}{kappa_crps_se}{0.022}
\setres{learnOJ}{CPT_post_m0}{kappa_prior_mean}{0.907}
\setres{learnOJ}{CPT_post_m0}{kappa_se}{0.020}
\setres{learnOJ}{CPT_post_m0}{mse}{61.61}
\setres{learnOJ}{CPT_post_m0}{n_subjects_seen}{0}
\setres{learnOJ}{CPT_post_m0}{r_ff}{0.173}
\setres{learnOJ}{CPT_post_m0}{r_ff_se}{0.002}
\setres{learnOJ}{CPT_post_m0}{r_prior}{0.9514}
\setres{learnOJ}{CPT_post_m0}{r_prior_se}{0.000}
\setres{learnOJ}{CPT_post_m0}{r_prob}{0.960}
\setres{learnOJ}{CPT_post_m0}{r_prob_se}{0.000}
\setres{learnOJ}{CPT_post_m0}{spread_penalty}{0.110}
\setres{learnOJ}{CPT_post_m0}{theta_mean}{0.70, 0.73, 1.08}
\setres{learnOJ}{CPT_post_m0}{theta_sd}{0.14, 0.18, 0.28}
\setres{learnOJ}{CPT_post_m0}{volume_ess}{58{,}937}
\setres{learnOJ}{CPT_post_m10}{ess}{10}
\setres{learnOJ}{CPT_post_m10}{kappa}{0.944}
\setres{learnOJ}{CPT_post_m10}{kappa_crps}{0.659}
\setres{learnOJ}{CPT_post_m10}{kappa_crps_se}{0.012}
\setres{learnOJ}{CPT_post_m10}{kappa_prior_mean}{0.953}
\setres{learnOJ}{CPT_post_m10}{kappa_se}{0.007}
\setres{learnOJ}{CPT_post_m10}{mse}{47.33}
\setres{learnOJ}{CPT_post_m10}{n_subjects_seen}{10}
\setres{learnOJ}{CPT_post_m10}{r_ff}{0.173}
\setres{learnOJ}{CPT_post_m10}{r_ff_se}{0.002}
\setres{learnOJ}{CPT_post_m10}{r_prior}{0.9979}
\setres{learnOJ}{CPT_post_m10}{r_prior_se}{0.000}
\setres{learnOJ}{CPT_post_m10}{r_prob}{0.998}
\setres{learnOJ}{CPT_post_m10}{r_prob_se}{0.000}
\setres{learnOJ}{CPT_post_m10}{spread_penalty}{0.009}
\setres{learnOJ}{CPT_post_m10}{theta_mean}{0.78, 0.32, 0.92}
\setres{learnOJ}{CPT_post_m10}{theta_sd}{0.06, 0.03, 0.09}
\setres{learnOJ}{CPT_post_m10}{volume_ess}{58{,}937}
\setres{learnOJ}{CPT_post_m100}{ess}{2}
\setres{learnOJ}{CPT_post_m100}{kappa}{0.981}
\setres{learnOJ}{CPT_post_m100}{kappa_crps}{0.721}
\setres{learnOJ}{CPT_post_m100}{kappa_crps_se}{0.011}
\setres{learnOJ}{CPT_post_m100}{kappa_prior_mean}{0.984}
\setres{learnOJ}{CPT_post_m100}{kappa_se}{0.004}
\setres{learnOJ}{CPT_post_m100}{mse}{43.76}
\setres{learnOJ}{CPT_post_m100}{n_subjects_seen}{100}
\setres{learnOJ}{CPT_post_m100}{r_ff}{0.173}
\setres{learnOJ}{CPT_post_m100}{r_ff_se}{0.002}
\setres{learnOJ}{CPT_post_m100}{r_prior}{0.9998}
\setres{learnOJ}{CPT_post_m100}{r_prior_se}{0.000}
\setres{learnOJ}{CPT_post_m100}{r_prob}{1.000}
\setres{learnOJ}{CPT_post_m100}{r_prob_se}{0.000}
\setres{learnOJ}{CPT_post_m100}{spread_penalty}{0.003}
\setres{learnOJ}{CPT_post_m100}{theta_mean}{0.78, 0.43, 1.02}
\setres{learnOJ}{CPT_post_m100}{theta_sd}{0.02, 0.00, 0.05}
\setres{learnOJ}{CPT_post_m100}{volume_ess}{58{,}937}
\setres{learnOJ}{CPT_post_m3}{ess}{45}
\setres{learnOJ}{CPT_post_m3}{kappa}{0.806}
\setres{learnOJ}{CPT_post_m3}{kappa_crps}{0.485}
\setres{learnOJ}{CPT_post_m3}{kappa_crps_se}{0.015}
\setres{learnOJ}{CPT_post_m3}{kappa_prior_mean}{0.826}
\setres{learnOJ}{CPT_post_m3}{kappa_se}{0.012}
\setres{learnOJ}{CPT_post_m3}{mse}{60.67}
\setres{learnOJ}{CPT_post_m3}{n_subjects_seen}{3}
\setres{learnOJ}{CPT_post_m3}{r_ff}{0.173}
\setres{learnOJ}{CPT_post_m3}{r_ff_se}{0.002}
\setres{learnOJ}{CPT_post_m3}{r_prior}{0.9905}
\setres{learnOJ}{CPT_post_m3}{r_prior_se}{0.000}
\setres{learnOJ}{CPT_post_m3}{r_prob}{0.992}
\setres{learnOJ}{CPT_post_m3}{r_prob_se}{0.000}
\setres{learnOJ}{CPT_post_m3}{spread_penalty}{0.019}
\setres{learnOJ}{CPT_post_m3}{theta_mean}{0.76, 0.25, 1.17}
\setres{learnOJ}{CPT_post_m3}{theta_sd}{0.08, 0.05, 0.16}
\setres{learnOJ}{CPT_post_m3}{volume_ess}{58{,}937}
\setres{learnOJ}{CPT_post_m30}{ess}{3}
\setres{learnOJ}{CPT_post_m30}{kappa}{0.972}
\setres{learnOJ}{CPT_post_m30}{kappa_crps}{0.697}
\setres{learnOJ}{CPT_post_m30}{kappa_crps_se}{0.011}
\setres{learnOJ}{CPT_post_m30}{kappa_prior_mean}{0.976}
\setres{learnOJ}{CPT_post_m30}{kappa_se}{0.006}
\setres{learnOJ}{CPT_post_m30}{mse}{44.65}
\setres{learnOJ}{CPT_post_m30}{n_subjects_seen}{30}
\setres{learnOJ}{CPT_post_m30}{r_ff}{0.173}
\setres{learnOJ}{CPT_post_m30}{r_ff_se}{0.002}
\setres{learnOJ}{CPT_post_m30}{r_prior}{0.9996}
\setres{learnOJ}{CPT_post_m30}{r_prior_se}{0.000}
\setres{learnOJ}{CPT_post_m30}{r_prob}{1.000}
\setres{learnOJ}{CPT_post_m30}{r_prob_se}{0.000}
\setres{learnOJ}{CPT_post_m30}{spread_penalty}{0.004}
\setres{learnOJ}{CPT_post_m30}{theta_mean}{0.81, 0.43, 1.00}
\setres{learnOJ}{CPT_post_m30}{theta_sd}{0.05, 0.01, 0.07}
\setres{learnOJ}{CPT_post_m30}{volume_ess}{58{,}937}
\setres{learnOJ}{CPT_post_mall}{ess}{2}
\setres{learnOJ}{CPT_post_mall}{kappa}{0.986}
\setres{learnOJ}{CPT_post_mall}{kappa_crps}{0.722}
\setres{learnOJ}{CPT_post_mall}{kappa_crps_se}{0.011}
\setres{learnOJ}{CPT_post_mall}{kappa_prior_mean}{0.989}
\setres{learnOJ}{CPT_post_mall}{kappa_se}{0.002}
\setres{learnOJ}{CPT_post_mall}{mse}{43.26}
\setres{learnOJ}{CPT_post_mall}{n_subjects_seen}{448}
\setres{learnOJ}{CPT_post_mall}{r_ff}{0.173}
\setres{learnOJ}{CPT_post_mall}{r_ff_se}{0.002}
\setres{learnOJ}{CPT_post_mall}{r_prior}{0.9998}
\setres{learnOJ}{CPT_post_mall}{r_prior_se}{0.000}
\setres{learnOJ}{CPT_post_mall}{r_prob}{1.000}
\setres{learnOJ}{CPT_post_mall}{r_prob_se}{0.000}
\setres{learnOJ}{CPT_post_mall}{spread_penalty}{0.003}
\setres{learnOJ}{CPT_post_mall}{theta_mean}{0.75, 0.43, 1.01}
\setres{learnOJ}{CPT_post_mall}{theta_sd}{0.04, 0.00, 0.06}
\setres{learnOJ}{CPT_post_mall}{volume_ess}{58{,}937}
\setres{learnOJ}{DA_post_m0}{ess}{439}
\setres{learnOJ}{DA_post_m0}{kappa}{0.485}
\setres{learnOJ}{DA_post_m0}{kappa_crps}{0.327}
\setres{learnOJ}{DA_post_m0}{kappa_crps_se}{0.039}
\setres{learnOJ}{DA_post_m0}{kappa_prior_mean}{0.635}
\setres{learnOJ}{DA_post_m0}{kappa_se}{0.051}
\setres{learnOJ}{DA_post_m0}{mse}{91.80}
\setres{learnOJ}{DA_post_m0}{n_subjects_seen}{0}
\setres{learnOJ}{DA_post_m0}{r_ff}{0.332}
\setres{learnOJ}{DA_post_m0}{r_ff_se}{0.005}
\setres{learnOJ}{DA_post_m0}{r_prior}{0.9520}
\setres{learnOJ}{DA_post_m0}{r_prior_se}{0.000}
\setres{learnOJ}{DA_post_m0}{r_prob}{0.968}
\setres{learnOJ}{DA_post_m0}{r_prob_se}{0.000}
\setres{learnOJ}{DA_post_m0}{spread_penalty}{0.150}
\setres{learnOJ}{DA_post_m0}{theta_mean}{0.58, 0.30}
\setres{learnOJ}{DA_post_m0}{theta_sd}{0.19, 0.45}
\setres{learnOJ}{DA_post_m0}{volume_ess}{518{,}446}
\setres{learnOJ}{DA_post_m10}{ess}{24}
\setres{learnOJ}{DA_post_m10}{kappa}{0.736}
\setres{learnOJ}{DA_post_m10}{kappa_crps}{0.401}
\setres{learnOJ}{DA_post_m10}{kappa_crps_se}{0.037}
\setres{learnOJ}{DA_post_m10}{kappa_prior_mean}{0.739}
\setres{learnOJ}{DA_post_m10}{kappa_se}{0.030}
\setres{learnOJ}{DA_post_m10}{mse}{67.48}
\setres{learnOJ}{DA_post_m10}{n_subjects_seen}{10}
\setres{learnOJ}{DA_post_m10}{r_ff}{0.332}
\setres{learnOJ}{DA_post_m10}{r_ff_se}{0.005}
\setres{learnOJ}{DA_post_m10}{r_prior}{0.9983}
\setres{learnOJ}{DA_post_m10}{r_prior_se}{0.000}
\setres{learnOJ}{DA_post_m10}{r_prob}{0.999}
\setres{learnOJ}{DA_post_m10}{r_prob_se}{0.000}
\setres{learnOJ}{DA_post_m10}{spread_penalty}{0.003}
\setres{learnOJ}{DA_post_m10}{theta_mean}{0.91, 0.50}
\setres{learnOJ}{DA_post_m10}{theta_sd}{0.05, 0.15}
\setres{learnOJ}{DA_post_m10}{volume_ess}{518{,}446}
\setres{learnOJ}{DA_post_m100}{ess}{1}
\setres{learnOJ}{DA_post_m100}{kappa}{0.758}
\setres{learnOJ}{DA_post_m100}{kappa_crps}{0.423}
\setres{learnOJ}{DA_post_m100}{kappa_crps_se}{0.035}
\setres{learnOJ}{DA_post_m100}{kappa_prior_mean}{0.758}
\setres{learnOJ}{DA_post_m100}{kappa_se}{0.026}
\setres{learnOJ}{DA_post_m100}{mse}{65.40}
\setres{learnOJ}{DA_post_m100}{n_subjects_seen}{100}
\setres{learnOJ}{DA_post_m100}{r_ff}{0.332}
\setres{learnOJ}{DA_post_m100}{r_ff_se}{0.005}
\setres{learnOJ}{DA_post_m100}{r_prior}{0.9999}
\setres{learnOJ}{DA_post_m100}{r_prior_se}{0.000}
\setres{learnOJ}{DA_post_m100}{r_prob}{1.000}
\setres{learnOJ}{DA_post_m100}{r_prob_se}{0.000}
\setres{learnOJ}{DA_post_m100}{spread_penalty}{0.001}
\setres{learnOJ}{DA_post_m100}{theta_mean}{0.99, 0.25}
\setres{learnOJ}{DA_post_m100}{theta_sd}{0.01, 0.01}
\setres{learnOJ}{DA_post_m100}{volume_ess}{518{,}446}
\setres{learnOJ}{DA_post_m3}{ess}{66}
\setres{learnOJ}{DA_post_m3}{kappa}{0.728}
\setres{learnOJ}{DA_post_m3}{kappa_crps}{0.399}
\setres{learnOJ}{DA_post_m3}{kappa_crps_se}{0.036}
\setres{learnOJ}{DA_post_m3}{kappa_prior_mean}{0.740}
\setres{learnOJ}{DA_post_m3}{kappa_se}{0.028}
\setres{learnOJ}{DA_post_m3}{mse}{68.32}
\setres{learnOJ}{DA_post_m3}{n_subjects_seen}{3}
\setres{learnOJ}{DA_post_m3}{r_ff}{0.332}
\setres{learnOJ}{DA_post_m3}{r_ff_se}{0.005}
\setres{learnOJ}{DA_post_m3}{r_prior}{0.9899}
\setres{learnOJ}{DA_post_m3}{r_prior_se}{0.000}
\setres{learnOJ}{DA_post_m3}{r_prob}{0.993}
\setres{learnOJ}{DA_post_m3}{r_prob_se}{0.000}
\setres{learnOJ}{DA_post_m3}{spread_penalty}{0.013}
\setres{learnOJ}{DA_post_m3}{theta_mean}{0.74, -0.10}
\setres{learnOJ}{DA_post_m3}{theta_sd}{0.15, 0.24}
\setres{learnOJ}{DA_post_m3}{volume_ess}{518{,}446}
\setres{learnOJ}{DA_post_m30}{ess}{7}
\setres{learnOJ}{DA_post_m30}{kappa}{0.755}
\setres{learnOJ}{DA_post_m30}{kappa_crps}{0.428}
\setres{learnOJ}{DA_post_m30}{kappa_crps_se}{0.035}
\setres{learnOJ}{DA_post_m30}{kappa_prior_mean}{0.756}
\setres{learnOJ}{DA_post_m30}{kappa_se}{0.026}
\setres{learnOJ}{DA_post_m30}{mse}{65.64}
\setres{learnOJ}{DA_post_m30}{n_subjects_seen}{30}
\setres{learnOJ}{DA_post_m30}{r_ff}{0.332}
\setres{learnOJ}{DA_post_m30}{r_ff_se}{0.005}
\setres{learnOJ}{DA_post_m30}{r_prior}{0.9994}
\setres{learnOJ}{DA_post_m30}{r_prior_se}{0.000}
\setres{learnOJ}{DA_post_m30}{r_prob}{1.000}
\setres{learnOJ}{DA_post_m30}{r_prob_se}{0.000}
\setres{learnOJ}{DA_post_m30}{spread_penalty}{0.001}
\setres{learnOJ}{DA_post_m30}{theta_mean}{0.96, 0.20}
\setres{learnOJ}{DA_post_m30}{theta_sd}{0.03, 0.07}
\setres{learnOJ}{DA_post_m30}{volume_ess}{518{,}446}
\setres{learnOJ}{DA_post_mall}{ess}{1}
\setres{learnOJ}{DA_post_mall}{kappa}{0.764}
\setres{learnOJ}{DA_post_mall}{kappa_crps}{0.420}
\setres{learnOJ}{DA_post_mall}{kappa_crps_se}{0.035}
\setres{learnOJ}{DA_post_mall}{kappa_prior_mean}{0.765}
\setres{learnOJ}{DA_post_mall}{kappa_se}{0.026}
\setres{learnOJ}{DA_post_mall}{mse}{64.76}
\setres{learnOJ}{DA_post_mall}{n_subjects_seen}{448}
\setres{learnOJ}{DA_post_mall}{r_ff}{0.332}
\setres{learnOJ}{DA_post_mall}{r_ff_se}{0.005}
\setres{learnOJ}{DA_post_mall}{r_prior}{0.9999}
\setres{learnOJ}{DA_post_mall}{r_prior_se}{0.000}
\setres{learnOJ}{DA_post_mall}{r_prob}{1.000}
\setres{learnOJ}{DA_post_mall}{r_prob_se}{0.000}
\setres{learnOJ}{DA_post_mall}{spread_penalty}{0.000}
\setres{learnOJ}{DA_post_mall}{theta_mean}{0.99, 0.35}
\setres{learnOJ}{DA_post_mall}{theta_sd}{0.00, 0.00}
\setres{learnOJ}{DA_post_mall}{volume_ess}{518{,}446}
\setres{learnOJ}{EU_post_m0}{ess}{98}
\setres{learnOJ}{EU_post_m0}{kappa}{0.700}
\setres{learnOJ}{EU_post_m0}{kappa_crps}{0.397}
\setres{learnOJ}{EU_post_m0}{kappa_crps_se}{0.036}
\setres{learnOJ}{EU_post_m0}{kappa_prior_mean}{0.736}
\setres{learnOJ}{EU_post_m0}{kappa_se}{0.031}
\setres{learnOJ}{EU_post_m0}{mse}{71.03}
\setres{learnOJ}{EU_post_m0}{n_subjects_seen}{0}
\setres{learnOJ}{EU_post_m0}{r_ff}{0.382}
\setres{learnOJ}{EU_post_m0}{r_ff_se}{0.005}
\setres{learnOJ}{EU_post_m0}{r_prior}{0.8459}
\setres{learnOJ}{EU_post_m0}{r_prior_se}{0.001}
\setres{learnOJ}{EU_post_m0}{r_prob}{0.905}
\setres{learnOJ}{EU_post_m0}{r_prob_se}{0.001}
\setres{learnOJ}{EU_post_m0}{spread_penalty}{0.036}
\setres{learnOJ}{EU_post_m0}{theta_mean}{0.70}
\setres{learnOJ}{EU_post_m0}{theta_sd}{0.14}
\setres{learnOJ}{EU_post_m0}{volume_ess}{160{,}713}
\setres{learnOJ}{EU_post_m10}{ess}{26}
\setres{learnOJ}{EU_post_m10}{kappa}{0.733}
\setres{learnOJ}{EU_post_m10}{kappa_crps}{0.383}
\setres{learnOJ}{EU_post_m10}{kappa_crps_se}{0.037}
\setres{learnOJ}{EU_post_m10}{kappa_prior_mean}{0.735}
\setres{learnOJ}{EU_post_m10}{kappa_se}{0.029}
\setres{learnOJ}{EU_post_m10}{mse}{67.82}
\setres{learnOJ}{EU_post_m10}{n_subjects_seen}{10}
\setres{learnOJ}{EU_post_m10}{r_ff}{0.382}
\setres{learnOJ}{EU_post_m10}{r_ff_se}{0.005}
\setres{learnOJ}{EU_post_m10}{r_prior}{0.9592}
\setres{learnOJ}{EU_post_m10}{r_prior_se}{0.000}
\setres{learnOJ}{EU_post_m10}{r_prob}{0.975}
\setres{learnOJ}{EU_post_m10}{r_prob_se}{0.000}
\setres{learnOJ}{EU_post_m10}{spread_penalty}{0.002}
\setres{learnOJ}{EU_post_m10}{theta_mean}{0.72}
\setres{learnOJ}{EU_post_m10}{theta_sd}{0.04}
\setres{learnOJ}{EU_post_m10}{volume_ess}{160{,}713}
\setres{learnOJ}{EU_post_m100}{ess}{10}
\setres{learnOJ}{EU_post_m100}{kappa}{0.745}
\setres{learnOJ}{EU_post_m100}{kappa_crps}{0.402}
\setres{learnOJ}{EU_post_m100}{kappa_crps_se}{0.036}
\setres{learnOJ}{EU_post_m100}{kappa_prior_mean}{0.745}
\setres{learnOJ}{EU_post_m100}{kappa_se}{0.026}
\setres{learnOJ}{EU_post_m100}{mse}{66.61}
\setres{learnOJ}{EU_post_m100}{n_subjects_seen}{100}
\setres{learnOJ}{EU_post_m100}{r_ff}{0.382}
\setres{learnOJ}{EU_post_m100}{r_ff_se}{0.005}
\setres{learnOJ}{EU_post_m100}{r_prior}{0.9861}
\setres{learnOJ}{EU_post_m100}{r_prior_se}{0.000}
\setres{learnOJ}{EU_post_m100}{r_prob}{0.991}
\setres{learnOJ}{EU_post_m100}{r_prob_se}{0.000}
\setres{learnOJ}{EU_post_m100}{spread_penalty}{0.000}
\setres{learnOJ}{EU_post_m100}{theta_mean}{0.86}
\setres{learnOJ}{EU_post_m100}{theta_sd}{0.01}
\setres{learnOJ}{EU_post_m100}{volume_ess}{160{,}713}
\setres{learnOJ}{EU_post_m3}{ess}{40}
\setres{learnOJ}{EU_post_m3}{kappa}{0.746}
\setres{learnOJ}{EU_post_m3}{kappa_crps}{0.418}
\setres{learnOJ}{EU_post_m3}{kappa_crps_se}{0.035}
\setres{learnOJ}{EU_post_m3}{kappa_prior_mean}{0.750}
\setres{learnOJ}{EU_post_m3}{kappa_se}{0.027}
\setres{learnOJ}{EU_post_m3}{mse}{66.52}
\setres{learnOJ}{EU_post_m3}{n_subjects_seen}{3}
\setres{learnOJ}{EU_post_m3}{r_ff}{0.382}
\setres{learnOJ}{EU_post_m3}{r_ff_se}{0.005}
\setres{learnOJ}{EU_post_m3}{r_prior}{0.9436}
\setres{learnOJ}{EU_post_m3}{r_prior_se}{0.000}
\setres{learnOJ}{EU_post_m3}{r_prob}{0.965}
\setres{learnOJ}{EU_post_m3}{r_prob_se}{0.000}
\setres{learnOJ}{EU_post_m3}{spread_penalty}{0.004}
\setres{learnOJ}{EU_post_m3}{theta_mean}{0.82}
\setres{learnOJ}{EU_post_m3}{theta_sd}{0.06}
\setres{learnOJ}{EU_post_m3}{volume_ess}{160{,}713}
\setres{learnOJ}{EU_post_m30}{ess}{18}
\setres{learnOJ}{EU_post_m30}{kappa}{0.747}
\setres{learnOJ}{EU_post_m30}{kappa_crps}{0.409}
\setres{learnOJ}{EU_post_m30}{kappa_crps_se}{0.036}
\setres{learnOJ}{EU_post_m30}{kappa_prior_mean}{0.748}
\setres{learnOJ}{EU_post_m30}{kappa_se}{0.026}
\setres{learnOJ}{EU_post_m30}{mse}{66.42}
\setres{learnOJ}{EU_post_m30}{n_subjects_seen}{30}
\setres{learnOJ}{EU_post_m30}{r_ff}{0.382}
\setres{learnOJ}{EU_post_m30}{r_ff_se}{0.005}
\setres{learnOJ}{EU_post_m30}{r_prior}{0.9756}
\setres{learnOJ}{EU_post_m30}{r_prior_se}{0.000}
\setres{learnOJ}{EU_post_m30}{r_prob}{0.985}
\setres{learnOJ}{EU_post_m30}{r_prob_se}{0.000}
\setres{learnOJ}{EU_post_m30}{spread_penalty}{0.001}
\setres{learnOJ}{EU_post_m30}{theta_mean}{0.87}
\setres{learnOJ}{EU_post_m30}{theta_sd}{0.03}
\setres{learnOJ}{EU_post_m30}{volume_ess}{160{,}713}
\setres{learnOJ}{EU_post_mall}{ess}{5}
\setres{learnOJ}{EU_post_mall}{kappa}{0.750}
\setres{learnOJ}{EU_post_mall}{kappa_crps}{0.406}
\setres{learnOJ}{EU_post_mall}{kappa_crps_se}{0.036}
\setres{learnOJ}{EU_post_mall}{kappa_prior_mean}{0.750}
\setres{learnOJ}{EU_post_mall}{kappa_se}{0.026}
\setres{learnOJ}{EU_post_mall}{mse}{66.18}
\setres{learnOJ}{EU_post_mall}{n_subjects_seen}{448}
\setres{learnOJ}{EU_post_mall}{r_ff}{0.382}
\setres{learnOJ}{EU_post_mall}{r_ff_se}{0.005}
\setres{learnOJ}{EU_post_mall}{r_prior}{0.9930}
\setres{learnOJ}{EU_post_mall}{r_prior_se}{0.000}
\setres{learnOJ}{EU_post_mall}{r_prob}{0.996}
\setres{learnOJ}{EU_post_mall}{r_prob_se}{0.000}
\setres{learnOJ}{EU_post_mall}{spread_penalty}{0.000}
\setres{learnOJ}{EU_post_mall}{theta_mean}{0.83}
\setres{learnOJ}{EU_post_mall}{theta_sd}{0.01}
\setres{learnOJ}{EU_post_mall}{volume_ess}{160{,}713}
\setres{learnOJ}{reference}{n_subjects}{448}
\setres{learnOJ}{reference}{sigma}{6.4}
\setres{learnUOJ}{CPT_post_m0}{ess}{8{,}000}
\setres{learnUOJ}{CPT_post_m0}{kappa}{0.219}
\setres{learnUOJ}{CPT_post_m0}{kappa_crps}{0.593}
\setres{learnUOJ}{CPT_post_m0}{kappa_crps_se}{0.015}
\setres{learnUOJ}{CPT_post_m0}{kappa_prior_mean}{0.908}
\setres{learnUOJ}{CPT_post_m0}{kappa_se}{0.048}
\setres{learnUOJ}{CPT_post_m0}{mse}{117.56}
\setres{learnUOJ}{CPT_post_m0}{n_subjects_seen}{0}
\setres{learnUOJ}{CPT_post_m0}{r_ff}{0.173}
\setres{learnUOJ}{CPT_post_m0}{r_ff_se}{0.002}
\setres{learnUOJ}{CPT_post_m0}{r_prior}{0.5323}
\setres{learnUOJ}{CPT_post_m0}{r_prior_se}{0.003}
\setres{learnUOJ}{CPT_post_m0}{r_prob}{0.613}
\setres{learnUOJ}{CPT_post_m0}{r_prob_se}{0.002}
\setres{learnUOJ}{CPT_post_m0}{spread_penalty}{0.688}
\setres{learnUOJ}{CPT_post_m0}{theta_mean}{0.50, 0.50, 1.55}
\setres{learnUOJ}{CPT_post_m0}{theta_sd}{0.29, 0.29, 0.84}
\setres{learnUOJ}{CPT_post_m0}{volume_ess}{58{,}937}
\setres{learnUOJ}{CPT_post_m10}{ess}{14}
\setres{learnUOJ}{CPT_post_m10}{kappa}{0.936}
\setres{learnUOJ}{CPT_post_m10}{kappa_crps}{0.648}
\setres{learnUOJ}{CPT_post_m10}{kappa_crps_se}{0.012}
\setres{learnUOJ}{CPT_post_m10}{kappa_prior_mean}{0.946}
\setres{learnUOJ}{CPT_post_m10}{kappa_se}{0.007}
\setres{learnUOJ}{CPT_post_m10}{mse}{48.08}
\setres{learnUOJ}{CPT_post_m10}{n_subjects_seen}{10}
\setres{learnUOJ}{CPT_post_m10}{r_ff}{0.173}
\setres{learnUOJ}{CPT_post_m10}{r_ff_se}{0.002}
\setres{learnUOJ}{CPT_post_m10}{r_prior}{0.9973}
\setres{learnUOJ}{CPT_post_m10}{r_prior_se}{0.000}
\setres{learnUOJ}{CPT_post_m10}{r_prob}{0.998}
\setres{learnUOJ}{CPT_post_m10}{r_prob_se}{0.000}
\setres{learnUOJ}{CPT_post_m10}{spread_penalty}{0.010}
\setres{learnUOJ}{CPT_post_m10}{theta_mean}{0.82, 0.31, 0.88}
\setres{learnUOJ}{CPT_post_m10}{theta_sd}{0.08, 0.03, 0.10}
\setres{learnUOJ}{CPT_post_m10}{volume_ess}{58{,}937}
\setres{learnUOJ}{CPT_post_m100}{ess}{2}
\setres{learnUOJ}{CPT_post_m100}{kappa}{0.981}
\setres{learnUOJ}{CPT_post_m100}{kappa_crps}{0.721}
\setres{learnUOJ}{CPT_post_m100}{kappa_crps_se}{0.011}
\setres{learnUOJ}{CPT_post_m100}{kappa_prior_mean}{0.984}
\setres{learnUOJ}{CPT_post_m100}{kappa_se}{0.004}
\setres{learnUOJ}{CPT_post_m100}{mse}{43.77}
\setres{learnUOJ}{CPT_post_m100}{n_subjects_seen}{100}
\setres{learnUOJ}{CPT_post_m100}{r_ff}{0.173}
\setres{learnUOJ}{CPT_post_m100}{r_ff_se}{0.002}
\setres{learnUOJ}{CPT_post_m100}{r_prior}{0.9998}
\setres{learnUOJ}{CPT_post_m100}{r_prior_se}{0.000}
\setres{learnUOJ}{CPT_post_m100}{r_prob}{1.000}
\setres{learnUOJ}{CPT_post_m100}{r_prob_se}{0.000}
\setres{learnUOJ}{CPT_post_m100}{spread_penalty}{0.003}
\setres{learnUOJ}{CPT_post_m100}{theta_mean}{0.78, 0.43, 1.02}
\setres{learnUOJ}{CPT_post_m100}{theta_sd}{0.03, 0.00, 0.05}
\setres{learnUOJ}{CPT_post_m100}{volume_ess}{58{,}937}
\setres{learnUOJ}{CPT_post_m3}{ess}{73}
\setres{learnUOJ}{CPT_post_m3}{kappa}{0.760}
\setres{learnUOJ}{CPT_post_m3}{kappa_crps}{0.429}
\setres{learnUOJ}{CPT_post_m3}{kappa_crps_se}{0.016}
\setres{learnUOJ}{CPT_post_m3}{kappa_prior_mean}{0.783}
\setres{learnUOJ}{CPT_post_m3}{kappa_se}{0.012}
\setres{learnUOJ}{CPT_post_m3}{mse}{65.21}
\setres{learnUOJ}{CPT_post_m3}{n_subjects_seen}{3}
\setres{learnUOJ}{CPT_post_m3}{r_ff}{0.173}
\setres{learnUOJ}{CPT_post_m3}{r_ff_se}{0.002}
\setres{learnUOJ}{CPT_post_m3}{r_prior}{0.9848}
\setres{learnUOJ}{CPT_post_m3}{r_prior_se}{0.000}
\setres{learnUOJ}{CPT_post_m3}{r_prob}{0.987}
\setres{learnUOJ}{CPT_post_m3}{r_prob_se}{0.000}
\setres{learnUOJ}{CPT_post_m3}{spread_penalty}{0.024}
\setres{learnUOJ}{CPT_post_m3}{theta_mean}{0.75, 0.22, 1.21}
\setres{learnUOJ}{CPT_post_m3}{theta_sd}{0.12, 0.05, 0.24}
\setres{learnUOJ}{CPT_post_m3}{volume_ess}{58{,}937}
\setres{learnUOJ}{CPT_post_m30}{ess}{3}
\setres{learnUOJ}{CPT_post_m30}{kappa}{0.971}
\setres{learnUOJ}{CPT_post_m30}{kappa_crps}{0.696}
\setres{learnUOJ}{CPT_post_m30}{kappa_crps_se}{0.011}
\setres{learnUOJ}{CPT_post_m30}{kappa_prior_mean}{0.975}
\setres{learnUOJ}{CPT_post_m30}{kappa_se}{0.006}
\setres{learnUOJ}{CPT_post_m30}{mse}{44.73}
\setres{learnUOJ}{CPT_post_m30}{n_subjects_seen}{30}
\setres{learnUOJ}{CPT_post_m30}{r_ff}{0.173}
\setres{learnUOJ}{CPT_post_m30}{r_ff_se}{0.002}
\setres{learnUOJ}{CPT_post_m30}{r_prior}{0.9995}
\setres{learnUOJ}{CPT_post_m30}{r_prior_se}{0.000}
\setres{learnUOJ}{CPT_post_m30}{r_prob}{1.000}
\setres{learnUOJ}{CPT_post_m30}{r_prob_se}{0.000}
\setres{learnUOJ}{CPT_post_m30}{spread_penalty}{0.004}
\setres{learnUOJ}{CPT_post_m30}{theta_mean}{0.83, 0.43, 0.98}
\setres{learnUOJ}{CPT_post_m30}{theta_sd}{0.05, 0.02, 0.08}
\setres{learnUOJ}{CPT_post_m30}{volume_ess}{58{,}937}
\setres{learnUOJ}{CPT_post_mall}{ess}{2}
\setres{learnUOJ}{CPT_post_mall}{kappa}{0.986}
\setres{learnUOJ}{CPT_post_mall}{kappa_crps}{0.722}
\setres{learnUOJ}{CPT_post_mall}{kappa_crps_se}{0.011}
\setres{learnUOJ}{CPT_post_mall}{kappa_prior_mean}{0.989}
\setres{learnUOJ}{CPT_post_mall}{kappa_se}{0.002}
\setres{learnUOJ}{CPT_post_mall}{mse}{43.26}
\setres{learnUOJ}{CPT_post_mall}{n_subjects_seen}{448}
\setres{learnUOJ}{CPT_post_mall}{r_ff}{0.173}
\setres{learnUOJ}{CPT_post_mall}{r_ff_se}{0.002}
\setres{learnUOJ}{CPT_post_mall}{r_prior}{0.9998}
\setres{learnUOJ}{CPT_post_mall}{r_prior_se}{0.000}
\setres{learnUOJ}{CPT_post_mall}{r_prob}{1.000}
\setres{learnUOJ}{CPT_post_mall}{r_prob_se}{0.000}
\setres{learnUOJ}{CPT_post_mall}{spread_penalty}{0.003}
\setres{learnUOJ}{CPT_post_mall}{theta_mean}{0.75, 0.43, 1.01}
\setres{learnUOJ}{CPT_post_mall}{theta_sd}{0.04, 0.00, 0.06}
\setres{learnUOJ}{CPT_post_mall}{volume_ess}{58{,}937}
\setres{learnUOJ}{DA_post_m0}{ess}{1{,}600}
\setres{learnUOJ}{DA_post_m0}{kappa}{0.091}
\setres{learnUOJ}{DA_post_m0}{kappa_crps}{0.233}
\setres{learnUOJ}{DA_post_m0}{kappa_crps_se}{0.042}
\setres{learnUOJ}{DA_post_m0}{kappa_prior_mean}{0.576}
\setres{learnUOJ}{DA_post_m0}{kappa_se}{0.079}
\setres{learnUOJ}{DA_post_m0}{mse}{129.96}
\setres{learnUOJ}{DA_post_m0}{n_subjects_seen}{0}
\setres{learnUOJ}{DA_post_m0}{r_ff}{0.332}
\setres{learnUOJ}{DA_post_m0}{r_ff_se}{0.005}
\setres{learnUOJ}{DA_post_m0}{r_prior}{0.8833}
\setres{learnUOJ}{DA_post_m0}{r_prior_se}{0.001}
\setres{learnUOJ}{DA_post_m0}{r_prob}{0.922}
\setres{learnUOJ}{DA_post_m0}{r_prob_se}{0.001}
\setres{learnUOJ}{DA_post_m0}{spread_penalty}{0.484}
\setres{learnUOJ}{DA_post_m0}{theta_mean}{0.50, 1.00}
\setres{learnUOJ}{DA_post_m0}{theta_sd}{0.29, 1.15}
\setres{learnUOJ}{DA_post_m0}{volume_ess}{518{,}446}
\setres{learnUOJ}{DA_post_m10}{ess}{14}
\setres{learnUOJ}{DA_post_m10}{kappa}{0.738}
\setres{learnUOJ}{DA_post_m10}{kappa_crps}{0.402}
\setres{learnUOJ}{DA_post_m10}{kappa_crps_se}{0.037}
\setres{learnUOJ}{DA_post_m10}{kappa_prior_mean}{0.740}
\setres{learnUOJ}{DA_post_m10}{kappa_se}{0.030}
\setres{learnUOJ}{DA_post_m10}{mse}{67.35}
\setres{learnUOJ}{DA_post_m10}{n_subjects_seen}{10}
\setres{learnUOJ}{DA_post_m10}{r_ff}{0.332}
\setres{learnUOJ}{DA_post_m10}{r_ff_se}{0.005}
\setres{learnUOJ}{DA_post_m10}{r_prior}{0.9991}
\setres{learnUOJ}{DA_post_m10}{r_prior_se}{0.000}
\setres{learnUOJ}{DA_post_m10}{r_prob}{0.999}
\setres{learnUOJ}{DA_post_m10}{r_prob_se}{0.000}
\setres{learnUOJ}{DA_post_m10}{spread_penalty}{0.003}
\setres{learnUOJ}{DA_post_m10}{theta_mean}{0.96, 0.63}
\setres{learnUOJ}{DA_post_m10}{theta_sd}{0.04, 0.15}
\setres{learnUOJ}{DA_post_m10}{volume_ess}{518{,}446}
\setres{learnUOJ}{DA_post_m100}{ess}{1}
\setres{learnUOJ}{DA_post_m100}{kappa}{0.758}
\setres{learnUOJ}{DA_post_m100}{kappa_crps}{0.423}
\setres{learnUOJ}{DA_post_m100}{kappa_crps_se}{0.035}
\setres{learnUOJ}{DA_post_m100}{kappa_prior_mean}{0.758}
\setres{learnUOJ}{DA_post_m100}{kappa_se}{0.026}
\setres{learnUOJ}{DA_post_m100}{mse}{65.40}
\setres{learnUOJ}{DA_post_m100}{n_subjects_seen}{100}
\setres{learnUOJ}{DA_post_m100}{r_ff}{0.332}
\setres{learnUOJ}{DA_post_m100}{r_ff_se}{0.005}
\setres{learnUOJ}{DA_post_m100}{r_prior}{0.9999}
\setres{learnUOJ}{DA_post_m100}{r_prior_se}{0.000}
\setres{learnUOJ}{DA_post_m100}{r_prob}{1.000}
\setres{learnUOJ}{DA_post_m100}{r_prob_se}{0.000}
\setres{learnUOJ}{DA_post_m100}{spread_penalty}{0.001}
\setres{learnUOJ}{DA_post_m100}{theta_mean}{0.99, 0.25}
\setres{learnUOJ}{DA_post_m100}{theta_sd}{0.00, 0.01}
\setres{learnUOJ}{DA_post_m100}{volume_ess}{518{,}446}
\setres{learnUOJ}{DA_post_m3}{ess}{21}
\setres{learnUOJ}{DA_post_m3}{kappa}{0.664}
\setres{learnUOJ}{DA_post_m3}{kappa_crps}{0.356}
\setres{learnUOJ}{DA_post_m3}{kappa_crps_se}{0.035}
\setres{learnUOJ}{DA_post_m3}{kappa_prior_mean}{0.729}
\setres{learnUOJ}{DA_post_m3}{kappa_se}{0.029}
\setres{learnUOJ}{DA_post_m3}{mse}{74.51}
\setres{learnUOJ}{DA_post_m3}{n_subjects_seen}{3}
\setres{learnUOJ}{DA_post_m3}{r_ff}{0.332}
\setres{learnUOJ}{DA_post_m3}{r_ff_se}{0.005}
\setres{learnUOJ}{DA_post_m3}{r_prior}{0.9758}
\setres{learnUOJ}{DA_post_m3}{r_prior_se}{0.000}
\setres{learnUOJ}{DA_post_m3}{r_prob}{0.984}
\setres{learnUOJ}{DA_post_m3}{r_prob_se}{0.000}
\setres{learnUOJ}{DA_post_m3}{spread_penalty}{0.065}
\setres{learnUOJ}{DA_post_m3}{theta_mean}{0.52, -0.39}
\setres{learnUOJ}{DA_post_m3}{theta_sd}{0.31, 0.43}
\setres{learnUOJ}{DA_post_m3}{volume_ess}{518{,}446}
\setres{learnUOJ}{DA_post_m30}{ess}{3}
\setres{learnUOJ}{DA_post_m30}{kappa}{0.757}
\setres{learnUOJ}{DA_post_m30}{kappa_crps}{0.431}
\setres{learnUOJ}{DA_post_m30}{kappa_crps_se}{0.035}
\setres{learnUOJ}{DA_post_m30}{kappa_prior_mean}{0.758}
\setres{learnUOJ}{DA_post_m30}{kappa_se}{0.026}
\setres{learnUOJ}{DA_post_m30}{mse}{65.49}
\setres{learnUOJ}{DA_post_m30}{n_subjects_seen}{30}
\setres{learnUOJ}{DA_post_m30}{r_ff}{0.332}
\setres{learnUOJ}{DA_post_m30}{r_ff_se}{0.005}
\setres{learnUOJ}{DA_post_m30}{r_prior}{0.9997}
\setres{learnUOJ}{DA_post_m30}{r_prior_se}{0.000}
\setres{learnUOJ}{DA_post_m30}{r_prob}{1.000}
\setres{learnUOJ}{DA_post_m30}{r_prob_se}{0.000}
\setres{learnUOJ}{DA_post_m30}{spread_penalty}{0.001}
\setres{learnUOJ}{DA_post_m30}{theta_mean}{0.98, 0.23}
\setres{learnUOJ}{DA_post_m30}{theta_sd}{0.02, 0.06}
\setres{learnUOJ}{DA_post_m30}{volume_ess}{518{,}446}
\setres{learnUOJ}{DA_post_mall}{ess}{1}
\setres{learnUOJ}{DA_post_mall}{kappa}{0.764}
\setres{learnUOJ}{DA_post_mall}{kappa_crps}{0.420}
\setres{learnUOJ}{DA_post_mall}{kappa_crps_se}{0.035}
\setres{learnUOJ}{DA_post_mall}{kappa_prior_mean}{0.765}
\setres{learnUOJ}{DA_post_mall}{kappa_se}{0.026}
\setres{learnUOJ}{DA_post_mall}{mse}{64.76}
\setres{learnUOJ}{DA_post_mall}{n_subjects_seen}{448}
\setres{learnUOJ}{DA_post_mall}{r_ff}{0.332}
\setres{learnUOJ}{DA_post_mall}{r_ff_se}{0.005}
\setres{learnUOJ}{DA_post_mall}{r_prior}{0.9999}
\setres{learnUOJ}{DA_post_mall}{r_prior_se}{0.000}
\setres{learnUOJ}{DA_post_mall}{r_prob}{1.000}
\setres{learnUOJ}{DA_post_mall}{r_prob_se}{0.000}
\setres{learnUOJ}{DA_post_mall}{spread_penalty}{0.000}
\setres{learnUOJ}{DA_post_mall}{theta_mean}{0.99, 0.35}
\setres{learnUOJ}{DA_post_mall}{theta_sd}{0.00, 0.00}
\setres{learnUOJ}{DA_post_mall}{volume_ess}{518{,}446}
\setres{learnUOJ}{EU_post_m0}{ess}{200}
\setres{learnUOJ}{EU_post_m0}{kappa}{0.408}
\setres{learnUOJ}{EU_post_m0}{kappa_crps}{0.309}
\setres{learnUOJ}{EU_post_m0}{kappa_crps_se}{0.040}
\setres{learnUOJ}{EU_post_m0}{kappa_prior_mean}{0.612}
\setres{learnUOJ}{EU_post_m0}{kappa_se}{0.057}
\setres{learnUOJ}{EU_post_m0}{mse}{99.30}
\setres{learnUOJ}{EU_post_m0}{n_subjects_seen}{0}
\setres{learnUOJ}{EU_post_m0}{r_ff}{0.382}
\setres{learnUOJ}{EU_post_m0}{r_ff_se}{0.005}
\setres{learnUOJ}{EU_post_m0}{r_prior}{0.6361}
\setres{learnUOJ}{EU_post_m0}{r_prior_se}{0.001}
\setres{learnUOJ}{EU_post_m0}{r_prob}{0.775}
\setres{learnUOJ}{EU_post_m0}{r_prob_se}{0.002}
\setres{learnUOJ}{EU_post_m0}{spread_penalty}{0.204}
\setres{learnUOJ}{EU_post_m0}{theta_mean}{0.51}
\setres{learnUOJ}{EU_post_m0}{theta_sd}{0.29}
\setres{learnUOJ}{EU_post_m0}{volume_ess}{160{,}713}
\setres{learnUOJ}{EU_post_m10}{ess}{27}
\setres{learnUOJ}{EU_post_m10}{kappa}{0.732}
\setres{learnUOJ}{EU_post_m10}{kappa_crps}{0.382}
\setres{learnUOJ}{EU_post_m10}{kappa_crps_se}{0.037}
\setres{learnUOJ}{EU_post_m10}{kappa_prior_mean}{0.734}
\setres{learnUOJ}{EU_post_m10}{kappa_se}{0.029}
\setres{learnUOJ}{EU_post_m10}{mse}{67.92}
\setres{learnUOJ}{EU_post_m10}{n_subjects_seen}{10}
\setres{learnUOJ}{EU_post_m10}{r_ff}{0.382}
\setres{learnUOJ}{EU_post_m10}{r_ff_se}{0.005}
\setres{learnUOJ}{EU_post_m10}{r_prior}{0.9581}
\setres{learnUOJ}{EU_post_m10}{r_prior_se}{0.000}
\setres{learnUOJ}{EU_post_m10}{r_prob}{0.974}
\setres{learnUOJ}{EU_post_m10}{r_prob_se}{0.000}
\setres{learnUOJ}{EU_post_m10}{spread_penalty}{0.003}
\setres{learnUOJ}{EU_post_m10}{theta_mean}{0.72}
\setres{learnUOJ}{EU_post_m10}{theta_sd}{0.04}
\setres{learnUOJ}{EU_post_m10}{volume_ess}{160{,}713}
\setres{learnUOJ}{EU_post_m100}{ess}{10}
\setres{learnUOJ}{EU_post_m100}{kappa}{0.745}
\setres{learnUOJ}{EU_post_m100}{kappa_crps}{0.402}
\setres{learnUOJ}{EU_post_m100}{kappa_crps_se}{0.036}
\setres{learnUOJ}{EU_post_m100}{kappa_prior_mean}{0.745}
\setres{learnUOJ}{EU_post_m100}{kappa_se}{0.026}
\setres{learnUOJ}{EU_post_m100}{mse}{66.66}
\setres{learnUOJ}{EU_post_m100}{n_subjects_seen}{100}
\setres{learnUOJ}{EU_post_m100}{r_ff}{0.382}
\setres{learnUOJ}{EU_post_m100}{r_ff_se}{0.005}
\setres{learnUOJ}{EU_post_m100}{r_prior}{0.9859}
\setres{learnUOJ}{EU_post_m100}{r_prior_se}{0.000}
\setres{learnUOJ}{EU_post_m100}{r_prob}{0.991}
\setres{learnUOJ}{EU_post_m100}{r_prob_se}{0.000}
\setres{learnUOJ}{EU_post_m100}{spread_penalty}{0.000}
\setres{learnUOJ}{EU_post_m100}{theta_mean}{0.87}
\setres{learnUOJ}{EU_post_m100}{theta_sd}{0.01}
\setres{learnUOJ}{EU_post_m100}{volume_ess}{160{,}713}
\setres{learnUOJ}{EU_post_m3}{ess}{48}
\setres{learnUOJ}{EU_post_m3}{kappa}{0.744}
\setres{learnUOJ}{EU_post_m3}{kappa_crps}{0.421}
\setres{learnUOJ}{EU_post_m3}{kappa_crps_se}{0.035}
\setres{learnUOJ}{EU_post_m3}{kappa_prior_mean}{0.749}
\setres{learnUOJ}{EU_post_m3}{kappa_se}{0.026}
\setres{learnUOJ}{EU_post_m3}{mse}{66.68}
\setres{learnUOJ}{EU_post_m3}{n_subjects_seen}{3}
\setres{learnUOJ}{EU_post_m3}{r_ff}{0.382}
\setres{learnUOJ}{EU_post_m3}{r_ff_se}{0.005}
\setres{learnUOJ}{EU_post_m3}{r_prior}{0.9351}
\setres{learnUOJ}{EU_post_m3}{r_prior_se}{0.000}
\setres{learnUOJ}{EU_post_m3}{r_prob}{0.960}
\setres{learnUOJ}{EU_post_m3}{r_prob_se}{0.000}
\setres{learnUOJ}{EU_post_m3}{spread_penalty}{0.005}
\setres{learnUOJ}{EU_post_m3}{theta_mean}{0.84}
\setres{learnUOJ}{EU_post_m3}{theta_sd}{0.07}
\setres{learnUOJ}{EU_post_m3}{volume_ess}{160{,}713}
\setres{learnUOJ}{EU_post_m30}{ess}{19}
\setres{learnUOJ}{EU_post_m30}{kappa}{0.746}
\setres{learnUOJ}{EU_post_m30}{kappa_crps}{0.409}
\setres{learnUOJ}{EU_post_m30}{kappa_crps_se}{0.036}
\setres{learnUOJ}{EU_post_m30}{kappa_prior_mean}{0.747}
\setres{learnUOJ}{EU_post_m30}{kappa_se}{0.026}
\setres{learnUOJ}{EU_post_m30}{mse}{66.52}
\setres{learnUOJ}{EU_post_m30}{n_subjects_seen}{30}
\setres{learnUOJ}{EU_post_m30}{r_ff}{0.382}
\setres{learnUOJ}{EU_post_m30}{r_ff_se}{0.005}
\setres{learnUOJ}{EU_post_m30}{r_prior}{0.9743}
\setres{learnUOJ}{EU_post_m30}{r_prior_se}{0.000}
\setres{learnUOJ}{EU_post_m30}{r_prob}{0.984}
\setres{learnUOJ}{EU_post_m30}{r_prob_se}{0.000}
\setres{learnUOJ}{EU_post_m30}{spread_penalty}{0.001}
\setres{learnUOJ}{EU_post_m30}{theta_mean}{0.88}
\setres{learnUOJ}{EU_post_m30}{theta_sd}{0.03}
\setres{learnUOJ}{EU_post_m30}{volume_ess}{160{,}713}
\setres{learnUOJ}{EU_post_mall}{ess}{5}
\setres{learnUOJ}{EU_post_mall}{kappa}{0.750}
\setres{learnUOJ}{EU_post_mall}{kappa_crps}{0.406}
\setres{learnUOJ}{EU_post_mall}{kappa_crps_se}{0.036}
\setres{learnUOJ}{EU_post_mall}{kappa_prior_mean}{0.750}
\setres{learnUOJ}{EU_post_mall}{kappa_se}{0.026}
\setres{learnUOJ}{EU_post_mall}{mse}{66.18}
\setres{learnUOJ}{EU_post_mall}{n_subjects_seen}{448}
\setres{learnUOJ}{EU_post_mall}{r_ff}{0.382}
\setres{learnUOJ}{EU_post_mall}{r_ff_se}{0.005}
\setres{learnUOJ}{EU_post_mall}{r_prior}{0.9930}
\setres{learnUOJ}{EU_post_mall}{r_prior_se}{0.000}
\setres{learnUOJ}{EU_post_mall}{r_prob}{0.996}
\setres{learnUOJ}{EU_post_mall}{r_prob_se}{0.000}
\setres{learnUOJ}{EU_post_mall}{spread_penalty}{0.000}
\setres{learnUOJ}{EU_post_mall}{theta_mean}{0.83}
\setres{learnUOJ}{EU_post_mall}{theta_sd}{0.01}
\setres{learnUOJ}{EU_post_mall}{volume_ess}{160{,}713}
\setres{learnUOJ}{reference}{n_subjects}{448}
\setres{learnUOJ}{reference}{sigma}{6.4}
\setres{lotL}{CPT_post_k0}{cov50}{0.24}
\setres{lotL}{CPT_post_k0}{cov50_mean}{0.240}
\setres{lotL}{CPT_post_k0}{cov90}{0.80}
\setres{lotL}{CPT_post_k0}{cov90_adj}{---}
\setres{lotL}{CPT_post_k0}{cov90_adj_mean}{---}
\setres{lotL}{CPT_post_k0}{cov90_mean}{0.800}
\setres{lotL}{CPT_post_k0}{kappa}{0.636}
\setres{lotL}{CPT_post_k0}{kappa_crps}{0.603}
\setres{lotL}{CPT_post_k0}{kappa_crps_mean}{0.606}
\setres{lotL}{CPT_post_k0}{kappa_mean}{0.636}
\setres{lotL}{CPT_post_k0}{logscore_next}{-2.99}
\setres{lotL}{CPT_post_k0}{logscore_next_mean}{-2.991}
\setres{lotL}{CPT_post_k0}{misfit_pct}{---}
\setres{lotL}{CPT_post_k0}{misfit_pct_mean}{---}
\setres{lotL}{CPT_post_k0}{n_seen}{0}
\setres{lotL}{CPT_post_k0}{surprise}{2.32}
\setres{lotL}{CPT_post_k0}{surprise_mean}{2.159}
\setres{lotL}{CPT_post_k0}{width90_adj_pct}{---}
\setres{lotL}{CPT_post_k0}{width90_adj_pct_mean}{---}
\setres{lotL}{CPT_post_k0}{width90_pct}{11.0}
\setres{lotL}{CPT_post_k0}{width90_pct_mean}{11.008}
\setres{lotL}{CPT_post_k1}{cov50}{0.42}
\setres{lotL}{CPT_post_k1}{cov50_mean}{0.365}
\setres{lotL}{CPT_post_k1}{cov90}{0.79}
\setres{lotL}{CPT_post_k1}{cov90_adj}{---}
\setres{lotL}{CPT_post_k1}{cov90_adj_mean}{---}
\setres{lotL}{CPT_post_k1}{cov90_mean}{0.769}
\setres{lotL}{CPT_post_k1}{kappa}{0.738}
\setres{lotL}{CPT_post_k1}{kappa_crps}{0.610}
\setres{lotL}{CPT_post_k1}{kappa_crps_mean}{0.612}
\setres{lotL}{CPT_post_k1}{kappa_mean}{0.718}
\setres{lotL}{CPT_post_k1}{logscore_next}{-2.98}
\setres{lotL}{CPT_post_k1}{logscore_next_mean}{-2.985}
\setres{lotL}{CPT_post_k1}{misfit_pct}{---}
\setres{lotL}{CPT_post_k1}{misfit_pct_mean}{---}
\setres{lotL}{CPT_post_k1}{n_seen}{1}
\setres{lotL}{CPT_post_k1}{surprise}{0.69}
\setres{lotL}{CPT_post_k1}{surprise_mean}{1.097}
\setres{lotL}{CPT_post_k1}{width90_adj_pct}{---}
\setres{lotL}{CPT_post_k1}{width90_adj_pct_mean}{---}
\setres{lotL}{CPT_post_k1}{width90_pct}{10.5}
\setres{lotL}{CPT_post_k1}{width90_pct_mean}{10.450}
\setres{lotL}{CPT_post_k10}{cov50}{0.03}
\setres{lotL}{CPT_post_k10}{cov50_mean}{0.090}
\setres{lotL}{CPT_post_k10}{cov90}{0.27}
\setres{lotL}{CPT_post_k10}{cov90_adj}{0.80}
\setres{lotL}{CPT_post_k10}{cov90_adj_mean}{0.807}
\setres{lotL}{CPT_post_k10}{cov90_mean}{0.237}
\setres{lotL}{CPT_post_k10}{kappa}{0.942}
\setres{lotL}{CPT_post_k10}{kappa_crps}{0.664}
\setres{lotL}{CPT_post_k10}{kappa_crps_mean}{0.641}
\setres{lotL}{CPT_post_k10}{kappa_mean}{0.945}
\setres{lotL}{CPT_post_k10}{logscore_next}{-2.98}
\setres{lotL}{CPT_post_k10}{logscore_next_mean}{-2.980}
\setres{lotL}{CPT_post_k10}{misfit_pct}{4.1}
\setres{lotL}{CPT_post_k10}{misfit_pct_mean}{4.189}
\setres{lotL}{CPT_post_k10}{n_seen}{10}
\setres{lotL}{CPT_post_k10}{surprise}{0.04}
\setres{lotL}{CPT_post_k10}{surprise_mean}{0.065}
\setres{lotL}{CPT_post_k10}{width90_adj_pct}{6.9}
\setres{lotL}{CPT_post_k10}{width90_adj_pct_mean}{7.111}
\setres{lotL}{CPT_post_k10}{width90_pct}{1.6}
\setres{lotL}{CPT_post_k10}{width90_pct_mean}{1.323}
\setres{lotL}{CPT_post_k11}{cov50}{0.00}
\setres{lotL}{CPT_post_k11}{cov50_mean}{0.086}
\setres{lotL}{CPT_post_k11}{cov90}{0.21}
\setres{lotL}{CPT_post_k11}{cov90_adj}{0.79}
\setres{lotL}{CPT_post_k11}{cov90_adj_mean}{0.786}
\setres{lotL}{CPT_post_k11}{cov90_mean}{0.189}
\setres{lotL}{CPT_post_k11}{kappa}{0.948}
\setres{lotL}{CPT_post_k11}{kappa_crps}{0.654}
\setres{lotL}{CPT_post_k11}{kappa_crps_mean}{0.651}
\setres{lotL}{CPT_post_k11}{kappa_mean}{0.945}
\setres{lotL}{CPT_post_k11}{logscore_next}{-2.98}
\setres{lotL}{CPT_post_k11}{logscore_next_mean}{-2.993}
\setres{lotL}{CPT_post_k11}{misfit_pct}{4.0}
\setres{lotL}{CPT_post_k11}{misfit_pct_mean}{4.225}
\setres{lotL}{CPT_post_k11}{n_seen}{11}
\setres{lotL}{CPT_post_k11}{surprise}{0.03}
\setres{lotL}{CPT_post_k11}{surprise_mean}{1.270}
\setres{lotL}{CPT_post_k11}{width90_adj_pct}{6.8}
\setres{lotL}{CPT_post_k11}{width90_adj_pct_mean}{7.148}
\setres{lotL}{CPT_post_k11}{width90_pct}{1.4}
\setres{lotL}{CPT_post_k11}{width90_pct_mean}{1.075}
\setres{lotL}{CPT_post_k12}{cov50}{0.00}
\setres{lotL}{CPT_post_k12}{cov50_mean}{0.073}
\setres{lotL}{CPT_post_k12}{cov90}{0.15}
\setres{lotL}{CPT_post_k12}{cov90_adj}{0.85}
\setres{lotL}{CPT_post_k12}{cov90_adj_mean}{0.819}
\setres{lotL}{CPT_post_k12}{cov90_mean}{0.181}
\setres{lotL}{CPT_post_k12}{kappa}{0.940}
\setres{lotL}{CPT_post_k12}{kappa_crps}{0.675}
\setres{lotL}{CPT_post_k12}{kappa_crps_mean}{0.644}
\setres{lotL}{CPT_post_k12}{kappa_mean}{0.942}
\setres{lotL}{CPT_post_k12}{logscore_next}{-2.98}
\setres{lotL}{CPT_post_k12}{logscore_next_mean}{-2.985}
\setres{lotL}{CPT_post_k12}{misfit_pct}{4.4}
\setres{lotL}{CPT_post_k12}{misfit_pct_mean}{4.516}
\setres{lotL}{CPT_post_k12}{n_seen}{12}
\setres{lotL}{CPT_post_k12}{surprise}{0.04}
\setres{lotL}{CPT_post_k12}{surprise_mean}{0.350}
\setres{lotL}{CPT_post_k12}{width90_adj_pct}{7.5}
\setres{lotL}{CPT_post_k12}{width90_adj_pct_mean}{7.580}
\setres{lotL}{CPT_post_k12}{width90_pct}{1.1}
\setres{lotL}{CPT_post_k12}{width90_pct_mean}{0.963}
\setres{lotL}{CPT_post_k13}{cov50}{0.08}
\setres{lotL}{CPT_post_k13}{cov50_mean}{0.117}
\setres{lotL}{CPT_post_k13}{cov90}{0.21}
\setres{lotL}{CPT_post_k13}{cov90_adj}{0.83}
\setres{lotL}{CPT_post_k13}{cov90_adj_mean}{0.817}
\setres{lotL}{CPT_post_k13}{cov90_mean}{0.208}
\setres{lotL}{CPT_post_k13}{kappa}{0.941}
\setres{lotL}{CPT_post_k13}{kappa_crps}{0.673}
\setres{lotL}{CPT_post_k13}{kappa_crps_mean}{0.641}
\setres{lotL}{CPT_post_k13}{kappa_mean}{0.937}
\setres{lotL}{CPT_post_k13}{logscore_next}{-2.98}
\setres{lotL}{CPT_post_k13}{logscore_next_mean}{-2.992}
\setres{lotL}{CPT_post_k13}{misfit_pct}{4.3}
\setres{lotL}{CPT_post_k13}{misfit_pct_mean}{4.430}
\setres{lotL}{CPT_post_k13}{n_seen}{13}
\setres{lotL}{CPT_post_k13}{surprise}{0.24}
\setres{lotL}{CPT_post_k13}{surprise_mean}{1.289}
\setres{lotL}{CPT_post_k13}{width90_adj_pct}{7.2}
\setres{lotL}{CPT_post_k13}{width90_adj_pct_mean}{7.472}
\setres{lotL}{CPT_post_k13}{width90_pct}{1.3}
\setres{lotL}{CPT_post_k13}{width90_pct_mean}{1.089}
\setres{lotL}{CPT_post_k14}{cov50}{0.00}
\setres{lotL}{CPT_post_k14}{cov50_mean}{0.073}
\setres{lotL}{CPT_post_k14}{cov90}{0.14}
\setres{lotL}{CPT_post_k14}{cov90_adj}{0.82}
\setres{lotL}{CPT_post_k14}{cov90_adj_mean}{0.773}
\setres{lotL}{CPT_post_k14}{cov90_mean}{0.195}
\setres{lotL}{CPT_post_k14}{kappa}{0.942}
\setres{lotL}{CPT_post_k14}{kappa_crps}{0.675}
\setres{lotL}{CPT_post_k14}{kappa_crps_mean}{0.632}
\setres{lotL}{CPT_post_k14}{kappa_mean}{0.939}
\setres{lotL}{CPT_post_k14}{logscore_next}{-2.98}
\setres{lotL}{CPT_post_k14}{logscore_next_mean}{-2.985}
\setres{lotL}{CPT_post_k14}{misfit_pct}{4.3}
\setres{lotL}{CPT_post_k14}{misfit_pct_mean}{4.477}
\setres{lotL}{CPT_post_k14}{n_seen}{14}
\setres{lotL}{CPT_post_k14}{surprise}{0.07}
\setres{lotL}{CPT_post_k14}{surprise_mean}{0.339}
\setres{lotL}{CPT_post_k14}{width90_adj_pct}{7.2}
\setres{lotL}{CPT_post_k14}{width90_adj_pct_mean}{7.515}
\setres{lotL}{CPT_post_k14}{width90_pct}{1.2}
\setres{lotL}{CPT_post_k14}{width90_pct_mean}{1.109}
\setres{lotL}{CPT_post_k15}{cov50}{0.00}
\setres{lotL}{CPT_post_k15}{cov50_mean}{0.070}
\setres{lotL}{CPT_post_k15}{cov90}{0.25}
\setres{lotL}{CPT_post_k15}{cov90_adj}{0.80}
\setres{lotL}{CPT_post_k15}{cov90_adj_mean}{0.790}
\setres{lotL}{CPT_post_k15}{cov90_mean}{0.205}
\setres{lotL}{CPT_post_k15}{kappa}{0.940}
\setres{lotL}{CPT_post_k15}{kappa_crps}{0.643}
\setres{lotL}{CPT_post_k15}{kappa_crps_mean}{0.615}
\setres{lotL}{CPT_post_k15}{kappa_mean}{0.934}
\setres{lotL}{CPT_post_k15}{logscore_next}{-2.98}
\setres{lotL}{CPT_post_k15}{logscore_next_mean}{-2.984}
\setres{lotL}{CPT_post_k15}{misfit_pct}{4.6}
\setres{lotL}{CPT_post_k15}{misfit_pct_mean}{4.551}
\setres{lotL}{CPT_post_k15}{n_seen}{15}
\setres{lotL}{CPT_post_k15}{surprise}{0.04}
\setres{lotL}{CPT_post_k15}{surprise_mean}{0.183}
\setres{lotL}{CPT_post_k15}{width90_adj_pct}{7.6}
\setres{lotL}{CPT_post_k15}{width90_adj_pct_mean}{7.631}
\setres{lotL}{CPT_post_k15}{width90_pct}{1.2}
\setres{lotL}{CPT_post_k15}{width90_pct_mean}{1.085}
\setres{lotL}{CPT_post_k16}{cov50}{0.00}
\setres{lotL}{CPT_post_k16}{cov50_mean}{0.056}
\setres{lotL}{CPT_post_k16}{cov90}{0.11}
\setres{lotL}{CPT_post_k16}{cov90_adj}{0.78}
\setres{lotL}{CPT_post_k16}{cov90_adj_mean}{0.800}
\setres{lotL}{CPT_post_k16}{cov90_mean}{0.167}
\setres{lotL}{CPT_post_k16}{kappa}{0.939}
\setres{lotL}{CPT_post_k16}{kappa_crps}{0.653}
\setres{lotL}{CPT_post_k16}{kappa_crps_mean}{0.616}
\setres{lotL}{CPT_post_k16}{kappa_mean}{0.936}
\setres{lotL}{CPT_post_k16}{logscore_next}{-2.98}
\setres{lotL}{CPT_post_k16}{logscore_next_mean}{-2.984}
\setres{lotL}{CPT_post_k16}{misfit_pct}{4.8}
\setres{lotL}{CPT_post_k16}{misfit_pct_mean}{4.639}
\setres{lotL}{CPT_post_k16}{n_seen}{16}
\setres{lotL}{CPT_post_k16}{surprise}{0.02}
\setres{lotL}{CPT_post_k16}{surprise_mean}{0.240}
\setres{lotL}{CPT_post_k16}{width90_adj_pct}{8.1}
\setres{lotL}{CPT_post_k16}{width90_adj_pct_mean}{7.753}
\setres{lotL}{CPT_post_k16}{width90_pct}{1.0}
\setres{lotL}{CPT_post_k16}{width90_pct_mean}{0.870}
\setres{lotL}{CPT_post_k17}{cov50}{0.00}
\setres{lotL}{CPT_post_k17}{cov50_mean}{0.062}
\setres{lotL}{CPT_post_k17}{cov90}{0.12}
\setres{lotL}{CPT_post_k17}{cov90_adj}{0.88}
\setres{lotL}{CPT_post_k17}{cov90_adj_mean}{0.831}
\setres{lotL}{CPT_post_k17}{cov90_mean}{0.144}
\setres{lotL}{CPT_post_k17}{kappa}{0.940}
\setres{lotL}{CPT_post_k17}{kappa_crps}{0.698}
\setres{lotL}{CPT_post_k17}{kappa_crps_mean}{0.635}
\setres{lotL}{CPT_post_k17}{kappa_mean}{0.936}
\setres{lotL}{CPT_post_k17}{logscore_next}{-2.99}
\setres{lotL}{CPT_post_k17}{logscore_next_mean}{-2.998}
\setres{lotL}{CPT_post_k17}{misfit_pct}{5.2}
\setres{lotL}{CPT_post_k17}{misfit_pct_mean}{4.803}
\setres{lotL}{CPT_post_k17}{n_seen}{17}
\setres{lotL}{CPT_post_k17}{surprise}{0.10}
\setres{lotL}{CPT_post_k17}{surprise_mean}{1.273}
\setres{lotL}{CPT_post_k17}{width90_adj_pct}{8.6}
\setres{lotL}{CPT_post_k17}{width90_adj_pct_mean}{8.006}
\setres{lotL}{CPT_post_k17}{width90_pct}{0.9}
\setres{lotL}{CPT_post_k17}{width90_pct_mean}{0.750}
\setres{lotL}{CPT_post_k18}{cov50}{0.00}
\setres{lotL}{CPT_post_k18}{cov50_mean}{0.050}
\setres{lotL}{CPT_post_k18}{cov90}{0.14}
\setres{lotL}{CPT_post_k18}{cov90_adj}{0.86}
\setres{lotL}{CPT_post_k18}{cov90_adj_mean}{0.836}
\setres{lotL}{CPT_post_k18}{cov90_mean}{0.157}
\setres{lotL}{CPT_post_k18}{kappa}{0.946}
\setres{lotL}{CPT_post_k18}{kappa_crps}{0.663}
\setres{lotL}{CPT_post_k18}{kappa_crps_mean}{0.609}
\setres{lotL}{CPT_post_k18}{kappa_mean}{0.927}
\setres{lotL}{CPT_post_k18}{logscore_next}{-2.98}
\setres{lotL}{CPT_post_k18}{logscore_next_mean}{-2.991}
\setres{lotL}{CPT_post_k18}{misfit_pct}{5.3}
\setres{lotL}{CPT_post_k18}{misfit_pct_mean}{4.944}
\setres{lotL}{CPT_post_k18}{n_seen}{18}
\setres{lotL}{CPT_post_k18}{surprise}{0.01}
\setres{lotL}{CPT_post_k18}{surprise_mean}{1.363}
\setres{lotL}{CPT_post_k18}{width90_adj_pct}{8.8}
\setres{lotL}{CPT_post_k18}{width90_adj_pct_mean}{8.236}
\setres{lotL}{CPT_post_k18}{width90_pct}{0.9}
\setres{lotL}{CPT_post_k18}{width90_pct_mean}{0.760}
\setres{lotL}{CPT_post_k19}{cov50}{0.00}
\setres{lotL}{CPT_post_k19}{cov50_mean}{0.050}
\setres{lotL}{CPT_post_k19}{cov90}{0.17}
\setres{lotL}{CPT_post_k19}{cov90_adj}{0.83}
\setres{lotL}{CPT_post_k19}{cov90_adj_mean}{0.858}
\setres{lotL}{CPT_post_k19}{cov90_mean}{0.142}
\setres{lotL}{CPT_post_k19}{kappa}{0.957}
\setres{lotL}{CPT_post_k19}{kappa_crps}{0.704}
\setres{lotL}{CPT_post_k19}{kappa_crps_mean}{0.629}
\setres{lotL}{CPT_post_k19}{kappa_mean}{0.929}
\setres{lotL}{CPT_post_k19}{logscore_next}{-2.98}
\setres{lotL}{CPT_post_k19}{logscore_next_mean}{-2.989}
\setres{lotL}{CPT_post_k19}{misfit_pct}{5.4}
\setres{lotL}{CPT_post_k19}{misfit_pct_mean}{5.056}
\setres{lotL}{CPT_post_k19}{n_seen}{19}
\setres{lotL}{CPT_post_k19}{surprise}{0.01}
\setres{lotL}{CPT_post_k19}{surprise_mean}{0.991}
\setres{lotL}{CPT_post_k19}{width90_adj_pct}{9.0}
\setres{lotL}{CPT_post_k19}{width90_adj_pct_mean}{8.403}
\setres{lotL}{CPT_post_k19}{width90_pct}{0.6}
\setres{lotL}{CPT_post_k19}{width90_pct_mean}{0.648}
\setres{lotL}{CPT_post_k2}{cov50}{0.26}
\setres{lotL}{CPT_post_k2}{cov50_mean}{0.250}
\setres{lotL}{CPT_post_k2}{cov90}{0.48}
\setres{lotL}{CPT_post_k2}{cov90_adj}{0.57}
\setres{lotL}{CPT_post_k2}{cov90_adj_mean}{0.572}
\setres{lotL}{CPT_post_k2}{cov90_mean}{0.524}
\setres{lotL}{CPT_post_k2}{kappa}{0.809}
\setres{lotL}{CPT_post_k2}{kappa_crps}{0.613}
\setres{lotL}{CPT_post_k2}{kappa_crps_mean}{0.606}
\setres{lotL}{CPT_post_k2}{kappa_mean}{0.791}
\setres{lotL}{CPT_post_k2}{logscore_next}{-2.98}
\setres{lotL}{CPT_post_k2}{logscore_next_mean}{-2.986}
\setres{lotL}{CPT_post_k2}{misfit_pct}{0.8}
\setres{lotL}{CPT_post_k2}{misfit_pct_mean}{1.367}
\setres{lotL}{CPT_post_k2}{n_seen}{2}
\setres{lotL}{CPT_post_k2}{surprise}{0.77}
\setres{lotL}{CPT_post_k2}{surprise_mean}{1.122}
\setres{lotL}{CPT_post_k2}{width90_adj_pct}{6.7}
\setres{lotL}{CPT_post_k2}{width90_adj_pct_mean}{7.406}
\setres{lotL}{CPT_post_k2}{width90_pct}{6.1}
\setres{lotL}{CPT_post_k2}{width90_pct_mean}{6.810}
\setres{lotL}{CPT_post_k20}{cov50}{0.00}
\setres{lotL}{CPT_post_k20}{cov50_mean}{0.060}
\setres{lotL}{CPT_post_k20}{cov90}{0.00}
\setres{lotL}{CPT_post_k20}{cov90_adj}{0.80}
\setres{lotL}{CPT_post_k20}{cov90_adj_mean}{0.850}
\setres{lotL}{CPT_post_k20}{cov90_mean}{0.120}
\setres{lotL}{CPT_post_k20}{kappa}{0.950}
\setres{lotL}{CPT_post_k20}{kappa_crps}{0.669}
\setres{lotL}{CPT_post_k20}{kappa_crps_mean}{0.600}
\setres{lotL}{CPT_post_k20}{kappa_mean}{0.921}
\setres{lotL}{CPT_post_k20}{logscore_next}{-2.98}
\setres{lotL}{CPT_post_k20}{logscore_next_mean}{-2.988}
\setres{lotL}{CPT_post_k20}{misfit_pct}{5.4}
\setres{lotL}{CPT_post_k20}{misfit_pct_mean}{5.181}
\setres{lotL}{CPT_post_k20}{n_seen}{20}
\setres{lotL}{CPT_post_k20}{surprise}{0.04}
\setres{lotL}{CPT_post_k20}{surprise_mean}{0.576}
\setres{lotL}{CPT_post_k20}{width90_adj_pct}{9.0}
\setres{lotL}{CPT_post_k20}{width90_adj_pct_mean}{8.617}
\setres{lotL}{CPT_post_k20}{width90_pct}{0.5}
\setres{lotL}{CPT_post_k20}{width90_pct_mean}{0.601}
\setres{lotL}{CPT_post_k21}{cov50}{0.00}
\setres{lotL}{CPT_post_k21}{cov50_mean}{0.075}
\setres{lotL}{CPT_post_k21}{cov90}{0.00}
\setres{lotL}{CPT_post_k21}{cov90_adj}{0.88}
\setres{lotL}{CPT_post_k21}{cov90_adj_mean}{0.863}
\setres{lotL}{CPT_post_k21}{cov90_mean}{0.163}
\setres{lotL}{CPT_post_k21}{kappa}{0.950}
\setres{lotL}{CPT_post_k21}{kappa_crps}{0.663}
\setres{lotL}{CPT_post_k21}{kappa_crps_mean}{0.586}
\setres{lotL}{CPT_post_k21}{kappa_mean}{0.919}
\setres{lotL}{CPT_post_k21}{logscore_next}{-2.98}
\setres{lotL}{CPT_post_k21}{logscore_next_mean}{-2.984}
\setres{lotL}{CPT_post_k21}{misfit_pct}{5.4}
\setres{lotL}{CPT_post_k21}{misfit_pct_mean}{5.154}
\setres{lotL}{CPT_post_k21}{n_seen}{21}
\setres{lotL}{CPT_post_k21}{surprise}{0.01}
\setres{lotL}{CPT_post_k21}{surprise_mean}{0.068}
\setres{lotL}{CPT_post_k21}{width90_adj_pct}{9.0}
\setres{lotL}{CPT_post_k21}{width90_adj_pct_mean}{8.564}
\setres{lotL}{CPT_post_k21}{width90_pct}{0.6}
\setres{lotL}{CPT_post_k21}{width90_pct_mean}{0.703}
\setres{lotL}{CPT_post_k22}{cov50}{0.00}
\setres{lotL}{CPT_post_k22}{cov50_mean}{0.067}
\setres{lotL}{CPT_post_k22}{cov90}{0.00}
\setres{lotL}{CPT_post_k22}{cov90_adj}{1.00}
\setres{lotL}{CPT_post_k22}{cov90_adj_mean}{0.833}
\setres{lotL}{CPT_post_k22}{cov90_mean}{0.150}
\setres{lotL}{CPT_post_k22}{kappa}{0.955}
\setres{lotL}{CPT_post_k22}{kappa_crps}{0.667}
\setres{lotL}{CPT_post_k22}{kappa_crps_mean}{0.575}
\setres{lotL}{CPT_post_k22}{kappa_mean}{0.907}
\setres{lotL}{CPT_post_k22}{logscore_next}{-2.98}
\setres{lotL}{CPT_post_k22}{logscore_next_mean}{-2.983}
\setres{lotL}{CPT_post_k22}{misfit_pct}{5.4}
\setres{lotL}{CPT_post_k22}{misfit_pct_mean}{5.121}
\setres{lotL}{CPT_post_k22}{n_seen}{22}
\setres{lotL}{CPT_post_k22}{surprise}{0.00}
\setres{lotL}{CPT_post_k22}{surprise_mean}{0.203}
\setres{lotL}{CPT_post_k22}{width90_adj_pct}{9.0}
\setres{lotL}{CPT_post_k22}{width90_adj_pct_mean}{8.507}
\setres{lotL}{CPT_post_k22}{width90_pct}{0.3}
\setres{lotL}{CPT_post_k22}{width90_pct_mean}{0.612}
\setres{lotL}{CPT_post_k23}{cov50}{0.00}
\setres{lotL}{CPT_post_k23}{cov50_mean}{0.050}
\setres{lotL}{CPT_post_k23}{cov90}{0.00}
\setres{lotL}{CPT_post_k23}{cov90_adj}{1.00}
\setres{lotL}{CPT_post_k23}{cov90_adj_mean}{0.825}
\setres{lotL}{CPT_post_k23}{cov90_mean}{0.125}
\setres{lotL}{CPT_post_k23}{kappa}{0.961}
\setres{lotL}{CPT_post_k23}{kappa_crps}{0.703}
\setres{lotL}{CPT_post_k23}{kappa_crps_mean}{0.642}
\setres{lotL}{CPT_post_k23}{kappa_mean}{0.897}
\setres{lotL}{CPT_post_k23}{logscore_next}{-2.99}
\setres{lotL}{CPT_post_k23}{logscore_next_mean}{-3.005}
\setres{lotL}{CPT_post_k23}{misfit_pct}{5.5}
\setres{lotL}{CPT_post_k23}{misfit_pct_mean}{5.162}
\setres{lotL}{CPT_post_k23}{n_seen}{23}
\setres{lotL}{CPT_post_k23}{surprise}{0.08}
\setres{lotL}{CPT_post_k23}{surprise_mean}{3.280}
\setres{lotL}{CPT_post_k23}{width90_adj_pct}{9.0}
\setres{lotL}{CPT_post_k23}{width90_adj_pct_mean}{8.579}
\setres{lotL}{CPT_post_k23}{width90_pct}{0.8}
\setres{lotL}{CPT_post_k23}{width90_pct_mean}{0.764}
\setres{lotL}{CPT_post_k24}{cov50}{0.00}
\setres{lotL}{CPT_post_k24}{cov50_mean}{0.100}
\setres{lotL}{CPT_post_k24}{cov90}{0.00}
\setres{lotL}{CPT_post_k24}{cov90_adj}{1.00}
\setres{lotL}{CPT_post_k24}{cov90_adj_mean}{0.950}
\setres{lotL}{CPT_post_k24}{cov90_mean}{0.100}
\setres{lotL}{CPT_post_k24}{kappa}{0.972}
\setres{lotL}{CPT_post_k24}{kappa_crps}{0.807}
\setres{lotL}{CPT_post_k24}{kappa_crps_mean}{0.747}
\setres{lotL}{CPT_post_k24}{kappa_mean}{0.879}
\setres{lotL}{CPT_post_k24}{logscore_next}{-2.98}
\setres{lotL}{CPT_post_k24}{logscore_next_mean}{-3.004}
\setres{lotL}{CPT_post_k24}{misfit_pct}{5.4}
\setres{lotL}{CPT_post_k24}{misfit_pct_mean}{5.356}
\setres{lotL}{CPT_post_k24}{n_seen}{24}
\setres{lotL}{CPT_post_k24}{surprise}{0.38}
\setres{lotL}{CPT_post_k24}{surprise_mean}{2.460}
\setres{lotL}{CPT_post_k24}{width90_adj_pct}{9.0}
\setres{lotL}{CPT_post_k24}{width90_adj_pct_mean}{8.911}
\setres{lotL}{CPT_post_k24}{width90_pct}{1.0}
\setres{lotL}{CPT_post_k24}{width90_pct_mean}{0.915}
\setres{lotL}{CPT_post_k25}{cov50}{---}
\setres{lotL}{CPT_post_k25}{cov50_mean}{---}
\setres{lotL}{CPT_post_k25}{cov90}{---}
\setres{lotL}{CPT_post_k25}{cov90_adj}{---}
\setres{lotL}{CPT_post_k25}{cov90_adj_mean}{---}
\setres{lotL}{CPT_post_k25}{cov90_mean}{---}
\setres{lotL}{CPT_post_k25}{kappa}{---}
\setres{lotL}{CPT_post_k25}{kappa_crps}{---}
\setres{lotL}{CPT_post_k25}{kappa_crps_mean}{---}
\setres{lotL}{CPT_post_k25}{kappa_mean}{---}
\setres{lotL}{CPT_post_k25}{logscore_next}{---}
\setres{lotL}{CPT_post_k25}{logscore_next_mean}{---}
\setres{lotL}{CPT_post_k25}{misfit_pct}{---}
\setres{lotL}{CPT_post_k25}{misfit_pct_mean}{---}
\setres{lotL}{CPT_post_k25}{n_seen}{25}
\setres{lotL}{CPT_post_k25}{surprise}{---}
\setres{lotL}{CPT_post_k25}{surprise_mean}{---}
\setres{lotL}{CPT_post_k25}{width90_adj_pct}{---}
\setres{lotL}{CPT_post_k25}{width90_adj_pct_mean}{---}
\setres{lotL}{CPT_post_k25}{width90_pct}{---}
\setres{lotL}{CPT_post_k25}{width90_pct_mean}{---}
\setres{lotL}{CPT_post_k3}{cov50}{0.14}
\setres{lotL}{CPT_post_k3}{cov50_mean}{0.175}
\setres{lotL}{CPT_post_k3}{cov90}{0.50}
\setres{lotL}{CPT_post_k3}{cov90_adj}{0.77}
\setres{lotL}{CPT_post_k3}{cov90_adj_mean}{0.723}
\setres{lotL}{CPT_post_k3}{cov90_mean}{0.489}
\setres{lotL}{CPT_post_k3}{kappa}{0.898}
\setres{lotL}{CPT_post_k3}{kappa_crps}{0.630}
\setres{lotL}{CPT_post_k3}{kappa_crps_mean}{0.623}
\setres{lotL}{CPT_post_k3}{kappa_mean}{0.863}
\setres{lotL}{CPT_post_k3}{logscore_next}{-2.99}
\setres{lotL}{CPT_post_k3}{logscore_next_mean}{-2.986}
\setres{lotL}{CPT_post_k3}{misfit_pct}{2.4}
\setres{lotL}{CPT_post_k3}{misfit_pct_mean}{3.043}
\setres{lotL}{CPT_post_k3}{n_seen}{3}
\setres{lotL}{CPT_post_k3}{surprise}{1.01}
\setres{lotL}{CPT_post_k3}{surprise_mean}{1.017}
\setres{lotL}{CPT_post_k3}{width90_adj_pct}{6.5}
\setres{lotL}{CPT_post_k3}{width90_adj_pct_mean}{7.522}
\setres{lotL}{CPT_post_k3}{width90_pct}{5.0}
\setres{lotL}{CPT_post_k3}{width90_pct_mean}{4.966}
\setres{lotL}{CPT_post_k4}{cov50}{0.14}
\setres{lotL}{CPT_post_k4}{cov50_mean}{0.164}
\setres{lotL}{CPT_post_k4}{cov90}{0.43}
\setres{lotL}{CPT_post_k4}{cov90_adj}{0.76}
\setres{lotL}{CPT_post_k4}{cov90_adj_mean}{0.724}
\setres{lotL}{CPT_post_k4}{cov90_mean}{0.436}
\setres{lotL}{CPT_post_k4}{kappa}{0.933}
\setres{lotL}{CPT_post_k4}{kappa_crps}{0.632}
\setres{lotL}{CPT_post_k4}{kappa_crps_mean}{0.641}
\setres{lotL}{CPT_post_k4}{kappa_mean}{0.917}
\setres{lotL}{CPT_post_k4}{logscore_next}{-2.98}
\setres{lotL}{CPT_post_k4}{logscore_next_mean}{-2.984}
\setres{lotL}{CPT_post_k4}{misfit_pct}{2.9}
\setres{lotL}{CPT_post_k4}{misfit_pct_mean}{3.317}
\setres{lotL}{CPT_post_k4}{n_seen}{4}
\setres{lotL}{CPT_post_k4}{surprise}{0.24}
\setres{lotL}{CPT_post_k4}{surprise_mean}{0.517}
\setres{lotL}{CPT_post_k4}{width90_adj_pct}{5.6}
\setres{lotL}{CPT_post_k4}{width90_adj_pct_mean}{6.757}
\setres{lotL}{CPT_post_k4}{width90_pct}{3.3}
\setres{lotL}{CPT_post_k4}{width90_pct_mean}{3.539}
\setres{lotL}{CPT_post_k5}{cov50}{0.17}
\setres{lotL}{CPT_post_k5}{cov50_mean}{0.188}
\setres{lotL}{CPT_post_k5}{cov90}{0.40}
\setres{lotL}{CPT_post_k5}{cov90_adj}{0.78}
\setres{lotL}{CPT_post_k5}{cov90_adj_mean}{0.760}
\setres{lotL}{CPT_post_k5}{cov90_mean}{0.398}
\setres{lotL}{CPT_post_k5}{kappa}{0.945}
\setres{lotL}{CPT_post_k5}{kappa_crps}{0.639}
\setres{lotL}{CPT_post_k5}{kappa_crps_mean}{0.648}
\setres{lotL}{CPT_post_k5}{kappa_mean}{0.937}
\setres{lotL}{CPT_post_k5}{logscore_next}{-2.98}
\setres{lotL}{CPT_post_k5}{logscore_next_mean}{-2.984}
\setres{lotL}{CPT_post_k5}{misfit_pct}{2.8}
\setres{lotL}{CPT_post_k5}{misfit_pct_mean}{3.512}
\setres{lotL}{CPT_post_k5}{n_seen}{5}
\setres{lotL}{CPT_post_k5}{surprise}{0.35}
\setres{lotL}{CPT_post_k5}{surprise_mean}{0.568}
\setres{lotL}{CPT_post_k5}{width90_adj_pct}{5.7}
\setres{lotL}{CPT_post_k5}{width90_adj_pct_mean}{6.550}
\setres{lotL}{CPT_post_k5}{width90_pct}{2.4}
\setres{lotL}{CPT_post_k5}{width90_pct_mean}{2.641}
\setres{lotL}{CPT_post_k6}{cov50}{0.13}
\setres{lotL}{CPT_post_k6}{cov50_mean}{0.150}
\setres{lotL}{CPT_post_k6}{cov90}{0.39}
\setres{lotL}{CPT_post_k6}{cov90_adj}{0.79}
\setres{lotL}{CPT_post_k6}{cov90_adj_mean}{0.747}
\setres{lotL}{CPT_post_k6}{cov90_mean}{0.363}
\setres{lotL}{CPT_post_k6}{kappa}{0.948}
\setres{lotL}{CPT_post_k6}{kappa_crps}{0.644}
\setres{lotL}{CPT_post_k6}{kappa_crps_mean}{0.655}
\setres{lotL}{CPT_post_k6}{kappa_mean}{0.940}
\setres{lotL}{CPT_post_k6}{logscore_next}{-2.98}
\setres{lotL}{CPT_post_k6}{logscore_next_mean}{-2.986}
\setres{lotL}{CPT_post_k6}{misfit_pct}{3.5}
\setres{lotL}{CPT_post_k6}{misfit_pct_mean}{3.541}
\setres{lotL}{CPT_post_k6}{n_seen}{6}
\setres{lotL}{CPT_post_k6}{surprise}{0.07}
\setres{lotL}{CPT_post_k6}{surprise_mean}{1.058}
\setres{lotL}{CPT_post_k6}{width90_adj_pct}{6.3}
\setres{lotL}{CPT_post_k6}{width90_adj_pct_mean}{6.347}
\setres{lotL}{CPT_post_k6}{width90_pct}{2.0}
\setres{lotL}{CPT_post_k6}{width90_pct_mean}{2.150}
\setres{lotL}{CPT_post_k7}{cov50}{0.11}
\setres{lotL}{CPT_post_k7}{cov50_mean}{0.156}
\setres{lotL}{CPT_post_k7}{cov90}{0.39}
\setres{lotL}{CPT_post_k7}{cov90_adj}{0.78}
\setres{lotL}{CPT_post_k7}{cov90_adj_mean}{0.758}
\setres{lotL}{CPT_post_k7}{cov90_mean}{0.353}
\setres{lotL}{CPT_post_k7}{kappa}{0.946}
\setres{lotL}{CPT_post_k7}{kappa_crps}{0.663}
\setres{lotL}{CPT_post_k7}{kappa_crps_mean}{0.661}
\setres{lotL}{CPT_post_k7}{kappa_mean}{0.946}
\setres{lotL}{CPT_post_k7}{logscore_next}{-2.98}
\setres{lotL}{CPT_post_k7}{logscore_next_mean}{-2.985}
\setres{lotL}{CPT_post_k7}{misfit_pct}{3.8}
\setres{lotL}{CPT_post_k7}{misfit_pct_mean}{3.750}
\setres{lotL}{CPT_post_k7}{n_seen}{7}
\setres{lotL}{CPT_post_k7}{surprise}{0.03}
\setres{lotL}{CPT_post_k7}{surprise_mean}{0.511}
\setres{lotL}{CPT_post_k7}{width90_adj_pct}{6.8}
\setres{lotL}{CPT_post_k7}{width90_adj_pct_mean}{6.577}
\setres{lotL}{CPT_post_k7}{width90_pct}{1.6}
\setres{lotL}{CPT_post_k7}{width90_pct_mean}{1.870}
\setres{lotL}{CPT_post_k8}{cov50}{0.09}
\setres{lotL}{CPT_post_k8}{cov50_mean}{0.138}
\setres{lotL}{CPT_post_k8}{cov90}{0.29}
\setres{lotL}{CPT_post_k8}{cov90_adj}{0.82}
\setres{lotL}{CPT_post_k8}{cov90_adj_mean}{0.791}
\setres{lotL}{CPT_post_k8}{cov90_mean}{0.329}
\setres{lotL}{CPT_post_k8}{kappa}{0.950}
\setres{lotL}{CPT_post_k8}{kappa_crps}{0.677}
\setres{lotL}{CPT_post_k8}{kappa_crps_mean}{0.663}
\setres{lotL}{CPT_post_k8}{kappa_mean}{0.948}
\setres{lotL}{CPT_post_k8}{logscore_next}{-2.98}
\setres{lotL}{CPT_post_k8}{logscore_next_mean}{-2.985}
\setres{lotL}{CPT_post_k8}{misfit_pct}{3.8}
\setres{lotL}{CPT_post_k8}{misfit_pct_mean}{3.915}
\setres{lotL}{CPT_post_k8}{n_seen}{8}
\setres{lotL}{CPT_post_k8}{surprise}{0.12}
\setres{lotL}{CPT_post_k8}{surprise_mean}{0.656}
\setres{lotL}{CPT_post_k8}{width90_adj_pct}{6.6}
\setres{lotL}{CPT_post_k8}{width90_adj_pct_mean}{6.774}
\setres{lotL}{CPT_post_k8}{width90_pct}{1.8}
\setres{lotL}{CPT_post_k8}{width90_pct_mean}{1.676}
\setres{lotL}{CPT_post_k9}{cov50}{0.06}
\setres{lotL}{CPT_post_k9}{cov50_mean}{0.106}
\setres{lotL}{CPT_post_k9}{cov90}{0.31}
\setres{lotL}{CPT_post_k9}{cov90_adj}{0.81}
\setres{lotL}{CPT_post_k9}{cov90_adj_mean}{0.800}
\setres{lotL}{CPT_post_k9}{cov90_mean}{0.297}
\setres{lotL}{CPT_post_k9}{kappa}{0.948}
\setres{lotL}{CPT_post_k9}{kappa_crps}{0.676}
\setres{lotL}{CPT_post_k9}{kappa_crps_mean}{0.645}
\setres{lotL}{CPT_post_k9}{kappa_mean}{0.943}
\setres{lotL}{CPT_post_k9}{logscore_next}{-2.98}
\setres{lotL}{CPT_post_k9}{logscore_next_mean}{-2.988}
\setres{lotL}{CPT_post_k9}{misfit_pct}{4.0}
\setres{lotL}{CPT_post_k9}{misfit_pct_mean}{4.049}
\setres{lotL}{CPT_post_k9}{n_seen}{9}
\setres{lotL}{CPT_post_k9}{surprise}{0.16}
\setres{lotL}{CPT_post_k9}{surprise_mean}{0.984}
\setres{lotL}{CPT_post_k9}{width90_adj_pct}{6.9}
\setres{lotL}{CPT_post_k9}{width90_adj_pct_mean}{6.944}
\setres{lotL}{CPT_post_k9}{width90_pct}{1.7}
\setres{lotL}{CPT_post_k9}{width90_pct_mean}{1.570}
\setres{lotL}{DA_post_k0}{cov50}{0.16}
\setres{lotL}{DA_post_k0}{cov50_mean}{0.160}
\setres{lotL}{DA_post_k0}{cov90}{0.44}
\setres{lotL}{DA_post_k0}{cov90_adj}{---}
\setres{lotL}{DA_post_k0}{cov90_adj_mean}{---}
\setres{lotL}{DA_post_k0}{cov90_mean}{0.440}
\setres{lotL}{DA_post_k0}{kappa}{0.017}
\setres{lotL}{DA_post_k0}{kappa_crps}{0.232}
\setres{lotL}{DA_post_k0}{kappa_crps_mean}{0.237}
\setres{lotL}{DA_post_k0}{kappa_mean}{0.017}
\setres{lotL}{DA_post_k0}{logscore_next}{-3.00}
\setres{lotL}{DA_post_k0}{logscore_next_mean}{-3.011}
\setres{lotL}{DA_post_k0}{misfit_pct}{---}
\setres{lotL}{DA_post_k0}{misfit_pct_mean}{---}
\setres{lotL}{DA_post_k0}{n_seen}{0}
\setres{lotL}{DA_post_k0}{surprise}{4.30}
\setres{lotL}{DA_post_k0}{surprise_mean}{4.866}
\setres{lotL}{DA_post_k0}{width90_adj_pct}{---}
\setres{lotL}{DA_post_k0}{width90_adj_pct_mean}{---}
\setres{lotL}{DA_post_k0}{width90_pct}{10.7}
\setres{lotL}{DA_post_k0}{width90_pct_mean}{10.655}
\setres{lotL}{DA_post_k1}{cov50}{0.08}
\setres{lotL}{DA_post_k1}{cov50_mean}{0.073}
\setres{lotL}{DA_post_k1}{cov90}{0.12}
\setres{lotL}{DA_post_k1}{cov90_adj}{---}
\setres{lotL}{DA_post_k1}{cov90_adj_mean}{---}
\setres{lotL}{DA_post_k1}{cov90_mean}{0.138}
\setres{lotL}{DA_post_k1}{kappa}{-1.399}
\setres{lotL}{DA_post_k1}{kappa_crps}{-0.759}
\setres{lotL}{DA_post_k1}{kappa_crps_mean}{-0.611}
\setres{lotL}{DA_post_k1}{kappa_mean}{-1.134}
\setres{lotL}{DA_post_k1}{logscore_next}{-3.01}
\setres{lotL}{DA_post_k1}{logscore_next_mean}{-3.064}
\setres{lotL}{DA_post_k1}{misfit_pct}{---}
\setres{lotL}{DA_post_k1}{misfit_pct_mean}{---}
\setres{lotL}{DA_post_k1}{n_seen}{1}
\setres{lotL}{DA_post_k1}{surprise}{1.79}
\setres{lotL}{DA_post_k1}{surprise_mean}{6.348}
\setres{lotL}{DA_post_k1}{width90_adj_pct}{---}
\setres{lotL}{DA_post_k1}{width90_adj_pct_mean}{---}
\setres{lotL}{DA_post_k1}{width90_pct}{3.4}
\setres{lotL}{DA_post_k1}{width90_pct_mean}{4.102}
\setres{lotL}{DA_post_k10}{cov50}{0.00}
\setres{lotL}{DA_post_k10}{cov50_mean}{0.010}
\setres{lotL}{DA_post_k10}{cov90}{0.00}
\setres{lotL}{DA_post_k10}{cov90_adj}{1.00}
\setres{lotL}{DA_post_k10}{cov90_adj_mean}{0.913}
\setres{lotL}{DA_post_k10}{cov90_mean}{0.020}
\setres{lotL}{DA_post_k10}{kappa}{0.560}
\setres{lotL}{DA_post_k10}{kappa_crps}{0.307}
\setres{lotL}{DA_post_k10}{kappa_crps_mean}{-0.042}
\setres{lotL}{DA_post_k10}{kappa_mean}{-0.101}
\setres{lotL}{DA_post_k10}{logscore_next}{-3.06}
\setres{lotL}{DA_post_k10}{logscore_next_mean}{-3.119}
\setres{lotL}{DA_post_k10}{misfit_pct}{16.4}
\setres{lotL}{DA_post_k10}{misfit_pct_mean}{16.752}
\setres{lotL}{DA_post_k10}{n_seen}{10}
\setres{lotL}{DA_post_k10}{surprise}{0.06}
\setres{lotL}{DA_post_k10}{surprise_mean}{5.411}
\setres{lotL}{DA_post_k10}{width90_adj_pct}{27.1}
\setres{lotL}{DA_post_k10}{width90_adj_pct_mean}{27.568}
\setres{lotL}{DA_post_k10}{width90_pct}{0.4}
\setres{lotL}{DA_post_k10}{width90_pct_mean}{0.506}
\setres{lotL}{DA_post_k11}{cov50}{0.00}
\setres{lotL}{DA_post_k11}{cov50_mean}{0.007}
\setres{lotL}{DA_post_k11}{cov90}{0.00}
\setres{lotL}{DA_post_k11}{cov90_adj}{1.00}
\setres{lotL}{DA_post_k11}{cov90_adj_mean}{0.925}
\setres{lotL}{DA_post_k11}{cov90_mean}{0.021}
\setres{lotL}{DA_post_k11}{kappa}{0.559}
\setres{lotL}{DA_post_k11}{kappa_crps}{0.301}
\setres{lotL}{DA_post_k11}{kappa_crps_mean}{0.009}
\setres{lotL}{DA_post_k11}{kappa_mean}{0.013}
\setres{lotL}{DA_post_k11}{logscore_next}{-3.09}
\setres{lotL}{DA_post_k11}{logscore_next_mean}{-3.450}
\setres{lotL}{DA_post_k11}{misfit_pct}{15.9}
\setres{lotL}{DA_post_k11}{misfit_pct_mean}{16.474}
\setres{lotL}{DA_post_k11}{n_seen}{11}
\setres{lotL}{DA_post_k11}{surprise}{1.07}
\setres{lotL}{DA_post_k11}{surprise_mean}{47.706}
\setres{lotL}{DA_post_k11}{width90_adj_pct}{26.2}
\setres{lotL}{DA_post_k11}{width90_adj_pct_mean}{27.114}
\setres{lotL}{DA_post_k11}{width90_pct}{0.5}
\setres{lotL}{DA_post_k11}{width90_pct_mean}{0.607}
\setres{lotL}{DA_post_k12}{cov50}{0.00}
\setres{lotL}{DA_post_k12}{cov50_mean}{0.012}
\setres{lotL}{DA_post_k12}{cov90}{0.00}
\setres{lotL}{DA_post_k12}{cov90_adj}{1.00}
\setres{lotL}{DA_post_k12}{cov90_adj_mean}{0.923}
\setres{lotL}{DA_post_k12}{cov90_mean}{0.023}
\setres{lotL}{DA_post_k12}{kappa}{0.250}
\setres{lotL}{DA_post_k12}{kappa_crps}{0.139}
\setres{lotL}{DA_post_k12}{kappa_crps_mean}{0.071}
\setres{lotL}{DA_post_k12}{kappa_mean}{0.179}
\setres{lotL}{DA_post_k12}{logscore_next}{-3.09}
\setres{lotL}{DA_post_k12}{logscore_next_mean}{-3.179}
\setres{lotL}{DA_post_k12}{misfit_pct}{15.9}
\setres{lotL}{DA_post_k12}{misfit_pct_mean}{16.461}
\setres{lotL}{DA_post_k12}{n_seen}{12}
\setres{lotL}{DA_post_k12}{surprise}{0.13}
\setres{lotL}{DA_post_k12}{surprise_mean}{1.873}
\setres{lotL}{DA_post_k12}{width90_adj_pct}{26.2}
\setres{lotL}{DA_post_k12}{width90_adj_pct_mean}{27.089}
\setres{lotL}{DA_post_k12}{width90_pct}{0.5}
\setres{lotL}{DA_post_k12}{width90_pct_mean}{0.490}
\setres{lotL}{DA_post_k13}{cov50}{0.00}
\setres{lotL}{DA_post_k13}{cov50_mean}{0.004}
\setres{lotL}{DA_post_k13}{cov90}{0.00}
\setres{lotL}{DA_post_k13}{cov90_adj}{1.00}
\setres{lotL}{DA_post_k13}{cov90_adj_mean}{0.958}
\setres{lotL}{DA_post_k13}{cov90_mean}{0.017}
\setres{lotL}{DA_post_k13}{kappa}{0.367}
\setres{lotL}{DA_post_k13}{kappa_crps}{0.225}
\setres{lotL}{DA_post_k13}{kappa_crps_mean}{0.093}
\setres{lotL}{DA_post_k13}{kappa_mean}{0.147}
\setres{lotL}{DA_post_k13}{logscore_next}{-3.12}
\setres{lotL}{DA_post_k13}{logscore_next_mean}{-3.305}
\setres{lotL}{DA_post_k13}{misfit_pct}{15.9}
\setres{lotL}{DA_post_k13}{misfit_pct_mean}{16.230}
\setres{lotL}{DA_post_k13}{n_seen}{13}
\setres{lotL}{DA_post_k13}{surprise}{0.73}
\setres{lotL}{DA_post_k13}{surprise_mean}{4.194}
\setres{lotL}{DA_post_k13}{width90_adj_pct}{26.2}
\setres{lotL}{DA_post_k13}{width90_adj_pct_mean}{26.709}
\setres{lotL}{DA_post_k13}{width90_pct}{0.4}
\setres{lotL}{DA_post_k13}{width90_pct_mean}{0.460}
\setres{lotL}{DA_post_k14}{cov50}{0.00}
\setres{lotL}{DA_post_k14}{cov50_mean}{0.005}
\setres{lotL}{DA_post_k14}{cov90}{0.00}
\setres{lotL}{DA_post_k14}{cov90_adj}{1.00}
\setres{lotL}{DA_post_k14}{cov90_adj_mean}{0.941}
\setres{lotL}{DA_post_k14}{cov90_mean}{0.023}
\setres{lotL}{DA_post_k14}{kappa}{0.296}
\setres{lotL}{DA_post_k14}{kappa_crps}{0.172}
\setres{lotL}{DA_post_k14}{kappa_crps_mean}{0.067}
\setres{lotL}{DA_post_k14}{kappa_mean}{0.121}
\setres{lotL}{DA_post_k14}{logscore_next}{-3.09}
\setres{lotL}{DA_post_k14}{logscore_next_mean}{-3.337}
\setres{lotL}{DA_post_k14}{misfit_pct}{15.6}
\setres{lotL}{DA_post_k14}{misfit_pct_mean}{16.136}
\setres{lotL}{DA_post_k14}{n_seen}{14}
\setres{lotL}{DA_post_k14}{surprise}{0.15}
\setres{lotL}{DA_post_k14}{surprise_mean}{15.275}
\setres{lotL}{DA_post_k14}{width90_adj_pct}{25.6}
\setres{lotL}{DA_post_k14}{width90_adj_pct_mean}{26.555}
\setres{lotL}{DA_post_k14}{width90_pct}{0.4}
\setres{lotL}{DA_post_k14}{width90_pct_mean}{0.516}
\setres{lotL}{DA_post_k15}{cov50}{0.00}
\setres{lotL}{DA_post_k15}{cov50_mean}{0.000}
\setres{lotL}{DA_post_k15}{cov90}{0.00}
\setres{lotL}{DA_post_k15}{cov90_adj}{1.00}
\setres{lotL}{DA_post_k15}{cov90_adj_mean}{0.955}
\setres{lotL}{DA_post_k15}{cov90_mean}{0.015}
\setres{lotL}{DA_post_k15}{kappa}{0.325}
\setres{lotL}{DA_post_k15}{kappa_crps}{0.214}
\setres{lotL}{DA_post_k15}{kappa_crps_mean}{0.036}
\setres{lotL}{DA_post_k15}{kappa_mean}{0.055}
\setres{lotL}{DA_post_k15}{logscore_next}{-3.06}
\setres{lotL}{DA_post_k15}{logscore_next_mean}{-3.159}
\setres{lotL}{DA_post_k15}{misfit_pct}{15.6}
\setres{lotL}{DA_post_k15}{misfit_pct_mean}{16.039}
\setres{lotL}{DA_post_k15}{n_seen}{15}
\setres{lotL}{DA_post_k15}{surprise}{0.24}
\setres{lotL}{DA_post_k15}{surprise_mean}{2.735}
\setres{lotL}{DA_post_k15}{width90_adj_pct}{25.7}
\setres{lotL}{DA_post_k15}{width90_adj_pct_mean}{26.396}
\setres{lotL}{DA_post_k15}{width90_pct}{0.4}
\setres{lotL}{DA_post_k15}{width90_pct_mean}{0.469}
\setres{lotL}{DA_post_k16}{cov50}{0.00}
\setres{lotL}{DA_post_k16}{cov50_mean}{0.017}
\setres{lotL}{DA_post_k16}{cov90}{0.00}
\setres{lotL}{DA_post_k16}{cov90_adj}{1.00}
\setres{lotL}{DA_post_k16}{cov90_adj_mean}{0.956}
\setres{lotL}{DA_post_k16}{cov90_mean}{0.022}
\setres{lotL}{DA_post_k16}{kappa}{0.342}
\setres{lotL}{DA_post_k16}{kappa_crps}{0.222}
\setres{lotL}{DA_post_k16}{kappa_crps_mean}{0.020}
\setres{lotL}{DA_post_k16}{kappa_mean}{0.040}
\setres{lotL}{DA_post_k16}{logscore_next}{-3.02}
\setres{lotL}{DA_post_k16}{logscore_next_mean}{-3.175}
\setres{lotL}{DA_post_k16}{misfit_pct}{15.4}
\setres{lotL}{DA_post_k16}{misfit_pct_mean}{16.011}
\setres{lotL}{DA_post_k16}{n_seen}{16}
\setres{lotL}{DA_post_k16}{surprise}{0.07}
\setres{lotL}{DA_post_k16}{surprise_mean}{4.783}
\setres{lotL}{DA_post_k16}{width90_adj_pct}{25.4}
\setres{lotL}{DA_post_k16}{width90_adj_pct_mean}{26.351}
\setres{lotL}{DA_post_k16}{width90_pct}{0.4}
\setres{lotL}{DA_post_k16}{width90_pct_mean}{0.518}
\setres{lotL}{DA_post_k17}{cov50}{0.00}
\setres{lotL}{DA_post_k17}{cov50_mean}{0.006}
\setres{lotL}{DA_post_k17}{cov90}{0.00}
\setres{lotL}{DA_post_k17}{cov90_adj}{1.00}
\setres{lotL}{DA_post_k17}{cov90_adj_mean}{0.944}
\setres{lotL}{DA_post_k17}{cov90_mean}{0.025}
\setres{lotL}{DA_post_k17}{kappa}{0.319}
\setres{lotL}{DA_post_k17}{kappa_crps}{0.175}
\setres{lotL}{DA_post_k17}{kappa_crps_mean}{0.039}
\setres{lotL}{DA_post_k17}{kappa_mean}{0.032}
\setres{lotL}{DA_post_k17}{logscore_next}{-3.07}
\setres{lotL}{DA_post_k17}{logscore_next_mean}{-3.241}
\setres{lotL}{DA_post_k17}{misfit_pct}{15.6}
\setres{lotL}{DA_post_k17}{misfit_pct_mean}{15.868}
\setres{lotL}{DA_post_k17}{n_seen}{17}
\setres{lotL}{DA_post_k17}{surprise}{0.41}
\setres{lotL}{DA_post_k17}{surprise_mean}{2.993}
\setres{lotL}{DA_post_k17}{width90_adj_pct}{25.7}
\setres{lotL}{DA_post_k17}{width90_adj_pct_mean}{26.124}
\setres{lotL}{DA_post_k17}{width90_pct}{0.5}
\setres{lotL}{DA_post_k17}{width90_pct_mean}{0.670}
\setres{lotL}{DA_post_k18}{cov50}{0.00}
\setres{lotL}{DA_post_k18}{cov50_mean}{0.007}
\setres{lotL}{DA_post_k18}{cov90}{0.00}
\setres{lotL}{DA_post_k18}{cov90_adj}{1.00}
\setres{lotL}{DA_post_k18}{cov90_adj_mean}{0.943}
\setres{lotL}{DA_post_k18}{cov90_mean}{0.014}
\setres{lotL}{DA_post_k18}{kappa}{0.347}
\setres{lotL}{DA_post_k18}{kappa_crps}{0.187}
\setres{lotL}{DA_post_k18}{kappa_crps_mean}{-0.053}
\setres{lotL}{DA_post_k18}{kappa_mean}{-0.387}
\setres{lotL}{DA_post_k18}{logscore_next}{-3.08}
\setres{lotL}{DA_post_k18}{logscore_next_mean}{-3.147}
\setres{lotL}{DA_post_k18}{misfit_pct}{15.4}
\setres{lotL}{DA_post_k18}{misfit_pct_mean}{15.836}
\setres{lotL}{DA_post_k18}{n_seen}{18}
\setres{lotL}{DA_post_k18}{surprise}{0.14}
\setres{lotL}{DA_post_k18}{surprise_mean}{4.749}
\setres{lotL}{DA_post_k18}{width90_adj_pct}{25.4}
\setres{lotL}{DA_post_k18}{width90_adj_pct_mean}{26.059}
\setres{lotL}{DA_post_k18}{width90_pct}{0.4}
\setres{lotL}{DA_post_k18}{width90_pct_mean}{0.480}
\setres{lotL}{DA_post_k19}{cov50}{0.00}
\setres{lotL}{DA_post_k19}{cov50_mean}{0.008}
\setres{lotL}{DA_post_k19}{cov90}{0.00}
\setres{lotL}{DA_post_k19}{cov90_adj}{1.00}
\setres{lotL}{DA_post_k19}{cov90_adj_mean}{0.967}
\setres{lotL}{DA_post_k19}{cov90_mean}{0.008}
\setres{lotL}{DA_post_k19}{kappa}{0.392}
\setres{lotL}{DA_post_k19}{kappa_crps}{0.256}
\setres{lotL}{DA_post_k19}{kappa_crps_mean}{0.085}
\setres{lotL}{DA_post_k19}{kappa_mean}{0.069}
\setres{lotL}{DA_post_k19}{logscore_next}{-3.10}
\setres{lotL}{DA_post_k19}{logscore_next_mean}{-3.268}
\setres{lotL}{DA_post_k19}{misfit_pct}{15.3}
\setres{lotL}{DA_post_k19}{misfit_pct_mean}{15.659}
\setres{lotL}{DA_post_k19}{n_seen}{19}
\setres{lotL}{DA_post_k19}{surprise}{0.21}
\setres{lotL}{DA_post_k19}{surprise_mean}{10.051}
\setres{lotL}{DA_post_k19}{width90_adj_pct}{25.1}
\setres{lotL}{DA_post_k19}{width90_adj_pct_mean}{25.775}
\setres{lotL}{DA_post_k19}{width90_pct}{0.4}
\setres{lotL}{DA_post_k19}{width90_pct_mean}{0.554}
\setres{lotL}{DA_post_k2}{cov50}{0.00}
\setres{lotL}{DA_post_k2}{cov50_mean}{0.033}
\setres{lotL}{DA_post_k2}{cov90}{0.13}
\setres{lotL}{DA_post_k2}{cov90_adj}{0.33}
\setres{lotL}{DA_post_k2}{cov90_adj_mean}{0.439}
\setres{lotL}{DA_post_k2}{cov90_mean}{0.113}
\setres{lotL}{DA_post_k2}{kappa}{-0.281}
\setres{lotL}{DA_post_k2}{kappa_crps}{-0.150}
\setres{lotL}{DA_post_k2}{kappa_crps_mean}{-0.371}
\setres{lotL}{DA_post_k2}{kappa_mean}{-0.710}
\setres{lotL}{DA_post_k2}{logscore_next}{-3.02}
\setres{lotL}{DA_post_k2}{logscore_next_mean}{-3.131}
\setres{lotL}{DA_post_k2}{misfit_pct}{9.1}
\setres{lotL}{DA_post_k2}{misfit_pct_mean}{9.382}
\setres{lotL}{DA_post_k2}{n_seen}{2}
\setres{lotL}{DA_post_k2}{surprise}{1.72}
\setres{lotL}{DA_post_k2}{surprise_mean}{5.662}
\setres{lotL}{DA_post_k2}{width90_adj_pct}{15.2}
\setres{lotL}{DA_post_k2}{width90_adj_pct_mean}{15.850}
\setres{lotL}{DA_post_k2}{width90_pct}{2.4}
\setres{lotL}{DA_post_k2}{width90_pct_mean}{2.561}
\setres{lotL}{DA_post_k20}{cov50}{0.00}
\setres{lotL}{DA_post_k20}{cov50_mean}{0.010}
\setres{lotL}{DA_post_k20}{cov90}{0.00}
\setres{lotL}{DA_post_k20}{cov90_adj}{1.00}
\setres{lotL}{DA_post_k20}{cov90_adj_mean}{0.960}
\setres{lotL}{DA_post_k20}{cov90_mean}{0.010}
\setres{lotL}{DA_post_k20}{kappa}{0.341}
\setres{lotL}{DA_post_k20}{kappa_crps}{0.243}
\setres{lotL}{DA_post_k20}{kappa_crps_mean}{0.009}
\setres{lotL}{DA_post_k20}{kappa_mean}{-0.169}
\setres{lotL}{DA_post_k20}{logscore_next}{-3.05}
\setres{lotL}{DA_post_k20}{logscore_next_mean}{-3.197}
\setres{lotL}{DA_post_k20}{misfit_pct}{15.2}
\setres{lotL}{DA_post_k20}{misfit_pct_mean}{15.411}
\setres{lotL}{DA_post_k20}{n_seen}{20}
\setres{lotL}{DA_post_k20}{surprise}{0.42}
\setres{lotL}{DA_post_k20}{surprise_mean}{3.724}
\setres{lotL}{DA_post_k20}{width90_adj_pct}{25.1}
\setres{lotL}{DA_post_k20}{width90_adj_pct_mean}{25.388}
\setres{lotL}{DA_post_k20}{width90_pct}{0.4}
\setres{lotL}{DA_post_k20}{width90_pct_mean}{0.701}
\setres{lotL}{DA_post_k21}{cov50}{0.00}
\setres{lotL}{DA_post_k21}{cov50_mean}{0.013}
\setres{lotL}{DA_post_k21}{cov90}{0.00}
\setres{lotL}{DA_post_k21}{cov90_adj}{1.00}
\setres{lotL}{DA_post_k21}{cov90_adj_mean}{0.963}
\setres{lotL}{DA_post_k21}{cov90_mean}{0.037}
\setres{lotL}{DA_post_k21}{kappa}{0.312}
\setres{lotL}{DA_post_k21}{kappa_crps}{0.295}
\setres{lotL}{DA_post_k21}{kappa_crps_mean}{0.050}
\setres{lotL}{DA_post_k21}{kappa_mean}{-0.117}
\setres{lotL}{DA_post_k21}{logscore_next}{-3.05}
\setres{lotL}{DA_post_k21}{logscore_next_mean}{-3.095}
\setres{lotL}{DA_post_k21}{misfit_pct}{15.0}
\setres{lotL}{DA_post_k21}{misfit_pct_mean}{15.193}
\setres{lotL}{DA_post_k21}{n_seen}{21}
\setres{lotL}{DA_post_k21}{surprise}{0.17}
\setres{lotL}{DA_post_k21}{surprise_mean}{3.098}
\setres{lotL}{DA_post_k21}{width90_adj_pct}{24.7}
\setres{lotL}{DA_post_k21}{width90_adj_pct_mean}{25.062}
\setres{lotL}{DA_post_k21}{width90_pct}{0.4}
\setres{lotL}{DA_post_k21}{width90_pct_mean}{0.921}
\setres{lotL}{DA_post_k22}{cov50}{0.00}
\setres{lotL}{DA_post_k22}{cov50_mean}{0.000}
\setres{lotL}{DA_post_k22}{cov90}{0.00}
\setres{lotL}{DA_post_k22}{cov90_adj}{1.00}
\setres{lotL}{DA_post_k22}{cov90_adj_mean}{0.983}
\setres{lotL}{DA_post_k22}{cov90_mean}{0.000}
\setres{lotL}{DA_post_k22}{kappa}{0.340}
\setres{lotL}{DA_post_k22}{kappa_crps}{0.352}
\setres{lotL}{DA_post_k22}{kappa_crps_mean}{0.169}
\setres{lotL}{DA_post_k22}{kappa_mean}{0.309}
\setres{lotL}{DA_post_k22}{logscore_next}{-3.02}
\setres{lotL}{DA_post_k22}{logscore_next_mean}{-3.071}
\setres{lotL}{DA_post_k22}{misfit_pct}{15.0}
\setres{lotL}{DA_post_k22}{misfit_pct_mean}{14.983}
\setres{lotL}{DA_post_k22}{n_seen}{22}
\setres{lotL}{DA_post_k22}{surprise}{0.07}
\setres{lotL}{DA_post_k22}{surprise_mean}{0.804}
\setres{lotL}{DA_post_k22}{width90_adj_pct}{24.6}
\setres{lotL}{DA_post_k22}{width90_adj_pct_mean}{24.663}
\setres{lotL}{DA_post_k22}{width90_pct}{0.5}
\setres{lotL}{DA_post_k22}{width90_pct_mean}{0.642}
\setres{lotL}{DA_post_k23}{cov50}{0.00}
\setres{lotL}{DA_post_k23}{cov50_mean}{0.000}
\setres{lotL}{DA_post_k23}{cov90}{0.00}
\setres{lotL}{DA_post_k23}{cov90_adj}{1.00}
\setres{lotL}{DA_post_k23}{cov90_adj_mean}{1.000}
\setres{lotL}{DA_post_k23}{cov90_mean}{0.000}
\setres{lotL}{DA_post_k23}{kappa}{0.386}
\setres{lotL}{DA_post_k23}{kappa_crps}{0.331}
\setres{lotL}{DA_post_k23}{kappa_crps_mean}{0.218}
\setres{lotL}{DA_post_k23}{kappa_mean}{0.304}
\setres{lotL}{DA_post_k23}{logscore_next}{-3.10}
\setres{lotL}{DA_post_k23}{logscore_next_mean}{-3.206}
\setres{lotL}{DA_post_k23}{misfit_pct}{15.0}
\setres{lotL}{DA_post_k23}{misfit_pct_mean}{14.918}
\setres{lotL}{DA_post_k23}{n_seen}{23}
\setres{lotL}{DA_post_k23}{surprise}{0.34}
\setres{lotL}{DA_post_k23}{surprise_mean}{5.515}
\setres{lotL}{DA_post_k23}{width90_adj_pct}{24.6}
\setres{lotL}{DA_post_k23}{width90_adj_pct_mean}{24.560}
\setres{lotL}{DA_post_k23}{width90_pct}{0.4}
\setres{lotL}{DA_post_k23}{width90_pct_mean}{0.604}
\setres{lotL}{DA_post_k24}{cov50}{0.00}
\setres{lotL}{DA_post_k24}{cov50_mean}{0.000}
\setres{lotL}{DA_post_k24}{cov90}{0.00}
\setres{lotL}{DA_post_k24}{cov90_adj}{1.00}
\setres{lotL}{DA_post_k24}{cov90_adj_mean}{1.000}
\setres{lotL}{DA_post_k24}{cov90_mean}{0.000}
\setres{lotL}{DA_post_k24}{kappa}{0.179}
\setres{lotL}{DA_post_k24}{kappa_crps}{0.436}
\setres{lotL}{DA_post_k24}{kappa_crps_mean}{0.366}
\setres{lotL}{DA_post_k24}{kappa_mean}{0.270}
\setres{lotL}{DA_post_k24}{logscore_next}{-3.21}
\setres{lotL}{DA_post_k24}{logscore_next_mean}{-3.548}
\setres{lotL}{DA_post_k24}{misfit_pct}{14.8}
\setres{lotL}{DA_post_k24}{misfit_pct_mean}{14.895}
\setres{lotL}{DA_post_k24}{n_seen}{24}
\setres{lotL}{DA_post_k24}{surprise}{0.86}
\setres{lotL}{DA_post_k24}{surprise_mean}{3.211}
\setres{lotL}{DA_post_k24}{width90_adj_pct}{24.3}
\setres{lotL}{DA_post_k24}{width90_adj_pct_mean}{24.514}
\setres{lotL}{DA_post_k24}{width90_pct}{0.4}
\setres{lotL}{DA_post_k24}{width90_pct_mean}{0.554}
\setres{lotL}{DA_post_k25}{cov50}{---}
\setres{lotL}{DA_post_k25}{cov50_mean}{---}
\setres{lotL}{DA_post_k25}{cov90}{---}
\setres{lotL}{DA_post_k25}{cov90_adj}{---}
\setres{lotL}{DA_post_k25}{cov90_adj_mean}{---}
\setres{lotL}{DA_post_k25}{cov90_mean}{---}
\setres{lotL}{DA_post_k25}{kappa}{---}
\setres{lotL}{DA_post_k25}{kappa_crps}{---}
\setres{lotL}{DA_post_k25}{kappa_crps_mean}{---}
\setres{lotL}{DA_post_k25}{kappa_mean}{---}
\setres{lotL}{DA_post_k25}{logscore_next}{---}
\setres{lotL}{DA_post_k25}{logscore_next_mean}{---}
\setres{lotL}{DA_post_k25}{misfit_pct}{---}
\setres{lotL}{DA_post_k25}{misfit_pct_mean}{---}
\setres{lotL}{DA_post_k25}{n_seen}{25}
\setres{lotL}{DA_post_k25}{surprise}{---}
\setres{lotL}{DA_post_k25}{surprise_mean}{---}
\setres{lotL}{DA_post_k25}{width90_adj_pct}{---}
\setres{lotL}{DA_post_k25}{width90_adj_pct_mean}{---}
\setres{lotL}{DA_post_k25}{width90_pct}{---}
\setres{lotL}{DA_post_k25}{width90_pct_mean}{---}
\setres{lotL}{DA_post_k3}{cov50}{0.00}
\setres{lotL}{DA_post_k3}{cov50_mean}{0.023}
\setres{lotL}{DA_post_k3}{cov90}{0.05}
\setres{lotL}{DA_post_k3}{cov90_adj}{0.80}
\setres{lotL}{DA_post_k3}{cov90_adj_mean}{0.650}
\setres{lotL}{DA_post_k3}{cov90_mean}{0.068}
\setres{lotL}{DA_post_k3}{kappa}{0.162}
\setres{lotL}{DA_post_k3}{kappa_crps}{-0.115}
\setres{lotL}{DA_post_k3}{kappa_crps_mean}{-0.096}
\setres{lotL}{DA_post_k3}{kappa_mean}{-0.145}
\setres{lotL}{DA_post_k3}{logscore_next}{-3.10}
\setres{lotL}{DA_post_k3}{logscore_next_mean}{-3.159}
\setres{lotL}{DA_post_k3}{misfit_pct}{12.7}
\setres{lotL}{DA_post_k3}{misfit_pct_mean}{12.627}
\setres{lotL}{DA_post_k3}{n_seen}{3}
\setres{lotL}{DA_post_k3}{surprise}{1.53}
\setres{lotL}{DA_post_k3}{surprise_mean}{9.409}
\setres{lotL}{DA_post_k3}{width90_adj_pct}{21.1}
\setres{lotL}{DA_post_k3}{width90_adj_pct_mean}{20.890}
\setres{lotL}{DA_post_k3}{width90_pct}{1.3}
\setres{lotL}{DA_post_k3}{width90_pct_mean}{1.744}
\setres{lotL}{DA_post_k4}{cov50}{0.00}
\setres{lotL}{DA_post_k4}{cov50_mean}{0.012}
\setres{lotL}{DA_post_k4}{cov90}{0.00}
\setres{lotL}{DA_post_k4}{cov90_adj}{0.83}
\setres{lotL}{DA_post_k4}{cov90_adj_mean}{0.698}
\setres{lotL}{DA_post_k4}{cov90_mean}{0.024}
\setres{lotL}{DA_post_k4}{kappa}{0.347}
\setres{lotL}{DA_post_k4}{kappa_crps}{0.082}
\setres{lotL}{DA_post_k4}{kappa_crps_mean}{-0.075}
\setres{lotL}{DA_post_k4}{kappa_mean}{-0.037}
\setres{lotL}{DA_post_k4}{logscore_next}{-3.11}
\setres{lotL}{DA_post_k4}{logscore_next_mean}{-3.176}
\setres{lotL}{DA_post_k4}{misfit_pct}{13.4}
\setres{lotL}{DA_post_k4}{misfit_pct_mean}{13.557}
\setres{lotL}{DA_post_k4}{n_seen}{4}
\setres{lotL}{DA_post_k4}{surprise}{0.65}
\setres{lotL}{DA_post_k4}{surprise_mean}{3.105}
\setres{lotL}{DA_post_k4}{width90_adj_pct}{22.1}
\setres{lotL}{DA_post_k4}{width90_adj_pct_mean}{22.334}
\setres{lotL}{DA_post_k4}{width90_pct}{0.7}
\setres{lotL}{DA_post_k4}{width90_pct_mean}{0.827}
\setres{lotL}{DA_post_k5}{cov50}{0.00}
\setres{lotL}{DA_post_k5}{cov50_mean}{0.005}
\setres{lotL}{DA_post_k5}{cov90}{0.00}
\setres{lotL}{DA_post_k5}{cov90_adj}{0.82}
\setres{lotL}{DA_post_k5}{cov90_adj_mean}{0.767}
\setres{lotL}{DA_post_k5}{cov90_mean}{0.017}
\setres{lotL}{DA_post_k5}{kappa}{0.286}
\setres{lotL}{DA_post_k5}{kappa_crps}{0.033}
\setres{lotL}{DA_post_k5}{kappa_crps_mean}{-0.118}
\setres{lotL}{DA_post_k5}{kappa_mean}{-0.107}
\setres{lotL}{DA_post_k5}{logscore_next}{-3.07}
\setres{lotL}{DA_post_k5}{logscore_next_mean}{-3.195}
\setres{lotL}{DA_post_k5}{misfit_pct}{15.1}
\setres{lotL}{DA_post_k5}{misfit_pct_mean}{15.168}
\setres{lotL}{DA_post_k5}{n_seen}{5}
\setres{lotL}{DA_post_k5}{surprise}{0.56}
\setres{lotL}{DA_post_k5}{surprise_mean}{14.558}
\setres{lotL}{DA_post_k5}{width90_adj_pct}{24.9}
\setres{lotL}{DA_post_k5}{width90_adj_pct_mean}{25.001}
\setres{lotL}{DA_post_k5}{width90_pct}{0.6}
\setres{lotL}{DA_post_k5}{width90_pct_mean}{1.078}
\setres{lotL}{DA_post_k6}{cov50}{0.00}
\setres{lotL}{DA_post_k6}{cov50_mean}{0.008}
\setres{lotL}{DA_post_k6}{cov90}{0.00}
\setres{lotL}{DA_post_k6}{cov90_adj}{0.92}
\setres{lotL}{DA_post_k6}{cov90_adj_mean}{0.832}
\setres{lotL}{DA_post_k6}{cov90_mean}{0.021}
\setres{lotL}{DA_post_k6}{kappa}{0.453}
\setres{lotL}{DA_post_k6}{kappa_crps}{0.231}
\setres{lotL}{DA_post_k6}{kappa_crps_mean}{-0.017}
\setres{lotL}{DA_post_k6}{kappa_mean}{0.026}
\setres{lotL}{DA_post_k6}{logscore_next}{-3.04}
\setres{lotL}{DA_post_k6}{logscore_next_mean}{-3.102}
\setres{lotL}{DA_post_k6}{misfit_pct}{14.9}
\setres{lotL}{DA_post_k6}{misfit_pct_mean}{15.307}
\setres{lotL}{DA_post_k6}{n_seen}{6}
\setres{lotL}{DA_post_k6}{surprise}{0.14}
\setres{lotL}{DA_post_k6}{surprise_mean}{5.634}
\setres{lotL}{DA_post_k6}{width90_adj_pct}{24.6}
\setres{lotL}{DA_post_k6}{width90_adj_pct_mean}{25.209}
\setres{lotL}{DA_post_k6}{width90_pct}{0.8}
\setres{lotL}{DA_post_k6}{width90_pct_mean}{0.843}
\setres{lotL}{DA_post_k7}{cov50}{0.00}
\setres{lotL}{DA_post_k7}{cov50_mean}{0.006}
\setres{lotL}{DA_post_k7}{cov90}{0.00}
\setres{lotL}{DA_post_k7}{cov90_adj}{0.94}
\setres{lotL}{DA_post_k7}{cov90_adj_mean}{0.867}
\setres{lotL}{DA_post_k7}{cov90_mean}{0.017}
\setres{lotL}{DA_post_k7}{kappa}{0.566}
\setres{lotL}{DA_post_k7}{kappa_crps}{0.263}
\setres{lotL}{DA_post_k7}{kappa_crps_mean}{0.026}
\setres{lotL}{DA_post_k7}{kappa_mean}{0.136}
\setres{lotL}{DA_post_k7}{logscore_next}{-3.06}
\setres{lotL}{DA_post_k7}{logscore_next_mean}{-3.198}
\setres{lotL}{DA_post_k7}{misfit_pct}{14.6}
\setres{lotL}{DA_post_k7}{misfit_pct_mean}{15.568}
\setres{lotL}{DA_post_k7}{n_seen}{7}
\setres{lotL}{DA_post_k7}{surprise}{0.07}
\setres{lotL}{DA_post_k7}{surprise_mean}{10.774}
\setres{lotL}{DA_post_k7}{width90_adj_pct}{24.0}
\setres{lotL}{DA_post_k7}{width90_adj_pct_mean}{25.630}
\setres{lotL}{DA_post_k7}{width90_pct}{0.4}
\setres{lotL}{DA_post_k7}{width90_pct_mean}{0.643}
\setres{lotL}{DA_post_k8}{cov50}{0.00}
\setres{lotL}{DA_post_k8}{cov50_mean}{0.012}
\setres{lotL}{DA_post_k8}{cov90}{0.00}
\setres{lotL}{DA_post_k8}{cov90_adj}{0.94}
\setres{lotL}{DA_post_k8}{cov90_adj_mean}{0.924}
\setres{lotL}{DA_post_k8}{cov90_mean}{0.018}
\setres{lotL}{DA_post_k8}{kappa}{0.565}
\setres{lotL}{DA_post_k8}{kappa_crps}{0.286}
\setres{lotL}{DA_post_k8}{kappa_crps_mean}{0.052}
\setres{lotL}{DA_post_k8}{kappa_mean}{0.171}
\setres{lotL}{DA_post_k8}{logscore_next}{-3.13}
\setres{lotL}{DA_post_k8}{logscore_next_mean}{-3.250}
\setres{lotL}{DA_post_k8}{misfit_pct}{15.5}
\setres{lotL}{DA_post_k8}{misfit_pct_mean}{16.364}
\setres{lotL}{DA_post_k8}{n_seen}{8}
\setres{lotL}{DA_post_k8}{surprise}{0.40}
\setres{lotL}{DA_post_k8}{surprise_mean}{23.289}
\setres{lotL}{DA_post_k8}{width90_adj_pct}{25.6}
\setres{lotL}{DA_post_k8}{width90_adj_pct_mean}{26.938}
\setres{lotL}{DA_post_k8}{width90_pct}{0.4}
\setres{lotL}{DA_post_k8}{width90_pct_mean}{0.631}
\setres{lotL}{DA_post_k9}{cov50}{0.00}
\setres{lotL}{DA_post_k9}{cov50_mean}{0.009}
\setres{lotL}{DA_post_k9}{cov90}{0.00}
\setres{lotL}{DA_post_k9}{cov90_adj}{0.94}
\setres{lotL}{DA_post_k9}{cov90_adj_mean}{0.912}
\setres{lotL}{DA_post_k9}{cov90_mean}{0.013}
\setres{lotL}{DA_post_k9}{kappa}{0.561}
\setres{lotL}{DA_post_k9}{kappa_crps}{0.304}
\setres{lotL}{DA_post_k9}{kappa_crps_mean}{-0.055}
\setres{lotL}{DA_post_k9}{kappa_mean}{-0.141}
\setres{lotL}{DA_post_k9}{logscore_next}{-3.07}
\setres{lotL}{DA_post_k9}{logscore_next_mean}{-3.215}
\setres{lotL}{DA_post_k9}{misfit_pct}{15.6}
\setres{lotL}{DA_post_k9}{misfit_pct_mean}{16.629}
\setres{lotL}{DA_post_k9}{n_seen}{9}
\setres{lotL}{DA_post_k9}{surprise}{0.89}
\setres{lotL}{DA_post_k9}{surprise_mean}{13.809}
\setres{lotL}{DA_post_k9}{width90_adj_pct}{25.7}
\setres{lotL}{DA_post_k9}{width90_adj_pct_mean}{27.369}
\setres{lotL}{DA_post_k9}{width90_pct}{0.5}
\setres{lotL}{DA_post_k9}{width90_pct_mean}{0.528}
\setres{lotL}{EU_post_k0}{cov50}{0.12}
\setres{lotL}{EU_post_k0}{cov50_mean}{0.120}
\setres{lotL}{EU_post_k0}{cov90}{0.20}
\setres{lotL}{EU_post_k0}{cov90_adj}{---}
\setres{lotL}{EU_post_k0}{cov90_adj_mean}{---}
\setres{lotL}{EU_post_k0}{cov90_mean}{0.200}
\setres{lotL}{EU_post_k0}{kappa}{0.517}
\setres{lotL}{EU_post_k0}{kappa_crps}{0.359}
\setres{lotL}{EU_post_k0}{kappa_crps_mean}{0.362}
\setres{lotL}{EU_post_k0}{kappa_mean}{0.517}
\setres{lotL}{EU_post_k0}{logscore_next}{-3.03}
\setres{lotL}{EU_post_k0}{logscore_next_mean}{-3.198}
\setres{lotL}{EU_post_k0}{misfit_pct}{---}
\setres{lotL}{EU_post_k0}{misfit_pct_mean}{---}
\setres{lotL}{EU_post_k0}{n_seen}{0}
\setres{lotL}{EU_post_k0}{surprise}{2.08}
\setres{lotL}{EU_post_k0}{surprise_mean}{2.642}
\setres{lotL}{EU_post_k0}{width90_adj_pct}{---}
\setres{lotL}{EU_post_k0}{width90_adj_pct_mean}{---}
\setres{lotL}{EU_post_k0}{width90_pct}{2.6}
\setres{lotL}{EU_post_k0}{width90_pct_mean}{2.640}
\setres{lotL}{EU_post_k1}{cov50}{0.00}
\setres{lotL}{EU_post_k1}{cov50_mean}{0.029}
\setres{lotL}{EU_post_k1}{cov90}{0.04}
\setres{lotL}{EU_post_k1}{cov90_adj}{---}
\setres{lotL}{EU_post_k1}{cov90_adj_mean}{---}
\setres{lotL}{EU_post_k1}{cov90_mean}{0.081}
\setres{lotL}{EU_post_k1}{kappa}{0.518}
\setres{lotL}{EU_post_k1}{kappa_crps}{0.295}
\setres{lotL}{EU_post_k1}{kappa_crps_mean}{0.186}
\setres{lotL}{EU_post_k1}{kappa_mean}{0.283}
\setres{lotL}{EU_post_k1}{logscore_next}{-3.01}
\setres{lotL}{EU_post_k1}{logscore_next_mean}{-3.149}
\setres{lotL}{EU_post_k1}{misfit_pct}{---}
\setres{lotL}{EU_post_k1}{misfit_pct_mean}{---}
\setres{lotL}{EU_post_k1}{n_seen}{1}
\setres{lotL}{EU_post_k1}{surprise}{0.18}
\setres{lotL}{EU_post_k1}{surprise_mean}{1.954}
\setres{lotL}{EU_post_k1}{width90_adj_pct}{---}
\setres{lotL}{EU_post_k1}{width90_adj_pct_mean}{---}
\setres{lotL}{EU_post_k1}{width90_pct}{0.9}
\setres{lotL}{EU_post_k1}{width90_pct_mean}{1.153}
\setres{lotL}{EU_post_k10}{cov50}{0.00}
\setres{lotL}{EU_post_k10}{cov50_mean}{0.000}
\setres{lotL}{EU_post_k10}{cov90}{0.00}
\setres{lotL}{EU_post_k10}{cov90_adj}{1.00}
\setres{lotL}{EU_post_k10}{cov90_adj_mean}{0.967}
\setres{lotL}{EU_post_k10}{cov90_mean}{0.007}
\setres{lotL}{EU_post_k10}{kappa}{0.554}
\setres{lotL}{EU_post_k10}{kappa_crps}{0.303}
\setres{lotL}{EU_post_k10}{kappa_crps_mean}{0.244}
\setres{lotL}{EU_post_k10}{kappa_mean}{0.441}
\setres{lotL}{EU_post_k10}{logscore_next}{-3.01}
\setres{lotL}{EU_post_k10}{logscore_next_mean}{-3.052}
\setres{lotL}{EU_post_k10}{misfit_pct}{15.0}
\setres{lotL}{EU_post_k10}{misfit_pct_mean}{15.576}
\setres{lotL}{EU_post_k10}{n_seen}{10}
\setres{lotL}{EU_post_k10}{surprise}{0.01}
\setres{lotL}{EU_post_k10}{surprise_mean}{0.032}
\setres{lotL}{EU_post_k10}{width90_adj_pct}{24.7}
\setres{lotL}{EU_post_k10}{width90_adj_pct_mean}{25.624}
\setres{lotL}{EU_post_k10}{width90_pct}{0.2}
\setres{lotL}{EU_post_k10}{width90_pct_mean}{0.212}
\setres{lotL}{EU_post_k11}{cov50}{0.00}
\setres{lotL}{EU_post_k11}{cov50_mean}{0.000}
\setres{lotL}{EU_post_k11}{cov90}{0.00}
\setres{lotL}{EU_post_k11}{cov90_adj}{1.00}
\setres{lotL}{EU_post_k11}{cov90_adj_mean}{0.968}
\setres{lotL}{EU_post_k11}{cov90_mean}{0.007}
\setres{lotL}{EU_post_k11}{kappa}{0.556}
\setres{lotL}{EU_post_k11}{kappa_crps}{0.290}
\setres{lotL}{EU_post_k11}{kappa_crps_mean}{0.245}
\setres{lotL}{EU_post_k11}{kappa_mean}{0.444}
\setres{lotL}{EU_post_k11}{logscore_next}{-3.06}
\setres{lotL}{EU_post_k11}{logscore_next_mean}{-3.434}
\setres{lotL}{EU_post_k11}{misfit_pct}{15.4}
\setres{lotL}{EU_post_k11}{misfit_pct_mean}{15.538}
\setres{lotL}{EU_post_k11}{n_seen}{11}
\setres{lotL}{EU_post_k11}{surprise}{0.06}
\setres{lotL}{EU_post_k11}{surprise_mean}{1.634}
\setres{lotL}{EU_post_k11}{width90_adj_pct}{25.4}
\setres{lotL}{EU_post_k11}{width90_adj_pct_mean}{25.561}
\setres{lotL}{EU_post_k11}{width90_pct}{0.2}
\setres{lotL}{EU_post_k11}{width90_pct_mean}{0.218}
\setres{lotL}{EU_post_k12}{cov50}{0.00}
\setres{lotL}{EU_post_k12}{cov50_mean}{0.000}
\setres{lotL}{EU_post_k12}{cov90}{0.00}
\setres{lotL}{EU_post_k12}{cov90_adj}{1.00}
\setres{lotL}{EU_post_k12}{cov90_adj_mean}{0.965}
\setres{lotL}{EU_post_k12}{cov90_mean}{0.008}
\setres{lotL}{EU_post_k12}{kappa}{0.477}
\setres{lotL}{EU_post_k12}{kappa_crps}{0.321}
\setres{lotL}{EU_post_k12}{kappa_crps_mean}{0.266}
\setres{lotL}{EU_post_k12}{kappa_mean}{0.455}
\setres{lotL}{EU_post_k12}{logscore_next}{-3.09}
\setres{lotL}{EU_post_k12}{logscore_next_mean}{-3.151}
\setres{lotL}{EU_post_k12}{misfit_pct}{15.3}
\setres{lotL}{EU_post_k12}{misfit_pct_mean}{15.187}
\setres{lotL}{EU_post_k12}{n_seen}{12}
\setres{lotL}{EU_post_k12}{surprise}{0.03}
\setres{lotL}{EU_post_k12}{surprise_mean}{0.334}
\setres{lotL}{EU_post_k12}{width90_adj_pct}{25.1}
\setres{lotL}{EU_post_k12}{width90_adj_pct_mean}{24.984}
\setres{lotL}{EU_post_k12}{width90_pct}{0.2}
\setres{lotL}{EU_post_k12}{width90_pct_mean}{0.214}
\setres{lotL}{EU_post_k13}{cov50}{0.00}
\setres{lotL}{EU_post_k13}{cov50_mean}{0.004}
\setres{lotL}{EU_post_k13}{cov90}{0.00}
\setres{lotL}{EU_post_k13}{cov90_adj}{1.00}
\setres{lotL}{EU_post_k13}{cov90_adj_mean}{0.971}
\setres{lotL}{EU_post_k13}{cov90_mean}{0.004}
\setres{lotL}{EU_post_k13}{kappa}{0.479}
\setres{lotL}{EU_post_k13}{kappa_crps}{0.318}
\setres{lotL}{EU_post_k13}{kappa_crps_mean}{0.260}
\setres{lotL}{EU_post_k13}{kappa_mean}{0.439}
\setres{lotL}{EU_post_k13}{logscore_next}{-3.11}
\setres{lotL}{EU_post_k13}{logscore_next_mean}{-3.259}
\setres{lotL}{EU_post_k13}{misfit_pct}{15.5}
\setres{lotL}{EU_post_k13}{misfit_pct_mean}{15.277}
\setres{lotL}{EU_post_k13}{n_seen}{13}
\setres{lotL}{EU_post_k13}{surprise}{0.03}
\setres{lotL}{EU_post_k13}{surprise_mean}{3.069}
\setres{lotL}{EU_post_k13}{width90_adj_pct}{25.5}
\setres{lotL}{EU_post_k13}{width90_adj_pct_mean}{25.133}
\setres{lotL}{EU_post_k13}{width90_pct}{0.2}
\setres{lotL}{EU_post_k13}{width90_pct_mean}{0.235}
\setres{lotL}{EU_post_k14}{cov50}{0.00}
\setres{lotL}{EU_post_k14}{cov50_mean}{0.000}
\setres{lotL}{EU_post_k14}{cov90}{0.00}
\setres{lotL}{EU_post_k14}{cov90_adj}{1.00}
\setres{lotL}{EU_post_k14}{cov90_adj_mean}{0.950}
\setres{lotL}{EU_post_k14}{cov90_mean}{0.000}
\setres{lotL}{EU_post_k14}{kappa}{0.452}
\setres{lotL}{EU_post_k14}{kappa_crps}{0.259}
\setres{lotL}{EU_post_k14}{kappa_crps_mean}{0.219}
\setres{lotL}{EU_post_k14}{kappa_mean}{0.384}
\setres{lotL}{EU_post_k14}{logscore_next}{-3.10}
\setres{lotL}{EU_post_k14}{logscore_next_mean}{-3.311}
\setres{lotL}{EU_post_k14}{misfit_pct}{15.5}
\setres{lotL}{EU_post_k14}{misfit_pct_mean}{15.319}
\setres{lotL}{EU_post_k14}{n_seen}{14}
\setres{lotL}{EU_post_k14}{surprise}{0.03}
\setres{lotL}{EU_post_k14}{surprise_mean}{3.564}
\setres{lotL}{EU_post_k14}{width90_adj_pct}{25.6}
\setres{lotL}{EU_post_k14}{width90_adj_pct_mean}{25.201}
\setres{lotL}{EU_post_k14}{width90_pct}{0.2}
\setres{lotL}{EU_post_k14}{width90_pct_mean}{0.195}
\setres{lotL}{EU_post_k15}{cov50}{0.00}
\setres{lotL}{EU_post_k15}{cov50_mean}{0.005}
\setres{lotL}{EU_post_k15}{cov90}{0.00}
\setres{lotL}{EU_post_k15}{cov90_adj}{1.00}
\setres{lotL}{EU_post_k15}{cov90_adj_mean}{0.960}
\setres{lotL}{EU_post_k15}{cov90_mean}{0.005}
\setres{lotL}{EU_post_k15}{kappa}{0.436}
\setres{lotL}{EU_post_k15}{kappa_crps}{0.264}
\setres{lotL}{EU_post_k15}{kappa_crps_mean}{0.238}
\setres{lotL}{EU_post_k15}{kappa_mean}{0.416}
\setres{lotL}{EU_post_k15}{logscore_next}{-3.05}
\setres{lotL}{EU_post_k15}{logscore_next_mean}{-3.151}
\setres{lotL}{EU_post_k15}{misfit_pct}{15.5}
\setres{lotL}{EU_post_k15}{misfit_pct_mean}{15.278}
\setres{lotL}{EU_post_k15}{n_seen}{15}
\setres{lotL}{EU_post_k15}{surprise}{0.02}
\setres{lotL}{EU_post_k15}{surprise_mean}{1.871}
\setres{lotL}{EU_post_k15}{width90_adj_pct}{25.4}
\setres{lotL}{EU_post_k15}{width90_adj_pct_mean}{25.134}
\setres{lotL}{EU_post_k15}{width90_pct}{0.2}
\setres{lotL}{EU_post_k15}{width90_pct_mean}{0.201}
\setres{lotL}{EU_post_k16}{cov50}{0.00}
\setres{lotL}{EU_post_k16}{cov50_mean}{0.006}
\setres{lotL}{EU_post_k16}{cov90}{0.00}
\setres{lotL}{EU_post_k16}{cov90_adj}{1.00}
\setres{lotL}{EU_post_k16}{cov90_adj_mean}{0.956}
\setres{lotL}{EU_post_k16}{cov90_mean}{0.006}
\setres{lotL}{EU_post_k16}{kappa}{0.466}
\setres{lotL}{EU_post_k16}{kappa_crps}{0.321}
\setres{lotL}{EU_post_k16}{kappa_crps_mean}{0.243}
\setres{lotL}{EU_post_k16}{kappa_mean}{0.414}
\setres{lotL}{EU_post_k16}{logscore_next}{-3.00}
\setres{lotL}{EU_post_k16}{logscore_next_mean}{-3.178}
\setres{lotL}{EU_post_k16}{misfit_pct}{15.4}
\setres{lotL}{EU_post_k16}{misfit_pct_mean}{15.367}
\setres{lotL}{EU_post_k16}{n_seen}{16}
\setres{lotL}{EU_post_k16}{surprise}{0.01}
\setres{lotL}{EU_post_k16}{surprise_mean}{0.138}
\setres{lotL}{EU_post_k16}{width90_adj_pct}{25.4}
\setres{lotL}{EU_post_k16}{width90_adj_pct_mean}{25.279}
\setres{lotL}{EU_post_k16}{width90_pct}{0.1}
\setres{lotL}{EU_post_k16}{width90_pct_mean}{0.194}
\setres{lotL}{EU_post_k17}{cov50}{0.00}
\setres{lotL}{EU_post_k17}{cov50_mean}{0.006}
\setres{lotL}{EU_post_k17}{cov90}{0.00}
\setres{lotL}{EU_post_k17}{cov90_adj}{1.00}
\setres{lotL}{EU_post_k17}{cov90_adj_mean}{0.963}
\setres{lotL}{EU_post_k17}{cov90_mean}{0.013}
\setres{lotL}{EU_post_k17}{kappa}{0.408}
\setres{lotL}{EU_post_k17}{kappa_crps}{0.272}
\setres{lotL}{EU_post_k17}{kappa_crps_mean}{0.237}
\setres{lotL}{EU_post_k17}{kappa_mean}{0.396}
\setres{lotL}{EU_post_k17}{logscore_next}{-3.10}
\setres{lotL}{EU_post_k17}{logscore_next_mean}{-3.253}
\setres{lotL}{EU_post_k17}{misfit_pct}{15.4}
\setres{lotL}{EU_post_k17}{misfit_pct_mean}{15.288}
\setres{lotL}{EU_post_k17}{n_seen}{17}
\setres{lotL}{EU_post_k17}{surprise}{0.06}
\setres{lotL}{EU_post_k17}{surprise_mean}{0.675}
\setres{lotL}{EU_post_k17}{width90_adj_pct}{25.3}
\setres{lotL}{EU_post_k17}{width90_adj_pct_mean}{25.150}
\setres{lotL}{EU_post_k17}{width90_pct}{0.2}
\setres{lotL}{EU_post_k17}{width90_pct_mean}{0.201}
\setres{lotL}{EU_post_k18}{cov50}{0.00}
\setres{lotL}{EU_post_k18}{cov50_mean}{0.000}
\setres{lotL}{EU_post_k18}{cov90}{0.00}
\setres{lotL}{EU_post_k18}{cov90_adj}{1.00}
\setres{lotL}{EU_post_k18}{cov90_adj_mean}{0.971}
\setres{lotL}{EU_post_k18}{cov90_mean}{0.000}
\setres{lotL}{EU_post_k18}{kappa}{0.398}
\setres{lotL}{EU_post_k18}{kappa_crps}{0.309}
\setres{lotL}{EU_post_k18}{kappa_crps_mean}{0.210}
\setres{lotL}{EU_post_k18}{kappa_mean}{0.358}
\setres{lotL}{EU_post_k18}{logscore_next}{-3.06}
\setres{lotL}{EU_post_k18}{logscore_next_mean}{-3.125}
\setres{lotL}{EU_post_k18}{misfit_pct}{15.4}
\setres{lotL}{EU_post_k18}{misfit_pct_mean}{15.305}
\setres{lotL}{EU_post_k18}{n_seen}{18}
\setres{lotL}{EU_post_k18}{surprise}{0.01}
\setres{lotL}{EU_post_k18}{surprise_mean}{1.059}
\setres{lotL}{EU_post_k18}{width90_adj_pct}{25.3}
\setres{lotL}{EU_post_k18}{width90_adj_pct_mean}{25.177}
\setres{lotL}{EU_post_k18}{width90_pct}{0.1}
\setres{lotL}{EU_post_k18}{width90_pct_mean}{0.175}
\setres{lotL}{EU_post_k19}{cov50}{0.00}
\setres{lotL}{EU_post_k19}{cov50_mean}{0.000}
\setres{lotL}{EU_post_k19}{cov90}{0.00}
\setres{lotL}{EU_post_k19}{cov90_adj}{1.00}
\setres{lotL}{EU_post_k19}{cov90_adj_mean}{0.983}
\setres{lotL}{EU_post_k19}{cov90_mean}{0.000}
\setres{lotL}{EU_post_k19}{kappa}{0.420}
\setres{lotL}{EU_post_k19}{kappa_crps}{0.314}
\setres{lotL}{EU_post_k19}{kappa_crps_mean}{0.218}
\setres{lotL}{EU_post_k19}{kappa_mean}{0.368}
\setres{lotL}{EU_post_k19}{logscore_next}{-3.12}
\setres{lotL}{EU_post_k19}{logscore_next_mean}{-3.257}
\setres{lotL}{EU_post_k19}{misfit_pct}{15.4}
\setres{lotL}{EU_post_k19}{misfit_pct_mean}{15.403}
\setres{lotL}{EU_post_k19}{n_seen}{19}
\setres{lotL}{EU_post_k19}{surprise}{0.01}
\setres{lotL}{EU_post_k19}{surprise_mean}{1.743}
\setres{lotL}{EU_post_k19}{width90_adj_pct}{25.3}
\setres{lotL}{EU_post_k19}{width90_adj_pct_mean}{25.339}
\setres{lotL}{EU_post_k19}{width90_pct}{0.1}
\setres{lotL}{EU_post_k19}{width90_pct_mean}{0.177}
\setres{lotL}{EU_post_k2}{cov50}{0.00}
\setres{lotL}{EU_post_k2}{cov50_mean}{0.013}
\setres{lotL}{EU_post_k2}{cov90}{0.00}
\setres{lotL}{EU_post_k2}{cov90_adj}{0.96}
\setres{lotL}{EU_post_k2}{cov90_adj_mean}{0.754}
\setres{lotL}{EU_post_k2}{cov90_mean}{0.033}
\setres{lotL}{EU_post_k2}{kappa}{0.545}
\setres{lotL}{EU_post_k2}{kappa_crps}{0.301}
\setres{lotL}{EU_post_k2}{kappa_crps_mean}{0.210}
\setres{lotL}{EU_post_k2}{kappa_mean}{0.371}
\setres{lotL}{EU_post_k2}{logscore_next}{-3.01}
\setres{lotL}{EU_post_k2}{logscore_next_mean}{-3.059}
\setres{lotL}{EU_post_k2}{misfit_pct}{13.9}
\setres{lotL}{EU_post_k2}{misfit_pct_mean}{13.043}
\setres{lotL}{EU_post_k2}{n_seen}{2}
\setres{lotL}{EU_post_k2}{surprise}{0.01}
\setres{lotL}{EU_post_k2}{surprise_mean}{3.577}
\setres{lotL}{EU_post_k2}{width90_adj_pct}{22.8}
\setres{lotL}{EU_post_k2}{width90_adj_pct_mean}{21.472}
\setres{lotL}{EU_post_k2}{width90_pct}{0.4}
\setres{lotL}{EU_post_k2}{width90_pct_mean}{0.621}
\setres{lotL}{EU_post_k20}{cov50}{0.00}
\setres{lotL}{EU_post_k20}{cov50_mean}{0.000}
\setres{lotL}{EU_post_k20}{cov90}{0.00}
\setres{lotL}{EU_post_k20}{cov90_adj}{1.00}
\setres{lotL}{EU_post_k20}{cov90_adj_mean}{0.980}
\setres{lotL}{EU_post_k20}{cov90_mean}{0.000}
\setres{lotL}{EU_post_k20}{kappa}{0.351}
\setres{lotL}{EU_post_k20}{kappa_crps}{0.302}
\setres{lotL}{EU_post_k20}{kappa_crps_mean}{0.214}
\setres{lotL}{EU_post_k20}{kappa_mean}{0.259}
\setres{lotL}{EU_post_k20}{logscore_next}{-3.07}
\setres{lotL}{EU_post_k20}{logscore_next_mean}{-3.199}
\setres{lotL}{EU_post_k20}{misfit_pct}{15.4}
\setres{lotL}{EU_post_k20}{misfit_pct_mean}{15.369}
\setres{lotL}{EU_post_k20}{n_seen}{20}
\setres{lotL}{EU_post_k20}{surprise}{0.03}
\setres{lotL}{EU_post_k20}{surprise_mean}{0.406}
\setres{lotL}{EU_post_k20}{width90_adj_pct}{25.3}
\setres{lotL}{EU_post_k20}{width90_adj_pct_mean}{25.282}
\setres{lotL}{EU_post_k20}{width90_pct}{0.1}
\setres{lotL}{EU_post_k20}{width90_pct_mean}{0.144}
\setres{lotL}{EU_post_k21}{cov50}{0.00}
\setres{lotL}{EU_post_k21}{cov50_mean}{0.000}
\setres{lotL}{EU_post_k21}{cov90}{0.00}
\setres{lotL}{EU_post_k21}{cov90_adj}{1.00}
\setres{lotL}{EU_post_k21}{cov90_adj_mean}{0.975}
\setres{lotL}{EU_post_k21}{cov90_mean}{0.000}
\setres{lotL}{EU_post_k21}{kappa}{0.210}
\setres{lotL}{EU_post_k21}{kappa_crps}{0.255}
\setres{lotL}{EU_post_k21}{kappa_crps_mean}{0.172}
\setres{lotL}{EU_post_k21}{kappa_mean}{0.248}
\setres{lotL}{EU_post_k21}{logscore_next}{-3.06}
\setres{lotL}{EU_post_k21}{logscore_next_mean}{-3.092}
\setres{lotL}{EU_post_k21}{misfit_pct}{15.4}
\setres{lotL}{EU_post_k21}{misfit_pct_mean}{15.353}
\setres{lotL}{EU_post_k21}{n_seen}{21}
\setres{lotL}{EU_post_k21}{surprise}{0.01}
\setres{lotL}{EU_post_k21}{surprise_mean}{0.111}
\setres{lotL}{EU_post_k21}{width90_adj_pct}{25.3}
\setres{lotL}{EU_post_k21}{width90_adj_pct_mean}{25.257}
\setres{lotL}{EU_post_k21}{width90_pct}{0.1}
\setres{lotL}{EU_post_k21}{width90_pct_mean}{0.157}
\setres{lotL}{EU_post_k22}{cov50}{0.00}
\setres{lotL}{EU_post_k22}{cov50_mean}{0.000}
\setres{lotL}{EU_post_k22}{cov90}{0.00}
\setres{lotL}{EU_post_k22}{cov90_adj}{1.00}
\setres{lotL}{EU_post_k22}{cov90_adj_mean}{0.983}
\setres{lotL}{EU_post_k22}{cov90_mean}{0.000}
\setres{lotL}{EU_post_k22}{kappa}{0.297}
\setres{lotL}{EU_post_k22}{kappa_crps}{0.340}
\setres{lotL}{EU_post_k22}{kappa_crps_mean}{0.165}
\setres{lotL}{EU_post_k22}{kappa_mean}{0.274}
\setres{lotL}{EU_post_k22}{logscore_next}{-3.04}
\setres{lotL}{EU_post_k22}{logscore_next_mean}{-3.079}
\setres{lotL}{EU_post_k22}{misfit_pct}{15.5}
\setres{lotL}{EU_post_k22}{misfit_pct_mean}{15.405}
\setres{lotL}{EU_post_k22}{n_seen}{22}
\setres{lotL}{EU_post_k22}{surprise}{0.01}
\setres{lotL}{EU_post_k22}{surprise_mean}{0.086}
\setres{lotL}{EU_post_k22}{width90_adj_pct}{25.5}
\setres{lotL}{EU_post_k22}{width90_adj_pct_mean}{25.342}
\setres{lotL}{EU_post_k22}{width90_pct}{0.1}
\setres{lotL}{EU_post_k22}{width90_pct_mean}{0.152}
\setres{lotL}{EU_post_k23}{cov50}{0.00}
\setres{lotL}{EU_post_k23}{cov50_mean}{0.000}
\setres{lotL}{EU_post_k23}{cov90}{0.00}
\setres{lotL}{EU_post_k23}{cov90_adj}{1.00}
\setres{lotL}{EU_post_k23}{cov90_adj_mean}{1.000}
\setres{lotL}{EU_post_k23}{cov90_mean}{0.000}
\setres{lotL}{EU_post_k23}{kappa}{0.367}
\setres{lotL}{EU_post_k23}{kappa_crps}{0.330}
\setres{lotL}{EU_post_k23}{kappa_crps_mean}{0.220}
\setres{lotL}{EU_post_k23}{kappa_mean}{0.272}
\setres{lotL}{EU_post_k23}{logscore_next}{-3.11}
\setres{lotL}{EU_post_k23}{logscore_next_mean}{-3.198}
\setres{lotL}{EU_post_k23}{misfit_pct}{15.4}
\setres{lotL}{EU_post_k23}{misfit_pct_mean}{15.371}
\setres{lotL}{EU_post_k23}{n_seen}{23}
\setres{lotL}{EU_post_k23}{surprise}{0.01}
\setres{lotL}{EU_post_k23}{surprise_mean}{4.847}
\setres{lotL}{EU_post_k23}{width90_adj_pct}{25.3}
\setres{lotL}{EU_post_k23}{width90_adj_pct_mean}{25.288}
\setres{lotL}{EU_post_k23}{width90_pct}{0.1}
\setres{lotL}{EU_post_k23}{width90_pct_mean}{0.228}
\setres{lotL}{EU_post_k24}{cov50}{0.00}
\setres{lotL}{EU_post_k24}{cov50_mean}{0.000}
\setres{lotL}{EU_post_k24}{cov90}{0.00}
\setres{lotL}{EU_post_k24}{cov90_adj}{1.00}
\setres{lotL}{EU_post_k24}{cov90_adj_mean}{1.000}
\setres{lotL}{EU_post_k24}{cov90_mean}{0.000}
\setres{lotL}{EU_post_k24}{kappa}{0.276}
\setres{lotL}{EU_post_k24}{kappa_crps}{0.449}
\setres{lotL}{EU_post_k24}{kappa_crps_mean}{0.297}
\setres{lotL}{EU_post_k24}{kappa_mean}{0.125}
\setres{lotL}{EU_post_k24}{logscore_next}{-3.23}
\setres{lotL}{EU_post_k24}{logscore_next_mean}{-3.538}
\setres{lotL}{EU_post_k24}{misfit_pct}{15.4}
\setres{lotL}{EU_post_k24}{misfit_pct_mean}{15.392}
\setres{lotL}{EU_post_k24}{n_seen}{24}
\setres{lotL}{EU_post_k24}{surprise}{0.05}
\setres{lotL}{EU_post_k24}{surprise_mean}{2.462}
\setres{lotL}{EU_post_k24}{width90_adj_pct}{25.3}
\setres{lotL}{EU_post_k24}{width90_adj_pct_mean}{25.322}
\setres{lotL}{EU_post_k24}{width90_pct}{0.1}
\setres{lotL}{EU_post_k24}{width90_pct_mean}{0.194}
\setres{lotL}{EU_post_k25}{cov50}{---}
\setres{lotL}{EU_post_k25}{cov50_mean}{---}
\setres{lotL}{EU_post_k25}{cov90}{---}
\setres{lotL}{EU_post_k25}{cov90_adj}{---}
\setres{lotL}{EU_post_k25}{cov90_adj_mean}{---}
\setres{lotL}{EU_post_k25}{cov90_mean}{---}
\setres{lotL}{EU_post_k25}{kappa}{---}
\setres{lotL}{EU_post_k25}{kappa_crps}{---}
\setres{lotL}{EU_post_k25}{kappa_crps_mean}{---}
\setres{lotL}{EU_post_k25}{kappa_mean}{---}
\setres{lotL}{EU_post_k25}{logscore_next}{---}
\setres{lotL}{EU_post_k25}{logscore_next_mean}{---}
\setres{lotL}{EU_post_k25}{misfit_pct}{---}
\setres{lotL}{EU_post_k25}{misfit_pct_mean}{---}
\setres{lotL}{EU_post_k25}{n_seen}{25}
\setres{lotL}{EU_post_k25}{surprise}{---}
\setres{lotL}{EU_post_k25}{surprise_mean}{---}
\setres{lotL}{EU_post_k25}{width90_adj_pct}{---}
\setres{lotL}{EU_post_k25}{width90_adj_pct_mean}{---}
\setres{lotL}{EU_post_k25}{width90_pct}{---}
\setres{lotL}{EU_post_k25}{width90_pct_mean}{---}
\setres{lotL}{EU_post_k3}{cov50}{0.00}
\setres{lotL}{EU_post_k3}{cov50_mean}{0.005}
\setres{lotL}{EU_post_k3}{cov90}{0.00}
\setres{lotL}{EU_post_k3}{cov90_adj}{0.89}
\setres{lotL}{EU_post_k3}{cov90_adj_mean}{0.825}
\setres{lotL}{EU_post_k3}{cov90_mean}{0.011}
\setres{lotL}{EU_post_k3}{kappa}{0.496}
\setres{lotL}{EU_post_k3}{kappa_crps}{0.284}
\setres{lotL}{EU_post_k3}{kappa_crps_mean}{0.118}
\setres{lotL}{EU_post_k3}{kappa_mean}{0.224}
\setres{lotL}{EU_post_k3}{logscore_next}{-3.14}
\setres{lotL}{EU_post_k3}{logscore_next_mean}{-3.279}
\setres{lotL}{EU_post_k3}{misfit_pct}{13.3}
\setres{lotL}{EU_post_k3}{misfit_pct_mean}{13.988}
\setres{lotL}{EU_post_k3}{n_seen}{3}
\setres{lotL}{EU_post_k3}{surprise}{0.07}
\setres{lotL}{EU_post_k3}{surprise_mean}{10.321}
\setres{lotL}{EU_post_k3}{width90_adj_pct}{21.9}
\setres{lotL}{EU_post_k3}{width90_adj_pct_mean}{23.019}
\setres{lotL}{EU_post_k3}{width90_pct}{0.2}
\setres{lotL}{EU_post_k3}{width90_pct_mean}{0.411}
\setres{lotL}{EU_post_k4}{cov50}{0.00}
\setres{lotL}{EU_post_k4}{cov50_mean}{0.005}
\setres{lotL}{EU_post_k4}{cov90}{0.00}
\setres{lotL}{EU_post_k4}{cov90_adj}{0.95}
\setres{lotL}{EU_post_k4}{cov90_adj_mean}{0.893}
\setres{lotL}{EU_post_k4}{cov90_mean}{0.014}
\setres{lotL}{EU_post_k4}{kappa}{0.509}
\setres{lotL}{EU_post_k4}{kappa_crps}{0.273}
\setres{lotL}{EU_post_k4}{kappa_crps_mean}{0.163}
\setres{lotL}{EU_post_k4}{kappa_mean}{0.301}
\setres{lotL}{EU_post_k4}{logscore_next}{-3.15}
\setres{lotL}{EU_post_k4}{logscore_next_mean}{-3.225}
\setres{lotL}{EU_post_k4}{misfit_pct}{14.8}
\setres{lotL}{EU_post_k4}{misfit_pct_mean}{14.747}
\setres{lotL}{EU_post_k4}{n_seen}{4}
\setres{lotL}{EU_post_k4}{surprise}{0.03}
\setres{lotL}{EU_post_k4}{surprise_mean}{0.182}
\setres{lotL}{EU_post_k4}{width90_adj_pct}{24.3}
\setres{lotL}{EU_post_k4}{width90_adj_pct_mean}{24.262}
\setres{lotL}{EU_post_k4}{width90_pct}{0.2}
\setres{lotL}{EU_post_k4}{width90_pct_mean}{0.294}
\setres{lotL}{EU_post_k5}{cov50}{0.00}
\setres{lotL}{EU_post_k5}{cov50_mean}{0.003}
\setres{lotL}{EU_post_k5}{cov90}{0.00}
\setres{lotL}{EU_post_k5}{cov90_adj}{0.97}
\setres{lotL}{EU_post_k5}{cov90_adj_mean}{0.920}
\setres{lotL}{EU_post_k5}{cov90_mean}{0.008}
\setres{lotL}{EU_post_k5}{kappa}{0.485}
\setres{lotL}{EU_post_k5}{kappa_crps}{0.250}
\setres{lotL}{EU_post_k5}{kappa_crps_mean}{0.145}
\setres{lotL}{EU_post_k5}{kappa_mean}{0.289}
\setres{lotL}{EU_post_k5}{logscore_next}{-3.12}
\setres{lotL}{EU_post_k5}{logscore_next_mean}{-3.145}
\setres{lotL}{EU_post_k5}{misfit_pct}{15.0}
\setres{lotL}{EU_post_k5}{misfit_pct_mean}{15.381}
\setres{lotL}{EU_post_k5}{n_seen}{5}
\setres{lotL}{EU_post_k5}{surprise}{0.07}
\setres{lotL}{EU_post_k5}{surprise_mean}{11.826}
\setres{lotL}{EU_post_k5}{width90_adj_pct}{24.6}
\setres{lotL}{EU_post_k5}{width90_adj_pct_mean}{25.304}
\setres{lotL}{EU_post_k5}{width90_pct}{0.2}
\setres{lotL}{EU_post_k5}{width90_pct_mean}{0.285}
\setres{lotL}{EU_post_k6}{cov50}{0.00}
\setres{lotL}{EU_post_k6}{cov50_mean}{0.011}
\setres{lotL}{EU_post_k6}{cov90}{0.00}
\setres{lotL}{EU_post_k6}{cov90_adj}{1.00}
\setres{lotL}{EU_post_k6}{cov90_adj_mean}{0.955}
\setres{lotL}{EU_post_k6}{cov90_mean}{0.016}
\setres{lotL}{EU_post_k6}{kappa}{0.535}
\setres{lotL}{EU_post_k6}{kappa_crps}{0.289}
\setres{lotL}{EU_post_k6}{kappa_crps_mean}{0.249}
\setres{lotL}{EU_post_k6}{kappa_mean}{0.450}
\setres{lotL}{EU_post_k6}{logscore_next}{-3.05}
\setres{lotL}{EU_post_k6}{logscore_next_mean}{-3.104}
\setres{lotL}{EU_post_k6}{misfit_pct}{15.3}
\setres{lotL}{EU_post_k6}{misfit_pct_mean}{15.394}
\setres{lotL}{EU_post_k6}{n_seen}{6}
\setres{lotL}{EU_post_k6}{surprise}{0.03}
\setres{lotL}{EU_post_k6}{surprise_mean}{5.181}
\setres{lotL}{EU_post_k6}{width90_adj_pct}{25.2}
\setres{lotL}{EU_post_k6}{width90_adj_pct_mean}{25.326}
\setres{lotL}{EU_post_k6}{width90_pct}{0.2}
\setres{lotL}{EU_post_k6}{width90_pct_mean}{0.291}
\setres{lotL}{EU_post_k7}{cov50}{0.00}
\setres{lotL}{EU_post_k7}{cov50_mean}{0.003}
\setres{lotL}{EU_post_k7}{cov90}{0.00}
\setres{lotL}{EU_post_k7}{cov90_adj}{0.94}
\setres{lotL}{EU_post_k7}{cov90_adj_mean}{0.944}
\setres{lotL}{EU_post_k7}{cov90_mean}{0.006}
\setres{lotL}{EU_post_k7}{kappa}{0.566}
\setres{lotL}{EU_post_k7}{kappa_crps}{0.301}
\setres{lotL}{EU_post_k7}{kappa_crps_mean}{0.251}
\setres{lotL}{EU_post_k7}{kappa_mean}{0.454}
\setres{lotL}{EU_post_k7}{logscore_next}{-3.01}
\setres{lotL}{EU_post_k7}{logscore_next_mean}{-3.201}
\setres{lotL}{EU_post_k7}{misfit_pct}{15.0}
\setres{lotL}{EU_post_k7}{misfit_pct_mean}{15.239}
\setres{lotL}{EU_post_k7}{n_seen}{7}
\setres{lotL}{EU_post_k7}{surprise}{0.02}
\setres{lotL}{EU_post_k7}{surprise_mean}{0.223}
\setres{lotL}{EU_post_k7}{width90_adj_pct}{24.7}
\setres{lotL}{EU_post_k7}{width90_adj_pct_mean}{25.069}
\setres{lotL}{EU_post_k7}{width90_pct}{0.2}
\setres{lotL}{EU_post_k7}{width90_pct_mean}{0.249}
\setres{lotL}{EU_post_k8}{cov50}{0.00}
\setres{lotL}{EU_post_k8}{cov50_mean}{0.003}
\setres{lotL}{EU_post_k8}{cov90}{0.00}
\setres{lotL}{EU_post_k8}{cov90_adj}{1.00}
\setres{lotL}{EU_post_k8}{cov90_adj_mean}{0.971}
\setres{lotL}{EU_post_k8}{cov90_mean}{0.006}
\setres{lotL}{EU_post_k8}{kappa}{0.560}
\setres{lotL}{EU_post_k8}{kappa_crps}{0.294}
\setres{lotL}{EU_post_k8}{kappa_crps_mean}{0.262}
\setres{lotL}{EU_post_k8}{kappa_mean}{0.458}
\setres{lotL}{EU_post_k8}{logscore_next}{-3.12}
\setres{lotL}{EU_post_k8}{logscore_next_mean}{-3.251}
\setres{lotL}{EU_post_k8}{misfit_pct}{15.7}
\setres{lotL}{EU_post_k8}{misfit_pct_mean}{15.814}
\setres{lotL}{EU_post_k8}{n_seen}{8}
\setres{lotL}{EU_post_k8}{surprise}{0.04}
\setres{lotL}{EU_post_k8}{surprise_mean}{3.835}
\setres{lotL}{EU_post_k8}{width90_adj_pct}{25.8}
\setres{lotL}{EU_post_k8}{width90_adj_pct_mean}{26.016}
\setres{lotL}{EU_post_k8}{width90_pct}{0.2}
\setres{lotL}{EU_post_k8}{width90_pct_mean}{0.257}
\setres{lotL}{EU_post_k9}{cov50}{0.00}
\setres{lotL}{EU_post_k9}{cov50_mean}{0.003}
\setres{lotL}{EU_post_k9}{cov90}{0.00}
\setres{lotL}{EU_post_k9}{cov90_adj}{1.00}
\setres{lotL}{EU_post_k9}{cov90_adj_mean}{0.966}
\setres{lotL}{EU_post_k9}{cov90_mean}{0.006}
\setres{lotL}{EU_post_k9}{kappa}{0.554}
\setres{lotL}{EU_post_k9}{kappa_crps}{0.321}
\setres{lotL}{EU_post_k9}{kappa_crps_mean}{0.248}
\setres{lotL}{EU_post_k9}{kappa_mean}{0.435}
\setres{lotL}{EU_post_k9}{logscore_next}{-3.05}
\setres{lotL}{EU_post_k9}{logscore_next_mean}{-3.189}
\setres{lotL}{EU_post_k9}{misfit_pct}{15.7}
\setres{lotL}{EU_post_k9}{misfit_pct_mean}{15.721}
\setres{lotL}{EU_post_k9}{n_seen}{9}
\setres{lotL}{EU_post_k9}{surprise}{0.03}
\setres{lotL}{EU_post_k9}{surprise_mean}{0.449}
\setres{lotL}{EU_post_k9}{width90_adj_pct}{25.9}
\setres{lotL}{EU_post_k9}{width90_adj_pct_mean}{25.863}
\setres{lotL}{EU_post_k9}{width90_pct}{0.2}
\setres{lotL}{EU_post_k9}{width90_pct_mean}{0.221}
\setres{lotL}{reference}{domain}{Bruhin2010\_gains}
\setres{lotL}{reference}{n_draws}{400}
\setres{lotL}{reference}{n_lotteries}{25}
\setres{lotL}{reference}{n_orders}{20}
\setres{lotL}{reference}{start}{lit}
\setres{lotUL}{CPT_post_k0}{cov50}{0.88}
\setres{lotUL}{CPT_post_k0}{cov50_mean}{0.880}
\setres{lotUL}{CPT_post_k0}{cov90}{1.00}
\setres{lotUL}{CPT_post_k0}{cov90_adj}{---}
\setres{lotUL}{CPT_post_k0}{cov90_adj_mean}{---}
\setres{lotUL}{CPT_post_k0}{cov90_mean}{1.000}
\setres{lotUL}{CPT_post_k0}{kappa}{-0.449}
\setres{lotUL}{CPT_post_k0}{kappa_crps}{0.488}
\setres{lotUL}{CPT_post_k0}{kappa_crps_mean}{0.489}
\setres{lotUL}{CPT_post_k0}{kappa_mean}{-0.449}
\setres{lotUL}{CPT_post_k0}{logscore_next}{-2.99}
\setres{lotUL}{CPT_post_k0}{logscore_next_mean}{-2.991}
\setres{lotUL}{CPT_post_k0}{misfit_pct}{---}
\setres{lotUL}{CPT_post_k0}{misfit_pct_mean}{---}
\setres{lotUL}{CPT_post_k0}{n_seen}{0}
\setres{lotUL}{CPT_post_k0}{surprise}{2.05}
\setres{lotUL}{CPT_post_k0}{surprise_mean}{2.224}
\setres{lotUL}{CPT_post_k0}{width90_adj_pct}{---}
\setres{lotUL}{CPT_post_k0}{width90_adj_pct_mean}{---}
\setres{lotUL}{CPT_post_k0}{width90_pct}{28.4}
\setres{lotUL}{CPT_post_k0}{width90_pct_mean}{28.385}
\setres{lotUL}{CPT_post_k1}{cov50}{0.60}
\setres{lotUL}{CPT_post_k1}{cov50_mean}{0.590}
\setres{lotUL}{CPT_post_k1}{cov90}{0.90}
\setres{lotUL}{CPT_post_k1}{cov90_adj}{---}
\setres{lotUL}{CPT_post_k1}{cov90_adj_mean}{---}
\setres{lotUL}{CPT_post_k1}{cov90_mean}{0.892}
\setres{lotUL}{CPT_post_k1}{kappa}{0.035}
\setres{lotUL}{CPT_post_k1}{kappa_crps}{0.457}
\setres{lotUL}{CPT_post_k1}{kappa_crps_mean}{0.458}
\setres{lotUL}{CPT_post_k1}{kappa_mean}{-0.028}
\setres{lotUL}{CPT_post_k1}{logscore_next}{-2.99}
\setres{lotUL}{CPT_post_k1}{logscore_next_mean}{-2.988}
\setres{lotUL}{CPT_post_k1}{misfit_pct}{---}
\setres{lotUL}{CPT_post_k1}{misfit_pct_mean}{---}
\setres{lotUL}{CPT_post_k1}{n_seen}{1}
\setres{lotUL}{CPT_post_k1}{surprise}{1.50}
\setres{lotUL}{CPT_post_k1}{surprise_mean}{1.704}
\setres{lotUL}{CPT_post_k1}{width90_adj_pct}{---}
\setres{lotUL}{CPT_post_k1}{width90_adj_pct_mean}{---}
\setres{lotUL}{CPT_post_k1}{width90_pct}{19.9}
\setres{lotUL}{CPT_post_k1}{width90_pct_mean}{22.045}
\setres{lotUL}{CPT_post_k10}{cov50}{0.07}
\setres{lotUL}{CPT_post_k10}{cov50_mean}{0.080}
\setres{lotUL}{CPT_post_k10}{cov90}{0.27}
\setres{lotUL}{CPT_post_k10}{cov90_adj}{0.80}
\setres{lotUL}{CPT_post_k10}{cov90_adj_mean}{0.793}
\setres{lotUL}{CPT_post_k10}{cov90_mean}{0.233}
\setres{lotUL}{CPT_post_k10}{kappa}{0.937}
\setres{lotUL}{CPT_post_k10}{kappa_crps}{0.671}
\setres{lotUL}{CPT_post_k10}{kappa_crps_mean}{0.635}
\setres{lotUL}{CPT_post_k10}{kappa_mean}{0.938}
\setres{lotUL}{CPT_post_k10}{logscore_next}{-2.98}
\setres{lotUL}{CPT_post_k10}{logscore_next_mean}{-2.980}
\setres{lotUL}{CPT_post_k10}{misfit_pct}{4.0}
\setres{lotUL}{CPT_post_k10}{misfit_pct_mean}{4.092}
\setres{lotUL}{CPT_post_k10}{n_seen}{10}
\setres{lotUL}{CPT_post_k10}{surprise}{0.03}
\setres{lotUL}{CPT_post_k10}{surprise_mean}{0.063}
\setres{lotUL}{CPT_post_k10}{width90_adj_pct}{6.8}
\setres{lotUL}{CPT_post_k10}{width90_adj_pct_mean}{6.966}
\setres{lotUL}{CPT_post_k10}{width90_pct}{1.5}
\setres{lotUL}{CPT_post_k10}{width90_pct_mean}{1.400}
\setres{lotUL}{CPT_post_k11}{cov50}{0.07}
\setres{lotUL}{CPT_post_k11}{cov50_mean}{0.100}
\setres{lotUL}{CPT_post_k11}{cov90}{0.21}
\setres{lotUL}{CPT_post_k11}{cov90_adj}{0.82}
\setres{lotUL}{CPT_post_k11}{cov90_adj_mean}{0.786}
\setres{lotUL}{CPT_post_k11}{cov90_mean}{0.207}
\setres{lotUL}{CPT_post_k11}{kappa}{0.935}
\setres{lotUL}{CPT_post_k11}{kappa_crps}{0.669}
\setres{lotUL}{CPT_post_k11}{kappa_crps_mean}{0.642}
\setres{lotUL}{CPT_post_k11}{kappa_mean}{0.939}
\setres{lotUL}{CPT_post_k11}{logscore_next}{-2.98}
\setres{lotUL}{CPT_post_k11}{logscore_next_mean}{-2.994}
\setres{lotUL}{CPT_post_k11}{misfit_pct}{4.0}
\setres{lotUL}{CPT_post_k11}{misfit_pct_mean}{4.127}
\setres{lotUL}{CPT_post_k11}{n_seen}{11}
\setres{lotUL}{CPT_post_k11}{surprise}{0.03}
\setres{lotUL}{CPT_post_k11}{surprise_mean}{1.446}
\setres{lotUL}{CPT_post_k11}{width90_adj_pct}{6.6}
\setres{lotUL}{CPT_post_k11}{width90_adj_pct_mean}{7.005}
\setres{lotUL}{CPT_post_k11}{width90_pct}{1.3}
\setres{lotUL}{CPT_post_k11}{width90_pct_mean}{1.155}
\setres{lotUL}{CPT_post_k12}{cov50}{0.00}
\setres{lotUL}{CPT_post_k12}{cov50_mean}{0.088}
\setres{lotUL}{CPT_post_k12}{cov90}{0.19}
\setres{lotUL}{CPT_post_k12}{cov90_adj}{0.85}
\setres{lotUL}{CPT_post_k12}{cov90_adj_mean}{0.827}
\setres{lotUL}{CPT_post_k12}{cov90_mean}{0.181}
\setres{lotUL}{CPT_post_k12}{kappa}{0.938}
\setres{lotUL}{CPT_post_k12}{kappa_crps}{0.677}
\setres{lotUL}{CPT_post_k12}{kappa_crps_mean}{0.641}
\setres{lotUL}{CPT_post_k12}{kappa_mean}{0.937}
\setres{lotUL}{CPT_post_k12}{logscore_next}{-2.98}
\setres{lotUL}{CPT_post_k12}{logscore_next_mean}{-2.984}
\setres{lotUL}{CPT_post_k12}{misfit_pct}{4.4}
\setres{lotUL}{CPT_post_k12}{misfit_pct_mean}{4.432}
\setres{lotUL}{CPT_post_k12}{n_seen}{12}
\setres{lotUL}{CPT_post_k12}{surprise}{0.04}
\setres{lotUL}{CPT_post_k12}{surprise_mean}{0.391}
\setres{lotUL}{CPT_post_k12}{width90_adj_pct}{7.3}
\setres{lotUL}{CPT_post_k12}{width90_adj_pct_mean}{7.450}
\setres{lotUL}{CPT_post_k12}{width90_pct}{1.3}
\setres{lotUL}{CPT_post_k12}{width90_pct_mean}{0.996}
\setres{lotUL}{CPT_post_k13}{cov50}{0.12}
\setres{lotUL}{CPT_post_k13}{cov50_mean}{0.125}
\setres{lotUL}{CPT_post_k13}{cov90}{0.21}
\setres{lotUL}{CPT_post_k13}{cov90_adj}{0.83}
\setres{lotUL}{CPT_post_k13}{cov90_adj_mean}{0.821}
\setres{lotUL}{CPT_post_k13}{cov90_mean}{0.204}
\setres{lotUL}{CPT_post_k13}{kappa}{0.935}
\setres{lotUL}{CPT_post_k13}{kappa_crps}{0.671}
\setres{lotUL}{CPT_post_k13}{kappa_crps_mean}{0.630}
\setres{lotUL}{CPT_post_k13}{kappa_mean}{0.933}
\setres{lotUL}{CPT_post_k13}{logscore_next}{-2.98}
\setres{lotUL}{CPT_post_k13}{logscore_next_mean}{-2.992}
\setres{lotUL}{CPT_post_k13}{misfit_pct}{4.3}
\setres{lotUL}{CPT_post_k13}{misfit_pct_mean}{4.349}
\setres{lotUL}{CPT_post_k13}{n_seen}{13}
\setres{lotUL}{CPT_post_k13}{surprise}{0.13}
\setres{lotUL}{CPT_post_k13}{surprise_mean}{1.366}
\setres{lotUL}{CPT_post_k13}{width90_adj_pct}{7.1}
\setres{lotUL}{CPT_post_k13}{width90_adj_pct_mean}{7.342}
\setres{lotUL}{CPT_post_k13}{width90_pct}{1.3}
\setres{lotUL}{CPT_post_k13}{width90_pct_mean}{1.124}
\setres{lotUL}{CPT_post_k14}{cov50}{0.00}
\setres{lotUL}{CPT_post_k14}{cov50_mean}{0.059}
\setres{lotUL}{CPT_post_k14}{cov90}{0.14}
\setres{lotUL}{CPT_post_k14}{cov90_adj}{0.82}
\setres{lotUL}{CPT_post_k14}{cov90_adj_mean}{0.773}
\setres{lotUL}{CPT_post_k14}{cov90_mean}{0.177}
\setres{lotUL}{CPT_post_k14}{kappa}{0.937}
\setres{lotUL}{CPT_post_k14}{kappa_crps}{0.678}
\setres{lotUL}{CPT_post_k14}{kappa_crps_mean}{0.630}
\setres{lotUL}{CPT_post_k14}{kappa_mean}{0.935}
\setres{lotUL}{CPT_post_k14}{logscore_next}{-2.98}
\setres{lotUL}{CPT_post_k14}{logscore_next_mean}{-2.985}
\setres{lotUL}{CPT_post_k14}{misfit_pct}{4.3}
\setres{lotUL}{CPT_post_k14}{misfit_pct_mean}{4.403}
\setres{lotUL}{CPT_post_k14}{n_seen}{14}
\setres{lotUL}{CPT_post_k14}{surprise}{0.06}
\setres{lotUL}{CPT_post_k14}{surprise_mean}{0.326}
\setres{lotUL}{CPT_post_k14}{width90_adj_pct}{7.2}
\setres{lotUL}{CPT_post_k14}{width90_adj_pct_mean}{7.383}
\setres{lotUL}{CPT_post_k14}{width90_pct}{1.1}
\setres{lotUL}{CPT_post_k14}{width90_pct_mean}{1.102}
\setres{lotUL}{CPT_post_k15}{cov50}{0.05}
\setres{lotUL}{CPT_post_k15}{cov50_mean}{0.095}
\setres{lotUL}{CPT_post_k15}{cov90}{0.15}
\setres{lotUL}{CPT_post_k15}{cov90_adj}{0.80}
\setres{lotUL}{CPT_post_k15}{cov90_adj_mean}{0.780}
\setres{lotUL}{CPT_post_k15}{cov90_mean}{0.185}
\setres{lotUL}{CPT_post_k15}{kappa}{0.934}
\setres{lotUL}{CPT_post_k15}{kappa_crps}{0.628}
\setres{lotUL}{CPT_post_k15}{kappa_crps_mean}{0.607}
\setres{lotUL}{CPT_post_k15}{kappa_mean}{0.929}
\setres{lotUL}{CPT_post_k15}{logscore_next}{-2.98}
\setres{lotUL}{CPT_post_k15}{logscore_next_mean}{-2.985}
\setres{lotUL}{CPT_post_k15}{misfit_pct}{4.6}
\setres{lotUL}{CPT_post_k15}{misfit_pct_mean}{4.463}
\setres{lotUL}{CPT_post_k15}{n_seen}{15}
\setres{lotUL}{CPT_post_k15}{surprise}{0.04}
\setres{lotUL}{CPT_post_k15}{surprise_mean}{0.221}
\setres{lotUL}{CPT_post_k15}{width90_adj_pct}{7.6}
\setres{lotUL}{CPT_post_k15}{width90_adj_pct_mean}{7.495}
\setres{lotUL}{CPT_post_k15}{width90_pct}{1.2}
\setres{lotUL}{CPT_post_k15}{width90_pct_mean}{1.056}
\setres{lotUL}{CPT_post_k16}{cov50}{0.00}
\setres{lotUL}{CPT_post_k16}{cov50_mean}{0.078}
\setres{lotUL}{CPT_post_k16}{cov90}{0.11}
\setres{lotUL}{CPT_post_k16}{cov90_adj}{0.78}
\setres{lotUL}{CPT_post_k16}{cov90_adj_mean}{0.783}
\setres{lotUL}{CPT_post_k16}{cov90_mean}{0.144}
\setres{lotUL}{CPT_post_k16}{kappa}{0.939}
\setres{lotUL}{CPT_post_k16}{kappa_crps}{0.660}
\setres{lotUL}{CPT_post_k16}{kappa_crps_mean}{0.618}
\setres{lotUL}{CPT_post_k16}{kappa_mean}{0.931}
\setres{lotUL}{CPT_post_k16}{logscore_next}{-2.98}
\setres{lotUL}{CPT_post_k16}{logscore_next_mean}{-2.984}
\setres{lotUL}{CPT_post_k16}{misfit_pct}{4.8}
\setres{lotUL}{CPT_post_k16}{misfit_pct_mean}{4.566}
\setres{lotUL}{CPT_post_k16}{n_seen}{16}
\setres{lotUL}{CPT_post_k16}{surprise}{0.02}
\setres{lotUL}{CPT_post_k16}{surprise_mean}{0.261}
\setres{lotUL}{CPT_post_k16}{width90_adj_pct}{8.1}
\setres{lotUL}{CPT_post_k16}{width90_adj_pct_mean}{7.648}
\setres{lotUL}{CPT_post_k16}{width90_pct}{1.0}
\setres{lotUL}{CPT_post_k16}{width90_pct_mean}{0.882}
\setres{lotUL}{CPT_post_k17}{cov50}{0.00}
\setres{lotUL}{CPT_post_k17}{cov50_mean}{0.069}
\setres{lotUL}{CPT_post_k17}{cov90}{0.12}
\setres{lotUL}{CPT_post_k17}{cov90_adj}{0.75}
\setres{lotUL}{CPT_post_k17}{cov90_adj_mean}{0.806}
\setres{lotUL}{CPT_post_k17}{cov90_mean}{0.131}
\setres{lotUL}{CPT_post_k17}{kappa}{0.935}
\setres{lotUL}{CPT_post_k17}{kappa_crps}{0.671}
\setres{lotUL}{CPT_post_k17}{kappa_crps_mean}{0.622}
\setres{lotUL}{CPT_post_k17}{kappa_mean}{0.930}
\setres{lotUL}{CPT_post_k17}{logscore_next}{-2.99}
\setres{lotUL}{CPT_post_k17}{logscore_next_mean}{-2.998}
\setres{lotUL}{CPT_post_k17}{misfit_pct}{5.2}
\setres{lotUL}{CPT_post_k17}{misfit_pct_mean}{4.728}
\setres{lotUL}{CPT_post_k17}{n_seen}{17}
\setres{lotUL}{CPT_post_k17}{surprise}{0.08}
\setres{lotUL}{CPT_post_k17}{surprise_mean}{1.407}
\setres{lotUL}{CPT_post_k17}{width90_adj_pct}{8.5}
\setres{lotUL}{CPT_post_k17}{width90_adj_pct_mean}{7.894}
\setres{lotUL}{CPT_post_k17}{width90_pct}{0.8}
\setres{lotUL}{CPT_post_k17}{width90_pct_mean}{0.751}
\setres{lotUL}{CPT_post_k18}{cov50}{0.00}
\setres{lotUL}{CPT_post_k18}{cov50_mean}{0.064}
\setres{lotUL}{CPT_post_k18}{cov90}{0.14}
\setres{lotUL}{CPT_post_k18}{cov90_adj}{0.86}
\setres{lotUL}{CPT_post_k18}{cov90_adj_mean}{0.829}
\setres{lotUL}{CPT_post_k18}{cov90_mean}{0.164}
\setres{lotUL}{CPT_post_k18}{kappa}{0.946}
\setres{lotUL}{CPT_post_k18}{kappa_crps}{0.670}
\setres{lotUL}{CPT_post_k18}{kappa_crps_mean}{0.611}
\setres{lotUL}{CPT_post_k18}{kappa_mean}{0.920}
\setres{lotUL}{CPT_post_k18}{logscore_next}{-2.98}
\setres{lotUL}{CPT_post_k18}{logscore_next_mean}{-2.991}
\setres{lotUL}{CPT_post_k18}{misfit_pct}{5.3}
\setres{lotUL}{CPT_post_k18}{misfit_pct_mean}{4.879}
\setres{lotUL}{CPT_post_k18}{n_seen}{18}
\setres{lotUL}{CPT_post_k18}{surprise}{0.02}
\setres{lotUL}{CPT_post_k18}{surprise_mean}{1.424}
\setres{lotUL}{CPT_post_k18}{width90_adj_pct}{8.8}
\setres{lotUL}{CPT_post_k18}{width90_adj_pct_mean}{8.140}
\setres{lotUL}{CPT_post_k18}{width90_pct}{0.9}
\setres{lotUL}{CPT_post_k18}{width90_pct_mean}{0.756}
\setres{lotUL}{CPT_post_k19}{cov50}{0.00}
\setres{lotUL}{CPT_post_k19}{cov50_mean}{0.067}
\setres{lotUL}{CPT_post_k19}{cov90}{0.17}
\setres{lotUL}{CPT_post_k19}{cov90_adj}{0.83}
\setres{lotUL}{CPT_post_k19}{cov90_adj_mean}{0.833}
\setres{lotUL}{CPT_post_k19}{cov90_mean}{0.142}
\setres{lotUL}{CPT_post_k19}{kappa}{0.957}
\setres{lotUL}{CPT_post_k19}{kappa_crps}{0.707}
\setres{lotUL}{CPT_post_k19}{kappa_crps_mean}{0.611}
\setres{lotUL}{CPT_post_k19}{kappa_mean}{0.921}
\setres{lotUL}{CPT_post_k19}{logscore_next}{-2.98}
\setres{lotUL}{CPT_post_k19}{logscore_next_mean}{-2.989}
\setres{lotUL}{CPT_post_k19}{misfit_pct}{5.4}
\setres{lotUL}{CPT_post_k19}{misfit_pct_mean}{4.997}
\setres{lotUL}{CPT_post_k19}{n_seen}{19}
\setres{lotUL}{CPT_post_k19}{surprise}{0.02}
\setres{lotUL}{CPT_post_k19}{surprise_mean}{1.059}
\setres{lotUL}{CPT_post_k19}{width90_adj_pct}{8.9}
\setres{lotUL}{CPT_post_k19}{width90_adj_pct_mean}{8.315}
\setres{lotUL}{CPT_post_k19}{width90_pct}{0.6}
\setres{lotUL}{CPT_post_k19}{width90_pct_mean}{0.653}
\setres{lotUL}{CPT_post_k2}{cov50}{0.20}
\setres{lotUL}{CPT_post_k2}{cov50_mean}{0.202}
\setres{lotUL}{CPT_post_k2}{cov90}{0.46}
\setres{lotUL}{CPT_post_k2}{cov90_adj}{0.48}
\setres{lotUL}{CPT_post_k2}{cov90_adj_mean}{0.535}
\setres{lotUL}{CPT_post_k2}{cov90_mean}{0.528}
\setres{lotUL}{CPT_post_k2}{kappa}{0.480}
\setres{lotUL}{CPT_post_k2}{kappa_crps}{0.443}
\setres{lotUL}{CPT_post_k2}{kappa_crps_mean}{0.425}
\setres{lotUL}{CPT_post_k2}{kappa_mean}{0.399}
\setres{lotUL}{CPT_post_k2}{logscore_next}{-2.99}
\setres{lotUL}{CPT_post_k2}{logscore_next_mean}{-2.987}
\setres{lotUL}{CPT_post_k2}{misfit_pct}{0.5}
\setres{lotUL}{CPT_post_k2}{misfit_pct_mean}{0.822}
\setres{lotUL}{CPT_post_k2}{n_seen}{2}
\setres{lotUL}{CPT_post_k2}{surprise}{1.31}
\setres{lotUL}{CPT_post_k2}{surprise_mean}{1.465}
\setres{lotUL}{CPT_post_k2}{width90_adj_pct}{9.8}
\setres{lotUL}{CPT_post_k2}{width90_adj_pct_mean}{11.841}
\setres{lotUL}{CPT_post_k2}{width90_pct}{9.3}
\setres{lotUL}{CPT_post_k2}{width90_pct_mean}{11.365}
\setres{lotUL}{CPT_post_k20}{cov50}{0.00}
\setres{lotUL}{CPT_post_k20}{cov50_mean}{0.070}
\setres{lotUL}{CPT_post_k20}{cov90}{0.00}
\setres{lotUL}{CPT_post_k20}{cov90_adj}{0.80}
\setres{lotUL}{CPT_post_k20}{cov90_adj_mean}{0.820}
\setres{lotUL}{CPT_post_k20}{cov90_mean}{0.140}
\setres{lotUL}{CPT_post_k20}{kappa}{0.950}
\setres{lotUL}{CPT_post_k20}{kappa_crps}{0.671}
\setres{lotUL}{CPT_post_k20}{kappa_crps_mean}{0.596}
\setres{lotUL}{CPT_post_k20}{kappa_mean}{0.915}
\setres{lotUL}{CPT_post_k20}{logscore_next}{-2.98}
\setres{lotUL}{CPT_post_k20}{logscore_next_mean}{-2.988}
\setres{lotUL}{CPT_post_k20}{misfit_pct}{5.4}
\setres{lotUL}{CPT_post_k20}{misfit_pct_mean}{5.130}
\setres{lotUL}{CPT_post_k20}{n_seen}{20}
\setres{lotUL}{CPT_post_k20}{surprise}{0.04}
\setres{lotUL}{CPT_post_k20}{surprise_mean}{0.566}
\setres{lotUL}{CPT_post_k20}{width90_adj_pct}{9.0}
\setres{lotUL}{CPT_post_k20}{width90_adj_pct_mean}{8.538}
\setres{lotUL}{CPT_post_k20}{width90_pct}{0.5}
\setres{lotUL}{CPT_post_k20}{width90_pct_mean}{0.668}
\setres{lotUL}{CPT_post_k21}{cov50}{0.00}
\setres{lotUL}{CPT_post_k21}{cov50_mean}{0.075}
\setres{lotUL}{CPT_post_k21}{cov90}{0.12}
\setres{lotUL}{CPT_post_k21}{cov90_adj}{0.75}
\setres{lotUL}{CPT_post_k21}{cov90_adj_mean}{0.825}
\setres{lotUL}{CPT_post_k21}{cov90_mean}{0.175}
\setres{lotUL}{CPT_post_k21}{kappa}{0.949}
\setres{lotUL}{CPT_post_k21}{kappa_crps}{0.654}
\setres{lotUL}{CPT_post_k21}{kappa_crps_mean}{0.587}
\setres{lotUL}{CPT_post_k21}{kappa_mean}{0.912}
\setres{lotUL}{CPT_post_k21}{logscore_next}{-2.98}
\setres{lotUL}{CPT_post_k21}{logscore_next_mean}{-2.984}
\setres{lotUL}{CPT_post_k21}{misfit_pct}{5.4}
\setres{lotUL}{CPT_post_k21}{misfit_pct_mean}{5.111}
\setres{lotUL}{CPT_post_k21}{n_seen}{21}
\setres{lotUL}{CPT_post_k21}{surprise}{0.01}
\setres{lotUL}{CPT_post_k21}{surprise_mean}{0.070}
\setres{lotUL}{CPT_post_k21}{width90_adj_pct}{9.0}
\setres{lotUL}{CPT_post_k21}{width90_adj_pct_mean}{8.501}
\setres{lotUL}{CPT_post_k21}{width90_pct}{0.6}
\setres{lotUL}{CPT_post_k21}{width90_pct_mean}{0.749}
\setres{lotUL}{CPT_post_k22}{cov50}{0.00}
\setres{lotUL}{CPT_post_k22}{cov50_mean}{0.067}
\setres{lotUL}{CPT_post_k22}{cov90}{0.00}
\setres{lotUL}{CPT_post_k22}{cov90_adj}{0.67}
\setres{lotUL}{CPT_post_k22}{cov90_adj_mean}{0.800}
\setres{lotUL}{CPT_post_k22}{cov90_mean}{0.167}
\setres{lotUL}{CPT_post_k22}{kappa}{0.955}
\setres{lotUL}{CPT_post_k22}{kappa_crps}{0.657}
\setres{lotUL}{CPT_post_k22}{kappa_crps_mean}{0.561}
\setres{lotUL}{CPT_post_k22}{kappa_mean}{0.899}
\setres{lotUL}{CPT_post_k22}{logscore_next}{-2.98}
\setres{lotUL}{CPT_post_k22}{logscore_next_mean}{-2.983}
\setres{lotUL}{CPT_post_k22}{misfit_pct}{5.4}
\setres{lotUL}{CPT_post_k22}{misfit_pct_mean}{5.078}
\setres{lotUL}{CPT_post_k22}{n_seen}{22}
\setres{lotUL}{CPT_post_k22}{surprise}{0.00}
\setres{lotUL}{CPT_post_k22}{surprise_mean}{0.225}
\setres{lotUL}{CPT_post_k22}{width90_adj_pct}{8.9}
\setres{lotUL}{CPT_post_k22}{width90_adj_pct_mean}{8.441}
\setres{lotUL}{CPT_post_k22}{width90_pct}{0.5}
\setres{lotUL}{CPT_post_k22}{width90_pct_mean}{0.659}
\setres{lotUL}{CPT_post_k23}{cov50}{0.00}
\setres{lotUL}{CPT_post_k23}{cov50_mean}{0.100}
\setres{lotUL}{CPT_post_k23}{cov90}{0.00}
\setres{lotUL}{CPT_post_k23}{cov90_adj}{1.00}
\setres{lotUL}{CPT_post_k23}{cov90_adj_mean}{0.800}
\setres{lotUL}{CPT_post_k23}{cov90_mean}{0.125}
\setres{lotUL}{CPT_post_k23}{kappa}{0.961}
\setres{lotUL}{CPT_post_k23}{kappa_crps}{0.710}
\setres{lotUL}{CPT_post_k23}{kappa_crps_mean}{0.642}
\setres{lotUL}{CPT_post_k23}{kappa_mean}{0.890}
\setres{lotUL}{CPT_post_k23}{logscore_next}{-2.99}
\setres{lotUL}{CPT_post_k23}{logscore_next_mean}{-3.005}
\setres{lotUL}{CPT_post_k23}{misfit_pct}{5.5}
\setres{lotUL}{CPT_post_k23}{misfit_pct_mean}{5.120}
\setres{lotUL}{CPT_post_k23}{n_seen}{23}
\setres{lotUL}{CPT_post_k23}{surprise}{0.09}
\setres{lotUL}{CPT_post_k23}{surprise_mean}{3.444}
\setres{lotUL}{CPT_post_k23}{width90_adj_pct}{9.0}
\setres{lotUL}{CPT_post_k23}{width90_adj_pct_mean}{8.524}
\setres{lotUL}{CPT_post_k23}{width90_pct}{0.8}
\setres{lotUL}{CPT_post_k23}{width90_pct_mean}{0.794}
\setres{lotUL}{CPT_post_k24}{cov50}{0.00}
\setres{lotUL}{CPT_post_k24}{cov50_mean}{0.100}
\setres{lotUL}{CPT_post_k24}{cov90}{0.00}
\setres{lotUL}{CPT_post_k24}{cov90_adj}{1.00}
\setres{lotUL}{CPT_post_k24}{cov90_adj_mean}{0.950}
\setres{lotUL}{CPT_post_k24}{cov90_mean}{0.100}
\setres{lotUL}{CPT_post_k24}{kappa}{0.972}
\setres{lotUL}{CPT_post_k24}{kappa_crps}{0.797}
\setres{lotUL}{CPT_post_k24}{kappa_crps_mean}{0.743}
\setres{lotUL}{CPT_post_k24}{kappa_mean}{0.876}
\setres{lotUL}{CPT_post_k24}{logscore_next}{-2.98}
\setres{lotUL}{CPT_post_k24}{logscore_next_mean}{-3.005}
\setres{lotUL}{CPT_post_k24}{misfit_pct}{5.4}
\setres{lotUL}{CPT_post_k24}{misfit_pct_mean}{5.334}
\setres{lotUL}{CPT_post_k24}{n_seen}{24}
\setres{lotUL}{CPT_post_k24}{surprise}{0.42}
\setres{lotUL}{CPT_post_k24}{surprise_mean}{2.543}
\setres{lotUL}{CPT_post_k24}{width90_adj_pct}{9.0}
\setres{lotUL}{CPT_post_k24}{width90_adj_pct_mean}{8.883}
\setres{lotUL}{CPT_post_k24}{width90_pct}{1.0}
\setres{lotUL}{CPT_post_k24}{width90_pct_mean}{0.915}
\setres{lotUL}{CPT_post_k25}{cov50}{---}
\setres{lotUL}{CPT_post_k25}{cov50_mean}{---}
\setres{lotUL}{CPT_post_k25}{cov90}{---}
\setres{lotUL}{CPT_post_k25}{cov90_adj}{---}
\setres{lotUL}{CPT_post_k25}{cov90_adj_mean}{---}
\setres{lotUL}{CPT_post_k25}{cov90_mean}{---}
\setres{lotUL}{CPT_post_k25}{kappa}{---}
\setres{lotUL}{CPT_post_k25}{kappa_crps}{---}
\setres{lotUL}{CPT_post_k25}{kappa_crps_mean}{---}
\setres{lotUL}{CPT_post_k25}{kappa_mean}{---}
\setres{lotUL}{CPT_post_k25}{logscore_next}{---}
\setres{lotUL}{CPT_post_k25}{logscore_next_mean}{---}
\setres{lotUL}{CPT_post_k25}{misfit_pct}{---}
\setres{lotUL}{CPT_post_k25}{misfit_pct_mean}{---}
\setres{lotUL}{CPT_post_k25}{n_seen}{25}
\setres{lotUL}{CPT_post_k25}{surprise}{---}
\setres{lotUL}{CPT_post_k25}{surprise_mean}{---}
\setres{lotUL}{CPT_post_k25}{width90_adj_pct}{---}
\setres{lotUL}{CPT_post_k25}{width90_adj_pct_mean}{---}
\setres{lotUL}{CPT_post_k25}{width90_pct}{---}
\setres{lotUL}{CPT_post_k25}{width90_pct_mean}{---}
\setres{lotUL}{CPT_post_k3}{cov50}{0.16}
\setres{lotUL}{CPT_post_k3}{cov50_mean}{0.202}
\setres{lotUL}{CPT_post_k3}{cov90}{0.52}
\setres{lotUL}{CPT_post_k3}{cov90_adj}{0.77}
\setres{lotUL}{CPT_post_k3}{cov90_adj_mean}{0.693}
\setres{lotUL}{CPT_post_k3}{cov90_mean}{0.541}
\setres{lotUL}{CPT_post_k3}{kappa}{0.796}
\setres{lotUL}{CPT_post_k3}{kappa_crps}{0.553}
\setres{lotUL}{CPT_post_k3}{kappa_crps_mean}{0.514}
\setres{lotUL}{CPT_post_k3}{kappa_mean}{0.630}
\setres{lotUL}{CPT_post_k3}{logscore_next}{-2.99}
\setres{lotUL}{CPT_post_k3}{logscore_next_mean}{-2.988}
\setres{lotUL}{CPT_post_k3}{misfit_pct}{2.3}
\setres{lotUL}{CPT_post_k3}{misfit_pct_mean}{2.767}
\setres{lotUL}{CPT_post_k3}{n_seen}{3}
\setres{lotUL}{CPT_post_k3}{surprise}{1.30}
\setres{lotUL}{CPT_post_k3}{surprise_mean}{1.328}
\setres{lotUL}{CPT_post_k3}{width90_adj_pct}{9.3}
\setres{lotUL}{CPT_post_k3}{width90_adj_pct_mean}{9.299}
\setres{lotUL}{CPT_post_k3}{width90_pct}{7.3}
\setres{lotUL}{CPT_post_k3}{width90_pct_mean}{7.369}
\setres{lotUL}{CPT_post_k4}{cov50}{0.17}
\setres{lotUL}{CPT_post_k4}{cov50_mean}{0.183}
\setres{lotUL}{CPT_post_k4}{cov90}{0.52}
\setres{lotUL}{CPT_post_k4}{cov90_adj}{0.79}
\setres{lotUL}{CPT_post_k4}{cov90_adj_mean}{0.736}
\setres{lotUL}{CPT_post_k4}{cov90_mean}{0.510}
\setres{lotUL}{CPT_post_k4}{kappa}{0.914}
\setres{lotUL}{CPT_post_k4}{kappa_crps}{0.620}
\setres{lotUL}{CPT_post_k4}{kappa_crps_mean}{0.614}
\setres{lotUL}{CPT_post_k4}{kappa_mean}{0.880}
\setres{lotUL}{CPT_post_k4}{logscore_next}{-2.98}
\setres{lotUL}{CPT_post_k4}{logscore_next_mean}{-2.984}
\setres{lotUL}{CPT_post_k4}{misfit_pct}{2.7}
\setres{lotUL}{CPT_post_k4}{misfit_pct_mean}{3.121}
\setres{lotUL}{CPT_post_k4}{n_seen}{4}
\setres{lotUL}{CPT_post_k4}{surprise}{0.34}
\setres{lotUL}{CPT_post_k4}{surprise_mean}{0.646}
\setres{lotUL}{CPT_post_k4}{width90_adj_pct}{6.6}
\setres{lotUL}{CPT_post_k4}{width90_adj_pct_mean}{7.422}
\setres{lotUL}{CPT_post_k4}{width90_pct}{4.5}
\setres{lotUL}{CPT_post_k4}{width90_pct_mean}{5.064}
\setres{lotUL}{CPT_post_k5}{cov50}{0.25}
\setres{lotUL}{CPT_post_k5}{cov50_mean}{0.230}
\setres{lotUL}{CPT_post_k5}{cov90}{0.50}
\setres{lotUL}{CPT_post_k5}{cov90_adj}{0.78}
\setres{lotUL}{CPT_post_k5}{cov90_adj_mean}{0.773}
\setres{lotUL}{CPT_post_k5}{cov90_mean}{0.475}
\setres{lotUL}{CPT_post_k5}{kappa}{0.929}
\setres{lotUL}{CPT_post_k5}{kappa_crps}{0.639}
\setres{lotUL}{CPT_post_k5}{kappa_crps_mean}{0.647}
\setres{lotUL}{CPT_post_k5}{kappa_mean}{0.925}
\setres{lotUL}{CPT_post_k5}{logscore_next}{-2.98}
\setres{lotUL}{CPT_post_k5}{logscore_next_mean}{-2.984}
\setres{lotUL}{CPT_post_k5}{misfit_pct}{2.6}
\setres{lotUL}{CPT_post_k5}{misfit_pct_mean}{3.285}
\setres{lotUL}{CPT_post_k5}{n_seen}{5}
\setres{lotUL}{CPT_post_k5}{surprise}{0.31}
\setres{lotUL}{CPT_post_k5}{surprise_mean}{0.728}
\setres{lotUL}{CPT_post_k5}{width90_adj_pct}{6.0}
\setres{lotUL}{CPT_post_k5}{width90_adj_pct_mean}{6.556}
\setres{lotUL}{CPT_post_k5}{width90_pct}{3.1}
\setres{lotUL}{CPT_post_k5}{width90_pct_mean}{3.380}
\setres{lotUL}{CPT_post_k6}{cov50}{0.18}
\setres{lotUL}{CPT_post_k6}{cov50_mean}{0.168}
\setres{lotUL}{CPT_post_k6}{cov90}{0.42}
\setres{lotUL}{CPT_post_k6}{cov90_adj}{0.76}
\setres{lotUL}{CPT_post_k6}{cov90_adj_mean}{0.739}
\setres{lotUL}{CPT_post_k6}{cov90_mean}{0.382}
\setres{lotUL}{CPT_post_k6}{kappa}{0.941}
\setres{lotUL}{CPT_post_k6}{kappa_crps}{0.636}
\setres{lotUL}{CPT_post_k6}{kappa_crps_mean}{0.646}
\setres{lotUL}{CPT_post_k6}{kappa_mean}{0.928}
\setres{lotUL}{CPT_post_k6}{logscore_next}{-2.98}
\setres{lotUL}{CPT_post_k6}{logscore_next_mean}{-2.986}
\setres{lotUL}{CPT_post_k6}{misfit_pct}{3.3}
\setres{lotUL}{CPT_post_k6}{misfit_pct_mean}{3.357}
\setres{lotUL}{CPT_post_k6}{n_seen}{6}
\setres{lotUL}{CPT_post_k6}{surprise}{0.14}
\setres{lotUL}{CPT_post_k6}{surprise_mean}{1.053}
\setres{lotUL}{CPT_post_k6}{width90_adj_pct}{6.1}
\setres{lotUL}{CPT_post_k6}{width90_adj_pct_mean}{6.219}
\setres{lotUL}{CPT_post_k6}{width90_pct}{2.4}
\setres{lotUL}{CPT_post_k6}{width90_pct_mean}{2.345}
\setres{lotUL}{CPT_post_k7}{cov50}{0.11}
\setres{lotUL}{CPT_post_k7}{cov50_mean}{0.142}
\setres{lotUL}{CPT_post_k7}{cov90}{0.44}
\setres{lotUL}{CPT_post_k7}{cov90_adj}{0.78}
\setres{lotUL}{CPT_post_k7}{cov90_adj_mean}{0.761}
\setres{lotUL}{CPT_post_k7}{cov90_mean}{0.400}
\setres{lotUL}{CPT_post_k7}{kappa}{0.941}
\setres{lotUL}{CPT_post_k7}{kappa_crps}{0.654}
\setres{lotUL}{CPT_post_k7}{kappa_crps_mean}{0.654}
\setres{lotUL}{CPT_post_k7}{kappa_mean}{0.939}
\setres{lotUL}{CPT_post_k7}{logscore_next}{-2.98}
\setres{lotUL}{CPT_post_k7}{logscore_next_mean}{-2.984}
\setres{lotUL}{CPT_post_k7}{misfit_pct}{3.5}
\setres{lotUL}{CPT_post_k7}{misfit_pct_mean}{3.613}
\setres{lotUL}{CPT_post_k7}{n_seen}{7}
\setres{lotUL}{CPT_post_k7}{surprise}{0.04}
\setres{lotUL}{CPT_post_k7}{surprise_mean}{0.460}
\setres{lotUL}{CPT_post_k7}{width90_adj_pct}{6.6}
\setres{lotUL}{CPT_post_k7}{width90_adj_pct_mean}{6.440}
\setres{lotUL}{CPT_post_k7}{width90_pct}{2.2}
\setres{lotUL}{CPT_post_k7}{width90_pct_mean}{2.102}
\setres{lotUL}{CPT_post_k8}{cov50}{0.09}
\setres{lotUL}{CPT_post_k8}{cov50_mean}{0.132}
\setres{lotUL}{CPT_post_k8}{cov90}{0.35}
\setres{lotUL}{CPT_post_k8}{cov90_adj}{0.82}
\setres{lotUL}{CPT_post_k8}{cov90_adj_mean}{0.776}
\setres{lotUL}{CPT_post_k8}{cov90_mean}{0.347}
\setres{lotUL}{CPT_post_k8}{kappa}{0.940}
\setres{lotUL}{CPT_post_k8}{kappa_crps}{0.673}
\setres{lotUL}{CPT_post_k8}{kappa_crps_mean}{0.658}
\setres{lotUL}{CPT_post_k8}{kappa_mean}{0.939}
\setres{lotUL}{CPT_post_k8}{logscore_next}{-2.98}
\setres{lotUL}{CPT_post_k8}{logscore_next_mean}{-2.985}
\setres{lotUL}{CPT_post_k8}{misfit_pct}{3.7}
\setres{lotUL}{CPT_post_k8}{misfit_pct_mean}{3.772}
\setres{lotUL}{CPT_post_k8}{n_seen}{8}
\setres{lotUL}{CPT_post_k8}{surprise}{0.15}
\setres{lotUL}{CPT_post_k8}{surprise_mean}{0.631}
\setres{lotUL}{CPT_post_k8}{width90_adj_pct}{6.4}
\setres{lotUL}{CPT_post_k8}{width90_adj_pct_mean}{6.599}
\setres{lotUL}{CPT_post_k8}{width90_pct}{1.8}
\setres{lotUL}{CPT_post_k8}{width90_pct_mean}{1.763}
\setres{lotUL}{CPT_post_k9}{cov50}{0.06}
\setres{lotUL}{CPT_post_k9}{cov50_mean}{0.097}
\setres{lotUL}{CPT_post_k9}{cov90}{0.34}
\setres{lotUL}{CPT_post_k9}{cov90_adj}{0.81}
\setres{lotUL}{CPT_post_k9}{cov90_adj_mean}{0.781}
\setres{lotUL}{CPT_post_k9}{cov90_mean}{0.300}
\setres{lotUL}{CPT_post_k9}{kappa}{0.943}
\setres{lotUL}{CPT_post_k9}{kappa_crps}{0.662}
\setres{lotUL}{CPT_post_k9}{kappa_crps_mean}{0.643}
\setres{lotUL}{CPT_post_k9}{kappa_mean}{0.937}
\setres{lotUL}{CPT_post_k9}{logscore_next}{-2.98}
\setres{lotUL}{CPT_post_k9}{logscore_next_mean}{-2.988}
\setres{lotUL}{CPT_post_k9}{misfit_pct}{3.9}
\setres{lotUL}{CPT_post_k9}{misfit_pct_mean}{3.957}
\setres{lotUL}{CPT_post_k9}{n_seen}{9}
\setres{lotUL}{CPT_post_k9}{surprise}{0.10}
\setres{lotUL}{CPT_post_k9}{surprise_mean}{1.025}
\setres{lotUL}{CPT_post_k9}{width90_adj_pct}{6.9}
\setres{lotUL}{CPT_post_k9}{width90_adj_pct_mean}{6.809}
\setres{lotUL}{CPT_post_k9}{width90_pct}{1.8}
\setres{lotUL}{CPT_post_k9}{width90_pct_mean}{1.667}
\setres{lotUL}{DA_post_k0}{cov50}{0.24}
\setres{lotUL}{DA_post_k0}{cov50_mean}{0.240}
\setres{lotUL}{DA_post_k0}{cov90}{0.68}
\setres{lotUL}{DA_post_k0}{cov90_adj}{---}
\setres{lotUL}{DA_post_k0}{cov90_adj_mean}{---}
\setres{lotUL}{DA_post_k0}{cov90_mean}{0.680}
\setres{lotUL}{DA_post_k0}{kappa}{-0.837}
\setres{lotUL}{DA_post_k0}{kappa_crps}{0.056}
\setres{lotUL}{DA_post_k0}{kappa_crps_mean}{0.052}
\setres{lotUL}{DA_post_k0}{kappa_mean}{-0.837}
\setres{lotUL}{DA_post_k0}{logscore_next}{-3.00}
\setres{lotUL}{DA_post_k0}{logscore_next_mean}{-2.998}
\setres{lotUL}{DA_post_k0}{misfit_pct}{---}
\setres{lotUL}{DA_post_k0}{misfit_pct_mean}{---}
\setres{lotUL}{DA_post_k0}{n_seen}{0}
\setres{lotUL}{DA_post_k0}{surprise}{3.51}
\setres{lotUL}{DA_post_k0}{surprise_mean}{3.420}
\setres{lotUL}{DA_post_k0}{width90_adj_pct}{---}
\setres{lotUL}{DA_post_k0}{width90_adj_pct_mean}{---}
\setres{lotUL}{DA_post_k0}{width90_pct}{25.5}
\setres{lotUL}{DA_post_k0}{width90_pct_mean}{25.485}
\setres{lotUL}{DA_post_k1}{cov50}{0.10}
\setres{lotUL}{DA_post_k1}{cov50_mean}{0.075}
\setres{lotUL}{DA_post_k1}{cov90}{0.17}
\setres{lotUL}{DA_post_k1}{cov90_adj}{---}
\setres{lotUL}{DA_post_k1}{cov90_adj_mean}{---}
\setres{lotUL}{DA_post_k1}{cov90_mean}{0.167}
\setres{lotUL}{DA_post_k1}{kappa}{-1.500}
\setres{lotUL}{DA_post_k1}{kappa_crps}{-0.837}
\setres{lotUL}{DA_post_k1}{kappa_crps_mean}{-0.664}
\setres{lotUL}{DA_post_k1}{kappa_mean}{-1.277}
\setres{lotUL}{DA_post_k1}{logscore_next}{-3.01}
\setres{lotUL}{DA_post_k1}{logscore_next_mean}{-3.070}
\setres{lotUL}{DA_post_k1}{misfit_pct}{---}
\setres{lotUL}{DA_post_k1}{misfit_pct_mean}{---}
\setres{lotUL}{DA_post_k1}{n_seen}{1}
\setres{lotUL}{DA_post_k1}{surprise}{1.96}
\setres{lotUL}{DA_post_k1}{surprise_mean}{6.979}
\setres{lotUL}{DA_post_k1}{width90_adj_pct}{---}
\setres{lotUL}{DA_post_k1}{width90_adj_pct_mean}{---}
\setres{lotUL}{DA_post_k1}{width90_pct}{4.2}
\setres{lotUL}{DA_post_k1}{width90_pct_mean}{4.543}
\setres{lotUL}{DA_post_k10}{cov50}{0.00}
\setres{lotUL}{DA_post_k10}{cov50_mean}{0.007}
\setres{lotUL}{DA_post_k10}{cov90}{0.00}
\setres{lotUL}{DA_post_k10}{cov90_adj}{1.00}
\setres{lotUL}{DA_post_k10}{cov90_adj_mean}{0.917}
\setres{lotUL}{DA_post_k10}{cov90_mean}{0.020}
\setres{lotUL}{DA_post_k10}{kappa}{0.559}
\setres{lotUL}{DA_post_k10}{kappa_crps}{0.305}
\setres{lotUL}{DA_post_k10}{kappa_crps_mean}{-0.049}
\setres{lotUL}{DA_post_k10}{kappa_mean}{-0.110}
\setres{lotUL}{DA_post_k10}{logscore_next}{-3.06}
\setres{lotUL}{DA_post_k10}{logscore_next_mean}{-3.120}
\setres{lotUL}{DA_post_k10}{misfit_pct}{16.4}
\setres{lotUL}{DA_post_k10}{misfit_pct_mean}{16.742}
\setres{lotUL}{DA_post_k10}{n_seen}{10}
\setres{lotUL}{DA_post_k10}{surprise}{0.08}
\setres{lotUL}{DA_post_k10}{surprise_mean}{5.385}
\setres{lotUL}{DA_post_k10}{width90_adj_pct}{27.0}
\setres{lotUL}{DA_post_k10}{width90_adj_pct_mean}{27.552}
\setres{lotUL}{DA_post_k10}{width90_pct}{0.3}
\setres{lotUL}{DA_post_k10}{width90_pct_mean}{0.455}
\setres{lotUL}{DA_post_k11}{cov50}{0.00}
\setres{lotUL}{DA_post_k11}{cov50_mean}{0.007}
\setres{lotUL}{DA_post_k11}{cov90}{0.00}
\setres{lotUL}{DA_post_k11}{cov90_adj}{1.00}
\setres{lotUL}{DA_post_k11}{cov90_adj_mean}{0.925}
\setres{lotUL}{DA_post_k11}{cov90_mean}{0.025}
\setres{lotUL}{DA_post_k11}{kappa}{0.559}
\setres{lotUL}{DA_post_k11}{kappa_crps}{0.307}
\setres{lotUL}{DA_post_k11}{kappa_crps_mean}{0.006}
\setres{lotUL}{DA_post_k11}{kappa_mean}{0.008}
\setres{lotUL}{DA_post_k11}{logscore_next}{-3.09}
\setres{lotUL}{DA_post_k11}{logscore_next_mean}{-3.450}
\setres{lotUL}{DA_post_k11}{misfit_pct}{15.9}
\setres{lotUL}{DA_post_k11}{misfit_pct_mean}{16.466}
\setres{lotUL}{DA_post_k11}{n_seen}{11}
\setres{lotUL}{DA_post_k11}{surprise}{1.55}
\setres{lotUL}{DA_post_k11}{surprise_mean}{47.782}
\setres{lotUL}{DA_post_k11}{width90_adj_pct}{26.1}
\setres{lotUL}{DA_post_k11}{width90_adj_pct_mean}{27.102}
\setres{lotUL}{DA_post_k11}{width90_pct}{0.5}
\setres{lotUL}{DA_post_k11}{width90_pct_mean}{0.569}
\setres{lotUL}{DA_post_k12}{cov50}{0.00}
\setres{lotUL}{DA_post_k12}{cov50_mean}{0.012}
\setres{lotUL}{DA_post_k12}{cov90}{0.00}
\setres{lotUL}{DA_post_k12}{cov90_adj}{1.00}
\setres{lotUL}{DA_post_k12}{cov90_adj_mean}{0.923}
\setres{lotUL}{DA_post_k12}{cov90_mean}{0.023}
\setres{lotUL}{DA_post_k12}{kappa}{0.247}
\setres{lotUL}{DA_post_k12}{kappa_crps}{0.160}
\setres{lotUL}{DA_post_k12}{kappa_crps_mean}{0.079}
\setres{lotUL}{DA_post_k12}{kappa_mean}{0.175}
\setres{lotUL}{DA_post_k12}{logscore_next}{-3.09}
\setres{lotUL}{DA_post_k12}{logscore_next_mean}{-3.180}
\setres{lotUL}{DA_post_k12}{misfit_pct}{15.9}
\setres{lotUL}{DA_post_k12}{misfit_pct_mean}{16.455}
\setres{lotUL}{DA_post_k12}{n_seen}{12}
\setres{lotUL}{DA_post_k12}{surprise}{0.12}
\setres{lotUL}{DA_post_k12}{surprise_mean}{2.033}
\setres{lotUL}{DA_post_k12}{width90_adj_pct}{26.2}
\setres{lotUL}{DA_post_k12}{width90_adj_pct_mean}{27.080}
\setres{lotUL}{DA_post_k12}{width90_pct}{0.5}
\setres{lotUL}{DA_post_k12}{width90_pct_mean}{0.474}
\setres{lotUL}{DA_post_k13}{cov50}{0.00}
\setres{lotUL}{DA_post_k13}{cov50_mean}{0.004}
\setres{lotUL}{DA_post_k13}{cov90}{0.00}
\setres{lotUL}{DA_post_k13}{cov90_adj}{1.00}
\setres{lotUL}{DA_post_k13}{cov90_adj_mean}{0.958}
\setres{lotUL}{DA_post_k13}{cov90_mean}{0.017}
\setres{lotUL}{DA_post_k13}{kappa}{0.357}
\setres{lotUL}{DA_post_k13}{kappa_crps}{0.228}
\setres{lotUL}{DA_post_k13}{kappa_crps_mean}{0.084}
\setres{lotUL}{DA_post_k13}{kappa_mean}{0.143}
\setres{lotUL}{DA_post_k13}{logscore_next}{-3.12}
\setres{lotUL}{DA_post_k13}{logscore_next_mean}{-3.305}
\setres{lotUL}{DA_post_k13}{misfit_pct}{15.9}
\setres{lotUL}{DA_post_k13}{misfit_pct_mean}{16.232}
\setres{lotUL}{DA_post_k13}{n_seen}{13}
\setres{lotUL}{DA_post_k13}{surprise}{0.66}
\setres{lotUL}{DA_post_k13}{surprise_mean}{3.847}
\setres{lotUL}{DA_post_k13}{width90_adj_pct}{26.2}
\setres{lotUL}{DA_post_k13}{width90_adj_pct_mean}{26.711}
\setres{lotUL}{DA_post_k13}{width90_pct}{0.4}
\setres{lotUL}{DA_post_k13}{width90_pct_mean}{0.467}
\setres{lotUL}{DA_post_k14}{cov50}{0.00}
\setres{lotUL}{DA_post_k14}{cov50_mean}{0.005}
\setres{lotUL}{DA_post_k14}{cov90}{0.00}
\setres{lotUL}{DA_post_k14}{cov90_adj}{1.00}
\setres{lotUL}{DA_post_k14}{cov90_adj_mean}{0.941}
\setres{lotUL}{DA_post_k14}{cov90_mean}{0.018}
\setres{lotUL}{DA_post_k14}{kappa}{0.311}
\setres{lotUL}{DA_post_k14}{kappa_crps}{0.184}
\setres{lotUL}{DA_post_k14}{kappa_crps_mean}{0.066}
\setres{lotUL}{DA_post_k14}{kappa_mean}{0.116}
\setres{lotUL}{DA_post_k14}{logscore_next}{-3.09}
\setres{lotUL}{DA_post_k14}{logscore_next_mean}{-3.338}
\setres{lotUL}{DA_post_k14}{misfit_pct}{15.5}
\setres{lotUL}{DA_post_k14}{misfit_pct_mean}{16.129}
\setres{lotUL}{DA_post_k14}{n_seen}{14}
\setres{lotUL}{DA_post_k14}{surprise}{0.12}
\setres{lotUL}{DA_post_k14}{surprise_mean}{15.228}
\setres{lotUL}{DA_post_k14}{width90_adj_pct}{25.6}
\setres{lotUL}{DA_post_k14}{width90_adj_pct_mean}{26.544}
\setres{lotUL}{DA_post_k14}{width90_pct}{0.4}
\setres{lotUL}{DA_post_k14}{width90_pct_mean}{0.468}
\setres{lotUL}{DA_post_k15}{cov50}{0.00}
\setres{lotUL}{DA_post_k15}{cov50_mean}{0.000}
\setres{lotUL}{DA_post_k15}{cov90}{0.00}
\setres{lotUL}{DA_post_k15}{cov90_adj}{1.00}
\setres{lotUL}{DA_post_k15}{cov90_adj_mean}{0.955}
\setres{lotUL}{DA_post_k15}{cov90_mean}{0.020}
\setres{lotUL}{DA_post_k15}{kappa}{0.321}
\setres{lotUL}{DA_post_k15}{kappa_crps}{0.177}
\setres{lotUL}{DA_post_k15}{kappa_crps_mean}{0.021}
\setres{lotUL}{DA_post_k15}{kappa_mean}{0.047}
\setres{lotUL}{DA_post_k15}{logscore_next}{-3.05}
\setres{lotUL}{DA_post_k15}{logscore_next_mean}{-3.159}
\setres{lotUL}{DA_post_k15}{misfit_pct}{15.6}
\setres{lotUL}{DA_post_k15}{misfit_pct_mean}{16.041}
\setres{lotUL}{DA_post_k15}{n_seen}{15}
\setres{lotUL}{DA_post_k15}{surprise}{0.17}
\setres{lotUL}{DA_post_k15}{surprise_mean}{2.792}
\setres{lotUL}{DA_post_k15}{width90_adj_pct}{25.7}
\setres{lotUL}{DA_post_k15}{width90_adj_pct_mean}{26.398}
\setres{lotUL}{DA_post_k15}{width90_pct}{0.4}
\setres{lotUL}{DA_post_k15}{width90_pct_mean}{0.442}
\setres{lotUL}{DA_post_k16}{cov50}{0.00}
\setres{lotUL}{DA_post_k16}{cov50_mean}{0.017}
\setres{lotUL}{DA_post_k16}{cov90}{0.00}
\setres{lotUL}{DA_post_k16}{cov90_adj}{1.00}
\setres{lotUL}{DA_post_k16}{cov90_adj_mean}{0.956}
\setres{lotUL}{DA_post_k16}{cov90_mean}{0.022}
\setres{lotUL}{DA_post_k16}{kappa}{0.345}
\setres{lotUL}{DA_post_k16}{kappa_crps}{0.223}
\setres{lotUL}{DA_post_k16}{kappa_crps_mean}{0.024}
\setres{lotUL}{DA_post_k16}{kappa_mean}{0.034}
\setres{lotUL}{DA_post_k16}{logscore_next}{-3.02}
\setres{lotUL}{DA_post_k16}{logscore_next_mean}{-3.175}
\setres{lotUL}{DA_post_k16}{misfit_pct}{15.4}
\setres{lotUL}{DA_post_k16}{misfit_pct_mean}{16.008}
\setres{lotUL}{DA_post_k16}{n_seen}{16}
\setres{lotUL}{DA_post_k16}{surprise}{0.07}
\setres{lotUL}{DA_post_k16}{surprise_mean}{4.792}
\setres{lotUL}{DA_post_k16}{width90_adj_pct}{25.3}
\setres{lotUL}{DA_post_k16}{width90_adj_pct_mean}{26.344}
\setres{lotUL}{DA_post_k16}{width90_pct}{0.4}
\setres{lotUL}{DA_post_k16}{width90_pct_mean}{0.481}
\setres{lotUL}{DA_post_k17}{cov50}{0.00}
\setres{lotUL}{DA_post_k17}{cov50_mean}{0.013}
\setres{lotUL}{DA_post_k17}{cov90}{0.00}
\setres{lotUL}{DA_post_k17}{cov90_adj}{1.00}
\setres{lotUL}{DA_post_k17}{cov90_adj_mean}{0.944}
\setres{lotUL}{DA_post_k17}{cov90_mean}{0.025}
\setres{lotUL}{DA_post_k17}{kappa}{0.318}
\setres{lotUL}{DA_post_k17}{kappa_crps}{0.205}
\setres{lotUL}{DA_post_k17}{kappa_crps_mean}{0.023}
\setres{lotUL}{DA_post_k17}{kappa_mean}{0.018}
\setres{lotUL}{DA_post_k17}{logscore_next}{-3.07}
\setres{lotUL}{DA_post_k17}{logscore_next_mean}{-3.241}
\setres{lotUL}{DA_post_k17}{misfit_pct}{15.6}
\setres{lotUL}{DA_post_k17}{misfit_pct_mean}{15.884}
\setres{lotUL}{DA_post_k17}{n_seen}{17}
\setres{lotUL}{DA_post_k17}{surprise}{0.20}
\setres{lotUL}{DA_post_k17}{surprise_mean}{2.975}
\setres{lotUL}{DA_post_k17}{width90_adj_pct}{25.7}
\setres{lotUL}{DA_post_k17}{width90_adj_pct_mean}{26.146}
\setres{lotUL}{DA_post_k17}{width90_pct}{0.3}
\setres{lotUL}{DA_post_k17}{width90_pct_mean}{0.564}
\setres{lotUL}{DA_post_k18}{cov50}{0.00}
\setres{lotUL}{DA_post_k18}{cov50_mean}{0.007}
\setres{lotUL}{DA_post_k18}{cov90}{0.00}
\setres{lotUL}{DA_post_k18}{cov90_adj}{1.00}
\setres{lotUL}{DA_post_k18}{cov90_adj_mean}{0.943}
\setres{lotUL}{DA_post_k18}{cov90_mean}{0.014}
\setres{lotUL}{DA_post_k18}{kappa}{0.337}
\setres{lotUL}{DA_post_k18}{kappa_crps}{0.217}
\setres{lotUL}{DA_post_k18}{kappa_crps_mean}{-0.058}
\setres{lotUL}{DA_post_k18}{kappa_mean}{-0.389}
\setres{lotUL}{DA_post_k18}{logscore_next}{-3.08}
\setres{lotUL}{DA_post_k18}{logscore_next_mean}{-3.148}
\setres{lotUL}{DA_post_k18}{misfit_pct}{15.4}
\setres{lotUL}{DA_post_k18}{misfit_pct_mean}{15.826}
\setres{lotUL}{DA_post_k18}{n_seen}{18}
\setres{lotUL}{DA_post_k18}{surprise}{0.07}
\setres{lotUL}{DA_post_k18}{surprise_mean}{4.930}
\setres{lotUL}{DA_post_k18}{width90_adj_pct}{25.4}
\setres{lotUL}{DA_post_k18}{width90_adj_pct_mean}{26.042}
\setres{lotUL}{DA_post_k18}{width90_pct}{0.3}
\setres{lotUL}{DA_post_k18}{width90_pct_mean}{0.413}
\setres{lotUL}{DA_post_k19}{cov50}{0.00}
\setres{lotUL}{DA_post_k19}{cov50_mean}{0.008}
\setres{lotUL}{DA_post_k19}{cov90}{0.00}
\setres{lotUL}{DA_post_k19}{cov90_adj}{1.00}
\setres{lotUL}{DA_post_k19}{cov90_adj_mean}{0.967}
\setres{lotUL}{DA_post_k19}{cov90_mean}{0.008}
\setres{lotUL}{DA_post_k19}{kappa}{0.383}
\setres{lotUL}{DA_post_k19}{kappa_crps}{0.255}
\setres{lotUL}{DA_post_k19}{kappa_crps_mean}{0.047}
\setres{lotUL}{DA_post_k19}{kappa_mean}{0.047}
\setres{lotUL}{DA_post_k19}{logscore_next}{-3.10}
\setres{lotUL}{DA_post_k19}{logscore_next_mean}{-3.269}
\setres{lotUL}{DA_post_k19}{misfit_pct}{15.3}
\setres{lotUL}{DA_post_k19}{misfit_pct_mean}{15.659}
\setres{lotUL}{DA_post_k19}{n_seen}{19}
\setres{lotUL}{DA_post_k19}{surprise}{0.15}
\setres{lotUL}{DA_post_k19}{surprise_mean}{10.903}
\setres{lotUL}{DA_post_k19}{width90_adj_pct}{25.1}
\setres{lotUL}{DA_post_k19}{width90_adj_pct_mean}{25.774}
\setres{lotUL}{DA_post_k19}{width90_pct}{0.4}
\setres{lotUL}{DA_post_k19}{width90_pct_mean}{0.484}
\setres{lotUL}{DA_post_k2}{cov50}{0.00}
\setres{lotUL}{DA_post_k2}{cov50_mean}{0.030}
\setres{lotUL}{DA_post_k2}{cov90}{0.11}
\setres{lotUL}{DA_post_k2}{cov90_adj}{0.33}
\setres{lotUL}{DA_post_k2}{cov90_adj_mean}{0.443}
\setres{lotUL}{DA_post_k2}{cov90_mean}{0.102}
\setres{lotUL}{DA_post_k2}{kappa}{-0.351}
\setres{lotUL}{DA_post_k2}{kappa_crps}{-0.121}
\setres{lotUL}{DA_post_k2}{kappa_crps_mean}{-0.348}
\setres{lotUL}{DA_post_k2}{kappa_mean}{-0.710}
\setres{lotUL}{DA_post_k2}{logscore_next}{-3.02}
\setres{lotUL}{DA_post_k2}{logscore_next_mean}{-3.126}
\setres{lotUL}{DA_post_k2}{misfit_pct}{8.1}
\setres{lotUL}{DA_post_k2}{misfit_pct_mean}{8.746}
\setres{lotUL}{DA_post_k2}{n_seen}{2}
\setres{lotUL}{DA_post_k2}{surprise}{0.87}
\setres{lotUL}{DA_post_k2}{surprise_mean}{5.354}
\setres{lotUL}{DA_post_k2}{width90_adj_pct}{13.8}
\setres{lotUL}{DA_post_k2}{width90_adj_pct_mean}{14.903}
\setres{lotUL}{DA_post_k2}{width90_pct}{1.7}
\setres{lotUL}{DA_post_k2}{width90_pct_mean}{2.447}
\setres{lotUL}{DA_post_k20}{cov50}{0.00}
\setres{lotUL}{DA_post_k20}{cov50_mean}{0.000}
\setres{lotUL}{DA_post_k20}{cov90}{0.00}
\setres{lotUL}{DA_post_k20}{cov90_adj}{1.00}
\setres{lotUL}{DA_post_k20}{cov90_adj_mean}{0.960}
\setres{lotUL}{DA_post_k20}{cov90_mean}{0.010}
\setres{lotUL}{DA_post_k20}{kappa}{0.343}
\setres{lotUL}{DA_post_k20}{kappa_crps}{0.232}
\setres{lotUL}{DA_post_k20}{kappa_crps_mean}{0.006}
\setres{lotUL}{DA_post_k20}{kappa_mean}{-0.193}
\setres{lotUL}{DA_post_k20}{logscore_next}{-3.05}
\setres{lotUL}{DA_post_k20}{logscore_next_mean}{-3.197}
\setres{lotUL}{DA_post_k20}{misfit_pct}{15.2}
\setres{lotUL}{DA_post_k20}{misfit_pct_mean}{15.417}
\setres{lotUL}{DA_post_k20}{n_seen}{20}
\setres{lotUL}{DA_post_k20}{surprise}{0.52}
\setres{lotUL}{DA_post_k20}{surprise_mean}{4.165}
\setres{lotUL}{DA_post_k20}{width90_adj_pct}{25.1}
\setres{lotUL}{DA_post_k20}{width90_adj_pct_mean}{25.395}
\setres{lotUL}{DA_post_k20}{width90_pct}{0.3}
\setres{lotUL}{DA_post_k20}{width90_pct_mean}{0.664}
\setres{lotUL}{DA_post_k21}{cov50}{0.00}
\setres{lotUL}{DA_post_k21}{cov50_mean}{0.013}
\setres{lotUL}{DA_post_k21}{cov90}{0.00}
\setres{lotUL}{DA_post_k21}{cov90_adj}{1.00}
\setres{lotUL}{DA_post_k21}{cov90_adj_mean}{0.963}
\setres{lotUL}{DA_post_k21}{cov90_mean}{0.037}
\setres{lotUL}{DA_post_k21}{kappa}{0.309}
\setres{lotUL}{DA_post_k21}{kappa_crps}{0.208}
\setres{lotUL}{DA_post_k21}{kappa_crps_mean}{0.042}
\setres{lotUL}{DA_post_k21}{kappa_mean}{-0.124}
\setres{lotUL}{DA_post_k21}{logscore_next}{-3.05}
\setres{lotUL}{DA_post_k21}{logscore_next_mean}{-3.094}
\setres{lotUL}{DA_post_k21}{misfit_pct}{14.9}
\setres{lotUL}{DA_post_k21}{misfit_pct_mean}{15.168}
\setres{lotUL}{DA_post_k21}{n_seen}{21}
\setres{lotUL}{DA_post_k21}{surprise}{0.16}
\setres{lotUL}{DA_post_k21}{surprise_mean}{3.049}
\setres{lotUL}{DA_post_k21}{width90_adj_pct}{24.6}
\setres{lotUL}{DA_post_k21}{width90_adj_pct_mean}{25.069}
\setres{lotUL}{DA_post_k21}{width90_pct}{0.4}
\setres{lotUL}{DA_post_k21}{width90_pct_mean}{0.982}
\setres{lotUL}{DA_post_k22}{cov50}{0.00}
\setres{lotUL}{DA_post_k22}{cov50_mean}{0.000}
\setres{lotUL}{DA_post_k22}{cov90}{0.00}
\setres{lotUL}{DA_post_k22}{cov90_adj}{1.00}
\setres{lotUL}{DA_post_k22}{cov90_adj_mean}{0.983}
\setres{lotUL}{DA_post_k22}{cov90_mean}{0.000}
\setres{lotUL}{DA_post_k22}{kappa}{0.341}
\setres{lotUL}{DA_post_k22}{kappa_crps}{0.309}
\setres{lotUL}{DA_post_k22}{kappa_crps_mean}{0.180}
\setres{lotUL}{DA_post_k22}{kappa_mean}{0.320}
\setres{lotUL}{DA_post_k22}{logscore_next}{-3.02}
\setres{lotUL}{DA_post_k22}{logscore_next_mean}{-3.069}
\setres{lotUL}{DA_post_k22}{misfit_pct}{15.0}
\setres{lotUL}{DA_post_k22}{misfit_pct_mean}{14.956}
\setres{lotUL}{DA_post_k22}{n_seen}{22}
\setres{lotUL}{DA_post_k22}{surprise}{0.04}
\setres{lotUL}{DA_post_k22}{surprise_mean}{0.639}
\setres{lotUL}{DA_post_k22}{width90_adj_pct}{24.6}
\setres{lotUL}{DA_post_k22}{width90_adj_pct_mean}{24.618}
\setres{lotUL}{DA_post_k22}{width90_pct}{0.4}
\setres{lotUL}{DA_post_k22}{width90_pct_mean}{0.543}
\setres{lotUL}{DA_post_k23}{cov50}{0.00}
\setres{lotUL}{DA_post_k23}{cov50_mean}{0.000}
\setres{lotUL}{DA_post_k23}{cov90}{0.00}
\setres{lotUL}{DA_post_k23}{cov90_adj}{1.00}
\setres{lotUL}{DA_post_k23}{cov90_adj_mean}{1.000}
\setres{lotUL}{DA_post_k23}{cov90_mean}{0.000}
\setres{lotUL}{DA_post_k23}{kappa}{0.394}
\setres{lotUL}{DA_post_k23}{kappa_crps}{0.294}
\setres{lotUL}{DA_post_k23}{kappa_crps_mean}{0.237}
\setres{lotUL}{DA_post_k23}{kappa_mean}{0.314}
\setres{lotUL}{DA_post_k23}{logscore_next}{-3.10}
\setres{lotUL}{DA_post_k23}{logscore_next_mean}{-3.206}
\setres{lotUL}{DA_post_k23}{misfit_pct}{14.9}
\setres{lotUL}{DA_post_k23}{misfit_pct_mean}{14.892}
\setres{lotUL}{DA_post_k23}{n_seen}{23}
\setres{lotUL}{DA_post_k23}{surprise}{0.22}
\setres{lotUL}{DA_post_k23}{surprise_mean}{5.572}
\setres{lotUL}{DA_post_k23}{width90_adj_pct}{24.6}
\setres{lotUL}{DA_post_k23}{width90_adj_pct_mean}{24.518}
\setres{lotUL}{DA_post_k23}{width90_pct}{0.4}
\setres{lotUL}{DA_post_k23}{width90_pct_mean}{0.525}
\setres{lotUL}{DA_post_k24}{cov50}{0.00}
\setres{lotUL}{DA_post_k24}{cov50_mean}{0.000}
\setres{lotUL}{DA_post_k24}{cov90}{0.00}
\setres{lotUL}{DA_post_k24}{cov90_adj}{1.00}
\setres{lotUL}{DA_post_k24}{cov90_adj_mean}{1.000}
\setres{lotUL}{DA_post_k24}{cov90_mean}{0.000}
\setres{lotUL}{DA_post_k24}{kappa}{0.249}
\setres{lotUL}{DA_post_k24}{kappa_crps}{0.424}
\setres{lotUL}{DA_post_k24}{kappa_crps_mean}{0.353}
\setres{lotUL}{DA_post_k24}{kappa_mean}{0.305}
\setres{lotUL}{DA_post_k24}{logscore_next}{-3.20}
\setres{lotUL}{DA_post_k24}{logscore_next_mean}{-3.544}
\setres{lotUL}{DA_post_k24}{misfit_pct}{14.8}
\setres{lotUL}{DA_post_k24}{misfit_pct_mean}{14.860}
\setres{lotUL}{DA_post_k24}{n_seen}{24}
\setres{lotUL}{DA_post_k24}{surprise}{0.27}
\setres{lotUL}{DA_post_k24}{surprise_mean}{2.482}
\setres{lotUL}{DA_post_k24}{width90_adj_pct}{24.3}
\setres{lotUL}{DA_post_k24}{width90_adj_pct_mean}{24.454}
\setres{lotUL}{DA_post_k24}{width90_pct}{0.2}
\setres{lotUL}{DA_post_k24}{width90_pct_mean}{0.408}
\setres{lotUL}{DA_post_k25}{cov50}{---}
\setres{lotUL}{DA_post_k25}{cov50_mean}{---}
\setres{lotUL}{DA_post_k25}{cov90}{---}
\setres{lotUL}{DA_post_k25}{cov90_adj}{---}
\setres{lotUL}{DA_post_k25}{cov90_adj_mean}{---}
\setres{lotUL}{DA_post_k25}{cov90_mean}{---}
\setres{lotUL}{DA_post_k25}{kappa}{---}
\setres{lotUL}{DA_post_k25}{kappa_crps}{---}
\setres{lotUL}{DA_post_k25}{kappa_crps_mean}{---}
\setres{lotUL}{DA_post_k25}{kappa_mean}{---}
\setres{lotUL}{DA_post_k25}{logscore_next}{---}
\setres{lotUL}{DA_post_k25}{logscore_next_mean}{---}
\setres{lotUL}{DA_post_k25}{misfit_pct}{---}
\setres{lotUL}{DA_post_k25}{misfit_pct_mean}{---}
\setres{lotUL}{DA_post_k25}{n_seen}{25}
\setres{lotUL}{DA_post_k25}{surprise}{---}
\setres{lotUL}{DA_post_k25}{surprise_mean}{---}
\setres{lotUL}{DA_post_k25}{width90_adj_pct}{---}
\setres{lotUL}{DA_post_k25}{width90_adj_pct_mean}{---}
\setres{lotUL}{DA_post_k25}{width90_pct}{---}
\setres{lotUL}{DA_post_k25}{width90_pct_mean}{---}
\setres{lotUL}{DA_post_k3}{cov50}{0.00}
\setres{lotUL}{DA_post_k3}{cov50_mean}{0.023}
\setres{lotUL}{DA_post_k3}{cov90}{0.05}
\setres{lotUL}{DA_post_k3}{cov90_adj}{0.57}
\setres{lotUL}{DA_post_k3}{cov90_adj_mean}{0.611}
\setres{lotUL}{DA_post_k3}{cov90_mean}{0.068}
\setres{lotUL}{DA_post_k3}{kappa}{-0.005}
\setres{lotUL}{DA_post_k3}{kappa_crps}{-0.132}
\setres{lotUL}{DA_post_k3}{kappa_crps_mean}{-0.147}
\setres{lotUL}{DA_post_k3}{kappa_mean}{-0.246}
\setres{lotUL}{DA_post_k3}{logscore_next}{-3.09}
\setres{lotUL}{DA_post_k3}{logscore_next_mean}{-3.157}
\setres{lotUL}{DA_post_k3}{misfit_pct}{12.5}
\setres{lotUL}{DA_post_k3}{misfit_pct_mean}{12.178}
\setres{lotUL}{DA_post_k3}{n_seen}{3}
\setres{lotUL}{DA_post_k3}{surprise}{1.54}
\setres{lotUL}{DA_post_k3}{surprise_mean}{9.053}
\setres{lotUL}{DA_post_k3}{width90_adj_pct}{20.6}
\setres{lotUL}{DA_post_k3}{width90_adj_pct_mean}{20.194}
\setres{lotUL}{DA_post_k3}{width90_pct}{1.2}
\setres{lotUL}{DA_post_k3}{width90_pct_mean}{1.713}
\setres{lotUL}{DA_post_k4}{cov50}{0.00}
\setres{lotUL}{DA_post_k4}{cov50_mean}{0.010}
\setres{lotUL}{DA_post_k4}{cov90}{0.00}
\setres{lotUL}{DA_post_k4}{cov90_adj}{0.83}
\setres{lotUL}{DA_post_k4}{cov90_adj_mean}{0.693}
\setres{lotUL}{DA_post_k4}{cov90_mean}{0.024}
\setres{lotUL}{DA_post_k4}{kappa}{0.323}
\setres{lotUL}{DA_post_k4}{kappa_crps}{0.039}
\setres{lotUL}{DA_post_k4}{kappa_crps_mean}{-0.085}
\setres{lotUL}{DA_post_k4}{kappa_mean}{-0.047}
\setres{lotUL}{DA_post_k4}{logscore_next}{-3.11}
\setres{lotUL}{DA_post_k4}{logscore_next_mean}{-3.177}
\setres{lotUL}{DA_post_k4}{misfit_pct}{13.4}
\setres{lotUL}{DA_post_k4}{misfit_pct_mean}{13.516}
\setres{lotUL}{DA_post_k4}{n_seen}{4}
\setres{lotUL}{DA_post_k4}{surprise}{0.46}
\setres{lotUL}{DA_post_k4}{surprise_mean}{2.700}
\setres{lotUL}{DA_post_k4}{width90_adj_pct}{22.0}
\setres{lotUL}{DA_post_k4}{width90_adj_pct_mean}{22.267}
\setres{lotUL}{DA_post_k4}{width90_pct}{0.5}
\setres{lotUL}{DA_post_k4}{width90_pct_mean}{0.755}
\setres{lotUL}{DA_post_k5}{cov50}{0.00}
\setres{lotUL}{DA_post_k5}{cov50_mean}{0.003}
\setres{lotUL}{DA_post_k5}{cov90}{0.00}
\setres{lotUL}{DA_post_k5}{cov90_adj}{0.80}
\setres{lotUL}{DA_post_k5}{cov90_adj_mean}{0.762}
\setres{lotUL}{DA_post_k5}{cov90_mean}{0.015}
\setres{lotUL}{DA_post_k5}{kappa}{0.249}
\setres{lotUL}{DA_post_k5}{kappa_crps}{0.007}
\setres{lotUL}{DA_post_k5}{kappa_crps_mean}{-0.136}
\setres{lotUL}{DA_post_k5}{kappa_mean}{-0.117}
\setres{lotUL}{DA_post_k5}{logscore_next}{-3.07}
\setres{lotUL}{DA_post_k5}{logscore_next_mean}{-3.198}
\setres{lotUL}{DA_post_k5}{misfit_pct}{15.1}
\setres{lotUL}{DA_post_k5}{misfit_pct_mean}{15.172}
\setres{lotUL}{DA_post_k5}{n_seen}{5}
\setres{lotUL}{DA_post_k5}{surprise}{0.38}
\setres{lotUL}{DA_post_k5}{surprise_mean}{15.138}
\setres{lotUL}{DA_post_k5}{width90_adj_pct}{24.8}
\setres{lotUL}{DA_post_k5}{width90_adj_pct_mean}{24.995}
\setres{lotUL}{DA_post_k5}{width90_pct}{0.5}
\setres{lotUL}{DA_post_k5}{width90_pct_mean}{0.934}
\setres{lotUL}{DA_post_k6}{cov50}{0.00}
\setres{lotUL}{DA_post_k6}{cov50_mean}{0.008}
\setres{lotUL}{DA_post_k6}{cov90}{0.00}
\setres{lotUL}{DA_post_k6}{cov90_adj}{0.92}
\setres{lotUL}{DA_post_k6}{cov90_adj_mean}{0.829}
\setres{lotUL}{DA_post_k6}{cov90_mean}{0.021}
\setres{lotUL}{DA_post_k6}{kappa}{0.431}
\setres{lotUL}{DA_post_k6}{kappa_crps}{0.196}
\setres{lotUL}{DA_post_k6}{kappa_crps_mean}{-0.037}
\setres{lotUL}{DA_post_k6}{kappa_mean}{0.016}
\setres{lotUL}{DA_post_k6}{logscore_next}{-3.04}
\setres{lotUL}{DA_post_k6}{logscore_next_mean}{-3.102}
\setres{lotUL}{DA_post_k6}{misfit_pct}{14.9}
\setres{lotUL}{DA_post_k6}{misfit_pct_mean}{15.285}
\setres{lotUL}{DA_post_k6}{n_seen}{6}
\setres{lotUL}{DA_post_k6}{surprise}{0.05}
\setres{lotUL}{DA_post_k6}{surprise_mean}{5.527}
\setres{lotUL}{DA_post_k6}{width90_adj_pct}{24.5}
\setres{lotUL}{DA_post_k6}{width90_adj_pct_mean}{25.169}
\setres{lotUL}{DA_post_k6}{width90_pct}{0.7}
\setres{lotUL}{DA_post_k6}{width90_pct_mean}{0.766}
\setres{lotUL}{DA_post_k7}{cov50}{0.00}
\setres{lotUL}{DA_post_k7}{cov50_mean}{0.003}
\setres{lotUL}{DA_post_k7}{cov90}{0.00}
\setres{lotUL}{DA_post_k7}{cov90_adj}{0.94}
\setres{lotUL}{DA_post_k7}{cov90_adj_mean}{0.864}
\setres{lotUL}{DA_post_k7}{cov90_mean}{0.019}
\setres{lotUL}{DA_post_k7}{kappa}{0.566}
\setres{lotUL}{DA_post_k7}{kappa_crps}{0.291}
\setres{lotUL}{DA_post_k7}{kappa_crps_mean}{0.015}
\setres{lotUL}{DA_post_k7}{kappa_mean}{0.131}
\setres{lotUL}{DA_post_k7}{logscore_next}{-3.06}
\setres{lotUL}{DA_post_k7}{logscore_next_mean}{-3.198}
\setres{lotUL}{DA_post_k7}{misfit_pct}{14.5}
\setres{lotUL}{DA_post_k7}{misfit_pct_mean}{15.553}
\setres{lotUL}{DA_post_k7}{n_seen}{7}
\setres{lotUL}{DA_post_k7}{surprise}{0.06}
\setres{lotUL}{DA_post_k7}{surprise_mean}{10.568}
\setres{lotUL}{DA_post_k7}{width90_adj_pct}{24.0}
\setres{lotUL}{DA_post_k7}{width90_adj_pct_mean}{25.602}
\setres{lotUL}{DA_post_k7}{width90_pct}{0.3}
\setres{lotUL}{DA_post_k7}{width90_pct_mean}{0.584}
\setres{lotUL}{DA_post_k8}{cov50}{0.00}
\setres{lotUL}{DA_post_k8}{cov50_mean}{0.012}
\setres{lotUL}{DA_post_k8}{cov90}{0.00}
\setres{lotUL}{DA_post_k8}{cov90_adj}{0.94}
\setres{lotUL}{DA_post_k8}{cov90_adj_mean}{0.924}
\setres{lotUL}{DA_post_k8}{cov90_mean}{0.015}
\setres{lotUL}{DA_post_k8}{kappa}{0.564}
\setres{lotUL}{DA_post_k8}{kappa_crps}{0.274}
\setres{lotUL}{DA_post_k8}{kappa_crps_mean}{0.037}
\setres{lotUL}{DA_post_k8}{kappa_mean}{0.160}
\setres{lotUL}{DA_post_k8}{logscore_next}{-3.13}
\setres{lotUL}{DA_post_k8}{logscore_next_mean}{-3.252}
\setres{lotUL}{DA_post_k8}{misfit_pct}{15.5}
\setres{lotUL}{DA_post_k8}{misfit_pct_mean}{16.358}
\setres{lotUL}{DA_post_k8}{n_seen}{8}
\setres{lotUL}{DA_post_k8}{surprise}{0.36}
\setres{lotUL}{DA_post_k8}{surprise_mean}{23.728}
\setres{lotUL}{DA_post_k8}{width90_adj_pct}{25.5}
\setres{lotUL}{DA_post_k8}{width90_adj_pct_mean}{26.926}
\setres{lotUL}{DA_post_k8}{width90_pct}{0.3}
\setres{lotUL}{DA_post_k8}{width90_pct_mean}{0.541}
\setres{lotUL}{DA_post_k9}{cov50}{0.00}
\setres{lotUL}{DA_post_k9}{cov50_mean}{0.006}
\setres{lotUL}{DA_post_k9}{cov90}{0.00}
\setres{lotUL}{DA_post_k9}{cov90_adj}{0.94}
\setres{lotUL}{DA_post_k9}{cov90_adj_mean}{0.912}
\setres{lotUL}{DA_post_k9}{cov90_mean}{0.009}
\setres{lotUL}{DA_post_k9}{kappa}{0.561}
\setres{lotUL}{DA_post_k9}{kappa_crps}{0.269}
\setres{lotUL}{DA_post_k9}{kappa_crps_mean}{-0.054}
\setres{lotUL}{DA_post_k9}{kappa_mean}{-0.142}
\setres{lotUL}{DA_post_k9}{logscore_next}{-3.07}
\setres{lotUL}{DA_post_k9}{logscore_next_mean}{-3.215}
\setres{lotUL}{DA_post_k9}{misfit_pct}{15.6}
\setres{lotUL}{DA_post_k9}{misfit_pct_mean}{16.609}
\setres{lotUL}{DA_post_k9}{n_seen}{9}
\setres{lotUL}{DA_post_k9}{surprise}{0.72}
\setres{lotUL}{DA_post_k9}{surprise_mean}{13.701}
\setres{lotUL}{DA_post_k9}{width90_adj_pct}{25.7}
\setres{lotUL}{DA_post_k9}{width90_adj_pct_mean}{27.335}
\setres{lotUL}{DA_post_k9}{width90_pct}{0.4}
\setres{lotUL}{DA_post_k9}{width90_pct_mean}{0.490}
\setres{lotUL}{EU_post_k0}{cov50}{0.24}
\setres{lotUL}{EU_post_k0}{cov50_mean}{0.240}
\setres{lotUL}{EU_post_k0}{cov90}{0.36}
\setres{lotUL}{EU_post_k0}{cov90_adj}{---}
\setres{lotUL}{EU_post_k0}{cov90_adj_mean}{---}
\setres{lotUL}{EU_post_k0}{cov90_mean}{0.360}
\setres{lotUL}{EU_post_k0}{kappa}{-0.190}
\setres{lotUL}{EU_post_k0}{kappa_crps}{0.231}
\setres{lotUL}{EU_post_k0}{kappa_crps_mean}{0.230}
\setres{lotUL}{EU_post_k0}{kappa_mean}{-0.190}
\setres{lotUL}{EU_post_k0}{logscore_next}{-3.02}
\setres{lotUL}{EU_post_k0}{logscore_next_mean}{-3.188}
\setres{lotUL}{EU_post_k0}{misfit_pct}{---}
\setres{lotUL}{EU_post_k0}{misfit_pct_mean}{---}
\setres{lotUL}{EU_post_k0}{n_seen}{0}
\setres{lotUL}{EU_post_k0}{surprise}{2.17}
\setres{lotUL}{EU_post_k0}{surprise_mean}{2.213}
\setres{lotUL}{EU_post_k0}{width90_adj_pct}{---}
\setres{lotUL}{EU_post_k0}{width90_adj_pct_mean}{---}
\setres{lotUL}{EU_post_k0}{width90_pct}{4.4}
\setres{lotUL}{EU_post_k0}{width90_pct_mean}{4.407}
\setres{lotUL}{EU_post_k1}{cov50}{0.00}
\setres{lotUL}{EU_post_k1}{cov50_mean}{0.021}
\setres{lotUL}{EU_post_k1}{cov90}{0.00}
\setres{lotUL}{EU_post_k1}{cov90_adj}{---}
\setres{lotUL}{EU_post_k1}{cov90_adj_mean}{---}
\setres{lotUL}{EU_post_k1}{cov90_mean}{0.065}
\setres{lotUL}{EU_post_k1}{kappa}{0.463}
\setres{lotUL}{EU_post_k1}{kappa_crps}{0.271}
\setres{lotUL}{EU_post_k1}{kappa_crps_mean}{0.026}
\setres{lotUL}{EU_post_k1}{kappa_mean}{-0.032}
\setres{lotUL}{EU_post_k1}{logscore_next}{-3.01}
\setres{lotUL}{EU_post_k1}{logscore_next_mean}{-3.151}
\setres{lotUL}{EU_post_k1}{misfit_pct}{---}
\setres{lotUL}{EU_post_k1}{misfit_pct_mean}{---}
\setres{lotUL}{EU_post_k1}{n_seen}{1}
\setres{lotUL}{EU_post_k1}{surprise}{0.18}
\setres{lotUL}{EU_post_k1}{surprise_mean}{2.533}
\setres{lotUL}{EU_post_k1}{width90_adj_pct}{---}
\setres{lotUL}{EU_post_k1}{width90_adj_pct_mean}{---}
\setres{lotUL}{EU_post_k1}{width90_pct}{0.6}
\setres{lotUL}{EU_post_k1}{width90_pct_mean}{1.057}
\setres{lotUL}{EU_post_k10}{cov50}{0.00}
\setres{lotUL}{EU_post_k10}{cov50_mean}{0.003}
\setres{lotUL}{EU_post_k10}{cov90}{0.00}
\setres{lotUL}{EU_post_k10}{cov90_adj}{1.00}
\setres{lotUL}{EU_post_k10}{cov90_adj_mean}{0.963}
\setres{lotUL}{EU_post_k10}{cov90_mean}{0.007}
\setres{lotUL}{EU_post_k10}{kappa}{0.538}
\setres{lotUL}{EU_post_k10}{kappa_crps}{0.313}
\setres{lotUL}{EU_post_k10}{kappa_crps_mean}{0.245}
\setres{lotUL}{EU_post_k10}{kappa_mean}{0.434}
\setres{lotUL}{EU_post_k10}{logscore_next}{-3.01}
\setres{lotUL}{EU_post_k10}{logscore_next_mean}{-3.053}
\setres{lotUL}{EU_post_k10}{misfit_pct}{15.0}
\setres{lotUL}{EU_post_k10}{misfit_pct_mean}{15.569}
\setres{lotUL}{EU_post_k10}{n_seen}{10}
\setres{lotUL}{EU_post_k10}{surprise}{0.01}
\setres{lotUL}{EU_post_k10}{surprise_mean}{0.031}
\setres{lotUL}{EU_post_k10}{width90_adj_pct}{24.7}
\setres{lotUL}{EU_post_k10}{width90_adj_pct_mean}{25.612}
\setres{lotUL}{EU_post_k10}{width90_pct}{0.2}
\setres{lotUL}{EU_post_k10}{width90_pct_mean}{0.200}
\setres{lotUL}{EU_post_k11}{cov50}{0.00}
\setres{lotUL}{EU_post_k11}{cov50_mean}{0.000}
\setres{lotUL}{EU_post_k11}{cov90}{0.00}
\setres{lotUL}{EU_post_k11}{cov90_adj}{1.00}
\setres{lotUL}{EU_post_k11}{cov90_adj_mean}{0.968}
\setres{lotUL}{EU_post_k11}{cov90_mean}{0.007}
\setres{lotUL}{EU_post_k11}{kappa}{0.542}
\setres{lotUL}{EU_post_k11}{kappa_crps}{0.296}
\setres{lotUL}{EU_post_k11}{kappa_crps_mean}{0.252}
\setres{lotUL}{EU_post_k11}{kappa_mean}{0.437}
\setres{lotUL}{EU_post_k11}{logscore_next}{-3.06}
\setres{lotUL}{EU_post_k11}{logscore_next_mean}{-3.435}
\setres{lotUL}{EU_post_k11}{misfit_pct}{15.4}
\setres{lotUL}{EU_post_k11}{misfit_pct_mean}{15.531}
\setres{lotUL}{EU_post_k11}{n_seen}{11}
\setres{lotUL}{EU_post_k11}{surprise}{0.06}
\setres{lotUL}{EU_post_k11}{surprise_mean}{1.688}
\setres{lotUL}{EU_post_k11}{width90_adj_pct}{25.4}
\setres{lotUL}{EU_post_k11}{width90_adj_pct_mean}{25.550}
\setres{lotUL}{EU_post_k11}{width90_pct}{0.2}
\setres{lotUL}{EU_post_k11}{width90_pct_mean}{0.206}
\setres{lotUL}{EU_post_k12}{cov50}{0.00}
\setres{lotUL}{EU_post_k12}{cov50_mean}{0.000}
\setres{lotUL}{EU_post_k12}{cov90}{0.00}
\setres{lotUL}{EU_post_k12}{cov90_adj}{1.00}
\setres{lotUL}{EU_post_k12}{cov90_adj_mean}{0.965}
\setres{lotUL}{EU_post_k12}{cov90_mean}{0.008}
\setres{lotUL}{EU_post_k12}{kappa}{0.473}
\setres{lotUL}{EU_post_k12}{kappa_crps}{0.311}
\setres{lotUL}{EU_post_k12}{kappa_crps_mean}{0.262}
\setres{lotUL}{EU_post_k12}{kappa_mean}{0.449}
\setres{lotUL}{EU_post_k12}{logscore_next}{-3.09}
\setres{lotUL}{EU_post_k12}{logscore_next_mean}{-3.152}
\setres{lotUL}{EU_post_k12}{misfit_pct}{15.3}
\setres{lotUL}{EU_post_k12}{misfit_pct_mean}{15.175}
\setres{lotUL}{EU_post_k12}{n_seen}{12}
\setres{lotUL}{EU_post_k12}{surprise}{0.03}
\setres{lotUL}{EU_post_k12}{surprise_mean}{0.342}
\setres{lotUL}{EU_post_k12}{width90_adj_pct}{25.1}
\setres{lotUL}{EU_post_k12}{width90_adj_pct_mean}{24.964}
\setres{lotUL}{EU_post_k12}{width90_pct}{0.2}
\setres{lotUL}{EU_post_k12}{width90_pct_mean}{0.201}
\setres{lotUL}{EU_post_k13}{cov50}{0.00}
\setres{lotUL}{EU_post_k13}{cov50_mean}{0.000}
\setres{lotUL}{EU_post_k13}{cov90}{0.00}
\setres{lotUL}{EU_post_k13}{cov90_adj}{1.00}
\setres{lotUL}{EU_post_k13}{cov90_adj_mean}{0.971}
\setres{lotUL}{EU_post_k13}{cov90_mean}{0.004}
\setres{lotUL}{EU_post_k13}{kappa}{0.474}
\setres{lotUL}{EU_post_k13}{kappa_crps}{0.294}
\setres{lotUL}{EU_post_k13}{kappa_crps_mean}{0.259}
\setres{lotUL}{EU_post_k13}{kappa_mean}{0.433}
\setres{lotUL}{EU_post_k13}{logscore_next}{-3.11}
\setres{lotUL}{EU_post_k13}{logscore_next_mean}{-3.259}
\setres{lotUL}{EU_post_k13}{misfit_pct}{15.5}
\setres{lotUL}{EU_post_k13}{misfit_pct_mean}{15.267}
\setres{lotUL}{EU_post_k13}{n_seen}{13}
\setres{lotUL}{EU_post_k13}{surprise}{0.03}
\setres{lotUL}{EU_post_k13}{surprise_mean}{3.044}
\setres{lotUL}{EU_post_k13}{width90_adj_pct}{25.5}
\setres{lotUL}{EU_post_k13}{width90_adj_pct_mean}{25.116}
\setres{lotUL}{EU_post_k13}{width90_pct}{0.2}
\setres{lotUL}{EU_post_k13}{width90_pct_mean}{0.233}
\setres{lotUL}{EU_post_k14}{cov50}{0.00}
\setres{lotUL}{EU_post_k14}{cov50_mean}{0.000}
\setres{lotUL}{EU_post_k14}{cov90}{0.00}
\setres{lotUL}{EU_post_k14}{cov90_adj}{1.00}
\setres{lotUL}{EU_post_k14}{cov90_adj_mean}{0.950}
\setres{lotUL}{EU_post_k14}{cov90_mean}{0.000}
\setres{lotUL}{EU_post_k14}{kappa}{0.452}
\setres{lotUL}{EU_post_k14}{kappa_crps}{0.247}
\setres{lotUL}{EU_post_k14}{kappa_crps_mean}{0.206}
\setres{lotUL}{EU_post_k14}{kappa_mean}{0.377}
\setres{lotUL}{EU_post_k14}{logscore_next}{-3.10}
\setres{lotUL}{EU_post_k14}{logscore_next_mean}{-3.312}
\setres{lotUL}{EU_post_k14}{misfit_pct}{15.5}
\setres{lotUL}{EU_post_k14}{misfit_pct_mean}{15.316}
\setres{lotUL}{EU_post_k14}{n_seen}{14}
\setres{lotUL}{EU_post_k14}{surprise}{0.02}
\setres{lotUL}{EU_post_k14}{surprise_mean}{3.700}
\setres{lotUL}{EU_post_k14}{width90_adj_pct}{25.6}
\setres{lotUL}{EU_post_k14}{width90_adj_pct_mean}{25.195}
\setres{lotUL}{EU_post_k14}{width90_pct}{0.2}
\setres{lotUL}{EU_post_k14}{width90_pct_mean}{0.189}
\setres{lotUL}{EU_post_k15}{cov50}{0.00}
\setres{lotUL}{EU_post_k15}{cov50_mean}{0.005}
\setres{lotUL}{EU_post_k15}{cov90}{0.00}
\setres{lotUL}{EU_post_k15}{cov90_adj}{1.00}
\setres{lotUL}{EU_post_k15}{cov90_adj_mean}{0.960}
\setres{lotUL}{EU_post_k15}{cov90_mean}{0.005}
\setres{lotUL}{EU_post_k15}{kappa}{0.436}
\setres{lotUL}{EU_post_k15}{kappa_crps}{0.282}
\setres{lotUL}{EU_post_k15}{kappa_crps_mean}{0.235}
\setres{lotUL}{EU_post_k15}{kappa_mean}{0.411}
\setres{lotUL}{EU_post_k15}{logscore_next}{-3.05}
\setres{lotUL}{EU_post_k15}{logscore_next_mean}{-3.151}
\setres{lotUL}{EU_post_k15}{misfit_pct}{15.5}
\setres{lotUL}{EU_post_k15}{misfit_pct_mean}{15.273}
\setres{lotUL}{EU_post_k15}{n_seen}{15}
\setres{lotUL}{EU_post_k15}{surprise}{0.02}
\setres{lotUL}{EU_post_k15}{surprise_mean}{2.001}
\setres{lotUL}{EU_post_k15}{width90_adj_pct}{25.4}
\setres{lotUL}{EU_post_k15}{width90_adj_pct_mean}{25.125}
\setres{lotUL}{EU_post_k15}{width90_pct}{0.2}
\setres{lotUL}{EU_post_k15}{width90_pct_mean}{0.196}
\setres{lotUL}{EU_post_k16}{cov50}{0.00}
\setres{lotUL}{EU_post_k16}{cov50_mean}{0.006}
\setres{lotUL}{EU_post_k16}{cov90}{0.00}
\setres{lotUL}{EU_post_k16}{cov90_adj}{1.00}
\setres{lotUL}{EU_post_k16}{cov90_adj_mean}{0.956}
\setres{lotUL}{EU_post_k16}{cov90_mean}{0.006}
\setres{lotUL}{EU_post_k16}{kappa}{0.466}
\setres{lotUL}{EU_post_k16}{kappa_crps}{0.281}
\setres{lotUL}{EU_post_k16}{kappa_crps_mean}{0.237}
\setres{lotUL}{EU_post_k16}{kappa_mean}{0.408}
\setres{lotUL}{EU_post_k16}{logscore_next}{-3.00}
\setres{lotUL}{EU_post_k16}{logscore_next_mean}{-3.178}
\setres{lotUL}{EU_post_k16}{misfit_pct}{15.4}
\setres{lotUL}{EU_post_k16}{misfit_pct_mean}{15.359}
\setres{lotUL}{EU_post_k16}{n_seen}{16}
\setres{lotUL}{EU_post_k16}{surprise}{0.01}
\setres{lotUL}{EU_post_k16}{surprise_mean}{0.140}
\setres{lotUL}{EU_post_k16}{width90_adj_pct}{25.4}
\setres{lotUL}{EU_post_k16}{width90_adj_pct_mean}{25.267}
\setres{lotUL}{EU_post_k16}{width90_pct}{0.1}
\setres{lotUL}{EU_post_k16}{width90_pct_mean}{0.191}
\setres{lotUL}{EU_post_k17}{cov50}{0.00}
\setres{lotUL}{EU_post_k17}{cov50_mean}{0.006}
\setres{lotUL}{EU_post_k17}{cov90}{0.00}
\setres{lotUL}{EU_post_k17}{cov90_adj}{1.00}
\setres{lotUL}{EU_post_k17}{cov90_adj_mean}{0.963}
\setres{lotUL}{EU_post_k17}{cov90_mean}{0.013}
\setres{lotUL}{EU_post_k17}{kappa}{0.408}
\setres{lotUL}{EU_post_k17}{kappa_crps}{0.302}
\setres{lotUL}{EU_post_k17}{kappa_crps_mean}{0.234}
\setres{lotUL}{EU_post_k17}{kappa_mean}{0.390}
\setres{lotUL}{EU_post_k17}{logscore_next}{-3.10}
\setres{lotUL}{EU_post_k17}{logscore_next_mean}{-3.253}
\setres{lotUL}{EU_post_k17}{misfit_pct}{15.4}
\setres{lotUL}{EU_post_k17}{misfit_pct_mean}{15.281}
\setres{lotUL}{EU_post_k17}{n_seen}{17}
\setres{lotUL}{EU_post_k17}{surprise}{0.04}
\setres{lotUL}{EU_post_k17}{surprise_mean}{0.738}
\setres{lotUL}{EU_post_k17}{width90_adj_pct}{25.3}
\setres{lotUL}{EU_post_k17}{width90_adj_pct_mean}{25.139}
\setres{lotUL}{EU_post_k17}{width90_pct}{0.1}
\setres{lotUL}{EU_post_k17}{width90_pct_mean}{0.198}
\setres{lotUL}{EU_post_k18}{cov50}{0.00}
\setres{lotUL}{EU_post_k18}{cov50_mean}{0.000}
\setres{lotUL}{EU_post_k18}{cov90}{0.00}
\setres{lotUL}{EU_post_k18}{cov90_adj}{1.00}
\setres{lotUL}{EU_post_k18}{cov90_adj_mean}{0.971}
\setres{lotUL}{EU_post_k18}{cov90_mean}{0.000}
\setres{lotUL}{EU_post_k18}{kappa}{0.398}
\setres{lotUL}{EU_post_k18}{kappa_crps}{0.260}
\setres{lotUL}{EU_post_k18}{kappa_crps_mean}{0.216}
\setres{lotUL}{EU_post_k18}{kappa_mean}{0.350}
\setres{lotUL}{EU_post_k18}{logscore_next}{-3.06}
\setres{lotUL}{EU_post_k18}{logscore_next_mean}{-3.125}
\setres{lotUL}{EU_post_k18}{misfit_pct}{15.4}
\setres{lotUL}{EU_post_k18}{misfit_pct_mean}{15.298}
\setres{lotUL}{EU_post_k18}{n_seen}{18}
\setres{lotUL}{EU_post_k18}{surprise}{0.01}
\setres{lotUL}{EU_post_k18}{surprise_mean}{1.110}
\setres{lotUL}{EU_post_k18}{width90_adj_pct}{25.3}
\setres{lotUL}{EU_post_k18}{width90_adj_pct_mean}{25.166}
\setres{lotUL}{EU_post_k18}{width90_pct}{0.1}
\setres{lotUL}{EU_post_k18}{width90_pct_mean}{0.172}
\setres{lotUL}{EU_post_k19}{cov50}{0.00}
\setres{lotUL}{EU_post_k19}{cov50_mean}{0.000}
\setres{lotUL}{EU_post_k19}{cov90}{0.00}
\setres{lotUL}{EU_post_k19}{cov90_adj}{1.00}
\setres{lotUL}{EU_post_k19}{cov90_adj_mean}{0.983}
\setres{lotUL}{EU_post_k19}{cov90_mean}{0.000}
\setres{lotUL}{EU_post_k19}{kappa}{0.418}
\setres{lotUL}{EU_post_k19}{kappa_crps}{0.318}
\setres{lotUL}{EU_post_k19}{kappa_crps_mean}{0.206}
\setres{lotUL}{EU_post_k19}{kappa_mean}{0.360}
\setres{lotUL}{EU_post_k19}{logscore_next}{-3.12}
\setres{lotUL}{EU_post_k19}{logscore_next_mean}{-3.258}
\setres{lotUL}{EU_post_k19}{misfit_pct}{15.4}
\setres{lotUL}{EU_post_k19}{misfit_pct_mean}{15.398}
\setres{lotUL}{EU_post_k19}{n_seen}{19}
\setres{lotUL}{EU_post_k19}{surprise}{0.01}
\setres{lotUL}{EU_post_k19}{surprise_mean}{1.848}
\setres{lotUL}{EU_post_k19}{width90_adj_pct}{25.3}
\setres{lotUL}{EU_post_k19}{width90_adj_pct_mean}{25.331}
\setres{lotUL}{EU_post_k19}{width90_pct}{0.1}
\setres{lotUL}{EU_post_k19}{width90_pct_mean}{0.173}
\setres{lotUL}{EU_post_k2}{cov50}{0.00}
\setres{lotUL}{EU_post_k2}{cov50_mean}{0.004}
\setres{lotUL}{EU_post_k2}{cov90}{0.00}
\setres{lotUL}{EU_post_k2}{cov90_adj}{0.93}
\setres{lotUL}{EU_post_k2}{cov90_adj_mean}{0.750}
\setres{lotUL}{EU_post_k2}{cov90_mean}{0.022}
\setres{lotUL}{EU_post_k2}{kappa}{0.514}
\setres{lotUL}{EU_post_k2}{kappa_crps}{0.291}
\setres{lotUL}{EU_post_k2}{kappa_crps_mean}{0.194}
\setres{lotUL}{EU_post_k2}{kappa_mean}{0.340}
\setres{lotUL}{EU_post_k2}{logscore_next}{-3.01}
\setres{lotUL}{EU_post_k2}{logscore_next_mean}{-3.057}
\setres{lotUL}{EU_post_k2}{misfit_pct}{13.6}
\setres{lotUL}{EU_post_k2}{misfit_pct_mean}{12.865}
\setres{lotUL}{EU_post_k2}{n_seen}{2}
\setres{lotUL}{EU_post_k2}{surprise}{0.01}
\setres{lotUL}{EU_post_k2}{surprise_mean}{3.390}
\setres{lotUL}{EU_post_k2}{width90_adj_pct}{22.4}
\setres{lotUL}{EU_post_k2}{width90_adj_pct_mean}{21.178}
\setres{lotUL}{EU_post_k2}{width90_pct}{0.3}
\setres{lotUL}{EU_post_k2}{width90_pct_mean}{0.520}
\setres{lotUL}{EU_post_k20}{cov50}{0.00}
\setres{lotUL}{EU_post_k20}{cov50_mean}{0.000}
\setres{lotUL}{EU_post_k20}{cov90}{0.00}
\setres{lotUL}{EU_post_k20}{cov90_adj}{1.00}
\setres{lotUL}{EU_post_k20}{cov90_adj_mean}{0.980}
\setres{lotUL}{EU_post_k20}{cov90_mean}{0.000}
\setres{lotUL}{EU_post_k20}{kappa}{0.351}
\setres{lotUL}{EU_post_k20}{kappa_crps}{0.279}
\setres{lotUL}{EU_post_k20}{kappa_crps_mean}{0.180}
\setres{lotUL}{EU_post_k20}{kappa_mean}{0.253}
\setres{lotUL}{EU_post_k20}{logscore_next}{-3.07}
\setres{lotUL}{EU_post_k20}{logscore_next_mean}{-3.199}
\setres{lotUL}{EU_post_k20}{misfit_pct}{15.4}
\setres{lotUL}{EU_post_k20}{misfit_pct_mean}{15.364}
\setres{lotUL}{EU_post_k20}{n_seen}{20}
\setres{lotUL}{EU_post_k20}{surprise}{0.02}
\setres{lotUL}{EU_post_k20}{surprise_mean}{0.371}
\setres{lotUL}{EU_post_k20}{width90_adj_pct}{25.3}
\setres{lotUL}{EU_post_k20}{width90_adj_pct_mean}{25.274}
\setres{lotUL}{EU_post_k20}{width90_pct}{0.1}
\setres{lotUL}{EU_post_k20}{width90_pct_mean}{0.139}
\setres{lotUL}{EU_post_k21}{cov50}{0.00}
\setres{lotUL}{EU_post_k21}{cov50_mean}{0.000}
\setres{lotUL}{EU_post_k21}{cov90}{0.00}
\setres{lotUL}{EU_post_k21}{cov90_adj}{1.00}
\setres{lotUL}{EU_post_k21}{cov90_adj_mean}{0.975}
\setres{lotUL}{EU_post_k21}{cov90_mean}{0.000}
\setres{lotUL}{EU_post_k21}{kappa}{0.210}
\setres{lotUL}{EU_post_k21}{kappa_crps}{0.274}
\setres{lotUL}{EU_post_k21}{kappa_crps_mean}{0.165}
\setres{lotUL}{EU_post_k21}{kappa_mean}{0.239}
\setres{lotUL}{EU_post_k21}{logscore_next}{-3.06}
\setres{lotUL}{EU_post_k21}{logscore_next_mean}{-3.092}
\setres{lotUL}{EU_post_k21}{misfit_pct}{15.4}
\setres{lotUL}{EU_post_k21}{misfit_pct_mean}{15.349}
\setres{lotUL}{EU_post_k21}{n_seen}{21}
\setres{lotUL}{EU_post_k21}{surprise}{0.01}
\setres{lotUL}{EU_post_k21}{surprise_mean}{0.107}
\setres{lotUL}{EU_post_k21}{width90_adj_pct}{25.3}
\setres{lotUL}{EU_post_k21}{width90_adj_pct_mean}{25.249}
\setres{lotUL}{EU_post_k21}{width90_pct}{0.1}
\setres{lotUL}{EU_post_k21}{width90_pct_mean}{0.148}
\setres{lotUL}{EU_post_k22}{cov50}{0.00}
\setres{lotUL}{EU_post_k22}{cov50_mean}{0.000}
\setres{lotUL}{EU_post_k22}{cov90}{0.00}
\setres{lotUL}{EU_post_k22}{cov90_adj}{1.00}
\setres{lotUL}{EU_post_k22}{cov90_adj_mean}{0.983}
\setres{lotUL}{EU_post_k22}{cov90_mean}{0.000}
\setres{lotUL}{EU_post_k22}{kappa}{0.297}
\setres{lotUL}{EU_post_k22}{kappa_crps}{0.294}
\setres{lotUL}{EU_post_k22}{kappa_crps_mean}{0.150}
\setres{lotUL}{EU_post_k22}{kappa_mean}{0.265}
\setres{lotUL}{EU_post_k22}{logscore_next}{-3.04}
\setres{lotUL}{EU_post_k22}{logscore_next_mean}{-3.078}
\setres{lotUL}{EU_post_k22}{misfit_pct}{15.5}
\setres{lotUL}{EU_post_k22}{misfit_pct_mean}{15.401}
\setres{lotUL}{EU_post_k22}{n_seen}{22}
\setres{lotUL}{EU_post_k22}{surprise}{0.01}
\setres{lotUL}{EU_post_k22}{surprise_mean}{0.084}
\setres{lotUL}{EU_post_k22}{width90_adj_pct}{25.4}
\setres{lotUL}{EU_post_k22}{width90_adj_pct_mean}{25.336}
\setres{lotUL}{EU_post_k22}{width90_pct}{0.1}
\setres{lotUL}{EU_post_k22}{width90_pct_mean}{0.147}
\setres{lotUL}{EU_post_k23}{cov50}{0.00}
\setres{lotUL}{EU_post_k23}{cov50_mean}{0.000}
\setres{lotUL}{EU_post_k23}{cov90}{0.00}
\setres{lotUL}{EU_post_k23}{cov90_adj}{1.00}
\setres{lotUL}{EU_post_k23}{cov90_adj_mean}{1.000}
\setres{lotUL}{EU_post_k23}{cov90_mean}{0.000}
\setres{lotUL}{EU_post_k23}{kappa}{0.367}
\setres{lotUL}{EU_post_k23}{kappa_crps}{0.392}
\setres{lotUL}{EU_post_k23}{kappa_crps_mean}{0.224}
\setres{lotUL}{EU_post_k23}{kappa_mean}{0.263}
\setres{lotUL}{EU_post_k23}{logscore_next}{-3.11}
\setres{lotUL}{EU_post_k23}{logscore_next_mean}{-3.200}
\setres{lotUL}{EU_post_k23}{misfit_pct}{15.4}
\setres{lotUL}{EU_post_k23}{misfit_pct_mean}{15.367}
\setres{lotUL}{EU_post_k23}{n_seen}{23}
\setres{lotUL}{EU_post_k23}{surprise}{0.01}
\setres{lotUL}{EU_post_k23}{surprise_mean}{5.246}
\setres{lotUL}{EU_post_k23}{width90_adj_pct}{25.3}
\setres{lotUL}{EU_post_k23}{width90_adj_pct_mean}{25.282}
\setres{lotUL}{EU_post_k23}{width90_pct}{0.1}
\setres{lotUL}{EU_post_k23}{width90_pct_mean}{0.218}
\setres{lotUL}{EU_post_k24}{cov50}{0.00}
\setres{lotUL}{EU_post_k24}{cov50_mean}{0.000}
\setres{lotUL}{EU_post_k24}{cov90}{0.00}
\setres{lotUL}{EU_post_k24}{cov90_adj}{1.00}
\setres{lotUL}{EU_post_k24}{cov90_adj_mean}{1.000}
\setres{lotUL}{EU_post_k24}{cov90_mean}{0.000}
\setres{lotUL}{EU_post_k24}{kappa}{0.244}
\setres{lotUL}{EU_post_k24}{kappa_crps}{0.429}
\setres{lotUL}{EU_post_k24}{kappa_crps_mean}{0.273}
\setres{lotUL}{EU_post_k24}{kappa_mean}{0.120}
\setres{lotUL}{EU_post_k24}{logscore_next}{-3.24}
\setres{lotUL}{EU_post_k24}{logscore_next_mean}{-3.539}
\setres{lotUL}{EU_post_k24}{misfit_pct}{15.4}
\setres{lotUL}{EU_post_k24}{misfit_pct_mean}{15.391}
\setres{lotUL}{EU_post_k24}{n_seen}{24}
\setres{lotUL}{EU_post_k24}{surprise}{0.05}
\setres{lotUL}{EU_post_k24}{surprise_mean}{2.642}
\setres{lotUL}{EU_post_k24}{width90_adj_pct}{25.3}
\setres{lotUL}{EU_post_k24}{width90_adj_pct_mean}{25.319}
\setres{lotUL}{EU_post_k24}{width90_pct}{0.1}
\setres{lotUL}{EU_post_k24}{width90_pct_mean}{0.182}
\setres{lotUL}{EU_post_k25}{cov50}{---}
\setres{lotUL}{EU_post_k25}{cov50_mean}{---}
\setres{lotUL}{EU_post_k25}{cov90}{---}
\setres{lotUL}{EU_post_k25}{cov90_adj}{---}
\setres{lotUL}{EU_post_k25}{cov90_adj_mean}{---}
\setres{lotUL}{EU_post_k25}{cov90_mean}{---}
\setres{lotUL}{EU_post_k25}{kappa}{---}
\setres{lotUL}{EU_post_k25}{kappa_crps}{---}
\setres{lotUL}{EU_post_k25}{kappa_crps_mean}{---}
\setres{lotUL}{EU_post_k25}{kappa_mean}{---}
\setres{lotUL}{EU_post_k25}{logscore_next}{---}
\setres{lotUL}{EU_post_k25}{logscore_next_mean}{---}
\setres{lotUL}{EU_post_k25}{misfit_pct}{---}
\setres{lotUL}{EU_post_k25}{misfit_pct_mean}{---}
\setres{lotUL}{EU_post_k25}{n_seen}{25}
\setres{lotUL}{EU_post_k25}{surprise}{---}
\setres{lotUL}{EU_post_k25}{surprise_mean}{---}
\setres{lotUL}{EU_post_k25}{width90_adj_pct}{---}
\setres{lotUL}{EU_post_k25}{width90_adj_pct_mean}{---}
\setres{lotUL}{EU_post_k25}{width90_pct}{---}
\setres{lotUL}{EU_post_k25}{width90_pct_mean}{---}
\setres{lotUL}{EU_post_k3}{cov50}{0.00}
\setres{lotUL}{EU_post_k3}{cov50_mean}{0.002}
\setres{lotUL}{EU_post_k3}{cov90}{0.00}
\setres{lotUL}{EU_post_k3}{cov90_adj}{0.89}
\setres{lotUL}{EU_post_k3}{cov90_adj_mean}{0.816}
\setres{lotUL}{EU_post_k3}{cov90_mean}{0.009}
\setres{lotUL}{EU_post_k3}{kappa}{0.481}
\setres{lotUL}{EU_post_k3}{kappa_crps}{0.266}
\setres{lotUL}{EU_post_k3}{kappa_crps_mean}{0.114}
\setres{lotUL}{EU_post_k3}{kappa_mean}{0.198}
\setres{lotUL}{EU_post_k3}{logscore_next}{-3.14}
\setres{lotUL}{EU_post_k3}{logscore_next_mean}{-3.280}
\setres{lotUL}{EU_post_k3}{misfit_pct}{13.2}
\setres{lotUL}{EU_post_k3}{misfit_pct_mean}{13.896}
\setres{lotUL}{EU_post_k3}{n_seen}{3}
\setres{lotUL}{EU_post_k3}{surprise}{0.07}
\setres{lotUL}{EU_post_k3}{surprise_mean}{10.802}
\setres{lotUL}{EU_post_k3}{width90_adj_pct}{21.8}
\setres{lotUL}{EU_post_k3}{width90_adj_pct_mean}{22.867}
\setres{lotUL}{EU_post_k3}{width90_pct}{0.2}
\setres{lotUL}{EU_post_k3}{width90_pct_mean}{0.351}
\setres{lotUL}{EU_post_k4}{cov50}{0.00}
\setres{lotUL}{EU_post_k4}{cov50_mean}{0.005}
\setres{lotUL}{EU_post_k4}{cov90}{0.00}
\setres{lotUL}{EU_post_k4}{cov90_adj}{0.95}
\setres{lotUL}{EU_post_k4}{cov90_adj_mean}{0.890}
\setres{lotUL}{EU_post_k4}{cov90_mean}{0.012}
\setres{lotUL}{EU_post_k4}{kappa}{0.491}
\setres{lotUL}{EU_post_k4}{kappa_crps}{0.252}
\setres{lotUL}{EU_post_k4}{kappa_crps_mean}{0.147}
\setres{lotUL}{EU_post_k4}{kappa_mean}{0.289}
\setres{lotUL}{EU_post_k4}{logscore_next}{-3.15}
\setres{lotUL}{EU_post_k4}{logscore_next_mean}{-3.225}
\setres{lotUL}{EU_post_k4}{misfit_pct}{14.6}
\setres{lotUL}{EU_post_k4}{misfit_pct_mean}{14.708}
\setres{lotUL}{EU_post_k4}{n_seen}{4}
\setres{lotUL}{EU_post_k4}{surprise}{0.03}
\setres{lotUL}{EU_post_k4}{surprise_mean}{0.168}
\setres{lotUL}{EU_post_k4}{width90_adj_pct}{24.1}
\setres{lotUL}{EU_post_k4}{width90_adj_pct_mean}{24.199}
\setres{lotUL}{EU_post_k4}{width90_pct}{0.2}
\setres{lotUL}{EU_post_k4}{width90_pct_mean}{0.262}
\setres{lotUL}{EU_post_k5}{cov50}{0.00}
\setres{lotUL}{EU_post_k5}{cov50_mean}{0.003}
\setres{lotUL}{EU_post_k5}{cov90}{0.00}
\setres{lotUL}{EU_post_k5}{cov90_adj}{0.97}
\setres{lotUL}{EU_post_k5}{cov90_adj_mean}{0.917}
\setres{lotUL}{EU_post_k5}{cov90_mean}{0.005}
\setres{lotUL}{EU_post_k5}{kappa}{0.475}
\setres{lotUL}{EU_post_k5}{kappa_crps}{0.266}
\setres{lotUL}{EU_post_k5}{kappa_crps_mean}{0.140}
\setres{lotUL}{EU_post_k5}{kappa_mean}{0.276}
\setres{lotUL}{EU_post_k5}{logscore_next}{-3.12}
\setres{lotUL}{EU_post_k5}{logscore_next_mean}{-3.148}
\setres{lotUL}{EU_post_k5}{misfit_pct}{14.8}
\setres{lotUL}{EU_post_k5}{misfit_pct_mean}{15.346}
\setres{lotUL}{EU_post_k5}{n_seen}{5}
\setres{lotUL}{EU_post_k5}{surprise}{0.03}
\setres{lotUL}{EU_post_k5}{surprise_mean}{12.235}
\setres{lotUL}{EU_post_k5}{width90_adj_pct}{24.4}
\setres{lotUL}{EU_post_k5}{width90_adj_pct_mean}{25.246}
\setres{lotUL}{EU_post_k5}{width90_pct}{0.2}
\setres{lotUL}{EU_post_k5}{width90_pct_mean}{0.245}
\setres{lotUL}{EU_post_k6}{cov50}{0.00}
\setres{lotUL}{EU_post_k6}{cov50_mean}{0.008}
\setres{lotUL}{EU_post_k6}{cov90}{0.00}
\setres{lotUL}{EU_post_k6}{cov90_adj}{0.97}
\setres{lotUL}{EU_post_k6}{cov90_adj_mean}{0.953}
\setres{lotUL}{EU_post_k6}{cov90_mean}{0.016}
\setres{lotUL}{EU_post_k6}{kappa}{0.508}
\setres{lotUL}{EU_post_k6}{kappa_crps}{0.278}
\setres{lotUL}{EU_post_k6}{kappa_crps_mean}{0.247}
\setres{lotUL}{EU_post_k6}{kappa_mean}{0.440}
\setres{lotUL}{EU_post_k6}{logscore_next}{-3.06}
\setres{lotUL}{EU_post_k6}{logscore_next_mean}{-3.108}
\setres{lotUL}{EU_post_k6}{misfit_pct}{15.2}
\setres{lotUL}{EU_post_k6}{misfit_pct_mean}{15.374}
\setres{lotUL}{EU_post_k6}{n_seen}{6}
\setres{lotUL}{EU_post_k6}{surprise}{0.02}
\setres{lotUL}{EU_post_k6}{surprise_mean}{5.698}
\setres{lotUL}{EU_post_k6}{width90_adj_pct}{25.1}
\setres{lotUL}{EU_post_k6}{width90_adj_pct_mean}{25.294}
\setres{lotUL}{EU_post_k6}{width90_pct}{0.2}
\setres{lotUL}{EU_post_k6}{width90_pct_mean}{0.267}
\setres{lotUL}{EU_post_k7}{cov50}{0.00}
\setres{lotUL}{EU_post_k7}{cov50_mean}{0.003}
\setres{lotUL}{EU_post_k7}{cov90}{0.00}
\setres{lotUL}{EU_post_k7}{cov90_adj}{0.94}
\setres{lotUL}{EU_post_k7}{cov90_adj_mean}{0.944}
\setres{lotUL}{EU_post_k7}{cov90_mean}{0.006}
\setres{lotUL}{EU_post_k7}{kappa}{0.566}
\setres{lotUL}{EU_post_k7}{kappa_crps}{0.303}
\setres{lotUL}{EU_post_k7}{kappa_crps_mean}{0.248}
\setres{lotUL}{EU_post_k7}{kappa_mean}{0.444}
\setres{lotUL}{EU_post_k7}{logscore_next}{-3.01}
\setres{lotUL}{EU_post_k7}{logscore_next_mean}{-3.201}
\setres{lotUL}{EU_post_k7}{misfit_pct}{15.0}
\setres{lotUL}{EU_post_k7}{misfit_pct_mean}{15.223}
\setres{lotUL}{EU_post_k7}{n_seen}{7}
\setres{lotUL}{EU_post_k7}{surprise}{0.01}
\setres{lotUL}{EU_post_k7}{surprise_mean}{0.212}
\setres{lotUL}{EU_post_k7}{width90_adj_pct}{24.6}
\setres{lotUL}{EU_post_k7}{width90_adj_pct_mean}{25.043}
\setres{lotUL}{EU_post_k7}{width90_pct}{0.2}
\setres{lotUL}{EU_post_k7}{width90_pct_mean}{0.218}
\setres{lotUL}{EU_post_k8}{cov50}{0.00}
\setres{lotUL}{EU_post_k8}{cov50_mean}{0.003}
\setres{lotUL}{EU_post_k8}{cov90}{0.00}
\setres{lotUL}{EU_post_k8}{cov90_adj}{1.00}
\setres{lotUL}{EU_post_k8}{cov90_adj_mean}{0.974}
\setres{lotUL}{EU_post_k8}{cov90_mean}{0.006}
\setres{lotUL}{EU_post_k8}{kappa}{0.545}
\setres{lotUL}{EU_post_k8}{kappa_crps}{0.289}
\setres{lotUL}{EU_post_k8}{kappa_crps_mean}{0.254}
\setres{lotUL}{EU_post_k8}{kappa_mean}{0.447}
\setres{lotUL}{EU_post_k8}{logscore_next}{-3.12}
\setres{lotUL}{EU_post_k8}{logscore_next_mean}{-3.252}
\setres{lotUL}{EU_post_k8}{misfit_pct}{15.7}
\setres{lotUL}{EU_post_k8}{misfit_pct_mean}{15.798}
\setres{lotUL}{EU_post_k8}{n_seen}{8}
\setres{lotUL}{EU_post_k8}{surprise}{0.03}
\setres{lotUL}{EU_post_k8}{surprise_mean}{4.067}
\setres{lotUL}{EU_post_k8}{width90_adj_pct}{25.8}
\setres{lotUL}{EU_post_k8}{width90_adj_pct_mean}{25.988}
\setres{lotUL}{EU_post_k8}{width90_pct}{0.2}
\setres{lotUL}{EU_post_k8}{width90_pct_mean}{0.221}
\setres{lotUL}{EU_post_k9}{cov50}{0.00}
\setres{lotUL}{EU_post_k9}{cov50_mean}{0.003}
\setres{lotUL}{EU_post_k9}{cov90}{0.00}
\setres{lotUL}{EU_post_k9}{cov90_adj}{1.00}
\setres{lotUL}{EU_post_k9}{cov90_adj_mean}{0.963}
\setres{lotUL}{EU_post_k9}{cov90_mean}{0.006}
\setres{lotUL}{EU_post_k9}{kappa}{0.548}
\setres{lotUL}{EU_post_k9}{kappa_crps}{0.311}
\setres{lotUL}{EU_post_k9}{kappa_crps_mean}{0.250}
\setres{lotUL}{EU_post_k9}{kappa_mean}{0.429}
\setres{lotUL}{EU_post_k9}{logscore_next}{-3.05}
\setres{lotUL}{EU_post_k9}{logscore_next_mean}{-3.190}
\setres{lotUL}{EU_post_k9}{misfit_pct}{15.7}
\setres{lotUL}{EU_post_k9}{misfit_pct_mean}{15.710}
\setres{lotUL}{EU_post_k9}{n_seen}{9}
\setres{lotUL}{EU_post_k9}{surprise}{0.03}
\setres{lotUL}{EU_post_k9}{surprise_mean}{0.470}
\setres{lotUL}{EU_post_k9}{width90_adj_pct}{25.8}
\setres{lotUL}{EU_post_k9}{width90_adj_pct_mean}{25.844}
\setres{lotUL}{EU_post_k9}{width90_pct}{0.2}
\setres{lotUL}{EU_post_k9}{width90_pct_mean}{0.211}
\setres{lotUL}{reference}{domain}{Bruhin2010\_gains}
\setres{lotUL}{reference}{n_draws}{400}
\setres{lotUL}{reference}{n_lotteries}{25}
\setres{lotUL}{reference}{n_orders}{20}
\setres{lotUL}{reference}{start}{uniform}
\setres{lotLJ}{CPT_post_k0}{cov50}{0.25}
\setres{lotLJ}{CPT_post_k0}{cov50_mean}{0.250}
\setres{lotLJ}{CPT_post_k0}{cov90}{0.70}
\setres{lotLJ}{CPT_post_k0}{cov90_adj}{---}
\setres{lotLJ}{CPT_post_k0}{cov90_adj_mean}{---}
\setres{lotLJ}{CPT_post_k0}{cov90_mean}{0.705}
\setres{lotLJ}{CPT_post_k0}{kappa}{0.794}
\setres{lotLJ}{CPT_post_k0}{kappa_crps}{0.653}
\setres{lotLJ}{CPT_post_k0}{kappa_crps_mean}{0.648}
\setres{lotLJ}{CPT_post_k0}{kappa_mean}{0.794}
\setres{lotLJ}{CPT_post_k0}{logscore_next}{-2.80}
\setres{lotLJ}{CPT_post_k0}{logscore_next_mean}{-2.799}
\setres{lotLJ}{CPT_post_k0}{misfit_pct}{---}
\setres{lotLJ}{CPT_post_k0}{misfit_pct_mean}{---}
\setres{lotLJ}{CPT_post_k0}{n_seen}{0}
\setres{lotLJ}{CPT_post_k0}{surprise}{2.50}
\setres{lotLJ}{CPT_post_k0}{surprise_mean}{2.265}
\setres{lotLJ}{CPT_post_k0}{width90_adj_pct}{---}
\setres{lotLJ}{CPT_post_k0}{width90_adj_pct_mean}{---}
\setres{lotLJ}{CPT_post_k0}{width90_pct}{11.0}
\setres{lotLJ}{CPT_post_k0}{width90_pct_mean}{10.969}
\setres{lotLJ}{CPT_post_k1}{cov50}{0.27}
\setres{lotLJ}{CPT_post_k1}{cov50_mean}{0.293}
\setres{lotLJ}{CPT_post_k1}{cov90}{0.73}
\setres{lotLJ}{CPT_post_k1}{cov90_adj}{---}
\setres{lotLJ}{CPT_post_k1}{cov90_adj_mean}{---}
\setres{lotLJ}{CPT_post_k1}{cov90_mean}{0.706}
\setres{lotLJ}{CPT_post_k1}{kappa}{0.821}
\setres{lotLJ}{CPT_post_k1}{kappa_crps}{0.635}
\setres{lotLJ}{CPT_post_k1}{kappa_crps_mean}{0.635}
\setres{lotLJ}{CPT_post_k1}{kappa_mean}{0.829}
\setres{lotLJ}{CPT_post_k1}{logscore_next}{-2.79}
\setres{lotLJ}{CPT_post_k1}{logscore_next_mean}{-2.798}
\setres{lotLJ}{CPT_post_k1}{misfit_pct}{---}
\setres{lotLJ}{CPT_post_k1}{misfit_pct_mean}{---}
\setres{lotLJ}{CPT_post_k1}{n_seen}{1}
\setres{lotLJ}{CPT_post_k1}{surprise}{1.21}
\setres{lotLJ}{CPT_post_k1}{surprise_mean}{1.477}
\setres{lotLJ}{CPT_post_k1}{width90_adj_pct}{---}
\setres{lotLJ}{CPT_post_k1}{width90_adj_pct_mean}{---}
\setres{lotLJ}{CPT_post_k1}{width90_pct}{9.5}
\setres{lotLJ}{CPT_post_k1}{width90_pct_mean}{9.585}
\setres{lotLJ}{CPT_post_k10}{cov50}{0.00}
\setres{lotLJ}{CPT_post_k10}{cov50_mean}{0.057}
\setres{lotLJ}{CPT_post_k10}{cov90}{0.13}
\setres{lotLJ}{CPT_post_k10}{cov90_adj}{0.85}
\setres{lotLJ}{CPT_post_k10}{cov90_adj_mean}{0.824}
\setres{lotLJ}{CPT_post_k10}{cov90_mean}{0.157}
\setres{lotLJ}{CPT_post_k10}{kappa}{0.961}
\setres{lotLJ}{CPT_post_k10}{kappa_crps}{0.670}
\setres{lotLJ}{CPT_post_k10}{kappa_crps_mean}{0.666}
\setres{lotLJ}{CPT_post_k10}{kappa_mean}{0.957}
\setres{lotLJ}{CPT_post_k10}{logscore_next}{-2.79}
\setres{lotLJ}{CPT_post_k10}{logscore_next_mean}{-2.801}
\setres{lotLJ}{CPT_post_k10}{misfit_pct}{5.3}
\setres{lotLJ}{CPT_post_k10}{misfit_pct_mean}{5.344}
\setres{lotLJ}{CPT_post_k10}{n_seen}{10}
\setres{lotLJ}{CPT_post_k10}{surprise}{0.03}
\setres{lotLJ}{CPT_post_k10}{surprise_mean}{0.135}
\setres{lotLJ}{CPT_post_k10}{width90_adj_pct}{8.8}
\setres{lotLJ}{CPT_post_k10}{width90_adj_pct_mean}{8.945}
\setres{lotLJ}{CPT_post_k10}{width90_pct}{0.9}
\setres{lotLJ}{CPT_post_k10}{width90_pct_mean}{1.001}
\setres{lotLJ}{CPT_post_k11}{cov50}{0.00}
\setres{lotLJ}{CPT_post_k11}{cov50_mean}{0.047}
\setres{lotLJ}{CPT_post_k11}{cov90}{0.12}
\setres{lotLJ}{CPT_post_k11}{cov90_adj}{0.88}
\setres{lotLJ}{CPT_post_k11}{cov90_adj_mean}{0.847}
\setres{lotLJ}{CPT_post_k11}{cov90_mean}{0.158}
\setres{lotLJ}{CPT_post_k11}{kappa}{0.966}
\setres{lotLJ}{CPT_post_k11}{kappa_crps}{0.669}
\setres{lotLJ}{CPT_post_k11}{kappa_crps_mean}{0.668}
\setres{lotLJ}{CPT_post_k11}{kappa_mean}{0.958}
\setres{lotLJ}{CPT_post_k11}{logscore_next}{-2.79}
\setres{lotLJ}{CPT_post_k11}{logscore_next_mean}{-2.804}
\setres{lotLJ}{CPT_post_k11}{misfit_pct}{5.4}
\setres{lotLJ}{CPT_post_k11}{misfit_pct_mean}{5.526}
\setres{lotLJ}{CPT_post_k11}{n_seen}{11}
\setres{lotLJ}{CPT_post_k11}{surprise}{0.11}
\setres{lotLJ}{CPT_post_k11}{surprise_mean}{0.164}
\setres{lotLJ}{CPT_post_k11}{width90_adj_pct}{9.0}
\setres{lotLJ}{CPT_post_k11}{width90_adj_pct_mean}{9.209}
\setres{lotLJ}{CPT_post_k11}{width90_pct}{0.9}
\setres{lotLJ}{CPT_post_k11}{width90_pct_mean}{1.003}
\setres{lotLJ}{CPT_post_k12}{cov50}{0.00}
\setres{lotLJ}{CPT_post_k12}{cov50_mean}{0.034}
\setres{lotLJ}{CPT_post_k12}{cov90}{0.20}
\setres{lotLJ}{CPT_post_k12}{cov90_adj}{0.88}
\setres{lotLJ}{CPT_post_k12}{cov90_adj_mean}{0.850}
\setres{lotLJ}{CPT_post_k12}{cov90_mean}{0.173}
\setres{lotLJ}{CPT_post_k12}{kappa}{0.963}
\setres{lotLJ}{CPT_post_k12}{kappa_crps}{0.664}
\setres{lotLJ}{CPT_post_k12}{kappa_crps_mean}{0.662}
\setres{lotLJ}{CPT_post_k12}{kappa_mean}{0.958}
\setres{lotLJ}{CPT_post_k12}{logscore_next}{-2.79}
\setres{lotLJ}{CPT_post_k12}{logscore_next_mean}{-2.810}
\setres{lotLJ}{CPT_post_k12}{misfit_pct}{5.5}
\setres{lotLJ}{CPT_post_k12}{misfit_pct_mean}{5.602}
\setres{lotLJ}{CPT_post_k12}{n_seen}{12}
\setres{lotLJ}{CPT_post_k12}{surprise}{0.17}
\setres{lotLJ}{CPT_post_k12}{surprise_mean}{1.596}
\setres{lotLJ}{CPT_post_k12}{width90_adj_pct}{9.2}
\setres{lotLJ}{CPT_post_k12}{width90_adj_pct_mean}{9.342}
\setres{lotLJ}{CPT_post_k12}{width90_pct}{1.1}
\setres{lotLJ}{CPT_post_k12}{width90_pct_mean}{1.067}
\setres{lotLJ}{CPT_post_k13}{cov50}{0.02}
\setres{lotLJ}{CPT_post_k13}{cov50_mean}{0.071}
\setres{lotLJ}{CPT_post_k13}{cov90}{0.16}
\setres{lotLJ}{CPT_post_k13}{cov90_adj}{0.85}
\setres{lotLJ}{CPT_post_k13}{cov90_adj_mean}{0.850}
\setres{lotLJ}{CPT_post_k13}{cov90_mean}{0.148}
\setres{lotLJ}{CPT_post_k13}{kappa}{0.959}
\setres{lotLJ}{CPT_post_k13}{kappa_crps}{0.662}
\setres{lotLJ}{CPT_post_k13}{kappa_crps_mean}{0.654}
\setres{lotLJ}{CPT_post_k13}{kappa_mean}{0.954}
\setres{lotLJ}{CPT_post_k13}{logscore_next}{-2.79}
\setres{lotLJ}{CPT_post_k13}{logscore_next_mean}{-2.802}
\setres{lotLJ}{CPT_post_k13}{misfit_pct}{5.7}
\setres{lotLJ}{CPT_post_k13}{misfit_pct_mean}{5.689}
\setres{lotLJ}{CPT_post_k13}{n_seen}{13}
\setres{lotLJ}{CPT_post_k13}{surprise}{0.03}
\setres{lotLJ}{CPT_post_k13}{surprise_mean}{1.130}
\setres{lotLJ}{CPT_post_k13}{width90_adj_pct}{9.7}
\setres{lotLJ}{CPT_post_k13}{width90_adj_pct_mean}{9.495}
\setres{lotLJ}{CPT_post_k13}{width90_pct}{1.0}
\setres{lotLJ}{CPT_post_k13}{width90_pct_mean}{0.983}
\setres{lotLJ}{CPT_post_k14}{cov50}{0.03}
\setres{lotLJ}{CPT_post_k14}{cov50_mean}{0.090}
\setres{lotLJ}{CPT_post_k14}{cov90}{0.15}
\setres{lotLJ}{CPT_post_k14}{cov90_adj}{0.87}
\setres{lotLJ}{CPT_post_k14}{cov90_adj_mean}{0.847}
\setres{lotLJ}{CPT_post_k14}{cov90_mean}{0.142}
\setres{lotLJ}{CPT_post_k14}{kappa}{0.957}
\setres{lotLJ}{CPT_post_k14}{kappa_crps}{0.669}
\setres{lotLJ}{CPT_post_k14}{kappa_crps_mean}{0.659}
\setres{lotLJ}{CPT_post_k14}{kappa_mean}{0.956}
\setres{lotLJ}{CPT_post_k14}{logscore_next}{-2.79}
\setres{lotLJ}{CPT_post_k14}{logscore_next_mean}{-2.805}
\setres{lotLJ}{CPT_post_k14}{misfit_pct}{5.7}
\setres{lotLJ}{CPT_post_k14}{misfit_pct_mean}{5.739}
\setres{lotLJ}{CPT_post_k14}{n_seen}{14}
\setres{lotLJ}{CPT_post_k14}{surprise}{0.02}
\setres{lotLJ}{CPT_post_k14}{surprise_mean}{0.672}
\setres{lotLJ}{CPT_post_k14}{width90_adj_pct}{9.5}
\setres{lotLJ}{CPT_post_k14}{width90_adj_pct_mean}{9.573}
\setres{lotLJ}{CPT_post_k14}{width90_pct}{1.0}
\setres{lotLJ}{CPT_post_k14}{width90_pct_mean}{0.867}
\setres{lotLJ}{CPT_post_k15}{cov50}{0.03}
\setres{lotLJ}{CPT_post_k15}{cov50_mean}{0.078}
\setres{lotLJ}{CPT_post_k15}{cov90}{0.09}
\setres{lotLJ}{CPT_post_k15}{cov90_adj}{0.86}
\setres{lotLJ}{CPT_post_k15}{cov90_adj_mean}{0.850}
\setres{lotLJ}{CPT_post_k15}{cov90_mean}{0.133}
\setres{lotLJ}{CPT_post_k15}{kappa}{0.960}
\setres{lotLJ}{CPT_post_k15}{kappa_crps}{0.666}
\setres{lotLJ}{CPT_post_k15}{kappa_crps_mean}{0.660}
\setres{lotLJ}{CPT_post_k15}{kappa_mean}{0.960}
\setres{lotLJ}{CPT_post_k15}{logscore_next}{-2.78}
\setres{lotLJ}{CPT_post_k15}{logscore_next_mean}{-2.791}
\setres{lotLJ}{CPT_post_k15}{misfit_pct}{5.7}
\setres{lotLJ}{CPT_post_k15}{misfit_pct_mean}{5.723}
\setres{lotLJ}{CPT_post_k15}{n_seen}{15}
\setres{lotLJ}{CPT_post_k15}{surprise}{0.01}
\setres{lotLJ}{CPT_post_k15}{surprise_mean}{0.231}
\setres{lotLJ}{CPT_post_k15}{width90_adj_pct}{9.4}
\setres{lotLJ}{CPT_post_k15}{width90_adj_pct_mean}{9.517}
\setres{lotLJ}{CPT_post_k15}{width90_pct}{1.0}
\setres{lotLJ}{CPT_post_k15}{width90_pct_mean}{0.815}
\setres{lotLJ}{CPT_post_k16}{cov50}{0.00}
\setres{lotLJ}{CPT_post_k16}{cov50_mean}{0.073}
\setres{lotLJ}{CPT_post_k16}{cov90}{0.11}
\setres{lotLJ}{CPT_post_k16}{cov90_adj}{0.88}
\setres{lotLJ}{CPT_post_k16}{cov90_adj_mean}{0.857}
\setres{lotLJ}{CPT_post_k16}{cov90_mean}{0.127}
\setres{lotLJ}{CPT_post_k16}{kappa}{0.961}
\setres{lotLJ}{CPT_post_k16}{kappa_crps}{0.665}
\setres{lotLJ}{CPT_post_k16}{kappa_crps_mean}{0.659}
\setres{lotLJ}{CPT_post_k16}{kappa_mean}{0.960}
\setres{lotLJ}{CPT_post_k16}{logscore_next}{-2.78}
\setres{lotLJ}{CPT_post_k16}{logscore_next_mean}{-2.787}
\setres{lotLJ}{CPT_post_k16}{misfit_pct}{5.7}
\setres{lotLJ}{CPT_post_k16}{misfit_pct_mean}{5.670}
\setres{lotLJ}{CPT_post_k16}{n_seen}{16}
\setres{lotLJ}{CPT_post_k16}{surprise}{0.00}
\setres{lotLJ}{CPT_post_k16}{surprise_mean}{0.071}
\setres{lotLJ}{CPT_post_k16}{width90_adj_pct}{9.4}
\setres{lotLJ}{CPT_post_k16}{width90_adj_pct_mean}{9.426}
\setres{lotLJ}{CPT_post_k16}{width90_pct}{1.0}
\setres{lotLJ}{CPT_post_k16}{width90_pct_mean}{0.703}
\setres{lotLJ}{CPT_post_k17}{cov50}{0.00}
\setres{lotLJ}{CPT_post_k17}{cov50_mean}{0.050}
\setres{lotLJ}{CPT_post_k17}{cov90}{0.07}
\setres{lotLJ}{CPT_post_k17}{cov90_adj}{0.87}
\setres{lotLJ}{CPT_post_k17}{cov90_adj_mean}{0.854}
\setres{lotLJ}{CPT_post_k17}{cov90_mean}{0.124}
\setres{lotLJ}{CPT_post_k17}{kappa}{0.963}
\setres{lotLJ}{CPT_post_k17}{kappa_crps}{0.673}
\setres{lotLJ}{CPT_post_k17}{kappa_crps_mean}{0.669}
\setres{lotLJ}{CPT_post_k17}{kappa_mean}{0.960}
\setres{lotLJ}{CPT_post_k17}{logscore_next}{-2.79}
\setres{lotLJ}{CPT_post_k17}{logscore_next_mean}{-2.803}
\setres{lotLJ}{CPT_post_k17}{misfit_pct}{5.8}
\setres{lotLJ}{CPT_post_k17}{misfit_pct_mean}{5.688}
\setres{lotLJ}{CPT_post_k17}{n_seen}{17}
\setres{lotLJ}{CPT_post_k17}{surprise}{0.01}
\setres{lotLJ}{CPT_post_k17}{surprise_mean}{1.529}
\setres{lotLJ}{CPT_post_k17}{width90_adj_pct}{9.6}
\setres{lotLJ}{CPT_post_k17}{width90_adj_pct_mean}{9.450}
\setres{lotLJ}{CPT_post_k17}{width90_pct}{0.6}
\setres{lotLJ}{CPT_post_k17}{width90_pct_mean}{0.665}
\setres{lotLJ}{CPT_post_k18}{cov50}{0.00}
\setres{lotLJ}{CPT_post_k18}{cov50_mean}{0.044}
\setres{lotLJ}{CPT_post_k18}{cov90}{0.06}
\setres{lotLJ}{CPT_post_k18}{cov90_adj}{0.87}
\setres{lotLJ}{CPT_post_k18}{cov90_adj_mean}{0.852}
\setres{lotLJ}{CPT_post_k18}{cov90_mean}{0.117}
\setres{lotLJ}{CPT_post_k18}{kappa}{0.960}
\setres{lotLJ}{CPT_post_k18}{kappa_crps}{0.679}
\setres{lotLJ}{CPT_post_k18}{kappa_crps_mean}{0.664}
\setres{lotLJ}{CPT_post_k18}{kappa_mean}{0.958}
\setres{lotLJ}{CPT_post_k18}{logscore_next}{-2.79}
\setres{lotLJ}{CPT_post_k18}{logscore_next_mean}{-2.795}
\setres{lotLJ}{CPT_post_k18}{misfit_pct}{5.8}
\setres{lotLJ}{CPT_post_k18}{misfit_pct_mean}{5.698}
\setres{lotLJ}{CPT_post_k18}{n_seen}{18}
\setres{lotLJ}{CPT_post_k18}{surprise}{0.00}
\setres{lotLJ}{CPT_post_k18}{surprise_mean}{0.140}
\setres{lotLJ}{CPT_post_k18}{width90_adj_pct}{9.6}
\setres{lotLJ}{CPT_post_k18}{width90_adj_pct_mean}{9.451}
\setres{lotLJ}{CPT_post_k18}{width90_pct}{0.2}
\setres{lotLJ}{CPT_post_k18}{width90_pct_mean}{0.583}
\setres{lotLJ}{CPT_post_k19}{cov50}{0.00}
\setres{lotLJ}{CPT_post_k19}{cov50_mean}{0.020}
\setres{lotLJ}{CPT_post_k19}{cov90}{0.08}
\setres{lotLJ}{CPT_post_k19}{cov90_adj}{0.86}
\setres{lotLJ}{CPT_post_k19}{cov90_adj_mean}{0.852}
\setres{lotLJ}{CPT_post_k19}{cov90_mean}{0.110}
\setres{lotLJ}{CPT_post_k19}{kappa}{0.961}
\setres{lotLJ}{CPT_post_k19}{kappa_crps}{0.665}
\setres{lotLJ}{CPT_post_k19}{kappa_crps_mean}{0.662}
\setres{lotLJ}{CPT_post_k19}{kappa_mean}{0.959}
\setres{lotLJ}{CPT_post_k19}{logscore_next}{-2.79}
\setres{lotLJ}{CPT_post_k19}{logscore_next_mean}{-2.838}
\setres{lotLJ}{CPT_post_k19}{misfit_pct}{5.9}
\setres{lotLJ}{CPT_post_k19}{misfit_pct_mean}{5.667}
\setres{lotLJ}{CPT_post_k19}{n_seen}{19}
\setres{lotLJ}{CPT_post_k19}{surprise}{0.01}
\setres{lotLJ}{CPT_post_k19}{surprise_mean}{4.471}
\setres{lotLJ}{CPT_post_k19}{width90_adj_pct}{9.8}
\setres{lotLJ}{CPT_post_k19}{width90_adj_pct_mean}{9.378}
\setres{lotLJ}{CPT_post_k19}{width90_pct}{0.5}
\setres{lotLJ}{CPT_post_k19}{width90_pct_mean}{0.583}
\setres{lotLJ}{CPT_post_k2}{cov50}{0.19}
\setres{lotLJ}{CPT_post_k2}{cov50_mean}{0.229}
\setres{lotLJ}{CPT_post_k2}{cov90}{0.63}
\setres{lotLJ}{CPT_post_k2}{cov90_adj}{0.69}
\setres{lotLJ}{CPT_post_k2}{cov90_adj_mean}{0.661}
\setres{lotLJ}{CPT_post_k2}{cov90_mean}{0.586}
\setres{lotLJ}{CPT_post_k2}{kappa}{0.872}
\setres{lotLJ}{CPT_post_k2}{kappa_crps}{0.644}
\setres{lotLJ}{CPT_post_k2}{kappa_crps_mean}{0.653}
\setres{lotLJ}{CPT_post_k2}{kappa_mean}{0.891}
\setres{lotLJ}{CPT_post_k2}{logscore_next}{-2.79}
\setres{lotLJ}{CPT_post_k2}{logscore_next_mean}{-2.791}
\setres{lotLJ}{CPT_post_k2}{misfit_pct}{1.1}
\setres{lotLJ}{CPT_post_k2}{misfit_pct_mean}{2.026}
\setres{lotLJ}{CPT_post_k2}{n_seen}{2}
\setres{lotLJ}{CPT_post_k2}{surprise}{0.66}
\setres{lotLJ}{CPT_post_k2}{surprise_mean}{0.951}
\setres{lotLJ}{CPT_post_k2}{width90_adj_pct}{6.8}
\setres{lotLJ}{CPT_post_k2}{width90_adj_pct_mean}{7.925}
\setres{lotLJ}{CPT_post_k2}{width90_pct}{6.9}
\setres{lotLJ}{CPT_post_k2}{width90_pct_mean}{6.895}
\setres{lotLJ}{CPT_post_k20}{cov50}{0.00}
\setres{lotLJ}{CPT_post_k20}{cov50_mean}{0.025}
\setres{lotLJ}{CPT_post_k20}{cov90}{0.00}
\setres{lotLJ}{CPT_post_k20}{cov90_adj}{0.83}
\setres{lotLJ}{CPT_post_k20}{cov90_adj_mean}{0.850}
\setres{lotLJ}{CPT_post_k20}{cov90_mean}{0.087}
\setres{lotLJ}{CPT_post_k20}{kappa}{0.961}
\setres{lotLJ}{CPT_post_k20}{kappa_crps}{0.685}
\setres{lotLJ}{CPT_post_k20}{kappa_crps_mean}{0.661}
\setres{lotLJ}{CPT_post_k20}{kappa_mean}{0.960}
\setres{lotLJ}{CPT_post_k20}{logscore_next}{-2.78}
\setres{lotLJ}{CPT_post_k20}{logscore_next_mean}{-2.797}
\setres{lotLJ}{CPT_post_k20}{misfit_pct}{6.0}
\setres{lotLJ}{CPT_post_k20}{misfit_pct_mean}{5.878}
\setres{lotLJ}{CPT_post_k20}{n_seen}{20}
\setres{lotLJ}{CPT_post_k20}{surprise}{0.00}
\setres{lotLJ}{CPT_post_k20}{surprise_mean}{1.377}
\setres{lotLJ}{CPT_post_k20}{width90_adj_pct}{9.9}
\setres{lotLJ}{CPT_post_k20}{width90_adj_pct_mean}{9.720}
\setres{lotLJ}{CPT_post_k20}{width90_pct}{0.0}
\setres{lotLJ}{CPT_post_k20}{width90_pct_mean}{0.506}
\setres{lotLJ}{CPT_post_k21}{cov50}{0.00}
\setres{lotLJ}{CPT_post_k21}{cov50_mean}{0.024}
\setres{lotLJ}{CPT_post_k21}{cov90}{0.00}
\setres{lotLJ}{CPT_post_k21}{cov90_adj}{0.83}
\setres{lotLJ}{CPT_post_k21}{cov90_adj_mean}{0.850}
\setres{lotLJ}{CPT_post_k21}{cov90_mean}{0.078}
\setres{lotLJ}{CPT_post_k21}{kappa}{0.962}
\setres{lotLJ}{CPT_post_k21}{kappa_crps}{0.645}
\setres{lotLJ}{CPT_post_k21}{kappa_crps_mean}{0.653}
\setres{lotLJ}{CPT_post_k21}{kappa_mean}{0.957}
\setres{lotLJ}{CPT_post_k21}{logscore_next}{-2.79}
\setres{lotLJ}{CPT_post_k21}{logscore_next_mean}{-2.808}
\setres{lotLJ}{CPT_post_k21}{misfit_pct}{6.0}
\setres{lotLJ}{CPT_post_k21}{misfit_pct_mean}{5.904}
\setres{lotLJ}{CPT_post_k21}{n_seen}{21}
\setres{lotLJ}{CPT_post_k21}{surprise}{0.01}
\setres{lotLJ}{CPT_post_k21}{surprise_mean}{0.180}
\setres{lotLJ}{CPT_post_k21}{width90_adj_pct}{9.9}
\setres{lotLJ}{CPT_post_k21}{width90_adj_pct_mean}{9.746}
\setres{lotLJ}{CPT_post_k21}{width90_pct}{0.0}
\setres{lotLJ}{CPT_post_k21}{width90_pct_mean}{0.406}
\setres{lotLJ}{CPT_post_k22}{cov50}{0.00}
\setres{lotLJ}{CPT_post_k22}{cov50_mean}{0.020}
\setres{lotLJ}{CPT_post_k22}{cov90}{0.00}
\setres{lotLJ}{CPT_post_k22}{cov90_adj}{0.84}
\setres{lotLJ}{CPT_post_k22}{cov90_adj_mean}{0.857}
\setres{lotLJ}{CPT_post_k22}{cov90_mean}{0.048}
\setres{lotLJ}{CPT_post_k22}{kappa}{0.962}
\setres{lotLJ}{CPT_post_k22}{kappa_crps}{0.659}
\setres{lotLJ}{CPT_post_k22}{kappa_crps_mean}{0.656}
\setres{lotLJ}{CPT_post_k22}{kappa_mean}{0.958}
\setres{lotLJ}{CPT_post_k22}{logscore_next}{-2.79}
\setres{lotLJ}{CPT_post_k22}{logscore_next_mean}{-2.807}
\setres{lotLJ}{CPT_post_k22}{misfit_pct}{5.9}
\setres{lotLJ}{CPT_post_k22}{misfit_pct_mean}{5.970}
\setres{lotLJ}{CPT_post_k22}{n_seen}{22}
\setres{lotLJ}{CPT_post_k22}{surprise}{0.00}
\setres{lotLJ}{CPT_post_k22}{surprise_mean}{0.469}
\setres{lotLJ}{CPT_post_k22}{width90_adj_pct}{9.8}
\setres{lotLJ}{CPT_post_k22}{width90_adj_pct_mean}{9.851}
\setres{lotLJ}{CPT_post_k22}{width90_pct}{0.0}
\setres{lotLJ}{CPT_post_k22}{width90_pct_mean}{0.274}
\setres{lotLJ}{CPT_post_k23}{cov50}{0.00}
\setres{lotLJ}{CPT_post_k23}{cov50_mean}{0.033}
\setres{lotLJ}{CPT_post_k23}{cov90}{0.00}
\setres{lotLJ}{CPT_post_k23}{cov90_adj}{0.86}
\setres{lotLJ}{CPT_post_k23}{cov90_adj_mean}{0.860}
\setres{lotLJ}{CPT_post_k23}{cov90_mean}{0.086}
\setres{lotLJ}{CPT_post_k23}{kappa}{0.962}
\setres{lotLJ}{CPT_post_k23}{kappa_crps}{0.682}
\setres{lotLJ}{CPT_post_k23}{kappa_crps_mean}{0.652}
\setres{lotLJ}{CPT_post_k23}{kappa_mean}{0.957}
\setres{lotLJ}{CPT_post_k23}{logscore_next}{-2.79}
\setres{lotLJ}{CPT_post_k23}{logscore_next_mean}{-2.815}
\setres{lotLJ}{CPT_post_k23}{misfit_pct}{5.9}
\setres{lotLJ}{CPT_post_k23}{misfit_pct_mean}{5.948}
\setres{lotLJ}{CPT_post_k23}{n_seen}{23}
\setres{lotLJ}{CPT_post_k23}{surprise}{0.00}
\setres{lotLJ}{CPT_post_k23}{surprise_mean}{1.828}
\setres{lotLJ}{CPT_post_k23}{width90_adj_pct}{9.7}
\setres{lotLJ}{CPT_post_k23}{width90_adj_pct_mean}{9.822}
\setres{lotLJ}{CPT_post_k23}{width90_pct}{0.0}
\setres{lotLJ}{CPT_post_k23}{width90_pct_mean}{0.386}
\setres{lotLJ}{CPT_post_k24}{cov50}{0.00}
\setres{lotLJ}{CPT_post_k24}{cov50_mean}{0.032}
\setres{lotLJ}{CPT_post_k24}{cov90}{0.00}
\setres{lotLJ}{CPT_post_k24}{cov90_adj}{0.85}
\setres{lotLJ}{CPT_post_k24}{cov90_adj_mean}{0.872}
\setres{lotLJ}{CPT_post_k24}{cov90_mean}{0.058}
\setres{lotLJ}{CPT_post_k24}{kappa}{0.962}
\setres{lotLJ}{CPT_post_k24}{kappa_crps}{0.669}
\setres{lotLJ}{CPT_post_k24}{kappa_crps_mean}{0.647}
\setres{lotLJ}{CPT_post_k24}{kappa_mean}{0.957}
\setres{lotLJ}{CPT_post_k24}{logscore_next}{-2.79}
\setres{lotLJ}{CPT_post_k24}{logscore_next_mean}{-2.810}
\setres{lotLJ}{CPT_post_k24}{misfit_pct}{6.2}
\setres{lotLJ}{CPT_post_k24}{misfit_pct_mean}{6.043}
\setres{lotLJ}{CPT_post_k24}{n_seen}{24}
\setres{lotLJ}{CPT_post_k24}{surprise}{0.00}
\setres{lotLJ}{CPT_post_k24}{surprise_mean}{0.506}
\setres{lotLJ}{CPT_post_k24}{width90_adj_pct}{10.2}
\setres{lotLJ}{CPT_post_k24}{width90_adj_pct_mean}{9.968}
\setres{lotLJ}{CPT_post_k24}{width90_pct}{0.0}
\setres{lotLJ}{CPT_post_k24}{width90_pct_mean}{0.322}
\setres{lotLJ}{CPT_post_k25}{cov50}{0.00}
\setres{lotLJ}{CPT_post_k25}{cov50_mean}{0.024}
\setres{lotLJ}{CPT_post_k25}{cov90}{0.00}
\setres{lotLJ}{CPT_post_k25}{cov90_adj}{0.87}
\setres{lotLJ}{CPT_post_k25}{cov90_adj_mean}{0.866}
\setres{lotLJ}{CPT_post_k25}{cov90_mean}{0.029}
\setres{lotLJ}{CPT_post_k25}{kappa}{0.960}
\setres{lotLJ}{CPT_post_k25}{kappa_crps}{0.682}
\setres{lotLJ}{CPT_post_k25}{kappa_crps_mean}{0.651}
\setres{lotLJ}{CPT_post_k25}{kappa_mean}{0.959}
\setres{lotLJ}{CPT_post_k25}{logscore_next}{-2.78}
\setres{lotLJ}{CPT_post_k25}{logscore_next_mean}{-2.790}
\setres{lotLJ}{CPT_post_k25}{misfit_pct}{6.1}
\setres{lotLJ}{CPT_post_k25}{misfit_pct_mean}{5.955}
\setres{lotLJ}{CPT_post_k25}{n_seen}{25}
\setres{lotLJ}{CPT_post_k25}{surprise}{0.00}
\setres{lotLJ}{CPT_post_k25}{surprise_mean}{0.033}
\setres{lotLJ}{CPT_post_k25}{width90_adj_pct}{10.0}
\setres{lotLJ}{CPT_post_k25}{width90_adj_pct_mean}{9.824}
\setres{lotLJ}{CPT_post_k25}{width90_pct}{0.0}
\setres{lotLJ}{CPT_post_k25}{width90_pct_mean}{0.230}
\setres{lotLJ}{CPT_post_k26}{cov50}{0.00}
\setres{lotLJ}{CPT_post_k26}{cov50_mean}{0.022}
\setres{lotLJ}{CPT_post_k26}{cov90}{0.00}
\setres{lotLJ}{CPT_post_k26}{cov90_adj}{0.83}
\setres{lotLJ}{CPT_post_k26}{cov90_adj_mean}{0.853}
\setres{lotLJ}{CPT_post_k26}{cov90_mean}{0.022}
\setres{lotLJ}{CPT_post_k26}{kappa}{0.961}
\setres{lotLJ}{CPT_post_k26}{kappa_crps}{0.663}
\setres{lotLJ}{CPT_post_k26}{kappa_crps_mean}{0.653}
\setres{lotLJ}{CPT_post_k26}{kappa_mean}{0.960}
\setres{lotLJ}{CPT_post_k26}{logscore_next}{-2.79}
\setres{lotLJ}{CPT_post_k26}{logscore_next_mean}{-2.805}
\setres{lotLJ}{CPT_post_k26}{misfit_pct}{6.0}
\setres{lotLJ}{CPT_post_k26}{misfit_pct_mean}{5.895}
\setres{lotLJ}{CPT_post_k26}{n_seen}{26}
\setres{lotLJ}{CPT_post_k26}{surprise}{0.00}
\setres{lotLJ}{CPT_post_k26}{surprise_mean}{1.759}
\setres{lotLJ}{CPT_post_k26}{width90_adj_pct}{9.8}
\setres{lotLJ}{CPT_post_k26}{width90_adj_pct_mean}{9.724}
\setres{lotLJ}{CPT_post_k26}{width90_pct}{0.0}
\setres{lotLJ}{CPT_post_k26}{width90_pct_mean}{0.151}
\setres{lotLJ}{CPT_post_k27}{cov50}{0.00}
\setres{lotLJ}{CPT_post_k27}{cov50_mean}{0.026}
\setres{lotLJ}{CPT_post_k27}{cov90}{0.00}
\setres{lotLJ}{CPT_post_k27}{cov90_adj}{0.85}
\setres{lotLJ}{CPT_post_k27}{cov90_adj_mean}{0.859}
\setres{lotLJ}{CPT_post_k27}{cov90_mean}{0.041}
\setres{lotLJ}{CPT_post_k27}{kappa}{0.960}
\setres{lotLJ}{CPT_post_k27}{kappa_crps}{0.671}
\setres{lotLJ}{CPT_post_k27}{kappa_crps_mean}{0.653}
\setres{lotLJ}{CPT_post_k27}{kappa_mean}{0.961}
\setres{lotLJ}{CPT_post_k27}{logscore_next}{-2.79}
\setres{lotLJ}{CPT_post_k27}{logscore_next_mean}{-2.791}
\setres{lotLJ}{CPT_post_k27}{misfit_pct}{5.9}
\setres{lotLJ}{CPT_post_k27}{misfit_pct_mean}{5.926}
\setres{lotLJ}{CPT_post_k27}{n_seen}{27}
\setres{lotLJ}{CPT_post_k27}{surprise}{0.00}
\setres{lotLJ}{CPT_post_k27}{surprise_mean}{0.238}
\setres{lotLJ}{CPT_post_k27}{width90_adj_pct}{9.8}
\setres{lotLJ}{CPT_post_k27}{width90_adj_pct_mean}{9.779}
\setres{lotLJ}{CPT_post_k27}{width90_pct}{0.0}
\setres{lotLJ}{CPT_post_k27}{width90_pct_mean}{0.265}
\setres{lotLJ}{CPT_post_k28}{cov50}{0.00}
\setres{lotLJ}{CPT_post_k28}{cov50_mean}{0.031}
\setres{lotLJ}{CPT_post_k28}{cov90}{0.00}
\setres{lotLJ}{CPT_post_k28}{cov90_adj}{0.84}
\setres{lotLJ}{CPT_post_k28}{cov90_adj_mean}{0.866}
\setres{lotLJ}{CPT_post_k28}{cov90_mean}{0.050}
\setres{lotLJ}{CPT_post_k28}{kappa}{0.958}
\setres{lotLJ}{CPT_post_k28}{kappa_crps}{0.677}
\setres{lotLJ}{CPT_post_k28}{kappa_crps_mean}{0.660}
\setres{lotLJ}{CPT_post_k28}{kappa_mean}{0.960}
\setres{lotLJ}{CPT_post_k28}{logscore_next}{-2.80}
\setres{lotLJ}{CPT_post_k28}{logscore_next_mean}{-2.807}
\setres{lotLJ}{CPT_post_k28}{misfit_pct}{5.8}
\setres{lotLJ}{CPT_post_k28}{misfit_pct_mean}{5.883}
\setres{lotLJ}{CPT_post_k28}{n_seen}{28}
\setres{lotLJ}{CPT_post_k28}{surprise}{0.00}
\setres{lotLJ}{CPT_post_k28}{surprise_mean}{0.032}
\setres{lotLJ}{CPT_post_k28}{width90_adj_pct}{9.6}
\setres{lotLJ}{CPT_post_k28}{width90_adj_pct_mean}{9.714}
\setres{lotLJ}{CPT_post_k28}{width90_pct}{0.0}
\setres{lotLJ}{CPT_post_k28}{width90_pct_mean}{0.269}
\setres{lotLJ}{CPT_post_k29}{cov50}{0.00}
\setres{lotLJ}{CPT_post_k29}{cov50_mean}{0.017}
\setres{lotLJ}{CPT_post_k29}{cov90}{0.00}
\setres{lotLJ}{CPT_post_k29}{cov90_adj}{0.87}
\setres{lotLJ}{CPT_post_k29}{cov90_adj_mean}{0.877}
\setres{lotLJ}{CPT_post_k29}{cov90_mean}{0.043}
\setres{lotLJ}{CPT_post_k29}{kappa}{0.959}
\setres{lotLJ}{CPT_post_k29}{kappa_crps}{0.714}
\setres{lotLJ}{CPT_post_k29}{kappa_crps_mean}{0.670}
\setres{lotLJ}{CPT_post_k29}{kappa_mean}{0.961}
\setres{lotLJ}{CPT_post_k29}{logscore_next}{-2.79}
\setres{lotLJ}{CPT_post_k29}{logscore_next_mean}{-2.798}
\setres{lotLJ}{CPT_post_k29}{misfit_pct}{6.0}
\setres{lotLJ}{CPT_post_k29}{misfit_pct_mean}{5.923}
\setres{lotLJ}{CPT_post_k29}{n_seen}{29}
\setres{lotLJ}{CPT_post_k29}{surprise}{0.00}
\setres{lotLJ}{CPT_post_k29}{surprise_mean}{0.111}
\setres{lotLJ}{CPT_post_k29}{width90_adj_pct}{9.8}
\setres{lotLJ}{CPT_post_k29}{width90_adj_pct_mean}{9.777}
\setres{lotLJ}{CPT_post_k29}{width90_pct}{0.0}
\setres{lotLJ}{CPT_post_k29}{width90_pct_mean}{0.227}
\setres{lotLJ}{CPT_post_k3}{cov50}{0.22}
\setres{lotLJ}{CPT_post_k3}{cov50_mean}{0.234}
\setres{lotLJ}{CPT_post_k3}{cov90}{0.56}
\setres{lotLJ}{CPT_post_k3}{cov90_adj}{0.80}
\setres{lotLJ}{CPT_post_k3}{cov90_adj_mean}{0.740}
\setres{lotLJ}{CPT_post_k3}{cov90_mean}{0.530}
\setres{lotLJ}{CPT_post_k3}{kappa}{0.932}
\setres{lotLJ}{CPT_post_k3}{kappa_crps}{0.674}
\setres{lotLJ}{CPT_post_k3}{kappa_crps_mean}{0.664}
\setres{lotLJ}{CPT_post_k3}{kappa_mean}{0.915}
\setres{lotLJ}{CPT_post_k3}{logscore_next}{-2.79}
\setres{lotLJ}{CPT_post_k3}{logscore_next_mean}{-2.793}
\setres{lotLJ}{CPT_post_k3}{misfit_pct}{3.0}
\setres{lotLJ}{CPT_post_k3}{misfit_pct_mean}{3.543}
\setres{lotLJ}{CPT_post_k3}{n_seen}{3}
\setres{lotLJ}{CPT_post_k3}{surprise}{0.90}
\setres{lotLJ}{CPT_post_k3}{surprise_mean}{1.300}
\setres{lotLJ}{CPT_post_k3}{width90_adj_pct}{8.5}
\setres{lotLJ}{CPT_post_k3}{width90_adj_pct_mean}{8.712}
\setres{lotLJ}{CPT_post_k3}{width90_pct}{5.2}
\setres{lotLJ}{CPT_post_k3}{width90_pct_mean}{5.620}
\setres{lotLJ}{CPT_post_k30}{cov50}{0.00}
\setres{lotLJ}{CPT_post_k30}{cov50_mean}{0.011}
\setres{lotLJ}{CPT_post_k30}{cov90}{0.00}
\setres{lotLJ}{CPT_post_k30}{cov90_adj}{0.89}
\setres{lotLJ}{CPT_post_k30}{cov90_adj_mean}{0.871}
\setres{lotLJ}{CPT_post_k30}{cov90_mean}{0.029}
\setres{lotLJ}{CPT_post_k30}{kappa}{0.959}
\setres{lotLJ}{CPT_post_k30}{kappa_crps}{0.689}
\setres{lotLJ}{CPT_post_k30}{kappa_crps_mean}{0.662}
\setres{lotLJ}{CPT_post_k30}{kappa_mean}{0.956}
\setres{lotLJ}{CPT_post_k30}{logscore_next}{-2.79}
\setres{lotLJ}{CPT_post_k30}{logscore_next_mean}{-2.802}
\setres{lotLJ}{CPT_post_k30}{misfit_pct}{6.1}
\setres{lotLJ}{CPT_post_k30}{misfit_pct_mean}{5.915}
\setres{lotLJ}{CPT_post_k30}{n_seen}{30}
\setres{lotLJ}{CPT_post_k30}{surprise}{0.00}
\setres{lotLJ}{CPT_post_k30}{surprise_mean}{0.035}
\setres{lotLJ}{CPT_post_k30}{width90_adj_pct}{10.0}
\setres{lotLJ}{CPT_post_k30}{width90_adj_pct_mean}{9.751}
\setres{lotLJ}{CPT_post_k30}{width90_pct}{0.0}
\setres{lotLJ}{CPT_post_k30}{width90_pct_mean}{0.152}
\setres{lotLJ}{CPT_post_k31}{cov50}{0.00}
\setres{lotLJ}{CPT_post_k31}{cov50_mean}{0.023}
\setres{lotLJ}{CPT_post_k31}{cov90}{0.00}
\setres{lotLJ}{CPT_post_k31}{cov90_adj}{0.88}
\setres{lotLJ}{CPT_post_k31}{cov90_adj_mean}{0.869}
\setres{lotLJ}{CPT_post_k31}{cov90_mean}{0.031}
\setres{lotLJ}{CPT_post_k31}{kappa}{0.953}
\setres{lotLJ}{CPT_post_k31}{kappa_crps}{0.654}
\setres{lotLJ}{CPT_post_k31}{kappa_crps_mean}{0.646}
\setres{lotLJ}{CPT_post_k31}{kappa_mean}{0.956}
\setres{lotLJ}{CPT_post_k31}{logscore_next}{-2.78}
\setres{lotLJ}{CPT_post_k31}{logscore_next_mean}{-2.816}
\setres{lotLJ}{CPT_post_k31}{misfit_pct}{6.1}
\setres{lotLJ}{CPT_post_k31}{misfit_pct_mean}{5.892}
\setres{lotLJ}{CPT_post_k31}{n_seen}{31}
\setres{lotLJ}{CPT_post_k31}{surprise}{0.00}
\setres{lotLJ}{CPT_post_k31}{surprise_mean}{1.876}
\setres{lotLJ}{CPT_post_k31}{width90_adj_pct}{10.0}
\setres{lotLJ}{CPT_post_k31}{width90_adj_pct_mean}{9.719}
\setres{lotLJ}{CPT_post_k31}{width90_pct}{0.0}
\setres{lotLJ}{CPT_post_k31}{width90_pct_mean}{0.172}
\setres{lotLJ}{CPT_post_k32}{cov50}{0.00}
\setres{lotLJ}{CPT_post_k32}{cov50_mean}{0.017}
\setres{lotLJ}{CPT_post_k32}{cov90}{0.00}
\setres{lotLJ}{CPT_post_k32}{cov90_adj}{0.92}
\setres{lotLJ}{CPT_post_k32}{cov90_adj_mean}{0.858}
\setres{lotLJ}{CPT_post_k32}{cov90_mean}{0.046}
\setres{lotLJ}{CPT_post_k32}{kappa}{0.956}
\setres{lotLJ}{CPT_post_k32}{kappa_crps}{0.678}
\setres{lotLJ}{CPT_post_k32}{kappa_crps_mean}{0.656}
\setres{lotLJ}{CPT_post_k32}{kappa_mean}{0.957}
\setres{lotLJ}{CPT_post_k32}{logscore_next}{-2.78}
\setres{lotLJ}{CPT_post_k32}{logscore_next_mean}{-2.800}
\setres{lotLJ}{CPT_post_k32}{misfit_pct}{6.0}
\setres{lotLJ}{CPT_post_k32}{misfit_pct_mean}{5.936}
\setres{lotLJ}{CPT_post_k32}{n_seen}{32}
\setres{lotLJ}{CPT_post_k32}{surprise}{0.00}
\setres{lotLJ}{CPT_post_k32}{surprise_mean}{0.385}
\setres{lotLJ}{CPT_post_k32}{width90_adj_pct}{10.0}
\setres{lotLJ}{CPT_post_k32}{width90_adj_pct_mean}{9.798}
\setres{lotLJ}{CPT_post_k32}{width90_pct}{0.0}
\setres{lotLJ}{CPT_post_k32}{width90_pct_mean}{0.315}
\setres{lotLJ}{CPT_post_k33}{cov50}{0.00}
\setres{lotLJ}{CPT_post_k33}{cov50_mean}{0.027}
\setres{lotLJ}{CPT_post_k33}{cov90}{0.00}
\setres{lotLJ}{CPT_post_k33}{cov90_adj}{0.91}
\setres{lotLJ}{CPT_post_k33}{cov90_adj_mean}{0.850}
\setres{lotLJ}{CPT_post_k33}{cov90_mean}{0.041}
\setres{lotLJ}{CPT_post_k33}{kappa}{0.960}
\setres{lotLJ}{CPT_post_k33}{kappa_crps}{0.659}
\setres{lotLJ}{CPT_post_k33}{kappa_crps_mean}{0.624}
\setres{lotLJ}{CPT_post_k33}{kappa_mean}{0.952}
\setres{lotLJ}{CPT_post_k33}{logscore_next}{-2.79}
\setres{lotLJ}{CPT_post_k33}{logscore_next_mean}{-2.797}
\setres{lotLJ}{CPT_post_k33}{misfit_pct}{6.0}
\setres{lotLJ}{CPT_post_k33}{misfit_pct_mean}{5.872}
\setres{lotLJ}{CPT_post_k33}{n_seen}{33}
\setres{lotLJ}{CPT_post_k33}{surprise}{0.00}
\setres{lotLJ}{CPT_post_k33}{surprise_mean}{0.297}
\setres{lotLJ}{CPT_post_k33}{width90_adj_pct}{9.8}
\setres{lotLJ}{CPT_post_k33}{width90_adj_pct_mean}{9.693}
\setres{lotLJ}{CPT_post_k33}{width90_pct}{0.0}
\setres{lotLJ}{CPT_post_k33}{width90_pct_mean}{0.247}
\setres{lotLJ}{CPT_post_k34}{cov50}{0.00}
\setres{lotLJ}{CPT_post_k34}{cov50_mean}{0.020}
\setres{lotLJ}{CPT_post_k34}{cov90}{0.00}
\setres{lotLJ}{CPT_post_k34}{cov90_adj}{0.90}
\setres{lotLJ}{CPT_post_k34}{cov90_adj_mean}{0.860}
\setres{lotLJ}{CPT_post_k34}{cov90_mean}{0.030}
\setres{lotLJ}{CPT_post_k34}{kappa}{0.956}
\setres{lotLJ}{CPT_post_k34}{kappa_crps}{0.667}
\setres{lotLJ}{CPT_post_k34}{kappa_crps_mean}{0.621}
\setres{lotLJ}{CPT_post_k34}{kappa_mean}{0.952}
\setres{lotLJ}{CPT_post_k34}{logscore_next}{-2.79}
\setres{lotLJ}{CPT_post_k34}{logscore_next_mean}{-2.808}
\setres{lotLJ}{CPT_post_k34}{misfit_pct}{6.0}
\setres{lotLJ}{CPT_post_k34}{misfit_pct_mean}{5.843}
\setres{lotLJ}{CPT_post_k34}{n_seen}{34}
\setres{lotLJ}{CPT_post_k34}{surprise}{0.01}
\setres{lotLJ}{CPT_post_k34}{surprise_mean}{1.339}
\setres{lotLJ}{CPT_post_k34}{width90_adj_pct}{9.8}
\setres{lotLJ}{CPT_post_k34}{width90_adj_pct_mean}{9.630}
\setres{lotLJ}{CPT_post_k34}{width90_pct}{0.0}
\setres{lotLJ}{CPT_post_k34}{width90_pct_mean}{0.147}
\setres{lotLJ}{CPT_post_k35}{cov50}{0.00}
\setres{lotLJ}{CPT_post_k35}{cov50_mean}{0.033}
\setres{lotLJ}{CPT_post_k35}{cov90}{0.00}
\setres{lotLJ}{CPT_post_k35}{cov90_adj}{0.89}
\setres{lotLJ}{CPT_post_k35}{cov90_adj_mean}{0.867}
\setres{lotLJ}{CPT_post_k35}{cov90_mean}{0.067}
\setres{lotLJ}{CPT_post_k35}{kappa}{0.942}
\setres{lotLJ}{CPT_post_k35}{kappa_crps}{0.641}
\setres{lotLJ}{CPT_post_k35}{kappa_crps_mean}{0.589}
\setres{lotLJ}{CPT_post_k35}{kappa_mean}{0.937}
\setres{lotLJ}{CPT_post_k35}{logscore_next}{-2.79}
\setres{lotLJ}{CPT_post_k35}{logscore_next_mean}{-2.806}
\setres{lotLJ}{CPT_post_k35}{misfit_pct}{5.9}
\setres{lotLJ}{CPT_post_k35}{misfit_pct_mean}{5.889}
\setres{lotLJ}{CPT_post_k35}{n_seen}{35}
\setres{lotLJ}{CPT_post_k35}{surprise}{0.00}
\setres{lotLJ}{CPT_post_k35}{surprise_mean}{0.542}
\setres{lotLJ}{CPT_post_k35}{width90_adj_pct}{9.8}
\setres{lotLJ}{CPT_post_k35}{width90_adj_pct_mean}{9.723}
\setres{lotLJ}{CPT_post_k35}{width90_pct}{0.0}
\setres{lotLJ}{CPT_post_k35}{width90_pct_mean}{0.309}
\setres{lotLJ}{CPT_post_k36}{cov50}{0.00}
\setres{lotLJ}{CPT_post_k36}{cov50_mean}{0.025}
\setres{lotLJ}{CPT_post_k36}{cov90}{0.00}
\setres{lotLJ}{CPT_post_k36}{cov90_adj}{0.88}
\setres{lotLJ}{CPT_post_k36}{cov90_adj_mean}{0.850}
\setres{lotLJ}{CPT_post_k36}{cov90_mean}{0.031}
\setres{lotLJ}{CPT_post_k36}{kappa}{0.961}
\setres{lotLJ}{CPT_post_k36}{kappa_crps}{0.639}
\setres{lotLJ}{CPT_post_k36}{kappa_crps_mean}{0.611}
\setres{lotLJ}{CPT_post_k36}{kappa_mean}{0.944}
\setres{lotLJ}{CPT_post_k36}{logscore_next}{-2.78}
\setres{lotLJ}{CPT_post_k36}{logscore_next_mean}{-2.807}
\setres{lotLJ}{CPT_post_k36}{misfit_pct}{5.9}
\setres{lotLJ}{CPT_post_k36}{misfit_pct_mean}{5.846}
\setres{lotLJ}{CPT_post_k36}{n_seen}{36}
\setres{lotLJ}{CPT_post_k36}{surprise}{0.00}
\setres{lotLJ}{CPT_post_k36}{surprise_mean}{0.387}
\setres{lotLJ}{CPT_post_k36}{width90_adj_pct}{9.7}
\setres{lotLJ}{CPT_post_k36}{width90_adj_pct_mean}{9.643}
\setres{lotLJ}{CPT_post_k36}{width90_pct}{0.0}
\setres{lotLJ}{CPT_post_k36}{width90_pct_mean}{0.244}
\setres{lotLJ}{CPT_post_k37}{cov50}{0.00}
\setres{lotLJ}{CPT_post_k37}{cov50_mean}{0.029}
\setres{lotLJ}{CPT_post_k37}{cov90}{0.00}
\setres{lotLJ}{CPT_post_k37}{cov90_adj}{0.86}
\setres{lotLJ}{CPT_post_k37}{cov90_adj_mean}{0.821}
\setres{lotLJ}{CPT_post_k37}{cov90_mean}{0.043}
\setres{lotLJ}{CPT_post_k37}{kappa}{0.954}
\setres{lotLJ}{CPT_post_k37}{kappa_crps}{0.633}
\setres{lotLJ}{CPT_post_k37}{kappa_crps_mean}{0.518}
\setres{lotLJ}{CPT_post_k37}{kappa_mean}{0.904}
\setres{lotLJ}{CPT_post_k37}{logscore_next}{-2.79}
\setres{lotLJ}{CPT_post_k37}{logscore_next_mean}{-2.800}
\setres{lotLJ}{CPT_post_k37}{misfit_pct}{5.9}
\setres{lotLJ}{CPT_post_k37}{misfit_pct_mean}{5.803}
\setres{lotLJ}{CPT_post_k37}{n_seen}{37}
\setres{lotLJ}{CPT_post_k37}{surprise}{0.00}
\setres{lotLJ}{CPT_post_k37}{surprise_mean}{0.107}
\setres{lotLJ}{CPT_post_k37}{width90_adj_pct}{9.7}
\setres{lotLJ}{CPT_post_k37}{width90_adj_pct_mean}{9.581}
\setres{lotLJ}{CPT_post_k37}{width90_pct}{0.0}
\setres{lotLJ}{CPT_post_k37}{width90_pct_mean}{0.310}
\setres{lotLJ}{CPT_post_k38}{cov50}{0.00}
\setres{lotLJ}{CPT_post_k38}{cov50_mean}{0.000}
\setres{lotLJ}{CPT_post_k38}{cov90}{0.00}
\setres{lotLJ}{CPT_post_k38}{cov90_adj}{0.83}
\setres{lotLJ}{CPT_post_k38}{cov90_adj_mean}{0.825}
\setres{lotLJ}{CPT_post_k38}{cov90_mean}{0.025}
\setres{lotLJ}{CPT_post_k38}{kappa}{0.952}
\setres{lotLJ}{CPT_post_k38}{kappa_crps}{0.609}
\setres{lotLJ}{CPT_post_k38}{kappa_crps_mean}{0.469}
\setres{lotLJ}{CPT_post_k38}{kappa_mean}{0.896}
\setres{lotLJ}{CPT_post_k38}{logscore_next}{-2.79}
\setres{lotLJ}{CPT_post_k38}{logscore_next_mean}{-2.802}
\setres{lotLJ}{CPT_post_k38}{misfit_pct}{5.8}
\setres{lotLJ}{CPT_post_k38}{misfit_pct_mean}{5.787}
\setres{lotLJ}{CPT_post_k38}{n_seen}{38}
\setres{lotLJ}{CPT_post_k38}{surprise}{0.00}
\setres{lotLJ}{CPT_post_k38}{surprise_mean}{0.021}
\setres{lotLJ}{CPT_post_k38}{width90_adj_pct}{9.6}
\setres{lotLJ}{CPT_post_k38}{width90_adj_pct_mean}{9.530}
\setres{lotLJ}{CPT_post_k38}{width90_pct}{0.0}
\setres{lotLJ}{CPT_post_k38}{width90_pct_mean}{0.146}
\setres{lotLJ}{CPT_post_k39}{cov50}{0.00}
\setres{lotLJ}{CPT_post_k39}{cov50_mean}{0.010}
\setres{lotLJ}{CPT_post_k39}{cov90}{0.00}
\setres{lotLJ}{CPT_post_k39}{cov90_adj}{0.90}
\setres{lotLJ}{CPT_post_k39}{cov90_adj_mean}{0.830}
\setres{lotLJ}{CPT_post_k39}{cov90_mean}{0.030}
\setres{lotLJ}{CPT_post_k39}{kappa}{0.951}
\setres{lotLJ}{CPT_post_k39}{kappa_crps}{0.599}
\setres{lotLJ}{CPT_post_k39}{kappa_crps_mean}{0.421}
\setres{lotLJ}{CPT_post_k39}{kappa_mean}{0.879}
\setres{lotLJ}{CPT_post_k39}{logscore_next}{-2.79}
\setres{lotLJ}{CPT_post_k39}{logscore_next_mean}{-2.805}
\setres{lotLJ}{CPT_post_k39}{misfit_pct}{5.9}
\setres{lotLJ}{CPT_post_k39}{misfit_pct_mean}{5.775}
\setres{lotLJ}{CPT_post_k39}{n_seen}{39}
\setres{lotLJ}{CPT_post_k39}{surprise}{0.00}
\setres{lotLJ}{CPT_post_k39}{surprise_mean}{0.356}
\setres{lotLJ}{CPT_post_k39}{width90_adj_pct}{9.7}
\setres{lotLJ}{CPT_post_k39}{width90_adj_pct_mean}{9.516}
\setres{lotLJ}{CPT_post_k39}{width90_pct}{0.0}
\setres{lotLJ}{CPT_post_k39}{width90_pct_mean}{0.188}
\setres{lotLJ}{CPT_post_k4}{cov50}{0.17}
\setres{lotLJ}{CPT_post_k4}{cov50_mean}{0.178}
\setres{lotLJ}{CPT_post_k4}{cov90}{0.39}
\setres{lotLJ}{CPT_post_k4}{cov90_adj}{0.76}
\setres{lotLJ}{CPT_post_k4}{cov90_adj_mean}{0.726}
\setres{lotLJ}{CPT_post_k4}{cov90_mean}{0.381}
\setres{lotLJ}{CPT_post_k4}{kappa}{0.957}
\setres{lotLJ}{CPT_post_k4}{kappa_crps}{0.679}
\setres{lotLJ}{CPT_post_k4}{kappa_crps_mean}{0.654}
\setres{lotLJ}{CPT_post_k4}{kappa_mean}{0.923}
\setres{lotLJ}{CPT_post_k4}{logscore_next}{-2.79}
\setres{lotLJ}{CPT_post_k4}{logscore_next_mean}{-2.791}
\setres{lotLJ}{CPT_post_k4}{misfit_pct}{4.1}
\setres{lotLJ}{CPT_post_k4}{misfit_pct_mean}{3.916}
\setres{lotLJ}{CPT_post_k4}{n_seen}{4}
\setres{lotLJ}{CPT_post_k4}{surprise}{0.14}
\setres{lotLJ}{CPT_post_k4}{surprise_mean}{0.771}
\setres{lotLJ}{CPT_post_k4}{width90_adj_pct}{7.8}
\setres{lotLJ}{CPT_post_k4}{width90_adj_pct_mean}{7.947}
\setres{lotLJ}{CPT_post_k4}{width90_pct}{3.5}
\setres{lotLJ}{CPT_post_k4}{width90_pct_mean}{3.461}
\setres{lotLJ}{CPT_post_k40}{cov50}{0.00}
\setres{lotLJ}{CPT_post_k40}{cov50_mean}{0.000}
\setres{lotLJ}{CPT_post_k40}{cov90}{0.00}
\setres{lotLJ}{CPT_post_k40}{cov90_adj}{1.00}
\setres{lotLJ}{CPT_post_k40}{cov90_adj_mean}{0.838}
\setres{lotLJ}{CPT_post_k40}{cov90_mean}{0.050}
\setres{lotLJ}{CPT_post_k40}{kappa}{0.944}
\setres{lotLJ}{CPT_post_k40}{kappa_crps}{0.622}
\setres{lotLJ}{CPT_post_k40}{kappa_crps_mean}{0.445}
\setres{lotLJ}{CPT_post_k40}{kappa_mean}{0.848}
\setres{lotLJ}{CPT_post_k40}{logscore_next}{-2.78}
\setres{lotLJ}{CPT_post_k40}{logscore_next_mean}{-2.818}
\setres{lotLJ}{CPT_post_k40}{misfit_pct}{5.9}
\setres{lotLJ}{CPT_post_k40}{misfit_pct_mean}{5.754}
\setres{lotLJ}{CPT_post_k40}{n_seen}{40}
\setres{lotLJ}{CPT_post_k40}{surprise}{0.01}
\setres{lotLJ}{CPT_post_k40}{surprise_mean}{1.173}
\setres{lotLJ}{CPT_post_k40}{width90_adj_pct}{9.7}
\setres{lotLJ}{CPT_post_k40}{width90_adj_pct_mean}{9.490}
\setres{lotLJ}{CPT_post_k40}{width90_pct}{0.0}
\setres{lotLJ}{CPT_post_k40}{width90_pct_mean}{0.278}
\setres{lotLJ}{CPT_post_k41}{cov50}{0.00}
\setres{lotLJ}{CPT_post_k41}{cov50_mean}{0.000}
\setres{lotLJ}{CPT_post_k41}{cov90}{0.00}
\setres{lotLJ}{CPT_post_k41}{cov90_adj}{1.00}
\setres{lotLJ}{CPT_post_k41}{cov90_adj_mean}{0.867}
\setres{lotLJ}{CPT_post_k41}{cov90_mean}{0.050}
\setres{lotLJ}{CPT_post_k41}{kappa}{0.955}
\setres{lotLJ}{CPT_post_k41}{kappa_crps}{0.645}
\setres{lotLJ}{CPT_post_k41}{kappa_crps_mean}{0.434}
\setres{lotLJ}{CPT_post_k41}{kappa_mean}{0.843}
\setres{lotLJ}{CPT_post_k41}{logscore_next}{-2.79}
\setres{lotLJ}{CPT_post_k41}{logscore_next_mean}{-2.816}
\setres{lotLJ}{CPT_post_k41}{misfit_pct}{5.8}
\setres{lotLJ}{CPT_post_k41}{misfit_pct_mean}{5.737}
\setres{lotLJ}{CPT_post_k41}{n_seen}{41}
\setres{lotLJ}{CPT_post_k41}{surprise}{0.01}
\setres{lotLJ}{CPT_post_k41}{surprise_mean}{1.348}
\setres{lotLJ}{CPT_post_k41}{width90_adj_pct}{9.6}
\setres{lotLJ}{CPT_post_k41}{width90_adj_pct_mean}{9.463}
\setres{lotLJ}{CPT_post_k41}{width90_pct}{0.0}
\setres{lotLJ}{CPT_post_k41}{width90_pct_mean}{0.289}
\setres{lotLJ}{CPT_post_k42}{cov50}{0.00}
\setres{lotLJ}{CPT_post_k42}{cov50_mean}{0.050}
\setres{lotLJ}{CPT_post_k42}{cov90}{0.00}
\setres{lotLJ}{CPT_post_k42}{cov90_adj}{1.00}
\setres{lotLJ}{CPT_post_k42}{cov90_adj_mean}{0.825}
\setres{lotLJ}{CPT_post_k42}{cov90_mean}{0.175}
\setres{lotLJ}{CPT_post_k42}{kappa}{0.956}
\setres{lotLJ}{CPT_post_k42}{kappa_crps}{0.512}
\setres{lotLJ}{CPT_post_k42}{kappa_crps_mean}{0.028}
\setres{lotLJ}{CPT_post_k42}{kappa_mean}{0.856}
\setres{lotLJ}{CPT_post_k42}{logscore_next}{-2.79}
\setres{lotLJ}{CPT_post_k42}{logscore_next_mean}{-2.796}
\setres{lotLJ}{CPT_post_k42}{misfit_pct}{5.8}
\setres{lotLJ}{CPT_post_k42}{misfit_pct_mean}{5.709}
\setres{lotLJ}{CPT_post_k42}{n_seen}{42}
\setres{lotLJ}{CPT_post_k42}{surprise}{0.02}
\setres{lotLJ}{CPT_post_k42}{surprise_mean}{0.279}
\setres{lotLJ}{CPT_post_k42}{width90_adj_pct}{9.5}
\setres{lotLJ}{CPT_post_k42}{width90_adj_pct_mean}{9.447}
\setres{lotLJ}{CPT_post_k42}{width90_pct}{0.7}
\setres{lotLJ}{CPT_post_k42}{width90_pct_mean}{0.585}
\setres{lotLJ}{CPT_post_k43}{cov50}{0.00}
\setres{lotLJ}{CPT_post_k43}{cov50_mean}{0.000}
\setres{lotLJ}{CPT_post_k43}{cov90}{0.00}
\setres{lotLJ}{CPT_post_k43}{cov90_adj}{1.00}
\setres{lotLJ}{CPT_post_k43}{cov90_adj_mean}{0.800}
\setres{lotLJ}{CPT_post_k43}{cov90_mean}{0.200}
\setres{lotLJ}{CPT_post_k43}{kappa}{0.953}
\setres{lotLJ}{CPT_post_k43}{kappa_crps}{0.539}
\setres{lotLJ}{CPT_post_k43}{kappa_crps_mean}{-0.384}
\setres{lotLJ}{CPT_post_k43}{kappa_mean}{-3.902}
\setres{lotLJ}{CPT_post_k43}{logscore_next}{-2.79}
\setres{lotLJ}{CPT_post_k43}{logscore_next_mean}{-2.800}
\setres{lotLJ}{CPT_post_k43}{misfit_pct}{5.8}
\setres{lotLJ}{CPT_post_k43}{misfit_pct_mean}{5.707}
\setres{lotLJ}{CPT_post_k43}{n_seen}{43}
\setres{lotLJ}{CPT_post_k43}{surprise}{0.04}
\setres{lotLJ}{CPT_post_k43}{surprise_mean}{0.360}
\setres{lotLJ}{CPT_post_k43}{width90_adj_pct}{9.5}
\setres{lotLJ}{CPT_post_k43}{width90_adj_pct_mean}{9.442}
\setres{lotLJ}{CPT_post_k43}{width90_pct}{0.6}
\setres{lotLJ}{CPT_post_k43}{width90_pct_mean}{0.602}
\setres{lotLJ}{CPT_post_k44}{cov50}{---}
\setres{lotLJ}{CPT_post_k44}{cov50_mean}{---}
\setres{lotLJ}{CPT_post_k44}{cov90}{---}
\setres{lotLJ}{CPT_post_k44}{cov90_adj}{---}
\setres{lotLJ}{CPT_post_k44}{cov90_adj_mean}{---}
\setres{lotLJ}{CPT_post_k44}{cov90_mean}{---}
\setres{lotLJ}{CPT_post_k44}{kappa}{---}
\setres{lotLJ}{CPT_post_k44}{kappa_crps}{---}
\setres{lotLJ}{CPT_post_k44}{kappa_crps_mean}{---}
\setres{lotLJ}{CPT_post_k44}{kappa_mean}{---}
\setres{lotLJ}{CPT_post_k44}{logscore_next}{---}
\setres{lotLJ}{CPT_post_k44}{logscore_next_mean}{---}
\setres{lotLJ}{CPT_post_k44}{misfit_pct}{---}
\setres{lotLJ}{CPT_post_k44}{misfit_pct_mean}{---}
\setres{lotLJ}{CPT_post_k44}{n_seen}{44}
\setres{lotLJ}{CPT_post_k44}{surprise}{---}
\setres{lotLJ}{CPT_post_k44}{surprise_mean}{---}
\setres{lotLJ}{CPT_post_k44}{width90_adj_pct}{---}
\setres{lotLJ}{CPT_post_k44}{width90_adj_pct_mean}{---}
\setres{lotLJ}{CPT_post_k44}{width90_pct}{---}
\setres{lotLJ}{CPT_post_k44}{width90_pct_mean}{---}
\setres{lotLJ}{CPT_post_k5}{cov50}{0.08}
\setres{lotLJ}{CPT_post_k5}{cov50_mean}{0.101}
\setres{lotLJ}{CPT_post_k5}{cov90}{0.27}
\setres{lotLJ}{CPT_post_k5}{cov90_adj}{0.77}
\setres{lotLJ}{CPT_post_k5}{cov90_adj_mean}{0.737}
\setres{lotLJ}{CPT_post_k5}{cov90_mean}{0.269}
\setres{lotLJ}{CPT_post_k5}{kappa}{0.947}
\setres{lotLJ}{CPT_post_k5}{kappa_crps}{0.662}
\setres{lotLJ}{CPT_post_k5}{kappa_crps_mean}{0.651}
\setres{lotLJ}{CPT_post_k5}{kappa_mean}{0.934}
\setres{lotLJ}{CPT_post_k5}{logscore_next}{-2.79}
\setres{lotLJ}{CPT_post_k5}{logscore_next_mean}{-2.812}
\setres{lotLJ}{CPT_post_k5}{misfit_pct}{4.0}
\setres{lotLJ}{CPT_post_k5}{misfit_pct_mean}{4.594}
\setres{lotLJ}{CPT_post_k5}{n_seen}{5}
\setres{lotLJ}{CPT_post_k5}{surprise}{0.54}
\setres{lotLJ}{CPT_post_k5}{surprise_mean}{2.371}
\setres{lotLJ}{CPT_post_k5}{width90_adj_pct}{7.4}
\setres{lotLJ}{CPT_post_k5}{width90_adj_pct_mean}{8.245}
\setres{lotLJ}{CPT_post_k5}{width90_pct}{2.9}
\setres{lotLJ}{CPT_post_k5}{width90_pct_mean}{2.481}
\setres{lotLJ}{CPT_post_k6}{cov50}{0.11}
\setres{lotLJ}{CPT_post_k6}{cov50_mean}{0.086}
\setres{lotLJ}{CPT_post_k6}{cov90}{0.25}
\setres{lotLJ}{CPT_post_k6}{cov90_adj}{0.80}
\setres{lotLJ}{CPT_post_k6}{cov90_adj_mean}{0.751}
\setres{lotLJ}{CPT_post_k6}{cov90_mean}{0.229}
\setres{lotLJ}{CPT_post_k6}{kappa}{0.953}
\setres{lotLJ}{CPT_post_k6}{kappa_crps}{0.659}
\setres{lotLJ}{CPT_post_k6}{kappa_crps_mean}{0.652}
\setres{lotLJ}{CPT_post_k6}{kappa_mean}{0.944}
\setres{lotLJ}{CPT_post_k6}{logscore_next}{-2.79}
\setres{lotLJ}{CPT_post_k6}{logscore_next_mean}{-2.803}
\setres{lotLJ}{CPT_post_k6}{misfit_pct}{5.7}
\setres{lotLJ}{CPT_post_k6}{misfit_pct_mean}{5.045}
\setres{lotLJ}{CPT_post_k6}{n_seen}{6}
\setres{lotLJ}{CPT_post_k6}{surprise}{0.33}
\setres{lotLJ}{CPT_post_k6}{surprise_mean}{1.678}
\setres{lotLJ}{CPT_post_k6}{width90_adj_pct}{10.0}
\setres{lotLJ}{CPT_post_k6}{width90_adj_pct_mean}{8.700}
\setres{lotLJ}{CPT_post_k6}{width90_pct}{1.8}
\setres{lotLJ}{CPT_post_k6}{width90_pct_mean}{1.901}
\setres{lotLJ}{CPT_post_k7}{cov50}{0.05}
\setres{lotLJ}{CPT_post_k7}{cov50_mean}{0.069}
\setres{lotLJ}{CPT_post_k7}{cov90}{0.16}
\setres{lotLJ}{CPT_post_k7}{cov90_adj}{0.84}
\setres{lotLJ}{CPT_post_k7}{cov90_adj_mean}{0.788}
\setres{lotLJ}{CPT_post_k7}{cov90_mean}{0.191}
\setres{lotLJ}{CPT_post_k7}{kappa}{0.959}
\setres{lotLJ}{CPT_post_k7}{kappa_crps}{0.684}
\setres{lotLJ}{CPT_post_k7}{kappa_crps_mean}{0.669}
\setres{lotLJ}{CPT_post_k7}{kappa_mean}{0.952}
\setres{lotLJ}{CPT_post_k7}{logscore_next}{-2.79}
\setres{lotLJ}{CPT_post_k7}{logscore_next_mean}{-2.796}
\setres{lotLJ}{CPT_post_k7}{misfit_pct}{5.3}
\setres{lotLJ}{CPT_post_k7}{misfit_pct_mean}{5.312}
\setres{lotLJ}{CPT_post_k7}{n_seen}{7}
\setres{lotLJ}{CPT_post_k7}{surprise}{0.11}
\setres{lotLJ}{CPT_post_k7}{surprise_mean}{1.385}
\setres{lotLJ}{CPT_post_k7}{width90_adj_pct}{9.1}
\setres{lotLJ}{CPT_post_k7}{width90_adj_pct_mean}{8.999}
\setres{lotLJ}{CPT_post_k7}{width90_pct}{1.1}
\setres{lotLJ}{CPT_post_k7}{width90_pct_mean}{1.488}
\setres{lotLJ}{CPT_post_k8}{cov50}{0.00}
\setres{lotLJ}{CPT_post_k8}{cov50_mean}{0.051}
\setres{lotLJ}{CPT_post_k8}{cov90}{0.15}
\setres{lotLJ}{CPT_post_k8}{cov90_adj}{0.82}
\setres{lotLJ}{CPT_post_k8}{cov90_adj_mean}{0.824}
\setres{lotLJ}{CPT_post_k8}{cov90_mean}{0.146}
\setres{lotLJ}{CPT_post_k8}{kappa}{0.958}
\setres{lotLJ}{CPT_post_k8}{kappa_crps}{0.680}
\setres{lotLJ}{CPT_post_k8}{kappa_crps_mean}{0.663}
\setres{lotLJ}{CPT_post_k8}{kappa_mean}{0.954}
\setres{lotLJ}{CPT_post_k8}{logscore_next}{-2.79}
\setres{lotLJ}{CPT_post_k8}{logscore_next_mean}{-2.794}
\setres{lotLJ}{CPT_post_k8}{misfit_pct}{5.5}
\setres{lotLJ}{CPT_post_k8}{misfit_pct_mean}{5.342}
\setres{lotLJ}{CPT_post_k8}{n_seen}{8}
\setres{lotLJ}{CPT_post_k8}{surprise}{0.02}
\setres{lotLJ}{CPT_post_k8}{surprise_mean}{0.664}
\setres{lotLJ}{CPT_post_k8}{width90_adj_pct}{9.1}
\setres{lotLJ}{CPT_post_k8}{width90_adj_pct_mean}{8.951}
\setres{lotLJ}{CPT_post_k8}{width90_pct}{1.1}
\setres{lotLJ}{CPT_post_k8}{width90_pct_mean}{1.066}
\setres{lotLJ}{CPT_post_k9}{cov50}{0.01}
\setres{lotLJ}{CPT_post_k9}{cov50_mean}{0.057}
\setres{lotLJ}{CPT_post_k9}{cov90}{0.14}
\setres{lotLJ}{CPT_post_k9}{cov90_adj}{0.84}
\setres{lotLJ}{CPT_post_k9}{cov90_adj_mean}{0.817}
\setres{lotLJ}{CPT_post_k9}{cov90_mean}{0.156}
\setres{lotLJ}{CPT_post_k9}{kappa}{0.962}
\setres{lotLJ}{CPT_post_k9}{kappa_crps}{0.671}
\setres{lotLJ}{CPT_post_k9}{kappa_crps_mean}{0.666}
\setres{lotLJ}{CPT_post_k9}{kappa_mean}{0.957}
\setres{lotLJ}{CPT_post_k9}{logscore_next}{-2.79}
\setres{lotLJ}{CPT_post_k9}{logscore_next_mean}{-2.799}
\setres{lotLJ}{CPT_post_k9}{misfit_pct}{5.4}
\setres{lotLJ}{CPT_post_k9}{misfit_pct_mean}{5.292}
\setres{lotLJ}{CPT_post_k9}{n_seen}{9}
\setres{lotLJ}{CPT_post_k9}{surprise}{0.04}
\setres{lotLJ}{CPT_post_k9}{surprise_mean}{1.330}
\setres{lotLJ}{CPT_post_k9}{width90_adj_pct}{8.9}
\setres{lotLJ}{CPT_post_k9}{width90_adj_pct_mean}{8.850}
\setres{lotLJ}{CPT_post_k9}{width90_pct}{1.0}
\setres{lotLJ}{CPT_post_k9}{width90_pct_mean}{1.095}
\setres{lotLJ}{DA_post_k0}{cov50}{0.09}
\setres{lotLJ}{DA_post_k0}{cov50_mean}{0.091}
\setres{lotLJ}{DA_post_k0}{cov90}{0.43}
\setres{lotLJ}{DA_post_k0}{cov90_adj}{---}
\setres{lotLJ}{DA_post_k0}{cov90_adj_mean}{---}
\setres{lotLJ}{DA_post_k0}{cov90_mean}{0.432}
\setres{lotLJ}{DA_post_k0}{kappa}{0.483}
\setres{lotLJ}{DA_post_k0}{kappa_crps}{0.322}
\setres{lotLJ}{DA_post_k0}{kappa_crps_mean}{0.320}
\setres{lotLJ}{DA_post_k0}{kappa_mean}{0.483}
\setres{lotLJ}{DA_post_k0}{logscore_next}{-2.81}
\setres{lotLJ}{DA_post_k0}{logscore_next_mean}{-2.810}
\setres{lotLJ}{DA_post_k0}{misfit_pct}{---}
\setres{lotLJ}{DA_post_k0}{misfit_pct_mean}{---}
\setres{lotLJ}{DA_post_k0}{n_seen}{0}
\setres{lotLJ}{DA_post_k0}{surprise}{3.89}
\setres{lotLJ}{DA_post_k0}{surprise_mean}{4.113}
\setres{lotLJ}{DA_post_k0}{width90_adj_pct}{---}
\setres{lotLJ}{DA_post_k0}{width90_adj_pct_mean}{---}
\setres{lotLJ}{DA_post_k0}{width90_pct}{10.5}
\setres{lotLJ}{DA_post_k0}{width90_pct_mean}{10.469}
\setres{lotLJ}{DA_post_k1}{cov50}{0.07}
\setres{lotLJ}{DA_post_k1}{cov50_mean}{0.072}
\setres{lotLJ}{DA_post_k1}{cov90}{0.14}
\setres{lotLJ}{DA_post_k1}{cov90_adj}{---}
\setres{lotLJ}{DA_post_k1}{cov90_adj_mean}{---}
\setres{lotLJ}{DA_post_k1}{cov90_mean}{0.153}
\setres{lotLJ}{DA_post_k1}{kappa}{0.221}
\setres{lotLJ}{DA_post_k1}{kappa_crps}{-0.069}
\setres{lotLJ}{DA_post_k1}{kappa_crps_mean}{-0.169}
\setres{lotLJ}{DA_post_k1}{kappa_mean}{0.084}
\setres{lotLJ}{DA_post_k1}{logscore_next}{-2.86}
\setres{lotLJ}{DA_post_k1}{logscore_next_mean}{-3.053}
\setres{lotLJ}{DA_post_k1}{misfit_pct}{---}
\setres{lotLJ}{DA_post_k1}{misfit_pct_mean}{---}
\setres{lotLJ}{DA_post_k1}{n_seen}{1}
\setres{lotLJ}{DA_post_k1}{surprise}{4.36}
\setres{lotLJ}{DA_post_k1}{surprise_mean}{31.892}
\setres{lotLJ}{DA_post_k1}{width90_adj_pct}{---}
\setres{lotLJ}{DA_post_k1}{width90_adj_pct_mean}{---}
\setres{lotLJ}{DA_post_k1}{width90_pct}{3.9}
\setres{lotLJ}{DA_post_k1}{width90_pct_mean}{4.256}
\setres{lotLJ}{DA_post_k10}{cov50}{0.00}
\setres{lotLJ}{DA_post_k10}{cov50_mean}{0.001}
\setres{lotLJ}{DA_post_k10}{cov90}{0.00}
\setres{lotLJ}{DA_post_k10}{cov90_adj}{0.88}
\setres{lotLJ}{DA_post_k10}{cov90_adj_mean}{0.840}
\setres{lotLJ}{DA_post_k10}{cov90_mean}{0.009}
\setres{lotLJ}{DA_post_k10}{kappa}{0.501}
\setres{lotLJ}{DA_post_k10}{kappa_crps}{0.142}
\setres{lotLJ}{DA_post_k10}{kappa_crps_mean}{-0.033}
\setres{lotLJ}{DA_post_k10}{kappa_mean}{0.165}
\setres{lotLJ}{DA_post_k10}{logscore_next}{-2.94}
\setres{lotLJ}{DA_post_k10}{logscore_next_mean}{-3.013}
\setres{lotLJ}{DA_post_k10}{misfit_pct}{16.3}
\setres{lotLJ}{DA_post_k10}{misfit_pct_mean}{17.026}
\setres{lotLJ}{DA_post_k10}{n_seen}{10}
\setres{lotLJ}{DA_post_k10}{surprise}{0.02}
\setres{lotLJ}{DA_post_k10}{surprise_mean}{13.723}
\setres{lotLJ}{DA_post_k10}{width90_adj_pct}{26.9}
\setres{lotLJ}{DA_post_k10}{width90_adj_pct_mean}{28.014}
\setres{lotLJ}{DA_post_k10}{width90_pct}{0.0}
\setres{lotLJ}{DA_post_k10}{width90_pct_mean}{0.373}
\setres{lotLJ}{DA_post_k11}{cov50}{0.00}
\setres{lotLJ}{DA_post_k11}{cov50_mean}{0.008}
\setres{lotLJ}{DA_post_k11}{cov90}{0.00}
\setres{lotLJ}{DA_post_k11}{cov90_adj}{0.91}
\setres{lotLJ}{DA_post_k11}{cov90_adj_mean}{0.888}
\setres{lotLJ}{DA_post_k11}{cov90_mean}{0.017}
\setres{lotLJ}{DA_post_k11}{kappa}{0.560}
\setres{lotLJ}{DA_post_k11}{kappa_crps}{0.246}
\setres{lotLJ}{DA_post_k11}{kappa_crps_mean}{0.038}
\setres{lotLJ}{DA_post_k11}{kappa_mean}{0.226}
\setres{lotLJ}{DA_post_k11}{logscore_next}{-2.89}
\setres{lotLJ}{DA_post_k11}{logscore_next_mean}{-3.004}
\setres{lotLJ}{DA_post_k11}{misfit_pct}{17.4}
\setres{lotLJ}{DA_post_k11}{misfit_pct_mean}{17.297}
\setres{lotLJ}{DA_post_k11}{n_seen}{11}
\setres{lotLJ}{DA_post_k11}{surprise}{0.01}
\setres{lotLJ}{DA_post_k11}{surprise_mean}{2.438}
\setres{lotLJ}{DA_post_k11}{width90_adj_pct}{28.6}
\setres{lotLJ}{DA_post_k11}{width90_adj_pct_mean}{28.464}
\setres{lotLJ}{DA_post_k11}{width90_pct}{0.1}
\setres{lotLJ}{DA_post_k11}{width90_pct_mean}{0.463}
\setres{lotLJ}{DA_post_k12}{cov50}{0.00}
\setres{lotLJ}{DA_post_k12}{cov50_mean}{0.003}
\setres{lotLJ}{DA_post_k12}{cov90}{0.00}
\setres{lotLJ}{DA_post_k12}{cov90_adj}{0.95}
\setres{lotLJ}{DA_post_k12}{cov90_adj_mean}{0.891}
\setres{lotLJ}{DA_post_k12}{cov90_mean}{0.011}
\setres{lotLJ}{DA_post_k12}{kappa}{0.567}
\setres{lotLJ}{DA_post_k12}{kappa_crps}{0.262}
\setres{lotLJ}{DA_post_k12}{kappa_crps_mean}{0.016}
\setres{lotLJ}{DA_post_k12}{kappa_mean}{0.210}
\setres{lotLJ}{DA_post_k12}{logscore_next}{-2.95}
\setres{lotLJ}{DA_post_k12}{logscore_next_mean}{-3.454}
\setres{lotLJ}{DA_post_k12}{misfit_pct}{17.7}
\setres{lotLJ}{DA_post_k12}{misfit_pct_mean}{17.401}
\setres{lotLJ}{DA_post_k12}{n_seen}{12}
\setres{lotLJ}{DA_post_k12}{surprise}{0.10}
\setres{lotLJ}{DA_post_k12}{surprise_mean}{9.967}
\setres{lotLJ}{DA_post_k12}{width90_adj_pct}{29.1}
\setres{lotLJ}{DA_post_k12}{width90_adj_pct_mean}{28.634}
\setres{lotLJ}{DA_post_k12}{width90_pct}{0.2}
\setres{lotLJ}{DA_post_k12}{width90_pct_mean}{0.449}
\setres{lotLJ}{DA_post_k13}{cov50}{0.00}
\setres{lotLJ}{DA_post_k13}{cov50_mean}{0.003}
\setres{lotLJ}{DA_post_k13}{cov90}{0.00}
\setres{lotLJ}{DA_post_k13}{cov90_adj}{0.95}
\setres{lotLJ}{DA_post_k13}{cov90_adj_mean}{0.890}
\setres{lotLJ}{DA_post_k13}{cov90_mean}{0.015}
\setres{lotLJ}{DA_post_k13}{kappa}{0.595}
\setres{lotLJ}{DA_post_k13}{kappa_crps}{0.258}
\setres{lotLJ}{DA_post_k13}{kappa_crps_mean}{0.006}
\setres{lotLJ}{DA_post_k13}{kappa_mean}{0.169}
\setres{lotLJ}{DA_post_k13}{logscore_next}{-2.89}
\setres{lotLJ}{DA_post_k13}{logscore_next_mean}{-3.116}
\setres{lotLJ}{DA_post_k13}{misfit_pct}{17.7}
\setres{lotLJ}{DA_post_k13}{misfit_pct_mean}{17.391}
\setres{lotLJ}{DA_post_k13}{n_seen}{13}
\setres{lotLJ}{DA_post_k13}{surprise}{0.28}
\setres{lotLJ}{DA_post_k13}{surprise_mean}{7.143}
\setres{lotLJ}{DA_post_k13}{width90_adj_pct}{29.2}
\setres{lotLJ}{DA_post_k13}{width90_adj_pct_mean}{28.618}
\setres{lotLJ}{DA_post_k13}{width90_pct}{0.4}
\setres{lotLJ}{DA_post_k13}{width90_pct_mean}{0.513}
\setres{lotLJ}{DA_post_k14}{cov50}{0.00}
\setres{lotLJ}{DA_post_k14}{cov50_mean}{0.003}
\setres{lotLJ}{DA_post_k14}{cov90}{0.00}
\setres{lotLJ}{DA_post_k14}{cov90_adj}{0.95}
\setres{lotLJ}{DA_post_k14}{cov90_adj_mean}{0.897}
\setres{lotLJ}{DA_post_k14}{cov90_mean}{0.008}
\setres{lotLJ}{DA_post_k14}{kappa}{0.593}
\setres{lotLJ}{DA_post_k14}{kappa_crps}{0.288}
\setres{lotLJ}{DA_post_k14}{kappa_crps_mean}{0.008}
\setres{lotLJ}{DA_post_k14}{kappa_mean}{0.171}
\setres{lotLJ}{DA_post_k14}{logscore_next}{-2.94}
\setres{lotLJ}{DA_post_k14}{logscore_next_mean}{-3.054}
\setres{lotLJ}{DA_post_k14}{misfit_pct}{17.6}
\setres{lotLJ}{DA_post_k14}{misfit_pct_mean}{17.504}
\setres{lotLJ}{DA_post_k14}{n_seen}{14}
\setres{lotLJ}{DA_post_k14}{surprise}{0.09}
\setres{lotLJ}{DA_post_k14}{surprise_mean}{5.435}
\setres{lotLJ}{DA_post_k14}{width90_adj_pct}{29.0}
\setres{lotLJ}{DA_post_k14}{width90_adj_pct_mean}{28.806}
\setres{lotLJ}{DA_post_k14}{width90_pct}{0.2}
\setres{lotLJ}{DA_post_k14}{width90_pct_mean}{0.426}
\setres{lotLJ}{DA_post_k15}{cov50}{0.00}
\setres{lotLJ}{DA_post_k15}{cov50_mean}{0.010}
\setres{lotLJ}{DA_post_k15}{cov90}{0.00}
\setres{lotLJ}{DA_post_k15}{cov90_adj}{0.95}
\setres{lotLJ}{DA_post_k15}{cov90_adj_mean}{0.891}
\setres{lotLJ}{DA_post_k15}{cov90_mean}{0.028}
\setres{lotLJ}{DA_post_k15}{kappa}{0.589}
\setres{lotLJ}{DA_post_k15}{kappa_crps}{0.277}
\setres{lotLJ}{DA_post_k15}{kappa_crps_mean}{0.023}
\setres{lotLJ}{DA_post_k15}{kappa_mean}{0.181}
\setres{lotLJ}{DA_post_k15}{logscore_next}{-2.91}
\setres{lotLJ}{DA_post_k15}{logscore_next_mean}{-3.033}
\setres{lotLJ}{DA_post_k15}{misfit_pct}{17.5}
\setres{lotLJ}{DA_post_k15}{misfit_pct_mean}{17.260}
\setres{lotLJ}{DA_post_k15}{n_seen}{15}
\setres{lotLJ}{DA_post_k15}{surprise}{0.03}
\setres{lotLJ}{DA_post_k15}{surprise_mean}{14.710}
\setres{lotLJ}{DA_post_k15}{width90_adj_pct}{28.8}
\setres{lotLJ}{DA_post_k15}{width90_adj_pct_mean}{28.406}
\setres{lotLJ}{DA_post_k15}{width90_pct}{0.4}
\setres{lotLJ}{DA_post_k15}{width90_pct_mean}{0.497}
\setres{lotLJ}{DA_post_k16}{cov50}{0.00}
\setres{lotLJ}{DA_post_k16}{cov50_mean}{0.009}
\setres{lotLJ}{DA_post_k16}{cov90}{0.00}
\setres{lotLJ}{DA_post_k16}{cov90_adj}{0.96}
\setres{lotLJ}{DA_post_k16}{cov90_adj_mean}{0.923}
\setres{lotLJ}{DA_post_k16}{cov90_mean}{0.023}
\setres{lotLJ}{DA_post_k16}{kappa}{0.642}
\setres{lotLJ}{DA_post_k16}{kappa_crps}{0.324}
\setres{lotLJ}{DA_post_k16}{kappa_crps_mean}{0.095}
\setres{lotLJ}{DA_post_k16}{kappa_mean}{0.283}
\setres{lotLJ}{DA_post_k16}{logscore_next}{-2.85}
\setres{lotLJ}{DA_post_k16}{logscore_next_mean}{-3.123}
\setres{lotLJ}{DA_post_k16}{misfit_pct}{17.5}
\setres{lotLJ}{DA_post_k16}{misfit_pct_mean}{17.241}
\setres{lotLJ}{DA_post_k16}{n_seen}{16}
\setres{lotLJ}{DA_post_k16}{surprise}{0.01}
\setres{lotLJ}{DA_post_k16}{surprise_mean}{17.810}
\setres{lotLJ}{DA_post_k16}{width90_adj_pct}{28.9}
\setres{lotLJ}{DA_post_k16}{width90_adj_pct_mean}{28.370}
\setres{lotLJ}{DA_post_k16}{width90_pct}{0.3}
\setres{lotLJ}{DA_post_k16}{width90_pct_mean}{0.421}
\setres{lotLJ}{DA_post_k17}{cov50}{0.00}
\setres{lotLJ}{DA_post_k17}{cov50_mean}{0.004}
\setres{lotLJ}{DA_post_k17}{cov90}{0.00}
\setres{lotLJ}{DA_post_k17}{cov90_adj}{0.96}
\setres{lotLJ}{DA_post_k17}{cov90_adj_mean}{0.926}
\setres{lotLJ}{DA_post_k17}{cov90_mean}{0.009}
\setres{lotLJ}{DA_post_k17}{kappa}{0.643}
\setres{lotLJ}{DA_post_k17}{kappa_crps}{0.346}
\setres{lotLJ}{DA_post_k17}{kappa_crps_mean}{0.106}
\setres{lotLJ}{DA_post_k17}{kappa_mean}{0.307}
\setres{lotLJ}{DA_post_k17}{logscore_next}{-2.88}
\setres{lotLJ}{DA_post_k17}{logscore_next_mean}{-3.165}
\setres{lotLJ}{DA_post_k17}{misfit_pct}{17.7}
\setres{lotLJ}{DA_post_k17}{misfit_pct_mean}{17.281}
\setres{lotLJ}{DA_post_k17}{n_seen}{17}
\setres{lotLJ}{DA_post_k17}{surprise}{0.04}
\setres{lotLJ}{DA_post_k17}{surprise_mean}{7.276}
\setres{lotLJ}{DA_post_k17}{width90_adj_pct}{29.2}
\setres{lotLJ}{DA_post_k17}{width90_adj_pct_mean}{28.433}
\setres{lotLJ}{DA_post_k17}{width90_pct}{0.2}
\setres{lotLJ}{DA_post_k17}{width90_pct_mean}{0.317}
\setres{lotLJ}{DA_post_k18}{cov50}{0.00}
\setres{lotLJ}{DA_post_k18}{cov50_mean}{0.006}
\setres{lotLJ}{DA_post_k18}{cov90}{0.00}
\setres{lotLJ}{DA_post_k18}{cov90_adj}{0.96}
\setres{lotLJ}{DA_post_k18}{cov90_adj_mean}{0.940}
\setres{lotLJ}{DA_post_k18}{cov90_mean}{0.010}
\setres{lotLJ}{DA_post_k18}{kappa}{0.563}
\setres{lotLJ}{DA_post_k18}{kappa_crps}{0.265}
\setres{lotLJ}{DA_post_k18}{kappa_crps_mean}{0.125}
\setres{lotLJ}{DA_post_k18}{kappa_mean}{0.355}
\setres{lotLJ}{DA_post_k18}{logscore_next}{-2.88}
\setres{lotLJ}{DA_post_k18}{logscore_next_mean}{-2.972}
\setres{lotLJ}{DA_post_k18}{misfit_pct}{17.5}
\setres{lotLJ}{DA_post_k18}{misfit_pct_mean}{17.291}
\setres{lotLJ}{DA_post_k18}{n_seen}{18}
\setres{lotLJ}{DA_post_k18}{surprise}{0.10}
\setres{lotLJ}{DA_post_k18}{surprise_mean}{3.664}
\setres{lotLJ}{DA_post_k18}{width90_adj_pct}{28.8}
\setres{lotLJ}{DA_post_k18}{width90_adj_pct_mean}{28.455}
\setres{lotLJ}{DA_post_k18}{width90_pct}{0.3}
\setres{lotLJ}{DA_post_k18}{width90_pct_mean}{0.455}
\setres{lotLJ}{DA_post_k19}{cov50}{0.00}
\setres{lotLJ}{DA_post_k19}{cov50_mean}{0.010}
\setres{lotLJ}{DA_post_k19}{cov90}{0.00}
\setres{lotLJ}{DA_post_k19}{cov90_adj}{0.94}
\setres{lotLJ}{DA_post_k19}{cov90_adj_mean}{0.932}
\setres{lotLJ}{DA_post_k19}{cov90_mean}{0.014}
\setres{lotLJ}{DA_post_k19}{kappa}{0.594}
\setres{lotLJ}{DA_post_k19}{kappa_crps}{0.303}
\setres{lotLJ}{DA_post_k19}{kappa_crps_mean}{0.116}
\setres{lotLJ}{DA_post_k19}{kappa_mean}{0.344}
\setres{lotLJ}{DA_post_k19}{logscore_next}{-2.90}
\setres{lotLJ}{DA_post_k19}{logscore_next_mean}{-3.204}
\setres{lotLJ}{DA_post_k19}{misfit_pct}{17.3}
\setres{lotLJ}{DA_post_k19}{misfit_pct_mean}{17.232}
\setres{lotLJ}{DA_post_k19}{n_seen}{19}
\setres{lotLJ}{DA_post_k19}{surprise}{0.30}
\setres{lotLJ}{DA_post_k19}{surprise_mean}{38.356}
\setres{lotLJ}{DA_post_k19}{width90_adj_pct}{28.5}
\setres{lotLJ}{DA_post_k19}{width90_adj_pct_mean}{28.356}
\setres{lotLJ}{DA_post_k19}{width90_pct}{0.1}
\setres{lotLJ}{DA_post_k19}{width90_pct_mean}{0.309}
\setres{lotLJ}{DA_post_k2}{cov50}{0.02}
\setres{lotLJ}{DA_post_k2}{cov50_mean}{0.031}
\setres{lotLJ}{DA_post_k2}{cov90}{0.08}
\setres{lotLJ}{DA_post_k2}{cov90_adj}{0.33}
\setres{lotLJ}{DA_post_k2}{cov90_adj_mean}{0.460}
\setres{lotLJ}{DA_post_k2}{cov90_mean}{0.079}
\setres{lotLJ}{DA_post_k2}{kappa}{0.525}
\setres{lotLJ}{DA_post_k2}{kappa_crps}{0.251}
\setres{lotLJ}{DA_post_k2}{kappa_crps_mean}{-0.030}
\setres{lotLJ}{DA_post_k2}{kappa_mean}{0.256}
\setres{lotLJ}{DA_post_k2}{logscore_next}{-2.84}
\setres{lotLJ}{DA_post_k2}{logscore_next_mean}{-2.866}
\setres{lotLJ}{DA_post_k2}{misfit_pct}{9.7}
\setres{lotLJ}{DA_post_k2}{misfit_pct_mean}{10.072}
\setres{lotLJ}{DA_post_k2}{n_seen}{2}
\setres{lotLJ}{DA_post_k2}{surprise}{0.41}
\setres{lotLJ}{DA_post_k2}{surprise_mean}{6.552}
\setres{lotLJ}{DA_post_k2}{width90_adj_pct}{16.1}
\setres{lotLJ}{DA_post_k2}{width90_adj_pct_mean}{16.986}
\setres{lotLJ}{DA_post_k2}{width90_pct}{1.8}
\setres{lotLJ}{DA_post_k2}{width90_pct_mean}{2.125}
\setres{lotLJ}{DA_post_k20}{cov50}{0.00}
\setres{lotLJ}{DA_post_k20}{cov50_mean}{0.000}
\setres{lotLJ}{DA_post_k20}{cov90}{0.00}
\setres{lotLJ}{DA_post_k20}{cov90_adj}{0.96}
\setres{lotLJ}{DA_post_k20}{cov90_adj_mean}{0.938}
\setres{lotLJ}{DA_post_k20}{cov90_mean}{0.012}
\setres{lotLJ}{DA_post_k20}{kappa}{0.586}
\setres{lotLJ}{DA_post_k20}{kappa_crps}{0.284}
\setres{lotLJ}{DA_post_k20}{kappa_crps_mean}{0.152}
\setres{lotLJ}{DA_post_k20}{kappa_mean}{0.351}
\setres{lotLJ}{DA_post_k20}{logscore_next}{-2.87}
\setres{lotLJ}{DA_post_k20}{logscore_next_mean}{-3.216}
\setres{lotLJ}{DA_post_k20}{misfit_pct}{17.1}
\setres{lotLJ}{DA_post_k20}{misfit_pct_mean}{17.006}
\setres{lotLJ}{DA_post_k20}{n_seen}{20}
\setres{lotLJ}{DA_post_k20}{surprise}{0.01}
\setres{lotLJ}{DA_post_k20}{surprise_mean}{6.887}
\setres{lotLJ}{DA_post_k20}{width90_adj_pct}{28.2}
\setres{lotLJ}{DA_post_k20}{width90_adj_pct_mean}{27.982}
\setres{lotLJ}{DA_post_k20}{width90_pct}{0.1}
\setres{lotLJ}{DA_post_k20}{width90_pct_mean}{0.368}
\setres{lotLJ}{DA_post_k21}{cov50}{0.00}
\setres{lotLJ}{DA_post_k21}{cov50_mean}{0.004}
\setres{lotLJ}{DA_post_k21}{cov90}{0.00}
\setres{lotLJ}{DA_post_k21}{cov90_adj}{0.96}
\setres{lotLJ}{DA_post_k21}{cov90_adj_mean}{0.930}
\setres{lotLJ}{DA_post_k21}{cov90_mean}{0.017}
\setres{lotLJ}{DA_post_k21}{kappa}{0.496}
\setres{lotLJ}{DA_post_k21}{kappa_crps}{0.250}
\setres{lotLJ}{DA_post_k21}{kappa_crps_mean}{0.130}
\setres{lotLJ}{DA_post_k21}{kappa_mean}{0.331}
\setres{lotLJ}{DA_post_k21}{logscore_next}{-2.87}
\setres{lotLJ}{DA_post_k21}{logscore_next_mean}{-3.161}
\setres{lotLJ}{DA_post_k21}{misfit_pct}{16.9}
\setres{lotLJ}{DA_post_k21}{misfit_pct_mean}{16.918}
\setres{lotLJ}{DA_post_k21}{n_seen}{21}
\setres{lotLJ}{DA_post_k21}{surprise}{0.02}
\setres{lotLJ}{DA_post_k21}{surprise_mean}{25.158}
\setres{lotLJ}{DA_post_k21}{width90_adj_pct}{27.7}
\setres{lotLJ}{DA_post_k21}{width90_adj_pct_mean}{27.836}
\setres{lotLJ}{DA_post_k21}{width90_pct}{0.4}
\setres{lotLJ}{DA_post_k21}{width90_pct_mean}{0.399}
\setres{lotLJ}{DA_post_k22}{cov50}{0.00}
\setres{lotLJ}{DA_post_k22}{cov50_mean}{0.000}
\setres{lotLJ}{DA_post_k22}{cov90}{0.00}
\setres{lotLJ}{DA_post_k22}{cov90_adj}{0.95}
\setres{lotLJ}{DA_post_k22}{cov90_adj_mean}{0.939}
\setres{lotLJ}{DA_post_k22}{cov90_mean}{0.014}
\setres{lotLJ}{DA_post_k22}{kappa}{0.532}
\setres{lotLJ}{DA_post_k22}{kappa_crps}{0.249}
\setres{lotLJ}{DA_post_k22}{kappa_crps_mean}{0.217}
\setres{lotLJ}{DA_post_k22}{kappa_mean}{0.514}
\setres{lotLJ}{DA_post_k22}{logscore_next}{-2.88}
\setres{lotLJ}{DA_post_k22}{logscore_next_mean}{-3.083}
\setres{lotLJ}{DA_post_k22}{misfit_pct}{16.5}
\setres{lotLJ}{DA_post_k22}{misfit_pct_mean}{16.782}
\setres{lotLJ}{DA_post_k22}{n_seen}{22}
\setres{lotLJ}{DA_post_k22}{surprise}{0.01}
\setres{lotLJ}{DA_post_k22}{surprise_mean}{2.191}
\setres{lotLJ}{DA_post_k22}{width90_adj_pct}{27.2}
\setres{lotLJ}{DA_post_k22}{width90_adj_pct_mean}{27.615}
\setres{lotLJ}{DA_post_k22}{width90_pct}{0.2}
\setres{lotLJ}{DA_post_k22}{width90_pct_mean}{0.415}
\setres{lotLJ}{DA_post_k23}{cov50}{0.00}
\setres{lotLJ}{DA_post_k23}{cov50_mean}{0.002}
\setres{lotLJ}{DA_post_k23}{cov90}{0.00}
\setres{lotLJ}{DA_post_k23}{cov90_adj}{0.95}
\setres{lotLJ}{DA_post_k23}{cov90_adj_mean}{0.940}
\setres{lotLJ}{DA_post_k23}{cov90_mean}{0.007}
\setres{lotLJ}{DA_post_k23}{kappa}{0.524}
\setres{lotLJ}{DA_post_k23}{kappa_crps}{0.266}
\setres{lotLJ}{DA_post_k23}{kappa_crps_mean}{0.241}
\setres{lotLJ}{DA_post_k23}{kappa_mean}{0.540}
\setres{lotLJ}{DA_post_k23}{logscore_next}{-2.86}
\setres{lotLJ}{DA_post_k23}{logscore_next_mean}{-2.903}
\setres{lotLJ}{DA_post_k23}{misfit_pct}{16.5}
\setres{lotLJ}{DA_post_k23}{misfit_pct_mean}{16.753}
\setres{lotLJ}{DA_post_k23}{n_seen}{23}
\setres{lotLJ}{DA_post_k23}{surprise}{0.01}
\setres{lotLJ}{DA_post_k23}{surprise_mean}{1.230}
\setres{lotLJ}{DA_post_k23}{width90_adj_pct}{27.1}
\setres{lotLJ}{DA_post_k23}{width90_adj_pct_mean}{27.562}
\setres{lotLJ}{DA_post_k23}{width90_pct}{0.0}
\setres{lotLJ}{DA_post_k23}{width90_pct_mean}{0.199}
\setres{lotLJ}{DA_post_k24}{cov50}{0.00}
\setres{lotLJ}{DA_post_k24}{cov50_mean}{0.003}
\setres{lotLJ}{DA_post_k24}{cov90}{0.00}
\setres{lotLJ}{DA_post_k24}{cov90_adj}{0.95}
\setres{lotLJ}{DA_post_k24}{cov90_adj_mean}{0.947}
\setres{lotLJ}{DA_post_k24}{cov90_mean}{0.005}
\setres{lotLJ}{DA_post_k24}{kappa}{0.512}
\setres{lotLJ}{DA_post_k24}{kappa_crps}{0.258}
\setres{lotLJ}{DA_post_k24}{kappa_crps_mean}{0.209}
\setres{lotLJ}{DA_post_k24}{kappa_mean}{0.451}
\setres{lotLJ}{DA_post_k24}{logscore_next}{-2.89}
\setres{lotLJ}{DA_post_k24}{logscore_next_mean}{-2.967}
\setres{lotLJ}{DA_post_k24}{misfit_pct}{16.5}
\setres{lotLJ}{DA_post_k24}{misfit_pct_mean}{16.757}
\setres{lotLJ}{DA_post_k24}{n_seen}{24}
\setres{lotLJ}{DA_post_k24}{surprise}{0.01}
\setres{lotLJ}{DA_post_k24}{surprise_mean}{1.877}
\setres{lotLJ}{DA_post_k24}{width90_adj_pct}{27.2}
\setres{lotLJ}{DA_post_k24}{width90_adj_pct_mean}{27.569}
\setres{lotLJ}{DA_post_k24}{width90_pct}{0.0}
\setres{lotLJ}{DA_post_k24}{width90_pct_mean}{0.174}
\setres{lotLJ}{DA_post_k25}{cov50}{0.00}
\setres{lotLJ}{DA_post_k25}{cov50_mean}{0.000}
\setres{lotLJ}{DA_post_k25}{cov90}{0.00}
\setres{lotLJ}{DA_post_k25}{cov90_adj}{0.95}
\setres{lotLJ}{DA_post_k25}{cov90_adj_mean}{0.950}
\setres{lotLJ}{DA_post_k25}{cov90_mean}{0.005}
\setres{lotLJ}{DA_post_k25}{kappa}{0.514}
\setres{lotLJ}{DA_post_k25}{kappa_crps}{0.289}
\setres{lotLJ}{DA_post_k25}{kappa_crps_mean}{0.214}
\setres{lotLJ}{DA_post_k25}{kappa_mean}{0.466}
\setres{lotLJ}{DA_post_k25}{logscore_next}{-2.85}
\setres{lotLJ}{DA_post_k25}{logscore_next_mean}{-2.912}
\setres{lotLJ}{DA_post_k25}{misfit_pct}{16.4}
\setres{lotLJ}{DA_post_k25}{misfit_pct_mean}{16.668}
\setres{lotLJ}{DA_post_k25}{n_seen}{25}
\setres{lotLJ}{DA_post_k25}{surprise}{0.02}
\setres{lotLJ}{DA_post_k25}{surprise_mean}{2.534}
\setres{lotLJ}{DA_post_k25}{width90_adj_pct}{27.0}
\setres{lotLJ}{DA_post_k25}{width90_adj_pct_mean}{27.421}
\setres{lotLJ}{DA_post_k25}{width90_pct}{0.0}
\setres{lotLJ}{DA_post_k25}{width90_pct_mean}{0.135}
\setres{lotLJ}{DA_post_k26}{cov50}{0.00}
\setres{lotLJ}{DA_post_k26}{cov50_mean}{0.000}
\setres{lotLJ}{DA_post_k26}{cov90}{0.00}
\setres{lotLJ}{DA_post_k26}{cov90_adj}{0.94}
\setres{lotLJ}{DA_post_k26}{cov90_adj_mean}{0.947}
\setres{lotLJ}{DA_post_k26}{cov90_mean}{0.006}
\setres{lotLJ}{DA_post_k26}{kappa}{0.542}
\setres{lotLJ}{DA_post_k26}{kappa_crps}{0.299}
\setres{lotLJ}{DA_post_k26}{kappa_crps_mean}{0.235}
\setres{lotLJ}{DA_post_k26}{kappa_mean}{0.505}
\setres{lotLJ}{DA_post_k26}{logscore_next}{-2.89}
\setres{lotLJ}{DA_post_k26}{logscore_next_mean}{-3.067}
\setres{lotLJ}{DA_post_k26}{misfit_pct}{16.3}
\setres{lotLJ}{DA_post_k26}{misfit_pct_mean}{16.526}
\setres{lotLJ}{DA_post_k26}{n_seen}{26}
\setres{lotLJ}{DA_post_k26}{surprise}{0.12}
\setres{lotLJ}{DA_post_k26}{surprise_mean}{14.525}
\setres{lotLJ}{DA_post_k26}{width90_adj_pct}{26.8}
\setres{lotLJ}{DA_post_k26}{width90_adj_pct_mean}{27.190}
\setres{lotLJ}{DA_post_k26}{width90_pct}{0.0}
\setres{lotLJ}{DA_post_k26}{width90_pct_mean}{0.189}
\setres{lotLJ}{DA_post_k27}{cov50}{0.00}
\setres{lotLJ}{DA_post_k27}{cov50_mean}{0.003}
\setres{lotLJ}{DA_post_k27}{cov90}{0.00}
\setres{lotLJ}{DA_post_k27}{cov90_adj}{0.94}
\setres{lotLJ}{DA_post_k27}{cov90_adj_mean}{0.947}
\setres{lotLJ}{DA_post_k27}{cov90_mean}{0.006}
\setres{lotLJ}{DA_post_k27}{kappa}{0.540}
\setres{lotLJ}{DA_post_k27}{kappa_crps}{0.316}
\setres{lotLJ}{DA_post_k27}{kappa_crps_mean}{0.273}
\setres{lotLJ}{DA_post_k27}{kappa_mean}{0.518}
\setres{lotLJ}{DA_post_k27}{logscore_next}{-2.84}
\setres{lotLJ}{DA_post_k27}{logscore_next_mean}{-2.912}
\setres{lotLJ}{DA_post_k27}{misfit_pct}{16.2}
\setres{lotLJ}{DA_post_k27}{misfit_pct_mean}{16.416}
\setres{lotLJ}{DA_post_k27}{n_seen}{27}
\setres{lotLJ}{DA_post_k27}{surprise}{0.00}
\setres{lotLJ}{DA_post_k27}{surprise_mean}{5.395}
\setres{lotLJ}{DA_post_k27}{width90_adj_pct}{26.6}
\setres{lotLJ}{DA_post_k27}{width90_adj_pct_mean}{27.009}
\setres{lotLJ}{DA_post_k27}{width90_pct}{0.0}
\setres{lotLJ}{DA_post_k27}{width90_pct_mean}{0.206}
\setres{lotLJ}{DA_post_k28}{cov50}{0.00}
\setres{lotLJ}{DA_post_k28}{cov50_mean}{0.003}
\setres{lotLJ}{DA_post_k28}{cov90}{0.00}
\setres{lotLJ}{DA_post_k28}{cov90_adj}{0.94}
\setres{lotLJ}{DA_post_k28}{cov90_adj_mean}{0.950}
\setres{lotLJ}{DA_post_k28}{cov90_mean}{0.003}
\setres{lotLJ}{DA_post_k28}{kappa}{0.557}
\setres{lotLJ}{DA_post_k28}{kappa_crps}{0.329}
\setres{lotLJ}{DA_post_k28}{kappa_crps_mean}{0.271}
\setres{lotLJ}{DA_post_k28}{kappa_mean}{0.518}
\setres{lotLJ}{DA_post_k28}{logscore_next}{-2.84}
\setres{lotLJ}{DA_post_k28}{logscore_next_mean}{-3.068}
\setres{lotLJ}{DA_post_k28}{misfit_pct}{16.1}
\setres{lotLJ}{DA_post_k28}{misfit_pct_mean}{16.324}
\setres{lotLJ}{DA_post_k28}{n_seen}{28}
\setres{lotLJ}{DA_post_k28}{surprise}{0.01}
\setres{lotLJ}{DA_post_k28}{surprise_mean}{5.532}
\setres{lotLJ}{DA_post_k28}{width90_adj_pct}{26.6}
\setres{lotLJ}{DA_post_k28}{width90_adj_pct_mean}{26.855}
\setres{lotLJ}{DA_post_k28}{width90_pct}{0.0}
\setres{lotLJ}{DA_post_k28}{width90_pct_mean}{0.119}
\setres{lotLJ}{DA_post_k29}{cov50}{0.00}
\setres{lotLJ}{DA_post_k29}{cov50_mean}{0.007}
\setres{lotLJ}{DA_post_k29}{cov90}{0.00}
\setres{lotLJ}{DA_post_k29}{cov90_adj}{0.93}
\setres{lotLJ}{DA_post_k29}{cov90_adj_mean}{0.947}
\setres{lotLJ}{DA_post_k29}{cov90_mean}{0.013}
\setres{lotLJ}{DA_post_k29}{kappa}{0.587}
\setres{lotLJ}{DA_post_k29}{kappa_crps}{0.371}
\setres{lotLJ}{DA_post_k29}{kappa_crps_mean}{0.298}
\setres{lotLJ}{DA_post_k29}{kappa_mean}{0.529}
\setres{lotLJ}{DA_post_k29}{logscore_next}{-2.96}
\setres{lotLJ}{DA_post_k29}{logscore_next_mean}{-3.293}
\setres{lotLJ}{DA_post_k29}{misfit_pct}{16.2}
\setres{lotLJ}{DA_post_k29}{misfit_pct_mean}{16.232}
\setres{lotLJ}{DA_post_k29}{n_seen}{29}
\setres{lotLJ}{DA_post_k29}{surprise}{0.03}
\setres{lotLJ}{DA_post_k29}{surprise_mean}{1.426}
\setres{lotLJ}{DA_post_k29}{width90_adj_pct}{26.6}
\setres{lotLJ}{DA_post_k29}{width90_adj_pct_mean}{26.706}
\setres{lotLJ}{DA_post_k29}{width90_pct}{0.0}
\setres{lotLJ}{DA_post_k29}{width90_pct_mean}{0.216}
\setres{lotLJ}{DA_post_k3}{cov50}{0.01}
\setres{lotLJ}{DA_post_k3}{cov50_mean}{0.026}
\setres{lotLJ}{DA_post_k3}{cov90}{0.05}
\setres{lotLJ}{DA_post_k3}{cov90_adj}{0.82}
\setres{lotLJ}{DA_post_k3}{cov90_adj_mean}{0.718}
\setres{lotLJ}{DA_post_k3}{cov90_mean}{0.052}
\setres{lotLJ}{DA_post_k3}{kappa}{0.642}
\setres{lotLJ}{DA_post_k3}{kappa_crps}{0.297}
\setres{lotLJ}{DA_post_k3}{kappa_crps_mean}{-0.067}
\setres{lotLJ}{DA_post_k3}{kappa_mean}{0.216}
\setres{lotLJ}{DA_post_k3}{logscore_next}{-2.92}
\setres{lotLJ}{DA_post_k3}{logscore_next_mean}{-3.010}
\setres{lotLJ}{DA_post_k3}{misfit_pct}{13.7}
\setres{lotLJ}{DA_post_k3}{misfit_pct_mean}{15.521}
\setres{lotLJ}{DA_post_k3}{n_seen}{3}
\setres{lotLJ}{DA_post_k3}{surprise}{4.72}
\setres{lotLJ}{DA_post_k3}{surprise_mean}{12.880}
\setres{lotLJ}{DA_post_k3}{width90_adj_pct}{22.6}
\setres{lotLJ}{DA_post_k3}{width90_adj_pct_mean}{25.616}
\setres{lotLJ}{DA_post_k3}{width90_pct}{1.0}
\setres{lotLJ}{DA_post_k3}{width90_pct_mean}{1.306}
\setres{lotLJ}{DA_post_k30}{cov50}{0.00}
\setres{lotLJ}{DA_post_k30}{cov50_mean}{0.000}
\setres{lotLJ}{DA_post_k30}{cov90}{0.00}
\setres{lotLJ}{DA_post_k30}{cov90_adj}{0.93}
\setres{lotLJ}{DA_post_k30}{cov90_adj_mean}{0.946}
\setres{lotLJ}{DA_post_k30}{cov90_mean}{0.011}
\setres{lotLJ}{DA_post_k30}{kappa}{0.576}
\setres{lotLJ}{DA_post_k30}{kappa_crps}{0.353}
\setres{lotLJ}{DA_post_k30}{kappa_crps_mean}{0.293}
\setres{lotLJ}{DA_post_k30}{kappa_mean}{0.519}
\setres{lotLJ}{DA_post_k30}{logscore_next}{-2.93}
\setres{lotLJ}{DA_post_k30}{logscore_next_mean}{-3.046}
\setres{lotLJ}{DA_post_k30}{misfit_pct}{16.3}
\setres{lotLJ}{DA_post_k30}{misfit_pct_mean}{16.201}
\setres{lotLJ}{DA_post_k30}{n_seen}{30}
\setres{lotLJ}{DA_post_k30}{surprise}{0.03}
\setres{lotLJ}{DA_post_k30}{surprise_mean}{1.393}
\setres{lotLJ}{DA_post_k30}{width90_adj_pct}{26.8}
\setres{lotLJ}{DA_post_k30}{width90_adj_pct_mean}{26.654}
\setres{lotLJ}{DA_post_k30}{width90_pct}{0.0}
\setres{lotLJ}{DA_post_k30}{width90_pct_mean}{0.211}
\setres{lotLJ}{DA_post_k31}{cov50}{0.00}
\setres{lotLJ}{DA_post_k31}{cov50_mean}{0.000}
\setres{lotLJ}{DA_post_k31}{cov90}{0.00}
\setres{lotLJ}{DA_post_k31}{cov90_adj}{0.92}
\setres{lotLJ}{DA_post_k31}{cov90_adj_mean}{0.946}
\setres{lotLJ}{DA_post_k31}{cov90_mean}{0.000}
\setres{lotLJ}{DA_post_k31}{kappa}{0.562}
\setres{lotLJ}{DA_post_k31}{kappa_crps}{0.374}
\setres{lotLJ}{DA_post_k31}{kappa_crps_mean}{0.291}
\setres{lotLJ}{DA_post_k31}{kappa_mean}{0.538}
\setres{lotLJ}{DA_post_k31}{logscore_next}{-2.85}
\setres{lotLJ}{DA_post_k31}{logscore_next_mean}{-2.951}
\setres{lotLJ}{DA_post_k31}{misfit_pct}{16.3}
\setres{lotLJ}{DA_post_k31}{misfit_pct_mean}{16.178}
\setres{lotLJ}{DA_post_k31}{n_seen}{31}
\setres{lotLJ}{DA_post_k31}{surprise}{0.01}
\setres{lotLJ}{DA_post_k31}{surprise_mean}{1.463}
\setres{lotLJ}{DA_post_k31}{width90_adj_pct}{26.8}
\setres{lotLJ}{DA_post_k31}{width90_adj_pct_mean}{26.615}
\setres{lotLJ}{DA_post_k31}{width90_pct}{0.0}
\setres{lotLJ}{DA_post_k31}{width90_pct_mean}{0.140}
\setres{lotLJ}{DA_post_k32}{cov50}{0.00}
\setres{lotLJ}{DA_post_k32}{cov50_mean}{0.000}
\setres{lotLJ}{DA_post_k32}{cov90}{0.00}
\setres{lotLJ}{DA_post_k32}{cov90_adj}{0.96}
\setres{lotLJ}{DA_post_k32}{cov90_adj_mean}{0.946}
\setres{lotLJ}{DA_post_k32}{cov90_mean}{0.000}
\setres{lotLJ}{DA_post_k32}{kappa}{0.572}
\setres{lotLJ}{DA_post_k32}{kappa_crps}{0.325}
\setres{lotLJ}{DA_post_k32}{kappa_crps_mean}{0.284}
\setres{lotLJ}{DA_post_k32}{kappa_mean}{0.517}
\setres{lotLJ}{DA_post_k32}{logscore_next}{-2.96}
\setres{lotLJ}{DA_post_k32}{logscore_next_mean}{-3.015}
\setres{lotLJ}{DA_post_k32}{misfit_pct}{16.2}
\setres{lotLJ}{DA_post_k32}{misfit_pct_mean}{16.210}
\setres{lotLJ}{DA_post_k32}{n_seen}{32}
\setres{lotLJ}{DA_post_k32}{surprise}{0.01}
\setres{lotLJ}{DA_post_k32}{surprise_mean}{0.707}
\setres{lotLJ}{DA_post_k32}{width90_adj_pct}{26.7}
\setres{lotLJ}{DA_post_k32}{width90_adj_pct_mean}{26.667}
\setres{lotLJ}{DA_post_k32}{width90_pct}{0.0}
\setres{lotLJ}{DA_post_k32}{width90_pct_mean}{0.148}
\setres{lotLJ}{DA_post_k33}{cov50}{0.00}
\setres{lotLJ}{DA_post_k33}{cov50_mean}{0.000}
\setres{lotLJ}{DA_post_k33}{cov90}{0.00}
\setres{lotLJ}{DA_post_k33}{cov90_adj}{1.00}
\setres{lotLJ}{DA_post_k33}{cov90_adj_mean}{0.945}
\setres{lotLJ}{DA_post_k33}{cov90_mean}{0.005}
\setres{lotLJ}{DA_post_k33}{kappa}{0.567}
\setres{lotLJ}{DA_post_k33}{kappa_crps}{0.317}
\setres{lotLJ}{DA_post_k33}{kappa_crps_mean}{0.269}
\setres{lotLJ}{DA_post_k33}{kappa_mean}{0.506}
\setres{lotLJ}{DA_post_k33}{logscore_next}{-2.86}
\setres{lotLJ}{DA_post_k33}{logscore_next_mean}{-2.945}
\setres{lotLJ}{DA_post_k33}{misfit_pct}{16.2}
\setres{lotLJ}{DA_post_k33}{misfit_pct_mean}{16.191}
\setres{lotLJ}{DA_post_k33}{n_seen}{33}
\setres{lotLJ}{DA_post_k33}{surprise}{0.00}
\setres{lotLJ}{DA_post_k33}{surprise_mean}{0.744}
\setres{lotLJ}{DA_post_k33}{width90_adj_pct}{26.6}
\setres{lotLJ}{DA_post_k33}{width90_adj_pct_mean}{26.635}
\setres{lotLJ}{DA_post_k33}{width90_pct}{0.0}
\setres{lotLJ}{DA_post_k33}{width90_pct_mean}{0.126}
\setres{lotLJ}{DA_post_k34}{cov50}{0.00}
\setres{lotLJ}{DA_post_k34}{cov50_mean}{0.000}
\setres{lotLJ}{DA_post_k34}{cov90}{0.00}
\setres{lotLJ}{DA_post_k34}{cov90_adj}{1.00}
\setres{lotLJ}{DA_post_k34}{cov90_adj_mean}{0.955}
\setres{lotLJ}{DA_post_k34}{cov90_mean}{0.000}
\setres{lotLJ}{DA_post_k34}{kappa}{0.574}
\setres{lotLJ}{DA_post_k34}{kappa_crps}{0.335}
\setres{lotLJ}{DA_post_k34}{kappa_crps_mean}{0.259}
\setres{lotLJ}{DA_post_k34}{kappa_mean}{0.495}
\setres{lotLJ}{DA_post_k34}{logscore_next}{-2.86}
\setres{lotLJ}{DA_post_k34}{logscore_next_mean}{-3.234}
\setres{lotLJ}{DA_post_k34}{misfit_pct}{16.3}
\setres{lotLJ}{DA_post_k34}{misfit_pct_mean}{16.226}
\setres{lotLJ}{DA_post_k34}{n_seen}{34}
\setres{lotLJ}{DA_post_k34}{surprise}{0.01}
\setres{lotLJ}{DA_post_k34}{surprise_mean}{13.443}
\setres{lotLJ}{DA_post_k34}{width90_adj_pct}{26.9}
\setres{lotLJ}{DA_post_k34}{width90_adj_pct_mean}{26.694}
\setres{lotLJ}{DA_post_k34}{width90_pct}{0.0}
\setres{lotLJ}{DA_post_k34}{width90_pct_mean}{0.104}
\setres{lotLJ}{DA_post_k35}{cov50}{0.00}
\setres{lotLJ}{DA_post_k35}{cov50_mean}{0.000}
\setres{lotLJ}{DA_post_k35}{cov90}{0.00}
\setres{lotLJ}{DA_post_k35}{cov90_adj}{1.00}
\setres{lotLJ}{DA_post_k35}{cov90_adj_mean}{0.961}
\setres{lotLJ}{DA_post_k35}{cov90_mean}{0.000}
\setres{lotLJ}{DA_post_k35}{kappa}{0.538}
\setres{lotLJ}{DA_post_k35}{kappa_crps}{0.297}
\setres{lotLJ}{DA_post_k35}{kappa_crps_mean}{0.252}
\setres{lotLJ}{DA_post_k35}{kappa_mean}{0.514}
\setres{lotLJ}{DA_post_k35}{logscore_next}{-2.84}
\setres{lotLJ}{DA_post_k35}{logscore_next_mean}{-3.029}
\setres{lotLJ}{DA_post_k35}{misfit_pct}{16.2}
\setres{lotLJ}{DA_post_k35}{misfit_pct_mean}{16.186}
\setres{lotLJ}{DA_post_k35}{n_seen}{35}
\setres{lotLJ}{DA_post_k35}{surprise}{0.00}
\setres{lotLJ}{DA_post_k35}{surprise_mean}{0.459}
\setres{lotLJ}{DA_post_k35}{width90_adj_pct}{26.7}
\setres{lotLJ}{DA_post_k35}{width90_adj_pct_mean}{26.630}
\setres{lotLJ}{DA_post_k35}{width90_pct}{0.0}
\setres{lotLJ}{DA_post_k35}{width90_pct_mean}{0.169}
\setres{lotLJ}{DA_post_k36}{cov50}{0.00}
\setres{lotLJ}{DA_post_k36}{cov50_mean}{0.000}
\setres{lotLJ}{DA_post_k36}{cov90}{0.00}
\setres{lotLJ}{DA_post_k36}{cov90_adj}{1.00}
\setres{lotLJ}{DA_post_k36}{cov90_adj_mean}{0.956}
\setres{lotLJ}{DA_post_k36}{cov90_mean}{0.013}
\setres{lotLJ}{DA_post_k36}{kappa}{0.593}
\setres{lotLJ}{DA_post_k36}{kappa_crps}{0.287}
\setres{lotLJ}{DA_post_k36}{kappa_crps_mean}{0.280}
\setres{lotLJ}{DA_post_k36}{kappa_mean}{0.528}
\setres{lotLJ}{DA_post_k36}{logscore_next}{-2.83}
\setres{lotLJ}{DA_post_k36}{logscore_next_mean}{-3.082}
\setres{lotLJ}{DA_post_k36}{misfit_pct}{16.3}
\setres{lotLJ}{DA_post_k36}{misfit_pct_mean}{16.230}
\setres{lotLJ}{DA_post_k36}{n_seen}{36}
\setres{lotLJ}{DA_post_k36}{surprise}{0.00}
\setres{lotLJ}{DA_post_k36}{surprise_mean}{1.398}
\setres{lotLJ}{DA_post_k36}{width90_adj_pct}{26.8}
\setres{lotLJ}{DA_post_k36}{width90_adj_pct_mean}{26.702}
\setres{lotLJ}{DA_post_k36}{width90_pct}{0.0}
\setres{lotLJ}{DA_post_k36}{width90_pct_mean}{0.198}
\setres{lotLJ}{DA_post_k37}{cov50}{0.00}
\setres{lotLJ}{DA_post_k37}{cov50_mean}{0.000}
\setres{lotLJ}{DA_post_k37}{cov90}{0.00}
\setres{lotLJ}{DA_post_k37}{cov90_adj}{1.00}
\setres{lotLJ}{DA_post_k37}{cov90_adj_mean}{0.950}
\setres{lotLJ}{DA_post_k37}{cov90_mean}{0.000}
\setres{lotLJ}{DA_post_k37}{kappa}{0.475}
\setres{lotLJ}{DA_post_k37}{kappa_crps}{0.315}
\setres{lotLJ}{DA_post_k37}{kappa_crps_mean}{0.193}
\setres{lotLJ}{DA_post_k37}{kappa_mean}{0.470}
\setres{lotLJ}{DA_post_k37}{logscore_next}{-2.86}
\setres{lotLJ}{DA_post_k37}{logscore_next_mean}{-3.075}
\setres{lotLJ}{DA_post_k37}{misfit_pct}{16.2}
\setres{lotLJ}{DA_post_k37}{misfit_pct_mean}{16.199}
\setres{lotLJ}{DA_post_k37}{n_seen}{37}
\setres{lotLJ}{DA_post_k37}{surprise}{0.00}
\setres{lotLJ}{DA_post_k37}{surprise_mean}{7.444}
\setres{lotLJ}{DA_post_k37}{width90_adj_pct}{26.7}
\setres{lotLJ}{DA_post_k37}{width90_adj_pct_mean}{26.648}
\setres{lotLJ}{DA_post_k37}{width90_pct}{0.0}
\setres{lotLJ}{DA_post_k37}{width90_pct_mean}{0.023}
\setres{lotLJ}{DA_post_k38}{cov50}{0.00}
\setres{lotLJ}{DA_post_k38}{cov50_mean}{0.000}
\setres{lotLJ}{DA_post_k38}{cov90}{0.00}
\setres{lotLJ}{DA_post_k38}{cov90_adj}{1.00}
\setres{lotLJ}{DA_post_k38}{cov90_adj_mean}{0.958}
\setres{lotLJ}{DA_post_k38}{cov90_mean}{0.000}
\setres{lotLJ}{DA_post_k38}{kappa}{0.457}
\setres{lotLJ}{DA_post_k38}{kappa_crps}{0.293}
\setres{lotLJ}{DA_post_k38}{kappa_crps_mean}{0.134}
\setres{lotLJ}{DA_post_k38}{kappa_mean}{0.408}
\setres{lotLJ}{DA_post_k38}{logscore_next}{-2.94}
\setres{lotLJ}{DA_post_k38}{logscore_next_mean}{-3.050}
\setres{lotLJ}{DA_post_k38}{misfit_pct}{16.2}
\setres{lotLJ}{DA_post_k38}{misfit_pct_mean}{16.252}
\setres{lotLJ}{DA_post_k38}{n_seen}{38}
\setres{lotLJ}{DA_post_k38}{surprise}{0.00}
\setres{lotLJ}{DA_post_k38}{surprise_mean}{0.977}
\setres{lotLJ}{DA_post_k38}{width90_adj_pct}{26.7}
\setres{lotLJ}{DA_post_k38}{width90_adj_pct_mean}{26.735}
\setres{lotLJ}{DA_post_k38}{width90_pct}{0.0}
\setres{lotLJ}{DA_post_k38}{width90_pct_mean}{0.049}
\setres{lotLJ}{DA_post_k39}{cov50}{0.00}
\setres{lotLJ}{DA_post_k39}{cov50_mean}{0.000}
\setres{lotLJ}{DA_post_k39}{cov90}{0.00}
\setres{lotLJ}{DA_post_k39}{cov90_adj}{1.00}
\setres{lotLJ}{DA_post_k39}{cov90_adj_mean}{0.960}
\setres{lotLJ}{DA_post_k39}{cov90_mean}{0.000}
\setres{lotLJ}{DA_post_k39}{kappa}{0.463}
\setres{lotLJ}{DA_post_k39}{kappa_crps}{0.231}
\setres{lotLJ}{DA_post_k39}{kappa_crps_mean}{0.076}
\setres{lotLJ}{DA_post_k39}{kappa_mean}{0.347}
\setres{lotLJ}{DA_post_k39}{logscore_next}{-2.86}
\setres{lotLJ}{DA_post_k39}{logscore_next_mean}{-3.073}
\setres{lotLJ}{DA_post_k39}{misfit_pct}{16.2}
\setres{lotLJ}{DA_post_k39}{misfit_pct_mean}{16.266}
\setres{lotLJ}{DA_post_k39}{n_seen}{39}
\setres{lotLJ}{DA_post_k39}{surprise}{0.00}
\setres{lotLJ}{DA_post_k39}{surprise_mean}{0.562}
\setres{lotLJ}{DA_post_k39}{width90_adj_pct}{26.7}
\setres{lotLJ}{DA_post_k39}{width90_adj_pct_mean}{26.761}
\setres{lotLJ}{DA_post_k39}{width90_pct}{0.0}
\setres{lotLJ}{DA_post_k39}{width90_pct_mean}{0.104}
\setres{lotLJ}{DA_post_k4}{cov50}{0.00}
\setres{lotLJ}{DA_post_k4}{cov50_mean}{0.005}
\setres{lotLJ}{DA_post_k4}{cov90}{0.03}
\setres{lotLJ}{DA_post_k4}{cov90_adj}{0.85}
\setres{lotLJ}{DA_post_k4}{cov90_adj_mean}{0.791}
\setres{lotLJ}{DA_post_k4}{cov90_mean}{0.036}
\setres{lotLJ}{DA_post_k4}{kappa}{0.439}
\setres{lotLJ}{DA_post_k4}{kappa_crps}{-0.015}
\setres{lotLJ}{DA_post_k4}{kappa_crps_mean}{-0.091}
\setres{lotLJ}{DA_post_k4}{kappa_mean}{0.222}
\setres{lotLJ}{DA_post_k4}{logscore_next}{-2.93}
\setres{lotLJ}{DA_post_k4}{logscore_next_mean}{-3.050}
\setres{lotLJ}{DA_post_k4}{misfit_pct}{15.1}
\setres{lotLJ}{DA_post_k4}{misfit_pct_mean}{17.597}
\setres{lotLJ}{DA_post_k4}{n_seen}{4}
\setres{lotLJ}{DA_post_k4}{surprise}{2.30}
\setres{lotLJ}{DA_post_k4}{surprise_mean}{14.248}
\setres{lotLJ}{DA_post_k4}{width90_adj_pct}{25.0}
\setres{lotLJ}{DA_post_k4}{width90_adj_pct_mean}{28.979}
\setres{lotLJ}{DA_post_k4}{width90_pct}{0.7}
\setres{lotLJ}{DA_post_k4}{width90_pct_mean}{0.837}
\setres{lotLJ}{DA_post_k40}{cov50}{0.00}
\setres{lotLJ}{DA_post_k40}{cov50_mean}{0.000}
\setres{lotLJ}{DA_post_k40}{cov90}{0.00}
\setres{lotLJ}{DA_post_k40}{cov90_adj}{1.00}
\setres{lotLJ}{DA_post_k40}{cov90_adj_mean}{0.963}
\setres{lotLJ}{DA_post_k40}{cov90_mean}{0.000}
\setres{lotLJ}{DA_post_k40}{kappa}{0.427}
\setres{lotLJ}{DA_post_k40}{kappa_crps}{0.312}
\setres{lotLJ}{DA_post_k40}{kappa_crps_mean}{0.117}
\setres{lotLJ}{DA_post_k40}{kappa_mean}{0.276}
\setres{lotLJ}{DA_post_k40}{logscore_next}{-2.84}
\setres{lotLJ}{DA_post_k40}{logscore_next_mean}{-3.034}
\setres{lotLJ}{DA_post_k40}{misfit_pct}{16.4}
\setres{lotLJ}{DA_post_k40}{misfit_pct_mean}{16.330}
\setres{lotLJ}{DA_post_k40}{n_seen}{40}
\setres{lotLJ}{DA_post_k40}{surprise}{0.00}
\setres{lotLJ}{DA_post_k40}{surprise_mean}{6.188}
\setres{lotLJ}{DA_post_k40}{width90_adj_pct}{27.0}
\setres{lotLJ}{DA_post_k40}{width90_adj_pct_mean}{26.864}
\setres{lotLJ}{DA_post_k40}{width90_pct}{0.0}
\setres{lotLJ}{DA_post_k40}{width90_pct_mean}{0.078}
\setres{lotLJ}{DA_post_k41}{cov50}{0.00}
\setres{lotLJ}{DA_post_k41}{cov50_mean}{0.000}
\setres{lotLJ}{DA_post_k41}{cov90}{0.00}
\setres{lotLJ}{DA_post_k41}{cov90_adj}{1.00}
\setres{lotLJ}{DA_post_k41}{cov90_adj_mean}{0.983}
\setres{lotLJ}{DA_post_k41}{cov90_mean}{0.000}
\setres{lotLJ}{DA_post_k41}{kappa}{0.432}
\setres{lotLJ}{DA_post_k41}{kappa_crps}{0.251}
\setres{lotLJ}{DA_post_k41}{kappa_crps_mean}{0.123}
\setres{lotLJ}{DA_post_k41}{kappa_mean}{0.320}
\setres{lotLJ}{DA_post_k41}{logscore_next}{-2.87}
\setres{lotLJ}{DA_post_k41}{logscore_next_mean}{-3.030}
\setres{lotLJ}{DA_post_k41}{misfit_pct}{16.3}
\setres{lotLJ}{DA_post_k41}{misfit_pct_mean}{16.281}
\setres{lotLJ}{DA_post_k41}{n_seen}{41}
\setres{lotLJ}{DA_post_k41}{surprise}{0.00}
\setres{lotLJ}{DA_post_k41}{surprise_mean}{0.757}
\setres{lotLJ}{DA_post_k41}{width90_adj_pct}{26.8}
\setres{lotLJ}{DA_post_k41}{width90_adj_pct_mean}{26.784}
\setres{lotLJ}{DA_post_k41}{width90_pct}{0.0}
\setres{lotLJ}{DA_post_k41}{width90_pct_mean}{0.093}
\setres{lotLJ}{DA_post_k42}{cov50}{0.00}
\setres{lotLJ}{DA_post_k42}{cov50_mean}{0.000}
\setres{lotLJ}{DA_post_k42}{cov90}{0.00}
\setres{lotLJ}{DA_post_k42}{cov90_adj}{1.00}
\setres{lotLJ}{DA_post_k42}{cov90_adj_mean}{0.975}
\setres{lotLJ}{DA_post_k42}{cov90_mean}{0.000}
\setres{lotLJ}{DA_post_k42}{kappa}{0.210}
\setres{lotLJ}{DA_post_k42}{kappa_crps}{0.237}
\setres{lotLJ}{DA_post_k42}{kappa_crps_mean}{-0.307}
\setres{lotLJ}{DA_post_k42}{kappa_mean}{0.009}
\setres{lotLJ}{DA_post_k42}{logscore_next}{-2.85}
\setres{lotLJ}{DA_post_k42}{logscore_next_mean}{-3.056}
\setres{lotLJ}{DA_post_k42}{misfit_pct}{16.2}
\setres{lotLJ}{DA_post_k42}{misfit_pct_mean}{16.210}
\setres{lotLJ}{DA_post_k42}{n_seen}{42}
\setres{lotLJ}{DA_post_k42}{surprise}{0.00}
\setres{lotLJ}{DA_post_k42}{surprise_mean}{6.344}
\setres{lotLJ}{DA_post_k42}{width90_adj_pct}{26.7}
\setres{lotLJ}{DA_post_k42}{width90_adj_pct_mean}{26.665}
\setres{lotLJ}{DA_post_k42}{width90_pct}{0.0}
\setres{lotLJ}{DA_post_k42}{width90_pct_mean}{0.032}
\setres{lotLJ}{DA_post_k43}{cov50}{0.00}
\setres{lotLJ}{DA_post_k43}{cov50_mean}{0.000}
\setres{lotLJ}{DA_post_k43}{cov90}{0.00}
\setres{lotLJ}{DA_post_k43}{cov90_adj}{1.00}
\setres{lotLJ}{DA_post_k43}{cov90_adj_mean}{0.950}
\setres{lotLJ}{DA_post_k43}{cov90_mean}{0.000}
\setres{lotLJ}{DA_post_k43}{kappa}{0.250}
\setres{lotLJ}{DA_post_k43}{kappa_crps}{0.238}
\setres{lotLJ}{DA_post_k43}{kappa_crps_mean}{-0.621}
\setres{lotLJ}{DA_post_k43}{kappa_mean}{-2.853}
\setres{lotLJ}{DA_post_k43}{logscore_next}{-2.91}
\setres{lotLJ}{DA_post_k43}{logscore_next_mean}{-2.957}
\setres{lotLJ}{DA_post_k43}{misfit_pct}{16.1}
\setres{lotLJ}{DA_post_k43}{misfit_pct_mean}{16.143}
\setres{lotLJ}{DA_post_k43}{n_seen}{43}
\setres{lotLJ}{DA_post_k43}{surprise}{0.00}
\setres{lotLJ}{DA_post_k43}{surprise_mean}{6.711}
\setres{lotLJ}{DA_post_k43}{width90_adj_pct}{26.6}
\setres{lotLJ}{DA_post_k43}{width90_adj_pct_mean}{26.555}
\setres{lotLJ}{DA_post_k43}{width90_pct}{0.0}
\setres{lotLJ}{DA_post_k43}{width90_pct_mean}{0.037}
\setres{lotLJ}{DA_post_k44}{cov50}{---}
\setres{lotLJ}{DA_post_k44}{cov50_mean}{---}
\setres{lotLJ}{DA_post_k44}{cov90}{---}
\setres{lotLJ}{DA_post_k44}{cov90_adj}{---}
\setres{lotLJ}{DA_post_k44}{cov90_adj_mean}{---}
\setres{lotLJ}{DA_post_k44}{cov90_mean}{---}
\setres{lotLJ}{DA_post_k44}{kappa}{---}
\setres{lotLJ}{DA_post_k44}{kappa_crps}{---}
\setres{lotLJ}{DA_post_k44}{kappa_crps_mean}{---}
\setres{lotLJ}{DA_post_k44}{kappa_mean}{---}
\setres{lotLJ}{DA_post_k44}{logscore_next}{---}
\setres{lotLJ}{DA_post_k44}{logscore_next_mean}{---}
\setres{lotLJ}{DA_post_k44}{misfit_pct}{---}
\setres{lotLJ}{DA_post_k44}{misfit_pct_mean}{---}
\setres{lotLJ}{DA_post_k44}{n_seen}{44}
\setres{lotLJ}{DA_post_k44}{surprise}{---}
\setres{lotLJ}{DA_post_k44}{surprise_mean}{---}
\setres{lotLJ}{DA_post_k44}{width90_adj_pct}{---}
\setres{lotLJ}{DA_post_k44}{width90_adj_pct_mean}{---}
\setres{lotLJ}{DA_post_k44}{width90_pct}{---}
\setres{lotLJ}{DA_post_k44}{width90_pct_mean}{---}
\setres{lotLJ}{DA_post_k5}{cov50}{0.00}
\setres{lotLJ}{DA_post_k5}{cov50_mean}{0.021}
\setres{lotLJ}{DA_post_k5}{cov90}{0.01}
\setres{lotLJ}{DA_post_k5}{cov90_adj}{0.88}
\setres{lotLJ}{DA_post_k5}{cov90_adj_mean}{0.841}
\setres{lotLJ}{DA_post_k5}{cov90_mean}{0.046}
\setres{lotLJ}{DA_post_k5}{kappa}{0.497}
\setres{lotLJ}{DA_post_k5}{kappa_crps}{0.038}
\setres{lotLJ}{DA_post_k5}{kappa_crps_mean}{-0.046}
\setres{lotLJ}{DA_post_k5}{kappa_mean}{0.213}
\setres{lotLJ}{DA_post_k5}{logscore_next}{-2.96}
\setres{lotLJ}{DA_post_k5}{logscore_next_mean}{-3.417}
\setres{lotLJ}{DA_post_k5}{misfit_pct}{16.5}
\setres{lotLJ}{DA_post_k5}{misfit_pct_mean}{18.187}
\setres{lotLJ}{DA_post_k5}{n_seen}{5}
\setres{lotLJ}{DA_post_k5}{surprise}{1.36}
\setres{lotLJ}{DA_post_k5}{surprise_mean}{27.194}
\setres{lotLJ}{DA_post_k5}{width90_adj_pct}{27.2}
\setres{lotLJ}{DA_post_k5}{width90_adj_pct_mean}{29.947}
\setres{lotLJ}{DA_post_k5}{width90_pct}{0.4}
\setres{lotLJ}{DA_post_k5}{width90_pct_mean}{0.809}
\setres{lotLJ}{DA_post_k6}{cov50}{0.00}
\setres{lotLJ}{DA_post_k6}{cov50_mean}{0.013}
\setres{lotLJ}{DA_post_k6}{cov90}{0.01}
\setres{lotLJ}{DA_post_k6}{cov90_adj}{0.92}
\setres{lotLJ}{DA_post_k6}{cov90_adj_mean}{0.883}
\setres{lotLJ}{DA_post_k6}{cov90_mean}{0.026}
\setres{lotLJ}{DA_post_k6}{kappa}{0.444}
\setres{lotLJ}{DA_post_k6}{kappa_crps}{0.193}
\setres{lotLJ}{DA_post_k6}{kappa_crps_mean}{-0.035}
\setres{lotLJ}{DA_post_k6}{kappa_mean}{0.131}
\setres{lotLJ}{DA_post_k6}{logscore_next}{-2.89}
\setres{lotLJ}{DA_post_k6}{logscore_next_mean}{-3.154}
\setres{lotLJ}{DA_post_k6}{misfit_pct}{17.0}
\setres{lotLJ}{DA_post_k6}{misfit_pct_mean}{17.829}
\setres{lotLJ}{DA_post_k6}{n_seen}{6}
\setres{lotLJ}{DA_post_k6}{surprise}{0.97}
\setres{lotLJ}{DA_post_k6}{surprise_mean}{10.158}
\setres{lotLJ}{DA_post_k6}{width90_adj_pct}{28.0}
\setres{lotLJ}{DA_post_k6}{width90_adj_pct_mean}{29.349}
\setres{lotLJ}{DA_post_k6}{width90_pct}{0.4}
\setres{lotLJ}{DA_post_k6}{width90_pct_mean}{0.733}
\setres{lotLJ}{DA_post_k7}{cov50}{0.00}
\setres{lotLJ}{DA_post_k7}{cov50_mean}{0.016}
\setres{lotLJ}{DA_post_k7}{cov90}{0.00}
\setres{lotLJ}{DA_post_k7}{cov90_adj}{0.92}
\setres{lotLJ}{DA_post_k7}{cov90_adj_mean}{0.891}
\setres{lotLJ}{DA_post_k7}{cov90_mean}{0.028}
\setres{lotLJ}{DA_post_k7}{kappa}{0.648}
\setres{lotLJ}{DA_post_k7}{kappa_crps}{0.346}
\setres{lotLJ}{DA_post_k7}{kappa_crps_mean}{0.059}
\setres{lotLJ}{DA_post_k7}{kappa_mean}{0.242}
\setres{lotLJ}{DA_post_k7}{logscore_next}{-2.95}
\setres{lotLJ}{DA_post_k7}{logscore_next_mean}{-3.355}
\setres{lotLJ}{DA_post_k7}{misfit_pct}{16.9}
\setres{lotLJ}{DA_post_k7}{misfit_pct_mean}{17.675}
\setres{lotLJ}{DA_post_k7}{n_seen}{7}
\setres{lotLJ}{DA_post_k7}{surprise}{0.25}
\setres{lotLJ}{DA_post_k7}{surprise_mean}{52.443}
\setres{lotLJ}{DA_post_k7}{width90_adj_pct}{27.8}
\setres{lotLJ}{DA_post_k7}{width90_adj_pct_mean}{29.090}
\setres{lotLJ}{DA_post_k7}{width90_pct}{0.5}
\setres{lotLJ}{DA_post_k7}{width90_pct_mean}{0.567}
\setres{lotLJ}{DA_post_k8}{cov50}{0.00}
\setres{lotLJ}{DA_post_k8}{cov50_mean}{0.006}
\setres{lotLJ}{DA_post_k8}{cov90}{0.00}
\setres{lotLJ}{DA_post_k8}{cov90_adj}{0.92}
\setres{lotLJ}{DA_post_k8}{cov90_adj_mean}{0.856}
\setres{lotLJ}{DA_post_k8}{cov90_mean}{0.021}
\setres{lotLJ}{DA_post_k8}{kappa}{0.612}
\setres{lotLJ}{DA_post_k8}{kappa_crps}{0.313}
\setres{lotLJ}{DA_post_k8}{kappa_crps_mean}{-0.025}
\setres{lotLJ}{DA_post_k8}{kappa_mean}{0.119}
\setres{lotLJ}{DA_post_k8}{logscore_next}{-2.90}
\setres{lotLJ}{DA_post_k8}{logscore_next_mean}{-2.996}
\setres{lotLJ}{DA_post_k8}{misfit_pct}{16.6}
\setres{lotLJ}{DA_post_k8}{misfit_pct_mean}{17.142}
\setres{lotLJ}{DA_post_k8}{n_seen}{8}
\setres{lotLJ}{DA_post_k8}{surprise}{0.14}
\setres{lotLJ}{DA_post_k8}{surprise_mean}{5.415}
\setres{lotLJ}{DA_post_k8}{width90_adj_pct}{27.3}
\setres{lotLJ}{DA_post_k8}{width90_adj_pct_mean}{28.210}
\setres{lotLJ}{DA_post_k8}{width90_pct}{0.2}
\setres{lotLJ}{DA_post_k8}{width90_pct_mean}{0.518}
\setres{lotLJ}{DA_post_k9}{cov50}{0.00}
\setres{lotLJ}{DA_post_k9}{cov50_mean}{0.006}
\setres{lotLJ}{DA_post_k9}{cov90}{0.00}
\setres{lotLJ}{DA_post_k9}{cov90_adj}{0.87}
\setres{lotLJ}{DA_post_k9}{cov90_adj_mean}{0.840}
\setres{lotLJ}{DA_post_k9}{cov90_mean}{0.007}
\setres{lotLJ}{DA_post_k9}{kappa}{0.594}
\setres{lotLJ}{DA_post_k9}{kappa_crps}{0.271}
\setres{lotLJ}{DA_post_k9}{kappa_crps_mean}{-0.019}
\setres{lotLJ}{DA_post_k9}{kappa_mean}{0.145}
\setres{lotLJ}{DA_post_k9}{logscore_next}{-2.84}
\setres{lotLJ}{DA_post_k9}{logscore_next_mean}{-3.093}
\setres{lotLJ}{DA_post_k9}{misfit_pct}{16.4}
\setres{lotLJ}{DA_post_k9}{misfit_pct_mean}{16.794}
\setres{lotLJ}{DA_post_k9}{n_seen}{9}
\setres{lotLJ}{DA_post_k9}{surprise}{0.16}
\setres{lotLJ}{DA_post_k9}{surprise_mean}{52.196}
\setres{lotLJ}{DA_post_k9}{width90_adj_pct}{27.1}
\setres{lotLJ}{DA_post_k9}{width90_adj_pct_mean}{27.636}
\setres{lotLJ}{DA_post_k9}{width90_pct}{0.0}
\setres{lotLJ}{DA_post_k9}{width90_pct_mean}{0.370}
\setres{lotLJ}{EU_post_k0}{cov50}{0.07}
\setres{lotLJ}{EU_post_k0}{cov50_mean}{0.068}
\setres{lotLJ}{EU_post_k0}{cov90}{0.18}
\setres{lotLJ}{EU_post_k0}{cov90_adj}{---}
\setres{lotLJ}{EU_post_k0}{cov90_adj_mean}{---}
\setres{lotLJ}{EU_post_k0}{cov90_mean}{0.182}
\setres{lotLJ}{EU_post_k0}{kappa}{0.697}
\setres{lotLJ}{EU_post_k0}{kappa_crps}{0.397}
\setres{lotLJ}{EU_post_k0}{kappa_crps_mean}{0.398}
\setres{lotLJ}{EU_post_k0}{kappa_mean}{0.697}
\setres{lotLJ}{EU_post_k0}{logscore_next}{-2.85}
\setres{lotLJ}{EU_post_k0}{logscore_next_mean}{-2.961}
\setres{lotLJ}{EU_post_k0}{misfit_pct}{---}
\setres{lotLJ}{EU_post_k0}{misfit_pct_mean}{---}
\setres{lotLJ}{EU_post_k0}{n_seen}{0}
\setres{lotLJ}{EU_post_k0}{surprise}{2.60}
\setres{lotLJ}{EU_post_k0}{surprise_mean}{3.332}
\setres{lotLJ}{EU_post_k0}{width90_adj_pct}{---}
\setres{lotLJ}{EU_post_k0}{width90_adj_pct_mean}{---}
\setres{lotLJ}{EU_post_k0}{width90_pct}{2.4}
\setres{lotLJ}{EU_post_k0}{width90_pct_mean}{2.395}
\setres{lotLJ}{EU_post_k1}{cov50}{0.00}
\setres{lotLJ}{EU_post_k1}{cov50_mean}{0.026}
\setres{lotLJ}{EU_post_k1}{cov90}{0.02}
\setres{lotLJ}{EU_post_k1}{cov90_adj}{---}
\setres{lotLJ}{EU_post_k1}{cov90_adj_mean}{---}
\setres{lotLJ}{EU_post_k1}{cov90_mean}{0.065}
\setres{lotLJ}{EU_post_k1}{kappa}{0.724}
\setres{lotLJ}{EU_post_k1}{kappa_crps}{0.399}
\setres{lotLJ}{EU_post_k1}{kappa_crps_mean}{0.312}
\setres{lotLJ}{EU_post_k1}{kappa_mean}{0.614}
\setres{lotLJ}{EU_post_k1}{logscore_next}{-2.87}
\setres{lotLJ}{EU_post_k1}{logscore_next_mean}{-3.175}
\setres{lotLJ}{EU_post_k1}{misfit_pct}{---}
\setres{lotLJ}{EU_post_k1}{misfit_pct_mean}{---}
\setres{lotLJ}{EU_post_k1}{n_seen}{1}
\setres{lotLJ}{EU_post_k1}{surprise}{0.81}
\setres{lotLJ}{EU_post_k1}{surprise_mean}{27.169}
\setres{lotLJ}{EU_post_k1}{width90_adj_pct}{---}
\setres{lotLJ}{EU_post_k1}{width90_adj_pct_mean}{---}
\setres{lotLJ}{EU_post_k1}{width90_pct}{0.5}
\setres{lotLJ}{EU_post_k1}{width90_pct_mean}{0.885}
\setres{lotLJ}{EU_post_k10}{cov50}{0.00}
\setres{lotLJ}{EU_post_k10}{cov50_mean}{0.003}
\setres{lotLJ}{EU_post_k10}{cov90}{0.00}
\setres{lotLJ}{EU_post_k10}{cov90_adj}{0.94}
\setres{lotLJ}{EU_post_k10}{cov90_adj_mean}{0.913}
\setres{lotLJ}{EU_post_k10}{cov90_mean}{0.007}
\setres{lotLJ}{EU_post_k10}{kappa}{0.692}
\setres{lotLJ}{EU_post_k10}{kappa_crps}{0.305}
\setres{lotLJ}{EU_post_k10}{kappa_crps_mean}{0.309}
\setres{lotLJ}{EU_post_k10}{kappa_mean}{0.606}
\setres{lotLJ}{EU_post_k10}{logscore_next}{-2.92}
\setres{lotLJ}{EU_post_k10}{logscore_next_mean}{-2.963}
\setres{lotLJ}{EU_post_k10}{misfit_pct}{16.3}
\setres{lotLJ}{EU_post_k10}{misfit_pct_mean}{16.161}
\setres{lotLJ}{EU_post_k10}{n_seen}{10}
\setres{lotLJ}{EU_post_k10}{surprise}{0.01}
\setres{lotLJ}{EU_post_k10}{surprise_mean}{0.799}
\setres{lotLJ}{EU_post_k10}{width90_adj_pct}{26.8}
\setres{lotLJ}{EU_post_k10}{width90_adj_pct_mean}{26.585}
\setres{lotLJ}{EU_post_k10}{width90_pct}{0.1}
\setres{lotLJ}{EU_post_k10}{width90_pct_mean}{0.131}
\setres{lotLJ}{EU_post_k11}{cov50}{0.00}
\setres{lotLJ}{EU_post_k11}{cov50_mean}{0.002}
\setres{lotLJ}{EU_post_k11}{cov90}{0.00}
\setres{lotLJ}{EU_post_k11}{cov90_adj}{0.97}
\setres{lotLJ}{EU_post_k11}{cov90_adj_mean}{0.936}
\setres{lotLJ}{EU_post_k11}{cov90_mean}{0.009}
\setres{lotLJ}{EU_post_k11}{kappa}{0.698}
\setres{lotLJ}{EU_post_k11}{kappa_crps}{0.310}
\setres{lotLJ}{EU_post_k11}{kappa_crps_mean}{0.316}
\setres{lotLJ}{EU_post_k11}{kappa_mean}{0.616}
\setres{lotLJ}{EU_post_k11}{logscore_next}{-2.88}
\setres{lotLJ}{EU_post_k11}{logscore_next_mean}{-2.991}
\setres{lotLJ}{EU_post_k11}{misfit_pct}{16.7}
\setres{lotLJ}{EU_post_k11}{misfit_pct_mean}{16.507}
\setres{lotLJ}{EU_post_k11}{n_seen}{11}
\setres{lotLJ}{EU_post_k11}{surprise}{0.00}
\setres{lotLJ}{EU_post_k11}{surprise_mean}{6.889}
\setres{lotLJ}{EU_post_k11}{width90_adj_pct}{27.4}
\setres{lotLJ}{EU_post_k11}{width90_adj_pct_mean}{27.154}
\setres{lotLJ}{EU_post_k11}{width90_pct}{0.1}
\setres{lotLJ}{EU_post_k11}{width90_pct_mean}{0.127}
\setres{lotLJ}{EU_post_k12}{cov50}{0.00}
\setres{lotLJ}{EU_post_k12}{cov50_mean}{0.000}
\setres{lotLJ}{EU_post_k12}{cov90}{0.00}
\setres{lotLJ}{EU_post_k12}{cov90_adj}{0.97}
\setres{lotLJ}{EU_post_k12}{cov90_adj_mean}{0.928}
\setres{lotLJ}{EU_post_k12}{cov90_mean}{0.006}
\setres{lotLJ}{EU_post_k12}{kappa}{0.671}
\setres{lotLJ}{EU_post_k12}{kappa_crps}{0.309}
\setres{lotLJ}{EU_post_k12}{kappa_crps_mean}{0.295}
\setres{lotLJ}{EU_post_k12}{kappa_mean}{0.580}
\setres{lotLJ}{EU_post_k12}{logscore_next}{-2.88}
\setres{lotLJ}{EU_post_k12}{logscore_next_mean}{-3.278}
\setres{lotLJ}{EU_post_k12}{misfit_pct}{16.5}
\setres{lotLJ}{EU_post_k12}{misfit_pct_mean}{16.526}
\setres{lotLJ}{EU_post_k12}{n_seen}{12}
\setres{lotLJ}{EU_post_k12}{surprise}{0.01}
\setres{lotLJ}{EU_post_k12}{surprise_mean}{2.204}
\setres{lotLJ}{EU_post_k12}{width90_adj_pct}{27.1}
\setres{lotLJ}{EU_post_k12}{width90_adj_pct_mean}{27.186}
\setres{lotLJ}{EU_post_k12}{width90_pct}{0.1}
\setres{lotLJ}{EU_post_k12}{width90_pct_mean}{0.125}
\setres{lotLJ}{EU_post_k13}{cov50}{0.00}
\setres{lotLJ}{EU_post_k13}{cov50_mean}{0.005}
\setres{lotLJ}{EU_post_k13}{cov90}{0.00}
\setres{lotLJ}{EU_post_k13}{cov90_adj}{0.95}
\setres{lotLJ}{EU_post_k13}{cov90_adj_mean}{0.923}
\setres{lotLJ}{EU_post_k13}{cov90_mean}{0.010}
\setres{lotLJ}{EU_post_k13}{kappa}{0.634}
\setres{lotLJ}{EU_post_k13}{kappa_crps}{0.305}
\setres{lotLJ}{EU_post_k13}{kappa_crps_mean}{0.281}
\setres{lotLJ}{EU_post_k13}{kappa_mean}{0.561}
\setres{lotLJ}{EU_post_k13}{logscore_next}{-2.88}
\setres{lotLJ}{EU_post_k13}{logscore_next_mean}{-3.080}
\setres{lotLJ}{EU_post_k13}{misfit_pct}{16.5}
\setres{lotLJ}{EU_post_k13}{misfit_pct_mean}{16.430}
\setres{lotLJ}{EU_post_k13}{n_seen}{13}
\setres{lotLJ}{EU_post_k13}{surprise}{0.01}
\setres{lotLJ}{EU_post_k13}{surprise_mean}{3.604}
\setres{lotLJ}{EU_post_k13}{width90_adj_pct}{27.2}
\setres{lotLJ}{EU_post_k13}{width90_adj_pct_mean}{27.028}
\setres{lotLJ}{EU_post_k13}{width90_pct}{0.1}
\setres{lotLJ}{EU_post_k13}{width90_pct_mean}{0.126}
\setres{lotLJ}{EU_post_k14}{cov50}{0.00}
\setres{lotLJ}{EU_post_k14}{cov50_mean}{0.002}
\setres{lotLJ}{EU_post_k14}{cov90}{0.00}
\setres{lotLJ}{EU_post_k14}{cov90_adj}{0.97}
\setres{lotLJ}{EU_post_k14}{cov90_adj_mean}{0.932}
\setres{lotLJ}{EU_post_k14}{cov90_mean}{0.008}
\setres{lotLJ}{EU_post_k14}{kappa}{0.653}
\setres{lotLJ}{EU_post_k14}{kappa_crps}{0.313}
\setres{lotLJ}{EU_post_k14}{kappa_crps_mean}{0.294}
\setres{lotLJ}{EU_post_k14}{kappa_mean}{0.579}
\setres{lotLJ}{EU_post_k14}{logscore_next}{-2.95}
\setres{lotLJ}{EU_post_k14}{logscore_next_mean}{-3.072}
\setres{lotLJ}{EU_post_k14}{misfit_pct}{16.8}
\setres{lotLJ}{EU_post_k14}{misfit_pct_mean}{16.443}
\setres{lotLJ}{EU_post_k14}{n_seen}{14}
\setres{lotLJ}{EU_post_k14}{surprise}{0.01}
\setres{lotLJ}{EU_post_k14}{surprise_mean}{0.111}
\setres{lotLJ}{EU_post_k14}{width90_adj_pct}{27.6}
\setres{lotLJ}{EU_post_k14}{width90_adj_pct_mean}{27.050}
\setres{lotLJ}{EU_post_k14}{width90_pct}{0.1}
\setres{lotLJ}{EU_post_k14}{width90_pct_mean}{0.120}
\setres{lotLJ}{EU_post_k15}{cov50}{0.00}
\setres{lotLJ}{EU_post_k15}{cov50_mean}{0.000}
\setres{lotLJ}{EU_post_k15}{cov90}{0.00}
\setres{lotLJ}{EU_post_k15}{cov90_adj}{0.97}
\setres{lotLJ}{EU_post_k15}{cov90_adj_mean}{0.945}
\setres{lotLJ}{EU_post_k15}{cov90_mean}{0.005}
\setres{lotLJ}{EU_post_k15}{kappa}{0.653}
\setres{lotLJ}{EU_post_k15}{kappa_crps}{0.317}
\setres{lotLJ}{EU_post_k15}{kappa_crps_mean}{0.297}
\setres{lotLJ}{EU_post_k15}{kappa_mean}{0.583}
\setres{lotLJ}{EU_post_k15}{logscore_next}{-2.86}
\setres{lotLJ}{EU_post_k15}{logscore_next_mean}{-2.953}
\setres{lotLJ}{EU_post_k15}{misfit_pct}{16.8}
\setres{lotLJ}{EU_post_k15}{misfit_pct_mean}{16.583}
\setres{lotLJ}{EU_post_k15}{n_seen}{15}
\setres{lotLJ}{EU_post_k15}{surprise}{0.00}
\setres{lotLJ}{EU_post_k15}{surprise_mean}{0.131}
\setres{lotLJ}{EU_post_k15}{width90_adj_pct}{27.7}
\setres{lotLJ}{EU_post_k15}{width90_adj_pct_mean}{27.280}
\setres{lotLJ}{EU_post_k15}{width90_pct}{0.1}
\setres{lotLJ}{EU_post_k15}{width90_pct_mean}{0.117}
\setres{lotLJ}{EU_post_k16}{cov50}{0.00}
\setres{lotLJ}{EU_post_k16}{cov50_mean}{0.000}
\setres{lotLJ}{EU_post_k16}{cov90}{0.00}
\setres{lotLJ}{EU_post_k16}{cov90_adj}{0.98}
\setres{lotLJ}{EU_post_k16}{cov90_adj_mean}{0.946}
\setres{lotLJ}{EU_post_k16}{cov90_mean}{0.007}
\setres{lotLJ}{EU_post_k16}{kappa}{0.655}
\setres{lotLJ}{EU_post_k16}{kappa_crps}{0.328}
\setres{lotLJ}{EU_post_k16}{kappa_crps_mean}{0.310}
\setres{lotLJ}{EU_post_k16}{kappa_mean}{0.586}
\setres{lotLJ}{EU_post_k16}{logscore_next}{-2.83}
\setres{lotLJ}{EU_post_k16}{logscore_next_mean}{-3.097}
\setres{lotLJ}{EU_post_k16}{misfit_pct}{16.8}
\setres{lotLJ}{EU_post_k16}{misfit_pct_mean}{16.627}
\setres{lotLJ}{EU_post_k16}{n_seen}{16}
\setres{lotLJ}{EU_post_k16}{surprise}{0.00}
\setres{lotLJ}{EU_post_k16}{surprise_mean}{10.418}
\setres{lotLJ}{EU_post_k16}{width90_adj_pct}{27.6}
\setres{lotLJ}{EU_post_k16}{width90_adj_pct_mean}{27.351}
\setres{lotLJ}{EU_post_k16}{width90_pct}{0.1}
\setres{lotLJ}{EU_post_k16}{width90_pct_mean}{0.126}
\setres{lotLJ}{EU_post_k17}{cov50}{0.00}
\setres{lotLJ}{EU_post_k17}{cov50_mean}{0.000}
\setres{lotLJ}{EU_post_k17}{cov90}{0.00}
\setres{lotLJ}{EU_post_k17}{cov90_adj}{0.96}
\setres{lotLJ}{EU_post_k17}{cov90_adj_mean}{0.943}
\setres{lotLJ}{EU_post_k17}{cov90_mean}{0.007}
\setres{lotLJ}{EU_post_k17}{kappa}{0.663}
\setres{lotLJ}{EU_post_k17}{kappa_crps}{0.327}
\setres{lotLJ}{EU_post_k17}{kappa_crps_mean}{0.310}
\setres{lotLJ}{EU_post_k17}{kappa_mean}{0.592}
\setres{lotLJ}{EU_post_k17}{logscore_next}{-2.90}
\setres{lotLJ}{EU_post_k17}{logscore_next_mean}{-3.125}
\setres{lotLJ}{EU_post_k17}{misfit_pct}{16.9}
\setres{lotLJ}{EU_post_k17}{misfit_pct_mean}{16.672}
\setres{lotLJ}{EU_post_k17}{n_seen}{17}
\setres{lotLJ}{EU_post_k17}{surprise}{0.01}
\setres{lotLJ}{EU_post_k17}{surprise_mean}{2.325}
\setres{lotLJ}{EU_post_k17}{width90_adj_pct}{27.8}
\setres{lotLJ}{EU_post_k17}{width90_adj_pct_mean}{27.425}
\setres{lotLJ}{EU_post_k17}{width90_pct}{0.1}
\setres{lotLJ}{EU_post_k17}{width90_pct_mean}{0.133}
\setres{lotLJ}{EU_post_k18}{cov50}{0.00}
\setres{lotLJ}{EU_post_k18}{cov50_mean}{0.002}
\setres{lotLJ}{EU_post_k18}{cov90}{0.00}
\setres{lotLJ}{EU_post_k18}{cov90_adj}{0.98}
\setres{lotLJ}{EU_post_k18}{cov90_adj_mean}{0.946}
\setres{lotLJ}{EU_post_k18}{cov90_mean}{0.010}
\setres{lotLJ}{EU_post_k18}{kappa}{0.653}
\setres{lotLJ}{EU_post_k18}{kappa_crps}{0.303}
\setres{lotLJ}{EU_post_k18}{kappa_crps_mean}{0.302}
\setres{lotLJ}{EU_post_k18}{kappa_mean}{0.586}
\setres{lotLJ}{EU_post_k18}{logscore_next}{-2.86}
\setres{lotLJ}{EU_post_k18}{logscore_next_mean}{-2.971}
\setres{lotLJ}{EU_post_k18}{misfit_pct}{16.7}
\setres{lotLJ}{EU_post_k18}{misfit_pct_mean}{16.648}
\setres{lotLJ}{EU_post_k18}{n_seen}{18}
\setres{lotLJ}{EU_post_k18}{surprise}{0.01}
\setres{lotLJ}{EU_post_k18}{surprise_mean}{0.395}
\setres{lotLJ}{EU_post_k18}{width90_adj_pct}{27.5}
\setres{lotLJ}{EU_post_k18}{width90_adj_pct_mean}{27.387}
\setres{lotLJ}{EU_post_k18}{width90_pct}{0.1}
\setres{lotLJ}{EU_post_k18}{width90_pct_mean}{0.138}
\setres{lotLJ}{EU_post_k19}{cov50}{0.00}
\setres{lotLJ}{EU_post_k19}{cov50_mean}{0.004}
\setres{lotLJ}{EU_post_k19}{cov90}{0.00}
\setres{lotLJ}{EU_post_k19}{cov90_adj}{0.98}
\setres{lotLJ}{EU_post_k19}{cov90_adj_mean}{0.954}
\setres{lotLJ}{EU_post_k19}{cov90_mean}{0.010}
\setres{lotLJ}{EU_post_k19}{kappa}{0.658}
\setres{lotLJ}{EU_post_k19}{kappa_crps}{0.322}
\setres{lotLJ}{EU_post_k19}{kappa_crps_mean}{0.310}
\setres{lotLJ}{EU_post_k19}{kappa_mean}{0.593}
\setres{lotLJ}{EU_post_k19}{logscore_next}{-2.88}
\setres{lotLJ}{EU_post_k19}{logscore_next_mean}{-3.244}
\setres{lotLJ}{EU_post_k19}{misfit_pct}{16.7}
\setres{lotLJ}{EU_post_k19}{misfit_pct_mean}{16.729}
\setres{lotLJ}{EU_post_k19}{n_seen}{19}
\setres{lotLJ}{EU_post_k19}{surprise}{0.02}
\setres{lotLJ}{EU_post_k19}{surprise_mean}{16.142}
\setres{lotLJ}{EU_post_k19}{width90_adj_pct}{27.4}
\setres{lotLJ}{EU_post_k19}{width90_adj_pct_mean}{27.521}
\setres{lotLJ}{EU_post_k19}{width90_pct}{0.1}
\setres{lotLJ}{EU_post_k19}{width90_pct_mean}{0.140}
\setres{lotLJ}{EU_post_k2}{cov50}{0.00}
\setres{lotLJ}{EU_post_k2}{cov50_mean}{0.020}
\setres{lotLJ}{EU_post_k2}{cov90}{0.02}
\setres{lotLJ}{EU_post_k2}{cov90_adj}{0.82}
\setres{lotLJ}{EU_post_k2}{cov90_adj_mean}{0.694}
\setres{lotLJ}{EU_post_k2}{cov90_mean}{0.043}
\setres{lotLJ}{EU_post_k2}{kappa}{0.733}
\setres{lotLJ}{EU_post_k2}{kappa_crps}{0.385}
\setres{lotLJ}{EU_post_k2}{kappa_crps_mean}{0.326}
\setres{lotLJ}{EU_post_k2}{kappa_mean}{0.634}
\setres{lotLJ}{EU_post_k2}{logscore_next}{-2.84}
\setres{lotLJ}{EU_post_k2}{logscore_next_mean}{-2.981}
\setres{lotLJ}{EU_post_k2}{misfit_pct}{13.2}
\setres{lotLJ}{EU_post_k2}{misfit_pct_mean}{13.840}
\setres{lotLJ}{EU_post_k2}{n_seen}{2}
\setres{lotLJ}{EU_post_k2}{surprise}{0.03}
\setres{lotLJ}{EU_post_k2}{surprise_mean}{2.324}
\setres{lotLJ}{EU_post_k2}{width90_adj_pct}{21.7}
\setres{lotLJ}{EU_post_k2}{width90_adj_pct_mean}{22.786}
\setres{lotLJ}{EU_post_k2}{width90_pct}{0.4}
\setres{lotLJ}{EU_post_k2}{width90_pct_mean}{0.658}
\setres{lotLJ}{EU_post_k20}{cov50}{0.00}
\setres{lotLJ}{EU_post_k20}{cov50_mean}{0.000}
\setres{lotLJ}{EU_post_k20}{cov90}{0.00}
\setres{lotLJ}{EU_post_k20}{cov90_adj}{0.96}
\setres{lotLJ}{EU_post_k20}{cov90_adj_mean}{0.948}
\setres{lotLJ}{EU_post_k20}{cov90_mean}{0.004}
\setres{lotLJ}{EU_post_k20}{kappa}{0.649}
\setres{lotLJ}{EU_post_k20}{kappa_crps}{0.297}
\setres{lotLJ}{EU_post_k20}{kappa_crps_mean}{0.293}
\setres{lotLJ}{EU_post_k20}{kappa_mean}{0.573}
\setres{lotLJ}{EU_post_k20}{logscore_next}{-2.83}
\setres{lotLJ}{EU_post_k20}{logscore_next_mean}{-3.191}
\setres{lotLJ}{EU_post_k20}{misfit_pct}{16.6}
\setres{lotLJ}{EU_post_k20}{misfit_pct_mean}{16.643}
\setres{lotLJ}{EU_post_k20}{n_seen}{20}
\setres{lotLJ}{EU_post_k20}{surprise}{0.00}
\setres{lotLJ}{EU_post_k20}{surprise_mean}{3.426}
\setres{lotLJ}{EU_post_k20}{width90_adj_pct}{27.2}
\setres{lotLJ}{EU_post_k20}{width90_adj_pct_mean}{27.379}
\setres{lotLJ}{EU_post_k20}{width90_pct}{0.1}
\setres{lotLJ}{EU_post_k20}{width90_pct_mean}{0.132}
\setres{lotLJ}{EU_post_k21}{cov50}{0.00}
\setres{lotLJ}{EU_post_k21}{cov50_mean}{0.002}
\setres{lotLJ}{EU_post_k21}{cov90}{0.00}
\setres{lotLJ}{EU_post_k21}{cov90_adj}{0.96}
\setres{lotLJ}{EU_post_k21}{cov90_adj_mean}{0.943}
\setres{lotLJ}{EU_post_k21}{cov90_mean}{0.007}
\setres{lotLJ}{EU_post_k21}{kappa}{0.532}
\setres{lotLJ}{EU_post_k21}{kappa_crps}{0.268}
\setres{lotLJ}{EU_post_k21}{kappa_crps_mean}{0.285}
\setres{lotLJ}{EU_post_k21}{kappa_mean}{0.549}
\setres{lotLJ}{EU_post_k21}{logscore_next}{-2.88}
\setres{lotLJ}{EU_post_k21}{logscore_next_mean}{-3.080}
\setres{lotLJ}{EU_post_k21}{misfit_pct}{16.4}
\setres{lotLJ}{EU_post_k21}{misfit_pct_mean}{16.513}
\setres{lotLJ}{EU_post_k21}{n_seen}{21}
\setres{lotLJ}{EU_post_k21}{surprise}{0.00}
\setres{lotLJ}{EU_post_k21}{surprise_mean}{13.199}
\setres{lotLJ}{EU_post_k21}{width90_adj_pct}{26.9}
\setres{lotLJ}{EU_post_k21}{width90_adj_pct_mean}{27.164}
\setres{lotLJ}{EU_post_k21}{width90_pct}{0.1}
\setres{lotLJ}{EU_post_k21}{width90_pct_mean}{0.126}
\setres{lotLJ}{EU_post_k22}{cov50}{0.00}
\setres{lotLJ}{EU_post_k22}{cov50_mean}{0.002}
\setres{lotLJ}{EU_post_k22}{cov90}{0.00}
\setres{lotLJ}{EU_post_k22}{cov90_adj}{0.95}
\setres{lotLJ}{EU_post_k22}{cov90_adj_mean}{0.948}
\setres{lotLJ}{EU_post_k22}{cov90_mean}{0.002}
\setres{lotLJ}{EU_post_k22}{kappa}{0.549}
\setres{lotLJ}{EU_post_k22}{kappa_crps}{0.298}
\setres{lotLJ}{EU_post_k22}{kappa_crps_mean}{0.290}
\setres{lotLJ}{EU_post_k22}{kappa_mean}{0.565}
\setres{lotLJ}{EU_post_k22}{logscore_next}{-2.89}
\setres{lotLJ}{EU_post_k22}{logscore_next_mean}{-3.055}
\setres{lotLJ}{EU_post_k22}{misfit_pct}{16.5}
\setres{lotLJ}{EU_post_k22}{misfit_pct_mean}{16.651}
\setres{lotLJ}{EU_post_k22}{n_seen}{22}
\setres{lotLJ}{EU_post_k22}{surprise}{0.02}
\setres{lotLJ}{EU_post_k22}{surprise_mean}{0.351}
\setres{lotLJ}{EU_post_k22}{width90_adj_pct}{27.2}
\setres{lotLJ}{EU_post_k22}{width90_adj_pct_mean}{27.391}
\setres{lotLJ}{EU_post_k22}{width90_pct}{0.1}
\setres{lotLJ}{EU_post_k22}{width90_pct_mean}{0.118}
\setres{lotLJ}{EU_post_k23}{cov50}{0.00}
\setres{lotLJ}{EU_post_k23}{cov50_mean}{0.002}
\setres{lotLJ}{EU_post_k23}{cov90}{0.00}
\setres{lotLJ}{EU_post_k23}{cov90_adj}{0.95}
\setres{lotLJ}{EU_post_k23}{cov90_adj_mean}{0.952}
\setres{lotLJ}{EU_post_k23}{cov90_mean}{0.002}
\setres{lotLJ}{EU_post_k23}{kappa}{0.541}
\setres{lotLJ}{EU_post_k23}{kappa_crps}{0.304}
\setres{lotLJ}{EU_post_k23}{kappa_crps_mean}{0.296}
\setres{lotLJ}{EU_post_k23}{kappa_mean}{0.569}
\setres{lotLJ}{EU_post_k23}{logscore_next}{-2.84}
\setres{lotLJ}{EU_post_k23}{logscore_next_mean}{-2.929}
\setres{lotLJ}{EU_post_k23}{misfit_pct}{16.8}
\setres{lotLJ}{EU_post_k23}{misfit_pct_mean}{16.739}
\setres{lotLJ}{EU_post_k23}{n_seen}{23}
\setres{lotLJ}{EU_post_k23}{surprise}{0.00}
\setres{lotLJ}{EU_post_k23}{surprise_mean}{0.193}
\setres{lotLJ}{EU_post_k23}{width90_adj_pct}{27.6}
\setres{lotLJ}{EU_post_k23}{width90_adj_pct_mean}{27.535}
\setres{lotLJ}{EU_post_k23}{width90_pct}{0.1}
\setres{lotLJ}{EU_post_k23}{width90_pct_mean}{0.115}
\setres{lotLJ}{EU_post_k24}{cov50}{0.00}
\setres{lotLJ}{EU_post_k24}{cov50_mean}{0.003}
\setres{lotLJ}{EU_post_k24}{cov90}{0.00}
\setres{lotLJ}{EU_post_k24}{cov90_adj}{0.95}
\setres{lotLJ}{EU_post_k24}{cov90_adj_mean}{0.955}
\setres{lotLJ}{EU_post_k24}{cov90_mean}{0.003}
\setres{lotLJ}{EU_post_k24}{kappa}{0.518}
\setres{lotLJ}{EU_post_k24}{kappa_crps}{0.281}
\setres{lotLJ}{EU_post_k24}{kappa_crps_mean}{0.283}
\setres{lotLJ}{EU_post_k24}{kappa_mean}{0.544}
\setres{lotLJ}{EU_post_k24}{logscore_next}{-2.93}
\setres{lotLJ}{EU_post_k24}{logscore_next_mean}{-2.984}
\setres{lotLJ}{EU_post_k24}{misfit_pct}{16.6}
\setres{lotLJ}{EU_post_k24}{misfit_pct_mean}{16.678}
\setres{lotLJ}{EU_post_k24}{n_seen}{24}
\setres{lotLJ}{EU_post_k24}{surprise}{0.02}
\setres{lotLJ}{EU_post_k24}{surprise_mean}{0.208}
\setres{lotLJ}{EU_post_k24}{width90_adj_pct}{27.3}
\setres{lotLJ}{EU_post_k24}{width90_adj_pct_mean}{27.435}
\setres{lotLJ}{EU_post_k24}{width90_pct}{0.1}
\setres{lotLJ}{EU_post_k24}{width90_pct_mean}{0.118}
\setres{lotLJ}{EU_post_k25}{cov50}{0.00}
\setres{lotLJ}{EU_post_k25}{cov50_mean}{0.000}
\setres{lotLJ}{EU_post_k25}{cov90}{0.00}
\setres{lotLJ}{EU_post_k25}{cov90_adj}{0.95}
\setres{lotLJ}{EU_post_k25}{cov90_adj_mean}{0.961}
\setres{lotLJ}{EU_post_k25}{cov90_mean}{0.003}
\setres{lotLJ}{EU_post_k25}{kappa}{0.488}
\setres{lotLJ}{EU_post_k25}{kappa_crps}{0.296}
\setres{lotLJ}{EU_post_k25}{kappa_crps_mean}{0.293}
\setres{lotLJ}{EU_post_k25}{kappa_mean}{0.550}
\setres{lotLJ}{EU_post_k25}{logscore_next}{-2.84}
\setres{lotLJ}{EU_post_k25}{logscore_next_mean}{-2.889}
\setres{lotLJ}{EU_post_k25}{misfit_pct}{16.5}
\setres{lotLJ}{EU_post_k25}{misfit_pct_mean}{16.652}
\setres{lotLJ}{EU_post_k25}{n_seen}{25}
\setres{lotLJ}{EU_post_k25}{surprise}{0.00}
\setres{lotLJ}{EU_post_k25}{surprise_mean}{0.131}
\setres{lotLJ}{EU_post_k25}{width90_adj_pct}{27.2}
\setres{lotLJ}{EU_post_k25}{width90_adj_pct_mean}{27.393}
\setres{lotLJ}{EU_post_k25}{width90_pct}{0.1}
\setres{lotLJ}{EU_post_k25}{width90_pct_mean}{0.111}
\setres{lotLJ}{EU_post_k26}{cov50}{0.00}
\setres{lotLJ}{EU_post_k26}{cov50_mean}{0.000}
\setres{lotLJ}{EU_post_k26}{cov90}{0.00}
\setres{lotLJ}{EU_post_k26}{cov90_adj}{0.94}
\setres{lotLJ}{EU_post_k26}{cov90_adj_mean}{0.958}
\setres{lotLJ}{EU_post_k26}{cov90_mean}{0.003}
\setres{lotLJ}{EU_post_k26}{kappa}{0.490}
\setres{lotLJ}{EU_post_k26}{kappa_crps}{0.300}
\setres{lotLJ}{EU_post_k26}{kappa_crps_mean}{0.292}
\setres{lotLJ}{EU_post_k26}{kappa_mean}{0.552}
\setres{lotLJ}{EU_post_k26}{logscore_next}{-2.90}
\setres{lotLJ}{EU_post_k26}{logscore_next_mean}{-3.090}
\setres{lotLJ}{EU_post_k26}{misfit_pct}{16.3}
\setres{lotLJ}{EU_post_k26}{misfit_pct_mean}{16.581}
\setres{lotLJ}{EU_post_k26}{n_seen}{26}
\setres{lotLJ}{EU_post_k26}{surprise}{0.02}
\setres{lotLJ}{EU_post_k26}{surprise_mean}{12.773}
\setres{lotLJ}{EU_post_k26}{width90_adj_pct}{26.9}
\setres{lotLJ}{EU_post_k26}{width90_adj_pct_mean}{27.275}
\setres{lotLJ}{EU_post_k26}{width90_pct}{0.1}
\setres{lotLJ}{EU_post_k26}{width90_pct_mean}{0.110}
\setres{lotLJ}{EU_post_k27}{cov50}{0.00}
\setres{lotLJ}{EU_post_k27}{cov50_mean}{0.000}
\setres{lotLJ}{EU_post_k27}{cov90}{0.00}
\setres{lotLJ}{EU_post_k27}{cov90_adj}{0.94}
\setres{lotLJ}{EU_post_k27}{cov90_adj_mean}{0.959}
\setres{lotLJ}{EU_post_k27}{cov90_mean}{0.000}
\setres{lotLJ}{EU_post_k27}{kappa}{0.599}
\setres{lotLJ}{EU_post_k27}{kappa_crps}{0.327}
\setres{lotLJ}{EU_post_k27}{kappa_crps_mean}{0.312}
\setres{lotLJ}{EU_post_k27}{kappa_mean}{0.574}
\setres{lotLJ}{EU_post_k27}{logscore_next}{-2.83}
\setres{lotLJ}{EU_post_k27}{logscore_next_mean}{-2.935}
\setres{lotLJ}{EU_post_k27}{misfit_pct}{16.3}
\setres{lotLJ}{EU_post_k27}{misfit_pct_mean}{16.488}
\setres{lotLJ}{EU_post_k27}{n_seen}{27}
\setres{lotLJ}{EU_post_k27}{surprise}{0.00}
\setres{lotLJ}{EU_post_k27}{surprise_mean}{9.298}
\setres{lotLJ}{EU_post_k27}{width90_adj_pct}{26.7}
\setres{lotLJ}{EU_post_k27}{width90_adj_pct_mean}{27.123}
\setres{lotLJ}{EU_post_k27}{width90_pct}{0.1}
\setres{lotLJ}{EU_post_k27}{width90_pct_mean}{0.111}
\setres{lotLJ}{EU_post_k28}{cov50}{0.00}
\setres{lotLJ}{EU_post_k28}{cov50_mean}{0.000}
\setres{lotLJ}{EU_post_k28}{cov90}{0.00}
\setres{lotLJ}{EU_post_k28}{cov90_adj}{0.97}
\setres{lotLJ}{EU_post_k28}{cov90_adj_mean}{0.959}
\setres{lotLJ}{EU_post_k28}{cov90_mean}{0.000}
\setres{lotLJ}{EU_post_k28}{kappa}{0.609}
\setres{lotLJ}{EU_post_k28}{kappa_crps}{0.365}
\setres{lotLJ}{EU_post_k28}{kappa_crps_mean}{0.318}
\setres{lotLJ}{EU_post_k28}{kappa_mean}{0.573}
\setres{lotLJ}{EU_post_k28}{logscore_next}{-2.85}
\setres{lotLJ}{EU_post_k28}{logscore_next_mean}{-3.069}
\setres{lotLJ}{EU_post_k28}{misfit_pct}{16.5}
\setres{lotLJ}{EU_post_k28}{misfit_pct_mean}{16.432}
\setres{lotLJ}{EU_post_k28}{n_seen}{28}
\setres{lotLJ}{EU_post_k28}{surprise}{0.01}
\setres{lotLJ}{EU_post_k28}{surprise_mean}{0.232}
\setres{lotLJ}{EU_post_k28}{width90_adj_pct}{27.1}
\setres{lotLJ}{EU_post_k28}{width90_adj_pct_mean}{27.032}
\setres{lotLJ}{EU_post_k28}{width90_pct}{0.1}
\setres{lotLJ}{EU_post_k28}{width90_pct_mean}{0.116}
\setres{lotLJ}{EU_post_k29}{cov50}{0.00}
\setres{lotLJ}{EU_post_k29}{cov50_mean}{0.000}
\setres{lotLJ}{EU_post_k29}{cov90}{0.00}
\setres{lotLJ}{EU_post_k29}{cov90_adj}{0.97}
\setres{lotLJ}{EU_post_k29}{cov90_adj_mean}{0.963}
\setres{lotLJ}{EU_post_k29}{cov90_mean}{0.000}
\setres{lotLJ}{EU_post_k29}{kappa}{0.614}
\setres{lotLJ}{EU_post_k29}{kappa_crps}{0.371}
\setres{lotLJ}{EU_post_k29}{kappa_crps_mean}{0.322}
\setres{lotLJ}{EU_post_k29}{kappa_mean}{0.581}
\setres{lotLJ}{EU_post_k29}{logscore_next}{-2.99}
\setres{lotLJ}{EU_post_k29}{logscore_next_mean}{-3.266}
\setres{lotLJ}{EU_post_k29}{misfit_pct}{16.5}
\setres{lotLJ}{EU_post_k29}{misfit_pct_mean}{16.439}
\setres{lotLJ}{EU_post_k29}{n_seen}{29}
\setres{lotLJ}{EU_post_k29}{surprise}{0.02}
\setres{lotLJ}{EU_post_k29}{surprise_mean}{0.279}
\setres{lotLJ}{EU_post_k29}{width90_adj_pct}{27.1}
\setres{lotLJ}{EU_post_k29}{width90_adj_pct_mean}{27.043}
\setres{lotLJ}{EU_post_k29}{width90_pct}{0.1}
\setres{lotLJ}{EU_post_k29}{width90_pct_mean}{0.118}
\setres{lotLJ}{EU_post_k3}{cov50}{0.00}
\setres{lotLJ}{EU_post_k3}{cov50_mean}{0.015}
\setres{lotLJ}{EU_post_k3}{cov90}{0.02}
\setres{lotLJ}{EU_post_k3}{cov90_adj}{0.96}
\setres{lotLJ}{EU_post_k3}{cov90_adj_mean}{0.861}
\setres{lotLJ}{EU_post_k3}{cov90_mean}{0.045}
\setres{lotLJ}{EU_post_k3}{kappa}{0.735}
\setres{lotLJ}{EU_post_k3}{kappa_crps}{0.394}
\setres{lotLJ}{EU_post_k3}{kappa_crps_mean}{0.324}
\setres{lotLJ}{EU_post_k3}{kappa_mean}{0.629}
\setres{lotLJ}{EU_post_k3}{logscore_next}{-2.94}
\setres{lotLJ}{EU_post_k3}{logscore_next_mean}{-3.042}
\setres{lotLJ}{EU_post_k3}{misfit_pct}{16.9}
\setres{lotLJ}{EU_post_k3}{misfit_pct_mean}{16.404}
\setres{lotLJ}{EU_post_k3}{n_seen}{3}
\setres{lotLJ}{EU_post_k3}{surprise}{0.11}
\setres{lotLJ}{EU_post_k3}{surprise_mean}{1.632}
\setres{lotLJ}{EU_post_k3}{width90_adj_pct}{27.8}
\setres{lotLJ}{EU_post_k3}{width90_adj_pct_mean}{26.999}
\setres{lotLJ}{EU_post_k3}{width90_pct}{0.3}
\setres{lotLJ}{EU_post_k3}{width90_pct_mean}{0.596}
\setres{lotLJ}{EU_post_k30}{cov50}{0.00}
\setres{lotLJ}{EU_post_k30}{cov50_mean}{0.000}
\setres{lotLJ}{EU_post_k30}{cov90}{0.00}
\setres{lotLJ}{EU_post_k30}{cov90_adj}{0.96}
\setres{lotLJ}{EU_post_k30}{cov90_adj_mean}{0.954}
\setres{lotLJ}{EU_post_k30}{cov90_mean}{0.004}
\setres{lotLJ}{EU_post_k30}{kappa}{0.620}
\setres{lotLJ}{EU_post_k30}{kappa_crps}{0.393}
\setres{lotLJ}{EU_post_k30}{kappa_crps_mean}{0.314}
\setres{lotLJ}{EU_post_k30}{kappa_mean}{0.571}
\setres{lotLJ}{EU_post_k30}{logscore_next}{-2.87}
\setres{lotLJ}{EU_post_k30}{logscore_next_mean}{-3.045}
\setres{lotLJ}{EU_post_k30}{misfit_pct}{16.4}
\setres{lotLJ}{EU_post_k30}{misfit_pct_mean}{16.372}
\setres{lotLJ}{EU_post_k30}{n_seen}{30}
\setres{lotLJ}{EU_post_k30}{surprise}{0.03}
\setres{lotLJ}{EU_post_k30}{surprise_mean}{0.375}
\setres{lotLJ}{EU_post_k30}{width90_adj_pct}{27.1}
\setres{lotLJ}{EU_post_k30}{width90_adj_pct_mean}{26.933}
\setres{lotLJ}{EU_post_k30}{width90_pct}{0.1}
\setres{lotLJ}{EU_post_k30}{width90_pct_mean}{0.117}
\setres{lotLJ}{EU_post_k31}{cov50}{0.00}
\setres{lotLJ}{EU_post_k31}{cov50_mean}{0.000}
\setres{lotLJ}{EU_post_k31}{cov90}{0.00}
\setres{lotLJ}{EU_post_k31}{cov90_adj}{1.00}
\setres{lotLJ}{EU_post_k31}{cov90_adj_mean}{0.954}
\setres{lotLJ}{EU_post_k31}{cov90_mean}{0.004}
\setres{lotLJ}{EU_post_k31}{kappa}{0.618}
\setres{lotLJ}{EU_post_k31}{kappa_crps}{0.371}
\setres{lotLJ}{EU_post_k31}{kappa_crps_mean}{0.302}
\setres{lotLJ}{EU_post_k31}{kappa_mean}{0.563}
\setres{lotLJ}{EU_post_k31}{logscore_next}{-2.86}
\setres{lotLJ}{EU_post_k31}{logscore_next_mean}{-2.977}
\setres{lotLJ}{EU_post_k31}{misfit_pct}{16.4}
\setres{lotLJ}{EU_post_k31}{misfit_pct_mean}{16.338}
\setres{lotLJ}{EU_post_k31}{n_seen}{31}
\setres{lotLJ}{EU_post_k31}{surprise}{0.00}
\setres{lotLJ}{EU_post_k31}{surprise_mean}{0.186}
\setres{lotLJ}{EU_post_k31}{width90_adj_pct}{26.9}
\setres{lotLJ}{EU_post_k31}{width90_adj_pct_mean}{26.876}
\setres{lotLJ}{EU_post_k31}{width90_pct}{0.1}
\setres{lotLJ}{EU_post_k31}{width90_pct_mean}{0.103}
\setres{lotLJ}{EU_post_k32}{cov50}{0.00}
\setres{lotLJ}{EU_post_k32}{cov50_mean}{0.000}
\setres{lotLJ}{EU_post_k32}{cov90}{0.00}
\setres{lotLJ}{EU_post_k32}{cov90_adj}{1.00}
\setres{lotLJ}{EU_post_k32}{cov90_adj_mean}{0.958}
\setres{lotLJ}{EU_post_k32}{cov90_mean}{0.004}
\setres{lotLJ}{EU_post_k32}{kappa}{0.594}
\setres{lotLJ}{EU_post_k32}{kappa_crps}{0.326}
\setres{lotLJ}{EU_post_k32}{kappa_crps_mean}{0.290}
\setres{lotLJ}{EU_post_k32}{kappa_mean}{0.542}
\setres{lotLJ}{EU_post_k32}{logscore_next}{-2.92}
\setres{lotLJ}{EU_post_k32}{logscore_next_mean}{-3.033}
\setres{lotLJ}{EU_post_k32}{misfit_pct}{16.4}
\setres{lotLJ}{EU_post_k32}{misfit_pct_mean}{16.363}
\setres{lotLJ}{EU_post_k32}{n_seen}{32}
\setres{lotLJ}{EU_post_k32}{surprise}{0.01}
\setres{lotLJ}{EU_post_k32}{surprise_mean}{0.289}
\setres{lotLJ}{EU_post_k32}{width90_adj_pct}{26.9}
\setres{lotLJ}{EU_post_k32}{width90_adj_pct_mean}{26.917}
\setres{lotLJ}{EU_post_k32}{width90_pct}{0.1}
\setres{lotLJ}{EU_post_k32}{width90_pct_mean}{0.094}
\setres{lotLJ}{EU_post_k33}{cov50}{0.00}
\setres{lotLJ}{EU_post_k33}{cov50_mean}{0.000}
\setres{lotLJ}{EU_post_k33}{cov90}{0.00}
\setres{lotLJ}{EU_post_k33}{cov90_adj}{1.00}
\setres{lotLJ}{EU_post_k33}{cov90_adj_mean}{0.959}
\setres{lotLJ}{EU_post_k33}{cov90_mean}{0.005}
\setres{lotLJ}{EU_post_k33}{kappa}{0.581}
\setres{lotLJ}{EU_post_k33}{kappa_crps}{0.313}
\setres{lotLJ}{EU_post_k33}{kappa_crps_mean}{0.277}
\setres{lotLJ}{EU_post_k33}{kappa_mean}{0.524}
\setres{lotLJ}{EU_post_k33}{logscore_next}{-2.86}
\setres{lotLJ}{EU_post_k33}{logscore_next_mean}{-2.955}
\setres{lotLJ}{EU_post_k33}{misfit_pct}{16.4}
\setres{lotLJ}{EU_post_k33}{misfit_pct_mean}{16.358}
\setres{lotLJ}{EU_post_k33}{n_seen}{33}
\setres{lotLJ}{EU_post_k33}{surprise}{0.00}
\setres{lotLJ}{EU_post_k33}{surprise_mean}{0.289}
\setres{lotLJ}{EU_post_k33}{width90_adj_pct}{27.0}
\setres{lotLJ}{EU_post_k33}{width90_adj_pct_mean}{26.910}
\setres{lotLJ}{EU_post_k33}{width90_pct}{0.1}
\setres{lotLJ}{EU_post_k33}{width90_pct_mean}{0.093}
\setres{lotLJ}{EU_post_k34}{cov50}{0.00}
\setres{lotLJ}{EU_post_k34}{cov50_mean}{0.000}
\setres{lotLJ}{EU_post_k34}{cov90}{0.00}
\setres{lotLJ}{EU_post_k34}{cov90_adj}{1.00}
\setres{lotLJ}{EU_post_k34}{cov90_adj_mean}{0.955}
\setres{lotLJ}{EU_post_k34}{cov90_mean}{0.005}
\setres{lotLJ}{EU_post_k34}{kappa}{0.584}
\setres{lotLJ}{EU_post_k34}{kappa_crps}{0.354}
\setres{lotLJ}{EU_post_k34}{kappa_crps_mean}{0.272}
\setres{lotLJ}{EU_post_k34}{kappa_mean}{0.512}
\setres{lotLJ}{EU_post_k34}{logscore_next}{-2.85}
\setres{lotLJ}{EU_post_k34}{logscore_next_mean}{-3.256}
\setres{lotLJ}{EU_post_k34}{misfit_pct}{16.4}
\setres{lotLJ}{EU_post_k34}{misfit_pct_mean}{16.385}
\setres{lotLJ}{EU_post_k34}{n_seen}{34}
\setres{lotLJ}{EU_post_k34}{surprise}{0.00}
\setres{lotLJ}{EU_post_k34}{surprise_mean}{14.069}
\setres{lotLJ}{EU_post_k34}{width90_adj_pct}{27.0}
\setres{lotLJ}{EU_post_k34}{width90_adj_pct_mean}{26.954}
\setres{lotLJ}{EU_post_k34}{width90_pct}{0.1}
\setres{lotLJ}{EU_post_k34}{width90_pct_mean}{0.092}
\setres{lotLJ}{EU_post_k35}{cov50}{0.00}
\setres{lotLJ}{EU_post_k35}{cov50_mean}{0.000}
\setres{lotLJ}{EU_post_k35}{cov90}{0.00}
\setres{lotLJ}{EU_post_k35}{cov90_adj}{1.00}
\setres{lotLJ}{EU_post_k35}{cov90_adj_mean}{0.961}
\setres{lotLJ}{EU_post_k35}{cov90_mean}{0.006}
\setres{lotLJ}{EU_post_k35}{kappa}{0.577}
\setres{lotLJ}{EU_post_k35}{kappa_crps}{0.280}
\setres{lotLJ}{EU_post_k35}{kappa_crps_mean}{0.238}
\setres{lotLJ}{EU_post_k35}{kappa_mean}{0.481}
\setres{lotLJ}{EU_post_k35}{logscore_next}{-2.84}
\setres{lotLJ}{EU_post_k35}{logscore_next_mean}{-3.029}
\setres{lotLJ}{EU_post_k35}{misfit_pct}{16.4}
\setres{lotLJ}{EU_post_k35}{misfit_pct_mean}{16.376}
\setres{lotLJ}{EU_post_k35}{n_seen}{35}
\setres{lotLJ}{EU_post_k35}{surprise}{0.00}
\setres{lotLJ}{EU_post_k35}{surprise_mean}{0.021}
\setres{lotLJ}{EU_post_k35}{width90_adj_pct}{27.0}
\setres{lotLJ}{EU_post_k35}{width90_adj_pct_mean}{26.939}
\setres{lotLJ}{EU_post_k35}{width90_pct}{0.1}
\setres{lotLJ}{EU_post_k35}{width90_pct_mean}{0.099}
\setres{lotLJ}{EU_post_k36}{cov50}{0.00}
\setres{lotLJ}{EU_post_k36}{cov50_mean}{0.000}
\setres{lotLJ}{EU_post_k36}{cov90}{0.00}
\setres{lotLJ}{EU_post_k36}{cov90_adj}{1.00}
\setres{lotLJ}{EU_post_k36}{cov90_adj_mean}{0.956}
\setres{lotLJ}{EU_post_k36}{cov90_mean}{0.006}
\setres{lotLJ}{EU_post_k36}{kappa}{0.587}
\setres{lotLJ}{EU_post_k36}{kappa_crps}{0.294}
\setres{lotLJ}{EU_post_k36}{kappa_crps_mean}{0.261}
\setres{lotLJ}{EU_post_k36}{kappa_mean}{0.502}
\setres{lotLJ}{EU_post_k36}{logscore_next}{-2.83}
\setres{lotLJ}{EU_post_k36}{logscore_next_mean}{-3.143}
\setres{lotLJ}{EU_post_k36}{misfit_pct}{16.5}
\setres{lotLJ}{EU_post_k36}{misfit_pct_mean}{16.422}
\setres{lotLJ}{EU_post_k36}{n_seen}{36}
\setres{lotLJ}{EU_post_k36}{surprise}{0.00}
\setres{lotLJ}{EU_post_k36}{surprise_mean}{2.382}
\setres{lotLJ}{EU_post_k36}{width90_adj_pct}{27.2}
\setres{lotLJ}{EU_post_k36}{width90_adj_pct_mean}{27.014}
\setres{lotLJ}{EU_post_k36}{width90_pct}{0.1}
\setres{lotLJ}{EU_post_k36}{width90_pct_mean}{0.107}
\setres{lotLJ}{EU_post_k37}{cov50}{0.00}
\setres{lotLJ}{EU_post_k37}{cov50_mean}{0.000}
\setres{lotLJ}{EU_post_k37}{cov90}{0.00}
\setres{lotLJ}{EU_post_k37}{cov90_adj}{1.00}
\setres{lotLJ}{EU_post_k37}{cov90_adj_mean}{0.950}
\setres{lotLJ}{EU_post_k37}{cov90_mean}{0.000}
\setres{lotLJ}{EU_post_k37}{kappa}{0.414}
\setres{lotLJ}{EU_post_k37}{kappa_crps}{0.268}
\setres{lotLJ}{EU_post_k37}{kappa_crps_mean}{0.185}
\setres{lotLJ}{EU_post_k37}{kappa_mean}{0.449}
\setres{lotLJ}{EU_post_k37}{logscore_next}{-2.85}
\setres{lotLJ}{EU_post_k37}{logscore_next_mean}{-3.141}
\setres{lotLJ}{EU_post_k37}{misfit_pct}{16.5}
\setres{lotLJ}{EU_post_k37}{misfit_pct_mean}{16.399}
\setres{lotLJ}{EU_post_k37}{n_seen}{37}
\setres{lotLJ}{EU_post_k37}{surprise}{0.02}
\setres{lotLJ}{EU_post_k37}{surprise_mean}{11.436}
\setres{lotLJ}{EU_post_k37}{width90_adj_pct}{27.1}
\setres{lotLJ}{EU_post_k37}{width90_adj_pct_mean}{26.976}
\setres{lotLJ}{EU_post_k37}{width90_pct}{0.1}
\setres{lotLJ}{EU_post_k37}{width90_pct_mean}{0.111}
\setres{lotLJ}{EU_post_k38}{cov50}{0.00}
\setres{lotLJ}{EU_post_k38}{cov50_mean}{0.000}
\setres{lotLJ}{EU_post_k38}{cov90}{0.00}
\setres{lotLJ}{EU_post_k38}{cov90_adj}{1.00}
\setres{lotLJ}{EU_post_k38}{cov90_adj_mean}{0.967}
\setres{lotLJ}{EU_post_k38}{cov90_mean}{0.000}
\setres{lotLJ}{EU_post_k38}{kappa}{0.397}
\setres{lotLJ}{EU_post_k38}{kappa_crps}{0.217}
\setres{lotLJ}{EU_post_k38}{kappa_crps_mean}{0.137}
\setres{lotLJ}{EU_post_k38}{kappa_mean}{0.411}
\setres{lotLJ}{EU_post_k38}{logscore_next}{-2.93}
\setres{lotLJ}{EU_post_k38}{logscore_next_mean}{-3.063}
\setres{lotLJ}{EU_post_k38}{misfit_pct}{16.5}
\setres{lotLJ}{EU_post_k38}{misfit_pct_mean}{16.420}
\setres{lotLJ}{EU_post_k38}{n_seen}{38}
\setres{lotLJ}{EU_post_k38}{surprise}{0.04}
\setres{lotLJ}{EU_post_k38}{surprise_mean}{0.178}
\setres{lotLJ}{EU_post_k38}{width90_adj_pct}{27.1}
\setres{lotLJ}{EU_post_k38}{width90_adj_pct_mean}{27.011}
\setres{lotLJ}{EU_post_k38}{width90_pct}{0.1}
\setres{lotLJ}{EU_post_k38}{width90_pct_mean}{0.114}
\setres{lotLJ}{EU_post_k39}{cov50}{0.00}
\setres{lotLJ}{EU_post_k39}{cov50_mean}{0.000}
\setres{lotLJ}{EU_post_k39}{cov90}{0.00}
\setres{lotLJ}{EU_post_k39}{cov90_adj}{1.00}
\setres{lotLJ}{EU_post_k39}{cov90_adj_mean}{0.970}
\setres{lotLJ}{EU_post_k39}{cov90_mean}{0.000}
\setres{lotLJ}{EU_post_k39}{kappa}{0.435}
\setres{lotLJ}{EU_post_k39}{kappa_crps}{0.190}
\setres{lotLJ}{EU_post_k39}{kappa_crps_mean}{0.079}
\setres{lotLJ}{EU_post_k39}{kappa_mean}{0.289}
\setres{lotLJ}{EU_post_k39}{logscore_next}{-2.85}
\setres{lotLJ}{EU_post_k39}{logscore_next_mean}{-3.067}
\setres{lotLJ}{EU_post_k39}{misfit_pct}{16.5}
\setres{lotLJ}{EU_post_k39}{misfit_pct_mean}{16.425}
\setres{lotLJ}{EU_post_k39}{n_seen}{39}
\setres{lotLJ}{EU_post_k39}{surprise}{0.00}
\setres{lotLJ}{EU_post_k39}{surprise_mean}{0.129}
\setres{lotLJ}{EU_post_k39}{width90_adj_pct}{27.1}
\setres{lotLJ}{EU_post_k39}{width90_adj_pct_mean}{27.020}
\setres{lotLJ}{EU_post_k39}{width90_pct}{0.1}
\setres{lotLJ}{EU_post_k39}{width90_pct_mean}{0.113}
\setres{lotLJ}{EU_post_k4}{cov50}{0.00}
\setres{lotLJ}{EU_post_k4}{cov50_mean}{0.014}
\setres{lotLJ}{EU_post_k4}{cov90}{0.01}
\setres{lotLJ}{EU_post_k4}{cov90_adj}{1.00}
\setres{lotLJ}{EU_post_k4}{cov90_adj_mean}{0.909}
\setres{lotLJ}{EU_post_k4}{cov90_mean}{0.036}
\setres{lotLJ}{EU_post_k4}{kappa}{0.732}
\setres{lotLJ}{EU_post_k4}{kappa_crps}{0.380}
\setres{lotLJ}{EU_post_k4}{kappa_crps_mean}{0.331}
\setres{lotLJ}{EU_post_k4}{kappa_mean}{0.644}
\setres{lotLJ}{EU_post_k4}{logscore_next}{-2.86}
\setres{lotLJ}{EU_post_k4}{logscore_next_mean}{-3.265}
\setres{lotLJ}{EU_post_k4}{misfit_pct}{16.8}
\setres{lotLJ}{EU_post_k4}{misfit_pct_mean}{16.903}
\setres{lotLJ}{EU_post_k4}{n_seen}{4}
\setres{lotLJ}{EU_post_k4}{surprise}{0.03}
\setres{lotLJ}{EU_post_k4}{surprise_mean}{2.937}
\setres{lotLJ}{EU_post_k4}{width90_adj_pct}{27.7}
\setres{lotLJ}{EU_post_k4}{width90_adj_pct_mean}{27.813}
\setres{lotLJ}{EU_post_k4}{width90_pct}{0.1}
\setres{lotLJ}{EU_post_k4}{width90_pct_mean}{0.429}
\setres{lotLJ}{EU_post_k40}{cov50}{0.00}
\setres{lotLJ}{EU_post_k40}{cov50_mean}{0.000}
\setres{lotLJ}{EU_post_k40}{cov90}{0.00}
\setres{lotLJ}{EU_post_k40}{cov90_adj}{1.00}
\setres{lotLJ}{EU_post_k40}{cov90_adj_mean}{0.975}
\setres{lotLJ}{EU_post_k40}{cov90_mean}{0.000}
\setres{lotLJ}{EU_post_k40}{kappa}{0.357}
\setres{lotLJ}{EU_post_k40}{kappa_crps}{0.278}
\setres{lotLJ}{EU_post_k40}{kappa_crps_mean}{0.096}
\setres{lotLJ}{EU_post_k40}{kappa_mean}{0.208}
\setres{lotLJ}{EU_post_k40}{logscore_next}{-2.83}
\setres{lotLJ}{EU_post_k40}{logscore_next_mean}{-3.128}
\setres{lotLJ}{EU_post_k40}{misfit_pct}{16.5}
\setres{lotLJ}{EU_post_k40}{misfit_pct_mean}{16.474}
\setres{lotLJ}{EU_post_k40}{n_seen}{40}
\setres{lotLJ}{EU_post_k40}{surprise}{0.01}
\setres{lotLJ}{EU_post_k40}{surprise_mean}{12.883}
\setres{lotLJ}{EU_post_k40}{width90_adj_pct}{27.2}
\setres{lotLJ}{EU_post_k40}{width90_adj_pct_mean}{27.101}
\setres{lotLJ}{EU_post_k40}{width90_pct}{0.1}
\setres{lotLJ}{EU_post_k40}{width90_pct_mean}{0.136}
\setres{lotLJ}{EU_post_k41}{cov50}{0.00}
\setres{lotLJ}{EU_post_k41}{cov50_mean}{0.000}
\setres{lotLJ}{EU_post_k41}{cov90}{0.00}
\setres{lotLJ}{EU_post_k41}{cov90_adj}{1.00}
\setres{lotLJ}{EU_post_k41}{cov90_adj_mean}{0.983}
\setres{lotLJ}{EU_post_k41}{cov90_mean}{0.000}
\setres{lotLJ}{EU_post_k41}{kappa}{0.418}
\setres{lotLJ}{EU_post_k41}{kappa_crps}{0.122}
\setres{lotLJ}{EU_post_k41}{kappa_crps_mean}{0.089}
\setres{lotLJ}{EU_post_k41}{kappa_mean}{0.236}
\setres{lotLJ}{EU_post_k41}{logscore_next}{-2.88}
\setres{lotLJ}{EU_post_k41}{logscore_next_mean}{-3.075}
\setres{lotLJ}{EU_post_k41}{misfit_pct}{16.4}
\setres{lotLJ}{EU_post_k41}{misfit_pct_mean}{16.451}
\setres{lotLJ}{EU_post_k41}{n_seen}{41}
\setres{lotLJ}{EU_post_k41}{surprise}{0.01}
\setres{lotLJ}{EU_post_k41}{surprise_mean}{2.112}
\setres{lotLJ}{EU_post_k41}{width90_adj_pct}{27.0}
\setres{lotLJ}{EU_post_k41}{width90_adj_pct_mean}{27.062}
\setres{lotLJ}{EU_post_k41}{width90_pct}{0.1}
\setres{lotLJ}{EU_post_k41}{width90_pct_mean}{0.144}
\setres{lotLJ}{EU_post_k42}{cov50}{0.00}
\setres{lotLJ}{EU_post_k42}{cov50_mean}{0.000}
\setres{lotLJ}{EU_post_k42}{cov90}{0.00}
\setres{lotLJ}{EU_post_k42}{cov90_adj}{1.00}
\setres{lotLJ}{EU_post_k42}{cov90_adj_mean}{0.975}
\setres{lotLJ}{EU_post_k42}{cov90_mean}{0.000}
\setres{lotLJ}{EU_post_k42}{kappa}{0.104}
\setres{lotLJ}{EU_post_k42}{kappa_crps}{0.146}
\setres{lotLJ}{EU_post_k42}{kappa_crps_mean}{-0.257}
\setres{lotLJ}{EU_post_k42}{kappa_mean}{-0.158}
\setres{lotLJ}{EU_post_k42}{logscore_next}{-2.85}
\setres{lotLJ}{EU_post_k42}{logscore_next_mean}{-3.093}
\setres{lotLJ}{EU_post_k42}{misfit_pct}{16.4}
\setres{lotLJ}{EU_post_k42}{misfit_pct_mean}{16.388}
\setres{lotLJ}{EU_post_k42}{n_seen}{42}
\setres{lotLJ}{EU_post_k42}{surprise}{0.01}
\setres{lotLJ}{EU_post_k42}{surprise_mean}{11.484}
\setres{lotLJ}{EU_post_k42}{width90_adj_pct}{27.0}
\setres{lotLJ}{EU_post_k42}{width90_adj_pct_mean}{26.959}
\setres{lotLJ}{EU_post_k42}{width90_pct}{0.2}
\setres{lotLJ}{EU_post_k42}{width90_pct_mean}{0.179}
\setres{lotLJ}{EU_post_k43}{cov50}{0.00}
\setres{lotLJ}{EU_post_k43}{cov50_mean}{0.000}
\setres{lotLJ}{EU_post_k43}{cov90}{0.00}
\setres{lotLJ}{EU_post_k43}{cov90_adj}{1.00}
\setres{lotLJ}{EU_post_k43}{cov90_adj_mean}{0.950}
\setres{lotLJ}{EU_post_k43}{cov90_mean}{0.000}
\setres{lotLJ}{EU_post_k43}{kappa}{0.088}
\setres{lotLJ}{EU_post_k43}{kappa_crps}{-0.161}
\setres{lotLJ}{EU_post_k43}{kappa_crps_mean}{-0.763}
\setres{lotLJ}{EU_post_k43}{kappa_mean}{-2.375}
\setres{lotLJ}{EU_post_k43}{logscore_next}{-2.89}
\setres{lotLJ}{EU_post_k43}{logscore_next_mean}{-3.005}
\setres{lotLJ}{EU_post_k43}{misfit_pct}{16.3}
\setres{lotLJ}{EU_post_k43}{misfit_pct_mean}{16.339}
\setres{lotLJ}{EU_post_k43}{n_seen}{43}
\setres{lotLJ}{EU_post_k43}{surprise}{0.02}
\setres{lotLJ}{EU_post_k43}{surprise_mean}{11.401}
\setres{lotLJ}{EU_post_k43}{width90_adj_pct}{26.9}
\setres{lotLJ}{EU_post_k43}{width90_adj_pct_mean}{26.879}
\setres{lotLJ}{EU_post_k43}{width90_pct}{0.1}
\setres{lotLJ}{EU_post_k43}{width90_pct_mean}{0.192}
\setres{lotLJ}{EU_post_k44}{cov50}{---}
\setres{lotLJ}{EU_post_k44}{cov50_mean}{---}
\setres{lotLJ}{EU_post_k44}{cov90}{---}
\setres{lotLJ}{EU_post_k44}{cov90_adj}{---}
\setres{lotLJ}{EU_post_k44}{cov90_adj_mean}{---}
\setres{lotLJ}{EU_post_k44}{cov90_mean}{---}
\setres{lotLJ}{EU_post_k44}{kappa}{---}
\setres{lotLJ}{EU_post_k44}{kappa_crps}{---}
\setres{lotLJ}{EU_post_k44}{kappa_crps_mean}{---}
\setres{lotLJ}{EU_post_k44}{kappa_mean}{---}
\setres{lotLJ}{EU_post_k44}{logscore_next}{---}
\setres{lotLJ}{EU_post_k44}{logscore_next_mean}{---}
\setres{lotLJ}{EU_post_k44}{misfit_pct}{---}
\setres{lotLJ}{EU_post_k44}{misfit_pct_mean}{---}
\setres{lotLJ}{EU_post_k44}{n_seen}{44}
\setres{lotLJ}{EU_post_k44}{surprise}{---}
\setres{lotLJ}{EU_post_k44}{surprise_mean}{---}
\setres{lotLJ}{EU_post_k44}{width90_adj_pct}{---}
\setres{lotLJ}{EU_post_k44}{width90_adj_pct_mean}{---}
\setres{lotLJ}{EU_post_k44}{width90_pct}{---}
\setres{lotLJ}{EU_post_k44}{width90_pct_mean}{---}
\setres{lotLJ}{EU_post_k5}{cov50}{0.00}
\setres{lotLJ}{EU_post_k5}{cov50_mean}{0.013}
\setres{lotLJ}{EU_post_k5}{cov90}{0.00}
\setres{lotLJ}{EU_post_k5}{cov90_adj}{0.99}
\setres{lotLJ}{EU_post_k5}{cov90_adj_mean}{0.924}
\setres{lotLJ}{EU_post_k5}{cov90_mean}{0.033}
\setres{lotLJ}{EU_post_k5}{kappa}{0.729}
\setres{lotLJ}{EU_post_k5}{kappa_crps}{0.384}
\setres{lotLJ}{EU_post_k5}{kappa_crps_mean}{0.326}
\setres{lotLJ}{EU_post_k5}{kappa_mean}{0.638}
\setres{lotLJ}{EU_post_k5}{logscore_next}{-2.95}
\setres{lotLJ}{EU_post_k5}{logscore_next_mean}{-3.364}
\setres{lotLJ}{EU_post_k5}{misfit_pct}{16.6}
\setres{lotLJ}{EU_post_k5}{misfit_pct_mean}{16.771}
\setres{lotLJ}{EU_post_k5}{n_seen}{5}
\setres{lotLJ}{EU_post_k5}{surprise}{0.14}
\setres{lotLJ}{EU_post_k5}{surprise_mean}{22.430}
\setres{lotLJ}{EU_post_k5}{width90_adj_pct}{27.3}
\setres{lotLJ}{EU_post_k5}{width90_adj_pct_mean}{27.594}
\setres{lotLJ}{EU_post_k5}{width90_pct}{0.1}
\setres{lotLJ}{EU_post_k5}{width90_pct_mean}{0.390}
\setres{lotLJ}{EU_post_k6}{cov50}{0.00}
\setres{lotLJ}{EU_post_k6}{cov50_mean}{0.005}
\setres{lotLJ}{EU_post_k6}{cov90}{0.00}
\setres{lotLJ}{EU_post_k6}{cov90_adj}{0.99}
\setres{lotLJ}{EU_post_k6}{cov90_adj_mean}{0.943}
\setres{lotLJ}{EU_post_k6}{cov90_mean}{0.021}
\setres{lotLJ}{EU_post_k6}{kappa}{0.710}
\setres{lotLJ}{EU_post_k6}{kappa_crps}{0.381}
\setres{lotLJ}{EU_post_k6}{kappa_crps_mean}{0.357}
\setres{lotLJ}{EU_post_k6}{kappa_mean}{0.663}
\setres{lotLJ}{EU_post_k6}{logscore_next}{-2.86}
\setres{lotLJ}{EU_post_k6}{logscore_next_mean}{-3.117}
\setres{lotLJ}{EU_post_k6}{misfit_pct}{16.8}
\setres{lotLJ}{EU_post_k6}{misfit_pct_mean}{16.619}
\setres{lotLJ}{EU_post_k6}{n_seen}{6}
\setres{lotLJ}{EU_post_k6}{surprise}{0.05}
\setres{lotLJ}{EU_post_k6}{surprise_mean}{13.272}
\setres{lotLJ}{EU_post_k6}{width90_adj_pct}{27.6}
\setres{lotLJ}{EU_post_k6}{width90_adj_pct_mean}{27.341}
\setres{lotLJ}{EU_post_k6}{width90_pct}{0.1}
\setres{lotLJ}{EU_post_k6}{width90_pct_mean}{0.249}
\setres{lotLJ}{EU_post_k7}{cov50}{0.00}
\setres{lotLJ}{EU_post_k7}{cov50_mean}{0.003}
\setres{lotLJ}{EU_post_k7}{cov90}{0.00}
\setres{lotLJ}{EU_post_k7}{cov90_adj}{0.95}
\setres{lotLJ}{EU_post_k7}{cov90_adj_mean}{0.919}
\setres{lotLJ}{EU_post_k7}{cov90_mean}{0.016}
\setres{lotLJ}{EU_post_k7}{kappa}{0.697}
\setres{lotLJ}{EU_post_k7}{kappa_crps}{0.369}
\setres{lotLJ}{EU_post_k7}{kappa_crps_mean}{0.328}
\setres{lotLJ}{EU_post_k7}{kappa_mean}{0.626}
\setres{lotLJ}{EU_post_k7}{logscore_next}{-2.84}
\setres{lotLJ}{EU_post_k7}{logscore_next_mean}{-3.518}
\setres{lotLJ}{EU_post_k7}{misfit_pct}{16.1}
\setres{lotLJ}{EU_post_k7}{misfit_pct_mean}{16.614}
\setres{lotLJ}{EU_post_k7}{n_seen}{7}
\setres{lotLJ}{EU_post_k7}{surprise}{0.05}
\setres{lotLJ}{EU_post_k7}{surprise_mean}{10.390}
\setres{lotLJ}{EU_post_k7}{width90_adj_pct}{26.5}
\setres{lotLJ}{EU_post_k7}{width90_adj_pct_mean}{27.334}
\setres{lotLJ}{EU_post_k7}{width90_pct}{0.1}
\setres{lotLJ}{EU_post_k7}{width90_pct_mean}{0.264}
\setres{lotLJ}{EU_post_k8}{cov50}{0.00}
\setres{lotLJ}{EU_post_k8}{cov50_mean}{0.001}
\setres{lotLJ}{EU_post_k8}{cov90}{0.00}
\setres{lotLJ}{EU_post_k8}{cov90_adj}{0.97}
\setres{lotLJ}{EU_post_k8}{cov90_adj_mean}{0.933}
\setres{lotLJ}{EU_post_k8}{cov90_mean}{0.013}
\setres{lotLJ}{EU_post_k8}{kappa}{0.707}
\setres{lotLJ}{EU_post_k8}{kappa_crps}{0.339}
\setres{lotLJ}{EU_post_k8}{kappa_crps_mean}{0.347}
\setres{lotLJ}{EU_post_k8}{kappa_mean}{0.645}
\setres{lotLJ}{EU_post_k8}{logscore_next}{-2.86}
\setres{lotLJ}{EU_post_k8}{logscore_next_mean}{-3.005}
\setres{lotLJ}{EU_post_k8}{misfit_pct}{16.6}
\setres{lotLJ}{EU_post_k8}{misfit_pct_mean}{16.442}
\setres{lotLJ}{EU_post_k8}{n_seen}{8}
\setres{lotLJ}{EU_post_k8}{surprise}{0.00}
\setres{lotLJ}{EU_post_k8}{surprise_mean}{8.502}
\setres{lotLJ}{EU_post_k8}{width90_adj_pct}{27.2}
\setres{lotLJ}{EU_post_k8}{width90_adj_pct_mean}{27.048}
\setres{lotLJ}{EU_post_k8}{width90_pct}{0.1}
\setres{lotLJ}{EU_post_k8}{width90_pct_mean}{0.147}
\setres{lotLJ}{EU_post_k9}{cov50}{0.00}
\setres{lotLJ}{EU_post_k9}{cov50_mean}{0.003}
\setres{lotLJ}{EU_post_k9}{cov90}{0.00}
\setres{lotLJ}{EU_post_k9}{cov90_adj}{0.97}
\setres{lotLJ}{EU_post_k9}{cov90_adj_mean}{0.920}
\setres{lotLJ}{EU_post_k9}{cov90_mean}{0.010}
\setres{lotLJ}{EU_post_k9}{kappa}{0.692}
\setres{lotLJ}{EU_post_k9}{kappa_crps}{0.324}
\setres{lotLJ}{EU_post_k9}{kappa_crps_mean}{0.317}
\setres{lotLJ}{EU_post_k9}{kappa_mean}{0.623}
\setres{lotLJ}{EU_post_k9}{logscore_next}{-2.85}
\setres{lotLJ}{EU_post_k9}{logscore_next_mean}{-3.028}
\setres{lotLJ}{EU_post_k9}{misfit_pct}{16.8}
\setres{lotLJ}{EU_post_k9}{misfit_pct_mean}{16.266}
\setres{lotLJ}{EU_post_k9}{n_seen}{9}
\setres{lotLJ}{EU_post_k9}{surprise}{0.00}
\setres{lotLJ}{EU_post_k9}{surprise_mean}{3.430}
\setres{lotLJ}{EU_post_k9}{width90_adj_pct}{27.6}
\setres{lotLJ}{EU_post_k9}{width90_adj_pct_mean}{26.758}
\setres{lotLJ}{EU_post_k9}{width90_pct}{0.1}
\setres{lotLJ}{EU_post_k9}{width90_pct_mean}{0.135}
\setres{lotLJ}{reference}{domain}{Zurich03\_Zurich06\_Beijing\_gains}
\setres{lotLJ}{reference}{n_draws}{400}
\setres{lotLJ}{reference}{n_lotteries}{44}
\setres{lotLJ}{reference}{n_orders}{20}
\setres{lotLJ}{reference}{start}{lit}
\setres{lotULJ}{CPT_post_k0}{cov50}{0.93}
\setres{lotULJ}{CPT_post_k0}{cov50_mean}{0.932}
\setres{lotULJ}{CPT_post_k0}{cov90}{1.00}
\setres{lotULJ}{CPT_post_k0}{cov90_adj}{---}
\setres{lotULJ}{CPT_post_k0}{cov90_adj_mean}{---}
\setres{lotULJ}{CPT_post_k0}{cov90_mean}{1.000}
\setres{lotULJ}{CPT_post_k0}{kappa}{0.221}
\setres{lotULJ}{CPT_post_k0}{kappa_crps}{0.590}
\setres{lotULJ}{CPT_post_k0}{kappa_crps_mean}{0.591}
\setres{lotULJ}{CPT_post_k0}{kappa_mean}{0.221}
\setres{lotULJ}{CPT_post_k0}{logscore_next}{-2.80}
\setres{lotULJ}{CPT_post_k0}{logscore_next_mean}{-2.798}
\setres{lotULJ}{CPT_post_k0}{misfit_pct}{---}
\setres{lotULJ}{CPT_post_k0}{misfit_pct_mean}{---}
\setres{lotULJ}{CPT_post_k0}{n_seen}{0}
\setres{lotULJ}{CPT_post_k0}{surprise}{2.40}
\setres{lotULJ}{CPT_post_k0}{surprise_mean}{2.323}
\setres{lotULJ}{CPT_post_k0}{width90_adj_pct}{---}
\setres{lotULJ}{CPT_post_k0}{width90_adj_pct_mean}{---}
\setres{lotULJ}{CPT_post_k0}{width90_pct}{28.2}
\setres{lotULJ}{CPT_post_k0}{width90_pct_mean}{28.191}
\setres{lotULJ}{CPT_post_k1}{cov50}{0.58}
\setres{lotULJ}{CPT_post_k1}{cov50_mean}{0.586}
\setres{lotULJ}{CPT_post_k1}{cov90}{0.90}
\setres{lotULJ}{CPT_post_k1}{cov90_adj}{---}
\setres{lotULJ}{CPT_post_k1}{cov90_adj_mean}{---}
\setres{lotULJ}{CPT_post_k1}{cov90_mean}{0.881}
\setres{lotULJ}{CPT_post_k1}{kappa}{0.512}
\setres{lotULJ}{CPT_post_k1}{kappa_crps}{0.555}
\setres{lotULJ}{CPT_post_k1}{kappa_crps_mean}{0.549}
\setres{lotULJ}{CPT_post_k1}{kappa_mean}{0.487}
\setres{lotULJ}{CPT_post_k1}{logscore_next}{-2.79}
\setres{lotULJ}{CPT_post_k1}{logscore_next_mean}{-2.799}
\setres{lotULJ}{CPT_post_k1}{misfit_pct}{---}
\setres{lotULJ}{CPT_post_k1}{misfit_pct_mean}{---}
\setres{lotULJ}{CPT_post_k1}{n_seen}{1}
\setres{lotULJ}{CPT_post_k1}{surprise}{1.45}
\setres{lotULJ}{CPT_post_k1}{surprise_mean}{1.884}
\setres{lotULJ}{CPT_post_k1}{width90_adj_pct}{---}
\setres{lotULJ}{CPT_post_k1}{width90_adj_pct_mean}{---}
\setres{lotULJ}{CPT_post_k1}{width90_pct}{18.4}
\setres{lotULJ}{CPT_post_k1}{width90_pct_mean}{19.523}
\setres{lotULJ}{CPT_post_k10}{cov50}{0.00}
\setres{lotULJ}{CPT_post_k10}{cov50_mean}{0.054}
\setres{lotULJ}{CPT_post_k10}{cov90}{0.15}
\setres{lotULJ}{CPT_post_k10}{cov90_adj}{0.84}
\setres{lotULJ}{CPT_post_k10}{cov90_adj_mean}{0.824}
\setres{lotULJ}{CPT_post_k10}{cov90_mean}{0.166}
\setres{lotULJ}{CPT_post_k10}{kappa}{0.958}
\setres{lotULJ}{CPT_post_k10}{kappa_crps}{0.670}
\setres{lotULJ}{CPT_post_k10}{kappa_crps_mean}{0.661}
\setres{lotULJ}{CPT_post_k10}{kappa_mean}{0.955}
\setres{lotULJ}{CPT_post_k10}{logscore_next}{-2.79}
\setres{lotULJ}{CPT_post_k10}{logscore_next_mean}{-2.802}
\setres{lotULJ}{CPT_post_k10}{misfit_pct}{5.3}
\setres{lotULJ}{CPT_post_k10}{misfit_pct_mean}{5.317}
\setres{lotULJ}{CPT_post_k10}{n_seen}{10}
\setres{lotULJ}{CPT_post_k10}{surprise}{0.03}
\setres{lotULJ}{CPT_post_k10}{surprise_mean}{0.147}
\setres{lotULJ}{CPT_post_k10}{width90_adj_pct}{8.8}
\setres{lotULJ}{CPT_post_k10}{width90_adj_pct_mean}{8.913}
\setres{lotULJ}{CPT_post_k10}{width90_pct}{0.9}
\setres{lotULJ}{CPT_post_k10}{width90_pct_mean}{1.049}
\setres{lotULJ}{CPT_post_k11}{cov50}{0.00}
\setres{lotULJ}{CPT_post_k11}{cov50_mean}{0.033}
\setres{lotULJ}{CPT_post_k11}{cov90}{0.14}
\setres{lotULJ}{CPT_post_k11}{cov90_adj}{0.86}
\setres{lotULJ}{CPT_post_k11}{cov90_adj_mean}{0.844}
\setres{lotULJ}{CPT_post_k11}{cov90_mean}{0.168}
\setres{lotULJ}{CPT_post_k11}{kappa}{0.964}
\setres{lotULJ}{CPT_post_k11}{kappa_crps}{0.666}
\setres{lotULJ}{CPT_post_k11}{kappa_crps_mean}{0.664}
\setres{lotULJ}{CPT_post_k11}{kappa_mean}{0.957}
\setres{lotULJ}{CPT_post_k11}{logscore_next}{-2.79}
\setres{lotULJ}{CPT_post_k11}{logscore_next_mean}{-2.804}
\setres{lotULJ}{CPT_post_k11}{misfit_pct}{5.3}
\setres{lotULJ}{CPT_post_k11}{misfit_pct_mean}{5.512}
\setres{lotULJ}{CPT_post_k11}{n_seen}{11}
\setres{lotULJ}{CPT_post_k11}{surprise}{0.07}
\setres{lotULJ}{CPT_post_k11}{surprise_mean}{0.187}
\setres{lotULJ}{CPT_post_k11}{width90_adj_pct}{8.9}
\setres{lotULJ}{CPT_post_k11}{width90_adj_pct_mean}{9.194}
\setres{lotULJ}{CPT_post_k11}{width90_pct}{1.0}
\setres{lotULJ}{CPT_post_k11}{width90_pct_mean}{1.055}
\setres{lotULJ}{CPT_post_k12}{cov50}{0.00}
\setres{lotULJ}{CPT_post_k12}{cov50_mean}{0.030}
\setres{lotULJ}{CPT_post_k12}{cov90}{0.20}
\setres{lotULJ}{CPT_post_k12}{cov90_adj}{0.86}
\setres{lotULJ}{CPT_post_k12}{cov90_adj_mean}{0.845}
\setres{lotULJ}{CPT_post_k12}{cov90_mean}{0.192}
\setres{lotULJ}{CPT_post_k12}{kappa}{0.961}
\setres{lotULJ}{CPT_post_k12}{kappa_crps}{0.664}
\setres{lotULJ}{CPT_post_k12}{kappa_crps_mean}{0.663}
\setres{lotULJ}{CPT_post_k12}{kappa_mean}{0.957}
\setres{lotULJ}{CPT_post_k12}{logscore_next}{-2.79}
\setres{lotULJ}{CPT_post_k12}{logscore_next_mean}{-2.810}
\setres{lotULJ}{CPT_post_k12}{misfit_pct}{5.5}
\setres{lotULJ}{CPT_post_k12}{misfit_pct_mean}{5.586}
\setres{lotULJ}{CPT_post_k12}{n_seen}{12}
\setres{lotULJ}{CPT_post_k12}{surprise}{0.20}
\setres{lotULJ}{CPT_post_k12}{surprise_mean}{1.631}
\setres{lotULJ}{CPT_post_k12}{width90_adj_pct}{9.2}
\setres{lotULJ}{CPT_post_k12}{width90_adj_pct_mean}{9.330}
\setres{lotULJ}{CPT_post_k12}{width90_pct}{1.1}
\setres{lotULJ}{CPT_post_k12}{width90_pct_mean}{1.113}
\setres{lotULJ}{CPT_post_k13}{cov50}{0.05}
\setres{lotULJ}{CPT_post_k13}{cov50_mean}{0.081}
\setres{lotULJ}{CPT_post_k13}{cov90}{0.16}
\setres{lotULJ}{CPT_post_k13}{cov90_adj}{0.87}
\setres{lotULJ}{CPT_post_k13}{cov90_adj_mean}{0.852}
\setres{lotULJ}{CPT_post_k13}{cov90_mean}{0.174}
\setres{lotULJ}{CPT_post_k13}{kappa}{0.958}
\setres{lotULJ}{CPT_post_k13}{kappa_crps}{0.667}
\setres{lotULJ}{CPT_post_k13}{kappa_crps_mean}{0.655}
\setres{lotULJ}{CPT_post_k13}{kappa_mean}{0.953}
\setres{lotULJ}{CPT_post_k13}{logscore_next}{-2.79}
\setres{lotULJ}{CPT_post_k13}{logscore_next_mean}{-2.802}
\setres{lotULJ}{CPT_post_k13}{misfit_pct}{5.8}
\setres{lotULJ}{CPT_post_k13}{misfit_pct_mean}{5.675}
\setres{lotULJ}{CPT_post_k13}{n_seen}{13}
\setres{lotULJ}{CPT_post_k13}{surprise}{0.04}
\setres{lotULJ}{CPT_post_k13}{surprise_mean}{1.147}
\setres{lotULJ}{CPT_post_k13}{width90_adj_pct}{9.7}
\setres{lotULJ}{CPT_post_k13}{width90_adj_pct_mean}{9.483}
\setres{lotULJ}{CPT_post_k13}{width90_pct}{1.0}
\setres{lotULJ}{CPT_post_k13}{width90_pct_mean}{1.055}
\setres{lotULJ}{CPT_post_k14}{cov50}{0.02}
\setres{lotULJ}{CPT_post_k14}{cov50_mean}{0.092}
\setres{lotULJ}{CPT_post_k14}{cov90}{0.12}
\setres{lotULJ}{CPT_post_k14}{cov90_adj}{0.85}
\setres{lotULJ}{CPT_post_k14}{cov90_adj_mean}{0.847}
\setres{lotULJ}{CPT_post_k14}{cov90_mean}{0.150}
\setres{lotULJ}{CPT_post_k14}{kappa}{0.956}
\setres{lotULJ}{CPT_post_k14}{kappa_crps}{0.668}
\setres{lotULJ}{CPT_post_k14}{kappa_crps_mean}{0.659}
\setres{lotULJ}{CPT_post_k14}{kappa_mean}{0.956}
\setres{lotULJ}{CPT_post_k14}{logscore_next}{-2.79}
\setres{lotULJ}{CPT_post_k14}{logscore_next_mean}{-2.804}
\setres{lotULJ}{CPT_post_k14}{misfit_pct}{5.7}
\setres{lotULJ}{CPT_post_k14}{misfit_pct_mean}{5.727}
\setres{lotULJ}{CPT_post_k14}{n_seen}{14}
\setres{lotULJ}{CPT_post_k14}{surprise}{0.02}
\setres{lotULJ}{CPT_post_k14}{surprise_mean}{0.631}
\setres{lotULJ}{CPT_post_k14}{width90_adj_pct}{9.5}
\setres{lotULJ}{CPT_post_k14}{width90_adj_pct_mean}{9.554}
\setres{lotULJ}{CPT_post_k14}{width90_pct}{1.0}
\setres{lotULJ}{CPT_post_k14}{width90_pct_mean}{0.855}
\setres{lotULJ}{CPT_post_k15}{cov50}{0.00}
\setres{lotULJ}{CPT_post_k15}{cov50_mean}{0.069}
\setres{lotULJ}{CPT_post_k15}{cov90}{0.07}
\setres{lotULJ}{CPT_post_k15}{cov90_adj}{0.86}
\setres{lotULJ}{CPT_post_k15}{cov90_adj_mean}{0.845}
\setres{lotULJ}{CPT_post_k15}{cov90_mean}{0.119}
\setres{lotULJ}{CPT_post_k15}{kappa}{0.958}
\setres{lotULJ}{CPT_post_k15}{kappa_crps}{0.671}
\setres{lotULJ}{CPT_post_k15}{kappa_crps_mean}{0.661}
\setres{lotULJ}{CPT_post_k15}{kappa_mean}{0.958}
\setres{lotULJ}{CPT_post_k15}{logscore_next}{-2.78}
\setres{lotULJ}{CPT_post_k15}{logscore_next_mean}{-2.791}
\setres{lotULJ}{CPT_post_k15}{misfit_pct}{5.7}
\setres{lotULJ}{CPT_post_k15}{misfit_pct_mean}{5.708}
\setres{lotULJ}{CPT_post_k15}{n_seen}{15}
\setres{lotULJ}{CPT_post_k15}{surprise}{0.01}
\setres{lotULJ}{CPT_post_k15}{surprise_mean}{0.243}
\setres{lotULJ}{CPT_post_k15}{width90_adj_pct}{9.4}
\setres{lotULJ}{CPT_post_k15}{width90_adj_pct_mean}{9.489}
\setres{lotULJ}{CPT_post_k15}{width90_pct}{0.9}
\setres{lotULJ}{CPT_post_k15}{width90_pct_mean}{0.715}
\setres{lotULJ}{CPT_post_k16}{cov50}{0.00}
\setres{lotULJ}{CPT_post_k16}{cov50_mean}{0.066}
\setres{lotULJ}{CPT_post_k16}{cov90}{0.09}
\setres{lotULJ}{CPT_post_k16}{cov90_adj}{0.86}
\setres{lotULJ}{CPT_post_k16}{cov90_adj_mean}{0.850}
\setres{lotULJ}{CPT_post_k16}{cov90_mean}{0.118}
\setres{lotULJ}{CPT_post_k16}{kappa}{0.960}
\setres{lotULJ}{CPT_post_k16}{kappa_crps}{0.661}
\setres{lotULJ}{CPT_post_k16}{kappa_crps_mean}{0.659}
\setres{lotULJ}{CPT_post_k16}{kappa_mean}{0.959}
\setres{lotULJ}{CPT_post_k16}{logscore_next}{-2.78}
\setres{lotULJ}{CPT_post_k16}{logscore_next_mean}{-2.787}
\setres{lotULJ}{CPT_post_k16}{misfit_pct}{5.7}
\setres{lotULJ}{CPT_post_k16}{misfit_pct_mean}{5.660}
\setres{lotULJ}{CPT_post_k16}{n_seen}{16}
\setres{lotULJ}{CPT_post_k16}{surprise}{0.00}
\setres{lotULJ}{CPT_post_k16}{surprise_mean}{0.079}
\setres{lotULJ}{CPT_post_k16}{width90_adj_pct}{9.4}
\setres{lotULJ}{CPT_post_k16}{width90_adj_pct_mean}{9.400}
\setres{lotULJ}{CPT_post_k16}{width90_pct}{0.8}
\setres{lotULJ}{CPT_post_k16}{width90_pct_mean}{0.659}
\setres{lotULJ}{CPT_post_k17}{cov50}{0.00}
\setres{lotULJ}{CPT_post_k17}{cov50_mean}{0.057}
\setres{lotULJ}{CPT_post_k17}{cov90}{0.07}
\setres{lotULJ}{CPT_post_k17}{cov90_adj}{0.87}
\setres{lotULJ}{CPT_post_k17}{cov90_adj_mean}{0.852}
\setres{lotULJ}{CPT_post_k17}{cov90_mean}{0.120}
\setres{lotULJ}{CPT_post_k17}{kappa}{0.961}
\setres{lotULJ}{CPT_post_k17}{kappa_crps}{0.673}
\setres{lotULJ}{CPT_post_k17}{kappa_crps_mean}{0.668}
\setres{lotULJ}{CPT_post_k17}{kappa_mean}{0.959}
\setres{lotULJ}{CPT_post_k17}{logscore_next}{-2.79}
\setres{lotULJ}{CPT_post_k17}{logscore_next_mean}{-2.804}
\setres{lotULJ}{CPT_post_k17}{misfit_pct}{5.8}
\setres{lotULJ}{CPT_post_k17}{misfit_pct_mean}{5.683}
\setres{lotULJ}{CPT_post_k17}{n_seen}{17}
\setres{lotULJ}{CPT_post_k17}{surprise}{0.01}
\setres{lotULJ}{CPT_post_k17}{surprise_mean}{1.536}
\setres{lotULJ}{CPT_post_k17}{width90_adj_pct}{9.6}
\setres{lotULJ}{CPT_post_k17}{width90_adj_pct_mean}{9.433}
\setres{lotULJ}{CPT_post_k17}{width90_pct}{0.6}
\setres{lotULJ}{CPT_post_k17}{width90_pct_mean}{0.646}
\setres{lotULJ}{CPT_post_k18}{cov50}{0.00}
\setres{lotULJ}{CPT_post_k18}{cov50_mean}{0.046}
\setres{lotULJ}{CPT_post_k18}{cov90}{0.06}
\setres{lotULJ}{CPT_post_k18}{cov90_adj}{0.87}
\setres{lotULJ}{CPT_post_k18}{cov90_adj_mean}{0.848}
\setres{lotULJ}{CPT_post_k18}{cov90_mean}{0.110}
\setres{lotULJ}{CPT_post_k18}{kappa}{0.960}
\setres{lotULJ}{CPT_post_k18}{kappa_crps}{0.677}
\setres{lotULJ}{CPT_post_k18}{kappa_crps_mean}{0.663}
\setres{lotULJ}{CPT_post_k18}{kappa_mean}{0.957}
\setres{lotULJ}{CPT_post_k18}{logscore_next}{-2.79}
\setres{lotULJ}{CPT_post_k18}{logscore_next_mean}{-2.795}
\setres{lotULJ}{CPT_post_k18}{misfit_pct}{5.8}
\setres{lotULJ}{CPT_post_k18}{misfit_pct_mean}{5.685}
\setres{lotULJ}{CPT_post_k18}{n_seen}{18}
\setres{lotULJ}{CPT_post_k18}{surprise}{0.00}
\setres{lotULJ}{CPT_post_k18}{surprise_mean}{0.141}
\setres{lotULJ}{CPT_post_k18}{width90_adj_pct}{9.6}
\setres{lotULJ}{CPT_post_k18}{width90_adj_pct_mean}{9.426}
\setres{lotULJ}{CPT_post_k18}{width90_pct}{0.2}
\setres{lotULJ}{CPT_post_k18}{width90_pct_mean}{0.537}
\setres{lotULJ}{CPT_post_k19}{cov50}{0.00}
\setres{lotULJ}{CPT_post_k19}{cov50_mean}{0.030}
\setres{lotULJ}{CPT_post_k19}{cov90}{0.04}
\setres{lotULJ}{CPT_post_k19}{cov90_adj}{0.84}
\setres{lotULJ}{CPT_post_k19}{cov90_adj_mean}{0.848}
\setres{lotULJ}{CPT_post_k19}{cov90_mean}{0.102}
\setres{lotULJ}{CPT_post_k19}{kappa}{0.959}
\setres{lotULJ}{CPT_post_k19}{kappa_crps}{0.657}
\setres{lotULJ}{CPT_post_k19}{kappa_crps_mean}{0.660}
\setres{lotULJ}{CPT_post_k19}{kappa_mean}{0.958}
\setres{lotULJ}{CPT_post_k19}{logscore_next}{-2.79}
\setres{lotULJ}{CPT_post_k19}{logscore_next_mean}{-2.839}
\setres{lotULJ}{CPT_post_k19}{misfit_pct}{5.9}
\setres{lotULJ}{CPT_post_k19}{misfit_pct_mean}{5.650}
\setres{lotULJ}{CPT_post_k19}{n_seen}{19}
\setres{lotULJ}{CPT_post_k19}{surprise}{0.01}
\setres{lotULJ}{CPT_post_k19}{surprise_mean}{4.561}
\setres{lotULJ}{CPT_post_k19}{width90_adj_pct}{9.8}
\setres{lotULJ}{CPT_post_k19}{width90_adj_pct_mean}{9.357}
\setres{lotULJ}{CPT_post_k19}{width90_pct}{0.2}
\setres{lotULJ}{CPT_post_k19}{width90_pct_mean}{0.528}
\setres{lotULJ}{CPT_post_k2}{cov50}{0.38}
\setres{lotULJ}{CPT_post_k2}{cov50_mean}{0.351}
\setres{lotULJ}{CPT_post_k2}{cov90}{0.71}
\setres{lotULJ}{CPT_post_k2}{cov90_adj}{0.77}
\setres{lotULJ}{CPT_post_k2}{cov90_adj_mean}{0.683}
\setres{lotULJ}{CPT_post_k2}{cov90_mean}{0.674}
\setres{lotULJ}{CPT_post_k2}{kappa}{0.666}
\setres{lotULJ}{CPT_post_k2}{kappa_crps}{0.519}
\setres{lotULJ}{CPT_post_k2}{kappa_crps_mean}{0.506}
\setres{lotULJ}{CPT_post_k2}{kappa_mean}{0.613}
\setres{lotULJ}{CPT_post_k2}{logscore_next}{-2.79}
\setres{lotULJ}{CPT_post_k2}{logscore_next_mean}{-2.791}
\setres{lotULJ}{CPT_post_k2}{misfit_pct}{0.4}
\setres{lotULJ}{CPT_post_k2}{misfit_pct_mean}{1.331}
\setres{lotULJ}{CPT_post_k2}{n_seen}{2}
\setres{lotULJ}{CPT_post_k2}{surprise}{1.10}
\setres{lotULJ}{CPT_post_k2}{surprise_mean}{1.265}
\setres{lotULJ}{CPT_post_k2}{width90_adj_pct}{10.9}
\setres{lotULJ}{CPT_post_k2}{width90_adj_pct_mean}{13.587}
\setres{lotULJ}{CPT_post_k2}{width90_pct}{10.5}
\setres{lotULJ}{CPT_post_k2}{width90_pct_mean}{12.646}
\setres{lotULJ}{CPT_post_k20}{cov50}{0.00}
\setres{lotULJ}{CPT_post_k20}{cov50_mean}{0.037}
\setres{lotULJ}{CPT_post_k20}{cov90}{0.04}
\setres{lotULJ}{CPT_post_k20}{cov90_adj}{0.83}
\setres{lotULJ}{CPT_post_k20}{cov90_adj_mean}{0.846}
\setres{lotULJ}{CPT_post_k20}{cov90_mean}{0.094}
\setres{lotULJ}{CPT_post_k20}{kappa}{0.961}
\setres{lotULJ}{CPT_post_k20}{kappa_crps}{0.683}
\setres{lotULJ}{CPT_post_k20}{kappa_crps_mean}{0.660}
\setres{lotULJ}{CPT_post_k20}{kappa_mean}{0.959}
\setres{lotULJ}{CPT_post_k20}{logscore_next}{-2.78}
\setres{lotULJ}{CPT_post_k20}{logscore_next_mean}{-2.797}
\setres{lotULJ}{CPT_post_k20}{misfit_pct}{6.0}
\setres{lotULJ}{CPT_post_k20}{misfit_pct_mean}{5.851}
\setres{lotULJ}{CPT_post_k20}{n_seen}{20}
\setres{lotULJ}{CPT_post_k20}{surprise}{0.00}
\setres{lotULJ}{CPT_post_k20}{surprise_mean}{1.389}
\setres{lotULJ}{CPT_post_k20}{width90_adj_pct}{9.9}
\setres{lotULJ}{CPT_post_k20}{width90_adj_pct_mean}{9.688}
\setres{lotULJ}{CPT_post_k20}{width90_pct}{0.0}
\setres{lotULJ}{CPT_post_k20}{width90_pct_mean}{0.516}
\setres{lotULJ}{CPT_post_k21}{cov50}{0.00}
\setres{lotULJ}{CPT_post_k21}{cov50_mean}{0.022}
\setres{lotULJ}{CPT_post_k21}{cov90}{0.00}
\setres{lotULJ}{CPT_post_k21}{cov90_adj}{0.83}
\setres{lotULJ}{CPT_post_k21}{cov90_adj_mean}{0.846}
\setres{lotULJ}{CPT_post_k21}{cov90_mean}{0.089}
\setres{lotULJ}{CPT_post_k21}{kappa}{0.960}
\setres{lotULJ}{CPT_post_k21}{kappa_crps}{0.643}
\setres{lotULJ}{CPT_post_k21}{kappa_crps_mean}{0.653}
\setres{lotULJ}{CPT_post_k21}{kappa_mean}{0.956}
\setres{lotULJ}{CPT_post_k21}{logscore_next}{-2.79}
\setres{lotULJ}{CPT_post_k21}{logscore_next_mean}{-2.808}
\setres{lotULJ}{CPT_post_k21}{misfit_pct}{6.0}
\setres{lotULJ}{CPT_post_k21}{misfit_pct_mean}{5.885}
\setres{lotULJ}{CPT_post_k21}{n_seen}{21}
\setres{lotULJ}{CPT_post_k21}{surprise}{0.01}
\setres{lotULJ}{CPT_post_k21}{surprise_mean}{0.174}
\setres{lotULJ}{CPT_post_k21}{width90_adj_pct}{9.9}
\setres{lotULJ}{CPT_post_k21}{width90_adj_pct_mean}{9.714}
\setres{lotULJ}{CPT_post_k21}{width90_pct}{0.0}
\setres{lotULJ}{CPT_post_k21}{width90_pct_mean}{0.424}
\setres{lotULJ}{CPT_post_k22}{cov50}{0.00}
\setres{lotULJ}{CPT_post_k22}{cov50_mean}{0.020}
\setres{lotULJ}{CPT_post_k22}{cov90}{0.00}
\setres{lotULJ}{CPT_post_k22}{cov90_adj}{0.84}
\setres{lotULJ}{CPT_post_k22}{cov90_adj_mean}{0.857}
\setres{lotULJ}{CPT_post_k22}{cov90_mean}{0.057}
\setres{lotULJ}{CPT_post_k22}{kappa}{0.960}
\setres{lotULJ}{CPT_post_k22}{kappa_crps}{0.654}
\setres{lotULJ}{CPT_post_k22}{kappa_crps_mean}{0.654}
\setres{lotULJ}{CPT_post_k22}{kappa_mean}{0.957}
\setres{lotULJ}{CPT_post_k22}{logscore_next}{-2.79}
\setres{lotULJ}{CPT_post_k22}{logscore_next_mean}{-2.807}
\setres{lotULJ}{CPT_post_k22}{misfit_pct}{5.9}
\setres{lotULJ}{CPT_post_k22}{misfit_pct_mean}{5.953}
\setres{lotULJ}{CPT_post_k22}{n_seen}{22}
\setres{lotULJ}{CPT_post_k22}{surprise}{0.00}
\setres{lotULJ}{CPT_post_k22}{surprise_mean}{0.388}
\setres{lotULJ}{CPT_post_k22}{width90_adj_pct}{9.8}
\setres{lotULJ}{CPT_post_k22}{width90_adj_pct_mean}{9.825}
\setres{lotULJ}{CPT_post_k22}{width90_pct}{0.0}
\setres{lotULJ}{CPT_post_k22}{width90_pct_mean}{0.285}
\setres{lotULJ}{CPT_post_k23}{cov50}{0.00}
\setres{lotULJ}{CPT_post_k23}{cov50_mean}{0.033}
\setres{lotULJ}{CPT_post_k23}{cov90}{0.00}
\setres{lotULJ}{CPT_post_k23}{cov90_adj}{0.86}
\setres{lotULJ}{CPT_post_k23}{cov90_adj_mean}{0.860}
\setres{lotULJ}{CPT_post_k23}{cov90_mean}{0.090}
\setres{lotULJ}{CPT_post_k23}{kappa}{0.960}
\setres{lotULJ}{CPT_post_k23}{kappa_crps}{0.674}
\setres{lotULJ}{CPT_post_k23}{kappa_crps_mean}{0.647}
\setres{lotULJ}{CPT_post_k23}{kappa_mean}{0.956}
\setres{lotULJ}{CPT_post_k23}{logscore_next}{-2.79}
\setres{lotULJ}{CPT_post_k23}{logscore_next_mean}{-2.816}
\setres{lotULJ}{CPT_post_k23}{misfit_pct}{5.9}
\setres{lotULJ}{CPT_post_k23}{misfit_pct_mean}{5.945}
\setres{lotULJ}{CPT_post_k23}{n_seen}{23}
\setres{lotULJ}{CPT_post_k23}{surprise}{0.00}
\setres{lotULJ}{CPT_post_k23}{surprise_mean}{1.848}
\setres{lotULJ}{CPT_post_k23}{width90_adj_pct}{9.7}
\setres{lotULJ}{CPT_post_k23}{width90_adj_pct_mean}{9.817}
\setres{lotULJ}{CPT_post_k23}{width90_pct}{0.0}
\setres{lotULJ}{CPT_post_k23}{width90_pct_mean}{0.395}
\setres{lotULJ}{CPT_post_k24}{cov50}{0.00}
\setres{lotULJ}{CPT_post_k24}{cov50_mean}{0.032}
\setres{lotULJ}{CPT_post_k24}{cov90}{0.00}
\setres{lotULJ}{CPT_post_k24}{cov90_adj}{0.85}
\setres{lotULJ}{CPT_post_k24}{cov90_adj_mean}{0.870}
\setres{lotULJ}{CPT_post_k24}{cov90_mean}{0.058}
\setres{lotULJ}{CPT_post_k24}{kappa}{0.961}
\setres{lotULJ}{CPT_post_k24}{kappa_crps}{0.674}
\setres{lotULJ}{CPT_post_k24}{kappa_crps_mean}{0.648}
\setres{lotULJ}{CPT_post_k24}{kappa_mean}{0.957}
\setres{lotULJ}{CPT_post_k24}{logscore_next}{-2.79}
\setres{lotULJ}{CPT_post_k24}{logscore_next_mean}{-2.810}
\setres{lotULJ}{CPT_post_k24}{misfit_pct}{6.2}
\setres{lotULJ}{CPT_post_k24}{misfit_pct_mean}{6.034}
\setres{lotULJ}{CPT_post_k24}{n_seen}{24}
\setres{lotULJ}{CPT_post_k24}{surprise}{0.00}
\setres{lotULJ}{CPT_post_k24}{surprise_mean}{0.489}
\setres{lotULJ}{CPT_post_k24}{width90_adj_pct}{10.2}
\setres{lotULJ}{CPT_post_k24}{width90_adj_pct_mean}{9.954}
\setres{lotULJ}{CPT_post_k24}{width90_pct}{0.0}
\setres{lotULJ}{CPT_post_k24}{width90_pct_mean}{0.322}
\setres{lotULJ}{CPT_post_k25}{cov50}{0.00}
\setres{lotULJ}{CPT_post_k25}{cov50_mean}{0.024}
\setres{lotULJ}{CPT_post_k25}{cov90}{0.00}
\setres{lotULJ}{CPT_post_k25}{cov90_adj}{0.87}
\setres{lotULJ}{CPT_post_k25}{cov90_adj_mean}{0.863}
\setres{lotULJ}{CPT_post_k25}{cov90_mean}{0.029}
\setres{lotULJ}{CPT_post_k25}{kappa}{0.960}
\setres{lotULJ}{CPT_post_k25}{kappa_crps}{0.670}
\setres{lotULJ}{CPT_post_k25}{kappa_crps_mean}{0.650}
\setres{lotULJ}{CPT_post_k25}{kappa_mean}{0.958}
\setres{lotULJ}{CPT_post_k25}{logscore_next}{-2.78}
\setres{lotULJ}{CPT_post_k25}{logscore_next_mean}{-2.790}
\setres{lotULJ}{CPT_post_k25}{misfit_pct}{6.1}
\setres{lotULJ}{CPT_post_k25}{misfit_pct_mean}{5.946}
\setres{lotULJ}{CPT_post_k25}{n_seen}{25}
\setres{lotULJ}{CPT_post_k25}{surprise}{0.00}
\setres{lotULJ}{CPT_post_k25}{surprise_mean}{0.057}
\setres{lotULJ}{CPT_post_k25}{width90_adj_pct}{10.0}
\setres{lotULJ}{CPT_post_k25}{width90_adj_pct_mean}{9.808}
\setres{lotULJ}{CPT_post_k25}{width90_pct}{0.0}
\setres{lotULJ}{CPT_post_k25}{width90_pct_mean}{0.230}
\setres{lotULJ}{CPT_post_k26}{cov50}{0.00}
\setres{lotULJ}{CPT_post_k26}{cov50_mean}{0.022}
\setres{lotULJ}{CPT_post_k26}{cov90}{0.00}
\setres{lotULJ}{CPT_post_k26}{cov90_adj}{0.83}
\setres{lotULJ}{CPT_post_k26}{cov90_adj_mean}{0.853}
\setres{lotULJ}{CPT_post_k26}{cov90_mean}{0.028}
\setres{lotULJ}{CPT_post_k26}{kappa}{0.960}
\setres{lotULJ}{CPT_post_k26}{kappa_crps}{0.670}
\setres{lotULJ}{CPT_post_k26}{kappa_crps_mean}{0.657}
\setres{lotULJ}{CPT_post_k26}{kappa_mean}{0.960}
\setres{lotULJ}{CPT_post_k26}{logscore_next}{-2.79}
\setres{lotULJ}{CPT_post_k26}{logscore_next_mean}{-2.806}
\setres{lotULJ}{CPT_post_k26}{misfit_pct}{6.0}
\setres{lotULJ}{CPT_post_k26}{misfit_pct_mean}{5.890}
\setres{lotULJ}{CPT_post_k26}{n_seen}{26}
\setres{lotULJ}{CPT_post_k26}{surprise}{0.00}
\setres{lotULJ}{CPT_post_k26}{surprise_mean}{1.801}
\setres{lotULJ}{CPT_post_k26}{width90_adj_pct}{9.8}
\setres{lotULJ}{CPT_post_k26}{width90_adj_pct_mean}{9.719}
\setres{lotULJ}{CPT_post_k26}{width90_pct}{0.0}
\setres{lotULJ}{CPT_post_k26}{width90_pct_mean}{0.224}
\setres{lotULJ}{CPT_post_k27}{cov50}{0.00}
\setres{lotULJ}{CPT_post_k27}{cov50_mean}{0.026}
\setres{lotULJ}{CPT_post_k27}{cov90}{0.00}
\setres{lotULJ}{CPT_post_k27}{cov90_adj}{0.85}
\setres{lotULJ}{CPT_post_k27}{cov90_adj_mean}{0.859}
\setres{lotULJ}{CPT_post_k27}{cov90_mean}{0.047}
\setres{lotULJ}{CPT_post_k27}{kappa}{0.960}
\setres{lotULJ}{CPT_post_k27}{kappa_crps}{0.661}
\setres{lotULJ}{CPT_post_k27}{kappa_crps_mean}{0.660}
\setres{lotULJ}{CPT_post_k27}{kappa_mean}{0.961}
\setres{lotULJ}{CPT_post_k27}{logscore_next}{-2.79}
\setres{lotULJ}{CPT_post_k27}{logscore_next_mean}{-2.791}
\setres{lotULJ}{CPT_post_k27}{misfit_pct}{5.9}
\setres{lotULJ}{CPT_post_k27}{misfit_pct_mean}{5.924}
\setres{lotULJ}{CPT_post_k27}{n_seen}{27}
\setres{lotULJ}{CPT_post_k27}{surprise}{0.00}
\setres{lotULJ}{CPT_post_k27}{surprise_mean}{0.230}
\setres{lotULJ}{CPT_post_k27}{width90_adj_pct}{9.8}
\setres{lotULJ}{CPT_post_k27}{width90_adj_pct_mean}{9.776}
\setres{lotULJ}{CPT_post_k27}{width90_pct}{0.0}
\setres{lotULJ}{CPT_post_k27}{width90_pct_mean}{0.285}
\setres{lotULJ}{CPT_post_k28}{cov50}{0.00}
\setres{lotULJ}{CPT_post_k28}{cov50_mean}{0.031}
\setres{lotULJ}{CPT_post_k28}{cov90}{0.00}
\setres{lotULJ}{CPT_post_k28}{cov90_adj}{0.84}
\setres{lotULJ}{CPT_post_k28}{cov90_adj_mean}{0.866}
\setres{lotULJ}{CPT_post_k28}{cov90_mean}{0.056}
\setres{lotULJ}{CPT_post_k28}{kappa}{0.958}
\setres{lotULJ}{CPT_post_k28}{kappa_crps}{0.690}
\setres{lotULJ}{CPT_post_k28}{kappa_crps_mean}{0.661}
\setres{lotULJ}{CPT_post_k28}{kappa_mean}{0.960}
\setres{lotULJ}{CPT_post_k28}{logscore_next}{-2.80}
\setres{lotULJ}{CPT_post_k28}{logscore_next_mean}{-2.808}
\setres{lotULJ}{CPT_post_k28}{misfit_pct}{5.8}
\setres{lotULJ}{CPT_post_k28}{misfit_pct_mean}{5.880}
\setres{lotULJ}{CPT_post_k28}{n_seen}{28}
\setres{lotULJ}{CPT_post_k28}{surprise}{0.00}
\setres{lotULJ}{CPT_post_k28}{surprise_mean}{0.036}
\setres{lotULJ}{CPT_post_k28}{width90_adj_pct}{9.6}
\setres{lotULJ}{CPT_post_k28}{width90_adj_pct_mean}{9.710}
\setres{lotULJ}{CPT_post_k28}{width90_pct}{0.0}
\setres{lotULJ}{CPT_post_k28}{width90_pct_mean}{0.279}
\setres{lotULJ}{CPT_post_k29}{cov50}{0.00}
\setres{lotULJ}{CPT_post_k29}{cov50_mean}{0.017}
\setres{lotULJ}{CPT_post_k29}{cov90}{0.00}
\setres{lotULJ}{CPT_post_k29}{cov90_adj}{0.87}
\setres{lotULJ}{CPT_post_k29}{cov90_adj_mean}{0.873}
\setres{lotULJ}{CPT_post_k29}{cov90_mean}{0.050}
\setres{lotULJ}{CPT_post_k29}{kappa}{0.959}
\setres{lotULJ}{CPT_post_k29}{kappa_crps}{0.706}
\setres{lotULJ}{CPT_post_k29}{kappa_crps_mean}{0.665}
\setres{lotULJ}{CPT_post_k29}{kappa_mean}{0.961}
\setres{lotULJ}{CPT_post_k29}{logscore_next}{-2.79}
\setres{lotULJ}{CPT_post_k29}{logscore_next_mean}{-2.798}
\setres{lotULJ}{CPT_post_k29}{misfit_pct}{6.0}
\setres{lotULJ}{CPT_post_k29}{misfit_pct_mean}{5.919}
\setres{lotULJ}{CPT_post_k29}{n_seen}{29}
\setres{lotULJ}{CPT_post_k29}{surprise}{0.00}
\setres{lotULJ}{CPT_post_k29}{surprise_mean}{0.106}
\setres{lotULJ}{CPT_post_k29}{width90_adj_pct}{9.8}
\setres{lotULJ}{CPT_post_k29}{width90_adj_pct_mean}{9.772}
\setres{lotULJ}{CPT_post_k29}{width90_pct}{0.0}
\setres{lotULJ}{CPT_post_k29}{width90_pct_mean}{0.237}
\setres{lotULJ}{CPT_post_k3}{cov50}{0.28}
\setres{lotULJ}{CPT_post_k3}{cov50_mean}{0.255}
\setres{lotULJ}{CPT_post_k3}{cov90}{0.51}
\setres{lotULJ}{CPT_post_k3}{cov90_adj}{0.70}
\setres{lotULJ}{CPT_post_k3}{cov90_adj_mean}{0.659}
\setres{lotULJ}{CPT_post_k3}{cov90_mean}{0.515}
\setres{lotULJ}{CPT_post_k3}{kappa}{0.802}
\setres{lotULJ}{CPT_post_k3}{kappa_crps}{0.502}
\setres{lotULJ}{CPT_post_k3}{kappa_crps_mean}{0.523}
\setres{lotULJ}{CPT_post_k3}{kappa_mean}{0.732}
\setres{lotULJ}{CPT_post_k3}{logscore_next}{-2.79}
\setres{lotULJ}{CPT_post_k3}{logscore_next_mean}{-2.796}
\setres{lotULJ}{CPT_post_k3}{misfit_pct}{1.3}
\setres{lotULJ}{CPT_post_k3}{misfit_pct_mean}{2.451}
\setres{lotULJ}{CPT_post_k3}{n_seen}{3}
\setres{lotULJ}{CPT_post_k3}{surprise}{1.51}
\setres{lotULJ}{CPT_post_k3}{surprise_mean}{1.900}
\setres{lotULJ}{CPT_post_k3}{width90_adj_pct}{8.4}
\setres{lotULJ}{CPT_post_k3}{width90_adj_pct_mean}{10.578}
\setres{lotULJ}{CPT_post_k3}{width90_pct}{6.7}
\setres{lotULJ}{CPT_post_k3}{width90_pct_mean}{8.753}
\setres{lotULJ}{CPT_post_k30}{cov50}{0.00}
\setres{lotULJ}{CPT_post_k30}{cov50_mean}{0.011}
\setres{lotULJ}{CPT_post_k30}{cov90}{0.00}
\setres{lotULJ}{CPT_post_k30}{cov90_adj}{0.89}
\setres{lotULJ}{CPT_post_k30}{cov90_adj_mean}{0.871}
\setres{lotULJ}{CPT_post_k30}{cov90_mean}{0.021}
\setres{lotULJ}{CPT_post_k30}{kappa}{0.959}
\setres{lotULJ}{CPT_post_k30}{kappa_crps}{0.681}
\setres{lotULJ}{CPT_post_k30}{kappa_crps_mean}{0.658}
\setres{lotULJ}{CPT_post_k30}{kappa_mean}{0.956}
\setres{lotULJ}{CPT_post_k30}{logscore_next}{-2.79}
\setres{lotULJ}{CPT_post_k30}{logscore_next_mean}{-2.802}
\setres{lotULJ}{CPT_post_k30}{misfit_pct}{6.1}
\setres{lotULJ}{CPT_post_k30}{misfit_pct_mean}{5.913}
\setres{lotULJ}{CPT_post_k30}{n_seen}{30}
\setres{lotULJ}{CPT_post_k30}{surprise}{0.00}
\setres{lotULJ}{CPT_post_k30}{surprise_mean}{0.037}
\setres{lotULJ}{CPT_post_k30}{width90_adj_pct}{10.0}
\setres{lotULJ}{CPT_post_k30}{width90_adj_pct_mean}{9.749}
\setres{lotULJ}{CPT_post_k30}{width90_pct}{0.0}
\setres{lotULJ}{CPT_post_k30}{width90_pct_mean}{0.124}
\setres{lotULJ}{CPT_post_k31}{cov50}{0.00}
\setres{lotULJ}{CPT_post_k31}{cov50_mean}{0.023}
\setres{lotULJ}{CPT_post_k31}{cov90}{0.00}
\setres{lotULJ}{CPT_post_k31}{cov90_adj}{0.88}
\setres{lotULJ}{CPT_post_k31}{cov90_adj_mean}{0.869}
\setres{lotULJ}{CPT_post_k31}{cov90_mean}{0.031}
\setres{lotULJ}{CPT_post_k31}{kappa}{0.953}
\setres{lotULJ}{CPT_post_k31}{kappa_crps}{0.652}
\setres{lotULJ}{CPT_post_k31}{kappa_crps_mean}{0.651}
\setres{lotULJ}{CPT_post_k31}{kappa_mean}{0.956}
\setres{lotULJ}{CPT_post_k31}{logscore_next}{-2.78}
\setres{lotULJ}{CPT_post_k31}{logscore_next_mean}{-2.817}
\setres{lotULJ}{CPT_post_k31}{misfit_pct}{6.1}
\setres{lotULJ}{CPT_post_k31}{misfit_pct_mean}{5.890}
\setres{lotULJ}{CPT_post_k31}{n_seen}{31}
\setres{lotULJ}{CPT_post_k31}{surprise}{0.00}
\setres{lotULJ}{CPT_post_k31}{surprise_mean}{1.858}
\setres{lotULJ}{CPT_post_k31}{width90_adj_pct}{10.0}
\setres{lotULJ}{CPT_post_k31}{width90_adj_pct_mean}{9.717}
\setres{lotULJ}{CPT_post_k31}{width90_pct}{0.0}
\setres{lotULJ}{CPT_post_k31}{width90_pct_mean}{0.172}
\setres{lotULJ}{CPT_post_k32}{cov50}{0.00}
\setres{lotULJ}{CPT_post_k32}{cov50_mean}{0.017}
\setres{lotULJ}{CPT_post_k32}{cov90}{0.00}
\setres{lotULJ}{CPT_post_k32}{cov90_adj}{0.92}
\setres{lotULJ}{CPT_post_k32}{cov90_adj_mean}{0.862}
\setres{lotULJ}{CPT_post_k32}{cov90_mean}{0.046}
\setres{lotULJ}{CPT_post_k32}{kappa}{0.957}
\setres{lotULJ}{CPT_post_k32}{kappa_crps}{0.675}
\setres{lotULJ}{CPT_post_k32}{kappa_crps_mean}{0.659}
\setres{lotULJ}{CPT_post_k32}{kappa_mean}{0.957}
\setres{lotULJ}{CPT_post_k32}{logscore_next}{-2.78}
\setres{lotULJ}{CPT_post_k32}{logscore_next_mean}{-2.799}
\setres{lotULJ}{CPT_post_k32}{misfit_pct}{6.0}
\setres{lotULJ}{CPT_post_k32}{misfit_pct_mean}{5.932}
\setres{lotULJ}{CPT_post_k32}{n_seen}{32}
\setres{lotULJ}{CPT_post_k32}{surprise}{0.00}
\setres{lotULJ}{CPT_post_k32}{surprise_mean}{0.372}
\setres{lotULJ}{CPT_post_k32}{width90_adj_pct}{10.0}
\setres{lotULJ}{CPT_post_k32}{width90_adj_pct_mean}{9.792}
\setres{lotULJ}{CPT_post_k32}{width90_pct}{0.0}
\setres{lotULJ}{CPT_post_k32}{width90_pct_mean}{0.315}
\setres{lotULJ}{CPT_post_k33}{cov50}{0.00}
\setres{lotULJ}{CPT_post_k33}{cov50_mean}{0.027}
\setres{lotULJ}{CPT_post_k33}{cov90}{0.00}
\setres{lotULJ}{CPT_post_k33}{cov90_adj}{0.91}
\setres{lotULJ}{CPT_post_k33}{cov90_adj_mean}{0.850}
\setres{lotULJ}{CPT_post_k33}{cov90_mean}{0.045}
\setres{lotULJ}{CPT_post_k33}{kappa}{0.960}
\setres{lotULJ}{CPT_post_k33}{kappa_crps}{0.656}
\setres{lotULJ}{CPT_post_k33}{kappa_crps_mean}{0.632}
\setres{lotULJ}{CPT_post_k33}{kappa_mean}{0.952}
\setres{lotULJ}{CPT_post_k33}{logscore_next}{-2.79}
\setres{lotULJ}{CPT_post_k33}{logscore_next_mean}{-2.797}
\setres{lotULJ}{CPT_post_k33}{misfit_pct}{5.9}
\setres{lotULJ}{CPT_post_k33}{misfit_pct_mean}{5.871}
\setres{lotULJ}{CPT_post_k33}{n_seen}{33}
\setres{lotULJ}{CPT_post_k33}{surprise}{0.00}
\setres{lotULJ}{CPT_post_k33}{surprise_mean}{0.287}
\setres{lotULJ}{CPT_post_k33}{width90_adj_pct}{9.8}
\setres{lotULJ}{CPT_post_k33}{width90_adj_pct_mean}{9.692}
\setres{lotULJ}{CPT_post_k33}{width90_pct}{0.0}
\setres{lotULJ}{CPT_post_k33}{width90_pct_mean}{0.294}
\setres{lotULJ}{CPT_post_k34}{cov50}{0.00}
\setres{lotULJ}{CPT_post_k34}{cov50_mean}{0.020}
\setres{lotULJ}{CPT_post_k34}{cov90}{0.00}
\setres{lotULJ}{CPT_post_k34}{cov90_adj}{0.90}
\setres{lotULJ}{CPT_post_k34}{cov90_adj_mean}{0.860}
\setres{lotULJ}{CPT_post_k34}{cov90_mean}{0.030}
\setres{lotULJ}{CPT_post_k34}{kappa}{0.956}
\setres{lotULJ}{CPT_post_k34}{kappa_crps}{0.646}
\setres{lotULJ}{CPT_post_k34}{kappa_crps_mean}{0.617}
\setres{lotULJ}{CPT_post_k34}{kappa_mean}{0.952}
\setres{lotULJ}{CPT_post_k34}{logscore_next}{-2.79}
\setres{lotULJ}{CPT_post_k34}{logscore_next_mean}{-2.808}
\setres{lotULJ}{CPT_post_k34}{misfit_pct}{6.0}
\setres{lotULJ}{CPT_post_k34}{misfit_pct_mean}{5.842}
\setres{lotULJ}{CPT_post_k34}{n_seen}{34}
\setres{lotULJ}{CPT_post_k34}{surprise}{0.01}
\setres{lotULJ}{CPT_post_k34}{surprise_mean}{1.353}
\setres{lotULJ}{CPT_post_k34}{width90_adj_pct}{9.8}
\setres{lotULJ}{CPT_post_k34}{width90_adj_pct_mean}{9.630}
\setres{lotULJ}{CPT_post_k34}{width90_pct}{0.0}
\setres{lotULJ}{CPT_post_k34}{width90_pct_mean}{0.147}
\setres{lotULJ}{CPT_post_k35}{cov50}{0.00}
\setres{lotULJ}{CPT_post_k35}{cov50_mean}{0.033}
\setres{lotULJ}{CPT_post_k35}{cov90}{0.00}
\setres{lotULJ}{CPT_post_k35}{cov90_adj}{0.89}
\setres{lotULJ}{CPT_post_k35}{cov90_adj_mean}{0.867}
\setres{lotULJ}{CPT_post_k35}{cov90_mean}{0.067}
\setres{lotULJ}{CPT_post_k35}{kappa}{0.942}
\setres{lotULJ}{CPT_post_k35}{kappa_crps}{0.663}
\setres{lotULJ}{CPT_post_k35}{kappa_crps_mean}{0.596}
\setres{lotULJ}{CPT_post_k35}{kappa_mean}{0.937}
\setres{lotULJ}{CPT_post_k35}{logscore_next}{-2.79}
\setres{lotULJ}{CPT_post_k35}{logscore_next_mean}{-2.806}
\setres{lotULJ}{CPT_post_k35}{misfit_pct}{5.9}
\setres{lotULJ}{CPT_post_k35}{misfit_pct_mean}{5.885}
\setres{lotULJ}{CPT_post_k35}{n_seen}{35}
\setres{lotULJ}{CPT_post_k35}{surprise}{0.00}
\setres{lotULJ}{CPT_post_k35}{surprise_mean}{0.510}
\setres{lotULJ}{CPT_post_k35}{width90_adj_pct}{9.8}
\setres{lotULJ}{CPT_post_k35}{width90_adj_pct_mean}{9.716}
\setres{lotULJ}{CPT_post_k35}{width90_pct}{0.0}
\setres{lotULJ}{CPT_post_k35}{width90_pct_mean}{0.309}
\setres{lotULJ}{CPT_post_k36}{cov50}{0.00}
\setres{lotULJ}{CPT_post_k36}{cov50_mean}{0.025}
\setres{lotULJ}{CPT_post_k36}{cov90}{0.00}
\setres{lotULJ}{CPT_post_k36}{cov90_adj}{0.88}
\setres{lotULJ}{CPT_post_k36}{cov90_adj_mean}{0.850}
\setres{lotULJ}{CPT_post_k36}{cov90_mean}{0.031}
\setres{lotULJ}{CPT_post_k36}{kappa}{0.961}
\setres{lotULJ}{CPT_post_k36}{kappa_crps}{0.632}
\setres{lotULJ}{CPT_post_k36}{kappa_crps_mean}{0.610}
\setres{lotULJ}{CPT_post_k36}{kappa_mean}{0.944}
\setres{lotULJ}{CPT_post_k36}{logscore_next}{-2.78}
\setres{lotULJ}{CPT_post_k36}{logscore_next_mean}{-2.807}
\setres{lotULJ}{CPT_post_k36}{misfit_pct}{5.9}
\setres{lotULJ}{CPT_post_k36}{misfit_pct_mean}{5.843}
\setres{lotULJ}{CPT_post_k36}{n_seen}{36}
\setres{lotULJ}{CPT_post_k36}{surprise}{0.00}
\setres{lotULJ}{CPT_post_k36}{surprise_mean}{0.396}
\setres{lotULJ}{CPT_post_k36}{width90_adj_pct}{9.7}
\setres{lotULJ}{CPT_post_k36}{width90_adj_pct_mean}{9.636}
\setres{lotULJ}{CPT_post_k36}{width90_pct}{0.0}
\setres{lotULJ}{CPT_post_k36}{width90_pct_mean}{0.244}
\setres{lotULJ}{CPT_post_k37}{cov50}{0.00}
\setres{lotULJ}{CPT_post_k37}{cov50_mean}{0.029}
\setres{lotULJ}{CPT_post_k37}{cov90}{0.00}
\setres{lotULJ}{CPT_post_k37}{cov90_adj}{0.86}
\setres{lotULJ}{CPT_post_k37}{cov90_adj_mean}{0.821}
\setres{lotULJ}{CPT_post_k37}{cov90_mean}{0.043}
\setres{lotULJ}{CPT_post_k37}{kappa}{0.954}
\setres{lotULJ}{CPT_post_k37}{kappa_crps}{0.633}
\setres{lotULJ}{CPT_post_k37}{kappa_crps_mean}{0.516}
\setres{lotULJ}{CPT_post_k37}{kappa_mean}{0.904}
\setres{lotULJ}{CPT_post_k37}{logscore_next}{-2.79}
\setres{lotULJ}{CPT_post_k37}{logscore_next_mean}{-2.800}
\setres{lotULJ}{CPT_post_k37}{misfit_pct}{5.9}
\setres{lotULJ}{CPT_post_k37}{misfit_pct_mean}{5.799}
\setres{lotULJ}{CPT_post_k37}{n_seen}{37}
\setres{lotULJ}{CPT_post_k37}{surprise}{0.00}
\setres{lotULJ}{CPT_post_k37}{surprise_mean}{0.099}
\setres{lotULJ}{CPT_post_k37}{width90_adj_pct}{9.7}
\setres{lotULJ}{CPT_post_k37}{width90_adj_pct_mean}{9.574}
\setres{lotULJ}{CPT_post_k37}{width90_pct}{0.0}
\setres{lotULJ}{CPT_post_k37}{width90_pct_mean}{0.310}
\setres{lotULJ}{CPT_post_k38}{cov50}{0.00}
\setres{lotULJ}{CPT_post_k38}{cov50_mean}{0.000}
\setres{lotULJ}{CPT_post_k38}{cov90}{0.00}
\setres{lotULJ}{CPT_post_k38}{cov90_adj}{0.83}
\setres{lotULJ}{CPT_post_k38}{cov90_adj_mean}{0.825}
\setres{lotULJ}{CPT_post_k38}{cov90_mean}{0.025}
\setres{lotULJ}{CPT_post_k38}{kappa}{0.952}
\setres{lotULJ}{CPT_post_k38}{kappa_crps}{0.607}
\setres{lotULJ}{CPT_post_k38}{kappa_crps_mean}{0.469}
\setres{lotULJ}{CPT_post_k38}{kappa_mean}{0.896}
\setres{lotULJ}{CPT_post_k38}{logscore_next}{-2.79}
\setres{lotULJ}{CPT_post_k38}{logscore_next_mean}{-2.802}
\setres{lotULJ}{CPT_post_k38}{misfit_pct}{5.8}
\setres{lotULJ}{CPT_post_k38}{misfit_pct_mean}{5.787}
\setres{lotULJ}{CPT_post_k38}{n_seen}{38}
\setres{lotULJ}{CPT_post_k38}{surprise}{0.00}
\setres{lotULJ}{CPT_post_k38}{surprise_mean}{0.024}
\setres{lotULJ}{CPT_post_k38}{width90_adj_pct}{9.6}
\setres{lotULJ}{CPT_post_k38}{width90_adj_pct_mean}{9.530}
\setres{lotULJ}{CPT_post_k38}{width90_pct}{0.0}
\setres{lotULJ}{CPT_post_k38}{width90_pct_mean}{0.146}
\setres{lotULJ}{CPT_post_k39}{cov50}{0.00}
\setres{lotULJ}{CPT_post_k39}{cov50_mean}{0.010}
\setres{lotULJ}{CPT_post_k39}{cov90}{0.00}
\setres{lotULJ}{CPT_post_k39}{cov90_adj}{0.90}
\setres{lotULJ}{CPT_post_k39}{cov90_adj_mean}{0.830}
\setres{lotULJ}{CPT_post_k39}{cov90_mean}{0.030}
\setres{lotULJ}{CPT_post_k39}{kappa}{0.950}
\setres{lotULJ}{CPT_post_k39}{kappa_crps}{0.599}
\setres{lotULJ}{CPT_post_k39}{kappa_crps_mean}{0.432}
\setres{lotULJ}{CPT_post_k39}{kappa_mean}{0.879}
\setres{lotULJ}{CPT_post_k39}{logscore_next}{-2.79}
\setres{lotULJ}{CPT_post_k39}{logscore_next_mean}{-2.805}
\setres{lotULJ}{CPT_post_k39}{misfit_pct}{5.9}
\setres{lotULJ}{CPT_post_k39}{misfit_pct_mean}{5.774}
\setres{lotULJ}{CPT_post_k39}{n_seen}{39}
\setres{lotULJ}{CPT_post_k39}{surprise}{0.00}
\setres{lotULJ}{CPT_post_k39}{surprise_mean}{0.388}
\setres{lotULJ}{CPT_post_k39}{width90_adj_pct}{9.7}
\setres{lotULJ}{CPT_post_k39}{width90_adj_pct_mean}{9.516}
\setres{lotULJ}{CPT_post_k39}{width90_pct}{0.0}
\setres{lotULJ}{CPT_post_k39}{width90_pct_mean}{0.188}
\setres{lotULJ}{CPT_post_k4}{cov50}{0.16}
\setres{lotULJ}{CPT_post_k4}{cov50_mean}{0.179}
\setres{lotULJ}{CPT_post_k4}{cov90}{0.41}
\setres{lotULJ}{CPT_post_k4}{cov90_adj}{0.78}
\setres{lotULJ}{CPT_post_k4}{cov90_adj_mean}{0.704}
\setres{lotULJ}{CPT_post_k4}{cov90_mean}{0.426}
\setres{lotULJ}{CPT_post_k4}{kappa}{0.938}
\setres{lotULJ}{CPT_post_k4}{kappa_crps}{0.651}
\setres{lotULJ}{CPT_post_k4}{kappa_crps_mean}{0.594}
\setres{lotULJ}{CPT_post_k4}{kappa_mean}{0.837}
\setres{lotULJ}{CPT_post_k4}{logscore_next}{-2.79}
\setres{lotULJ}{CPT_post_k4}{logscore_next_mean}{-2.791}
\setres{lotULJ}{CPT_post_k4}{misfit_pct}{3.8}
\setres{lotULJ}{CPT_post_k4}{misfit_pct_mean}{3.730}
\setres{lotULJ}{CPT_post_k4}{n_seen}{4}
\setres{lotULJ}{CPT_post_k4}{surprise}{0.23}
\setres{lotULJ}{CPT_post_k4}{surprise_mean}{0.862}
\setres{lotULJ}{CPT_post_k4}{width90_adj_pct}{8.6}
\setres{lotULJ}{CPT_post_k4}{width90_adj_pct_mean}{8.599}
\setres{lotULJ}{CPT_post_k4}{width90_pct}{4.0}
\setres{lotULJ}{CPT_post_k4}{width90_pct_mean}{4.777}
\setres{lotULJ}{CPT_post_k40}{cov50}{0.00}
\setres{lotULJ}{CPT_post_k40}{cov50_mean}{0.000}
\setres{lotULJ}{CPT_post_k40}{cov90}{0.00}
\setres{lotULJ}{CPT_post_k40}{cov90_adj}{1.00}
\setres{lotULJ}{CPT_post_k40}{cov90_adj_mean}{0.838}
\setres{lotULJ}{CPT_post_k40}{cov90_mean}{0.050}
\setres{lotULJ}{CPT_post_k40}{kappa}{0.945}
\setres{lotULJ}{CPT_post_k40}{kappa_crps}{0.636}
\setres{lotULJ}{CPT_post_k40}{kappa_crps_mean}{0.444}
\setres{lotULJ}{CPT_post_k40}{kappa_mean}{0.849}
\setres{lotULJ}{CPT_post_k40}{logscore_next}{-2.78}
\setres{lotULJ}{CPT_post_k40}{logscore_next_mean}{-2.818}
\setres{lotULJ}{CPT_post_k40}{misfit_pct}{5.9}
\setres{lotULJ}{CPT_post_k40}{misfit_pct_mean}{5.751}
\setres{lotULJ}{CPT_post_k40}{n_seen}{40}
\setres{lotULJ}{CPT_post_k40}{surprise}{0.01}
\setres{lotULJ}{CPT_post_k40}{surprise_mean}{1.170}
\setres{lotULJ}{CPT_post_k40}{width90_adj_pct}{9.7}
\setres{lotULJ}{CPT_post_k40}{width90_adj_pct_mean}{9.487}
\setres{lotULJ}{CPT_post_k40}{width90_pct}{0.0}
\setres{lotULJ}{CPT_post_k40}{width90_pct_mean}{0.279}
\setres{lotULJ}{CPT_post_k41}{cov50}{0.00}
\setres{lotULJ}{CPT_post_k41}{cov50_mean}{0.000}
\setres{lotULJ}{CPT_post_k41}{cov90}{0.00}
\setres{lotULJ}{CPT_post_k41}{cov90_adj}{1.00}
\setres{lotULJ}{CPT_post_k41}{cov90_adj_mean}{0.867}
\setres{lotULJ}{CPT_post_k41}{cov90_mean}{0.050}
\setres{lotULJ}{CPT_post_k41}{kappa}{0.955}
\setres{lotULJ}{CPT_post_k41}{kappa_crps}{0.645}
\setres{lotULJ}{CPT_post_k41}{kappa_crps_mean}{0.419}
\setres{lotULJ}{CPT_post_k41}{kappa_mean}{0.843}
\setres{lotULJ}{CPT_post_k41}{logscore_next}{-2.79}
\setres{lotULJ}{CPT_post_k41}{logscore_next_mean}{-2.816}
\setres{lotULJ}{CPT_post_k41}{misfit_pct}{5.8}
\setres{lotULJ}{CPT_post_k41}{misfit_pct_mean}{5.737}
\setres{lotULJ}{CPT_post_k41}{n_seen}{41}
\setres{lotULJ}{CPT_post_k41}{surprise}{0.01}
\setres{lotULJ}{CPT_post_k41}{surprise_mean}{1.321}
\setres{lotULJ}{CPT_post_k41}{width90_adj_pct}{9.6}
\setres{lotULJ}{CPT_post_k41}{width90_adj_pct_mean}{9.463}
\setres{lotULJ}{CPT_post_k41}{width90_pct}{0.0}
\setres{lotULJ}{CPT_post_k41}{width90_pct_mean}{0.290}
\setres{lotULJ}{CPT_post_k42}{cov50}{0.00}
\setres{lotULJ}{CPT_post_k42}{cov50_mean}{0.050}
\setres{lotULJ}{CPT_post_k42}{cov90}{0.00}
\setres{lotULJ}{CPT_post_k42}{cov90_adj}{1.00}
\setres{lotULJ}{CPT_post_k42}{cov90_adj_mean}{0.825}
\setres{lotULJ}{CPT_post_k42}{cov90_mean}{0.175}
\setres{lotULJ}{CPT_post_k42}{kappa}{0.956}
\setres{lotULJ}{CPT_post_k42}{kappa_crps}{0.526}
\setres{lotULJ}{CPT_post_k42}{kappa_crps_mean}{0.020}
\setres{lotULJ}{CPT_post_k42}{kappa_mean}{0.856}
\setres{lotULJ}{CPT_post_k42}{logscore_next}{-2.79}
\setres{lotULJ}{CPT_post_k42}{logscore_next_mean}{-2.796}
\setres{lotULJ}{CPT_post_k42}{misfit_pct}{5.8}
\setres{lotULJ}{CPT_post_k42}{misfit_pct_mean}{5.708}
\setres{lotULJ}{CPT_post_k42}{n_seen}{42}
\setres{lotULJ}{CPT_post_k42}{surprise}{0.02}
\setres{lotULJ}{CPT_post_k42}{surprise_mean}{0.284}
\setres{lotULJ}{CPT_post_k42}{width90_adj_pct}{9.5}
\setres{lotULJ}{CPT_post_k42}{width90_adj_pct_mean}{9.450}
\setres{lotULJ}{CPT_post_k42}{width90_pct}{0.7}
\setres{lotULJ}{CPT_post_k42}{width90_pct_mean}{0.585}
\setres{lotULJ}{CPT_post_k43}{cov50}{0.00}
\setres{lotULJ}{CPT_post_k43}{cov50_mean}{0.000}
\setres{lotULJ}{CPT_post_k43}{cov90}{0.00}
\setres{lotULJ}{CPT_post_k43}{cov90_adj}{1.00}
\setres{lotULJ}{CPT_post_k43}{cov90_adj_mean}{0.800}
\setres{lotULJ}{CPT_post_k43}{cov90_mean}{0.200}
\setres{lotULJ}{CPT_post_k43}{kappa}{0.953}
\setres{lotULJ}{CPT_post_k43}{kappa_crps}{0.556}
\setres{lotULJ}{CPT_post_k43}{kappa_crps_mean}{-0.366}
\setres{lotULJ}{CPT_post_k43}{kappa_mean}{-3.887}
\setres{lotULJ}{CPT_post_k43}{logscore_next}{-2.79}
\setres{lotULJ}{CPT_post_k43}{logscore_next_mean}{-2.800}
\setres{lotULJ}{CPT_post_k43}{misfit_pct}{5.8}
\setres{lotULJ}{CPT_post_k43}{misfit_pct_mean}{5.706}
\setres{lotULJ}{CPT_post_k43}{n_seen}{43}
\setres{lotULJ}{CPT_post_k43}{surprise}{0.05}
\setres{lotULJ}{CPT_post_k43}{surprise_mean}{0.376}
\setres{lotULJ}{CPT_post_k43}{width90_adj_pct}{9.5}
\setres{lotULJ}{CPT_post_k43}{width90_adj_pct_mean}{9.445}
\setres{lotULJ}{CPT_post_k43}{width90_pct}{0.6}
\setres{lotULJ}{CPT_post_k43}{width90_pct_mean}{0.602}
\setres{lotULJ}{CPT_post_k44}{cov50}{---}
\setres{lotULJ}{CPT_post_k44}{cov50_mean}{---}
\setres{lotULJ}{CPT_post_k44}{cov90}{---}
\setres{lotULJ}{CPT_post_k44}{cov90_adj}{---}
\setres{lotULJ}{CPT_post_k44}{cov90_adj_mean}{---}
\setres{lotULJ}{CPT_post_k44}{cov90_mean}{---}
\setres{lotULJ}{CPT_post_k44}{kappa}{---}
\setres{lotULJ}{CPT_post_k44}{kappa_crps}{---}
\setres{lotULJ}{CPT_post_k44}{kappa_crps_mean}{---}
\setres{lotULJ}{CPT_post_k44}{kappa_mean}{---}
\setres{lotULJ}{CPT_post_k44}{logscore_next}{---}
\setres{lotULJ}{CPT_post_k44}{logscore_next_mean}{---}
\setres{lotULJ}{CPT_post_k44}{misfit_pct}{---}
\setres{lotULJ}{CPT_post_k44}{misfit_pct_mean}{---}
\setres{lotULJ}{CPT_post_k44}{n_seen}{44}
\setres{lotULJ}{CPT_post_k44}{surprise}{---}
\setres{lotULJ}{CPT_post_k44}{surprise_mean}{---}
\setres{lotULJ}{CPT_post_k44}{width90_adj_pct}{---}
\setres{lotULJ}{CPT_post_k44}{width90_adj_pct_mean}{---}
\setres{lotULJ}{CPT_post_k44}{width90_pct}{---}
\setres{lotULJ}{CPT_post_k44}{width90_pct_mean}{---}
\setres{lotULJ}{CPT_post_k5}{cov50}{0.18}
\setres{lotULJ}{CPT_post_k5}{cov50_mean}{0.156}
\setres{lotULJ}{CPT_post_k5}{cov90}{0.31}
\setres{lotULJ}{CPT_post_k5}{cov90_adj}{0.77}
\setres{lotULJ}{CPT_post_k5}{cov90_adj_mean}{0.715}
\setres{lotULJ}{CPT_post_k5}{cov90_mean}{0.310}
\setres{lotULJ}{CPT_post_k5}{kappa}{0.938}
\setres{lotULJ}{CPT_post_k5}{kappa_crps}{0.617}
\setres{lotULJ}{CPT_post_k5}{kappa_crps_mean}{0.610}
\setres{lotULJ}{CPT_post_k5}{kappa_mean}{0.876}
\setres{lotULJ}{CPT_post_k5}{logscore_next}{-2.80}
\setres{lotULJ}{CPT_post_k5}{logscore_next_mean}{-2.814}
\setres{lotULJ}{CPT_post_k5}{misfit_pct}{3.8}
\setres{lotULJ}{CPT_post_k5}{misfit_pct_mean}{4.366}
\setres{lotULJ}{CPT_post_k5}{n_seen}{5}
\setres{lotULJ}{CPT_post_k5}{surprise}{0.70}
\setres{lotULJ}{CPT_post_k5}{surprise_mean}{2.667}
\setres{lotULJ}{CPT_post_k5}{width90_adj_pct}{7.9}
\setres{lotULJ}{CPT_post_k5}{width90_adj_pct_mean}{8.301}
\setres{lotULJ}{CPT_post_k5}{width90_pct}{3.4}
\setres{lotULJ}{CPT_post_k5}{width90_pct_mean}{3.145}
\setres{lotULJ}{CPT_post_k6}{cov50}{0.08}
\setres{lotULJ}{CPT_post_k6}{cov50_mean}{0.092}
\setres{lotULJ}{CPT_post_k6}{cov90}{0.26}
\setres{lotULJ}{CPT_post_k6}{cov90_adj}{0.76}
\setres{lotULJ}{CPT_post_k6}{cov90_adj_mean}{0.736}
\setres{lotULJ}{CPT_post_k6}{cov90_mean}{0.266}
\setres{lotULJ}{CPT_post_k6}{kappa}{0.945}
\setres{lotULJ}{CPT_post_k6}{kappa_crps}{0.629}
\setres{lotULJ}{CPT_post_k6}{kappa_crps_mean}{0.633}
\setres{lotULJ}{CPT_post_k6}{kappa_mean}{0.931}
\setres{lotULJ}{CPT_post_k6}{logscore_next}{-2.79}
\setres{lotULJ}{CPT_post_k6}{logscore_next_mean}{-2.804}
\setres{lotULJ}{CPT_post_k6}{misfit_pct}{5.1}
\setres{lotULJ}{CPT_post_k6}{misfit_pct_mean}{4.887}
\setres{lotULJ}{CPT_post_k6}{n_seen}{6}
\setres{lotULJ}{CPT_post_k6}{surprise}{0.37}
\setres{lotULJ}{CPT_post_k6}{surprise_mean}{1.808}
\setres{lotULJ}{CPT_post_k6}{width90_adj_pct}{9.3}
\setres{lotULJ}{CPT_post_k6}{width90_adj_pct_mean}{8.626}
\setres{lotULJ}{CPT_post_k6}{width90_pct}{2.2}
\setres{lotULJ}{CPT_post_k6}{width90_pct_mean}{2.266}
\setres{lotULJ}{CPT_post_k7}{cov50}{0.12}
\setres{lotULJ}{CPT_post_k7}{cov50_mean}{0.100}
\setres{lotULJ}{CPT_post_k7}{cov90}{0.20}
\setres{lotULJ}{CPT_post_k7}{cov90_adj}{0.82}
\setres{lotULJ}{CPT_post_k7}{cov90_adj_mean}{0.782}
\setres{lotULJ}{CPT_post_k7}{cov90_mean}{0.214}
\setres{lotULJ}{CPT_post_k7}{kappa}{0.953}
\setres{lotULJ}{CPT_post_k7}{kappa_crps}{0.675}
\setres{lotULJ}{CPT_post_k7}{kappa_crps_mean}{0.656}
\setres{lotULJ}{CPT_post_k7}{kappa_mean}{0.944}
\setres{lotULJ}{CPT_post_k7}{logscore_next}{-2.79}
\setres{lotULJ}{CPT_post_k7}{logscore_next_mean}{-2.798}
\setres{lotULJ}{CPT_post_k7}{misfit_pct}{4.9}
\setres{lotULJ}{CPT_post_k7}{misfit_pct_mean}{5.245}
\setres{lotULJ}{CPT_post_k7}{n_seen}{7}
\setres{lotULJ}{CPT_post_k7}{surprise}{0.12}
\setres{lotULJ}{CPT_post_k7}{surprise_mean}{1.722}
\setres{lotULJ}{CPT_post_k7}{width90_adj_pct}{9.0}
\setres{lotULJ}{CPT_post_k7}{width90_adj_pct_mean}{9.027}
\setres{lotULJ}{CPT_post_k7}{width90_pct}{1.4}
\setres{lotULJ}{CPT_post_k7}{width90_pct_mean}{1.716}
\setres{lotULJ}{CPT_post_k8}{cov50}{0.00}
\setres{lotULJ}{CPT_post_k8}{cov50_mean}{0.072}
\setres{lotULJ}{CPT_post_k8}{cov90}{0.15}
\setres{lotULJ}{CPT_post_k8}{cov90_adj}{0.82}
\setres{lotULJ}{CPT_post_k8}{cov90_adj_mean}{0.822}
\setres{lotULJ}{CPT_post_k8}{cov90_mean}{0.167}
\setres{lotULJ}{CPT_post_k8}{kappa}{0.955}
\setres{lotULJ}{CPT_post_k8}{kappa_crps}{0.655}
\setres{lotULJ}{CPT_post_k8}{kappa_crps_mean}{0.657}
\setres{lotULJ}{CPT_post_k8}{kappa_mean}{0.949}
\setres{lotULJ}{CPT_post_k8}{logscore_next}{-2.79}
\setres{lotULJ}{CPT_post_k8}{logscore_next_mean}{-2.794}
\setres{lotULJ}{CPT_post_k8}{misfit_pct}{5.2}
\setres{lotULJ}{CPT_post_k8}{misfit_pct_mean}{5.285}
\setres{lotULJ}{CPT_post_k8}{n_seen}{8}
\setres{lotULJ}{CPT_post_k8}{surprise}{0.03}
\setres{lotULJ}{CPT_post_k8}{surprise_mean}{0.746}
\setres{lotULJ}{CPT_post_k8}{width90_adj_pct}{9.1}
\setres{lotULJ}{CPT_post_k8}{width90_adj_pct_mean}{9.030}
\setres{lotULJ}{CPT_post_k8}{width90_pct}{1.0}
\setres{lotULJ}{CPT_post_k8}{width90_pct_mean}{1.360}
\setres{lotULJ}{CPT_post_k9}{cov50}{0.01}
\setres{lotULJ}{CPT_post_k9}{cov50_mean}{0.064}
\setres{lotULJ}{CPT_post_k9}{cov90}{0.14}
\setres{lotULJ}{CPT_post_k9}{cov90_adj}{0.84}
\setres{lotULJ}{CPT_post_k9}{cov90_adj_mean}{0.821}
\setres{lotULJ}{CPT_post_k9}{cov90_mean}{0.163}
\setres{lotULJ}{CPT_post_k9}{kappa}{0.961}
\setres{lotULJ}{CPT_post_k9}{kappa_crps}{0.671}
\setres{lotULJ}{CPT_post_k9}{kappa_crps_mean}{0.666}
\setres{lotULJ}{CPT_post_k9}{kappa_mean}{0.955}
\setres{lotULJ}{CPT_post_k9}{logscore_next}{-2.79}
\setres{lotULJ}{CPT_post_k9}{logscore_next_mean}{-2.799}
\setres{lotULJ}{CPT_post_k9}{misfit_pct}{5.4}
\setres{lotULJ}{CPT_post_k9}{misfit_pct_mean}{5.272}
\setres{lotULJ}{CPT_post_k9}{n_seen}{9}
\setres{lotULJ}{CPT_post_k9}{surprise}{0.04}
\setres{lotULJ}{CPT_post_k9}{surprise_mean}{1.373}
\setres{lotULJ}{CPT_post_k9}{width90_adj_pct}{8.9}
\setres{lotULJ}{CPT_post_k9}{width90_adj_pct_mean}{8.839}
\setres{lotULJ}{CPT_post_k9}{width90_pct}{1.0}
\setres{lotULJ}{CPT_post_k9}{width90_pct_mean}{1.150}
\setres{lotULJ}{DA_post_k0}{cov50}{0.20}
\setres{lotULJ}{DA_post_k0}{cov50_mean}{0.205}
\setres{lotULJ}{DA_post_k0}{cov90}{0.66}
\setres{lotULJ}{DA_post_k0}{cov90_adj}{---}
\setres{lotULJ}{DA_post_k0}{cov90_adj_mean}{---}
\setres{lotULJ}{DA_post_k0}{cov90_mean}{0.659}
\setres{lotULJ}{DA_post_k0}{kappa}{0.110}
\setres{lotULJ}{DA_post_k0}{kappa_crps}{0.224}
\setres{lotULJ}{DA_post_k0}{kappa_crps_mean}{0.225}
\setres{lotULJ}{DA_post_k0}{kappa_mean}{0.110}
\setres{lotULJ}{DA_post_k0}{logscore_next}{-2.81}
\setres{lotULJ}{DA_post_k0}{logscore_next_mean}{-2.805}
\setres{lotULJ}{DA_post_k0}{misfit_pct}{---}
\setres{lotULJ}{DA_post_k0}{misfit_pct_mean}{---}
\setres{lotULJ}{DA_post_k0}{n_seen}{0}
\setres{lotULJ}{DA_post_k0}{surprise}{3.68}
\setres{lotULJ}{DA_post_k0}{surprise_mean}{3.440}
\setres{lotULJ}{DA_post_k0}{width90_adj_pct}{---}
\setres{lotULJ}{DA_post_k0}{width90_adj_pct_mean}{---}
\setres{lotULJ}{DA_post_k0}{width90_pct}{25.3}
\setres{lotULJ}{DA_post_k0}{width90_pct_mean}{25.283}
\setres{lotULJ}{DA_post_k1}{cov50}{0.09}
\setres{lotULJ}{DA_post_k1}{cov50_mean}{0.094}
\setres{lotULJ}{DA_post_k1}{cov90}{0.21}
\setres{lotULJ}{DA_post_k1}{cov90_adj}{---}
\setres{lotULJ}{DA_post_k1}{cov90_adj_mean}{---}
\setres{lotULJ}{DA_post_k1}{cov90_mean}{0.203}
\setres{lotULJ}{DA_post_k1}{kappa}{0.282}
\setres{lotULJ}{DA_post_k1}{kappa_crps}{-0.030}
\setres{lotULJ}{DA_post_k1}{kappa_crps_mean}{-0.167}
\setres{lotULJ}{DA_post_k1}{kappa_mean}{0.035}
\setres{lotULJ}{DA_post_k1}{logscore_next}{-2.86}
\setres{lotULJ}{DA_post_k1}{logscore_next_mean}{-3.049}
\setres{lotULJ}{DA_post_k1}{misfit_pct}{---}
\setres{lotULJ}{DA_post_k1}{misfit_pct_mean}{---}
\setres{lotULJ}{DA_post_k1}{n_seen}{1}
\setres{lotULJ}{DA_post_k1}{surprise}{2.77}
\setres{lotULJ}{DA_post_k1}{surprise_mean}{31.163}
\setres{lotULJ}{DA_post_k1}{width90_adj_pct}{---}
\setres{lotULJ}{DA_post_k1}{width90_adj_pct_mean}{---}
\setres{lotULJ}{DA_post_k1}{width90_pct}{5.1}
\setres{lotULJ}{DA_post_k1}{width90_pct_mean}{5.644}
\setres{lotULJ}{DA_post_k10}{cov50}{0.00}
\setres{lotULJ}{DA_post_k10}{cov50_mean}{0.001}
\setres{lotULJ}{DA_post_k10}{cov90}{0.00}
\setres{lotULJ}{DA_post_k10}{cov90_adj}{0.88}
\setres{lotULJ}{DA_post_k10}{cov90_adj_mean}{0.840}
\setres{lotULJ}{DA_post_k10}{cov90_mean}{0.009}
\setres{lotULJ}{DA_post_k10}{kappa}{0.481}
\setres{lotULJ}{DA_post_k10}{kappa_crps}{0.138}
\setres{lotULJ}{DA_post_k10}{kappa_crps_mean}{-0.034}
\setres{lotULJ}{DA_post_k10}{kappa_mean}{0.163}
\setres{lotULJ}{DA_post_k10}{logscore_next}{-2.94}
\setres{lotULJ}{DA_post_k10}{logscore_next_mean}{-3.016}
\setres{lotULJ}{DA_post_k10}{misfit_pct}{16.3}
\setres{lotULJ}{DA_post_k10}{misfit_pct_mean}{17.049}
\setres{lotULJ}{DA_post_k10}{n_seen}{10}
\setres{lotULJ}{DA_post_k10}{surprise}{0.02}
\setres{lotULJ}{DA_post_k10}{surprise_mean}{13.731}
\setres{lotULJ}{DA_post_k10}{width90_adj_pct}{26.9}
\setres{lotULJ}{DA_post_k10}{width90_adj_pct_mean}{28.050}
\setres{lotULJ}{DA_post_k10}{width90_pct}{0.0}
\setres{lotULJ}{DA_post_k10}{width90_pct_mean}{0.290}
\setres{lotULJ}{DA_post_k11}{cov50}{0.00}
\setres{lotULJ}{DA_post_k11}{cov50_mean}{0.002}
\setres{lotULJ}{DA_post_k11}{cov90}{0.00}
\setres{lotULJ}{DA_post_k11}{cov90_adj}{0.91}
\setres{lotULJ}{DA_post_k11}{cov90_adj_mean}{0.891}
\setres{lotULJ}{DA_post_k11}{cov90_mean}{0.015}
\setres{lotULJ}{DA_post_k11}{kappa}{0.532}
\setres{lotULJ}{DA_post_k11}{kappa_crps}{0.234}
\setres{lotULJ}{DA_post_k11}{kappa_crps_mean}{0.031}
\setres{lotULJ}{DA_post_k11}{kappa_mean}{0.219}
\setres{lotULJ}{DA_post_k11}{logscore_next}{-2.89}
\setres{lotULJ}{DA_post_k11}{logscore_next_mean}{-3.003}
\setres{lotULJ}{DA_post_k11}{misfit_pct}{17.4}
\setres{lotULJ}{DA_post_k11}{misfit_pct_mean}{17.354}
\setres{lotULJ}{DA_post_k11}{n_seen}{11}
\setres{lotULJ}{DA_post_k11}{surprise}{0.01}
\setres{lotULJ}{DA_post_k11}{surprise_mean}{2.759}
\setres{lotULJ}{DA_post_k11}{width90_adj_pct}{28.6}
\setres{lotULJ}{DA_post_k11}{width90_adj_pct_mean}{28.556}
\setres{lotULJ}{DA_post_k11}{width90_pct}{0.1}
\setres{lotULJ}{DA_post_k11}{width90_pct_mean}{0.404}
\setres{lotULJ}{DA_post_k12}{cov50}{0.00}
\setres{lotULJ}{DA_post_k12}{cov50_mean}{0.005}
\setres{lotULJ}{DA_post_k12}{cov90}{0.00}
\setres{lotULJ}{DA_post_k12}{cov90_adj}{0.95}
\setres{lotULJ}{DA_post_k12}{cov90_adj_mean}{0.892}
\setres{lotULJ}{DA_post_k12}{cov90_mean}{0.009}
\setres{lotULJ}{DA_post_k12}{kappa}{0.567}
\setres{lotULJ}{DA_post_k12}{kappa_crps}{0.230}
\setres{lotULJ}{DA_post_k12}{kappa_crps_mean}{0.011}
\setres{lotULJ}{DA_post_k12}{kappa_mean}{0.206}
\setres{lotULJ}{DA_post_k12}{logscore_next}{-2.95}
\setres{lotULJ}{DA_post_k12}{logscore_next_mean}{-3.455}
\setres{lotULJ}{DA_post_k12}{misfit_pct}{17.8}
\setres{lotULJ}{DA_post_k12}{misfit_pct_mean}{17.482}
\setres{lotULJ}{DA_post_k12}{n_seen}{12}
\setres{lotULJ}{DA_post_k12}{surprise}{0.08}
\setres{lotULJ}{DA_post_k12}{surprise_mean}{10.219}
\setres{lotULJ}{DA_post_k12}{width90_adj_pct}{29.3}
\setres{lotULJ}{DA_post_k12}{width90_adj_pct_mean}{28.767}
\setres{lotULJ}{DA_post_k12}{width90_pct}{0.1}
\setres{lotULJ}{DA_post_k12}{width90_pct_mean}{0.381}
\setres{lotULJ}{DA_post_k13}{cov50}{0.00}
\setres{lotULJ}{DA_post_k13}{cov50_mean}{0.003}
\setres{lotULJ}{DA_post_k13}{cov90}{0.00}
\setres{lotULJ}{DA_post_k13}{cov90_adj}{0.95}
\setres{lotULJ}{DA_post_k13}{cov90_adj_mean}{0.889}
\setres{lotULJ}{DA_post_k13}{cov90_mean}{0.015}
\setres{lotULJ}{DA_post_k13}{kappa}{0.594}
\setres{lotULJ}{DA_post_k13}{kappa_crps}{0.267}
\setres{lotULJ}{DA_post_k13}{kappa_crps_mean}{-0.002}
\setres{lotULJ}{DA_post_k13}{kappa_mean}{0.165}
\setres{lotULJ}{DA_post_k13}{logscore_next}{-2.89}
\setres{lotULJ}{DA_post_k13}{logscore_next_mean}{-3.116}
\setres{lotULJ}{DA_post_k13}{misfit_pct}{17.7}
\setres{lotULJ}{DA_post_k13}{misfit_pct_mean}{17.471}
\setres{lotULJ}{DA_post_k13}{n_seen}{13}
\setres{lotULJ}{DA_post_k13}{surprise}{0.30}
\setres{lotULJ}{DA_post_k13}{surprise_mean}{7.247}
\setres{lotULJ}{DA_post_k13}{width90_adj_pct}{29.2}
\setres{lotULJ}{DA_post_k13}{width90_adj_pct_mean}{28.751}
\setres{lotULJ}{DA_post_k13}{width90_pct}{0.4}
\setres{lotULJ}{DA_post_k13}{width90_pct_mean}{0.523}
\setres{lotULJ}{DA_post_k14}{cov50}{0.00}
\setres{lotULJ}{DA_post_k14}{cov50_mean}{0.005}
\setres{lotULJ}{DA_post_k14}{cov90}{0.00}
\setres{lotULJ}{DA_post_k14}{cov90_adj}{0.95}
\setres{lotULJ}{DA_post_k14}{cov90_adj_mean}{0.898}
\setres{lotULJ}{DA_post_k14}{cov90_mean}{0.007}
\setres{lotULJ}{DA_post_k14}{kappa}{0.576}
\setres{lotULJ}{DA_post_k14}{kappa_crps}{0.246}
\setres{lotULJ}{DA_post_k14}{kappa_crps_mean}{0.000}
\setres{lotULJ}{DA_post_k14}{kappa_mean}{0.169}
\setres{lotULJ}{DA_post_k14}{logscore_next}{-2.94}
\setres{lotULJ}{DA_post_k14}{logscore_next_mean}{-3.055}
\setres{lotULJ}{DA_post_k14}{misfit_pct}{17.7}
\setres{lotULJ}{DA_post_k14}{misfit_pct_mean}{17.566}
\setres{lotULJ}{DA_post_k14}{n_seen}{14}
\setres{lotULJ}{DA_post_k14}{surprise}{0.09}
\setres{lotULJ}{DA_post_k14}{surprise_mean}{5.510}
\setres{lotULJ}{DA_post_k14}{width90_adj_pct}{29.2}
\setres{lotULJ}{DA_post_k14}{width90_adj_pct_mean}{28.908}
\setres{lotULJ}{DA_post_k14}{width90_pct}{0.3}
\setres{lotULJ}{DA_post_k14}{width90_pct_mean}{0.488}
\setres{lotULJ}{DA_post_k15}{cov50}{0.00}
\setres{lotULJ}{DA_post_k15}{cov50_mean}{0.010}
\setres{lotULJ}{DA_post_k15}{cov90}{0.00}
\setres{lotULJ}{DA_post_k15}{cov90_adj}{0.95}
\setres{lotULJ}{DA_post_k15}{cov90_adj_mean}{0.891}
\setres{lotULJ}{DA_post_k15}{cov90_mean}{0.028}
\setres{lotULJ}{DA_post_k15}{kappa}{0.587}
\setres{lotULJ}{DA_post_k15}{kappa_crps}{0.268}
\setres{lotULJ}{DA_post_k15}{kappa_crps_mean}{0.016}
\setres{lotULJ}{DA_post_k15}{kappa_mean}{0.180}
\setres{lotULJ}{DA_post_k15}{logscore_next}{-2.91}
\setres{lotULJ}{DA_post_k15}{logscore_next_mean}{-3.033}
\setres{lotULJ}{DA_post_k15}{misfit_pct}{17.5}
\setres{lotULJ}{DA_post_k15}{misfit_pct_mean}{17.288}
\setres{lotULJ}{DA_post_k15}{n_seen}{15}
\setres{lotULJ}{DA_post_k15}{surprise}{0.03}
\setres{lotULJ}{DA_post_k15}{surprise_mean}{14.735}
\setres{lotULJ}{DA_post_k15}{width90_adj_pct}{28.8}
\setres{lotULJ}{DA_post_k15}{width90_adj_pct_mean}{28.451}
\setres{lotULJ}{DA_post_k15}{width90_pct}{0.4}
\setres{lotULJ}{DA_post_k15}{width90_pct_mean}{0.513}
\setres{lotULJ}{DA_post_k16}{cov50}{0.00}
\setres{lotULJ}{DA_post_k16}{cov50_mean}{0.007}
\setres{lotULJ}{DA_post_k16}{cov90}{0.00}
\setres{lotULJ}{DA_post_k16}{cov90_adj}{0.96}
\setres{lotULJ}{DA_post_k16}{cov90_adj_mean}{0.923}
\setres{lotULJ}{DA_post_k16}{cov90_mean}{0.020}
\setres{lotULJ}{DA_post_k16}{kappa}{0.621}
\setres{lotULJ}{DA_post_k16}{kappa_crps}{0.309}
\setres{lotULJ}{DA_post_k16}{kappa_crps_mean}{0.085}
\setres{lotULJ}{DA_post_k16}{kappa_mean}{0.280}
\setres{lotULJ}{DA_post_k16}{logscore_next}{-2.85}
\setres{lotULJ}{DA_post_k16}{logscore_next_mean}{-3.123}
\setres{lotULJ}{DA_post_k16}{misfit_pct}{17.5}
\setres{lotULJ}{DA_post_k16}{misfit_pct_mean}{17.282}
\setres{lotULJ}{DA_post_k16}{n_seen}{16}
\setres{lotULJ}{DA_post_k16}{surprise}{0.01}
\setres{lotULJ}{DA_post_k16}{surprise_mean}{17.997}
\setres{lotULJ}{DA_post_k16}{width90_adj_pct}{28.8}
\setres{lotULJ}{DA_post_k16}{width90_adj_pct_mean}{28.437}
\setres{lotULJ}{DA_post_k16}{width90_pct}{0.2}
\setres{lotULJ}{DA_post_k16}{width90_pct_mean}{0.392}
\setres{lotULJ}{DA_post_k17}{cov50}{0.00}
\setres{lotULJ}{DA_post_k17}{cov50_mean}{0.002}
\setres{lotULJ}{DA_post_k17}{cov90}{0.00}
\setres{lotULJ}{DA_post_k17}{cov90_adj}{0.96}
\setres{lotULJ}{DA_post_k17}{cov90_adj_mean}{0.926}
\setres{lotULJ}{DA_post_k17}{cov90_mean}{0.006}
\setres{lotULJ}{DA_post_k17}{kappa}{0.628}
\setres{lotULJ}{DA_post_k17}{kappa_crps}{0.316}
\setres{lotULJ}{DA_post_k17}{kappa_crps_mean}{0.099}
\setres{lotULJ}{DA_post_k17}{kappa_mean}{0.304}
\setres{lotULJ}{DA_post_k17}{logscore_next}{-2.87}
\setres{lotULJ}{DA_post_k17}{logscore_next_mean}{-3.164}
\setres{lotULJ}{DA_post_k17}{misfit_pct}{17.7}
\setres{lotULJ}{DA_post_k17}{misfit_pct_mean}{17.312}
\setres{lotULJ}{DA_post_k17}{n_seen}{17}
\setres{lotULJ}{DA_post_k17}{surprise}{0.03}
\setres{lotULJ}{DA_post_k17}{surprise_mean}{7.585}
\setres{lotULJ}{DA_post_k17}{width90_adj_pct}{29.2}
\setres{lotULJ}{DA_post_k17}{width90_adj_pct_mean}{28.484}
\setres{lotULJ}{DA_post_k17}{width90_pct}{0.1}
\setres{lotULJ}{DA_post_k17}{width90_pct_mean}{0.336}
\setres{lotULJ}{DA_post_k18}{cov50}{0.00}
\setres{lotULJ}{DA_post_k18}{cov50_mean}{0.006}
\setres{lotULJ}{DA_post_k18}{cov90}{0.00}
\setres{lotULJ}{DA_post_k18}{cov90_adj}{0.96}
\setres{lotULJ}{DA_post_k18}{cov90_adj_mean}{0.940}
\setres{lotULJ}{DA_post_k18}{cov90_mean}{0.010}
\setres{lotULJ}{DA_post_k18}{kappa}{0.553}
\setres{lotULJ}{DA_post_k18}{kappa_crps}{0.263}
\setres{lotULJ}{DA_post_k18}{kappa_crps_mean}{0.129}
\setres{lotULJ}{DA_post_k18}{kappa_mean}{0.353}
\setres{lotULJ}{DA_post_k18}{logscore_next}{-2.88}
\setres{lotULJ}{DA_post_k18}{logscore_next_mean}{-2.972}
\setres{lotULJ}{DA_post_k18}{misfit_pct}{17.5}
\setres{lotULJ}{DA_post_k18}{misfit_pct_mean}{17.328}
\setres{lotULJ}{DA_post_k18}{n_seen}{18}
\setres{lotULJ}{DA_post_k18}{surprise}{0.10}
\setres{lotULJ}{DA_post_k18}{surprise_mean}{3.638}
\setres{lotULJ}{DA_post_k18}{width90_adj_pct}{28.8}
\setres{lotULJ}{DA_post_k18}{width90_adj_pct_mean}{28.516}
\setres{lotULJ}{DA_post_k18}{width90_pct}{0.3}
\setres{lotULJ}{DA_post_k18}{width90_pct_mean}{0.402}
\setres{lotULJ}{DA_post_k19}{cov50}{0.00}
\setres{lotULJ}{DA_post_k19}{cov50_mean}{0.010}
\setres{lotULJ}{DA_post_k19}{cov90}{0.00}
\setres{lotULJ}{DA_post_k19}{cov90_adj}{0.94}
\setres{lotULJ}{DA_post_k19}{cov90_adj_mean}{0.932}
\setres{lotULJ}{DA_post_k19}{cov90_mean}{0.012}
\setres{lotULJ}{DA_post_k19}{kappa}{0.593}
\setres{lotULJ}{DA_post_k19}{kappa_crps}{0.287}
\setres{lotULJ}{DA_post_k19}{kappa_crps_mean}{0.110}
\setres{lotULJ}{DA_post_k19}{kappa_mean}{0.341}
\setres{lotULJ}{DA_post_k19}{logscore_next}{-2.90}
\setres{lotULJ}{DA_post_k19}{logscore_next_mean}{-3.205}
\setres{lotULJ}{DA_post_k19}{misfit_pct}{17.5}
\setres{lotULJ}{DA_post_k19}{misfit_pct_mean}{17.244}
\setres{lotULJ}{DA_post_k19}{n_seen}{19}
\setres{lotULJ}{DA_post_k19}{surprise}{0.26}
\setres{lotULJ}{DA_post_k19}{surprise_mean}{38.693}
\setres{lotULJ}{DA_post_k19}{width90_adj_pct}{28.7}
\setres{lotULJ}{DA_post_k19}{width90_adj_pct_mean}{28.376}
\setres{lotULJ}{DA_post_k19}{width90_pct}{0.1}
\setres{lotULJ}{DA_post_k19}{width90_pct_mean}{0.357}
\setres{lotULJ}{DA_post_k2}{cov50}{0.04}
\setres{lotULJ}{DA_post_k2}{cov50_mean}{0.045}
\setres{lotULJ}{DA_post_k2}{cov90}{0.10}
\setres{lotULJ}{DA_post_k2}{cov90_adj}{0.33}
\setres{lotULJ}{DA_post_k2}{cov90_adj_mean}{0.471}
\setres{lotULJ}{DA_post_k2}{cov90_mean}{0.105}
\setres{lotULJ}{DA_post_k2}{kappa}{0.466}
\setres{lotULJ}{DA_post_k2}{kappa_crps}{0.255}
\setres{lotULJ}{DA_post_k2}{kappa_crps_mean}{-0.042}
\setres{lotULJ}{DA_post_k2}{kappa_mean}{0.229}
\setres{lotULJ}{DA_post_k2}{logscore_next}{-2.83}
\setres{lotULJ}{DA_post_k2}{logscore_next_mean}{-2.861}
\setres{lotULJ}{DA_post_k2}{misfit_pct}{8.6}
\setres{lotULJ}{DA_post_k2}{misfit_pct_mean}{9.824}
\setres{lotULJ}{DA_post_k2}{n_seen}{2}
\setres{lotULJ}{DA_post_k2}{surprise}{0.52}
\setres{lotULJ}{DA_post_k2}{surprise_mean}{5.634}
\setres{lotULJ}{DA_post_k2}{width90_adj_pct}{14.3}
\setres{lotULJ}{DA_post_k2}{width90_adj_pct_mean}{16.802}
\setres{lotULJ}{DA_post_k2}{width90_pct}{2.1}
\setres{lotULJ}{DA_post_k2}{width90_pct_mean}{2.500}
\setres{lotULJ}{DA_post_k20}{cov50}{0.00}
\setres{lotULJ}{DA_post_k20}{cov50_mean}{0.000}
\setres{lotULJ}{DA_post_k20}{cov90}{0.00}
\setres{lotULJ}{DA_post_k20}{cov90_adj}{0.96}
\setres{lotULJ}{DA_post_k20}{cov90_adj_mean}{0.938}
\setres{lotULJ}{DA_post_k20}{cov90_mean}{0.008}
\setres{lotULJ}{DA_post_k20}{kappa}{0.586}
\setres{lotULJ}{DA_post_k20}{kappa_crps}{0.274}
\setres{lotULJ}{DA_post_k20}{kappa_crps_mean}{0.145}
\setres{lotULJ}{DA_post_k20}{kappa_mean}{0.350}
\setres{lotULJ}{DA_post_k20}{logscore_next}{-2.87}
\setres{lotULJ}{DA_post_k20}{logscore_next_mean}{-3.217}
\setres{lotULJ}{DA_post_k20}{misfit_pct}{17.2}
\setres{lotULJ}{DA_post_k20}{misfit_pct_mean}{17.010}
\setres{lotULJ}{DA_post_k20}{n_seen}{20}
\setres{lotULJ}{DA_post_k20}{surprise}{0.01}
\setres{lotULJ}{DA_post_k20}{surprise_mean}{6.731}
\setres{lotULJ}{DA_post_k20}{width90_adj_pct}{28.2}
\setres{lotULJ}{DA_post_k20}{width90_adj_pct_mean}{27.989}
\setres{lotULJ}{DA_post_k20}{width90_pct}{0.0}
\setres{lotULJ}{DA_post_k20}{width90_pct_mean}{0.337}
\setres{lotULJ}{DA_post_k21}{cov50}{0.00}
\setres{lotULJ}{DA_post_k21}{cov50_mean}{0.002}
\setres{lotULJ}{DA_post_k21}{cov90}{0.00}
\setres{lotULJ}{DA_post_k21}{cov90_adj}{0.96}
\setres{lotULJ}{DA_post_k21}{cov90_adj_mean}{0.928}
\setres{lotULJ}{DA_post_k21}{cov90_mean}{0.013}
\setres{lotULJ}{DA_post_k21}{kappa}{0.496}
\setres{lotULJ}{DA_post_k21}{kappa_crps}{0.255}
\setres{lotULJ}{DA_post_k21}{kappa_crps_mean}{0.126}
\setres{lotULJ}{DA_post_k21}{kappa_mean}{0.330}
\setres{lotULJ}{DA_post_k21}{logscore_next}{-2.87}
\setres{lotULJ}{DA_post_k21}{logscore_next_mean}{-3.161}
\setres{lotULJ}{DA_post_k21}{misfit_pct}{16.9}
\setres{lotULJ}{DA_post_k21}{misfit_pct_mean}{16.923}
\setres{lotULJ}{DA_post_k21}{n_seen}{21}
\setres{lotULJ}{DA_post_k21}{surprise}{0.01}
\setres{lotULJ}{DA_post_k21}{surprise_mean}{25.153}
\setres{lotULJ}{DA_post_k21}{width90_adj_pct}{27.8}
\setres{lotULJ}{DA_post_k21}{width90_adj_pct_mean}{27.844}
\setres{lotULJ}{DA_post_k21}{width90_pct}{0.2}
\setres{lotULJ}{DA_post_k21}{width90_pct_mean}{0.322}
\setres{lotULJ}{DA_post_k22}{cov50}{0.00}
\setres{lotULJ}{DA_post_k22}{cov50_mean}{0.005}
\setres{lotULJ}{DA_post_k22}{cov90}{0.00}
\setres{lotULJ}{DA_post_k22}{cov90_adj}{0.95}
\setres{lotULJ}{DA_post_k22}{cov90_adj_mean}{0.934}
\setres{lotULJ}{DA_post_k22}{cov90_mean}{0.014}
\setres{lotULJ}{DA_post_k22}{kappa}{0.532}
\setres{lotULJ}{DA_post_k22}{kappa_crps}{0.262}
\setres{lotULJ}{DA_post_k22}{kappa_crps_mean}{0.213}
\setres{lotULJ}{DA_post_k22}{kappa_mean}{0.514}
\setres{lotULJ}{DA_post_k22}{logscore_next}{-2.88}
\setres{lotULJ}{DA_post_k22}{logscore_next_mean}{-3.083}
\setres{lotULJ}{DA_post_k22}{misfit_pct}{16.5}
\setres{lotULJ}{DA_post_k22}{misfit_pct_mean}{16.788}
\setres{lotULJ}{DA_post_k22}{n_seen}{22}
\setres{lotULJ}{DA_post_k22}{surprise}{0.01}
\setres{lotULJ}{DA_post_k22}{surprise_mean}{2.232}
\setres{lotULJ}{DA_post_k22}{width90_adj_pct}{27.2}
\setres{lotULJ}{DA_post_k22}{width90_adj_pct_mean}{27.626}
\setres{lotULJ}{DA_post_k22}{width90_pct}{0.0}
\setres{lotULJ}{DA_post_k22}{width90_pct_mean}{0.367}
\setres{lotULJ}{DA_post_k23}{cov50}{0.00}
\setres{lotULJ}{DA_post_k23}{cov50_mean}{0.002}
\setres{lotULJ}{DA_post_k23}{cov90}{0.00}
\setres{lotULJ}{DA_post_k23}{cov90_adj}{0.95}
\setres{lotULJ}{DA_post_k23}{cov90_adj_mean}{0.940}
\setres{lotULJ}{DA_post_k23}{cov90_mean}{0.007}
\setres{lotULJ}{DA_post_k23}{kappa}{0.524}
\setres{lotULJ}{DA_post_k23}{kappa_crps}{0.252}
\setres{lotULJ}{DA_post_k23}{kappa_crps_mean}{0.234}
\setres{lotULJ}{DA_post_k23}{kappa_mean}{0.540}
\setres{lotULJ}{DA_post_k23}{logscore_next}{-2.86}
\setres{lotULJ}{DA_post_k23}{logscore_next_mean}{-2.902}
\setres{lotULJ}{DA_post_k23}{misfit_pct}{16.5}
\setres{lotULJ}{DA_post_k23}{misfit_pct_mean}{16.753}
\setres{lotULJ}{DA_post_k23}{n_seen}{23}
\setres{lotULJ}{DA_post_k23}{surprise}{0.01}
\setres{lotULJ}{DA_post_k23}{surprise_mean}{1.181}
\setres{lotULJ}{DA_post_k23}{width90_adj_pct}{27.1}
\setres{lotULJ}{DA_post_k23}{width90_adj_pct_mean}{27.562}
\setres{lotULJ}{DA_post_k23}{width90_pct}{0.0}
\setres{lotULJ}{DA_post_k23}{width90_pct_mean}{0.194}
\setres{lotULJ}{DA_post_k24}{cov50}{0.00}
\setres{lotULJ}{DA_post_k24}{cov50_mean}{0.003}
\setres{lotULJ}{DA_post_k24}{cov90}{0.00}
\setres{lotULJ}{DA_post_k24}{cov90_adj}{0.95}
\setres{lotULJ}{DA_post_k24}{cov90_adj_mean}{0.947}
\setres{lotULJ}{DA_post_k24}{cov90_mean}{0.005}
\setres{lotULJ}{DA_post_k24}{kappa}{0.512}
\setres{lotULJ}{DA_post_k24}{kappa_crps}{0.276}
\setres{lotULJ}{DA_post_k24}{kappa_crps_mean}{0.207}
\setres{lotULJ}{DA_post_k24}{kappa_mean}{0.451}
\setres{lotULJ}{DA_post_k24}{logscore_next}{-2.89}
\setres{lotULJ}{DA_post_k24}{logscore_next_mean}{-2.967}
\setres{lotULJ}{DA_post_k24}{misfit_pct}{16.6}
\setres{lotULJ}{DA_post_k24}{misfit_pct_mean}{16.758}
\setres{lotULJ}{DA_post_k24}{n_seen}{24}
\setres{lotULJ}{DA_post_k24}{surprise}{0.01}
\setres{lotULJ}{DA_post_k24}{surprise_mean}{1.938}
\setres{lotULJ}{DA_post_k24}{width90_adj_pct}{27.2}
\setres{lotULJ}{DA_post_k24}{width90_adj_pct_mean}{27.570}
\setres{lotULJ}{DA_post_k24}{width90_pct}{0.0}
\setres{lotULJ}{DA_post_k24}{width90_pct_mean}{0.164}
\setres{lotULJ}{DA_post_k25}{cov50}{0.00}
\setres{lotULJ}{DA_post_k25}{cov50_mean}{0.000}
\setres{lotULJ}{DA_post_k25}{cov90}{0.00}
\setres{lotULJ}{DA_post_k25}{cov90_adj}{0.95}
\setres{lotULJ}{DA_post_k25}{cov90_adj_mean}{0.950}
\setres{lotULJ}{DA_post_k25}{cov90_mean}{0.005}
\setres{lotULJ}{DA_post_k25}{kappa}{0.514}
\setres{lotULJ}{DA_post_k25}{kappa_crps}{0.305}
\setres{lotULJ}{DA_post_k25}{kappa_crps_mean}{0.215}
\setres{lotULJ}{DA_post_k25}{kappa_mean}{0.466}
\setres{lotULJ}{DA_post_k25}{logscore_next}{-2.85}
\setres{lotULJ}{DA_post_k25}{logscore_next_mean}{-2.912}
\setres{lotULJ}{DA_post_k25}{misfit_pct}{16.4}
\setres{lotULJ}{DA_post_k25}{misfit_pct_mean}{16.670}
\setres{lotULJ}{DA_post_k25}{n_seen}{25}
\setres{lotULJ}{DA_post_k25}{surprise}{0.02}
\setres{lotULJ}{DA_post_k25}{surprise_mean}{2.592}
\setres{lotULJ}{DA_post_k25}{width90_adj_pct}{27.0}
\setres{lotULJ}{DA_post_k25}{width90_adj_pct_mean}{27.425}
\setres{lotULJ}{DA_post_k25}{width90_pct}{0.0}
\setres{lotULJ}{DA_post_k25}{width90_pct_mean}{0.158}
\setres{lotULJ}{DA_post_k26}{cov50}{0.00}
\setres{lotULJ}{DA_post_k26}{cov50_mean}{0.000}
\setres{lotULJ}{DA_post_k26}{cov90}{0.00}
\setres{lotULJ}{DA_post_k26}{cov90_adj}{0.94}
\setres{lotULJ}{DA_post_k26}{cov90_adj_mean}{0.947}
\setres{lotULJ}{DA_post_k26}{cov90_mean}{0.006}
\setres{lotULJ}{DA_post_k26}{kappa}{0.541}
\setres{lotULJ}{DA_post_k26}{kappa_crps}{0.321}
\setres{lotULJ}{DA_post_k26}{kappa_crps_mean}{0.237}
\setres{lotULJ}{DA_post_k26}{kappa_mean}{0.504}
\setres{lotULJ}{DA_post_k26}{logscore_next}{-2.89}
\setres{lotULJ}{DA_post_k26}{logscore_next_mean}{-3.067}
\setres{lotULJ}{DA_post_k26}{misfit_pct}{16.3}
\setres{lotULJ}{DA_post_k26}{misfit_pct_mean}{16.527}
\setres{lotULJ}{DA_post_k26}{n_seen}{26}
\setres{lotULJ}{DA_post_k26}{surprise}{0.12}
\setres{lotULJ}{DA_post_k26}{surprise_mean}{14.789}
\setres{lotULJ}{DA_post_k26}{width90_adj_pct}{26.8}
\setres{lotULJ}{DA_post_k26}{width90_adj_pct_mean}{27.192}
\setres{lotULJ}{DA_post_k26}{width90_pct}{0.0}
\setres{lotULJ}{DA_post_k26}{width90_pct_mean}{0.189}
\setres{lotULJ}{DA_post_k27}{cov50}{0.00}
\setres{lotULJ}{DA_post_k27}{cov50_mean}{0.003}
\setres{lotULJ}{DA_post_k27}{cov90}{0.00}
\setres{lotULJ}{DA_post_k27}{cov90_adj}{0.94}
\setres{lotULJ}{DA_post_k27}{cov90_adj_mean}{0.944}
\setres{lotULJ}{DA_post_k27}{cov90_mean}{0.003}
\setres{lotULJ}{DA_post_k27}{kappa}{0.540}
\setres{lotULJ}{DA_post_k27}{kappa_crps}{0.327}
\setres{lotULJ}{DA_post_k27}{kappa_crps_mean}{0.271}
\setres{lotULJ}{DA_post_k27}{kappa_mean}{0.518}
\setres{lotULJ}{DA_post_k27}{logscore_next}{-2.84}
\setres{lotULJ}{DA_post_k27}{logscore_next_mean}{-2.912}
\setres{lotULJ}{DA_post_k27}{misfit_pct}{16.1}
\setres{lotULJ}{DA_post_k27}{misfit_pct_mean}{16.415}
\setres{lotULJ}{DA_post_k27}{n_seen}{27}
\setres{lotULJ}{DA_post_k27}{surprise}{0.00}
\setres{lotULJ}{DA_post_k27}{surprise_mean}{5.431}
\setres{lotULJ}{DA_post_k27}{width90_adj_pct}{26.5}
\setres{lotULJ}{DA_post_k27}{width90_adj_pct_mean}{27.007}
\setres{lotULJ}{DA_post_k27}{width90_pct}{0.0}
\setres{lotULJ}{DA_post_k27}{width90_pct_mean}{0.180}
\setres{lotULJ}{DA_post_k28}{cov50}{0.00}
\setres{lotULJ}{DA_post_k28}{cov50_mean}{0.003}
\setres{lotULJ}{DA_post_k28}{cov90}{0.00}
\setres{lotULJ}{DA_post_k28}{cov90_adj}{0.94}
\setres{lotULJ}{DA_post_k28}{cov90_adj_mean}{0.950}
\setres{lotULJ}{DA_post_k28}{cov90_mean}{0.003}
\setres{lotULJ}{DA_post_k28}{kappa}{0.557}
\setres{lotULJ}{DA_post_k28}{kappa_crps}{0.328}
\setres{lotULJ}{DA_post_k28}{kappa_crps_mean}{0.273}
\setres{lotULJ}{DA_post_k28}{kappa_mean}{0.518}
\setres{lotULJ}{DA_post_k28}{logscore_next}{-2.84}
\setres{lotULJ}{DA_post_k28}{logscore_next_mean}{-3.068}
\setres{lotULJ}{DA_post_k28}{misfit_pct}{16.1}
\setres{lotULJ}{DA_post_k28}{misfit_pct_mean}{16.323}
\setres{lotULJ}{DA_post_k28}{n_seen}{28}
\setres{lotULJ}{DA_post_k28}{surprise}{0.01}
\setres{lotULJ}{DA_post_k28}{surprise_mean}{5.712}
\setres{lotULJ}{DA_post_k28}{width90_adj_pct}{26.5}
\setres{lotULJ}{DA_post_k28}{width90_adj_pct_mean}{26.853}
\setres{lotULJ}{DA_post_k28}{width90_pct}{0.0}
\setres{lotULJ}{DA_post_k28}{width90_pct_mean}{0.086}
\setres{lotULJ}{DA_post_k29}{cov50}{0.00}
\setres{lotULJ}{DA_post_k29}{cov50_mean}{0.003}
\setres{lotULJ}{DA_post_k29}{cov90}{0.00}
\setres{lotULJ}{DA_post_k29}{cov90_adj}{0.93}
\setres{lotULJ}{DA_post_k29}{cov90_adj_mean}{0.947}
\setres{lotULJ}{DA_post_k29}{cov90_mean}{0.007}
\setres{lotULJ}{DA_post_k29}{kappa}{0.587}
\setres{lotULJ}{DA_post_k29}{kappa_crps}{0.366}
\setres{lotULJ}{DA_post_k29}{kappa_crps_mean}{0.305}
\setres{lotULJ}{DA_post_k29}{kappa_mean}{0.529}
\setres{lotULJ}{DA_post_k29}{logscore_next}{-2.96}
\setres{lotULJ}{DA_post_k29}{logscore_next_mean}{-3.293}
\setres{lotULJ}{DA_post_k29}{misfit_pct}{16.2}
\setres{lotULJ}{DA_post_k29}{misfit_pct_mean}{16.231}
\setres{lotULJ}{DA_post_k29}{n_seen}{29}
\setres{lotULJ}{DA_post_k29}{surprise}{0.03}
\setres{lotULJ}{DA_post_k29}{surprise_mean}{1.486}
\setres{lotULJ}{DA_post_k29}{width90_adj_pct}{26.6}
\setres{lotULJ}{DA_post_k29}{width90_adj_pct_mean}{26.702}
\setres{lotULJ}{DA_post_k29}{width90_pct}{0.0}
\setres{lotULJ}{DA_post_k29}{width90_pct_mean}{0.167}
\setres{lotULJ}{DA_post_k3}{cov50}{0.02}
\setres{lotULJ}{DA_post_k3}{cov50_mean}{0.018}
\setres{lotULJ}{DA_post_k3}{cov90}{0.05}
\setres{lotULJ}{DA_post_k3}{cov90_adj}{0.83}
\setres{lotULJ}{DA_post_k3}{cov90_adj_mean}{0.732}
\setres{lotULJ}{DA_post_k3}{cov90_mean}{0.052}
\setres{lotULJ}{DA_post_k3}{kappa}{0.621}
\setres{lotULJ}{DA_post_k3}{kappa_crps}{0.263}
\setres{lotULJ}{DA_post_k3}{kappa_crps_mean}{-0.095}
\setres{lotULJ}{DA_post_k3}{kappa_mean}{0.191}
\setres{lotULJ}{DA_post_k3}{logscore_next}{-2.91}
\setres{lotULJ}{DA_post_k3}{logscore_next_mean}{-3.006}
\setres{lotULJ}{DA_post_k3}{misfit_pct}{14.3}
\setres{lotULJ}{DA_post_k3}{misfit_pct_mean}{15.764}
\setres{lotULJ}{DA_post_k3}{n_seen}{3}
\setres{lotULJ}{DA_post_k3}{surprise}{2.04}
\setres{lotULJ}{DA_post_k3}{surprise_mean}{12.000}
\setres{lotULJ}{DA_post_k3}{width90_adj_pct}{23.6}
\setres{lotULJ}{DA_post_k3}{width90_adj_pct_mean}{26.037}
\setres{lotULJ}{DA_post_k3}{width90_pct}{0.9}
\setres{lotULJ}{DA_post_k3}{width90_pct_mean}{1.277}
\setres{lotULJ}{DA_post_k30}{cov50}{0.00}
\setres{lotULJ}{DA_post_k30}{cov50_mean}{0.000}
\setres{lotULJ}{DA_post_k30}{cov90}{0.00}
\setres{lotULJ}{DA_post_k30}{cov90_adj}{0.93}
\setres{lotULJ}{DA_post_k30}{cov90_adj_mean}{0.946}
\setres{lotULJ}{DA_post_k30}{cov90_mean}{0.007}
\setres{lotULJ}{DA_post_k30}{kappa}{0.576}
\setres{lotULJ}{DA_post_k30}{kappa_crps}{0.363}
\setres{lotULJ}{DA_post_k30}{kappa_crps_mean}{0.298}
\setres{lotULJ}{DA_post_k30}{kappa_mean}{0.520}
\setres{lotULJ}{DA_post_k30}{logscore_next}{-2.93}
\setres{lotULJ}{DA_post_k30}{logscore_next_mean}{-3.045}
\setres{lotULJ}{DA_post_k30}{misfit_pct}{16.3}
\setres{lotULJ}{DA_post_k30}{misfit_pct_mean}{16.196}
\setres{lotULJ}{DA_post_k30}{n_seen}{30}
\setres{lotULJ}{DA_post_k30}{surprise}{0.03}
\setres{lotULJ}{DA_post_k30}{surprise_mean}{1.376}
\setres{lotULJ}{DA_post_k30}{width90_adj_pct}{26.8}
\setres{lotULJ}{DA_post_k30}{width90_adj_pct_mean}{26.645}
\setres{lotULJ}{DA_post_k30}{width90_pct}{0.0}
\setres{lotULJ}{DA_post_k30}{width90_pct_mean}{0.206}
\setres{lotULJ}{DA_post_k31}{cov50}{0.00}
\setres{lotULJ}{DA_post_k31}{cov50_mean}{0.000}
\setres{lotULJ}{DA_post_k31}{cov90}{0.00}
\setres{lotULJ}{DA_post_k31}{cov90_adj}{0.92}
\setres{lotULJ}{DA_post_k31}{cov90_adj_mean}{0.946}
\setres{lotULJ}{DA_post_k31}{cov90_mean}{0.000}
\setres{lotULJ}{DA_post_k31}{kappa}{0.562}
\setres{lotULJ}{DA_post_k31}{kappa_crps}{0.365}
\setres{lotULJ}{DA_post_k31}{kappa_crps_mean}{0.295}
\setres{lotULJ}{DA_post_k31}{kappa_mean}{0.538}
\setres{lotULJ}{DA_post_k31}{logscore_next}{-2.85}
\setres{lotULJ}{DA_post_k31}{logscore_next_mean}{-2.951}
\setres{lotULJ}{DA_post_k31}{misfit_pct}{16.3}
\setres{lotULJ}{DA_post_k31}{misfit_pct_mean}{16.178}
\setres{lotULJ}{DA_post_k31}{n_seen}{31}
\setres{lotULJ}{DA_post_k31}{surprise}{0.01}
\setres{lotULJ}{DA_post_k31}{surprise_mean}{1.561}
\setres{lotULJ}{DA_post_k31}{width90_adj_pct}{26.8}
\setres{lotULJ}{DA_post_k31}{width90_adj_pct_mean}{26.615}
\setres{lotULJ}{DA_post_k31}{width90_pct}{0.0}
\setres{lotULJ}{DA_post_k31}{width90_pct_mean}{0.130}
\setres{lotULJ}{DA_post_k32}{cov50}{0.00}
\setres{lotULJ}{DA_post_k32}{cov50_mean}{0.000}
\setres{lotULJ}{DA_post_k32}{cov90}{0.00}
\setres{lotULJ}{DA_post_k32}{cov90_adj}{0.96}
\setres{lotULJ}{DA_post_k32}{cov90_adj_mean}{0.946}
\setres{lotULJ}{DA_post_k32}{cov90_mean}{0.000}
\setres{lotULJ}{DA_post_k32}{kappa}{0.572}
\setres{lotULJ}{DA_post_k32}{kappa_crps}{0.336}
\setres{lotULJ}{DA_post_k32}{kappa_crps_mean}{0.294}
\setres{lotULJ}{DA_post_k32}{kappa_mean}{0.517}
\setres{lotULJ}{DA_post_k32}{logscore_next}{-2.97}
\setres{lotULJ}{DA_post_k32}{logscore_next_mean}{-3.015}
\setres{lotULJ}{DA_post_k32}{misfit_pct}{16.2}
\setres{lotULJ}{DA_post_k32}{misfit_pct_mean}{16.211}
\setres{lotULJ}{DA_post_k32}{n_seen}{32}
\setres{lotULJ}{DA_post_k32}{surprise}{0.01}
\setres{lotULJ}{DA_post_k32}{surprise_mean}{0.704}
\setres{lotULJ}{DA_post_k32}{width90_adj_pct}{26.7}
\setres{lotULJ}{DA_post_k32}{width90_adj_pct_mean}{26.670}
\setres{lotULJ}{DA_post_k32}{width90_pct}{0.0}
\setres{lotULJ}{DA_post_k32}{width90_pct_mean}{0.151}
\setres{lotULJ}{DA_post_k33}{cov50}{0.00}
\setres{lotULJ}{DA_post_k33}{cov50_mean}{0.000}
\setres{lotULJ}{DA_post_k33}{cov90}{0.00}
\setres{lotULJ}{DA_post_k33}{cov90_adj}{1.00}
\setres{lotULJ}{DA_post_k33}{cov90_adj_mean}{0.945}
\setres{lotULJ}{DA_post_k33}{cov90_mean}{0.005}
\setres{lotULJ}{DA_post_k33}{kappa}{0.567}
\setres{lotULJ}{DA_post_k33}{kappa_crps}{0.348}
\setres{lotULJ}{DA_post_k33}{kappa_crps_mean}{0.267}
\setres{lotULJ}{DA_post_k33}{kappa_mean}{0.506}
\setres{lotULJ}{DA_post_k33}{logscore_next}{-2.86}
\setres{lotULJ}{DA_post_k33}{logscore_next_mean}{-2.944}
\setres{lotULJ}{DA_post_k33}{misfit_pct}{16.2}
\setres{lotULJ}{DA_post_k33}{misfit_pct_mean}{16.193}
\setres{lotULJ}{DA_post_k33}{n_seen}{33}
\setres{lotULJ}{DA_post_k33}{surprise}{0.00}
\setres{lotULJ}{DA_post_k33}{surprise_mean}{0.683}
\setres{lotULJ}{DA_post_k33}{width90_adj_pct}{26.6}
\setres{lotULJ}{DA_post_k33}{width90_adj_pct_mean}{26.639}
\setres{lotULJ}{DA_post_k33}{width90_pct}{0.0}
\setres{lotULJ}{DA_post_k33}{width90_pct_mean}{0.123}
\setres{lotULJ}{DA_post_k34}{cov50}{0.00}
\setres{lotULJ}{DA_post_k34}{cov50_mean}{0.000}
\setres{lotULJ}{DA_post_k34}{cov90}{0.00}
\setres{lotULJ}{DA_post_k34}{cov90_adj}{1.00}
\setres{lotULJ}{DA_post_k34}{cov90_adj_mean}{0.955}
\setres{lotULJ}{DA_post_k34}{cov90_mean}{0.000}
\setres{lotULJ}{DA_post_k34}{kappa}{0.574}
\setres{lotULJ}{DA_post_k34}{kappa_crps}{0.355}
\setres{lotULJ}{DA_post_k34}{kappa_crps_mean}{0.254}
\setres{lotULJ}{DA_post_k34}{kappa_mean}{0.494}
\setres{lotULJ}{DA_post_k34}{logscore_next}{-2.86}
\setres{lotULJ}{DA_post_k34}{logscore_next_mean}{-3.234}
\setres{lotULJ}{DA_post_k34}{misfit_pct}{16.3}
\setres{lotULJ}{DA_post_k34}{misfit_pct_mean}{16.230}
\setres{lotULJ}{DA_post_k34}{n_seen}{34}
\setres{lotULJ}{DA_post_k34}{surprise}{0.00}
\setres{lotULJ}{DA_post_k34}{surprise_mean}{12.910}
\setres{lotULJ}{DA_post_k34}{width90_adj_pct}{26.9}
\setres{lotULJ}{DA_post_k34}{width90_adj_pct_mean}{26.700}
\setres{lotULJ}{DA_post_k34}{width90_pct}{0.0}
\setres{lotULJ}{DA_post_k34}{width90_pct_mean}{0.099}
\setres{lotULJ}{DA_post_k35}{cov50}{0.00}
\setres{lotULJ}{DA_post_k35}{cov50_mean}{0.000}
\setres{lotULJ}{DA_post_k35}{cov90}{0.00}
\setres{lotULJ}{DA_post_k35}{cov90_adj}{1.00}
\setres{lotULJ}{DA_post_k35}{cov90_adj_mean}{0.961}
\setres{lotULJ}{DA_post_k35}{cov90_mean}{0.000}
\setres{lotULJ}{DA_post_k35}{kappa}{0.538}
\setres{lotULJ}{DA_post_k35}{kappa_crps}{0.299}
\setres{lotULJ}{DA_post_k35}{kappa_crps_mean}{0.248}
\setres{lotULJ}{DA_post_k35}{kappa_mean}{0.514}
\setres{lotULJ}{DA_post_k35}{logscore_next}{-2.84}
\setres{lotULJ}{DA_post_k35}{logscore_next_mean}{-3.029}
\setres{lotULJ}{DA_post_k35}{misfit_pct}{16.2}
\setres{lotULJ}{DA_post_k35}{misfit_pct_mean}{16.186}
\setres{lotULJ}{DA_post_k35}{n_seen}{35}
\setres{lotULJ}{DA_post_k35}{surprise}{0.00}
\setres{lotULJ}{DA_post_k35}{surprise_mean}{0.456}
\setres{lotULJ}{DA_post_k35}{width90_adj_pct}{26.7}
\setres{lotULJ}{DA_post_k35}{width90_adj_pct_mean}{26.629}
\setres{lotULJ}{DA_post_k35}{width90_pct}{0.0}
\setres{lotULJ}{DA_post_k35}{width90_pct_mean}{0.160}
\setres{lotULJ}{DA_post_k36}{cov50}{0.00}
\setres{lotULJ}{DA_post_k36}{cov50_mean}{0.000}
\setres{lotULJ}{DA_post_k36}{cov90}{0.00}
\setres{lotULJ}{DA_post_k36}{cov90_adj}{1.00}
\setres{lotULJ}{DA_post_k36}{cov90_adj_mean}{0.956}
\setres{lotULJ}{DA_post_k36}{cov90_mean}{0.000}
\setres{lotULJ}{DA_post_k36}{kappa}{0.593}
\setres{lotULJ}{DA_post_k36}{kappa_crps}{0.300}
\setres{lotULJ}{DA_post_k36}{kappa_crps_mean}{0.282}
\setres{lotULJ}{DA_post_k36}{kappa_mean}{0.528}
\setres{lotULJ}{DA_post_k36}{logscore_next}{-2.83}
\setres{lotULJ}{DA_post_k36}{logscore_next_mean}{-3.082}
\setres{lotULJ}{DA_post_k36}{misfit_pct}{16.3}
\setres{lotULJ}{DA_post_k36}{misfit_pct_mean}{16.230}
\setres{lotULJ}{DA_post_k36}{n_seen}{36}
\setres{lotULJ}{DA_post_k36}{surprise}{0.00}
\setres{lotULJ}{DA_post_k36}{surprise_mean}{1.301}
\setres{lotULJ}{DA_post_k36}{width90_adj_pct}{26.8}
\setres{lotULJ}{DA_post_k36}{width90_adj_pct_mean}{26.701}
\setres{lotULJ}{DA_post_k36}{width90_pct}{0.0}
\setres{lotULJ}{DA_post_k36}{width90_pct_mean}{0.121}
\setres{lotULJ}{DA_post_k37}{cov50}{0.00}
\setres{lotULJ}{DA_post_k37}{cov50_mean}{0.000}
\setres{lotULJ}{DA_post_k37}{cov90}{0.00}
\setres{lotULJ}{DA_post_k37}{cov90_adj}{1.00}
\setres{lotULJ}{DA_post_k37}{cov90_adj_mean}{0.950}
\setres{lotULJ}{DA_post_k37}{cov90_mean}{0.000}
\setres{lotULJ}{DA_post_k37}{kappa}{0.475}
\setres{lotULJ}{DA_post_k37}{kappa_crps}{0.305}
\setres{lotULJ}{DA_post_k37}{kappa_crps_mean}{0.208}
\setres{lotULJ}{DA_post_k37}{kappa_mean}{0.470}
\setres{lotULJ}{DA_post_k37}{logscore_next}{-2.86}
\setres{lotULJ}{DA_post_k37}{logscore_next_mean}{-3.075}
\setres{lotULJ}{DA_post_k37}{misfit_pct}{16.2}
\setres{lotULJ}{DA_post_k37}{misfit_pct_mean}{16.200}
\setres{lotULJ}{DA_post_k37}{n_seen}{37}
\setres{lotULJ}{DA_post_k37}{surprise}{0.00}
\setres{lotULJ}{DA_post_k37}{surprise_mean}{7.342}
\setres{lotULJ}{DA_post_k37}{width90_adj_pct}{26.7}
\setres{lotULJ}{DA_post_k37}{width90_adj_pct_mean}{26.649}
\setres{lotULJ}{DA_post_k37}{width90_pct}{0.0}
\setres{lotULJ}{DA_post_k37}{width90_pct_mean}{0.014}
\setres{lotULJ}{DA_post_k38}{cov50}{0.00}
\setres{lotULJ}{DA_post_k38}{cov50_mean}{0.000}
\setres{lotULJ}{DA_post_k38}{cov90}{0.00}
\setres{lotULJ}{DA_post_k38}{cov90_adj}{1.00}
\setres{lotULJ}{DA_post_k38}{cov90_adj_mean}{0.958}
\setres{lotULJ}{DA_post_k38}{cov90_mean}{0.000}
\setres{lotULJ}{DA_post_k38}{kappa}{0.457}
\setres{lotULJ}{DA_post_k38}{kappa_crps}{0.295}
\setres{lotULJ}{DA_post_k38}{kappa_crps_mean}{0.158}
\setres{lotULJ}{DA_post_k38}{kappa_mean}{0.408}
\setres{lotULJ}{DA_post_k38}{logscore_next}{-2.94}
\setres{lotULJ}{DA_post_k38}{logscore_next_mean}{-3.051}
\setres{lotULJ}{DA_post_k38}{misfit_pct}{16.2}
\setres{lotULJ}{DA_post_k38}{misfit_pct_mean}{16.253}
\setres{lotULJ}{DA_post_k38}{n_seen}{38}
\setres{lotULJ}{DA_post_k38}{surprise}{0.00}
\setres{lotULJ}{DA_post_k38}{surprise_mean}{0.958}
\setres{lotULJ}{DA_post_k38}{width90_adj_pct}{26.7}
\setres{lotULJ}{DA_post_k38}{width90_adj_pct_mean}{26.737}
\setres{lotULJ}{DA_post_k38}{width90_pct}{0.0}
\setres{lotULJ}{DA_post_k38}{width90_pct_mean}{0.034}
\setres{lotULJ}{DA_post_k39}{cov50}{0.00}
\setres{lotULJ}{DA_post_k39}{cov50_mean}{0.000}
\setres{lotULJ}{DA_post_k39}{cov90}{0.00}
\setres{lotULJ}{DA_post_k39}{cov90_adj}{1.00}
\setres{lotULJ}{DA_post_k39}{cov90_adj_mean}{0.960}
\setres{lotULJ}{DA_post_k39}{cov90_mean}{0.000}
\setres{lotULJ}{DA_post_k39}{kappa}{0.463}
\setres{lotULJ}{DA_post_k39}{kappa_crps}{0.226}
\setres{lotULJ}{DA_post_k39}{kappa_crps_mean}{0.087}
\setres{lotULJ}{DA_post_k39}{kappa_mean}{0.347}
\setres{lotULJ}{DA_post_k39}{logscore_next}{-2.86}
\setres{lotULJ}{DA_post_k39}{logscore_next_mean}{-3.073}
\setres{lotULJ}{DA_post_k39}{misfit_pct}{16.2}
\setres{lotULJ}{DA_post_k39}{misfit_pct_mean}{16.266}
\setres{lotULJ}{DA_post_k39}{n_seen}{39}
\setres{lotULJ}{DA_post_k39}{surprise}{0.00}
\setres{lotULJ}{DA_post_k39}{surprise_mean}{0.548}
\setres{lotULJ}{DA_post_k39}{width90_adj_pct}{26.7}
\setres{lotULJ}{DA_post_k39}{width90_adj_pct_mean}{26.760}
\setres{lotULJ}{DA_post_k39}{width90_pct}{0.0}
\setres{lotULJ}{DA_post_k39}{width90_pct_mean}{0.095}
\setres{lotULJ}{DA_post_k4}{cov50}{0.00}
\setres{lotULJ}{DA_post_k4}{cov50_mean}{0.013}
\setres{lotULJ}{DA_post_k4}{cov90}{0.03}
\setres{lotULJ}{DA_post_k4}{cov90_adj}{0.86}
\setres{lotULJ}{DA_post_k4}{cov90_adj_mean}{0.792}
\setres{lotULJ}{DA_post_k4}{cov90_mean}{0.036}
\setres{lotULJ}{DA_post_k4}{kappa}{0.354}
\setres{lotULJ}{DA_post_k4}{kappa_crps}{-0.033}
\setres{lotULJ}{DA_post_k4}{kappa_crps_mean}{-0.119}
\setres{lotULJ}{DA_post_k4}{kappa_mean}{0.196}
\setres{lotULJ}{DA_post_k4}{logscore_next}{-2.93}
\setres{lotULJ}{DA_post_k4}{logscore_next_mean}{-3.051}
\setres{lotULJ}{DA_post_k4}{misfit_pct}{15.2}
\setres{lotULJ}{DA_post_k4}{misfit_pct_mean}{17.619}
\setres{lotULJ}{DA_post_k4}{n_seen}{4}
\setres{lotULJ}{DA_post_k4}{surprise}{1.75}
\setres{lotULJ}{DA_post_k4}{surprise_mean}{14.521}
\setres{lotULJ}{DA_post_k4}{width90_adj_pct}{25.0}
\setres{lotULJ}{DA_post_k4}{width90_adj_pct_mean}{29.013}
\setres{lotULJ}{DA_post_k4}{width90_pct}{0.5}
\setres{lotULJ}{DA_post_k4}{width90_pct_mean}{0.710}
\setres{lotULJ}{DA_post_k40}{cov50}{0.00}
\setres{lotULJ}{DA_post_k40}{cov50_mean}{0.000}
\setres{lotULJ}{DA_post_k40}{cov90}{0.00}
\setres{lotULJ}{DA_post_k40}{cov90_adj}{1.00}
\setres{lotULJ}{DA_post_k40}{cov90_adj_mean}{0.963}
\setres{lotULJ}{DA_post_k40}{cov90_mean}{0.000}
\setres{lotULJ}{DA_post_k40}{kappa}{0.427}
\setres{lotULJ}{DA_post_k40}{kappa_crps}{0.289}
\setres{lotULJ}{DA_post_k40}{kappa_crps_mean}{0.096}
\setres{lotULJ}{DA_post_k40}{kappa_mean}{0.276}
\setres{lotULJ}{DA_post_k40}{logscore_next}{-2.84}
\setres{lotULJ}{DA_post_k40}{logscore_next_mean}{-3.033}
\setres{lotULJ}{DA_post_k40}{misfit_pct}{16.4}
\setres{lotULJ}{DA_post_k40}{misfit_pct_mean}{16.330}
\setres{lotULJ}{DA_post_k40}{n_seen}{40}
\setres{lotULJ}{DA_post_k40}{surprise}{0.00}
\setres{lotULJ}{DA_post_k40}{surprise_mean}{6.115}
\setres{lotULJ}{DA_post_k40}{width90_adj_pct}{27.0}
\setres{lotULJ}{DA_post_k40}{width90_adj_pct_mean}{26.863}
\setres{lotULJ}{DA_post_k40}{width90_pct}{0.0}
\setres{lotULJ}{DA_post_k40}{width90_pct_mean}{0.048}
\setres{lotULJ}{DA_post_k41}{cov50}{0.00}
\setres{lotULJ}{DA_post_k41}{cov50_mean}{0.000}
\setres{lotULJ}{DA_post_k41}{cov90}{0.00}
\setres{lotULJ}{DA_post_k41}{cov90_adj}{1.00}
\setres{lotULJ}{DA_post_k41}{cov90_adj_mean}{0.983}
\setres{lotULJ}{DA_post_k41}{cov90_mean}{0.000}
\setres{lotULJ}{DA_post_k41}{kappa}{0.432}
\setres{lotULJ}{DA_post_k41}{kappa_crps}{0.205}
\setres{lotULJ}{DA_post_k41}{kappa_crps_mean}{0.104}
\setres{lotULJ}{DA_post_k41}{kappa_mean}{0.320}
\setres{lotULJ}{DA_post_k41}{logscore_next}{-2.87}
\setres{lotULJ}{DA_post_k41}{logscore_next_mean}{-3.030}
\setres{lotULJ}{DA_post_k41}{misfit_pct}{16.3}
\setres{lotULJ}{DA_post_k41}{misfit_pct_mean}{16.282}
\setres{lotULJ}{DA_post_k41}{n_seen}{41}
\setres{lotULJ}{DA_post_k41}{surprise}{0.00}
\setres{lotULJ}{DA_post_k41}{surprise_mean}{0.765}
\setres{lotULJ}{DA_post_k41}{width90_adj_pct}{26.8}
\setres{lotULJ}{DA_post_k41}{width90_adj_pct_mean}{26.784}
\setres{lotULJ}{DA_post_k41}{width90_pct}{0.0}
\setres{lotULJ}{DA_post_k41}{width90_pct_mean}{0.060}
\setres{lotULJ}{DA_post_k42}{cov50}{0.00}
\setres{lotULJ}{DA_post_k42}{cov50_mean}{0.000}
\setres{lotULJ}{DA_post_k42}{cov90}{0.00}
\setres{lotULJ}{DA_post_k42}{cov90_adj}{1.00}
\setres{lotULJ}{DA_post_k42}{cov90_adj_mean}{0.975}
\setres{lotULJ}{DA_post_k42}{cov90_mean}{0.000}
\setres{lotULJ}{DA_post_k42}{kappa}{0.210}
\setres{lotULJ}{DA_post_k42}{kappa_crps}{0.171}
\setres{lotULJ}{DA_post_k42}{kappa_crps_mean}{-0.302}
\setres{lotULJ}{DA_post_k42}{kappa_mean}{0.009}
\setres{lotULJ}{DA_post_k42}{logscore_next}{-2.85}
\setres{lotULJ}{DA_post_k42}{logscore_next_mean}{-3.057}
\setres{lotULJ}{DA_post_k42}{misfit_pct}{16.2}
\setres{lotULJ}{DA_post_k42}{misfit_pct_mean}{16.210}
\setres{lotULJ}{DA_post_k42}{n_seen}{42}
\setres{lotULJ}{DA_post_k42}{surprise}{0.00}
\setres{lotULJ}{DA_post_k42}{surprise_mean}{6.361}
\setres{lotULJ}{DA_post_k42}{width90_adj_pct}{26.7}
\setres{lotULJ}{DA_post_k42}{width90_adj_pct_mean}{26.665}
\setres{lotULJ}{DA_post_k42}{width90_pct}{0.0}
\setres{lotULJ}{DA_post_k42}{width90_pct_mean}{0.032}
\setres{lotULJ}{DA_post_k43}{cov50}{0.00}
\setres{lotULJ}{DA_post_k43}{cov50_mean}{0.000}
\setres{lotULJ}{DA_post_k43}{cov90}{0.00}
\setres{lotULJ}{DA_post_k43}{cov90_adj}{1.00}
\setres{lotULJ}{DA_post_k43}{cov90_adj_mean}{0.950}
\setres{lotULJ}{DA_post_k43}{cov90_mean}{0.000}
\setres{lotULJ}{DA_post_k43}{kappa}{0.250}
\setres{lotULJ}{DA_post_k43}{kappa_crps}{0.219}
\setres{lotULJ}{DA_post_k43}{kappa_crps_mean}{-0.637}
\setres{lotULJ}{DA_post_k43}{kappa_mean}{-2.853}
\setres{lotULJ}{DA_post_k43}{logscore_next}{-2.91}
\setres{lotULJ}{DA_post_k43}{logscore_next_mean}{-2.958}
\setres{lotULJ}{DA_post_k43}{misfit_pct}{16.1}
\setres{lotULJ}{DA_post_k43}{misfit_pct_mean}{16.143}
\setres{lotULJ}{DA_post_k43}{n_seen}{43}
\setres{lotULJ}{DA_post_k43}{surprise}{0.00}
\setres{lotULJ}{DA_post_k43}{surprise_mean}{6.726}
\setres{lotULJ}{DA_post_k43}{width90_adj_pct}{26.6}
\setres{lotULJ}{DA_post_k43}{width90_adj_pct_mean}{26.555}
\setres{lotULJ}{DA_post_k43}{width90_pct}{0.0}
\setres{lotULJ}{DA_post_k43}{width90_pct_mean}{0.037}
\setres{lotULJ}{DA_post_k44}{cov50}{---}
\setres{lotULJ}{DA_post_k44}{cov50_mean}{---}
\setres{lotULJ}{DA_post_k44}{cov90}{---}
\setres{lotULJ}{DA_post_k44}{cov90_adj}{---}
\setres{lotULJ}{DA_post_k44}{cov90_adj_mean}{---}
\setres{lotULJ}{DA_post_k44}{cov90_mean}{---}
\setres{lotULJ}{DA_post_k44}{kappa}{---}
\setres{lotULJ}{DA_post_k44}{kappa_crps}{---}
\setres{lotULJ}{DA_post_k44}{kappa_crps_mean}{---}
\setres{lotULJ}{DA_post_k44}{kappa_mean}{---}
\setres{lotULJ}{DA_post_k44}{logscore_next}{---}
\setres{lotULJ}{DA_post_k44}{logscore_next_mean}{---}
\setres{lotULJ}{DA_post_k44}{misfit_pct}{---}
\setres{lotULJ}{DA_post_k44}{misfit_pct_mean}{---}
\setres{lotULJ}{DA_post_k44}{n_seen}{44}
\setres{lotULJ}{DA_post_k44}{surprise}{---}
\setres{lotULJ}{DA_post_k44}{surprise_mean}{---}
\setres{lotULJ}{DA_post_k44}{width90_adj_pct}{---}
\setres{lotULJ}{DA_post_k44}{width90_adj_pct_mean}{---}
\setres{lotULJ}{DA_post_k44}{width90_pct}{---}
\setres{lotULJ}{DA_post_k44}{width90_pct_mean}{---}
\setres{lotULJ}{DA_post_k5}{cov50}{0.00}
\setres{lotULJ}{DA_post_k5}{cov50_mean}{0.018}
\setres{lotULJ}{DA_post_k5}{cov90}{0.01}
\setres{lotULJ}{DA_post_k5}{cov90_adj}{0.88}
\setres{lotULJ}{DA_post_k5}{cov90_adj_mean}{0.840}
\setres{lotULJ}{DA_post_k5}{cov90_mean}{0.041}
\setres{lotULJ}{DA_post_k5}{kappa}{0.476}
\setres{lotULJ}{DA_post_k5}{kappa_crps}{0.033}
\setres{lotULJ}{DA_post_k5}{kappa_crps_mean}{-0.048}
\setres{lotULJ}{DA_post_k5}{kappa_mean}{0.201}
\setres{lotULJ}{DA_post_k5}{logscore_next}{-2.97}
\setres{lotULJ}{DA_post_k5}{logscore_next_mean}{-3.422}
\setres{lotULJ}{DA_post_k5}{misfit_pct}{16.5}
\setres{lotULJ}{DA_post_k5}{misfit_pct_mean}{18.419}
\setres{lotULJ}{DA_post_k5}{n_seen}{5}
\setres{lotULJ}{DA_post_k5}{surprise}{1.48}
\setres{lotULJ}{DA_post_k5}{surprise_mean}{26.727}
\setres{lotULJ}{DA_post_k5}{width90_adj_pct}{27.2}
\setres{lotULJ}{DA_post_k5}{width90_adj_pct_mean}{30.329}
\setres{lotULJ}{DA_post_k5}{width90_pct}{0.4}
\setres{lotULJ}{DA_post_k5}{width90_pct_mean}{0.838}
\setres{lotULJ}{DA_post_k6}{cov50}{0.00}
\setres{lotULJ}{DA_post_k6}{cov50_mean}{0.011}
\setres{lotULJ}{DA_post_k6}{cov90}{0.03}
\setres{lotULJ}{DA_post_k6}{cov90_adj}{0.92}
\setres{lotULJ}{DA_post_k6}{cov90_adj_mean}{0.884}
\setres{lotULJ}{DA_post_k6}{cov90_mean}{0.028}
\setres{lotULJ}{DA_post_k6}{kappa}{0.444}
\setres{lotULJ}{DA_post_k6}{kappa_crps}{0.164}
\setres{lotULJ}{DA_post_k6}{kappa_crps_mean}{-0.051}
\setres{lotULJ}{DA_post_k6}{kappa_mean}{0.119}
\setres{lotULJ}{DA_post_k6}{logscore_next}{-2.90}
\setres{lotULJ}{DA_post_k6}{logscore_next_mean}{-3.159}
\setres{lotULJ}{DA_post_k6}{misfit_pct}{17.0}
\setres{lotULJ}{DA_post_k6}{misfit_pct_mean}{17.877}
\setres{lotULJ}{DA_post_k6}{n_seen}{6}
\setres{lotULJ}{DA_post_k6}{surprise}{0.99}
\setres{lotULJ}{DA_post_k6}{surprise_mean}{10.834}
\setres{lotULJ}{DA_post_k6}{width90_adj_pct}{28.0}
\setres{lotULJ}{DA_post_k6}{width90_adj_pct_mean}{29.427}
\setres{lotULJ}{DA_post_k6}{width90_pct}{0.4}
\setres{lotULJ}{DA_post_k6}{width90_pct_mean}{0.720}
\setres{lotULJ}{DA_post_k7}{cov50}{0.00}
\setres{lotULJ}{DA_post_k7}{cov50_mean}{0.011}
\setres{lotULJ}{DA_post_k7}{cov90}{0.00}
\setres{lotULJ}{DA_post_k7}{cov90_adj}{0.92}
\setres{lotULJ}{DA_post_k7}{cov90_adj_mean}{0.893}
\setres{lotULJ}{DA_post_k7}{cov90_mean}{0.026}
\setres{lotULJ}{DA_post_k7}{kappa}{0.627}
\setres{lotULJ}{DA_post_k7}{kappa_crps}{0.313}
\setres{lotULJ}{DA_post_k7}{kappa_crps_mean}{0.053}
\setres{lotULJ}{DA_post_k7}{kappa_mean}{0.239}
\setres{lotULJ}{DA_post_k7}{logscore_next}{-2.95}
\setres{lotULJ}{DA_post_k7}{logscore_next_mean}{-3.352}
\setres{lotULJ}{DA_post_k7}{misfit_pct}{16.9}
\setres{lotULJ}{DA_post_k7}{misfit_pct_mean}{17.696}
\setres{lotULJ}{DA_post_k7}{n_seen}{7}
\setres{lotULJ}{DA_post_k7}{surprise}{0.32}
\setres{lotULJ}{DA_post_k7}{surprise_mean}{52.050}
\setres{lotULJ}{DA_post_k7}{width90_adj_pct}{27.8}
\setres{lotULJ}{DA_post_k7}{width90_adj_pct_mean}{29.126}
\setres{lotULJ}{DA_post_k7}{width90_pct}{0.5}
\setres{lotULJ}{DA_post_k7}{width90_pct_mean}{0.651}
\setres{lotULJ}{DA_post_k8}{cov50}{0.00}
\setres{lotULJ}{DA_post_k8}{cov50_mean}{0.004}
\setres{lotULJ}{DA_post_k8}{cov90}{0.00}
\setres{lotULJ}{DA_post_k8}{cov90_adj}{0.92}
\setres{lotULJ}{DA_post_k8}{cov90_adj_mean}{0.856}
\setres{lotULJ}{DA_post_k8}{cov90_mean}{0.019}
\setres{lotULJ}{DA_post_k8}{kappa}{0.610}
\setres{lotULJ}{DA_post_k8}{kappa_crps}{0.288}
\setres{lotULJ}{DA_post_k8}{kappa_crps_mean}{-0.016}
\setres{lotULJ}{DA_post_k8}{kappa_mean}{0.118}
\setres{lotULJ}{DA_post_k8}{logscore_next}{-2.91}
\setres{lotULJ}{DA_post_k8}{logscore_next_mean}{-2.995}
\setres{lotULJ}{DA_post_k8}{misfit_pct}{16.6}
\setres{lotULJ}{DA_post_k8}{misfit_pct_mean}{17.152}
\setres{lotULJ}{DA_post_k8}{n_seen}{8}
\setres{lotULJ}{DA_post_k8}{surprise}{0.15}
\setres{lotULJ}{DA_post_k8}{surprise_mean}{5.435}
\setres{lotULJ}{DA_post_k8}{width90_adj_pct}{27.3}
\setres{lotULJ}{DA_post_k8}{width90_adj_pct_mean}{28.229}
\setres{lotULJ}{DA_post_k8}{width90_pct}{0.2}
\setres{lotULJ}{DA_post_k8}{width90_pct_mean}{0.542}
\setres{lotULJ}{DA_post_k9}{cov50}{0.00}
\setres{lotULJ}{DA_post_k9}{cov50_mean}{0.007}
\setres{lotULJ}{DA_post_k9}{cov90}{0.00}
\setres{lotULJ}{DA_post_k9}{cov90_adj}{0.87}
\setres{lotULJ}{DA_post_k9}{cov90_adj_mean}{0.839}
\setres{lotULJ}{DA_post_k9}{cov90_mean}{0.010}
\setres{lotULJ}{DA_post_k9}{kappa}{0.593}
\setres{lotULJ}{DA_post_k9}{kappa_crps}{0.285}
\setres{lotULJ}{DA_post_k9}{kappa_crps_mean}{-0.012}
\setres{lotULJ}{DA_post_k9}{kappa_mean}{0.144}
\setres{lotULJ}{DA_post_k9}{logscore_next}{-2.84}
\setres{lotULJ}{DA_post_k9}{logscore_next_mean}{-3.091}
\setres{lotULJ}{DA_post_k9}{misfit_pct}{16.4}
\setres{lotULJ}{DA_post_k9}{misfit_pct_mean}{16.799}
\setres{lotULJ}{DA_post_k9}{n_seen}{9}
\setres{lotULJ}{DA_post_k9}{surprise}{0.16}
\setres{lotULJ}{DA_post_k9}{surprise_mean}{52.086}
\setres{lotULJ}{DA_post_k9}{width90_adj_pct}{27.1}
\setres{lotULJ}{DA_post_k9}{width90_adj_pct_mean}{27.644}
\setres{lotULJ}{DA_post_k9}{width90_pct}{0.0}
\setres{lotULJ}{DA_post_k9}{width90_pct_mean}{0.316}
\setres{lotULJ}{EU_post_k0}{cov50}{0.16}
\setres{lotULJ}{EU_post_k0}{cov50_mean}{0.159}
\setres{lotULJ}{EU_post_k0}{cov90}{0.30}
\setres{lotULJ}{EU_post_k0}{cov90_adj}{---}
\setres{lotULJ}{EU_post_k0}{cov90_adj_mean}{---}
\setres{lotULJ}{EU_post_k0}{cov90_mean}{0.295}
\setres{lotULJ}{EU_post_k0}{kappa}{0.408}
\setres{lotULJ}{EU_post_k0}{kappa_crps}{0.324}
\setres{lotULJ}{EU_post_k0}{kappa_crps_mean}{0.325}
\setres{lotULJ}{EU_post_k0}{kappa_mean}{0.408}
\setres{lotULJ}{EU_post_k0}{logscore_next}{-2.84}
\setres{lotULJ}{EU_post_k0}{logscore_next_mean}{-2.951}
\setres{lotULJ}{EU_post_k0}{misfit_pct}{---}
\setres{lotULJ}{EU_post_k0}{misfit_pct_mean}{---}
\setres{lotULJ}{EU_post_k0}{n_seen}{0}
\setres{lotULJ}{EU_post_k0}{surprise}{2.81}
\setres{lotULJ}{EU_post_k0}{surprise_mean}{2.419}
\setres{lotULJ}{EU_post_k0}{width90_adj_pct}{---}
\setres{lotULJ}{EU_post_k0}{width90_adj_pct_mean}{---}
\setres{lotULJ}{EU_post_k0}{width90_pct}{4.4}
\setres{lotULJ}{EU_post_k0}{width90_pct_mean}{4.407}
\setres{lotULJ}{EU_post_k1}{cov50}{0.00}
\setres{lotULJ}{EU_post_k1}{cov50_mean}{0.026}
\setres{lotULJ}{EU_post_k1}{cov90}{0.00}
\setres{lotULJ}{EU_post_k1}{cov90_adj}{---}
\setres{lotULJ}{EU_post_k1}{cov90_adj_mean}{---}
\setres{lotULJ}{EU_post_k1}{cov90_mean}{0.066}
\setres{lotULJ}{EU_post_k1}{kappa}{0.723}
\setres{lotULJ}{EU_post_k1}{kappa_crps}{0.390}
\setres{lotULJ}{EU_post_k1}{kappa_crps_mean}{0.265}
\setres{lotULJ}{EU_post_k1}{kappa_mean}{0.535}
\setres{lotULJ}{EU_post_k1}{logscore_next}{-2.86}
\setres{lotULJ}{EU_post_k1}{logscore_next_mean}{-3.177}
\setres{lotULJ}{EU_post_k1}{misfit_pct}{---}
\setres{lotULJ}{EU_post_k1}{misfit_pct_mean}{---}
\setres{lotULJ}{EU_post_k1}{n_seen}{1}
\setres{lotULJ}{EU_post_k1}{surprise}{0.42}
\setres{lotULJ}{EU_post_k1}{surprise_mean}{28.194}
\setres{lotULJ}{EU_post_k1}{width90_adj_pct}{---}
\setres{lotULJ}{EU_post_k1}{width90_adj_pct_mean}{---}
\setres{lotULJ}{EU_post_k1}{width90_pct}{0.3}
\setres{lotULJ}{EU_post_k1}{width90_pct_mean}{1.086}
\setres{lotULJ}{EU_post_k10}{cov50}{0.00}
\setres{lotULJ}{EU_post_k10}{cov50_mean}{0.003}
\setres{lotULJ}{EU_post_k10}{cov90}{0.00}
\setres{lotULJ}{EU_post_k10}{cov90_adj}{0.94}
\setres{lotULJ}{EU_post_k10}{cov90_adj_mean}{0.913}
\setres{lotULJ}{EU_post_k10}{cov90_mean}{0.007}
\setres{lotULJ}{EU_post_k10}{kappa}{0.692}
\setres{lotULJ}{EU_post_k10}{kappa_crps}{0.311}
\setres{lotULJ}{EU_post_k10}{kappa_crps_mean}{0.306}
\setres{lotULJ}{EU_post_k10}{kappa_mean}{0.606}
\setres{lotULJ}{EU_post_k10}{logscore_next}{-2.91}
\setres{lotULJ}{EU_post_k10}{logscore_next_mean}{-2.963}
\setres{lotULJ}{EU_post_k10}{misfit_pct}{16.3}
\setres{lotULJ}{EU_post_k10}{misfit_pct_mean}{16.160}
\setres{lotULJ}{EU_post_k10}{n_seen}{10}
\setres{lotULJ}{EU_post_k10}{surprise}{0.00}
\setres{lotULJ}{EU_post_k10}{surprise_mean}{0.837}
\setres{lotULJ}{EU_post_k10}{width90_adj_pct}{26.8}
\setres{lotULJ}{EU_post_k10}{width90_adj_pct_mean}{26.583}
\setres{lotULJ}{EU_post_k10}{width90_pct}{0.1}
\setres{lotULJ}{EU_post_k10}{width90_pct_mean}{0.121}
\setres{lotULJ}{EU_post_k11}{cov50}{0.00}
\setres{lotULJ}{EU_post_k11}{cov50_mean}{0.000}
\setres{lotULJ}{EU_post_k11}{cov90}{0.00}
\setres{lotULJ}{EU_post_k11}{cov90_adj}{0.97}
\setres{lotULJ}{EU_post_k11}{cov90_adj_mean}{0.935}
\setres{lotULJ}{EU_post_k11}{cov90_mean}{0.008}
\setres{lotULJ}{EU_post_k11}{kappa}{0.698}
\setres{lotULJ}{EU_post_k11}{kappa_crps}{0.320}
\setres{lotULJ}{EU_post_k11}{kappa_crps_mean}{0.322}
\setres{lotULJ}{EU_post_k11}{kappa_mean}{0.615}
\setres{lotULJ}{EU_post_k11}{logscore_next}{-2.88}
\setres{lotULJ}{EU_post_k11}{logscore_next_mean}{-2.991}
\setres{lotULJ}{EU_post_k11}{misfit_pct}{16.7}
\setres{lotULJ}{EU_post_k11}{misfit_pct_mean}{16.505}
\setres{lotULJ}{EU_post_k11}{n_seen}{11}
\setres{lotULJ}{EU_post_k11}{surprise}{0.00}
\setres{lotULJ}{EU_post_k11}{surprise_mean}{7.022}
\setres{lotULJ}{EU_post_k11}{width90_adj_pct}{27.4}
\setres{lotULJ}{EU_post_k11}{width90_adj_pct_mean}{27.151}
\setres{lotULJ}{EU_post_k11}{width90_pct}{0.1}
\setres{lotULJ}{EU_post_k11}{width90_pct_mean}{0.116}
\setres{lotULJ}{EU_post_k12}{cov50}{0.00}
\setres{lotULJ}{EU_post_k12}{cov50_mean}{0.000}
\setres{lotULJ}{EU_post_k12}{cov90}{0.00}
\setres{lotULJ}{EU_post_k12}{cov90_adj}{0.97}
\setres{lotULJ}{EU_post_k12}{cov90_adj_mean}{0.928}
\setres{lotULJ}{EU_post_k12}{cov90_mean}{0.005}
\setres{lotULJ}{EU_post_k12}{kappa}{0.671}
\setres{lotULJ}{EU_post_k12}{kappa_crps}{0.304}
\setres{lotULJ}{EU_post_k12}{kappa_crps_mean}{0.285}
\setres{lotULJ}{EU_post_k12}{kappa_mean}{0.579}
\setres{lotULJ}{EU_post_k12}{logscore_next}{-2.88}
\setres{lotULJ}{EU_post_k12}{logscore_next_mean}{-3.278}
\setres{lotULJ}{EU_post_k12}{misfit_pct}{16.5}
\setres{lotULJ}{EU_post_k12}{misfit_pct_mean}{16.523}
\setres{lotULJ}{EU_post_k12}{n_seen}{12}
\setres{lotULJ}{EU_post_k12}{surprise}{0.00}
\setres{lotULJ}{EU_post_k12}{surprise_mean}{2.198}
\setres{lotULJ}{EU_post_k12}{width90_adj_pct}{27.1}
\setres{lotULJ}{EU_post_k12}{width90_adj_pct_mean}{27.181}
\setres{lotULJ}{EU_post_k12}{width90_pct}{0.1}
\setres{lotULJ}{EU_post_k12}{width90_pct_mean}{0.117}
\setres{lotULJ}{EU_post_k13}{cov50}{0.00}
\setres{lotULJ}{EU_post_k13}{cov50_mean}{0.005}
\setres{lotULJ}{EU_post_k13}{cov90}{0.00}
\setres{lotULJ}{EU_post_k13}{cov90_adj}{0.95}
\setres{lotULJ}{EU_post_k13}{cov90_adj_mean}{0.923}
\setres{lotULJ}{EU_post_k13}{cov90_mean}{0.008}
\setres{lotULJ}{EU_post_k13}{kappa}{0.634}
\setres{lotULJ}{EU_post_k13}{kappa_crps}{0.286}
\setres{lotULJ}{EU_post_k13}{kappa_crps_mean}{0.274}
\setres{lotULJ}{EU_post_k13}{kappa_mean}{0.560}
\setres{lotULJ}{EU_post_k13}{logscore_next}{-2.88}
\setres{lotULJ}{EU_post_k13}{logscore_next_mean}{-3.078}
\setres{lotULJ}{EU_post_k13}{misfit_pct}{16.6}
\setres{lotULJ}{EU_post_k13}{misfit_pct_mean}{16.428}
\setres{lotULJ}{EU_post_k13}{n_seen}{13}
\setres{lotULJ}{EU_post_k13}{surprise}{0.00}
\setres{lotULJ}{EU_post_k13}{surprise_mean}{3.395}
\setres{lotULJ}{EU_post_k13}{width90_adj_pct}{27.2}
\setres{lotULJ}{EU_post_k13}{width90_adj_pct_mean}{27.025}
\setres{lotULJ}{EU_post_k13}{width90_pct}{0.1}
\setres{lotULJ}{EU_post_k13}{width90_pct_mean}{0.118}
\setres{lotULJ}{EU_post_k14}{cov50}{0.00}
\setres{lotULJ}{EU_post_k14}{cov50_mean}{0.000}
\setres{lotULJ}{EU_post_k14}{cov90}{0.00}
\setres{lotULJ}{EU_post_k14}{cov90_adj}{0.97}
\setres{lotULJ}{EU_post_k14}{cov90_adj_mean}{0.932}
\setres{lotULJ}{EU_post_k14}{cov90_mean}{0.007}
\setres{lotULJ}{EU_post_k14}{kappa}{0.652}
\setres{lotULJ}{EU_post_k14}{kappa_crps}{0.333}
\setres{lotULJ}{EU_post_k14}{kappa_crps_mean}{0.300}
\setres{lotULJ}{EU_post_k14}{kappa_mean}{0.578}
\setres{lotULJ}{EU_post_k14}{logscore_next}{-2.95}
\setres{lotULJ}{EU_post_k14}{logscore_next_mean}{-3.072}
\setres{lotULJ}{EU_post_k14}{misfit_pct}{16.8}
\setres{lotULJ}{EU_post_k14}{misfit_pct_mean}{16.441}
\setres{lotULJ}{EU_post_k14}{n_seen}{14}
\setres{lotULJ}{EU_post_k14}{surprise}{0.00}
\setres{lotULJ}{EU_post_k14}{surprise_mean}{0.059}
\setres{lotULJ}{EU_post_k14}{width90_adj_pct}{27.6}
\setres{lotULJ}{EU_post_k14}{width90_adj_pct_mean}{27.046}
\setres{lotULJ}{EU_post_k14}{width90_pct}{0.1}
\setres{lotULJ}{EU_post_k14}{width90_pct_mean}{0.111}
\setres{lotULJ}{EU_post_k15}{cov50}{0.00}
\setres{lotULJ}{EU_post_k15}{cov50_mean}{0.000}
\setres{lotULJ}{EU_post_k15}{cov90}{0.00}
\setres{lotULJ}{EU_post_k15}{cov90_adj}{0.97}
\setres{lotULJ}{EU_post_k15}{cov90_adj_mean}{0.945}
\setres{lotULJ}{EU_post_k15}{cov90_mean}{0.005}
\setres{lotULJ}{EU_post_k15}{kappa}{0.653}
\setres{lotULJ}{EU_post_k15}{kappa_crps}{0.308}
\setres{lotULJ}{EU_post_k15}{kappa_crps_mean}{0.293}
\setres{lotULJ}{EU_post_k15}{kappa_mean}{0.582}
\setres{lotULJ}{EU_post_k15}{logscore_next}{-2.86}
\setres{lotULJ}{EU_post_k15}{logscore_next_mean}{-2.953}
\setres{lotULJ}{EU_post_k15}{misfit_pct}{16.8}
\setres{lotULJ}{EU_post_k15}{misfit_pct_mean}{16.581}
\setres{lotULJ}{EU_post_k15}{n_seen}{15}
\setres{lotULJ}{EU_post_k15}{surprise}{0.00}
\setres{lotULJ}{EU_post_k15}{surprise_mean}{0.094}
\setres{lotULJ}{EU_post_k15}{width90_adj_pct}{27.7}
\setres{lotULJ}{EU_post_k15}{width90_adj_pct_mean}{27.276}
\setres{lotULJ}{EU_post_k15}{width90_pct}{0.1}
\setres{lotULJ}{EU_post_k15}{width90_pct_mean}{0.103}
\setres{lotULJ}{EU_post_k16}{cov50}{0.00}
\setres{lotULJ}{EU_post_k16}{cov50_mean}{0.000}
\setres{lotULJ}{EU_post_k16}{cov90}{0.00}
\setres{lotULJ}{EU_post_k16}{cov90_adj}{0.98}
\setres{lotULJ}{EU_post_k16}{cov90_adj_mean}{0.946}
\setres{lotULJ}{EU_post_k16}{cov90_mean}{0.005}
\setres{lotULJ}{EU_post_k16}{kappa}{0.655}
\setres{lotULJ}{EU_post_k16}{kappa_crps}{0.331}
\setres{lotULJ}{EU_post_k16}{kappa_crps_mean}{0.300}
\setres{lotULJ}{EU_post_k16}{kappa_mean}{0.585}
\setres{lotULJ}{EU_post_k16}{logscore_next}{-2.83}
\setres{lotULJ}{EU_post_k16}{logscore_next_mean}{-3.099}
\setres{lotULJ}{EU_post_k16}{misfit_pct}{16.8}
\setres{lotULJ}{EU_post_k16}{misfit_pct_mean}{16.625}
\setres{lotULJ}{EU_post_k16}{n_seen}{16}
\setres{lotULJ}{EU_post_k16}{surprise}{0.00}
\setres{lotULJ}{EU_post_k16}{surprise_mean}{10.725}
\setres{lotULJ}{EU_post_k16}{width90_adj_pct}{27.6}
\setres{lotULJ}{EU_post_k16}{width90_adj_pct_mean}{27.349}
\setres{lotULJ}{EU_post_k16}{width90_pct}{0.1}
\setres{lotULJ}{EU_post_k16}{width90_pct_mean}{0.112}
\setres{lotULJ}{EU_post_k17}{cov50}{0.00}
\setres{lotULJ}{EU_post_k17}{cov50_mean}{0.000}
\setres{lotULJ}{EU_post_k17}{cov90}{0.00}
\setres{lotULJ}{EU_post_k17}{cov90_adj}{0.96}
\setres{lotULJ}{EU_post_k17}{cov90_adj_mean}{0.943}
\setres{lotULJ}{EU_post_k17}{cov90_mean}{0.006}
\setres{lotULJ}{EU_post_k17}{kappa}{0.663}
\setres{lotULJ}{EU_post_k17}{kappa_crps}{0.340}
\setres{lotULJ}{EU_post_k17}{kappa_crps_mean}{0.305}
\setres{lotULJ}{EU_post_k17}{kappa_mean}{0.591}
\setres{lotULJ}{EU_post_k17}{logscore_next}{-2.90}
\setres{lotULJ}{EU_post_k17}{logscore_next_mean}{-3.125}
\setres{lotULJ}{EU_post_k17}{misfit_pct}{16.9}
\setres{lotULJ}{EU_post_k17}{misfit_pct_mean}{16.672}
\setres{lotULJ}{EU_post_k17}{n_seen}{17}
\setres{lotULJ}{EU_post_k17}{surprise}{0.00}
\setres{lotULJ}{EU_post_k17}{surprise_mean}{2.327}
\setres{lotULJ}{EU_post_k17}{width90_adj_pct}{27.8}
\setres{lotULJ}{EU_post_k17}{width90_adj_pct_mean}{27.426}
\setres{lotULJ}{EU_post_k17}{width90_pct}{0.1}
\setres{lotULJ}{EU_post_k17}{width90_pct_mean}{0.123}
\setres{lotULJ}{EU_post_k18}{cov50}{0.00}
\setres{lotULJ}{EU_post_k18}{cov50_mean}{0.000}
\setres{lotULJ}{EU_post_k18}{cov90}{0.00}
\setres{lotULJ}{EU_post_k18}{cov90_adj}{0.98}
\setres{lotULJ}{EU_post_k18}{cov90_adj_mean}{0.946}
\setres{lotULJ}{EU_post_k18}{cov90_mean}{0.008}
\setres{lotULJ}{EU_post_k18}{kappa}{0.653}
\setres{lotULJ}{EU_post_k18}{kappa_crps}{0.315}
\setres{lotULJ}{EU_post_k18}{kappa_crps_mean}{0.305}
\setres{lotULJ}{EU_post_k18}{kappa_mean}{0.585}
\setres{lotULJ}{EU_post_k18}{logscore_next}{-2.86}
\setres{lotULJ}{EU_post_k18}{logscore_next_mean}{-2.971}
\setres{lotULJ}{EU_post_k18}{misfit_pct}{16.7}
\setres{lotULJ}{EU_post_k18}{misfit_pct_mean}{16.649}
\setres{lotULJ}{EU_post_k18}{n_seen}{18}
\setres{lotULJ}{EU_post_k18}{surprise}{0.00}
\setres{lotULJ}{EU_post_k18}{surprise_mean}{0.407}
\setres{lotULJ}{EU_post_k18}{width90_adj_pct}{27.5}
\setres{lotULJ}{EU_post_k18}{width90_adj_pct_mean}{27.389}
\setres{lotULJ}{EU_post_k18}{width90_pct}{0.1}
\setres{lotULJ}{EU_post_k18}{width90_pct_mean}{0.132}
\setres{lotULJ}{EU_post_k19}{cov50}{0.00}
\setres{lotULJ}{EU_post_k19}{cov50_mean}{0.004}
\setres{lotULJ}{EU_post_k19}{cov90}{0.00}
\setres{lotULJ}{EU_post_k19}{cov90_adj}{0.98}
\setres{lotULJ}{EU_post_k19}{cov90_adj_mean}{0.954}
\setres{lotULJ}{EU_post_k19}{cov90_mean}{0.010}
\setres{lotULJ}{EU_post_k19}{kappa}{0.657}
\setres{lotULJ}{EU_post_k19}{kappa_crps}{0.328}
\setres{lotULJ}{EU_post_k19}{kappa_crps_mean}{0.313}
\setres{lotULJ}{EU_post_k19}{kappa_mean}{0.592}
\setres{lotULJ}{EU_post_k19}{logscore_next}{-2.88}
\setres{lotULJ}{EU_post_k19}{logscore_next_mean}{-3.243}
\setres{lotULJ}{EU_post_k19}{misfit_pct}{16.7}
\setres{lotULJ}{EU_post_k19}{misfit_pct_mean}{16.731}
\setres{lotULJ}{EU_post_k19}{n_seen}{19}
\setres{lotULJ}{EU_post_k19}{surprise}{0.02}
\setres{lotULJ}{EU_post_k19}{surprise_mean}{16.316}
\setres{lotULJ}{EU_post_k19}{width90_adj_pct}{27.4}
\setres{lotULJ}{EU_post_k19}{width90_adj_pct_mean}{27.523}
\setres{lotULJ}{EU_post_k19}{width90_pct}{0.1}
\setres{lotULJ}{EU_post_k19}{width90_pct_mean}{0.135}
\setres{lotULJ}{EU_post_k2}{cov50}{0.00}
\setres{lotULJ}{EU_post_k2}{cov50_mean}{0.006}
\setres{lotULJ}{EU_post_k2}{cov90}{0.00}
\setres{lotULJ}{EU_post_k2}{cov90_adj}{0.82}
\setres{lotULJ}{EU_post_k2}{cov90_adj_mean}{0.683}
\setres{lotULJ}{EU_post_k2}{cov90_mean}{0.027}
\setres{lotULJ}{EU_post_k2}{kappa}{0.728}
\setres{lotULJ}{EU_post_k2}{kappa_crps}{0.392}
\setres{lotULJ}{EU_post_k2}{kappa_crps_mean}{0.269}
\setres{lotULJ}{EU_post_k2}{kappa_mean}{0.557}
\setres{lotULJ}{EU_post_k2}{logscore_next}{-2.84}
\setres{lotULJ}{EU_post_k2}{logscore_next_mean}{-2.982}
\setres{lotULJ}{EU_post_k2}{misfit_pct}{13.0}
\setres{lotULJ}{EU_post_k2}{misfit_pct_mean}{13.575}
\setres{lotULJ}{EU_post_k2}{n_seen}{2}
\setres{lotULJ}{EU_post_k2}{surprise}{0.03}
\setres{lotULJ}{EU_post_k2}{surprise_mean}{2.521}
\setres{lotULJ}{EU_post_k2}{width90_adj_pct}{21.4}
\setres{lotULJ}{EU_post_k2}{width90_adj_pct_mean}{22.344}
\setres{lotULJ}{EU_post_k2}{width90_pct}{0.3}
\setres{lotULJ}{EU_post_k2}{width90_pct_mean}{0.476}
\setres{lotULJ}{EU_post_k20}{cov50}{0.00}
\setres{lotULJ}{EU_post_k20}{cov50_mean}{0.000}
\setres{lotULJ}{EU_post_k20}{cov90}{0.00}
\setres{lotULJ}{EU_post_k20}{cov90_adj}{0.96}
\setres{lotULJ}{EU_post_k20}{cov90_adj_mean}{0.948}
\setres{lotULJ}{EU_post_k20}{cov90_mean}{0.004}
\setres{lotULJ}{EU_post_k20}{kappa}{0.648}
\setres{lotULJ}{EU_post_k20}{kappa_crps}{0.295}
\setres{lotULJ}{EU_post_k20}{kappa_crps_mean}{0.297}
\setres{lotULJ}{EU_post_k20}{kappa_mean}{0.572}
\setres{lotULJ}{EU_post_k20}{logscore_next}{-2.83}
\setres{lotULJ}{EU_post_k20}{logscore_next_mean}{-3.192}
\setres{lotULJ}{EU_post_k20}{misfit_pct}{16.6}
\setres{lotULJ}{EU_post_k20}{misfit_pct_mean}{16.644}
\setres{lotULJ}{EU_post_k20}{n_seen}{20}
\setres{lotULJ}{EU_post_k20}{surprise}{0.00}
\setres{lotULJ}{EU_post_k20}{surprise_mean}{3.622}
\setres{lotULJ}{EU_post_k20}{width90_adj_pct}{27.2}
\setres{lotULJ}{EU_post_k20}{width90_adj_pct_mean}{27.381}
\setres{lotULJ}{EU_post_k20}{width90_pct}{0.1}
\setres{lotULJ}{EU_post_k20}{width90_pct_mean}{0.123}
\setres{lotULJ}{EU_post_k21}{cov50}{0.00}
\setres{lotULJ}{EU_post_k21}{cov50_mean}{0.002}
\setres{lotULJ}{EU_post_k21}{cov90}{0.00}
\setres{lotULJ}{EU_post_k21}{cov90_adj}{0.96}
\setres{lotULJ}{EU_post_k21}{cov90_adj_mean}{0.943}
\setres{lotULJ}{EU_post_k21}{cov90_mean}{0.007}
\setres{lotULJ}{EU_post_k21}{kappa}{0.531}
\setres{lotULJ}{EU_post_k21}{kappa_crps}{0.262}
\setres{lotULJ}{EU_post_k21}{kappa_crps_mean}{0.274}
\setres{lotULJ}{EU_post_k21}{kappa_mean}{0.549}
\setres{lotULJ}{EU_post_k21}{logscore_next}{-2.88}
\setres{lotULJ}{EU_post_k21}{logscore_next_mean}{-3.082}
\setres{lotULJ}{EU_post_k21}{misfit_pct}{16.4}
\setres{lotULJ}{EU_post_k21}{misfit_pct_mean}{16.515}
\setres{lotULJ}{EU_post_k21}{n_seen}{21}
\setres{lotULJ}{EU_post_k21}{surprise}{0.00}
\setres{lotULJ}{EU_post_k21}{surprise_mean}{13.542}
\setres{lotULJ}{EU_post_k21}{width90_adj_pct}{26.9}
\setres{lotULJ}{EU_post_k21}{width90_adj_pct_mean}{27.167}
\setres{lotULJ}{EU_post_k21}{width90_pct}{0.1}
\setres{lotULJ}{EU_post_k21}{width90_pct_mean}{0.120}
\setres{lotULJ}{EU_post_k22}{cov50}{0.00}
\setres{lotULJ}{EU_post_k22}{cov50_mean}{0.002}
\setres{lotULJ}{EU_post_k22}{cov90}{0.00}
\setres{lotULJ}{EU_post_k22}{cov90_adj}{0.95}
\setres{lotULJ}{EU_post_k22}{cov90_adj_mean}{0.948}
\setres{lotULJ}{EU_post_k22}{cov90_mean}{0.002}
\setres{lotULJ}{EU_post_k22}{kappa}{0.549}
\setres{lotULJ}{EU_post_k22}{kappa_crps}{0.299}
\setres{lotULJ}{EU_post_k22}{kappa_crps_mean}{0.292}
\setres{lotULJ}{EU_post_k22}{kappa_mean}{0.564}
\setres{lotULJ}{EU_post_k22}{logscore_next}{-2.89}
\setres{lotULJ}{EU_post_k22}{logscore_next_mean}{-3.055}
\setres{lotULJ}{EU_post_k22}{misfit_pct}{16.5}
\setres{lotULJ}{EU_post_k22}{misfit_pct_mean}{16.652}
\setres{lotULJ}{EU_post_k22}{n_seen}{22}
\setres{lotULJ}{EU_post_k22}{surprise}{0.01}
\setres{lotULJ}{EU_post_k22}{surprise_mean}{0.348}
\setres{lotULJ}{EU_post_k22}{width90_adj_pct}{27.2}
\setres{lotULJ}{EU_post_k22}{width90_adj_pct_mean}{27.394}
\setres{lotULJ}{EU_post_k22}{width90_pct}{0.1}
\setres{lotULJ}{EU_post_k22}{width90_pct_mean}{0.112}
\setres{lotULJ}{EU_post_k23}{cov50}{0.00}
\setres{lotULJ}{EU_post_k23}{cov50_mean}{0.002}
\setres{lotULJ}{EU_post_k23}{cov90}{0.00}
\setres{lotULJ}{EU_post_k23}{cov90_adj}{0.95}
\setres{lotULJ}{EU_post_k23}{cov90_adj_mean}{0.952}
\setres{lotULJ}{EU_post_k23}{cov90_mean}{0.002}
\setres{lotULJ}{EU_post_k23}{kappa}{0.541}
\setres{lotULJ}{EU_post_k23}{kappa_crps}{0.305}
\setres{lotULJ}{EU_post_k23}{kappa_crps_mean}{0.295}
\setres{lotULJ}{EU_post_k23}{kappa_mean}{0.568}
\setres{lotULJ}{EU_post_k23}{logscore_next}{-2.84}
\setres{lotULJ}{EU_post_k23}{logscore_next_mean}{-2.929}
\setres{lotULJ}{EU_post_k23}{misfit_pct}{16.8}
\setres{lotULJ}{EU_post_k23}{misfit_pct_mean}{16.740}
\setres{lotULJ}{EU_post_k23}{n_seen}{23}
\setres{lotULJ}{EU_post_k23}{surprise}{0.00}
\setres{lotULJ}{EU_post_k23}{surprise_mean}{0.142}
\setres{lotULJ}{EU_post_k23}{width90_adj_pct}{27.6}
\setres{lotULJ}{EU_post_k23}{width90_adj_pct_mean}{27.538}
\setres{lotULJ}{EU_post_k23}{width90_pct}{0.1}
\setres{lotULJ}{EU_post_k23}{width90_pct_mean}{0.109}
\setres{lotULJ}{EU_post_k24}{cov50}{0.00}
\setres{lotULJ}{EU_post_k24}{cov50_mean}{0.000}
\setres{lotULJ}{EU_post_k24}{cov90}{0.00}
\setres{lotULJ}{EU_post_k24}{cov90_adj}{0.95}
\setres{lotULJ}{EU_post_k24}{cov90_adj_mean}{0.955}
\setres{lotULJ}{EU_post_k24}{cov90_mean}{0.003}
\setres{lotULJ}{EU_post_k24}{kappa}{0.511}
\setres{lotULJ}{EU_post_k24}{kappa_crps}{0.295}
\setres{lotULJ}{EU_post_k24}{kappa_crps_mean}{0.287}
\setres{lotULJ}{EU_post_k24}{kappa_mean}{0.543}
\setres{lotULJ}{EU_post_k24}{logscore_next}{-2.93}
\setres{lotULJ}{EU_post_k24}{logscore_next_mean}{-2.984}
\setres{lotULJ}{EU_post_k24}{misfit_pct}{16.6}
\setres{lotULJ}{EU_post_k24}{misfit_pct_mean}{16.681}
\setres{lotULJ}{EU_post_k24}{n_seen}{24}
\setres{lotULJ}{EU_post_k24}{surprise}{0.00}
\setres{lotULJ}{EU_post_k24}{surprise_mean}{0.206}
\setres{lotULJ}{EU_post_k24}{width90_adj_pct}{27.3}
\setres{lotULJ}{EU_post_k24}{width90_adj_pct_mean}{27.440}
\setres{lotULJ}{EU_post_k24}{width90_pct}{0.1}
\setres{lotULJ}{EU_post_k24}{width90_pct_mean}{0.113}
\setres{lotULJ}{EU_post_k25}{cov50}{0.00}
\setres{lotULJ}{EU_post_k25}{cov50_mean}{0.003}
\setres{lotULJ}{EU_post_k25}{cov90}{0.00}
\setres{lotULJ}{EU_post_k25}{cov90_adj}{0.95}
\setres{lotULJ}{EU_post_k25}{cov90_adj_mean}{0.961}
\setres{lotULJ}{EU_post_k25}{cov90_mean}{0.003}
\setres{lotULJ}{EU_post_k25}{kappa}{0.488}
\setres{lotULJ}{EU_post_k25}{kappa_crps}{0.279}
\setres{lotULJ}{EU_post_k25}{kappa_crps_mean}{0.286}
\setres{lotULJ}{EU_post_k25}{kappa_mean}{0.549}
\setres{lotULJ}{EU_post_k25}{logscore_next}{-2.84}
\setres{lotULJ}{EU_post_k25}{logscore_next_mean}{-2.889}
\setres{lotULJ}{EU_post_k25}{misfit_pct}{16.5}
\setres{lotULJ}{EU_post_k25}{misfit_pct_mean}{16.654}
\setres{lotULJ}{EU_post_k25}{n_seen}{25}
\setres{lotULJ}{EU_post_k25}{surprise}{0.00}
\setres{lotULJ}{EU_post_k25}{surprise_mean}{0.132}
\setres{lotULJ}{EU_post_k25}{width90_adj_pct}{27.2}
\setres{lotULJ}{EU_post_k25}{width90_adj_pct_mean}{27.396}
\setres{lotULJ}{EU_post_k25}{width90_pct}{0.1}
\setres{lotULJ}{EU_post_k25}{width90_pct_mean}{0.107}
\setres{lotULJ}{EU_post_k26}{cov50}{0.00}
\setres{lotULJ}{EU_post_k26}{cov50_mean}{0.003}
\setres{lotULJ}{EU_post_k26}{cov90}{0.00}
\setres{lotULJ}{EU_post_k26}{cov90_adj}{0.94}
\setres{lotULJ}{EU_post_k26}{cov90_adj_mean}{0.958}
\setres{lotULJ}{EU_post_k26}{cov90_mean}{0.003}
\setres{lotULJ}{EU_post_k26}{kappa}{0.490}
\setres{lotULJ}{EU_post_k26}{kappa_crps}{0.308}
\setres{lotULJ}{EU_post_k26}{kappa_crps_mean}{0.289}
\setres{lotULJ}{EU_post_k26}{kappa_mean}{0.551}
\setres{lotULJ}{EU_post_k26}{logscore_next}{-2.90}
\setres{lotULJ}{EU_post_k26}{logscore_next_mean}{-3.092}
\setres{lotULJ}{EU_post_k26}{misfit_pct}{16.3}
\setres{lotULJ}{EU_post_k26}{misfit_pct_mean}{16.583}
\setres{lotULJ}{EU_post_k26}{n_seen}{26}
\setres{lotULJ}{EU_post_k26}{surprise}{0.02}
\setres{lotULJ}{EU_post_k26}{surprise_mean}{13.356}
\setres{lotULJ}{EU_post_k26}{width90_adj_pct}{26.9}
\setres{lotULJ}{EU_post_k26}{width90_adj_pct_mean}{27.279}
\setres{lotULJ}{EU_post_k26}{width90_pct}{0.1}
\setres{lotULJ}{EU_post_k26}{width90_pct_mean}{0.105}
\setres{lotULJ}{EU_post_k27}{cov50}{0.00}
\setres{lotULJ}{EU_post_k27}{cov50_mean}{0.000}
\setres{lotULJ}{EU_post_k27}{cov90}{0.00}
\setres{lotULJ}{EU_post_k27}{cov90_adj}{0.94}
\setres{lotULJ}{EU_post_k27}{cov90_adj_mean}{0.959}
\setres{lotULJ}{EU_post_k27}{cov90_mean}{0.000}
\setres{lotULJ}{EU_post_k27}{kappa}{0.598}
\setres{lotULJ}{EU_post_k27}{kappa_crps}{0.317}
\setres{lotULJ}{EU_post_k27}{kappa_crps_mean}{0.317}
\setres{lotULJ}{EU_post_k27}{kappa_mean}{0.573}
\setres{lotULJ}{EU_post_k27}{logscore_next}{-2.83}
\setres{lotULJ}{EU_post_k27}{logscore_next_mean}{-2.937}
\setres{lotULJ}{EU_post_k27}{misfit_pct}{16.3}
\setres{lotULJ}{EU_post_k27}{misfit_pct_mean}{16.488}
\setres{lotULJ}{EU_post_k27}{n_seen}{27}
\setres{lotULJ}{EU_post_k27}{surprise}{0.00}
\setres{lotULJ}{EU_post_k27}{surprise_mean}{9.650}
\setres{lotULJ}{EU_post_k27}{width90_adj_pct}{26.7}
\setres{lotULJ}{EU_post_k27}{width90_adj_pct_mean}{27.124}
\setres{lotULJ}{EU_post_k27}{width90_pct}{0.1}
\setres{lotULJ}{EU_post_k27}{width90_pct_mean}{0.110}
\setres{lotULJ}{EU_post_k28}{cov50}{0.00}
\setres{lotULJ}{EU_post_k28}{cov50_mean}{0.000}
\setres{lotULJ}{EU_post_k28}{cov90}{0.00}
\setres{lotULJ}{EU_post_k28}{cov90_adj}{0.97}
\setres{lotULJ}{EU_post_k28}{cov90_adj_mean}{0.959}
\setres{lotULJ}{EU_post_k28}{cov90_mean}{0.000}
\setres{lotULJ}{EU_post_k28}{kappa}{0.609}
\setres{lotULJ}{EU_post_k28}{kappa_crps}{0.350}
\setres{lotULJ}{EU_post_k28}{kappa_crps_mean}{0.316}
\setres{lotULJ}{EU_post_k28}{kappa_mean}{0.573}
\setres{lotULJ}{EU_post_k28}{logscore_next}{-2.85}
\setres{lotULJ}{EU_post_k28}{logscore_next_mean}{-3.069}
\setres{lotULJ}{EU_post_k28}{misfit_pct}{16.5}
\setres{lotULJ}{EU_post_k28}{misfit_pct_mean}{16.433}
\setres{lotULJ}{EU_post_k28}{n_seen}{28}
\setres{lotULJ}{EU_post_k28}{surprise}{0.01}
\setres{lotULJ}{EU_post_k28}{surprise_mean}{0.218}
\setres{lotULJ}{EU_post_k28}{width90_adj_pct}{27.1}
\setres{lotULJ}{EU_post_k28}{width90_adj_pct_mean}{27.033}
\setres{lotULJ}{EU_post_k28}{width90_pct}{0.1}
\setres{lotULJ}{EU_post_k28}{width90_pct_mean}{0.117}
\setres{lotULJ}{EU_post_k29}{cov50}{0.00}
\setres{lotULJ}{EU_post_k29}{cov50_mean}{0.000}
\setres{lotULJ}{EU_post_k29}{cov90}{0.00}
\setres{lotULJ}{EU_post_k29}{cov90_adj}{0.97}
\setres{lotULJ}{EU_post_k29}{cov90_adj_mean}{0.963}
\setres{lotULJ}{EU_post_k29}{cov90_mean}{0.000}
\setres{lotULJ}{EU_post_k29}{kappa}{0.614}
\setres{lotULJ}{EU_post_k29}{kappa_crps}{0.368}
\setres{lotULJ}{EU_post_k29}{kappa_crps_mean}{0.332}
\setres{lotULJ}{EU_post_k29}{kappa_mean}{0.580}
\setres{lotULJ}{EU_post_k29}{logscore_next}{-2.99}
\setres{lotULJ}{EU_post_k29}{logscore_next_mean}{-3.266}
\setres{lotULJ}{EU_post_k29}{misfit_pct}{16.5}
\setres{lotULJ}{EU_post_k29}{misfit_pct_mean}{16.440}
\setres{lotULJ}{EU_post_k29}{n_seen}{29}
\setres{lotULJ}{EU_post_k29}{surprise}{0.02}
\setres{lotULJ}{EU_post_k29}{surprise_mean}{0.261}
\setres{lotULJ}{EU_post_k29}{width90_adj_pct}{27.1}
\setres{lotULJ}{EU_post_k29}{width90_adj_pct_mean}{27.044}
\setres{lotULJ}{EU_post_k29}{width90_pct}{0.1}
\setres{lotULJ}{EU_post_k29}{width90_pct_mean}{0.119}
\setres{lotULJ}{EU_post_k3}{cov50}{0.00}
\setres{lotULJ}{EU_post_k3}{cov50_mean}{0.011}
\setres{lotULJ}{EU_post_k3}{cov90}{0.00}
\setres{lotULJ}{EU_post_k3}{cov90_adj}{0.89}
\setres{lotULJ}{EU_post_k3}{cov90_adj_mean}{0.838}
\setres{lotULJ}{EU_post_k3}{cov90_mean}{0.027}
\setres{lotULJ}{EU_post_k3}{kappa}{0.729}
\setres{lotULJ}{EU_post_k3}{kappa_crps}{0.390}
\setres{lotULJ}{EU_post_k3}{kappa_crps_mean}{0.262}
\setres{lotULJ}{EU_post_k3}{kappa_mean}{0.555}
\setres{lotULJ}{EU_post_k3}{logscore_next}{-2.94}
\setres{lotULJ}{EU_post_k3}{logscore_next_mean}{-3.040}
\setres{lotULJ}{EU_post_k3}{misfit_pct}{16.6}
\setres{lotULJ}{EU_post_k3}{misfit_pct_mean}{16.360}
\setres{lotULJ}{EU_post_k3}{n_seen}{3}
\setres{lotULJ}{EU_post_k3}{surprise}{0.05}
\setres{lotULJ}{EU_post_k3}{surprise_mean}{1.483}
\setres{lotULJ}{EU_post_k3}{width90_adj_pct}{27.4}
\setres{lotULJ}{EU_post_k3}{width90_adj_pct_mean}{26.923}
\setres{lotULJ}{EU_post_k3}{width90_pct}{0.3}
\setres{lotULJ}{EU_post_k3}{width90_pct_mean}{0.446}
\setres{lotULJ}{EU_post_k30}{cov50}{0.00}
\setres{lotULJ}{EU_post_k30}{cov50_mean}{0.000}
\setres{lotULJ}{EU_post_k30}{cov90}{0.00}
\setres{lotULJ}{EU_post_k30}{cov90_adj}{0.96}
\setres{lotULJ}{EU_post_k30}{cov90_adj_mean}{0.954}
\setres{lotULJ}{EU_post_k30}{cov90_mean}{0.004}
\setres{lotULJ}{EU_post_k30}{kappa}{0.620}
\setres{lotULJ}{EU_post_k30}{kappa_crps}{0.374}
\setres{lotULJ}{EU_post_k30}{kappa_crps_mean}{0.308}
\setres{lotULJ}{EU_post_k30}{kappa_mean}{0.570}
\setres{lotULJ}{EU_post_k30}{logscore_next}{-2.87}
\setres{lotULJ}{EU_post_k30}{logscore_next_mean}{-3.044}
\setres{lotULJ}{EU_post_k30}{misfit_pct}{16.4}
\setres{lotULJ}{EU_post_k30}{misfit_pct_mean}{16.373}
\setres{lotULJ}{EU_post_k30}{n_seen}{30}
\setres{lotULJ}{EU_post_k30}{surprise}{0.03}
\setres{lotULJ}{EU_post_k30}{surprise_mean}{0.277}
\setres{lotULJ}{EU_post_k30}{width90_adj_pct}{27.1}
\setres{lotULJ}{EU_post_k30}{width90_adj_pct_mean}{26.934}
\setres{lotULJ}{EU_post_k30}{width90_pct}{0.1}
\setres{lotULJ}{EU_post_k30}{width90_pct_mean}{0.116}
\setres{lotULJ}{EU_post_k31}{cov50}{0.00}
\setres{lotULJ}{EU_post_k31}{cov50_mean}{0.000}
\setres{lotULJ}{EU_post_k31}{cov90}{0.00}
\setres{lotULJ}{EU_post_k31}{cov90_adj}{1.00}
\setres{lotULJ}{EU_post_k31}{cov90_adj_mean}{0.954}
\setres{lotULJ}{EU_post_k31}{cov90_mean}{0.004}
\setres{lotULJ}{EU_post_k31}{kappa}{0.618}
\setres{lotULJ}{EU_post_k31}{kappa_crps}{0.369}
\setres{lotULJ}{EU_post_k31}{kappa_crps_mean}{0.304}
\setres{lotULJ}{EU_post_k31}{kappa_mean}{0.563}
\setres{lotULJ}{EU_post_k31}{logscore_next}{-2.86}
\setres{lotULJ}{EU_post_k31}{logscore_next_mean}{-2.977}
\setres{lotULJ}{EU_post_k31}{misfit_pct}{16.4}
\setres{lotULJ}{EU_post_k31}{misfit_pct_mean}{16.338}
\setres{lotULJ}{EU_post_k31}{n_seen}{31}
\setres{lotULJ}{EU_post_k31}{surprise}{0.00}
\setres{lotULJ}{EU_post_k31}{surprise_mean}{0.121}
\setres{lotULJ}{EU_post_k31}{width90_adj_pct}{26.9}
\setres{lotULJ}{EU_post_k31}{width90_adj_pct_mean}{26.876}
\setres{lotULJ}{EU_post_k31}{width90_pct}{0.1}
\setres{lotULJ}{EU_post_k31}{width90_pct_mean}{0.098}
\setres{lotULJ}{EU_post_k32}{cov50}{0.00}
\setres{lotULJ}{EU_post_k32}{cov50_mean}{0.000}
\setres{lotULJ}{EU_post_k32}{cov90}{0.00}
\setres{lotULJ}{EU_post_k32}{cov90_adj}{1.00}
\setres{lotULJ}{EU_post_k32}{cov90_adj_mean}{0.958}
\setres{lotULJ}{EU_post_k32}{cov90_mean}{0.004}
\setres{lotULJ}{EU_post_k32}{kappa}{0.594}
\setres{lotULJ}{EU_post_k32}{kappa_crps}{0.329}
\setres{lotULJ}{EU_post_k32}{kappa_crps_mean}{0.299}
\setres{lotULJ}{EU_post_k32}{kappa_mean}{0.542}
\setres{lotULJ}{EU_post_k32}{logscore_next}{-2.92}
\setres{lotULJ}{EU_post_k32}{logscore_next_mean}{-3.033}
\setres{lotULJ}{EU_post_k32}{misfit_pct}{16.4}
\setres{lotULJ}{EU_post_k32}{misfit_pct_mean}{16.362}
\setres{lotULJ}{EU_post_k32}{n_seen}{32}
\setres{lotULJ}{EU_post_k32}{surprise}{0.01}
\setres{lotULJ}{EU_post_k32}{surprise_mean}{0.290}
\setres{lotULJ}{EU_post_k32}{width90_adj_pct}{26.9}
\setres{lotULJ}{EU_post_k32}{width90_adj_pct_mean}{26.916}
\setres{lotULJ}{EU_post_k32}{width90_pct}{0.1}
\setres{lotULJ}{EU_post_k32}{width90_pct_mean}{0.091}
\setres{lotULJ}{EU_post_k33}{cov50}{0.00}
\setres{lotULJ}{EU_post_k33}{cov50_mean}{0.000}
\setres{lotULJ}{EU_post_k33}{cov90}{0.00}
\setres{lotULJ}{EU_post_k33}{cov90_adj}{1.00}
\setres{lotULJ}{EU_post_k33}{cov90_adj_mean}{0.959}
\setres{lotULJ}{EU_post_k33}{cov90_mean}{0.005}
\setres{lotULJ}{EU_post_k33}{kappa}{0.581}
\setres{lotULJ}{EU_post_k33}{kappa_crps}{0.315}
\setres{lotULJ}{EU_post_k33}{kappa_crps_mean}{0.269}
\setres{lotULJ}{EU_post_k33}{kappa_mean}{0.524}
\setres{lotULJ}{EU_post_k33}{logscore_next}{-2.86}
\setres{lotULJ}{EU_post_k33}{logscore_next_mean}{-2.955}
\setres{lotULJ}{EU_post_k33}{misfit_pct}{16.4}
\setres{lotULJ}{EU_post_k33}{misfit_pct_mean}{16.358}
\setres{lotULJ}{EU_post_k33}{n_seen}{33}
\setres{lotULJ}{EU_post_k33}{surprise}{0.00}
\setres{lotULJ}{EU_post_k33}{surprise_mean}{0.290}
\setres{lotULJ}{EU_post_k33}{width90_adj_pct}{27.0}
\setres{lotULJ}{EU_post_k33}{width90_adj_pct_mean}{26.909}
\setres{lotULJ}{EU_post_k33}{width90_pct}{0.1}
\setres{lotULJ}{EU_post_k33}{width90_pct_mean}{0.090}
\setres{lotULJ}{EU_post_k34}{cov50}{0.00}
\setres{lotULJ}{EU_post_k34}{cov50_mean}{0.000}
\setres{lotULJ}{EU_post_k34}{cov90}{0.00}
\setres{lotULJ}{EU_post_k34}{cov90_adj}{1.00}
\setres{lotULJ}{EU_post_k34}{cov90_adj_mean}{0.955}
\setres{lotULJ}{EU_post_k34}{cov90_mean}{0.005}
\setres{lotULJ}{EU_post_k34}{kappa}{0.584}
\setres{lotULJ}{EU_post_k34}{kappa_crps}{0.342}
\setres{lotULJ}{EU_post_k34}{kappa_crps_mean}{0.262}
\setres{lotULJ}{EU_post_k34}{kappa_mean}{0.512}
\setres{lotULJ}{EU_post_k34}{logscore_next}{-2.85}
\setres{lotULJ}{EU_post_k34}{logscore_next_mean}{-3.258}
\setres{lotULJ}{EU_post_k34}{misfit_pct}{16.4}
\setres{lotULJ}{EU_post_k34}{misfit_pct_mean}{16.385}
\setres{lotULJ}{EU_post_k34}{n_seen}{34}
\setres{lotULJ}{EU_post_k34}{surprise}{0.00}
\setres{lotULJ}{EU_post_k34}{surprise_mean}{14.409}
\setres{lotULJ}{EU_post_k34}{width90_adj_pct}{27.0}
\setres{lotULJ}{EU_post_k34}{width90_adj_pct_mean}{26.954}
\setres{lotULJ}{EU_post_k34}{width90_pct}{0.1}
\setres{lotULJ}{EU_post_k34}{width90_pct_mean}{0.090}
\setres{lotULJ}{EU_post_k35}{cov50}{0.00}
\setres{lotULJ}{EU_post_k35}{cov50_mean}{0.000}
\setres{lotULJ}{EU_post_k35}{cov90}{0.00}
\setres{lotULJ}{EU_post_k35}{cov90_adj}{1.00}
\setres{lotULJ}{EU_post_k35}{cov90_adj_mean}{0.961}
\setres{lotULJ}{EU_post_k35}{cov90_mean}{0.006}
\setres{lotULJ}{EU_post_k35}{kappa}{0.577}
\setres{lotULJ}{EU_post_k35}{kappa_crps}{0.276}
\setres{lotULJ}{EU_post_k35}{kappa_crps_mean}{0.232}
\setres{lotULJ}{EU_post_k35}{kappa_mean}{0.481}
\setres{lotULJ}{EU_post_k35}{logscore_next}{-2.84}
\setres{lotULJ}{EU_post_k35}{logscore_next_mean}{-3.029}
\setres{lotULJ}{EU_post_k35}{misfit_pct}{16.4}
\setres{lotULJ}{EU_post_k35}{misfit_pct_mean}{16.376}
\setres{lotULJ}{EU_post_k35}{n_seen}{35}
\setres{lotULJ}{EU_post_k35}{surprise}{0.00}
\setres{lotULJ}{EU_post_k35}{surprise_mean}{0.021}
\setres{lotULJ}{EU_post_k35}{width90_adj_pct}{27.0}
\setres{lotULJ}{EU_post_k35}{width90_adj_pct_mean}{26.938}
\setres{lotULJ}{EU_post_k35}{width90_pct}{0.1}
\setres{lotULJ}{EU_post_k35}{width90_pct_mean}{0.096}
\setres{lotULJ}{EU_post_k36}{cov50}{0.00}
\setres{lotULJ}{EU_post_k36}{cov50_mean}{0.000}
\setres{lotULJ}{EU_post_k36}{cov90}{0.00}
\setres{lotULJ}{EU_post_k36}{cov90_adj}{1.00}
\setres{lotULJ}{EU_post_k36}{cov90_adj_mean}{0.956}
\setres{lotULJ}{EU_post_k36}{cov90_mean}{0.006}
\setres{lotULJ}{EU_post_k36}{kappa}{0.587}
\setres{lotULJ}{EU_post_k36}{kappa_crps}{0.311}
\setres{lotULJ}{EU_post_k36}{kappa_crps_mean}{0.265}
\setres{lotULJ}{EU_post_k36}{kappa_mean}{0.502}
\setres{lotULJ}{EU_post_k36}{logscore_next}{-2.83}
\setres{lotULJ}{EU_post_k36}{logscore_next_mean}{-3.143}
\setres{lotULJ}{EU_post_k36}{misfit_pct}{16.5}
\setres{lotULJ}{EU_post_k36}{misfit_pct_mean}{16.421}
\setres{lotULJ}{EU_post_k36}{n_seen}{36}
\setres{lotULJ}{EU_post_k36}{surprise}{0.00}
\setres{lotULJ}{EU_post_k36}{surprise_mean}{2.390}
\setres{lotULJ}{EU_post_k36}{width90_adj_pct}{27.2}
\setres{lotULJ}{EU_post_k36}{width90_adj_pct_mean}{27.013}
\setres{lotULJ}{EU_post_k36}{width90_pct}{0.1}
\setres{lotULJ}{EU_post_k36}{width90_pct_mean}{0.104}
\setres{lotULJ}{EU_post_k37}{cov50}{0.00}
\setres{lotULJ}{EU_post_k37}{cov50_mean}{0.000}
\setres{lotULJ}{EU_post_k37}{cov90}{0.00}
\setres{lotULJ}{EU_post_k37}{cov90_adj}{1.00}
\setres{lotULJ}{EU_post_k37}{cov90_adj_mean}{0.950}
\setres{lotULJ}{EU_post_k37}{cov90_mean}{0.000}
\setres{lotULJ}{EU_post_k37}{kappa}{0.414}
\setres{lotULJ}{EU_post_k37}{kappa_crps}{0.241}
\setres{lotULJ}{EU_post_k37}{kappa_crps_mean}{0.182}
\setres{lotULJ}{EU_post_k37}{kappa_mean}{0.449}
\setres{lotULJ}{EU_post_k37}{logscore_next}{-2.85}
\setres{lotULJ}{EU_post_k37}{logscore_next_mean}{-3.143}
\setres{lotULJ}{EU_post_k37}{misfit_pct}{16.5}
\setres{lotULJ}{EU_post_k37}{misfit_pct_mean}{16.398}
\setres{lotULJ}{EU_post_k37}{n_seen}{37}
\setres{lotULJ}{EU_post_k37}{surprise}{0.02}
\setres{lotULJ}{EU_post_k37}{surprise_mean}{11.786}
\setres{lotULJ}{EU_post_k37}{width90_adj_pct}{27.1}
\setres{lotULJ}{EU_post_k37}{width90_adj_pct_mean}{26.975}
\setres{lotULJ}{EU_post_k37}{width90_pct}{0.1}
\setres{lotULJ}{EU_post_k37}{width90_pct_mean}{0.108}
\setres{lotULJ}{EU_post_k38}{cov50}{0.00}
\setres{lotULJ}{EU_post_k38}{cov50_mean}{0.000}
\setres{lotULJ}{EU_post_k38}{cov90}{0.00}
\setres{lotULJ}{EU_post_k38}{cov90_adj}{1.00}
\setres{lotULJ}{EU_post_k38}{cov90_adj_mean}{0.967}
\setres{lotULJ}{EU_post_k38}{cov90_mean}{0.000}
\setres{lotULJ}{EU_post_k38}{kappa}{0.396}
\setres{lotULJ}{EU_post_k38}{kappa_crps}{0.201}
\setres{lotULJ}{EU_post_k38}{kappa_crps_mean}{0.127}
\setres{lotULJ}{EU_post_k38}{kappa_mean}{0.411}
\setres{lotULJ}{EU_post_k38}{logscore_next}{-2.93}
\setres{lotULJ}{EU_post_k38}{logscore_next_mean}{-3.063}
\setres{lotULJ}{EU_post_k38}{misfit_pct}{16.5}
\setres{lotULJ}{EU_post_k38}{misfit_pct_mean}{16.419}
\setres{lotULJ}{EU_post_k38}{n_seen}{38}
\setres{lotULJ}{EU_post_k38}{surprise}{0.04}
\setres{lotULJ}{EU_post_k38}{surprise_mean}{0.178}
\setres{lotULJ}{EU_post_k38}{width90_adj_pct}{27.1}
\setres{lotULJ}{EU_post_k38}{width90_adj_pct_mean}{27.010}
\setres{lotULJ}{EU_post_k38}{width90_pct}{0.1}
\setres{lotULJ}{EU_post_k38}{width90_pct_mean}{0.111}
\setres{lotULJ}{EU_post_k39}{cov50}{0.00}
\setres{lotULJ}{EU_post_k39}{cov50_mean}{0.000}
\setres{lotULJ}{EU_post_k39}{cov90}{0.00}
\setres{lotULJ}{EU_post_k39}{cov90_adj}{1.00}
\setres{lotULJ}{EU_post_k39}{cov90_adj_mean}{0.970}
\setres{lotULJ}{EU_post_k39}{cov90_mean}{0.000}
\setres{lotULJ}{EU_post_k39}{kappa}{0.435}
\setres{lotULJ}{EU_post_k39}{kappa_crps}{0.185}
\setres{lotULJ}{EU_post_k39}{kappa_crps_mean}{0.074}
\setres{lotULJ}{EU_post_k39}{kappa_mean}{0.289}
\setres{lotULJ}{EU_post_k39}{logscore_next}{-2.85}
\setres{lotULJ}{EU_post_k39}{logscore_next_mean}{-3.067}
\setres{lotULJ}{EU_post_k39}{misfit_pct}{16.5}
\setres{lotULJ}{EU_post_k39}{misfit_pct_mean}{16.425}
\setres{lotULJ}{EU_post_k39}{n_seen}{39}
\setres{lotULJ}{EU_post_k39}{surprise}{0.00}
\setres{lotULJ}{EU_post_k39}{surprise_mean}{0.125}
\setres{lotULJ}{EU_post_k39}{width90_adj_pct}{27.1}
\setres{lotULJ}{EU_post_k39}{width90_adj_pct_mean}{27.019}
\setres{lotULJ}{EU_post_k39}{width90_pct}{0.1}
\setres{lotULJ}{EU_post_k39}{width90_pct_mean}{0.110}
\setres{lotULJ}{EU_post_k4}{cov50}{0.00}
\setres{lotULJ}{EU_post_k4}{cov50_mean}{0.014}
\setres{lotULJ}{EU_post_k4}{cov90}{0.00}
\setres{lotULJ}{EU_post_k4}{cov90_adj}{0.99}
\setres{lotULJ}{EU_post_k4}{cov90_adj_mean}{0.904}
\setres{lotULJ}{EU_post_k4}{cov90_mean}{0.039}
\setres{lotULJ}{EU_post_k4}{kappa}{0.728}
\setres{lotULJ}{EU_post_k4}{kappa_crps}{0.392}
\setres{lotULJ}{EU_post_k4}{kappa_crps_mean}{0.322}
\setres{lotULJ}{EU_post_k4}{kappa_mean}{0.625}
\setres{lotULJ}{EU_post_k4}{logscore_next}{-2.86}
\setres{lotULJ}{EU_post_k4}{logscore_next_mean}{-3.265}
\setres{lotULJ}{EU_post_k4}{misfit_pct}{16.8}
\setres{lotULJ}{EU_post_k4}{misfit_pct_mean}{16.851}
\setres{lotULJ}{EU_post_k4}{n_seen}{4}
\setres{lotULJ}{EU_post_k4}{surprise}{0.01}
\setres{lotULJ}{EU_post_k4}{surprise_mean}{3.177}
\setres{lotULJ}{EU_post_k4}{width90_adj_pct}{27.6}
\setres{lotULJ}{EU_post_k4}{width90_adj_pct_mean}{27.741}
\setres{lotULJ}{EU_post_k4}{width90_pct}{0.1}
\setres{lotULJ}{EU_post_k4}{width90_pct_mean}{0.533}
\setres{lotULJ}{EU_post_k40}{cov50}{0.00}
\setres{lotULJ}{EU_post_k40}{cov50_mean}{0.000}
\setres{lotULJ}{EU_post_k40}{cov90}{0.00}
\setres{lotULJ}{EU_post_k40}{cov90_adj}{1.00}
\setres{lotULJ}{EU_post_k40}{cov90_adj_mean}{0.975}
\setres{lotULJ}{EU_post_k40}{cov90_mean}{0.000}
\setres{lotULJ}{EU_post_k40}{kappa}{0.357}
\setres{lotULJ}{EU_post_k40}{kappa_crps}{0.242}
\setres{lotULJ}{EU_post_k40}{kappa_crps_mean}{0.085}
\setres{lotULJ}{EU_post_k40}{kappa_mean}{0.208}
\setres{lotULJ}{EU_post_k40}{logscore_next}{-2.83}
\setres{lotULJ}{EU_post_k40}{logscore_next_mean}{-3.130}
\setres{lotULJ}{EU_post_k40}{misfit_pct}{16.5}
\setres{lotULJ}{EU_post_k40}{misfit_pct_mean}{16.474}
\setres{lotULJ}{EU_post_k40}{n_seen}{40}
\setres{lotULJ}{EU_post_k40}{surprise}{0.01}
\setres{lotULJ}{EU_post_k40}{surprise_mean}{13.237}
\setres{lotULJ}{EU_post_k40}{width90_adj_pct}{27.2}
\setres{lotULJ}{EU_post_k40}{width90_adj_pct_mean}{27.100}
\setres{lotULJ}{EU_post_k40}{width90_pct}{0.1}
\setres{lotULJ}{EU_post_k40}{width90_pct_mean}{0.134}
\setres{lotULJ}{EU_post_k41}{cov50}{0.00}
\setres{lotULJ}{EU_post_k41}{cov50_mean}{0.000}
\setres{lotULJ}{EU_post_k41}{cov90}{0.00}
\setres{lotULJ}{EU_post_k41}{cov90_adj}{1.00}
\setres{lotULJ}{EU_post_k41}{cov90_adj_mean}{0.983}
\setres{lotULJ}{EU_post_k41}{cov90_mean}{0.000}
\setres{lotULJ}{EU_post_k41}{kappa}{0.418}
\setres{lotULJ}{EU_post_k41}{kappa_crps}{0.261}
\setres{lotULJ}{EU_post_k41}{kappa_crps_mean}{0.108}
\setres{lotULJ}{EU_post_k41}{kappa_mean}{0.236}
\setres{lotULJ}{EU_post_k41}{logscore_next}{-2.88}
\setres{lotULJ}{EU_post_k41}{logscore_next_mean}{-3.075}
\setres{lotULJ}{EU_post_k41}{misfit_pct}{16.4}
\setres{lotULJ}{EU_post_k41}{misfit_pct_mean}{16.450}
\setres{lotULJ}{EU_post_k41}{n_seen}{41}
\setres{lotULJ}{EU_post_k41}{surprise}{0.01}
\setres{lotULJ}{EU_post_k41}{surprise_mean}{2.118}
\setres{lotULJ}{EU_post_k41}{width90_adj_pct}{27.0}
\setres{lotULJ}{EU_post_k41}{width90_adj_pct_mean}{27.061}
\setres{lotULJ}{EU_post_k41}{width90_pct}{0.1}
\setres{lotULJ}{EU_post_k41}{width90_pct_mean}{0.143}
\setres{lotULJ}{EU_post_k42}{cov50}{0.00}
\setres{lotULJ}{EU_post_k42}{cov50_mean}{0.000}
\setres{lotULJ}{EU_post_k42}{cov90}{0.00}
\setres{lotULJ}{EU_post_k42}{cov90_adj}{1.00}
\setres{lotULJ}{EU_post_k42}{cov90_adj_mean}{0.975}
\setres{lotULJ}{EU_post_k42}{cov90_mean}{0.000}
\setres{lotULJ}{EU_post_k42}{kappa}{0.104}
\setres{lotULJ}{EU_post_k42}{kappa_crps}{0.095}
\setres{lotULJ}{EU_post_k42}{kappa_crps_mean}{-0.352}
\setres{lotULJ}{EU_post_k42}{kappa_mean}{-0.158}
\setres{lotULJ}{EU_post_k42}{logscore_next}{-2.85}
\setres{lotULJ}{EU_post_k42}{logscore_next_mean}{-3.095}
\setres{lotULJ}{EU_post_k42}{misfit_pct}{16.4}
\setres{lotULJ}{EU_post_k42}{misfit_pct_mean}{16.388}
\setres{lotULJ}{EU_post_k42}{n_seen}{42}
\setres{lotULJ}{EU_post_k42}{surprise}{0.01}
\setres{lotULJ}{EU_post_k42}{surprise_mean}{11.831}
\setres{lotULJ}{EU_post_k42}{width90_adj_pct}{27.0}
\setres{lotULJ}{EU_post_k42}{width90_adj_pct_mean}{26.958}
\setres{lotULJ}{EU_post_k42}{width90_pct}{0.2}
\setres{lotULJ}{EU_post_k42}{width90_pct_mean}{0.177}
\setres{lotULJ}{EU_post_k43}{cov50}{0.00}
\setres{lotULJ}{EU_post_k43}{cov50_mean}{0.000}
\setres{lotULJ}{EU_post_k43}{cov90}{0.00}
\setres{lotULJ}{EU_post_k43}{cov90_adj}{1.00}
\setres{lotULJ}{EU_post_k43}{cov90_adj_mean}{0.950}
\setres{lotULJ}{EU_post_k43}{cov90_mean}{0.000}
\setres{lotULJ}{EU_post_k43}{kappa}{0.088}
\setres{lotULJ}{EU_post_k43}{kappa_crps}{0.108}
\setres{lotULJ}{EU_post_k43}{kappa_crps_mean}{-0.775}
\setres{lotULJ}{EU_post_k43}{kappa_mean}{-2.370}
\setres{lotULJ}{EU_post_k43}{logscore_next}{-2.89}
\setres{lotULJ}{EU_post_k43}{logscore_next_mean}{-3.007}
\setres{lotULJ}{EU_post_k43}{misfit_pct}{16.3}
\setres{lotULJ}{EU_post_k43}{misfit_pct_mean}{16.339}
\setres{lotULJ}{EU_post_k43}{n_seen}{43}
\setres{lotULJ}{EU_post_k43}{surprise}{0.02}
\setres{lotULJ}{EU_post_k43}{surprise_mean}{11.746}
\setres{lotULJ}{EU_post_k43}{width90_adj_pct}{26.9}
\setres{lotULJ}{EU_post_k43}{width90_adj_pct_mean}{26.878}
\setres{lotULJ}{EU_post_k43}{width90_pct}{0.1}
\setres{lotULJ}{EU_post_k43}{width90_pct_mean}{0.191}
\setres{lotULJ}{EU_post_k44}{cov50}{---}
\setres{lotULJ}{EU_post_k44}{cov50_mean}{---}
\setres{lotULJ}{EU_post_k44}{cov90}{---}
\setres{lotULJ}{EU_post_k44}{cov90_adj}{---}
\setres{lotULJ}{EU_post_k44}{cov90_adj_mean}{---}
\setres{lotULJ}{EU_post_k44}{cov90_mean}{---}
\setres{lotULJ}{EU_post_k44}{kappa}{---}
\setres{lotULJ}{EU_post_k44}{kappa_crps}{---}
\setres{lotULJ}{EU_post_k44}{kappa_crps_mean}{---}
\setres{lotULJ}{EU_post_k44}{kappa_mean}{---}
\setres{lotULJ}{EU_post_k44}{logscore_next}{---}
\setres{lotULJ}{EU_post_k44}{logscore_next_mean}{---}
\setres{lotULJ}{EU_post_k44}{misfit_pct}{---}
\setres{lotULJ}{EU_post_k44}{misfit_pct_mean}{---}
\setres{lotULJ}{EU_post_k44}{n_seen}{44}
\setres{lotULJ}{EU_post_k44}{surprise}{---}
\setres{lotULJ}{EU_post_k44}{surprise_mean}{---}
\setres{lotULJ}{EU_post_k44}{width90_adj_pct}{---}
\setres{lotULJ}{EU_post_k44}{width90_adj_pct_mean}{---}
\setres{lotULJ}{EU_post_k44}{width90_pct}{---}
\setres{lotULJ}{EU_post_k44}{width90_pct_mean}{---}
\setres{lotULJ}{EU_post_k5}{cov50}{0.00}
\setres{lotULJ}{EU_post_k5}{cov50_mean}{0.008}
\setres{lotULJ}{EU_post_k5}{cov90}{0.00}
\setres{lotULJ}{EU_post_k5}{cov90_adj}{0.99}
\setres{lotULJ}{EU_post_k5}{cov90_adj_mean}{0.924}
\setres{lotULJ}{EU_post_k5}{cov90_mean}{0.031}
\setres{lotULJ}{EU_post_k5}{kappa}{0.726}
\setres{lotULJ}{EU_post_k5}{kappa_crps}{0.384}
\setres{lotULJ}{EU_post_k5}{kappa_crps_mean}{0.322}
\setres{lotULJ}{EU_post_k5}{kappa_mean}{0.634}
\setres{lotULJ}{EU_post_k5}{logscore_next}{-2.95}
\setres{lotULJ}{EU_post_k5}{logscore_next_mean}{-3.354}
\setres{lotULJ}{EU_post_k5}{misfit_pct}{16.6}
\setres{lotULJ}{EU_post_k5}{misfit_pct_mean}{16.755}
\setres{lotULJ}{EU_post_k5}{n_seen}{5}
\setres{lotULJ}{EU_post_k5}{surprise}{0.15}
\setres{lotULJ}{EU_post_k5}{surprise_mean}{21.438}
\setres{lotULJ}{EU_post_k5}{width90_adj_pct}{27.3}
\setres{lotULJ}{EU_post_k5}{width90_adj_pct_mean}{27.569}
\setres{lotULJ}{EU_post_k5}{width90_pct}{0.1}
\setres{lotULJ}{EU_post_k5}{width90_pct_mean}{0.386}
\setres{lotULJ}{EU_post_k6}{cov50}{0.00}
\setres{lotULJ}{EU_post_k6}{cov50_mean}{0.005}
\setres{lotULJ}{EU_post_k6}{cov90}{0.00}
\setres{lotULJ}{EU_post_k6}{cov90_adj}{0.99}
\setres{lotULJ}{EU_post_k6}{cov90_adj_mean}{0.942}
\setres{lotULJ}{EU_post_k6}{cov90_mean}{0.020}
\setres{lotULJ}{EU_post_k6}{kappa}{0.699}
\setres{lotULJ}{EU_post_k6}{kappa_crps}{0.372}
\setres{lotULJ}{EU_post_k6}{kappa_crps_mean}{0.356}
\setres{lotULJ}{EU_post_k6}{kappa_mean}{0.660}
\setres{lotULJ}{EU_post_k6}{logscore_next}{-2.86}
\setres{lotULJ}{EU_post_k6}{logscore_next_mean}{-3.121}
\setres{lotULJ}{EU_post_k6}{misfit_pct}{16.8}
\setres{lotULJ}{EU_post_k6}{misfit_pct_mean}{16.601}
\setres{lotULJ}{EU_post_k6}{n_seen}{6}
\setres{lotULJ}{EU_post_k6}{surprise}{0.01}
\setres{lotULJ}{EU_post_k6}{surprise_mean}{13.821}
\setres{lotULJ}{EU_post_k6}{width90_adj_pct}{27.6}
\setres{lotULJ}{EU_post_k6}{width90_adj_pct_mean}{27.315}
\setres{lotULJ}{EU_post_k6}{width90_pct}{0.1}
\setres{lotULJ}{EU_post_k6}{width90_pct_mean}{0.279}
\setres{lotULJ}{EU_post_k7}{cov50}{0.00}
\setres{lotULJ}{EU_post_k7}{cov50_mean}{0.005}
\setres{lotULJ}{EU_post_k7}{cov90}{0.00}
\setres{lotULJ}{EU_post_k7}{cov90_adj}{0.95}
\setres{lotULJ}{EU_post_k7}{cov90_adj_mean}{0.920}
\setres{lotULJ}{EU_post_k7}{cov90_mean}{0.018}
\setres{lotULJ}{EU_post_k7}{kappa}{0.694}
\setres{lotULJ}{EU_post_k7}{kappa_crps}{0.377}
\setres{lotULJ}{EU_post_k7}{kappa_crps_mean}{0.329}
\setres{lotULJ}{EU_post_k7}{kappa_mean}{0.619}
\setres{lotULJ}{EU_post_k7}{logscore_next}{-2.84}
\setres{lotULJ}{EU_post_k7}{logscore_next_mean}{-3.516}
\setres{lotULJ}{EU_post_k7}{misfit_pct}{16.1}
\setres{lotULJ}{EU_post_k7}{misfit_pct_mean}{16.613}
\setres{lotULJ}{EU_post_k7}{n_seen}{7}
\setres{lotULJ}{EU_post_k7}{surprise}{0.01}
\setres{lotULJ}{EU_post_k7}{surprise_mean}{10.145}
\setres{lotULJ}{EU_post_k7}{width90_adj_pct}{26.5}
\setres{lotULJ}{EU_post_k7}{width90_adj_pct_mean}{27.335}
\setres{lotULJ}{EU_post_k7}{width90_pct}{0.1}
\setres{lotULJ}{EU_post_k7}{width90_pct_mean}{0.293}
\setres{lotULJ}{EU_post_k8}{cov50}{0.00}
\setres{lotULJ}{EU_post_k8}{cov50_mean}{0.003}
\setres{lotULJ}{EU_post_k8}{cov90}{0.00}
\setres{lotULJ}{EU_post_k8}{cov90_adj}{0.97}
\setres{lotULJ}{EU_post_k8}{cov90_adj_mean}{0.933}
\setres{lotULJ}{EU_post_k8}{cov90_mean}{0.014}
\setres{lotULJ}{EU_post_k8}{kappa}{0.707}
\setres{lotULJ}{EU_post_k8}{kappa_crps}{0.343}
\setres{lotULJ}{EU_post_k8}{kappa_crps_mean}{0.342}
\setres{lotULJ}{EU_post_k8}{kappa_mean}{0.644}
\setres{lotULJ}{EU_post_k8}{logscore_next}{-2.86}
\setres{lotULJ}{EU_post_k8}{logscore_next_mean}{-3.005}
\setres{lotULJ}{EU_post_k8}{misfit_pct}{16.6}
\setres{lotULJ}{EU_post_k8}{misfit_pct_mean}{16.438}
\setres{lotULJ}{EU_post_k8}{n_seen}{8}
\setres{lotULJ}{EU_post_k8}{surprise}{0.00}
\setres{lotULJ}{EU_post_k8}{surprise_mean}{8.611}
\setres{lotULJ}{EU_post_k8}{width90_adj_pct}{27.2}
\setres{lotULJ}{EU_post_k8}{width90_adj_pct_mean}{27.041}
\setres{lotULJ}{EU_post_k8}{width90_pct}{0.1}
\setres{lotULJ}{EU_post_k8}{width90_pct_mean}{0.134}
\setres{lotULJ}{EU_post_k9}{cov50}{0.00}
\setres{lotULJ}{EU_post_k9}{cov50_mean}{0.003}
\setres{lotULJ}{EU_post_k9}{cov90}{0.00}
\setres{lotULJ}{EU_post_k9}{cov90_adj}{0.97}
\setres{lotULJ}{EU_post_k9}{cov90_adj_mean}{0.920}
\setres{lotULJ}{EU_post_k9}{cov90_mean}{0.011}
\setres{lotULJ}{EU_post_k9}{kappa}{0.692}
\setres{lotULJ}{EU_post_k9}{kappa_crps}{0.328}
\setres{lotULJ}{EU_post_k9}{kappa_crps_mean}{0.317}
\setres{lotULJ}{EU_post_k9}{kappa_mean}{0.622}
\setres{lotULJ}{EU_post_k9}{logscore_next}{-2.85}
\setres{lotULJ}{EU_post_k9}{logscore_next_mean}{-3.030}
\setres{lotULJ}{EU_post_k9}{misfit_pct}{16.8}
\setres{lotULJ}{EU_post_k9}{misfit_pct_mean}{16.264}
\setres{lotULJ}{EU_post_k9}{n_seen}{9}
\setres{lotULJ}{EU_post_k9}{surprise}{0.00}
\setres{lotULJ}{EU_post_k9}{surprise_mean}{3.749}
\setres{lotULJ}{EU_post_k9}{width90_adj_pct}{27.6}
\setres{lotULJ}{EU_post_k9}{width90_adj_pct_mean}{26.755}
\setres{lotULJ}{EU_post_k9}{width90_pct}{0.1}
\setres{lotULJ}{EU_post_k9}{width90_pct_mean}{0.126}
\setres{lotULJ}{reference}{domain}{Zurich03\_Zurich06\_Beijing\_gains}
\setres{lotULJ}{reference}{n_draws}{400}
\setres{lotULJ}{reference}{n_lotteries}{44}
\setres{lotULJ}{reference}{n_orders}{20}
\setres{lotULJ}{reference}{start}{uniform}
\setres{crpsfit}{CPT}{det_pct}{97}
\setres{crpsfit}{CPT}{gap_lit}{0.323}
\setres{crpsfit}{CPT}{gap_uniform}{1.469}
\setres{crpsfit}{CPT}{share_lit}{67}
\setres{crpsfit}{CPT}{share_uniform}{-51}
\setres{crpsfit}{CPT_crpsfit}{ess}{33}
\setres{crpsfit}{CPT_crpsfit}{kappa}{0.480}
\setres{crpsfit}{CPT_crpsfit}{kappa_crps}{0.890}
\setres{crpsfit}{CPT_crpsfit}{kappa_crps_se}{0.015}
\setres{crpsfit}{CPT_crpsfit}{r_cop}{0.570}
\setres{crpsfit}{CPT_crpsfit}{r_dist}{0.166}
\setres{crpsfit}{CPT_crpsfit}{r_form}{0.018}
\setres{crpsfit}{CPT_crpsfit}{r_pi}{0.136}
\setres{crpsfit}{CPT_crpsfit}{theta_mean}{0.93, 0.39, 1.14}
\setres{crpsfit}{CPT_crpsfit}{theta_sd}{0.10, 0.21, 0.76}
\setres{crpsfit}{CPT_deterministic}{kappa}{0.972}
\setres{crpsfit}{CPT_deterministic}{kappa_crps}{0.287}
\setres{crpsfit}{CPT_deterministic}{kappa_crps_se}{0.043}
\setres{crpsfit}{CPT_deterministic}{r_dist}{0.359}
\setres{crpsfit}{CPT_deterministic}{r_form}{0.018}
\setres{crpsfit}{CPT_deterministic}{r_pi}{0.321}
\setres{crpsfit}{CPT_lit}{kappa}{0.649}
\setres{crpsfit}{CPT_lit}{kappa_crps}{0.661}
\setres{crpsfit}{CPT_lit}{kappa_crps_se}{0.042}
\setres{crpsfit}{CPT_lit}{r_cop}{0.418}
\setres{crpsfit}{CPT_lit}{r_dist}{0.374}
\setres{crpsfit}{CPT_lit}{r_form}{0.018}
\setres{crpsfit}{CPT_lit}{r_pi}{0.342}
\setres{crpsfit}{CPT_types}{kappa}{0.629}
\setres{crpsfit}{CPT_types}{kappa_crps}{0.696}
\setres{crpsfit}{CPT_types}{kappa_crps_se}{0.022}
\setres{crpsfit}{CPT_types}{n_atoms}{4}
\setres{crpsfit}{CPT_types}{r_dist}{0.264}
\setres{crpsfit}{CPT_types}{samples}{ZH06, BJ05}
\setres{crpsfit}{CPT_types_bj05}{kappa}{0.359}
\setres{crpsfit}{CPT_types_bj05}{kappa_crps}{0.102}
\setres{crpsfit}{CPT_types_bj05}{kappa_crps_se}{0.048}
\setres{crpsfit}{CPT_types_bj05}{n_atoms}{2}
\setres{crpsfit}{CPT_types_bj05}{r_dist}{0.310}
\setres{crpsfit}{CPT_types_bj05}{samples}{BJ05}
\setres{crpsfit}{CPT_types_fit}{kappa}{0.693}
\setres{crpsfit}{CPT_types_fit}{kappa_crps}{0.651}
\setres{crpsfit}{CPT_types_fit}{kappa_crps_se}{0.025}
\setres{crpsfit}{CPT_types_fit}{n_atoms}{4}
\setres{crpsfit}{CPT_types_fit}{r_dist}{0.261}
\setres{crpsfit}{CPT_types_fit}{samples}{ZH06, BJ05}
\setres{crpsfit}{CPT_types_zh06}{kappa}{0.899}
\setres{crpsfit}{CPT_types_zh06}{kappa_crps}{0.574}
\setres{crpsfit}{CPT_types_zh06}{kappa_crps_se}{0.035}
\setres{crpsfit}{CPT_types_zh06}{n_atoms}{2}
\setres{crpsfit}{CPT_types_zh06}{r_dist}{0.349}
\setres{crpsfit}{CPT_types_zh06}{samples}{ZH06}
\setres{crpsfit}{CPT_uniform}{kappa}{-0.496}
\setres{crpsfit}{CPT_uniform}{kappa_crps}{0.700}
\setres{crpsfit}{CPT_uniform}{kappa_crps_se}{0.018}
\setres{crpsfit}{CPT_uniform}{r_cop}{0.458}
\setres{crpsfit}{CPT_uniform}{r_dist}{0.181}
\setres{crpsfit}{CPT_uniform}{r_form}{0.018}
\setres{crpsfit}{CPT_uniform}{r_pi}{0.151}
\setres{crpsfit}{DA}{det_pct}{63}
\setres{crpsfit}{DA}{gap_cpt_lit}{0.621}
\setres{crpsfit}{DA}{gap_lit}{0.597}
\setres{crpsfit}{DA}{gap_uniform}{1.454}
\setres{crpsfit}{DA}{share_lit}{4}
\setres{crpsfit}{DA}{share_uniform}{-132}
\setres{crpsfit}{DA_crpsfit}{ess}{28}
\setres{crpsfit}{DA_crpsfit}{kappa}{-0.350}
\setres{crpsfit}{DA_crpsfit}{kappa_crps}{0.532}
\setres{crpsfit}{DA_crpsfit}{kappa_crps_se}{0.046}
\setres{crpsfit}{DA_crpsfit}{r_cop}{0.754}
\setres{crpsfit}{DA_crpsfit}{r_dist}{0.287}
\setres{crpsfit}{DA_crpsfit}{r_form}{0.187}
\setres{crpsfit}{DA_crpsfit}{r_pi}{0.099}
\setres{crpsfit}{DA_crpsfit}{theta_mean}{0.80, 0.45}
\setres{crpsfit}{DA_crpsfit}{theta_sd}{0.36, 1.40}
\setres{crpsfit}{DA_deterministic}{kappa}{0.625}
\setres{crpsfit}{DA_deterministic}{kappa_crps}{-0.143}
\setres{crpsfit}{DA_deterministic}{kappa_crps_se}{0.092}
\setres{crpsfit}{DA_deterministic}{r_dist}{0.557}
\setres{crpsfit}{DA_deterministic}{r_form}{0.187}
\setres{crpsfit}{DA_deterministic}{r_pi}{0.387}
\setres{crpsfit}{DA_lit}{kappa}{0.028}
\setres{crpsfit}{DA_lit}{kappa_crps}{0.048}
\setres{crpsfit}{DA_lit}{kappa_crps_se}{0.083}
\setres{crpsfit}{DA_lit}{r_cop}{0.683}
\setres{crpsfit}{DA_lit}{r_dist}{0.498}
\setres{crpsfit}{DA_lit}{r_form}{0.187}
\setres{crpsfit}{DA_lit}{r_pi}{0.290}
\setres{crpsfit}{DA_types}{kappa}{0.277}
\setres{crpsfit}{DA_types}{kappa_crps}{0.298}
\setres{crpsfit}{DA_types}{kappa_crps_se}{0.069}
\setres{crpsfit}{DA_types}{n_atoms}{4}
\setres{crpsfit}{DA_types}{r_dist}{0.449}
\setres{crpsfit}{DA_types}{samples}{ZH06, BJ05}
\setres{crpsfit}{DA_uniform}{kappa}{-0.829}
\setres{crpsfit}{DA_uniform}{kappa_crps}{-0.147}
\setres{crpsfit}{DA_uniform}{kappa_crps_se}{0.088}
\setres{crpsfit}{DA_uniform}{r_cop}{0.656}
\setres{crpsfit}{DA_uniform}{r_dist}{0.457}
\setres{crpsfit}{DA_uniform}{r_form}{0.187}
\setres{crpsfit}{DA_uniform}{r_pi}{0.255}
\setres{crpsfit}{EU}{det_pct}{62}
\setres{crpsfit}{EU}{gap_lit}{0.085}
\setres{crpsfit}{EU}{gap_uniform}{0.826}
\setres{crpsfit}{EU}{share_lit}{86}
\setres{crpsfit}{EU}{share_uniform}{-34}
\setres{crpsfit}{EU_crpsfit}{ess}{5}
\setres{crpsfit}{EU_crpsfit}{kappa}{0.106}
\setres{crpsfit}{EU_crpsfit}{kappa_crps}{0.189}
\setres{crpsfit}{EU_crpsfit}{kappa_crps_se}{0.076}
\setres{crpsfit}{EU_crpsfit}{r_cop}{0.870}
\setres{crpsfit}{EU_crpsfit}{r_dist}{0.475}
\setres{crpsfit}{EU_crpsfit}{r_form}{0.456}
\setres{crpsfit}{EU_crpsfit}{r_pi}{0.019}
\setres{crpsfit}{EU_crpsfit}{theta_mean}{0.73}
\setres{crpsfit}{EU_crpsfit}{theta_sd}{0.32}
\setres{crpsfit}{EU_deterministic}{kappa}{0.616}
\setres{crpsfit}{EU_deterministic}{kappa_crps}{-0.153}
\setres{crpsfit}{EU_deterministic}{kappa_crps_se}{0.096}
\setres{crpsfit}{EU_deterministic}{r_dist}{0.571}
\setres{crpsfit}{EU_deterministic}{r_form}{0.456}
\setres{crpsfit}{EU_deterministic}{r_pi}{0.115}
\setres{crpsfit}{EU_lit}{kappa}{0.530}
\setres{crpsfit}{EU_lit}{kappa_crps}{0.049}
\setres{crpsfit}{EU_lit}{kappa_crps_se}{0.086}
\setres{crpsfit}{EU_lit}{r_cop}{0.993}
\setres{crpsfit}{EU_lit}{r_dist}{0.532}
\setres{crpsfit}{EU_lit}{r_form}{0.456}
\setres{crpsfit}{EU_lit}{r_pi}{0.076}
\setres{crpsfit}{EU_types}{kappa}{0.337}
\setres{crpsfit}{EU_types}{kappa_crps}{-0.064}
\setres{crpsfit}{EU_types}{kappa_crps_se}{0.091}
\setres{crpsfit}{EU_types}{n_atoms}{4}
\setres{crpsfit}{EU_types}{r_dist}{0.524}
\setres{crpsfit}{EU_types}{samples}{ZH06, BJ05}
\setres{crpsfit}{EU_uniform}{kappa}{-0.211}
\setres{crpsfit}{EU_uniform}{kappa_crps}{-0.012}
\setres{crpsfit}{EU_uniform}{kappa_crps_se}{0.085}
\setres{crpsfit}{EU_uniform}{r_cop}{0.994}
\setres{crpsfit}{EU_uniform}{r_dist}{0.504}
\setres{crpsfit}{EU_uniform}{r_form}{0.456}
\setres{crpsfit}{EU_uniform}{r_pi}{0.046}
\setres{crpsfit}{EV_deterministic}{kappa}{0.493}
\setres{crpsfit}{EV_deterministic}{kappa_crps}{-0.183}
\setres{crpsfit}{EV_deterministic}{kappa_crps_se}{0.099}
\setres{crpsfit}{EV_deterministic}{r_dist}{0.554}
\setres{crpsfit}{reference}{n_draws}{1{,}000}
\setres{crpsfit}{reference}{n_folds}{10}
\setres{crpsfit}{reference}{sample}{Bruhin2010\_gains}
\setres{crpsfit}{subjects}{n_boot}{1{,}000}
\setres{crpsfit}{subjects}{n_complete_profiles}{166}
\setres{crpsfit}{subjects}{n_subjects}{179}
\setres{crpsfit}{subjects}{r_cop}{0.127}
\setres{crpsfit}{subjects}{r_dist}{0.185}
\setres{crpsfit}{subjects}{r_dist_se}{0.009}
\setres{crpsfitR}{CPT_crpsfit}{r_ff}{0.140}
\setres{crpsfitR}{CPT_crpsfit}{r_ff_se}{0.001}
\setres{crpsfitR}{CPT_crpsfit}{r_prior}{0.9953}
\setres{crpsfitR}{CPT_crpsfit}{r_prior_se}{0.000}
\setres{crpsfitR}{CPT_crpsfit}{r_prob}{0.996}
\setres{crpsfitR}{CPT_crpsfit}{r_prob_se}{0.000}
\setres{crpsfitR}{CPT_crpsfit}{volume_ess}{55{,}652}
\setres{crpsfitR}{DA_crpsfit}{r_ff}{0.281}
\setres{crpsfitR}{DA_crpsfit}{r_ff_se}{0.004}
\setres{crpsfitR}{DA_crpsfit}{r_prior}{0.9941}
\setres{crpsfitR}{DA_crpsfit}{r_prior_se}{0.000}
\setres{crpsfitR}{DA_crpsfit}{r_prob}{0.996}
\setres{crpsfitR}{DA_crpsfit}{r_prob_se}{0.000}
\setres{crpsfitR}{DA_crpsfit}{volume_ess}{436{,}452}
\setres{crpsfitR}{EU_crpsfit}{r_ff}{0.336}
\setres{crpsfitR}{EU_crpsfit}{r_ff_se}{0.005}
\setres{crpsfitR}{EU_crpsfit}{r_prior}{0.9719}
\setres{crpsfitR}{EU_crpsfit}{r_prior_se}{0.000}
\setres{crpsfitR}{EU_crpsfit}{r_prob}{0.981}
\setres{crpsfitR}{EU_crpsfit}{r_prob_se}{0.000}
\setres{crpsfitR}{EU_crpsfit}{volume_ess}{172{,}086}
\setres{crpsfitP}{CPT}{alpha_lit_mean}{0.700}
\setres{crpsfitP}{CPT}{alpha_lit_q05}{0.449}
\setres{crpsfitP}{CPT}{alpha_lit_q95}{0.901}
\setres{crpsfitP}{CPT}{alpha_lit_sd}{0.138}
\setres{crpsfitP}{CPT}{alpha_mean}{0.935}
\setres{crpsfitP}{CPT}{alpha_q05}{0.777}
\setres{crpsfitP}{CPT}{alpha_q50}{0.975}
\setres{crpsfitP}{CPT}{alpha_q95}{0.975}
\setres{crpsfitP}{CPT}{alpha_sd}{0.10}
\setres{crpsfitP}{CPT}{delta_lit_mean}{1.084}
\setres{crpsfitP}{CPT}{delta_lit_q05}{0.696}
\setres{crpsfitP}{CPT}{delta_lit_q95}{1.590}
\setres{crpsfitP}{CPT}{delta_lit_sd}{0.276}
\setres{crpsfitP}{CPT}{delta_mean}{1.136}
\setres{crpsfitP}{CPT}{delta_q05}{0.318}
\setres{crpsfitP}{CPT}{delta_q50}{0.897}
\setres{crpsfitP}{CPT}{delta_q95}{2.927}
\setres{crpsfitP}{CPT}{delta_sd}{0.76}
\setres{crpsfitP}{CPT}{gamma_lit_mean}{0.728}
\setres{crpsfitP}{CPT}{gamma_lit_q05}{0.397}
\setres{crpsfitP}{CPT}{gamma_lit_q95}{0.959}
\setres{crpsfitP}{CPT}{gamma_lit_sd}{0.174}
\setres{crpsfitP}{CPT}{gamma_mean}{0.393}
\setres{crpsfitP}{CPT}{gamma_q05}{0.035}
\setres{crpsfitP}{CPT}{gamma_q50}{0.431}
\setres{crpsfitP}{CPT}{gamma_q95}{0.777}
\setres{crpsfitP}{CPT}{gamma_sd}{0.21}
\setres{crpsfitP}{DA}{alpha_lit_mean}{0.582}
\setres{crpsfitP}{DA}{alpha_lit_q05}{0.258}
\setres{crpsfitP}{DA}{alpha_lit_q95}{0.873}
\setres{crpsfitP}{DA}{alpha_lit_sd}{0.186}
\setres{crpsfitP}{DA}{alpha_mean}{0.796}
\setres{crpsfitP}{DA}{alpha_q05}{0.022}
\setres{crpsfitP}{DA}{alpha_q50}{0.988}
\setres{crpsfitP}{DA}{alpha_q95}{0.988}
\setres{crpsfitP}{DA}{alpha_sd}{0.36}
\setres{crpsfitP}{DA}{eta_lit_mean}{0.307}
\setres{crpsfitP}{DA}{eta_lit_q05}{-0.432}
\setres{crpsfitP}{DA}{eta_lit_q95}{1.048}
\setres{crpsfitP}{DA}{eta_lit_sd}{0.446}
\setres{crpsfitP}{DA}{eta_mean}{0.451}
\setres{crpsfitP}{DA}{eta_q05}{-0.950}
\setres{crpsfitP}{DA}{eta_q50}{-0.050}
\setres{crpsfitP}{DA}{eta_q95}{2.950}
\setres{crpsfitP}{DA}{eta_sd}{1.40}
\setres{crpsfitP}{EU}{alpha_lit_mean}{0.701}
\setres{crpsfitP}{EU}{alpha_lit_q05}{0.453}
\setres{crpsfitP}{EU}{alpha_lit_q95}{0.902}
\setres{crpsfitP}{EU}{alpha_lit_sd}{0.138}
\setres{crpsfitP}{EU}{alpha_mean}{0.734}
\setres{crpsfitP}{EU}{alpha_q05}{0.062}
\setres{crpsfitP}{EU}{alpha_q50}{0.925}
\setres{crpsfitP}{EU}{alpha_q95}{0.997}
\setres{crpsfitP}{EU}{alpha_sd}{0.32}
\setres{crpsfitP}{reference}{sample}{Bruhin2010\_gains}
\setres{crpsN}{CPT}{c_range}{0.062}
\setres{crpsN}{CPT}{c_range_full}{0.062}
\setres{crpsN}{CPT}{c_range_pct}{6.2}
\setres{crpsN}{CPT}{c_rms}{0.156}
\setres{crpsN}{CPT}{c_rms_full}{0.156}
\setres{crpsN}{CPT}{c_rms_pct}{15.6}
\setres{crpsN}{CPT}{sigma_range_rel}{0.062}
\setres{crpsN}{CPT}{sigma_rms_rel}{0.156}
\setres{crpsN}{CPT_crpsfit}{kappa_crps_none}{0.890}
\setres{crpsN}{CPT_crpsfit}{kappa_crps_none_se}{0.015}
\setres{crpsN}{CPT_crpsfit}{kappa_crps_range}{0.964}
\setres{crpsN}{CPT_crpsfit}{kappa_crps_range_se}{0.005}
\setres{crpsN}{CPT_crpsfit}{kappa_crps_rms}{0.944}
\setres{crpsN}{CPT_crpsfit}{kappa_crps_rms_se}{0.005}
\setres{crpsN}{CPT_crpsfit}{r_dist_none}{0.166}
\setres{crpsN}{CPT_crpsfit}{r_dist_range}{0.179}
\setres{crpsN}{CPT_crpsfit}{r_dist_rms}{0.173}
\setres{crpsN}{CPT_deterministic}{kappa_crps_none}{0.287}
\setres{crpsN}{CPT_deterministic}{kappa_crps_none_se}{0.043}
\setres{crpsN}{CPT_deterministic}{kappa_crps_range}{0.710}
\setres{crpsN}{CPT_deterministic}{kappa_crps_range_se}{0.023}
\setres{crpsN}{CPT_deterministic}{kappa_crps_rms}{0.904}
\setres{crpsN}{CPT_deterministic}{kappa_crps_rms_se}{0.009}
\setres{crpsN}{CPT_deterministic}{r_dist_none}{0.359}
\setres{crpsN}{CPT_deterministic}{r_dist_range}{0.266}
\setres{crpsN}{CPT_deterministic}{r_dist_rms}{0.167}
\setres{crpsN}{CPT_lit}{kappa_crps_none}{0.661}
\setres{crpsN}{CPT_lit}{kappa_crps_none_se}{0.042}
\setres{crpsN}{CPT_lit}{kappa_crps_range}{0.732}
\setres{crpsN}{CPT_lit}{kappa_crps_range_se}{0.035}
\setres{crpsN}{CPT_lit}{kappa_crps_rms}{0.794}
\setres{crpsN}{CPT_lit}{kappa_crps_rms_se}{0.023}
\setres{crpsN}{CPT_lit}{r_dist_none}{0.374}
\setres{crpsN}{CPT_lit}{r_dist_range}{0.345}
\setres{crpsN}{CPT_lit}{r_dist_rms}{0.272}
\setres{crpsN}{CPT_types}{kappa_crps_none}{0.696}
\setres{crpsN}{CPT_types}{kappa_crps_none_se}{0.022}
\setres{crpsN}{CPT_types}{kappa_crps_range}{0.825}
\setres{crpsN}{CPT_types}{kappa_crps_range_se}{0.016}
\setres{crpsN}{CPT_types}{kappa_crps_rms}{0.860}
\setres{crpsN}{CPT_types}{kappa_crps_rms_se}{0.012}
\setres{crpsN}{CPT_types}{r_dist_none}{0.264}
\setres{crpsN}{CPT_types}{r_dist_range}{0.227}
\setres{crpsN}{CPT_types}{r_dist_rms}{0.168}
\setres{crpsN}{CPT_types_bj05}{kappa_crps_none}{0.102}
\setres{crpsN}{CPT_types_bj05}{kappa_crps_none_se}{0.048}
\setres{crpsN}{CPT_types_bj05}{kappa_crps_range}{0.401}
\setres{crpsN}{CPT_types_bj05}{kappa_crps_range_se}{0.036}
\setres{crpsN}{CPT_types_bj05}{kappa_crps_rms}{0.570}
\setres{crpsN}{CPT_types_bj05}{kappa_crps_rms_se}{0.027}
\setres{crpsN}{CPT_types_bj05}{r_dist_none}{0.310}
\setres{crpsN}{CPT_types_bj05}{r_dist_range}{0.245}
\setres{crpsN}{CPT_types_bj05}{r_dist_rms}{0.175}
\setres{crpsN}{CPT_types_fit}{kappa_crps_none}{0.651}
\setres{crpsN}{CPT_types_fit}{kappa_crps_none_se}{0.025}
\setres{crpsN}{CPT_types_fit}{kappa_crps_range}{0.785}
\setres{crpsN}{CPT_types_fit}{kappa_crps_range_se}{0.020}
\setres{crpsN}{CPT_types_fit}{kappa_crps_rms}{0.835}
\setres{crpsN}{CPT_types_fit}{kappa_crps_rms_se}{0.015}
\setres{crpsN}{CPT_types_fit}{r_dist_none}{0.261}
\setres{crpsN}{CPT_types_fit}{r_dist_range}{0.223}
\setres{crpsN}{CPT_types_fit}{r_dist_rms}{0.160}
\setres{crpsN}{CPT_types_zh06}{kappa_crps_none}{0.574}
\setres{crpsN}{CPT_types_zh06}{kappa_crps_none_se}{0.035}
\setres{crpsN}{CPT_types_zh06}{kappa_crps_range}{0.832}
\setres{crpsN}{CPT_types_zh06}{kappa_crps_range_se}{0.020}
\setres{crpsN}{CPT_types_zh06}{kappa_crps_rms}{0.934}
\setres{crpsN}{CPT_types_zh06}{kappa_crps_rms_se}{0.007}
\setres{crpsN}{CPT_types_zh06}{r_dist_none}{0.349}
\setres{crpsN}{CPT_types_zh06}{r_dist_range}{0.285}
\setres{crpsN}{CPT_types_zh06}{r_dist_rms}{0.199}
\setres{crpsN}{CPT_uniform}{kappa_crps_none}{0.700}
\setres{crpsN}{CPT_uniform}{kappa_crps_none_se}{0.018}
\setres{crpsN}{CPT_uniform}{kappa_crps_range}{0.710}
\setres{crpsN}{CPT_uniform}{kappa_crps_range_se}{0.018}
\setres{crpsN}{CPT_uniform}{kappa_crps_rms}{0.675}
\setres{crpsN}{CPT_uniform}{kappa_crps_rms_se}{0.018}
\setres{crpsN}{CPT_uniform}{r_dist_none}{0.181}
\setres{crpsN}{CPT_uniform}{r_dist_range}{0.172}
\setres{crpsN}{CPT_uniform}{r_dist_rms}{0.149}
\setres{crpsN}{DA}{c_range}{0.086}
\setres{crpsN}{DA}{c_range_full}{0.086}
\setres{crpsN}{DA}{c_range_pct}{8.6}
\setres{crpsN}{DA}{c_rms}{0.220}
\setres{crpsN}{DA}{c_rms_full}{0.220}
\setres{crpsN}{DA}{c_rms_pct}{22.0}
\setres{crpsN}{DA}{sigma_range_rel}{0.086}
\setres{crpsN}{DA}{sigma_rms_rel}{0.220}
\setres{crpsN}{DA_crpsfit}{kappa_crps_none}{0.532}
\setres{crpsN}{DA_crpsfit}{kappa_crps_none_se}{0.046}
\setres{crpsN}{DA_crpsfit}{kappa_crps_range}{0.635}
\setres{crpsN}{DA_crpsfit}{kappa_crps_range_se}{0.036}
\setres{crpsN}{DA_crpsfit}{kappa_crps_rms}{0.593}
\setres{crpsN}{DA_crpsfit}{kappa_crps_rms_se}{0.029}
\setres{crpsN}{DA_crpsfit}{r_dist_none}{0.287}
\setres{crpsN}{DA_crpsfit}{r_dist_range}{0.278}
\setres{crpsN}{DA_crpsfit}{r_dist_rms}{0.246}
\setres{crpsN}{DA_deterministic}{kappa_crps_none}{-0.143}
\setres{crpsN}{DA_deterministic}{kappa_crps_none_se}{0.092}
\setres{crpsN}{DA_deterministic}{kappa_crps_range}{0.401}
\setres{crpsN}{DA_deterministic}{kappa_crps_range_se}{0.060}
\setres{crpsN}{DA_deterministic}{kappa_crps_rms}{0.557}
\setres{crpsN}{DA_deterministic}{kappa_crps_rms_se}{0.037}
\setres{crpsN}{DA_deterministic}{r_dist_none}{0.557}
\setres{crpsN}{DA_deterministic}{r_dist_range}{0.432}
\setres{crpsN}{DA_deterministic}{r_dist_rms}{0.306}
\setres{crpsN}{DA_lit}{kappa_crps_none}{0.048}
\setres{crpsN}{DA_lit}{kappa_crps_none_se}{0.083}
\setres{crpsN}{DA_lit}{kappa_crps_range}{0.267}
\setres{crpsN}{DA_lit}{kappa_crps_range_se}{0.068}
\setres{crpsN}{DA_lit}{kappa_crps_rms}{0.389}
\setres{crpsN}{DA_lit}{kappa_crps_rms_se}{0.048}
\setres{crpsN}{DA_lit}{r_dist_none}{0.498}
\setres{crpsN}{DA_lit}{r_dist_range}{0.439}
\setres{crpsN}{DA_lit}{r_dist_rms}{0.339}
\setres{crpsN}{DA_types}{kappa_crps_none}{0.298}
\setres{crpsN}{DA_types}{kappa_crps_none_se}{0.069}
\setres{crpsN}{DA_types}{kappa_crps_range}{0.530}
\setres{crpsN}{DA_types}{kappa_crps_range_se}{0.051}
\setres{crpsN}{DA_types}{kappa_crps_rms}{0.568}
\setres{crpsN}{DA_types}{kappa_crps_rms_se}{0.035}
\setres{crpsN}{DA_types}{r_dist_none}{0.449}
\setres{crpsN}{DA_types}{r_dist_range}{0.382}
\setres{crpsN}{DA_types}{r_dist_rms}{0.288}
\setres{crpsN}{DA_uniform}{kappa_crps_none}{-0.147}
\setres{crpsN}{DA_uniform}{kappa_crps_none_se}{0.088}
\setres{crpsN}{DA_uniform}{kappa_crps_range}{0.046}
\setres{crpsN}{DA_uniform}{kappa_crps_range_se}{0.076}
\setres{crpsN}{DA_uniform}{kappa_crps_rms}{0.183}
\setres{crpsN}{DA_uniform}{kappa_crps_rms_se}{0.059}
\setres{crpsN}{DA_uniform}{r_dist_none}{0.457}
\setres{crpsN}{DA_uniform}{r_dist_range}{0.412}
\setres{crpsN}{DA_uniform}{r_dist_rms}{0.333}
\setres{crpsN}{EU}{c_range}{0.135}
\setres{crpsN}{EU}{c_range_full}{0.135}
\setres{crpsN}{EU}{c_range_pct}{13.5}
\setres{crpsN}{EU}{c_rms}{0.226}
\setres{crpsN}{EU}{c_rms_full}{0.226}
\setres{crpsN}{EU}{c_rms_pct}{22.6}
\setres{crpsN}{EU}{sigma_range_rel}{0.135}
\setres{crpsN}{EU}{sigma_rms_rel}{0.226}
\setres{crpsN}{EU_crpsfit}{kappa_crps_none}{0.189}
\setres{crpsN}{EU_crpsfit}{kappa_crps_none_se}{0.076}
\setres{crpsN}{EU_crpsfit}{kappa_crps_range}{0.545}
\setres{crpsN}{EU_crpsfit}{kappa_crps_range_se}{0.045}
\setres{crpsN}{EU_crpsfit}{kappa_crps_rms}{0.538}
\setres{crpsN}{EU_crpsfit}{kappa_crps_rms_se}{0.036}
\setres{crpsN}{EU_crpsfit}{r_dist_none}{0.475}
\setres{crpsN}{EU_crpsfit}{r_dist_range}{0.348}
\setres{crpsN}{EU_crpsfit}{r_dist_rms}{0.289}
\setres{crpsN}{EU_deterministic}{kappa_crps_none}{-0.153}
\setres{crpsN}{EU_deterministic}{kappa_crps_none_se}{0.096}
\setres{crpsN}{EU_deterministic}{kappa_crps_range}{0.493}
\setres{crpsN}{EU_deterministic}{kappa_crps_range_se}{0.051}
\setres{crpsN}{EU_deterministic}{kappa_crps_rms}{0.520}
\setres{crpsN}{EU_deterministic}{kappa_crps_rms_se}{0.039}
\setres{crpsN}{EU_deterministic}{r_dist_none}{0.571}
\setres{crpsN}{EU_deterministic}{r_dist_range}{0.391}
\setres{crpsN}{EU_deterministic}{r_dist_rms}{0.316}
\setres{crpsN}{EU_lit}{kappa_crps_none}{0.049}
\setres{crpsN}{EU_lit}{kappa_crps_none_se}{0.086}
\setres{crpsN}{EU_lit}{kappa_crps_range}{0.491}
\setres{crpsN}{EU_lit}{kappa_crps_range_se}{0.051}
\setres{crpsN}{EU_lit}{kappa_crps_rms}{0.511}
\setres{crpsN}{EU_lit}{kappa_crps_rms_se}{0.040}
\setres{crpsN}{EU_lit}{r_dist_none}{0.532}
\setres{crpsN}{EU_lit}{r_dist_range}{0.391}
\setres{crpsN}{EU_lit}{r_dist_rms}{0.320}
\setres{crpsN}{EU_types}{kappa_crps_none}{-0.064}
\setres{crpsN}{EU_types}{kappa_crps_none_se}{0.091}
\setres{crpsN}{EU_types}{kappa_crps_range}{0.506}
\setres{crpsN}{EU_types}{kappa_crps_range_se}{0.050}
\setres{crpsN}{EU_types}{kappa_crps_rms}{0.523}
\setres{crpsN}{EU_types}{kappa_crps_rms_se}{0.038}
\setres{crpsN}{EU_types}{r_dist_none}{0.524}
\setres{crpsN}{EU_types}{r_dist_range}{0.359}
\setres{crpsN}{EU_types}{r_dist_rms}{0.291}
\setres{crpsN}{EU_uniform}{kappa_crps_none}{-0.012}
\setres{crpsN}{EU_uniform}{kappa_crps_none_se}{0.085}
\setres{crpsN}{EU_uniform}{kappa_crps_range}{0.332}
\setres{crpsN}{EU_uniform}{kappa_crps_range_se}{0.059}
\setres{crpsN}{EU_uniform}{kappa_crps_rms}{0.354}
\setres{crpsN}{EU_uniform}{kappa_crps_rms_se}{0.048}
\setres{crpsN}{EU_uniform}{r_dist_none}{0.504}
\setres{crpsN}{EU_uniform}{r_dist_range}{0.395}
\setres{crpsN}{EU_uniform}{r_dist_rms}{0.334}
\setres{crpsN}{EV_deterministic}{kappa_crps_none}{-0.183}
\setres{crpsN}{EV_deterministic}{kappa_crps_none_se}{0.099}
\setres{crpsN}{EV_deterministic}{kappa_crps_range}{0.456}
\setres{crpsN}{EV_deterministic}{kappa_crps_range_se}{0.054}
\setres{crpsN}{EV_deterministic}{kappa_crps_rms}{0.492}
\setres{crpsN}{EV_deterministic}{kappa_crps_rms_se}{0.041}
\setres{crpsN}{EV_deterministic}{r_dist_none}{0.554}
\setres{crpsN}{EV_deterministic}{r_dist_range}{0.374}
\setres{crpsN}{EV_deterministic}{r_dist_rms}{0.300}
\setres{crpsN}{reference}{calibrations}{none, range, rms}
\setres{crpsN}{reference}{n_folds}{10}
\setres{crpsN}{reference}{sample}{Bruhin2010\_gains}
\setres{crpsN}{reference}{tails_sd}{6.000}
\setres{crpsN}{subjects}{n_boot}{1{,}000}
\setres{crpsN}{subjects}{n_subjects}{179}
\setres{crpsN}{subjects}{r_dist}{0.185}
\setres{crpsN}{subjects}{r_dist_se}{0.009}
\setres{typesfit}{CPT_BJ05}{alpha_1}{0.480}
\setres{typesfit}{CPT_BJ05}{alpha_2}{0.381}
\setres{typesfit}{CPT_BJ05}{c}{0.060}
\setres{typesfit}{CPT_BJ05}{delta_1}{1.188}
\setres{typesfit}{CPT_BJ05}{delta_2}{1.478}
\setres{typesfit}{CPT_BJ05}{gamma_1}{0.134}
\setres{typesfit}{CPT_BJ05}{gamma_2}{0.480}
\setres{typesfit}{CPT_BJ05}{loglik}{-6604.314}
\setres{typesfit}{CPT_BJ05}{n_subjects}{151}
\setres{typesfit}{CPT_BJ05}{share_1}{1}
\setres{typesfit}{CPT_BJ05}{share_2}{0}
\setres{typesfit}{CPT_BJ05}{types}{0.50: (0.48, 0.13, 1.19); 0.50: (0.38, 0.48, 1.48)}
\setres{typesfit}{CPT_BJ05_published}{types}{0.20: (1.08, 0.91, 0.89); 0.80: (0.38, 0.24, 1.33)}
\setres{typesfit}{CPT_ZH03_published}{types}{0.18: (0.98, 0.95, 0.91); 0.82: (1.06, 0.41, 0.85)}
\setres{typesfit}{CPT_ZH06}{alpha_1}{0.876}
\setres{typesfit}{CPT_ZH06}{alpha_2}{0.876}
\setres{typesfit}{CPT_ZH06}{c}{0.038}
\setres{typesfit}{CPT_ZH06}{delta_1}{1.042}
\setres{typesfit}{CPT_ZH06}{delta_2}{0.607}
\setres{typesfit}{CPT_ZH06}{gamma_1}{0.530}
\setres{typesfit}{CPT_ZH06}{gamma_2}{0.332}
\setres{typesfit}{CPT_ZH06}{loglik}{-13364.843}
\setres{typesfit}{CPT_ZH06}{n_subjects}{118}
\setres{typesfit}{CPT_ZH06}{share_1}{1}
\setres{typesfit}{CPT_ZH06}{share_2}{0}
\setres{typesfit}{CPT_ZH06}{types}{0.71: (0.88, 0.53, 1.04); 0.29: (0.88, 0.33, 0.61)}
\setres{typesfit}{CPT_ZH06_published}{types}{0.22: (0.99, 0.94, 0.91); 0.78: (0.90, 0.42, 0.86)}
\setres{typesfit}{DA_BJ05}{alpha_1}{0.988}
\setres{typesfit}{DA_BJ05}{alpha_2}{0.022}
\setres{typesfit}{DA_BJ05}{c}{0.179}
\setres{typesfit}{DA_BJ05}{eta_1}{0.250}
\setres{typesfit}{DA_BJ05}{eta_2}{-0.850}
\setres{typesfit}{DA_BJ05}{loglik}{-2090.718}
\setres{typesfit}{DA_BJ05}{n_subjects}{151}
\setres{typesfit}{DA_BJ05}{share_1}{1}
\setres{typesfit}{DA_BJ05}{share_2}{0}
\setres{typesfit}{DA_BJ05}{types}{0.81: (0.99, 0.25); 0.19: (0.02, -0.85)}
\setres{typesfit}{DA_ZH06}{alpha_1}{0.988}
\setres{typesfit}{DA_ZH06}{alpha_2}{0.988}
\setres{typesfit}{DA_ZH06}{c}{0.076}
\setres{typesfit}{DA_ZH06}{eta_1}{-0.050}
\setres{typesfit}{DA_ZH06}{eta_2}{2.050}
\setres{typesfit}{DA_ZH06}{loglik}{-6619.351}
\setres{typesfit}{DA_ZH06}{n_subjects}{118}
\setres{typesfit}{DA_ZH06}{share_1}{1}
\setres{typesfit}{DA_ZH06}{share_2}{0}
\setres{typesfit}{DA_ZH06}{types}{0.74: (0.99, -0.05); 0.26: (0.99, 2.05)}
\setres{typesfit}{EU_BJ05}{alpha_1}{0.998}
\setres{typesfit}{EU_BJ05}{alpha_2}{0.998}
\setres{typesfit}{EU_BJ05}{c}{0.291}
\setres{typesfit}{EU_BJ05}{loglik}{-821.943}
\setres{typesfit}{EU_BJ05}{n_subjects}{151}
\setres{typesfit}{EU_BJ05}{share_1}{1}
\setres{typesfit}{EU_BJ05}{share_2}{0}
\setres{typesfit}{EU_BJ05}{types}{0.60: (1.00); 0.40: (1.00)}
\setres{typesfit}{EU_ZH06}{alpha_1}{0.998}
\setres{typesfit}{EU_ZH06}{alpha_2}{0.057}
\setres{typesfit}{EU_ZH06}{c}{0.133}
\setres{typesfit}{EU_ZH06}{loglik}{-2329.285}
\setres{typesfit}{EU_ZH06}{n_subjects}{118}
\setres{typesfit}{EU_ZH06}{share_1}{1}
\setres{typesfit}{EU_ZH06}{share_2}{0}
\setres{typesfit}{EU_ZH06}{types}{0.88: (1.00); 0.12: (0.06)}
\setres{typesfit}{reference}{K}{2.000}
\setres{typesfit}{reference}{samples}{Zurich 2006, Beijing 2005 (gains)}
\setres{csG}{CPT_crpsfit}{kappa_crps}{0.890}
\setres{csG}{CPT_crpsfit}{kappa_crps_se}{0.015}
\setres{csG}{CPT_crpsfit}{r_dist}{0.166}
\setres{csG}{CPT_deterministic}{kappa_crps}{0.287}
\setres{csG}{CPT_deterministic}{kappa_crps_se}{0.043}
\setres{csG}{CPT_deterministic}{r_dist}{0.359}
\setres{csG}{CPT_types}{kappa_crps}{0.696}
\setres{csG}{CPT_types}{kappa_crps_se}{0.022}
\setres{csG}{CPT_types}{r_dist}{0.264}
\setres{csG}{CPT_uniform}{kappa_crps}{0.700}
\setres{csG}{CPT_uniform}{kappa_crps_se}{0.018}
\setres{csG}{CPT_uniform}{r_dist}{0.181}
\setres{csG}{DA_crpsfit}{kappa_crps}{0.532}
\setres{csG}{DA_crpsfit}{kappa_crps_se}{0.046}
\setres{csG}{DA_crpsfit}{r_dist}{0.287}
\setres{csG}{DA_deterministic}{kappa_crps}{-0.143}
\setres{csG}{DA_deterministic}{kappa_crps_se}{0.092}
\setres{csG}{DA_deterministic}{r_dist}{0.557}
\setres{csG}{DA_types}{kappa_crps}{0.298}
\setres{csG}{DA_types}{kappa_crps_se}{0.069}
\setres{csG}{DA_types}{r_dist}{0.449}
\setres{csG}{DA_uniform}{kappa_crps}{-0.147}
\setres{csG}{DA_uniform}{kappa_crps_se}{0.088}
\setres{csG}{DA_uniform}{r_dist}{0.457}
\setres{csG}{EU_crpsfit}{kappa_crps}{0.189}
\setres{csG}{EU_crpsfit}{kappa_crps_se}{0.076}
\setres{csG}{EU_crpsfit}{r_dist}{0.475}
\setres{csG}{EU_deterministic}{kappa_crps}{-0.153}
\setres{csG}{EU_deterministic}{kappa_crps_se}{0.096}
\setres{csG}{EU_deterministic}{r_dist}{0.571}
\setres{csG}{EU_types}{kappa_crps}{-0.064}
\setres{csG}{EU_types}{kappa_crps_se}{0.091}
\setres{csG}{EU_types}{r_dist}{0.524}
\setres{csG}{EU_uniform}{kappa_crps}{-0.012}
\setres{csG}{EU_uniform}{kappa_crps_se}{0.085}
\setres{csG}{EU_uniform}{r_dist}{0.504}
\setres{csG}{EV_deterministic}{kappa_crps}{-0.183}
\setres{csG}{EV_deterministic}{kappa_crps_se}{0.099}
\setres{csG}{EV_deterministic}{r_dist}{0.554}
\setres{csL}{CPT_crpsfit}{kappa_crps}{0.944}
\setres{csL}{CPT_crpsfit}{kappa_crps_se}{0.012}
\setres{csL}{CPT_crpsfit}{r_dist}{0.213}
\setres{csL}{CPT_deterministic}{kappa_crps}{0.173}
\setres{csL}{CPT_deterministic}{kappa_crps_se}{0.053}
\setres{csL}{CPT_deterministic}{r_dist}{0.349}
\setres{csL}{CPT_types}{kappa_crps}{0.634}
\setres{csL}{CPT_types}{kappa_crps_se}{0.033}
\setres{csL}{CPT_types}{r_dist}{0.273}
\setres{csL}{CPT_uniform}{kappa_crps}{0.822}
\setres{csL}{CPT_uniform}{kappa_crps_se}{0.010}
\setres{csL}{CPT_uniform}{r_dist}{0.131}
\setres{csL}{DA_crpsfit}{kappa_crps}{0.600}
\setres{csL}{DA_crpsfit}{kappa_crps_se}{0.046}
\setres{csL}{DA_crpsfit}{r_dist}{0.251}
\setres{csL}{DA_deterministic}{kappa_crps}{-0.106}
\setres{csL}{DA_deterministic}{kappa_crps_se}{0.065}
\setres{csL}{DA_deterministic}{r_dist}{0.385}
\setres{csL}{DA_types}{kappa_crps}{0.419}
\setres{csL}{DA_types}{kappa_crps_se}{0.067}
\setres{csL}{DA_types}{r_dist}{0.417}
\setres{csL}{DA_uniform}{kappa_crps}{0.407}
\setres{csL}{DA_uniform}{kappa_crps_se}{0.060}
\setres{csL}{DA_uniform}{r_dist}{0.343}
\setres{csL}{EU_crpsfit}{kappa_crps}{0.254}
\setres{csL}{EU_crpsfit}{kappa_crps_se}{0.079}
\setres{csL}{EU_crpsfit}{r_dist}{0.418}
\setres{csL}{EU_deterministic}{kappa_crps}{-0.125}
\setres{csL}{EU_deterministic}{kappa_crps_se}{0.105}
\setres{csL}{EU_deterministic}{r_dist}{0.551}
\setres{csL}{EU_types}{kappa_crps}{0.241}
\setres{csL}{EU_types}{kappa_crps_se}{0.073}
\setres{csL}{EU_types}{r_dist}{0.384}
\setres{csL}{EU_uniform}{kappa_crps}{0.211}
\setres{csL}{EU_uniform}{kappa_crps_se}{0.081}
\setres{csL}{EU_uniform}{r_dist}{0.411}
\setres{csL}{EV_deterministic}{kappa_crps}{-0.223}
\setres{csL}{EV_deterministic}{kappa_crps_se}{0.115}
\setres{csL}{EV_deterministic}{r_dist}{0.554}
\setres{csA}{CPT_crpsfit}{kappa_crps}{0.917}
\setres{csA}{CPT_crpsfit}{kappa_crps_se}{0.009}
\setres{csA}{CPT_crpsfit}{r_dist}{0.189}
\setres{csA}{CPT_deterministic}{kappa_crps}{0.230}
\setres{csA}{CPT_deterministic}{kappa_crps_se}{0.044}
\setres{csA}{CPT_deterministic}{r_dist}{0.354}
\setres{csA}{CPT_types}{kappa_crps}{0.665}
\setres{csA}{CPT_types}{kappa_crps_se}{0.021}
\setres{csA}{CPT_types}{r_dist}{0.269}
\setres{csA}{CPT_uniform}{kappa_crps}{0.761}
\setres{csA}{CPT_uniform}{kappa_crps_se}{0.011}
\setres{csA}{CPT_uniform}{r_dist}{0.156}
\setres{csA}{DA_crpsfit}{kappa_crps}{0.566}
\setres{csA}{DA_crpsfit}{kappa_crps_se}{0.043}
\setres{csA}{DA_crpsfit}{r_dist}{0.269}
\setres{csA}{DA_deterministic}{kappa_crps}{-0.125}
\setres{csA}{DA_deterministic}{kappa_crps_se}{0.072}
\setres{csA}{DA_deterministic}{r_dist}{0.471}
\setres{csA}{DA_types}{kappa_crps}{0.359}
\setres{csA}{DA_types}{kappa_crps_se}{0.064}
\setres{csA}{DA_types}{r_dist}{0.433}
\setres{csA}{DA_uniform}{kappa_crps}{0.130}
\setres{csA}{DA_uniform}{kappa_crps_se}{0.068}
\setres{csA}{DA_uniform}{r_dist}{0.400}
\setres{csA}{EU_crpsfit}{kappa_crps}{0.222}
\setres{csA}{EU_crpsfit}{kappa_crps_se}{0.073}
\setres{csA}{EU_crpsfit}{r_dist}{0.447}
\setres{csA}{EU_deterministic}{kappa_crps}{-0.139}
\setres{csA}{EU_deterministic}{kappa_crps_se}{0.095}
\setres{csA}{EU_deterministic}{r_dist}{0.561}
\setres{csA}{EU_types}{kappa_crps}{0.088}
\setres{csA}{EU_types}{kappa_crps_se}{0.078}
\setres{csA}{EU_types}{r_dist}{0.454}
\setres{csA}{EU_uniform}{kappa_crps}{0.100}
\setres{csA}{EU_uniform}{kappa_crps_se}{0.076}
\setres{csA}{EU_uniform}{r_dist}{0.458}
\setres{csA}{EV_deterministic}{kappa_crps}{-0.203}
\setres{csA}{EV_deterministic}{kappa_crps_se}{0.102}
\setres{csA}{EV_deterministic}{r_dist}{0.554}
\setres{cuG}{CPT_crpsfit}{kappa_crps}{0.845}
\setres{cuG}{CPT_crpsfit}{kappa_crps_se}{0.025}
\setres{cuG}{CPT_crpsfit}{r_dist}{0.207}
\setres{cuG}{CPT_deterministic}{kappa_crps}{0.188}
\setres{cuG}{CPT_deterministic}{kappa_crps_se}{0.046}
\setres{cuG}{CPT_deterministic}{r_dist}{0.343}
\setres{cuG}{CPT_types}{kappa_crps}{0.689}
\setres{cuG}{CPT_types}{kappa_crps_se}{0.023}
\setres{cuG}{CPT_types}{r_dist}{0.258}
\setres{cuG}{CPT_uniform}{kappa_crps}{0.694}
\setres{cuG}{CPT_uniform}{kappa_crps_se}{0.021}
\setres{cuG}{CPT_uniform}{r_dist}{0.116}
\setres{cuG}{DA_crpsfit}{kappa_crps}{0.448}
\setres{cuG}{DA_crpsfit}{kappa_crps_se}{0.051}
\setres{cuG}{DA_crpsfit}{r_dist}{0.348}
\setres{cuG}{DA_deterministic}{kappa_crps}{-0.128}
\setres{cuG}{DA_deterministic}{kappa_crps_se}{0.094}
\setres{cuG}{DA_deterministic}{r_dist}{0.563}
\setres{cuG}{DA_types}{kappa_crps}{0.198}
\setres{cuG}{DA_types}{kappa_crps_se}{0.076}
\setres{cuG}{DA_types}{r_dist}{0.451}
\setres{cuG}{DA_uniform}{kappa_crps}{0.455}
\setres{cuG}{DA_uniform}{kappa_crps_se}{0.052}
\setres{cuG}{DA_uniform}{r_dist}{0.357}
\setres{cuG}{EU_crpsfit}{kappa_crps}{0.319}
\setres{cuG}{EU_crpsfit}{kappa_crps_se}{0.063}
\setres{cuG}{EU_crpsfit}{r_dist}{0.410}
\setres{cuG}{EU_deterministic}{kappa_crps}{-0.202}
\setres{cuG}{EU_deterministic}{kappa_crps_se}{0.100}
\setres{cuG}{EU_deterministic}{r_dist}{0.552}
\setres{cuG}{EU_types}{kappa_crps}{-0.136}
\setres{cuG}{EU_types}{kappa_crps_se}{0.096}
\setres{cuG}{EU_types}{r_dist}{0.531}
\setres{cuG}{EU_uniform}{kappa_crps}{0.331}
\setres{cuG}{EU_uniform}{kappa_crps_se}{0.063}
\setres{cuG}{EU_uniform}{r_dist}{0.409}
\setres{cuG}{EV_deterministic}{kappa_crps}{-0.183}
\setres{cuG}{EV_deterministic}{kappa_crps_se}{0.099}
\setres{cuG}{EV_deterministic}{r_dist}{0.554}
\setres{cuL}{CPT_crpsfit}{kappa_crps}{0.822}
\setres{cuL}{CPT_crpsfit}{kappa_crps_se}{0.028}
\setres{cuL}{CPT_crpsfit}{r_dist}{0.207}
\setres{cuL}{CPT_deterministic}{kappa_crps}{0.111}
\setres{cuL}{CPT_deterministic}{kappa_crps_se}{0.059}
\setres{cuL}{CPT_deterministic}{r_dist}{0.343}
\setres{cuL}{CPT_types}{kappa_crps}{0.444}
\setres{cuL}{CPT_types}{kappa_crps_se}{0.048}
\setres{cuL}{CPT_types}{r_dist}{0.254}
\setres{cuL}{CPT_uniform}{kappa_crps}{0.822}
\setres{cuL}{CPT_uniform}{kappa_crps_se}{0.010}
\setres{cuL}{CPT_uniform}{r_dist}{0.131}
\setres{cuL}{DA_crpsfit}{kappa_crps}{0.446}
\setres{cuL}{DA_crpsfit}{kappa_crps_se}{0.060}
\setres{cuL}{DA_crpsfit}{r_dist}{0.339}
\setres{cuL}{DA_deterministic}{kappa_crps}{-0.228}
\setres{cuL}{DA_deterministic}{kappa_crps_se}{0.112}
\setres{cuL}{DA_deterministic}{r_dist}{0.566}
\setres{cuL}{DA_types}{kappa_crps}{0.305}
\setres{cuL}{DA_types}{kappa_crps_se}{0.078}
\setres{cuL}{DA_types}{r_dist}{0.443}
\setres{cuL}{DA_uniform}{kappa_crps}{0.407}
\setres{cuL}{DA_uniform}{kappa_crps_se}{0.060}
\setres{cuL}{DA_uniform}{r_dist}{0.343}
\setres{cuL}{EU_crpsfit}{kappa_crps}{0.238}
\setres{cuL}{EU_crpsfit}{kappa_crps_se}{0.079}
\setres{cuL}{EU_crpsfit}{r_dist}{0.412}
\setres{cuL}{EU_deterministic}{kappa_crps}{-0.194}
\setres{cuL}{EU_deterministic}{kappa_crps_se}{0.113}
\setres{cuL}{EU_deterministic}{r_dist}{0.553}
\setres{cuL}{EU_types}{kappa_crps}{-0.054}
\setres{cuL}{EU_types}{kappa_crps_se}{0.105}
\setres{cuL}{EU_types}{r_dist}{0.532}
\setres{cuL}{EU_uniform}{kappa_crps}{0.211}
\setres{cuL}{EU_uniform}{kappa_crps_se}{0.081}
\setres{cuL}{EU_uniform}{r_dist}{0.411}
\setres{cuL}{EV_deterministic}{kappa_crps}{-0.223}
\setres{cuL}{EV_deterministic}{kappa_crps_se}{0.115}
\setres{cuL}{EV_deterministic}{r_dist}{0.554}
\setres{cuA}{CPT_crpsfit}{kappa_crps}{0.834}
\setres{cuA}{CPT_crpsfit}{kappa_crps_se}{0.023}
\setres{cuA}{CPT_crpsfit}{r_dist}{0.207}
\setres{cuA}{CPT_deterministic}{kappa_crps}{0.149}
\setres{cuA}{CPT_deterministic}{kappa_crps_se}{0.049}
\setres{cuA}{CPT_deterministic}{r_dist}{0.343}
\setres{cuA}{CPT_types}{kappa_crps}{0.566}
\setres{cuA}{CPT_types}{kappa_crps_se}{0.032}
\setres{cuA}{CPT_types}{r_dist}{0.256}
\setres{cuA}{CPT_uniform}{kappa_crps}{0.758}
\setres{cuA}{CPT_uniform}{kappa_crps_se}{0.012}
\setres{cuA}{CPT_uniform}{r_dist}{0.124}
\setres{cuA}{DA_crpsfit}{kappa_crps}{0.447}
\setres{cuA}{DA_crpsfit}{kappa_crps_se}{0.052}
\setres{cuA}{DA_crpsfit}{r_dist}{0.343}
\setres{cuA}{DA_deterministic}{kappa_crps}{-0.178}
\setres{cuA}{DA_deterministic}{kappa_crps_se}{0.097}
\setres{cuA}{DA_deterministic}{r_dist}{0.565}
\setres{cuA}{DA_types}{kappa_crps}{0.252}
\setres{cuA}{DA_types}{kappa_crps_se}{0.073}
\setres{cuA}{DA_types}{r_dist}{0.447}
\setres{cuA}{DA_uniform}{kappa_crps}{0.431}
\setres{cuA}{DA_uniform}{kappa_crps_se}{0.053}
\setres{cuA}{DA_uniform}{r_dist}{0.350}
\setres{cuA}{EU_crpsfit}{kappa_crps}{0.278}
\setres{cuA}{EU_crpsfit}{kappa_crps_se}{0.067}
\setres{cuA}{EU_crpsfit}{r_dist}{0.411}
\setres{cuA}{EU_deterministic}{kappa_crps}{-0.198}
\setres{cuA}{EU_deterministic}{kappa_crps_se}{0.101}
\setres{cuA}{EU_deterministic}{r_dist}{0.552}
\setres{cuA}{EU_types}{kappa_crps}{-0.095}
\setres{cuA}{EU_types}{kappa_crps_se}{0.096}
\setres{cuA}{EU_types}{r_dist}{0.532}
\setres{cuA}{EU_uniform}{kappa_crps}{0.271}
\setres{cuA}{EU_uniform}{kappa_crps_se}{0.068}
\setres{cuA}{EU_uniform}{r_dist}{0.410}
\setres{cuA}{EV_deterministic}{kappa_crps}{-0.203}
\setres{cuA}{EV_deterministic}{kappa_crps_se}{0.102}
\setres{cuA}{EV_deterministic}{r_dist}{0.554}
\setres{csSG}{subjects}{n_boot}{1{,}000}
\setres{csSG}{subjects}{n_subjects}{179}
\setres{csSG}{subjects}{r_dist}{0.185}
\setres{csSG}{subjects}{r_dist_se}{0.009}
\setres{csSL}{subjects}{n_boot}{1{,}000}
\setres{csSL}{subjects}{n_subjects}{179}
\setres{csSL}{subjects}{r_dist}{0.172}
\setres{csSL}{subjects}{r_dist_se}{0.010}
\setres{indcov}{all_three}{between}{0.0138}
\setres{indcov}{all_three}{between_p}{0.07}
\setres{indcov}{all_three}{between_ref_median}{0.0114}
\setres{indcov}{all_three}{between_ref_p95}{0.0140}
\setres{indcov}{all_three}{cells}{8.000}
\setres{indcov}{all_three}{fit_gain_crps_pct}{1.9}
\setres{indcov}{all_three}{fit_gain_crps_pct_se}{1.1}
\setres{indcov}{all_three}{fit_gain_pct}{1.3}
\setres{indcov}{all_three}{fit_gain_pct_se}{1.9}
\setres{indcov}{all_three}{fit_p}{0.23}
\setres{indcov}{all_three}{fit_p_crps}{0.04}
\setres{indcov}{all_three}{lookup_gain_crps_pct}{-1.5}
\setres{indcov}{all_three}{lookup_gain_crps_pct_se}{0.7}
\setres{indcov}{all_three}{lookup_gain_pct}{-1.1}
\setres{indcov}{all_three}{lookup_gain_pct_se}{1.3}
\setres{indcov}{all_three}{lookup_p}{0.79}
\setres{indcov}{all_three}{lookup_p_crps}{0.98}
\setres{indcov}{all_three}{r_star_max}{0.250}
\setres{indcov}{all_three}{r_star_min}{0.139}
\setres{indcov}{female}{between}{0.0041}
\setres{indcov}{female}{between_p}{0.01}
\setres{indcov}{female}{between_ref_median}{0.0015}
\setres{indcov}{female}{between_ref_p95}{0.0027}
\setres{indcov}{female}{cells}{2.000}
\setres{indcov}{female}{fit_gain_crps_pct}{2.7}
\setres{indcov}{female}{fit_gain_crps_pct_se}{1.2}
\setres{indcov}{female}{fit_gain_pct}{1.4}
\setres{indcov}{female}{fit_gain_pct_se}{2.4}
\setres{indcov}{female}{fit_p}{0.25}
\setres{indcov}{female}{fit_p_crps}{0.01}
\setres{indcov}{female}{lookup_gain_crps_pct}{0.1}
\setres{indcov}{female}{lookup_gain_crps_pct_se}{0.5}
\setres{indcov}{female}{lookup_gain_pct}{0.0}
\setres{indcov}{female}{lookup_gain_pct_se}{0.8}
\setres{indcov}{female}{lookup_p}{0.51}
\setres{indcov}{female}{lookup_p_crps}{0.41}
\setres{indcov}{female}{r_star_max}{0.208}
\setres{indcov}{female}{r_star_min}{0.153}
\setres{indcov}{female_x_highincome}{between}{0.0077}
\setres{indcov}{female_x_highincome}{between_p}{0.00}
\setres{indcov}{female_x_highincome}{between_ref_median}{0.0049}
\setres{indcov}{female_x_highincome}{between_ref_p95}{0.0064}
\setres{indcov}{female_x_highincome}{cells}{4.000}
\setres{indcov}{female_x_highincome}{fit_gain_crps_pct}{1.9}
\setres{indcov}{female_x_highincome}{fit_gain_crps_pct_se}{1.2}
\setres{indcov}{female_x_highincome}{fit_gain_pct}{2.5}
\setres{indcov}{female_x_highincome}{fit_gain_pct_se}{2.6}
\setres{indcov}{female_x_highincome}{fit_p}{0.17}
\setres{indcov}{female_x_highincome}{fit_p_crps}{0.05}
\setres{indcov}{female_x_highincome}{lookup_gain_crps_pct}{-1.4}
\setres{indcov}{female_x_highincome}{lookup_gain_crps_pct_se}{0.9}
\setres{indcov}{female_x_highincome}{lookup_gain_pct}{-0.5}
\setres{indcov}{female_x_highincome}{lookup_gain_pct_se}{1.9}
\setres{indcov}{female_x_highincome}{lookup_p}{0.61}
\setres{indcov}{female_x_highincome}{lookup_p_crps}{0.94}
\setres{indcov}{female_x_highincome}{r_star_max}{0.216}
\setres{indcov}{female_x_highincome}{r_star_min}{0.154}
\setres{indcov}{highincome}{between}{0.0014}
\setres{indcov}{highincome}{between_p}{0.68}
\setres{indcov}{highincome}{between_ref_median}{0.0016}
\setres{indcov}{highincome}{between_ref_p95}{0.0025}
\setres{indcov}{highincome}{cells}{2.000}
\setres{indcov}{highincome}{fit_gain_crps_pct}{0.7}
\setres{indcov}{highincome}{fit_gain_crps_pct_se}{0.9}
\setres{indcov}{highincome}{fit_gain_pct}{1.1}
\setres{indcov}{highincome}{fit_gain_pct_se}{1.6}
\setres{indcov}{highincome}{fit_p}{0.25}
\setres{indcov}{highincome}{fit_p_crps}{0.23}
\setres{indcov}{highincome}{lookup_gain_crps_pct}{-1.2}
\setres{indcov}{highincome}{lookup_gain_crps_pct_se}{0.3}
\setres{indcov}{highincome}{lookup_gain_pct}{-1.4}
\setres{indcov}{highincome}{lookup_gain_pct_se}{0.5}
\setres{indcov}{highincome}{lookup_p}{1.00}
\setres{indcov}{highincome}{lookup_p_crps}{1.00}
\setres{indcov}{highincome}{r_star_max}{0.182}
\setres{indcov}{highincome}{r_star_min}{0.180}
\setres{indcov}{reference}{family}{CPT}
\setres{indcov}{reference}{min_cell}{8.000}
\setres{indcov}{reference}{n_boot}{2{,}000}
\setres{indcov}{reference}{n_draws}{1{,}500}
\setres{indcov}{reference}{n_folds}{10}
\setres{indcov}{reference}{n_split}{200}
\setres{indcov}{reference}{n_subjects}{166}
\setres{indcov}{reference}{r_star_pooled}{0.180}
\setres{indcov}{semester}{between}{0.0014}
\setres{indcov}{semester}{between_p}{0.74}
\setres{indcov}{semester}{between_ref_median}{0.0016}
\setres{indcov}{semester}{between_ref_p95}{0.0025}
\setres{indcov}{semester}{cells}{2.000}
\setres{indcov}{semester}{fit_gain_crps_pct}{0.5}
\setres{indcov}{semester}{fit_gain_crps_pct_se}{1.0}
\setres{indcov}{semester}{fit_gain_pct}{0.8}
\setres{indcov}{semester}{fit_gain_pct_se}{1.7}
\setres{indcov}{semester}{fit_p}{0.30}
\setres{indcov}{semester}{fit_p_crps}{0.31}
\setres{indcov}{semester}{lookup_gain_crps_pct}{-0.9}
\setres{indcov}{semester}{lookup_gain_crps_pct_se}{0.2}
\setres{indcov}{semester}{lookup_gain_pct}{-0.3}
\setres{indcov}{semester}{lookup_gain_pct_se}{0.7}
\setres{indcov}{semester}{lookup_p}{0.66}
\setres{indcov}{semester}{lookup_p_crps}{1.00}
\setres{indcov}{semester}{r_star_max}{0.182}
\setres{indcov}{semester}{r_star_min}{0.181}
\setres{inddec}{CPT}{cov50}{0.32}
\setres{inddec}{CPT}{cov90}{0.64}
\setres{inddec}{CPT}{err_crps_loc_pct}{-14.6}
\setres{inddec}{CPT}{err_crps_pct}{-9.9}
\setres{inddec}{CPT}{err_crps_shape_pct}{-13.6}
\setres{inddec}{CPT}{kappa_ind_crps}{0.58}
\setres{inddec}{CPT}{kappa_ind_crps_run}{0.58}
\setres{inddec}{CPT}{kappa_ind_sq}{0.87}
\setres{inddec}{CPT}{kappa_loc}{0.86}
\setres{inddec}{CPT}{kappa_loc_se}{0.08}
\setres{inddec}{CPT}{kappa_shape}{0.80}
\setres{inddec}{CPT}{kappa_shape_se}{0.05}
\setres{inddec}{CPT}{noise_c_pct}{6.3}
\setres{inddec}{CPT}{ratio}{2.24}
\setres{inddec}{CPT}{resid_pct}{15.6}
\setres{inddec}{CPT}{rmse_pct}{14.8}
\setres{inddec}{CPT}{sd_par_pct}{2.0}
\setres{inddec}{CPT}{sd_pct}{6.6}
\setres{inddec}{CPT}{share_loc_pct}{42}
\setres{inddec}{CPT}{share_spread_pct}{58}
\setres{inddec}{CPT}{spread_first_pct}{67}
\setres{inddec}{CPT}{spread_last_pct}{48}
\setres{inddec}{DA}{cov50}{0.29}
\setres{inddec}{DA}{cov90}{0.56}
\setres{inddec}{DA}{err_crps_loc_pct}{24.2}
\setres{inddec}{DA}{err_crps_pct}{31.7}
\setres{inddec}{DA}{err_crps_shape_pct}{-16.5}
\setres{inddec}{DA}{kappa_ind_crps}{-1.87}
\setres{inddec}{DA}{kappa_ind_crps_run}{-1.87}
\setres{inddec}{DA}{kappa_ind_sq}{-1.13}
\setres{inddec}{DA}{kappa_loc}{-1.43}
\setres{inddec}{DA}{kappa_loc_se}{0.33}
\setres{inddec}{DA}{kappa_shape}{0.97}
\setres{inddec}{DA}{kappa_shape_se}{0.03}
\setres{inddec}{DA}{noise_c_pct}{8.6}
\setres{inddec}{DA}{ratio}{2.40}
\setres{inddec}{DA}{resid_pct}{22.0}
\setres{inddec}{DA}{rmse_pct}{21.5}
\setres{inddec}{DA}{sd_par_pct}{2.0}
\setres{inddec}{DA}{sd_pct}{9.0}
\setres{inddec}{DA}{share_loc_pct}{92}
\setres{inddec}{DA}{share_spread_pct}{8}
\setres{inddec}{DA}{spread_first_pct}{15}
\setres{inddec}{DA}{spread_last_pct}{1}
\setres{inddec}{EU}{cov50}{0.41}
\setres{inddec}{EU}{cov90}{0.75}
\setres{inddec}{EU}{err_crps_loc_pct}{27.9}
\setres{inddec}{EU}{err_crps_pct}{28.3}
\setres{inddec}{EU}{err_crps_shape_pct}{-16.7}
\setres{inddec}{EU}{kappa_ind_crps}{-1.67}
\setres{inddec}{EU}{kappa_ind_crps_run}{-1.67}
\setres{inddec}{EU}{kappa_ind_sq}{-1.27}
\setres{inddec}{EU}{kappa_loc}{-1.64}
\setres{inddec}{EU}{kappa_loc_se}{0.29}
\setres{inddec}{EU}{kappa_shape}{0.99}
\setres{inddec}{EU}{kappa_shape_se}{0.01}
\setres{inddec}{EU}{noise_c_pct}{13.7}
\setres{inddec}{EU}{ratio}{1.62}
\setres{inddec}{EU}{resid_pct}{22.6}
\setres{inddec}{EU}{rmse_pct}{22.6}
\setres{inddec}{EU}{sd_par_pct}{1.5}
\setres{inddec}{EU}{sd_pct}{14.0}
\setres{inddec}{EU}{share_loc_pct}{99}
\setres{inddec}{EU}{share_spread_pct}{1}
\setres{inddec}{EU}{spread_first_pct}{1}
\setres{inddec}{EU}{spread_last_pct}{1}
\setres{inddec}{pop}{cov50}{0.51}
\setres{inddec}{pop}{cov90}{0.90}
\setres{inddec}{pop}{kappa_ind_crps}{0.00}
\setres{inddec}{pop}{ratio}{1.01}
\setres{inddec}{pop}{rmse_pct}{17.7}
\setres{inddec}{pop}{sd_pct}{17.5}
\setres{inddec}{reference}{crps_pop}{3.073}
\setres{inddec}{reference}{crps_rf}{2.552}
\setres{inddec}{reference}{k_full}{24}
\setres{inddec}{reference}{n_boot}{200}
\setres{inddec}{reference}{n_folds}{10}
\setres{inddec}{reference}{n_lotteries}{25}
\setres{inddec}{reference}{n_subjects}{166}
\setres{inddec}{reference}{rf_gain_crps_pct_refit}{17.0}
\setres{inddec}{reference}{rf_gain_pct_refit}{28.6}
\setres{inddec}{reference}{sample}{Bruhin2010\_gains}
\setres{inddec}{rf}{cov50}{0.60}
\setres{inddec}{rf}{cov90}{0.91}
\setres{inddec}{rf}{kappa_ind_crps}{1.00}
\setres{inddec}{rf}{ratio}{1.02}
\setres{inddec}{rf}{rmse_pct}{14.5}
\setres{inddec}{rf}{sd_pct}{14.2}
\setres{ind}{CPT_crpsfit_k0}{err_crps_pct}{6.2}
\setres{ind}{CPT_crpsfit_k0}{err_pct}{4.3}
\setres{ind}{CPT_crpsfit_k0}{gain_crps_pct}{-6.2}
\setres{ind}{CPT_crpsfit_k0}{gain_pct}{-4.3}
\setres{ind}{CPT_crpsfit_k0}{k}{0}
\setres{ind}{CPT_crpsfit_k1}{err_crps_pct}{11.3}
\setres{ind}{CPT_crpsfit_k1}{err_pct}{14.2}
\setres{ind}{CPT_crpsfit_k1}{gain_crps_pct}{-11.3}
\setres{ind}{CPT_crpsfit_k1}{gain_pct}{-14.2}
\setres{ind}{CPT_crpsfit_k1}{k}{1}
\setres{ind}{CPT_crpsfit_k1}{kappa_ind}{2.14}
\setres{ind}{CPT_crpsfit_k1}{kappa_ind_crps}{2.16}
\setres{ind}{CPT_crpsfit_k1}{kappa_ind_crps_nn}{11.073}
\setres{ind}{CPT_crpsfit_k1}{kappa_ind_nn}{941.287}
\setres{ind}{CPT_crpsfit_k16}{err_crps_pct}{-7.2}
\setres{ind}{CPT_crpsfit_k16}{err_pct}{-17.5}
\setres{ind}{CPT_crpsfit_k16}{gain_crps_pct}{7.2}
\setres{ind}{CPT_crpsfit_k16}{gain_pct}{17.5}
\setres{ind}{CPT_crpsfit_k16}{k}{16}
\setres{ind}{CPT_crpsfit_k16}{kappa_ind}{0.73}
\setres{ind}{CPT_crpsfit_k16}{kappa_ind_crps}{0.58}
\setres{ind}{CPT_crpsfit_k16}{kappa_ind_crps_nn}{0.510}
\setres{ind}{CPT_crpsfit_k16}{kappa_ind_nn}{0.871}
\setres{ind}{CPT_crpsfit_k2}{err_crps_pct}{9.9}
\setres{ind}{CPT_crpsfit_k2}{err_pct}{13.2}
\setres{ind}{CPT_crpsfit_k2}{gain_crps_pct}{-9.9}
\setres{ind}{CPT_crpsfit_k2}{gain_pct}{-13.2}
\setres{ind}{CPT_crpsfit_k2}{k}{2}
\setres{ind}{CPT_crpsfit_k2}{kappa_ind}{3.12}
\setres{ind}{CPT_crpsfit_k2}{kappa_ind_crps}{2.40}
\setres{ind}{CPT_crpsfit_k2}{kappa_ind_crps_nn}{-2.093}
\setres{ind}{CPT_crpsfit_k2}{kappa_ind_nn}{-1.544}
\setres{ind}{CPT_crpsfit_k24}{err_crps_pct}{-9.4}
\setres{ind}{CPT_crpsfit_k24}{err_pct}{-21.9}
\setres{ind}{CPT_crpsfit_k24}{gain0_crps_pct}{45.558}
\setres{ind}{CPT_crpsfit_k24}{gain0_pct}{63.727}
\setres{ind}{CPT_crpsfit_k24}{gain_crps_pct}{9.4}
\setres{ind}{CPT_crpsfit_k24}{gain_crps_pct_se}{2.9}
\setres{ind}{CPT_crpsfit_k24}{gain_pct}{21.9}
\setres{ind}{CPT_crpsfit_k24}{gain_pct_se}{6.5}
\setres{ind}{CPT_crpsfit_k24}{k}{24}
\setres{ind}{CPT_crpsfit_k24}{kappa_ind}{0.77}
\setres{ind}{CPT_crpsfit_k24}{kappa_ind0}{0.954}
\setres{ind}{CPT_crpsfit_k24}{kappa_ind0_crps}{0.909}
\setres{ind}{CPT_crpsfit_k24}{kappa_ind_crps}{0.55}
\setres{ind}{CPT_crpsfit_k24}{kappa_ind_crps_nn}{0.660}
\setres{ind}{CPT_crpsfit_k24}{kappa_ind_crps_se}{0.13}
\setres{ind}{CPT_crpsfit_k24}{kappa_ind_nn}{1.104}
\setres{ind}{CPT_crpsfit_k24}{kappa_ind_se}{0.18}
\setres{ind}{CPT_crpsfit_k4}{err_crps_pct}{4.8}
\setres{ind}{CPT_crpsfit_k4}{err_pct}{0.1}
\setres{ind}{CPT_crpsfit_k4}{gain_crps_pct}{-4.8}
\setres{ind}{CPT_crpsfit_k4}{gain_pct}{-0.1}
\setres{ind}{CPT_crpsfit_k4}{k}{4}
\setres{ind}{CPT_crpsfit_k4}{kappa_ind}{-0.01}
\setres{ind}{CPT_crpsfit_k4}{kappa_ind_crps}{-9.91}
\setres{ind}{CPT_crpsfit_k4}{kappa_ind_crps_nn}{-0.561}
\setres{ind}{CPT_crpsfit_k4}{kappa_ind_nn}{-0.004}
\setres{ind}{CPT_crpsfit_k8}{err_crps_pct}{-1.1}
\setres{ind}{CPT_crpsfit_k8}{err_pct}{-10.2}
\setres{ind}{CPT_crpsfit_k8}{gain_crps_pct}{1.1}
\setres{ind}{CPT_crpsfit_k8}{gain_pct}{10.2}
\setres{ind}{CPT_crpsfit_k8}{k}{8}
\setres{ind}{CPT_crpsfit_k8}{kappa_ind}{0.70}
\setres{ind}{CPT_crpsfit_k8}{kappa_ind_crps}{0.20}
\setres{ind}{CPT_crpsfit_k8}{kappa_ind_crps_nn}{0.093}
\setres{ind}{CPT_crpsfit_k8}{kappa_ind_nn}{0.542}
\setres{ind}{CPT_fit_k1}{err_crps_pct}{118.5}
\setres{ind}{CPT_fit_k1}{err_pct}{114.2}
\setres{ind}{CPT_fit_k1}{gain_crps_pct}{-118.5}
\setres{ind}{CPT_fit_k1}{gain_pct}{-114.2}
\setres{ind}{CPT_fit_k1}{k}{1}
\setres{ind}{CPT_fit_k1}{kappa_ind}{17.20}
\setres{ind}{CPT_fit_k1}{kappa_ind_crps}{22.71}
\setres{ind}{CPT_fit_k1}{kappa_ind_crps_nn}{116.207}
\setres{ind}{CPT_fit_k1}{kappa_ind_nn}{7575.414}
\setres{ind}{CPT_fit_k16}{err_crps_pct}{17.4}
\setres{ind}{CPT_fit_k16}{err_pct}{-19.0}
\setres{ind}{CPT_fit_k16}{gain_crps_pct}{-17.4}
\setres{ind}{CPT_fit_k16}{gain_pct}{19.0}
\setres{ind}{CPT_fit_k16}{k}{16}
\setres{ind}{CPT_fit_k16}{kappa_ind}{0.79}
\setres{ind}{CPT_fit_k16}{kappa_ind_crps}{-1.41}
\setres{ind}{CPT_fit_k16}{kappa_ind_crps_nn}{-1.237}
\setres{ind}{CPT_fit_k16}{kappa_ind_nn}{0.944}
\setres{ind}{CPT_fit_k2}{err_crps_pct}{88.8}
\setres{ind}{CPT_fit_k2}{err_pct}{80.5}
\setres{ind}{CPT_fit_k2}{gain_crps_pct}{-88.8}
\setres{ind}{CPT_fit_k2}{gain_pct}{-80.5}
\setres{ind}{CPT_fit_k2}{k}{2}
\setres{ind}{CPT_fit_k2}{kappa_ind}{18.98}
\setres{ind}{CPT_fit_k2}{kappa_ind_crps}{21.51}
\setres{ind}{CPT_fit_k2}{kappa_ind_crps_nn}{-18.784}
\setres{ind}{CPT_fit_k2}{kappa_ind_nn}{-9.387}
\setres{ind}{CPT_fit_k24}{err_crps_pct}{16.2}
\setres{ind}{CPT_fit_k24}{err_pct}{-23.8}
\setres{ind}{CPT_fit_k24}{gain0_crps_pct}{30.195}
\setres{ind}{CPT_fit_k24}{gain0_pct}{64.630}
\setres{ind}{CPT_fit_k24}{gain_crps_pct}{-16.2}
\setres{ind}{CPT_fit_k24}{gain_crps_pct_se}{3.7}
\setres{ind}{CPT_fit_k24}{gain_pct}{23.8}
\setres{ind}{CPT_fit_k24}{gain_pct_se}{5.3}
\setres{ind}{CPT_fit_k24}{k}{24}
\setres{ind}{CPT_fit_k24}{kappa_ind}{0.83}
\setres{ind}{CPT_fit_k24}{kappa_ind0}{0.967}
\setres{ind}{CPT_fit_k24}{kappa_ind0_crps}{0.602}
\setres{ind}{CPT_fit_k24}{kappa_ind_crps}{-0.95}
\setres{ind}{CPT_fit_k24}{kappa_ind_crps_nn}{-1.141}
\setres{ind}{CPT_fit_k24}{kappa_ind_crps_se}{0.33}
\setres{ind}{CPT_fit_k24}{kappa_ind_nn}{1.202}
\setres{ind}{CPT_fit_k24}{kappa_ind_se}{0.15}
\setres{ind}{CPT_fit_k4}{err_crps_pct}{52.3}
\setres{ind}{CPT_fit_k4}{err_pct}{32.2}
\setres{ind}{CPT_fit_k4}{gain_crps_pct}{-52.3}
\setres{ind}{CPT_fit_k4}{gain_pct}{-32.2}
\setres{ind}{CPT_fit_k4}{k}{4}
\setres{ind}{CPT_fit_k4}{kappa_ind}{-5.04}
\setres{ind}{CPT_fit_k4}{kappa_ind_crps}{-107.38}
\setres{ind}{CPT_fit_k4}{kappa_ind_crps_nn}{-6.078}
\setres{ind}{CPT_fit_k4}{kappa_ind_nn}{-2.231}
\setres{ind}{CPT_fit_k8}{err_crps_pct}{27.6}
\setres{ind}{CPT_fit_k8}{err_pct}{-10.3}
\setres{ind}{CPT_fit_k8}{gain_crps_pct}{-27.6}
\setres{ind}{CPT_fit_k8}{gain_pct}{10.3}
\setres{ind}{CPT_fit_k8}{k}{8}
\setres{ind}{CPT_fit_k8}{kappa_ind}{0.70}
\setres{ind}{CPT_fit_k8}{kappa_ind_crps}{-4.77}
\setres{ind}{CPT_fit_k8}{kappa_ind_crps_nn}{-2.243}
\setres{ind}{CPT_fit_k8}{kappa_ind_nn}{0.547}
\setres{ind}{CPT_genderfit_k0}{err_crps_pct}{3.5}
\setres{ind}{CPT_genderfit_k0}{err_pct}{2.8}
\setres{ind}{CPT_genderfit_k0}{gain0_crps_pct}{37.807}
\setres{ind}{CPT_genderfit_k0}{gain0_pct}{52.260}
\setres{ind}{CPT_genderfit_k0}{gain_crps_pct}{-3.5}
\setres{ind}{CPT_genderfit_k0}{gain_crps_pct_se}{1.0}
\setres{ind}{CPT_genderfit_k0}{gain_pct}{-2.8}
\setres{ind}{CPT_genderfit_k0}{gain_pct_se}{2.0}
\setres{ind}{CPT_genderfit_k0}{k}{0}
\setres{ind}{CPT_genderfit_k0}{kappa_ind}{-0.10}
\setres{ind}{CPT_genderfit_k0}{kappa_ind0}{0.782}
\setres{ind}{CPT_genderfit_k0}{kappa_ind0_crps}{0.754}
\setres{ind}{CPT_genderfit_k0}{kappa_ind_crps}{-0.21}
\setres{ind}{CPT_genderfit_k0}{kappa_pop}{0.976}
\setres{ind}{CPT_genderfit_k0}{kappa_pop_crps}{0.947}
\setres{ind}{CPT_genderfit_k0}{r_dist}{0.204}
\setres{ind}{CPT_genderfit_k0}{r_ff}{0.140}
\setres{ind}{CPT_genderfit_k0}{r_prior}{0.9924}
\setres{ind}{CPT_genderfit_k0}{r_prob}{0.993}
\setres{ind}{CPT_genderlit_k0}{err_crps_pct}{21.7}
\setres{ind}{CPT_genderlit_k0}{err_pct}{16.3}
\setres{ind}{CPT_genderlit_k0}{gain0_crps_pct}{26.919}
\setres{ind}{CPT_genderlit_k0}{gain0_pct}{45.985}
\setres{ind}{CPT_genderlit_k0}{gain_crps_pct}{-21.7}
\setres{ind}{CPT_genderlit_k0}{gain_crps_pct_se}{1.8}
\setres{ind}{CPT_genderlit_k0}{gain_pct}{-16.3}
\setres{ind}{CPT_genderlit_k0}{gain_pct_se}{2.4}
\setres{ind}{CPT_genderlit_k0}{k}{0}
\setres{ind}{CPT_genderlit_k0}{kappa_ind}{-0.57}
\setres{ind}{CPT_genderlit_k0}{kappa_ind0}{0.688}
\setres{ind}{CPT_genderlit_k0}{kappa_ind0_crps}{0.537}
\setres{ind}{CPT_genderlit_k0}{kappa_ind_crps}{-1.28}
\setres{ind}{CPT_genderlit_k0}{kappa_pop}{0.859}
\setres{ind}{CPT_genderlit_k0}{kappa_pop_crps}{0.674}
\setres{ind}{CPT_genderlit_k0}{r_dist}{0.391}
\setres{ind}{CPT_genderlit_k0}{r_ff}{0.140}
\setres{ind}{CPT_genderlit_k0}{r_prior}{0.9502}
\setres{ind}{CPT_genderlit_k0}{r_prob}{0.957}
\setres{ind}{CPT_lit_k0}{err_crps_pct}{17.4}
\setres{ind}{CPT_lit_k0}{err_pct}{17.0}
\setres{ind}{CPT_lit_k0}{gain_crps_pct}{-17.4}
\setres{ind}{CPT_lit_k0}{gain_pct}{-17.0}
\setres{ind}{CPT_lit_k0}{k}{0}
\setres{ind}{CPT_lit_k1}{err_crps_pct}{15.1}
\setres{ind}{CPT_lit_k1}{err_pct}{14.2}
\setres{ind}{CPT_lit_k1}{gain_crps_pct}{-15.1}
\setres{ind}{CPT_lit_k1}{gain_pct}{-14.2}
\setres{ind}{CPT_lit_k1}{k}{1}
\setres{ind}{CPT_lit_k1}{kappa_ind}{2.14}
\setres{ind}{CPT_lit_k1}{kappa_ind_crps}{2.90}
\setres{ind}{CPT_lit_k1}{kappa_ind_crps_nn}{14.818}
\setres{ind}{CPT_lit_k1}{kappa_ind_nn}{942.642}
\setres{ind}{CPT_lit_k16}{err_crps_pct}{-8.6}
\setres{ind}{CPT_lit_k16}{err_pct}{-21.6}
\setres{ind}{CPT_lit_k16}{gain_crps_pct}{8.6}
\setres{ind}{CPT_lit_k16}{gain_pct}{21.6}
\setres{ind}{CPT_lit_k16}{k}{16}
\setres{ind}{CPT_lit_k16}{kappa_ind}{0.90}
\setres{ind}{CPT_lit_k16}{kappa_ind_crps}{0.69}
\setres{ind}{CPT_lit_k16}{kappa_ind_crps_nn}{0.609}
\setres{ind}{CPT_lit_k16}{kappa_ind_nn}{1.075}
\setres{ind}{CPT_lit_k2}{err_crps_pct}{8.4}
\setres{ind}{CPT_lit_k2}{err_pct}{4.1}
\setres{ind}{CPT_lit_k2}{gain_crps_pct}{-8.4}
\setres{ind}{CPT_lit_k2}{gain_pct}{-4.1}
\setres{ind}{CPT_lit_k2}{k}{2}
\setres{ind}{CPT_lit_k2}{kappa_ind}{0.96}
\setres{ind}{CPT_lit_k2}{kappa_ind_crps}{2.04}
\setres{ind}{CPT_lit_k2}{kappa_ind_crps_nn}{-1.778}
\setres{ind}{CPT_lit_k2}{kappa_ind_nn}{-0.473}
\setres{ind}{CPT_lit_k24}{err_crps_pct}{-10.3}
\setres{ind}{CPT_lit_k24}{err_pct}{-26.0}
\setres{ind}{CPT_lit_k24}{gain0_crps_pct}{46.143}
\setres{ind}{CPT_lit_k24}{gain0_pct}{65.639}
\setres{ind}{CPT_lit_k24}{gain_crps_pct}{10.3}
\setres{ind}{CPT_lit_k24}{gain_crps_pct_se}{2.9}
\setres{ind}{CPT_lit_k24}{gain_pct}{26.0}
\setres{ind}{CPT_lit_k24}{gain_pct_se}{5.2}
\setres{ind}{CPT_lit_k24}{k}{24}
\setres{ind}{CPT_lit_k24}{kappa_ind}{0.91}
\setres{ind}{CPT_lit_k24}{kappa_ind0}{0.982}
\setres{ind}{CPT_lit_k24}{kappa_ind0_crps}{0.921}
\setres{ind}{CPT_lit_k24}{kappa_ind_crps}{0.61}
\setres{ind}{CPT_lit_k24}{kappa_ind_crps_nn}{0.729}
\setres{ind}{CPT_lit_k24}{kappa_ind_crps_se}{0.12}
\setres{ind}{CPT_lit_k24}{kappa_ind_nn}{1.311}
\setres{ind}{CPT_lit_k24}{kappa_ind_se}{0.13}
\setres{ind}{CPT_lit_k4}{err_crps_pct}{3.9}
\setres{ind}{CPT_lit_k4}{err_pct}{-1.2}
\setres{ind}{CPT_lit_k4}{gain_crps_pct}{-3.9}
\setres{ind}{CPT_lit_k4}{gain_pct}{1.2}
\setres{ind}{CPT_lit_k4}{k}{4}
\setres{ind}{CPT_lit_k4}{kappa_ind}{0.19}
\setres{ind}{CPT_lit_k4}{kappa_ind_crps}{-7.94}
\setres{ind}{CPT_lit_k4}{kappa_ind_crps_nn}{-0.449}
\setres{ind}{CPT_lit_k4}{kappa_ind_nn}{0.086}
\setres{ind}{CPT_lit_k8}{err_crps_pct}{-3.0}
\setres{ind}{CPT_lit_k8}{err_pct}{-16.2}
\setres{ind}{CPT_lit_k8}{gain_crps_pct}{3.0}
\setres{ind}{CPT_lit_k8}{gain_pct}{16.2}
\setres{ind}{CPT_lit_k8}{k}{8}
\setres{ind}{CPT_lit_k8}{kappa_ind}{1.11}
\setres{ind}{CPT_lit_k8}{kappa_ind_crps}{0.51}
\setres{ind}{CPT_lit_k8}{kappa_ind_crps_nn}{0.240}
\setres{ind}{CPT_lit_k8}{kappa_ind_nn}{0.863}
\setres{ind}{CPT_uniform_k0}{err_crps_pct}{18.5}
\setres{ind}{CPT_uniform_k0}{err_pct}{21.1}
\setres{ind}{CPT_uniform_k0}{gain_crps_pct}{-18.5}
\setres{ind}{CPT_uniform_k0}{gain_pct}{-21.1}
\setres{ind}{CPT_uniform_k0}{k}{0}
\setres{ind}{CPT_uniform_k1}{err_crps_pct}{22.9}
\setres{ind}{CPT_uniform_k1}{err_pct}{28.6}
\setres{ind}{CPT_uniform_k1}{gain_crps_pct}{-22.9}
\setres{ind}{CPT_uniform_k1}{gain_pct}{-28.6}
\setres{ind}{CPT_uniform_k1}{k}{1}
\setres{ind}{CPT_uniform_k1}{kappa_ind}{4.31}
\setres{ind}{CPT_uniform_k1}{kappa_ind_crps}{4.39}
\setres{ind}{CPT_uniform_k1}{kappa_ind_crps_nn}{22.435}
\setres{ind}{CPT_uniform_k1}{kappa_ind_nn}{1896.972}
\setres{ind}{CPT_uniform_k16}{err_crps_pct}{-8.0}
\setres{ind}{CPT_uniform_k16}{err_pct}{-20.2}
\setres{ind}{CPT_uniform_k16}{gain_crps_pct}{8.0}
\setres{ind}{CPT_uniform_k16}{gain_pct}{20.2}
\setres{ind}{CPT_uniform_k16}{k}{16}
\setres{ind}{CPT_uniform_k16}{kappa_ind}{0.84}
\setres{ind}{CPT_uniform_k16}{kappa_ind_crps}{0.64}
\setres{ind}{CPT_uniform_k16}{kappa_ind_crps_nn}{0.565}
\setres{ind}{CPT_uniform_k16}{kappa_ind_nn}{1.004}
\setres{ind}{CPT_uniform_k2}{err_crps_pct}{21.5}
\setres{ind}{CPT_uniform_k2}{err_pct}{26.1}
\setres{ind}{CPT_uniform_k2}{gain_crps_pct}{-21.5}
\setres{ind}{CPT_uniform_k2}{gain_pct}{-26.1}
\setres{ind}{CPT_uniform_k2}{k}{2}
\setres{ind}{CPT_uniform_k2}{kappa_ind}{6.17}
\setres{ind}{CPT_uniform_k2}{kappa_ind_crps}{5.21}
\setres{ind}{CPT_uniform_k2}{kappa_ind_crps_nn}{-4.550}
\setres{ind}{CPT_uniform_k2}{kappa_ind_nn}{-3.050}
\setres{ind}{CPT_uniform_k24}{err_crps_pct}{-9.9}
\setres{ind}{CPT_uniform_k24}{err_pct}{-24.9}
\setres{ind}{CPT_uniform_k24}{gain0_crps_pct}{45.866}
\setres{ind}{CPT_uniform_k24}{gain0_pct}{65.123}
\setres{ind}{CPT_uniform_k24}{gain_crps_pct}{9.9}
\setres{ind}{CPT_uniform_k24}{gain_crps_pct_se}{2.9}
\setres{ind}{CPT_uniform_k24}{gain_pct}{24.9}
\setres{ind}{CPT_uniform_k24}{gain_pct_se}{5.3}
\setres{ind}{CPT_uniform_k24}{k}{24}
\setres{ind}{CPT_uniform_k24}{kappa_ind}{0.87}
\setres{ind}{CPT_uniform_k24}{kappa_ind0}{0.975}
\setres{ind}{CPT_uniform_k24}{kappa_ind0_crps}{0.915}
\setres{ind}{CPT_uniform_k24}{kappa_ind_crps}{0.58}
\setres{ind}{CPT_uniform_k24}{kappa_ind_crps_nn}{0.696}
\setres{ind}{CPT_uniform_k24}{kappa_ind_crps_se}{0.12}
\setres{ind}{CPT_uniform_k24}{kappa_ind_nn}{1.255}
\setres{ind}{CPT_uniform_k24}{kappa_ind_se}{0.14}
\setres{ind}{CPT_uniform_k24}{n_ind_r}{40}
\setres{ind}{CPT_uniform_k24}{r_dist}{0.305}
\setres{ind}{CPT_uniform_k24}{r_ff}{0.140}
\setres{ind}{CPT_uniform_k24}{r_prior}{0.9987}
\setres{ind}{CPT_uniform_k24}{r_prior_sd}{0.001}
\setres{ind}{CPT_uniform_k24}{r_prob}{0.999}
\setres{ind}{CPT_uniform_k4}{err_crps_pct}{13.0}
\setres{ind}{CPT_uniform_k4}{err_pct}{15.5}
\setres{ind}{CPT_uniform_k4}{gain_crps_pct}{-13.0}
\setres{ind}{CPT_uniform_k4}{gain_pct}{-15.5}
\setres{ind}{CPT_uniform_k4}{k}{4}
\setres{ind}{CPT_uniform_k4}{kappa_ind}{-2.43}
\setres{ind}{CPT_uniform_k4}{kappa_ind_crps}{-26.65}
\setres{ind}{CPT_uniform_k4}{kappa_ind_crps_nn}{-1.508}
\setres{ind}{CPT_uniform_k4}{kappa_ind_nn}{-1.078}
\setres{ind}{CPT_uniform_k8}{err_crps_pct}{0.0}
\setres{ind}{CPT_uniform_k8}{err_pct}{-11.5}
\setres{ind}{CPT_uniform_k8}{gain_crps_pct}{0.0}
\setres{ind}{CPT_uniform_k8}{gain_pct}{11.5}
\setres{ind}{CPT_uniform_k8}{k}{8}
\setres{ind}{CPT_uniform_k8}{kappa_ind}{0.79}
\setres{ind}{CPT_uniform_k8}{kappa_ind_crps}{-0.01}
\setres{ind}{CPT_uniform_k8}{kappa_ind_crps_nn}{-0.003}
\setres{ind}{CPT_uniform_k8}{kappa_ind_nn}{0.613}
\setres{ind}{DA_crpsfit_k0}{err_crps_pct}{26.4}
\setres{ind}{DA_crpsfit_k0}{err_pct}{27.1}
\setres{ind}{DA_crpsfit_k0}{gain_crps_pct}{-26.4}
\setres{ind}{DA_crpsfit_k0}{gain_pct}{-27.1}
\setres{ind}{DA_crpsfit_k0}{k}{0}
\setres{ind}{DA_crpsfit_k1}{err_crps_pct}{53.6}
\setres{ind}{DA_crpsfit_k1}{err_pct}{86.8}
\setres{ind}{DA_crpsfit_k1}{gain_crps_pct}{-53.6}
\setres{ind}{DA_crpsfit_k1}{gain_pct}{-86.8}
\setres{ind}{DA_crpsfit_k1}{k}{1}
\setres{ind}{DA_crpsfit_k1}{kappa_ind}{13.08}
\setres{ind}{DA_crpsfit_k1}{kappa_ind_crps}{10.27}
\setres{ind}{DA_crpsfit_k1}{kappa_ind_crps_nn}{52.520}
\setres{ind}{DA_crpsfit_k1}{kappa_ind_nn}{5758.122}
\setres{ind}{DA_crpsfit_k16}{err_crps_pct}{32.0}
\setres{ind}{DA_crpsfit_k16}{err_pct}{35.7}
\setres{ind}{DA_crpsfit_k16}{gain_crps_pct}{-32.0}
\setres{ind}{DA_crpsfit_k16}{gain_pct}{-35.7}
\setres{ind}{DA_crpsfit_k16}{k}{16}
\setres{ind}{DA_crpsfit_k16}{kappa_ind}{-1.48}
\setres{ind}{DA_crpsfit_k16}{kappa_ind_crps}{-2.58}
\setres{ind}{DA_crpsfit_k16}{kappa_ind_crps_nn}{-2.267}
\setres{ind}{DA_crpsfit_k16}{kappa_ind_nn}{-1.772}
\setres{ind}{DA_crpsfit_k2}{err_crps_pct}{47.1}
\setres{ind}{DA_crpsfit_k2}{err_pct}{71.5}
\setres{ind}{DA_crpsfit_k2}{gain_crps_pct}{-47.1}
\setres{ind}{DA_crpsfit_k2}{gain_pct}{-71.5}
\setres{ind}{DA_crpsfit_k2}{k}{2}
\setres{ind}{DA_crpsfit_k2}{kappa_ind}{16.86}
\setres{ind}{DA_crpsfit_k2}{kappa_ind_crps}{11.42}
\setres{ind}{DA_crpsfit_k2}{kappa_ind_crps_nn}{-9.973}
\setres{ind}{DA_crpsfit_k2}{kappa_ind_nn}{-8.338}
\setres{ind}{DA_crpsfit_k24}{err_crps_pct}{30.8}
\setres{ind}{DA_crpsfit_k24}{err_pct}{32.7}
\setres{ind}{DA_crpsfit_k24}{gain0_crps_pct}{21.417}
\setres{ind}{DA_crpsfit_k24}{gain0_pct}{38.382}
\setres{ind}{DA_crpsfit_k24}{gain_crps_pct}{-30.8}
\setres{ind}{DA_crpsfit_k24}{gain_crps_pct_se}{4.4}
\setres{ind}{DA_crpsfit_k24}{gain_pct}{-32.7}
\setres{ind}{DA_crpsfit_k24}{gain_pct_se}{10.2}
\setres{ind}{DA_crpsfit_k24}{k}{24}
\setres{ind}{DA_crpsfit_k24}{kappa_ind}{-1.14}
\setres{ind}{DA_crpsfit_k24}{kappa_ind0}{0.574}
\setres{ind}{DA_crpsfit_k24}{kappa_ind0_crps}{0.427}
\setres{ind}{DA_crpsfit_k24}{kappa_ind_crps}{-1.82}
\setres{ind}{DA_crpsfit_k24}{kappa_ind_crps_nn}{-2.171}
\setres{ind}{DA_crpsfit_k24}{kappa_ind_crps_se}{0.41}
\setres{ind}{DA_crpsfit_k24}{kappa_ind_nn}{-1.647}
\setres{ind}{DA_crpsfit_k24}{kappa_ind_se}{0.50}
\setres{ind}{DA_crpsfit_k4}{err_crps_pct}{46.1}
\setres{ind}{DA_crpsfit_k4}{err_pct}{70.5}
\setres{ind}{DA_crpsfit_k4}{gain_crps_pct}{-46.1}
\setres{ind}{DA_crpsfit_k4}{gain_pct}{-70.5}
\setres{ind}{DA_crpsfit_k4}{k}{4}
\setres{ind}{DA_crpsfit_k4}{kappa_ind}{-11.04}
\setres{ind}{DA_crpsfit_k4}{kappa_ind_crps}{-94.53}
\setres{ind}{DA_crpsfit_k4}{kappa_ind_crps_nn}{-5.350}
\setres{ind}{DA_crpsfit_k4}{kappa_ind_nn}{-4.890}
\setres{ind}{DA_crpsfit_k8}{err_crps_pct}{38.8}
\setres{ind}{DA_crpsfit_k8}{err_pct}{49.0}
\setres{ind}{DA_crpsfit_k8}{gain_crps_pct}{-38.8}
\setres{ind}{DA_crpsfit_k8}{gain_pct}{-49.0}
\setres{ind}{DA_crpsfit_k8}{k}{8}
\setres{ind}{DA_crpsfit_k8}{kappa_ind}{-3.35}
\setres{ind}{DA_crpsfit_k8}{kappa_ind_crps}{-6.71}
\setres{ind}{DA_crpsfit_k8}{kappa_ind_crps_nn}{-3.152}
\setres{ind}{DA_crpsfit_k8}{kappa_ind_nn}{-2.607}
\setres{ind}{DA_fit_k1}{err_crps_pct}{139.7}
\setres{ind}{DA_fit_k1}{err_pct}{182.6}
\setres{ind}{DA_fit_k1}{gain_crps_pct}{-139.7}
\setres{ind}{DA_fit_k1}{gain_pct}{-182.6}
\setres{ind}{DA_fit_k1}{k}{1}
\setres{ind}{DA_fit_k1}{kappa_ind}{27.52}
\setres{ind}{DA_fit_k1}{kappa_ind_crps}{26.78}
\setres{ind}{DA_fit_k1}{kappa_ind_crps_nn}{137.003}
\setres{ind}{DA_fit_k1}{kappa_ind_nn}{12116.572}
\setres{ind}{DA_fit_k16}{err_crps_pct}{71.4}
\setres{ind}{DA_fit_k16}{err_pct}{43.0}
\setres{ind}{DA_fit_k16}{gain_crps_pct}{-71.4}
\setres{ind}{DA_fit_k16}{gain_pct}{-43.0}
\setres{ind}{DA_fit_k16}{k}{16}
\setres{ind}{DA_fit_k16}{kappa_ind}{-1.79}
\setres{ind}{DA_fit_k16}{kappa_ind_crps}{-5.76}
\setres{ind}{DA_fit_k16}{kappa_ind_crps_nn}{-5.068}
\setres{ind}{DA_fit_k16}{kappa_ind_nn}{-2.136}
\setres{ind}{DA_fit_k2}{err_crps_pct}{119.8}
\setres{ind}{DA_fit_k2}{err_pct}{133.0}
\setres{ind}{DA_fit_k2}{gain_crps_pct}{-119.8}
\setres{ind}{DA_fit_k2}{gain_pct}{-133.0}
\setres{ind}{DA_fit_k2}{k}{2}
\setres{ind}{DA_fit_k2}{kappa_ind}{31.39}
\setres{ind}{DA_fit_k2}{kappa_ind_crps}{29.04}
\setres{ind}{DA_fit_k2}{kappa_ind_crps_nn}{-25.356}
\setres{ind}{DA_fit_k2}{kappa_ind_nn}{-15.521}
\setres{ind}{DA_fit_k24}{err_crps_pct}{67.1}
\setres{ind}{DA_fit_k24}{err_pct}{34.0}
\setres{ind}{DA_fit_k24}{gain0_crps_pct}{-0.378}
\setres{ind}{DA_fit_k24}{gain0_pct}{37.753}
\setres{ind}{DA_fit_k24}{gain_crps_pct}{-67.1}
\setres{ind}{DA_fit_k24}{gain_crps_pct_se}{6.0}
\setres{ind}{DA_fit_k24}{gain_pct}{-34.0}
\setres{ind}{DA_fit_k24}{gain_pct_se}{11.8}
\setres{ind}{DA_fit_k24}{k}{24}
\setres{ind}{DA_fit_k24}{kappa_ind}{-1.19}
\setres{ind}{DA_fit_k24}{kappa_ind0}{0.565}
\setres{ind}{DA_fit_k24}{kappa_ind0_crps}{-0.008}
\setres{ind}{DA_fit_k24}{kappa_ind_crps}{-3.95}
\setres{ind}{DA_fit_k24}{kappa_ind_crps_nn}{-4.726}
\setres{ind}{DA_fit_k24}{kappa_ind_crps_se}{0.73}
\setres{ind}{DA_fit_k24}{kappa_ind_nn}{-1.715}
\setres{ind}{DA_fit_k24}{kappa_ind_se}{0.57}
\setres{ind}{DA_fit_k4}{err_crps_pct}{101.3}
\setres{ind}{DA_fit_k4}{err_pct}{98.0}
\setres{ind}{DA_fit_k4}{gain_crps_pct}{-101.3}
\setres{ind}{DA_fit_k4}{gain_pct}{-98.0}
\setres{ind}{DA_fit_k4}{k}{4}
\setres{ind}{DA_fit_k4}{kappa_ind}{-15.34}
\setres{ind}{DA_fit_k4}{kappa_ind_crps}{-207.93}
\setres{ind}{DA_fit_k4}{kappa_ind_crps_nn}{-11.769}
\setres{ind}{DA_fit_k4}{kappa_ind_nn}{-6.791}
\setres{ind}{DA_fit_k8}{err_crps_pct}{85.5}
\setres{ind}{DA_fit_k8}{err_pct}{65.2}
\setres{ind}{DA_fit_k8}{gain_crps_pct}{-85.5}
\setres{ind}{DA_fit_k8}{gain_pct}{-65.2}
\setres{ind}{DA_fit_k8}{k}{8}
\setres{ind}{DA_fit_k8}{kappa_ind}{-4.45}
\setres{ind}{DA_fit_k8}{kappa_ind_crps}{-14.77}
\setres{ind}{DA_fit_k8}{kappa_ind_crps_nn}{-6.940}
\setres{ind}{DA_fit_k8}{kappa_ind_nn}{-3.467}
\setres{ind}{DA_genderfit_k0}{err_crps_pct}{26.1}
\setres{ind}{DA_genderfit_k0}{err_pct}{25.0}
\setres{ind}{DA_genderfit_k0}{gain0_crps_pct}{24.281}
\setres{ind}{DA_genderfit_k0}{gain0_pct}{41.958}
\setres{ind}{DA_genderfit_k0}{gain_crps_pct}{-26.1}
\setres{ind}{DA_genderfit_k0}{gain_crps_pct_se}{2.2}
\setres{ind}{DA_genderfit_k0}{gain_pct}{-25.0}
\setres{ind}{DA_genderfit_k0}{gain_pct_se}{4.0}
\setres{ind}{DA_genderfit_k0}{k}{0}
\setres{ind}{DA_genderfit_k0}{kappa_ind}{-0.87}
\setres{ind}{DA_genderfit_k0}{kappa_ind0}{0.628}
\setres{ind}{DA_genderfit_k0}{kappa_ind0_crps}{0.484}
\setres{ind}{DA_genderfit_k0}{kappa_ind_crps}{-1.53}
\setres{ind}{DA_genderfit_k0}{kappa_pop}{0.783}
\setres{ind}{DA_genderfit_k0}{kappa_pop_crps}{0.608}
\setres{ind}{DA_genderfit_k0}{r_dist}{0.310}
\setres{ind}{DA_genderfit_k0}{r_ff}{0.281}
\setres{ind}{DA_genderfit_k0}{r_prior}{0.9962}
\setres{ind}{DA_genderfit_k0}{r_prob}{0.997}
\setres{ind}{DA_lit_k0}{err_crps_pct}{48.3}
\setres{ind}{DA_lit_k0}{err_pct}{72.0}
\setres{ind}{DA_lit_k0}{gain_crps_pct}{-48.3}
\setres{ind}{DA_lit_k0}{gain_pct}{-72.0}
\setres{ind}{DA_lit_k0}{k}{0}
\setres{ind}{DA_lit_k1}{err_crps_pct}{52.1}
\setres{ind}{DA_lit_k1}{err_pct}{81.3}
\setres{ind}{DA_lit_k1}{gain_crps_pct}{-52.1}
\setres{ind}{DA_lit_k1}{gain_pct}{-81.3}
\setres{ind}{DA_lit_k1}{k}{1}
\setres{ind}{DA_lit_k1}{kappa_ind}{12.26}
\setres{ind}{DA_lit_k1}{kappa_ind_crps}{9.99}
\setres{ind}{DA_lit_k1}{kappa_ind_crps_nn}{51.132}
\setres{ind}{DA_lit_k1}{kappa_ind_nn}{5397.472}
\setres{ind}{DA_lit_k16}{err_crps_pct}{34.8}
\setres{ind}{DA_lit_k16}{err_pct}{39.3}
\setres{ind}{DA_lit_k16}{gain_crps_pct}{-34.8}
\setres{ind}{DA_lit_k16}{gain_pct}{-39.3}
\setres{ind}{DA_lit_k16}{k}{16}
\setres{ind}{DA_lit_k16}{kappa_ind}{-1.63}
\setres{ind}{DA_lit_k16}{kappa_ind_crps}{-2.81}
\setres{ind}{DA_lit_k16}{kappa_ind_crps_nn}{-2.467}
\setres{ind}{DA_lit_k16}{kappa_ind_nn}{-1.952}
\setres{ind}{DA_lit_k2}{err_crps_pct}{48.5}
\setres{ind}{DA_lit_k2}{err_pct}{73.7}
\setres{ind}{DA_lit_k2}{gain_crps_pct}{-48.5}
\setres{ind}{DA_lit_k2}{gain_pct}{-73.7}
\setres{ind}{DA_lit_k2}{k}{2}
\setres{ind}{DA_lit_k2}{kappa_ind}{17.39}
\setres{ind}{DA_lit_k2}{kappa_ind_crps}{11.76}
\setres{ind}{DA_lit_k2}{kappa_ind_crps_nn}{-10.270}
\setres{ind}{DA_lit_k2}{kappa_ind_nn}{-8.600}
\setres{ind}{DA_lit_k24}{err_crps_pct}{31.7}
\setres{ind}{DA_lit_k24}{err_pct}{30.6}
\setres{ind}{DA_lit_k24}{gain0_crps_pct}{20.871}
\setres{ind}{DA_lit_k24}{gain0_pct}{39.342}
\setres{ind}{DA_lit_k24}{gain_crps_pct}{-31.7}
\setres{ind}{DA_lit_k24}{gain_crps_pct_se}{4.7}
\setres{ind}{DA_lit_k24}{gain_pct}{-30.6}
\setres{ind}{DA_lit_k24}{gain_pct_se}{11.5}
\setres{ind}{DA_lit_k24}{k}{24}
\setres{ind}{DA_lit_k24}{kappa_ind}{-1.07}
\setres{ind}{DA_lit_k24}{kappa_ind0}{0.589}
\setres{ind}{DA_lit_k24}{kappa_ind0_crps}{0.416}
\setres{ind}{DA_lit_k24}{kappa_ind_crps}{-1.87}
\setres{ind}{DA_lit_k24}{kappa_ind_crps_nn}{-2.235}
\setres{ind}{DA_lit_k24}{kappa_ind_crps_se}{0.44}
\setres{ind}{DA_lit_k24}{kappa_ind_nn}{-1.543}
\setres{ind}{DA_lit_k24}{kappa_ind_se}{0.54}
\setres{ind}{DA_lit_k4}{err_crps_pct}{45.2}
\setres{ind}{DA_lit_k4}{err_pct}{61.1}
\setres{ind}{DA_lit_k4}{gain_crps_pct}{-45.2}
\setres{ind}{DA_lit_k4}{gain_pct}{-61.1}
\setres{ind}{DA_lit_k4}{k}{4}
\setres{ind}{DA_lit_k4}{kappa_ind}{-9.56}
\setres{ind}{DA_lit_k4}{kappa_ind_crps}{-92.86}
\setres{ind}{DA_lit_k4}{kappa_ind_crps_nn}{-5.256}
\setres{ind}{DA_lit_k4}{kappa_ind_nn}{-4.234}
\setres{ind}{DA_lit_k8}{err_crps_pct}{41.1}
\setres{ind}{DA_lit_k8}{err_pct}{49.1}
\setres{ind}{DA_lit_k8}{gain_crps_pct}{-41.1}
\setres{ind}{DA_lit_k8}{gain_pct}{-49.1}
\setres{ind}{DA_lit_k8}{k}{8}
\setres{ind}{DA_lit_k8}{kappa_ind}{-3.35}
\setres{ind}{DA_lit_k8}{kappa_ind_crps}{-7.11}
\setres{ind}{DA_lit_k8}{kappa_ind_crps_nn}{-3.339}
\setres{ind}{DA_lit_k8}{kappa_ind_nn}{-2.609}
\setres{ind}{DA_uniform_k0}{err_crps_pct}{59.6}
\setres{ind}{DA_uniform_k0}{err_pct}{98.3}
\setres{ind}{DA_uniform_k0}{gain_crps_pct}{-59.6}
\setres{ind}{DA_uniform_k0}{gain_pct}{-98.3}
\setres{ind}{DA_uniform_k0}{k}{0}
\setres{ind}{DA_uniform_k1}{err_crps_pct}{61.1}
\setres{ind}{DA_uniform_k1}{err_pct}{98.0}
\setres{ind}{DA_uniform_k1}{gain_crps_pct}{-61.1}
\setres{ind}{DA_uniform_k1}{gain_pct}{-98.0}
\setres{ind}{DA_uniform_k1}{k}{1}
\setres{ind}{DA_uniform_k1}{kappa_ind}{14.77}
\setres{ind}{DA_uniform_k1}{kappa_ind_crps}{11.72}
\setres{ind}{DA_uniform_k1}{kappa_ind_crps_nn}{59.957}
\setres{ind}{DA_uniform_k1}{kappa_ind_nn}{6502.251}
\setres{ind}{DA_uniform_k16}{err_crps_pct}{34.7}
\setres{ind}{DA_uniform_k16}{err_pct}{41.2}
\setres{ind}{DA_uniform_k16}{gain_crps_pct}{-34.7}
\setres{ind}{DA_uniform_k16}{gain_pct}{-41.2}
\setres{ind}{DA_uniform_k16}{k}{16}
\setres{ind}{DA_uniform_k16}{kappa_ind}{-1.71}
\setres{ind}{DA_uniform_k16}{kappa_ind_crps}{-2.80}
\setres{ind}{DA_uniform_k16}{kappa_ind_crps_nn}{-2.462}
\setres{ind}{DA_uniform_k16}{kappa_ind_nn}{-2.044}
\setres{ind}{DA_uniform_k2}{err_crps_pct}{54.6}
\setres{ind}{DA_uniform_k2}{err_pct}{86.2}
\setres{ind}{DA_uniform_k2}{gain_crps_pct}{-54.6}
\setres{ind}{DA_uniform_k2}{gain_pct}{-86.2}
\setres{ind}{DA_uniform_k2}{k}{2}
\setres{ind}{DA_uniform_k2}{kappa_ind}{20.35}
\setres{ind}{DA_uniform_k2}{kappa_ind_crps}{13.23}
\setres{ind}{DA_uniform_k2}{kappa_ind_crps_nn}{-11.549}
\setres{ind}{DA_uniform_k2}{kappa_ind_nn}{-10.062}
\setres{ind}{DA_uniform_k24}{err_crps_pct}{31.7}
\setres{ind}{DA_uniform_k24}{err_pct}{32.3}
\setres{ind}{DA_uniform_k24}{gain0_crps_pct}{20.892}
\setres{ind}{DA_uniform_k24}{gain0_pct}{38.546}
\setres{ind}{DA_uniform_k24}{gain_crps_pct}{-31.7}
\setres{ind}{DA_uniform_k24}{gain_crps_pct_se}{4.7}
\setres{ind}{DA_uniform_k24}{gain_pct}{-32.3}
\setres{ind}{DA_uniform_k24}{gain_pct_se}{11.6}
\setres{ind}{DA_uniform_k24}{k}{24}
\setres{ind}{DA_uniform_k24}{kappa_ind}{-1.13}
\setres{ind}{DA_uniform_k24}{kappa_ind0}{0.577}
\setres{ind}{DA_uniform_k24}{kappa_ind0_crps}{0.417}
\setres{ind}{DA_uniform_k24}{kappa_ind_crps}{-1.87}
\setres{ind}{DA_uniform_k24}{kappa_ind_crps_nn}{-2.232}
\setres{ind}{DA_uniform_k24}{kappa_ind_crps_se}{0.44}
\setres{ind}{DA_uniform_k24}{kappa_ind_nn}{-1.629}
\setres{ind}{DA_uniform_k24}{kappa_ind_se}{0.55}
\setres{ind}{DA_uniform_k24}{n_ind_r}{40}
\setres{ind}{DA_uniform_k24}{r_dist}{0.480}
\setres{ind}{DA_uniform_k24}{r_ff}{0.281}
\setres{ind}{DA_uniform_k24}{r_prior}{0.9964}
\setres{ind}{DA_uniform_k24}{r_prior_sd}{0.005}
\setres{ind}{DA_uniform_k24}{r_prob}{0.997}
\setres{ind}{DA_uniform_k4}{err_crps_pct}{49.7}
\setres{ind}{DA_uniform_k4}{err_pct}{72.3}
\setres{ind}{DA_uniform_k4}{gain_crps_pct}{-49.7}
\setres{ind}{DA_uniform_k4}{gain_pct}{-72.3}
\setres{ind}{DA_uniform_k4}{k}{4}
\setres{ind}{DA_uniform_k4}{kappa_ind}{-11.32}
\setres{ind}{DA_uniform_k4}{kappa_ind_crps}{-102.02}
\setres{ind}{DA_uniform_k4}{kappa_ind_crps_nn}{-5.774}
\setres{ind}{DA_uniform_k4}{kappa_ind_nn}{-5.010}
\setres{ind}{DA_uniform_k8}{err_crps_pct}{43.5}
\setres{ind}{DA_uniform_k8}{err_pct}{53.7}
\setres{ind}{DA_uniform_k8}{gain_crps_pct}{-43.5}
\setres{ind}{DA_uniform_k8}{gain_pct}{-53.7}
\setres{ind}{DA_uniform_k8}{k}{8}
\setres{ind}{DA_uniform_k8}{kappa_ind}{-3.66}
\setres{ind}{DA_uniform_k8}{kappa_ind_crps}{-7.51}
\setres{ind}{DA_uniform_k8}{kappa_ind_crps_nn}{-3.531}
\setres{ind}{DA_uniform_k8}{kappa_ind_nn}{-2.854}
\setres{ind}{EU_crpsfit_k0}{err_crps_pct}{31.0}
\setres{ind}{EU_crpsfit_k0}{err_pct}{40.1}
\setres{ind}{EU_crpsfit_k0}{gain_crps_pct}{-31.0}
\setres{ind}{EU_crpsfit_k0}{gain_pct}{-40.1}
\setres{ind}{EU_crpsfit_k0}{k}{0}
\setres{ind}{EU_crpsfit_k1}{err_crps_pct}{32.3}
\setres{ind}{EU_crpsfit_k1}{err_pct}{46.1}
\setres{ind}{EU_crpsfit_k1}{gain_crps_pct}{-32.3}
\setres{ind}{EU_crpsfit_k1}{gain_pct}{-46.1}
\setres{ind}{EU_crpsfit_k1}{k}{1}
\setres{ind}{EU_crpsfit_k1}{kappa_ind}{6.95}
\setres{ind}{EU_crpsfit_k1}{kappa_ind_crps}{6.20}
\setres{ind}{EU_crpsfit_k1}{kappa_ind_crps_nn}{31.717}
\setres{ind}{EU_crpsfit_k1}{kappa_ind_nn}{3059.859}
\setres{ind}{EU_crpsfit_k16}{err_crps_pct}{28.9}
\setres{ind}{EU_crpsfit_k16}{err_pct}{38.9}
\setres{ind}{EU_crpsfit_k16}{gain_crps_pct}{-28.9}
\setres{ind}{EU_crpsfit_k16}{gain_pct}{-38.9}
\setres{ind}{EU_crpsfit_k16}{k}{16}
\setres{ind}{EU_crpsfit_k16}{kappa_ind}{-1.62}
\setres{ind}{EU_crpsfit_k16}{kappa_ind_crps}{-2.33}
\setres{ind}{EU_crpsfit_k16}{kappa_ind_crps_nn}{-2.047}
\setres{ind}{EU_crpsfit_k16}{kappa_ind_nn}{-1.933}
\setres{ind}{EU_crpsfit_k2}{err_crps_pct}{30.4}
\setres{ind}{EU_crpsfit_k2}{err_pct}{41.0}
\setres{ind}{EU_crpsfit_k2}{gain_crps_pct}{-30.4}
\setres{ind}{EU_crpsfit_k2}{gain_pct}{-41.0}
\setres{ind}{EU_crpsfit_k2}{k}{2}
\setres{ind}{EU_crpsfit_k2}{kappa_ind}{9.68}
\setres{ind}{EU_crpsfit_k2}{kappa_ind_crps}{7.36}
\setres{ind}{EU_crpsfit_k2}{kappa_ind_crps_nn}{-6.428}
\setres{ind}{EU_crpsfit_k2}{kappa_ind_nn}{-4.787}
\setres{ind}{EU_crpsfit_k24}{err_crps_pct}{28.1}
\setres{ind}{EU_crpsfit_k24}{err_pct}{37.3}
\setres{ind}{EU_crpsfit_k24}{gain0_crps_pct}{23.057}
\setres{ind}{EU_crpsfit_k24}{gain0_pct}{36.247}
\setres{ind}{EU_crpsfit_k24}{gain_crps_pct}{-28.1}
\setres{ind}{EU_crpsfit_k24}{gain_crps_pct_se}{3.2}
\setres{ind}{EU_crpsfit_k24}{gain_pct}{-37.3}
\setres{ind}{EU_crpsfit_k24}{gain_pct_se}{7.5}
\setres{ind}{EU_crpsfit_k24}{k}{24}
\setres{ind}{EU_crpsfit_k24}{kappa_ind}{-1.31}
\setres{ind}{EU_crpsfit_k24}{kappa_ind0}{0.542}
\setres{ind}{EU_crpsfit_k24}{kappa_ind0_crps}{0.460}
\setres{ind}{EU_crpsfit_k24}{kappa_ind_crps}{-1.65}
\setres{ind}{EU_crpsfit_k24}{kappa_ind_crps_nn}{-1.978}
\setres{ind}{EU_crpsfit_k24}{kappa_ind_crps_se}{0.33}
\setres{ind}{EU_crpsfit_k24}{kappa_ind_nn}{-1.879}
\setres{ind}{EU_crpsfit_k24}{kappa_ind_se}{0.43}
\setres{ind}{EU_crpsfit_k4}{err_crps_pct}{31.6}
\setres{ind}{EU_crpsfit_k4}{err_pct}{45.1}
\setres{ind}{EU_crpsfit_k4}{gain_crps_pct}{-31.6}
\setres{ind}{EU_crpsfit_k4}{gain_pct}{-45.1}
\setres{ind}{EU_crpsfit_k4}{k}{4}
\setres{ind}{EU_crpsfit_k4}{kappa_ind}{-7.06}
\setres{ind}{EU_crpsfit_k4}{kappa_ind_crps}{-64.80}
\setres{ind}{EU_crpsfit_k4}{kappa_ind_crps_nn}{-3.668}
\setres{ind}{EU_crpsfit_k4}{kappa_ind_nn}{-3.127}
\setres{ind}{EU_crpsfit_k8}{err_crps_pct}{30.2}
\setres{ind}{EU_crpsfit_k8}{err_pct}{41.5}
\setres{ind}{EU_crpsfit_k8}{gain_crps_pct}{-30.2}
\setres{ind}{EU_crpsfit_k8}{gain_pct}{-41.5}
\setres{ind}{EU_crpsfit_k8}{k}{8}
\setres{ind}{EU_crpsfit_k8}{kappa_ind}{-2.83}
\setres{ind}{EU_crpsfit_k8}{kappa_ind_crps}{-5.22}
\setres{ind}{EU_crpsfit_k8}{kappa_ind_crps_nn}{-2.452}
\setres{ind}{EU_crpsfit_k8}{kappa_ind_nn}{-2.203}
\setres{ind}{EU_fit_k1}{err_crps_pct}{110.1}
\setres{ind}{EU_fit_k1}{err_pct}{124.5}
\setres{ind}{EU_fit_k1}{gain_crps_pct}{-110.1}
\setres{ind}{EU_fit_k1}{gain_pct}{-124.5}
\setres{ind}{EU_fit_k1}{k}{1}
\setres{ind}{EU_fit_k1}{kappa_ind}{18.76}
\setres{ind}{EU_fit_k1}{kappa_ind_crps}{21.10}
\setres{ind}{EU_fit_k1}{kappa_ind_crps_nn}{107.931}
\setres{ind}{EU_fit_k1}{kappa_ind_nn}{8262.776}
\setres{ind}{EU_fit_k16}{err_crps_pct}{74.1}
\setres{ind}{EU_fit_k16}{err_pct}{46.8}
\setres{ind}{EU_fit_k16}{gain_crps_pct}{-74.1}
\setres{ind}{EU_fit_k16}{gain_pct}{-46.8}
\setres{ind}{EU_fit_k16}{k}{16}
\setres{ind}{EU_fit_k16}{kappa_ind}{-1.94}
\setres{ind}{EU_fit_k16}{kappa_ind_crps}{-5.98}
\setres{ind}{EU_fit_k16}{kappa_ind_crps_nn}{-5.254}
\setres{ind}{EU_fit_k16}{kappa_ind_nn}{-2.322}
\setres{ind}{EU_fit_k2}{err_crps_pct}{95.3}
\setres{ind}{EU_fit_k2}{err_pct}{87.9}
\setres{ind}{EU_fit_k2}{gain_crps_pct}{-95.3}
\setres{ind}{EU_fit_k2}{gain_pct}{-87.9}
\setres{ind}{EU_fit_k2}{k}{2}
\setres{ind}{EU_fit_k2}{kappa_ind}{20.74}
\setres{ind}{EU_fit_k2}{kappa_ind_crps}{23.10}
\setres{ind}{EU_fit_k2}{kappa_ind_crps_nn}{-20.167}
\setres{ind}{EU_fit_k2}{kappa_ind_nn}{-10.257}
\setres{ind}{EU_fit_k24}{err_crps_pct}{72.4}
\setres{ind}{EU_fit_k24}{err_pct}{41.8}
\setres{ind}{EU_fit_k24}{gain0_crps_pct}{-3.572}
\setres{ind}{EU_fit_k24}{gain0_pct}{34.123}
\setres{ind}{EU_fit_k24}{gain_crps_pct}{-72.4}
\setres{ind}{EU_fit_k24}{gain_crps_pct_se}{5.2}
\setres{ind}{EU_fit_k24}{gain_pct}{-41.8}
\setres{ind}{EU_fit_k24}{gain_pct_se}{8.3}
\setres{ind}{EU_fit_k24}{k}{24}
\setres{ind}{EU_fit_k24}{kappa_ind}{-1.47}
\setres{ind}{EU_fit_k24}{kappa_ind0}{0.511}
\setres{ind}{EU_fit_k24}{kappa_ind0_crps}{-0.071}
\setres{ind}{EU_fit_k24}{kappa_ind_crps}{-4.27}
\setres{ind}{EU_fit_k24}{kappa_ind_crps_nn}{-5.101}
\setres{ind}{EU_fit_k24}{kappa_ind_crps_se}{0.70}
\setres{ind}{EU_fit_k24}{kappa_ind_nn}{-2.109}
\setres{ind}{EU_fit_k24}{kappa_ind_se}{0.48}
\setres{ind}{EU_fit_k4}{err_crps_pct}{88.2}
\setres{ind}{EU_fit_k4}{err_pct}{77.4}
\setres{ind}{EU_fit_k4}{gain_crps_pct}{-88.2}
\setres{ind}{EU_fit_k4}{gain_pct}{-77.4}
\setres{ind}{EU_fit_k4}{k}{4}
\setres{ind}{EU_fit_k4}{kappa_ind}{-12.12}
\setres{ind}{EU_fit_k4}{kappa_ind_crps}{-181.06}
\setres{ind}{EU_fit_k4}{kappa_ind_crps_nn}{-10.248}
\setres{ind}{EU_fit_k4}{kappa_ind_nn}{-5.365}
\setres{ind}{EU_fit_k8}{err_crps_pct}{82.1}
\setres{ind}{EU_fit_k8}{err_pct}{63.3}
\setres{ind}{EU_fit_k8}{gain_crps_pct}{-82.1}
\setres{ind}{EU_fit_k8}{gain_pct}{-63.3}
\setres{ind}{EU_fit_k8}{k}{8}
\setres{ind}{EU_fit_k8}{kappa_ind}{-4.32}
\setres{ind}{EU_fit_k8}{kappa_ind_crps}{-14.18}
\setres{ind}{EU_fit_k8}{kappa_ind_crps_nn}{-6.663}
\setres{ind}{EU_fit_k8}{kappa_ind_nn}{-3.366}
\setres{ind}{EU_genderfit_k0}{err_crps_pct}{30.8}
\setres{ind}{EU_genderfit_k0}{err_pct}{39.7}
\setres{ind}{EU_genderfit_k0}{gain0_crps_pct}{21.454}
\setres{ind}{EU_genderfit_k0}{gain0_pct}{35.126}
\setres{ind}{EU_genderfit_k0}{gain_crps_pct}{-30.8}
\setres{ind}{EU_genderfit_k0}{gain_crps_pct_se}{2.6}
\setres{ind}{EU_genderfit_k0}{gain_pct}{-39.7}
\setres{ind}{EU_genderfit_k0}{gain_pct_se}{4.3}
\setres{ind}{EU_genderfit_k0}{k}{0}
\setres{ind}{EU_genderfit_k0}{kappa_ind}{-1.39}
\setres{ind}{EU_genderfit_k0}{kappa_ind0}{0.526}
\setres{ind}{EU_genderfit_k0}{kappa_ind0_crps}{0.428}
\setres{ind}{EU_genderfit_k0}{kappa_ind_crps}{-1.81}
\setres{ind}{EU_genderfit_k0}{kappa_pop}{0.656}
\setres{ind}{EU_genderfit_k0}{kappa_pop_crps}{0.537}
\setres{ind}{EU_genderfit_k0}{r_dist}{0.370}
\setres{ind}{EU_genderfit_k0}{r_ff}{0.336}
\setres{ind}{EU_genderfit_k0}{r_prior}{0.9696}
\setres{ind}{EU_genderfit_k0}{r_prob}{0.980}
\setres{ind}{EU_lit_k0}{err_crps_pct}{33.5}
\setres{ind}{EU_lit_k0}{err_pct}{44.6}
\setres{ind}{EU_lit_k0}{gain_crps_pct}{-33.5}
\setres{ind}{EU_lit_k0}{gain_pct}{-44.6}
\setres{ind}{EU_lit_k0}{k}{0}
\setres{ind}{EU_lit_k1}{err_crps_pct}{33.2}
\setres{ind}{EU_lit_k1}{err_pct}{44.2}
\setres{ind}{EU_lit_k1}{gain_crps_pct}{-33.2}
\setres{ind}{EU_lit_k1}{gain_pct}{-44.2}
\setres{ind}{EU_lit_k1}{k}{1}
\setres{ind}{EU_lit_k1}{kappa_ind}{6.66}
\setres{ind}{EU_lit_k1}{kappa_ind_crps}{6.36}
\setres{ind}{EU_lit_k1}{kappa_ind_crps_nn}{32.541}
\setres{ind}{EU_lit_k1}{kappa_ind_nn}{2934.729}
\setres{ind}{EU_lit_k16}{err_crps_pct}{27.3}
\setres{ind}{EU_lit_k16}{err_pct}{31.1}
\setres{ind}{EU_lit_k16}{gain_crps_pct}{-27.3}
\setres{ind}{EU_lit_k16}{gain_pct}{-31.1}
\setres{ind}{EU_lit_k16}{k}{16}
\setres{ind}{EU_lit_k16}{kappa_ind}{-1.29}
\setres{ind}{EU_lit_k16}{kappa_ind_crps}{-2.20}
\setres{ind}{EU_lit_k16}{kappa_ind_crps_nn}{-1.936}
\setres{ind}{EU_lit_k16}{kappa_ind_nn}{-1.546}
\setres{ind}{EU_lit_k2}{err_crps_pct}{31.8}
\setres{ind}{EU_lit_k2}{err_pct}{41.1}
\setres{ind}{EU_lit_k2}{gain_crps_pct}{-31.8}
\setres{ind}{EU_lit_k2}{gain_pct}{-41.1}
\setres{ind}{EU_lit_k2}{k}{2}
\setres{ind}{EU_lit_k2}{kappa_ind}{9.70}
\setres{ind}{EU_lit_k2}{kappa_ind_crps}{7.70}
\setres{ind}{EU_lit_k2}{kappa_ind_crps_nn}{-6.721}
\setres{ind}{EU_lit_k2}{kappa_ind_nn}{-4.797}
\setres{ind}{EU_lit_k24}{err_crps_pct}{26.8}
\setres{ind}{EU_lit_k24}{err_pct}{29.9}
\setres{ind}{EU_lit_k24}{gain0_crps_pct}{23.807}
\setres{ind}{EU_lit_k24}{gain0_pct}{39.674}
\setres{ind}{EU_lit_k24}{gain_crps_pct}{-26.8}
\setres{ind}{EU_lit_k24}{gain_crps_pct_se}{2.9}
\setres{ind}{EU_lit_k24}{gain_pct}{-29.9}
\setres{ind}{EU_lit_k24}{gain_pct_se}{5.8}
\setres{ind}{EU_lit_k24}{k}{24}
\setres{ind}{EU_lit_k24}{kappa_ind}{-1.05}
\setres{ind}{EU_lit_k24}{kappa_ind0}{0.594}
\setres{ind}{EU_lit_k24}{kappa_ind0_crps}{0.475}
\setres{ind}{EU_lit_k24}{kappa_ind_crps}{-1.58}
\setres{ind}{EU_lit_k24}{kappa_ind_crps_nn}{-1.890}
\setres{ind}{EU_lit_k24}{kappa_ind_crps_se}{0.30}
\setres{ind}{EU_lit_k24}{kappa_ind_nn}{-1.507}
\setres{ind}{EU_lit_k24}{kappa_ind_se}{0.34}
\setres{ind}{EU_lit_k4}{err_crps_pct}{31.3}
\setres{ind}{EU_lit_k4}{err_pct}{40.6}
\setres{ind}{EU_lit_k4}{gain_crps_pct}{-31.3}
\setres{ind}{EU_lit_k4}{gain_pct}{-40.6}
\setres{ind}{EU_lit_k4}{k}{4}
\setres{ind}{EU_lit_k4}{kappa_ind}{-6.36}
\setres{ind}{EU_lit_k4}{kappa_ind_crps}{-64.26}
\setres{ind}{EU_lit_k4}{kappa_ind_crps_nn}{-3.637}
\setres{ind}{EU_lit_k4}{kappa_ind_nn}{-2.816}
\setres{ind}{EU_lit_k8}{err_crps_pct}{29.1}
\setres{ind}{EU_lit_k8}{err_pct}{33.8}
\setres{ind}{EU_lit_k8}{gain_crps_pct}{-29.1}
\setres{ind}{EU_lit_k8}{gain_pct}{-33.8}
\setres{ind}{EU_lit_k8}{k}{8}
\setres{ind}{EU_lit_k8}{kappa_ind}{-2.31}
\setres{ind}{EU_lit_k8}{kappa_ind_crps}{-5.03}
\setres{ind}{EU_lit_k8}{kappa_ind_crps_nn}{-2.362}
\setres{ind}{EU_lit_k8}{kappa_ind_nn}{-1.797}
\setres{ind}{EU_uniform_k0}{err_crps_pct}{43.1}
\setres{ind}{EU_uniform_k0}{err_pct}{78.7}
\setres{ind}{EU_uniform_k0}{gain_crps_pct}{-43.1}
\setres{ind}{EU_uniform_k0}{gain_pct}{-78.7}
\setres{ind}{EU_uniform_k0}{k}{0}
\setres{ind}{EU_uniform_k1}{err_crps_pct}{41.1}
\setres{ind}{EU_uniform_k1}{err_pct}{73.1}
\setres{ind}{EU_uniform_k1}{gain_crps_pct}{-41.1}
\setres{ind}{EU_uniform_k1}{gain_pct}{-73.1}
\setres{ind}{EU_uniform_k1}{k}{1}
\setres{ind}{EU_uniform_k1}{kappa_ind}{11.01}
\setres{ind}{EU_uniform_k1}{kappa_ind_crps}{7.88}
\setres{ind}{EU_uniform_k1}{kappa_ind_crps_nn}{40.304}
\setres{ind}{EU_uniform_k1}{kappa_ind_nn}{4850.312}
\setres{ind}{EU_uniform_k16}{err_crps_pct}{29.5}
\setres{ind}{EU_uniform_k16}{err_pct}{39.4}
\setres{ind}{EU_uniform_k16}{gain_crps_pct}{-29.5}
\setres{ind}{EU_uniform_k16}{gain_pct}{-39.4}
\setres{ind}{EU_uniform_k16}{k}{16}
\setres{ind}{EU_uniform_k16}{kappa_ind}{-1.64}
\setres{ind}{EU_uniform_k16}{kappa_ind_crps}{-2.38}
\setres{ind}{EU_uniform_k16}{kappa_ind_crps_nn}{-2.096}
\setres{ind}{EU_uniform_k16}{kappa_ind_nn}{-1.957}
\setres{ind}{EU_uniform_k2}{err_crps_pct}{36.9}
\setres{ind}{EU_uniform_k2}{err_pct}{59.1}
\setres{ind}{EU_uniform_k2}{gain_crps_pct}{-36.9}
\setres{ind}{EU_uniform_k2}{gain_pct}{-59.1}
\setres{ind}{EU_uniform_k2}{k}{2}
\setres{ind}{EU_uniform_k2}{kappa_ind}{13.95}
\setres{ind}{EU_uniform_k2}{kappa_ind_crps}{8.94}
\setres{ind}{EU_uniform_k2}{kappa_ind_crps_nn}{-7.806}
\setres{ind}{EU_uniform_k2}{kappa_ind_nn}{-6.897}
\setres{ind}{EU_uniform_k24}{err_crps_pct}{28.3}
\setres{ind}{EU_uniform_k24}{err_pct}{36.3}
\setres{ind}{EU_uniform_k24}{gain0_crps_pct}{22.910}
\setres{ind}{EU_uniform_k24}{gain0_pct}{36.720}
\setres{ind}{EU_uniform_k24}{gain_crps_pct}{-28.3}
\setres{ind}{EU_uniform_k24}{gain_crps_pct_se}{3.2}
\setres{ind}{EU_uniform_k24}{gain_pct}{-36.3}
\setres{ind}{EU_uniform_k24}{gain_pct_se}{7.3}
\setres{ind}{EU_uniform_k24}{k}{24}
\setres{ind}{EU_uniform_k24}{kappa_ind}{-1.27}
\setres{ind}{EU_uniform_k24}{kappa_ind0}{0.550}
\setres{ind}{EU_uniform_k24}{kappa_ind0_crps}{0.457}
\setres{ind}{EU_uniform_k24}{kappa_ind_crps}{-1.67}
\setres{ind}{EU_uniform_k24}{kappa_ind_crps_nn}{-1.996}
\setres{ind}{EU_uniform_k24}{kappa_ind_crps_se}{0.33}
\setres{ind}{EU_uniform_k24}{kappa_ind_nn}{-1.828}
\setres{ind}{EU_uniform_k24}{kappa_ind_se}{0.42}
\setres{ind}{EU_uniform_k24}{n_ind_r}{40}
\setres{ind}{EU_uniform_k24}{r_dist}{0.409}
\setres{ind}{EU_uniform_k24}{r_ff}{0.336}
\setres{ind}{EU_uniform_k24}{r_prior}{0.9062}
\setres{ind}{EU_uniform_k24}{r_prior_sd}{0.057}
\setres{ind}{EU_uniform_k24}{r_prob}{0.938}
\setres{ind}{EU_uniform_k4}{err_crps_pct}{34.8}
\setres{ind}{EU_uniform_k4}{err_pct}{52.0}
\setres{ind}{EU_uniform_k4}{gain_crps_pct}{-34.8}
\setres{ind}{EU_uniform_k4}{gain_pct}{-52.0}
\setres{ind}{EU_uniform_k4}{k}{4}
\setres{ind}{EU_uniform_k4}{kappa_ind}{-8.14}
\setres{ind}{EU_uniform_k4}{kappa_ind_crps}{-71.49}
\setres{ind}{EU_uniform_k4}{kappa_ind_crps_nn}{-4.046}
\setres{ind}{EU_uniform_k4}{kappa_ind_nn}{-3.605}
\setres{ind}{EU_uniform_k8}{err_crps_pct}{32.2}
\setres{ind}{EU_uniform_k8}{err_pct}{45.3}
\setres{ind}{EU_uniform_k8}{gain_crps_pct}{-32.2}
\setres{ind}{EU_uniform_k8}{gain_pct}{-45.3}
\setres{ind}{EU_uniform_k8}{k}{8}
\setres{ind}{EU_uniform_k8}{kappa_ind}{-3.09}
\setres{ind}{EU_uniform_k8}{kappa_ind_crps}{-5.57}
\setres{ind}{EU_uniform_k8}{kappa_ind_crps_nn}{-2.618}
\setres{ind}{EU_uniform_k8}{kappa_ind_nn}{-2.407}
\setres{ind}{nn_k1}{err_crps_pct}{1.0}
\setres{ind}{nn_k1}{err_pct}{0.0}
\setres{ind}{nn_k1}{gain_crps_pct}{-1.0}
\setres{ind}{nn_k1}{gain_pct}{0.0}
\setres{ind}{nn_k1}{k}{1}
\setres{ind}{nn_k16}{err_crps_pct}{-14.1}
\setres{ind}{nn_k16}{err_pct}{-20.1}
\setres{ind}{nn_k16}{gain_crps_pct}{14.1}
\setres{ind}{nn_k16}{gain_pct}{20.1}
\setres{ind}{nn_k16}{k}{16}
\setres{ind}{nn_k2}{err_crps_pct}{-4.7}
\setres{ind}{nn_k2}{err_pct}{-8.6}
\setres{ind}{nn_k2}{gain_crps_pct}{4.7}
\setres{ind}{nn_k2}{gain_pct}{8.6}
\setres{ind}{nn_k2}{k}{2}
\setres{ind}{nn_k24}{err_crps_pct}{-14.2}
\setres{ind}{nn_k24}{err_pct}{-19.8}
\setres{ind}{nn_k24}{gain0_crps_pct}{48.458}
\setres{ind}{nn_k24}{gain0_pct}{62.773}
\setres{ind}{nn_k24}{gain_crps_pct}{14.2}
\setres{ind}{nn_k24}{gain_crps_pct_se}{1.7}
\setres{ind}{nn_k24}{gain_pct}{19.8}
\setres{ind}{nn_k24}{gain_pct_se}{3.7}
\setres{ind}{nn_k24}{k}{24}
\setres{ind}{nn_k24}{kappa_ind0}{0.939}
\setres{ind}{nn_k24}{kappa_ind0_crps}{0.967}
\setres{ind}{nn_k4}{err_crps_pct}{-8.6}
\setres{ind}{nn_k4}{err_pct}{-14.4}
\setres{ind}{nn_k4}{gain_crps_pct}{8.6}
\setres{ind}{nn_k4}{gain_pct}{14.4}
\setres{ind}{nn_k4}{k}{4}
\setres{ind}{nn_k8}{err_crps_pct}{-12.3}
\setres{ind}{nn_k8}{err_pct}{-18.8}
\setres{ind}{nn_k8}{gain_crps_pct}{12.3}
\setres{ind}{nn_k8}{gain_pct}{18.8}
\setres{ind}{nn_k8}{k}{8}
\setres{ind}{reference}{ceiling}{random forest}
\setres{ind}{reference}{e_mid_sq}{135.616}
\setres{ind}{reference}{e_pop_sq}{62.982}
\setres{ind}{reference}{female_share}{0.440}
\setres{ind}{reference}{gain0_pop_crps_pct}{39.930}
\setres{ind}{reference}{gain0_pop_pct}{53.559}
\setres{ind}{reference}{k_full}{24}
\setres{ind}{reference}{k_nn}{15}
\setres{ind}{reference}{kappa_ind0_pop}{0.802}
\setres{ind}{reference}{kappa_ind0_pop_crps}{0.797}
\setres{ind}{reference}{ks}{0.000, 1.000, 2.000, 4.000, 8.000, 16.000, 24.000}
\setres{ind}{reference}{n_boot}{1{,}000}
\setres{ind}{reference}{n_female}{73}
\setres{ind}{reference}{n_folds}{10}
\setres{ind}{reference}{n_lotteries}{25}
\setres{ind}{reference}{n_subjects}{166}
\setres{ind}{reference}{n_trees}{200}
\setres{ind}{reference}{noise_c_CPT}{0.063}
\setres{ind}{reference}{noise_c_DA}{0.086}
\setres{ind}{reference}{noise_c_EU}{0.137}
\setres{ind}{reference}{r0_crps}{1.665}
\setres{ind}{reference}{r0_sq}{2.153}
\setres{ind}{reference}{r_dist_pop}{0.180}
\setres{ind}{reference}{r_star_ind}{0.455}
\setres{ind}{reference}{r_star_noise_CPT}{0.359}
\setres{ind}{reference}{r_star_noise_DA}{0.329}
\setres{ind}{reference}{r_star_noise_EU}{0.273}
\setres{ind}{reference}{s_pop_crps}{3.073}
\setres{ind}{reference}{s_unif_crps}{5.116}
\setres{ind}{reference}{sample}{Bruhin2010\_gains}
\setres{ind}{reference}{share_within_crps}{0.203}
\setres{ind}{reference}{share_within_sq}{0.198}
\setres{ind}{rf_k1}{err_crps_pct}{5.2}
\setres{ind}{rf_k1}{err_pct}{6.6}
\setres{ind}{rf_k1}{gain_crps_pct}{-5.2}
\setres{ind}{rf_k1}{gain_pct}{-6.6}
\setres{ind}{rf_k1}{k}{1}
\setres{ind}{rf_k16}{err_crps_pct}{-12.4}
\setres{ind}{rf_k16}{err_pct}{-24.1}
\setres{ind}{rf_k16}{gain_crps_pct}{12.4}
\setres{ind}{rf_k16}{gain_pct}{24.1}
\setres{ind}{rf_k16}{k}{16}
\setres{ind}{rf_k2}{err_crps_pct}{4.1}
\setres{ind}{rf_k2}{err_pct}{4.2}
\setres{ind}{rf_k2}{gain_crps_pct}{-4.1}
\setres{ind}{rf_k2}{gain_pct}{-4.2}
\setres{ind}{rf_k2}{k}{2}
\setres{ind}{rf_k24}{err_crps_pct}{-17.0}
\setres{ind}{rf_k24}{err_pct}{-28.6}
\setres{ind}{rf_k24}{gain0_crps_pct}{50.128}
\setres{ind}{rf_k24}{gain0_pct}{66.819}
\setres{ind}{rf_k24}{gain_crps_pct}{17.0}
\setres{ind}{rf_k24}{gain_crps_pct_se}{2.0}
\setres{ind}{rf_k24}{gain_pct}{28.6}
\setres{ind}{rf_k24}{gain_pct_se}{4.3}
\setres{ind}{rf_k24}{k}{24}
\setres{ind}{rf_k24}{kappa_ind0}{1.000}
\setres{ind}{rf_k24}{kappa_ind0_crps}{1.000}
\setres{ind}{rf_k4}{err_crps_pct}{-0.5}
\setres{ind}{rf_k4}{err_pct}{-6.4}
\setres{ind}{rf_k4}{gain_crps_pct}{0.5}
\setres{ind}{rf_k4}{gain_pct}{6.4}
\setres{ind}{rf_k4}{k}{4}
\setres{ind}{rf_k8}{err_crps_pct}{-5.8}
\setres{ind}{rf_k8}{err_pct}{-14.7}
\setres{ind}{rf_k8}{gain_crps_pct}{5.8}
\setres{ind}{rf_k8}{gain_pct}{14.7}
\setres{ind}{rf_k8}{k}{8}
\setres{shared15}{reference}{n_lotteries}{15}
\setres{shared15}{reference}{n_subjects_2003}{179}
\setres{shared15}{reference}{n_subjects_2006}{118}
\setres{shared15}{z03__CPT_crpsfit}{kappa}{0.587}
\setres{shared15}{z03__CPT_crpsfit}{kappa_crps}{0.977}
\setres{shared15}{z03__CPT_crpsfit}{kappa_crps_se}{0.006}
\setres{shared15}{z03__CPT_crpsfit}{r_cop}{0.501}
\setres{shared15}{z03__CPT_crpsfit}{r_dist}{0.206}
\setres{shared15}{z03__CPT_crpsfit}{r_form}{0.019}
\setres{shared15}{z03__CPT_crpsfit}{r_pi}{0.175}
\setres{shared15}{z03__CPT_crpsfit}{theta_mean}{0.93, 0.50, 1.09}
\setres{shared15}{z03__CPT_crpsfit}{theta_sd}{0.13, 0.31, 0.68}
\setres{shared15}{z03__CPT_deterministic}{kappa}{0.999}
\setres{shared15}{z03__CPT_deterministic}{kappa_crps}{0.447}
\setres{shared15}{z03__CPT_deterministic}{kappa_crps_se}{0.036}
\setres{shared15}{z03__CPT_deterministic}{r_dist}{0.383}
\setres{shared15}{z03__CPT_deterministic}{r_form}{0.019}
\setres{shared15}{z03__CPT_deterministic}{r_pi}{0.343}
\setres{shared15}{z03__CPT_lit}{kappa}{0.788}
\setres{shared15}{z03__CPT_lit}{kappa_crps}{0.699}
\setres{shared15}{z03__CPT_lit}{kappa_crps_se}{0.045}
\setres{shared15}{z03__CPT_lit}{r_cop}{0.462}
\setres{shared15}{z03__CPT_lit}{r_dist}{0.428}
\setres{shared15}{z03__CPT_lit}{r_form}{0.019}
\setres{shared15}{z03__CPT_lit}{r_pi}{0.393}
\setres{shared15}{z03__CPT_uniform}{kappa}{0.488}
\setres{shared15}{z03__CPT_uniform}{kappa_crps}{0.904}
\setres{shared15}{z03__CPT_uniform}{kappa_crps_se}{0.010}
\setres{shared15}{z03__CPT_uniform}{r_cop}{0.509}
\setres{shared15}{z03__CPT_uniform}{r_dist}{0.182}
\setres{shared15}{z03__CPT_uniform}{r_form}{0.019}
\setres{shared15}{z03__CPT_uniform}{r_pi}{0.152}
\setres{shared15}{z03__DA_crpsfit}{kappa}{0.324}
\setres{shared15}{z03__DA_crpsfit}{kappa_crps}{0.575}
\setres{shared15}{z03__DA_crpsfit}{kappa_crps_se}{0.044}
\setres{shared15}{z03__DA_crpsfit}{r_cop}{0.970}
\setres{shared15}{z03__DA_crpsfit}{r_dist}{0.364}
\setres{shared15}{z03__DA_crpsfit}{r_form}{0.229}
\setres{shared15}{z03__DA_crpsfit}{r_pi}{0.135}
\setres{shared15}{z03__DA_crpsfit}{theta_mean}{0.75, 0.63}
\setres{shared15}{z03__DA_crpsfit}{theta_sd}{0.39, 1.46}
\setres{shared15}{z03__DA_deterministic}{kappa}{0.806}
\setres{shared15}{z03__DA_deterministic}{kappa_crps}{0.031}
\setres{shared15}{z03__DA_deterministic}{kappa_crps_se}{0.075}
\setres{shared15}{z03__DA_deterministic}{r_dist}{0.519}
\setres{shared15}{z03__DA_deterministic}{r_form}{0.229}
\setres{shared15}{z03__DA_deterministic}{r_pi}{0.389}
\setres{shared15}{z03__DA_lit}{kappa}{0.557}
\setres{shared15}{z03__DA_lit}{kappa_crps}{0.166}
\setres{shared15}{z03__DA_lit}{kappa_crps_se}{0.085}
\setres{shared15}{z03__DA_lit}{r_cop}{0.872}
\setres{shared15}{z03__DA_lit}{r_dist}{0.567}
\setres{shared15}{z03__DA_lit}{r_form}{0.229}
\setres{shared15}{z03__DA_lit}{r_pi}{0.333}
\setres{shared15}{z03__DA_uniform}{kappa}{0.415}
\setres{shared15}{z03__DA_uniform}{kappa_crps}{0.192}
\setres{shared15}{z03__DA_uniform}{kappa_crps_se}{0.078}
\setres{shared15}{z03__DA_uniform}{r_cop}{0.862}
\setres{shared15}{z03__DA_uniform}{r_dist}{0.506}
\setres{shared15}{z03__DA_uniform}{r_form}{0.229}
\setres{shared15}{z03__DA_uniform}{r_pi}{0.273}
\setres{shared15}{z03__EU_crpsfit}{kappa}{0.594}
\setres{shared15}{z03__EU_crpsfit}{kappa_crps}{0.132}
\setres{shared15}{z03__EU_crpsfit}{kappa_crps_se}{0.090}
\setres{shared15}{z03__EU_crpsfit}{r_cop}{0.781}
\setres{shared15}{z03__EU_crpsfit}{r_dist}{0.591}
\setres{shared15}{z03__EU_crpsfit}{r_form}{0.590}
\setres{shared15}{z03__EU_crpsfit}{r_pi}{0.001}
\setres{shared15}{z03__EU_crpsfit}{theta_mean}{0.64}
\setres{shared15}{z03__EU_crpsfit}{theta_sd}{0.44}
\setres{shared15}{z03__EU_deterministic}{kappa}{0.641}
\setres{shared15}{z03__EU_deterministic}{kappa_crps}{-0.233}
\setres{shared15}{z03__EU_deterministic}{kappa_crps_se}{0.098}
\setres{shared15}{z03__EU_deterministic}{r_dist}{0.599}
\setres{shared15}{z03__EU_deterministic}{r_form}{0.590}
\setres{shared15}{z03__EU_deterministic}{r_pi}{0.254}
\setres{shared15}{z03__EU_lit}{kappa}{0.608}
\setres{shared15}{z03__EU_lit}{kappa_crps}{0.030}
\setres{shared15}{z03__EU_lit}{kappa_crps_se}{0.097}
\setres{shared15}{z03__EU_lit}{r_cop}{1.002}
\setres{shared15}{z03__EU_lit}{r_dist}{0.618}
\setres{shared15}{z03__EU_lit}{r_form}{0.590}
\setres{shared15}{z03__EU_lit}{r_pi}{0.028}
\setres{shared15}{z03__EU_uniform}{kappa}{0.587}
\setres{shared15}{z03__EU_uniform}{kappa_crps}{0.054}
\setres{shared15}{z03__EU_uniform}{kappa_crps_se}{0.094}
\setres{shared15}{z03__EU_uniform}{r_cop}{1.012}
\setres{shared15}{z03__EU_uniform}{r_dist}{0.607}
\setres{shared15}{z03__EU_uniform}{r_form}{0.590}
\setres{shared15}{z03__EU_uniform}{r_pi}{0.017}
\setres{shared15}{z03__EV_deterministic}{kappa}{0.624}
\setres{shared15}{z03__EV_deterministic}{kappa_crps}{-0.011}
\setres{shared15}{z03__EV_deterministic}{kappa_crps_se}{0.099}
\setres{shared15}{z03__EV_deterministic}{r_dist}{0.627}
\setres{shared15}{z03__subjects}{n_boot}{1{,}000}
\setres{shared15}{z03__subjects}{n_complete_profiles}{168}
\setres{shared15}{z03__subjects}{n_subjects}{179}
\setres{shared15}{z03__subjects}{r_cop}{0.142}
\setres{shared15}{z03__subjects}{r_dist}{0.212}
\setres{shared15}{z03__subjects}{r_dist_se}{0.012}
\setres{shared15}{z06__CPT_crpsfit}{kappa}{0.796}
\setres{shared15}{z06__CPT_crpsfit}{kappa_crps}{0.986}
\setres{shared15}{z06__CPT_crpsfit}{kappa_crps_se}{0.002}
\setres{shared15}{z06__CPT_crpsfit}{r_cop}{0.505}
\setres{shared15}{z06__CPT_crpsfit}{r_dist}{0.263}
\setres{shared15}{z06__CPT_crpsfit}{r_form}{0.019}
\setres{shared15}{z06__CPT_crpsfit}{r_pi}{0.230}
\setres{shared15}{z06__CPT_crpsfit}{theta_mean}{0.91, 0.55, 0.92}
\setres{shared15}{z06__CPT_crpsfit}{theta_sd}{0.17, 0.28, 0.46}
\setres{shared15}{z06__CPT_deterministic}{kappa}{1.000}
\setres{shared15}{z06__CPT_deterministic}{kappa_crps}{0.646}
\setres{shared15}{z06__CPT_deterministic}{kappa_crps_se}{0.021}
\setres{shared15}{z06__CPT_deterministic}{r_dist}{0.438}
\setres{shared15}{z06__CPT_deterministic}{r_form}{0.019}
\setres{shared15}{z06__CPT_deterministic}{r_pi}{0.396}
\setres{shared15}{z06__CPT_from03_dist}{kappa}{0.667}
\setres{shared15}{z06__CPT_from03_dist}{kappa_crps}{0.965}
\setres{shared15}{z06__CPT_from03_dist}{kappa_crps_se}{0.009}
\setres{shared15}{z06__CPT_from03_dist}{r_cop}{0.476}
\setres{shared15}{z06__CPT_from03_dist}{r_dist}{0.206}
\setres{shared15}{z06__CPT_from03_dist}{r_form}{0.019}
\setres{shared15}{z06__CPT_from03_dist}{r_pi}{0.175}
\setres{shared15}{z06__CPT_from03_value}{kappa}{0.977}
\setres{shared15}{z06__CPT_from03_value}{kappa_crps}{0.558}
\setres{shared15}{z06__CPT_from03_value}{kappa_crps_se}{0.018}
\setres{shared15}{z06__CPT_from03_value}{r_dist}{0.383}
\setres{shared15}{z06__CPT_from03_value}{r_form}{0.019}
\setres{shared15}{z06__CPT_from03_value}{r_pi}{0.343}
\setres{shared15}{z06__CPT_lit}{kappa}{0.907}
\setres{shared15}{z06__CPT_lit}{kappa_crps}{0.828}
\setres{shared15}{z06__CPT_lit}{kappa_crps_se}{0.032}
\setres{shared15}{z06__CPT_lit}{r_cop}{0.463}
\setres{shared15}{z06__CPT_lit}{r_dist}{0.428}
\setres{shared15}{z06__CPT_lit}{r_form}{0.019}
\setres{shared15}{z06__CPT_lit}{r_pi}{0.393}
\setres{shared15}{z06__CPT_uniform}{kappa}{0.585}
\setres{shared15}{z06__CPT_uniform}{kappa_crps}{0.871}
\setres{shared15}{z06__CPT_uniform}{kappa_crps_se}{0.012}
\setres{shared15}{z06__CPT_uniform}{r_cop}{0.511}
\setres{shared15}{z06__CPT_uniform}{r_dist}{0.182}
\setres{shared15}{z06__CPT_uniform}{r_form}{0.019}
\setres{shared15}{z06__CPT_uniform}{r_pi}{0.152}
\setres{shared15}{z06__DA_crpsfit}{kappa}{0.639}
\setres{shared15}{z06__DA_crpsfit}{kappa_crps}{0.683}
\setres{shared15}{z06__DA_crpsfit}{kappa_crps_se}{0.041}
\setres{shared15}{z06__DA_crpsfit}{r_cop}{0.977}
\setres{shared15}{z06__DA_crpsfit}{r_dist}{0.404}
\setres{shared15}{z06__DA_crpsfit}{r_form}{0.229}
\setres{shared15}{z06__DA_crpsfit}{r_pi}{0.176}
\setres{shared15}{z06__DA_crpsfit}{theta_mean}{0.79, 0.64}
\setres{shared15}{z06__DA_crpsfit}{theta_sd}{0.37, 1.32}
\setres{shared15}{z06__DA_deterministic}{kappa}{0.899}
\setres{shared15}{z06__DA_deterministic}{kappa_crps}{0.367}
\setres{shared15}{z06__DA_deterministic}{kappa_crps_se}{0.061}
\setres{shared15}{z06__DA_deterministic}{r_dist}{0.556}
\setres{shared15}{z06__DA_deterministic}{r_form}{0.229}
\setres{shared15}{z06__DA_deterministic}{r_pi}{0.396}
\setres{shared15}{z06__DA_from03_dist}{kappa}{0.587}
\setres{shared15}{z06__DA_from03_dist}{kappa_crps}{0.682}
\setres{shared15}{z06__DA_from03_dist}{kappa_crps_se}{0.038}
\setres{shared15}{z06__DA_from03_dist}{r_cop}{0.954}
\setres{shared15}{z06__DA_from03_dist}{r_dist}{0.364}
\setres{shared15}{z06__DA_from03_dist}{r_form}{0.229}
\setres{shared15}{z06__DA_from03_dist}{r_pi}{0.135}
\setres{shared15}{z06__DA_from03_value}{kappa}{0.875}
\setres{shared15}{z06__DA_from03_value}{kappa_crps}{0.201}
\setres{shared15}{z06__DA_from03_value}{kappa_crps_se}{0.060}
\setres{shared15}{z06__DA_from03_value}{r_dist}{0.519}
\setres{shared15}{z06__DA_from03_value}{r_form}{0.229}
\setres{shared15}{z06__DA_from03_value}{r_pi}{0.389}
\setres{shared15}{z06__DA_lit}{kappa}{0.800}
\setres{shared15}{z06__DA_lit}{kappa_crps}{0.455}
\setres{shared15}{z06__DA_lit}{kappa_crps_se}{0.065}
\setres{shared15}{z06__DA_lit}{r_cop}{0.872}
\setres{shared15}{z06__DA_lit}{r_dist}{0.567}
\setres{shared15}{z06__DA_lit}{r_form}{0.229}
\setres{shared15}{z06__DA_lit}{r_pi}{0.333}
\setres{shared15}{z06__DA_uniform}{kappa}{0.693}
\setres{shared15}{z06__DA_uniform}{kappa_crps}{0.448}
\setres{shared15}{z06__DA_uniform}{kappa_crps_se}{0.058}
\setres{shared15}{z06__DA_uniform}{r_cop}{0.862}
\setres{shared15}{z06__DA_uniform}{r_dist}{0.506}
\setres{shared15}{z06__DA_uniform}{r_form}{0.229}
\setres{shared15}{z06__DA_uniform}{r_pi}{0.273}
\setres{shared15}{z06__EU_crpsfit}{kappa}{0.820}
\setres{shared15}{z06__EU_crpsfit}{kappa_crps}{0.425}
\setres{shared15}{z06__EU_crpsfit}{kappa_crps_se}{0.069}
\setres{shared15}{z06__EU_crpsfit}{r_cop}{0.827}
\setres{shared15}{z06__EU_crpsfit}{r_dist}{0.591}
\setres{shared15}{z06__EU_crpsfit}{r_form}{0.590}
\setres{shared15}{z06__EU_crpsfit}{r_pi}{0.001}
\setres{shared15}{z06__EU_crpsfit}{theta_mean}{0.61}
\setres{shared15}{z06__EU_crpsfit}{theta_sd}{0.44}
\setres{shared15}{z06__EU_deterministic}{kappa}{0.832}
\setres{shared15}{z06__EU_deterministic}{kappa_crps}{0.297}
\setres{shared15}{z06__EU_deterministic}{kappa_crps_se}{0.077}
\setres{shared15}{z06__EU_deterministic}{r_dist}{0.628}
\setres{shared15}{z06__EU_deterministic}{r_form}{0.590}
\setres{shared15}{z06__EU_deterministic}{r_pi}{0.038}
\setres{shared15}{z06__EU_from03_dist}{kappa}{0.820}
\setres{shared15}{z06__EU_from03_dist}{kappa_crps}{0.424}
\setres{shared15}{z06__EU_from03_dist}{kappa_crps_se}{0.070}
\setres{shared15}{z06__EU_from03_dist}{r_cop}{0.747}
\setres{shared15}{z06__EU_from03_dist}{r_dist}{0.591}
\setres{shared15}{z06__EU_from03_dist}{r_form}{0.590}
\setres{shared15}{z06__EU_from03_dist}{r_pi}{0.001}
\setres{shared15}{z06__EU_from03_value}{kappa}{0.801}
\setres{shared15}{z06__EU_from03_value}{kappa_crps}{0.017}
\setres{shared15}{z06__EU_from03_value}{kappa_crps_se}{0.078}
\setres{shared15}{z06__EU_from03_value}{r_dist}{0.599}
\setres{shared15}{z06__EU_from03_value}{r_form}{0.590}
\setres{shared15}{z06__EU_from03_value}{r_pi}{0.254}
\setres{shared15}{z06__EU_lit}{kappa}{0.830}
\setres{shared15}{z06__EU_lit}{kappa_crps}{0.352}
\setres{shared15}{z06__EU_lit}{kappa_crps_se}{0.075}
\setres{shared15}{z06__EU_lit}{r_cop}{1.002}
\setres{shared15}{z06__EU_lit}{r_dist}{0.618}
\setres{shared15}{z06__EU_lit}{r_form}{0.590}
\setres{shared15}{z06__EU_lit}{r_pi}{0.028}
\setres{shared15}{z06__EU_uniform}{kappa}{0.820}
\setres{shared15}{z06__EU_uniform}{kappa_crps}{0.374}
\setres{shared15}{z06__EU_uniform}{kappa_crps_se}{0.072}
\setres{shared15}{z06__EU_uniform}{r_cop}{1.012}
\setres{shared15}{z06__EU_uniform}{r_dist}{0.607}
\setres{shared15}{z06__EU_uniform}{r_form}{0.590}
\setres{shared15}{z06__EU_uniform}{r_pi}{0.017}
\setres{shared15}{z06__EV_deterministic}{kappa}{0.833}
\setres{shared15}{z06__EV_deterministic}{kappa_crps}{0.293}
\setres{shared15}{z06__EV_deterministic}{kappa_crps_se}{0.077}
\setres{shared15}{z06__EV_deterministic}{r_dist}{0.627}
\setres{shared15}{z06__subjects}{n_boot}{1{,}000}
\setres{shared15}{z06__subjects}{n_complete_profiles}{105}
\setres{shared15}{z06__subjects}{n_subjects}{118}
\setres{shared15}{z06__subjects}{r_cop}{0.060}
\setres{shared15}{z06__subjects}{r_dist}{0.269}
\setres{shared15}{z06__subjects}{r_dist_se}{0.012}
\setres{extM}{lottery_150_0_0p10}{CPT_lit__kappa_25}{0.641}
\setres{extM}{lottery_150_0_0p10}{CPT_lit__kappa_lottery}{0.928}
\setres{extM}{lottery_150_0_0p10}{CPT_lit__kappa_union}{0.779}
\setres{extM}{lottery_150_0_0p10}{CPT_lit__r_ff_25}{0.140}
\setres{extM}{lottery_150_0_0p10}{CPT_lit__r_ff_26}{0.155}
\setres{extM}{lottery_150_0_0p10}{CPT_lit__r_prior_25}{0.949}
\setres{extM}{lottery_150_0_0p10}{CPT_lit__r_prior_26}{0.949}
\setres{extM}{lottery_150_0_0p10}{CPT_lit__r_prob_25}{0.956}
\setres{extM}{lottery_150_0_0p10}{CPT_lit__r_prob_26}{0.957}
\setres{extM}{lottery_150_0_0p10}{CPT_uniform__kappa_25}{-0.495}
\setres{extM}{lottery_150_0_0p10}{CPT_uniform__kappa_lottery}{0.751}
\setres{extM}{lottery_150_0_0p10}{CPT_uniform__kappa_union}{0.106}
\setres{extM}{lottery_150_0_0p10}{CPT_uniform__r_ff_25}{0.140}
\setres{extM}{lottery_150_0_0p10}{CPT_uniform__r_ff_26}{0.155}
\setres{extM}{lottery_150_0_0p10}{CPT_uniform__r_prior_25}{0.515}
\setres{extM}{lottery_150_0_0p10}{CPT_uniform__r_prior_26}{0.543}
\setres{extM}{lottery_150_0_0p10}{CPT_uniform__r_prob_25}{0.583}
\setres{extM}{lottery_150_0_0p10}{CPT_uniform__r_prob_26}{0.614}
\setres{extM}{lottery_150_0_0p10}{DA_lit__kappa_25}{0.020}
\setres{extM}{lottery_150_0_0p10}{DA_lit__kappa_lottery}{0.836}
\setres{extM}{lottery_150_0_0p10}{DA_lit__kappa_union}{0.413}
\setres{extM}{lottery_150_0_0p10}{DA_lit__r_ff_25}{0.281}
\setres{extM}{lottery_150_0_0p10}{DA_lit__r_ff_26}{0.299}
\setres{extM}{lottery_150_0_0p10}{DA_lit__r_prior_25}{0.938}
\setres{extM}{lottery_150_0_0p10}{DA_lit__r_prior_26}{0.951}
\setres{extM}{lottery_150_0_0p10}{DA_lit__r_prob_25}{0.956}
\setres{extM}{lottery_150_0_0p10}{DA_lit__r_prob_26}{0.966}
\setres{extM}{lottery_150_0_0p10}{DA_uniform__kappa_25}{-0.827}
\setres{extM}{lottery_150_0_0p10}{DA_uniform__kappa_lottery}{0.797}
\setres{extM}{lottery_150_0_0p10}{DA_uniform__kappa_union}{-0.044}
\setres{extM}{lottery_150_0_0p10}{DA_uniform__r_ff_25}{0.281}
\setres{extM}{lottery_150_0_0p10}{DA_uniform__r_ff_26}{0.299}
\setres{extM}{lottery_150_0_0p10}{DA_uniform__r_prior_25}{0.867}
\setres{extM}{lottery_150_0_0p10}{DA_uniform__r_prior_26}{0.894}
\setres{extM}{lottery_150_0_0p10}{DA_uniform__r_prob_25}{0.904}
\setres{extM}{lottery_150_0_0p10}{DA_uniform__r_prob_26}{0.926}
\setres{extM}{lottery_150_0_0p10}{EU_lit__kappa_25}{0.521}
\setres{extM}{lottery_150_0_0p10}{EU_lit__kappa_lottery}{0.874}
\setres{extM}{lottery_150_0_0p10}{EU_lit__kappa_union}{0.691}
\setres{extM}{lottery_150_0_0p10}{EU_lit__r_ff_25}{0.336}
\setres{extM}{lottery_150_0_0p10}{EU_lit__r_ff_26}{0.350}
\setres{extM}{lottery_150_0_0p10}{EU_lit__r_prior_25}{0.814}
\setres{extM}{lottery_150_0_0p10}{EU_lit__r_prior_26}{0.835}
\setres{extM}{lottery_150_0_0p10}{EU_lit__r_prob_25}{0.877}
\setres{extM}{lottery_150_0_0p10}{EU_lit__r_prob_26}{0.893}
\setres{extM}{lottery_150_0_0p10}{EU_uniform__kappa_25}{-0.209}
\setres{extM}{lottery_150_0_0p10}{EU_uniform__kappa_lottery}{0.843}
\setres{extM}{lottery_150_0_0p10}{EU_uniform__kappa_union}{0.298}
\setres{extM}{lottery_150_0_0p10}{EU_uniform__r_ff_25}{0.336}
\setres{extM}{lottery_150_0_0p10}{EU_uniform__r_ff_26}{0.350}
\setres{extM}{lottery_150_0_0p10}{EU_uniform__r_prior_25}{0.578}
\setres{extM}{lottery_150_0_0p10}{EU_uniform__r_prior_26}{0.633}
\setres{extM}{lottery_150_0_0p10}{EU_uniform__r_prob_25}{0.720}
\setres{extM}{lottery_150_0_0p10}{EU_uniform__r_prob_26}{0.761}
\setres{extM}{lottery_150_0_0p10}{headroom_share}{0.482}
\setres{extM}{lottery_150_0_0p10}{n_obs}{114}
\setres{extM}{lottery_150_0_0p10}{p}{0.10}
\setres{extM}{lottery_150_0_0p10}{z1}{150}
\setres{extM}{lottery_150_0_0p10}{z2}{0}
\setres{extM}{lottery_20_0_0p50}{CPT_lit__kappa_25}{0.641}
\setres{extM}{lottery_20_0_0p50}{CPT_lit__kappa_lottery}{-3.665}
\setres{extM}{lottery_20_0_0p50}{CPT_lit__kappa_union}{0.639}
\setres{extM}{lottery_20_0_0p50}{CPT_lit__r_ff_25}{0.140}
\setres{extM}{lottery_20_0_0p50}{CPT_lit__r_ff_26}{0.142}
\setres{extM}{lottery_20_0_0p50}{CPT_lit__r_prior_25}{0.949}
\setres{extM}{lottery_20_0_0p50}{CPT_lit__r_prior_26}{0.949}
\setres{extM}{lottery_20_0_0p50}{CPT_lit__r_prob_25}{0.956}
\setres{extM}{lottery_20_0_0p50}{CPT_lit__r_prob_26}{0.956}
\setres{extM}{lottery_20_0_0p50}{CPT_uniform__kappa_25}{-0.495}
\setres{extM}{lottery_20_0_0p50}{CPT_uniform__kappa_lottery}{-22.869}
\setres{extM}{lottery_20_0_0p50}{CPT_uniform__kappa_union}{-0.505}
\setres{extM}{lottery_20_0_0p50}{CPT_uniform__r_ff_25}{0.140}
\setres{extM}{lottery_20_0_0p50}{CPT_uniform__r_ff_26}{0.142}
\setres{extM}{lottery_20_0_0p50}{CPT_uniform__r_prior_25}{0.515}
\setres{extM}{lottery_20_0_0p50}{CPT_uniform__r_prior_26}{0.515}
\setres{extM}{lottery_20_0_0p50}{CPT_uniform__r_prob_25}{0.583}
\setres{extM}{lottery_20_0_0p50}{CPT_uniform__r_prob_26}{0.584}
\setres{extM}{lottery_20_0_0p50}{DA_lit__kappa_25}{0.020}
\setres{extM}{lottery_20_0_0p50}{DA_lit__kappa_lottery}{-16.894}
\setres{extM}{lottery_20_0_0p50}{DA_lit__kappa_union}{0.012}
\setres{extM}{lottery_20_0_0p50}{DA_lit__r_ff_25}{0.281}
\setres{extM}{lottery_20_0_0p50}{DA_lit__r_ff_26}{0.290}
\setres{extM}{lottery_20_0_0p50}{DA_lit__r_prior_25}{0.938}
\setres{extM}{lottery_20_0_0p50}{DA_lit__r_prior_26}{0.938}
\setres{extM}{lottery_20_0_0p50}{DA_lit__r_prob_25}{0.956}
\setres{extM}{lottery_20_0_0p50}{DA_lit__r_prob_26}{0.956}
\setres{extM}{lottery_20_0_0p50}{DA_uniform__kappa_25}{-0.827}
\setres{extM}{lottery_20_0_0p50}{DA_uniform__kappa_lottery}{-34.006}
\setres{extM}{lottery_20_0_0p50}{DA_uniform__kappa_union}{-0.842}
\setres{extM}{lottery_20_0_0p50}{DA_uniform__r_ff_25}{0.281}
\setres{extM}{lottery_20_0_0p50}{DA_uniform__r_ff_26}{0.290}
\setres{extM}{lottery_20_0_0p50}{DA_uniform__r_prior_25}{0.867}
\setres{extM}{lottery_20_0_0p50}{DA_uniform__r_prior_26}{0.867}
\setres{extM}{lottery_20_0_0p50}{DA_uniform__r_prob_25}{0.904}
\setres{extM}{lottery_20_0_0p50}{DA_uniform__r_prob_26}{0.905}
\setres{extM}{lottery_20_0_0p50}{EU_lit__kappa_25}{0.521}
\setres{extM}{lottery_20_0_0p50}{EU_lit__kappa_lottery}{-2.986}
\setres{extM}{lottery_20_0_0p50}{EU_lit__kappa_union}{0.520}
\setres{extM}{lottery_20_0_0p50}{EU_lit__r_ff_25}{0.336}
\setres{extM}{lottery_20_0_0p50}{EU_lit__r_ff_26}{0.346}
\setres{extM}{lottery_20_0_0p50}{EU_lit__r_prior_25}{0.814}
\setres{extM}{lottery_20_0_0p50}{EU_lit__r_prior_26}{0.814}
\setres{extM}{lottery_20_0_0p50}{EU_lit__r_prob_25}{0.877}
\setres{extM}{lottery_20_0_0p50}{EU_lit__r_prob_26}{0.878}
\setres{extM}{lottery_20_0_0p50}{EU_uniform__kappa_25}{-0.209}
\setres{extM}{lottery_20_0_0p50}{EU_uniform__kappa_lottery}{-20.673}
\setres{extM}{lottery_20_0_0p50}{EU_uniform__kappa_union}{-0.218}
\setres{extM}{lottery_20_0_0p50}{EU_uniform__r_ff_25}{0.336}
\setres{extM}{lottery_20_0_0p50}{EU_uniform__r_ff_26}{0.346}
\setres{extM}{lottery_20_0_0p50}{EU_uniform__r_prior_25}{0.578}
\setres{extM}{lottery_20_0_0p50}{EU_uniform__r_prior_26}{0.577}
\setres{extM}{lottery_20_0_0p50}{EU_uniform__r_prob_25}{0.720}
\setres{extM}{lottery_20_0_0p50}{EU_uniform__r_prob_26}{0.724}
\setres{extM}{lottery_20_0_0p50}{headroom_share}{0.000}
\setres{extM}{lottery_20_0_0p50}{n_obs}{117}
\setres{extM}{lottery_20_0_0p50}{p}{0.50}
\setres{extM}{lottery_20_0_0p50}{z1}{20}
\setres{extM}{lottery_20_0_0p50}{z2}{0}
\setres{extM}{lottery_40_0_0p05}{CPT_lit__kappa_25}{0.641}
\setres{extM}{lottery_40_0_0p05}{CPT_lit__kappa_lottery}{0.808}
\setres{extM}{lottery_40_0_0p05}{CPT_lit__kappa_union}{0.650}
\setres{extM}{lottery_40_0_0p05}{CPT_lit__r_ff_25}{0.140}
\setres{extM}{lottery_40_0_0p05}{CPT_lit__r_ff_26}{0.147}
\setres{extM}{lottery_40_0_0p05}{CPT_lit__r_prior_25}{0.949}
\setres{extM}{lottery_40_0_0p05}{CPT_lit__r_prior_26}{0.950}
\setres{extM}{lottery_40_0_0p05}{CPT_lit__r_prob_25}{0.956}
\setres{extM}{lottery_40_0_0p05}{CPT_lit__r_prob_26}{0.957}
\setres{extM}{lottery_40_0_0p05}{CPT_uniform__kappa_25}{-0.495}
\setres{extM}{lottery_40_0_0p05}{CPT_uniform__kappa_lottery}{0.705}
\setres{extM}{lottery_40_0_0p05}{CPT_uniform__kappa_union}{-0.430}
\setres{extM}{lottery_40_0_0p05}{CPT_uniform__r_ff_25}{0.140}
\setres{extM}{lottery_40_0_0p05}{CPT_uniform__r_ff_26}{0.147}
\setres{extM}{lottery_40_0_0p05}{CPT_uniform__r_prior_25}{0.515}
\setres{extM}{lottery_40_0_0p05}{CPT_uniform__r_prior_26}{0.520}
\setres{extM}{lottery_40_0_0p05}{CPT_uniform__r_prob_25}{0.583}
\setres{extM}{lottery_40_0_0p05}{CPT_uniform__r_prob_26}{0.591}
\setres{extM}{lottery_40_0_0p05}{DA_lit__kappa_25}{0.020}
\setres{extM}{lottery_40_0_0p05}{DA_lit__kappa_lottery}{0.691}
\setres{extM}{lottery_40_0_0p05}{DA_lit__kappa_union}{0.056}
\setres{extM}{lottery_40_0_0p05}{DA_lit__r_ff_25}{0.281}
\setres{extM}{lottery_40_0_0p05}{DA_lit__r_ff_26}{0.301}
\setres{extM}{lottery_40_0_0p05}{DA_lit__r_prior_25}{0.938}
\setres{extM}{lottery_40_0_0p05}{DA_lit__r_prior_26}{0.941}
\setres{extM}{lottery_40_0_0p05}{DA_lit__r_prob_25}{0.956}
\setres{extM}{lottery_40_0_0p05}{DA_lit__r_prob_26}{0.959}
\setres{extM}{lottery_40_0_0p05}{DA_uniform__kappa_25}{-0.827}
\setres{extM}{lottery_40_0_0p05}{DA_uniform__kappa_lottery}{0.682}
\setres{extM}{lottery_40_0_0p05}{DA_uniform__kappa_union}{-0.744}
\setres{extM}{lottery_40_0_0p05}{DA_uniform__r_ff_25}{0.281}
\setres{extM}{lottery_40_0_0p05}{DA_uniform__r_ff_26}{0.301}
\setres{extM}{lottery_40_0_0p05}{DA_uniform__r_prior_25}{0.867}
\setres{extM}{lottery_40_0_0p05}{DA_uniform__r_prior_26}{0.872}
\setres{extM}{lottery_40_0_0p05}{DA_uniform__r_prob_25}{0.904}
\setres{extM}{lottery_40_0_0p05}{DA_uniform__r_prob_26}{0.910}
\setres{extM}{lottery_40_0_0p05}{EU_lit__kappa_25}{0.521}
\setres{extM}{lottery_40_0_0p05}{EU_lit__kappa_lottery}{0.716}
\setres{extM}{lottery_40_0_0p05}{EU_lit__kappa_union}{0.532}
\setres{extM}{lottery_40_0_0p05}{EU_lit__r_ff_25}{0.336}
\setres{extM}{lottery_40_0_0p05}{EU_lit__r_ff_26}{0.355}
\setres{extM}{lottery_40_0_0p05}{EU_lit__r_prior_25}{0.814}
\setres{extM}{lottery_40_0_0p05}{EU_lit__r_prior_26}{0.817}
\setres{extM}{lottery_40_0_0p05}{EU_lit__r_prob_25}{0.877}
\setres{extM}{lottery_40_0_0p05}{EU_lit__r_prob_26}{0.882}
\setres{extM}{lottery_40_0_0p05}{EU_uniform__kappa_25}{-0.209}
\setres{extM}{lottery_40_0_0p05}{EU_uniform__kappa_lottery}{0.699}
\setres{extM}{lottery_40_0_0p05}{EU_uniform__kappa_union}{-0.160}
\setres{extM}{lottery_40_0_0p05}{EU_uniform__r_ff_25}{0.336}
\setres{extM}{lottery_40_0_0p05}{EU_uniform__r_ff_26}{0.355}
\setres{extM}{lottery_40_0_0p05}{EU_uniform__r_prior_25}{0.578}
\setres{extM}{lottery_40_0_0p05}{EU_uniform__r_prior_26}{0.584}
\setres{extM}{lottery_40_0_0p05}{EU_uniform__r_prob_25}{0.720}
\setres{extM}{lottery_40_0_0p05}{EU_uniform__r_prob_26}{0.732}
\setres{extM}{lottery_40_0_0p05}{headroom_share}{0.055}
\setres{extM}{lottery_40_0_0p05}{n_obs}{116}
\setres{extM}{lottery_40_0_0p05}{p}{0.05}
\setres{extM}{lottery_40_0_0p05}{z1}{40}
\setres{extM}{lottery_40_0_0p05}{z2}{0}
\setres{extM}{lottery_40_0_0p25}{CPT_lit__kappa_25}{0.641}
\setres{extM}{lottery_40_0_0p25}{CPT_lit__kappa_lottery}{0.749}
\setres{extM}{lottery_40_0_0p25}{CPT_lit__kappa_union}{0.644}
\setres{extM}{lottery_40_0_0p25}{CPT_lit__r_ff_25}{0.140}
\setres{extM}{lottery_40_0_0p25}{CPT_lit__r_ff_26}{0.147}
\setres{extM}{lottery_40_0_0p25}{CPT_lit__r_prior_25}{0.949}
\setres{extM}{lottery_40_0_0p25}{CPT_lit__r_prior_26}{0.948}
\setres{extM}{lottery_40_0_0p25}{CPT_lit__r_prob_25}{0.956}
\setres{extM}{lottery_40_0_0p25}{CPT_lit__r_prob_26}{0.955}
\setres{extM}{lottery_40_0_0p25}{CPT_uniform__kappa_25}{-0.495}
\setres{extM}{lottery_40_0_0p25}{CPT_uniform__kappa_lottery}{0.163}
\setres{extM}{lottery_40_0_0p25}{CPT_uniform__kappa_union}{-0.476}
\setres{extM}{lottery_40_0_0p25}{CPT_uniform__r_ff_25}{0.140}
\setres{extM}{lottery_40_0_0p25}{CPT_uniform__r_ff_26}{0.147}
\setres{extM}{lottery_40_0_0p25}{CPT_uniform__r_prior_25}{0.515}
\setres{extM}{lottery_40_0_0p25}{CPT_uniform__r_prior_26}{0.514}
\setres{extM}{lottery_40_0_0p25}{CPT_uniform__r_prob_25}{0.583}
\setres{extM}{lottery_40_0_0p25}{CPT_uniform__r_prob_26}{0.585}
\setres{extM}{lottery_40_0_0p25}{DA_lit__kappa_25}{0.020}
\setres{extM}{lottery_40_0_0p25}{DA_lit__kappa_lottery}{0.203}
\setres{extM}{lottery_40_0_0p25}{DA_lit__kappa_union}{0.025}
\setres{extM}{lottery_40_0_0p25}{DA_lit__r_ff_25}{0.281}
\setres{extM}{lottery_40_0_0p25}{DA_lit__r_ff_26}{0.299}
\setres{extM}{lottery_40_0_0p25}{DA_lit__r_prior_25}{0.938}
\setres{extM}{lottery_40_0_0p25}{DA_lit__r_prior_26}{0.939}
\setres{extM}{lottery_40_0_0p25}{DA_lit__r_prob_25}{0.956}
\setres{extM}{lottery_40_0_0p25}{DA_lit__r_prob_26}{0.957}
\setres{extM}{lottery_40_0_0p25}{DA_uniform__kappa_25}{-0.827}
\setres{extM}{lottery_40_0_0p25}{DA_uniform__kappa_lottery}{-0.123}
\setres{extM}{lottery_40_0_0p25}{DA_uniform__kappa_union}{-0.807}
\setres{extM}{lottery_40_0_0p25}{DA_uniform__r_ff_25}{0.281}
\setres{extM}{lottery_40_0_0p25}{DA_uniform__r_ff_26}{0.299}
\setres{extM}{lottery_40_0_0p25}{DA_uniform__r_prior_25}{0.867}
\setres{extM}{lottery_40_0_0p25}{DA_uniform__r_prior_26}{0.869}
\setres{extM}{lottery_40_0_0p25}{DA_uniform__r_prob_25}{0.904}
\setres{extM}{lottery_40_0_0p25}{DA_uniform__r_prob_26}{0.908}
\setres{extM}{lottery_40_0_0p25}{EU_lit__kappa_25}{0.521}
\setres{extM}{lottery_40_0_0p25}{EU_lit__kappa_lottery}{0.581}
\setres{extM}{lottery_40_0_0p25}{EU_lit__kappa_union}{0.523}
\setres{extM}{lottery_40_0_0p25}{EU_lit__r_ff_25}{0.336}
\setres{extM}{lottery_40_0_0p25}{EU_lit__r_ff_26}{0.351}
\setres{extM}{lottery_40_0_0p25}{EU_lit__r_prior_25}{0.814}
\setres{extM}{lottery_40_0_0p25}{EU_lit__r_prior_26}{0.814}
\setres{extM}{lottery_40_0_0p25}{EU_lit__r_prob_25}{0.877}
\setres{extM}{lottery_40_0_0p25}{EU_lit__r_prob_26}{0.879}
\setres{extM}{lottery_40_0_0p25}{EU_uniform__kappa_25}{-0.209}
\setres{extM}{lottery_40_0_0p25}{EU_uniform__kappa_lottery}{0.201}
\setres{extM}{lottery_40_0_0p25}{EU_uniform__kappa_union}{-0.197}
\setres{extM}{lottery_40_0_0p25}{EU_uniform__r_ff_25}{0.336}
\setres{extM}{lottery_40_0_0p25}{EU_uniform__r_ff_26}{0.351}
\setres{extM}{lottery_40_0_0p25}{EU_uniform__r_prior_25}{0.578}
\setres{extM}{lottery_40_0_0p25}{EU_uniform__r_prior_26}{0.580}
\setres{extM}{lottery_40_0_0p25}{EU_uniform__r_prob_25}{0.720}
\setres{extM}{lottery_40_0_0p25}{EU_uniform__r_prob_26}{0.728}
\setres{extM}{lottery_40_0_0p25}{headroom_share}{0.029}
\setres{extM}{lottery_40_0_0p25}{n_obs}{116}
\setres{extM}{lottery_40_0_0p25}{p}{0.25}
\setres{extM}{lottery_40_0_0p25}{z1}{40}
\setres{extM}{lottery_40_0_0p25}{z2}{0}
\setres{extM}{lottery_50_0_0p95}{CPT_lit__kappa_25}{0.641}
\setres{extM}{lottery_50_0_0p95}{CPT_lit__kappa_lottery}{0.676}
\setres{extM}{lottery_50_0_0p95}{CPT_lit__kappa_union}{0.642}
\setres{extM}{lottery_50_0_0p95}{CPT_lit__r_ff_25}{0.140}
\setres{extM}{lottery_50_0_0p95}{CPT_lit__r_ff_26}{0.149}
\setres{extM}{lottery_50_0_0p95}{CPT_lit__r_prior_25}{0.949}
\setres{extM}{lottery_50_0_0p95}{CPT_lit__r_prior_26}{0.952}
\setres{extM}{lottery_50_0_0p95}{CPT_lit__r_prob_25}{0.956}
\setres{extM}{lottery_50_0_0p95}{CPT_lit__r_prob_26}{0.959}
\setres{extM}{lottery_50_0_0p95}{CPT_uniform__kappa_25}{-0.495}
\setres{extM}{lottery_50_0_0p95}{CPT_uniform__kappa_lottery}{-0.971}
\setres{extM}{lottery_50_0_0p95}{CPT_uniform__kappa_union}{-0.519}
\setres{extM}{lottery_50_0_0p95}{CPT_uniform__r_ff_25}{0.140}
\setres{extM}{lottery_50_0_0p95}{CPT_uniform__r_ff_26}{0.149}
\setres{extM}{lottery_50_0_0p95}{CPT_uniform__r_prior_25}{0.515}
\setres{extM}{lottery_50_0_0p95}{CPT_uniform__r_prior_26}{0.511}
\setres{extM}{lottery_50_0_0p95}{CPT_uniform__r_prob_25}{0.583}
\setres{extM}{lottery_50_0_0p95}{CPT_uniform__r_prob_26}{0.584}
\setres{extM}{lottery_50_0_0p95}{DA_lit__kappa_25}{0.020}
\setres{extM}{lottery_50_0_0p95}{DA_lit__kappa_lottery}{0.561}
\setres{extM}{lottery_50_0_0p95}{DA_lit__kappa_union}{0.046}
\setres{extM}{lottery_50_0_0p95}{DA_lit__r_ff_25}{0.281}
\setres{extM}{lottery_50_0_0p95}{DA_lit__r_ff_26}{0.312}
\setres{extM}{lottery_50_0_0p95}{DA_lit__r_prior_25}{0.938}
\setres{extM}{lottery_50_0_0p95}{DA_lit__r_prior_26}{0.940}
\setres{extM}{lottery_50_0_0p95}{DA_lit__r_prob_25}{0.956}
\setres{extM}{lottery_50_0_0p95}{DA_lit__r_prob_26}{0.958}
\setres{extM}{lottery_50_0_0p95}{DA_uniform__kappa_25}{-0.827}
\setres{extM}{lottery_50_0_0p95}{DA_uniform__kappa_lottery}{0.049}
\setres{extM}{lottery_50_0_0p95}{DA_uniform__kappa_union}{-0.783}
\setres{extM}{lottery_50_0_0p95}{DA_uniform__r_ff_25}{0.281}
\setres{extM}{lottery_50_0_0p95}{DA_uniform__r_ff_26}{0.312}
\setres{extM}{lottery_50_0_0p95}{DA_uniform__r_prior_25}{0.867}
\setres{extM}{lottery_50_0_0p95}{DA_uniform__r_prior_26}{0.855}
\setres{extM}{lottery_50_0_0p95}{DA_uniform__r_prob_25}{0.904}
\setres{extM}{lottery_50_0_0p95}{DA_uniform__r_prob_26}{0.900}
\setres{extM}{lottery_50_0_0p95}{EU_lit__kappa_25}{0.521}
\setres{extM}{lottery_50_0_0p95}{EU_lit__kappa_lottery}{0.368}
\setres{extM}{lottery_50_0_0p95}{EU_lit__kappa_union}{0.514}
\setres{extM}{lottery_50_0_0p95}{EU_lit__r_ff_25}{0.336}
\setres{extM}{lottery_50_0_0p95}{EU_lit__r_ff_26}{0.364}
\setres{extM}{lottery_50_0_0p95}{EU_lit__r_prior_25}{0.814}
\setres{extM}{lottery_50_0_0p95}{EU_lit__r_prior_26}{0.827}
\setres{extM}{lottery_50_0_0p95}{EU_lit__r_prob_25}{0.877}
\setres{extM}{lottery_50_0_0p95}{EU_lit__r_prob_26}{0.890}
\setres{extM}{lottery_50_0_0p95}{EU_uniform__kappa_25}{-0.209}
\setres{extM}{lottery_50_0_0p95}{EU_uniform__kappa_lottery}{0.260}
\setres{extM}{lottery_50_0_0p95}{EU_uniform__kappa_union}{-0.186}
\setres{extM}{lottery_50_0_0p95}{EU_uniform__r_ff_25}{0.336}
\setres{extM}{lottery_50_0_0p95}{EU_uniform__r_ff_26}{0.364}
\setres{extM}{lottery_50_0_0p95}{EU_uniform__r_prior_25}{0.578}
\setres{extM}{lottery_50_0_0p95}{EU_uniform__r_prior_26}{0.580}
\setres{extM}{lottery_50_0_0p95}{EU_uniform__r_prob_25}{0.720}
\setres{extM}{lottery_50_0_0p95}{EU_uniform__r_prob_26}{0.733}
\setres{extM}{lottery_50_0_0p95}{headroom_share}{0.049}
\setres{extM}{lottery_50_0_0p95}{n_obs}{118}
\setres{extM}{lottery_50_0_0p95}{p}{0.95}
\setres{extM}{lottery_50_0_0p95}{z1}{50}
\setres{extM}{lottery_50_0_0p95}{z2}{0}
\setres{extM}{reference}{headroom_25}{320786.066}
\setres{extM}{reference}{n_candidates}{5}
\setres{extC}{joint__CPT_lit_k10}{coverage}{0.90}
\setres{extC}{joint__CPT_lit_k10}{k}{10}
\setres{extC}{joint__CPT_lit_k10}{m}{20}
\setres{extC}{joint__CPT_lit_k10}{width_median}{10.6}
\setres{extC}{joint__CPT_lit_k10}{width_q25}{9.0}
\setres{extC}{joint__CPT_lit_k10}{width_q75}{12.3}
\setres{extC}{joint__CPT_lit_k15}{coverage}{0.89}
\setres{extC}{joint__CPT_lit_k15}{k}{15}
\setres{extC}{joint__CPT_lit_k15}{m}{20}
\setres{extC}{joint__CPT_lit_k15}{width_median}{10.8}
\setres{extC}{joint__CPT_lit_k15}{width_q25}{9.1}
\setres{extC}{joint__CPT_lit_k15}{width_q75}{12.1}
\setres{extC}{joint__CPT_lit_k2}{coverage}{0.90}
\setres{extC}{joint__CPT_lit_k2}{k}{2}
\setres{extC}{joint__CPT_lit_k2}{m}{20}
\setres{extC}{joint__CPT_lit_k2}{width_median}{11.9}
\setres{extC}{joint__CPT_lit_k2}{width_q25}{10.3}
\setres{extC}{joint__CPT_lit_k2}{width_q75}{15.4}
\setres{extC}{joint__CPT_lit_k20}{coverage}{0.89}
\setres{extC}{joint__CPT_lit_k20}{k}{20}
\setres{extC}{joint__CPT_lit_k20}{m}{20}
\setres{extC}{joint__CPT_lit_k20}{width_median}{11.2}
\setres{extC}{joint__CPT_lit_k20}{width_q25}{9.4}
\setres{extC}{joint__CPT_lit_k20}{width_q75}{12.1}
\setres{extC}{joint__CPT_lit_k3}{coverage}{0.91}
\setres{extC}{joint__CPT_lit_k3}{k}{3}
\setres{extC}{joint__CPT_lit_k3}{m}{20}
\setres{extC}{joint__CPT_lit_k3}{width_median}{11.5}
\setres{extC}{joint__CPT_lit_k3}{width_q25}{9.5}
\setres{extC}{joint__CPT_lit_k3}{width_q75}{13.3}
\setres{extC}{joint__CPT_lit_k5}{coverage}{0.91}
\setres{extC}{joint__CPT_lit_k5}{k}{5}
\setres{extC}{joint__CPT_lit_k5}{m}{20}
\setres{extC}{joint__CPT_lit_k5}{width_median}{11.0}
\setres{extC}{joint__CPT_lit_k5}{width_q25}{9.6}
\setres{extC}{joint__CPT_lit_k5}{width_q75}{12.3}
\setres{extC}{joint__CPT_lit_k8}{coverage}{0.90}
\setres{extC}{joint__CPT_lit_k8}{k}{8}
\setres{extC}{joint__CPT_lit_k8}{m}{20}
\setres{extC}{joint__CPT_lit_k8}{width_median}{10.8}
\setres{extC}{joint__CPT_lit_k8}{width_q25}{9.2}
\setres{extC}{joint__CPT_lit_k8}{width_q75}{12.3}
\setres{extC}{joint__CPT_uniform_k10}{coverage}{0.91}
\setres{extC}{joint__CPT_uniform_k10}{k}{10}
\setres{extC}{joint__CPT_uniform_k10}{m}{20}
\setres{extC}{joint__CPT_uniform_k10}{width_median}{10.8}
\setres{extC}{joint__CPT_uniform_k10}{width_q25}{9.0}
\setres{extC}{joint__CPT_uniform_k10}{width_q75}{12.3}
\setres{extC}{joint__CPT_uniform_k15}{coverage}{0.89}
\setres{extC}{joint__CPT_uniform_k15}{k}{15}
\setres{extC}{joint__CPT_uniform_k15}{m}{20}
\setres{extC}{joint__CPT_uniform_k15}{width_median}{10.8}
\setres{extC}{joint__CPT_uniform_k15}{width_q25}{9.0}
\setres{extC}{joint__CPT_uniform_k15}{width_q75}{12.1}
\setres{extC}{joint__CPT_uniform_k2}{coverage}{0.91}
\setres{extC}{joint__CPT_uniform_k2}{k}{2}
\setres{extC}{joint__CPT_uniform_k2}{m}{20}
\setres{extC}{joint__CPT_uniform_k2}{width_median}{18.8}
\setres{extC}{joint__CPT_uniform_k2}{width_q25}{14.7}
\setres{extC}{joint__CPT_uniform_k2}{width_q75}{25.4}
\setres{extC}{joint__CPT_uniform_k20}{coverage}{0.89}
\setres{extC}{joint__CPT_uniform_k20}{k}{20}
\setres{extC}{joint__CPT_uniform_k20}{m}{20}
\setres{extC}{joint__CPT_uniform_k20}{width_median}{11.4}
\setres{extC}{joint__CPT_uniform_k20}{width_q25}{9.3}
\setres{extC}{joint__CPT_uniform_k20}{width_q75}{12.0}
\setres{extC}{joint__CPT_uniform_k3}{coverage}{0.90}
\setres{extC}{joint__CPT_uniform_k3}{k}{3}
\setres{extC}{joint__CPT_uniform_k3}{m}{20}
\setres{extC}{joint__CPT_uniform_k3}{width_median}{15.2}
\setres{extC}{joint__CPT_uniform_k3}{width_q25}{11.1}
\setres{extC}{joint__CPT_uniform_k3}{width_q75}{23.2}
\setres{extC}{joint__CPT_uniform_k5}{coverage}{0.90}
\setres{extC}{joint__CPT_uniform_k5}{k}{5}
\setres{extC}{joint__CPT_uniform_k5}{m}{20}
\setres{extC}{joint__CPT_uniform_k5}{width_median}{11.5}
\setres{extC}{joint__CPT_uniform_k5}{width_q25}{9.8}
\setres{extC}{joint__CPT_uniform_k5}{width_q75}{13.2}
\setres{extC}{joint__CPT_uniform_k8}{coverage}{0.90}
\setres{extC}{joint__CPT_uniform_k8}{k}{8}
\setres{extC}{joint__CPT_uniform_k8}{m}{20}
\setres{extC}{joint__CPT_uniform_k8}{width_median}{11.1}
\setres{extC}{joint__CPT_uniform_k8}{width_q25}{9.3}
\setres{extC}{joint__CPT_uniform_k8}{width_q75}{12.5}
\setres{extC}{joint__DA_lit_k10}{coverage}{0.91}
\setres{extC}{joint__DA_lit_k10}{k}{10}
\setres{extC}{joint__DA_lit_k10}{m}{20}
\setres{extC}{joint__DA_lit_k10}{width_median}{27.3}
\setres{extC}{joint__DA_lit_k10}{width_q25}{24.8}
\setres{extC}{joint__DA_lit_k10}{width_q75}{33.1}
\setres{extC}{joint__DA_lit_k15}{coverage}{0.90}
\setres{extC}{joint__DA_lit_k15}{k}{15}
\setres{extC}{joint__DA_lit_k15}{m}{20}
\setres{extC}{joint__DA_lit_k15}{width_median}{25.6}
\setres{extC}{joint__DA_lit_k15}{width_q25}{24.0}
\setres{extC}{joint__DA_lit_k15}{width_q75}{28.3}
\setres{extC}{joint__DA_lit_k2}{coverage}{0.92}
\setres{extC}{joint__DA_lit_k2}{k}{2}
\setres{extC}{joint__DA_lit_k2}{m}{20}
\setres{extC}{joint__DA_lit_k2}{width_median}{31.1}
\setres{extC}{joint__DA_lit_k2}{width_q25}{26.7}
\setres{extC}{joint__DA_lit_k2}{width_q75}{36.9}
\setres{extC}{joint__DA_lit_k20}{coverage}{0.91}
\setres{extC}{joint__DA_lit_k20}{k}{20}
\setres{extC}{joint__DA_lit_k20}{m}{20}
\setres{extC}{joint__DA_lit_k20}{width_median}{25.5}
\setres{extC}{joint__DA_lit_k20}{width_q25}{24.0}
\setres{extC}{joint__DA_lit_k20}{width_q75}{27.6}
\setres{extC}{joint__DA_lit_k3}{coverage}{0.91}
\setres{extC}{joint__DA_lit_k3}{k}{3}
\setres{extC}{joint__DA_lit_k3}{m}{20}
\setres{extC}{joint__DA_lit_k3}{width_median}{29.6}
\setres{extC}{joint__DA_lit_k3}{width_q25}{26.3}
\setres{extC}{joint__DA_lit_k3}{width_q75}{35.6}
\setres{extC}{joint__DA_lit_k5}{coverage}{0.91}
\setres{extC}{joint__DA_lit_k5}{k}{5}
\setres{extC}{joint__DA_lit_k5}{m}{20}
\setres{extC}{joint__DA_lit_k5}{width_median}{28.5}
\setres{extC}{joint__DA_lit_k5}{width_q25}{25.7}
\setres{extC}{joint__DA_lit_k5}{width_q75}{35.3}
\setres{extC}{joint__DA_lit_k8}{coverage}{0.91}
\setres{extC}{joint__DA_lit_k8}{k}{8}
\setres{extC}{joint__DA_lit_k8}{m}{20}
\setres{extC}{joint__DA_lit_k8}{width_median}{27.7}
\setres{extC}{joint__DA_lit_k8}{width_q25}{24.8}
\setres{extC}{joint__DA_lit_k8}{width_q75}{33.6}
\setres{extC}{joint__DA_uniform_k10}{coverage}{0.91}
\setres{extC}{joint__DA_uniform_k10}{k}{10}
\setres{extC}{joint__DA_uniform_k10}{m}{20}
\setres{extC}{joint__DA_uniform_k10}{width_median}{27.3}
\setres{extC}{joint__DA_uniform_k10}{width_q25}{24.8}
\setres{extC}{joint__DA_uniform_k10}{width_q75}{33.6}
\setres{extC}{joint__DA_uniform_k15}{coverage}{0.90}
\setres{extC}{joint__DA_uniform_k15}{k}{15}
\setres{extC}{joint__DA_uniform_k15}{m}{20}
\setres{extC}{joint__DA_uniform_k15}{width_median}{25.6}
\setres{extC}{joint__DA_uniform_k15}{width_q25}{24.0}
\setres{extC}{joint__DA_uniform_k15}{width_q75}{28.3}
\setres{extC}{joint__DA_uniform_k2}{coverage}{0.91}
\setres{extC}{joint__DA_uniform_k2}{k}{2}
\setres{extC}{joint__DA_uniform_k2}{m}{20}
\setres{extC}{joint__DA_uniform_k2}{width_median}{31.6}
\setres{extC}{joint__DA_uniform_k2}{width_q25}{26.9}
\setres{extC}{joint__DA_uniform_k2}{width_q75}{36.9}
\setres{extC}{joint__DA_uniform_k20}{coverage}{0.91}
\setres{extC}{joint__DA_uniform_k20}{k}{20}
\setres{extC}{joint__DA_uniform_k20}{m}{20}
\setres{extC}{joint__DA_uniform_k20}{width_median}{25.5}
\setres{extC}{joint__DA_uniform_k20}{width_q25}{24.0}
\setres{extC}{joint__DA_uniform_k20}{width_q75}{27.6}
\setres{extC}{joint__DA_uniform_k3}{coverage}{0.91}
\setres{extC}{joint__DA_uniform_k3}{k}{3}
\setres{extC}{joint__DA_uniform_k3}{m}{20}
\setres{extC}{joint__DA_uniform_k3}{width_median}{29.2}
\setres{extC}{joint__DA_uniform_k3}{width_q25}{26.4}
\setres{extC}{joint__DA_uniform_k3}{width_q75}{35.8}
\setres{extC}{joint__DA_uniform_k5}{coverage}{0.91}
\setres{extC}{joint__DA_uniform_k5}{k}{5}
\setres{extC}{joint__DA_uniform_k5}{m}{20}
\setres{extC}{joint__DA_uniform_k5}{width_median}{28.7}
\setres{extC}{joint__DA_uniform_k5}{width_q25}{25.7}
\setres{extC}{joint__DA_uniform_k5}{width_q75}{35.5}
\setres{extC}{joint__DA_uniform_k8}{coverage}{0.91}
\setres{extC}{joint__DA_uniform_k8}{k}{8}
\setres{extC}{joint__DA_uniform_k8}{m}{20}
\setres{extC}{joint__DA_uniform_k8}{width_median}{27.7}
\setres{extC}{joint__DA_uniform_k8}{width_q25}{24.8}
\setres{extC}{joint__DA_uniform_k8}{width_q75}{33.5}
\setres{extC}{joint__EU_lit_k10}{coverage}{0.89}
\setres{extC}{joint__EU_lit_k10}{k}{10}
\setres{extC}{joint__EU_lit_k10}{m}{20}
\setres{extC}{joint__EU_lit_k10}{width_median}{26.2}
\setres{extC}{joint__EU_lit_k10}{width_q25}{23.6}
\setres{extC}{joint__EU_lit_k10}{width_q75}{27.0}
\setres{extC}{joint__EU_lit_k15}{coverage}{0.90}
\setres{extC}{joint__EU_lit_k15}{k}{15}
\setres{extC}{joint__EU_lit_k15}{m}{20}
\setres{extC}{joint__EU_lit_k15}{width_median}{25.6}
\setres{extC}{joint__EU_lit_k15}{width_q25}{24.2}
\setres{extC}{joint__EU_lit_k15}{width_q75}{26.4}
\setres{extC}{joint__EU_lit_k2}{coverage}{0.91}
\setres{extC}{joint__EU_lit_k2}{k}{2}
\setres{extC}{joint__EU_lit_k2}{m}{20}
\setres{extC}{joint__EU_lit_k2}{width_median}{26.4}
\setres{extC}{joint__EU_lit_k2}{width_q25}{24.9}
\setres{extC}{joint__EU_lit_k2}{width_q75}{28.5}
\setres{extC}{joint__EU_lit_k20}{coverage}{0.90}
\setres{extC}{joint__EU_lit_k20}{k}{20}
\setres{extC}{joint__EU_lit_k20}{m}{20}
\setres{extC}{joint__EU_lit_k20}{width_median}{25.6}
\setres{extC}{joint__EU_lit_k20}{width_q25}{24.1}
\setres{extC}{joint__EU_lit_k20}{width_q75}{26.4}
\setres{extC}{joint__EU_lit_k3}{coverage}{0.91}
\setres{extC}{joint__EU_lit_k3}{k}{3}
\setres{extC}{joint__EU_lit_k3}{m}{20}
\setres{extC}{joint__EU_lit_k3}{width_median}{26.3}
\setres{extC}{joint__EU_lit_k3}{width_q25}{25.0}
\setres{extC}{joint__EU_lit_k3}{width_q75}{29.1}
\setres{extC}{joint__EU_lit_k5}{coverage}{0.91}
\setres{extC}{joint__EU_lit_k5}{k}{5}
\setres{extC}{joint__EU_lit_k5}{m}{20}
\setres{extC}{joint__EU_lit_k5}{width_median}{26.3}
\setres{extC}{joint__EU_lit_k5}{width_q25}{23.9}
\setres{extC}{joint__EU_lit_k5}{width_q75}{27.9}
\setres{extC}{joint__EU_lit_k8}{coverage}{0.90}
\setres{extC}{joint__EU_lit_k8}{k}{8}
\setres{extC}{joint__EU_lit_k8}{m}{20}
\setres{extC}{joint__EU_lit_k8}{width_median}{26.3}
\setres{extC}{joint__EU_lit_k8}{width_q25}{23.8}
\setres{extC}{joint__EU_lit_k8}{width_q75}{27.2}
\setres{extC}{joint__EU_uniform_k10}{coverage}{0.89}
\setres{extC}{joint__EU_uniform_k10}{k}{10}
\setres{extC}{joint__EU_uniform_k10}{m}{20}
\setres{extC}{joint__EU_uniform_k10}{width_median}{26.2}
\setres{extC}{joint__EU_uniform_k10}{width_q25}{23.5}
\setres{extC}{joint__EU_uniform_k10}{width_q75}{27.0}
\setres{extC}{joint__EU_uniform_k15}{coverage}{0.90}
\setres{extC}{joint__EU_uniform_k15}{k}{15}
\setres{extC}{joint__EU_uniform_k15}{m}{20}
\setres{extC}{joint__EU_uniform_k15}{width_median}{25.6}
\setres{extC}{joint__EU_uniform_k15}{width_q25}{24.2}
\setres{extC}{joint__EU_uniform_k15}{width_q75}{26.4}
\setres{extC}{joint__EU_uniform_k2}{coverage}{0.91}
\setres{extC}{joint__EU_uniform_k2}{k}{2}
\setres{extC}{joint__EU_uniform_k2}{m}{20}
\setres{extC}{joint__EU_uniform_k2}{width_median}{26.3}
\setres{extC}{joint__EU_uniform_k2}{width_q25}{25.3}
\setres{extC}{joint__EU_uniform_k2}{width_q75}{33.2}
\setres{extC}{joint__EU_uniform_k20}{coverage}{0.90}
\setres{extC}{joint__EU_uniform_k20}{k}{20}
\setres{extC}{joint__EU_uniform_k20}{m}{20}
\setres{extC}{joint__EU_uniform_k20}{width_median}{25.6}
\setres{extC}{joint__EU_uniform_k20}{width_q25}{24.1}
\setres{extC}{joint__EU_uniform_k20}{width_q75}{26.3}
\setres{extC}{joint__EU_uniform_k3}{coverage}{0.91}
\setres{extC}{joint__EU_uniform_k3}{k}{3}
\setres{extC}{joint__EU_uniform_k3}{m}{20}
\setres{extC}{joint__EU_uniform_k3}{width_median}{26.3}
\setres{extC}{joint__EU_uniform_k3}{width_q25}{24.9}
\setres{extC}{joint__EU_uniform_k3}{width_q75}{29.8}
\setres{extC}{joint__EU_uniform_k5}{coverage}{0.91}
\setres{extC}{joint__EU_uniform_k5}{k}{5}
\setres{extC}{joint__EU_uniform_k5}{m}{20}
\setres{extC}{joint__EU_uniform_k5}{width_median}{26.2}
\setres{extC}{joint__EU_uniform_k5}{width_q25}{23.9}
\setres{extC}{joint__EU_uniform_k5}{width_q75}{28.6}
\setres{extC}{joint__EU_uniform_k8}{coverage}{0.90}
\setres{extC}{joint__EU_uniform_k8}{k}{8}
\setres{extC}{joint__EU_uniform_k8}{m}{20}
\setres{extC}{joint__EU_uniform_k8}{width_median}{26.2}
\setres{extC}{joint__EU_uniform_k8}{width_q25}{23.8}
\setres{extC}{joint__EU_uniform_k8}{width_q75}{27.2}
\setres{extC}{reference}{alpha}{0.1}
\setres{extC}{reference}{n_orders}{200}
\setres{extC}{zurich03__CPT_lit_k10}{coverage}{0.89}
\setres{extC}{zurich03__CPT_lit_k10}{k}{10}
\setres{extC}{zurich03__CPT_lit_k10}{m}{10}
\setres{extC}{zurich03__CPT_lit_k10}{width_median}{8.5}
\setres{extC}{zurich03__CPT_lit_k10}{width_q25}{7.5}
\setres{extC}{zurich03__CPT_lit_k10}{width_q75}{11.8}
\setres{extC}{zurich03__CPT_lit_k12}{coverage}{0.88}
\setres{extC}{zurich03__CPT_lit_k12}{k}{12}
\setres{extC}{zurich03__CPT_lit_k12}{m}{10}
\setres{extC}{zurich03__CPT_lit_k12}{width_median}{8.9}
\setres{extC}{zurich03__CPT_lit_k12}{width_q25}{7.5}
\setres{extC}{zurich03__CPT_lit_k12}{width_q75}{12.1}
\setres{extC}{zurich03__CPT_lit_k2}{coverage}{0.92}
\setres{extC}{zurich03__CPT_lit_k2}{k}{2}
\setres{extC}{zurich03__CPT_lit_k2}{m}{10}
\setres{extC}{zurich03__CPT_lit_k2}{width_median}{13.0}
\setres{extC}{zurich03__CPT_lit_k2}{width_q25}{9.7}
\setres{extC}{zurich03__CPT_lit_k2}{width_q75}{15.9}
\setres{extC}{zurich03__CPT_lit_k3}{coverage}{0.91}
\setres{extC}{zurich03__CPT_lit_k3}{k}{3}
\setres{extC}{zurich03__CPT_lit_k3}{m}{10}
\setres{extC}{zurich03__CPT_lit_k3}{width_median}{10.6}
\setres{extC}{zurich03__CPT_lit_k3}{width_q25}{8.3}
\setres{extC}{zurich03__CPT_lit_k3}{width_q75}{13.6}
\setres{extC}{zurich03__CPT_lit_k4}{coverage}{0.91}
\setres{extC}{zurich03__CPT_lit_k4}{k}{4}
\setres{extC}{zurich03__CPT_lit_k4}{m}{10}
\setres{extC}{zurich03__CPT_lit_k4}{width_median}{9.7}
\setres{extC}{zurich03__CPT_lit_k4}{width_q25}{8.0}
\setres{extC}{zurich03__CPT_lit_k4}{width_q75}{12.7}
\setres{extC}{zurich03__CPT_lit_k5}{coverage}{0.90}
\setres{extC}{zurich03__CPT_lit_k5}{k}{5}
\setres{extC}{zurich03__CPT_lit_k5}{m}{10}
\setres{extC}{zurich03__CPT_lit_k5}{width_median}{9.2}
\setres{extC}{zurich03__CPT_lit_k5}{width_q25}{7.9}
\setres{extC}{zurich03__CPT_lit_k5}{width_q75}{11.2}
\setres{extC}{zurich03__CPT_lit_k6}{coverage}{0.90}
\setres{extC}{zurich03__CPT_lit_k6}{k}{6}
\setres{extC}{zurich03__CPT_lit_k6}{m}{10}
\setres{extC}{zurich03__CPT_lit_k6}{width_median}{8.6}
\setres{extC}{zurich03__CPT_lit_k6}{width_q25}{7.5}
\setres{extC}{zurich03__CPT_lit_k6}{width_q75}{10.9}
\setres{extC}{zurich03__CPT_lit_k8}{coverage}{0.90}
\setres{extC}{zurich03__CPT_lit_k8}{k}{8}
\setres{extC}{zurich03__CPT_lit_k8}{m}{10}
\setres{extC}{zurich03__CPT_lit_k8}{width_median}{8.5}
\setres{extC}{zurich03__CPT_lit_k8}{width_q25}{7.4}
\setres{extC}{zurich03__CPT_lit_k8}{width_q75}{11.5}
\setres{extC}{zurich03__CPT_uniform_k10}{coverage}{0.89}
\setres{extC}{zurich03__CPT_uniform_k10}{k}{10}
\setres{extC}{zurich03__CPT_uniform_k10}{m}{10}
\setres{extC}{zurich03__CPT_uniform_k10}{width_median}{8.2}
\setres{extC}{zurich03__CPT_uniform_k10}{width_q25}{7.0}
\setres{extC}{zurich03__CPT_uniform_k10}{width_q75}{11.8}
\setres{extC}{zurich03__CPT_uniform_k12}{coverage}{0.88}
\setres{extC}{zurich03__CPT_uniform_k12}{k}{12}
\setres{extC}{zurich03__CPT_uniform_k12}{m}{10}
\setres{extC}{zurich03__CPT_uniform_k12}{width_median}{8.7}
\setres{extC}{zurich03__CPT_uniform_k12}{width_q25}{7.1}
\setres{extC}{zurich03__CPT_uniform_k12}{width_q75}{12.1}
\setres{extC}{zurich03__CPT_uniform_k2}{coverage}{0.91}
\setres{extC}{zurich03__CPT_uniform_k2}{k}{2}
\setres{extC}{zurich03__CPT_uniform_k2}{m}{10}
\setres{extC}{zurich03__CPT_uniform_k2}{width_median}{20.5}
\setres{extC}{zurich03__CPT_uniform_k2}{width_q25}{15.8}
\setres{extC}{zurich03__CPT_uniform_k2}{width_q75}{27.2}
\setres{extC}{zurich03__CPT_uniform_k3}{coverage}{0.92}
\setres{extC}{zurich03__CPT_uniform_k3}{k}{3}
\setres{extC}{zurich03__CPT_uniform_k3}{m}{10}
\setres{extC}{zurich03__CPT_uniform_k3}{width_median}{13.5}
\setres{extC}{zurich03__CPT_uniform_k3}{width_q25}{9.3}
\setres{extC}{zurich03__CPT_uniform_k3}{width_q75}{19.4}
\setres{extC}{zurich03__CPT_uniform_k4}{coverage}{0.91}
\setres{extC}{zurich03__CPT_uniform_k4}{k}{4}
\setres{extC}{zurich03__CPT_uniform_k4}{m}{10}
\setres{extC}{zurich03__CPT_uniform_k4}{width_median}{9.5}
\setres{extC}{zurich03__CPT_uniform_k4}{width_q25}{7.7}
\setres{extC}{zurich03__CPT_uniform_k4}{width_q75}{13.4}
\setres{extC}{zurich03__CPT_uniform_k5}{coverage}{0.90}
\setres{extC}{zurich03__CPT_uniform_k5}{k}{5}
\setres{extC}{zurich03__CPT_uniform_k5}{m}{10}
\setres{extC}{zurich03__CPT_uniform_k5}{width_median}{8.8}
\setres{extC}{zurich03__CPT_uniform_k5}{width_q25}{7.4}
\setres{extC}{zurich03__CPT_uniform_k5}{width_q75}{11.3}
\setres{extC}{zurich03__CPT_uniform_k6}{coverage}{0.90}
\setres{extC}{zurich03__CPT_uniform_k6}{k}{6}
\setres{extC}{zurich03__CPT_uniform_k6}{m}{10}
\setres{extC}{zurich03__CPT_uniform_k6}{width_median}{8.1}
\setres{extC}{zurich03__CPT_uniform_k6}{width_q25}{7.1}
\setres{extC}{zurich03__CPT_uniform_k6}{width_q75}{10.5}
\setres{extC}{zurich03__CPT_uniform_k8}{coverage}{0.90}
\setres{extC}{zurich03__CPT_uniform_k8}{k}{8}
\setres{extC}{zurich03__CPT_uniform_k8}{m}{10}
\setres{extC}{zurich03__CPT_uniform_k8}{width_median}{8.1}
\setres{extC}{zurich03__CPT_uniform_k8}{width_q25}{6.9}
\setres{extC}{zurich03__CPT_uniform_k8}{width_q75}{11.4}
\setres{extC}{zurich03__DA_lit_k10}{coverage}{0.90}
\setres{extC}{zurich03__DA_lit_k10}{k}{10}
\setres{extC}{zurich03__DA_lit_k10}{m}{10}
\setres{extC}{zurich03__DA_lit_k10}{width_median}{24.1}
\setres{extC}{zurich03__DA_lit_k10}{width_q25}{21.5}
\setres{extC}{zurich03__DA_lit_k10}{width_q75}{31.6}
\setres{extC}{zurich03__DA_lit_k12}{coverage}{0.89}
\setres{extC}{zurich03__DA_lit_k12}{k}{12}
\setres{extC}{zurich03__DA_lit_k12}{m}{10}
\setres{extC}{zurich03__DA_lit_k12}{width_median}{23.8}
\setres{extC}{zurich03__DA_lit_k12}{width_q25}{21.2}
\setres{extC}{zurich03__DA_lit_k12}{width_q75}{29.0}
\setres{extC}{zurich03__DA_lit_k2}{coverage}{0.92}
\setres{extC}{zurich03__DA_lit_k2}{k}{2}
\setres{extC}{zurich03__DA_lit_k2}{m}{10}
\setres{extC}{zurich03__DA_lit_k2}{width_median}{28.2}
\setres{extC}{zurich03__DA_lit_k2}{width_q25}{24.2}
\setres{extC}{zurich03__DA_lit_k2}{width_q75}{37.3}
\setres{extC}{zurich03__DA_lit_k3}{coverage}{0.92}
\setres{extC}{zurich03__DA_lit_k3}{k}{3}
\setres{extC}{zurich03__DA_lit_k3}{m}{10}
\setres{extC}{zurich03__DA_lit_k3}{width_median}{27.0}
\setres{extC}{zurich03__DA_lit_k3}{width_q25}{23.3}
\setres{extC}{zurich03__DA_lit_k3}{width_q75}{35.1}
\setres{extC}{zurich03__DA_lit_k4}{coverage}{0.93}
\setres{extC}{zurich03__DA_lit_k4}{k}{4}
\setres{extC}{zurich03__DA_lit_k4}{m}{10}
\setres{extC}{zurich03__DA_lit_k4}{width_median}{26.4}
\setres{extC}{zurich03__DA_lit_k4}{width_q25}{23.3}
\setres{extC}{zurich03__DA_lit_k4}{width_q75}{34.8}
\setres{extC}{zurich03__DA_lit_k5}{coverage}{0.92}
\setres{extC}{zurich03__DA_lit_k5}{k}{5}
\setres{extC}{zurich03__DA_lit_k5}{m}{10}
\setres{extC}{zurich03__DA_lit_k5}{width_median}{25.2}
\setres{extC}{zurich03__DA_lit_k5}{width_q25}{22.0}
\setres{extC}{zurich03__DA_lit_k5}{width_q75}{33.9}
\setres{extC}{zurich03__DA_lit_k6}{coverage}{0.91}
\setres{extC}{zurich03__DA_lit_k6}{k}{6}
\setres{extC}{zurich03__DA_lit_k6}{m}{10}
\setres{extC}{zurich03__DA_lit_k6}{width_median}{24.7}
\setres{extC}{zurich03__DA_lit_k6}{width_q25}{21.6}
\setres{extC}{zurich03__DA_lit_k6}{width_q75}{33.8}
\setres{extC}{zurich03__DA_lit_k8}{coverage}{0.90}
\setres{extC}{zurich03__DA_lit_k8}{k}{8}
\setres{extC}{zurich03__DA_lit_k8}{m}{10}
\setres{extC}{zurich03__DA_lit_k8}{width_median}{24.3}
\setres{extC}{zurich03__DA_lit_k8}{width_q25}{21.8}
\setres{extC}{zurich03__DA_lit_k8}{width_q75}{34.0}
\setres{extC}{zurich03__DA_uniform_k10}{coverage}{0.90}
\setres{extC}{zurich03__DA_uniform_k10}{k}{10}
\setres{extC}{zurich03__DA_uniform_k10}{m}{10}
\setres{extC}{zurich03__DA_uniform_k10}{width_median}{24.1}
\setres{extC}{zurich03__DA_uniform_k10}{width_q25}{21.5}
\setres{extC}{zurich03__DA_uniform_k10}{width_q75}{31.9}
\setres{extC}{zurich03__DA_uniform_k12}{coverage}{0.89}
\setres{extC}{zurich03__DA_uniform_k12}{k}{12}
\setres{extC}{zurich03__DA_uniform_k12}{m}{10}
\setres{extC}{zurich03__DA_uniform_k12}{width_median}{23.8}
\setres{extC}{zurich03__DA_uniform_k12}{width_q25}{21.3}
\setres{extC}{zurich03__DA_uniform_k12}{width_q75}{29.1}
\setres{extC}{zurich03__DA_uniform_k2}{coverage}{0.92}
\setres{extC}{zurich03__DA_uniform_k2}{k}{2}
\setres{extC}{zurich03__DA_uniform_k2}{m}{10}
\setres{extC}{zurich03__DA_uniform_k2}{width_median}{28.8}
\setres{extC}{zurich03__DA_uniform_k2}{width_q25}{24.3}
\setres{extC}{zurich03__DA_uniform_k2}{width_q75}{37.3}
\setres{extC}{zurich03__DA_uniform_k3}{coverage}{0.92}
\setres{extC}{zurich03__DA_uniform_k3}{k}{3}
\setres{extC}{zurich03__DA_uniform_k3}{m}{10}
\setres{extC}{zurich03__DA_uniform_k3}{width_median}{27.8}
\setres{extC}{zurich03__DA_uniform_k3}{width_q25}{23.1}
\setres{extC}{zurich03__DA_uniform_k3}{width_q75}{35.3}
\setres{extC}{zurich03__DA_uniform_k4}{coverage}{0.92}
\setres{extC}{zurich03__DA_uniform_k4}{k}{4}
\setres{extC}{zurich03__DA_uniform_k4}{m}{10}
\setres{extC}{zurich03__DA_uniform_k4}{width_median}{26.2}
\setres{extC}{zurich03__DA_uniform_k4}{width_q25}{23.0}
\setres{extC}{zurich03__DA_uniform_k4}{width_q75}{35.0}
\setres{extC}{zurich03__DA_uniform_k5}{coverage}{0.91}
\setres{extC}{zurich03__DA_uniform_k5}{k}{5}
\setres{extC}{zurich03__DA_uniform_k5}{m}{10}
\setres{extC}{zurich03__DA_uniform_k5}{width_median}{25.0}
\setres{extC}{zurich03__DA_uniform_k5}{width_q25}{21.8}
\setres{extC}{zurich03__DA_uniform_k5}{width_q75}{34.0}
\setres{extC}{zurich03__DA_uniform_k6}{coverage}{0.91}
\setres{extC}{zurich03__DA_uniform_k6}{k}{6}
\setres{extC}{zurich03__DA_uniform_k6}{m}{10}
\setres{extC}{zurich03__DA_uniform_k6}{width_median}{24.5}
\setres{extC}{zurich03__DA_uniform_k6}{width_q25}{21.5}
\setres{extC}{zurich03__DA_uniform_k6}{width_q75}{34.5}
\setres{extC}{zurich03__DA_uniform_k8}{coverage}{0.90}
\setres{extC}{zurich03__DA_uniform_k8}{k}{8}
\setres{extC}{zurich03__DA_uniform_k8}{m}{10}
\setres{extC}{zurich03__DA_uniform_k8}{width_median}{24.3}
\setres{extC}{zurich03__DA_uniform_k8}{width_q25}{21.8}
\setres{extC}{zurich03__DA_uniform_k8}{width_q75}{34.0}
\setres{extC}{zurich03__EU_lit_k10}{coverage}{0.93}
\setres{extC}{zurich03__EU_lit_k10}{k}{10}
\setres{extC}{zurich03__EU_lit_k10}{m}{10}
\setres{extC}{zurich03__EU_lit_k10}{width_median}{22.2}
\setres{extC}{zurich03__EU_lit_k10}{width_q25}{21.5}
\setres{extC}{zurich03__EU_lit_k10}{width_q75}{25.9}
\setres{extC}{zurich03__EU_lit_k12}{coverage}{0.93}
\setres{extC}{zurich03__EU_lit_k12}{k}{12}
\setres{extC}{zurich03__EU_lit_k12}{m}{10}
\setres{extC}{zurich03__EU_lit_k12}{width_median}{22.2}
\setres{extC}{zurich03__EU_lit_k12}{width_q25}{21.6}
\setres{extC}{zurich03__EU_lit_k12}{width_q75}{25.7}
\setres{extC}{zurich03__EU_lit_k2}{coverage}{0.92}
\setres{extC}{zurich03__EU_lit_k2}{k}{2}
\setres{extC}{zurich03__EU_lit_k2}{m}{10}
\setres{extC}{zurich03__EU_lit_k2}{width_median}{23.0}
\setres{extC}{zurich03__EU_lit_k2}{width_q25}{22.1}
\setres{extC}{zurich03__EU_lit_k2}{width_q75}{28.5}
\setres{extC}{zurich03__EU_lit_k3}{coverage}{0.90}
\setres{extC}{zurich03__EU_lit_k3}{k}{3}
\setres{extC}{zurich03__EU_lit_k3}{m}{10}
\setres{extC}{zurich03__EU_lit_k3}{width_median}{22.5}
\setres{extC}{zurich03__EU_lit_k3}{width_q25}{21.8}
\setres{extC}{zurich03__EU_lit_k3}{width_q75}{26.2}
\setres{extC}{zurich03__EU_lit_k4}{coverage}{0.91}
\setres{extC}{zurich03__EU_lit_k4}{k}{4}
\setres{extC}{zurich03__EU_lit_k4}{m}{10}
\setres{extC}{zurich03__EU_lit_k4}{width_median}{22.5}
\setres{extC}{zurich03__EU_lit_k4}{width_q25}{21.8}
\setres{extC}{zurich03__EU_lit_k4}{width_q75}{27.2}
\setres{extC}{zurich03__EU_lit_k5}{coverage}{0.90}
\setres{extC}{zurich03__EU_lit_k5}{k}{5}
\setres{extC}{zurich03__EU_lit_k5}{m}{10}
\setres{extC}{zurich03__EU_lit_k5}{width_median}{22.4}
\setres{extC}{zurich03__EU_lit_k5}{width_q25}{21.8}
\setres{extC}{zurich03__EU_lit_k5}{width_q75}{26.1}
\setres{extC}{zurich03__EU_lit_k6}{coverage}{0.91}
\setres{extC}{zurich03__EU_lit_k6}{k}{6}
\setres{extC}{zurich03__EU_lit_k6}{m}{10}
\setres{extC}{zurich03__EU_lit_k6}{width_median}{22.4}
\setres{extC}{zurich03__EU_lit_k6}{width_q25}{21.7}
\setres{extC}{zurich03__EU_lit_k6}{width_q75}{26.0}
\setres{extC}{zurich03__EU_lit_k8}{coverage}{0.92}
\setres{extC}{zurich03__EU_lit_k8}{k}{8}
\setres{extC}{zurich03__EU_lit_k8}{m}{10}
\setres{extC}{zurich03__EU_lit_k8}{width_median}{22.3}
\setres{extC}{zurich03__EU_lit_k8}{width_q25}{21.5}
\setres{extC}{zurich03__EU_lit_k8}{width_q75}{25.8}
\setres{extC}{zurich03__EU_uniform_k10}{coverage}{0.92}
\setres{extC}{zurich03__EU_uniform_k10}{k}{10}
\setres{extC}{zurich03__EU_uniform_k10}{m}{10}
\setres{extC}{zurich03__EU_uniform_k10}{width_median}{22.2}
\setres{extC}{zurich03__EU_uniform_k10}{width_q25}{21.6}
\setres{extC}{zurich03__EU_uniform_k10}{width_q75}{25.9}
\setres{extC}{zurich03__EU_uniform_k12}{coverage}{0.93}
\setres{extC}{zurich03__EU_uniform_k12}{k}{12}
\setres{extC}{zurich03__EU_uniform_k12}{m}{10}
\setres{extC}{zurich03__EU_uniform_k12}{width_median}{22.2}
\setres{extC}{zurich03__EU_uniform_k12}{width_q25}{21.6}
\setres{extC}{zurich03__EU_uniform_k12}{width_q75}{25.7}
\setres{extC}{zurich03__EU_uniform_k2}{coverage}{0.91}
\setres{extC}{zurich03__EU_uniform_k2}{k}{2}
\setres{extC}{zurich03__EU_uniform_k2}{m}{10}
\setres{extC}{zurich03__EU_uniform_k2}{width_median}{22.5}
\setres{extC}{zurich03__EU_uniform_k2}{width_q25}{22.2}
\setres{extC}{zurich03__EU_uniform_k2}{width_q75}{35.2}
\setres{extC}{zurich03__EU_uniform_k3}{coverage}{0.90}
\setres{extC}{zurich03__EU_uniform_k3}{k}{3}
\setres{extC}{zurich03__EU_uniform_k3}{m}{10}
\setres{extC}{zurich03__EU_uniform_k3}{width_median}{22.5}
\setres{extC}{zurich03__EU_uniform_k3}{width_q25}{21.8}
\setres{extC}{zurich03__EU_uniform_k3}{width_q75}{28.2}
\setres{extC}{zurich03__EU_uniform_k4}{coverage}{0.91}
\setres{extC}{zurich03__EU_uniform_k4}{k}{4}
\setres{extC}{zurich03__EU_uniform_k4}{m}{10}
\setres{extC}{zurich03__EU_uniform_k4}{width_median}{22.5}
\setres{extC}{zurich03__EU_uniform_k4}{width_q25}{21.9}
\setres{extC}{zurich03__EU_uniform_k4}{width_q75}{28.1}
\setres{extC}{zurich03__EU_uniform_k5}{coverage}{0.91}
\setres{extC}{zurich03__EU_uniform_k5}{k}{5}
\setres{extC}{zurich03__EU_uniform_k5}{m}{10}
\setres{extC}{zurich03__EU_uniform_k5}{width_median}{22.5}
\setres{extC}{zurich03__EU_uniform_k5}{width_q25}{21.8}
\setres{extC}{zurich03__EU_uniform_k5}{width_q75}{26.1}
\setres{extC}{zurich03__EU_uniform_k6}{coverage}{0.91}
\setres{extC}{zurich03__EU_uniform_k6}{k}{6}
\setres{extC}{zurich03__EU_uniform_k6}{m}{10}
\setres{extC}{zurich03__EU_uniform_k6}{width_median}{22.4}
\setres{extC}{zurich03__EU_uniform_k6}{width_q25}{21.6}
\setres{extC}{zurich03__EU_uniform_k6}{width_q75}{26.0}
\setres{extC}{zurich03__EU_uniform_k8}{coverage}{0.92}
\setres{extC}{zurich03__EU_uniform_k8}{k}{8}
\setres{extC}{zurich03__EU_uniform_k8}{m}{10}
\setres{extC}{zurich03__EU_uniform_k8}{width_median}{22.3}
\setres{extC}{zurich03__EU_uniform_k8}{width_q25}{21.5}
\setres{extC}{zurich03__EU_uniform_k8}{width_q75}{25.8}
\setres{extS}{CPT_after03}{misfit_pct}{5.3}
\setres{extS}{CPT_after0306}{misfit_pct}{4.9}
\setres{extS}{DA_after03}{misfit_pct}{14.9}
\setres{extS}{DA_after0306}{misfit_pct}{14.0}
\setres{extS}{EU_after03}{misfit_pct}{15.4}
\setres{extS}{EU_after0306}{misfit_pct}{14.6}
\setres{extS}{reference}{start}{lit}
\setres{sampP}{CPT_fit}{kappa_crps_median}{0.735}
\setres{sampP}{CPT_lit}{kappa_crps_median}{0.568}
\setres{sampP}{CPT_s1}{alpha_sd}{0.29}
\setres{sampP}{CPT_s1}{delta_sd}{0.68}
\setres{sampP}{CPT_s1}{gamma_sd}{0.22}
\setres{sampP}{CPT_s1}{kappa_crps_mean}{0.500}
\setres{sampP}{CPT_s1}{kappa_crps_median}{0.551}
\setres{sampP}{CPT_s1}{s}{1}
\setres{sampP}{CPT_s1}{theta_sd}{0.29, 0.22, 0.68}
\setres{sampP}{CPT_s12}{alpha_sd}{0.15}
\setres{sampP}{CPT_s12}{delta_sd}{0.32}
\setres{sampP}{CPT_s12}{gamma_sd}{0.13}
\setres{sampP}{CPT_s12}{kappa_crps_mean}{0.668}
\setres{sampP}{CPT_s12}{kappa_crps_median}{0.736}
\setres{sampP}{CPT_s12}{s}{12}
\setres{sampP}{CPT_s12}{theta_sd}{0.15, 0.13, 0.32}
\setres{sampP}{CPT_s2}{alpha_sd}{0.27}
\setres{sampP}{CPT_s2}{delta_sd}{0.57}
\setres{sampP}{CPT_s2}{gamma_sd}{0.19}
\setres{sampP}{CPT_s2}{kappa_crps_mean}{0.586}
\setres{sampP}{CPT_s2}{kappa_crps_median}{0.638}
\setres{sampP}{CPT_s2}{s}{2}
\setres{sampP}{CPT_s2}{theta_sd}{0.27, 0.19, 0.57}
\setres{sampP}{CPT_s20}{alpha_sd}{0.12}
\setres{sampP}{CPT_s20}{delta_sd}{0.28}
\setres{sampP}{CPT_s20}{gamma_sd}{0.15}
\setres{sampP}{CPT_s20}{kappa_crps_mean}{0.676}
\setres{sampP}{CPT_s20}{kappa_crps_median}{0.737}
\setres{sampP}{CPT_s20}{s}{20}
\setres{sampP}{CPT_s20}{theta_sd}{0.12, 0.15, 0.28}
\setres{sampP}{CPT_s29}{alpha_sd}{0.10}
\setres{sampP}{CPT_s29}{delta_sd}{0.25}
\setres{sampP}{CPT_s29}{gamma_sd}{0.14}
\setres{sampP}{CPT_s29}{kappa_crps_mean}{0.677}
\setres{sampP}{CPT_s29}{kappa_crps_median}{0.735}
\setres{sampP}{CPT_s29}{s}{29}
\setres{sampP}{CPT_s29}{theta_sd}{0.10, 0.14, 0.25}
\setres{sampP}{CPT_s3}{alpha_sd}{0.24}
\setres{sampP}{CPT_s3}{delta_sd}{0.50}
\setres{sampP}{CPT_s3}{gamma_sd}{0.17}
\setres{sampP}{CPT_s3}{kappa_crps_mean}{0.619}
\setres{sampP}{CPT_s3}{kappa_crps_median}{0.678}
\setres{sampP}{CPT_s3}{s}{3}
\setres{sampP}{CPT_s3}{theta_sd}{0.24, 0.17, 0.50}
\setres{sampP}{CPT_s5}{alpha_sd}{0.20}
\setres{sampP}{CPT_s5}{delta_sd}{0.43}
\setres{sampP}{CPT_s5}{gamma_sd}{0.15}
\setres{sampP}{CPT_s5}{kappa_crps_mean}{0.647}
\setres{sampP}{CPT_s5}{kappa_crps_median}{0.703}
\setres{sampP}{CPT_s5}{s}{5}
\setres{sampP}{CPT_s5}{theta_sd}{0.20, 0.15, 0.43}
\setres{sampP}{CPT_s8}{alpha_sd}{0.17}
\setres{sampP}{CPT_s8}{delta_sd}{0.37}
\setres{sampP}{CPT_s8}{gamma_sd}{0.14}
\setres{sampP}{CPT_s8}{kappa_crps_mean}{0.663}
\setres{sampP}{CPT_s8}{kappa_crps_median}{0.724}
\setres{sampP}{CPT_s8}{s}{8}
\setres{sampP}{CPT_s8}{theta_sd}{0.17, 0.14, 0.37}
\setres{sampP}{DA_fit}{kappa_crps_median}{0.578}
\setres{sampP}{DA_lit}{kappa_crps_median}{0.023}
\setres{sampP}{DA_s1}{alpha_sd}{0.31}
\setres{sampP}{DA_s1}{eta_sd}{0.90}
\setres{sampP}{DA_s1}{kappa_crps_mean}{0.246}
\setres{sampP}{DA_s1}{kappa_crps_median}{0.335}
\setres{sampP}{DA_s1}{s}{1}
\setres{sampP}{DA_s1}{theta_sd}{0.31, 0.90}
\setres{sampP}{DA_s12}{alpha_sd}{0.15}
\setres{sampP}{DA_s12}{eta_sd}{0.44}
\setres{sampP}{DA_s12}{kappa_crps_mean}{0.476}
\setres{sampP}{DA_s12}{kappa_crps_median}{0.578}
\setres{sampP}{DA_s12}{s}{12}
\setres{sampP}{DA_s12}{theta_sd}{0.15, 0.44}
\setres{sampP}{DA_s2}{alpha_sd}{0.28}
\setres{sampP}{DA_s2}{eta_sd}{0.78}
\setres{sampP}{DA_s2}{kappa_crps_mean}{0.383}
\setres{sampP}{DA_s2}{kappa_crps_median}{0.486}
\setres{sampP}{DA_s2}{s}{2}
\setres{sampP}{DA_s2}{theta_sd}{0.28, 0.78}
\setres{sampP}{DA_s20}{alpha_sd}{0.12}
\setres{sampP}{DA_s20}{eta_sd}{0.37}
\setres{sampP}{DA_s20}{kappa_crps_mean}{0.481}
\setres{sampP}{DA_s20}{kappa_crps_median}{0.579}
\setres{sampP}{DA_s20}{s}{20}
\setres{sampP}{DA_s20}{theta_sd}{0.12, 0.37}
\setres{sampP}{DA_s29}{alpha_sd}{0.10}
\setres{sampP}{DA_s29}{eta_sd}{0.33}
\setres{sampP}{DA_s29}{kappa_crps_mean}{0.482}
\setres{sampP}{DA_s29}{kappa_crps_median}{0.578}
\setres{sampP}{DA_s29}{s}{29}
\setres{sampP}{DA_s29}{theta_sd}{0.10, 0.33}
\setres{sampP}{DA_s3}{alpha_sd}{0.25}
\setres{sampP}{DA_s3}{eta_sd}{0.71}
\setres{sampP}{DA_s3}{kappa_crps_mean}{0.432}
\setres{sampP}{DA_s3}{kappa_crps_median}{0.531}
\setres{sampP}{DA_s3}{s}{3}
\setres{sampP}{DA_s3}{theta_sd}{0.25, 0.71}
\setres{sampP}{DA_s5}{alpha_sd}{0.21}
\setres{sampP}{DA_s5}{eta_sd}{0.60}
\setres{sampP}{DA_s5}{kappa_crps_mean}{0.462}
\setres{sampP}{DA_s5}{kappa_crps_median}{0.560}
\setres{sampP}{DA_s5}{s}{5}
\setres{sampP}{DA_s5}{theta_sd}{0.21, 0.60}
\setres{sampP}{DA_s8}{alpha_sd}{0.18}
\setres{sampP}{DA_s8}{eta_sd}{0.51}
\setres{sampP}{DA_s8}{kappa_crps_mean}{0.474}
\setres{sampP}{DA_s8}{kappa_crps_median}{0.578}
\setres{sampP}{DA_s8}{s}{8}
\setres{sampP}{DA_s8}{theta_sd}{0.18, 0.51}
\setres{sampP}{EU_fit}{kappa_crps_median}{0.491}
\setres{sampP}{EU_lit}{kappa_crps_median}{0.356}
\setres{sampP}{EU_s1}{alpha_sd}{0.28}
\setres{sampP}{EU_s1}{kappa_crps_mean}{0.321}
\setres{sampP}{EU_s1}{kappa_crps_median}{0.425}
\setres{sampP}{EU_s1}{s}{1}
\setres{sampP}{EU_s1}{theta_sd}{0.28}
\setres{sampP}{EU_s12}{alpha_sd}{0.15}
\setres{sampP}{EU_s12}{kappa_crps_mean}{0.417}
\setres{sampP}{EU_s12}{kappa_crps_median}{0.500}
\setres{sampP}{EU_s12}{s}{12}
\setres{sampP}{EU_s12}{theta_sd}{0.15}
\setres{sampP}{EU_s2}{alpha_sd}{0.26}
\setres{sampP}{EU_s2}{kappa_crps_mean}{0.385}
\setres{sampP}{EU_s2}{kappa_crps_median}{0.486}
\setres{sampP}{EU_s2}{s}{2}
\setres{sampP}{EU_s2}{theta_sd}{0.26}
\setres{sampP}{EU_s20}{alpha_sd}{0.13}
\setres{sampP}{EU_s20}{kappa_crps_mean}{0.418}
\setres{sampP}{EU_s20}{kappa_crps_median}{0.496}
\setres{sampP}{EU_s20}{s}{20}
\setres{sampP}{EU_s20}{theta_sd}{0.13}
\setres{sampP}{EU_s29}{alpha_sd}{0.12}
\setres{sampP}{EU_s29}{kappa_crps_mean}{0.418}
\setres{sampP}{EU_s29}{kappa_crps_median}{0.491}
\setres{sampP}{EU_s29}{s}{29}
\setres{sampP}{EU_s29}{theta_sd}{0.12}
\setres{sampP}{EU_s3}{alpha_sd}{0.23}
\setres{sampP}{EU_s3}{kappa_crps_mean}{0.403}
\setres{sampP}{EU_s3}{kappa_crps_median}{0.503}
\setres{sampP}{EU_s3}{s}{3}
\setres{sampP}{EU_s3}{theta_sd}{0.23}
\setres{sampP}{EU_s5}{alpha_sd}{0.20}
\setres{sampP}{EU_s5}{kappa_crps_mean}{0.415}
\setres{sampP}{EU_s5}{kappa_crps_median}{0.511}
\setres{sampP}{EU_s5}{s}{5}
\setres{sampP}{EU_s5}{theta_sd}{0.20}
\setres{sampP}{EU_s8}{alpha_sd}{0.18}
\setres{sampP}{EU_s8}{kappa_crps_mean}{0.417}
\setres{sampP}{EU_s8}{kappa_crps_median}{0.505}
\setres{sampP}{EU_s8}{s}{8}
\setres{sampP}{EU_s8}{theta_sd}{0.18}
\setres{sampP}{reference}{n_orders}{10}
\setres{sampP}{reference}{s_grid}{1.000, 2.000, 3.000, 5.000, 8.000, 12.000, 20.000, 29.000}
\setres{sampP}{reference}{score}{CRPS of the country mean}
\setres{sampP}{reference}{start}{uniform}
\setres{ctyP}{CPT_lit}{kappa_crps_median}{0.489}
\setres{ctyP}{CPT_s1}{kappa_crps_max}{0.954}
\setres{ctyP}{CPT_s1}{kappa_crps_median}{0.581}
\setres{ctyP}{CPT_s1}{kappa_crps_min}{-0.308}
\setres{ctyP}{CPT_s1}{kappa_crps_q25}{0.462}
\setres{ctyP}{CPT_s1}{kappa_crps_q75}{0.783}
\setres{ctyP}{CPT_s1}{s}{1}
\setres{ctyP}{CPT_s10}{kappa_crps_max}{0.925}
\setres{ctyP}{CPT_s10}{kappa_crps_median}{0.897}
\setres{ctyP}{CPT_s10}{kappa_crps_min}{-0.381}
\setres{ctyP}{CPT_s10}{kappa_crps_q25}{0.820}
\setres{ctyP}{CPT_s10}{kappa_crps_q75}{0.911}
\setres{ctyP}{CPT_s10}{s}{10}
\setres{ctyP}{CPT_s11}{kappa_crps_max}{0.937}
\setres{ctyP}{CPT_s11}{kappa_crps_median}{0.868}
\setres{ctyP}{CPT_s11}{kappa_crps_min}{0.646}
\setres{ctyP}{CPT_s11}{kappa_crps_q25}{0.737}
\setres{ctyP}{CPT_s11}{kappa_crps_q75}{0.883}
\setres{ctyP}{CPT_s11}{s}{11}
\setres{ctyP}{CPT_s12}{kappa_crps_max}{0.895}
\setres{ctyP}{CPT_s12}{kappa_crps_median}{0.733}
\setres{ctyP}{CPT_s12}{kappa_crps_min}{-3.296}
\setres{ctyP}{CPT_s12}{kappa_crps_q25}{0.638}
\setres{ctyP}{CPT_s12}{kappa_crps_q75}{0.840}
\setres{ctyP}{CPT_s12}{s}{12}
\setres{ctyP}{CPT_s13}{kappa_crps_max}{0.948}
\setres{ctyP}{CPT_s13}{kappa_crps_median}{0.897}
\setres{ctyP}{CPT_s13}{kappa_crps_min}{0.672}
\setres{ctyP}{CPT_s13}{kappa_crps_q25}{0.835}
\setres{ctyP}{CPT_s13}{kappa_crps_q75}{0.925}
\setres{ctyP}{CPT_s13}{s}{13}
\setres{ctyP}{CPT_s14}{kappa_crps_max}{0.945}
\setres{ctyP}{CPT_s14}{kappa_crps_median}{0.858}
\setres{ctyP}{CPT_s14}{kappa_crps_min}{0.747}
\setres{ctyP}{CPT_s14}{kappa_crps_q25}{0.838}
\setres{ctyP}{CPT_s14}{kappa_crps_q75}{0.900}
\setres{ctyP}{CPT_s14}{s}{14}
\setres{ctyP}{CPT_s15}{kappa_crps_max}{0.937}
\setres{ctyP}{CPT_s15}{kappa_crps_median}{0.761}
\setres{ctyP}{CPT_s15}{kappa_crps_min}{-0.124}
\setres{ctyP}{CPT_s15}{kappa_crps_q25}{0.559}
\setres{ctyP}{CPT_s15}{kappa_crps_q75}{0.915}
\setres{ctyP}{CPT_s15}{s}{15}
\setres{ctyP}{CPT_s16}{kappa_crps_max}{0.973}
\setres{ctyP}{CPT_s16}{kappa_crps_median}{0.896}
\setres{ctyP}{CPT_s16}{kappa_crps_min}{0.770}
\setres{ctyP}{CPT_s16}{kappa_crps_q25}{0.856}
\setres{ctyP}{CPT_s16}{kappa_crps_q75}{0.928}
\setres{ctyP}{CPT_s16}{s}{16}
\setres{ctyP}{CPT_s17}{kappa_crps_max}{0.911}
\setres{ctyP}{CPT_s17}{kappa_crps_median}{0.849}
\setres{ctyP}{CPT_s17}{kappa_crps_min}{-3.277}
\setres{ctyP}{CPT_s17}{kappa_crps_q25}{0.780}
\setres{ctyP}{CPT_s17}{kappa_crps_q75}{0.871}
\setres{ctyP}{CPT_s17}{s}{17}
\setres{ctyP}{CPT_s18}{kappa_crps_max}{0.968}
\setres{ctyP}{CPT_s18}{kappa_crps_median}{0.924}
\setres{ctyP}{CPT_s18}{kappa_crps_min}{0.819}
\setres{ctyP}{CPT_s18}{kappa_crps_q25}{0.901}
\setres{ctyP}{CPT_s18}{kappa_crps_q75}{0.939}
\setres{ctyP}{CPT_s18}{s}{18}
\setres{ctyP}{CPT_s19}{kappa_crps_max}{0.925}
\setres{ctyP}{CPT_s19}{kappa_crps_median}{0.874}
\setres{ctyP}{CPT_s19}{kappa_crps_min}{-0.022}
\setres{ctyP}{CPT_s19}{kappa_crps_q25}{0.768}
\setres{ctyP}{CPT_s19}{kappa_crps_q75}{0.916}
\setres{ctyP}{CPT_s19}{s}{19}
\setres{ctyP}{CPT_s2}{kappa_crps_max}{0.919}
\setres{ctyP}{CPT_s2}{kappa_crps_median}{0.807}
\setres{ctyP}{CPT_s2}{kappa_crps_min}{0.444}
\setres{ctyP}{CPT_s2}{kappa_crps_q25}{0.754}
\setres{ctyP}{CPT_s2}{kappa_crps_q75}{0.872}
\setres{ctyP}{CPT_s2}{s}{2}
\setres{ctyP}{CPT_s20}{kappa_crps_max}{0.936}
\setres{ctyP}{CPT_s20}{kappa_crps_median}{0.887}
\setres{ctyP}{CPT_s20}{kappa_crps_min}{0.623}
\setres{ctyP}{CPT_s20}{kappa_crps_q25}{0.808}
\setres{ctyP}{CPT_s20}{kappa_crps_q75}{0.907}
\setres{ctyP}{CPT_s20}{s}{20}
\setres{ctyP}{CPT_s21}{kappa_crps_max}{0.932}
\setres{ctyP}{CPT_s21}{kappa_crps_median}{0.855}
\setres{ctyP}{CPT_s21}{kappa_crps_min}{-0.339}
\setres{ctyP}{CPT_s21}{kappa_crps_q25}{0.593}
\setres{ctyP}{CPT_s21}{kappa_crps_q75}{0.916}
\setres{ctyP}{CPT_s21}{s}{21}
\setres{ctyP}{CPT_s22}{kappa_crps_max}{0.944}
\setres{ctyP}{CPT_s22}{kappa_crps_median}{0.775}
\setres{ctyP}{CPT_s22}{kappa_crps_min}{-3.160}
\setres{ctyP}{CPT_s22}{kappa_crps_q25}{-0.355}
\setres{ctyP}{CPT_s22}{kappa_crps_q75}{0.904}
\setres{ctyP}{CPT_s22}{s}{22}
\setres{ctyP}{CPT_s23}{kappa_crps_max}{0.961}
\setres{ctyP}{CPT_s23}{kappa_crps_median}{0.811}
\setres{ctyP}{CPT_s23}{kappa_crps_min}{-0.257}
\setres{ctyP}{CPT_s23}{kappa_crps_q25}{0.721}
\setres{ctyP}{CPT_s23}{kappa_crps_q75}{0.874}
\setres{ctyP}{CPT_s23}{s}{23}
\setres{ctyP}{CPT_s24}{kappa_crps_max}{0.945}
\setres{ctyP}{CPT_s24}{kappa_crps_median}{0.899}
\setres{ctyP}{CPT_s24}{kappa_crps_min}{0.486}
\setres{ctyP}{CPT_s24}{kappa_crps_q25}{0.672}
\setres{ctyP}{CPT_s24}{kappa_crps_q75}{0.933}
\setres{ctyP}{CPT_s24}{s}{24}
\setres{ctyP}{CPT_s25}{kappa_crps_max}{0.945}
\setres{ctyP}{CPT_s25}{kappa_crps_median}{0.807}
\setres{ctyP}{CPT_s25}{kappa_crps_min}{-3.073}
\setres{ctyP}{CPT_s25}{kappa_crps_q25}{0.680}
\setres{ctyP}{CPT_s25}{kappa_crps_q75}{0.899}
\setres{ctyP}{CPT_s25}{s}{25}
\setres{ctyP}{CPT_s26}{kappa_crps_max}{0.929}
\setres{ctyP}{CPT_s26}{kappa_crps_median}{0.878}
\setres{ctyP}{CPT_s26}{kappa_crps_min}{0.508}
\setres{ctyP}{CPT_s26}{kappa_crps_q25}{0.844}
\setres{ctyP}{CPT_s26}{kappa_crps_q75}{0.906}
\setres{ctyP}{CPT_s26}{s}{26}
\setres{ctyP}{CPT_s27}{kappa_crps_max}{0.904}
\setres{ctyP}{CPT_s27}{kappa_crps_median}{0.812}
\setres{ctyP}{CPT_s27}{kappa_crps_min}{-0.226}
\setres{ctyP}{CPT_s27}{kappa_crps_q25}{0.640}
\setres{ctyP}{CPT_s27}{kappa_crps_q75}{0.876}
\setres{ctyP}{CPT_s27}{s}{27}
\setres{ctyP}{CPT_s28}{kappa_crps_max}{0.942}
\setres{ctyP}{CPT_s28}{kappa_crps_median}{0.879}
\setres{ctyP}{CPT_s28}{kappa_crps_min}{0.418}
\setres{ctyP}{CPT_s28}{kappa_crps_q25}{0.755}
\setres{ctyP}{CPT_s28}{kappa_crps_q75}{0.905}
\setres{ctyP}{CPT_s28}{s}{28}
\setres{ctyP}{CPT_s29}{kappa_crps_max}{0.959}
\setres{ctyP}{CPT_s29}{kappa_crps_median}{0.897}
\setres{ctyP}{CPT_s29}{kappa_crps_min}{0.396}
\setres{ctyP}{CPT_s29}{kappa_crps_q25}{0.876}
\setres{ctyP}{CPT_s29}{kappa_crps_q75}{0.955}
\setres{ctyP}{CPT_s29}{s}{29}
\setres{ctyP}{CPT_s3}{kappa_crps_max}{0.938}
\setres{ctyP}{CPT_s3}{kappa_crps_median}{0.798}
\setres{ctyP}{CPT_s3}{kappa_crps_min}{-4.855}
\setres{ctyP}{CPT_s3}{kappa_crps_q25}{0.428}
\setres{ctyP}{CPT_s3}{kappa_crps_q75}{0.846}
\setres{ctyP}{CPT_s3}{s}{3}
\setres{ctyP}{CPT_s4}{kappa_crps_max}{0.926}
\setres{ctyP}{CPT_s4}{kappa_crps_median}{0.856}
\setres{ctyP}{CPT_s4}{kappa_crps_min}{0.252}
\setres{ctyP}{CPT_s4}{kappa_crps_q25}{0.701}
\setres{ctyP}{CPT_s4}{kappa_crps_q75}{0.871}
\setres{ctyP}{CPT_s4}{s}{4}
\setres{ctyP}{CPT_s5}{kappa_crps_max}{0.930}
\setres{ctyP}{CPT_s5}{kappa_crps_median}{0.856}
\setres{ctyP}{CPT_s5}{kappa_crps_min}{0.679}
\setres{ctyP}{CPT_s5}{kappa_crps_q25}{0.769}
\setres{ctyP}{CPT_s5}{kappa_crps_q75}{0.884}
\setres{ctyP}{CPT_s5}{s}{5}
\setres{ctyP}{CPT_s6}{kappa_crps_max}{0.938}
\setres{ctyP}{CPT_s6}{kappa_crps_median}{0.791}
\setres{ctyP}{CPT_s6}{kappa_crps_min}{0.499}
\setres{ctyP}{CPT_s6}{kappa_crps_q25}{0.762}
\setres{ctyP}{CPT_s6}{kappa_crps_q75}{0.898}
\setres{ctyP}{CPT_s6}{s}{6}
\setres{ctyP}{CPT_s7}{kappa_crps_max}{0.915}
\setres{ctyP}{CPT_s7}{kappa_crps_median}{0.550}
\setres{ctyP}{CPT_s7}{kappa_crps_min}{-2.697}
\setres{ctyP}{CPT_s7}{kappa_crps_q25}{0.374}
\setres{ctyP}{CPT_s7}{kappa_crps_q75}{0.676}
\setres{ctyP}{CPT_s7}{s}{7}
\setres{ctyP}{CPT_s8}{kappa_crps_max}{0.923}
\setres{ctyP}{CPT_s8}{kappa_crps_median}{0.822}
\setres{ctyP}{CPT_s8}{kappa_crps_min}{0.088}
\setres{ctyP}{CPT_s8}{kappa_crps_q25}{0.722}
\setres{ctyP}{CPT_s8}{kappa_crps_q75}{0.884}
\setres{ctyP}{CPT_s8}{s}{8}
\setres{ctyP}{CPT_s9}{kappa_crps_max}{0.932}
\setres{ctyP}{CPT_s9}{kappa_crps_median}{0.843}
\setres{ctyP}{CPT_s9}{kappa_crps_min}{-4.367}
\setres{ctyP}{CPT_s9}{kappa_crps_q25}{0.812}
\setres{ctyP}{CPT_s9}{kappa_crps_q75}{0.877}
\setres{ctyP}{CPT_s9}{s}{9}
\setres{ctyP}{CPT_typical}{order_index}{7}
\setres{ctyP}{CPT_within}{kappa_crps_median}{0.858}
\setres{ctyP}{DA_lit}{kappa_crps_median}{-0.302}
\setres{ctyP}{DA_s1}{kappa_crps_max}{0.914}
\setres{ctyP}{DA_s1}{kappa_crps_median}{0.624}
\setres{ctyP}{DA_s1}{kappa_crps_min}{-0.124}
\setres{ctyP}{DA_s1}{kappa_crps_q25}{0.451}
\setres{ctyP}{DA_s1}{kappa_crps_q75}{0.796}
\setres{ctyP}{DA_s1}{s}{1}
\setres{ctyP}{DA_s10}{kappa_crps_max}{0.913}
\setres{ctyP}{DA_s10}{kappa_crps_median}{0.797}
\setres{ctyP}{DA_s10}{kappa_crps_min}{-0.016}
\setres{ctyP}{DA_s10}{kappa_crps_q25}{0.764}
\setres{ctyP}{DA_s10}{kappa_crps_q75}{0.853}
\setres{ctyP}{DA_s10}{s}{10}
\setres{ctyP}{DA_s11}{kappa_crps_max}{0.896}
\setres{ctyP}{DA_s11}{kappa_crps_median}{0.810}
\setres{ctyP}{DA_s11}{kappa_crps_min}{0.744}
\setres{ctyP}{DA_s11}{kappa_crps_q25}{0.757}
\setres{ctyP}{DA_s11}{kappa_crps_q75}{0.830}
\setres{ctyP}{DA_s11}{s}{11}
\setres{ctyP}{DA_s12}{kappa_crps_max}{0.830}
\setres{ctyP}{DA_s12}{kappa_crps_median}{0.804}
\setres{ctyP}{DA_s12}{kappa_crps_min}{-2.595}
\setres{ctyP}{DA_s12}{kappa_crps_q25}{0.701}
\setres{ctyP}{DA_s12}{kappa_crps_q75}{0.821}
\setres{ctyP}{DA_s12}{s}{12}
\setres{ctyP}{DA_s13}{kappa_crps_max}{0.890}
\setres{ctyP}{DA_s13}{kappa_crps_median}{0.817}
\setres{ctyP}{DA_s13}{kappa_crps_min}{0.489}
\setres{ctyP}{DA_s13}{kappa_crps_q25}{0.801}
\setres{ctyP}{DA_s13}{kappa_crps_q75}{0.872}
\setres{ctyP}{DA_s13}{s}{13}
\setres{ctyP}{DA_s14}{kappa_crps_max}{0.899}
\setres{ctyP}{DA_s14}{kappa_crps_median}{0.802}
\setres{ctyP}{DA_s14}{kappa_crps_min}{0.581}
\setres{ctyP}{DA_s14}{kappa_crps_q25}{0.770}
\setres{ctyP}{DA_s14}{kappa_crps_q75}{0.850}
\setres{ctyP}{DA_s14}{s}{14}
\setres{ctyP}{DA_s15}{kappa_crps_max}{0.957}
\setres{ctyP}{DA_s15}{kappa_crps_median}{0.744}
\setres{ctyP}{DA_s15}{kappa_crps_min}{0.173}
\setres{ctyP}{DA_s15}{kappa_crps_q25}{0.508}
\setres{ctyP}{DA_s15}{kappa_crps_q75}{0.873}
\setres{ctyP}{DA_s15}{s}{15}
\setres{ctyP}{DA_s16}{kappa_crps_max}{0.943}
\setres{ctyP}{DA_s16}{kappa_crps_median}{0.841}
\setres{ctyP}{DA_s16}{kappa_crps_min}{0.549}
\setres{ctyP}{DA_s16}{kappa_crps_q25}{0.714}
\setres{ctyP}{DA_s16}{kappa_crps_q75}{0.890}
\setres{ctyP}{DA_s16}{s}{16}
\setres{ctyP}{DA_s17}{kappa_crps_max}{0.892}
\setres{ctyP}{DA_s17}{kappa_crps_median}{0.860}
\setres{ctyP}{DA_s17}{kappa_crps_min}{-3.519}
\setres{ctyP}{DA_s17}{kappa_crps_q25}{0.759}
\setres{ctyP}{DA_s17}{kappa_crps_q75}{0.868}
\setres{ctyP}{DA_s17}{s}{17}
\setres{ctyP}{DA_s18}{kappa_crps_max}{0.864}
\setres{ctyP}{DA_s18}{kappa_crps_median}{0.852}
\setres{ctyP}{DA_s18}{kappa_crps_min}{0.827}
\setres{ctyP}{DA_s18}{kappa_crps_q25}{0.847}
\setres{ctyP}{DA_s18}{kappa_crps_q75}{0.856}
\setres{ctyP}{DA_s18}{s}{18}
\setres{ctyP}{DA_s19}{kappa_crps_max}{0.956}
\setres{ctyP}{DA_s19}{kappa_crps_median}{0.845}
\setres{ctyP}{DA_s19}{kappa_crps_min}{0.322}
\setres{ctyP}{DA_s19}{kappa_crps_q25}{0.770}
\setres{ctyP}{DA_s19}{kappa_crps_q75}{0.879}
\setres{ctyP}{DA_s19}{s}{19}
\setres{ctyP}{DA_s2}{kappa_crps_max}{0.860}
\setres{ctyP}{DA_s2}{kappa_crps_median}{0.763}
\setres{ctyP}{DA_s2}{kappa_crps_min}{0.244}
\setres{ctyP}{DA_s2}{kappa_crps_q25}{0.603}
\setres{ctyP}{DA_s2}{kappa_crps_q75}{0.791}
\setres{ctyP}{DA_s2}{s}{2}
\setres{ctyP}{DA_s20}{kappa_crps_max}{0.967}
\setres{ctyP}{DA_s20}{kappa_crps_median}{0.857}
\setres{ctyP}{DA_s20}{kappa_crps_min}{0.595}
\setres{ctyP}{DA_s20}{kappa_crps_q25}{0.777}
\setres{ctyP}{DA_s20}{kappa_crps_q75}{0.893}
\setres{ctyP}{DA_s20}{s}{20}
\setres{ctyP}{DA_s21}{kappa_crps_max}{0.898}
\setres{ctyP}{DA_s21}{kappa_crps_median}{0.754}
\setres{ctyP}{DA_s21}{kappa_crps_min}{0.084}
\setres{ctyP}{DA_s21}{kappa_crps_q25}{0.576}
\setres{ctyP}{DA_s21}{kappa_crps_q75}{0.872}
\setres{ctyP}{DA_s21}{s}{21}
\setres{ctyP}{DA_s22}{kappa_crps_max}{0.873}
\setres{ctyP}{DA_s22}{kappa_crps_median}{0.825}
\setres{ctyP}{DA_s22}{kappa_crps_min}{-3.169}
\setres{ctyP}{DA_s22}{kappa_crps_q25}{-0.525}
\setres{ctyP}{DA_s22}{kappa_crps_q75}{0.847}
\setres{ctyP}{DA_s22}{s}{22}
\setres{ctyP}{DA_s23}{kappa_crps_max}{0.966}
\setres{ctyP}{DA_s23}{kappa_crps_median}{0.749}
\setres{ctyP}{DA_s23}{kappa_crps_min}{0.320}
\setres{ctyP}{DA_s23}{kappa_crps_q25}{0.504}
\setres{ctyP}{DA_s23}{kappa_crps_q75}{0.805}
\setres{ctyP}{DA_s23}{s}{23}
\setres{ctyP}{DA_s24}{kappa_crps_max}{0.971}
\setres{ctyP}{DA_s24}{kappa_crps_median}{0.803}
\setres{ctyP}{DA_s24}{kappa_crps_min}{0.406}
\setres{ctyP}{DA_s24}{kappa_crps_q25}{0.517}
\setres{ctyP}{DA_s24}{kappa_crps_q75}{0.864}
\setres{ctyP}{DA_s24}{s}{24}
\setres{ctyP}{DA_s25}{kappa_crps_max}{0.846}
\setres{ctyP}{DA_s25}{kappa_crps_median}{0.760}
\setres{ctyP}{DA_s25}{kappa_crps_min}{-3.522}
\setres{ctyP}{DA_s25}{kappa_crps_q25}{0.590}
\setres{ctyP}{DA_s25}{kappa_crps_q75}{0.834}
\setres{ctyP}{DA_s25}{s}{25}
\setres{ctyP}{DA_s26}{kappa_crps_max}{0.860}
\setres{ctyP}{DA_s26}{kappa_crps_median}{0.802}
\setres{ctyP}{DA_s26}{kappa_crps_min}{0.480}
\setres{ctyP}{DA_s26}{kappa_crps_q25}{0.778}
\setres{ctyP}{DA_s26}{kappa_crps_q75}{0.853}
\setres{ctyP}{DA_s26}{s}{26}
\setres{ctyP}{DA_s27}{kappa_crps_max}{0.856}
\setres{ctyP}{DA_s27}{kappa_crps_median}{0.635}
\setres{ctyP}{DA_s27}{kappa_crps_min}{0.232}
\setres{ctyP}{DA_s27}{kappa_crps_q25}{0.594}
\setres{ctyP}{DA_s27}{kappa_crps_q75}{0.777}
\setres{ctyP}{DA_s27}{s}{27}
\setres{ctyP}{DA_s28}{kappa_crps_max}{0.881}
\setres{ctyP}{DA_s28}{kappa_crps_median}{0.804}
\setres{ctyP}{DA_s28}{kappa_crps_min}{0.353}
\setres{ctyP}{DA_s28}{kappa_crps_q25}{0.658}
\setres{ctyP}{DA_s28}{kappa_crps_q75}{0.842}
\setres{ctyP}{DA_s28}{s}{28}
\setres{ctyP}{DA_s29}{kappa_crps_max}{0.879}
\setres{ctyP}{DA_s29}{kappa_crps_median}{0.852}
\setres{ctyP}{DA_s29}{kappa_crps_min}{0.593}
\setres{ctyP}{DA_s29}{kappa_crps_q25}{0.806}
\setres{ctyP}{DA_s29}{kappa_crps_q75}{0.867}
\setres{ctyP}{DA_s29}{s}{29}
\setres{ctyP}{DA_s3}{kappa_crps_max}{0.872}
\setres{ctyP}{DA_s3}{kappa_crps_median}{0.750}
\setres{ctyP}{DA_s3}{kappa_crps_min}{-5.826}
\setres{ctyP}{DA_s3}{kappa_crps_q25}{0.486}
\setres{ctyP}{DA_s3}{kappa_crps_q75}{0.862}
\setres{ctyP}{DA_s3}{s}{3}
\setres{ctyP}{DA_s4}{kappa_crps_max}{0.866}
\setres{ctyP}{DA_s4}{kappa_crps_median}{0.771}
\setres{ctyP}{DA_s4}{kappa_crps_min}{0.516}
\setres{ctyP}{DA_s4}{kappa_crps_q25}{0.725}
\setres{ctyP}{DA_s4}{kappa_crps_q75}{0.780}
\setres{ctyP}{DA_s4}{s}{4}
\setres{ctyP}{DA_s5}{kappa_crps_max}{0.925}
\setres{ctyP}{DA_s5}{kappa_crps_median}{0.837}
\setres{ctyP}{DA_s5}{kappa_crps_min}{0.568}
\setres{ctyP}{DA_s5}{kappa_crps_q25}{0.800}
\setres{ctyP}{DA_s5}{kappa_crps_q75}{0.844}
\setres{ctyP}{DA_s5}{s}{5}
\setres{ctyP}{DA_s6}{kappa_crps_max}{0.892}
\setres{ctyP}{DA_s6}{kappa_crps_median}{0.797}
\setres{ctyP}{DA_s6}{kappa_crps_min}{0.561}
\setres{ctyP}{DA_s6}{kappa_crps_q25}{0.758}
\setres{ctyP}{DA_s6}{kappa_crps_q75}{0.836}
\setres{ctyP}{DA_s6}{s}{6}
\setres{ctyP}{DA_s7}{kappa_crps_max}{0.868}
\setres{ctyP}{DA_s7}{kappa_crps_median}{0.600}
\setres{ctyP}{DA_s7}{kappa_crps_min}{-2.692}
\setres{ctyP}{DA_s7}{kappa_crps_q25}{0.408}
\setres{ctyP}{DA_s7}{kappa_crps_q75}{0.817}
\setres{ctyP}{DA_s7}{s}{7}
\setres{ctyP}{DA_s8}{kappa_crps_max}{0.862}
\setres{ctyP}{DA_s8}{kappa_crps_median}{0.719}
\setres{ctyP}{DA_s8}{kappa_crps_min}{0.468}
\setres{ctyP}{DA_s8}{kappa_crps_q25}{0.676}
\setres{ctyP}{DA_s8}{kappa_crps_q75}{0.759}
\setres{ctyP}{DA_s8}{s}{8}
\setres{ctyP}{DA_s9}{kappa_crps_max}{0.898}
\setres{ctyP}{DA_s9}{kappa_crps_median}{0.765}
\setres{ctyP}{DA_s9}{kappa_crps_min}{-5.308}
\setres{ctyP}{DA_s9}{kappa_crps_q25}{0.569}
\setres{ctyP}{DA_s9}{kappa_crps_q75}{0.813}
\setres{ctyP}{DA_s9}{s}{9}
\setres{ctyP}{DA_typical}{order_index}{0}
\setres{ctyP}{DA_within}{kappa_crps_median}{0.776}
\setres{ctyP}{EU_lit}{kappa_crps_median}{0.060}
\setres{ctyP}{EU_s1}{kappa_crps_max}{0.780}
\setres{ctyP}{EU_s1}{kappa_crps_median}{0.241}
\setres{ctyP}{EU_s1}{kappa_crps_min}{-0.648}
\setres{ctyP}{EU_s1}{kappa_crps_q25}{-0.498}
\setres{ctyP}{EU_s1}{kappa_crps_q75}{0.470}
\setres{ctyP}{EU_s1}{s}{1}
\setres{ctyP}{EU_s10}{kappa_crps_max}{0.728}
\setres{ctyP}{EU_s10}{kappa_crps_median}{0.509}
\setres{ctyP}{EU_s10}{kappa_crps_min}{-1.558}
\setres{ctyP}{EU_s10}{kappa_crps_q25}{0.232}
\setres{ctyP}{EU_s10}{kappa_crps_q75}{0.645}
\setres{ctyP}{EU_s10}{s}{10}
\setres{ctyP}{EU_s11}{kappa_crps_max}{0.776}
\setres{ctyP}{EU_s11}{kappa_crps_median}{0.537}
\setres{ctyP}{EU_s11}{kappa_crps_min}{0.035}
\setres{ctyP}{EU_s11}{kappa_crps_q25}{0.337}
\setres{ctyP}{EU_s11}{kappa_crps_q75}{0.730}
\setres{ctyP}{EU_s11}{s}{11}
\setres{ctyP}{EU_s12}{kappa_crps_max}{0.669}
\setres{ctyP}{EU_s12}{kappa_crps_median}{0.396}
\setres{ctyP}{EU_s12}{kappa_crps_min}{-7.534}
\setres{ctyP}{EU_s12}{kappa_crps_q25}{0.187}
\setres{ctyP}{EU_s12}{kappa_crps_q75}{0.484}
\setres{ctyP}{EU_s12}{s}{12}
\setres{ctyP}{EU_s13}{kappa_crps_max}{0.770}
\setres{ctyP}{EU_s13}{kappa_crps_median}{0.521}
\setres{ctyP}{EU_s13}{kappa_crps_min}{-0.465}
\setres{ctyP}{EU_s13}{kappa_crps_q25}{0.253}
\setres{ctyP}{EU_s13}{kappa_crps_q75}{0.608}
\setres{ctyP}{EU_s13}{s}{13}
\setres{ctyP}{EU_s14}{kappa_crps_max}{0.730}
\setres{ctyP}{EU_s14}{kappa_crps_median}{0.426}
\setres{ctyP}{EU_s14}{kappa_crps_min}{-0.620}
\setres{ctyP}{EU_s14}{kappa_crps_q25}{0.303}
\setres{ctyP}{EU_s14}{kappa_crps_q75}{0.647}
\setres{ctyP}{EU_s14}{s}{14}
\setres{ctyP}{EU_s15}{kappa_crps_max}{0.757}
\setres{ctyP}{EU_s15}{kappa_crps_median}{0.197}
\setres{ctyP}{EU_s15}{kappa_crps_min}{-1.509}
\setres{ctyP}{EU_s15}{kappa_crps_q25}{-0.940}
\setres{ctyP}{EU_s15}{kappa_crps_q75}{0.491}
\setres{ctyP}{EU_s15}{s}{15}
\setres{ctyP}{EU_s16}{kappa_crps_max}{0.775}
\setres{ctyP}{EU_s16}{kappa_crps_median}{0.550}
\setres{ctyP}{EU_s16}{kappa_crps_min}{-0.485}
\setres{ctyP}{EU_s16}{kappa_crps_q25}{0.194}
\setres{ctyP}{EU_s16}{kappa_crps_q75}{0.686}
\setres{ctyP}{EU_s16}{s}{16}
\setres{ctyP}{EU_s17}{kappa_crps_max}{0.722}
\setres{ctyP}{EU_s17}{kappa_crps_median}{0.653}
\setres{ctyP}{EU_s17}{kappa_crps_min}{-7.669}
\setres{ctyP}{EU_s17}{kappa_crps_q25}{0.243}
\setres{ctyP}{EU_s17}{kappa_crps_q75}{0.663}
\setres{ctyP}{EU_s17}{s}{17}
\setres{ctyP}{EU_s18}{kappa_crps_max}{0.760}
\setres{ctyP}{EU_s18}{kappa_crps_median}{0.696}
\setres{ctyP}{EU_s18}{kappa_crps_min}{0.261}
\setres{ctyP}{EU_s18}{kappa_crps_q25}{0.660}
\setres{ctyP}{EU_s18}{kappa_crps_q75}{0.731}
\setres{ctyP}{EU_s18}{s}{18}
\setres{ctyP}{EU_s19}{kappa_crps_max}{0.716}
\setres{ctyP}{EU_s19}{kappa_crps_median}{0.438}
\setres{ctyP}{EU_s19}{kappa_crps_min}{-1.521}
\setres{ctyP}{EU_s19}{kappa_crps_q25}{0.007}
\setres{ctyP}{EU_s19}{kappa_crps_q75}{0.547}
\setres{ctyP}{EU_s19}{s}{19}
\setres{ctyP}{EU_s2}{kappa_crps_max}{0.652}
\setres{ctyP}{EU_s2}{kappa_crps_median}{0.444}
\setres{ctyP}{EU_s2}{kappa_crps_min}{-0.463}
\setres{ctyP}{EU_s2}{kappa_crps_q25}{0.144}
\setres{ctyP}{EU_s2}{kappa_crps_q75}{0.650}
\setres{ctyP}{EU_s2}{s}{2}
\setres{ctyP}{EU_s20}{kappa_crps_max}{0.726}
\setres{ctyP}{EU_s20}{kappa_crps_median}{0.337}
\setres{ctyP}{EU_s20}{kappa_crps_min}{-0.939}
\setres{ctyP}{EU_s20}{kappa_crps_q25}{0.078}
\setres{ctyP}{EU_s20}{kappa_crps_q75}{0.419}
\setres{ctyP}{EU_s20}{s}{20}
\setres{ctyP}{EU_s21}{kappa_crps_max}{0.763}
\setres{ctyP}{EU_s21}{kappa_crps_median}{0.290}
\setres{ctyP}{EU_s21}{kappa_crps_min}{-1.533}
\setres{ctyP}{EU_s21}{kappa_crps_q25}{-0.492}
\setres{ctyP}{EU_s21}{kappa_crps_q75}{0.467}
\setres{ctyP}{EU_s21}{s}{21}
\setres{ctyP}{EU_s22}{kappa_crps_max}{0.655}
\setres{ctyP}{EU_s22}{kappa_crps_median}{0.271}
\setres{ctyP}{EU_s22}{kappa_crps_min}{-7.618}
\setres{ctyP}{EU_s22}{kappa_crps_q25}{-2.257}
\setres{ctyP}{EU_s22}{kappa_crps_q75}{0.404}
\setres{ctyP}{EU_s22}{s}{22}
\setres{ctyP}{EU_s23}{kappa_crps_max}{0.661}
\setres{ctyP}{EU_s23}{kappa_crps_median}{0.187}
\setres{ctyP}{EU_s23}{kappa_crps_min}{-1.526}
\setres{ctyP}{EU_s23}{kappa_crps_q25}{-0.476}
\setres{ctyP}{EU_s23}{kappa_crps_q75}{0.400}
\setres{ctyP}{EU_s23}{s}{23}
\setres{ctyP}{EU_s24}{kappa_crps_max}{0.763}
\setres{ctyP}{EU_s24}{kappa_crps_median}{0.307}
\setres{ctyP}{EU_s24}{kappa_crps_min}{-0.522}
\setres{ctyP}{EU_s24}{kappa_crps_q25}{-0.426}
\setres{ctyP}{EU_s24}{kappa_crps_q75}{0.572}
\setres{ctyP}{EU_s24}{s}{24}
\setres{ctyP}{EU_s25}{kappa_crps_max}{0.594}
\setres{ctyP}{EU_s25}{kappa_crps_median}{0.172}
\setres{ctyP}{EU_s25}{kappa_crps_min}{-7.519}
\setres{ctyP}{EU_s25}{kappa_crps_q25}{-0.054}
\setres{ctyP}{EU_s25}{kappa_crps_q75}{0.242}
\setres{ctyP}{EU_s25}{s}{25}
\setres{ctyP}{EU_s26}{kappa_crps_max}{0.772}
\setres{ctyP}{EU_s26}{kappa_crps_median}{0.643}
\setres{ctyP}{EU_s26}{kappa_crps_min}{-0.397}
\setres{ctyP}{EU_s26}{kappa_crps_q25}{0.338}
\setres{ctyP}{EU_s26}{kappa_crps_q75}{0.659}
\setres{ctyP}{EU_s26}{s}{26}
\setres{ctyP}{EU_s27}{kappa_crps_max}{0.724}
\setres{ctyP}{EU_s27}{kappa_crps_median}{-0.555}
\setres{ctyP}{EU_s27}{kappa_crps_min}{-1.524}
\setres{ctyP}{EU_s27}{kappa_crps_q25}{-0.704}
\setres{ctyP}{EU_s27}{kappa_crps_q75}{0.641}
\setres{ctyP}{EU_s27}{s}{27}
\setres{ctyP}{EU_s28}{kappa_crps_max}{0.659}
\setres{ctyP}{EU_s28}{kappa_crps_median}{0.298}
\setres{ctyP}{EU_s28}{kappa_crps_min}{-0.930}
\setres{ctyP}{EU_s28}{kappa_crps_q25}{-0.496}
\setres{ctyP}{EU_s28}{kappa_crps_q75}{0.338}
\setres{ctyP}{EU_s28}{s}{28}
\setres{ctyP}{EU_s29}{kappa_crps_max}{0.779}
\setres{ctyP}{EU_s29}{kappa_crps_median}{0.473}
\setres{ctyP}{EU_s29}{kappa_crps_min}{-0.934}
\setres{ctyP}{EU_s29}{kappa_crps_q25}{0.360}
\setres{ctyP}{EU_s29}{kappa_crps_q75}{0.688}
\setres{ctyP}{EU_s29}{s}{29}
\setres{ctyP}{EU_s3}{kappa_crps_max}{0.727}
\setres{ctyP}{EU_s3}{kappa_crps_median}{0.342}
\setres{ctyP}{EU_s3}{kappa_crps_min}{-7.584}
\setres{ctyP}{EU_s3}{kappa_crps_q25}{-0.746}
\setres{ctyP}{EU_s3}{kappa_crps_q75}{0.592}
\setres{ctyP}{EU_s3}{s}{3}
\setres{ctyP}{EU_s4}{kappa_crps_max}{0.750}
\setres{ctyP}{EU_s4}{kappa_crps_median}{0.397}
\setres{ctyP}{EU_s4}{kappa_crps_min}{0.044}
\setres{ctyP}{EU_s4}{kappa_crps_q25}{0.344}
\setres{ctyP}{EU_s4}{kappa_crps_q75}{0.469}
\setres{ctyP}{EU_s4}{s}{4}
\setres{ctyP}{EU_s5}{kappa_crps_max}{0.728}
\setres{ctyP}{EU_s5}{kappa_crps_median}{0.409}
\setres{ctyP}{EU_s5}{kappa_crps_min}{-0.619}
\setres{ctyP}{EU_s5}{kappa_crps_q25}{0.338}
\setres{ctyP}{EU_s5}{kappa_crps_q75}{0.572}
\setres{ctyP}{EU_s5}{s}{5}
\setres{ctyP}{EU_s6}{kappa_crps_max}{0.653}
\setres{ctyP}{EU_s6}{kappa_crps_median}{0.364}
\setres{ctyP}{EU_s6}{kappa_crps_min}{-0.923}
\setres{ctyP}{EU_s6}{kappa_crps_q25}{0.105}
\setres{ctyP}{EU_s6}{kappa_crps_q75}{0.452}
\setres{ctyP}{EU_s6}{s}{6}
\setres{ctyP}{EU_s7}{kappa_crps_max}{0.644}
\setres{ctyP}{EU_s7}{kappa_crps_median}{-0.204}
\setres{ctyP}{EU_s7}{kappa_crps_min}{-7.556}
\setres{ctyP}{EU_s7}{kappa_crps_q25}{-0.839}
\setres{ctyP}{EU_s7}{kappa_crps_q75}{0.239}
\setres{ctyP}{EU_s7}{s}{7}
\setres{ctyP}{EU_s8}{kappa_crps_max}{0.767}
\setres{ctyP}{EU_s8}{kappa_crps_median}{0.469}
\setres{ctyP}{EU_s8}{kappa_crps_min}{-0.948}
\setres{ctyP}{EU_s8}{kappa_crps_q25}{0.220}
\setres{ctyP}{EU_s8}{kappa_crps_q75}{0.604}
\setres{ctyP}{EU_s8}{s}{8}
\setres{ctyP}{EU_s9}{kappa_crps_max}{0.763}
\setres{ctyP}{EU_s9}{kappa_crps_median}{0.346}
\setres{ctyP}{EU_s9}{kappa_crps_min}{-7.470}
\setres{ctyP}{EU_s9}{kappa_crps_q25}{-0.458}
\setres{ctyP}{EU_s9}{kappa_crps_q75}{0.663}
\setres{ctyP}{EU_s9}{s}{9}
\setres{ctyP}{EU_typical}{order_index}{7}
\setres{ctyP}{EU_within}{kappa_crps_median}{0.395}
\setres{ctyP}{reference}{n_countries}{30}
\setres{ctyP}{reference}{n_folds}{10}
\setres{ctyP}{reference}{n_iter}{300}
\setres{ctyP}{reference}{n_orders}{8}
\setres{learnUZ}{CPT_post_m0}{ess}{8{,}000}
\setres{learnUZ}{CPT_post_m0}{kappa}{0.580}
\setres{learnUZ}{CPT_post_m0}{kappa_prior_mean}{0.946}
\setres{learnUZ}{CPT_post_m0}{kappa_se}{0.013}
\setres{learnUZ}{CPT_post_m0}{mse}{66.44}
\setres{learnUZ}{CPT_post_m0}{n_subjects_seen}{0}
\setres{learnUZ}{CPT_post_m0}{spread_penalty}{0.366}
\setres{learnUZ}{CPT_post_m0}{theta_mean}{0.50, 0.50, 1.55}
\setres{learnUZ}{CPT_post_m0}{theta_sd}{0.29, 0.29, 0.84}
\setres{learnUZ}{CPT_post_m10}{ess}{135}
\setres{learnUZ}{CPT_post_m10}{kappa}{0.963}
\setres{learnUZ}{CPT_post_m10}{kappa_prior_mean}{0.970}
\setres{learnUZ}{CPT_post_m10}{kappa_se}{0.007}
\setres{learnUZ}{CPT_post_m10}{mse}{25.26}
\setres{learnUZ}{CPT_post_m10}{n_subjects_seen}{10}
\setres{learnUZ}{CPT_post_m10}{spread_penalty}{0.007}
\setres{learnUZ}{CPT_post_m10}{theta_mean}{0.65, 0.56, 1.29}
\setres{learnUZ}{CPT_post_m10}{theta_sd}{0.22, 0.05, 0.21}
\setres{learnUZ}{CPT_post_m100}{ess}{3}
\setres{learnUZ}{CPT_post_m100}{kappa}{0.982}
\setres{learnUZ}{CPT_post_m100}{kappa_prior_mean}{0.984}
\setres{learnUZ}{CPT_post_m100}{kappa_se}{0.006}
\setres{learnUZ}{CPT_post_m100}{mse}{23.19}
\setres{learnUZ}{CPT_post_m100}{n_subjects_seen}{100}
\setres{learnUZ}{CPT_post_m100}{spread_penalty}{0.002}
\setres{learnUZ}{CPT_post_m100}{theta_mean}{0.93, 0.43, 0.91}
\setres{learnUZ}{CPT_post_m100}{theta_sd}{0.06, 0.01, 0.03}
\setres{learnUZ}{CPT_post_m3}{ess}{618}
\setres{learnUZ}{CPT_post_m3}{kappa}{0.928}
\setres{learnUZ}{CPT_post_m3}{kappa_prior_mean}{0.950}
\setres{learnUZ}{CPT_post_m3}{kappa_se}{0.008}
\setres{learnUZ}{CPT_post_m3}{mse}{29.03}
\setres{learnUZ}{CPT_post_m3}{n_subjects_seen}{3}
\setres{learnUZ}{CPT_post_m3}{spread_penalty}{0.021}
\setres{learnUZ}{CPT_post_m3}{theta_mean}{0.54, 0.53, 1.51}
\setres{learnUZ}{CPT_post_m3}{theta_sd}{0.27, 0.09, 0.37}
\setres{learnUZ}{CPT_post_m30}{ess}{15}
\setres{learnUZ}{CPT_post_m30}{kappa}{0.943}
\setres{learnUZ}{CPT_post_m30}{kappa_prior_mean}{0.946}
\setres{learnUZ}{CPT_post_m30}{kappa_se}{0.009}
\setres{learnUZ}{CPT_post_m30}{mse}{27.40}
\setres{learnUZ}{CPT_post_m30}{n_subjects_seen}{30}
\setres{learnUZ}{CPT_post_m30}{spread_penalty}{0.003}
\setres{learnUZ}{CPT_post_m30}{theta_mean}{0.81, 0.43, 1.16}
\setres{learnUZ}{CPT_post_m30}{theta_sd}{0.14, 0.02, 0.11}
\setres{learnUZ}{CPT_post_mall}{ess}{3}
\setres{learnUZ}{CPT_post_mall}{kappa}{0.979}
\setres{learnUZ}{CPT_post_mall}{kappa_prior_mean}{0.981}
\setres{learnUZ}{CPT_post_mall}{kappa_se}{0.006}
\setres{learnUZ}{CPT_post_mall}{mse}{23.59}
\setres{learnUZ}{CPT_post_mall}{n_subjects_seen}{179}
\setres{learnUZ}{CPT_post_mall}{spread_penalty}{0.002}
\setres{learnUZ}{CPT_post_mall}{theta_mean}{0.94, 0.42, 0.90}
\setres{learnUZ}{CPT_post_mall}{theta_sd}{0.05, 0.02, 0.01}
\setres{learnUZ}{DA_post_m0}{ess}{1{,}600}
\setres{learnUZ}{DA_post_m0}{kappa}{0.679}
\setres{learnUZ}{DA_post_m0}{kappa_prior_mean}{0.862}
\setres{learnUZ}{DA_post_m0}{kappa_se}{0.038}
\setres{learnUZ}{DA_post_m0}{mse}{55.83}
\setres{learnUZ}{DA_post_m0}{n_subjects_seen}{0}
\setres{learnUZ}{DA_post_m0}{spread_penalty}{0.184}
\setres{learnUZ}{DA_post_m0}{theta_mean}{0.50, 1.00}
\setres{learnUZ}{DA_post_m0}{theta_sd}{0.29, 1.15}
\setres{learnUZ}{DA_post_m10}{ess}{86}
\setres{learnUZ}{DA_post_m10}{kappa}{0.801}
\setres{learnUZ}{DA_post_m10}{kappa_prior_mean}{0.807}
\setres{learnUZ}{DA_post_m10}{kappa_se}{0.030}
\setres{learnUZ}{DA_post_m10}{mse}{42.68}
\setres{learnUZ}{DA_post_m10}{n_subjects_seen}{10}
\setres{learnUZ}{DA_post_m10}{spread_penalty}{0.006}
\setres{learnUZ}{DA_post_m10}{theta_mean}{0.72, -0.39}
\setres{learnUZ}{DA_post_m10}{theta_sd}{0.21, 0.12}
\setres{learnUZ}{DA_post_m100}{ess}{5}
\setres{learnUZ}{DA_post_m100}{kappa}{0.827}
\setres{learnUZ}{DA_post_m100}{kappa_prior_mean}{0.828}
\setres{learnUZ}{DA_post_m100}{kappa_se}{0.030}
\setres{learnUZ}{DA_post_m100}{mse}{39.91}
\setres{learnUZ}{DA_post_m100}{n_subjects_seen}{100}
\setres{learnUZ}{DA_post_m100}{spread_penalty}{0.001}
\setres{learnUZ}{DA_post_m100}{theta_mean}{0.97, -0.13}
\setres{learnUZ}{DA_post_m100}{theta_sd}{0.03, 0.05}
\setres{learnUZ}{DA_post_m3}{ess}{182}
\setres{learnUZ}{DA_post_m3}{kappa}{0.778}
\setres{learnUZ}{DA_post_m3}{kappa_prior_mean}{0.797}
\setres{learnUZ}{DA_post_m3}{kappa_se}{0.031}
\setres{learnUZ}{DA_post_m3}{mse}{45.20}
\setres{learnUZ}{DA_post_m3}{n_subjects_seen}{3}
\setres{learnUZ}{DA_post_m3}{spread_penalty}{0.019}
\setres{learnUZ}{DA_post_m3}{theta_mean}{0.62, -0.46}
\setres{learnUZ}{DA_post_m3}{theta_sd}{0.26, 0.18}
\setres{learnUZ}{DA_post_m30}{ess}{17}
\setres{learnUZ}{DA_post_m30}{kappa}{0.763}
\setres{learnUZ}{DA_post_m30}{kappa_prior_mean}{0.766}
\setres{learnUZ}{DA_post_m30}{kappa_se}{0.031}
\setres{learnUZ}{DA_post_m30}{mse}{46.79}
\setres{learnUZ}{DA_post_m30}{n_subjects_seen}{30}
\setres{learnUZ}{DA_post_m30}{spread_penalty}{0.003}
\setres{learnUZ}{DA_post_m30}{theta_mean}{0.89, -0.50}
\setres{learnUZ}{DA_post_m30}{theta_sd}{0.11, 0.06}
\setres{learnUZ}{DA_post_mall}{ess}{2}
\setres{learnUZ}{DA_post_mall}{kappa}{0.826}
\setres{learnUZ}{DA_post_mall}{kappa_prior_mean}{0.826}
\setres{learnUZ}{DA_post_mall}{kappa_se}{0.030}
\setres{learnUZ}{DA_post_mall}{mse}{40.02}
\setres{learnUZ}{DA_post_mall}{n_subjects_seen}{179}
\setres{learnUZ}{DA_post_mall}{spread_penalty}{0.000}
\setres{learnUZ}{DA_post_mall}{theta_mean}{0.98, -0.14}
\setres{learnUZ}{DA_post_mall}{theta_sd}{0.02, 0.03}
\setres{learnUZ}{EU_post_m0}{ess}{200}
\setres{learnUZ}{EU_post_m0}{kappa}{0.820}
\setres{learnUZ}{EU_post_m0}{kappa_prior_mean}{0.827}
\setres{learnUZ}{EU_post_m0}{kappa_se}{0.032}
\setres{learnUZ}{EU_post_m0}{mse}{40.61}
\setres{learnUZ}{EU_post_m0}{n_subjects_seen}{0}
\setres{learnUZ}{EU_post_m0}{spread_penalty}{0.007}
\setres{learnUZ}{EU_post_m0}{theta_mean}{0.51}
\setres{learnUZ}{EU_post_m0}{theta_sd}{0.29}
\setres{learnUZ}{EU_post_m10}{ess}{33}
\setres{learnUZ}{EU_post_m10}{kappa}{0.833}
\setres{learnUZ}{EU_post_m10}{kappa_prior_mean}{0.833}
\setres{learnUZ}{EU_post_m10}{kappa_se}{0.031}
\setres{learnUZ}{EU_post_m10}{mse}{39.26}
\setres{learnUZ}{EU_post_m10}{n_subjects_seen}{10}
\setres{learnUZ}{EU_post_m10}{spread_penalty}{0.000}
\setres{learnUZ}{EU_post_m10}{theta_mean}{0.92}
\setres{learnUZ}{EU_post_m10}{theta_sd}{0.07}
\setres{learnUZ}{EU_post_m100}{ess}{7}
\setres{learnUZ}{EU_post_m100}{kappa}{0.833}
\setres{learnUZ}{EU_post_m100}{kappa_prior_mean}{0.833}
\setres{learnUZ}{EU_post_m100}{kappa_se}{0.030}
\setres{learnUZ}{EU_post_m100}{mse}{39.24}
\setres{learnUZ}{EU_post_m100}{n_subjects_seen}{100}
\setres{learnUZ}{EU_post_m100}{spread_penalty}{0.000}
\setres{learnUZ}{EU_post_m100}{theta_mean}{0.98}
\setres{learnUZ}{EU_post_m100}{theta_sd}{0.02}
\setres{learnUZ}{EU_post_m3}{ess}{77}
\setres{learnUZ}{EU_post_m3}{kappa}{0.831}
\setres{learnUZ}{EU_post_m3}{kappa_prior_mean}{0.833}
\setres{learnUZ}{EU_post_m3}{kappa_se}{0.031}
\setres{learnUZ}{EU_post_m3}{mse}{39.42}
\setres{learnUZ}{EU_post_m3}{n_subjects_seen}{3}
\setres{learnUZ}{EU_post_m3}{spread_penalty}{0.002}
\setres{learnUZ}{EU_post_m3}{theta_mean}{0.82}
\setres{learnUZ}{EU_post_m3}{theta_sd}{0.16}
\setres{learnUZ}{EU_post_m30}{ess}{9}
\setres{learnUZ}{EU_post_m30}{kappa}{0.833}
\setres{learnUZ}{EU_post_m30}{kappa_prior_mean}{0.833}
\setres{learnUZ}{EU_post_m30}{kappa_se}{0.030}
\setres{learnUZ}{EU_post_m30}{mse}{39.24}
\setres{learnUZ}{EU_post_m30}{n_subjects_seen}{30}
\setres{learnUZ}{EU_post_m30}{spread_penalty}{0.000}
\setres{learnUZ}{EU_post_m30}{theta_mean}{0.98}
\setres{learnUZ}{EU_post_m30}{theta_sd}{0.02}
\setres{learnUZ}{EU_post_mall}{ess}{4}
\setres{learnUZ}{EU_post_mall}{kappa}{0.833}
\setres{learnUZ}{EU_post_mall}{kappa_prior_mean}{0.833}
\setres{learnUZ}{EU_post_mall}{kappa_se}{0.030}
\setres{learnUZ}{EU_post_mall}{mse}{39.24}
\setres{learnUZ}{EU_post_mall}{n_subjects_seen}{179}
\setres{learnUZ}{EU_post_mall}{spread_penalty}{0.000}
\setres{learnUZ}{EU_post_mall}{theta_mean}{0.99}
\setres{learnUZ}{EU_post_mall}{theta_sd}{0.01}
\setres{learnUZ}{reference}{n_source_subjects}{179}
\setres{learnUZ}{reference}{sigma}{6.6}
\setres{samp}{CPT_samples}{ess}{13}
\setres{samp}{CPT_samples}{kappa_max}{0.892}
\setres{samp}{CPT_samples}{kappa_mean}{0.631}
\setres{samp}{CPT_samples}{kappa_mean_se}{0.103}
\setres{samp}{CPT_samples}{kappa_median}{0.840}
\setres{samp}{CPT_samples}{kappa_min}{-4.386}
\setres{samp}{CPT_samples}{kappa_prior_mean_median}{0.941}
\setres{samp}{CPT_samples}{n_countries}{30}
\setres{samp}{CPT_samples}{n_pos}{29}
\setres{samp}{CPT_samples}{r_ff}{0.199}
\setres{samp}{CPT_samples}{r_prior}{0.9988}
\setres{samp}{CPT_samples}{r_prob}{0.999}
\setres{samp}{CPT_samples}{r_prob_se}{0.000}
\setres{samp}{CPT_samples}{theta_mean}{0.936, 0.631, 0.974}
\setres{samp}{CPT_samples}{theta_sd}{0.064, 0.128, 0.178}
\setres{samp}{CPT_samples_u}{ess}{13}
\setres{samp}{CPT_samples_u}{kappa_max}{0.845}
\setres{samp}{CPT_samples_u}{kappa_mean}{0.553}
\setres{samp}{CPT_samples_u}{kappa_mean_se}{0.109}
\setres{samp}{CPT_samples_u}{kappa_median}{0.778}
\setres{samp}{CPT_samples_u}{kappa_min}{-4.820}
\setres{samp}{CPT_samples_u}{kappa_prior_mean_median}{0.940}
\setres{samp}{CPT_samples_u}{n_countries}{30}
\setres{samp}{CPT_samples_u}{n_pos}{29}
\setres{samp}{CPT_samples_u}{r_ff}{0.199}
\setres{samp}{CPT_samples_u}{r_prior}{0.9987}
\setres{samp}{CPT_samples_u}{r_prob}{0.999}
\setres{samp}{CPT_samples_u}{r_prob_se}{0.000}
\setres{samp}{CPT_samples_u}{theta_mean}{0.930, 0.623, 0.989}
\setres{samp}{CPT_samples_u}{theta_sd}{0.101, 0.135, 0.249}
\setres{samp}{DA_samples}{ess}{6}
\setres{samp}{DA_samples}{kappa_max}{0.869}
\setres{samp}{DA_samples}{kappa_mean}{0.274}
\setres{samp}{DA_samples}{kappa_mean_se}{0.209}
\setres{samp}{DA_samples}{kappa_median}{0.767}
\setres{samp}{DA_samples}{kappa_min}{-10.362}
\setres{samp}{DA_samples}{kappa_prior_mean_median}{0.865}
\setres{samp}{DA_samples}{n_countries}{30}
\setres{samp}{DA_samples}{n_pos}{28}
\setres{samp}{DA_samples}{r_ff}{0.314}
\setres{samp}{DA_samples}{r_prior}{0.9988}
\setres{samp}{DA_samples}{r_prob}{0.999}
\setres{samp}{DA_samples}{r_prob_se}{0.000}
\setres{samp}{DA_samples}{theta_mean}{0.974, 0.163}
\setres{samp}{DA_samples}{theta_sd}{0.079, 0.225}
\setres{samp}{DA_samples_u}{ess}{6}
\setres{samp}{DA_samples_u}{kappa_max}{0.818}
\setres{samp}{DA_samples_u}{kappa_mean}{0.196}
\setres{samp}{DA_samples_u}{kappa_mean_se}{0.216}
\setres{samp}{DA_samples_u}{kappa_median}{0.710}
\setres{samp}{DA_samples_u}{kappa_min}{-10.854}
\setres{samp}{DA_samples_u}{kappa_prior_mean_median}{0.873}
\setres{samp}{DA_samples_u}{n_countries}{30}
\setres{samp}{DA_samples_u}{n_pos}{27}
\setres{samp}{DA_samples_u}{r_ff}{0.314}
\setres{samp}{DA_samples_u}{r_prior}{0.9988}
\setres{samp}{DA_samples_u}{r_prob}{0.999}
\setres{samp}{DA_samples_u}{r_prob_se}{0.000}
\setres{samp}{DA_samples_u}{theta_mean}{0.972, 0.185}
\setres{samp}{DA_samples_u}{theta_sd}{0.100, 0.329}
\setres{samp}{EU_samples}{ess}{4}
\setres{samp}{EU_samples}{kappa_max}{0.923}
\setres{samp}{EU_samples}{kappa_mean}{0.340}
\setres{samp}{EU_samples}{kappa_mean_se}{0.207}
\setres{samp}{EU_samples}{kappa_median}{0.807}
\setres{samp}{EU_samples}{kappa_min}{-10.013}
\setres{samp}{EU_samples}{kappa_prior_mean_median}{0.840}
\setres{samp}{EU_samples}{n_countries}{30}
\setres{samp}{EU_samples}{n_pos}{29}
\setres{samp}{EU_samples}{r_ff}{0.367}
\setres{samp}{EU_samples}{r_prior}{0.9873}
\setres{samp}{EU_samples}{r_prob}{0.992}
\setres{samp}{EU_samples}{r_prob_se}{0.000}
\setres{samp}{EU_samples}{theta_mean}{0.906}
\setres{samp}{EU_samples}{theta_sd}{0.090}
\setres{samp}{EU_samples_u}{ess}{4}
\setres{samp}{EU_samples_u}{kappa_max}{0.876}
\setres{samp}{EU_samples_u}{kappa_mean}{0.259}
\setres{samp}{EU_samples_u}{kappa_mean_se}{0.215}
\setres{samp}{EU_samples_u}{kappa_median}{0.755}
\setres{samp}{EU_samples_u}{kappa_min}{-10.592}
\setres{samp}{EU_samples_u}{kappa_prior_mean_median}{0.847}
\setres{samp}{EU_samples_u}{n_countries}{30}
\setres{samp}{EU_samples_u}{n_pos}{28}
\setres{samp}{EU_samples_u}{r_ff}{0.367}
\setres{samp}{EU_samples_u}{r_prior}{0.9885}
\setres{samp}{EU_samples_u}{r_prob}{0.993}
\setres{samp}{EU_samples_u}{r_prob_se}{0.000}
\setres{samp}{EU_samples_u}{theta_mean}{0.900}
\setres{samp}{EU_samples_u}{theta_sd}{0.118}
\setres{samp}{reference}{alpha}{1.0}
\setres{samp}{reference}{n_countries}{30}
\setres{disp}{CPT}{alpha_lit_sd}{0.138}
\setres{disp}{CPT}{alpha_max}{0.975}
\setres{disp}{CPT}{alpha_mean}{0.936}
\setres{disp}{CPT}{alpha_min}{0.832}
\setres{disp}{CPT}{alpha_sd_countries}{0.038}
\setres{disp}{CPT}{alpha_se_mean}{0.030}
\setres{disp}{CPT}{alpha_tau}{0.024}
\setres{disp}{CPT}{delta_lit_sd}{0.275}
\setres{disp}{CPT}{delta_max}{1.377}
\setres{disp}{CPT}{delta_mean}{0.986}
\setres{disp}{CPT}{delta_min}{0.753}
\setres{disp}{CPT}{delta_sd_countries}{0.181}
\setres{disp}{CPT}{delta_se_mean}{0.051}
\setres{disp}{CPT}{delta_tau}{0.174}
\setres{disp}{CPT}{gamma_lit_sd}{0.175}
\setres{disp}{CPT}{gamma_max}{0.793}
\setres{disp}{CPT}{gamma_mean}{0.631}
\setres{disp}{CPT}{gamma_min}{0.187}
\setres{disp}{CPT}{gamma_sd_countries}{0.128}
\setres{disp}{CPT}{gamma_se_mean}{0.028}
\setres{disp}{CPT}{gamma_tau}{0.125}
\setres{disp}{CPT}{n_countries}{30}
\setres{disp}{DA}{alpha_lit_sd}{0.187}
\setres{disp}{DA}{alpha_max}{0.988}
\setres{disp}{DA}{alpha_mean}{0.983}
\setres{disp}{DA}{alpha_min}{0.954}
\setres{disp}{DA}{alpha_sd_countries}{0.008}
\setres{disp}{DA}{alpha_se_mean}{0.011}
\setres{disp}{DA}{alpha_tau}{0.000}
\setres{disp}{DA}{eta_lit_sd}{0.447}
\setres{disp}{DA}{eta_max}{0.601}
\setres{disp}{DA}{eta_mean}{0.158}
\setres{disp}{DA}{eta_min}{-0.251}
\setres{disp}{DA}{eta_sd_countries}{0.218}
\setres{disp}{DA}{eta_se_mean}{0.036}
\setres{disp}{DA}{eta_tau}{0.215}
\setres{disp}{DA}{n_countries}{30}
\setres{disp}{EU}{alpha_lit_sd}{0.138}
\setres{disp}{EU}{alpha_max}{0.997}
\setres{disp}{EU}{alpha_mean}{0.907}
\setres{disp}{EU}{alpha_min}{0.770}
\setres{disp}{EU}{alpha_sd_countries}{0.076}
\setres{disp}{EU}{alpha_se_mean}{0.015}
\setres{disp}{EU}{alpha_tau}{0.075}
\setres{disp}{EU}{n_countries}{30}
\setres{disp}{reference}{n_countries}{30}
\setres{pop}{CPT_fitted}{H_F}{24.68}
\setres{pop}{CPT_fitted}{H_L}{47.83}
\setres{pop}{CPT_fitted}{ess}{45}
\setres{pop}{CPT_fitted}{kappa_crps}{0.957}
\setres{pop}{CPT_fitted}{kappa_crps_se}{0.008}
\setres{pop}{CPT_fitted}{kappa_draw}{0.881}
\setres{pop}{CPT_fitted}{kappa_draw_se}{0.003}
\setres{pop}{CPT_fitted}{kappa_mean}{0.992}
\setres{pop}{CPT_fitted}{kappa_mean_se}{0.005}
\setres{pop}{CPT_fitted}{r_stoch}{0.786}
\setres{pop}{CPT_fitted}{s_crps}{2.22}
\setres{pop}{CPT_fitted}{theta_mean}{0.89, 0.49, 0.94}
\setres{pop}{CPT_fitted}{theta_sd}{0.18, 0.22, 0.32}
\setres{pop}{CPT_from_zurich03}{H_F}{24.78}
\setres{pop}{CPT_from_zurich03}{H_L}{47.83}
\setres{pop}{CPT_from_zurich03}{ess}{40}
\setres{pop}{CPT_from_zurich03}{kappa_crps}{0.944}
\setres{pop}{CPT_from_zurich03}{kappa_crps_se}{0.013}
\setres{pop}{CPT_from_zurich03}{kappa_draw}{0.804}
\setres{pop}{CPT_from_zurich03}{kappa_draw_se}{0.003}
\setres{pop}{CPT_from_zurich03}{kappa_mean}{0.973}
\setres{pop}{CPT_from_zurich03}{kappa_mean_se}{0.008}
\setres{pop}{CPT_from_zurich03}{r_stoch}{0.785}
\setres{pop}{CPT_from_zurich03}{s_crps}{2.25}
\setres{pop}{CPT_from_zurich03}{theta_mean}{0.93, 0.45, 1.00}
\setres{pop}{CPT_from_zurich03}{theta_sd}{0.15, 0.22, 0.50}
\setres{pop}{CPT_from_zurich03_types1}{H_F}{21.21}
\setres{pop}{CPT_from_zurich03_types1}{H_L}{47.83}
\setres{pop}{CPT_from_zurich03_types1}{ess}{1}
\setres{pop}{CPT_from_zurich03_types1}{kappa_crps}{0.790}
\setres{pop}{CPT_from_zurich03_types1}{kappa_crps_se}{0.017}
\setres{pop}{CPT_from_zurich03_types1}{kappa_draw}{0.982}
\setres{pop}{CPT_from_zurich03_types1}{kappa_draw_se}{0.007}
\setres{pop}{CPT_from_zurich03_types1}{kappa_mean}{0.982}
\setres{pop}{CPT_from_zurich03_types1}{kappa_mean_se}{0.007}
\setres{pop}{CPT_from_zurich03_types1}{r_stoch}{0.830}
\setres{pop}{CPT_from_zurich03_types1}{s_crps}{2.65}
\setres{pop}{CPT_from_zurich03_types1}{shares}{1.000}
\setres{pop}{CPT_from_zurich03_types1}{theta_mean}{0.98, 0.43, 0.90}
\setres{pop}{CPT_from_zurich03_types1}{theta_sd}{0.00, 0.00, 0.00}
\setres{pop}{CPT_from_zurich03_types1}{types}{1.00: (0.98, 0.43, 0.90)}
\setres{pop}{CPT_from_zurich03_types2}{H_F}{21.93}
\setres{pop}{CPT_from_zurich03_types2}{H_L}{47.83}
\setres{pop}{CPT_from_zurich03_types2}{ess}{2}
\setres{pop}{CPT_from_zurich03_types2}{kappa_crps}{0.849}
\setres{pop}{CPT_from_zurich03_types2}{kappa_crps_se}{0.016}
\setres{pop}{CPT_from_zurich03_types2}{kappa_draw}{0.937}
\setres{pop}{CPT_from_zurich03_types2}{kappa_draw_se}{0.006}
\setres{pop}{CPT_from_zurich03_types2}{kappa_mean}{0.975}
\setres{pop}{CPT_from_zurich03_types2}{kappa_mean_se}{0.008}
\setres{pop}{CPT_from_zurich03_types2}{r_stoch}{0.822}
\setres{pop}{CPT_from_zurich03_types2}{s_crps}{2.50}
\setres{pop}{CPT_from_zurich03_types2}{shares}{0.604, 0.396}
\setres{pop}{CPT_from_zurich03_types2}{theta_mean}{0.98, 0.42, 0.96}
\setres{pop}{CPT_from_zurich03_types2}{theta_sd}{0.00, 0.07, 0.28}
\setres{pop}{CPT_from_zurich03_types2}{types}{0.60: (0.98, 0.48, 1.19); 0.40: (0.98, 0.33, 0.61)}
\setres{pop}{CPT_from_zurich03_types3}{H_F}{22.20}
\setres{pop}{CPT_from_zurich03_types3}{H_L}{47.83}
\setres{pop}{CPT_from_zurich03_types3}{ess}{3}
\setres{pop}{CPT_from_zurich03_types3}{kappa_crps}{0.908}
\setres{pop}{CPT_from_zurich03_types3}{kappa_crps_se}{0.014}
\setres{pop}{CPT_from_zurich03_types3}{kappa_draw}{0.881}
\setres{pop}{CPT_from_zurich03_types3}{kappa_draw_se}{0.004}
\setres{pop}{CPT_from_zurich03_types3}{kappa_mean}{0.974}
\setres{pop}{CPT_from_zurich03_types3}{kappa_mean_se}{0.008}
\setres{pop}{CPT_from_zurich03_types3}{r_stoch}{0.819}
\setres{pop}{CPT_from_zurich03_types3}{s_crps}{2.35}
\setres{pop}{CPT_from_zurich03_types3}{shares}{0.438, 0.379, 0.184}
\setres{pop}{CPT_from_zurich03_types3}{theta_mean}{0.98, 0.43, 0.96}
\setres{pop}{CPT_from_zurich03_types3}{theta_sd}{0.00, 0.13, 0.32}
\setres{pop}{CPT_from_zurich03_types3}{types}{0.44: (0.98, 0.58, 1.04); 0.38: (0.98, 0.33, 0.61); 0.18: (0.98, 0.28, 1.48)}
\setres{pop}{CPT_from_zurich03_types5}{H_F}{22.59}
\setres{pop}{CPT_from_zurich03_types5}{H_L}{47.83}
\setres{pop}{CPT_from_zurich03_types5}{ess}{4}
\setres{pop}{CPT_from_zurich03_types5}{kappa_crps}{0.939}
\setres{pop}{CPT_from_zurich03_types5}{kappa_crps_se}{0.013}
\setres{pop}{CPT_from_zurich03_types5}{kappa_draw}{0.849}
\setres{pop}{CPT_from_zurich03_types5}{kappa_draw_se}{0.002}
\setres{pop}{CPT_from_zurich03_types5}{kappa_mean}{0.975}
\setres{pop}{CPT_from_zurich03_types5}{kappa_mean_se}{0.008}
\setres{pop}{CPT_from_zurich03_types5}{r_stoch}{0.814}
\setres{pop}{CPT_from_zurich03_types5}{s_crps}{2.27}
\setres{pop}{CPT_from_zurich03_types5}{shares}{0.327, 0.310, 0.232, 0.097, 0.034}
\setres{pop}{CPT_from_zurich03_types5}{theta_mean}{0.98, 0.43, 0.95}
\setres{pop}{CPT_from_zurich03_types5}{theta_sd}{0.00, 0.18, 0.45}
\setres{pop}{CPT_from_zurich03_types5}{types}{0.33: (0.98, 0.33, 1.19); 0.31: (0.98, 0.68, 0.90); 0.23: (0.98, 0.38, 0.61); 0.10: (0.98, 0.13, 0.46); 0.03: (0.98, 0.38, 2.93)}
\setres{pop}{CPT_literature}{H_F}{24.53}
\setres{pop}{CPT_literature}{H_L}{47.83}
\setres{pop}{CPT_literature}{ess}{732}
\setres{pop}{CPT_literature}{kappa_crps}{0.851}
\setres{pop}{CPT_literature}{kappa_crps_se}{0.030}
\setres{pop}{CPT_literature}{kappa_draw}{0.904}
\setres{pop}{CPT_literature}{kappa_draw_se}{0.016}
\setres{pop}{CPT_literature}{kappa_mean}{0.954}
\setres{pop}{CPT_literature}{kappa_mean_se}{0.013}
\setres{pop}{CPT_literature}{r_stoch}{0.788}
\setres{pop}{CPT_literature}{s_crps}{2.49}
\setres{pop}{CPT_literature}{theta_mean}{0.70, 0.73, 1.08}
\setres{pop}{CPT_literature}{theta_sd}{0.14, 0.18, 0.28}
\setres{pop}{CPT_types1}{H_F}{21.23}
\setres{pop}{CPT_types1}{H_L}{47.83}
\setres{pop}{CPT_types1}{ess}{1}
\setres{pop}{CPT_types1}{kappa_crps}{0.800}
\setres{pop}{CPT_types1}{kappa_crps_se}{0.016}
\setres{pop}{CPT_types1}{kappa_draw}{0.985}
\setres{pop}{CPT_types1}{kappa_draw_se}{0.006}
\setres{pop}{CPT_types1}{kappa_mean}{0.985}
\setres{pop}{CPT_types1}{kappa_mean_se}{0.006}
\setres{pop}{CPT_types1}{r_stoch}{0.830}
\setres{pop}{CPT_types1}{s_crps}{2.63}
\setres{pop}{CPT_types1}{shares}{1.000}
\setres{pop}{CPT_types1}{theta_mean}{0.93, 0.43, 0.90}
\setres{pop}{CPT_types1}{theta_sd}{0.00, 0.00, 0.00}
\setres{pop}{CPT_types1}{types}{1.00: (0.93, 0.43, 0.90)}
\setres{pop}{CPT_types2}{H_F}{21.82}
\setres{pop}{CPT_types2}{H_L}{47.83}
\setres{pop}{CPT_types2}{ess}{2}
\setres{pop}{CPT_types2}{kappa_crps}{0.888}
\setres{pop}{CPT_types2}{kappa_crps_se}{0.011}
\setres{pop}{CPT_types2}{kappa_draw}{0.964}
\setres{pop}{CPT_types2}{kappa_draw_se}{0.003}
\setres{pop}{CPT_types2}{kappa_mean}{0.993}
\setres{pop}{CPT_types2}{kappa_mean_se}{0.004}
\setres{pop}{CPT_types2}{r_stoch}{0.823}
\setres{pop}{CPT_types2}{s_crps}{2.40}
\setres{pop}{CPT_types2}{shares}{0.656, 0.344}
\setres{pop}{CPT_types2}{theta_mean}{0.98, 0.46, 0.89}
\setres{pop}{CPT_types2}{theta_sd}{0.00, 0.09, 0.21}
\setres{pop}{CPT_types2}{types}{0.66: (0.98, 0.53, 1.04); 0.34: (0.98, 0.33, 0.61)}
\setres{pop}{CPT_types3}{H_F}{22.24}
\setres{pop}{CPT_types3}{H_L}{47.83}
\setres{pop}{CPT_types3}{ess}{3}
\setres{pop}{CPT_types3}{kappa_crps}{0.940}
\setres{pop}{CPT_types3}{kappa_crps_se}{0.009}
\setres{pop}{CPT_types3}{kappa_draw}{0.931}
\setres{pop}{CPT_types3}{kappa_draw_se}{0.002}
\setres{pop}{CPT_types3}{kappa_mean}{0.994}
\setres{pop}{CPT_types3}{kappa_mean_se}{0.004}
\setres{pop}{CPT_types3}{r_stoch}{0.818}
\setres{pop}{CPT_types3}{s_crps}{2.26}
\setres{pop}{CPT_types3}{shares}{0.387, 0.322, 0.290}
\setres{pop}{CPT_types3}{theta_mean}{0.96, 0.48, 0.87}
\setres{pop}{CPT_types3}{theta_sd}{0.02, 0.17, 0.18}
\setres{pop}{CPT_types3}{types}{0.39: (0.98, 0.38, 1.04); 0.32: (0.98, 0.73, 0.90); 0.29: (0.93, 0.33, 0.61)}
\setres{pop}{CPT_types5}{H_F}{22.79}
\setres{pop}{CPT_types5}{H_L}{47.83}
\setres{pop}{CPT_types5}{ess}{5}
\setres{pop}{CPT_types5}{kappa_crps}{0.948}
\setres{pop}{CPT_types5}{kappa_crps_se}{0.009}
\setres{pop}{CPT_types5}{kappa_draw}{0.913}
\setres{pop}{CPT_types5}{kappa_draw_se}{0.002}
\setres{pop}{CPT_types5}{kappa_mean}{0.992}
\setres{pop}{CPT_types5}{kappa_mean_se}{0.005}
\setres{pop}{CPT_types5}{r_stoch}{0.812}
\setres{pop}{CPT_types5}{s_crps}{2.24}
\setres{pop}{CPT_types5}{shares}{0.264, 0.249, 0.202, 0.183, 0.102}
\setres{pop}{CPT_types5}{theta_mean}{0.96, 0.48, 0.89}
\setres{pop}{CPT_types5}{theta_sd}{0.02, 0.20, 0.23}
\setres{pop}{CPT_types5}{types}{0.26: (0.98, 0.53, 0.75); 0.25: (0.98, 0.38, 1.19); 0.20: (0.93, 0.83, 1.04); 0.18: (0.98, 0.28, 0.75); 0.10: (0.93, 0.28, 0.46)}
\setres{pop}{CPT_uniform}{H_F}{26.67}
\setres{pop}{CPT_uniform}{H_L}{47.83}
\setres{pop}{CPT_uniform}{ess}{8{,}000}
\setres{pop}{CPT_uniform}{kappa_crps}{0.867}
\setres{pop}{CPT_uniform}{kappa_crps_se}{0.013}
\setres{pop}{CPT_uniform}{kappa_draw}{0.597}
\setres{pop}{CPT_uniform}{kappa_draw_se}{0.013}
\setres{pop}{CPT_uniform}{kappa_mean}{0.949}
\setres{pop}{CPT_uniform}{kappa_mean_se}{0.010}
\setres{pop}{CPT_uniform}{r_stoch}{0.756}
\setres{pop}{CPT_uniform}{s_crps}{2.45}
\setres{pop}{CPT_uniform}{theta_mean}{0.50, 0.50, 1.55}
\setres{pop}{CPT_uniform}{theta_sd}{0.29, 0.29, 0.84}
\setres{pop}{DA_fitted}{H_F}{23.36}
\setres{pop}{DA_fitted}{H_L}{47.83}
\setres{pop}{DA_fitted}{ess}{11}
\setres{pop}{DA_fitted}{kappa_crps}{0.620}
\setres{pop}{DA_fitted}{kappa_crps_se}{0.057}
\setres{pop}{DA_fitted}{kappa_draw}{0.798}
\setres{pop}{DA_fitted}{kappa_draw_se}{0.033}
\setres{pop}{DA_fitted}{kappa_mean}{0.851}
\setres{pop}{DA_fitted}{kappa_mean_se}{0.030}
\setres{pop}{DA_fitted}{r_stoch}{0.804}
\setres{pop}{DA_fitted}{s_crps}{3.09}
\setres{pop}{DA_fitted}{theta_mean}{0.92, 0.47}
\setres{pop}{DA_fitted}{theta_sd}{0.20, 0.91}
\setres{pop}{DA_from_zurich03}{H_F}{23.49}
\setres{pop}{DA_from_zurich03}{H_L}{47.83}
\setres{pop}{DA_from_zurich03}{ess}{9}
\setres{pop}{DA_from_zurich03}{kappa_crps}{0.670}
\setres{pop}{DA_from_zurich03}{kappa_crps_se}{0.050}
\setres{pop}{DA_from_zurich03}{kappa_draw}{0.744}
\setres{pop}{DA_from_zurich03}{kappa_draw_se}{0.034}
\setres{pop}{DA_from_zurich03}{kappa_mean}{0.878}
\setres{pop}{DA_from_zurich03}{kappa_mean_se}{0.026}
\setres{pop}{DA_from_zurich03}{r_stoch}{0.803}
\setres{pop}{DA_from_zurich03}{s_crps}{2.97}
\setres{pop}{DA_from_zurich03}{theta_mean}{0.91, 0.53}
\setres{pop}{DA_from_zurich03}{theta_sd}{0.24, 1.13}
\setres{pop}{DA_from_zurich03_types1}{H_F}{21.21}
\setres{pop}{DA_from_zurich03_types1}{H_L}{47.83}
\setres{pop}{DA_from_zurich03_types1}{ess}{1}
\setres{pop}{DA_from_zurich03_types1}{kappa_crps}{0.508}
\setres{pop}{DA_from_zurich03_types1}{kappa_crps_se}{0.066}
\setres{pop}{DA_from_zurich03_types1}{kappa_draw}{0.834}
\setres{pop}{DA_from_zurich03_types1}{kappa_draw_se}{0.032}
\setres{pop}{DA_from_zurich03_types1}{kappa_mean}{0.834}
\setres{pop}{DA_from_zurich03_types1}{kappa_mean_se}{0.032}
\setres{pop}{DA_from_zurich03_types1}{r_stoch}{0.830}
\setres{pop}{DA_from_zurich03_types1}{s_crps}{3.38}
\setres{pop}{DA_from_zurich03_types1}{shares}{1.000}
\setres{pop}{DA_from_zurich03_types1}{theta_mean}{0.99, 0.25}
\setres{pop}{DA_from_zurich03_types1}{theta_sd}{0.00, 0.00}
\setres{pop}{DA_from_zurich03_types1}{types}{1.00: (0.99, 0.25)}
\setres{pop}{DA_from_zurich03_types2}{H_F}{21.87}
\setres{pop}{DA_from_zurich03_types2}{H_L}{47.83}
\setres{pop}{DA_from_zurich03_types2}{ess}{2}
\setres{pop}{DA_from_zurich03_types2}{kappa_crps}{0.617}
\setres{pop}{DA_from_zurich03_types2}{kappa_crps_se}{0.056}
\setres{pop}{DA_from_zurich03_types2}{kappa_draw}{0.791}
\setres{pop}{DA_from_zurich03_types2}{kappa_draw_se}{0.033}
\setres{pop}{DA_from_zurich03_types2}{kappa_mean}{0.855}
\setres{pop}{DA_from_zurich03_types2}{kappa_mean_se}{0.029}
\setres{pop}{DA_from_zurich03_types2}{r_stoch}{0.823}
\setres{pop}{DA_from_zurich03_types2}{s_crps}{3.10}
\setres{pop}{DA_from_zurich03_types2}{shares}{0.705, 0.295}
\setres{pop}{DA_from_zurich03_types2}{theta_mean}{0.99, 0.62}
\setres{pop}{DA_from_zurich03_types2}{theta_sd}{0.00, 1.19}
\setres{pop}{DA_from_zurich03_types2}{types}{0.70: (0.99, -0.15); 0.30: (0.99, 2.45)}
\setres{pop}{DA_from_zurich03_types3}{H_F}{21.91}
\setres{pop}{DA_from_zurich03_types3}{H_L}{47.83}
\setres{pop}{DA_from_zurich03_types3}{ess}{2}
\setres{pop}{DA_from_zurich03_types3}{kappa_crps}{0.651}
\setres{pop}{DA_from_zurich03_types3}{kappa_crps_se}{0.052}
\setres{pop}{DA_from_zurich03_types3}{kappa_draw}{0.748}
\setres{pop}{DA_from_zurich03_types3}{kappa_draw_se}{0.034}
\setres{pop}{DA_from_zurich03_types3}{kappa_mean}{0.876}
\setres{pop}{DA_from_zurich03_types3}{kappa_mean_se}{0.026}
\setres{pop}{DA_from_zurich03_types3}{r_stoch}{0.822}
\setres{pop}{DA_from_zurich03_types3}{s_crps}{3.02}
\setres{pop}{DA_from_zurich03_types3}{shares}{0.676, 0.277, 0.047}
\setres{pop}{DA_from_zurich03_types3}{theta_mean}{0.95, 0.66}
\setres{pop}{DA_from_zurich03_types3}{theta_sd}{0.19, 1.25}
\setres{pop}{DA_from_zurich03_types3}{types}{0.68: (0.99, -0.05); 0.28: (0.99, 2.65); 0.05: (0.10, -0.95)}
\setres{pop}{DA_from_zurich03_types5}{H_F}{22.63}
\setres{pop}{DA_from_zurich03_types5}{H_L}{47.83}
\setres{pop}{DA_from_zurich03_types5}{ess}{4}
\setres{pop}{DA_from_zurich03_types5}{kappa_crps}{0.656}
\setres{pop}{DA_from_zurich03_types5}{kappa_crps_se}{0.052}
\setres{pop}{DA_from_zurich03_types5}{kappa_draw}{0.759}
\setres{pop}{DA_from_zurich03_types5}{kappa_draw_se}{0.033}
\setres{pop}{DA_from_zurich03_types5}{kappa_mean}{0.873}
\setres{pop}{DA_from_zurich03_types5}{kappa_mean_se}{0.027}
\setres{pop}{DA_from_zurich03_types5}{r_stoch}{0.814}
\setres{pop}{DA_from_zurich03_types5}{s_crps}{3.00}
\setres{pop}{DA_from_zurich03_types5}{shares}{0.380, 0.222, 0.189, 0.170, 0.039}
\setres{pop}{DA_from_zurich03_types5}{theta_mean}{0.95, 0.57}
\setres{pop}{DA_from_zurich03_types5}{theta_sd}{0.17, 1.17}
\setres{pop}{DA_from_zurich03_types5}{types}{0.38: (0.99, 0.05); 0.22: (0.99, -0.35); 0.19: (0.99, 0.85); 0.17: (0.99, 2.95); 0.04: (0.10, -0.95)}
\setres{pop}{DA_literature}{H_F}{23.69}
\setres{pop}{DA_literature}{H_L}{47.83}
\setres{pop}{DA_literature}{ess}{439}
\setres{pop}{DA_literature}{kappa_crps}{0.531}
\setres{pop}{DA_literature}{kappa_crps_se}{0.061}
\setres{pop}{DA_literature}{kappa_draw}{0.796}
\setres{pop}{DA_literature}{kappa_draw_se}{0.034}
\setres{pop}{DA_literature}{kappa_mean}{0.833}
\setres{pop}{DA_literature}{kappa_mean_se}{0.032}
\setres{pop}{DA_literature}{r_stoch}{0.800}
\setres{pop}{DA_literature}{s_crps}{3.33}
\setres{pop}{DA_literature}{theta_mean}{0.58, 0.30}
\setres{pop}{DA_literature}{theta_sd}{0.19, 0.45}
\setres{pop}{DA_types1}{H_F}{21.23}
\setres{pop}{DA_types1}{H_L}{47.83}
\setres{pop}{DA_types1}{ess}{1}
\setres{pop}{DA_types1}{kappa_crps}{0.508}
\setres{pop}{DA_types1}{kappa_crps_se}{0.067}
\setres{pop}{DA_types1}{kappa_draw}{0.834}
\setres{pop}{DA_types1}{kappa_draw_se}{0.032}
\setres{pop}{DA_types1}{kappa_mean}{0.834}
\setres{pop}{DA_types1}{kappa_mean_se}{0.032}
\setres{pop}{DA_types1}{r_stoch}{0.830}
\setres{pop}{DA_types1}{s_crps}{3.39}
\setres{pop}{DA_types1}{shares}{1.000}
\setres{pop}{DA_types1}{theta_mean}{0.99, 0.25}
\setres{pop}{DA_types1}{theta_sd}{0.00, 0.00}
\setres{pop}{DA_types1}{types}{1.00: (0.99, 0.25)}
\setres{pop}{DA_types2}{H_F}{21.82}
\setres{pop}{DA_types2}{H_L}{47.83}
\setres{pop}{DA_types2}{ess}{2}
\setres{pop}{DA_types2}{kappa_crps}{0.590}
\setres{pop}{DA_types2}{kappa_crps_se}{0.060}
\setres{pop}{DA_types2}{kappa_draw}{0.809}
\setres{pop}{DA_types2}{kappa_draw_se}{0.033}
\setres{pop}{DA_types2}{kappa_mean}{0.844}
\setres{pop}{DA_types2}{kappa_mean_se}{0.031}
\setres{pop}{DA_types2}{r_stoch}{0.823}
\setres{pop}{DA_types2}{s_crps}{3.17}
\setres{pop}{DA_types2}{shares}{0.651, 0.349}
\setres{pop}{DA_types2}{theta_mean}{0.99, 0.47}
\setres{pop}{DA_types2}{theta_sd}{0.00, 0.72}
\setres{pop}{DA_types2}{types}{0.65: (0.99, -0.05); 0.35: (0.99, 1.45)}
\setres{pop}{DA_types3}{H_F}{22.08}
\setres{pop}{DA_types3}{H_L}{47.83}
\setres{pop}{DA_types3}{ess}{2}
\setres{pop}{DA_types3}{kappa_crps}{0.604}
\setres{pop}{DA_types3}{kappa_crps_se}{0.058}
\setres{pop}{DA_types3}{kappa_draw}{0.808}
\setres{pop}{DA_types3}{kappa_draw_se}{0.033}
\setres{pop}{DA_types3}{kappa_mean}{0.847}
\setres{pop}{DA_types3}{kappa_mean_se}{0.030}
\setres{pop}{DA_types3}{r_stoch}{0.820}
\setres{pop}{DA_types3}{s_crps}{3.14}
\setres{pop}{DA_types3}{shares}{0.510, 0.394, 0.096}
\setres{pop}{DA_types3}{theta_mean}{0.99, 0.46}
\setres{pop}{DA_types3}{theta_sd}{0.00, 0.89}
\setres{pop}{DA_types3}{types}{0.51: (0.99, -0.15); 0.39: (0.99, 0.65); 0.10: (0.99, 2.95)}
\setres{pop}{DA_types5}{H_F}{22.62}
\setres{pop}{DA_types5}{H_L}{47.83}
\setres{pop}{DA_types5}{ess}{4}
\setres{pop}{DA_types5}{kappa_crps}{0.616}
\setres{pop}{DA_types5}{kappa_crps_se}{0.057}
\setres{pop}{DA_types5}{kappa_draw}{0.801}
\setres{pop}{DA_types5}{kappa_draw_se}{0.033}
\setres{pop}{DA_types5}{kappa_mean}{0.850}
\setres{pop}{DA_types5}{kappa_mean_se}{0.030}
\setres{pop}{DA_types5}{r_stoch}{0.814}
\setres{pop}{DA_types5}{s_crps}{3.11}
\setres{pop}{DA_types5}{shares}{0.371, 0.313, 0.173, 0.093, 0.050}
\setres{pop}{DA_types5}{theta_mean}{0.99, 0.46}
\setres{pop}{DA_types5}{theta_sd}{0.00, 0.90}
\setres{pop}{DA_types5}{types}{0.37: (0.99, 0.05); 0.31: (0.99, 0.75); 0.17: (0.99, -0.25); 0.09: (0.99, 2.95); 0.05: (0.99, -0.55)}
\setres{pop}{DA_uniform}{H_F}{24.32}
\setres{pop}{DA_uniform}{H_L}{47.83}
\setres{pop}{DA_uniform}{ess}{1{,}600}
\setres{pop}{DA_uniform}{kappa_crps}{0.521}
\setres{pop}{DA_uniform}{kappa_crps_se}{0.054}
\setres{pop}{DA_uniform}{kappa_draw}{0.690}
\setres{pop}{DA_uniform}{kappa_draw_se}{0.039}
\setres{pop}{DA_uniform}{kappa_mean}{0.853}
\setres{pop}{DA_uniform}{kappa_mean_se}{0.029}
\setres{pop}{DA_uniform}{r_stoch}{0.791}
\setres{pop}{DA_uniform}{s_crps}{3.35}
\setres{pop}{DA_uniform}{theta_mean}{0.50, 1.00}
\setres{pop}{DA_uniform}{theta_sd}{0.29, 1.15}
\setres{pop}{EU_fitted}{H_F}{22.22}
\setres{pop}{EU_fitted}{H_L}{47.83}
\setres{pop}{EU_fitted}{ess}{4}
\setres{pop}{EU_fitted}{kappa_crps}{0.523}
\setres{pop}{EU_fitted}{kappa_crps_se}{0.066}
\setres{pop}{EU_fitted}{kappa_draw}{0.829}
\setres{pop}{EU_fitted}{kappa_draw_se}{0.032}
\setres{pop}{EU_fitted}{kappa_mean}{0.833}
\setres{pop}{EU_fitted}{kappa_mean_se}{0.031}
\setres{pop}{EU_fitted}{r_stoch}{0.819}
\setres{pop}{EU_fitted}{s_crps}{3.35}
\setres{pop}{EU_fitted}{theta_mean}{0.80}
\setres{pop}{EU_fitted}{theta_sd}{0.26}
\setres{pop}{EU_from_zurich03}{H_F}{22.12}
\setres{pop}{EU_from_zurich03}{H_L}{47.83}
\setres{pop}{EU_from_zurich03}{ess}{2}
\setres{pop}{EU_from_zurich03}{kappa_crps}{0.528}
\setres{pop}{EU_from_zurich03}{kappa_crps_se}{0.066}
\setres{pop}{EU_from_zurich03}{kappa_draw}{0.828}
\setres{pop}{EU_from_zurich03}{kappa_draw_se}{0.032}
\setres{pop}{EU_from_zurich03}{kappa_mean}{0.834}
\setres{pop}{EU_from_zurich03}{kappa_mean_se}{0.031}
\setres{pop}{EU_from_zurich03}{r_stoch}{0.820}
\setres{pop}{EU_from_zurich03}{s_crps}{3.33}
\setres{pop}{EU_from_zurich03}{theta_mean}{0.82}
\setres{pop}{EU_from_zurich03}{theta_sd}{0.28}
\setres{pop}{EU_from_zurich03_types1}{H_F}{21.21}
\setres{pop}{EU_from_zurich03_types1}{H_L}{47.83}
\setres{pop}{EU_from_zurich03_types1}{ess}{1}
\setres{pop}{EU_from_zurich03_types1}{kappa_crps}{0.503}
\setres{pop}{EU_from_zurich03_types1}{kappa_crps_se}{0.068}
\setres{pop}{EU_from_zurich03_types1}{kappa_draw}{0.832}
\setres{pop}{EU_from_zurich03_types1}{kappa_draw_se}{0.031}
\setres{pop}{EU_from_zurich03_types1}{kappa_mean}{0.832}
\setres{pop}{EU_from_zurich03_types1}{kappa_mean_se}{0.031}
\setres{pop}{EU_from_zurich03_types1}{r_stoch}{0.830}
\setres{pop}{EU_from_zurich03_types1}{s_crps}{3.40}
\setres{pop}{EU_from_zurich03_types1}{shares}{1.000}
\setres{pop}{EU_from_zurich03_types1}{theta_mean}{0.92}
\setres{pop}{EU_from_zurich03_types1}{theta_sd}{0.00}
\setres{pop}{EU_from_zurich03_types1}{types}{1.00: (0.92)}
\setres{pop}{EU_from_zurich03_types2}{H_F}{21.81}
\setres{pop}{EU_from_zurich03_types2}{H_L}{47.83}
\setres{pop}{EU_from_zurich03_types2}{ess}{2}
\setres{pop}{EU_from_zurich03_types2}{kappa_crps}{0.521}
\setres{pop}{EU_from_zurich03_types2}{kappa_crps_se}{0.066}
\setres{pop}{EU_from_zurich03_types2}{kappa_draw}{0.830}
\setres{pop}{EU_from_zurich03_types2}{kappa_draw_se}{0.031}
\setres{pop}{EU_from_zurich03_types2}{kappa_mean}{0.834}
\setres{pop}{EU_from_zurich03_types2}{kappa_mean_se}{0.031}
\setres{pop}{EU_from_zurich03_types2}{r_stoch}{0.824}
\setres{pop}{EU_from_zurich03_types2}{s_crps}{3.35}
\setres{pop}{EU_from_zurich03_types2}{shares}{0.744, 0.256}
\setres{pop}{EU_from_zurich03_types2}{theta_mean}{0.86}
\setres{pop}{EU_from_zurich03_types2}{theta_sd}{0.24}
\setres{pop}{EU_from_zurich03_types2}{types}{0.74: (1.00); 0.26: (0.46)}
\setres{pop}{EU_from_zurich03_types3}{H_F}{21.95}
\setres{pop}{EU_from_zurich03_types3}{H_L}{47.83}
\setres{pop}{EU_from_zurich03_types3}{ess}{2}
\setres{pop}{EU_from_zurich03_types3}{kappa_crps}{0.526}
\setres{pop}{EU_from_zurich03_types3}{kappa_crps_se}{0.066}
\setres{pop}{EU_from_zurich03_types3}{kappa_draw}{0.828}
\setres{pop}{EU_from_zurich03_types3}{kappa_draw_se}{0.032}
\setres{pop}{EU_from_zurich03_types3}{kappa_mean}{0.834}
\setres{pop}{EU_from_zurich03_types3}{kappa_mean_se}{0.031}
\setres{pop}{EU_from_zurich03_types3}{r_stoch}{0.822}
\setres{pop}{EU_from_zurich03_types3}{s_crps}{3.34}
\setres{pop}{EU_from_zurich03_types3}{shares}{0.674, 0.254, 0.072}
\setres{pop}{EU_from_zurich03_types3}{theta_mean}{0.82}
\setres{pop}{EU_from_zurich03_types3}{theta_sd}{0.29}
\setres{pop}{EU_from_zurich03_types3}{types}{0.67: (1.00); 0.25: (0.59); 0.07: (0.01)}
\setres{pop}{EU_from_zurich03_types5}{H_F}{22.09}
\setres{pop}{EU_from_zurich03_types5}{H_L}{47.83}
\setres{pop}{EU_from_zurich03_types5}{ess}{2}
\setres{pop}{EU_from_zurich03_types5}{kappa_crps}{0.524}
\setres{pop}{EU_from_zurich03_types5}{kappa_crps_se}{0.066}
\setres{pop}{EU_from_zurich03_types5}{kappa_draw}{0.829}
\setres{pop}{EU_from_zurich03_types5}{kappa_draw_se}{0.032}
\setres{pop}{EU_from_zurich03_types5}{kappa_mean}{0.834}
\setres{pop}{EU_from_zurich03_types5}{kappa_mean_se}{0.031}
\setres{pop}{EU_from_zurich03_types5}{r_stoch}{0.820}
\setres{pop}{EU_from_zurich03_types5}{s_crps}{3.34}
\setres{pop}{EU_from_zurich03_types5}{shares}{0.652, 0.156, 0.080, 0.069, 0.042}
\setres{pop}{EU_from_zurich03_types5}{theta_mean}{0.82}
\setres{pop}{EU_from_zurich03_types5}{theta_sd}{0.28}
\setres{pop}{EU_from_zurich03_types5}{types}{0.65: (1.00); 0.16: (0.54); 0.08: (0.69); 0.07: (0.01); 0.04: (0.70)}
\setres{pop}{EU_literature}{H_F}{21.99}
\setres{pop}{EU_literature}{H_L}{47.83}
\setres{pop}{EU_literature}{ess}{98}
\setres{pop}{EU_literature}{kappa_crps}{0.504}
\setres{pop}{EU_literature}{kappa_crps_se}{0.067}
\setres{pop}{EU_literature}{kappa_draw}{0.829}
\setres{pop}{EU_literature}{kappa_draw_se}{0.032}
\setres{pop}{EU_literature}{kappa_mean}{0.831}
\setres{pop}{EU_literature}{kappa_mean_se}{0.032}
\setres{pop}{EU_literature}{r_stoch}{0.821}
\setres{pop}{EU_literature}{s_crps}{3.40}
\setres{pop}{EU_literature}{theta_mean}{0.70}
\setres{pop}{EU_literature}{theta_sd}{0.14}
\setres{pop}{EU_types1}{H_F}{21.23}
\setres{pop}{EU_types1}{H_L}{47.83}
\setres{pop}{EU_types1}{ess}{1}
\setres{pop}{EU_types1}{kappa_crps}{0.503}
\setres{pop}{EU_types1}{kappa_crps_se}{0.068}
\setres{pop}{EU_types1}{kappa_draw}{0.832}
\setres{pop}{EU_types1}{kappa_draw_se}{0.031}
\setres{pop}{EU_types1}{kappa_mean}{0.832}
\setres{pop}{EU_types1}{kappa_mean_se}{0.031}
\setres{pop}{EU_types1}{r_stoch}{0.830}
\setres{pop}{EU_types1}{s_crps}{3.40}
\setres{pop}{EU_types1}{shares}{1.000}
\setres{pop}{EU_types1}{theta_mean}{0.85}
\setres{pop}{EU_types1}{theta_sd}{0.00}
\setres{pop}{EU_types1}{types}{1.00: (0.85)}
\setres{pop}{EU_types2}{H_F}{21.81}
\setres{pop}{EU_types2}{H_L}{47.83}
\setres{pop}{EU_types2}{ess}{2}
\setres{pop}{EU_types2}{kappa_crps}{0.518}
\setres{pop}{EU_types2}{kappa_crps_se}{0.067}
\setres{pop}{EU_types2}{kappa_draw}{0.830}
\setres{pop}{EU_types2}{kappa_draw_se}{0.032}
\setres{pop}{EU_types2}{kappa_mean}{0.833}
\setres{pop}{EU_types2}{kappa_mean_se}{0.031}
\setres{pop}{EU_types2}{r_stoch}{0.824}
\setres{pop}{EU_types2}{s_crps}{3.36}
\setres{pop}{EU_types2}{shares}{0.668, 0.332}
\setres{pop}{EU_types2}{theta_mean}{0.84}
\setres{pop}{EU_types2}{theta_sd}{0.23}
\setres{pop}{EU_types2}{types}{0.67: (1.00); 0.33: (0.52)}
\setres{pop}{EU_types3}{H_F}{21.95}
\setres{pop}{EU_types3}{H_L}{47.83}
\setres{pop}{EU_types3}{ess}{2}
\setres{pop}{EU_types3}{kappa_crps}{0.525}
\setres{pop}{EU_types3}{kappa_crps_se}{0.066}
\setres{pop}{EU_types3}{kappa_draw}{0.829}
\setres{pop}{EU_types3}{kappa_draw_se}{0.032}
\setres{pop}{EU_types3}{kappa_mean}{0.834}
\setres{pop}{EU_types3}{kappa_mean_se}{0.031}
\setres{pop}{EU_types3}{r_stoch}{0.822}
\setres{pop}{EU_types3}{s_crps}{3.34}
\setres{pop}{EU_types3}{shares}{0.651, 0.291, 0.058}
\setres{pop}{EU_types3}{theta_mean}{0.82}
\setres{pop}{EU_types3}{theta_sd}{0.27}
\setres{pop}{EU_types3}{types}{0.65: (1.00); 0.29: (0.59); 0.06: (0.01)}
\setres{pop}{EU_types5}{H_F}{22.21}
\setres{pop}{EU_types5}{H_L}{47.83}
\setres{pop}{EU_types5}{ess}{3}
\setres{pop}{EU_types5}{kappa_crps}{0.522}
\setres{pop}{EU_types5}{kappa_crps_se}{0.066}
\setres{pop}{EU_types5}{kappa_draw}{0.829}
\setres{pop}{EU_types5}{kappa_draw_se}{0.032}
\setres{pop}{EU_types5}{kappa_mean}{0.833}
\setres{pop}{EU_types5}{kappa_mean_se}{0.031}
\setres{pop}{EU_types5}{r_stoch}{0.819}
\setres{pop}{EU_types5}{s_crps}{3.35}
\setres{pop}{EU_types5}{shares}{0.529, 0.145, 0.141, 0.138, 0.048}
\setres{pop}{EU_types5}{theta_mean}{0.81}
\setres{pop}{EU_types5}{theta_sd}{0.26}
\setres{pop}{EU_types5}{types}{0.53: (1.00); 0.14: (0.64); 0.14: (0.52); 0.14: (0.82); 0.05: (0.01)}
\setres{pop}{EU_uniform}{H_F}{22.65}
\setres{pop}{EU_uniform}{H_L}{47.83}
\setres{pop}{EU_uniform}{ess}{200}
\setres{pop}{EU_uniform}{kappa_crps}{0.494}
\setres{pop}{EU_uniform}{kappa_crps_se}{0.066}
\setres{pop}{EU_uniform}{kappa_draw}{0.820}
\setres{pop}{EU_uniform}{kappa_draw_se}{0.033}
\setres{pop}{EU_uniform}{kappa_mean}{0.826}
\setres{pop}{EU_uniform}{kappa_mean_se}{0.033}
\setres{pop}{EU_uniform}{r_stoch}{0.813}
\setres{pop}{EU_uniform}{s_crps}{3.42}
\setres{pop}{EU_uniform}{theta_mean}{0.51}
\setres{pop}{EU_uniform}{theta_sd}{0.29}
\setres{pop}{MIX_from_zurich03_mix_eu_cpt}{H_F}{21.72}
\setres{pop}{MIX_from_zurich03_mix_eu_cpt}{H_L}{47.83}
\setres{pop}{MIX_from_zurich03_mix_eu_cpt}{ess}{2}
\setres{pop}{MIX_from_zurich03_mix_eu_cpt}{eu_alpha}{1.00}
\setres{pop}{MIX_from_zurich03_mix_eu_cpt}{eu_share}{0.21}
\setres{pop}{MIX_from_zurich03_mix_eu_cpt}{kappa_crps}{0.849}
\setres{pop}{MIX_from_zurich03_mix_eu_cpt}{kappa_crps_se}{0.016}
\setres{pop}{MIX_from_zurich03_mix_eu_cpt}{kappa_draw}{0.905}
\setres{pop}{MIX_from_zurich03_mix_eu_cpt}{kappa_draw_se}{0.004}
\setres{pop}{MIX_from_zurich03_mix_eu_cpt}{kappa_mean}{0.976}
\setres{pop}{MIX_from_zurich03_mix_eu_cpt}{kappa_mean_se}{0.008}
\setres{pop}{MIX_from_zurich03_mix_eu_cpt}{r_stoch}{0.825}
\setres{pop}{MIX_from_zurich03_mix_eu_cpt}{s_crps}{2.50}
\setres{pop}{MIX_from_zurich03_mix_eu_cpt}{shares}{0.213, 0.787}
\setres{pop}{MIX_from_zurich03_mix_eu_cpt}{types}{0.79: CPT (0.98, 0.33, 0.90); 0.21: EU (1.00)}
\setres{pop}{MIX_from_zurich03_mix_eu_cpt2}{H_F}{22.24}
\setres{pop}{MIX_from_zurich03_mix_eu_cpt2}{H_L}{47.83}
\setres{pop}{MIX_from_zurich03_mix_eu_cpt2}{ess}{3}
\setres{pop}{MIX_from_zurich03_mix_eu_cpt2}{eu_alpha}{0.97}
\setres{pop}{MIX_from_zurich03_mix_eu_cpt2}{eu_share}{0.22}
\setres{pop}{MIX_from_zurich03_mix_eu_cpt2}{kappa_crps}{0.941}
\setres{pop}{MIX_from_zurich03_mix_eu_cpt2}{kappa_crps_se}{0.013}
\setres{pop}{MIX_from_zurich03_mix_eu_cpt2}{kappa_draw}{0.865}
\setres{pop}{MIX_from_zurich03_mix_eu_cpt2}{kappa_draw_se}{0.003}
\setres{pop}{MIX_from_zurich03_mix_eu_cpt2}{kappa_mean}{0.985}
\setres{pop}{MIX_from_zurich03_mix_eu_cpt2}{kappa_mean_se}{0.007}
\setres{pop}{MIX_from_zurich03_mix_eu_cpt2}{r_stoch}{0.818}
\setres{pop}{MIX_from_zurich03_mix_eu_cpt2}{s_crps}{2.26}
\setres{pop}{MIX_from_zurich03_mix_eu_cpt2}{shares}{0.223, 0.374, 0.403}
\setres{pop}{MIX_from_zurich03_mix_eu_cpt2}{types}{0.40: CPT (0.98, 0.33, 0.61); 0.37: CPT (0.98, 0.38, 1.33); 0.22: EU (0.97)}
\setres{pop}{MIX_mix_eu_cpt}{H_F}{21.77}
\setres{pop}{MIX_mix_eu_cpt}{H_L}{47.83}
\setres{pop}{MIX_mix_eu_cpt}{ess}{2}
\setres{pop}{MIX_mix_eu_cpt}{eu_alpha}{0.97}
\setres{pop}{MIX_mix_eu_cpt}{eu_share}{0.23}
\setres{pop}{MIX_mix_eu_cpt}{kappa_crps}{0.901}
\setres{pop}{MIX_mix_eu_cpt}{kappa_crps_se}{0.009}
\setres{pop}{MIX_mix_eu_cpt}{kappa_draw}{0.937}
\setres{pop}{MIX_mix_eu_cpt}{kappa_draw_se}{0.004}
\setres{pop}{MIX_mix_eu_cpt}{kappa_mean}{0.994}
\setres{pop}{MIX_mix_eu_cpt}{kappa_mean_se}{0.003}
\setres{pop}{MIX_mix_eu_cpt}{r_stoch}{0.824}
\setres{pop}{MIX_mix_eu_cpt}{s_crps}{2.37}
\setres{pop}{MIX_mix_eu_cpt}{shares}{0.228, 0.772}
\setres{pop}{MIX_mix_eu_cpt}{types}{0.77: CPT (0.88, 0.38, 0.90); 0.23: EU (0.97)}
\setres{pop}{MIX_mix_eu_cpt2}{H_F}{22.30}
\setres{pop}{MIX_mix_eu_cpt2}{H_L}{47.83}
\setres{pop}{MIX_mix_eu_cpt2}{ess}{3}
\setres{pop}{MIX_mix_eu_cpt2}{eu_alpha}{0.92}
\setres{pop}{MIX_mix_eu_cpt2}{eu_share}{0.22}
\setres{pop}{MIX_mix_eu_cpt2}{kappa_crps}{0.940}
\setres{pop}{MIX_mix_eu_cpt2}{kappa_crps_se}{0.009}
\setres{pop}{MIX_mix_eu_cpt2}{kappa_draw}{0.916}
\setres{pop}{MIX_mix_eu_cpt2}{kappa_draw_se}{0.003}
\setres{pop}{MIX_mix_eu_cpt2}{kappa_mean}{0.993}
\setres{pop}{MIX_mix_eu_cpt2}{kappa_mean_se}{0.004}
\setres{pop}{MIX_mix_eu_cpt2}{r_stoch}{0.818}
\setres{pop}{MIX_mix_eu_cpt2}{s_crps}{2.26}
\setres{pop}{MIX_mix_eu_cpt2}{shares}{0.222, 0.474, 0.304}
\setres{pop}{MIX_mix_eu_cpt2}{types}{0.47: CPT (0.98, 0.43, 1.04); 0.30: CPT (0.93, 0.33, 0.61); 0.22: EU (0.92)}
\setres{pop}{reference}{CPT_noise_c}{0.041}
\setres{pop}{reference}{DA_noise_c}{0.041}
\setres{pop}{reference}{EU_noise_c}{0.041}
\setres{pop}{reference}{MIX_noise_c}{0.041}
\setres{pop}{reference}{mse_base}{128.78}
\setres{pop}{reference}{mse_best}{21.33}
\setres{pop}{reference}{n_lotteries}{15}
\setres{pop}{reference}{n_subjects}{118}
\setres{pop}{reference}{s_crps_base}{4.70}
\setres{pop}{reference}{s_crps_best}{2.11}
\setres{pop}{reference}{target}{Zurich2006\_shared}
\setres{popL}{CPT_fitted}{H_F}{28.20}
\setres{popL}{CPT_fitted}{H_L}{47.83}
\setres{popL}{CPT_fitted}{ess}{43}
\setres{popL}{CPT_fitted}{kappa_crps}{0.969}
\setres{popL}{CPT_fitted}{kappa_crps_se}{0.008}
\setres{popL}{CPT_fitted}{kappa_draw}{0.845}
\setres{popL}{CPT_fitted}{kappa_draw_se}{0.008}
\setres{popL}{CPT_fitted}{kappa_mean}{0.996}
\setres{popL}{CPT_fitted}{kappa_mean_se}{0.004}
\setres{popL}{CPT_fitted}{r_stoch}{0.730}
\setres{popL}{CPT_fitted}{s_crps}{2.38}
\setres{popL}{CPT_fitted}{theta_mean}{0.84, 0.52, 1.30}
\setres{popL}{CPT_fitted}{theta_sd}{0.23, 0.24, 0.51}
\setres{popL}{CPT_from_zurich03}{H_F}{28.34}
\setres{popL}{CPT_from_zurich03}{H_L}{47.83}
\setres{popL}{CPT_from_zurich03}{ess}{34}
\setres{popL}{CPT_from_zurich03}{kappa_crps}{0.953}
\setres{popL}{CPT_from_zurich03}{kappa_crps_se}{0.012}
\setres{popL}{CPT_from_zurich03}{kappa_draw}{0.761}
\setres{popL}{CPT_from_zurich03}{kappa_draw_se}{0.007}
\setres{popL}{CPT_from_zurich03}{kappa_mean}{0.981}
\setres{popL}{CPT_from_zurich03}{kappa_mean_se}{0.009}
\setres{popL}{CPT_from_zurich03}{r_stoch}{0.727}
\setres{popL}{CPT_from_zurich03}{s_crps}{2.42}
\setres{popL}{CPT_from_zurich03}{theta_mean}{0.85, 0.45, 1.34}
\setres{popL}{CPT_from_zurich03}{theta_sd}{0.22, 0.23, 0.65}
\setres{popL}{CPT_from_zurich03_types1}{H_F}{24.91}
\setres{popL}{CPT_from_zurich03_types1}{H_L}{47.83}
\setres{popL}{CPT_from_zurich03_types1}{ess}{1}
\setres{popL}{CPT_from_zurich03_types1}{kappa_crps}{0.755}
\setres{popL}{CPT_from_zurich03_types1}{kappa_crps_se}{0.021}
\setres{popL}{CPT_from_zurich03_types1}{kappa_draw}{0.973}
\setres{popL}{CPT_from_zurich03_types1}{kappa_draw_se}{0.010}
\setres{popL}{CPT_from_zurich03_types1}{kappa_mean}{0.973}
\setres{popL}{CPT_from_zurich03_types1}{kappa_mean_se}{0.010}
\setres{popL}{CPT_from_zurich03_types1}{r_stoch}{0.783}
\setres{popL}{CPT_from_zurich03_types1}{s_crps}{2.91}
\setres{popL}{CPT_from_zurich03_types1}{shares}{1.000}
\setres{popL}{CPT_from_zurich03_types1}{theta_mean}{0.88, 0.38, 1.19}
\setres{popL}{CPT_from_zurich03_types1}{theta_sd}{0.00, 0.00, 0.00}
\setres{popL}{CPT_from_zurich03_types1}{types}{1.00: (0.88, 0.38, 1.19)}
\setres{popL}{CPT_from_zurich03_types2}{H_F}{25.51}
\setres{popL}{CPT_from_zurich03_types2}{H_L}{47.83}
\setres{popL}{CPT_from_zurich03_types2}{ess}{2}
\setres{popL}{CPT_from_zurich03_types2}{kappa_crps}{0.856}
\setres{popL}{CPT_from_zurich03_types2}{kappa_crps_se}{0.017}
\setres{popL}{CPT_from_zurich03_types2}{kappa_draw}{0.927}
\setres{popL}{CPT_from_zurich03_types2}{kappa_draw_se}{0.006}
\setres{popL}{CPT_from_zurich03_types2}{kappa_mean}{0.979}
\setres{popL}{CPT_from_zurich03_types2}{kappa_mean_se}{0.009}
\setres{popL}{CPT_from_zurich03_types2}{r_stoch}{0.774}
\setres{popL}{CPT_from_zurich03_types2}{s_crps}{2.65}
\setres{popL}{CPT_from_zurich03_types2}{shares}{0.741, 0.259}
\setres{popL}{CPT_from_zurich03_types2}{theta_mean}{0.98, 0.41, 1.20}
\setres{popL}{CPT_from_zurich03_types2}{theta_sd}{0.00, 0.04, 0.51}
\setres{popL}{CPT_from_zurich03_types2}{types}{0.74: (0.98, 0.43, 0.90); 0.26: (0.98, 0.33, 2.06)}
\setres{popL}{CPT_from_zurich03_types3}{H_F}{25.86}
\setres{popL}{CPT_from_zurich03_types3}{H_L}{47.83}
\setres{popL}{CPT_from_zurich03_types3}{ess}{2}
\setres{popL}{CPT_from_zurich03_types3}{kappa_crps}{0.939}
\setres{popL}{CPT_from_zurich03_types3}{kappa_crps_se}{0.012}
\setres{popL}{CPT_from_zurich03_types3}{kappa_draw}{0.839}
\setres{popL}{CPT_from_zurich03_types3}{kappa_draw_se}{0.003}
\setres{popL}{CPT_from_zurich03_types3}{kappa_mean}{0.985}
\setres{popL}{CPT_from_zurich03_types3}{kappa_mean_se}{0.009}
\setres{popL}{CPT_from_zurich03_types3}{r_stoch}{0.769}
\setres{popL}{CPT_from_zurich03_types3}{s_crps}{2.45}
\setres{popL}{CPT_from_zurich03_types3}{shares}{0.555, 0.243, 0.203}
\setres{popL}{CPT_from_zurich03_types3}{theta_mean}{0.98, 0.43, 1.21}
\setres{popL}{CPT_from_zurich03_types3}{theta_sd}{0.00, 0.16, 0.52}
\setres{popL}{CPT_from_zurich03_types3}{types}{0.55: (0.98, 0.58, 1.04); 0.24: (0.98, 0.23, 0.75); 0.20: (0.98, 0.28, 2.20)}
\setres{popL}{CPT_from_zurich03_types5}{H_F}{26.25}
\setres{popL}{CPT_from_zurich03_types5}{H_L}{47.83}
\setres{popL}{CPT_from_zurich03_types5}{ess}{3}
\setres{popL}{CPT_from_zurich03_types5}{kappa_crps}{0.940}
\setres{popL}{CPT_from_zurich03_types5}{kappa_crps_se}{0.012}
\setres{popL}{CPT_from_zurich03_types5}{kappa_draw}{0.821}
\setres{popL}{CPT_from_zurich03_types5}{kappa_draw_se}{0.004}
\setres{popL}{CPT_from_zurich03_types5}{kappa_mean}{0.982}
\setres{popL}{CPT_from_zurich03_types5}{kappa_mean_se}{0.008}
\setres{popL}{CPT_from_zurich03_types5}{r_stoch}{0.763}
\setres{popL}{CPT_from_zurich03_types5}{s_crps}{2.45}
\setres{popL}{CPT_from_zurich03_types5}{shares}{0.442, 0.230, 0.207, 0.069, 0.052}
\setres{popL}{CPT_from_zurich03_types5}{theta_mean}{0.90, 0.42, 1.29}
\setres{popL}{CPT_from_zurich03_types5}{theta_sd}{0.04, 0.16, 0.56}
\setres{popL}{CPT_from_zurich03_types5}{types}{0.44: (0.88, 0.58, 1.04); 0.23: (0.88, 0.23, 1.04); 0.21: (0.98, 0.43, 1.77); 0.07: (0.98, 0.18, 2.93); 0.05: (0.88, 0.18, 0.46)}
\setres{popL}{CPT_literature}{H_F}{27.97}
\setres{popL}{CPT_literature}{H_L}{47.83}
\setres{popL}{CPT_literature}{ess}{1{,}104}
\setres{popL}{CPT_literature}{kappa_crps}{0.847}
\setres{popL}{CPT_literature}{kappa_crps_se}{0.028}
\setres{popL}{CPT_literature}{kappa_draw}{0.879}
\setres{popL}{CPT_literature}{kappa_draw_se}{0.019}
\setres{popL}{CPT_literature}{kappa_mean}{0.947}
\setres{popL}{CPT_literature}{kappa_mean_se}{0.014}
\setres{popL}{CPT_literature}{r_stoch}{0.734}
\setres{popL}{CPT_literature}{s_crps}{2.68}
\setres{popL}{CPT_literature}{theta_mean}{0.67, 0.72, 1.08}
\setres{popL}{CPT_literature}{theta_sd}{0.20, 0.18, 0.28}
\setres{popL}{CPT_types1}{H_F}{24.93}
\setres{popL}{CPT_types1}{H_L}{47.83}
\setres{popL}{CPT_types1}{ess}{1}
\setres{popL}{CPT_types1}{kappa_crps}{0.856}
\setres{popL}{CPT_types1}{kappa_crps_se}{0.015}
\setres{popL}{CPT_types1}{kappa_draw}{0.997}
\setres{popL}{CPT_types1}{kappa_draw_se}{0.002}
\setres{popL}{CPT_types1}{kappa_mean}{0.997}
\setres{popL}{CPT_types1}{kappa_mean_se}{0.002}
\setres{popL}{CPT_types1}{r_stoch}{0.783}
\setres{popL}{CPT_types1}{s_crps}{2.66}
\setres{popL}{CPT_types1}{shares}{1.000}
\setres{popL}{CPT_types1}{theta_mean}{0.88, 0.48, 1.19}
\setres{popL}{CPT_types1}{theta_sd}{0.00, 0.00, 0.00}
\setres{popL}{CPT_types1}{types}{1.00: (0.88, 0.48, 1.19)}
\setres{popL}{CPT_types2}{H_F}{25.56}
\setres{popL}{CPT_types2}{H_L}{47.83}
\setres{popL}{CPT_types2}{ess}{2}
\setres{popL}{CPT_types2}{kappa_crps}{0.901}
\setres{popL}{CPT_types2}{kappa_crps_se}{0.012}
\setres{popL}{CPT_types2}{kappa_draw}{0.957}
\setres{popL}{CPT_types2}{kappa_draw_se}{0.003}
\setres{popL}{CPT_types2}{kappa_mean}{0.995}
\setres{popL}{CPT_types2}{kappa_mean_se}{0.003}
\setres{popL}{CPT_types2}{r_stoch}{0.773}
\setres{popL}{CPT_types2}{s_crps}{2.54}
\setres{popL}{CPT_types2}{shares}{0.567, 0.433}
\setres{popL}{CPT_types2}{theta_mean}{0.90, 0.48, 1.17}
\setres{popL}{CPT_types2}{theta_sd}{0.02, 0.17, 0.14}
\setres{popL}{CPT_types2}{types}{0.57: (0.93, 0.63, 1.04); 0.43: (0.88, 0.28, 1.33)}
\setres{popL}{CPT_types3}{H_F}{25.94}
\setres{popL}{CPT_types3}{H_L}{47.83}
\setres{popL}{CPT_types3}{ess}{3}
\setres{popL}{CPT_types3}{kappa_crps}{0.952}
\setres{popL}{CPT_types3}{kappa_crps_se}{0.008}
\setres{popL}{CPT_types3}{kappa_draw}{0.904}
\setres{popL}{CPT_types3}{kappa_draw_se}{0.004}
\setres{popL}{CPT_types3}{kappa_mean}{0.997}
\setres{popL}{CPT_types3}{kappa_mean_se}{0.004}
\setres{popL}{CPT_types3}{r_stoch}{0.768}
\setres{popL}{CPT_types3}{s_crps}{2.42}
\setres{popL}{CPT_types3}{shares}{0.413, 0.303, 0.284}
\setres{popL}{CPT_types3}{theta_mean}{0.95, 0.51, 1.16}
\setres{popL}{CPT_types3}{theta_sd}{0.05, 0.18, 0.30}
\setres{popL}{CPT_types3}{types}{0.41: (0.98, 0.73, 1.04); 0.30: (0.88, 0.33, 0.90); 0.28: (0.98, 0.38, 1.62)}
\setres{popL}{CPT_types5}{H_F}{26.45}
\setres{popL}{CPT_types5}{H_L}{47.83}
\setres{popL}{CPT_types5}{ess}{4}
\setres{popL}{CPT_types5}{kappa_crps}{0.956}
\setres{popL}{CPT_types5}{kappa_crps_se}{0.008}
\setres{popL}{CPT_types5}{kappa_draw}{0.890}
\setres{popL}{CPT_types5}{kappa_draw_se}{0.005}
\setres{popL}{CPT_types5}{kappa_mean}{0.997}
\setres{popL}{CPT_types5}{kappa_mean_se}{0.004}
\setres{popL}{CPT_types5}{r_stoch}{0.760}
\setres{popL}{CPT_types5}{s_crps}{2.41}
\setres{popL}{CPT_types5}{shares}{0.324, 0.258, 0.173, 0.129, 0.117}
\setres{popL}{CPT_types5}{theta_mean}{0.90, 0.51, 1.22}
\setres{popL}{CPT_types5}{theta_sd}{0.08, 0.21, 0.31}
\setres{popL}{CPT_types5}{types}{0.32: (0.98, 0.48, 1.04); 0.26: (0.83, 0.83, 1.19); 0.17: (0.98, 0.43, 1.77); 0.13: (0.88, 0.33, 0.75); 0.12: (0.78, 0.18, 1.48)}
\setres{popL}{CPT_uniform}{H_F}{29.73}
\setres{popL}{CPT_uniform}{H_L}{47.83}
\setres{popL}{CPT_uniform}{ess}{8{,}000}
\setres{popL}{CPT_uniform}{kappa_crps}{0.918}
\setres{popL}{CPT_uniform}{kappa_crps_se}{0.012}
\setres{popL}{CPT_uniform}{kappa_draw}{0.527}
\setres{popL}{CPT_uniform}{kappa_draw_se}{0.020}
\setres{popL}{CPT_uniform}{kappa_mean}{0.960}
\setres{popL}{CPT_uniform}{kappa_mean_se}{0.013}
\setres{popL}{CPT_uniform}{r_stoch}{0.701}
\setres{popL}{CPT_uniform}{s_crps}{2.50}
\setres{popL}{CPT_uniform}{theta_mean}{0.50, 0.50, 1.55}
\setres{popL}{CPT_uniform}{theta_sd}{0.29, 0.29, 0.84}
\setres{popL}{DA_fitted}{H_F}{27.20}
\setres{popL}{DA_fitted}{H_L}{47.83}
\setres{popL}{DA_fitted}{ess}{15}
\setres{popL}{DA_fitted}{kappa_crps}{0.662}
\setres{popL}{DA_fitted}{kappa_crps_se}{0.055}
\setres{popL}{DA_fitted}{kappa_draw}{0.752}
\setres{popL}{DA_fitted}{kappa_draw_se}{0.046}
\setres{popL}{DA_fitted}{kappa_mean}{0.826}
\setres{popL}{DA_fitted}{kappa_mean_se}{0.040}
\setres{popL}{DA_fitted}{r_stoch}{0.747}
\setres{popL}{DA_fitted}{s_crps}{3.14}
\setres{popL}{DA_fitted}{theta_mean}{0.85, 0.54}
\setres{popL}{DA_fitted}{theta_sd}{0.26, 1.03}
\setres{popL}{DA_from_zurich03}{H_F}{27.15}
\setres{popL}{DA_from_zurich03}{H_L}{47.83}
\setres{popL}{DA_from_zurich03}{ess}{8}
\setres{popL}{DA_from_zurich03}{kappa_crps}{0.701}
\setres{popL}{DA_from_zurich03}{kappa_crps_se}{0.049}
\setres{popL}{DA_from_zurich03}{kappa_draw}{0.685}
\setres{popL}{DA_from_zurich03}{kappa_draw_se}{0.047}
\setres{popL}{DA_from_zurich03}{kappa_mean}{0.858}
\setres{popL}{DA_from_zurich03}{kappa_mean_se}{0.035}
\setres{popL}{DA_from_zurich03}{r_stoch}{0.748}
\setres{popL}{DA_from_zurich03}{s_crps}{3.04}
\setres{popL}{DA_from_zurich03}{theta_mean}{0.87, 0.61}
\setres{popL}{DA_from_zurich03}{theta_sd}{0.27, 1.21}
\setres{popL}{DA_from_zurich03_types1}{H_F}{24.91}
\setres{popL}{DA_from_zurich03_types1}{H_L}{47.83}
\setres{popL}{DA_from_zurich03_types1}{ess}{1}
\setres{popL}{DA_from_zurich03_types1}{kappa_crps}{0.568}
\setres{popL}{DA_from_zurich03_types1}{kappa_crps_se}{0.066}
\setres{popL}{DA_from_zurich03_types1}{kappa_draw}{0.802}
\setres{popL}{DA_from_zurich03_types1}{kappa_draw_se}{0.044}
\setres{popL}{DA_from_zurich03_types1}{kappa_mean}{0.802}
\setres{popL}{DA_from_zurich03_types1}{kappa_mean_se}{0.044}
\setres{popL}{DA_from_zurich03_types1}{r_stoch}{0.783}
\setres{popL}{DA_from_zurich03_types1}{s_crps}{3.37}
\setres{popL}{DA_from_zurich03_types1}{shares}{1.000}
\setres{popL}{DA_from_zurich03_types1}{theta_mean}{0.99, 0.15}
\setres{popL}{DA_from_zurich03_types1}{theta_sd}{0.00, 0.00}
\setres{popL}{DA_from_zurich03_types1}{types}{1.00: (0.99, 0.15)}
\setres{popL}{DA_from_zurich03_types2}{H_F}{25.50}
\setres{popL}{DA_from_zurich03_types2}{H_L}{47.83}
\setres{popL}{DA_from_zurich03_types2}{ess}{2}
\setres{popL}{DA_from_zurich03_types2}{kappa_crps}{0.655}
\setres{popL}{DA_from_zurich03_types2}{kappa_crps_se}{0.055}
\setres{popL}{DA_from_zurich03_types2}{kappa_draw}{0.752}
\setres{popL}{DA_from_zurich03_types2}{kappa_draw_se}{0.046}
\setres{popL}{DA_from_zurich03_types2}{kappa_mean}{0.824}
\setres{popL}{DA_from_zurich03_types2}{kappa_mean_se}{0.041}
\setres{popL}{DA_from_zurich03_types2}{r_stoch}{0.774}
\setres{popL}{DA_from_zurich03_types2}{s_crps}{3.15}
\setres{popL}{DA_from_zurich03_types2}{shares}{0.762, 0.238}
\setres{popL}{DA_from_zurich03_types2}{theta_mean}{0.99, 0.59}
\setres{popL}{DA_from_zurich03_types2}{theta_sd}{0.00, 1.32}
\setres{popL}{DA_from_zurich03_types2}{types}{0.76: (0.99, -0.15); 0.24: (0.99, 2.95)}
\setres{popL}{DA_from_zurich03_types3}{H_F}{25.55}
\setres{popL}{DA_from_zurich03_types3}{H_L}{47.83}
\setres{popL}{DA_from_zurich03_types3}{ess}{2}
\setres{popL}{DA_from_zurich03_types3}{kappa_crps}{0.691}
\setres{popL}{DA_from_zurich03_types3}{kappa_crps_se}{0.051}
\setres{popL}{DA_from_zurich03_types3}{kappa_draw}{0.685}
\setres{popL}{DA_from_zurich03_types3}{kappa_draw_se}{0.048}
\setres{popL}{DA_from_zurich03_types3}{kappa_mean}{0.858}
\setres{popL}{DA_from_zurich03_types3}{kappa_mean_se}{0.035}
\setres{popL}{DA_from_zurich03_types3}{r_stoch}{0.774}
\setres{popL}{DA_from_zurich03_types3}{s_crps}{3.06}
\setres{popL}{DA_from_zurich03_types3}{shares}{0.733, 0.223, 0.045}
\setres{popL}{DA_from_zurich03_types3}{theta_mean}{0.99, 0.58}
\setres{popL}{DA_from_zurich03_types3}{theta_sd}{0.00, 1.28}
\setres{popL}{DA_from_zurich03_types3}{types}{0.73: (0.99, -0.05); 0.22: (0.99, 2.95); 0.04: (0.99, -0.95)}
\setres{popL}{DA_from_zurich03_types5}{H_F}{26.29}
\setres{popL}{DA_from_zurich03_types5}{H_L}{47.83}
\setres{popL}{DA_from_zurich03_types5}{ess}{3}
\setres{popL}{DA_from_zurich03_types5}{kappa_crps}{0.700}
\setres{popL}{DA_from_zurich03_types5}{kappa_crps_se}{0.050}
\setres{popL}{DA_from_zurich03_types5}{kappa_draw}{0.697}
\setres{popL}{DA_from_zurich03_types5}{kappa_draw_se}{0.047}
\setres{popL}{DA_from_zurich03_types5}{kappa_mean}{0.856}
\setres{popL}{DA_from_zurich03_types5}{kappa_mean_se}{0.036}
\setres{popL}{DA_from_zurich03_types5}{r_stoch}{0.762}
\setres{popL}{DA_from_zurich03_types5}{s_crps}{3.04}
\setres{popL}{DA_from_zurich03_types5}{shares}{0.445, 0.202, 0.171, 0.143, 0.039}
\setres{popL}{DA_from_zurich03_types5}{theta_mean}{0.95, 0.51}
\setres{popL}{DA_from_zurich03_types5}{theta_sd}{0.07, 1.13}
\setres{popL}{DA_from_zurich03_types5}{types}{0.44: (0.99, 0.05); 0.20: (0.99, -0.45); 0.17: (0.79, 1.15); 0.14: (0.99, 2.95); 0.04: (0.99, -0.95)}
\setres{popL}{DA_literature}{H_F}{27.22}
\setres{popL}{DA_literature}{H_L}{47.83}
\setres{popL}{DA_literature}{ess}{439}
\setres{popL}{DA_literature}{kappa_crps}{0.576}
\setres{popL}{DA_literature}{kappa_crps_se}{0.060}
\setres{popL}{DA_literature}{kappa_draw}{0.768}
\setres{popL}{DA_literature}{kappa_draw_se}{0.043}
\setres{popL}{DA_literature}{kappa_mean}{0.810}
\setres{popL}{DA_literature}{kappa_mean_se}{0.040}
\setres{popL}{DA_literature}{r_stoch}{0.747}
\setres{popL}{DA_literature}{s_crps}{3.35}
\setres{popL}{DA_literature}{theta_mean}{0.42, 0.30}
\setres{popL}{DA_literature}{theta_sd}{0.19, 0.45}
\setres{popL}{DA_types1}{H_F}{24.93}
\setres{popL}{DA_types1}{H_L}{47.83}
\setres{popL}{DA_types1}{ess}{1}
\setres{popL}{DA_types1}{kappa_crps}{0.566}
\setres{popL}{DA_types1}{kappa_crps_se}{0.066}
\setres{popL}{DA_types1}{kappa_draw}{0.801}
\setres{popL}{DA_types1}{kappa_draw_se}{0.044}
\setres{popL}{DA_types1}{kappa_mean}{0.801}
\setres{popL}{DA_types1}{kappa_mean_se}{0.044}
\setres{popL}{DA_types1}{r_stoch}{0.783}
\setres{popL}{DA_types1}{s_crps}{3.37}
\setres{popL}{DA_types1}{shares}{1.000}
\setres{popL}{DA_types1}{theta_mean}{0.99, 0.15}
\setres{popL}{DA_types1}{theta_sd}{0.00, 0.00}
\setres{popL}{DA_types1}{types}{1.00: (0.99, 0.15)}
\setres{popL}{DA_types2}{H_F}{25.50}
\setres{popL}{DA_types2}{H_L}{47.83}
\setres{popL}{DA_types2}{ess}{2}
\setres{popL}{DA_types2}{kappa_crps}{0.635}
\setres{popL}{DA_types2}{kappa_crps_se}{0.058}
\setres{popL}{DA_types2}{kappa_draw}{0.776}
\setres{popL}{DA_types2}{kappa_draw_se}{0.045}
\setres{popL}{DA_types2}{kappa_mean}{0.813}
\setres{popL}{DA_types2}{kappa_mean_se}{0.042}
\setres{popL}{DA_types2}{r_stoch}{0.774}
\setres{popL}{DA_types2}{s_crps}{3.20}
\setres{popL}{DA_types2}{shares}{0.681, 0.319}
\setres{popL}{DA_types2}{theta_mean}{0.90, 0.42}
\setres{popL}{DA_types2}{theta_sd}{0.13, 0.84}
\setres{popL}{DA_types2}{types}{0.68: (0.99, -0.15); 0.32: (0.72, 1.65)}
\setres{popL}{DA_types3}{H_F}{25.56}
\setres{popL}{DA_types3}{H_L}{47.83}
\setres{popL}{DA_types3}{ess}{2}
\setres{popL}{DA_types3}{kappa_crps}{0.652}
\setres{popL}{DA_types3}{kappa_crps_se}{0.056}
\setres{popL}{DA_types3}{kappa_draw}{0.763}
\setres{popL}{DA_types3}{kappa_draw_se}{0.046}
\setres{popL}{DA_types3}{kappa_mean}{0.819}
\setres{popL}{DA_types3}{kappa_mean_se}{0.042}
\setres{popL}{DA_types3}{r_stoch}{0.773}
\setres{popL}{DA_types3}{s_crps}{3.16}
\setres{popL}{DA_types3}{shares}{0.695, 0.280, 0.025}
\setres{popL}{DA_types3}{theta_mean}{0.88, 0.52}
\setres{popL}{DA_types3}{theta_sd}{0.17, 0.96}
\setres{popL}{DA_types3}{types}{0.69: (0.99, -0.05); 0.28: (0.62, 2.05); 0.03: (0.99, -0.85)}
\setres{popL}{DA_types5}{H_F}{26.23}
\setres{popL}{DA_types5}{H_L}{47.83}
\setres{popL}{DA_types5}{ess}{3}
\setres{popL}{DA_types5}{kappa_crps}{0.660}
\setres{popL}{DA_types5}{kappa_crps_se}{0.055}
\setres{popL}{DA_types5}{kappa_draw}{0.758}
\setres{popL}{DA_types5}{kappa_draw_se}{0.046}
\setres{popL}{DA_types5}{kappa_mean}{0.823}
\setres{popL}{DA_types5}{kappa_mean_se}{0.041}
\setres{popL}{DA_types5}{r_stoch}{0.763}
\setres{popL}{DA_types5}{s_crps}{3.14}
\setres{popL}{DA_types5}{shares}{0.360, 0.337, 0.218, 0.060, 0.025}
\setres{popL}{DA_types5}{theta_mean}{0.91, 0.46}
\setres{popL}{DA_types5}{theta_sd}{0.14, 0.94}
\setres{popL}{DA_types5}{types}{0.36: (0.99, 0.15); 0.34: (0.99, -0.25); 0.22: (0.64, 1.55); 0.06: (0.99, 2.95); 0.03: (0.99, -0.85)}
\setres{popL}{DA_uniform}{H_F}{27.95}
\setres{popL}{DA_uniform}{H_L}{47.83}
\setres{popL}{DA_uniform}{ess}{1{,}600}
\setres{popL}{DA_uniform}{kappa_crps}{0.681}
\setres{popL}{DA_uniform}{kappa_crps_se}{0.047}
\setres{popL}{DA_uniform}{kappa_draw}{0.657}
\setres{popL}{DA_uniform}{kappa_draw_se}{0.049}
\setres{popL}{DA_uniform}{kappa_mean}{0.865}
\setres{popL}{DA_uniform}{kappa_mean_se}{0.034}
\setres{popL}{DA_uniform}{r_stoch}{0.734}
\setres{popL}{DA_uniform}{s_crps}{3.09}
\setres{popL}{DA_uniform}{theta_mean}{0.50, 1.00}
\setres{popL}{DA_uniform}{theta_sd}{0.29, 1.15}
\setres{popL}{EU_fitted}{H_F}{25.55}
\setres{popL}{EU_fitted}{H_L}{47.83}
\setres{popL}{EU_fitted}{ess}{2}
\setres{popL}{EU_fitted}{kappa_crps}{0.565}
\setres{popL}{EU_fitted}{kappa_crps_se}{0.065}
\setres{popL}{EU_fitted}{kappa_draw}{0.803}
\setres{popL}{EU_fitted}{kappa_draw_se}{0.042}
\setres{popL}{EU_fitted}{kappa_mean}{0.806}
\setres{popL}{EU_fitted}{kappa_mean_se}{0.042}
\setres{popL}{EU_fitted}{r_stoch}{0.774}
\setres{popL}{EU_fitted}{s_crps}{3.37}
\setres{popL}{EU_fitted}{theta_mean}{0.90}
\setres{popL}{EU_fitted}{theta_sd}{0.20}
\setres{popL}{EU_from_zurich03}{H_F}{25.58}
\setres{popL}{EU_from_zurich03}{H_L}{47.83}
\setres{popL}{EU_from_zurich03}{ess}{2}
\setres{popL}{EU_from_zurich03}{kappa_crps}{0.569}
\setres{popL}{EU_from_zurich03}{kappa_crps_se}{0.064}
\setres{popL}{EU_from_zurich03}{kappa_draw}{0.800}
\setres{popL}{EU_from_zurich03}{kappa_draw_se}{0.042}
\setres{popL}{EU_from_zurich03}{kappa_mean}{0.806}
\setres{popL}{EU_from_zurich03}{kappa_mean_se}{0.041}
\setres{popL}{EU_from_zurich03}{r_stoch}{0.773}
\setres{popL}{EU_from_zurich03}{s_crps}{3.37}
\setres{popL}{EU_from_zurich03}{theta_mean}{0.87}
\setres{popL}{EU_from_zurich03}{theta_sd}{0.27}
\setres{popL}{EU_from_zurich03_types1}{H_F}{24.91}
\setres{popL}{EU_from_zurich03_types1}{H_L}{47.83}
\setres{popL}{EU_from_zurich03_types1}{ess}{1}
\setres{popL}{EU_from_zurich03_types1}{kappa_crps}{0.568}
\setres{popL}{EU_from_zurich03_types1}{kappa_crps_se}{0.065}
\setres{popL}{EU_from_zurich03_types1}{kappa_draw}{0.806}
\setres{popL}{EU_from_zurich03_types1}{kappa_draw_se}{0.042}
\setres{popL}{EU_from_zurich03_types1}{kappa_mean}{0.806}
\setres{popL}{EU_from_zurich03_types1}{kappa_mean_se}{0.042}
\setres{popL}{EU_from_zurich03_types1}{r_stoch}{0.783}
\setres{popL}{EU_from_zurich03_types1}{s_crps}{3.37}
\setres{popL}{EU_from_zurich03_types1}{shares}{1.000}
\setres{popL}{EU_from_zurich03_types1}{theta_mean}{1.00}
\setres{popL}{EU_from_zurich03_types1}{theta_sd}{0.00}
\setres{popL}{EU_from_zurich03_types1}{types}{1.00: (1.00)}
\setres{popL}{EU_from_zurich03_types2}{H_F}{25.26}
\setres{popL}{EU_from_zurich03_types2}{H_L}{47.83}
\setres{popL}{EU_from_zurich03_types2}{ess}{1}
\setres{popL}{EU_from_zurich03_types2}{kappa_crps}{0.573}
\setres{popL}{EU_from_zurich03_types2}{kappa_crps_se}{0.064}
\setres{popL}{EU_from_zurich03_types2}{kappa_draw}{0.801}
\setres{popL}{EU_from_zurich03_types2}{kappa_draw_se}{0.042}
\setres{popL}{EU_from_zurich03_types2}{kappa_mean}{0.807}
\setres{popL}{EU_from_zurich03_types2}{kappa_mean_se}{0.042}
\setres{popL}{EU_from_zurich03_types2}{r_stoch}{0.778}
\setres{popL}{EU_from_zurich03_types2}{s_crps}{3.35}
\setres{popL}{EU_from_zurich03_types2}{shares}{0.904, 0.096}
\setres{popL}{EU_from_zurich03_types2}{theta_mean}{0.90}
\setres{popL}{EU_from_zurich03_types2}{theta_sd}{0.29}
\setres{popL}{EU_from_zurich03_types2}{types}{0.90: (1.00); 0.10: (0.01)}
\setres{popL}{EU_from_zurich03_types3}{H_F}{25.45}
\setres{popL}{EU_from_zurich03_types3}{H_L}{47.83}
\setres{popL}{EU_from_zurich03_types3}{ess}{2}
\setres{popL}{EU_from_zurich03_types3}{kappa_crps}{0.571}
\setres{popL}{EU_from_zurich03_types3}{kappa_crps_se}{0.064}
\setres{popL}{EU_from_zurich03_types3}{kappa_draw}{0.801}
\setres{popL}{EU_from_zurich03_types3}{kappa_draw_se}{0.042}
\setres{popL}{EU_from_zurich03_types3}{kappa_mean}{0.806}
\setres{popL}{EU_from_zurich03_types3}{kappa_mean_se}{0.042}
\setres{popL}{EU_from_zurich03_types3}{r_stoch}{0.775}
\setres{popL}{EU_from_zurich03_types3}{s_crps}{3.36}
\setres{popL}{EU_from_zurich03_types3}{shares}{0.798, 0.139, 0.063}
\setres{popL}{EU_from_zurich03_types3}{theta_mean}{0.88}
\setres{popL}{EU_from_zurich03_types3}{theta_sd}{0.27}
\setres{popL}{EU_from_zurich03_types3}{types}{0.80: (1.00); 0.14: (0.57); 0.06: (0.01)}
\setres{popL}{EU_from_zurich03_types5}{H_F}{25.73}
\setres{popL}{EU_from_zurich03_types5}{H_L}{47.83}
\setres{popL}{EU_from_zurich03_types5}{ess}{2}
\setres{popL}{EU_from_zurich03_types5}{kappa_crps}{0.566}
\setres{popL}{EU_from_zurich03_types5}{kappa_crps_se}{0.064}
\setres{popL}{EU_from_zurich03_types5}{kappa_draw}{0.798}
\setres{popL}{EU_from_zurich03_types5}{kappa_draw_se}{0.042}
\setres{popL}{EU_from_zurich03_types5}{kappa_mean}{0.806}
\setres{popL}{EU_from_zurich03_types5}{kappa_mean_se}{0.041}
\setres{popL}{EU_from_zurich03_types5}{r_stoch}{0.771}
\setres{popL}{EU_from_zurich03_types5}{s_crps}{3.37}
\setres{popL}{EU_from_zurich03_types5}{shares}{0.439, 0.081, 0.071, 0.062}
\setres{popL}{EU_from_zurich03_types5}{theta_mean}{0.81}
\setres{popL}{EU_from_zurich03_types5}{theta_sd}{0.31}
\setres{popL}{EU_from_zurich03_types5}{types}{0.44: (1.00); 0.08: (0.66); 0.07: (0.50); 0.06: (0.01)}
\setres{popL}{EU_literature}{H_F}{25.82}
\setres{popL}{EU_literature}{H_L}{47.83}
\setres{popL}{EU_literature}{ess}{145}
\setres{popL}{EU_literature}{kappa_crps}{0.517}
\setres{popL}{EU_literature}{kappa_crps_se}{0.066}
\setres{popL}{EU_literature}{kappa_draw}{0.796}
\setres{popL}{EU_literature}{kappa_draw_se}{0.041}
\setres{popL}{EU_literature}{kappa_mean}{0.799}
\setres{popL}{EU_literature}{kappa_mean_se}{0.041}
\setres{popL}{EU_literature}{r_stoch}{0.769}
\setres{popL}{EU_literature}{s_crps}{3.49}
\setres{popL}{EU_literature}{theta_mean}{0.67}
\setres{popL}{EU_literature}{theta_sd}{0.20}
\setres{popL}{EU_types1}{H_F}{24.93}
\setres{popL}{EU_types1}{H_L}{47.83}
\setres{popL}{EU_types1}{ess}{1}
\setres{popL}{EU_types1}{kappa_crps}{0.569}
\setres{popL}{EU_types1}{kappa_crps_se}{0.065}
\setres{popL}{EU_types1}{kappa_draw}{0.806}
\setres{popL}{EU_types1}{kappa_draw_se}{0.042}
\setres{popL}{EU_types1}{kappa_mean}{0.806}
\setres{popL}{EU_types1}{kappa_mean_se}{0.042}
\setres{popL}{EU_types1}{r_stoch}{0.783}
\setres{popL}{EU_types1}{s_crps}{3.37}
\setres{popL}{EU_types1}{shares}{1.000}
\setres{popL}{EU_types1}{theta_mean}{1.00}
\setres{popL}{EU_types1}{theta_sd}{0.00}
\setres{popL}{EU_types1}{types}{1.00: (1.00)}
\setres{popL}{EU_types2}{H_F}{25.29}
\setres{popL}{EU_types2}{H_L}{47.83}
\setres{popL}{EU_types2}{ess}{1}
\setres{popL}{EU_types2}{kappa_crps}{0.568}
\setres{popL}{EU_types2}{kappa_crps_se}{0.065}
\setres{popL}{EU_types2}{kappa_draw}{0.803}
\setres{popL}{EU_types2}{kappa_draw_se}{0.042}
\setres{popL}{EU_types2}{kappa_mean}{0.806}
\setres{popL}{EU_types2}{kappa_mean_se}{0.042}
\setres{popL}{EU_types2}{r_stoch}{0.777}
\setres{popL}{EU_types2}{s_crps}{3.37}
\setres{popL}{EU_types2}{shares}{0.859, 0.141}
\setres{popL}{EU_types2}{theta_mean}{0.92}
\setres{popL}{EU_types2}{theta_sd}{0.19}
\setres{popL}{EU_types2}{types}{0.86: (1.00); 0.14: (0.46)}
\setres{popL}{EU_types3}{H_F}{25.40}
\setres{popL}{EU_types3}{H_L}{47.83}
\setres{popL}{EU_types3}{ess}{2}
\setres{popL}{EU_types3}{kappa_crps}{0.570}
\setres{popL}{EU_types3}{kappa_crps_se}{0.065}
\setres{popL}{EU_types3}{kappa_draw}{0.802}
\setres{popL}{EU_types3}{kappa_draw_se}{0.042}
\setres{popL}{EU_types3}{kappa_mean}{0.806}
\setres{popL}{EU_types3}{kappa_mean_se}{0.042}
\setres{popL}{EU_types3}{r_stoch}{0.776}
\setres{popL}{EU_types3}{s_crps}{3.36}
\setres{popL}{EU_types3}{shares}{0.800, 0.159, 0.041}
\setres{popL}{EU_types3}{theta_mean}{0.90}
\setres{popL}{EU_types3}{theta_sd}{0.23}
\setres{popL}{EU_types3}{types}{0.80: (1.00); 0.16: (0.62); 0.04: (0.01)}
\setres{popL}{EU_types5}{H_F}{25.53}
\setres{popL}{EU_types5}{H_L}{47.83}
\setres{popL}{EU_types5}{ess}{2}
\setres{popL}{EU_types5}{kappa_crps}{0.563}
\setres{popL}{EU_types5}{kappa_crps_se}{0.065}
\setres{popL}{EU_types5}{kappa_draw}{0.802}
\setres{popL}{EU_types5}{kappa_draw_se}{0.042}
\setres{popL}{EU_types5}{kappa_mean}{0.806}
\setres{popL}{EU_types5}{kappa_mean_se}{0.042}
\setres{popL}{EU_types5}{r_stoch}{0.774}
\setres{popL}{EU_types5}{s_crps}{3.38}
\setres{popL}{EU_types5}{shares}{0.786, 0.106, 0.058, 0.041, 0.009}
\setres{popL}{EU_types5}{theta_mean}{0.90}
\setres{popL}{EU_types5}{theta_sd}{0.20}
\setres{popL}{EU_types5}{types}{0.79: (1.00); 0.11: (0.65); 0.06: (0.38); 0.04: (0.69); 0.01: (0.01)}
\setres{popL}{EU_uniform}{H_F}{26.14}
\setres{popL}{EU_uniform}{H_L}{47.83}
\setres{popL}{EU_uniform}{ess}{200}
\setres{popL}{EU_uniform}{kappa_crps}{0.478}
\setres{popL}{EU_uniform}{kappa_crps_se}{0.067}
\setres{popL}{EU_uniform}{kappa_draw}{0.786}
\setres{popL}{EU_uniform}{kappa_draw_se}{0.041}
\setres{popL}{EU_uniform}{kappa_mean}{0.793}
\setres{popL}{EU_uniform}{kappa_mean_se}{0.040}
\setres{popL}{EU_uniform}{r_stoch}{0.764}
\setres{popL}{EU_uniform}{s_crps}{3.59}
\setres{popL}{EU_uniform}{theta_mean}{0.51}
\setres{popL}{EU_uniform}{theta_sd}{0.29}
\setres{popL}{MIX_from_zurich03_mix_eu_cpt}{H_F}{25.47}
\setres{popL}{MIX_from_zurich03_mix_eu_cpt}{H_L}{47.83}
\setres{popL}{MIX_from_zurich03_mix_eu_cpt}{ess}{2}
\setres{popL}{MIX_from_zurich03_mix_eu_cpt}{eu_alpha}{0.92}
\setres{popL}{MIX_from_zurich03_mix_eu_cpt}{eu_share}{0.26}
\setres{popL}{MIX_from_zurich03_mix_eu_cpt}{kappa_crps}{0.900}
\setres{popL}{MIX_from_zurich03_mix_eu_cpt}{kappa_crps_se}{0.012}
\setres{popL}{MIX_from_zurich03_mix_eu_cpt}{kappa_draw}{0.911}
\setres{popL}{MIX_from_zurich03_mix_eu_cpt}{kappa_draw_se}{0.002}
\setres{popL}{MIX_from_zurich03_mix_eu_cpt}{kappa_mean}{0.997}
\setres{popL}{MIX_from_zurich03_mix_eu_cpt}{kappa_mean_se}{0.004}
\setres{popL}{MIX_from_zurich03_mix_eu_cpt}{r_stoch}{0.775}
\setres{popL}{MIX_from_zurich03_mix_eu_cpt}{s_crps}{2.55}
\setres{popL}{MIX_from_zurich03_mix_eu_cpt}{shares}{0.259, 0.741}
\setres{popL}{MIX_from_zurich03_mix_eu_cpt}{types}{0.74: CPT (0.98, 0.33, 1.19); 0.26: EU (0.92)}
\setres{popL}{MIX_from_zurich03_mix_eu_cpt2}{H_F}{25.94}
\setres{popL}{MIX_from_zurich03_mix_eu_cpt2}{H_L}{47.83}
\setres{popL}{MIX_from_zurich03_mix_eu_cpt2}{ess}{3}
\setres{popL}{MIX_from_zurich03_mix_eu_cpt2}{eu_alpha}{1.00}
\setres{popL}{MIX_from_zurich03_mix_eu_cpt2}{eu_share}{0.26}
\setres{popL}{MIX_from_zurich03_mix_eu_cpt2}{kappa_crps}{0.949}
\setres{popL}{MIX_from_zurich03_mix_eu_cpt2}{kappa_crps_se}{0.010}
\setres{popL}{MIX_from_zurich03_mix_eu_cpt2}{kappa_draw}{0.832}
\setres{popL}{MIX_from_zurich03_mix_eu_cpt2}{kappa_draw_se}{0.006}
\setres{popL}{MIX_from_zurich03_mix_eu_cpt2}{kappa_mean}{0.992}
\setres{popL}{MIX_from_zurich03_mix_eu_cpt2}{kappa_mean_se}{0.006}
\setres{popL}{MIX_from_zurich03_mix_eu_cpt2}{r_stoch}{0.768}
\setres{popL}{MIX_from_zurich03_mix_eu_cpt2}{s_crps}{2.43}
\setres{popL}{MIX_from_zurich03_mix_eu_cpt2}{shares}{0.260, 0.298, 0.442}
\setres{popL}{MIX_from_zurich03_mix_eu_cpt2}{types}{0.44: CPT (0.83, 0.33, 0.90); 0.30: CPT (0.98, 0.33, 1.91); 0.26: EU (1.00)}
\setres{popL}{MIX_mix_eu_cpt}{H_F}{25.52}
\setres{popL}{MIX_mix_eu_cpt}{H_L}{47.83}
\setres{popL}{MIX_mix_eu_cpt}{ess}{2}
\setres{popL}{MIX_mix_eu_cpt}{eu_alpha}{1.00}
\setres{popL}{MIX_mix_eu_cpt}{eu_share}{0.28}
\setres{popL}{MIX_mix_eu_cpt}{kappa_crps}{0.929}
\setres{popL}{MIX_mix_eu_cpt}{kappa_crps_se}{0.009}
\setres{popL}{MIX_mix_eu_cpt}{kappa_draw}{0.928}
\setres{popL}{MIX_mix_eu_cpt}{kappa_draw_se}{0.005}
\setres{popL}{MIX_mix_eu_cpt}{kappa_mean}{0.999}
\setres{popL}{MIX_mix_eu_cpt}{kappa_mean_se}{0.002}
\setres{popL}{MIX_mix_eu_cpt}{r_stoch}{0.774}
\setres{popL}{MIX_mix_eu_cpt}{s_crps}{2.48}
\setres{popL}{MIX_mix_eu_cpt}{shares}{0.281, 0.719}
\setres{popL}{MIX_mix_eu_cpt}{types}{0.72: CPT (0.93, 0.38, 1.19); 0.28: EU (1.00)}
\setres{popL}{MIX_mix_eu_cpt2}{H_F}{26.04}
\setres{popL}{MIX_mix_eu_cpt2}{H_L}{47.83}
\setres{popL}{MIX_mix_eu_cpt2}{ess}{3}
\setres{popL}{MIX_mix_eu_cpt2}{eu_alpha}{1.00}
\setres{popL}{MIX_mix_eu_cpt2}{eu_share}{0.27}
\setres{popL}{MIX_mix_eu_cpt2}{kappa_crps}{0.959}
\setres{popL}{MIX_mix_eu_cpt2}{kappa_crps_se}{0.007}
\setres{popL}{MIX_mix_eu_cpt2}{kappa_draw}{0.892}
\setres{popL}{MIX_mix_eu_cpt2}{kappa_draw_se}{0.007}
\setres{popL}{MIX_mix_eu_cpt2}{kappa_mean}{0.998}
\setres{popL}{MIX_mix_eu_cpt2}{kappa_mean_se}{0.002}
\setres{popL}{MIX_mix_eu_cpt2}{r_stoch}{0.766}
\setres{popL}{MIX_mix_eu_cpt2}{s_crps}{2.40}
\setres{popL}{MIX_mix_eu_cpt2}{shares}{0.269, 0.332, 0.399}
\setres{popL}{MIX_mix_eu_cpt2}{types}{0.40: CPT (0.93, 0.38, 0.90); 0.33: CPT (0.93, 0.38, 1.62); 0.27: EU (1.00)}
\setres{popL}{reference}{CPT_noise_c}{0.052}
\setres{popL}{reference}{DA_noise_c}{0.052}
\setres{popL}{reference}{EU_noise_c}{0.052}
\setres{popL}{reference}{MIX_noise_c}{0.052}
\setres{popL}{reference}{mse_base}{129.04}
\setres{popL}{reference}{mse_best}{30.21}
\setres{popL}{reference}{n_lotteries}{15}
\setres{popL}{reference}{n_subjects}{118}
\setres{popL}{reference}{s_crps_base}{4.77}
\setres{popL}{reference}{s_crps_best}{2.30}
\setres{popL}{reference}{target}{Zurich2006\_shared\_losses}
\setres{popZ03}{CPT_fitted}{H_F}{50.46}
\setres{popZ03}{CPT_fitted}{H_L}{81.16}
\setres{popZ03}{CPT_fitted}{ess}{57}
\setres{popZ03}{CPT_fitted}{kappa_crps}{0.910}
\setres{popZ03}{CPT_fitted}{kappa_crps_se}{0.012}
\setres{popZ03}{CPT_fitted}{kappa_draw}{0.599}
\setres{popZ03}{CPT_fitted}{kappa_draw_se}{0.024}
\setres{popZ03}{CPT_fitted}{kappa_mean}{0.944}
\setres{popZ03}{CPT_fitted}{kappa_mean_se}{0.018}
\setres{popZ03}{CPT_fitted}{r_stoch}{0.707}
\setres{popZ03}{CPT_fitted}{s_crps}{3.24}
\setres{popZ03}{CPT_fitted}{theta_mean}{0.92, 0.45, 1.04}
\setres{popZ03}{CPT_fitted}{theta_sd}{0.13, 0.20, 0.49}
\setres{popZ03}{CPT_from_beijing}{H_F}{50.77}
\setres{popZ03}{CPT_from_beijing}{H_L}{81.16}
\setres{popZ03}{CPT_from_beijing}{ess}{59}
\setres{popZ03}{CPT_from_beijing}{kappa_crps}{0.415}
\setres{popZ03}{CPT_from_beijing}{kappa_crps_se}{0.025}
\setres{popZ03}{CPT_from_beijing}{kappa_draw}{-0.788}
\setres{popZ03}{CPT_from_beijing}{kappa_draw_se}{0.158}
\setres{popZ03}{CPT_from_beijing}{kappa_mean}{0.452}
\setres{popZ03}{CPT_from_beijing}{kappa_mean_se}{0.071}
\setres{popZ03}{CPT_from_beijing}{r_stoch}{0.703}
\setres{popZ03}{CPT_from_beijing}{s_crps}{4.25}
\setres{popZ03}{CPT_from_beijing}{theta_mean}{0.50, 0.34, 1.40}
\setres{popZ03}{CPT_from_beijing}{theta_sd}{0.43, 0.24, 0.60}
\setres{popZ03}{CPT_from_countries}{H_F}{50.52}
\setres{popZ03}{CPT_from_countries}{H_L}{81.16}
\setres{popZ03}{CPT_from_countries}{ess}{41}
\setres{popZ03}{CPT_from_countries}{kappa_crps}{0.864}
\setres{popZ03}{CPT_from_countries}{kappa_crps_se}{0.020}
\setres{popZ03}{CPT_from_countries}{kappa_draw}{0.220}
\setres{popZ03}{CPT_from_countries}{kappa_draw_se}{0.057}
\setres{popZ03}{CPT_from_countries}{kappa_mean}{0.879}
\setres{popZ03}{CPT_from_countries}{kappa_mean_se}{0.025}
\setres{popZ03}{CPT_from_countries}{r_stoch}{0.706}
\setres{popZ03}{CPT_from_countries}{s_crps}{3.34}
\setres{popZ03}{CPT_from_countries}{theta_mean}{0.89, 0.66, 1.35}
\setres{popZ03}{CPT_from_countries}{theta_sd}{0.19, 0.33, 0.89}
\setres{popZ03}{CPT_from_zurich06}{H_F}{50.44}
\setres{popZ03}{CPT_from_zurich06}{H_L}{81.16}
\setres{popZ03}{CPT_from_zurich06}{ess}{54}
\setres{popZ03}{CPT_from_zurich06}{kappa_crps}{0.929}
\setres{popZ03}{CPT_from_zurich06}{kappa_crps_se}{0.009}
\setres{popZ03}{CPT_from_zurich06}{kappa_draw}{0.707}
\setres{popZ03}{CPT_from_zurich06}{kappa_draw_se}{0.025}
\setres{popZ03}{CPT_from_zurich06}{kappa_mean}{0.983}
\setres{popZ03}{CPT_from_zurich06}{kappa_mean_se}{0.009}
\setres{popZ03}{CPT_from_zurich06}{r_stoch}{0.707}
\setres{popZ03}{CPT_from_zurich06}{s_crps}{3.20}
\setres{popZ03}{CPT_from_zurich06}{theta_mean}{0.87, 0.49, 0.95}
\setres{popZ03}{CPT_from_zurich06}{theta_sd}{0.15, 0.22, 0.32}
\setres{popZ03}{CPT_literature}{H_F}{50.37}
\setres{popZ03}{CPT_literature}{H_L}{81.16}
\setres{popZ03}{CPT_literature}{ess}{732}
\setres{popZ03}{CPT_literature}{kappa_crps}{0.735}
\setres{popZ03}{CPT_literature}{kappa_crps_se}{0.035}
\setres{popZ03}{CPT_literature}{kappa_draw}{0.642}
\setres{popZ03}{CPT_literature}{kappa_draw_se}{0.052}
\setres{popZ03}{CPT_literature}{kappa_mean}{0.852}
\setres{popZ03}{CPT_literature}{kappa_mean_se}{0.036}
\setres{popZ03}{CPT_literature}{r_stoch}{0.708}
\setres{popZ03}{CPT_literature}{s_crps}{3.60}
\setres{popZ03}{CPT_literature}{theta_mean}{0.70, 0.73, 1.08}
\setres{popZ03}{CPT_literature}{theta_sd}{0.14, 0.18, 0.28}
\setres{popZ03}{CPT_types1}{H_F}{46.93}
\setres{popZ03}{CPT_types1}{H_L}{81.16}
\setres{popZ03}{CPT_types1}{ess}{1}
\setres{popZ03}{CPT_types1}{kappa_crps}{0.734}
\setres{popZ03}{CPT_types1}{kappa_crps_se}{0.022}
\setres{popZ03}{CPT_types1}{kappa_draw}{0.935}
\setres{popZ03}{CPT_types1}{kappa_draw_se}{0.020}
\setres{popZ03}{CPT_types1}{kappa_mean}{0.935}
\setres{popZ03}{CPT_types1}{kappa_mean_se}{0.020}
\setres{popZ03}{CPT_types1}{r_stoch}{0.746}
\setres{popZ03}{CPT_types1}{s_crps}{3.60}
\setres{popZ03}{CPT_types1}{shares}{1.000}
\setres{popZ03}{CPT_types1}{theta_mean}{0.98, 0.43, 0.90}
\setres{popZ03}{CPT_types1}{theta_sd}{0.00, 0.00, 0.00}
\setres{popZ03}{CPT_types1}{types}{1.00: (0.98, 0.43, 0.90)}
\setres{popZ03}{CPT_types2}{H_F}{47.63}
\setres{popZ03}{CPT_types2}{H_L}{81.16}
\setres{popZ03}{CPT_types2}{ess}{2}
\setres{popZ03}{CPT_types2}{kappa_crps}{0.830}
\setres{popZ03}{CPT_types2}{kappa_crps_se}{0.017}
\setres{popZ03}{CPT_types2}{kappa_draw}{0.790}
\setres{popZ03}{CPT_types2}{kappa_draw_se}{0.019}
\setres{popZ03}{CPT_types2}{kappa_mean}{0.921}
\setres{popZ03}{CPT_types2}{kappa_mean_se}{0.021}
\setres{popZ03}{CPT_types2}{r_stoch}{0.738}
\setres{popZ03}{CPT_types2}{s_crps}{3.41}
\setres{popZ03}{CPT_types2}{shares}{0.636, 0.364}
\setres{popZ03}{CPT_types2}{theta_mean}{0.98, 0.43, 0.98}
\setres{popZ03}{CPT_types2}{theta_sd}{0.00, 0.07, 0.28}
\setres{popZ03}{CPT_types2}{types}{0.64: (0.98, 0.48, 1.19); 0.36: (0.98, 0.33, 0.61)}
\setres{popZ03}{CPT_types3}{H_F}{48.05}
\setres{popZ03}{CPT_types3}{H_L}{81.16}
\setres{popZ03}{CPT_types3}{ess}{3}
\setres{popZ03}{CPT_types3}{kappa_crps}{0.877}
\setres{popZ03}{CPT_types3}{kappa_crps_se}{0.014}
\setres{popZ03}{CPT_types3}{kappa_draw}{0.733}
\setres{popZ03}{CPT_types3}{kappa_draw_se}{0.019}
\setres{popZ03}{CPT_types3}{kappa_mean}{0.932}
\setres{popZ03}{CPT_types3}{kappa_mean_se}{0.019}
\setres{popZ03}{CPT_types3}{r_stoch}{0.734}
\setres{popZ03}{CPT_types3}{s_crps}{3.31}
\setres{popZ03}{CPT_types3}{shares}{0.467, 0.275, 0.258}
\setres{popZ03}{CPT_types3}{theta_mean}{0.96, 0.43, 0.94}
\setres{popZ03}{CPT_types3}{theta_sd}{0.02, 0.14, 0.27}
\setres{popZ03}{CPT_types3}{types}{0.47: (0.98, 0.58, 0.90); 0.28: (0.98, 0.33, 1.33); 0.26: (0.93, 0.28, 0.61)}
\setres{popZ03}{CPT_types5}{H_F}{48.41}
\setres{popZ03}{CPT_types5}{H_L}{81.16}
\setres{popZ03}{CPT_types5}{ess}{4}
\setres{popZ03}{CPT_types5}{kappa_crps}{0.887}
\setres{popZ03}{CPT_types5}{kappa_crps_se}{0.014}
\setres{popZ03}{CPT_types5}{kappa_draw}{0.695}
\setres{popZ03}{CPT_types5}{kappa_draw_se}{0.020}
\setres{popZ03}{CPT_types5}{kappa_mean}{0.933}
\setres{popZ03}{CPT_types5}{kappa_mean_se}{0.020}
\setres{popZ03}{CPT_types5}{r_stoch}{0.730}
\setres{popZ03}{CPT_types5}{s_crps}{3.29}
\setres{popZ03}{CPT_types5}{shares}{0.325, 0.286, 0.216, 0.123, 0.050}
\setres{popZ03}{CPT_types5}{theta_mean}{0.96, 0.43, 1.01}
\setres{popZ03}{CPT_types5}{theta_sd}{0.09, 0.15, 0.50}
\setres{popZ03}{CPT_types5}{types}{0.32: (0.98, 0.63, 1.04); 0.29: (0.98, 0.43, 0.75); 0.22: (0.98, 0.28, 1.19); 0.12: (0.98, 0.23, 0.46); 0.05: (0.58, 0.33, 2.93)}
\setres{popZ03}{CPT_uniform}{H_F}{52.57}
\setres{popZ03}{CPT_uniform}{H_L}{81.16}
\setres{popZ03}{CPT_uniform}{ess}{8{,}000}
\setres{popZ03}{CPT_uniform}{kappa_crps}{0.713}
\setres{popZ03}{CPT_uniform}{kappa_crps_se}{0.019}
\setres{popZ03}{CPT_uniform}{kappa_draw}{-0.455}
\setres{popZ03}{CPT_uniform}{kappa_draw_se}{0.131}
\setres{popZ03}{CPT_uniform}{kappa_mean}{0.806}
\setres{popZ03}{CPT_uniform}{kappa_mean_se}{0.039}
\setres{popZ03}{CPT_uniform}{r_stoch}{0.681}
\setres{popZ03}{CPT_uniform}{s_crps}{3.65}
\setres{popZ03}{CPT_uniform}{theta_mean}{0.50, 0.50, 1.55}
\setres{popZ03}{CPT_uniform}{theta_sd}{0.29, 0.29, 0.84}
\setres{popZ03}{DA_fitted}{H_F}{49.27}
\setres{popZ03}{DA_fitted}{H_L}{81.16}
\setres{popZ03}{DA_fitted}{ess}{11}
\setres{popZ03}{DA_fitted}{kappa_crps}{0.542}
\setres{popZ03}{DA_fitted}{kappa_crps_se}{0.050}
\setres{popZ03}{DA_fitted}{kappa_draw}{0.098}
\setres{popZ03}{DA_fitted}{kappa_draw_se}{0.097}
\setres{popZ03}{DA_fitted}{kappa_mean}{0.691}
\setres{popZ03}{DA_fitted}{kappa_mean_se}{0.054}
\setres{popZ03}{DA_fitted}{r_stoch}{0.721}
\setres{popZ03}{DA_fitted}{s_crps}{4.00}
\setres{popZ03}{DA_fitted}{theta_mean}{0.93, 0.57}
\setres{popZ03}{DA_fitted}{theta_sd}{0.19, 1.14}
\setres{popZ03}{DA_from_beijing}{H_F}{49.44}
\setres{popZ03}{DA_from_beijing}{H_L}{81.16}
\setres{popZ03}{DA_from_beijing}{ess}{11}
\setres{popZ03}{DA_from_beijing}{kappa_crps}{0.485}
\setres{popZ03}{DA_from_beijing}{kappa_crps_se}{0.047}
\setres{popZ03}{DA_from_beijing}{kappa_draw}{-0.667}
\setres{popZ03}{DA_from_beijing}{kappa_draw_se}{0.167}
\setres{popZ03}{DA_from_beijing}{kappa_mean}{0.708}
\setres{popZ03}{DA_from_beijing}{kappa_mean_se}{0.060}
\setres{popZ03}{DA_from_beijing}{r_stoch}{0.719}
\setres{popZ03}{DA_from_beijing}{s_crps}{4.11}
\setres{popZ03}{DA_from_beijing}{theta_mean}{0.65, 0.18}
\setres{popZ03}{DA_from_beijing}{theta_sd}{0.45, 1.01}
\setres{popZ03}{DA_from_countries}{H_F}{49.46}
\setres{popZ03}{DA_from_countries}{H_L}{81.16}
\setres{popZ03}{DA_from_countries}{ess}{22}
\setres{popZ03}{DA_from_countries}{kappa_crps}{0.517}
\setres{popZ03}{DA_from_countries}{kappa_crps_se}{0.050}
\setres{popZ03}{DA_from_countries}{kappa_draw}{0.044}
\setres{popZ03}{DA_from_countries}{kappa_draw_se}{0.093}
\setres{popZ03}{DA_from_countries}{kappa_mean}{0.641}
\setres{popZ03}{DA_from_countries}{kappa_mean_se}{0.053}
\setres{popZ03}{DA_from_countries}{r_stoch}{0.719}
\setres{popZ03}{DA_from_countries}{s_crps}{4.04}
\setres{popZ03}{DA_from_countries}{theta_mean}{0.89, 0.28}
\setres{popZ03}{DA_from_countries}{theta_sd}{0.25, 1.03}
\setres{popZ03}{DA_from_zurich06}{H_F}{49.15}
\setres{popZ03}{DA_from_zurich06}{H_L}{81.16}
\setres{popZ03}{DA_from_zurich06}{ess}{14}
\setres{popZ03}{DA_from_zurich06}{kappa_crps}{0.497}
\setres{popZ03}{DA_from_zurich06}{kappa_crps_se}{0.055}
\setres{popZ03}{DA_from_zurich06}{kappa_draw}{0.178}
\setres{popZ03}{DA_from_zurich06}{kappa_draw_se}{0.095}
\setres{popZ03}{DA_from_zurich06}{kappa_mean}{0.686}
\setres{popZ03}{DA_from_zurich06}{kappa_mean_se}{0.057}
\setres{popZ03}{DA_from_zurich06}{r_stoch}{0.722}
\setres{popZ03}{DA_from_zurich06}{s_crps}{4.09}
\setres{popZ03}{DA_from_zurich06}{theta_mean}{0.90, 0.42}
\setres{popZ03}{DA_from_zurich06}{theta_sd}{0.25, 0.91}
\setres{popZ03}{DA_literature}{H_F}{49.67}
\setres{popZ03}{DA_literature}{H_L}{81.16}
\setres{popZ03}{DA_literature}{ess}{439}
\setres{popZ03}{DA_literature}{kappa_crps}{0.203}
\setres{popZ03}{DA_literature}{kappa_crps_se}{0.075}
\setres{popZ03}{DA_literature}{kappa_draw}{0.010}
\setres{popZ03}{DA_literature}{kappa_draw_se}{0.140}
\setres{popZ03}{DA_literature}{kappa_mean}{0.357}
\setres{popZ03}{DA_literature}{kappa_mean_se}{0.113}
\setres{popZ03}{DA_literature}{r_stoch}{0.716}
\setres{popZ03}{DA_literature}{s_crps}{4.69}
\setres{popZ03}{DA_literature}{theta_mean}{0.58, 0.30}
\setres{popZ03}{DA_literature}{theta_sd}{0.19, 0.45}
\setres{popZ03}{DA_types1}{H_F}{46.93}
\setres{popZ03}{DA_types1}{H_L}{81.16}
\setres{popZ03}{DA_types1}{ess}{1}
\setres{popZ03}{DA_types1}{kappa_crps}{0.312}
\setres{popZ03}{DA_types1}{kappa_crps_se}{0.069}
\setres{popZ03}{DA_types1}{kappa_draw}{0.610}
\setres{popZ03}{DA_types1}{kappa_draw_se}{0.063}
\setres{popZ03}{DA_types1}{kappa_mean}{0.610}
\setres{popZ03}{DA_types1}{kappa_mean_se}{0.063}
\setres{popZ03}{DA_types1}{r_stoch}{0.746}
\setres{popZ03}{DA_types1}{s_crps}{4.46}
\setres{popZ03}{DA_types1}{shares}{1.000}
\setres{popZ03}{DA_types1}{theta_mean}{0.99, 0.25}
\setres{popZ03}{DA_types1}{theta_sd}{0.00, 0.00}
\setres{popZ03}{DA_types1}{types}{1.00: (0.99, 0.25)}
\setres{popZ03}{DA_types2}{H_F}{47.60}
\setres{popZ03}{DA_types2}{H_L}{81.16}
\setres{popZ03}{DA_types2}{ess}{2}
\setres{popZ03}{DA_types2}{kappa_crps}{0.459}
\setres{popZ03}{DA_types2}{kappa_crps_se}{0.058}
\setres{popZ03}{DA_types2}{kappa_draw}{0.283}
\setres{popZ03}{DA_types2}{kappa_draw_se}{0.087}
\setres{popZ03}{DA_types2}{kappa_mean}{0.649}
\setres{popZ03}{DA_types2}{kappa_mean_se}{0.061}
\setres{popZ03}{DA_types2}{r_stoch}{0.739}
\setres{popZ03}{DA_types2}{s_crps}{4.16}
\setres{popZ03}{DA_types2}{shares}{0.713, 0.287}
\setres{popZ03}{DA_types2}{theta_mean}{0.99, 0.71}
\setres{popZ03}{DA_types2}{theta_sd}{0.00, 1.36}
\setres{popZ03}{DA_types2}{types}{0.71: (0.99, -0.15); 0.29: (0.99, 2.85)}
\setres{popZ03}{DA_types3}{H_F}{47.89}
\setres{popZ03}{DA_types3}{H_L}{81.16}
\setres{popZ03}{DA_types3}{ess}{2}
\setres{popZ03}{DA_types3}{kappa_crps}{0.518}
\setres{popZ03}{DA_types3}{kappa_crps_se}{0.052}
\setres{popZ03}{DA_types3}{kappa_draw}{0.169}
\setres{popZ03}{DA_types3}{kappa_draw_se}{0.093}
\setres{popZ03}{DA_types3}{kappa_mean}{0.664}
\setres{popZ03}{DA_types3}{kappa_mean_se}{0.057}
\setres{popZ03}{DA_types3}{r_stoch}{0.736}
\setres{popZ03}{DA_types3}{s_crps}{4.04}
\setres{popZ03}{DA_types3}{shares}{0.628, 0.246, 0.126}
\setres{popZ03}{DA_types3}{theta_mean}{0.99, 0.67}
\setres{popZ03}{DA_types3}{theta_sd}{0.00, 1.32}
\setres{popZ03}{DA_types3}{types}{0.63: (0.99, 0.05); 0.25: (0.99, 2.95); 0.13: (0.99, -0.65)}
\setres{popZ03}{DA_types5}{H_F}{48.31}
\setres{popZ03}{DA_types5}{H_L}{81.16}
\setres{popZ03}{DA_types5}{ess}{4}
\setres{popZ03}{DA_types5}{kappa_crps}{0.535}
\setres{popZ03}{DA_types5}{kappa_crps_se}{0.051}
\setres{popZ03}{DA_types5}{kappa_draw}{0.107}
\setres{popZ03}{DA_types5}{kappa_draw_se}{0.096}
\setres{popZ03}{DA_types5}{kappa_mean}{0.674}
\setres{popZ03}{DA_types5}{kappa_mean_se}{0.055}
\setres{popZ03}{DA_types5}{r_stoch}{0.731}
\setres{popZ03}{DA_types5}{s_crps}{4.01}
\setres{popZ03}{DA_types5}{shares}{0.386, 0.295, 0.207, 0.067, 0.045}
\setres{popZ03}{DA_types5}{theta_mean}{0.95, 0.54}
\setres{popZ03}{DA_types5}{theta_sd}{0.14, 1.07}
\setres{popZ03}{DA_types5}{types}{0.39: (0.99, 0.25); 0.30: (0.99, -0.25); 0.21: (0.99, 2.15); 0.07: (0.86, -0.85); 0.04: (0.32, 2.95)}
\setres{popZ03}{DA_uniform}{H_F}{50.62}
\setres{popZ03}{DA_uniform}{H_L}{81.16}
\setres{popZ03}{DA_uniform}{ess}{1{,}600}
\setres{popZ03}{DA_uniform}{kappa_crps}{0.026}
\setres{popZ03}{DA_uniform}{kappa_crps_se}{0.081}
\setres{popZ03}{DA_uniform}{kappa_draw}{-0.847}
\setres{popZ03}{DA_uniform}{kappa_draw_se}{0.219}
\setres{popZ03}{DA_uniform}{kappa_mean}{0.102}
\setres{popZ03}{DA_uniform}{kappa_mean_se}{0.146}
\setres{popZ03}{DA_uniform}{r_stoch}{0.705}
\setres{popZ03}{DA_uniform}{s_crps}{5.05}
\setres{popZ03}{DA_uniform}{theta_mean}{0.50, 1.00}
\setres{popZ03}{DA_uniform}{theta_sd}{0.29, 1.15}
\setres{popZ03}{EU_fitted}{H_F}{48.12}
\setres{popZ03}{EU_fitted}{H_L}{81.16}
\setres{popZ03}{EU_fitted}{ess}{2}
\setres{popZ03}{EU_fitted}{kappa_crps}{0.388}
\setres{popZ03}{EU_fitted}{kappa_crps_se}{0.063}
\setres{popZ03}{EU_fitted}{kappa_draw}{0.327}
\setres{popZ03}{EU_fitted}{kappa_draw_se}{0.084}
\setres{popZ03}{EU_fitted}{kappa_mean}{0.649}
\setres{popZ03}{EU_fitted}{kappa_mean_se}{0.061}
\setres{popZ03}{EU_fitted}{r_stoch}{0.733}
\setres{popZ03}{EU_fitted}{s_crps}{4.31}
\setres{popZ03}{EU_fitted}{theta_mean}{0.82}
\setres{popZ03}{EU_fitted}{theta_sd}{0.27}
\setres{popZ03}{EU_from_beijing}{H_F}{47.66}
\setres{popZ03}{EU_from_beijing}{H_L}{81.16}
\setres{popZ03}{EU_from_beijing}{ess}{2}
\setres{popZ03}{EU_from_beijing}{kappa_crps}{0.311}
\setres{popZ03}{EU_from_beijing}{kappa_crps_se}{0.062}
\setres{popZ03}{EU_from_beijing}{kappa_draw}{-0.503}
\setres{popZ03}{EU_from_beijing}{kappa_draw_se}{0.161}
\setres{popZ03}{EU_from_beijing}{kappa_mean}{0.611}
\setres{popZ03}{EU_from_beijing}{kappa_mean_se}{0.074}
\setres{popZ03}{EU_from_beijing}{r_stoch}{0.738}
\setres{popZ03}{EU_from_beijing}{s_crps}{4.47}
\setres{popZ03}{EU_from_beijing}{theta_mean}{0.67}
\setres{popZ03}{EU_from_beijing}{theta_sd}{0.46}
\setres{popZ03}{EU_from_countries}{H_F}{47.88}
\setres{popZ03}{EU_from_countries}{H_L}{81.16}
\setres{popZ03}{EU_from_countries}{ess}{2}
\setres{popZ03}{EU_from_countries}{kappa_crps}{0.364}
\setres{popZ03}{EU_from_countries}{kappa_crps_se}{0.065}
\setres{popZ03}{EU_from_countries}{kappa_draw}{0.405}
\setres{popZ03}{EU_from_countries}{kappa_draw_se}{0.076}
\setres{popZ03}{EU_from_countries}{kappa_mean}{0.628}
\setres{popZ03}{EU_from_countries}{kappa_mean_se}{0.060}
\setres{popZ03}{EU_from_countries}{r_stoch}{0.736}
\setres{popZ03}{EU_from_countries}{s_crps}{4.36}
\setres{popZ03}{EU_from_countries}{theta_mean}{0.87}
\setres{popZ03}{EU_from_countries}{theta_sd}{0.22}
\setres{popZ03}{EU_from_zurich06}{H_F}{48.13}
\setres{popZ03}{EU_from_zurich06}{H_L}{81.16}
\setres{popZ03}{EU_from_zurich06}{ess}{3}
\setres{popZ03}{EU_from_zurich06}{kappa_crps}{0.392}
\setres{popZ03}{EU_from_zurich06}{kappa_crps_se}{0.062}
\setres{popZ03}{EU_from_zurich06}{kappa_draw}{0.232}
\setres{popZ03}{EU_from_zurich06}{kappa_draw_se}{0.091}
\setres{popZ03}{EU_from_zurich06}{kappa_mean}{0.672}
\setres{popZ03}{EU_from_zurich06}{kappa_mean_se}{0.058}
\setres{popZ03}{EU_from_zurich06}{r_stoch}{0.733}
\setres{popZ03}{EU_from_zurich06}{s_crps}{4.30}
\setres{popZ03}{EU_from_zurich06}{theta_mean}{0.82}
\setres{popZ03}{EU_from_zurich06}{theta_sd}{0.30}
\setres{popZ03}{EU_literature}{H_F}{48.08}
\setres{popZ03}{EU_literature}{H_L}{81.16}
\setres{popZ03}{EU_literature}{ess}{98}
\setres{popZ03}{EU_literature}{kappa_crps}{0.308}
\setres{popZ03}{EU_literature}{kappa_crps_se}{0.069}
\setres{popZ03}{EU_literature}{kappa_draw}{0.522}
\setres{popZ03}{EU_literature}{kappa_draw_se}{0.078}
\setres{popZ03}{EU_literature}{kappa_mean}{0.609}
\setres{popZ03}{EU_literature}{kappa_mean_se}{0.071}
\setres{popZ03}{EU_literature}{r_stoch}{0.734}
\setres{popZ03}{EU_literature}{s_crps}{4.47}
\setres{popZ03}{EU_literature}{theta_mean}{0.70}
\setres{popZ03}{EU_literature}{theta_sd}{0.14}
\setres{popZ03}{EU_types1}{H_F}{46.93}
\setres{popZ03}{EU_types1}{H_L}{81.16}
\setres{popZ03}{EU_types1}{ess}{1}
\setres{popZ03}{EU_types1}{kappa_crps}{0.240}
\setres{popZ03}{EU_types1}{kappa_crps_se}{0.073}
\setres{popZ03}{EU_types1}{kappa_draw}{0.501}
\setres{popZ03}{EU_types1}{kappa_draw_se}{0.068}
\setres{popZ03}{EU_types1}{kappa_mean}{0.501}
\setres{popZ03}{EU_types1}{kappa_mean_se}{0.068}
\setres{popZ03}{EU_types1}{r_stoch}{0.746}
\setres{popZ03}{EU_types1}{s_crps}{4.61}
\setres{popZ03}{EU_types1}{shares}{1.000}
\setres{popZ03}{EU_types1}{theta_mean}{0.99}
\setres{popZ03}{EU_types1}{theta_sd}{0.00}
\setres{popZ03}{EU_types1}{types}{1.00: (0.99)}
\setres{popZ03}{EU_types2}{H_F}{47.48}
\setres{popZ03}{EU_types2}{H_L}{81.16}
\setres{popZ03}{EU_types2}{ess}{1}
\setres{popZ03}{EU_types2}{kappa_crps}{0.363}
\setres{popZ03}{EU_types2}{kappa_crps_se}{0.065}
\setres{popZ03}{EU_types2}{kappa_draw}{0.294}
\setres{popZ03}{EU_types2}{kappa_draw_se}{0.084}
\setres{popZ03}{EU_types2}{kappa_mean}{0.639}
\setres{popZ03}{EU_types2}{kappa_mean_se}{0.060}
\setres{popZ03}{EU_types2}{r_stoch}{0.740}
\setres{popZ03}{EU_types2}{s_crps}{4.36}
\setres{popZ03}{EU_types2}{shares}{0.794, 0.206}
\setres{popZ03}{EU_types2}{theta_mean}{0.86}
\setres{popZ03}{EU_types2}{theta_sd}{0.27}
\setres{popZ03}{EU_types2}{types}{0.79: (1.00); 0.21: (0.33)}
\setres{popZ03}{EU_types3}{H_F}{47.69}
\setres{popZ03}{EU_types3}{H_L}{81.16}
\setres{popZ03}{EU_types3}{ess}{2}
\setres{popZ03}{EU_types3}{kappa_crps}{0.383}
\setres{popZ03}{EU_types3}{kappa_crps_se}{0.064}
\setres{popZ03}{EU_types3}{kappa_draw}{0.331}
\setres{popZ03}{EU_types3}{kappa_draw_se}{0.083}
\setres{popZ03}{EU_types3}{kappa_mean}{0.644}
\setres{popZ03}{EU_types3}{kappa_mean_se}{0.061}
\setres{popZ03}{EU_types3}{r_stoch}{0.738}
\setres{popZ03}{EU_types3}{s_crps}{4.32}
\setres{popZ03}{EU_types3}{shares}{0.739, 0.210, 0.051}
\setres{popZ03}{EU_types3}{theta_mean}{0.85}
\setres{popZ03}{EU_types3}{theta_sd}{0.27}
\setres{popZ03}{EU_types3}{types}{0.74: (1.00); 0.21: (0.50); 0.05: (0.09)}
\setres{popZ03}{EU_types5}{H_F}{47.99}
\setres{popZ03}{EU_types5}{H_L}{81.16}
\setres{popZ03}{EU_types5}{ess}{2}
\setres{popZ03}{EU_types5}{kappa_crps}{0.385}
\setres{popZ03}{EU_types5}{kappa_crps_se}{0.064}
\setres{popZ03}{EU_types5}{kappa_draw}{0.334}
\setres{popZ03}{EU_types5}{kappa_draw_se}{0.084}
\setres{popZ03}{EU_types5}{kappa_mean}{0.645}
\setres{popZ03}{EU_types5}{kappa_mean_se}{0.061}
\setres{popZ03}{EU_types5}{r_stoch}{0.735}
\setres{popZ03}{EU_types5}{s_crps}{4.31}
\setres{popZ03}{EU_types5}{shares}{0.647, 0.146, 0.128, 0.045, 0.034}
\setres{popZ03}{EU_types5}{theta_mean}{0.83}
\setres{popZ03}{EU_types5}{theta_sd}{0.26}
\setres{popZ03}{EU_types5}{types}{0.65: (1.00); 0.15: (0.72); 0.13: (0.50); 0.04: (0.23); 0.03: (0.07)}
\setres{popZ03}{EU_uniform}{H_F}{49.01}
\setres{popZ03}{EU_uniform}{H_L}{81.16}
\setres{popZ03}{EU_uniform}{ess}{200}
\setres{popZ03}{EU_uniform}{kappa_crps}{0.187}
\setres{popZ03}{EU_uniform}{kappa_crps_se}{0.075}
\setres{popZ03}{EU_uniform}{kappa_draw}{-0.198}
\setres{popZ03}{EU_uniform}{kappa_draw_se}{0.159}
\setres{popZ03}{EU_uniform}{kappa_mean}{0.296}
\setres{popZ03}{EU_uniform}{kappa_mean_se}{0.121}
\setres{popZ03}{EU_uniform}{r_stoch}{0.724}
\setres{popZ03}{EU_uniform}{s_crps}{4.72}
\setres{popZ03}{EU_uniform}{theta_mean}{0.51}
\setres{popZ03}{EU_uniform}{theta_sd}{0.29}
\setres{popZ03}{reference}{CPT_noise_c}{0.062}
\setres{popZ03}{reference}{DA_noise_c}{0.062}
\setres{popZ03}{reference}{EU_noise_c}{0.062}
\setres{popZ03}{reference}{mse_base}{133.78}
\setres{popZ03}{reference}{mse_best}{61.60}
\setres{popZ03}{reference}{n_lotteries}{25}
\setres{popZ03}{reference}{n_subjects}{179}
\setres{popZ03}{reference}{s_crps_base}{5.10}
\setres{popZ03}{reference}{s_crps_best}{3.06}
\setres{popZ03}{reference}{target}{Bruhin2010\_gains}
\setres{popC}{CPT_fitted}{H_F}{32.53}
\setres{popC}{CPT_fitted}{H_L}{39.00}
\setres{popC}{CPT_fitted}{ess}{41}
\setres{popC}{CPT_fitted}{kappa_crps}{0.896}
\setres{popC}{CPT_fitted}{kappa_crps_se}{0.004}
\setres{popC}{CPT_fitted}{kappa_draw}{0.079}
\setres{popC}{CPT_fitted}{kappa_draw_se}{0.019}
\setres{popC}{CPT_fitted}{kappa_mean}{0.960}
\setres{popC}{CPT_fitted}{kappa_mean_se}{0.003}
\setres{popC}{CPT_fitted}{r_stoch}{0.370}
\setres{popC}{CPT_fitted}{s_crps}{2.16}
\setres{popC}{CPT_fitted}{theta_mean}{0.89, 0.66, 1.35}
\setres{popC}{CPT_fitted}{theta_sd}{0.19, 0.33, 0.89}
\setres{popC}{CPT_from_beijing}{H_F}{32.29}
\setres{popC}{CPT_from_beijing}{H_L}{39.00}
\setres{popC}{CPT_from_beijing}{ess}{59}
\setres{popC}{CPT_from_beijing}{kappa_crps}{-0.417}
\setres{popC}{CPT_from_beijing}{kappa_crps_se}{0.031}
\setres{popC}{CPT_from_beijing}{kappa_draw}{-3.096}
\setres{popC}{CPT_from_beijing}{kappa_draw_se}{0.092}
\setres{popC}{CPT_from_beijing}{kappa_mean}{-0.978}
\setres{popC}{CPT_from_beijing}{kappa_mean_se}{0.056}
\setres{popC}{CPT_from_beijing}{r_stoch}{0.381}
\setres{popC}{CPT_from_beijing}{s_crps}{3.06}
\setres{popC}{CPT_from_beijing}{theta_mean}{0.50, 0.34, 1.40}
\setres{popC}{CPT_from_beijing}{theta_sd}{0.43, 0.24, 0.60}
\setres{popC}{CPT_from_zurich03}{H_F}{31.92}
\setres{popC}{CPT_from_zurich03}{H_L}{39.00}
\setres{popC}{CPT_from_zurich03}{ess}{57}
\setres{popC}{CPT_from_zurich03}{kappa_crps}{0.731}
\setres{popC}{CPT_from_zurich03}{kappa_crps_se}{0.008}
\setres{popC}{CPT_from_zurich03}{kappa_draw}{0.374}
\setres{popC}{CPT_from_zurich03}{kappa_draw_se}{0.010}
\setres{popC}{CPT_from_zurich03}{kappa_mean}{0.890}
\setres{popC}{CPT_from_zurich03}{kappa_mean_se}{0.006}
\setres{popC}{CPT_from_zurich03}{r_stoch}{0.397}
\setres{popC}{CPT_from_zurich03}{s_crps}{2.27}
\setres{popC}{CPT_from_zurich03}{theta_mean}{0.92, 0.45, 1.04}
\setres{popC}{CPT_from_zurich03}{theta_sd}{0.13, 0.20, 0.49}
\setres{popC}{CPT_from_zurich06}{H_F}{31.91}
\setres{popC}{CPT_from_zurich06}{H_L}{39.00}
\setres{popC}{CPT_from_zurich06}{ess}{54}
\setres{popC}{CPT_from_zurich06}{kappa_crps}{0.672}
\setres{popC}{CPT_from_zurich06}{kappa_crps_se}{0.012}
\setres{popC}{CPT_from_zurich06}{kappa_draw}{0.389}
\setres{popC}{CPT_from_zurich06}{kappa_draw_se}{0.016}
\setres{popC}{CPT_from_zurich06}{kappa_mean}{0.823}
\setres{popC}{CPT_from_zurich06}{kappa_mean_se}{0.011}
\setres{popC}{CPT_from_zurich06}{r_stoch}{0.397}
\setres{popC}{CPT_from_zurich06}{s_crps}{2.31}
\setres{popC}{CPT_from_zurich06}{theta_mean}{0.87, 0.49, 0.95}
\setres{popC}{CPT_from_zurich06}{theta_sd}{0.15, 0.22, 0.32}
\setres{popC}{CPT_literature}{H_F}{31.69}
\setres{popC}{CPT_literature}{H_L}{39.00}
\setres{popC}{CPT_literature}{ess}{732}
\setres{popC}{CPT_literature}{kappa_crps}{0.629}
\setres{popC}{CPT_literature}{kappa_crps_se}{0.017}
\setres{popC}{CPT_literature}{kappa_draw}{0.357}
\setres{popC}{CPT_literature}{kappa_draw_se}{0.024}
\setres{popC}{CPT_literature}{kappa_mean}{0.704}
\setres{popC}{CPT_literature}{kappa_mean_se}{0.019}
\setres{popC}{CPT_literature}{r_stoch}{0.407}
\setres{popC}{CPT_literature}{s_crps}{2.34}
\setres{popC}{CPT_literature}{theta_mean}{0.70, 0.73, 1.08}
\setres{popC}{CPT_literature}{theta_sd}{0.14, 0.18, 0.28}
\setres{popC}{CPT_types1}{H_F}{30.25}
\setres{popC}{CPT_types1}{H_L}{39.00}
\setres{popC}{CPT_types1}{ess}{1}
\setres{popC}{CPT_types1}{kappa_crps}{0.738}
\setres{popC}{CPT_types1}{kappa_crps_se}{0.009}
\setres{popC}{CPT_types1}{kappa_draw}{0.960}
\setres{popC}{CPT_types1}{kappa_draw_se}{0.003}
\setres{popC}{CPT_types1}{kappa_mean}{0.960}
\setres{popC}{CPT_types1}{kappa_mean_se}{0.003}
\setres{popC}{CPT_types1}{r_stoch}{0.465}
\setres{popC}{CPT_types1}{s_crps}{2.26}
\setres{popC}{CPT_types1}{shares}{1.000}
\setres{popC}{CPT_types1}{theta_mean}{0.93, 0.58, 1.04}
\setres{popC}{CPT_types1}{theta_sd}{0.00, 0.00, 0.00}
\setres{popC}{CPT_types1}{types}{1.00: (0.93, 0.58, 1.04)}
\setres{popC}{CPT_types2}{H_F}{31.01}
\setres{popC}{CPT_types2}{H_L}{39.00}
\setres{popC}{CPT_types2}{ess}{2}
\setres{popC}{CPT_types2}{kappa_crps}{0.829}
\setres{popC}{CPT_types2}{kappa_crps_se}{0.006}
\setres{popC}{CPT_types2}{kappa_draw}{0.529}
\setres{popC}{CPT_types2}{kappa_draw_se}{0.009}
\setres{popC}{CPT_types2}{kappa_mean}{0.956}
\setres{popC}{CPT_types2}{kappa_mean_se}{0.003}
\setres{popC}{CPT_types2}{r_stoch}{0.435}
\setres{popC}{CPT_types2}{s_crps}{2.20}
\setres{popC}{CPT_types2}{shares}{0.551, 0.449}
\setres{popC}{CPT_types2}{theta_mean}{0.98, 0.61, 1.09}
\setres{popC}{CPT_types2}{theta_sd}{0.00, 0.07, 0.43}
\setres{popC}{CPT_types2}{types}{0.55: (0.98, 0.68, 1.48); 0.45: (0.98, 0.53, 0.61)}
\setres{popC}{CPT_types3}{H_F}{31.09}
\setres{popC}{CPT_types3}{H_L}{39.00}
\setres{popC}{CPT_types3}{ess}{2}
\setres{popC}{CPT_types3}{kappa_crps}{0.885}
\setres{popC}{CPT_types3}{kappa_crps_se}{0.004}
\setres{popC}{CPT_types3}{kappa_draw}{0.287}
\setres{popC}{CPT_types3}{kappa_draw_se}{0.014}
\setres{popC}{CPT_types3}{kappa_mean}{0.961}
\setres{popC}{CPT_types3}{kappa_mean_se}{0.003}
\setres{popC}{CPT_types3}{r_stoch}{0.432}
\setres{popC}{CPT_types3}{s_crps}{2.16}
\setres{popC}{CPT_types3}{shares}{0.614, 0.234, 0.152}
\setres{popC}{CPT_types3}{theta_mean}{0.97, 0.65, 1.13}
\setres{popC}{CPT_types3}{theta_sd}{0.02, 0.23, 0.62}
\setres{popC}{CPT_types3}{types}{0.61: (0.98, 0.83, 1.04); 0.23: (0.98, 0.38, 0.46); 0.15: (0.93, 0.33, 2.49)}
\setres{popC}{CPT_types5}{H_F}{31.72}
\setres{popC}{CPT_types5}{H_L}{39.00}
\setres{popC}{CPT_types5}{ess}{4}
\setres{popC}{CPT_types5}{kappa_crps}{0.891}
\setres{popC}{CPT_types5}{kappa_crps_se}{0.004}
\setres{popC}{CPT_types5}{kappa_draw}{0.165}
\setres{popC}{CPT_types5}{kappa_draw_se}{0.016}
\setres{popC}{CPT_types5}{kappa_mean}{0.961}
\setres{popC}{CPT_types5}{kappa_mean_se}{0.003}
\setres{popC}{CPT_types5}{r_stoch}{0.406}
\setres{popC}{CPT_types5}{s_crps}{2.16}
\setres{popC}{CPT_types5}{shares}{0.392, 0.308, 0.111, 0.104, 0.085}
\setres{popC}{CPT_types5}{theta_mean}{0.94, 0.67, 1.19}
\setres{popC}{CPT_types5}{theta_sd}{0.05, 0.28, 0.69}
\setres{popC}{CPT_types5}{types}{0.39: (0.98, 0.83, 0.75); 0.31: (0.88, 0.88, 1.62); 0.11: (0.98, 0.33, 0.32); 0.10: (0.98, 0.08, 1.04); 0.08: (0.98, 0.33, 2.93)}
\setres{popC}{CPT_uniform}{H_F}{33.08}
\setres{popC}{CPT_uniform}{H_L}{39.00}
\setres{popC}{CPT_uniform}{ess}{8{,}000}
\setres{popC}{CPT_uniform}{kappa_crps}{0.158}
\setres{popC}{CPT_uniform}{kappa_crps_se}{0.022}
\setres{popC}{CPT_uniform}{kappa_draw}{-2.011}
\setres{popC}{CPT_uniform}{kappa_draw_se}{0.069}
\setres{popC}{CPT_uniform}{kappa_mean}{-0.009}
\setres{popC}{CPT_uniform}{kappa_mean_se}{0.036}
\setres{popC}{CPT_uniform}{r_stoch}{0.345}
\setres{popC}{CPT_uniform}{s_crps}{2.66}
\setres{popC}{CPT_uniform}{theta_mean}{0.50, 0.50, 1.55}
\setres{popC}{CPT_uniform}{theta_sd}{0.29, 0.29, 0.84}
\setres{popC}{DA_fitted}{H_F}{32.05}
\setres{popC}{DA_fitted}{H_L}{39.00}
\setres{popC}{DA_fitted}{ess}{22}
\setres{popC}{DA_fitted}{kappa_crps}{0.845}
\setres{popC}{DA_fitted}{kappa_crps_se}{0.007}
\setres{popC}{DA_fitted}{kappa_draw}{-0.153}
\setres{popC}{DA_fitted}{kappa_draw_se}{0.030}
\setres{popC}{DA_fitted}{kappa_mean}{0.834}
\setres{popC}{DA_fitted}{kappa_mean_se}{0.010}
\setres{popC}{DA_fitted}{r_stoch}{0.391}
\setres{popC}{DA_fitted}{s_crps}{2.19}
\setres{popC}{DA_fitted}{theta_mean}{0.89, 0.28}
\setres{popC}{DA_fitted}{theta_sd}{0.25, 1.03}
\setres{popC}{DA_from_beijing}{H_F}{32.17}
\setres{popC}{DA_from_beijing}{H_L}{39.00}
\setres{popC}{DA_from_beijing}{ess}{11}
\setres{popC}{DA_from_beijing}{kappa_crps}{0.464}
\setres{popC}{DA_from_beijing}{kappa_crps_se}{0.017}
\setres{popC}{DA_from_beijing}{kappa_draw}{-2.150}
\setres{popC}{DA_from_beijing}{kappa_draw_se}{0.075}
\setres{popC}{DA_from_beijing}{kappa_mean}{0.349}
\setres{popC}{DA_from_beijing}{kappa_mean_se}{0.029}
\setres{popC}{DA_from_beijing}{r_stoch}{0.386}
\setres{popC}{DA_from_beijing}{s_crps}{2.45}
\setres{popC}{DA_from_beijing}{theta_mean}{0.65, 0.18}
\setres{popC}{DA_from_beijing}{theta_sd}{0.45, 1.01}
\setres{popC}{DA_from_zurich03}{H_F}{31.92}
\setres{popC}{DA_from_zurich03}{H_L}{39.00}
\setres{popC}{DA_from_zurich03}{ess}{11}
\setres{popC}{DA_from_zurich03}{kappa_crps}{0.827}
\setres{popC}{DA_from_zurich03}{kappa_crps_se}{0.009}
\setres{popC}{DA_from_zurich03}{kappa_draw}{-0.213}
\setres{popC}{DA_from_zurich03}{kappa_draw_se}{0.033}
\setres{popC}{DA_from_zurich03}{kappa_mean}{0.829}
\setres{popC}{DA_from_zurich03}{kappa_mean_se}{0.013}
\setres{popC}{DA_from_zurich03}{r_stoch}{0.397}
\setres{popC}{DA_from_zurich03}{s_crps}{2.20}
\setres{popC}{DA_from_zurich03}{theta_mean}{0.93, 0.57}
\setres{popC}{DA_from_zurich03}{theta_sd}{0.19, 1.14}
\setres{popC}{DA_from_zurich06}{H_F}{31.75}
\setres{popC}{DA_from_zurich06}{H_L}{39.00}
\setres{popC}{DA_from_zurich06}{ess}{14}
\setres{popC}{DA_from_zurich06}{kappa_crps}{0.785}
\setres{popC}{DA_from_zurich06}{kappa_crps_se}{0.011}
\setres{popC}{DA_from_zurich06}{kappa_draw}{-0.179}
\setres{popC}{DA_from_zurich06}{kappa_draw_se}{0.034}
\setres{popC}{DA_from_zurich06}{kappa_mean}{0.779}
\setres{popC}{DA_from_zurich06}{kappa_mean_se}{0.015}
\setres{popC}{DA_from_zurich06}{r_stoch}{0.404}
\setres{popC}{DA_from_zurich06}{s_crps}{2.23}
\setres{popC}{DA_from_zurich06}{theta_mean}{0.90, 0.42}
\setres{popC}{DA_from_zurich06}{theta_sd}{0.25, 0.91}
\setres{popC}{DA_literature}{H_F}{31.73}
\setres{popC}{DA_literature}{H_L}{39.00}
\setres{popC}{DA_literature}{ess}{439}
\setres{popC}{DA_literature}{kappa_crps}{-0.153}
\setres{popC}{DA_literature}{kappa_crps_se}{0.038}
\setres{popC}{DA_literature}{kappa_draw}{-0.889}
\setres{popC}{DA_literature}{kappa_draw_se}{0.062}
\setres{popC}{DA_literature}{kappa_mean}{-0.283}
\setres{popC}{DA_literature}{kappa_mean_se}{0.052}
\setres{popC}{DA_literature}{r_stoch}{0.405}
\setres{popC}{DA_literature}{s_crps}{2.88}
\setres{popC}{DA_literature}{theta_mean}{0.58, 0.30}
\setres{popC}{DA_literature}{theta_sd}{0.19, 0.45}
\setres{popC}{DA_types1}{H_F}{30.25}
\setres{popC}{DA_types1}{H_L}{39.00}
\setres{popC}{DA_types1}{ess}{1}
\setres{popC}{DA_types1}{kappa_crps}{0.706}
\setres{popC}{DA_types1}{kappa_crps_se}{0.014}
\setres{popC}{DA_types1}{kappa_draw}{0.771}
\setres{popC}{DA_types1}{kappa_draw_se}{0.013}
\setres{popC}{DA_types1}{kappa_mean}{0.771}
\setres{popC}{DA_types1}{kappa_mean_se}{0.013}
\setres{popC}{DA_types1}{r_stoch}{0.465}
\setres{popC}{DA_types1}{s_crps}{2.29}
\setres{popC}{DA_types1}{shares}{1.000}
\setres{popC}{DA_types1}{theta_mean}{0.99, 0.05}
\setres{popC}{DA_types1}{theta_sd}{0.00, 0.00}
\setres{popC}{DA_types1}{types}{1.00: (0.99, 0.05)}
\setres{popC}{DA_types2}{H_F}{31.00}
\setres{popC}{DA_types2}{H_L}{39.00}
\setres{popC}{DA_types2}{ess}{2}
\setres{popC}{DA_types2}{kappa_crps}{0.810}
\setres{popC}{DA_types2}{kappa_crps_se}{0.009}
\setres{popC}{DA_types2}{kappa_draw}{0.309}
\setres{popC}{DA_types2}{kappa_draw_se}{0.021}
\setres{popC}{DA_types2}{kappa_mean}{0.817}
\setres{popC}{DA_types2}{kappa_mean_se}{0.011}
\setres{popC}{DA_types2}{r_stoch}{0.435}
\setres{popC}{DA_types2}{s_crps}{2.21}
\setres{popC}{DA_types2}{shares}{0.576, 0.424}
\setres{popC}{DA_types2}{theta_mean}{0.99, 0.18}
\setres{popC}{DA_types2}{theta_sd}{0.00, 0.54}
\setres{popC}{DA_types2}{types}{0.58: (0.99, 0.65); 0.42: (0.99, -0.45)}
\setres{popC}{DA_types3}{H_F}{31.08}
\setres{popC}{DA_types3}{H_L}{39.00}
\setres{popC}{DA_types3}{ess}{2}
\setres{popC}{DA_types3}{kappa_crps}{0.837}
\setres{popC}{DA_types3}{kappa_crps_se}{0.008}
\setres{popC}{DA_types3}{kappa_draw}{0.075}
\setres{popC}{DA_types3}{kappa_draw_se}{0.025}
\setres{popC}{DA_types3}{kappa_mean}{0.834}
\setres{popC}{DA_types3}{kappa_mean_se}{0.010}
\setres{popC}{DA_types3}{r_stoch}{0.432}
\setres{popC}{DA_types3}{s_crps}{2.20}
\setres{popC}{DA_types3}{shares}{0.624, 0.193, 0.182}
\setres{popC}{DA_types3}{theta_mean}{0.99, 0.37}
\setres{popC}{DA_types3}{theta_sd}{0.00, 1.01}
\setres{popC}{DA_types3}{types}{0.62: (0.99, 0.05); 0.19: (0.99, 2.35); 0.18: (0.99, -0.65)}
\setres{popC}{DA_types5}{H_F}{31.65}
\setres{popC}{DA_types5}{H_L}{39.00}
\setres{popC}{DA_types5}{ess}{4}
\setres{popC}{DA_types5}{kappa_crps}{0.850}
\setres{popC}{DA_types5}{kappa_crps_se}{0.007}
\setres{popC}{DA_types5}{kappa_draw}{-0.022}
\setres{popC}{DA_types5}{kappa_draw_se}{0.027}
\setres{popC}{DA_types5}{kappa_mean}{0.840}
\setres{popC}{DA_types5}{kappa_mean_se}{0.010}
\setres{popC}{DA_types5}{r_stoch}{0.408}
\setres{popC}{DA_types5}{s_crps}{2.19}
\setres{popC}{DA_types5}{shares}{0.393, 0.310, 0.122, 0.106, 0.069}
\setres{popC}{DA_types5}{theta_mean}{0.93, 0.33}
\setres{popC}{DA_types5}{theta_sd}{0.19, 1.08}
\setres{popC}{DA_types5}{types}{0.39: (0.99, -0.15); 0.31: (0.99, 0.55); 0.12: (0.99, 2.95); 0.11: (0.99, -0.75); 0.07: (0.22, -0.85)}
\setres{popC}{DA_uniform}{H_F}{32.38}
\setres{popC}{DA_uniform}{H_L}{39.00}
\setres{popC}{DA_uniform}{ess}{1{,}600}
\setres{popC}{DA_uniform}{kappa_crps}{-0.743}
\setres{popC}{DA_uniform}{kappa_crps_se}{0.046}
\setres{popC}{DA_uniform}{kappa_draw}{-2.893}
\setres{popC}{DA_uniform}{kappa_draw_se}{0.103}
\setres{popC}{DA_uniform}{kappa_mean}{-1.075}
\setres{popC}{DA_uniform}{kappa_mean_se}{0.070}
\setres{popC}{DA_uniform}{r_stoch}{0.377}
\setres{popC}{DA_uniform}{s_crps}{3.29}
\setres{popC}{DA_uniform}{theta_mean}{0.50, 1.00}
\setres{popC}{DA_uniform}{theta_sd}{0.29, 1.15}
\setres{popC}{EU_fitted}{H_F}{31.00}
\setres{popC}{EU_fitted}{H_L}{39.00}
\setres{popC}{EU_fitted}{ess}{2}
\setres{popC}{EU_fitted}{kappa_crps}{0.732}
\setres{popC}{EU_fitted}{kappa_crps_se}{0.013}
\setres{popC}{EU_fitted}{kappa_draw}{0.388}
\setres{popC}{EU_fitted}{kappa_draw_se}{0.022}
\setres{popC}{EU_fitted}{kappa_mean}{0.762}
\setres{popC}{EU_fitted}{kappa_mean_se}{0.015}
\setres{popC}{EU_fitted}{r_stoch}{0.435}
\setres{popC}{EU_fitted}{s_crps}{2.27}
\setres{popC}{EU_fitted}{theta_mean}{0.87}
\setres{popC}{EU_fitted}{theta_sd}{0.22}
\setres{popC}{EU_from_beijing}{H_F}{30.92}
\setres{popC}{EU_from_beijing}{H_L}{39.00}
\setres{popC}{EU_from_beijing}{ess}{2}
\setres{popC}{EU_from_beijing}{kappa_crps}{0.407}
\setres{popC}{EU_from_beijing}{kappa_crps_se}{0.020}
\setres{popC}{EU_from_beijing}{kappa_draw}{-1.889}
\setres{popC}{EU_from_beijing}{kappa_draw_se}{0.072}
\setres{popC}{EU_from_beijing}{kappa_mean}{0.243}
\setres{popC}{EU_from_beijing}{kappa_mean_se}{0.033}
\setres{popC}{EU_from_beijing}{r_stoch}{0.439}
\setres{popC}{EU_from_beijing}{s_crps}{2.49}
\setres{popC}{EU_from_beijing}{theta_mean}{0.67}
\setres{popC}{EU_from_beijing}{theta_sd}{0.46}
\setres{popC}{EU_from_zurich03}{H_F}{31.18}
\setres{popC}{EU_from_zurich03}{H_L}{39.00}
\setres{popC}{EU_from_zurich03}{ess}{2}
\setres{popC}{EU_from_zurich03}{kappa_crps}{0.711}
\setres{popC}{EU_from_zurich03}{kappa_crps_se}{0.014}
\setres{popC}{EU_from_zurich03}{kappa_draw}{0.108}
\setres{popC}{EU_from_zurich03}{kappa_draw_se}{0.029}
\setres{popC}{EU_from_zurich03}{kappa_mean}{0.714}
\setres{popC}{EU_from_zurich03}{kappa_mean_se}{0.018}
\setres{popC}{EU_from_zurich03}{r_stoch}{0.428}
\setres{popC}{EU_from_zurich03}{s_crps}{2.28}
\setres{popC}{EU_from_zurich03}{theta_mean}{0.82}
\setres{popC}{EU_from_zurich03}{theta_sd}{0.27}
\setres{popC}{EU_from_zurich06}{H_F}{31.11}
\setres{popC}{EU_from_zurich06}{H_L}{39.00}
\setres{popC}{EU_from_zurich06}{ess}{3}
\setres{popC}{EU_from_zurich06}{kappa_crps}{0.715}
\setres{popC}{EU_from_zurich06}{kappa_crps_se}{0.014}
\setres{popC}{EU_from_zurich06}{kappa_draw}{-0.156}
\setres{popC}{EU_from_zurich06}{kappa_draw_se}{0.035}
\setres{popC}{EU_from_zurich06}{kappa_mean}{0.721}
\setres{popC}{EU_from_zurich06}{kappa_mean_se}{0.018}
\setres{popC}{EU_from_zurich06}{r_stoch}{0.431}
\setres{popC}{EU_from_zurich06}{s_crps}{2.28}
\setres{popC}{EU_from_zurich06}{theta_mean}{0.82}
\setres{popC}{EU_from_zurich06}{theta_sd}{0.30}
\setres{popC}{EU_literature}{H_F}{30.90}
\setres{popC}{EU_literature}{H_L}{39.00}
\setres{popC}{EU_literature}{ess}{98}
\setres{popC}{EU_literature}{kappa_crps}{0.460}
\setres{popC}{EU_literature}{kappa_crps_se}{0.023}
\setres{popC}{EU_literature}{kappa_draw}{0.344}
\setres{popC}{EU_literature}{kappa_draw_se}{0.030}
\setres{popC}{EU_literature}{kappa_mean}{0.486}
\setres{popC}{EU_literature}{kappa_mean_se}{0.027}
\setres{popC}{EU_literature}{r_stoch}{0.439}
\setres{popC}{EU_literature}{s_crps}{2.46}
\setres{popC}{EU_literature}{theta_mean}{0.70}
\setres{popC}{EU_literature}{theta_sd}{0.14}
\setres{popC}{EU_types1}{H_F}{30.25}
\setres{popC}{EU_types1}{H_L}{39.00}
\setres{popC}{EU_types1}{ess}{1}
\setres{popC}{EU_types1}{kappa_crps}{0.689}
\setres{popC}{EU_types1}{kappa_crps_se}{0.014}
\setres{popC}{EU_types1}{kappa_draw}{0.752}
\setres{popC}{EU_types1}{kappa_draw_se}{0.014}
\setres{popC}{EU_types1}{kappa_mean}{0.752}
\setres{popC}{EU_types1}{kappa_mean_se}{0.014}
\setres{popC}{EU_types1}{r_stoch}{0.465}
\setres{popC}{EU_types1}{s_crps}{2.30}
\setres{popC}{EU_types1}{shares}{1.000}
\setres{popC}{EU_types1}{theta_mean}{1.00}
\setres{popC}{EU_types1}{theta_sd}{0.00}
\setres{popC}{EU_types1}{types}{1.00: (1.00)}
\setres{popC}{EU_types2}{H_F}{30.82}
\setres{popC}{EU_types2}{H_L}{39.00}
\setres{popC}{EU_types2}{ess}{1}
\setres{popC}{EU_types2}{kappa_crps}{0.734}
\setres{popC}{EU_types2}{kappa_crps_se}{0.013}
\setres{popC}{EU_types2}{kappa_draw}{0.514}
\setres{popC}{EU_types2}{kappa_draw_se}{0.019}
\setres{popC}{EU_types2}{kappa_mean}{0.765}
\setres{popC}{EU_types2}{kappa_mean_se}{0.014}
\setres{popC}{EU_types2}{r_stoch}{0.442}
\setres{popC}{EU_types2}{s_crps}{2.27}
\setres{popC}{EU_types2}{shares}{0.810, 0.190}
\setres{popC}{EU_types2}{theta_mean}{0.90}
\setres{popC}{EU_types2}{theta_sd}{0.20}
\setres{popC}{EU_types2}{types}{0.81: (1.00); 0.19: (0.48)}
\setres{popC}{EU_types3}{H_F}{30.78}
\setres{popC}{EU_types3}{H_L}{39.00}
\setres{popC}{EU_types3}{ess}{2}
\setres{popC}{EU_types3}{kappa_crps}{0.734}
\setres{popC}{EU_types3}{kappa_crps_se}{0.013}
\setres{popC}{EU_types3}{kappa_draw}{0.387}
\setres{popC}{EU_types3}{kappa_draw_se}{0.022}
\setres{popC}{EU_types3}{kappa_mean}{0.768}
\setres{popC}{EU_types3}{kappa_mean_se}{0.014}
\setres{popC}{EU_types3}{r_stoch}{0.444}
\setres{popC}{EU_types3}{s_crps}{2.27}
\setres{popC}{EU_types3}{shares}{0.765, 0.204, 0.030}
\setres{popC}{EU_types3}{theta_mean}{0.88}
\setres{popC}{EU_types3}{theta_sd}{0.22}
\setres{popC}{EU_types3}{types}{0.77: (1.00); 0.20: (0.57); 0.03: (0.08)}
\setres{popC}{EU_types5}{H_F}{31.03}
\setres{popC}{EU_types5}{H_L}{39.00}
\setres{popC}{EU_types5}{ess}{2}
\setres{popC}{EU_types5}{kappa_crps}{0.734}
\setres{popC}{EU_types5}{kappa_crps_se}{0.013}
\setres{popC}{EU_types5}{kappa_draw}{0.384}
\setres{popC}{EU_types5}{kappa_draw_se}{0.022}
\setres{popC}{EU_types5}{kappa_mean}{0.762}
\setres{popC}{EU_types5}{kappa_mean_se}{0.015}
\setres{popC}{EU_types5}{r_stoch}{0.434}
\setres{popC}{EU_types5}{s_crps}{2.27}
\setres{popC}{EU_types5}{shares}{0.724, 0.155, 0.075, 0.025, 0.022}
\setres{popC}{EU_types5}{theta_mean}{0.88}
\setres{popC}{EU_types5}{theta_sd}{0.22}
\setres{popC}{EU_types5}{types}{0.72: (1.00); 0.15: (0.69); 0.07: (0.51); 0.02: (0.33); 0.02: (0.06)}
\setres{popC}{EU_uniform}{H_F}{31.67}
\setres{popC}{EU_uniform}{H_L}{39.00}
\setres{popC}{EU_uniform}{ess}{200}
\setres{popC}{EU_uniform}{kappa_crps}{-0.198}
\setres{popC}{EU_uniform}{kappa_crps_se}{0.038}
\setres{popC}{EU_uniform}{kappa_draw}{-1.407}
\setres{popC}{EU_uniform}{kappa_draw_se}{0.072}
\setres{popC}{EU_uniform}{kappa_mean}{-0.466}
\setres{popC}{EU_uniform}{kappa_mean_se}{0.056}
\setres{popC}{EU_uniform}{r_stoch}{0.407}
\setres{popC}{EU_uniform}{s_crps}{2.91}
\setres{popC}{EU_uniform}{theta_mean}{0.51}
\setres{popC}{EU_uniform}{theta_sd}{0.29}
\setres{popC}{reference}{CPT_noise_c}{0.130}
\setres{popC}{reference}{DA_noise_c}{0.130}
\setres{popC}{reference}{EU_noise_c}{0.130}
\setres{popC}{reference}{mse_base}{24.41}
\setres{popC}{reference}{mse_best}{16.04}
\setres{popC}{reference}{n_lotteries}{14}
\setres{popC}{reference}{n_subjects}{2{,}939}
\setres{popC}{reference}{s_crps_base}{2.77}
\setres{popC}{reference}{s_crps_best}{2.08}
\setres{popC}{reference}{target}{LV2019\_pooled}
\setres{types}{K1}{K}{1.000}
\setres{types}{K1}{bic_per_subject}{25.91}
\setres{types}{K1}{crps_test}{2.909}
\setres{types}{K1}{heldout_loglik_per_subject}{-13.18}
\setres{types}{K1}{kappa_crps_test}{0.567}
\setres{types}{K1}{loglik_per_subject}{-12.95}
\setres{types}{K1}{shares}{1.000}
\setres{types}{K1}{types}{1.00: (0.98, 0.63, 0.90)}
\setres{types}{K2}{K}{2.000}
\setres{types}{K2}{bic_per_subject}{21.03}
\setres{types}{K2}{crps_test}{2.451}
\setres{types}{K2}{heldout_loglik_per_subject}{-10.54}
\setres{types}{K2}{kappa_crps_test}{0.808}
\setres{types}{K2}{loglik_per_subject}{-10.51}
\setres{types}{K2}{shares}{0.669, 0.331}
\setres{types}{K2}{types}{0.67: (0.93, 0.68, 1.33); 0.33: (0.98, 0.53, 0.46)}
\setres{types}{K3}{K}{3.000}
\setres{types}{K3}{bic_per_subject}{18.71}
\setres{types}{K3}{crps_test}{2.379}
\setres{types}{K3}{heldout_loglik_per_subject}{-9.47}
\setres{types}{K3}{kappa_crps_test}{0.846}
\setres{types}{K3}{loglik_per_subject}{-9.34}
\setres{types}{K3}{shares}{0.596, 0.269, 0.135}
\setres{types}{K3}{types}{0.60: (0.98, 0.83, 1.04); 0.27: (0.98, 0.43, 0.46); 0.13: (0.98, 0.28, 2.35)}
\setres{types}{K4}{K}{4.000}
\setres{types}{K4}{bic_per_subject}{17.98}
\setres{types}{K4}{crps_test}{2.352}
\setres{types}{K4}{heldout_loglik_per_subject}{-9.22}
\setres{types}{K4}{kappa_crps_test}{0.860}
\setres{types}{K4}{loglik_per_subject}{-8.97}
\setres{types}{K4}{shares}{0.438, 0.307, 0.129, 0.126}
\setres{types}{K4}{types}{0.44: (0.98, 0.88, 1.19); 0.31: (0.98, 0.78, 0.61); 0.13: (0.98, 0.23, 2.20); 0.13: (0.88, 0.18, 0.46)}
\setres{types}{K5}{K}{5.000}
\setres{types}{K5}{bic_per_subject}{17.28}
\setres{types}{K5}{crps_test}{2.295}
\setres{types}{K5}{heldout_loglik_per_subject}{-8.79}
\setres{types}{K5}{kappa_crps_test}{0.890}
\setres{types}{K5}{loglik_per_subject}{-8.61}
\setres{types}{K5}{shares}{0.430, 0.303, 0.097, 0.092, 0.078}
\setres{types}{K5}{types}{0.43: (0.98, 0.88, 1.19); 0.30: (0.98, 0.78, 0.61); 0.10: (0.98, 0.23, 0.32); 0.09: (0.98, 0.38, 2.93); 0.08: (0.98, 0.03, 1.04)}
\setres{types}{K8}{K}{8.000}
\setres{types}{K8}{bic_per_subject}{16.89}
\setres{types}{K8}{crps_test}{2.268}
\setres{types}{K8}{heldout_loglik_per_subject}{-8.55}
\setres{types}{K8}{kappa_crps_test}{0.904}
\setres{types}{K8}{loglik_per_subject}{-8.40}
\setres{types}{K8}{shares}{0.228, 0.216, 0.133, 0.125, 0.079, 0.079, 0.070, 0.070}
\setres{types}{K8}{types}{0.23: (0.98, 0.98, 1.04); 0.22: (0.98, 0.78, 0.75); 0.13: (0.98, 0.78, 0.46); 0.12: (0.98, 0.58, 1.33); 0.08: (0.98, 0.18, 0.32); 0.08: (0.68, 0.98, 2.93); 0.07: (0.98, 0.23, 2.93); 0.07: (0.98, 0.03, 0.90)}
\setres{types}{one_per_subject}{loglik_per_subject}{-5.46}
\setres{types}{reference}{crps_base_test}{3.985}
\setres{types}{reference}{crps_floor_test}{2.086}
\setres{types}{reference}{n_subjects}{2{,}939}
\setres{types}{unrestricted}{crps_test}{2.227}
\setres{types}{unrestricted}{ess}{43}
\setres{types}{unrestricted}{heldout_loglik_per_subject}{-8.33}
\setres{types}{unrestricted}{kappa_crps_test}{0.926}
\setres{types}{unrestricted}{loglik_per_subject}{-8.18}
\setres{cty_Australia}{CPT_deterministic}{kappa}{0.980}
\setres{cty_Australia}{CPT_deterministic}{kappa_se}{0.011}
\setres{cty_Australia}{CPT_deterministic}{mse}{10.91}
\setres{cty_Australia}{CPT_deterministic}{nrmse_pct}{19.2}
\setres{cty_Australia}{CPT_deterministic}{r_ff}{---}
\setres{cty_Australia}{CPT_deterministic}{r_ff_se}{---}
\setres{cty_Australia}{CPT_deterministic}{r_prior}{---}
\setres{cty_Australia}{CPT_deterministic}{r_prior_se}{---}
\setres{cty_Australia}{CPT_deterministic}{r_prob}{---}
\setres{cty_Australia}{CPT_deterministic}{r_prob_se}{---}
\setres{cty_Australia}{CPT_deterministic}{theta_star}{1.30, 0.80, 0.50}
\setres{cty_Australia}{CPT_deterministic}{volume_ess}{---}
\setres{cty_Australia}{CPT_lit}{kappa}{0.644}
\setres{cty_Australia}{CPT_lit}{kappa_prior_mean}{0.884}
\setres{cty_Australia}{CPT_lit}{kappa_prior_mean_se}{0.032}
\setres{cty_Australia}{CPT_lit}{kappa_se}{0.049}
\setres{cty_Australia}{CPT_lit}{mse}{14.96}
\setres{cty_Australia}{CPT_lit}{mse_prior_mean}{12.06}
\setres{cty_Australia}{CPT_lit}{nrmse_pct}{19.0}
\setres{cty_Australia}{CPT_lit}{pred_variance}{2.90}
\setres{cty_Australia}{CPT_lit}{r_ff}{---}
\setres{cty_Australia}{CPT_lit}{r_ff_se}{---}
\setres{cty_Australia}{CPT_lit}{r_prior}{---}
\setres{cty_Australia}{CPT_lit}{r_prior_se}{---}
\setres{cty_Australia}{CPT_lit}{r_prob}{---}
\setres{cty_Australia}{CPT_lit}{r_prob_se}{---}
\setres{cty_Australia}{CPT_lit}{spread_penalty}{0.240}
\setres{cty_Australia}{CPT_lit}{spread_penalty_se}{0.021}
\setres{cty_Australia}{CPT_lit}{volume_ess}{---}
\setres{cty_Australia}{CPT_transfer_others}{kappa}{0.900}
\setres{cty_Australia}{CPT_transfer_others}{kappa_prior_mean}{0.900}
\setres{cty_Australia}{CPT_transfer_others}{kappa_prior_mean_se}{0.040}
\setres{cty_Australia}{CPT_transfer_others}{kappa_se}{0.040}
\setres{cty_Australia}{CPT_transfer_others}{mse}{11.87}
\setres{cty_Australia}{CPT_transfer_others}{mse_prior_mean}{11.87}
\setres{cty_Australia}{CPT_transfer_others}{nrmse_pct}{18.5}
\setres{cty_Australia}{CPT_transfer_others}{pred_variance}{0.00}
\setres{cty_Australia}{CPT_transfer_others}{r_ff}{---}
\setres{cty_Australia}{CPT_transfer_others}{r_ff_se}{---}
\setres{cty_Australia}{CPT_transfer_others}{r_prior}{---}
\setres{cty_Australia}{CPT_transfer_others}{r_prior_se}{---}
\setres{cty_Australia}{CPT_transfer_others}{r_prob}{---}
\setres{cty_Australia}{CPT_transfer_others}{r_prob_se}{---}
\setres{cty_Australia}{CPT_transfer_others}{spread_penalty}{0.000}
\setres{cty_Australia}{CPT_transfer_others}{spread_penalty_se}{0.000}
\setres{cty_Australia}{CPT_transfer_others}{theta_star}{1.00, 0.60, 0.90}
\setres{cty_Australia}{CPT_transfer_others}{transfer}{others}
\setres{cty_Australia}{CPT_transfer_others}{volume_ess}{---}
\setres{cty_Australia}{CPT_uniform}{kappa}{-1.021}
\setres{cty_Australia}{CPT_uniform}{kappa_prior_mean}{0.417}
\setres{cty_Australia}{CPT_uniform}{kappa_prior_mean_se}{0.116}
\setres{cty_Australia}{CPT_uniform}{kappa_se}{0.219}
\setres{cty_Australia}{CPT_uniform}{mse}{35.07}
\setres{cty_Australia}{CPT_uniform}{mse_prior_mean}{17.71}
\setres{cty_Australia}{CPT_uniform}{nrmse_pct}{22.9}
\setres{cty_Australia}{CPT_uniform}{pred_variance}{17.37}
\setres{cty_Australia}{CPT_uniform}{r_ff}{---}
\setres{cty_Australia}{CPT_uniform}{r_ff_se}{---}
\setres{cty_Australia}{CPT_uniform}{r_prior}{---}
\setres{cty_Australia}{CPT_uniform}{r_prior_se}{---}
\setres{cty_Australia}{CPT_uniform}{r_prob}{---}
\setres{cty_Australia}{CPT_uniform}{r_prob_se}{---}
\setres{cty_Australia}{CPT_uniform}{spread_penalty}{1.438}
\setres{cty_Australia}{CPT_uniform}{spread_penalty_se}{0.124}
\setres{cty_Australia}{CPT_uniform}{volume_ess}{---}
\setres{cty_Australia}{DA_deterministic}{kappa}{0.958}
\setres{cty_Australia}{DA_deterministic}{kappa_se}{0.014}
\setres{cty_Australia}{DA_deterministic}{mse}{11.17}
\setres{cty_Australia}{DA_deterministic}{nrmse_pct}{19.3}
\setres{cty_Australia}{DA_deterministic}{r_ff}{---}
\setres{cty_Australia}{DA_deterministic}{r_ff_se}{---}
\setres{cty_Australia}{DA_deterministic}{r_prior}{---}
\setres{cty_Australia}{DA_deterministic}{r_prior_se}{---}
\setres{cty_Australia}{DA_deterministic}{r_prob}{---}
\setres{cty_Australia}{DA_deterministic}{r_prob_se}{---}
\setres{cty_Australia}{DA_deterministic}{theta_star}{1.73, 2.32}
\setres{cty_Australia}{DA_deterministic}{volume_ess}{---}
\setres{cty_Australia}{DA_lit}{kappa}{0.035}
\setres{cty_Australia}{DA_lit}{kappa_prior_mean}{0.428}
\setres{cty_Australia}{DA_lit}{kappa_prior_mean_se}{0.145}
\setres{cty_Australia}{DA_lit}{kappa_se}{0.175}
\setres{cty_Australia}{DA_lit}{mse}{22.31}
\setres{cty_Australia}{DA_lit}{mse_prior_mean}{17.57}
\setres{cty_Australia}{DA_lit}{nrmse_pct}{22.3}
\setres{cty_Australia}{DA_lit}{pred_variance}{4.75}
\setres{cty_Australia}{DA_lit}{r_ff}{---}
\setres{cty_Australia}{DA_lit}{r_ff_se}{---}
\setres{cty_Australia}{DA_lit}{r_prior}{---}
\setres{cty_Australia}{DA_lit}{r_prior_se}{---}
\setres{cty_Australia}{DA_lit}{r_prob}{---}
\setres{cty_Australia}{DA_lit}{r_prob_se}{---}
\setres{cty_Australia}{DA_lit}{spread_penalty}{0.393}
\setres{cty_Australia}{DA_lit}{spread_penalty_se}{0.034}
\setres{cty_Australia}{DA_lit}{volume_ess}{---}
\setres{cty_Australia}{DA_transfer_others}{kappa}{0.919}
\setres{cty_Australia}{DA_transfer_others}{kappa_prior_mean}{0.919}
\setres{cty_Australia}{DA_transfer_others}{kappa_prior_mean_se}{0.046}
\setres{cty_Australia}{DA_transfer_others}{kappa_se}{0.046}
\setres{cty_Australia}{DA_transfer_others}{mse}{11.64}
\setres{cty_Australia}{DA_transfer_others}{mse_prior_mean}{11.64}
\setres{cty_Australia}{DA_transfer_others}{nrmse_pct}{18.9}
\setres{cty_Australia}{DA_transfer_others}{pred_variance}{0.00}
\setres{cty_Australia}{DA_transfer_others}{r_ff}{---}
\setres{cty_Australia}{DA_transfer_others}{r_ff_se}{---}
\setres{cty_Australia}{DA_transfer_others}{r_prior}{---}
\setres{cty_Australia}{DA_transfer_others}{r_prior_se}{---}
\setres{cty_Australia}{DA_transfer_others}{r_prob}{---}
\setres{cty_Australia}{DA_transfer_others}{r_prob_se}{---}
\setres{cty_Australia}{DA_transfer_others}{spread_penalty}{0.000}
\setres{cty_Australia}{DA_transfer_others}{spread_penalty_se}{0.000}
\setres{cty_Australia}{DA_transfer_others}{theta_star}{1.69, 1.43}
\setres{cty_Australia}{DA_transfer_others}{transfer}{others}
\setres{cty_Australia}{DA_transfer_others}{volume_ess}{---}
\setres{cty_Australia}{DA_uniform}{kappa}{-1.235}
\setres{cty_Australia}{DA_uniform}{kappa_prior_mean}{-0.104}
\setres{cty_Australia}{DA_uniform}{kappa_prior_mean_se}{0.241}
\setres{cty_Australia}{DA_uniform}{kappa_se}{0.324}
\setres{cty_Australia}{DA_uniform}{mse}{37.65}
\setres{cty_Australia}{DA_uniform}{mse_prior_mean}{24.00}
\setres{cty_Australia}{DA_uniform}{nrmse_pct}{28.1}
\setres{cty_Australia}{DA_uniform}{pred_variance}{13.65}
\setres{cty_Australia}{DA_uniform}{r_ff}{---}
\setres{cty_Australia}{DA_uniform}{r_ff_se}{---}
\setres{cty_Australia}{DA_uniform}{r_prior}{---}
\setres{cty_Australia}{DA_uniform}{r_prior_se}{---}
\setres{cty_Australia}{DA_uniform}{r_prob}{---}
\setres{cty_Australia}{DA_uniform}{r_prob_se}{---}
\setres{cty_Australia}{DA_uniform}{spread_penalty}{1.130}
\setres{cty_Australia}{DA_uniform}{spread_penalty_se}{0.098}
\setres{cty_Australia}{DA_uniform}{volume_ess}{---}
\setres{cty_Australia}{EU_deterministic}{kappa}{0.825}
\setres{cty_Australia}{EU_deterministic}{kappa_se}{0.040}
\setres{cty_Australia}{EU_deterministic}{mse}{12.78}
\setres{cty_Australia}{EU_deterministic}{nrmse_pct}{19.9}
\setres{cty_Australia}{EU_deterministic}{r_ff}{---}
\setres{cty_Australia}{EU_deterministic}{r_ff_se}{---}
\setres{cty_Australia}{EU_deterministic}{r_prior}{---}
\setres{cty_Australia}{EU_deterministic}{r_prior_se}{---}
\setres{cty_Australia}{EU_deterministic}{r_prob}{---}
\setres{cty_Australia}{EU_deterministic}{r_prob_se}{---}
\setres{cty_Australia}{EU_deterministic}{theta_star}{0.81}
\setres{cty_Australia}{EU_deterministic}{volume_ess}{---}
\setres{cty_Australia}{EU_lit}{kappa}{0.696}
\setres{cty_Australia}{EU_lit}{kappa_prior_mean}{0.796}
\setres{cty_Australia}{EU_lit}{kappa_prior_mean_se}{0.050}
\setres{cty_Australia}{EU_lit}{kappa_se}{0.058}
\setres{cty_Australia}{EU_lit}{mse}{14.34}
\setres{cty_Australia}{EU_lit}{mse_prior_mean}{13.13}
\setres{cty_Australia}{EU_lit}{nrmse_pct}{19.6}
\setres{cty_Australia}{EU_lit}{pred_variance}{1.20}
\setres{cty_Australia}{EU_lit}{r_ff}{---}
\setres{cty_Australia}{EU_lit}{r_ff_se}{---}
\setres{cty_Australia}{EU_lit}{r_prior}{---}
\setres{cty_Australia}{EU_lit}{r_prior_se}{---}
\setres{cty_Australia}{EU_lit}{r_prob}{---}
\setres{cty_Australia}{EU_lit}{r_prob_se}{---}
\setres{cty_Australia}{EU_lit}{spread_penalty}{0.100}
\setres{cty_Australia}{EU_lit}{spread_penalty_se}{0.009}
\setres{cty_Australia}{EU_lit}{volume_ess}{---}
\setres{cty_Australia}{EU_transfer_others}{kappa}{0.795}
\setres{cty_Australia}{EU_transfer_others}{kappa_prior_mean}{0.795}
\setres{cty_Australia}{EU_transfer_others}{kappa_prior_mean_se}{0.065}
\setres{cty_Australia}{EU_transfer_others}{kappa_se}{0.065}
\setres{cty_Australia}{EU_transfer_others}{mse}{13.14}
\setres{cty_Australia}{EU_transfer_others}{mse_prior_mean}{13.14}
\setres{cty_Australia}{EU_transfer_others}{nrmse_pct}{18.9}
\setres{cty_Australia}{EU_transfer_others}{pred_variance}{0.00}
\setres{cty_Australia}{EU_transfer_others}{r_ff}{---}
\setres{cty_Australia}{EU_transfer_others}{r_ff_se}{---}
\setres{cty_Australia}{EU_transfer_others}{r_prior}{---}
\setres{cty_Australia}{EU_transfer_others}{r_prior_se}{---}
\setres{cty_Australia}{EU_transfer_others}{r_prob}{---}
\setres{cty_Australia}{EU_transfer_others}{r_prob_se}{---}
\setres{cty_Australia}{EU_transfer_others}{spread_penalty}{0.000}
\setres{cty_Australia}{EU_transfer_others}{spread_penalty_se}{0.000}
\setres{cty_Australia}{EU_transfer_others}{theta_star}{0.95}
\setres{cty_Australia}{EU_transfer_others}{transfer}{others}
\setres{cty_Australia}{EU_transfer_others}{volume_ess}{---}
\setres{cty_Australia}{EU_uniform}{kappa}{-0.382}
\setres{cty_Australia}{EU_uniform}{kappa_prior_mean}{0.297}
\setres{cty_Australia}{EU_uniform}{kappa_prior_mean_se}{0.161}
\setres{cty_Australia}{EU_uniform}{kappa_se}{0.213}
\setres{cty_Australia}{EU_uniform}{mse}{27.35}
\setres{cty_Australia}{EU_uniform}{mse_prior_mean}{19.16}
\setres{cty_Australia}{EU_uniform}{nrmse_pct}{23.9}
\setres{cty_Australia}{EU_uniform}{pred_variance}{8.20}
\setres{cty_Australia}{EU_uniform}{r_ff}{---}
\setres{cty_Australia}{EU_uniform}{r_ff_se}{---}
\setres{cty_Australia}{EU_uniform}{r_prior}{---}
\setres{cty_Australia}{EU_uniform}{r_prior_se}{---}
\setres{cty_Australia}{EU_uniform}{r_prob}{---}
\setres{cty_Australia}{EU_uniform}{r_prob_se}{---}
\setres{cty_Australia}{EU_uniform}{spread_penalty}{0.679}
\setres{cty_Australia}{EU_uniform}{spread_penalty_se}{0.059}
\setres{cty_Australia}{EU_uniform}{volume_ess}{---}
\setres{cty_Australia}{EV_deterministic}{kappa}{0.755}
\setres{cty_Australia}{EV_deterministic}{kappa_se}{0.073}
\setres{cty_Australia}{EV_deterministic}{mse}{13.63}
\setres{cty_Australia}{EV_deterministic}{nrmse_pct}{18.6}
\setres{cty_Australia}{EV_deterministic}{r_ff}{---}
\setres{cty_Australia}{EV_deterministic}{r_ff_se}{---}
\setres{cty_Australia}{EV_deterministic}{r_prior}{---}
\setres{cty_Australia}{EV_deterministic}{r_prior_se}{---}
\setres{cty_Australia}{EV_deterministic}{r_prob}{---}
\setres{cty_Australia}{EV_deterministic}{r_prob_se}{---}
\setres{cty_Australia}{EV_deterministic}{volume_ess}{---}
\setres{cty_Australia}{reference}{country_name}{Australia}
\setres{cty_Australia}{reference}{lookup_nrmse_pct}{18.3}
\setres{cty_Australia}{reference}{midpoint_mse}{22.74}
\setres{cty_Australia}{reference}{midpoint_nrmse_pct}{26.4}
\setres{cty_Australia}{reference}{missing_lotteries}{none}
\setres{cty_Australia}{reference}{n_lotteries}{14}
\setres{cty_Australia}{reference}{n_obs}{854}
\setres{cty_Australia}{reference}{n_subjects}{61}
\setres{cty_Australia}{reference}{oracle_mse}{10.66}
\setres{cty_Australia}{reference}{sample}{LV2019\_Australia}
\setres{cty_Belgium}{CPT_deterministic}{kappa}{0.975}
\setres{cty_Belgium}{CPT_deterministic}{kappa_se}{0.011}
\setres{cty_Belgium}{CPT_deterministic}{mse}{8.89}
\setres{cty_Belgium}{CPT_deterministic}{nrmse_pct}{17.0}
\setres{cty_Belgium}{CPT_deterministic}{r_ff}{---}
\setres{cty_Belgium}{CPT_deterministic}{r_ff_se}{---}
\setres{cty_Belgium}{CPT_deterministic}{r_prior}{---}
\setres{cty_Belgium}{CPT_deterministic}{r_prior_se}{---}
\setres{cty_Belgium}{CPT_deterministic}{r_prob}{---}
\setres{cty_Belgium}{CPT_deterministic}{r_prob_se}{---}
\setres{cty_Belgium}{CPT_deterministic}{theta_star}{1.00, 0.70, 0.80}
\setres{cty_Belgium}{CPT_deterministic}{volume_ess}{---}
\setres{cty_Belgium}{CPT_lit}{kappa}{0.619}
\setres{cty_Belgium}{CPT_lit}{kappa_prior_mean}{0.875}
\setres{cty_Belgium}{CPT_lit}{kappa_prior_mean_se}{0.038}
\setres{cty_Belgium}{CPT_lit}{kappa_se}{0.046}
\setres{cty_Belgium}{CPT_lit}{mse}{12.93}
\setres{cty_Belgium}{CPT_lit}{mse_prior_mean}{10.03}
\setres{cty_Belgium}{CPT_lit}{nrmse_pct}{17.9}
\setres{cty_Belgium}{CPT_lit}{pred_variance}{2.90}
\setres{cty_Belgium}{CPT_lit}{r_ff}{---}
\setres{cty_Belgium}{CPT_lit}{r_ff_se}{---}
\setres{cty_Belgium}{CPT_lit}{r_prior}{---}
\setres{cty_Belgium}{CPT_lit}{r_prior_se}{---}
\setres{cty_Belgium}{CPT_lit}{r_prob}{---}
\setres{cty_Belgium}{CPT_lit}{r_prob_se}{---}
\setres{cty_Belgium}{CPT_lit}{spread_penalty}{0.256}
\setres{cty_Belgium}{CPT_lit}{spread_penalty_se}{0.015}
\setres{cty_Belgium}{CPT_lit}{volume_ess}{---}
\setres{cty_Belgium}{CPT_transfer_others}{kappa}{0.948}
\setres{cty_Belgium}{CPT_transfer_others}{kappa_prior_mean}{0.948}
\setres{cty_Belgium}{CPT_transfer_others}{kappa_prior_mean_se}{0.013}
\setres{cty_Belgium}{CPT_transfer_others}{kappa_se}{0.013}
\setres{cty_Belgium}{CPT_transfer_others}{mse}{9.20}
\setres{cty_Belgium}{CPT_transfer_others}{mse_prior_mean}{9.20}
\setres{cty_Belgium}{CPT_transfer_others}{nrmse_pct}{17.2}
\setres{cty_Belgium}{CPT_transfer_others}{pred_variance}{0.00}
\setres{cty_Belgium}{CPT_transfer_others}{r_ff}{---}
\setres{cty_Belgium}{CPT_transfer_others}{r_ff_se}{---}
\setres{cty_Belgium}{CPT_transfer_others}{r_prior}{---}
\setres{cty_Belgium}{CPT_transfer_others}{r_prior_se}{---}
\setres{cty_Belgium}{CPT_transfer_others}{r_prob}{---}
\setres{cty_Belgium}{CPT_transfer_others}{r_prob_se}{---}
\setres{cty_Belgium}{CPT_transfer_others}{spread_penalty}{0.000}
\setres{cty_Belgium}{CPT_transfer_others}{spread_penalty_se}{0.000}
\setres{cty_Belgium}{CPT_transfer_others}{theta_star}{1.10, 0.60, 0.80}
\setres{cty_Belgium}{CPT_transfer_others}{transfer}{others}
\setres{cty_Belgium}{CPT_transfer_others}{volume_ess}{---}
\setres{cty_Belgium}{CPT_uniform}{kappa}{-1.227}
\setres{cty_Belgium}{CPT_uniform}{kappa_prior_mean}{0.304}
\setres{cty_Belgium}{CPT_uniform}{kappa_prior_mean_se}{0.095}
\setres{cty_Belgium}{CPT_uniform}{kappa_se}{0.154}
\setres{cty_Belgium}{CPT_uniform}{mse}{33.86}
\setres{cty_Belgium}{CPT_uniform}{mse_prior_mean}{16.50}
\setres{cty_Belgium}{CPT_uniform}{nrmse_pct}{22.3}
\setres{cty_Belgium}{CPT_uniform}{pred_variance}{17.37}
\setres{cty_Belgium}{CPT_uniform}{r_ff}{---}
\setres{cty_Belgium}{CPT_uniform}{r_ff_se}{---}
\setres{cty_Belgium}{CPT_uniform}{r_prior}{---}
\setres{cty_Belgium}{CPT_uniform}{r_prior_se}{---}
\setres{cty_Belgium}{CPT_uniform}{r_prob}{---}
\setres{cty_Belgium}{CPT_uniform}{r_prob_se}{---}
\setres{cty_Belgium}{CPT_uniform}{spread_penalty}{1.532}
\setres{cty_Belgium}{CPT_uniform}{spread_penalty_se}{0.092}
\setres{cty_Belgium}{CPT_uniform}{volume_ess}{---}
\setres{cty_Belgium}{DA_deterministic}{kappa}{0.943}
\setres{cty_Belgium}{DA_deterministic}{kappa_se}{0.017}
\setres{cty_Belgium}{DA_deterministic}{mse}{9.26}
\setres{cty_Belgium}{DA_deterministic}{nrmse_pct}{17.0}
\setres{cty_Belgium}{DA_deterministic}{r_ff}{---}
\setres{cty_Belgium}{DA_deterministic}{r_ff_se}{---}
\setres{cty_Belgium}{DA_deterministic}{r_prior}{---}
\setres{cty_Belgium}{DA_deterministic}{r_prior_se}{---}
\setres{cty_Belgium}{DA_deterministic}{r_prob}{---}
\setres{cty_Belgium}{DA_deterministic}{r_prob_se}{---}
\setres{cty_Belgium}{DA_deterministic}{theta_star}{1.37, 0.95}
\setres{cty_Belgium}{DA_deterministic}{volume_ess}{---}
\setres{cty_Belgium}{DA_lit}{kappa}{-0.155}
\setres{cty_Belgium}{DA_lit}{kappa_prior_mean}{0.264}
\setres{cty_Belgium}{DA_lit}{kappa_prior_mean_se}{0.119}
\setres{cty_Belgium}{DA_lit}{kappa_se}{0.136}
\setres{cty_Belgium}{DA_lit}{mse}{21.70}
\setres{cty_Belgium}{DA_lit}{mse_prior_mean}{16.95}
\setres{cty_Belgium}{DA_lit}{nrmse_pct}{21.8}
\setres{cty_Belgium}{DA_lit}{pred_variance}{4.75}
\setres{cty_Belgium}{DA_lit}{r_ff}{---}
\setres{cty_Belgium}{DA_lit}{r_ff_se}{---}
\setres{cty_Belgium}{DA_lit}{r_prior}{---}
\setres{cty_Belgium}{DA_lit}{r_prior_se}{---}
\setres{cty_Belgium}{DA_lit}{r_prob}{---}
\setres{cty_Belgium}{DA_lit}{r_prob_se}{---}
\setres{cty_Belgium}{DA_lit}{spread_penalty}{0.419}
\setres{cty_Belgium}{DA_lit}{spread_penalty_se}{0.025}
\setres{cty_Belgium}{DA_lit}{volume_ess}{---}
\setres{cty_Belgium}{DA_transfer_others}{kappa}{0.928}
\setres{cty_Belgium}{DA_transfer_others}{kappa_prior_mean}{0.928}
\setres{cty_Belgium}{DA_transfer_others}{kappa_prior_mean_se}{0.021}
\setres{cty_Belgium}{DA_transfer_others}{kappa_se}{0.021}
\setres{cty_Belgium}{DA_transfer_others}{mse}{9.43}
\setres{cty_Belgium}{DA_transfer_others}{mse_prior_mean}{9.43}
\setres{cty_Belgium}{DA_transfer_others}{nrmse_pct}{17.2}
\setres{cty_Belgium}{DA_transfer_others}{pred_variance}{0.00}
\setres{cty_Belgium}{DA_transfer_others}{r_ff}{---}
\setres{cty_Belgium}{DA_transfer_others}{r_ff_se}{---}
\setres{cty_Belgium}{DA_transfer_others}{r_prior}{---}
\setres{cty_Belgium}{DA_transfer_others}{r_prior_se}{---}
\setres{cty_Belgium}{DA_transfer_others}{r_prob}{---}
\setres{cty_Belgium}{DA_transfer_others}{r_prob_se}{---}
\setres{cty_Belgium}{DA_transfer_others}{spread_penalty}{0.000}
\setres{cty_Belgium}{DA_transfer_others}{spread_penalty_se}{0.000}
\setres{cty_Belgium}{DA_transfer_others}{theta_star}{1.71, 1.47}
\setres{cty_Belgium}{DA_transfer_others}{transfer}{others}
\setres{cty_Belgium}{DA_transfer_others}{volume_ess}{---}
\setres{cty_Belgium}{DA_uniform}{kappa}{-1.625}
\setres{cty_Belgium}{DA_uniform}{kappa_prior_mean}{-0.422}
\setres{cty_Belgium}{DA_uniform}{kappa_prior_mean_se}{0.182}
\setres{cty_Belgium}{DA_uniform}{kappa_se}{0.232}
\setres{cty_Belgium}{DA_uniform}{mse}{38.37}
\setres{cty_Belgium}{DA_uniform}{mse_prior_mean}{24.73}
\setres{cty_Belgium}{DA_uniform}{nrmse_pct}{28.7}
\setres{cty_Belgium}{DA_uniform}{pred_variance}{13.65}
\setres{cty_Belgium}{DA_uniform}{r_ff}{---}
\setres{cty_Belgium}{DA_uniform}{r_ff_se}{---}
\setres{cty_Belgium}{DA_uniform}{r_prior}{---}
\setres{cty_Belgium}{DA_uniform}{r_prior_se}{---}
\setres{cty_Belgium}{DA_uniform}{r_prob}{---}
\setres{cty_Belgium}{DA_uniform}{r_prob_se}{---}
\setres{cty_Belgium}{DA_uniform}{spread_penalty}{1.204}
\setres{cty_Belgium}{DA_uniform}{spread_penalty_se}{0.072}
\setres{cty_Belgium}{DA_uniform}{volume_ess}{---}
\setres{cty_Belgium}{EU_deterministic}{kappa}{0.898}
\setres{cty_Belgium}{EU_deterministic}{kappa_se}{0.027}
\setres{cty_Belgium}{EU_deterministic}{mse}{9.77}
\setres{cty_Belgium}{EU_deterministic}{nrmse_pct}{17.8}
\setres{cty_Belgium}{EU_deterministic}{r_ff}{---}
\setres{cty_Belgium}{EU_deterministic}{r_ff_se}{---}
\setres{cty_Belgium}{EU_deterministic}{r_prior}{---}
\setres{cty_Belgium}{EU_deterministic}{r_prior_se}{---}
\setres{cty_Belgium}{EU_deterministic}{r_prob}{---}
\setres{cty_Belgium}{EU_deterministic}{r_prob_se}{---}
\setres{cty_Belgium}{EU_deterministic}{theta_star}{0.87}
\setres{cty_Belgium}{EU_deterministic}{volume_ess}{---}
\setres{cty_Belgium}{EU_lit}{kappa}{0.679}
\setres{cty_Belgium}{EU_lit}{kappa_prior_mean}{0.785}
\setres{cty_Belgium}{EU_lit}{kappa_prior_mean_se}{0.047}
\setres{cty_Belgium}{EU_lit}{kappa_se}{0.052}
\setres{cty_Belgium}{EU_lit}{mse}{12.25}
\setres{cty_Belgium}{EU_lit}{mse_prior_mean}{11.04}
\setres{cty_Belgium}{EU_lit}{nrmse_pct}{19.2}
\setres{cty_Belgium}{EU_lit}{pred_variance}{1.20}
\setres{cty_Belgium}{EU_lit}{r_ff}{---}
\setres{cty_Belgium}{EU_lit}{r_ff_se}{---}
\setres{cty_Belgium}{EU_lit}{r_prior}{---}
\setres{cty_Belgium}{EU_lit}{r_prior_se}{---}
\setres{cty_Belgium}{EU_lit}{r_prob}{---}
\setres{cty_Belgium}{EU_lit}{r_prob_se}{---}
\setres{cty_Belgium}{EU_lit}{spread_penalty}{0.106}
\setres{cty_Belgium}{EU_lit}{spread_penalty_se}{0.006}
\setres{cty_Belgium}{EU_lit}{volume_ess}{---}
\setres{cty_Belgium}{EU_transfer_others}{kappa}{0.885}
\setres{cty_Belgium}{EU_transfer_others}{kappa_prior_mean}{0.885}
\setres{cty_Belgium}{EU_transfer_others}{kappa_prior_mean_se}{0.036}
\setres{cty_Belgium}{EU_transfer_others}{kappa_se}{0.036}
\setres{cty_Belgium}{EU_transfer_others}{mse}{9.91}
\setres{cty_Belgium}{EU_transfer_others}{mse_prior_mean}{9.91}
\setres{cty_Belgium}{EU_transfer_others}{nrmse_pct}{17.8}
\setres{cty_Belgium}{EU_transfer_others}{pred_variance}{0.00}
\setres{cty_Belgium}{EU_transfer_others}{r_ff}{---}
\setres{cty_Belgium}{EU_transfer_others}{r_ff_se}{---}
\setres{cty_Belgium}{EU_transfer_others}{r_prior}{---}
\setres{cty_Belgium}{EU_transfer_others}{r_prior_se}{---}
\setres{cty_Belgium}{EU_transfer_others}{r_prob}{---}
\setres{cty_Belgium}{EU_transfer_others}{r_prob_se}{---}
\setres{cty_Belgium}{EU_transfer_others}{spread_penalty}{0.000}
\setres{cty_Belgium}{EU_transfer_others}{spread_penalty_se}{0.000}
\setres{cty_Belgium}{EU_transfer_others}{theta_star}{0.95}
\setres{cty_Belgium}{EU_transfer_others}{transfer}{others}
\setres{cty_Belgium}{EU_transfer_others}{volume_ess}{---}
\setres{cty_Belgium}{EU_uniform}{kappa}{-0.591}
\setres{cty_Belgium}{EU_uniform}{kappa_prior_mean}{0.132}
\setres{cty_Belgium}{EU_uniform}{kappa_prior_mean_se}{0.129}
\setres{cty_Belgium}{EU_uniform}{kappa_se}{0.159}
\setres{cty_Belgium}{EU_uniform}{mse}{26.64}
\setres{cty_Belgium}{EU_uniform}{mse_prior_mean}{18.45}
\setres{cty_Belgium}{EU_uniform}{nrmse_pct}{22.8}
\setres{cty_Belgium}{EU_uniform}{pred_variance}{8.20}
\setres{cty_Belgium}{EU_uniform}{r_ff}{---}
\setres{cty_Belgium}{EU_uniform}{r_ff_se}{---}
\setres{cty_Belgium}{EU_uniform}{r_prior}{---}
\setres{cty_Belgium}{EU_uniform}{r_prior_se}{---}
\setres{cty_Belgium}{EU_uniform}{r_prob}{---}
\setres{cty_Belgium}{EU_uniform}{r_prob_se}{---}
\setres{cty_Belgium}{EU_uniform}{spread_penalty}{0.723}
\setres{cty_Belgium}{EU_uniform}{spread_penalty_se}{0.043}
\setres{cty_Belgium}{EU_uniform}{volume_ess}{---}
\setres{cty_Belgium}{EV_deterministic}{kappa}{0.859}
\setres{cty_Belgium}{EV_deterministic}{kappa_se}{0.044}
\setres{cty_Belgium}{EV_deterministic}{mse}{10.21}
\setres{cty_Belgium}{EV_deterministic}{nrmse_pct}{17.4}
\setres{cty_Belgium}{EV_deterministic}{r_ff}{---}
\setres{cty_Belgium}{EV_deterministic}{r_ff_se}{---}
\setres{cty_Belgium}{EV_deterministic}{r_prior}{---}
\setres{cty_Belgium}{EV_deterministic}{r_prior_se}{---}
\setres{cty_Belgium}{EV_deterministic}{r_prob}{---}
\setres{cty_Belgium}{EV_deterministic}{r_prob_se}{---}
\setres{cty_Belgium}{EV_deterministic}{volume_ess}{---}
\setres{cty_Belgium}{reference}{country_name}{Belgium}
\setres{cty_Belgium}{reference}{lookup_nrmse_pct}{16.0}
\setres{cty_Belgium}{reference}{midpoint_mse}{19.95}
\setres{cty_Belgium}{reference}{midpoint_nrmse_pct}{22.5}
\setres{cty_Belgium}{reference}{missing_lotteries}{none}
\setres{cty_Belgium}{reference}{n_lotteries}{14}
\setres{cty_Belgium}{reference}{n_obs}{1{,}274}
\setres{cty_Belgium}{reference}{n_subjects}{91}
\setres{cty_Belgium}{reference}{oracle_mse}{8.61}
\setres{cty_Belgium}{reference}{sample}{LV2019\_Belgium}
\setres{cty_Brazil}{CPT_deterministic}{kappa}{0.957}
\setres{cty_Brazil}{CPT_deterministic}{kappa_se}{0.015}
\setres{cty_Brazil}{CPT_deterministic}{mse}{10.95}
\setres{cty_Brazil}{CPT_deterministic}{nrmse_pct}{17.7}
\setres{cty_Brazil}{CPT_deterministic}{r_ff}{---}
\setres{cty_Brazil}{CPT_deterministic}{r_ff_se}{---}
\setres{cty_Brazil}{CPT_deterministic}{r_prior}{---}
\setres{cty_Brazil}{CPT_deterministic}{r_prior_se}{---}
\setres{cty_Brazil}{CPT_deterministic}{r_prob}{---}
\setres{cty_Brazil}{CPT_deterministic}{r_prob_se}{---}
\setres{cty_Brazil}{CPT_deterministic}{theta_star}{1.10, 0.70, 0.80}
\setres{cty_Brazil}{CPT_deterministic}{volume_ess}{---}
\setres{cty_Brazil}{CPT_lit}{kappa}{0.448}
\setres{cty_Brazil}{CPT_lit}{kappa_prior_mean}{0.731}
\setres{cty_Brazil}{CPT_lit}{kappa_prior_mean_se}{0.073}
\setres{cty_Brazil}{CPT_lit}{kappa_se}{0.078}
\setres{cty_Brazil}{CPT_lit}{mse}{16.17}
\setres{cty_Brazil}{CPT_lit}{mse_prior_mean}{13.27}
\setres{cty_Brazil}{CPT_lit}{nrmse_pct}{19.8}
\setres{cty_Brazil}{CPT_lit}{pred_variance}{2.90}
\setres{cty_Brazil}{CPT_lit}{r_ff}{---}
\setres{cty_Brazil}{CPT_lit}{r_ff_se}{---}
\setres{cty_Brazil}{CPT_lit}{r_prior}{---}
\setres{cty_Brazil}{CPT_lit}{r_prior_se}{---}
\setres{cty_Brazil}{CPT_lit}{r_prob}{---}
\setres{cty_Brazil}{CPT_lit}{r_prob_se}{---}
\setres{cty_Brazil}{CPT_lit}{spread_penalty}{0.283}
\setres{cty_Brazil}{CPT_lit}{spread_penalty_se}{0.014}
\setres{cty_Brazil}{CPT_lit}{volume_ess}{---}
\setres{cty_Brazil}{CPT_transfer_others}{kappa}{0.976}
\setres{cty_Brazil}{CPT_transfer_others}{kappa_prior_mean}{0.976}
\setres{cty_Brazil}{CPT_transfer_others}{kappa_prior_mean_se}{0.015}
\setres{cty_Brazil}{CPT_transfer_others}{kappa_se}{0.015}
\setres{cty_Brazil}{CPT_transfer_others}{mse}{10.76}
\setres{cty_Brazil}{CPT_transfer_others}{mse_prior_mean}{10.76}
\setres{cty_Brazil}{CPT_transfer_others}{nrmse_pct}{17.9}
\setres{cty_Brazil}{CPT_transfer_others}{pred_variance}{0.00}
\setres{cty_Brazil}{CPT_transfer_others}{r_ff}{---}
\setres{cty_Brazil}{CPT_transfer_others}{r_ff_se}{---}
\setres{cty_Brazil}{CPT_transfer_others}{r_prior}{---}
\setres{cty_Brazil}{CPT_transfer_others}{r_prior_se}{---}
\setres{cty_Brazil}{CPT_transfer_others}{r_prob}{---}
\setres{cty_Brazil}{CPT_transfer_others}{r_prob_se}{---}
\setres{cty_Brazil}{CPT_transfer_others}{spread_penalty}{0.000}
\setres{cty_Brazil}{CPT_transfer_others}{spread_penalty_se}{0.000}
\setres{cty_Brazil}{CPT_transfer_others}{theta_star}{1.00, 0.60, 0.90}
\setres{cty_Brazil}{CPT_transfer_others}{transfer}{others}
\setres{cty_Brazil}{CPT_transfer_others}{volume_ess}{---}
\setres{cty_Brazil}{CPT_uniform}{kappa}{-1.707}
\setres{cty_Brazil}{CPT_uniform}{kappa_prior_mean}{-0.010}
\setres{cty_Brazil}{CPT_uniform}{kappa_prior_mean_se}{0.133}
\setres{cty_Brazil}{CPT_uniform}{kappa_se}{0.175}
\setres{cty_Brazil}{CPT_uniform}{mse}{38.22}
\setres{cty_Brazil}{CPT_uniform}{mse_prior_mean}{20.85}
\setres{cty_Brazil}{CPT_uniform}{nrmse_pct}{24.7}
\setres{cty_Brazil}{CPT_uniform}{pred_variance}{17.37}
\setres{cty_Brazil}{CPT_uniform}{r_ff}{---}
\setres{cty_Brazil}{CPT_uniform}{r_ff_se}{---}
\setres{cty_Brazil}{CPT_uniform}{r_prior}{---}
\setres{cty_Brazil}{CPT_uniform}{r_prior_se}{---}
\setres{cty_Brazil}{CPT_uniform}{r_prob}{---}
\setres{cty_Brazil}{CPT_uniform}{r_prob_se}{---}
\setres{cty_Brazil}{CPT_uniform}{spread_penalty}{1.696}
\setres{cty_Brazil}{CPT_uniform}{spread_penalty_se}{0.086}
\setres{cty_Brazil}{CPT_uniform}{volume_ess}{---}
\setres{cty_Brazil}{DA_deterministic}{kappa}{0.916}
\setres{cty_Brazil}{DA_deterministic}{kappa_se}{0.021}
\setres{cty_Brazil}{DA_deterministic}{mse}{11.37}
\setres{cty_Brazil}{DA_deterministic}{nrmse_pct}{17.8}
\setres{cty_Brazil}{DA_deterministic}{r_ff}{---}
\setres{cty_Brazil}{DA_deterministic}{r_ff_se}{---}
\setres{cty_Brazil}{DA_deterministic}{r_prior}{---}
\setres{cty_Brazil}{DA_deterministic}{r_prior_se}{---}
\setres{cty_Brazil}{DA_deterministic}{r_prob}{---}
\setres{cty_Brazil}{DA_deterministic}{r_prob_se}{---}
\setres{cty_Brazil}{DA_deterministic}{theta_star}{1.44, 0.78}
\setres{cty_Brazil}{DA_deterministic}{volume_ess}{---}
\setres{cty_Brazil}{DA_lit}{kappa}{-0.621}
\setres{cty_Brazil}{DA_lit}{kappa_prior_mean}{-0.157}
\setres{cty_Brazil}{DA_lit}{kappa_prior_mean_se}{0.167}
\setres{cty_Brazil}{DA_lit}{kappa_se}{0.179}
\setres{cty_Brazil}{DA_lit}{mse}{27.10}
\setres{cty_Brazil}{DA_lit}{mse_prior_mean}{22.36}
\setres{cty_Brazil}{DA_lit}{nrmse_pct}{26.0}
\setres{cty_Brazil}{DA_lit}{pred_variance}{4.75}
\setres{cty_Brazil}{DA_lit}{r_ff}{---}
\setres{cty_Brazil}{DA_lit}{r_ff_se}{---}
\setres{cty_Brazil}{DA_lit}{r_prior}{---}
\setres{cty_Brazil}{DA_lit}{r_prior_se}{---}
\setres{cty_Brazil}{DA_lit}{r_prob}{---}
\setres{cty_Brazil}{DA_lit}{r_prob_se}{---}
\setres{cty_Brazil}{DA_lit}{spread_penalty}{0.464}
\setres{cty_Brazil}{DA_lit}{spread_penalty_se}{0.024}
\setres{cty_Brazil}{DA_lit}{volume_ess}{---}
\setres{cty_Brazil}{DA_transfer_others}{kappa}{0.930}
\setres{cty_Brazil}{DA_transfer_others}{kappa_prior_mean}{0.930}
\setres{cty_Brazil}{DA_transfer_others}{kappa_prior_mean_se}{0.025}
\setres{cty_Brazil}{DA_transfer_others}{kappa_se}{0.025}
\setres{cty_Brazil}{DA_transfer_others}{mse}{11.23}
\setres{cty_Brazil}{DA_transfer_others}{mse_prior_mean}{11.23}
\setres{cty_Brazil}{DA_transfer_others}{nrmse_pct}{17.4}
\setres{cty_Brazil}{DA_transfer_others}{pred_variance}{0.00}
\setres{cty_Brazil}{DA_transfer_others}{r_ff}{---}
\setres{cty_Brazil}{DA_transfer_others}{r_ff_se}{---}
\setres{cty_Brazil}{DA_transfer_others}{r_prior}{---}
\setres{cty_Brazil}{DA_transfer_others}{r_prior_se}{---}
\setres{cty_Brazil}{DA_transfer_others}{r_prob}{---}
\setres{cty_Brazil}{DA_transfer_others}{r_prob_se}{---}
\setres{cty_Brazil}{DA_transfer_others}{spread_penalty}{0.000}
\setres{cty_Brazil}{DA_transfer_others}{spread_penalty_se}{0.000}
\setres{cty_Brazil}{DA_transfer_others}{theta_star}{1.70, 1.47}
\setres{cty_Brazil}{DA_transfer_others}{transfer}{others}
\setres{cty_Brazil}{DA_transfer_others}{volume_ess}{---}
\setres{cty_Brazil}{DA_uniform}{kappa}{-2.371}
\setres{cty_Brazil}{DA_uniform}{kappa_prior_mean}{-1.038}
\setres{cty_Brazil}{DA_uniform}{kappa_prior_mean_se}{0.233}
\setres{cty_Brazil}{DA_uniform}{kappa_se}{0.270}
\setres{cty_Brazil}{DA_uniform}{mse}{45.02}
\setres{cty_Brazil}{DA_uniform}{mse_prior_mean}{31.37}
\setres{cty_Brazil}{DA_uniform}{nrmse_pct}{31.5}
\setres{cty_Brazil}{DA_uniform}{pred_variance}{13.65}
\setres{cty_Brazil}{DA_uniform}{r_ff}{---}
\setres{cty_Brazil}{DA_uniform}{r_ff_se}{---}
\setres{cty_Brazil}{DA_uniform}{r_prior}{---}
\setres{cty_Brazil}{DA_uniform}{r_prior_se}{---}
\setres{cty_Brazil}{DA_uniform}{r_prob}{---}
\setres{cty_Brazil}{DA_uniform}{r_prob_se}{---}
\setres{cty_Brazil}{DA_uniform}{spread_penalty}{1.333}
\setres{cty_Brazil}{DA_uniform}{spread_penalty_se}{0.068}
\setres{cty_Brazil}{DA_uniform}{volume_ess}{---}
\setres{cty_Brazil}{EU_deterministic}{kappa}{0.885}
\setres{cty_Brazil}{EU_deterministic}{kappa_se}{0.027}
\setres{cty_Brazil}{EU_deterministic}{mse}{11.69}
\setres{cty_Brazil}{EU_deterministic}{nrmse_pct}{18.8}
\setres{cty_Brazil}{EU_deterministic}{r_ff}{---}
\setres{cty_Brazil}{EU_deterministic}{r_ff_se}{---}
\setres{cty_Brazil}{EU_deterministic}{r_prior}{---}
\setres{cty_Brazil}{EU_deterministic}{r_prior_se}{---}
\setres{cty_Brazil}{EU_deterministic}{r_prob}{---}
\setres{cty_Brazil}{EU_deterministic}{r_prob_se}{---}
\setres{cty_Brazil}{EU_deterministic}{theta_star}{0.98}
\setres{cty_Brazil}{EU_deterministic}{volume_ess}{---}
\setres{cty_Brazil}{EU_lit}{kappa}{0.489}
\setres{cty_Brazil}{EU_lit}{kappa_prior_mean}{0.607}
\setres{cty_Brazil}{EU_lit}{kappa_prior_mean_se}{0.082}
\setres{cty_Brazil}{EU_lit}{kappa_se}{0.085}
\setres{cty_Brazil}{EU_lit}{mse}{15.74}
\setres{cty_Brazil}{EU_lit}{mse_prior_mean}{14.54}
\setres{cty_Brazil}{EU_lit}{nrmse_pct}{21.1}
\setres{cty_Brazil}{EU_lit}{pred_variance}{1.20}
\setres{cty_Brazil}{EU_lit}{r_ff}{---}
\setres{cty_Brazil}{EU_lit}{r_ff_se}{---}
\setres{cty_Brazil}{EU_lit}{r_prior}{---}
\setres{cty_Brazil}{EU_lit}{r_prior_se}{---}
\setres{cty_Brazil}{EU_lit}{r_prob}{---}
\setres{cty_Brazil}{EU_lit}{r_prob_se}{---}
\setres{cty_Brazil}{EU_lit}{spread_penalty}{0.118}
\setres{cty_Brazil}{EU_lit}{spread_penalty_se}{0.006}
\setres{cty_Brazil}{EU_lit}{volume_ess}{---}
\setres{cty_Brazil}{EU_transfer_others}{kappa}{0.903}
\setres{cty_Brazil}{EU_transfer_others}{kappa_prior_mean}{0.903}
\setres{cty_Brazil}{EU_transfer_others}{kappa_prior_mean_se}{0.025}
\setres{cty_Brazil}{EU_transfer_others}{kappa_se}{0.025}
\setres{cty_Brazil}{EU_transfer_others}{mse}{11.50}
\setres{cty_Brazil}{EU_transfer_others}{mse_prior_mean}{11.50}
\setres{cty_Brazil}{EU_transfer_others}{nrmse_pct}{18.7}
\setres{cty_Brazil}{EU_transfer_others}{pred_variance}{0.00}
\setres{cty_Brazil}{EU_transfer_others}{r_ff}{---}
\setres{cty_Brazil}{EU_transfer_others}{r_ff_se}{---}
\setres{cty_Brazil}{EU_transfer_others}{r_prior}{---}
\setres{cty_Brazil}{EU_transfer_others}{r_prior_se}{---}
\setres{cty_Brazil}{EU_transfer_others}{r_prob}{---}
\setres{cty_Brazil}{EU_transfer_others}{r_prob_se}{---}
\setres{cty_Brazil}{EU_transfer_others}{spread_penalty}{0.000}
\setres{cty_Brazil}{EU_transfer_others}{spread_penalty_se}{0.000}
\setres{cty_Brazil}{EU_transfer_others}{theta_star}{0.94}
\setres{cty_Brazil}{EU_transfer_others}{transfer}{others}
\setres{cty_Brazil}{EU_transfer_others}{volume_ess}{---}
\setres{cty_Brazil}{EU_uniform}{kappa}{-1.118}
\setres{cty_Brazil}{EU_uniform}{kappa_prior_mean}{-0.317}
\setres{cty_Brazil}{EU_uniform}{kappa_prior_mean_se}{0.177}
\setres{cty_Brazil}{EU_uniform}{kappa_se}{0.198}
\setres{cty_Brazil}{EU_uniform}{mse}{32.19}
\setres{cty_Brazil}{EU_uniform}{mse_prior_mean}{23.99}
\setres{cty_Brazil}{EU_uniform}{nrmse_pct}{26.7}
\setres{cty_Brazil}{EU_uniform}{pred_variance}{8.20}
\setres{cty_Brazil}{EU_uniform}{r_ff}{---}
\setres{cty_Brazil}{EU_uniform}{r_ff_se}{---}
\setres{cty_Brazil}{EU_uniform}{r_prior}{---}
\setres{cty_Brazil}{EU_uniform}{r_prior_se}{---}
\setres{cty_Brazil}{EU_uniform}{r_prob}{---}
\setres{cty_Brazil}{EU_uniform}{r_prob_se}{---}
\setres{cty_Brazil}{EU_uniform}{spread_penalty}{0.801}
\setres{cty_Brazil}{EU_uniform}{spread_penalty_se}{0.041}
\setres{cty_Brazil}{EU_uniform}{volume_ess}{---}
\setres{cty_Brazil}{EV_deterministic}{kappa}{0.906}
\setres{cty_Brazil}{EV_deterministic}{kappa_se}{0.029}
\setres{cty_Brazil}{EV_deterministic}{mse}{11.47}
\setres{cty_Brazil}{EV_deterministic}{nrmse_pct}{18.6}
\setres{cty_Brazil}{EV_deterministic}{r_ff}{---}
\setres{cty_Brazil}{EV_deterministic}{r_ff_se}{---}
\setres{cty_Brazil}{EV_deterministic}{r_prior}{---}
\setres{cty_Brazil}{EV_deterministic}{r_prior_se}{---}
\setres{cty_Brazil}{EV_deterministic}{r_prob}{---}
\setres{cty_Brazil}{EV_deterministic}{r_prob_se}{---}
\setres{cty_Brazil}{EV_deterministic}{volume_ess}{---}
\setres{cty_Brazil}{reference}{country_name}{Brazil}
\setres{cty_Brazil}{reference}{lookup_nrmse_pct}{17.4}
\setres{cty_Brazil}{reference}{midpoint_mse}{20.75}
\setres{cty_Brazil}{reference}{midpoint_nrmse_pct}{24.0}
\setres{cty_Brazil}{reference}{missing_lotteries}{none}
\setres{cty_Brazil}{reference}{n_lotteries}{14}
\setres{cty_Brazil}{reference}{n_obs}{1{,}176}
\setres{cty_Brazil}{reference}{n_subjects}{84}
\setres{cty_Brazil}{reference}{oracle_mse}{10.51}
\setres{cty_Brazil}{reference}{sample}{LV2019\_Brazil}
\setres{cty_Cambodia}{CPT_deterministic}{kappa}{0.983}
\setres{cty_Cambodia}{CPT_deterministic}{kappa_se}{0.034}
\setres{cty_Cambodia}{CPT_deterministic}{mse}{17.83}
\setres{cty_Cambodia}{CPT_deterministic}{nrmse_pct}{23.9}
\setres{cty_Cambodia}{CPT_deterministic}{r_ff}{---}
\setres{cty_Cambodia}{CPT_deterministic}{r_ff_se}{---}
\setres{cty_Cambodia}{CPT_deterministic}{r_prior}{---}
\setres{cty_Cambodia}{CPT_deterministic}{r_prior_se}{---}
\setres{cty_Cambodia}{CPT_deterministic}{r_prob}{---}
\setres{cty_Cambodia}{CPT_deterministic}{r_prob_se}{---}
\setres{cty_Cambodia}{CPT_deterministic}{theta_star}{1.10, 0.50, 0.80}
\setres{cty_Cambodia}{CPT_deterministic}{volume_ess}{---}
\setres{cty_Cambodia}{CPT_lit}{kappa}{-0.055}
\setres{cty_Cambodia}{CPT_lit}{kappa_prior_mean}{0.452}
\setres{cty_Cambodia}{CPT_lit}{kappa_prior_mean_se}{0.216}
\setres{cty_Cambodia}{CPT_lit}{kappa_se}{0.297}
\setres{cty_Cambodia}{CPT_lit}{mse}{23.77}
\setres{cty_Cambodia}{CPT_lit}{mse_prior_mean}{20.87}
\setres{cty_Cambodia}{CPT_lit}{nrmse_pct}{26.2}
\setres{cty_Cambodia}{CPT_lit}{pred_variance}{2.90}
\setres{cty_Cambodia}{CPT_lit}{r_ff}{---}
\setres{cty_Cambodia}{CPT_lit}{r_ff_se}{---}
\setres{cty_Cambodia}{CPT_lit}{r_prior}{---}
\setres{cty_Cambodia}{CPT_lit}{r_prior_se}{---}
\setres{cty_Cambodia}{CPT_lit}{r_prob}{---}
\setres{cty_Cambodia}{CPT_lit}{r_prob_se}{---}
\setres{cty_Cambodia}{CPT_lit}{spread_penalty}{0.507}
\setres{cty_Cambodia}{CPT_lit}{spread_penalty_se}{0.089}
\setres{cty_Cambodia}{CPT_lit}{volume_ess}{---}
\setres{cty_Cambodia}{CPT_transfer_others}{kappa}{0.950}
\setres{cty_Cambodia}{CPT_transfer_others}{kappa_prior_mean}{0.950}
\setres{cty_Cambodia}{CPT_transfer_others}{kappa_prior_mean_se}{0.055}
\setres{cty_Cambodia}{CPT_transfer_others}{kappa_se}{0.055}
\setres{cty_Cambodia}{CPT_transfer_others}{mse}{18.02}
\setres{cty_Cambodia}{CPT_transfer_others}{mse_prior_mean}{18.02}
\setres{cty_Cambodia}{CPT_transfer_others}{nrmse_pct}{23.8}
\setres{cty_Cambodia}{CPT_transfer_others}{pred_variance}{0.00}
\setres{cty_Cambodia}{CPT_transfer_others}{r_ff}{---}
\setres{cty_Cambodia}{CPT_transfer_others}{r_ff_se}{---}
\setres{cty_Cambodia}{CPT_transfer_others}{r_prior}{---}
\setres{cty_Cambodia}{CPT_transfer_others}{r_prior_se}{---}
\setres{cty_Cambodia}{CPT_transfer_others}{r_prob}{---}
\setres{cty_Cambodia}{CPT_transfer_others}{r_prob_se}{---}
\setres{cty_Cambodia}{CPT_transfer_others}{spread_penalty}{0.000}
\setres{cty_Cambodia}{CPT_transfer_others}{spread_penalty_se}{0.000}
\setres{cty_Cambodia}{CPT_transfer_others}{theta_star}{1.00, 0.60, 0.90}
\setres{cty_Cambodia}{CPT_transfer_others}{transfer}{others}
\setres{cty_Cambodia}{CPT_transfer_others}{volume_ess}{---}
\setres{cty_Cambodia}{CPT_uniform}{kappa}{-3.443}
\setres{cty_Cambodia}{CPT_uniform}{kappa_prior_mean}{-0.403}
\setres{cty_Cambodia}{CPT_uniform}{kappa_prior_mean_se}{0.354}
\setres{cty_Cambodia}{CPT_uniform}{kappa_se}{0.825}
\setres{cty_Cambodia}{CPT_uniform}{mse}{43.12}
\setres{cty_Cambodia}{CPT_uniform}{mse_prior_mean}{25.75}
\setres{cty_Cambodia}{CPT_uniform}{nrmse_pct}{28.3}
\setres{cty_Cambodia}{CPT_uniform}{pred_variance}{17.37}
\setres{cty_Cambodia}{CPT_uniform}{r_ff}{---}
\setres{cty_Cambodia}{CPT_uniform}{r_ff_se}{---}
\setres{cty_Cambodia}{CPT_uniform}{r_prior}{---}
\setres{cty_Cambodia}{CPT_uniform}{r_prior_se}{---}
\setres{cty_Cambodia}{CPT_uniform}{r_prob}{---}
\setres{cty_Cambodia}{CPT_uniform}{r_prob_se}{---}
\setres{cty_Cambodia}{CPT_uniform}{spread_penalty}{3.040}
\setres{cty_Cambodia}{CPT_uniform}{spread_penalty_se}{0.535}
\setres{cty_Cambodia}{CPT_uniform}{volume_ess}{---}
\setres{cty_Cambodia}{DA_deterministic}{kappa}{0.659}
\setres{cty_Cambodia}{DA_deterministic}{kappa_se}{0.142}
\setres{cty_Cambodia}{DA_deterministic}{mse}{19.69}
\setres{cty_Cambodia}{DA_deterministic}{nrmse_pct}{24.3}
\setres{cty_Cambodia}{DA_deterministic}{r_ff}{---}
\setres{cty_Cambodia}{DA_deterministic}{r_ff_se}{---}
\setres{cty_Cambodia}{DA_deterministic}{r_prior}{---}
\setres{cty_Cambodia}{DA_deterministic}{r_prior_se}{---}
\setres{cty_Cambodia}{DA_deterministic}{r_prob}{---}
\setres{cty_Cambodia}{DA_deterministic}{r_prob_se}{---}
\setres{cty_Cambodia}{DA_deterministic}{theta_star}{2.24, 3.00}
\setres{cty_Cambodia}{DA_deterministic}{volume_ess}{---}
\setres{cty_Cambodia}{DA_lit}{kappa}{-2.017}
\setres{cty_Cambodia}{DA_lit}{kappa_prior_mean}{-1.186}
\setres{cty_Cambodia}{DA_lit}{kappa_prior_mean_se}{0.595}
\setres{cty_Cambodia}{DA_lit}{kappa_se}{0.727}
\setres{cty_Cambodia}{DA_lit}{mse}{34.97}
\setres{cty_Cambodia}{DA_lit}{mse_prior_mean}{30.22}
\setres{cty_Cambodia}{DA_lit}{nrmse_pct}{31.1}
\setres{cty_Cambodia}{DA_lit}{pred_variance}{4.75}
\setres{cty_Cambodia}{DA_lit}{r_ff}{---}
\setres{cty_Cambodia}{DA_lit}{r_ff_se}{---}
\setres{cty_Cambodia}{DA_lit}{r_prior}{---}
\setres{cty_Cambodia}{DA_lit}{r_prior_se}{---}
\setres{cty_Cambodia}{DA_lit}{r_prob}{---}
\setres{cty_Cambodia}{DA_lit}{r_prob_se}{---}
\setres{cty_Cambodia}{DA_lit}{spread_penalty}{0.831}
\setres{cty_Cambodia}{DA_lit}{spread_penalty_se}{0.146}
\setres{cty_Cambodia}{DA_lit}{volume_ess}{---}
\setres{cty_Cambodia}{DA_transfer_others}{kappa}{0.638}
\setres{cty_Cambodia}{DA_transfer_others}{kappa_prior_mean}{0.638}
\setres{cty_Cambodia}{DA_transfer_others}{kappa_prior_mean_se}{0.166}
\setres{cty_Cambodia}{DA_transfer_others}{kappa_se}{0.166}
\setres{cty_Cambodia}{DA_transfer_others}{mse}{19.81}
\setres{cty_Cambodia}{DA_transfer_others}{mse_prior_mean}{19.81}
\setres{cty_Cambodia}{DA_transfer_others}{nrmse_pct}{25.0}
\setres{cty_Cambodia}{DA_transfer_others}{pred_variance}{0.00}
\setres{cty_Cambodia}{DA_transfer_others}{r_ff}{---}
\setres{cty_Cambodia}{DA_transfer_others}{r_ff_se}{---}
\setres{cty_Cambodia}{DA_transfer_others}{r_prior}{---}
\setres{cty_Cambodia}{DA_transfer_others}{r_prior_se}{---}
\setres{cty_Cambodia}{DA_transfer_others}{r_prob}{---}
\setres{cty_Cambodia}{DA_transfer_others}{r_prob_se}{---}
\setres{cty_Cambodia}{DA_transfer_others}{spread_penalty}{0.000}
\setres{cty_Cambodia}{DA_transfer_others}{spread_penalty_se}{0.000}
\setres{cty_Cambodia}{DA_transfer_others}{theta_star}{1.68, 1.41}
\setres{cty_Cambodia}{DA_transfer_others}{transfer}{others}
\setres{cty_Cambodia}{DA_transfer_others}{volume_ess}{---}
\setres{cty_Cambodia}{DA_uniform}{kappa}{-4.801}
\setres{cty_Cambodia}{DA_uniform}{kappa_prior_mean}{-2.412}
\setres{cty_Cambodia}{DA_uniform}{kappa_prior_mean_se}{0.805}
\setres{cty_Cambodia}{DA_uniform}{kappa_se}{1.185}
\setres{cty_Cambodia}{DA_uniform}{mse}{50.87}
\setres{cty_Cambodia}{DA_uniform}{mse_prior_mean}{37.23}
\setres{cty_Cambodia}{DA_uniform}{nrmse_pct}{33.2}
\setres{cty_Cambodia}{DA_uniform}{pred_variance}{13.65}
\setres{cty_Cambodia}{DA_uniform}{r_ff}{---}
\setres{cty_Cambodia}{DA_uniform}{r_ff_se}{---}
\setres{cty_Cambodia}{DA_uniform}{r_prior}{---}
\setres{cty_Cambodia}{DA_uniform}{r_prior_se}{---}
\setres{cty_Cambodia}{DA_uniform}{r_prob}{---}
\setres{cty_Cambodia}{DA_uniform}{r_prob_se}{---}
\setres{cty_Cambodia}{DA_uniform}{spread_penalty}{2.389}
\setres{cty_Cambodia}{DA_uniform}{spread_penalty_se}{0.420}
\setres{cty_Cambodia}{DA_uniform}{volume_ess}{---}
\setres{cty_Cambodia}{EU_deterministic}{kappa}{0.394}
\setres{cty_Cambodia}{EU_deterministic}{kappa_se}{0.270}
\setres{cty_Cambodia}{EU_deterministic}{mse}{21.20}
\setres{cty_Cambodia}{EU_deterministic}{nrmse_pct}{25.2}
\setres{cty_Cambodia}{EU_deterministic}{r_ff}{---}
\setres{cty_Cambodia}{EU_deterministic}{r_ff_se}{---}
\setres{cty_Cambodia}{EU_deterministic}{r_prior}{---}
\setres{cty_Cambodia}{EU_deterministic}{r_prior_se}{---}
\setres{cty_Cambodia}{EU_deterministic}{r_prob}{---}
\setres{cty_Cambodia}{EU_deterministic}{r_prob_se}{---}
\setres{cty_Cambodia}{EU_deterministic}{theta_star}{0.96}
\setres{cty_Cambodia}{EU_deterministic}{volume_ess}{---}
\setres{cty_Cambodia}{EU_lit}{kappa}{-0.246}
\setres{cty_Cambodia}{EU_lit}{kappa_prior_mean}{-0.035}
\setres{cty_Cambodia}{EU_lit}{kappa_prior_mean_se}{0.356}
\setres{cty_Cambodia}{EU_lit}{kappa_se}{0.391}
\setres{cty_Cambodia}{EU_lit}{mse}{24.86}
\setres{cty_Cambodia}{EU_lit}{mse_prior_mean}{23.65}
\setres{cty_Cambodia}{EU_lit}{nrmse_pct}{26.6}
\setres{cty_Cambodia}{EU_lit}{pred_variance}{1.20}
\setres{cty_Cambodia}{EU_lit}{r_ff}{---}
\setres{cty_Cambodia}{EU_lit}{r_ff_se}{---}
\setres{cty_Cambodia}{EU_lit}{r_prior}{---}
\setres{cty_Cambodia}{EU_lit}{r_prior_se}{---}
\setres{cty_Cambodia}{EU_lit}{r_prob}{---}
\setres{cty_Cambodia}{EU_lit}{r_prob_se}{---}
\setres{cty_Cambodia}{EU_lit}{spread_penalty}{0.211}
\setres{cty_Cambodia}{EU_lit}{spread_penalty_se}{0.037}
\setres{cty_Cambodia}{EU_lit}{volume_ess}{---}
\setres{cty_Cambodia}{EU_transfer_others}{kappa}{0.408}
\setres{cty_Cambodia}{EU_transfer_others}{kappa_prior_mean}{0.408}
\setres{cty_Cambodia}{EU_transfer_others}{kappa_prior_mean_se}{0.256}
\setres{cty_Cambodia}{EU_transfer_others}{kappa_se}{0.256}
\setres{cty_Cambodia}{EU_transfer_others}{mse}{21.12}
\setres{cty_Cambodia}{EU_transfer_others}{mse_prior_mean}{21.12}
\setres{cty_Cambodia}{EU_transfer_others}{nrmse_pct}{25.1}
\setres{cty_Cambodia}{EU_transfer_others}{pred_variance}{0.00}
\setres{cty_Cambodia}{EU_transfer_others}{r_ff}{---}
\setres{cty_Cambodia}{EU_transfer_others}{r_ff_se}{---}
\setres{cty_Cambodia}{EU_transfer_others}{r_prior}{---}
\setres{cty_Cambodia}{EU_transfer_others}{r_prior_se}{---}
\setres{cty_Cambodia}{EU_transfer_others}{r_prob}{---}
\setres{cty_Cambodia}{EU_transfer_others}{r_prob_se}{---}
\setres{cty_Cambodia}{EU_transfer_others}{spread_penalty}{0.000}
\setres{cty_Cambodia}{EU_transfer_others}{spread_penalty_se}{0.000}
\setres{cty_Cambodia}{EU_transfer_others}{theta_star}{0.94}
\setres{cty_Cambodia}{EU_transfer_others}{transfer}{others}
\setres{cty_Cambodia}{EU_transfer_others}{volume_ess}{---}
\setres{cty_Cambodia}{EU_uniform}{kappa}{-2.851}
\setres{cty_Cambodia}{EU_uniform}{kappa_prior_mean}{-1.416}
\setres{cty_Cambodia}{EU_uniform}{kappa_prior_mean_se}{0.628}
\setres{cty_Cambodia}{EU_uniform}{kappa_se}{0.859}
\setres{cty_Cambodia}{EU_uniform}{mse}{39.74}
\setres{cty_Cambodia}{EU_uniform}{mse_prior_mean}{31.54}
\setres{cty_Cambodia}{EU_uniform}{nrmse_pct}{31.4}
\setres{cty_Cambodia}{EU_uniform}{pred_variance}{8.20}
\setres{cty_Cambodia}{EU_uniform}{r_ff}{---}
\setres{cty_Cambodia}{EU_uniform}{r_ff_se}{---}
\setres{cty_Cambodia}{EU_uniform}{r_prior}{---}
\setres{cty_Cambodia}{EU_uniform}{r_prior_se}{---}
\setres{cty_Cambodia}{EU_uniform}{r_prob}{---}
\setres{cty_Cambodia}{EU_uniform}{r_prob_se}{---}
\setres{cty_Cambodia}{EU_uniform}{spread_penalty}{1.435}
\setres{cty_Cambodia}{EU_uniform}{spread_penalty_se}{0.252}
\setres{cty_Cambodia}{EU_uniform}{volume_ess}{---}
\setres{cty_Cambodia}{EV_deterministic}{kappa}{0.404}
\setres{cty_Cambodia}{EV_deterministic}{kappa_se}{0.267}
\setres{cty_Cambodia}{EV_deterministic}{mse}{21.14}
\setres{cty_Cambodia}{EV_deterministic}{nrmse_pct}{25.3}
\setres{cty_Cambodia}{EV_deterministic}{r_ff}{---}
\setres{cty_Cambodia}{EV_deterministic}{r_ff_se}{---}
\setres{cty_Cambodia}{EV_deterministic}{r_prior}{---}
\setres{cty_Cambodia}{EV_deterministic}{r_prior_se}{---}
\setres{cty_Cambodia}{EV_deterministic}{r_prob}{---}
\setres{cty_Cambodia}{EV_deterministic}{r_prob_se}{---}
\setres{cty_Cambodia}{EV_deterministic}{volume_ess}{---}
\setres{cty_Cambodia}{reference}{country_name}{Cambodia}
\setres{cty_Cambodia}{reference}{lookup_nrmse_pct}{23.9}
\setres{cty_Cambodia}{reference}{midpoint_mse}{23.45}
\setres{cty_Cambodia}{reference}{midpoint_nrmse_pct}{26.7}
\setres{cty_Cambodia}{reference}{missing_lotteries}{none}
\setres{cty_Cambodia}{reference}{n_lotteries}{14}
\setres{cty_Cambodia}{reference}{n_obs}{1{,}120}
\setres{cty_Cambodia}{reference}{n_subjects}{80}
\setres{cty_Cambodia}{reference}{oracle_mse}{17.74}
\setres{cty_Cambodia}{reference}{sample}{LV2019\_Cambodia}
\setres{cty_Chile}{CPT_deterministic}{kappa}{0.954}
\setres{cty_Chile}{CPT_deterministic}{kappa_se}{0.020}
\setres{cty_Chile}{CPT_deterministic}{mse}{14.41}
\setres{cty_Chile}{CPT_deterministic}{nrmse_pct}{21.8}
\setres{cty_Chile}{CPT_deterministic}{r_ff}{---}
\setres{cty_Chile}{CPT_deterministic}{r_ff_se}{---}
\setres{cty_Chile}{CPT_deterministic}{r_prior}{---}
\setres{cty_Chile}{CPT_deterministic}{r_prior_se}{---}
\setres{cty_Chile}{CPT_deterministic}{r_prob}{---}
\setres{cty_Chile}{CPT_deterministic}{r_prob_se}{---}
\setres{cty_Chile}{CPT_deterministic}{theta_star}{1.10, 0.60, 0.70}
\setres{cty_Chile}{CPT_deterministic}{volume_ess}{---}
\setres{cty_Chile}{CPT_lit}{kappa}{0.504}
\setres{cty_Chile}{CPT_lit}{kappa_prior_mean}{0.856}
\setres{cty_Chile}{CPT_lit}{kappa_prior_mean_se}{0.055}
\setres{cty_Chile}{CPT_lit}{kappa_se}{0.090}
\setres{cty_Chile}{CPT_lit}{mse}{18.12}
\setres{cty_Chile}{CPT_lit}{mse_prior_mean}{15.22}
\setres{cty_Chile}{CPT_lit}{nrmse_pct}{22.3}
\setres{cty_Chile}{CPT_lit}{pred_variance}{2.90}
\setres{cty_Chile}{CPT_lit}{r_ff}{---}
\setres{cty_Chile}{CPT_lit}{r_ff_se}{---}
\setres{cty_Chile}{CPT_lit}{r_prior}{---}
\setres{cty_Chile}{CPT_lit}{r_prior_se}{---}
\setres{cty_Chile}{CPT_lit}{r_prob}{---}
\setres{cty_Chile}{CPT_lit}{r_prob_se}{---}
\setres{cty_Chile}{CPT_lit}{spread_penalty}{0.352}
\setres{cty_Chile}{CPT_lit}{spread_penalty_se}{0.038}
\setres{cty_Chile}{CPT_lit}{volume_ess}{---}
\setres{cty_Chile}{CPT_transfer_others}{kappa}{0.930}
\setres{cty_Chile}{CPT_transfer_others}{kappa_prior_mean}{0.930}
\setres{cty_Chile}{CPT_transfer_others}{kappa_prior_mean_se}{0.043}
\setres{cty_Chile}{CPT_transfer_others}{kappa_se}{0.043}
\setres{cty_Chile}{CPT_transfer_others}{mse}{14.61}
\setres{cty_Chile}{CPT_transfer_others}{mse_prior_mean}{14.61}
\setres{cty_Chile}{CPT_transfer_others}{nrmse_pct}{21.0}
\setres{cty_Chile}{CPT_transfer_others}{pred_variance}{0.00}
\setres{cty_Chile}{CPT_transfer_others}{r_ff}{---}
\setres{cty_Chile}{CPT_transfer_others}{r_ff_se}{---}
\setres{cty_Chile}{CPT_transfer_others}{r_prior}{---}
\setres{cty_Chile}{CPT_transfer_others}{r_prior_se}{---}
\setres{cty_Chile}{CPT_transfer_others}{r_prob}{---}
\setres{cty_Chile}{CPT_transfer_others}{r_prob_se}{---}
\setres{cty_Chile}{CPT_transfer_others}{spread_penalty}{0.000}
\setres{cty_Chile}{CPT_transfer_others}{spread_penalty_se}{0.000}
\setres{cty_Chile}{CPT_transfer_others}{theta_star}{1.00, 0.60, 0.90}
\setres{cty_Chile}{CPT_transfer_others}{transfer}{others}
\setres{cty_Chile}{CPT_transfer_others}{volume_ess}{---}
\setres{cty_Chile}{CPT_uniform}{kappa}{-1.735}
\setres{cty_Chile}{CPT_uniform}{kappa_prior_mean}{0.374}
\setres{cty_Chile}{CPT_uniform}{kappa_prior_mean_se}{0.143}
\setres{cty_Chile}{CPT_uniform}{kappa_se}{0.335}
\setres{cty_Chile}{CPT_uniform}{mse}{36.56}
\setres{cty_Chile}{CPT_uniform}{mse_prior_mean}{19.19}
\setres{cty_Chile}{CPT_uniform}{nrmse_pct}{25.4}
\setres{cty_Chile}{CPT_uniform}{pred_variance}{17.37}
\setres{cty_Chile}{CPT_uniform}{r_ff}{---}
\setres{cty_Chile}{CPT_uniform}{r_ff_se}{---}
\setres{cty_Chile}{CPT_uniform}{r_prior}{---}
\setres{cty_Chile}{CPT_uniform}{r_prior_se}{---}
\setres{cty_Chile}{CPT_uniform}{r_prob}{---}
\setres{cty_Chile}{CPT_uniform}{r_prob_se}{---}
\setres{cty_Chile}{CPT_uniform}{spread_penalty}{2.109}
\setres{cty_Chile}{CPT_uniform}{spread_penalty_se}{0.225}
\setres{cty_Chile}{CPT_uniform}{volume_ess}{---}
\setres{cty_Chile}{DA_deterministic}{kappa}{0.813}
\setres{cty_Chile}{DA_deterministic}{kappa_se}{0.057}
\setres{cty_Chile}{DA_deterministic}{mse}{15.58}
\setres{cty_Chile}{DA_deterministic}{nrmse_pct}{21.8}
\setres{cty_Chile}{DA_deterministic}{r_ff}{---}
\setres{cty_Chile}{DA_deterministic}{r_ff_se}{---}
\setres{cty_Chile}{DA_deterministic}{r_prior}{---}
\setres{cty_Chile}{DA_deterministic}{r_prior_se}{---}
\setres{cty_Chile}{DA_deterministic}{r_prob}{---}
\setres{cty_Chile}{DA_deterministic}{r_prob_se}{---}
\setres{cty_Chile}{DA_deterministic}{theta_star}{1.75, 2.27}
\setres{cty_Chile}{DA_deterministic}{volume_ess}{---}
\setres{cty_Chile}{DA_lit}{kappa}{-0.504}
\setres{cty_Chile}{DA_lit}{kappa_prior_mean}{0.073}
\setres{cty_Chile}{DA_lit}{kappa_prior_mean_se}{0.220}
\setres{cty_Chile}{DA_lit}{kappa_se}{0.277}
\setres{cty_Chile}{DA_lit}{mse}{26.42}
\setres{cty_Chile}{DA_lit}{mse_prior_mean}{21.67}
\setres{cty_Chile}{DA_lit}{nrmse_pct}{26.2}
\setres{cty_Chile}{DA_lit}{pred_variance}{4.75}
\setres{cty_Chile}{DA_lit}{r_ff}{---}
\setres{cty_Chile}{DA_lit}{r_ff_se}{---}
\setres{cty_Chile}{DA_lit}{r_prior}{---}
\setres{cty_Chile}{DA_lit}{r_prior_se}{---}
\setres{cty_Chile}{DA_lit}{r_prob}{---}
\setres{cty_Chile}{DA_lit}{r_prob_se}{---}
\setres{cty_Chile}{DA_lit}{spread_penalty}{0.577}
\setres{cty_Chile}{DA_lit}{spread_penalty_se}{0.062}
\setres{cty_Chile}{DA_lit}{volume_ess}{---}
\setres{cty_Chile}{DA_transfer_others}{kappa}{0.757}
\setres{cty_Chile}{DA_transfer_others}{kappa_prior_mean}{0.757}
\setres{cty_Chile}{DA_transfer_others}{kappa_prior_mean_se}{0.072}
\setres{cty_Chile}{DA_transfer_others}{kappa_se}{0.072}
\setres{cty_Chile}{DA_transfer_others}{mse}{16.04}
\setres{cty_Chile}{DA_transfer_others}{mse_prior_mean}{16.04}
\setres{cty_Chile}{DA_transfer_others}{nrmse_pct}{21.0}
\setres{cty_Chile}{DA_transfer_others}{pred_variance}{0.00}
\setres{cty_Chile}{DA_transfer_others}{r_ff}{---}
\setres{cty_Chile}{DA_transfer_others}{r_ff_se}{---}
\setres{cty_Chile}{DA_transfer_others}{r_prior}{---}
\setres{cty_Chile}{DA_transfer_others}{r_prior_se}{---}
\setres{cty_Chile}{DA_transfer_others}{r_prob}{---}
\setres{cty_Chile}{DA_transfer_others}{r_prob_se}{---}
\setres{cty_Chile}{DA_transfer_others}{spread_penalty}{0.000}
\setres{cty_Chile}{DA_transfer_others}{spread_penalty_se}{0.000}
\setres{cty_Chile}{DA_transfer_others}{theta_star}{1.69, 1.42}
\setres{cty_Chile}{DA_transfer_others}{transfer}{others}
\setres{cty_Chile}{DA_transfer_others}{volume_ess}{---}
\setres{cty_Chile}{DA_uniform}{kappa}{-2.295}
\setres{cty_Chile}{DA_uniform}{kappa_prior_mean}{-0.638}
\setres{cty_Chile}{DA_uniform}{kappa_prior_mean_se}{0.335}
\setres{cty_Chile}{DA_uniform}{kappa_se}{0.493}
\setres{cty_Chile}{DA_uniform}{mse}{41.17}
\setres{cty_Chile}{DA_uniform}{mse_prior_mean}{27.52}
\setres{cty_Chile}{DA_uniform}{nrmse_pct}{28.1}
\setres{cty_Chile}{DA_uniform}{pred_variance}{13.65}
\setres{cty_Chile}{DA_uniform}{r_ff}{---}
\setres{cty_Chile}{DA_uniform}{r_ff_se}{---}
\setres{cty_Chile}{DA_uniform}{r_prior}{---}
\setres{cty_Chile}{DA_uniform}{r_prior_se}{---}
\setres{cty_Chile}{DA_uniform}{r_prob}{---}
\setres{cty_Chile}{DA_uniform}{r_prob_se}{---}
\setres{cty_Chile}{DA_uniform}{spread_penalty}{1.658}
\setres{cty_Chile}{DA_uniform}{spread_penalty_se}{0.177}
\setres{cty_Chile}{DA_uniform}{volume_ess}{---}
\setres{cty_Chile}{EU_deterministic}{kappa}{0.667}
\setres{cty_Chile}{EU_deterministic}{kappa_se}{0.094}
\setres{cty_Chile}{EU_deterministic}{mse}{16.78}
\setres{cty_Chile}{EU_deterministic}{nrmse_pct}{23.4}
\setres{cty_Chile}{EU_deterministic}{r_ff}{---}
\setres{cty_Chile}{EU_deterministic}{r_ff_se}{---}
\setres{cty_Chile}{EU_deterministic}{r_prior}{---}
\setres{cty_Chile}{EU_deterministic}{r_prior_se}{---}
\setres{cty_Chile}{EU_deterministic}{r_prob}{---}
\setres{cty_Chile}{EU_deterministic}{r_prob_se}{---}
\setres{cty_Chile}{EU_deterministic}{theta_star}{0.82}
\setres{cty_Chile}{EU_deterministic}{volume_ess}{---}
\setres{cty_Chile}{EU_lit}{kappa}{0.454}
\setres{cty_Chile}{EU_lit}{kappa_prior_mean}{0.600}
\setres{cty_Chile}{EU_lit}{kappa_prior_mean_se}{0.108}
\setres{cty_Chile}{EU_lit}{kappa_se}{0.123}
\setres{cty_Chile}{EU_lit}{mse}{18.54}
\setres{cty_Chile}{EU_lit}{mse_prior_mean}{17.33}
\setres{cty_Chile}{EU_lit}{nrmse_pct}{24.3}
\setres{cty_Chile}{EU_lit}{pred_variance}{1.20}
\setres{cty_Chile}{EU_lit}{r_ff}{---}
\setres{cty_Chile}{EU_lit}{r_ff_se}{---}
\setres{cty_Chile}{EU_lit}{r_prior}{---}
\setres{cty_Chile}{EU_lit}{r_prior_se}{---}
\setres{cty_Chile}{EU_lit}{r_prob}{---}
\setres{cty_Chile}{EU_lit}{r_prob_se}{---}
\setres{cty_Chile}{EU_lit}{spread_penalty}{0.146}
\setres{cty_Chile}{EU_lit}{spread_penalty_se}{0.016}
\setres{cty_Chile}{EU_lit}{volume_ess}{---}
\setres{cty_Chile}{EU_transfer_others}{kappa}{0.613}
\setres{cty_Chile}{EU_transfer_others}{kappa_prior_mean}{0.613}
\setres{cty_Chile}{EU_transfer_others}{kappa_prior_mean_se}{0.104}
\setres{cty_Chile}{EU_transfer_others}{kappa_se}{0.104}
\setres{cty_Chile}{EU_transfer_others}{mse}{17.22}
\setres{cty_Chile}{EU_transfer_others}{mse_prior_mean}{17.22}
\setres{cty_Chile}{EU_transfer_others}{nrmse_pct}{21.8}
\setres{cty_Chile}{EU_transfer_others}{pred_variance}{0.00}
\setres{cty_Chile}{EU_transfer_others}{r_ff}{---}
\setres{cty_Chile}{EU_transfer_others}{r_ff_se}{---}
\setres{cty_Chile}{EU_transfer_others}{r_prior}{---}
\setres{cty_Chile}{EU_transfer_others}{r_prior_se}{---}
\setres{cty_Chile}{EU_transfer_others}{r_prob}{---}
\setres{cty_Chile}{EU_transfer_others}{r_prob_se}{---}
\setres{cty_Chile}{EU_transfer_others}{spread_penalty}{0.000}
\setres{cty_Chile}{EU_transfer_others}{spread_penalty_se}{0.000}
\setres{cty_Chile}{EU_transfer_others}{theta_star}{0.95}
\setres{cty_Chile}{EU_transfer_others}{transfer}{others}
\setres{cty_Chile}{EU_transfer_others}{volume_ess}{---}
\setres{cty_Chile}{EU_uniform}{kappa}{-1.059}
\setres{cty_Chile}{EU_uniform}{kappa_prior_mean}{-0.064}
\setres{cty_Chile}{EU_uniform}{kappa_prior_mean_se}{0.237}
\setres{cty_Chile}{EU_uniform}{kappa_se}{0.334}
\setres{cty_Chile}{EU_uniform}{mse}{30.99}
\setres{cty_Chile}{EU_uniform}{mse_prior_mean}{22.79}
\setres{cty_Chile}{EU_uniform}{nrmse_pct}{26.1}
\setres{cty_Chile}{EU_uniform}{pred_variance}{8.20}
\setres{cty_Chile}{EU_uniform}{r_ff}{---}
\setres{cty_Chile}{EU_uniform}{r_ff_se}{---}
\setres{cty_Chile}{EU_uniform}{r_prior}{---}
\setres{cty_Chile}{EU_uniform}{r_prior_se}{---}
\setres{cty_Chile}{EU_uniform}{r_prob}{---}
\setres{cty_Chile}{EU_uniform}{r_prob_se}{---}
\setres{cty_Chile}{EU_uniform}{spread_penalty}{0.996}
\setres{cty_Chile}{EU_uniform}{spread_penalty_se}{0.106}
\setres{cty_Chile}{EU_uniform}{volume_ess}{---}
\setres{cty_Chile}{EV_deterministic}{kappa}{0.562}
\setres{cty_Chile}{EV_deterministic}{kappa_se}{0.117}
\setres{cty_Chile}{EV_deterministic}{mse}{17.65}
\setres{cty_Chile}{EV_deterministic}{nrmse_pct}{21.9}
\setres{cty_Chile}{EV_deterministic}{r_ff}{---}
\setres{cty_Chile}{EV_deterministic}{r_ff_se}{---}
\setres{cty_Chile}{EV_deterministic}{r_prior}{---}
\setres{cty_Chile}{EV_deterministic}{r_prior_se}{---}
\setres{cty_Chile}{EV_deterministic}{r_prob}{---}
\setres{cty_Chile}{EV_deterministic}{r_prob_se}{---}
\setres{cty_Chile}{EV_deterministic}{volume_ess}{---}
\setres{cty_Chile}{reference}{country_name}{Chile}
\setres{cty_Chile}{reference}{lookup_nrmse_pct}{20.5}
\setres{cty_Chile}{reference}{midpoint_mse}{22.27}
\setres{cty_Chile}{reference}{midpoint_nrmse_pct}{25.6}
\setres{cty_Chile}{reference}{missing_lotteries}{none}
\setres{cty_Chile}{reference}{n_lotteries}{14}
\setres{cty_Chile}{reference}{n_obs}{1{,}344}
\setres{cty_Chile}{reference}{n_subjects}{96}
\setres{cty_Chile}{reference}{oracle_mse}{14.04}
\setres{cty_Chile}{reference}{sample}{LV2019\_Chile}
\setres{cty_China}{CPT_deterministic}{kappa}{0.964}
\setres{cty_China}{CPT_deterministic}{kappa_se}{0.008}
\setres{cty_China}{CPT_deterministic}{mse}{8.11}
\setres{cty_China}{CPT_deterministic}{nrmse_pct}{15.5}
\setres{cty_China}{CPT_deterministic}{r_ff}{---}
\setres{cty_China}{CPT_deterministic}{r_ff_se}{---}
\setres{cty_China}{CPT_deterministic}{r_prior}{---}
\setres{cty_China}{CPT_deterministic}{r_prior_se}{---}
\setres{cty_China}{CPT_deterministic}{r_prob}{---}
\setres{cty_China}{CPT_deterministic}{r_prob_se}{---}
\setres{cty_China}{CPT_deterministic}{theta_star}{1.00, 0.70, 0.90}
\setres{cty_China}{CPT_deterministic}{volume_ess}{---}
\setres{cty_China}{CPT_lit}{kappa}{0.555}
\setres{cty_China}{CPT_lit}{kappa_prior_mean}{0.829}
\setres{cty_China}{CPT_lit}{kappa_prior_mean_se}{0.028}
\setres{cty_China}{CPT_lit}{kappa_se}{0.035}
\setres{cty_China}{CPT_lit}{mse}{12.44}
\setres{cty_China}{CPT_lit}{mse_prior_mean}{9.55}
\setres{cty_China}{CPT_lit}{nrmse_pct}{17.3}
\setres{cty_China}{CPT_lit}{pred_variance}{2.90}
\setres{cty_China}{CPT_lit}{r_ff}{---}
\setres{cty_China}{CPT_lit}{r_ff_se}{---}
\setres{cty_China}{CPT_lit}{r_prior}{---}
\setres{cty_China}{CPT_lit}{r_prior_se}{---}
\setres{cty_China}{CPT_lit}{r_prob}{---}
\setres{cty_China}{CPT_lit}{r_prob_se}{---}
\setres{cty_China}{CPT_lit}{spread_penalty}{0.274}
\setres{cty_China}{CPT_lit}{spread_penalty_se}{0.012}
\setres{cty_China}{CPT_lit}{volume_ess}{---}
\setres{cty_China}{CPT_transfer_others}{kappa}{0.965}
\setres{cty_China}{CPT_transfer_others}{kappa_prior_mean}{0.965}
\setres{cty_China}{CPT_transfer_others}{kappa_prior_mean_se}{0.006}
\setres{cty_China}{CPT_transfer_others}{kappa_se}{0.006}
\setres{cty_China}{CPT_transfer_others}{mse}{8.11}
\setres{cty_China}{CPT_transfer_others}{mse_prior_mean}{8.11}
\setres{cty_China}{CPT_transfer_others}{nrmse_pct}{15.0}
\setres{cty_China}{CPT_transfer_others}{pred_variance}{0.00}
\setres{cty_China}{CPT_transfer_others}{r_ff}{---}
\setres{cty_China}{CPT_transfer_others}{r_ff_se}{---}
\setres{cty_China}{CPT_transfer_others}{r_prior}{---}
\setres{cty_China}{CPT_transfer_others}{r_prior_se}{---}
\setres{cty_China}{CPT_transfer_others}{r_prob}{---}
\setres{cty_China}{CPT_transfer_others}{r_prob_se}{---}
\setres{cty_China}{CPT_transfer_others}{spread_penalty}{0.000}
\setres{cty_China}{CPT_transfer_others}{spread_penalty_se}{0.000}
\setres{cty_China}{CPT_transfer_others}{theta_star}{1.10, 0.60, 0.80}
\setres{cty_China}{CPT_transfer_others}{transfer}{others}
\setres{cty_China}{CPT_transfer_others}{volume_ess}{---}
\setres{cty_China}{CPT_uniform}{kappa}{-1.446}
\setres{cty_China}{CPT_uniform}{kappa_prior_mean}{0.196}
\setres{cty_China}{CPT_uniform}{kappa_prior_mean_se}{0.062}
\setres{cty_China}{CPT_uniform}{kappa_se}{0.111}
\setres{cty_China}{CPT_uniform}{mse}{33.61}
\setres{cty_China}{CPT_uniform}{mse_prior_mean}{16.24}
\setres{cty_China}{CPT_uniform}{nrmse_pct}{21.6}
\setres{cty_China}{CPT_uniform}{pred_variance}{17.37}
\setres{cty_China}{CPT_uniform}{r_ff}{---}
\setres{cty_China}{CPT_uniform}{r_ff_se}{---}
\setres{cty_China}{CPT_uniform}{r_prior}{---}
\setres{cty_China}{CPT_uniform}{r_prior_se}{---}
\setres{cty_China}{CPT_uniform}{r_prob}{---}
\setres{cty_China}{CPT_uniform}{r_prob_se}{---}
\setres{cty_China}{CPT_uniform}{spread_penalty}{1.642}
\setres{cty_China}{CPT_uniform}{spread_penalty_se}{0.070}
\setres{cty_China}{CPT_uniform}{volume_ess}{---}
\setres{cty_China}{DA_deterministic}{kappa}{0.936}
\setres{cty_China}{DA_deterministic}{kappa_se}{0.012}
\setres{cty_China}{DA_deterministic}{mse}{8.41}
\setres{cty_China}{DA_deterministic}{nrmse_pct}{15.6}
\setres{cty_China}{DA_deterministic}{r_ff}{---}
\setres{cty_China}{DA_deterministic}{r_ff_se}{---}
\setres{cty_China}{DA_deterministic}{r_prior}{---}
\setres{cty_China}{DA_deterministic}{r_prior_se}{---}
\setres{cty_China}{DA_deterministic}{r_prob}{---}
\setres{cty_China}{DA_deterministic}{r_prob_se}{---}
\setres{cty_China}{DA_deterministic}{theta_star}{1.46, 1.06}
\setres{cty_China}{DA_deterministic}{volume_ess}{---}
\setres{cty_China}{DA_lit}{kappa}{-0.346}
\setres{cty_China}{DA_lit}{kappa_prior_mean}{0.103}
\setres{cty_China}{DA_lit}{kappa_prior_mean_se}{0.081}
\setres{cty_China}{DA_lit}{kappa_se}{0.095}
\setres{cty_China}{DA_lit}{mse}{21.97}
\setres{cty_China}{DA_lit}{mse_prior_mean}{17.22}
\setres{cty_China}{DA_lit}{nrmse_pct}{22.9}
\setres{cty_China}{DA_lit}{pred_variance}{4.75}
\setres{cty_China}{DA_lit}{r_ff}{---}
\setres{cty_China}{DA_lit}{r_ff_se}{---}
\setres{cty_China}{DA_lit}{r_prior}{---}
\setres{cty_China}{DA_lit}{r_prior_se}{---}
\setres{cty_China}{DA_lit}{r_prob}{---}
\setres{cty_China}{DA_lit}{r_prob_se}{---}
\setres{cty_China}{DA_lit}{spread_penalty}{0.449}
\setres{cty_China}{DA_lit}{spread_penalty_se}{0.019}
\setres{cty_China}{DA_lit}{volume_ess}{---}
\setres{cty_China}{DA_transfer_others}{kappa}{0.933}
\setres{cty_China}{DA_transfer_others}{kappa_prior_mean}{0.933}
\setres{cty_China}{DA_transfer_others}{kappa_prior_mean_se}{0.012}
\setres{cty_China}{DA_transfer_others}{kappa_se}{0.012}
\setres{cty_China}{DA_transfer_others}{mse}{8.45}
\setres{cty_China}{DA_transfer_others}{mse_prior_mean}{8.45}
\setres{cty_China}{DA_transfer_others}{nrmse_pct}{15.3}
\setres{cty_China}{DA_transfer_others}{pred_variance}{0.00}
\setres{cty_China}{DA_transfer_others}{r_ff}{---}
\setres{cty_China}{DA_transfer_others}{r_ff_se}{---}
\setres{cty_China}{DA_transfer_others}{r_prior}{---}
\setres{cty_China}{DA_transfer_others}{r_prior_se}{---}
\setres{cty_China}{DA_transfer_others}{r_prob}{---}
\setres{cty_China}{DA_transfer_others}{r_prob_se}{---}
\setres{cty_China}{DA_transfer_others}{spread_penalty}{0.000}
\setres{cty_China}{DA_transfer_others}{spread_penalty_se}{0.000}
\setres{cty_China}{DA_transfer_others}{theta_star}{1.71, 1.48}
\setres{cty_China}{DA_transfer_others}{transfer}{others}
\setres{cty_China}{DA_transfer_others}{volume_ess}{---}
\setres{cty_China}{DA_uniform}{kappa}{-1.947}
\setres{cty_China}{DA_uniform}{kappa_prior_mean}{-0.657}
\setres{cty_China}{DA_uniform}{kappa_prior_mean_se}{0.121}
\setres{cty_China}{DA_uniform}{kappa_se}{0.162}
\setres{cty_China}{DA_uniform}{mse}{38.92}
\setres{cty_China}{DA_uniform}{mse_prior_mean}{25.27}
\setres{cty_China}{DA_uniform}{nrmse_pct}{28.2}
\setres{cty_China}{DA_uniform}{pred_variance}{13.65}
\setres{cty_China}{DA_uniform}{r_ff}{---}
\setres{cty_China}{DA_uniform}{r_ff_se}{---}
\setres{cty_China}{DA_uniform}{r_prior}{---}
\setres{cty_China}{DA_uniform}{r_prior_se}{---}
\setres{cty_China}{DA_uniform}{r_prob}{---}
\setres{cty_China}{DA_uniform}{r_prob_se}{---}
\setres{cty_China}{DA_uniform}{spread_penalty}{1.290}
\setres{cty_China}{DA_uniform}{spread_penalty_se}{0.055}
\setres{cty_China}{DA_uniform}{volume_ess}{---}
\setres{cty_China}{EU_deterministic}{kappa}{0.882}
\setres{cty_China}{EU_deterministic}{kappa_se}{0.021}
\setres{cty_China}{EU_deterministic}{mse}{8.98}
\setres{cty_China}{EU_deterministic}{nrmse_pct}{16.7}
\setres{cty_China}{EU_deterministic}{r_ff}{---}
\setres{cty_China}{EU_deterministic}{r_ff_se}{---}
\setres{cty_China}{EU_deterministic}{r_prior}{---}
\setres{cty_China}{EU_deterministic}{r_prior_se}{---}
\setres{cty_China}{EU_deterministic}{r_prob}{---}
\setres{cty_China}{EU_deterministic}{r_prob_se}{---}
\setres{cty_China}{EU_deterministic}{theta_star}{0.90}
\setres{cty_China}{EU_deterministic}{volume_ess}{---}
\setres{cty_China}{EU_lit}{kappa}{0.601}
\setres{cty_China}{EU_lit}{kappa_prior_mean}{0.715}
\setres{cty_China}{EU_lit}{kappa_prior_mean_se}{0.037}
\setres{cty_China}{EU_lit}{kappa_se}{0.041}
\setres{cty_China}{EU_lit}{mse}{11.95}
\setres{cty_China}{EU_lit}{mse_prior_mean}{10.75}
\setres{cty_China}{EU_lit}{nrmse_pct}{18.3}
\setres{cty_China}{EU_lit}{pred_variance}{1.20}
\setres{cty_China}{EU_lit}{r_ff}{---}
\setres{cty_China}{EU_lit}{r_ff_se}{---}
\setres{cty_China}{EU_lit}{r_prior}{---}
\setres{cty_China}{EU_lit}{r_prior_se}{---}
\setres{cty_China}{EU_lit}{r_prob}{---}
\setres{cty_China}{EU_lit}{r_prob_se}{---}
\setres{cty_China}{EU_lit}{spread_penalty}{0.114}
\setres{cty_China}{EU_lit}{spread_penalty_se}{0.005}
\setres{cty_China}{EU_lit}{volume_ess}{---}
\setres{cty_China}{EU_transfer_others}{kappa}{0.880}
\setres{cty_China}{EU_transfer_others}{kappa_prior_mean}{0.880}
\setres{cty_China}{EU_transfer_others}{kappa_prior_mean_se}{0.023}
\setres{cty_China}{EU_transfer_others}{kappa_se}{0.023}
\setres{cty_China}{EU_transfer_others}{mse}{9.00}
\setres{cty_China}{EU_transfer_others}{mse_prior_mean}{9.00}
\setres{cty_China}{EU_transfer_others}{nrmse_pct}{17.0}
\setres{cty_China}{EU_transfer_others}{pred_variance}{0.00}
\setres{cty_China}{EU_transfer_others}{r_ff}{---}
\setres{cty_China}{EU_transfer_others}{r_ff_se}{---}
\setres{cty_China}{EU_transfer_others}{r_prior}{---}
\setres{cty_China}{EU_transfer_others}{r_prior_se}{---}
\setres{cty_China}{EU_transfer_others}{r_prob}{---}
\setres{cty_China}{EU_transfer_others}{r_prob_se}{---}
\setres{cty_China}{EU_transfer_others}{spread_penalty}{0.000}
\setres{cty_China}{EU_transfer_others}{spread_penalty_se}{0.000}
\setres{cty_China}{EU_transfer_others}{theta_star}{0.95}
\setres{cty_China}{EU_transfer_others}{transfer}{others}
\setres{cty_China}{EU_transfer_others}{volume_ess}{---}
\setres{cty_China}{EU_uniform}{kappa}{-0.819}
\setres{cty_China}{EU_uniform}{kappa_prior_mean}{-0.044}
\setres{cty_China}{EU_uniform}{kappa_prior_mean_se}{0.088}
\setres{cty_China}{EU_uniform}{kappa_se}{0.113}
\setres{cty_China}{EU_uniform}{mse}{26.98}
\setres{cty_China}{EU_uniform}{mse_prior_mean}{18.78}
\setres{cty_China}{EU_uniform}{nrmse_pct}{23.8}
\setres{cty_China}{EU_uniform}{pred_variance}{8.20}
\setres{cty_China}{EU_uniform}{r_ff}{---}
\setres{cty_China}{EU_uniform}{r_ff_se}{---}
\setres{cty_China}{EU_uniform}{r_prior}{---}
\setres{cty_China}{EU_uniform}{r_prior_se}{---}
\setres{cty_China}{EU_uniform}{r_prob}{---}
\setres{cty_China}{EU_uniform}{r_prob_se}{---}
\setres{cty_China}{EU_uniform}{spread_penalty}{0.775}
\setres{cty_China}{EU_uniform}{spread_penalty_se}{0.033}
\setres{cty_China}{EU_uniform}{volume_ess}{---}
\setres{cty_China}{EV_deterministic}{kappa}{0.862}
\setres{cty_China}{EV_deterministic}{kappa_se}{0.026}
\setres{cty_China}{EV_deterministic}{mse}{9.19}
\setres{cty_China}{EV_deterministic}{nrmse_pct}{17.0}
\setres{cty_China}{EV_deterministic}{r_ff}{---}
\setres{cty_China}{EV_deterministic}{r_ff_se}{---}
\setres{cty_China}{EV_deterministic}{r_prior}{---}
\setres{cty_China}{EV_deterministic}{r_prior_se}{---}
\setres{cty_China}{EV_deterministic}{r_prob}{---}
\setres{cty_China}{EV_deterministic}{r_prob_se}{---}
\setres{cty_China}{EV_deterministic}{volume_ess}{---}
\setres{cty_China}{reference}{country_name}{China}
\setres{cty_China}{reference}{lookup_nrmse_pct}{14.9}
\setres{cty_China}{reference}{midpoint_mse}{18.31}
\setres{cty_China}{reference}{midpoint_nrmse_pct}{21.4}
\setres{cty_China}{reference}{missing_lotteries}{none}
\setres{cty_China}{reference}{n_lotteries}{14}
\setres{cty_China}{reference}{n_obs}{2{,}855}
\setres{cty_China}{reference}{n_subjects}{204}
\setres{cty_China}{reference}{oracle_mse}{7.73}
\setres{cty_China}{reference}{sample}{LV2019\_China}
\setres{cty_Colombia}{CPT_deterministic}{kappa}{0.948}
\setres{cty_Colombia}{CPT_deterministic}{kappa_se}{0.019}
\setres{cty_Colombia}{CPT_deterministic}{mse}{16.01}
\setres{cty_Colombia}{CPT_deterministic}{nrmse_pct}{21.4}
\setres{cty_Colombia}{CPT_deterministic}{r_ff}{---}
\setres{cty_Colombia}{CPT_deterministic}{r_ff_se}{---}
\setres{cty_Colombia}{CPT_deterministic}{r_prior}{---}
\setres{cty_Colombia}{CPT_deterministic}{r_prior_se}{---}
\setres{cty_Colombia}{CPT_deterministic}{r_prob}{---}
\setres{cty_Colombia}{CPT_deterministic}{r_prob_se}{---}
\setres{cty_Colombia}{CPT_deterministic}{theta_star}{1.10, 0.70, 0.70}
\setres{cty_Colombia}{CPT_deterministic}{volume_ess}{---}
\setres{cty_Colombia}{CPT_lit}{kappa}{0.347}
\setres{cty_Colombia}{CPT_lit}{kappa_prior_mean}{0.756}
\setres{cty_Colombia}{CPT_lit}{kappa_prior_mean_se}{0.091}
\setres{cty_Colombia}{CPT_lit}{kappa_se}{0.132}
\setres{cty_Colombia}{CPT_lit}{mse}{20.69}
\setres{cty_Colombia}{CPT_lit}{mse_prior_mean}{17.50}
\setres{cty_Colombia}{CPT_lit}{nrmse_pct}{22.4}
\setres{cty_Colombia}{CPT_lit}{pred_variance}{3.19}
\setres{cty_Colombia}{CPT_lit}{r_ff}{---}
\setres{cty_Colombia}{CPT_lit}{r_ff_se}{---}
\setres{cty_Colombia}{CPT_lit}{r_prior}{---}
\setres{cty_Colombia}{CPT_lit}{r_prior_se}{---}
\setres{cty_Colombia}{CPT_lit}{r_prob}{---}
\setres{cty_Colombia}{CPT_lit}{r_prob_se}{---}
\setres{cty_Colombia}{CPT_lit}{spread_penalty}{0.409}
\setres{cty_Colombia}{CPT_lit}{spread_penalty_se}{0.048}
\setres{cty_Colombia}{CPT_lit}{volume_ess}{---}
\setres{cty_Colombia}{CPT_transfer_others}{kappa}{0.976}
\setres{cty_Colombia}{CPT_transfer_others}{kappa_prior_mean}{0.976}
\setres{cty_Colombia}{CPT_transfer_others}{kappa_prior_mean_se}{0.016}
\setres{cty_Colombia}{CPT_transfer_others}{kappa_se}{0.016}
\setres{cty_Colombia}{CPT_transfer_others}{mse}{15.79}
\setres{cty_Colombia}{CPT_transfer_others}{mse_prior_mean}{15.79}
\setres{cty_Colombia}{CPT_transfer_others}{nrmse_pct}{21.1}
\setres{cty_Colombia}{CPT_transfer_others}{pred_variance}{0.00}
\setres{cty_Colombia}{CPT_transfer_others}{r_ff}{---}
\setres{cty_Colombia}{CPT_transfer_others}{r_ff_se}{---}
\setres{cty_Colombia}{CPT_transfer_others}{r_prior}{---}
\setres{cty_Colombia}{CPT_transfer_others}{r_prior_se}{---}
\setres{cty_Colombia}{CPT_transfer_others}{r_prob}{---}
\setres{cty_Colombia}{CPT_transfer_others}{r_prob_se}{---}
\setres{cty_Colombia}{CPT_transfer_others}{spread_penalty}{0.000}
\setres{cty_Colombia}{CPT_transfer_others}{spread_penalty_se}{0.000}
\setres{cty_Colombia}{CPT_transfer_others}{theta_star}{1.00, 0.60, 0.90}
\setres{cty_Colombia}{CPT_transfer_others}{transfer}{others}
\setres{cty_Colombia}{CPT_transfer_others}{volume_ess}{---}
\setres{cty_Colombia}{CPT_uniform}{kappa}{-2.540}
\setres{cty_Colombia}{CPT_uniform}{kappa_prior_mean}{-0.098}
\setres{cty_Colombia}{CPT_uniform}{kappa_prior_mean_se}{0.201}
\setres{cty_Colombia}{CPT_uniform}{kappa_se}{0.453}
\setres{cty_Colombia}{CPT_uniform}{mse}{43.19}
\setres{cty_Colombia}{CPT_uniform}{mse_prior_mean}{24.16}
\setres{cty_Colombia}{CPT_uniform}{nrmse_pct}{27.3}
\setres{cty_Colombia}{CPT_uniform}{pred_variance}{19.03}
\setres{cty_Colombia}{CPT_uniform}{r_ff}{---}
\setres{cty_Colombia}{CPT_uniform}{r_ff_se}{---}
\setres{cty_Colombia}{CPT_uniform}{r_prior}{---}
\setres{cty_Colombia}{CPT_uniform}{r_prior_se}{---}
\setres{cty_Colombia}{CPT_uniform}{r_prob}{---}
\setres{cty_Colombia}{CPT_uniform}{r_prob_se}{---}
\setres{cty_Colombia}{CPT_uniform}{spread_penalty}{2.442}
\setres{cty_Colombia}{CPT_uniform}{spread_penalty_se}{0.285}
\setres{cty_Colombia}{CPT_uniform}{volume_ess}{---}
\setres{cty_Colombia}{DA_deterministic}{kappa}{0.880}
\setres{cty_Colombia}{DA_deterministic}{kappa_se}{0.045}
\setres{cty_Colombia}{DA_deterministic}{mse}{16.54}
\setres{cty_Colombia}{DA_deterministic}{nrmse_pct}{21.4}
\setres{cty_Colombia}{DA_deterministic}{r_ff}{---}
\setres{cty_Colombia}{DA_deterministic}{r_ff_se}{---}
\setres{cty_Colombia}{DA_deterministic}{r_prior}{---}
\setres{cty_Colombia}{DA_deterministic}{r_prior_se}{---}
\setres{cty_Colombia}{DA_deterministic}{r_prob}{---}
\setres{cty_Colombia}{DA_deterministic}{r_prob_se}{---}
\setres{cty_Colombia}{DA_deterministic}{theta_star}{1.54, 1.30}
\setres{cty_Colombia}{DA_deterministic}{volume_ess}{---}
\setres{cty_Colombia}{DA_lit}{kappa}{-1.026}
\setres{cty_Colombia}{DA_lit}{kappa_prior_mean}{-0.328}
\setres{cty_Colombia}{DA_lit}{kappa_prior_mean_se}{0.289}
\setres{cty_Colombia}{DA_lit}{kappa_se}{0.362}
\setres{cty_Colombia}{DA_lit}{mse}{31.39}
\setres{cty_Colombia}{DA_lit}{mse_prior_mean}{25.95}
\setres{cty_Colombia}{DA_lit}{nrmse_pct}{27.4}
\setres{cty_Colombia}{DA_lit}{pred_variance}{5.44}
\setres{cty_Colombia}{DA_lit}{r_ff}{---}
\setres{cty_Colombia}{DA_lit}{r_ff_se}{---}
\setres{cty_Colombia}{DA_lit}{r_prior}{---}
\setres{cty_Colombia}{DA_lit}{r_prior_se}{---}
\setres{cty_Colombia}{DA_lit}{r_prob}{---}
\setres{cty_Colombia}{DA_lit}{r_prob_se}{---}
\setres{cty_Colombia}{DA_lit}{spread_penalty}{0.698}
\setres{cty_Colombia}{DA_lit}{spread_penalty_se}{0.081}
\setres{cty_Colombia}{DA_lit}{volume_ess}{---}
\setres{cty_Colombia}{DA_transfer_others}{kappa}{0.895}
\setres{cty_Colombia}{DA_transfer_others}{kappa_prior_mean}{0.895}
\setres{cty_Colombia}{DA_transfer_others}{kappa_prior_mean_se}{0.041}
\setres{cty_Colombia}{DA_transfer_others}{kappa_se}{0.041}
\setres{cty_Colombia}{DA_transfer_others}{mse}{16.42}
\setres{cty_Colombia}{DA_transfer_others}{mse_prior_mean}{16.42}
\setres{cty_Colombia}{DA_transfer_others}{nrmse_pct}{21.5}
\setres{cty_Colombia}{DA_transfer_others}{pred_variance}{0.00}
\setres{cty_Colombia}{DA_transfer_others}{r_ff}{---}
\setres{cty_Colombia}{DA_transfer_others}{r_ff_se}{---}
\setres{cty_Colombia}{DA_transfer_others}{r_prior}{---}
\setres{cty_Colombia}{DA_transfer_others}{r_prior_se}{---}
\setres{cty_Colombia}{DA_transfer_others}{r_prob}{---}
\setres{cty_Colombia}{DA_transfer_others}{r_prob_se}{---}
\setres{cty_Colombia}{DA_transfer_others}{spread_penalty}{0.000}
\setres{cty_Colombia}{DA_transfer_others}{spread_penalty_se}{0.000}
\setres{cty_Colombia}{DA_transfer_others}{theta_star}{1.70, 1.45}
\setres{cty_Colombia}{DA_transfer_others}{transfer}{others}
\setres{cty_Colombia}{DA_transfer_others}{volume_ess}{---}
\setres{cty_Colombia}{DA_uniform}{kappa}{-3.431}
\setres{cty_Colombia}{DA_uniform}{kappa_prior_mean}{-1.456}
\setres{cty_Colombia}{DA_uniform}{kappa_prior_mean_se}{0.433}
\setres{cty_Colombia}{DA_uniform}{kappa_se}{0.643}
\setres{cty_Colombia}{DA_uniform}{mse}{50.13}
\setres{cty_Colombia}{DA_uniform}{mse_prior_mean}{34.74}
\setres{cty_Colombia}{DA_uniform}{nrmse_pct}{32.0}
\setres{cty_Colombia}{DA_uniform}{pred_variance}{15.39}
\setres{cty_Colombia}{DA_uniform}{r_ff}{---}
\setres{cty_Colombia}{DA_uniform}{r_ff_se}{---}
\setres{cty_Colombia}{DA_uniform}{r_prior}{---}
\setres{cty_Colombia}{DA_uniform}{r_prior_se}{---}
\setres{cty_Colombia}{DA_uniform}{r_prob}{---}
\setres{cty_Colombia}{DA_uniform}{r_prob_se}{---}
\setres{cty_Colombia}{DA_uniform}{spread_penalty}{1.975}
\setres{cty_Colombia}{DA_uniform}{spread_penalty_se}{0.230}
\setres{cty_Colombia}{DA_uniform}{volume_ess}{---}
\setres{cty_Colombia}{EU_deterministic}{kappa}{0.783}
\setres{cty_Colombia}{EU_deterministic}{kappa_se}{0.081}
\setres{cty_Colombia}{EU_deterministic}{mse}{17.30}
\setres{cty_Colombia}{EU_deterministic}{nrmse_pct}{21.9}
\setres{cty_Colombia}{EU_deterministic}{r_ff}{---}
\setres{cty_Colombia}{EU_deterministic}{r_ff_se}{---}
\setres{cty_Colombia}{EU_deterministic}{r_prior}{---}
\setres{cty_Colombia}{EU_deterministic}{r_prior_se}{---}
\setres{cty_Colombia}{EU_deterministic}{r_prob}{---}
\setres{cty_Colombia}{EU_deterministic}{r_prob_se}{---}
\setres{cty_Colombia}{EU_deterministic}{theta_star}{0.89}
\setres{cty_Colombia}{EU_deterministic}{volume_ess}{---}
\setres{cty_Colombia}{EU_lit}{kappa}{0.392}
\setres{cty_Colombia}{EU_lit}{kappa_prior_mean}{0.571}
\setres{cty_Colombia}{EU_lit}{kappa_prior_mean_se}{0.137}
\setres{cty_Colombia}{EU_lit}{kappa_se}{0.157}
\setres{cty_Colombia}{EU_lit}{mse}{20.35}
\setres{cty_Colombia}{EU_lit}{mse_prior_mean}{18.94}
\setres{cty_Colombia}{EU_lit}{nrmse_pct}{23.2}
\setres{cty_Colombia}{EU_lit}{pred_variance}{1.40}
\setres{cty_Colombia}{EU_lit}{r_ff}{---}
\setres{cty_Colombia}{EU_lit}{r_ff_se}{---}
\setres{cty_Colombia}{EU_lit}{r_prior}{---}
\setres{cty_Colombia}{EU_lit}{r_prior_se}{---}
\setres{cty_Colombia}{EU_lit}{r_prob}{---}
\setres{cty_Colombia}{EU_lit}{r_prob_se}{---}
\setres{cty_Colombia}{EU_lit}{spread_penalty}{0.180}
\setres{cty_Colombia}{EU_lit}{spread_penalty_se}{0.021}
\setres{cty_Colombia}{EU_lit}{volume_ess}{---}
\setres{cty_Colombia}{EU_transfer_others}{kappa}{0.787}
\setres{cty_Colombia}{EU_transfer_others}{kappa_prior_mean}{0.787}
\setres{cty_Colombia}{EU_transfer_others}{kappa_prior_mean_se}{0.076}
\setres{cty_Colombia}{EU_transfer_others}{kappa_se}{0.076}
\setres{cty_Colombia}{EU_transfer_others}{mse}{17.27}
\setres{cty_Colombia}{EU_transfer_others}{mse_prior_mean}{17.27}
\setres{cty_Colombia}{EU_transfer_others}{nrmse_pct}{22.3}
\setres{cty_Colombia}{EU_transfer_others}{pred_variance}{0.00}
\setres{cty_Colombia}{EU_transfer_others}{r_ff}{---}
\setres{cty_Colombia}{EU_transfer_others}{r_ff_se}{---}
\setres{cty_Colombia}{EU_transfer_others}{r_prior}{---}
\setres{cty_Colombia}{EU_transfer_others}{r_prior_se}{---}
\setres{cty_Colombia}{EU_transfer_others}{r_prob}{---}
\setres{cty_Colombia}{EU_transfer_others}{r_prob_se}{---}
\setres{cty_Colombia}{EU_transfer_others}{spread_penalty}{0.000}
\setres{cty_Colombia}{EU_transfer_others}{spread_penalty_se}{0.000}
\setres{cty_Colombia}{EU_transfer_others}{theta_star}{0.95}
\setres{cty_Colombia}{EU_transfer_others}{transfer}{others}
\setres{cty_Colombia}{EU_transfer_others}{volume_ess}{---}
\setres{cty_Colombia}{EU_uniform}{kappa}{-1.757}
\setres{cty_Colombia}{EU_uniform}{kappa_prior_mean}{-0.532}
\setres{cty_Colombia}{EU_uniform}{kappa_prior_mean_se}{0.311}
\setres{cty_Colombia}{EU_uniform}{kappa_se}{0.441}
\setres{cty_Colombia}{EU_uniform}{mse}{37.09}
\setres{cty_Colombia}{EU_uniform}{mse_prior_mean}{27.54}
\setres{cty_Colombia}{EU_uniform}{nrmse_pct}{28.3}
\setres{cty_Colombia}{EU_uniform}{pred_variance}{9.55}
\setres{cty_Colombia}{EU_uniform}{r_ff}{---}
\setres{cty_Colombia}{EU_uniform}{r_ff_se}{---}
\setres{cty_Colombia}{EU_uniform}{r_prior}{---}
\setres{cty_Colombia}{EU_uniform}{r_prior_se}{---}
\setres{cty_Colombia}{EU_uniform}{r_prob}{---}
\setres{cty_Colombia}{EU_uniform}{r_prob_se}{---}
\setres{cty_Colombia}{EU_uniform}{spread_penalty}{1.225}
\setres{cty_Colombia}{EU_uniform}{spread_penalty_se}{0.143}
\setres{cty_Colombia}{EU_uniform}{volume_ess}{---}
\setres{cty_Colombia}{EV_deterministic}{kappa}{0.753}
\setres{cty_Colombia}{EV_deterministic}{kappa_se}{0.084}
\setres{cty_Colombia}{EV_deterministic}{mse}{17.53}
\setres{cty_Colombia}{EV_deterministic}{nrmse_pct}{22.6}
\setres{cty_Colombia}{EV_deterministic}{r_ff}{---}
\setres{cty_Colombia}{EV_deterministic}{r_ff_se}{---}
\setres{cty_Colombia}{EV_deterministic}{r_prior}{---}
\setres{cty_Colombia}{EV_deterministic}{r_prior_se}{---}
\setres{cty_Colombia}{EV_deterministic}{r_prob}{---}
\setres{cty_Colombia}{EV_deterministic}{r_prob_se}{---}
\setres{cty_Colombia}{EV_deterministic}{volume_ess}{---}
\setres{cty_Colombia}{reference}{country_name}{Colombia}
\setres{cty_Colombia}{reference}{lookup_nrmse_pct}{21.2}
\setres{cty_Colombia}{reference}{midpoint_mse}{23.40}
\setres{cty_Colombia}{reference}{midpoint_nrmse_pct}{26.4}
\setres{cty_Colombia}{reference}{missing_lotteries}{8.000, 14.000}
\setres{cty_Colombia}{reference}{n_lotteries}{12}
\setres{cty_Colombia}{reference}{n_obs}{1{,}536}
\setres{cty_Colombia}{reference}{n_subjects}{128}
\setres{cty_Colombia}{reference}{oracle_mse}{15.60}
\setres{cty_Colombia}{reference}{sample}{LV2019\_Colombia}
\setres{cty_Costa_Rica}{CPT_deterministic}{kappa}{0.977}
\setres{cty_Costa_Rica}{CPT_deterministic}{kappa_se}{0.016}
\setres{cty_Costa_Rica}{CPT_deterministic}{mse}{17.72}
\setres{cty_Costa_Rica}{CPT_deterministic}{nrmse_pct}{23.6}
\setres{cty_Costa_Rica}{CPT_deterministic}{r_ff}{---}
\setres{cty_Costa_Rica}{CPT_deterministic}{r_ff_se}{---}
\setres{cty_Costa_Rica}{CPT_deterministic}{r_prior}{---}
\setres{cty_Costa_Rica}{CPT_deterministic}{r_prior_se}{---}
\setres{cty_Costa_Rica}{CPT_deterministic}{r_prob}{---}
\setres{cty_Costa_Rica}{CPT_deterministic}{r_prob_se}{---}
\setres{cty_Costa_Rica}{CPT_deterministic}{theta_star}{1.00, 0.70, 0.80}
\setres{cty_Costa_Rica}{CPT_deterministic}{volume_ess}{---}
\setres{cty_Costa_Rica}{CPT_lit}{kappa}{0.592}
\setres{cty_Costa_Rica}{CPT_lit}{kappa_prior_mean}{0.898}
\setres{cty_Costa_Rica}{CPT_lit}{kappa_prior_mean_se}{0.047}
\setres{cty_Costa_Rica}{CPT_lit}{kappa_se}{0.072}
\setres{cty_Costa_Rica}{CPT_lit}{mse}{21.37}
\setres{cty_Costa_Rica}{CPT_lit}{mse_prior_mean}{18.47}
\setres{cty_Costa_Rica}{CPT_lit}{nrmse_pct}{25.0}
\setres{cty_Costa_Rica}{CPT_lit}{pred_variance}{2.90}
\setres{cty_Costa_Rica}{CPT_lit}{r_ff}{---}
\setres{cty_Costa_Rica}{CPT_lit}{r_ff_se}{---}
\setres{cty_Costa_Rica}{CPT_lit}{r_prior}{---}
\setres{cty_Costa_Rica}{CPT_lit}{r_prior_se}{---}
\setres{cty_Costa_Rica}{CPT_lit}{r_prob}{---}
\setres{cty_Costa_Rica}{CPT_lit}{r_prob_se}{---}
\setres{cty_Costa_Rica}{CPT_lit}{spread_penalty}{0.306}
\setres{cty_Costa_Rica}{CPT_lit}{spread_penalty_se}{0.031}
\setres{cty_Costa_Rica}{CPT_lit}{volume_ess}{---}
\setres{cty_Costa_Rica}{CPT_transfer_others}{kappa}{0.967}
\setres{cty_Costa_Rica}{CPT_transfer_others}{kappa_prior_mean}{0.967}
\setres{cty_Costa_Rica}{CPT_transfer_others}{kappa_prior_mean_se}{0.022}
\setres{cty_Costa_Rica}{CPT_transfer_others}{kappa_se}{0.022}
\setres{cty_Costa_Rica}{CPT_transfer_others}{mse}{17.82}
\setres{cty_Costa_Rica}{CPT_transfer_others}{mse_prior_mean}{17.82}
\setres{cty_Costa_Rica}{CPT_transfer_others}{nrmse_pct}{22.5}
\setres{cty_Costa_Rica}{CPT_transfer_others}{pred_variance}{0.00}
\setres{cty_Costa_Rica}{CPT_transfer_others}{r_ff}{---}
\setres{cty_Costa_Rica}{CPT_transfer_others}{r_ff_se}{---}
\setres{cty_Costa_Rica}{CPT_transfer_others}{r_prior}{---}
\setres{cty_Costa_Rica}{CPT_transfer_others}{r_prior_se}{---}
\setres{cty_Costa_Rica}{CPT_transfer_others}{r_prob}{---}
\setres{cty_Costa_Rica}{CPT_transfer_others}{r_prob_se}{---}
\setres{cty_Costa_Rica}{CPT_transfer_others}{spread_penalty}{0.000}
\setres{cty_Costa_Rica}{CPT_transfer_others}{spread_penalty_se}{0.000}
\setres{cty_Costa_Rica}{CPT_transfer_others}{theta_star}{1.10, 0.60, 0.80}
\setres{cty_Costa_Rica}{CPT_transfer_others}{transfer}{others}
\setres{cty_Costa_Rica}{CPT_transfer_others}{volume_ess}{---}
\setres{cty_Costa_Rica}{CPT_uniform}{kappa}{-1.493}
\setres{cty_Costa_Rica}{CPT_uniform}{kappa_prior_mean}{0.341}
\setres{cty_Costa_Rica}{CPT_uniform}{kappa_prior_mean_se}{0.116}
\setres{cty_Costa_Rica}{CPT_uniform}{kappa_se}{0.258}
\setres{cty_Costa_Rica}{CPT_uniform}{mse}{41.12}
\setres{cty_Costa_Rica}{CPT_uniform}{mse_prior_mean}{23.75}
\setres{cty_Costa_Rica}{CPT_uniform}{nrmse_pct}{28.3}
\setres{cty_Costa_Rica}{CPT_uniform}{pred_variance}{17.37}
\setres{cty_Costa_Rica}{CPT_uniform}{r_ff}{---}
\setres{cty_Costa_Rica}{CPT_uniform}{r_ff_se}{---}
\setres{cty_Costa_Rica}{CPT_uniform}{r_prior}{---}
\setres{cty_Costa_Rica}{CPT_uniform}{r_prior_se}{---}
\setres{cty_Costa_Rica}{CPT_uniform}{r_prob}{---}
\setres{cty_Costa_Rica}{CPT_uniform}{r_prob_se}{---}
\setres{cty_Costa_Rica}{CPT_uniform}{spread_penalty}{1.833}
\setres{cty_Costa_Rica}{CPT_uniform}{spread_penalty_se}{0.188}
\setres{cty_Costa_Rica}{CPT_uniform}{volume_ess}{---}
\setres{cty_Costa_Rica}{DA_deterministic}{kappa}{0.887}
\setres{cty_Costa_Rica}{DA_deterministic}{kappa_se}{0.044}
\setres{cty_Costa_Rica}{DA_deterministic}{mse}{18.58}
\setres{cty_Costa_Rica}{DA_deterministic}{nrmse_pct}{24.1}
\setres{cty_Costa_Rica}{DA_deterministic}{r_ff}{---}
\setres{cty_Costa_Rica}{DA_deterministic}{r_ff_se}{---}
\setres{cty_Costa_Rica}{DA_deterministic}{r_prior}{---}
\setres{cty_Costa_Rica}{DA_deterministic}{r_prior_se}{---}
\setres{cty_Costa_Rica}{DA_deterministic}{r_prob}{---}
\setres{cty_Costa_Rica}{DA_deterministic}{r_prob_se}{---}
\setres{cty_Costa_Rica}{DA_deterministic}{theta_star}{1.43, 1.18}
\setres{cty_Costa_Rica}{DA_deterministic}{volume_ess}{---}
\setres{cty_Costa_Rica}{DA_lit}{kappa}{-0.319}
\setres{cty_Costa_Rica}{DA_lit}{kappa_prior_mean}{0.182}
\setres{cty_Costa_Rica}{DA_lit}{kappa_prior_mean_se}{0.167}
\setres{cty_Costa_Rica}{DA_lit}{kappa_se}{0.210}
\setres{cty_Costa_Rica}{DA_lit}{mse}{30.00}
\setres{cty_Costa_Rica}{DA_lit}{mse_prior_mean}{25.25}
\setres{cty_Costa_Rica}{DA_lit}{nrmse_pct}{28.6}
\setres{cty_Costa_Rica}{DA_lit}{pred_variance}{4.75}
\setres{cty_Costa_Rica}{DA_lit}{r_ff}{---}
\setres{cty_Costa_Rica}{DA_lit}{r_ff_se}{---}
\setres{cty_Costa_Rica}{DA_lit}{r_prior}{---}
\setres{cty_Costa_Rica}{DA_lit}{r_prior_se}{---}
\setres{cty_Costa_Rica}{DA_lit}{r_prob}{---}
\setres{cty_Costa_Rica}{DA_lit}{r_prob_se}{---}
\setres{cty_Costa_Rica}{DA_lit}{spread_penalty}{0.501}
\setres{cty_Costa_Rica}{DA_lit}{spread_penalty_se}{0.051}
\setres{cty_Costa_Rica}{DA_lit}{volume_ess}{---}
\setres{cty_Costa_Rica}{DA_transfer_others}{kappa}{0.876}
\setres{cty_Costa_Rica}{DA_transfer_others}{kappa_prior_mean}{0.876}
\setres{cty_Costa_Rica}{DA_transfer_others}{kappa_prior_mean_se}{0.051}
\setres{cty_Costa_Rica}{DA_transfer_others}{kappa_se}{0.051}
\setres{cty_Costa_Rica}{DA_transfer_others}{mse}{18.68}
\setres{cty_Costa_Rica}{DA_transfer_others}{mse_prior_mean}{18.68}
\setres{cty_Costa_Rica}{DA_transfer_others}{nrmse_pct}{22.9}
\setres{cty_Costa_Rica}{DA_transfer_others}{pred_variance}{0.00}
\setres{cty_Costa_Rica}{DA_transfer_others}{r_ff}{---}
\setres{cty_Costa_Rica}{DA_transfer_others}{r_ff_se}{---}
\setres{cty_Costa_Rica}{DA_transfer_others}{r_prior}{---}
\setres{cty_Costa_Rica}{DA_transfer_others}{r_prior_se}{---}
\setres{cty_Costa_Rica}{DA_transfer_others}{r_prob}{---}
\setres{cty_Costa_Rica}{DA_transfer_others}{r_prob_se}{---}
\setres{cty_Costa_Rica}{DA_transfer_others}{spread_penalty}{0.000}
\setres{cty_Costa_Rica}{DA_transfer_others}{spread_penalty_se}{0.000}
\setres{cty_Costa_Rica}{DA_transfer_others}{theta_star}{1.70, 1.46}
\setres{cty_Costa_Rica}{DA_transfer_others}{transfer}{others}
\setres{cty_Costa_Rica}{DA_transfer_others}{volume_ess}{---}
\setres{cty_Costa_Rica}{DA_uniform}{kappa}{-1.997}
\setres{cty_Costa_Rica}{DA_uniform}{kappa_prior_mean}{-0.557}
\setres{cty_Costa_Rica}{DA_uniform}{kappa_prior_mean_se}{0.249}
\setres{cty_Costa_Rica}{DA_uniform}{kappa_se}{0.370}
\setres{cty_Costa_Rica}{DA_uniform}{mse}{45.90}
\setres{cty_Costa_Rica}{DA_uniform}{mse_prior_mean}{32.25}
\setres{cty_Costa_Rica}{DA_uniform}{nrmse_pct}{30.9}
\setres{cty_Costa_Rica}{DA_uniform}{pred_variance}{13.65}
\setres{cty_Costa_Rica}{DA_uniform}{r_ff}{---}
\setres{cty_Costa_Rica}{DA_uniform}{r_ff_se}{---}
\setres{cty_Costa_Rica}{DA_uniform}{r_prior}{---}
\setres{cty_Costa_Rica}{DA_uniform}{r_prior_se}{---}
\setres{cty_Costa_Rica}{DA_uniform}{r_prob}{---}
\setres{cty_Costa_Rica}{DA_uniform}{r_prob_se}{---}
\setres{cty_Costa_Rica}{DA_uniform}{spread_penalty}{1.441}
\setres{cty_Costa_Rica}{DA_uniform}{spread_penalty_se}{0.148}
\setres{cty_Costa_Rica}{DA_uniform}{volume_ess}{---}
\setres{cty_Costa_Rica}{EU_deterministic}{kappa}{0.833}
\setres{cty_Costa_Rica}{EU_deterministic}{kappa_se}{0.069}
\setres{cty_Costa_Rica}{EU_deterministic}{mse}{19.08}
\setres{cty_Costa_Rica}{EU_deterministic}{nrmse_pct}{25.2}
\setres{cty_Costa_Rica}{EU_deterministic}{r_ff}{---}
\setres{cty_Costa_Rica}{EU_deterministic}{r_ff_se}{---}
\setres{cty_Costa_Rica}{EU_deterministic}{r_prior}{---}
\setres{cty_Costa_Rica}{EU_deterministic}{r_prior_se}{---}
\setres{cty_Costa_Rica}{EU_deterministic}{r_prob}{---}
\setres{cty_Costa_Rica}{EU_deterministic}{r_prob_se}{---}
\setres{cty_Costa_Rica}{EU_deterministic}{theta_star}{0.85}
\setres{cty_Costa_Rica}{EU_deterministic}{volume_ess}{---}
\setres{cty_Costa_Rica}{EU_lit}{kappa}{0.612}
\setres{cty_Costa_Rica}{EU_lit}{kappa_prior_mean}{0.739}
\setres{cty_Costa_Rica}{EU_lit}{kappa_prior_mean_se}{0.084}
\setres{cty_Costa_Rica}{EU_lit}{kappa_se}{0.096}
\setres{cty_Costa_Rica}{EU_lit}{mse}{21.18}
\setres{cty_Costa_Rica}{EU_lit}{mse_prior_mean}{19.97}
\setres{cty_Costa_Rica}{EU_lit}{nrmse_pct}{26.4}
\setres{cty_Costa_Rica}{EU_lit}{pred_variance}{1.20}
\setres{cty_Costa_Rica}{EU_lit}{r_ff}{---}
\setres{cty_Costa_Rica}{EU_lit}{r_ff_se}{---}
\setres{cty_Costa_Rica}{EU_lit}{r_prior}{---}
\setres{cty_Costa_Rica}{EU_lit}{r_prior_se}{---}
\setres{cty_Costa_Rica}{EU_lit}{r_prob}{---}
\setres{cty_Costa_Rica}{EU_lit}{r_prob_se}{---}
\setres{cty_Costa_Rica}{EU_lit}{spread_penalty}{0.127}
\setres{cty_Costa_Rica}{EU_lit}{spread_penalty_se}{0.013}
\setres{cty_Costa_Rica}{EU_lit}{volume_ess}{---}
\setres{cty_Costa_Rica}{EU_transfer_others}{kappa}{0.808}
\setres{cty_Costa_Rica}{EU_transfer_others}{kappa_prior_mean}{0.808}
\setres{cty_Costa_Rica}{EU_transfer_others}{kappa_prior_mean_se}{0.081}
\setres{cty_Costa_Rica}{EU_transfer_others}{kappa_se}{0.081}
\setres{cty_Costa_Rica}{EU_transfer_others}{mse}{19.32}
\setres{cty_Costa_Rica}{EU_transfer_others}{mse_prior_mean}{19.32}
\setres{cty_Costa_Rica}{EU_transfer_others}{nrmse_pct}{24.2}
\setres{cty_Costa_Rica}{EU_transfer_others}{pred_variance}{0.00}
\setres{cty_Costa_Rica}{EU_transfer_others}{r_ff}{---}
\setres{cty_Costa_Rica}{EU_transfer_others}{r_ff_se}{---}
\setres{cty_Costa_Rica}{EU_transfer_others}{r_prior}{---}
\setres{cty_Costa_Rica}{EU_transfer_others}{r_prior_se}{---}
\setres{cty_Costa_Rica}{EU_transfer_others}{r_prob}{---}
\setres{cty_Costa_Rica}{EU_transfer_others}{r_prob_se}{---}
\setres{cty_Costa_Rica}{EU_transfer_others}{spread_penalty}{0.000}
\setres{cty_Costa_Rica}{EU_transfer_others}{spread_penalty_se}{0.000}
\setres{cty_Costa_Rica}{EU_transfer_others}{theta_star}{0.95}
\setres{cty_Costa_Rica}{EU_transfer_others}{transfer}{others}
\setres{cty_Costa_Rica}{EU_transfer_others}{volume_ess}{---}
\setres{cty_Costa_Rica}{EU_uniform}{kappa}{-0.823}
\setres{cty_Costa_Rica}{EU_uniform}{kappa_prior_mean}{0.043}
\setres{cty_Costa_Rica}{EU_uniform}{kappa_prior_mean_se}{0.180}
\setres{cty_Costa_Rica}{EU_uniform}{kappa_se}{0.255}
\setres{cty_Costa_Rica}{EU_uniform}{mse}{34.77}
\setres{cty_Costa_Rica}{EU_uniform}{mse_prior_mean}{26.58}
\setres{cty_Costa_Rica}{EU_uniform}{nrmse_pct}{29.1}
\setres{cty_Costa_Rica}{EU_uniform}{pred_variance}{8.20}
\setres{cty_Costa_Rica}{EU_uniform}{r_ff}{---}
\setres{cty_Costa_Rica}{EU_uniform}{r_ff_se}{---}
\setres{cty_Costa_Rica}{EU_uniform}{r_prior}{---}
\setres{cty_Costa_Rica}{EU_uniform}{r_prior_se}{---}
\setres{cty_Costa_Rica}{EU_uniform}{r_prob}{---}
\setres{cty_Costa_Rica}{EU_uniform}{r_prob_se}{---}
\setres{cty_Costa_Rica}{EU_uniform}{spread_penalty}{0.865}
\setres{cty_Costa_Rica}{EU_uniform}{spread_penalty_se}{0.089}
\setres{cty_Costa_Rica}{EU_uniform}{volume_ess}{---}
\setres{cty_Costa_Rica}{EV_deterministic}{kappa}{0.770}
\setres{cty_Costa_Rica}{EV_deterministic}{kappa_se}{0.088}
\setres{cty_Costa_Rica}{EV_deterministic}{mse}{19.68}
\setres{cty_Costa_Rica}{EV_deterministic}{nrmse_pct}{24.1}
\setres{cty_Costa_Rica}{EV_deterministic}{r_ff}{---}
\setres{cty_Costa_Rica}{EV_deterministic}{r_ff_se}{---}
\setres{cty_Costa_Rica}{EV_deterministic}{r_prior}{---}
\setres{cty_Costa_Rica}{EV_deterministic}{r_prior_se}{---}
\setres{cty_Costa_Rica}{EV_deterministic}{r_prob}{---}
\setres{cty_Costa_Rica}{EV_deterministic}{r_prob_se}{---}
\setres{cty_Costa_Rica}{EV_deterministic}{volume_ess}{---}
\setres{cty_Costa_Rica}{reference}{country_name}{Costa Rica}
\setres{cty_Costa_Rica}{reference}{lookup_nrmse_pct}{22.9}
\setres{cty_Costa_Rica}{reference}{midpoint_mse}{26.98}
\setres{cty_Costa_Rica}{reference}{midpoint_nrmse_pct}{28.5}
\setres{cty_Costa_Rica}{reference}{missing_lotteries}{none}
\setres{cty_Costa_Rica}{reference}{n_lotteries}{14}
\setres{cty_Costa_Rica}{reference}{n_obs}{1{,}484}
\setres{cty_Costa_Rica}{reference}{n_subjects}{106}
\setres{cty_Costa_Rica}{reference}{oracle_mse}{17.50}
\setres{cty_Costa_Rica}{reference}{sample}{LV2019\_Costa\_Rica}
\setres{cty_Czech_Republic}{CPT_deterministic}{kappa}{0.986}
\setres{cty_Czech_Republic}{CPT_deterministic}{kappa_se}{0.007}
\setres{cty_Czech_Republic}{CPT_deterministic}{mse}{6.24}
\setres{cty_Czech_Republic}{CPT_deterministic}{nrmse_pct}{14.3}
\setres{cty_Czech_Republic}{CPT_deterministic}{r_ff}{---}
\setres{cty_Czech_Republic}{CPT_deterministic}{r_ff_se}{---}
\setres{cty_Czech_Republic}{CPT_deterministic}{r_prior}{---}
\setres{cty_Czech_Republic}{CPT_deterministic}{r_prior_se}{---}
\setres{cty_Czech_Republic}{CPT_deterministic}{r_prob}{---}
\setres{cty_Czech_Republic}{CPT_deterministic}{r_prob_se}{---}
\setres{cty_Czech_Republic}{CPT_deterministic}{theta_star}{0.90, 0.70, 0.90}
\setres{cty_Czech_Republic}{CPT_deterministic}{volume_ess}{---}
\setres{cty_Czech_Republic}{CPT_lit}{kappa}{0.653}
\setres{cty_Czech_Republic}{CPT_lit}{kappa_prior_mean}{0.905}
\setres{cty_Czech_Republic}{CPT_lit}{kappa_prior_mean_se}{0.029}
\setres{cty_Czech_Republic}{CPT_lit}{kappa_se}{0.035}
\setres{cty_Czech_Republic}{CPT_lit}{mse}{10.07}
\setres{cty_Czech_Republic}{CPT_lit}{mse_prior_mean}{7.17}
\setres{cty_Czech_Republic}{CPT_lit}{nrmse_pct}{15.3}
\setres{cty_Czech_Republic}{CPT_lit}{pred_variance}{2.90}
\setres{cty_Czech_Republic}{CPT_lit}{r_ff}{---}
\setres{cty_Czech_Republic}{CPT_lit}{r_ff_se}{---}
\setres{cty_Czech_Republic}{CPT_lit}{r_prior}{---}
\setres{cty_Czech_Republic}{CPT_lit}{r_prior_se}{---}
\setres{cty_Czech_Republic}{CPT_lit}{r_prob}{---}
\setres{cty_Czech_Republic}{CPT_lit}{r_prob_se}{---}
\setres{cty_Czech_Republic}{CPT_lit}{spread_penalty}{0.252}
\setres{cty_Czech_Republic}{CPT_lit}{spread_penalty_se}{0.012}
\setres{cty_Czech_Republic}{CPT_lit}{volume_ess}{---}
\setres{cty_Czech_Republic}{CPT_transfer_others}{kappa}{0.944}
\setres{cty_Czech_Republic}{CPT_transfer_others}{kappa_prior_mean}{0.944}
\setres{cty_Czech_Republic}{CPT_transfer_others}{kappa_prior_mean_se}{0.012}
\setres{cty_Czech_Republic}{CPT_transfer_others}{kappa_se}{0.012}
\setres{cty_Czech_Republic}{CPT_transfer_others}{mse}{6.72}
\setres{cty_Czech_Republic}{CPT_transfer_others}{mse_prior_mean}{6.72}
\setres{cty_Czech_Republic}{CPT_transfer_others}{nrmse_pct}{14.5}
\setres{cty_Czech_Republic}{CPT_transfer_others}{pred_variance}{0.00}
\setres{cty_Czech_Republic}{CPT_transfer_others}{r_ff}{---}
\setres{cty_Czech_Republic}{CPT_transfer_others}{r_ff_se}{---}
\setres{cty_Czech_Republic}{CPT_transfer_others}{r_prior}{---}
\setres{cty_Czech_Republic}{CPT_transfer_others}{r_prior_se}{---}
\setres{cty_Czech_Republic}{CPT_transfer_others}{r_prob}{---}
\setres{cty_Czech_Republic}{CPT_transfer_others}{r_prob_se}{---}
\setres{cty_Czech_Republic}{CPT_transfer_others}{spread_penalty}{0.000}
\setres{cty_Czech_Republic}{CPT_transfer_others}{spread_penalty_se}{0.000}
\setres{cty_Czech_Republic}{CPT_transfer_others}{theta_star}{1.10, 0.60, 0.80}
\setres{cty_Czech_Republic}{CPT_transfer_others}{transfer}{others}
\setres{cty_Czech_Republic}{CPT_transfer_others}{volume_ess}{---}
\setres{cty_Czech_Republic}{CPT_uniform}{kappa}{-1.133}
\setres{cty_Czech_Republic}{CPT_uniform}{kappa_prior_mean}{0.378}
\setres{cty_Czech_Republic}{CPT_uniform}{kappa_prior_mean_se}{0.077}
\setres{cty_Czech_Republic}{CPT_uniform}{kappa_se}{0.122}
\setres{cty_Czech_Republic}{CPT_uniform}{mse}{30.59}
\setres{cty_Czech_Republic}{CPT_uniform}{mse_prior_mean}{13.23}
\setres{cty_Czech_Republic}{CPT_uniform}{nrmse_pct}{19.7}
\setres{cty_Czech_Republic}{CPT_uniform}{pred_variance}{17.37}
\setres{cty_Czech_Republic}{CPT_uniform}{r_ff}{---}
\setres{cty_Czech_Republic}{CPT_uniform}{r_ff_se}{---}
\setres{cty_Czech_Republic}{CPT_uniform}{r_prior}{---}
\setres{cty_Czech_Republic}{CPT_uniform}{r_prior_se}{---}
\setres{cty_Czech_Republic}{CPT_uniform}{r_prob}{---}
\setres{cty_Czech_Republic}{CPT_uniform}{r_prob_se}{---}
\setres{cty_Czech_Republic}{CPT_uniform}{spread_penalty}{1.511}
\setres{cty_Czech_Republic}{CPT_uniform}{spread_penalty_se}{0.073}
\setres{cty_Czech_Republic}{CPT_uniform}{volume_ess}{---}
\setres{cty_Czech_Republic}{DA_deterministic}{kappa}{0.933}
\setres{cty_Czech_Republic}{DA_deterministic}{kappa_se}{0.015}
\setres{cty_Czech_Republic}{DA_deterministic}{mse}{6.85}
\setres{cty_Czech_Republic}{DA_deterministic}{nrmse_pct}{15.6}
\setres{cty_Czech_Republic}{DA_deterministic}{r_ff}{---}
\setres{cty_Czech_Republic}{DA_deterministic}{r_ff_se}{---}
\setres{cty_Czech_Republic}{DA_deterministic}{r_prior}{---}
\setres{cty_Czech_Republic}{DA_deterministic}{r_prior_se}{---}
\setres{cty_Czech_Republic}{DA_deterministic}{r_prob}{---}
\setres{cty_Czech_Republic}{DA_deterministic}{r_prob_se}{---}
\setres{cty_Czech_Republic}{DA_deterministic}{theta_star}{1.16, 0.53}
\setres{cty_Czech_Republic}{DA_deterministic}{volume_ess}{---}
\setres{cty_Czech_Republic}{DA_lit}{kappa}{-0.113}
\setres{cty_Czech_Republic}{DA_lit}{kappa_prior_mean}{0.300}
\setres{cty_Czech_Republic}{DA_lit}{kappa_prior_mean_se}{0.093}
\setres{cty_Czech_Republic}{DA_lit}{kappa_se}{0.106}
\setres{cty_Czech_Republic}{DA_lit}{mse}{18.87}
\setres{cty_Czech_Republic}{DA_lit}{mse_prior_mean}{14.12}
\setres{cty_Czech_Republic}{DA_lit}{nrmse_pct}{19.5}
\setres{cty_Czech_Republic}{DA_lit}{pred_variance}{4.75}
\setres{cty_Czech_Republic}{DA_lit}{r_ff}{---}
\setres{cty_Czech_Republic}{DA_lit}{r_ff_se}{---}
\setres{cty_Czech_Republic}{DA_lit}{r_prior}{---}
\setres{cty_Czech_Republic}{DA_lit}{r_prior_se}{---}
\setres{cty_Czech_Republic}{DA_lit}{r_prob}{---}
\setres{cty_Czech_Republic}{DA_lit}{r_prob_se}{---}
\setres{cty_Czech_Republic}{DA_lit}{spread_penalty}{0.413}
\setres{cty_Czech_Republic}{DA_lit}{spread_penalty_se}{0.020}
\setres{cty_Czech_Republic}{DA_lit}{volume_ess}{---}
\setres{cty_Czech_Republic}{DA_transfer_others}{kappa}{0.895}
\setres{cty_Czech_Republic}{DA_transfer_others}{kappa_prior_mean}{0.895}
\setres{cty_Czech_Republic}{DA_transfer_others}{kappa_prior_mean_se}{0.020}
\setres{cty_Czech_Republic}{DA_transfer_others}{kappa_se}{0.020}
\setres{cty_Czech_Republic}{DA_transfer_others}{mse}{7.29}
\setres{cty_Czech_Republic}{DA_transfer_others}{mse_prior_mean}{7.29}
\setres{cty_Czech_Republic}{DA_transfer_others}{nrmse_pct}{14.9}
\setres{cty_Czech_Republic}{DA_transfer_others}{pred_variance}{0.00}
\setres{cty_Czech_Republic}{DA_transfer_others}{r_ff}{---}
\setres{cty_Czech_Republic}{DA_transfer_others}{r_ff_se}{---}
\setres{cty_Czech_Republic}{DA_transfer_others}{r_prior}{---}
\setres{cty_Czech_Republic}{DA_transfer_others}{r_prior_se}{---}
\setres{cty_Czech_Republic}{DA_transfer_others}{r_prob}{---}
\setres{cty_Czech_Republic}{DA_transfer_others}{r_prob_se}{---}
\setres{cty_Czech_Republic}{DA_transfer_others}{spread_penalty}{0.000}
\setres{cty_Czech_Republic}{DA_transfer_others}{spread_penalty_se}{0.000}
\setres{cty_Czech_Republic}{DA_transfer_others}{theta_star}{1.72, 1.50}
\setres{cty_Czech_Republic}{DA_transfer_others}{transfer}{others}
\setres{cty_Czech_Republic}{DA_transfer_others}{volume_ess}{---}
\setres{cty_Czech_Republic}{DA_uniform}{kappa}{-1.546}
\setres{cty_Czech_Republic}{DA_uniform}{kappa_prior_mean}{-0.359}
\setres{cty_Czech_Republic}{DA_uniform}{kappa_prior_mean_se}{0.142}
\setres{cty_Czech_Republic}{DA_uniform}{kappa_se}{0.181}
\setres{cty_Czech_Republic}{DA_uniform}{mse}{35.34}
\setres{cty_Czech_Republic}{DA_uniform}{mse_prior_mean}{21.70}
\setres{cty_Czech_Republic}{DA_uniform}{nrmse_pct}{25.1}
\setres{cty_Czech_Republic}{DA_uniform}{pred_variance}{13.65}
\setres{cty_Czech_Republic}{DA_uniform}{r_ff}{---}
\setres{cty_Czech_Republic}{DA_uniform}{r_ff_se}{---}
\setres{cty_Czech_Republic}{DA_uniform}{r_prior}{---}
\setres{cty_Czech_Republic}{DA_uniform}{r_prior_se}{---}
\setres{cty_Czech_Republic}{DA_uniform}{r_prob}{---}
\setres{cty_Czech_Republic}{DA_uniform}{r_prob_se}{---}
\setres{cty_Czech_Republic}{DA_uniform}{spread_penalty}{1.187}
\setres{cty_Czech_Republic}{DA_uniform}{spread_penalty_se}{0.058}
\setres{cty_Czech_Republic}{DA_uniform}{volume_ess}{---}
\setres{cty_Czech_Republic}{EU_deterministic}{kappa}{0.915}
\setres{cty_Czech_Republic}{EU_deterministic}{kappa_se}{0.020}
\setres{cty_Czech_Republic}{EU_deterministic}{mse}{7.06}
\setres{cty_Czech_Republic}{EU_deterministic}{nrmse_pct}{15.3}
\setres{cty_Czech_Republic}{EU_deterministic}{r_ff}{---}
\setres{cty_Czech_Republic}{EU_deterministic}{r_ff_se}{---}
\setres{cty_Czech_Republic}{EU_deterministic}{r_prior}{---}
\setres{cty_Czech_Republic}{EU_deterministic}{r_prior_se}{---}
\setres{cty_Czech_Republic}{EU_deterministic}{r_prob}{---}
\setres{cty_Czech_Republic}{EU_deterministic}{r_prob_se}{---}
\setres{cty_Czech_Republic}{EU_deterministic}{theta_star}{0.86}
\setres{cty_Czech_Republic}{EU_deterministic}{volume_ess}{---}
\setres{cty_Czech_Republic}{EU_lit}{kappa}{0.705}
\setres{cty_Czech_Republic}{EU_lit}{kappa_prior_mean}{0.810}
\setres{cty_Czech_Republic}{EU_lit}{kappa_prior_mean_se}{0.036}
\setres{cty_Czech_Republic}{EU_lit}{kappa_se}{0.040}
\setres{cty_Czech_Republic}{EU_lit}{mse}{9.47}
\setres{cty_Czech_Republic}{EU_lit}{mse_prior_mean}{8.26}
\setres{cty_Czech_Republic}{EU_lit}{nrmse_pct}{15.6}
\setres{cty_Czech_Republic}{EU_lit}{pred_variance}{1.20}
\setres{cty_Czech_Republic}{EU_lit}{r_ff}{---}
\setres{cty_Czech_Republic}{EU_lit}{r_ff_se}{---}
\setres{cty_Czech_Republic}{EU_lit}{r_prior}{---}
\setres{cty_Czech_Republic}{EU_lit}{r_prior_se}{---}
\setres{cty_Czech_Republic}{EU_lit}{r_prob}{---}
\setres{cty_Czech_Republic}{EU_lit}{r_prob_se}{---}
\setres{cty_Czech_Republic}{EU_lit}{spread_penalty}{0.105}
\setres{cty_Czech_Republic}{EU_lit}{spread_penalty_se}{0.005}
\setres{cty_Czech_Republic}{EU_lit}{volume_ess}{---}
\setres{cty_Czech_Republic}{EU_transfer_others}{kappa}{0.896}
\setres{cty_Czech_Republic}{EU_transfer_others}{kappa_prior_mean}{0.896}
\setres{cty_Czech_Republic}{EU_transfer_others}{kappa_prior_mean_se}{0.029}
\setres{cty_Czech_Republic}{EU_transfer_others}{kappa_se}{0.029}
\setres{cty_Czech_Republic}{EU_transfer_others}{mse}{7.27}
\setres{cty_Czech_Republic}{EU_transfer_others}{mse_prior_mean}{7.27}
\setres{cty_Czech_Republic}{EU_transfer_others}{nrmse_pct}{14.5}
\setres{cty_Czech_Republic}{EU_transfer_others}{pred_variance}{0.00}
\setres{cty_Czech_Republic}{EU_transfer_others}{r_ff}{---}
\setres{cty_Czech_Republic}{EU_transfer_others}{r_ff_se}{---}
\setres{cty_Czech_Republic}{EU_transfer_others}{r_prior}{---}
\setres{cty_Czech_Republic}{EU_transfer_others}{r_prior_se}{---}
\setres{cty_Czech_Republic}{EU_transfer_others}{r_prob}{---}
\setres{cty_Czech_Republic}{EU_transfer_others}{r_prob_se}{---}
\setres{cty_Czech_Republic}{EU_transfer_others}{spread_penalty}{0.000}
\setres{cty_Czech_Republic}{EU_transfer_others}{spread_penalty_se}{0.000}
\setres{cty_Czech_Republic}{EU_transfer_others}{theta_star}{0.95}
\setres{cty_Czech_Republic}{EU_transfer_others}{transfer}{others}
\setres{cty_Czech_Republic}{EU_transfer_others}{volume_ess}{---}
\setres{cty_Czech_Republic}{EU_uniform}{kappa}{-0.522}
\setres{cty_Czech_Republic}{EU_uniform}{kappa_prior_mean}{0.191}
\setres{cty_Czech_Republic}{EU_uniform}{kappa_prior_mean_se}{0.101}
\setres{cty_Czech_Republic}{EU_uniform}{kappa_se}{0.124}
\setres{cty_Czech_Republic}{EU_uniform}{mse}{23.57}
\setres{cty_Czech_Republic}{EU_uniform}{mse_prior_mean}{15.38}
\setres{cty_Czech_Republic}{EU_uniform}{nrmse_pct}{20.5}
\setres{cty_Czech_Republic}{EU_uniform}{pred_variance}{8.20}
\setres{cty_Czech_Republic}{EU_uniform}{r_ff}{---}
\setres{cty_Czech_Republic}{EU_uniform}{r_ff_se}{---}
\setres{cty_Czech_Republic}{EU_uniform}{r_prior}{---}
\setres{cty_Czech_Republic}{EU_uniform}{r_prior_se}{---}
\setres{cty_Czech_Republic}{EU_uniform}{r_prob}{---}
\setres{cty_Czech_Republic}{EU_uniform}{r_prob_se}{---}
\setres{cty_Czech_Republic}{EU_uniform}{spread_penalty}{0.713}
\setres{cty_Czech_Republic}{EU_uniform}{spread_penalty_se}{0.035}
\setres{cty_Czech_Republic}{EU_uniform}{volume_ess}{---}
\setres{cty_Czech_Republic}{EV_deterministic}{kappa}{0.868}
\setres{cty_Czech_Republic}{EV_deterministic}{kappa_se}{0.036}
\setres{cty_Czech_Republic}{EV_deterministic}{mse}{7.59}
\setres{cty_Czech_Republic}{EV_deterministic}{nrmse_pct}{14.0}
\setres{cty_Czech_Republic}{EV_deterministic}{r_ff}{---}
\setres{cty_Czech_Republic}{EV_deterministic}{r_ff_se}{---}
\setres{cty_Czech_Republic}{EV_deterministic}{r_prior}{---}
\setres{cty_Czech_Republic}{EV_deterministic}{r_prior_se}{---}
\setres{cty_Czech_Republic}{EV_deterministic}{r_prob}{---}
\setres{cty_Czech_Republic}{EV_deterministic}{r_prob_se}{---}
\setres{cty_Czech_Republic}{EV_deterministic}{volume_ess}{---}
\setres{cty_Czech_Republic}{reference}{country_name}{Czech Republic}
\setres{cty_Czech_Republic}{reference}{lookup_nrmse_pct}{12.9}
\setres{cty_Czech_Republic}{reference}{midpoint_mse}{17.57}
\setres{cty_Czech_Republic}{reference}{midpoint_nrmse_pct}{20.6}
\setres{cty_Czech_Republic}{reference}{missing_lotteries}{none}
\setres{cty_Czech_Republic}{reference}{n_lotteries}{14}
\setres{cty_Czech_Republic}{reference}{n_obs}{1{,}386}
\setres{cty_Czech_Republic}{reference}{n_subjects}{99}
\setres{cty_Czech_Republic}{reference}{oracle_mse}{6.08}
\setres{cty_Czech_Republic}{reference}{sample}{LV2019\_Czech\_Republic}
\setres{cty_Ethiopia}{CPT_deterministic}{kappa}{0.924}
\setres{cty_Ethiopia}{CPT_deterministic}{kappa_se}{0.020}
\setres{cty_Ethiopia}{CPT_deterministic}{mse}{18.14}
\setres{cty_Ethiopia}{CPT_deterministic}{nrmse_pct}{21.9}
\setres{cty_Ethiopia}{CPT_deterministic}{r_ff}{---}
\setres{cty_Ethiopia}{CPT_deterministic}{r_ff_se}{---}
\setres{cty_Ethiopia}{CPT_deterministic}{r_prior}{---}
\setres{cty_Ethiopia}{CPT_deterministic}{r_prior_se}{---}
\setres{cty_Ethiopia}{CPT_deterministic}{r_prob}{---}
\setres{cty_Ethiopia}{CPT_deterministic}{r_prob_se}{---}
\setres{cty_Ethiopia}{CPT_deterministic}{theta_star}{1.20, 0.60, 0.90}
\setres{cty_Ethiopia}{CPT_deterministic}{volume_ess}{---}
\setres{cty_Ethiopia}{CPT_lit}{kappa}{-0.309}
\setres{cty_Ethiopia}{CPT_lit}{kappa_prior_mean}{0.068}
\setres{cty_Ethiopia}{CPT_lit}{kappa_prior_mean_se}{0.159}
\setres{cty_Ethiopia}{CPT_lit}{kappa_se}{0.176}
\setres{cty_Ethiopia}{CPT_lit}{mse}{27.63}
\setres{cty_Ethiopia}{CPT_lit}{mse_prior_mean}{24.73}
\setres{cty_Ethiopia}{CPT_lit}{nrmse_pct}{27.4}
\setres{cty_Ethiopia}{CPT_lit}{pred_variance}{2.90}
\setres{cty_Ethiopia}{CPT_lit}{r_ff}{---}
\setres{cty_Ethiopia}{CPT_lit}{r_ff_se}{---}
\setres{cty_Ethiopia}{CPT_lit}{r_prior}{---}
\setres{cty_Ethiopia}{CPT_lit}{r_prior_se}{---}
\setres{cty_Ethiopia}{CPT_lit}{r_prob}{---}
\setres{cty_Ethiopia}{CPT_lit}{r_prob_se}{---}
\setres{cty_Ethiopia}{CPT_lit}{spread_penalty}{0.377}
\setres{cty_Ethiopia}{CPT_lit}{spread_penalty_se}{0.038}
\setres{cty_Ethiopia}{CPT_lit}{volume_ess}{---}
\setres{cty_Ethiopia}{CPT_transfer_others}{kappa}{0.785}
\setres{cty_Ethiopia}{CPT_transfer_others}{kappa_prior_mean}{0.785}
\setres{cty_Ethiopia}{CPT_transfer_others}{kappa_prior_mean_se}{0.070}
\setres{cty_Ethiopia}{CPT_transfer_others}{kappa_se}{0.070}
\setres{cty_Ethiopia}{CPT_transfer_others}{mse}{19.21}
\setres{cty_Ethiopia}{CPT_transfer_others}{mse_prior_mean}{19.21}
\setres{cty_Ethiopia}{CPT_transfer_others}{nrmse_pct}{23.0}
\setres{cty_Ethiopia}{CPT_transfer_others}{pred_variance}{0.00}
\setres{cty_Ethiopia}{CPT_transfer_others}{r_ff}{---}
\setres{cty_Ethiopia}{CPT_transfer_others}{r_ff_se}{---}
\setres{cty_Ethiopia}{CPT_transfer_others}{r_prior}{---}
\setres{cty_Ethiopia}{CPT_transfer_others}{r_prior_se}{---}
\setres{cty_Ethiopia}{CPT_transfer_others}{r_prob}{---}
\setres{cty_Ethiopia}{CPT_transfer_others}{r_prob_se}{---}
\setres{cty_Ethiopia}{CPT_transfer_others}{spread_penalty}{0.000}
\setres{cty_Ethiopia}{CPT_transfer_others}{spread_penalty_se}{0.000}
\setres{cty_Ethiopia}{CPT_transfer_others}{theta_star}{1.00, 0.60, 0.90}
\setres{cty_Ethiopia}{CPT_transfer_others}{transfer}{others}
\setres{cty_Ethiopia}{CPT_transfer_others}{volume_ess}{---}
\setres{cty_Ethiopia}{CPT_uniform}{kappa}{-3.286}
\setres{cty_Ethiopia}{CPT_uniform}{kappa_prior_mean}{-1.030}
\setres{cty_Ethiopia}{CPT_uniform}{kappa_prior_mean_se}{0.214}
\setres{cty_Ethiopia}{CPT_uniform}{kappa_se}{0.378}
\setres{cty_Ethiopia}{CPT_uniform}{mse}{50.55}
\setres{cty_Ethiopia}{CPT_uniform}{mse_prior_mean}{33.19}
\setres{cty_Ethiopia}{CPT_uniform}{nrmse_pct}{30.9}
\setres{cty_Ethiopia}{CPT_uniform}{pred_variance}{17.37}
\setres{cty_Ethiopia}{CPT_uniform}{r_ff}{---}
\setres{cty_Ethiopia}{CPT_uniform}{r_ff_se}{---}
\setres{cty_Ethiopia}{CPT_uniform}{r_prior}{---}
\setres{cty_Ethiopia}{CPT_uniform}{r_prior_se}{---}
\setres{cty_Ethiopia}{CPT_uniform}{r_prob}{---}
\setres{cty_Ethiopia}{CPT_uniform}{r_prob_se}{---}
\setres{cty_Ethiopia}{CPT_uniform}{spread_penalty}{2.256}
\setres{cty_Ethiopia}{CPT_uniform}{spread_penalty_se}{0.229}
\setres{cty_Ethiopia}{CPT_uniform}{volume_ess}{---}
\setres{cty_Ethiopia}{DA_deterministic}{kappa}{0.793}
\setres{cty_Ethiopia}{DA_deterministic}{kappa_se}{0.051}
\setres{cty_Ethiopia}{DA_deterministic}{mse}{19.15}
\setres{cty_Ethiopia}{DA_deterministic}{nrmse_pct}{21.7}
\setres{cty_Ethiopia}{DA_deterministic}{r_ff}{---}
\setres{cty_Ethiopia}{DA_deterministic}{r_ff_se}{---}
\setres{cty_Ethiopia}{DA_deterministic}{r_prior}{---}
\setres{cty_Ethiopia}{DA_deterministic}{r_prior_se}{---}
\setres{cty_Ethiopia}{DA_deterministic}{r_prob}{---}
\setres{cty_Ethiopia}{DA_deterministic}{r_prob_se}{---}
\setres{cty_Ethiopia}{DA_deterministic}{theta_star}{2.23, 1.62}
\setres{cty_Ethiopia}{DA_deterministic}{volume_ess}{---}
\setres{cty_Ethiopia}{DA_lit}{kappa}{-2.214}
\setres{cty_Ethiopia}{DA_lit}{kappa_prior_mean}{-1.597}
\setres{cty_Ethiopia}{DA_lit}{kappa_prior_mean_se}{0.323}
\setres{cty_Ethiopia}{DA_lit}{kappa_se}{0.364}
\setres{cty_Ethiopia}{DA_lit}{mse}{42.30}
\setres{cty_Ethiopia}{DA_lit}{mse_prior_mean}{37.55}
\setres{cty_Ethiopia}{DA_lit}{nrmse_pct}{35.5}
\setres{cty_Ethiopia}{DA_lit}{pred_variance}{4.75}
\setres{cty_Ethiopia}{DA_lit}{r_ff}{---}
\setres{cty_Ethiopia}{DA_lit}{r_ff_se}{---}
\setres{cty_Ethiopia}{DA_lit}{r_prior}{---}
\setres{cty_Ethiopia}{DA_lit}{r_prior_se}{---}
\setres{cty_Ethiopia}{DA_lit}{r_prob}{---}
\setres{cty_Ethiopia}{DA_lit}{r_prob_se}{---}
\setres{cty_Ethiopia}{DA_lit}{spread_penalty}{0.617}
\setres{cty_Ethiopia}{DA_lit}{spread_penalty_se}{0.063}
\setres{cty_Ethiopia}{DA_lit}{volume_ess}{---}
\setres{cty_Ethiopia}{DA_transfer_others}{kappa}{0.608}
\setres{cty_Ethiopia}{DA_transfer_others}{kappa_prior_mean}{0.608}
\setres{cty_Ethiopia}{DA_transfer_others}{kappa_prior_mean_se}{0.099}
\setres{cty_Ethiopia}{DA_transfer_others}{kappa_se}{0.099}
\setres{cty_Ethiopia}{DA_transfer_others}{mse}{20.57}
\setres{cty_Ethiopia}{DA_transfer_others}{mse_prior_mean}{20.57}
\setres{cty_Ethiopia}{DA_transfer_others}{nrmse_pct}{22.2}
\setres{cty_Ethiopia}{DA_transfer_others}{pred_variance}{0.00}
\setres{cty_Ethiopia}{DA_transfer_others}{r_ff}{---}
\setres{cty_Ethiopia}{DA_transfer_others}{r_ff_se}{---}
\setres{cty_Ethiopia}{DA_transfer_others}{r_prior}{---}
\setres{cty_Ethiopia}{DA_transfer_others}{r_prior_se}{---}
\setres{cty_Ethiopia}{DA_transfer_others}{r_prob}{---}
\setres{cty_Ethiopia}{DA_transfer_others}{r_prob_se}{---}
\setres{cty_Ethiopia}{DA_transfer_others}{spread_penalty}{0.000}
\setres{cty_Ethiopia}{DA_transfer_others}{spread_penalty_se}{0.000}
\setres{cty_Ethiopia}{DA_transfer_others}{theta_star}{1.67, 1.45}
\setres{cty_Ethiopia}{DA_transfer_others}{transfer}{others}
\setres{cty_Ethiopia}{DA_transfer_others}{volume_ess}{---}
\setres{cty_Ethiopia}{DA_uniform}{kappa}{-4.729}
\setres{cty_Ethiopia}{DA_uniform}{kappa_prior_mean}{-2.956}
\setres{cty_Ethiopia}{DA_uniform}{kappa_prior_mean_se}{0.416}
\setres{cty_Ethiopia}{DA_uniform}{kappa_se}{0.553}
\setres{cty_Ethiopia}{DA_uniform}{mse}{61.66}
\setres{cty_Ethiopia}{DA_uniform}{mse_prior_mean}{48.01}
\setres{cty_Ethiopia}{DA_uniform}{nrmse_pct}{38.2}
\setres{cty_Ethiopia}{DA_uniform}{pred_variance}{13.65}
\setres{cty_Ethiopia}{DA_uniform}{r_ff}{---}
\setres{cty_Ethiopia}{DA_uniform}{r_ff_se}{---}
\setres{cty_Ethiopia}{DA_uniform}{r_prior}{---}
\setres{cty_Ethiopia}{DA_uniform}{r_prior_se}{---}
\setres{cty_Ethiopia}{DA_uniform}{r_prob}{---}
\setres{cty_Ethiopia}{DA_uniform}{r_prob_se}{---}
\setres{cty_Ethiopia}{DA_uniform}{spread_penalty}{1.773}
\setres{cty_Ethiopia}{DA_uniform}{spread_penalty_se}{0.180}
\setres{cty_Ethiopia}{DA_uniform}{volume_ess}{---}
\setres{cty_Ethiopia}{EU_deterministic}{kappa}{0.716}
\setres{cty_Ethiopia}{EU_deterministic}{kappa_se}{0.079}
\setres{cty_Ethiopia}{EU_deterministic}{mse}{19.74}
\setres{cty_Ethiopia}{EU_deterministic}{nrmse_pct}{21.9}
\setres{cty_Ethiopia}{EU_deterministic}{r_ff}{---}
\setres{cty_Ethiopia}{EU_deterministic}{r_ff_se}{---}
\setres{cty_Ethiopia}{EU_deterministic}{r_prior}{---}
\setres{cty_Ethiopia}{EU_deterministic}{r_prior_se}{---}
\setres{cty_Ethiopia}{EU_deterministic}{r_prob}{---}
\setres{cty_Ethiopia}{EU_deterministic}{r_prob_se}{---}
\setres{cty_Ethiopia}{EU_deterministic}{theta_star}{1.21}
\setres{cty_Ethiopia}{EU_deterministic}{volume_ess}{---}
\setres{cty_Ethiopia}{EU_lit}{kappa}{-0.388}
\setres{cty_Ethiopia}{EU_lit}{kappa_prior_mean}{-0.232}
\setres{cty_Ethiopia}{EU_lit}{kappa_prior_mean_se}{0.202}
\setres{cty_Ethiopia}{EU_lit}{kappa_se}{0.212}
\setres{cty_Ethiopia}{EU_lit}{mse}{28.24}
\setres{cty_Ethiopia}{EU_lit}{mse_prior_mean}{27.04}
\setres{cty_Ethiopia}{EU_lit}{nrmse_pct}{28.3}
\setres{cty_Ethiopia}{EU_lit}{pred_variance}{1.20}
\setres{cty_Ethiopia}{EU_lit}{r_ff}{---}
\setres{cty_Ethiopia}{EU_lit}{r_ff_se}{---}
\setres{cty_Ethiopia}{EU_lit}{r_prior}{---}
\setres{cty_Ethiopia}{EU_lit}{r_prior_se}{---}
\setres{cty_Ethiopia}{EU_lit}{r_prob}{---}
\setres{cty_Ethiopia}{EU_lit}{r_prob_se}{---}
\setres{cty_Ethiopia}{EU_lit}{spread_penalty}{0.156}
\setres{cty_Ethiopia}{EU_lit}{spread_penalty_se}{0.016}
\setres{cty_Ethiopia}{EU_lit}{volume_ess}{---}
\setres{cty_Ethiopia}{EU_transfer_others}{kappa}{0.529}
\setres{cty_Ethiopia}{EU_transfer_others}{kappa_prior_mean}{0.529}
\setres{cty_Ethiopia}{EU_transfer_others}{kappa_prior_mean_se}{0.110}
\setres{cty_Ethiopia}{EU_transfer_others}{kappa_se}{0.110}
\setres{cty_Ethiopia}{EU_transfer_others}{mse}{21.18}
\setres{cty_Ethiopia}{EU_transfer_others}{mse_prior_mean}{21.18}
\setres{cty_Ethiopia}{EU_transfer_others}{nrmse_pct}{22.3}
\setres{cty_Ethiopia}{EU_transfer_others}{pred_variance}{0.00}
\setres{cty_Ethiopia}{EU_transfer_others}{r_ff}{---}
\setres{cty_Ethiopia}{EU_transfer_others}{r_ff_se}{---}
\setres{cty_Ethiopia}{EU_transfer_others}{r_prior}{---}
\setres{cty_Ethiopia}{EU_transfer_others}{r_prior_se}{---}
\setres{cty_Ethiopia}{EU_transfer_others}{r_prob}{---}
\setres{cty_Ethiopia}{EU_transfer_others}{r_prob_se}{---}
\setres{cty_Ethiopia}{EU_transfer_others}{spread_penalty}{0.000}
\setres{cty_Ethiopia}{EU_transfer_others}{spread_penalty_se}{0.000}
\setres{cty_Ethiopia}{EU_transfer_others}{theta_star}{0.93}
\setres{cty_Ethiopia}{EU_transfer_others}{transfer}{others}
\setres{cty_Ethiopia}{EU_transfer_others}{volume_ess}{---}
\setres{cty_Ethiopia}{EU_uniform}{kappa}{-2.908}
\setres{cty_Ethiopia}{EU_uniform}{kappa_prior_mean}{-1.844}
\setres{cty_Ethiopia}{EU_uniform}{kappa_prior_mean_se}{0.336}
\setres{cty_Ethiopia}{EU_uniform}{kappa_se}{0.413}
\setres{cty_Ethiopia}{EU_uniform}{mse}{47.64}
\setres{cty_Ethiopia}{EU_uniform}{mse_prior_mean}{39.45}
\setres{cty_Ethiopia}{EU_uniform}{nrmse_pct}{36.0}
\setres{cty_Ethiopia}{EU_uniform}{pred_variance}{8.20}
\setres{cty_Ethiopia}{EU_uniform}{r_ff}{---}
\setres{cty_Ethiopia}{EU_uniform}{r_ff_se}{---}
\setres{cty_Ethiopia}{EU_uniform}{r_prior}{---}
\setres{cty_Ethiopia}{EU_uniform}{r_prior_se}{---}
\setres{cty_Ethiopia}{EU_uniform}{r_prob}{---}
\setres{cty_Ethiopia}{EU_uniform}{r_prob_se}{---}
\setres{cty_Ethiopia}{EU_uniform}{spread_penalty}{1.065}
\setres{cty_Ethiopia}{EU_uniform}{spread_penalty_se}{0.108}
\setres{cty_Ethiopia}{EU_uniform}{volume_ess}{---}
\setres{cty_Ethiopia}{EV_deterministic}{kappa}{0.622}
\setres{cty_Ethiopia}{EV_deterministic}{kappa_se}{0.095}
\setres{cty_Ethiopia}{EV_deterministic}{mse}{20.47}
\setres{cty_Ethiopia}{EV_deterministic}{nrmse_pct}{21.8}
\setres{cty_Ethiopia}{EV_deterministic}{r_ff}{---}
\setres{cty_Ethiopia}{EV_deterministic}{r_ff_se}{---}
\setres{cty_Ethiopia}{EV_deterministic}{r_prior}{---}
\setres{cty_Ethiopia}{EV_deterministic}{r_prior_se}{---}
\setres{cty_Ethiopia}{EV_deterministic}{r_prob}{---}
\setres{cty_Ethiopia}{EV_deterministic}{r_prob_se}{---}
\setres{cty_Ethiopia}{EV_deterministic}{volume_ess}{---}
\setres{cty_Ethiopia}{reference}{country_name}{Ethiopia}
\setres{cty_Ethiopia}{reference}{lookup_nrmse_pct}{21.5}
\setres{cty_Ethiopia}{reference}{midpoint_mse}{25.25}
\setres{cty_Ethiopia}{reference}{midpoint_nrmse_pct}{28.0}
\setres{cty_Ethiopia}{reference}{missing_lotteries}{none}
\setres{cty_Ethiopia}{reference}{n_lotteries}{14}
\setres{cty_Ethiopia}{reference}{n_obs}{1{,}960}
\setres{cty_Ethiopia}{reference}{n_subjects}{140}
\setres{cty_Ethiopia}{reference}{oracle_mse}{17.56}
\setres{cty_Ethiopia}{reference}{sample}{LV2019\_Ethiopia}
\setres{cty_France}{CPT_deterministic}{kappa}{0.967}
\setres{cty_France}{CPT_deterministic}{kappa_se}{0.012}
\setres{cty_France}{CPT_deterministic}{mse}{12.35}
\setres{cty_France}{CPT_deterministic}{nrmse_pct}{19.6}
\setres{cty_France}{CPT_deterministic}{r_ff}{---}
\setres{cty_France}{CPT_deterministic}{r_ff_se}{---}
\setres{cty_France}{CPT_deterministic}{r_prior}{---}
\setres{cty_France}{CPT_deterministic}{r_prior_se}{---}
\setres{cty_France}{CPT_deterministic}{r_prob}{---}
\setres{cty_France}{CPT_deterministic}{r_prob_se}{---}
\setres{cty_France}{CPT_deterministic}{theta_star}{1.40, 0.70, 0.50}
\setres{cty_France}{CPT_deterministic}{volume_ess}{---}
\setres{cty_France}{CPT_lit}{kappa}{0.514}
\setres{cty_France}{CPT_lit}{kappa_prior_mean}{0.807}
\setres{cty_France}{CPT_lit}{kappa_prior_mean_se}{0.058}
\setres{cty_France}{CPT_lit}{kappa_se}{0.077}
\setres{cty_France}{CPT_lit}{mse}{16.85}
\setres{cty_France}{CPT_lit}{mse_prior_mean}{13.95}
\setres{cty_France}{CPT_lit}{nrmse_pct}{21.0}
\setres{cty_France}{CPT_lit}{pred_variance}{2.90}
\setres{cty_France}{CPT_lit}{r_ff}{---}
\setres{cty_France}{CPT_lit}{r_ff_se}{---}
\setres{cty_France}{CPT_lit}{r_prior}{---}
\setres{cty_France}{CPT_lit}{r_prior_se}{---}
\setres{cty_France}{CPT_lit}{r_prob}{---}
\setres{cty_France}{CPT_lit}{r_prob_se}{---}
\setres{cty_France}{CPT_lit}{spread_penalty}{0.292}
\setres{cty_France}{CPT_lit}{spread_penalty_se}{0.024}
\setres{cty_France}{CPT_lit}{volume_ess}{---}
\setres{cty_France}{CPT_transfer_others}{kappa}{0.942}
\setres{cty_France}{CPT_transfer_others}{kappa_prior_mean}{0.942}
\setres{cty_France}{CPT_transfer_others}{kappa_prior_mean_se}{0.023}
\setres{cty_France}{CPT_transfer_others}{kappa_se}{0.023}
\setres{cty_France}{CPT_transfer_others}{mse}{12.61}
\setres{cty_France}{CPT_transfer_others}{mse_prior_mean}{12.61}
\setres{cty_France}{CPT_transfer_others}{nrmse_pct}{19.4}
\setres{cty_France}{CPT_transfer_others}{pred_variance}{0.00}
\setres{cty_France}{CPT_transfer_others}{r_ff}{---}
\setres{cty_France}{CPT_transfer_others}{r_ff_se}{---}
\setres{cty_France}{CPT_transfer_others}{r_prior}{---}
\setres{cty_France}{CPT_transfer_others}{r_prior_se}{---}
\setres{cty_France}{CPT_transfer_others}{r_prob}{---}
\setres{cty_France}{CPT_transfer_others}{r_prob_se}{---}
\setres{cty_France}{CPT_transfer_others}{spread_penalty}{0.000}
\setres{cty_France}{CPT_transfer_others}{spread_penalty_se}{0.000}
\setres{cty_France}{CPT_transfer_others}{theta_star}{1.00, 0.60, 0.90}
\setres{cty_France}{CPT_transfer_others}{transfer}{others}
\setres{cty_France}{CPT_transfer_others}{volume_ess}{---}
\setres{cty_France}{CPT_uniform}{kappa}{-1.550}
\setres{cty_France}{CPT_uniform}{kappa_prior_mean}{0.200}
\setres{cty_France}{CPT_uniform}{kappa_prior_mean_se}{0.141}
\setres{cty_France}{CPT_uniform}{kappa_se}{0.259}
\setres{cty_France}{CPT_uniform}{mse}{37.33}
\setres{cty_France}{CPT_uniform}{mse_prior_mean}{19.96}
\setres{cty_France}{CPT_uniform}{nrmse_pct}{26.4}
\setres{cty_France}{CPT_uniform}{pred_variance}{17.37}
\setres{cty_France}{CPT_uniform}{r_ff}{---}
\setres{cty_France}{CPT_uniform}{r_ff_se}{---}
\setres{cty_France}{CPT_uniform}{r_prior}{---}
\setres{cty_France}{CPT_uniform}{r_prior_se}{---}
\setres{cty_France}{CPT_uniform}{r_prob}{---}
\setres{cty_France}{CPT_uniform}{r_prob_se}{---}
\setres{cty_France}{CPT_uniform}{spread_penalty}{1.750}
\setres{cty_France}{CPT_uniform}{spread_penalty_se}{0.144}
\setres{cty_France}{CPT_uniform}{volume_ess}{---}
\setres{cty_France}{DA_deterministic}{kappa}{0.924}
\setres{cty_France}{DA_deterministic}{kappa_se}{0.022}
\setres{cty_France}{DA_deterministic}{mse}{12.78}
\setres{cty_France}{DA_deterministic}{nrmse_pct}{20.1}
\setres{cty_France}{DA_deterministic}{r_ff}{---}
\setres{cty_France}{DA_deterministic}{r_ff_se}{---}
\setres{cty_France}{DA_deterministic}{r_prior}{---}
\setres{cty_France}{DA_deterministic}{r_prior_se}{---}
\setres{cty_France}{DA_deterministic}{r_prob}{---}
\setres{cty_France}{DA_deterministic}{r_prob_se}{---}
\setres{cty_France}{DA_deterministic}{theta_star}{1.83, 2.22}
\setres{cty_France}{DA_deterministic}{volume_ess}{---}
\setres{cty_France}{DA_lit}{kappa}{-0.385}
\setres{cty_France}{DA_lit}{kappa_prior_mean}{0.094}
\setres{cty_France}{DA_lit}{kappa_prior_mean_se}{0.187}
\setres{cty_France}{DA_lit}{kappa_se}{0.220}
\setres{cty_France}{DA_lit}{mse}{25.76}
\setres{cty_France}{DA_lit}{mse_prior_mean}{21.02}
\setres{cty_France}{DA_lit}{nrmse_pct}{25.3}
\setres{cty_France}{DA_lit}{pred_variance}{4.75}
\setres{cty_France}{DA_lit}{r_ff}{---}
\setres{cty_France}{DA_lit}{r_ff_se}{---}
\setres{cty_France}{DA_lit}{r_prior}{---}
\setres{cty_France}{DA_lit}{r_prior_se}{---}
\setres{cty_France}{DA_lit}{r_prob}{---}
\setres{cty_France}{DA_lit}{r_prob_se}{---}
\setres{cty_France}{DA_lit}{spread_penalty}{0.479}
\setres{cty_France}{DA_lit}{spread_penalty_se}{0.039}
\setres{cty_France}{DA_lit}{volume_ess}{---}
\setres{cty_France}{DA_transfer_others}{kappa}{0.914}
\setres{cty_France}{DA_transfer_others}{kappa_prior_mean}{0.914}
\setres{cty_France}{DA_transfer_others}{kappa_prior_mean_se}{0.031}
\setres{cty_France}{DA_transfer_others}{kappa_se}{0.031}
\setres{cty_France}{DA_transfer_others}{mse}{12.87}
\setres{cty_France}{DA_transfer_others}{mse_prior_mean}{12.87}
\setres{cty_France}{DA_transfer_others}{nrmse_pct}{20.4}
\setres{cty_France}{DA_transfer_others}{pred_variance}{0.00}
\setres{cty_France}{DA_transfer_others}{r_ff}{---}
\setres{cty_France}{DA_transfer_others}{r_ff_se}{---}
\setres{cty_France}{DA_transfer_others}{r_prior}{---}
\setres{cty_France}{DA_transfer_others}{r_prior_se}{---}
\setres{cty_France}{DA_transfer_others}{r_prob}{---}
\setres{cty_France}{DA_transfer_others}{r_prob_se}{---}
\setres{cty_France}{DA_transfer_others}{spread_penalty}{0.000}
\setres{cty_France}{DA_transfer_others}{spread_penalty_se}{0.000}
\setres{cty_France}{DA_transfer_others}{theta_star}{1.69, 1.42}
\setres{cty_France}{DA_transfer_others}{transfer}{others}
\setres{cty_France}{DA_transfer_others}{volume_ess}{---}
\setres{cty_France}{DA_uniform}{kappa}{-2.010}
\setres{cty_France}{DA_uniform}{kappa_prior_mean}{-0.634}
\setres{cty_France}{DA_uniform}{kappa_prior_mean_se}{0.283}
\setres{cty_France}{DA_uniform}{kappa_se}{0.379}
\setres{cty_France}{DA_uniform}{mse}{41.88}
\setres{cty_France}{DA_uniform}{mse_prior_mean}{28.24}
\setres{cty_France}{DA_uniform}{nrmse_pct}{28.6}
\setres{cty_France}{DA_uniform}{pred_variance}{13.65}
\setres{cty_France}{DA_uniform}{r_ff}{---}
\setres{cty_France}{DA_uniform}{r_ff_se}{---}
\setres{cty_France}{DA_uniform}{r_prior}{---}
\setres{cty_France}{DA_uniform}{r_prior_se}{---}
\setres{cty_France}{DA_uniform}{r_prob}{---}
\setres{cty_France}{DA_uniform}{r_prob_se}{---}
\setres{cty_France}{DA_uniform}{spread_penalty}{1.376}
\setres{cty_France}{DA_uniform}{spread_penalty_se}{0.113}
\setres{cty_France}{DA_uniform}{volume_ess}{---}
\setres{cty_France}{EU_deterministic}{kappa}{0.788}
\setres{cty_France}{EU_deterministic}{kappa_se}{0.051}
\setres{cty_France}{EU_deterministic}{mse}{14.13}
\setres{cty_France}{EU_deterministic}{nrmse_pct}{20.9}
\setres{cty_France}{EU_deterministic}{r_ff}{---}
\setres{cty_France}{EU_deterministic}{r_ff_se}{---}
\setres{cty_France}{EU_deterministic}{r_prior}{---}
\setres{cty_France}{EU_deterministic}{r_prior_se}{---}
\setres{cty_France}{EU_deterministic}{r_prob}{---}
\setres{cty_France}{EU_deterministic}{r_prob_se}{---}
\setres{cty_France}{EU_deterministic}{theta_star}{0.87}
\setres{cty_France}{EU_deterministic}{volume_ess}{---}
\setres{cty_France}{EU_lit}{kappa}{0.540}
\setres{cty_France}{EU_lit}{kappa_prior_mean}{0.661}
\setres{cty_France}{EU_lit}{kappa_prior_mean_se}{0.082}
\setres{cty_France}{EU_lit}{kappa_se}{0.091}
\setres{cty_France}{EU_lit}{mse}{16.59}
\setres{cty_France}{EU_lit}{mse_prior_mean}{15.39}
\setres{cty_France}{EU_lit}{nrmse_pct}{20.4}
\setres{cty_France}{EU_lit}{pred_variance}{1.20}
\setres{cty_France}{EU_lit}{r_ff}{---}
\setres{cty_France}{EU_lit}{r_ff_se}{---}
\setres{cty_France}{EU_lit}{r_prior}{---}
\setres{cty_France}{EU_lit}{r_prior_se}{---}
\setres{cty_France}{EU_lit}{r_prob}{---}
\setres{cty_France}{EU_lit}{r_prob_se}{---}
\setres{cty_France}{EU_lit}{spread_penalty}{0.121}
\setres{cty_France}{EU_lit}{spread_penalty_se}{0.010}
\setres{cty_France}{EU_lit}{volume_ess}{---}
\setres{cty_France}{EU_transfer_others}{kappa}{0.782}
\setres{cty_France}{EU_transfer_others}{kappa_prior_mean}{0.782}
\setres{cty_France}{EU_transfer_others}{kappa_prior_mean_se}{0.055}
\setres{cty_France}{EU_transfer_others}{kappa_se}{0.055}
\setres{cty_France}{EU_transfer_others}{mse}{14.19}
\setres{cty_France}{EU_transfer_others}{mse_prior_mean}{14.19}
\setres{cty_France}{EU_transfer_others}{nrmse_pct}{21.5}
\setres{cty_France}{EU_transfer_others}{pred_variance}{0.00}
\setres{cty_France}{EU_transfer_others}{r_ff}{---}
\setres{cty_France}{EU_transfer_others}{r_ff_se}{---}
\setres{cty_France}{EU_transfer_others}{r_prior}{---}
\setres{cty_France}{EU_transfer_others}{r_prior_se}{---}
\setres{cty_France}{EU_transfer_others}{r_prob}{---}
\setres{cty_France}{EU_transfer_others}{r_prob_se}{---}
\setres{cty_France}{EU_transfer_others}{spread_penalty}{0.000}
\setres{cty_France}{EU_transfer_others}{spread_penalty_se}{0.000}
\setres{cty_France}{EU_transfer_others}{theta_star}{0.95}
\setres{cty_France}{EU_transfer_others}{transfer}{others}
\setres{cty_France}{EU_transfer_others}{volume_ess}{---}
\setres{cty_France}{EU_uniform}{kappa}{-0.896}
\setres{cty_France}{EU_uniform}{kappa_prior_mean}{-0.069}
\setres{cty_France}{EU_uniform}{kappa_prior_mean_se}{0.204}
\setres{cty_France}{EU_uniform}{kappa_se}{0.262}
\setres{cty_France}{EU_uniform}{mse}{30.83}
\setres{cty_France}{EU_uniform}{mse_prior_mean}{22.64}
\setres{cty_France}{EU_uniform}{nrmse_pct}{26.6}
\setres{cty_France}{EU_uniform}{pred_variance}{8.20}
\setres{cty_France}{EU_uniform}{r_ff}{---}
\setres{cty_France}{EU_uniform}{r_ff_se}{---}
\setres{cty_France}{EU_uniform}{r_prior}{---}
\setres{cty_France}{EU_uniform}{r_prior_se}{---}
\setres{cty_France}{EU_uniform}{r_prob}{---}
\setres{cty_France}{EU_uniform}{r_prob_se}{---}
\setres{cty_France}{EU_uniform}{spread_penalty}{0.826}
\setres{cty_France}{EU_uniform}{spread_penalty_se}{0.068}
\setres{cty_France}{EU_uniform}{volume_ess}{---}
\setres{cty_France}{EV_deterministic}{kappa}{0.754}
\setres{cty_France}{EV_deterministic}{kappa_se}{0.065}
\setres{cty_France}{EV_deterministic}{mse}{14.47}
\setres{cty_France}{EV_deterministic}{nrmse_pct}{21.2}
\setres{cty_France}{EV_deterministic}{r_ff}{---}
\setres{cty_France}{EV_deterministic}{r_ff_se}{---}
\setres{cty_France}{EV_deterministic}{r_prior}{---}
\setres{cty_France}{EV_deterministic}{r_prior_se}{---}
\setres{cty_France}{EV_deterministic}{r_prob}{---}
\setres{cty_France}{EV_deterministic}{r_prob_se}{---}
\setres{cty_France}{EV_deterministic}{volume_ess}{---}
\setres{cty_France}{reference}{country_name}{France}
\setres{cty_France}{reference}{lookup_nrmse_pct}{18.4}
\setres{cty_France}{reference}{midpoint_mse}{21.95}
\setres{cty_France}{reference}{midpoint_nrmse_pct}{26.3}
\setres{cty_France}{reference}{missing_lotteries}{none}
\setres{cty_France}{reference}{n_lotteries}{14}
\setres{cty_France}{reference}{n_obs}{1{,}302}
\setres{cty_France}{reference}{n_subjects}{93}
\setres{cty_France}{reference}{oracle_mse}{12.03}
\setres{cty_France}{reference}{sample}{LV2019\_France}
\setres{cty_Germany}{CPT_deterministic}{kappa}{0.976}
\setres{cty_Germany}{CPT_deterministic}{kappa_se}{0.010}
\setres{cty_Germany}{CPT_deterministic}{mse}{11.27}
\setres{cty_Germany}{CPT_deterministic}{nrmse_pct}{19.0}
\setres{cty_Germany}{CPT_deterministic}{r_ff}{---}
\setres{cty_Germany}{CPT_deterministic}{r_ff_se}{---}
\setres{cty_Germany}{CPT_deterministic}{r_prior}{---}
\setres{cty_Germany}{CPT_deterministic}{r_prior_se}{---}
\setres{cty_Germany}{CPT_deterministic}{r_prob}{---}
\setres{cty_Germany}{CPT_deterministic}{r_prob_se}{---}
\setres{cty_Germany}{CPT_deterministic}{theta_star}{1.10, 0.80, 0.60}
\setres{cty_Germany}{CPT_deterministic}{volume_ess}{---}
\setres{cty_Germany}{CPT_lit}{kappa}{0.719}
\setres{cty_Germany}{CPT_lit}{kappa_prior_mean}{0.948}
\setres{cty_Germany}{CPT_lit}{kappa_prior_mean_se}{0.013}
\setres{cty_Germany}{CPT_lit}{kappa_se}{0.022}
\setres{cty_Germany}{CPT_lit}{mse}{14.52}
\setres{cty_Germany}{CPT_lit}{mse_prior_mean}{11.62}
\setres{cty_Germany}{CPT_lit}{nrmse_pct}{19.8}
\setres{cty_Germany}{CPT_lit}{pred_variance}{2.90}
\setres{cty_Germany}{CPT_lit}{r_ff}{---}
\setres{cty_Germany}{CPT_lit}{r_ff_se}{---}
\setres{cty_Germany}{CPT_lit}{r_prior}{---}
\setres{cty_Germany}{CPT_lit}{r_prior_se}{---}
\setres{cty_Germany}{CPT_lit}{r_prob}{---}
\setres{cty_Germany}{CPT_lit}{r_prob_se}{---}
\setres{cty_Germany}{CPT_lit}{spread_penalty}{0.229}
\setres{cty_Germany}{CPT_lit}{spread_penalty_se}{0.013}
\setres{cty_Germany}{CPT_lit}{volume_ess}{---}
\setres{cty_Germany}{CPT_transfer_others}{kappa}{0.885}
\setres{cty_Germany}{CPT_transfer_others}{kappa_prior_mean}{0.885}
\setres{cty_Germany}{CPT_transfer_others}{kappa_prior_mean_se}{0.027}
\setres{cty_Germany}{CPT_transfer_others}{kappa_se}{0.027}
\setres{cty_Germany}{CPT_transfer_others}{mse}{12.42}
\setres{cty_Germany}{CPT_transfer_others}{mse_prior_mean}{12.42}
\setres{cty_Germany}{CPT_transfer_others}{nrmse_pct}{18.4}
\setres{cty_Germany}{CPT_transfer_others}{pred_variance}{0.00}
\setres{cty_Germany}{CPT_transfer_others}{r_ff}{---}
\setres{cty_Germany}{CPT_transfer_others}{r_ff_se}{---}
\setres{cty_Germany}{CPT_transfer_others}{r_prior}{---}
\setres{cty_Germany}{CPT_transfer_others}{r_prior_se}{---}
\setres{cty_Germany}{CPT_transfer_others}{r_prob}{---}
\setres{cty_Germany}{CPT_transfer_others}{r_prob_se}{---}
\setres{cty_Germany}{CPT_transfer_others}{spread_penalty}{0.000}
\setres{cty_Germany}{CPT_transfer_others}{spread_penalty_se}{0.000}
\setres{cty_Germany}{CPT_transfer_others}{theta_star}{1.10, 0.60, 0.80}
\setres{cty_Germany}{CPT_transfer_others}{transfer}{others}
\setres{cty_Germany}{CPT_transfer_others}{volume_ess}{---}
\setres{cty_Germany}{CPT_uniform}{kappa}{-0.816}
\setres{cty_Germany}{CPT_uniform}{kappa_prior_mean}{0.553}
\setres{cty_Germany}{CPT_uniform}{kappa_prior_mean_se}{0.066}
\setres{cty_Germany}{CPT_uniform}{kappa_se}{0.118}
\setres{cty_Germany}{CPT_uniform}{mse}{33.98}
\setres{cty_Germany}{CPT_uniform}{mse_prior_mean}{16.62}
\setres{cty_Germany}{CPT_uniform}{nrmse_pct}{24.2}
\setres{cty_Germany}{CPT_uniform}{pred_variance}{17.36}
\setres{cty_Germany}{CPT_uniform}{r_ff}{---}
\setres{cty_Germany}{CPT_uniform}{r_ff_se}{---}
\setres{cty_Germany}{CPT_uniform}{r_prior}{---}
\setres{cty_Germany}{CPT_uniform}{r_prior_se}{---}
\setres{cty_Germany}{CPT_uniform}{r_prob}{---}
\setres{cty_Germany}{CPT_uniform}{r_prob_se}{---}
\setres{cty_Germany}{CPT_uniform}{spread_penalty}{1.370}
\setres{cty_Germany}{CPT_uniform}{spread_penalty_se}{0.076}
\setres{cty_Germany}{CPT_uniform}{volume_ess}{---}
\setres{cty_Germany}{DA_deterministic}{kappa}{0.946}
\setres{cty_Germany}{DA_deterministic}{kappa_se}{0.013}
\setres{cty_Germany}{DA_deterministic}{mse}{11.64}
\setres{cty_Germany}{DA_deterministic}{nrmse_pct}{19.4}
\setres{cty_Germany}{DA_deterministic}{r_ff}{---}
\setres{cty_Germany}{DA_deterministic}{r_ff_se}{---}
\setres{cty_Germany}{DA_deterministic}{r_prior}{---}
\setres{cty_Germany}{DA_deterministic}{r_prior_se}{---}
\setres{cty_Germany}{DA_deterministic}{r_prob}{---}
\setres{cty_Germany}{DA_deterministic}{r_prob_se}{---}
\setres{cty_Germany}{DA_deterministic}{theta_star}{1.35, 1.31}
\setres{cty_Germany}{DA_deterministic}{volume_ess}{---}
\setres{cty_Germany}{DA_lit}{kappa}{0.198}
\setres{cty_Germany}{DA_lit}{kappa_prior_mean}{0.572}
\setres{cty_Germany}{DA_lit}{kappa_prior_mean_se}{0.076}
\setres{cty_Germany}{DA_lit}{kappa_se}{0.092}
\setres{cty_Germany}{DA_lit}{mse}{21.13}
\setres{cty_Germany}{DA_lit}{mse_prior_mean}{16.38}
\setres{cty_Germany}{DA_lit}{nrmse_pct}{22.6}
\setres{cty_Germany}{DA_lit}{pred_variance}{4.75}
\setres{cty_Germany}{DA_lit}{r_ff}{---}
\setres{cty_Germany}{DA_lit}{r_ff_se}{---}
\setres{cty_Germany}{DA_lit}{r_prior}{---}
\setres{cty_Germany}{DA_lit}{r_prior_se}{---}
\setres{cty_Germany}{DA_lit}{r_prob}{---}
\setres{cty_Germany}{DA_lit}{r_prob_se}{---}
\setres{cty_Germany}{DA_lit}{spread_penalty}{0.374}
\setres{cty_Germany}{DA_lit}{spread_penalty_se}{0.021}
\setres{cty_Germany}{DA_lit}{volume_ess}{---}
\setres{cty_Germany}{DA_transfer_others}{kappa}{0.866}
\setres{cty_Germany}{DA_transfer_others}{kappa_prior_mean}{0.866}
\setres{cty_Germany}{DA_transfer_others}{kappa_prior_mean_se}{0.036}
\setres{cty_Germany}{DA_transfer_others}{kappa_se}{0.036}
\setres{cty_Germany}{DA_transfer_others}{mse}{12.65}
\setres{cty_Germany}{DA_transfer_others}{mse_prior_mean}{12.65}
\setres{cty_Germany}{DA_transfer_others}{nrmse_pct}{17.7}
\setres{cty_Germany}{DA_transfer_others}{pred_variance}{0.00}
\setres{cty_Germany}{DA_transfer_others}{r_ff}{---}
\setres{cty_Germany}{DA_transfer_others}{r_ff_se}{---}
\setres{cty_Germany}{DA_transfer_others}{r_prior}{---}
\setres{cty_Germany}{DA_transfer_others}{r_prior_se}{---}
\setres{cty_Germany}{DA_transfer_others}{r_prob}{---}
\setres{cty_Germany}{DA_transfer_others}{r_prob_se}{---}
\setres{cty_Germany}{DA_transfer_others}{spread_penalty}{0.000}
\setres{cty_Germany}{DA_transfer_others}{spread_penalty_se}{0.000}
\setres{cty_Germany}{DA_transfer_others}{theta_star}{1.71, 1.46}
\setres{cty_Germany}{DA_transfer_others}{transfer}{others}
\setres{cty_Germany}{DA_transfer_others}{volume_ess}{---}
\setres{cty_Germany}{DA_uniform}{kappa}{-0.993}
\setres{cty_Germany}{DA_uniform}{kappa_prior_mean}{0.084}
\setres{cty_Germany}{DA_uniform}{kappa_prior_mean_se}{0.131}
\setres{cty_Germany}{DA_uniform}{kappa_se}{0.176}
\setres{cty_Germany}{DA_uniform}{mse}{36.22}
\setres{cty_Germany}{DA_uniform}{mse_prior_mean}{22.58}
\setres{cty_Germany}{DA_uniform}{nrmse_pct}{27.6}
\setres{cty_Germany}{DA_uniform}{pred_variance}{13.65}
\setres{cty_Germany}{DA_uniform}{r_ff}{---}
\setres{cty_Germany}{DA_uniform}{r_ff_se}{---}
\setres{cty_Germany}{DA_uniform}{r_prior}{---}
\setres{cty_Germany}{DA_uniform}{r_prior_se}{---}
\setres{cty_Germany}{DA_uniform}{r_prob}{---}
\setres{cty_Germany}{DA_uniform}{r_prob_se}{---}
\setres{cty_Germany}{DA_uniform}{spread_penalty}{1.077}
\setres{cty_Germany}{DA_uniform}{spread_penalty_se}{0.060}
\setres{cty_Germany}{DA_uniform}{volume_ess}{---}
\setres{cty_Germany}{EU_deterministic}{kappa}{0.881}
\setres{cty_Germany}{EU_deterministic}{kappa_se}{0.026}
\setres{cty_Germany}{EU_deterministic}{mse}{12.47}
\setres{cty_Germany}{EU_deterministic}{nrmse_pct}{19.6}
\setres{cty_Germany}{EU_deterministic}{r_ff}{---}
\setres{cty_Germany}{EU_deterministic}{r_ff_se}{---}
\setres{cty_Germany}{EU_deterministic}{r_prior}{---}
\setres{cty_Germany}{EU_deterministic}{r_prior_se}{---}
\setres{cty_Germany}{EU_deterministic}{r_prob}{---}
\setres{cty_Germany}{EU_deterministic}{r_prob_se}{---}
\setres{cty_Germany}{EU_deterministic}{theta_star}{0.77}
\setres{cty_Germany}{EU_deterministic}{volume_ess}{---}
\setres{cty_Germany}{EU_lit}{kappa}{0.778}
\setres{cty_Germany}{EU_lit}{kappa_prior_mean}{0.873}
\setres{cty_Germany}{EU_lit}{kappa_prior_mean_se}{0.025}
\setres{cty_Germany}{EU_lit}{kappa_se}{0.030}
\setres{cty_Germany}{EU_lit}{mse}{13.77}
\setres{cty_Germany}{EU_lit}{mse_prior_mean}{12.56}
\setres{cty_Germany}{EU_lit}{nrmse_pct}{20.3}
\setres{cty_Germany}{EU_lit}{pred_variance}{1.20}
\setres{cty_Germany}{EU_lit}{r_ff}{---}
\setres{cty_Germany}{EU_lit}{r_ff_se}{---}
\setres{cty_Germany}{EU_lit}{r_prior}{---}
\setres{cty_Germany}{EU_lit}{r_prior_se}{---}
\setres{cty_Germany}{EU_lit}{r_prob}{---}
\setres{cty_Germany}{EU_lit}{r_prob_se}{---}
\setres{cty_Germany}{EU_lit}{spread_penalty}{0.095}
\setres{cty_Germany}{EU_lit}{spread_penalty_se}{0.005}
\setres{cty_Germany}{EU_lit}{volume_ess}{---}
\setres{cty_Germany}{EU_transfer_others}{kappa}{0.801}
\setres{cty_Germany}{EU_transfer_others}{kappa_prior_mean}{0.801}
\setres{cty_Germany}{EU_transfer_others}{kappa_prior_mean_se}{0.050}
\setres{cty_Germany}{EU_transfer_others}{kappa_se}{0.050}
\setres{cty_Germany}{EU_transfer_others}{mse}{13.48}
\setres{cty_Germany}{EU_transfer_others}{mse_prior_mean}{13.48}
\setres{cty_Germany}{EU_transfer_others}{nrmse_pct}{17.1}
\setres{cty_Germany}{EU_transfer_others}{pred_variance}{0.00}
\setres{cty_Germany}{EU_transfer_others}{r_ff}{---}
\setres{cty_Germany}{EU_transfer_others}{r_ff_se}{---}
\setres{cty_Germany}{EU_transfer_others}{r_prior}{---}
\setres{cty_Germany}{EU_transfer_others}{r_prior_se}{---}
\setres{cty_Germany}{EU_transfer_others}{r_prob}{---}
\setres{cty_Germany}{EU_transfer_others}{r_prob_se}{---}
\setres{cty_Germany}{EU_transfer_others}{spread_penalty}{0.000}
\setres{cty_Germany}{EU_transfer_others}{spread_penalty_se}{0.000}
\setres{cty_Germany}{EU_transfer_others}{theta_star}{0.95}
\setres{cty_Germany}{EU_transfer_others}{transfer}{others}
\setres{cty_Germany}{EU_transfer_others}{volume_ess}{---}
\setres{cty_Germany}{EU_uniform}{kappa}{-0.176}
\setres{cty_Germany}{EU_uniform}{kappa_prior_mean}{0.471}
\setres{cty_Germany}{EU_uniform}{kappa_prior_mean_se}{0.085}
\setres{cty_Germany}{EU_uniform}{kappa_se}{0.113}
\setres{cty_Germany}{EU_uniform}{mse}{25.87}
\setres{cty_Germany}{EU_uniform}{mse_prior_mean}{17.67}
\setres{cty_Germany}{EU_uniform}{nrmse_pct}{24.1}
\setres{cty_Germany}{EU_uniform}{pred_variance}{8.19}
\setres{cty_Germany}{EU_uniform}{r_ff}{---}
\setres{cty_Germany}{EU_uniform}{r_ff_se}{---}
\setres{cty_Germany}{EU_uniform}{r_prior}{---}
\setres{cty_Germany}{EU_uniform}{r_prior_se}{---}
\setres{cty_Germany}{EU_uniform}{r_prob}{---}
\setres{cty_Germany}{EU_uniform}{r_prob_se}{---}
\setres{cty_Germany}{EU_uniform}{spread_penalty}{0.647}
\setres{cty_Germany}{EU_uniform}{spread_penalty_se}{0.036}
\setres{cty_Germany}{EU_uniform}{volume_ess}{---}
\setres{cty_Germany}{EV_deterministic}{kappa}{0.758}
\setres{cty_Germany}{EV_deterministic}{kappa_se}{0.056}
\setres{cty_Germany}{EV_deterministic}{mse}{14.03}
\setres{cty_Germany}{EV_deterministic}{nrmse_pct}{17.0}
\setres{cty_Germany}{EV_deterministic}{r_ff}{---}
\setres{cty_Germany}{EV_deterministic}{r_ff_se}{---}
\setres{cty_Germany}{EV_deterministic}{r_prior}{---}
\setres{cty_Germany}{EV_deterministic}{r_prior_se}{---}
\setres{cty_Germany}{EV_deterministic}{r_prob}{---}
\setres{cty_Germany}{EV_deterministic}{r_prob_se}{---}
\setres{cty_Germany}{EV_deterministic}{volume_ess}{---}
\setres{cty_Germany}{reference}{country_name}{Germany}
\setres{cty_Germany}{reference}{lookup_nrmse_pct}{16.1}
\setres{cty_Germany}{reference}{midpoint_mse}{23.64}
\setres{cty_Germany}{reference}{midpoint_nrmse_pct}{25.9}
\setres{cty_Germany}{reference}{missing_lotteries}{none}
\setres{cty_Germany}{reference}{n_lotteries}{14}
\setres{cty_Germany}{reference}{n_obs}{1{,}819}
\setres{cty_Germany}{reference}{n_subjects}{130}
\setres{cty_Germany}{reference}{oracle_mse}{10.96}
\setres{cty_Germany}{reference}{sample}{LV2019\_Germany}
\setres{cty_Guatemala}{CPT_deterministic}{kappa}{0.963}
\setres{cty_Guatemala}{CPT_deterministic}{kappa_se}{0.018}
\setres{cty_Guatemala}{CPT_deterministic}{mse}{22.37}
\setres{cty_Guatemala}{CPT_deterministic}{nrmse_pct}{26.0}
\setres{cty_Guatemala}{CPT_deterministic}{r_ff}{---}
\setres{cty_Guatemala}{CPT_deterministic}{r_ff_se}{---}
\setres{cty_Guatemala}{CPT_deterministic}{r_prior}{---}
\setres{cty_Guatemala}{CPT_deterministic}{r_prior_se}{---}
\setres{cty_Guatemala}{CPT_deterministic}{r_prob}{---}
\setres{cty_Guatemala}{CPT_deterministic}{r_prob_se}{---}
\setres{cty_Guatemala}{CPT_deterministic}{theta_star}{1.40, 0.90, 0.60}
\setres{cty_Guatemala}{CPT_deterministic}{volume_ess}{---}
\setres{cty_Guatemala}{CPT_lit}{kappa}{0.467}
\setres{cty_Guatemala}{CPT_lit}{kappa_prior_mean}{0.698}
\setres{cty_Guatemala}{CPT_lit}{kappa_prior_mean_se}{0.083}
\setres{cty_Guatemala}{CPT_lit}{kappa_se}{0.093}
\setres{cty_Guatemala}{CPT_lit}{mse}{28.59}
\setres{cty_Guatemala}{CPT_lit}{mse_prior_mean}{25.69}
\setres{cty_Guatemala}{CPT_lit}{nrmse_pct}{28.4}
\setres{cty_Guatemala}{CPT_lit}{pred_variance}{2.90}
\setres{cty_Guatemala}{CPT_lit}{r_ff}{---}
\setres{cty_Guatemala}{CPT_lit}{r_ff_se}{---}
\setres{cty_Guatemala}{CPT_lit}{r_prior}{---}
\setres{cty_Guatemala}{CPT_lit}{r_prior_se}{---}
\setres{cty_Guatemala}{CPT_lit}{r_prob}{---}
\setres{cty_Guatemala}{CPT_lit}{r_prob_se}{---}
\setres{cty_Guatemala}{CPT_lit}{spread_penalty}{0.231}
\setres{cty_Guatemala}{CPT_lit}{spread_penalty_se}{0.022}
\setres{cty_Guatemala}{CPT_lit}{volume_ess}{---}
\setres{cty_Guatemala}{CPT_transfer_others}{kappa}{0.911}
\setres{cty_Guatemala}{CPT_transfer_others}{kappa_prior_mean}{0.911}
\setres{cty_Guatemala}{CPT_transfer_others}{kappa_prior_mean_se}{0.025}
\setres{cty_Guatemala}{CPT_transfer_others}{kappa_se}{0.025}
\setres{cty_Guatemala}{CPT_transfer_others}{mse}{23.02}
\setres{cty_Guatemala}{CPT_transfer_others}{mse_prior_mean}{23.02}
\setres{cty_Guatemala}{CPT_transfer_others}{nrmse_pct}{26.2}
\setres{cty_Guatemala}{CPT_transfer_others}{pred_variance}{0.00}
\setres{cty_Guatemala}{CPT_transfer_others}{r_ff}{---}
\setres{cty_Guatemala}{CPT_transfer_others}{r_ff_se}{---}
\setres{cty_Guatemala}{CPT_transfer_others}{r_prior}{---}
\setres{cty_Guatemala}{CPT_transfer_others}{r_prior_se}{---}
\setres{cty_Guatemala}{CPT_transfer_others}{r_prob}{---}
\setres{cty_Guatemala}{CPT_transfer_others}{r_prob_se}{---}
\setres{cty_Guatemala}{CPT_transfer_others}{spread_penalty}{0.000}
\setres{cty_Guatemala}{CPT_transfer_others}{spread_penalty_se}{0.000}
\setres{cty_Guatemala}{CPT_transfer_others}{theta_star}{1.00, 0.60, 0.90}
\setres{cty_Guatemala}{CPT_transfer_others}{transfer}{others}
\setres{cty_Guatemala}{CPT_transfer_others}{volume_ess}{---}
\setres{cty_Guatemala}{CPT_uniform}{kappa}{-1.407}
\setres{cty_Guatemala}{CPT_uniform}{kappa_prior_mean}{-0.022}
\setres{cty_Guatemala}{CPT_uniform}{kappa_prior_mean_se}{0.150}
\setres{cty_Guatemala}{CPT_uniform}{kappa_se}{0.236}
\setres{cty_Guatemala}{CPT_uniform}{mse}{52.09}
\setres{cty_Guatemala}{CPT_uniform}{mse_prior_mean}{34.73}
\setres{cty_Guatemala}{CPT_uniform}{nrmse_pct}{31.6}
\setres{cty_Guatemala}{CPT_uniform}{pred_variance}{17.37}
\setres{cty_Guatemala}{CPT_uniform}{r_ff}{---}
\setres{cty_Guatemala}{CPT_uniform}{r_ff_se}{---}
\setres{cty_Guatemala}{CPT_uniform}{r_prior}{---}
\setres{cty_Guatemala}{CPT_uniform}{r_prior_se}{---}
\setres{cty_Guatemala}{CPT_uniform}{r_prob}{---}
\setres{cty_Guatemala}{CPT_uniform}{r_prob_se}{---}
\setres{cty_Guatemala}{CPT_uniform}{spread_penalty}{1.384}
\setres{cty_Guatemala}{CPT_uniform}{spread_penalty_se}{0.132}
\setres{cty_Guatemala}{CPT_uniform}{volume_ess}{---}
\setres{cty_Guatemala}{DA_deterministic}{kappa}{0.966}
\setres{cty_Guatemala}{DA_deterministic}{kappa_se}{0.018}
\setres{cty_Guatemala}{DA_deterministic}{mse}{22.33}
\setres{cty_Guatemala}{DA_deterministic}{nrmse_pct}{25.9}
\setres{cty_Guatemala}{DA_deterministic}{r_ff}{---}
\setres{cty_Guatemala}{DA_deterministic}{r_ff_se}{---}
\setres{cty_Guatemala}{DA_deterministic}{r_prior}{---}
\setres{cty_Guatemala}{DA_deterministic}{r_prior_se}{---}
\setres{cty_Guatemala}{DA_deterministic}{r_prob}{---}
\setres{cty_Guatemala}{DA_deterministic}{r_prob_se}{---}
\setres{cty_Guatemala}{DA_deterministic}{theta_star}{1.67, 1.23}
\setres{cty_Guatemala}{DA_deterministic}{volume_ess}{---}
\setres{cty_Guatemala}{DA_lit}{kappa}{-0.379}
\setres{cty_Guatemala}{DA_lit}{kappa_prior_mean}{0.000}
\setres{cty_Guatemala}{DA_lit}{kappa_prior_mean_se}{0.196}
\setres{cty_Guatemala}{DA_lit}{kappa_se}{0.219}
\setres{cty_Guatemala}{DA_lit}{mse}{39.20}
\setres{cty_Guatemala}{DA_lit}{mse_prior_mean}{34.45}
\setres{cty_Guatemala}{DA_lit}{nrmse_pct}{32.6}
\setres{cty_Guatemala}{DA_lit}{pred_variance}{4.75}
\setres{cty_Guatemala}{DA_lit}{r_ff}{---}
\setres{cty_Guatemala}{DA_lit}{r_ff_se}{---}
\setres{cty_Guatemala}{DA_lit}{r_prior}{---}
\setres{cty_Guatemala}{DA_lit}{r_prior_se}{---}
\setres{cty_Guatemala}{DA_lit}{r_prob}{---}
\setres{cty_Guatemala}{DA_lit}{r_prob_se}{---}
\setres{cty_Guatemala}{DA_lit}{spread_penalty}{0.379}
\setres{cty_Guatemala}{DA_lit}{spread_penalty_se}{0.036}
\setres{cty_Guatemala}{DA_lit}{volume_ess}{---}
\setres{cty_Guatemala}{DA_transfer_others}{kappa}{0.973}
\setres{cty_Guatemala}{DA_transfer_others}{kappa_prior_mean}{0.973}
\setres{cty_Guatemala}{DA_transfer_others}{kappa_prior_mean_se}{0.022}
\setres{cty_Guatemala}{DA_transfer_others}{kappa_se}{0.022}
\setres{cty_Guatemala}{DA_transfer_others}{mse}{22.24}
\setres{cty_Guatemala}{DA_transfer_others}{mse_prior_mean}{22.24}
\setres{cty_Guatemala}{DA_transfer_others}{nrmse_pct}{25.8}
\setres{cty_Guatemala}{DA_transfer_others}{pred_variance}{0.00}
\setres{cty_Guatemala}{DA_transfer_others}{r_ff}{---}
\setres{cty_Guatemala}{DA_transfer_others}{r_ff_se}{---}
\setres{cty_Guatemala}{DA_transfer_others}{r_prior}{---}
\setres{cty_Guatemala}{DA_transfer_others}{r_prior_se}{---}
\setres{cty_Guatemala}{DA_transfer_others}{r_prob}{---}
\setres{cty_Guatemala}{DA_transfer_others}{r_prob_se}{---}
\setres{cty_Guatemala}{DA_transfer_others}{spread_penalty}{0.000}
\setres{cty_Guatemala}{DA_transfer_others}{spread_penalty_se}{0.000}
\setres{cty_Guatemala}{DA_transfer_others}{theta_star}{1.70, 1.46}
\setres{cty_Guatemala}{DA_transfer_others}{transfer}{others}
\setres{cty_Guatemala}{DA_transfer_others}{volume_ess}{---}
\setres{cty_Guatemala}{DA_uniform}{kappa}{-1.871}
\setres{cty_Guatemala}{DA_uniform}{kappa_prior_mean}{-0.782}
\setres{cty_Guatemala}{DA_uniform}{kappa_prior_mean_se}{0.278}
\setres{cty_Guatemala}{DA_uniform}{kappa_se}{0.350}
\setres{cty_Guatemala}{DA_uniform}{mse}{57.91}
\setres{cty_Guatemala}{DA_uniform}{mse_prior_mean}{44.26}
\setres{cty_Guatemala}{DA_uniform}{nrmse_pct}{36.3}
\setres{cty_Guatemala}{DA_uniform}{pred_variance}{13.65}
\setres{cty_Guatemala}{DA_uniform}{r_ff}{---}
\setres{cty_Guatemala}{DA_uniform}{r_ff_se}{---}
\setres{cty_Guatemala}{DA_uniform}{r_prior}{---}
\setres{cty_Guatemala}{DA_uniform}{r_prior_se}{---}
\setres{cty_Guatemala}{DA_uniform}{r_prob}{---}
\setres{cty_Guatemala}{DA_uniform}{r_prob_se}{---}
\setres{cty_Guatemala}{DA_uniform}{spread_penalty}{1.088}
\setres{cty_Guatemala}{DA_uniform}{spread_penalty_se}{0.104}
\setres{cty_Guatemala}{DA_uniform}{volume_ess}{---}
\setres{cty_Guatemala}{EU_deterministic}{kappa}{0.907}
\setres{cty_Guatemala}{EU_deterministic}{kappa_se}{0.039}
\setres{cty_Guatemala}{EU_deterministic}{mse}{23.07}
\setres{cty_Guatemala}{EU_deterministic}{nrmse_pct}{25.7}
\setres{cty_Guatemala}{EU_deterministic}{r_ff}{---}
\setres{cty_Guatemala}{EU_deterministic}{r_ff_se}{---}
\setres{cty_Guatemala}{EU_deterministic}{r_prior}{---}
\setres{cty_Guatemala}{EU_deterministic}{r_prior_se}{---}
\setres{cty_Guatemala}{EU_deterministic}{r_prob}{---}
\setres{cty_Guatemala}{EU_deterministic}{r_prob_se}{---}
\setres{cty_Guatemala}{EU_deterministic}{theta_star}{0.99}
\setres{cty_Guatemala}{EU_deterministic}{volume_ess}{---}
\setres{cty_Guatemala}{EU_lit}{kappa}{0.556}
\setres{cty_Guatemala}{EU_lit}{kappa_prior_mean}{0.652}
\setres{cty_Guatemala}{EU_lit}{kappa_prior_mean_se}{0.098}
\setres{cty_Guatemala}{EU_lit}{kappa_se}{0.104}
\setres{cty_Guatemala}{EU_lit}{mse}{27.47}
\setres{cty_Guatemala}{EU_lit}{mse_prior_mean}{26.27}
\setres{cty_Guatemala}{EU_lit}{nrmse_pct}{28.3}
\setres{cty_Guatemala}{EU_lit}{pred_variance}{1.20}
\setres{cty_Guatemala}{EU_lit}{r_ff}{---}
\setres{cty_Guatemala}{EU_lit}{r_ff_se}{---}
\setres{cty_Guatemala}{EU_lit}{r_prior}{---}
\setres{cty_Guatemala}{EU_lit}{r_prior_se}{---}
\setres{cty_Guatemala}{EU_lit}{r_prob}{---}
\setres{cty_Guatemala}{EU_lit}{r_prob_se}{---}
\setres{cty_Guatemala}{EU_lit}{spread_penalty}{0.096}
\setres{cty_Guatemala}{EU_lit}{spread_penalty_se}{0.009}
\setres{cty_Guatemala}{EU_lit}{volume_ess}{---}
\setres{cty_Guatemala}{EU_transfer_others}{kappa}{0.911}
\setres{cty_Guatemala}{EU_transfer_others}{kappa_prior_mean}{0.911}
\setres{cty_Guatemala}{EU_transfer_others}{kappa_prior_mean_se}{0.037}
\setres{cty_Guatemala}{EU_transfer_others}{kappa_se}{0.037}
\setres{cty_Guatemala}{EU_transfer_others}{mse}{23.02}
\setres{cty_Guatemala}{EU_transfer_others}{mse_prior_mean}{23.02}
\setres{cty_Guatemala}{EU_transfer_others}{nrmse_pct}{25.6}
\setres{cty_Guatemala}{EU_transfer_others}{pred_variance}{0.00}
\setres{cty_Guatemala}{EU_transfer_others}{r_ff}{---}
\setres{cty_Guatemala}{EU_transfer_others}{r_ff_se}{---}
\setres{cty_Guatemala}{EU_transfer_others}{r_prior}{---}
\setres{cty_Guatemala}{EU_transfer_others}{r_prior_se}{---}
\setres{cty_Guatemala}{EU_transfer_others}{r_prob}{---}
\setres{cty_Guatemala}{EU_transfer_others}{r_prob_se}{---}
\setres{cty_Guatemala}{EU_transfer_others}{spread_penalty}{0.000}
\setres{cty_Guatemala}{EU_transfer_others}{spread_penalty_se}{0.000}
\setres{cty_Guatemala}{EU_transfer_others}{theta_star}{0.94}
\setres{cty_Guatemala}{EU_transfer_others}{transfer}{others}
\setres{cty_Guatemala}{EU_transfer_others}{volume_ess}{---}
\setres{cty_Guatemala}{EU_uniform}{kappa}{-0.825}
\setres{cty_Guatemala}{EU_uniform}{kappa_prior_mean}{-0.171}
\setres{cty_Guatemala}{EU_uniform}{kappa_prior_mean_se}{0.209}
\setres{cty_Guatemala}{EU_uniform}{kappa_se}{0.250}
\setres{cty_Guatemala}{EU_uniform}{mse}{44.79}
\setres{cty_Guatemala}{EU_uniform}{mse_prior_mean}{36.60}
\setres{cty_Guatemala}{EU_uniform}{nrmse_pct}{33.7}
\setres{cty_Guatemala}{EU_uniform}{pred_variance}{8.20}
\setres{cty_Guatemala}{EU_uniform}{r_ff}{---}
\setres{cty_Guatemala}{EU_uniform}{r_ff_se}{---}
\setres{cty_Guatemala}{EU_uniform}{r_prior}{---}
\setres{cty_Guatemala}{EU_uniform}{r_prior_se}{---}
\setres{cty_Guatemala}{EU_uniform}{r_prob}{---}
\setres{cty_Guatemala}{EU_uniform}{r_prob_se}{---}
\setres{cty_Guatemala}{EU_uniform}{spread_penalty}{0.654}
\setres{cty_Guatemala}{EU_uniform}{spread_penalty_se}{0.062}
\setres{cty_Guatemala}{EU_uniform}{volume_ess}{---}
\setres{cty_Guatemala}{EV_deterministic}{kappa}{0.915}
\setres{cty_Guatemala}{EV_deterministic}{kappa_se}{0.042}
\setres{cty_Guatemala}{EV_deterministic}{mse}{22.97}
\setres{cty_Guatemala}{EV_deterministic}{nrmse_pct}{25.7}
\setres{cty_Guatemala}{EV_deterministic}{r_ff}{---}
\setres{cty_Guatemala}{EV_deterministic}{r_ff_se}{---}
\setres{cty_Guatemala}{EV_deterministic}{r_prior}{---}
\setres{cty_Guatemala}{EV_deterministic}{r_prior_se}{---}
\setres{cty_Guatemala}{EV_deterministic}{r_prob}{---}
\setres{cty_Guatemala}{EV_deterministic}{r_prob_se}{---}
\setres{cty_Guatemala}{EV_deterministic}{volume_ess}{---}
\setres{cty_Guatemala}{reference}{country_name}{Guatemala}
\setres{cty_Guatemala}{reference}{lookup_nrmse_pct}{25.8}
\setres{cty_Guatemala}{reference}{midpoint_mse}{34.45}
\setres{cty_Guatemala}{reference}{midpoint_nrmse_pct}{32.9}
\setres{cty_Guatemala}{reference}{missing_lotteries}{none}
\setres{cty_Guatemala}{reference}{n_lotteries}{14}
\setres{cty_Guatemala}{reference}{n_obs}{1{,}176}
\setres{cty_Guatemala}{reference}{n_subjects}{84}
\setres{cty_Guatemala}{reference}{oracle_mse}{21.91}
\setres{cty_Guatemala}{reference}{sample}{LV2019\_Guatemala}
\setres{cty_India}{CPT_deterministic}{kappa}{0.958}
\setres{cty_India}{CPT_deterministic}{kappa_se}{0.026}
\setres{cty_India}{CPT_deterministic}{mse}{19.53}
\setres{cty_India}{CPT_deterministic}{nrmse_pct}{24.9}
\setres{cty_India}{CPT_deterministic}{r_ff}{---}
\setres{cty_India}{CPT_deterministic}{r_ff_se}{---}
\setres{cty_India}{CPT_deterministic}{r_prior}{---}
\setres{cty_India}{CPT_deterministic}{r_prior_se}{---}
\setres{cty_India}{CPT_deterministic}{r_prob}{---}
\setres{cty_India}{CPT_deterministic}{r_prob_se}{---}
\setres{cty_India}{CPT_deterministic}{theta_star}{1.10, 0.50, 0.60}
\setres{cty_India}{CPT_deterministic}{volume_ess}{---}
\setres{cty_India}{CPT_lit}{kappa}{0.269}
\setres{cty_India}{CPT_lit}{kappa_prior_mean}{0.695}
\setres{cty_India}{CPT_lit}{kappa_prior_mean_se}{0.122}
\setres{cty_India}{CPT_lit}{kappa_se}{0.184}
\setres{cty_India}{CPT_lit}{mse}{24.22}
\setres{cty_India}{CPT_lit}{mse_prior_mean}{21.32}
\setres{cty_India}{CPT_lit}{nrmse_pct}{26.7}
\setres{cty_India}{CPT_lit}{pred_variance}{2.90}
\setres{cty_India}{CPT_lit}{r_ff}{---}
\setres{cty_India}{CPT_lit}{r_ff_se}{---}
\setres{cty_India}{CPT_lit}{r_prior}{---}
\setres{cty_India}{CPT_lit}{r_prior_se}{---}
\setres{cty_India}{CPT_lit}{r_prob}{---}
\setres{cty_India}{CPT_lit}{r_prob_se}{---}
\setres{cty_India}{CPT_lit}{spread_penalty}{0.426}
\setres{cty_India}{CPT_lit}{spread_penalty_se}{0.069}
\setres{cty_India}{CPT_lit}{volume_ess}{---}
\setres{cty_India}{CPT_transfer_others}{kappa}{0.735}
\setres{cty_India}{CPT_transfer_others}{kappa_prior_mean}{0.735}
\setres{cty_India}{CPT_transfer_others}{kappa_prior_mean_se}{0.110}
\setres{cty_India}{CPT_transfer_others}{kappa_se}{0.110}
\setres{cty_India}{CPT_transfer_others}{mse}{21.05}
\setres{cty_India}{CPT_transfer_others}{mse_prior_mean}{21.05}
\setres{cty_India}{CPT_transfer_others}{nrmse_pct}{23.6}
\setres{cty_India}{CPT_transfer_others}{pred_variance}{0.00}
\setres{cty_India}{CPT_transfer_others}{r_ff}{---}
\setres{cty_India}{CPT_transfer_others}{r_ff_se}{---}
\setres{cty_India}{CPT_transfer_others}{r_prior}{---}
\setres{cty_India}{CPT_transfer_others}{r_prior_se}{---}
\setres{cty_India}{CPT_transfer_others}{r_prob}{---}
\setres{cty_India}{CPT_transfer_others}{r_prob_se}{---}
\setres{cty_India}{CPT_transfer_others}{spread_penalty}{0.000}
\setres{cty_India}{CPT_transfer_others}{spread_penalty_se}{0.000}
\setres{cty_India}{CPT_transfer_others}{theta_star}{1.00, 0.60, 0.90}
\setres{cty_India}{CPT_transfer_others}{transfer}{others}
\setres{cty_India}{CPT_transfer_others}{volume_ess}{---}
\setres{cty_India}{CPT_uniform}{kappa}{-2.104}
\setres{cty_India}{CPT_uniform}{kappa_prior_mean}{0.448}
\setres{cty_India}{CPT_uniform}{kappa_prior_mean_se}{0.202}
\setres{cty_India}{CPT_uniform}{kappa_se}{0.596}
\setres{cty_India}{CPT_uniform}{mse}{40.37}
\setres{cty_India}{CPT_uniform}{mse_prior_mean}{23.00}
\setres{cty_India}{CPT_uniform}{nrmse_pct}{27.6}
\setres{cty_India}{CPT_uniform}{pred_variance}{17.37}
\setres{cty_India}{CPT_uniform}{r_ff}{---}
\setres{cty_India}{CPT_uniform}{r_ff_se}{---}
\setres{cty_India}{CPT_uniform}{r_prior}{---}
\setres{cty_India}{CPT_uniform}{r_prior_se}{---}
\setres{cty_India}{CPT_uniform}{r_prob}{---}
\setres{cty_India}{CPT_uniform}{r_prob_se}{---}
\setres{cty_India}{CPT_uniform}{spread_penalty}{2.551}
\setres{cty_India}{CPT_uniform}{spread_penalty_se}{0.412}
\setres{cty_India}{CPT_uniform}{volume_ess}{---}
\setres{cty_India}{DA_deterministic}{kappa}{0.668}
\setres{cty_India}{DA_deterministic}{kappa_se}{0.127}
\setres{cty_India}{DA_deterministic}{mse}{21.51}
\setres{cty_India}{DA_deterministic}{nrmse_pct}{26.1}
\setres{cty_India}{DA_deterministic}{r_ff}{---}
\setres{cty_India}{DA_deterministic}{r_ff_se}{---}
\setres{cty_India}{DA_deterministic}{r_prior}{---}
\setres{cty_India}{DA_deterministic}{r_prior_se}{---}
\setres{cty_India}{DA_deterministic}{r_prob}{---}
\setres{cty_India}{DA_deterministic}{r_prob_se}{---}
\setres{cty_India}{DA_deterministic}{theta_star}{1.82, 3.00}
\setres{cty_India}{DA_deterministic}{volume_ess}{---}
\setres{cty_India}{DA_lit}{kappa}{-0.839}
\setres{cty_India}{DA_lit}{kappa_prior_mean}{-0.141}
\setres{cty_India}{DA_lit}{kappa_prior_mean_se}{0.372}
\setres{cty_India}{DA_lit}{kappa_se}{0.484}
\setres{cty_India}{DA_lit}{mse}{31.76}
\setres{cty_India}{DA_lit}{mse_prior_mean}{27.01}
\setres{cty_India}{DA_lit}{nrmse_pct}{28.8}
\setres{cty_India}{DA_lit}{pred_variance}{4.75}
\setres{cty_India}{DA_lit}{r_ff}{---}
\setres{cty_India}{DA_lit}{r_ff_se}{---}
\setres{cty_India}{DA_lit}{r_prior}{---}
\setres{cty_India}{DA_lit}{r_prior_se}{---}
\setres{cty_India}{DA_lit}{r_prob}{---}
\setres{cty_India}{DA_lit}{r_prob_se}{---}
\setres{cty_India}{DA_lit}{spread_penalty}{0.698}
\setres{cty_India}{DA_lit}{spread_penalty_se}{0.113}
\setres{cty_India}{DA_lit}{volume_ess}{---}
\setres{cty_India}{DA_transfer_others}{kappa}{0.434}
\setres{cty_India}{DA_transfer_others}{kappa_prior_mean}{0.434}
\setres{cty_India}{DA_transfer_others}{kappa_prior_mean_se}{0.167}
\setres{cty_India}{DA_transfer_others}{kappa_se}{0.167}
\setres{cty_India}{DA_transfer_others}{mse}{23.09}
\setres{cty_India}{DA_transfer_others}{mse_prior_mean}{23.09}
\setres{cty_India}{DA_transfer_others}{nrmse_pct}{24.9}
\setres{cty_India}{DA_transfer_others}{pred_variance}{0.00}
\setres{cty_India}{DA_transfer_others}{r_ff}{---}
\setres{cty_India}{DA_transfer_others}{r_ff_se}{---}
\setres{cty_India}{DA_transfer_others}{r_prior}{---}
\setres{cty_India}{DA_transfer_others}{r_prior_se}{---}
\setres{cty_India}{DA_transfer_others}{r_prob}{---}
\setres{cty_India}{DA_transfer_others}{r_prob_se}{---}
\setres{cty_India}{DA_transfer_others}{spread_penalty}{0.000}
\setres{cty_India}{DA_transfer_others}{spread_penalty_se}{0.000}
\setres{cty_India}{DA_transfer_others}{theta_star}{1.67, 1.37}
\setres{cty_India}{DA_transfer_others}{transfer}{others}
\setres{cty_India}{DA_transfer_others}{volume_ess}{---}
\setres{cty_India}{DA_uniform}{kappa}{-2.703}
\setres{cty_India}{DA_uniform}{kappa_prior_mean}{-0.698}
\setres{cty_India}{DA_uniform}{kappa_prior_mean_se}{0.534}
\setres{cty_India}{DA_uniform}{kappa_se}{0.849}
\setres{cty_India}{DA_uniform}{mse}{44.45}
\setres{cty_India}{DA_uniform}{mse_prior_mean}{30.80}
\setres{cty_India}{DA_uniform}{nrmse_pct}{30.7}
\setres{cty_India}{DA_uniform}{pred_variance}{13.65}
\setres{cty_India}{DA_uniform}{r_ff}{---}
\setres{cty_India}{DA_uniform}{r_ff_se}{---}
\setres{cty_India}{DA_uniform}{r_prior}{---}
\setres{cty_India}{DA_uniform}{r_prior_se}{---}
\setres{cty_India}{DA_uniform}{r_prob}{---}
\setres{cty_India}{DA_uniform}{r_prob_se}{---}
\setres{cty_India}{DA_uniform}{spread_penalty}{2.005}
\setres{cty_India}{DA_uniform}{spread_penalty_se}{0.323}
\setres{cty_India}{DA_uniform}{volume_ess}{---}
\setres{cty_India}{EU_deterministic}{kappa}{0.233}
\setres{cty_India}{EU_deterministic}{kappa_se}{0.225}
\setres{cty_India}{EU_deterministic}{mse}{24.46}
\setres{cty_India}{EU_deterministic}{nrmse_pct}{27.7}
\setres{cty_India}{EU_deterministic}{r_ff}{---}
\setres{cty_India}{EU_deterministic}{r_ff_se}{---}
\setres{cty_India}{EU_deterministic}{r_prior}{---}
\setres{cty_India}{EU_deterministic}{r_prior_se}{---}
\setres{cty_India}{EU_deterministic}{r_prob}{---}
\setres{cty_India}{EU_deterministic}{r_prob_se}{---}
\setres{cty_India}{EU_deterministic}{theta_star}{0.77}
\setres{cty_India}{EU_deterministic}{volume_ess}{---}
\setres{cty_India}{EU_lit}{kappa}{0.074}
\setres{cty_India}{EU_lit}{kappa_prior_mean}{0.251}
\setres{cty_India}{EU_lit}{kappa_prior_mean_se}{0.232}
\setres{cty_India}{EU_lit}{kappa_se}{0.258}
\setres{cty_India}{EU_lit}{mse}{25.55}
\setres{cty_India}{EU_lit}{mse_prior_mean}{24.34}
\setres{cty_India}{EU_lit}{nrmse_pct}{28.0}
\setres{cty_India}{EU_lit}{pred_variance}{1.20}
\setres{cty_India}{EU_lit}{r_ff}{---}
\setres{cty_India}{EU_lit}{r_ff_se}{---}
\setres{cty_India}{EU_lit}{r_prior}{---}
\setres{cty_India}{EU_lit}{r_prior_se}{---}
\setres{cty_India}{EU_lit}{r_prob}{---}
\setres{cty_India}{EU_lit}{r_prob_se}{---}
\setres{cty_India}{EU_lit}{spread_penalty}{0.177}
\setres{cty_India}{EU_lit}{spread_penalty_se}{0.029}
\setres{cty_India}{EU_lit}{volume_ess}{---}
\setres{cty_India}{EU_transfer_others}{kappa}{0.111}
\setres{cty_India}{EU_transfer_others}{kappa_prior_mean}{0.111}
\setres{cty_India}{EU_transfer_others}{kappa_prior_mean_se}{0.227}
\setres{cty_India}{EU_transfer_others}{kappa_se}{0.227}
\setres{cty_India}{EU_transfer_others}{mse}{25.29}
\setres{cty_India}{EU_transfer_others}{mse_prior_mean}{25.29}
\setres{cty_India}{EU_transfer_others}{nrmse_pct}{27.3}
\setres{cty_India}{EU_transfer_others}{pred_variance}{0.00}
\setres{cty_India}{EU_transfer_others}{r_ff}{---}
\setres{cty_India}{EU_transfer_others}{r_ff_se}{---}
\setres{cty_India}{EU_transfer_others}{r_prior}{---}
\setres{cty_India}{EU_transfer_others}{r_prior_se}{---}
\setres{cty_India}{EU_transfer_others}{r_prob}{---}
\setres{cty_India}{EU_transfer_others}{r_prob_se}{---}
\setres{cty_India}{EU_transfer_others}{spread_penalty}{0.000}
\setres{cty_India}{EU_transfer_others}{spread_penalty_se}{0.000}
\setres{cty_India}{EU_transfer_others}{theta_star}{0.95}
\setres{cty_India}{EU_transfer_others}{transfer}{others}
\setres{cty_India}{EU_transfer_others}{volume_ess}{---}
\setres{cty_India}{EU_uniform}{kappa}{-1.464}
\setres{cty_India}{EU_uniform}{kappa_prior_mean}{-0.259}
\setres{cty_India}{EU_uniform}{kappa_prior_mean_se}{0.398}
\setres{cty_India}{EU_uniform}{kappa_se}{0.589}
\setres{cty_India}{EU_uniform}{mse}{36.01}
\setres{cty_India}{EU_uniform}{mse_prior_mean}{27.82}
\setres{cty_India}{EU_uniform}{nrmse_pct}{29.2}
\setres{cty_India}{EU_uniform}{pred_variance}{8.20}
\setres{cty_India}{EU_uniform}{r_ff}{---}
\setres{cty_India}{EU_uniform}{r_ff_se}{---}
\setres{cty_India}{EU_uniform}{r_prior}{---}
\setres{cty_India}{EU_uniform}{r_prior_se}{---}
\setres{cty_India}{EU_uniform}{r_prob}{---}
\setres{cty_India}{EU_uniform}{r_prob_se}{---}
\setres{cty_India}{EU_uniform}{spread_penalty}{1.204}
\setres{cty_India}{EU_uniform}{spread_penalty_se}{0.194}
\setres{cty_India}{EU_uniform}{volume_ess}{---}
\setres{cty_India}{EV_deterministic}{kappa}{0.030}
\setres{cty_India}{EV_deterministic}{kappa_se}{0.241}
\setres{cty_India}{EV_deterministic}{mse}{25.84}
\setres{cty_India}{EV_deterministic}{nrmse_pct}{26.8}
\setres{cty_India}{EV_deterministic}{r_ff}{---}
\setres{cty_India}{EV_deterministic}{r_ff_se}{---}
\setres{cty_India}{EV_deterministic}{r_prior}{---}
\setres{cty_India}{EV_deterministic}{r_prior_se}{---}
\setres{cty_India}{EV_deterministic}{r_prob}{---}
\setres{cty_India}{EV_deterministic}{r_prob_se}{---}
\setres{cty_India}{EV_deterministic}{volume_ess}{---}
\setres{cty_India}{reference}{country_name}{India}
\setres{cty_India}{reference}{lookup_nrmse_pct}{23.5}
\setres{cty_India}{reference}{midpoint_mse}{26.05}
\setres{cty_India}{reference}{midpoint_nrmse_pct}{28.8}
\setres{cty_India}{reference}{missing_lotteries}{none}
\setres{cty_India}{reference}{n_lotteries}{14}
\setres{cty_India}{reference}{n_obs}{1{,}246}
\setres{cty_India}{reference}{n_subjects}{89}
\setres{cty_India}{reference}{oracle_mse}{19.24}
\setres{cty_India}{reference}{sample}{LV2019\_India}
\setres{cty_Japan}{CPT_deterministic}{kappa}{0.990}
\setres{cty_Japan}{CPT_deterministic}{kappa_se}{0.009}
\setres{cty_Japan}{CPT_deterministic}{mse}{8.66}
\setres{cty_Japan}{CPT_deterministic}{nrmse_pct}{16.3}
\setres{cty_Japan}{CPT_deterministic}{r_ff}{---}
\setres{cty_Japan}{CPT_deterministic}{r_ff_se}{---}
\setres{cty_Japan}{CPT_deterministic}{r_prior}{---}
\setres{cty_Japan}{CPT_deterministic}{r_prior_se}{---}
\setres{cty_Japan}{CPT_deterministic}{r_prob}{---}
\setres{cty_Japan}{CPT_deterministic}{r_prob_se}{---}
\setres{cty_Japan}{CPT_deterministic}{theta_star}{0.80, 0.70, 1.10}
\setres{cty_Japan}{CPT_deterministic}{volume_ess}{---}
\setres{cty_Japan}{CPT_lit}{kappa}{0.654}
\setres{cty_Japan}{CPT_lit}{kappa_prior_mean}{0.913}
\setres{cty_Japan}{CPT_lit}{kappa_prior_mean_se}{0.039}
\setres{cty_Japan}{CPT_lit}{kappa_se}{0.046}
\setres{cty_Japan}{CPT_lit}{mse}{12.43}
\setres{cty_Japan}{CPT_lit}{mse_prior_mean}{9.53}
\setres{cty_Japan}{CPT_lit}{nrmse_pct}{19.0}
\setres{cty_Japan}{CPT_lit}{pred_variance}{2.90}
\setres{cty_Japan}{CPT_lit}{r_ff}{---}
\setres{cty_Japan}{CPT_lit}{r_ff_se}{---}
\setres{cty_Japan}{CPT_lit}{r_prior}{---}
\setres{cty_Japan}{CPT_lit}{r_prior_se}{---}
\setres{cty_Japan}{CPT_lit}{r_prob}{---}
\setres{cty_Japan}{CPT_lit}{r_prob_se}{---}
\setres{cty_Japan}{CPT_lit}{spread_penalty}{0.259}
\setres{cty_Japan}{CPT_lit}{spread_penalty_se}{0.015}
\setres{cty_Japan}{CPT_lit}{volume_ess}{---}
\setres{cty_Japan}{CPT_transfer_others}{kappa}{0.956}
\setres{cty_Japan}{CPT_transfer_others}{kappa_prior_mean}{0.956}
\setres{cty_Japan}{CPT_transfer_others}{kappa_prior_mean_se}{0.015}
\setres{cty_Japan}{CPT_transfer_others}{kappa_se}{0.015}
\setres{cty_Japan}{CPT_transfer_others}{mse}{9.05}
\setres{cty_Japan}{CPT_transfer_others}{mse_prior_mean}{9.05}
\setres{cty_Japan}{CPT_transfer_others}{nrmse_pct}{15.9}
\setres{cty_Japan}{CPT_transfer_others}{pred_variance}{0.00}
\setres{cty_Japan}{CPT_transfer_others}{r_ff}{---}
\setres{cty_Japan}{CPT_transfer_others}{r_ff_se}{---}
\setres{cty_Japan}{CPT_transfer_others}{r_prior}{---}
\setres{cty_Japan}{CPT_transfer_others}{r_prior_se}{---}
\setres{cty_Japan}{CPT_transfer_others}{r_prob}{---}
\setres{cty_Japan}{CPT_transfer_others}{r_prob_se}{---}
\setres{cty_Japan}{CPT_transfer_others}{spread_penalty}{0.000}
\setres{cty_Japan}{CPT_transfer_others}{spread_penalty_se}{0.000}
\setres{cty_Japan}{CPT_transfer_others}{theta_star}{1.10, 0.60, 0.80}
\setres{cty_Japan}{CPT_transfer_others}{transfer}{others}
\setres{cty_Japan}{CPT_transfer_others}{volume_ess}{---}
\setres{cty_Japan}{CPT_uniform}{kappa}{-1.184}
\setres{cty_Japan}{CPT_uniform}{kappa_prior_mean}{0.368}
\setres{cty_Japan}{CPT_uniform}{kappa_prior_mean_se}{0.100}
\setres{cty_Japan}{CPT_uniform}{kappa_se}{0.151}
\setres{cty_Japan}{CPT_uniform}{mse}{32.99}
\setres{cty_Japan}{CPT_uniform}{mse_prior_mean}{15.63}
\setres{cty_Japan}{CPT_uniform}{nrmse_pct}{21.1}
\setres{cty_Japan}{CPT_uniform}{pred_variance}{17.37}
\setres{cty_Japan}{CPT_uniform}{r_ff}{---}
\setres{cty_Japan}{CPT_uniform}{r_ff_se}{---}
\setres{cty_Japan}{CPT_uniform}{r_prior}{---}
\setres{cty_Japan}{CPT_uniform}{r_prior_se}{---}
\setres{cty_Japan}{CPT_uniform}{r_prob}{---}
\setres{cty_Japan}{CPT_uniform}{r_prob_se}{---}
\setres{cty_Japan}{CPT_uniform}{spread_penalty}{1.551}
\setres{cty_Japan}{CPT_uniform}{spread_penalty_se}{0.092}
\setres{cty_Japan}{CPT_uniform}{volume_ess}{---}
\setres{cty_Japan}{DA_deterministic}{kappa}{0.928}
\setres{cty_Japan}{DA_deterministic}{kappa_se}{0.020}
\setres{cty_Japan}{DA_deterministic}{mse}{9.36}
\setres{cty_Japan}{DA_deterministic}{nrmse_pct}{16.1}
\setres{cty_Japan}{DA_deterministic}{r_ff}{---}
\setres{cty_Japan}{DA_deterministic}{r_ff_se}{---}
\setres{cty_Japan}{DA_deterministic}{r_prior}{---}
\setres{cty_Japan}{DA_deterministic}{r_prior_se}{---}
\setres{cty_Japan}{DA_deterministic}{r_prob}{---}
\setres{cty_Japan}{DA_deterministic}{r_prob_se}{---}
\setres{cty_Japan}{DA_deterministic}{theta_star}{1.13, 0.47}
\setres{cty_Japan}{DA_deterministic}{volume_ess}{---}
\setres{cty_Japan}{DA_lit}{kappa}{-0.131}
\setres{cty_Japan}{DA_lit}{kappa_prior_mean}{0.294}
\setres{cty_Japan}{DA_lit}{kappa_prior_mean_se}{0.118}
\setres{cty_Japan}{DA_lit}{kappa_se}{0.133}
\setres{cty_Japan}{DA_lit}{mse}{21.21}
\setres{cty_Japan}{DA_lit}{mse_prior_mean}{16.46}
\setres{cty_Japan}{DA_lit}{nrmse_pct}{22.8}
\setres{cty_Japan}{DA_lit}{pred_variance}{4.75}
\setres{cty_Japan}{DA_lit}{r_ff}{---}
\setres{cty_Japan}{DA_lit}{r_ff_se}{---}
\setres{cty_Japan}{DA_lit}{r_prior}{---}
\setres{cty_Japan}{DA_lit}{r_prior_se}{---}
\setres{cty_Japan}{DA_lit}{r_prob}{---}
\setres{cty_Japan}{DA_lit}{r_prob_se}{---}
\setres{cty_Japan}{DA_lit}{spread_penalty}{0.424}
\setres{cty_Japan}{DA_lit}{spread_penalty_se}{0.025}
\setres{cty_Japan}{DA_lit}{volume_ess}{---}
\setres{cty_Japan}{DA_transfer_others}{kappa}{0.898}
\setres{cty_Japan}{DA_transfer_others}{kappa_prior_mean}{0.898}
\setres{cty_Japan}{DA_transfer_others}{kappa_prior_mean_se}{0.026}
\setres{cty_Japan}{DA_transfer_others}{kappa_se}{0.026}
\setres{cty_Japan}{DA_transfer_others}{mse}{9.70}
\setres{cty_Japan}{DA_transfer_others}{mse_prior_mean}{9.70}
\setres{cty_Japan}{DA_transfer_others}{nrmse_pct}{15.4}
\setres{cty_Japan}{DA_transfer_others}{pred_variance}{0.00}
\setres{cty_Japan}{DA_transfer_others}{r_ff}{---}
\setres{cty_Japan}{DA_transfer_others}{r_ff_se}{---}
\setres{cty_Japan}{DA_transfer_others}{r_prior}{---}
\setres{cty_Japan}{DA_transfer_others}{r_prior_se}{---}
\setres{cty_Japan}{DA_transfer_others}{r_prob}{---}
\setres{cty_Japan}{DA_transfer_others}{r_prob_se}{---}
\setres{cty_Japan}{DA_transfer_others}{spread_penalty}{0.000}
\setres{cty_Japan}{DA_transfer_others}{spread_penalty_se}{0.000}
\setres{cty_Japan}{DA_transfer_others}{theta_star}{1.72, 1.49}
\setres{cty_Japan}{DA_transfer_others}{transfer}{others}
\setres{cty_Japan}{DA_transfer_others}{volume_ess}{---}
\setres{cty_Japan}{DA_uniform}{kappa}{-1.616}
\setres{cty_Japan}{DA_uniform}{kappa_prior_mean}{-0.396}
\setres{cty_Japan}{DA_uniform}{kappa_prior_mean_se}{0.180}
\setres{cty_Japan}{DA_uniform}{kappa_se}{0.225}
\setres{cty_Japan}{DA_uniform}{mse}{37.83}
\setres{cty_Japan}{DA_uniform}{mse_prior_mean}{24.18}
\setres{cty_Japan}{DA_uniform}{nrmse_pct}{26.5}
\setres{cty_Japan}{DA_uniform}{pred_variance}{13.65}
\setres{cty_Japan}{DA_uniform}{r_ff}{---}
\setres{cty_Japan}{DA_uniform}{r_ff_se}{---}
\setres{cty_Japan}{DA_uniform}{r_prior}{---}
\setres{cty_Japan}{DA_uniform}{r_prior_se}{---}
\setres{cty_Japan}{DA_uniform}{r_prob}{---}
\setres{cty_Japan}{DA_uniform}{r_prob_se}{---}
\setres{cty_Japan}{DA_uniform}{spread_penalty}{1.219}
\setres{cty_Japan}{DA_uniform}{spread_penalty_se}{0.073}
\setres{cty_Japan}{DA_uniform}{volume_ess}{---}
\setres{cty_Japan}{EU_deterministic}{kappa}{0.914}
\setres{cty_Japan}{EU_deterministic}{kappa_se}{0.026}
\setres{cty_Japan}{EU_deterministic}{mse}{9.51}
\setres{cty_Japan}{EU_deterministic}{nrmse_pct}{16.5}
\setres{cty_Japan}{EU_deterministic}{r_ff}{---}
\setres{cty_Japan}{EU_deterministic}{r_ff_se}{---}
\setres{cty_Japan}{EU_deterministic}{r_prior}{---}
\setres{cty_Japan}{EU_deterministic}{r_prior_se}{---}
\setres{cty_Japan}{EU_deterministic}{r_prob}{---}
\setres{cty_Japan}{EU_deterministic}{r_prob_se}{---}
\setres{cty_Japan}{EU_deterministic}{theta_star}{0.86}
\setres{cty_Japan}{EU_deterministic}{volume_ess}{---}
\setres{cty_Japan}{EU_lit}{kappa}{0.708}
\setres{cty_Japan}{EU_lit}{kappa_prior_mean}{0.816}
\setres{cty_Japan}{EU_lit}{kappa_prior_mean_se}{0.046}
\setres{cty_Japan}{EU_lit}{kappa_se}{0.050}
\setres{cty_Japan}{EU_lit}{mse}{11.82}
\setres{cty_Japan}{EU_lit}{mse_prior_mean}{10.61}
\setres{cty_Japan}{EU_lit}{nrmse_pct}{18.3}
\setres{cty_Japan}{EU_lit}{pred_variance}{1.20}
\setres{cty_Japan}{EU_lit}{r_ff}{---}
\setres{cty_Japan}{EU_lit}{r_ff_se}{---}
\setres{cty_Japan}{EU_lit}{r_prior}{---}
\setres{cty_Japan}{EU_lit}{r_prior_se}{---}
\setres{cty_Japan}{EU_lit}{r_prob}{---}
\setres{cty_Japan}{EU_lit}{r_prob_se}{---}
\setres{cty_Japan}{EU_lit}{spread_penalty}{0.108}
\setres{cty_Japan}{EU_lit}{spread_penalty_se}{0.006}
\setres{cty_Japan}{EU_lit}{volume_ess}{---}
\setres{cty_Japan}{EU_transfer_others}{kappa}{0.902}
\setres{cty_Japan}{EU_transfer_others}{kappa_prior_mean}{0.902}
\setres{cty_Japan}{EU_transfer_others}{kappa_prior_mean_se}{0.037}
\setres{cty_Japan}{EU_transfer_others}{kappa_se}{0.037}
\setres{cty_Japan}{EU_transfer_others}{mse}{9.65}
\setres{cty_Japan}{EU_transfer_others}{mse_prior_mean}{9.65}
\setres{cty_Japan}{EU_transfer_others}{nrmse_pct}{15.2}
\setres{cty_Japan}{EU_transfer_others}{pred_variance}{0.00}
\setres{cty_Japan}{EU_transfer_others}{r_ff}{---}
\setres{cty_Japan}{EU_transfer_others}{r_ff_se}{---}
\setres{cty_Japan}{EU_transfer_others}{r_prior}{---}
\setres{cty_Japan}{EU_transfer_others}{r_prior_se}{---}
\setres{cty_Japan}{EU_transfer_others}{r_prob}{---}
\setres{cty_Japan}{EU_transfer_others}{r_prob_se}{---}
\setres{cty_Japan}{EU_transfer_others}{spread_penalty}{0.000}
\setres{cty_Japan}{EU_transfer_others}{spread_penalty_se}{0.000}
\setres{cty_Japan}{EU_transfer_others}{theta_star}{0.95}
\setres{cty_Japan}{EU_transfer_others}{transfer}{others}
\setres{cty_Japan}{EU_transfer_others}{volume_ess}{---}
\setres{cty_Japan}{EU_uniform}{kappa}{-0.551}
\setres{cty_Japan}{EU_uniform}{kappa_prior_mean}{0.181}
\setres{cty_Japan}{EU_uniform}{kappa_prior_mean_se}{0.128}
\setres{cty_Japan}{EU_uniform}{kappa_se}{0.155}
\setres{cty_Japan}{EU_uniform}{mse}{25.91}
\setres{cty_Japan}{EU_uniform}{mse_prior_mean}{17.72}
\setres{cty_Japan}{EU_uniform}{nrmse_pct}{22.5}
\setres{cty_Japan}{EU_uniform}{pred_variance}{8.20}
\setres{cty_Japan}{EU_uniform}{r_ff}{---}
\setres{cty_Japan}{EU_uniform}{r_ff_se}{---}
\setres{cty_Japan}{EU_uniform}{r_prior}{---}
\setres{cty_Japan}{EU_uniform}{r_prior_se}{---}
\setres{cty_Japan}{EU_uniform}{r_prob}{---}
\setres{cty_Japan}{EU_uniform}{r_prob_se}{---}
\setres{cty_Japan}{EU_uniform}{spread_penalty}{0.732}
\setres{cty_Japan}{EU_uniform}{spread_penalty_se}{0.044}
\setres{cty_Japan}{EU_uniform}{volume_ess}{---}
\setres{cty_Japan}{EV_deterministic}{kappa}{0.874}
\setres{cty_Japan}{EV_deterministic}{kappa_se}{0.047}
\setres{cty_Japan}{EV_deterministic}{mse}{9.96}
\setres{cty_Japan}{EV_deterministic}{nrmse_pct}{14.9}
\setres{cty_Japan}{EV_deterministic}{r_ff}{---}
\setres{cty_Japan}{EV_deterministic}{r_ff_se}{---}
\setres{cty_Japan}{EV_deterministic}{r_prior}{---}
\setres{cty_Japan}{EV_deterministic}{r_prior_se}{---}
\setres{cty_Japan}{EV_deterministic}{r_prob}{---}
\setres{cty_Japan}{EV_deterministic}{r_prob_se}{---}
\setres{cty_Japan}{EV_deterministic}{volume_ess}{---}
\setres{cty_Japan}{reference}{country_name}{Japan}
\setres{cty_Japan}{reference}{lookup_nrmse_pct}{14.4}
\setres{cty_Japan}{reference}{midpoint_mse}{19.75}
\setres{cty_Japan}{reference}{midpoint_nrmse_pct}{21.7}
\setres{cty_Japan}{reference}{missing_lotteries}{none}
\setres{cty_Japan}{reference}{n_lotteries}{14}
\setres{cty_Japan}{reference}{n_obs}{1{,}176}
\setres{cty_Japan}{reference}{n_subjects}{84}
\setres{cty_Japan}{reference}{oracle_mse}{8.55}
\setres{cty_Japan}{reference}{sample}{LV2019\_Japan}
\setres{cty_Kyrgyzstan}{CPT_deterministic}{kappa}{0.975}
\setres{cty_Kyrgyzstan}{CPT_deterministic}{kappa_se}{0.014}
\setres{cty_Kyrgyzstan}{CPT_deterministic}{mse}{13.82}
\setres{cty_Kyrgyzstan}{CPT_deterministic}{nrmse_pct}{20.2}
\setres{cty_Kyrgyzstan}{CPT_deterministic}{r_ff}{---}
\setres{cty_Kyrgyzstan}{CPT_deterministic}{r_ff_se}{---}
\setres{cty_Kyrgyzstan}{CPT_deterministic}{r_prior}{---}
\setres{cty_Kyrgyzstan}{CPT_deterministic}{r_prior_se}{---}
\setres{cty_Kyrgyzstan}{CPT_deterministic}{r_prob}{---}
\setres{cty_Kyrgyzstan}{CPT_deterministic}{r_prob_se}{---}
\setres{cty_Kyrgyzstan}{CPT_deterministic}{theta_star}{1.00, 0.70, 0.90}
\setres{cty_Kyrgyzstan}{CPT_deterministic}{volume_ess}{---}
\setres{cty_Kyrgyzstan}{CPT_lit}{kappa}{0.441}
\setres{cty_Kyrgyzstan}{CPT_lit}{kappa_prior_mean}{0.751}
\setres{cty_Kyrgyzstan}{CPT_lit}{kappa_prior_mean_se}{0.074}
\setres{cty_Kyrgyzstan}{CPT_lit}{kappa_se}{0.089}
\setres{cty_Kyrgyzstan}{CPT_lit}{mse}{18.82}
\setres{cty_Kyrgyzstan}{CPT_lit}{mse_prior_mean}{15.92}
\setres{cty_Kyrgyzstan}{CPT_lit}{nrmse_pct}{22.6}
\setres{cty_Kyrgyzstan}{CPT_lit}{pred_variance}{2.90}
\setres{cty_Kyrgyzstan}{CPT_lit}{r_ff}{---}
\setres{cty_Kyrgyzstan}{CPT_lit}{r_ff_se}{---}
\setres{cty_Kyrgyzstan}{CPT_lit}{r_prior}{---}
\setres{cty_Kyrgyzstan}{CPT_lit}{r_prior_se}{---}
\setres{cty_Kyrgyzstan}{CPT_lit}{r_prob}{---}
\setres{cty_Kyrgyzstan}{CPT_lit}{r_prob_se}{---}
\setres{cty_Kyrgyzstan}{CPT_lit}{spread_penalty}{0.309}
\setres{cty_Kyrgyzstan}{CPT_lit}{spread_penalty_se}{0.030}
\setres{cty_Kyrgyzstan}{CPT_lit}{volume_ess}{---}
\setres{cty_Kyrgyzstan}{CPT_transfer_others}{kappa}{0.987}
\setres{cty_Kyrgyzstan}{CPT_transfer_others}{kappa_prior_mean}{0.987}
\setres{cty_Kyrgyzstan}{CPT_transfer_others}{kappa_prior_mean_se}{0.016}
\setres{cty_Kyrgyzstan}{CPT_transfer_others}{kappa_se}{0.016}
\setres{cty_Kyrgyzstan}{CPT_transfer_others}{mse}{13.70}
\setres{cty_Kyrgyzstan}{CPT_transfer_others}{mse_prior_mean}{13.70}
\setres{cty_Kyrgyzstan}{CPT_transfer_others}{nrmse_pct}{20.2}
\setres{cty_Kyrgyzstan}{CPT_transfer_others}{pred_variance}{0.00}
\setres{cty_Kyrgyzstan}{CPT_transfer_others}{r_ff}{---}
\setres{cty_Kyrgyzstan}{CPT_transfer_others}{r_ff_se}{---}
\setres{cty_Kyrgyzstan}{CPT_transfer_others}{r_prior}{---}
\setres{cty_Kyrgyzstan}{CPT_transfer_others}{r_prior_se}{---}
\setres{cty_Kyrgyzstan}{CPT_transfer_others}{r_prob}{---}
\setres{cty_Kyrgyzstan}{CPT_transfer_others}{r_prob_se}{---}
\setres{cty_Kyrgyzstan}{CPT_transfer_others}{spread_penalty}{0.000}
\setres{cty_Kyrgyzstan}{CPT_transfer_others}{spread_penalty_se}{0.000}
\setres{cty_Kyrgyzstan}{CPT_transfer_others}{theta_star}{1.10, 0.60, 0.80}
\setres{cty_Kyrgyzstan}{CPT_transfer_others}{transfer}{others}
\setres{cty_Kyrgyzstan}{CPT_transfer_others}{volume_ess}{---}
\setres{cty_Kyrgyzstan}{CPT_uniform}{kappa}{-1.844}
\setres{cty_Kyrgyzstan}{CPT_uniform}{kappa_prior_mean}{0.010}
\setres{cty_Kyrgyzstan}{CPT_uniform}{kappa_prior_mean_se}{0.136}
\setres{cty_Kyrgyzstan}{CPT_uniform}{kappa_se}{0.262}
\setres{cty_Kyrgyzstan}{CPT_uniform}{mse}{40.23}
\setres{cty_Kyrgyzstan}{CPT_uniform}{mse_prior_mean}{22.86}
\setres{cty_Kyrgyzstan}{CPT_uniform}{nrmse_pct}{27.3}
\setres{cty_Kyrgyzstan}{CPT_uniform}{pred_variance}{17.37}
\setres{cty_Kyrgyzstan}{CPT_uniform}{r_ff}{---}
\setres{cty_Kyrgyzstan}{CPT_uniform}{r_ff_se}{---}
\setres{cty_Kyrgyzstan}{CPT_uniform}{r_prior}{---}
\setres{cty_Kyrgyzstan}{CPT_uniform}{r_prior_se}{---}
\setres{cty_Kyrgyzstan}{CPT_uniform}{r_prob}{---}
\setres{cty_Kyrgyzstan}{CPT_uniform}{r_prob_se}{---}
\setres{cty_Kyrgyzstan}{CPT_uniform}{spread_penalty}{1.853}
\setres{cty_Kyrgyzstan}{CPT_uniform}{spread_penalty_se}{0.178}
\setres{cty_Kyrgyzstan}{CPT_uniform}{volume_ess}{---}
\setres{cty_Kyrgyzstan}{DA_deterministic}{kappa}{0.897}
\setres{cty_Kyrgyzstan}{DA_deterministic}{kappa_se}{0.037}
\setres{cty_Kyrgyzstan}{DA_deterministic}{mse}{14.55}
\setres{cty_Kyrgyzstan}{DA_deterministic}{nrmse_pct}{20.2}
\setres{cty_Kyrgyzstan}{DA_deterministic}{r_ff}{---}
\setres{cty_Kyrgyzstan}{DA_deterministic}{r_ff_se}{---}
\setres{cty_Kyrgyzstan}{DA_deterministic}{r_prior}{---}
\setres{cty_Kyrgyzstan}{DA_deterministic}{r_prior_se}{---}
\setres{cty_Kyrgyzstan}{DA_deterministic}{r_prob}{---}
\setres{cty_Kyrgyzstan}{DA_deterministic}{r_prob_se}{---}
\setres{cty_Kyrgyzstan}{DA_deterministic}{theta_star}{1.41, 0.76}
\setres{cty_Kyrgyzstan}{DA_deterministic}{volume_ess}{---}
\setres{cty_Kyrgyzstan}{DA_lit}{kappa}{-0.706}
\setres{cty_Kyrgyzstan}{DA_lit}{kappa_prior_mean}{-0.199}
\setres{cty_Kyrgyzstan}{DA_lit}{kappa_prior_mean_se}{0.195}
\setres{cty_Kyrgyzstan}{DA_lit}{kappa_se}{0.230}
\setres{cty_Kyrgyzstan}{DA_lit}{mse}{29.56}
\setres{cty_Kyrgyzstan}{DA_lit}{mse_prior_mean}{24.82}
\setres{cty_Kyrgyzstan}{DA_lit}{nrmse_pct}{27.7}
\setres{cty_Kyrgyzstan}{DA_lit}{pred_variance}{4.75}
\setres{cty_Kyrgyzstan}{DA_lit}{r_ff}{---}
\setres{cty_Kyrgyzstan}{DA_lit}{r_ff_se}{---}
\setres{cty_Kyrgyzstan}{DA_lit}{r_prior}{---}
\setres{cty_Kyrgyzstan}{DA_lit}{r_prior_se}{---}
\setres{cty_Kyrgyzstan}{DA_lit}{r_prob}{---}
\setres{cty_Kyrgyzstan}{DA_lit}{r_prob_se}{---}
\setres{cty_Kyrgyzstan}{DA_lit}{spread_penalty}{0.507}
\setres{cty_Kyrgyzstan}{DA_lit}{spread_penalty_se}{0.049}
\setres{cty_Kyrgyzstan}{DA_lit}{volume_ess}{---}
\setres{cty_Kyrgyzstan}{DA_transfer_others}{kappa}{0.906}
\setres{cty_Kyrgyzstan}{DA_transfer_others}{kappa_prior_mean}{0.906}
\setres{cty_Kyrgyzstan}{DA_transfer_others}{kappa_prior_mean_se}{0.035}
\setres{cty_Kyrgyzstan}{DA_transfer_others}{kappa_se}{0.035}
\setres{cty_Kyrgyzstan}{DA_transfer_others}{mse}{14.47}
\setres{cty_Kyrgyzstan}{DA_transfer_others}{mse_prior_mean}{14.47}
\setres{cty_Kyrgyzstan}{DA_transfer_others}{nrmse_pct}{20.2}
\setres{cty_Kyrgyzstan}{DA_transfer_others}{pred_variance}{0.00}
\setres{cty_Kyrgyzstan}{DA_transfer_others}{r_ff}{---}
\setres{cty_Kyrgyzstan}{DA_transfer_others}{r_ff_se}{---}
\setres{cty_Kyrgyzstan}{DA_transfer_others}{r_prior}{---}
\setres{cty_Kyrgyzstan}{DA_transfer_others}{r_prior_se}{---}
\setres{cty_Kyrgyzstan}{DA_transfer_others}{r_prob}{---}
\setres{cty_Kyrgyzstan}{DA_transfer_others}{r_prob_se}{---}
\setres{cty_Kyrgyzstan}{DA_transfer_others}{spread_penalty}{0.000}
\setres{cty_Kyrgyzstan}{DA_transfer_others}{spread_penalty_se}{0.000}
\setres{cty_Kyrgyzstan}{DA_transfer_others}{theta_star}{1.70, 1.48}
\setres{cty_Kyrgyzstan}{DA_transfer_others}{transfer}{others}
\setres{cty_Kyrgyzstan}{DA_transfer_others}{volume_ess}{---}
\setres{cty_Kyrgyzstan}{DA_uniform}{kappa}{-2.579}
\setres{cty_Kyrgyzstan}{DA_uniform}{kappa_prior_mean}{-1.123}
\setres{cty_Kyrgyzstan}{DA_uniform}{kappa_prior_mean_se}{0.274}
\setres{cty_Kyrgyzstan}{DA_uniform}{kappa_se}{0.382}
\setres{cty_Kyrgyzstan}{DA_uniform}{mse}{47.12}
\setres{cty_Kyrgyzstan}{DA_uniform}{mse_prior_mean}{33.47}
\setres{cty_Kyrgyzstan}{DA_uniform}{nrmse_pct}{30.9}
\setres{cty_Kyrgyzstan}{DA_uniform}{pred_variance}{13.65}
\setres{cty_Kyrgyzstan}{DA_uniform}{r_ff}{---}
\setres{cty_Kyrgyzstan}{DA_uniform}{r_ff_se}{---}
\setres{cty_Kyrgyzstan}{DA_uniform}{r_prior}{---}
\setres{cty_Kyrgyzstan}{DA_uniform}{r_prior_se}{---}
\setres{cty_Kyrgyzstan}{DA_uniform}{r_prob}{---}
\setres{cty_Kyrgyzstan}{DA_uniform}{r_prob_se}{---}
\setres{cty_Kyrgyzstan}{DA_uniform}{spread_penalty}{1.457}
\setres{cty_Kyrgyzstan}{DA_uniform}{spread_penalty_se}{0.140}
\setres{cty_Kyrgyzstan}{DA_uniform}{volume_ess}{---}
\setres{cty_Kyrgyzstan}{EU_deterministic}{kappa}{0.866}
\setres{cty_Kyrgyzstan}{EU_deterministic}{kappa_se}{0.051}
\setres{cty_Kyrgyzstan}{EU_deterministic}{mse}{14.83}
\setres{cty_Kyrgyzstan}{EU_deterministic}{nrmse_pct}{20.9}
\setres{cty_Kyrgyzstan}{EU_deterministic}{r_ff}{---}
\setres{cty_Kyrgyzstan}{EU_deterministic}{r_ff_se}{---}
\setres{cty_Kyrgyzstan}{EU_deterministic}{r_prior}{---}
\setres{cty_Kyrgyzstan}{EU_deterministic}{r_prior_se}{---}
\setres{cty_Kyrgyzstan}{EU_deterministic}{r_prob}{---}
\setres{cty_Kyrgyzstan}{EU_deterministic}{r_prob_se}{---}
\setres{cty_Kyrgyzstan}{EU_deterministic}{theta_star}{0.96}
\setres{cty_Kyrgyzstan}{EU_deterministic}{volume_ess}{---}
\setres{cty_Kyrgyzstan}{EU_lit}{kappa}{0.466}
\setres{cty_Kyrgyzstan}{EU_lit}{kappa_prior_mean}{0.595}
\setres{cty_Kyrgyzstan}{EU_lit}{kappa_prior_mean_se}{0.100}
\setres{cty_Kyrgyzstan}{EU_lit}{kappa_se}{0.109}
\setres{cty_Kyrgyzstan}{EU_lit}{mse}{18.58}
\setres{cty_Kyrgyzstan}{EU_lit}{mse_prior_mean}{17.38}
\setres{cty_Kyrgyzstan}{EU_lit}{nrmse_pct}{23.6}
\setres{cty_Kyrgyzstan}{EU_lit}{pred_variance}{1.20}
\setres{cty_Kyrgyzstan}{EU_lit}{r_ff}{---}
\setres{cty_Kyrgyzstan}{EU_lit}{r_ff_se}{---}
\setres{cty_Kyrgyzstan}{EU_lit}{r_prior}{---}
\setres{cty_Kyrgyzstan}{EU_lit}{r_prior_se}{---}
\setres{cty_Kyrgyzstan}{EU_lit}{r_prob}{---}
\setres{cty_Kyrgyzstan}{EU_lit}{r_prob_se}{---}
\setres{cty_Kyrgyzstan}{EU_lit}{spread_penalty}{0.129}
\setres{cty_Kyrgyzstan}{EU_lit}{spread_penalty_se}{0.012}
\setres{cty_Kyrgyzstan}{EU_lit}{volume_ess}{---}
\setres{cty_Kyrgyzstan}{EU_transfer_others}{kappa}{0.880}
\setres{cty_Kyrgyzstan}{EU_transfer_others}{kappa_prior_mean}{0.880}
\setres{cty_Kyrgyzstan}{EU_transfer_others}{kappa_prior_mean_se}{0.050}
\setres{cty_Kyrgyzstan}{EU_transfer_others}{kappa_se}{0.050}
\setres{cty_Kyrgyzstan}{EU_transfer_others}{mse}{14.71}
\setres{cty_Kyrgyzstan}{EU_transfer_others}{mse_prior_mean}{14.71}
\setres{cty_Kyrgyzstan}{EU_transfer_others}{nrmse_pct}{21.1}
\setres{cty_Kyrgyzstan}{EU_transfer_others}{pred_variance}{0.00}
\setres{cty_Kyrgyzstan}{EU_transfer_others}{r_ff}{---}
\setres{cty_Kyrgyzstan}{EU_transfer_others}{r_ff_se}{---}
\setres{cty_Kyrgyzstan}{EU_transfer_others}{r_prior}{---}
\setres{cty_Kyrgyzstan}{EU_transfer_others}{r_prior_se}{---}
\setres{cty_Kyrgyzstan}{EU_transfer_others}{r_prob}{---}
\setres{cty_Kyrgyzstan}{EU_transfer_others}{r_prob_se}{---}
\setres{cty_Kyrgyzstan}{EU_transfer_others}{spread_penalty}{0.000}
\setres{cty_Kyrgyzstan}{EU_transfer_others}{spread_penalty_se}{0.000}
\setres{cty_Kyrgyzstan}{EU_transfer_others}{theta_star}{0.94}
\setres{cty_Kyrgyzstan}{EU_transfer_others}{transfer}{others}
\setres{cty_Kyrgyzstan}{EU_transfer_others}{volume_ess}{---}
\setres{cty_Kyrgyzstan}{EU_uniform}{kappa}{-1.235}
\setres{cty_Kyrgyzstan}{EU_uniform}{kappa_prior_mean}{-0.360}
\setres{cty_Kyrgyzstan}{EU_uniform}{kappa_prior_mean_se}{0.207}
\setres{cty_Kyrgyzstan}{EU_uniform}{kappa_se}{0.270}
\setres{cty_Kyrgyzstan}{EU_uniform}{mse}{34.52}
\setres{cty_Kyrgyzstan}{EU_uniform}{mse_prior_mean}{26.33}
\setres{cty_Kyrgyzstan}{EU_uniform}{nrmse_pct}{28.4}
\setres{cty_Kyrgyzstan}{EU_uniform}{pred_variance}{8.20}
\setres{cty_Kyrgyzstan}{EU_uniform}{r_ff}{---}
\setres{cty_Kyrgyzstan}{EU_uniform}{r_ff_se}{---}
\setres{cty_Kyrgyzstan}{EU_uniform}{r_prior}{---}
\setres{cty_Kyrgyzstan}{EU_uniform}{r_prior_se}{---}
\setres{cty_Kyrgyzstan}{EU_uniform}{r_prob}{---}
\setres{cty_Kyrgyzstan}{EU_uniform}{r_prob_se}{---}
\setres{cty_Kyrgyzstan}{EU_uniform}{spread_penalty}{0.875}
\setres{cty_Kyrgyzstan}{EU_uniform}{spread_penalty_se}{0.084}
\setres{cty_Kyrgyzstan}{EU_uniform}{volume_ess}{---}
\setres{cty_Kyrgyzstan}{EV_deterministic}{kappa}{0.877}
\setres{cty_Kyrgyzstan}{EV_deterministic}{kappa_se}{0.053}
\setres{cty_Kyrgyzstan}{EV_deterministic}{mse}{14.73}
\setres{cty_Kyrgyzstan}{EV_deterministic}{nrmse_pct}{20.4}
\setres{cty_Kyrgyzstan}{EV_deterministic}{r_ff}{---}
\setres{cty_Kyrgyzstan}{EV_deterministic}{r_ff_se}{---}
\setres{cty_Kyrgyzstan}{EV_deterministic}{r_prior}{---}
\setres{cty_Kyrgyzstan}{EV_deterministic}{r_prior_se}{---}
\setres{cty_Kyrgyzstan}{EV_deterministic}{r_prob}{---}
\setres{cty_Kyrgyzstan}{EV_deterministic}{r_prob_se}{---}
\setres{cty_Kyrgyzstan}{EV_deterministic}{volume_ess}{---}
\setres{cty_Kyrgyzstan}{reference}{country_name}{Kyrgyzstan}
\setres{cty_Kyrgyzstan}{reference}{lookup_nrmse_pct}{19.0}
\setres{cty_Kyrgyzstan}{reference}{midpoint_mse}{22.95}
\setres{cty_Kyrgyzstan}{reference}{midpoint_nrmse_pct}{27.0}
\setres{cty_Kyrgyzstan}{reference}{missing_lotteries}{none}
\setres{cty_Kyrgyzstan}{reference}{n_lotteries}{14}
\setres{cty_Kyrgyzstan}{reference}{n_obs}{1{,}358}
\setres{cty_Kyrgyzstan}{reference}{n_subjects}{97}
\setres{cty_Kyrgyzstan}{reference}{oracle_mse}{13.58}
\setres{cty_Kyrgyzstan}{reference}{sample}{LV2019\_Kyrgyzstan}
\setres{cty_Malaysia}{CPT_deterministic}{kappa}{0.978}
\setres{cty_Malaysia}{CPT_deterministic}{kappa_se}{0.016}
\setres{cty_Malaysia}{CPT_deterministic}{mse}{12.94}
\setres{cty_Malaysia}{CPT_deterministic}{nrmse_pct}{18.7}
\setres{cty_Malaysia}{CPT_deterministic}{r_ff}{---}
\setres{cty_Malaysia}{CPT_deterministic}{r_ff_se}{---}
\setres{cty_Malaysia}{CPT_deterministic}{r_prior}{---}
\setres{cty_Malaysia}{CPT_deterministic}{r_prior_se}{---}
\setres{cty_Malaysia}{CPT_deterministic}{r_prob}{---}
\setres{cty_Malaysia}{CPT_deterministic}{r_prob_se}{---}
\setres{cty_Malaysia}{CPT_deterministic}{theta_star}{1.20, 0.70, 0.80}
\setres{cty_Malaysia}{CPT_deterministic}{volume_ess}{---}
\setres{cty_Malaysia}{CPT_lit}{kappa}{0.320}
\setres{cty_Malaysia}{CPT_lit}{kappa_prior_mean}{0.632}
\setres{cty_Malaysia}{CPT_lit}{kappa_prior_mean_se}{0.117}
\setres{cty_Malaysia}{CPT_lit}{kappa_se}{0.133}
\setres{cty_Malaysia}{CPT_lit}{mse}{19.06}
\setres{cty_Malaysia}{CPT_lit}{mse_prior_mean}{16.16}
\setres{cty_Malaysia}{CPT_lit}{nrmse_pct}{22.5}
\setres{cty_Malaysia}{CPT_lit}{pred_variance}{2.90}
\setres{cty_Malaysia}{CPT_lit}{r_ff}{---}
\setres{cty_Malaysia}{CPT_lit}{r_ff_se}{---}
\setres{cty_Malaysia}{CPT_lit}{r_prior}{---}
\setres{cty_Malaysia}{CPT_lit}{r_prior_se}{---}
\setres{cty_Malaysia}{CPT_lit}{r_prob}{---}
\setres{cty_Malaysia}{CPT_lit}{r_prob_se}{---}
\setres{cty_Malaysia}{CPT_lit}{spread_penalty}{0.312}
\setres{cty_Malaysia}{CPT_lit}{spread_penalty_se}{0.033}
\setres{cty_Malaysia}{CPT_lit}{volume_ess}{---}
\setres{cty_Malaysia}{CPT_transfer_others}{kappa}{0.977}
\setres{cty_Malaysia}{CPT_transfer_others}{kappa_prior_mean}{0.977}
\setres{cty_Malaysia}{CPT_transfer_others}{kappa_prior_mean_se}{0.028}
\setres{cty_Malaysia}{CPT_transfer_others}{kappa_se}{0.028}
\setres{cty_Malaysia}{CPT_transfer_others}{mse}{12.95}
\setres{cty_Malaysia}{CPT_transfer_others}{mse_prior_mean}{12.95}
\setres{cty_Malaysia}{CPT_transfer_others}{nrmse_pct}{19.2}
\setres{cty_Malaysia}{CPT_transfer_others}{pred_variance}{0.00}
\setres{cty_Malaysia}{CPT_transfer_others}{r_ff}{---}
\setres{cty_Malaysia}{CPT_transfer_others}{r_ff_se}{---}
\setres{cty_Malaysia}{CPT_transfer_others}{r_prior}{---}
\setres{cty_Malaysia}{CPT_transfer_others}{r_prior_se}{---}
\setres{cty_Malaysia}{CPT_transfer_others}{r_prob}{---}
\setres{cty_Malaysia}{CPT_transfer_others}{r_prob_se}{---}
\setres{cty_Malaysia}{CPT_transfer_others}{spread_penalty}{0.000}
\setres{cty_Malaysia}{CPT_transfer_others}{spread_penalty_se}{0.000}
\setres{cty_Malaysia}{CPT_transfer_others}{theta_star}{1.00, 0.60, 0.90}
\setres{cty_Malaysia}{CPT_transfer_others}{transfer}{others}
\setres{cty_Malaysia}{CPT_transfer_others}{volume_ess}{---}
\setres{cty_Malaysia}{CPT_uniform}{kappa}{-2.073}
\setres{cty_Malaysia}{CPT_uniform}{kappa_prior_mean}{-0.206}
\setres{cty_Malaysia}{CPT_uniform}{kappa_prior_mean_se}{0.197}
\setres{cty_Malaysia}{CPT_uniform}{kappa_se}{0.329}
\setres{cty_Malaysia}{CPT_uniform}{mse}{41.31}
\setres{cty_Malaysia}{CPT_uniform}{mse_prior_mean}{23.94}
\setres{cty_Malaysia}{CPT_uniform}{nrmse_pct}{26.8}
\setres{cty_Malaysia}{CPT_uniform}{pred_variance}{17.37}
\setres{cty_Malaysia}{CPT_uniform}{r_ff}{---}
\setres{cty_Malaysia}{CPT_uniform}{r_ff_se}{---}
\setres{cty_Malaysia}{CPT_uniform}{r_prior}{---}
\setres{cty_Malaysia}{CPT_uniform}{r_prior_se}{---}
\setres{cty_Malaysia}{CPT_uniform}{r_prob}{---}
\setres{cty_Malaysia}{CPT_uniform}{r_prob_se}{---}
\setres{cty_Malaysia}{CPT_uniform}{spread_penalty}{1.868}
\setres{cty_Malaysia}{CPT_uniform}{spread_penalty_se}{0.199}
\setres{cty_Malaysia}{CPT_uniform}{volume_ess}{---}
\setres{cty_Malaysia}{DA_deterministic}{kappa}{0.918}
\setres{cty_Malaysia}{DA_deterministic}{kappa_se}{0.037}
\setres{cty_Malaysia}{DA_deterministic}{mse}{13.50}
\setres{cty_Malaysia}{DA_deterministic}{nrmse_pct}{18.6}
\setres{cty_Malaysia}{DA_deterministic}{r_ff}{---}
\setres{cty_Malaysia}{DA_deterministic}{r_ff_se}{---}
\setres{cty_Malaysia}{DA_deterministic}{r_prior}{---}
\setres{cty_Malaysia}{DA_deterministic}{r_prior_se}{---}
\setres{cty_Malaysia}{DA_deterministic}{r_prob}{---}
\setres{cty_Malaysia}{DA_deterministic}{r_prob_se}{---}
\setres{cty_Malaysia}{DA_deterministic}{theta_star}{1.65, 1.06}
\setres{cty_Malaysia}{DA_deterministic}{volume_ess}{---}
\setres{cty_Malaysia}{DA_lit}{kappa}{-0.947}
\setres{cty_Malaysia}{DA_lit}{kappa_prior_mean}{-0.436}
\setres{cty_Malaysia}{DA_lit}{kappa_prior_mean_se}{0.274}
\setres{cty_Malaysia}{DA_lit}{kappa_se}{0.309}
\setres{cty_Malaysia}{DA_lit}{mse}{30.84}
\setres{cty_Malaysia}{DA_lit}{mse_prior_mean}{26.09}
\setres{cty_Malaysia}{DA_lit}{nrmse_pct}{28.5}
\setres{cty_Malaysia}{DA_lit}{pred_variance}{4.75}
\setres{cty_Malaysia}{DA_lit}{r_ff}{---}
\setres{cty_Malaysia}{DA_lit}{r_ff_se}{---}
\setres{cty_Malaysia}{DA_lit}{r_prior}{---}
\setres{cty_Malaysia}{DA_lit}{r_prior_se}{---}
\setres{cty_Malaysia}{DA_lit}{r_prob}{---}
\setres{cty_Malaysia}{DA_lit}{r_prob_se}{---}
\setres{cty_Malaysia}{DA_lit}{spread_penalty}{0.511}
\setres{cty_Malaysia}{DA_lit}{spread_penalty_se}{0.054}
\setres{cty_Malaysia}{DA_lit}{volume_ess}{---}
\setres{cty_Malaysia}{DA_transfer_others}{kappa}{0.922}
\setres{cty_Malaysia}{DA_transfer_others}{kappa_prior_mean}{0.922}
\setres{cty_Malaysia}{DA_transfer_others}{kappa_prior_mean_se}{0.047}
\setres{cty_Malaysia}{DA_transfer_others}{kappa_se}{0.047}
\setres{cty_Malaysia}{DA_transfer_others}{mse}{13.46}
\setres{cty_Malaysia}{DA_transfer_others}{mse_prior_mean}{13.46}
\setres{cty_Malaysia}{DA_transfer_others}{nrmse_pct}{18.7}
\setres{cty_Malaysia}{DA_transfer_others}{pred_variance}{0.00}
\setres{cty_Malaysia}{DA_transfer_others}{r_ff}{---}
\setres{cty_Malaysia}{DA_transfer_others}{r_ff_se}{---}
\setres{cty_Malaysia}{DA_transfer_others}{r_prior}{---}
\setres{cty_Malaysia}{DA_transfer_others}{r_prior_se}{---}
\setres{cty_Malaysia}{DA_transfer_others}{r_prob}{---}
\setres{cty_Malaysia}{DA_transfer_others}{r_prob_se}{---}
\setres{cty_Malaysia}{DA_transfer_others}{spread_penalty}{0.000}
\setres{cty_Malaysia}{DA_transfer_others}{spread_penalty_se}{0.000}
\setres{cty_Malaysia}{DA_transfer_others}{theta_star}{1.70, 1.46}
\setres{cty_Malaysia}{DA_transfer_others}{transfer}{others}
\setres{cty_Malaysia}{DA_transfer_others}{volume_ess}{---}
\setres{cty_Malaysia}{DA_uniform}{kappa}{-2.901}
\setres{cty_Malaysia}{DA_uniform}{kappa_prior_mean}{-1.433}
\setres{cty_Malaysia}{DA_uniform}{kappa_prior_mean_se}{0.376}
\setres{cty_Malaysia}{DA_uniform}{kappa_se}{0.487}
\setres{cty_Malaysia}{DA_uniform}{mse}{49.00}
\setres{cty_Malaysia}{DA_uniform}{mse_prior_mean}{35.35}
\setres{cty_Malaysia}{DA_uniform}{nrmse_pct}{31.9}
\setres{cty_Malaysia}{DA_uniform}{pred_variance}{13.65}
\setres{cty_Malaysia}{DA_uniform}{r_ff}{---}
\setres{cty_Malaysia}{DA_uniform}{r_ff_se}{---}
\setres{cty_Malaysia}{DA_uniform}{r_prior}{---}
\setres{cty_Malaysia}{DA_uniform}{r_prior_se}{---}
\setres{cty_Malaysia}{DA_uniform}{r_prob}{---}
\setres{cty_Malaysia}{DA_uniform}{r_prob_se}{---}
\setres{cty_Malaysia}{DA_uniform}{spread_penalty}{1.468}
\setres{cty_Malaysia}{DA_uniform}{spread_penalty_se}{0.156}
\setres{cty_Malaysia}{DA_uniform}{volume_ess}{---}
\setres{cty_Malaysia}{EU_deterministic}{kappa}{0.872}
\setres{cty_Malaysia}{EU_deterministic}{kappa_se}{0.061}
\setres{cty_Malaysia}{EU_deterministic}{mse}{13.93}
\setres{cty_Malaysia}{EU_deterministic}{nrmse_pct}{19.0}
\setres{cty_Malaysia}{EU_deterministic}{r_ff}{---}
\setres{cty_Malaysia}{EU_deterministic}{r_ff_se}{---}
\setres{cty_Malaysia}{EU_deterministic}{r_prior}{---}
\setres{cty_Malaysia}{EU_deterministic}{r_prior_se}{---}
\setres{cty_Malaysia}{EU_deterministic}{r_prob}{---}
\setres{cty_Malaysia}{EU_deterministic}{r_prob_se}{---}
\setres{cty_Malaysia}{EU_deterministic}{theta_star}{1.02}
\setres{cty_Malaysia}{EU_deterministic}{volume_ess}{---}
\setres{cty_Malaysia}{EU_lit}{kappa}{0.340}
\setres{cty_Malaysia}{EU_lit}{kappa_prior_mean}{0.470}
\setres{cty_Malaysia}{EU_lit}{kappa_prior_mean_se}{0.146}
\setres{cty_Malaysia}{EU_lit}{kappa_se}{0.155}
\setres{cty_Malaysia}{EU_lit}{mse}{18.87}
\setres{cty_Malaysia}{EU_lit}{mse_prior_mean}{17.67}
\setres{cty_Malaysia}{EU_lit}{nrmse_pct}{23.5}
\setres{cty_Malaysia}{EU_lit}{pred_variance}{1.20}
\setres{cty_Malaysia}{EU_lit}{r_ff}{---}
\setres{cty_Malaysia}{EU_lit}{r_ff_se}{---}
\setres{cty_Malaysia}{EU_lit}{r_prior}{---}
\setres{cty_Malaysia}{EU_lit}{r_prior_se}{---}
\setres{cty_Malaysia}{EU_lit}{r_prob}{---}
\setres{cty_Malaysia}{EU_lit}{r_prob_se}{---}
\setres{cty_Malaysia}{EU_lit}{spread_penalty}{0.130}
\setres{cty_Malaysia}{EU_lit}{spread_penalty_se}{0.014}
\setres{cty_Malaysia}{EU_lit}{volume_ess}{---}
\setres{cty_Malaysia}{EU_transfer_others}{kappa}{0.868}
\setres{cty_Malaysia}{EU_transfer_others}{kappa_prior_mean}{0.868}
\setres{cty_Malaysia}{EU_transfer_others}{kappa_prior_mean_se}{0.060}
\setres{cty_Malaysia}{EU_transfer_others}{kappa_se}{0.060}
\setres{cty_Malaysia}{EU_transfer_others}{mse}{13.96}
\setres{cty_Malaysia}{EU_transfer_others}{mse_prior_mean}{13.96}
\setres{cty_Malaysia}{EU_transfer_others}{nrmse_pct}{18.9}
\setres{cty_Malaysia}{EU_transfer_others}{pred_variance}{0.00}
\setres{cty_Malaysia}{EU_transfer_others}{r_ff}{---}
\setres{cty_Malaysia}{EU_transfer_others}{r_ff_se}{---}
\setres{cty_Malaysia}{EU_transfer_others}{r_prior}{---}
\setres{cty_Malaysia}{EU_transfer_others}{r_prior_se}{---}
\setres{cty_Malaysia}{EU_transfer_others}{r_prob}{---}
\setres{cty_Malaysia}{EU_transfer_others}{r_prob_se}{---}
\setres{cty_Malaysia}{EU_transfer_others}{spread_penalty}{0.000}
\setres{cty_Malaysia}{EU_transfer_others}{spread_penalty_se}{0.000}
\setres{cty_Malaysia}{EU_transfer_others}{theta_star}{0.94}
\setres{cty_Malaysia}{EU_transfer_others}{transfer}{others}
\setres{cty_Malaysia}{EU_transfer_others}{volume_ess}{---}
\setres{cty_Malaysia}{EU_uniform}{kappa}{-1.508}
\setres{cty_Malaysia}{EU_uniform}{kappa_prior_mean}{-0.626}
\setres{cty_Malaysia}{EU_uniform}{kappa_prior_mean_se}{0.289}
\setres{cty_Malaysia}{EU_uniform}{kappa_se}{0.353}
\setres{cty_Malaysia}{EU_uniform}{mse}{36.05}
\setres{cty_Malaysia}{EU_uniform}{mse_prior_mean}{27.86}
\setres{cty_Malaysia}{EU_uniform}{nrmse_pct}{28.8}
\setres{cty_Malaysia}{EU_uniform}{pred_variance}{8.20}
\setres{cty_Malaysia}{EU_uniform}{r_ff}{---}
\setres{cty_Malaysia}{EU_uniform}{r_ff_se}{---}
\setres{cty_Malaysia}{EU_uniform}{r_prior}{---}
\setres{cty_Malaysia}{EU_uniform}{r_prior_se}{---}
\setres{cty_Malaysia}{EU_uniform}{r_prob}{---}
\setres{cty_Malaysia}{EU_uniform}{r_prob_se}{---}
\setres{cty_Malaysia}{EU_uniform}{spread_penalty}{0.882}
\setres{cty_Malaysia}{EU_uniform}{spread_penalty_se}{0.094}
\setres{cty_Malaysia}{EU_uniform}{volume_ess}{---}
\setres{cty_Malaysia}{EV_deterministic}{kappa}{0.884}
\setres{cty_Malaysia}{EV_deterministic}{kappa_se}{0.059}
\setres{cty_Malaysia}{EV_deterministic}{mse}{13.81}
\setres{cty_Malaysia}{EV_deterministic}{nrmse_pct}{18.8}
\setres{cty_Malaysia}{EV_deterministic}{r_ff}{---}
\setres{cty_Malaysia}{EV_deterministic}{r_ff_se}{---}
\setres{cty_Malaysia}{EV_deterministic}{r_prior}{---}
\setres{cty_Malaysia}{EV_deterministic}{r_prior_se}{---}
\setres{cty_Malaysia}{EV_deterministic}{r_prob}{---}
\setres{cty_Malaysia}{EV_deterministic}{r_prob_se}{---}
\setres{cty_Malaysia}{EV_deterministic}{volume_ess}{---}
\setres{cty_Malaysia}{reference}{country_name}{Malaysia}
\setres{cty_Malaysia}{reference}{lookup_nrmse_pct}{18.6}
\setres{cty_Malaysia}{reference}{midpoint_mse}{22.03}
\setres{cty_Malaysia}{reference}{midpoint_nrmse_pct}{26.1}
\setres{cty_Malaysia}{reference}{missing_lotteries}{none}
\setres{cty_Malaysia}{reference}{n_lotteries}{14}
\setres{cty_Malaysia}{reference}{n_obs}{896}
\setres{cty_Malaysia}{reference}{n_subjects}{64}
\setres{cty_Malaysia}{reference}{oracle_mse}{12.74}
\setres{cty_Malaysia}{reference}{sample}{LV2019\_Malaysia}
\setres{cty_Nicaragua}{CPT_deterministic}{kappa}{0.997}
\setres{cty_Nicaragua}{CPT_deterministic}{kappa_se}{0.031}
\setres{cty_Nicaragua}{CPT_deterministic}{mse}{34.59}
\setres{cty_Nicaragua}{CPT_deterministic}{nrmse_pct}{33.2}
\setres{cty_Nicaragua}{CPT_deterministic}{r_ff}{---}
\setres{cty_Nicaragua}{CPT_deterministic}{r_ff_se}{---}
\setres{cty_Nicaragua}{CPT_deterministic}{r_prior}{---}
\setres{cty_Nicaragua}{CPT_deterministic}{r_prior_se}{---}
\setres{cty_Nicaragua}{CPT_deterministic}{r_prob}{---}
\setres{cty_Nicaragua}{CPT_deterministic}{r_prob_se}{---}
\setres{cty_Nicaragua}{CPT_deterministic}{theta_star}{1.10, 0.50, 1.10}
\setres{cty_Nicaragua}{CPT_deterministic}{volume_ess}{---}
\setres{cty_Nicaragua}{CPT_lit}{kappa}{-0.758}
\setres{cty_Nicaragua}{CPT_lit}{kappa_prior_mean}{-0.295}
\setres{cty_Nicaragua}{CPT_lit}{kappa_prior_mean_se}{0.257}
\setres{cty_Nicaragua}{CPT_lit}{kappa_se}{0.305}
\setres{cty_Nicaragua}{CPT_lit}{mse}{45.59}
\setres{cty_Nicaragua}{CPT_lit}{mse_prior_mean}{42.69}
\setres{cty_Nicaragua}{CPT_lit}{nrmse_pct}{36.4}
\setres{cty_Nicaragua}{CPT_lit}{pred_variance}{2.90}
\setres{cty_Nicaragua}{CPT_lit}{r_ff}{---}
\setres{cty_Nicaragua}{CPT_lit}{r_ff_se}{---}
\setres{cty_Nicaragua}{CPT_lit}{r_prior}{---}
\setres{cty_Nicaragua}{CPT_lit}{r_prior_se}{---}
\setres{cty_Nicaragua}{CPT_lit}{r_prob}{---}
\setres{cty_Nicaragua}{CPT_lit}{r_prob_se}{---}
\setres{cty_Nicaragua}{CPT_lit}{spread_penalty}{0.463}
\setres{cty_Nicaragua}{CPT_lit}{spread_penalty_se}{0.074}
\setres{cty_Nicaragua}{CPT_lit}{volume_ess}{---}
\setres{cty_Nicaragua}{CPT_transfer_others}{kappa}{0.708}
\setres{cty_Nicaragua}{CPT_transfer_others}{kappa_prior_mean}{0.708}
\setres{cty_Nicaragua}{CPT_transfer_others}{kappa_prior_mean_se}{0.108}
\setres{cty_Nicaragua}{CPT_transfer_others}{kappa_se}{0.108}
\setres{cty_Nicaragua}{CPT_transfer_others}{mse}{36.40}
\setres{cty_Nicaragua}{CPT_transfer_others}{mse_prior_mean}{36.40}
\setres{cty_Nicaragua}{CPT_transfer_others}{nrmse_pct}{33.5}
\setres{cty_Nicaragua}{CPT_transfer_others}{pred_variance}{0.00}
\setres{cty_Nicaragua}{CPT_transfer_others}{r_ff}{---}
\setres{cty_Nicaragua}{CPT_transfer_others}{r_ff_se}{---}
\setres{cty_Nicaragua}{CPT_transfer_others}{r_prior}{---}
\setres{cty_Nicaragua}{CPT_transfer_others}{r_prior_se}{---}
\setres{cty_Nicaragua}{CPT_transfer_others}{r_prob}{---}
\setres{cty_Nicaragua}{CPT_transfer_others}{r_prob_se}{---}
\setres{cty_Nicaragua}{CPT_transfer_others}{spread_penalty}{0.000}
\setres{cty_Nicaragua}{CPT_transfer_others}{spread_penalty_se}{0.000}
\setres{cty_Nicaragua}{CPT_transfer_others}{theta_star}{1.00, 0.60, 0.90}
\setres{cty_Nicaragua}{CPT_transfer_others}{transfer}{others}
\setres{cty_Nicaragua}{CPT_transfer_others}{volume_ess}{---}
\setres{cty_Nicaragua}{CPT_uniform}{kappa}{-4.352}
\setres{cty_Nicaragua}{CPT_uniform}{kappa_prior_mean}{-1.582}
\setres{cty_Nicaragua}{CPT_uniform}{kappa_prior_mean_se}{0.346}
\setres{cty_Nicaragua}{CPT_uniform}{kappa_se}{0.725}
\setres{cty_Nicaragua}{CPT_uniform}{mse}{68.11}
\setres{cty_Nicaragua}{CPT_uniform}{mse_prior_mean}{50.75}
\setres{cty_Nicaragua}{CPT_uniform}{nrmse_pct}{38.9}
\setres{cty_Nicaragua}{CPT_uniform}{pred_variance}{17.37}
\setres{cty_Nicaragua}{CPT_uniform}{r_ff}{---}
\setres{cty_Nicaragua}{CPT_uniform}{r_ff_se}{---}
\setres{cty_Nicaragua}{CPT_uniform}{r_prior}{---}
\setres{cty_Nicaragua}{CPT_uniform}{r_prior_se}{---}
\setres{cty_Nicaragua}{CPT_uniform}{r_prob}{---}
\setres{cty_Nicaragua}{CPT_uniform}{r_prob_se}{---}
\setres{cty_Nicaragua}{CPT_uniform}{spread_penalty}{2.771}
\setres{cty_Nicaragua}{CPT_uniform}{spread_penalty_se}{0.445}
\setres{cty_Nicaragua}{CPT_uniform}{volume_ess}{---}
\setres{cty_Nicaragua}{DA_deterministic}{kappa}{0.667}
\setres{cty_Nicaragua}{DA_deterministic}{kappa_se}{0.115}
\setres{cty_Nicaragua}{DA_deterministic}{mse}{36.65}
\setres{cty_Nicaragua}{DA_deterministic}{nrmse_pct}{32.9}
\setres{cty_Nicaragua}{DA_deterministic}{r_ff}{---}
\setres{cty_Nicaragua}{DA_deterministic}{r_ff_se}{---}
\setres{cty_Nicaragua}{DA_deterministic}{r_prior}{---}
\setres{cty_Nicaragua}{DA_deterministic}{r_prior_se}{---}
\setres{cty_Nicaragua}{DA_deterministic}{r_prob}{---}
\setres{cty_Nicaragua}{DA_deterministic}{r_prob_se}{---}
\setres{cty_Nicaragua}{DA_deterministic}{theta_star}{2.15, 1.25}
\setres{cty_Nicaragua}{DA_deterministic}{volume_ess}{---}
\setres{cty_Nicaragua}{DA_lit}{kappa}{-3.319}
\setres{cty_Nicaragua}{DA_lit}{kappa_prior_mean}{-2.561}
\setres{cty_Nicaragua}{DA_lit}{kappa_prior_mean_se}{0.574}
\setres{cty_Nicaragua}{DA_lit}{kappa_se}{0.675}
\setres{cty_Nicaragua}{DA_lit}{mse}{61.63}
\setres{cty_Nicaragua}{DA_lit}{mse_prior_mean}{56.89}
\setres{cty_Nicaragua}{DA_lit}{nrmse_pct}{42.2}
\setres{cty_Nicaragua}{DA_lit}{pred_variance}{4.75}
\setres{cty_Nicaragua}{DA_lit}{r_ff}{---}
\setres{cty_Nicaragua}{DA_lit}{r_ff_se}{---}
\setres{cty_Nicaragua}{DA_lit}{r_prior}{---}
\setres{cty_Nicaragua}{DA_lit}{r_prior_se}{---}
\setres{cty_Nicaragua}{DA_lit}{r_prob}{---}
\setres{cty_Nicaragua}{DA_lit}{r_prob_se}{---}
\setres{cty_Nicaragua}{DA_lit}{spread_penalty}{0.758}
\setres{cty_Nicaragua}{DA_lit}{spread_penalty_se}{0.122}
\setres{cty_Nicaragua}{DA_lit}{volume_ess}{---}
\setres{cty_Nicaragua}{DA_transfer_others}{kappa}{0.355}
\setres{cty_Nicaragua}{DA_transfer_others}{kappa_prior_mean}{0.355}
\setres{cty_Nicaragua}{DA_transfer_others}{kappa_prior_mean_se}{0.175}
\setres{cty_Nicaragua}{DA_transfer_others}{kappa_se}{0.175}
\setres{cty_Nicaragua}{DA_transfer_others}{mse}{38.61}
\setres{cty_Nicaragua}{DA_transfer_others}{mse_prior_mean}{38.61}
\setres{cty_Nicaragua}{DA_transfer_others}{nrmse_pct}{33.5}
\setres{cty_Nicaragua}{DA_transfer_others}{pred_variance}{0.00}
\setres{cty_Nicaragua}{DA_transfer_others}{r_ff}{---}
\setres{cty_Nicaragua}{DA_transfer_others}{r_ff_se}{---}
\setres{cty_Nicaragua}{DA_transfer_others}{r_prior}{---}
\setres{cty_Nicaragua}{DA_transfer_others}{r_prior_se}{---}
\setres{cty_Nicaragua}{DA_transfer_others}{r_prob}{---}
\setres{cty_Nicaragua}{DA_transfer_others}{r_prob_se}{---}
\setres{cty_Nicaragua}{DA_transfer_others}{spread_penalty}{0.000}
\setres{cty_Nicaragua}{DA_transfer_others}{spread_penalty_se}{0.000}
\setres{cty_Nicaragua}{DA_transfer_others}{theta_star}{1.68, 1.46}
\setres{cty_Nicaragua}{DA_transfer_others}{transfer}{others}
\setres{cty_Nicaragua}{DA_transfer_others}{volume_ess}{---}
\setres{cty_Nicaragua}{DA_uniform}{kappa}{-6.411}
\setres{cty_Nicaragua}{DA_uniform}{kappa_prior_mean}{-4.234}
\setres{cty_Nicaragua}{DA_uniform}{kappa_prior_mean_se}{0.743}
\setres{cty_Nicaragua}{DA_uniform}{kappa_se}{1.054}
\setres{cty_Nicaragua}{DA_uniform}{mse}{81.02}
\setres{cty_Nicaragua}{DA_uniform}{mse_prior_mean}{67.37}
\setres{cty_Nicaragua}{DA_uniform}{nrmse_pct}{45.9}
\setres{cty_Nicaragua}{DA_uniform}{pred_variance}{13.65}
\setres{cty_Nicaragua}{DA_uniform}{r_ff}{---}
\setres{cty_Nicaragua}{DA_uniform}{r_ff_se}{---}
\setres{cty_Nicaragua}{DA_uniform}{r_prior}{---}
\setres{cty_Nicaragua}{DA_uniform}{r_prior_se}{---}
\setres{cty_Nicaragua}{DA_uniform}{r_prob}{---}
\setres{cty_Nicaragua}{DA_uniform}{r_prob_se}{---}
\setres{cty_Nicaragua}{DA_uniform}{spread_penalty}{2.178}
\setres{cty_Nicaragua}{DA_uniform}{spread_penalty_se}{0.350}
\setres{cty_Nicaragua}{DA_uniform}{volume_ess}{---}
\setres{cty_Nicaragua}{EU_deterministic}{kappa}{0.640}
\setres{cty_Nicaragua}{EU_deterministic}{kappa_se}{0.165}
\setres{cty_Nicaragua}{EU_deterministic}{mse}{36.83}
\setres{cty_Nicaragua}{EU_deterministic}{nrmse_pct}{33.2}
\setres{cty_Nicaragua}{EU_deterministic}{r_ff}{---}
\setres{cty_Nicaragua}{EU_deterministic}{r_ff_se}{---}
\setres{cty_Nicaragua}{EU_deterministic}{r_prior}{---}
\setres{cty_Nicaragua}{EU_deterministic}{r_prior_se}{---}
\setres{cty_Nicaragua}{EU_deterministic}{r_prob}{---}
\setres{cty_Nicaragua}{EU_deterministic}{r_prob_se}{---}
\setres{cty_Nicaragua}{EU_deterministic}{theta_star}{1.28}
\setres{cty_Nicaragua}{EU_deterministic}{volume_ess}{---}
\setres{cty_Nicaragua}{EU_lit}{kappa}{-0.956}
\setres{cty_Nicaragua}{EU_lit}{kappa_prior_mean}{-0.764}
\setres{cty_Nicaragua}{EU_lit}{kappa_prior_mean_se}{0.360}
\setres{cty_Nicaragua}{EU_lit}{kappa_se}{0.383}
\setres{cty_Nicaragua}{EU_lit}{mse}{46.83}
\setres{cty_Nicaragua}{EU_lit}{mse_prior_mean}{45.62}
\setres{cty_Nicaragua}{EU_lit}{nrmse_pct}{37.2}
\setres{cty_Nicaragua}{EU_lit}{pred_variance}{1.20}
\setres{cty_Nicaragua}{EU_lit}{r_ff}{---}
\setres{cty_Nicaragua}{EU_lit}{r_ff_se}{---}
\setres{cty_Nicaragua}{EU_lit}{r_prior}{---}
\setres{cty_Nicaragua}{EU_lit}{r_prior_se}{---}
\setres{cty_Nicaragua}{EU_lit}{r_prob}{---}
\setres{cty_Nicaragua}{EU_lit}{r_prob_se}{---}
\setres{cty_Nicaragua}{EU_lit}{spread_penalty}{0.192}
\setres{cty_Nicaragua}{EU_lit}{spread_penalty_se}{0.031}
\setres{cty_Nicaragua}{EU_lit}{volume_ess}{---}
\setres{cty_Nicaragua}{EU_transfer_others}{kappa}{0.300}
\setres{cty_Nicaragua}{EU_transfer_others}{kappa_prior_mean}{0.300}
\setres{cty_Nicaragua}{EU_transfer_others}{kappa_prior_mean_se}{0.209}
\setres{cty_Nicaragua}{EU_transfer_others}{kappa_se}{0.209}
\setres{cty_Nicaragua}{EU_transfer_others}{mse}{38.96}
\setres{cty_Nicaragua}{EU_transfer_others}{mse_prior_mean}{38.96}
\setres{cty_Nicaragua}{EU_transfer_others}{nrmse_pct}{34.0}
\setres{cty_Nicaragua}{EU_transfer_others}{pred_variance}{0.00}
\setres{cty_Nicaragua}{EU_transfer_others}{r_ff}{---}
\setres{cty_Nicaragua}{EU_transfer_others}{r_ff_se}{---}
\setres{cty_Nicaragua}{EU_transfer_others}{r_prior}{---}
\setres{cty_Nicaragua}{EU_transfer_others}{r_prior_se}{---}
\setres{cty_Nicaragua}{EU_transfer_others}{r_prob}{---}
\setres{cty_Nicaragua}{EU_transfer_others}{r_prob_se}{---}
\setres{cty_Nicaragua}{EU_transfer_others}{spread_penalty}{0.000}
\setres{cty_Nicaragua}{EU_transfer_others}{spread_penalty_se}{0.000}
\setres{cty_Nicaragua}{EU_transfer_others}{theta_star}{0.93}
\setres{cty_Nicaragua}{EU_transfer_others}{transfer}{others}
\setres{cty_Nicaragua}{EU_transfer_others}{volume_ess}{---}
\setres{cty_Nicaragua}{EU_uniform}{kappa}{-4.148}
\setres{cty_Nicaragua}{EU_uniform}{kappa_prior_mean}{-2.840}
\setres{cty_Nicaragua}{EU_uniform}{kappa_prior_mean_se}{0.599}
\setres{cty_Nicaragua}{EU_uniform}{kappa_se}{0.779}
\setres{cty_Nicaragua}{EU_uniform}{mse}{66.83}
\setres{cty_Nicaragua}{EU_uniform}{mse_prior_mean}{58.63}
\setres{cty_Nicaragua}{EU_uniform}{nrmse_pct}{42.6}
\setres{cty_Nicaragua}{EU_uniform}{pred_variance}{8.20}
\setres{cty_Nicaragua}{EU_uniform}{r_ff}{---}
\setres{cty_Nicaragua}{EU_uniform}{r_ff_se}{---}
\setres{cty_Nicaragua}{EU_uniform}{r_prior}{---}
\setres{cty_Nicaragua}{EU_uniform}{r_prior_se}{---}
\setres{cty_Nicaragua}{EU_uniform}{r_prob}{---}
\setres{cty_Nicaragua}{EU_uniform}{r_prob_se}{---}
\setres{cty_Nicaragua}{EU_uniform}{spread_penalty}{1.308}
\setres{cty_Nicaragua}{EU_uniform}{spread_penalty_se}{0.210}
\setres{cty_Nicaragua}{EU_uniform}{volume_ess}{---}
\setres{cty_Nicaragua}{EV_deterministic}{kappa}{0.444}
\setres{cty_Nicaragua}{EV_deterministic}{kappa_se}{0.199}
\setres{cty_Nicaragua}{EV_deterministic}{mse}{38.05}
\setres{cty_Nicaragua}{EV_deterministic}{nrmse_pct}{33.4}
\setres{cty_Nicaragua}{EV_deterministic}{r_ff}{---}
\setres{cty_Nicaragua}{EV_deterministic}{r_ff_se}{---}
\setres{cty_Nicaragua}{EV_deterministic}{r_prior}{---}
\setres{cty_Nicaragua}{EV_deterministic}{r_prior_se}{---}
\setres{cty_Nicaragua}{EV_deterministic}{r_prob}{---}
\setres{cty_Nicaragua}{EV_deterministic}{r_prob_se}{---}
\setres{cty_Nicaragua}{EV_deterministic}{volume_ess}{---}
\setres{cty_Nicaragua}{reference}{country_name}{Nicaragua}
\setres{cty_Nicaragua}{reference}{lookup_nrmse_pct}{32.8}
\setres{cty_Nicaragua}{reference}{midpoint_mse}{40.84}
\setres{cty_Nicaragua}{reference}{midpoint_nrmse_pct}{37.0}
\setres{cty_Nicaragua}{reference}{missing_lotteries}{none}
\setres{cty_Nicaragua}{reference}{n_lotteries}{14}
\setres{cty_Nicaragua}{reference}{n_obs}{1{,}680}
\setres{cty_Nicaragua}{reference}{n_subjects}{120}
\setres{cty_Nicaragua}{reference}{oracle_mse}{34.57}
\setres{cty_Nicaragua}{reference}{sample}{LV2019\_Nicaragua}
\setres{cty_Nigeria}{CPT_deterministic}{kappa}{0.790}
\setres{cty_Nigeria}{CPT_deterministic}{kappa_se}{0.178}
\setres{cty_Nigeria}{CPT_deterministic}{mse}{35.65}
\setres{cty_Nigeria}{CPT_deterministic}{nrmse_pct}{34.7}
\setres{cty_Nigeria}{CPT_deterministic}{r_ff}{---}
\setres{cty_Nigeria}{CPT_deterministic}{r_ff_se}{---}
\setres{cty_Nigeria}{CPT_deterministic}{r_prior}{---}
\setres{cty_Nigeria}{CPT_deterministic}{r_prior_se}{---}
\setres{cty_Nigeria}{CPT_deterministic}{r_prob}{---}
\setres{cty_Nigeria}{CPT_deterministic}{r_prob_se}{---}
\setres{cty_Nigeria}{CPT_deterministic}{theta_star}{1.10, 0.20, 1.00}
\setres{cty_Nigeria}{CPT_deterministic}{volume_ess}{---}
\setres{cty_Nigeria}{CPT_lit}{kappa}{-11.792}
\setres{cty_Nigeria}{CPT_lit}{kappa_prior_mean}{-8.974}
\setres{cty_Nigeria}{CPT_lit}{kappa_prior_mean_se}{4.867}
\setres{cty_Nigeria}{CPT_lit}{kappa_se}{6.163}
\setres{cty_Nigeria}{CPT_lit}{mse}{48.59}
\setres{cty_Nigeria}{CPT_lit}{mse_prior_mean}{45.69}
\setres{cty_Nigeria}{CPT_lit}{nrmse_pct}{36.0}
\setres{cty_Nigeria}{CPT_lit}{pred_variance}{2.90}
\setres{cty_Nigeria}{CPT_lit}{r_ff}{---}
\setres{cty_Nigeria}{CPT_lit}{r_ff_se}{---}
\setres{cty_Nigeria}{CPT_lit}{r_prior}{---}
\setres{cty_Nigeria}{CPT_lit}{r_prior_se}{---}
\setres{cty_Nigeria}{CPT_lit}{r_prob}{---}
\setres{cty_Nigeria}{CPT_lit}{r_prob_se}{---}
\setres{cty_Nigeria}{CPT_lit}{spread_penalty}{2.818}
\setres{cty_Nigeria}{CPT_lit}{spread_penalty_se}{1.321}
\setres{cty_Nigeria}{CPT_lit}{volume_ess}{---}
\setres{cty_Nigeria}{CPT_transfer_others}{kappa}{-3.982}
\setres{cty_Nigeria}{CPT_transfer_others}{kappa_prior_mean}{-3.982}
\setres{cty_Nigeria}{CPT_transfer_others}{kappa_prior_mean_se}{2.996}
\setres{cty_Nigeria}{CPT_transfer_others}{kappa_se}{2.996}
\setres{cty_Nigeria}{CPT_transfer_others}{mse}{40.56}
\setres{cty_Nigeria}{CPT_transfer_others}{mse_prior_mean}{40.56}
\setres{cty_Nigeria}{CPT_transfer_others}{nrmse_pct}{34.7}
\setres{cty_Nigeria}{CPT_transfer_others}{pred_variance}{0.00}
\setres{cty_Nigeria}{CPT_transfer_others}{r_ff}{---}
\setres{cty_Nigeria}{CPT_transfer_others}{r_ff_se}{---}
\setres{cty_Nigeria}{CPT_transfer_others}{r_prior}{---}
\setres{cty_Nigeria}{CPT_transfer_others}{r_prior_se}{---}
\setres{cty_Nigeria}{CPT_transfer_others}{r_prob}{---}
\setres{cty_Nigeria}{CPT_transfer_others}{r_prob_se}{---}
\setres{cty_Nigeria}{CPT_transfer_others}{spread_penalty}{0.000}
\setres{cty_Nigeria}{CPT_transfer_others}{spread_penalty_se}{0.000}
\setres{cty_Nigeria}{CPT_transfer_others}{theta_star}{1.10, 0.70, 0.80}
\setres{cty_Nigeria}{CPT_transfer_others}{transfer}{others}
\setres{cty_Nigeria}{CPT_transfer_others}{volume_ess}{---}
\setres{cty_Nigeria}{CPT_uniform}{kappa}{-28.796}
\setres{cty_Nigeria}{CPT_uniform}{kappa_prior_mean}{-11.913}
\setres{cty_Nigeria}{CPT_uniform}{kappa_prior_mean_se}{5.197}
\setres{cty_Nigeria}{CPT_uniform}{kappa_se}{13.025}
\setres{cty_Nigeria}{CPT_uniform}{mse}{66.08}
\setres{cty_Nigeria}{CPT_uniform}{mse_prior_mean}{48.72}
\setres{cty_Nigeria}{CPT_uniform}{nrmse_pct}{38.1}
\setres{cty_Nigeria}{CPT_uniform}{pred_variance}{17.37}
\setres{cty_Nigeria}{CPT_uniform}{r_ff}{---}
\setres{cty_Nigeria}{CPT_uniform}{r_ff_se}{---}
\setres{cty_Nigeria}{CPT_uniform}{r_prior}{---}
\setres{cty_Nigeria}{CPT_uniform}{r_prior_se}{---}
\setres{cty_Nigeria}{CPT_uniform}{r_prob}{---}
\setres{cty_Nigeria}{CPT_uniform}{r_prob_se}{---}
\setres{cty_Nigeria}{CPT_uniform}{spread_penalty}{16.882}
\setres{cty_Nigeria}{CPT_uniform}{spread_penalty_se}{7.913}
\setres{cty_Nigeria}{CPT_uniform}{volume_ess}{---}
\setres{cty_Nigeria}{DA_deterministic}{kappa}{-4.797}
\setres{cty_Nigeria}{DA_deterministic}{kappa_se}{3.586}
\setres{cty_Nigeria}{DA_deterministic}{mse}{41.40}
\setres{cty_Nigeria}{DA_deterministic}{nrmse_pct}{35.7}
\setres{cty_Nigeria}{DA_deterministic}{r_ff}{---}
\setres{cty_Nigeria}{DA_deterministic}{r_ff_se}{---}
\setres{cty_Nigeria}{DA_deterministic}{r_prior}{---}
\setres{cty_Nigeria}{DA_deterministic}{r_prior_se}{---}
\setres{cty_Nigeria}{DA_deterministic}{r_prob}{---}
\setres{cty_Nigeria}{DA_deterministic}{r_prob_se}{---}
\setres{cty_Nigeria}{DA_deterministic}{theta_star}{2.80, 3.00}
\setres{cty_Nigeria}{DA_deterministic}{volume_ess}{---}
\setres{cty_Nigeria}{DA_lit}{kappa}{-26.492}
\setres{cty_Nigeria}{DA_lit}{kappa_prior_mean}{-21.875}
\setres{cty_Nigeria}{DA_lit}{kappa_prior_mean_se}{10.421}
\setres{cty_Nigeria}{DA_lit}{kappa_se}{12.563}
\setres{cty_Nigeria}{DA_lit}{mse}{63.71}
\setres{cty_Nigeria}{DA_lit}{mse_prior_mean}{58.96}
\setres{cty_Nigeria}{DA_lit}{nrmse_pct}{39.7}
\setres{cty_Nigeria}{DA_lit}{pred_variance}{4.75}
\setres{cty_Nigeria}{DA_lit}{r_ff}{---}
\setres{cty_Nigeria}{DA_lit}{r_ff_se}{---}
\setres{cty_Nigeria}{DA_lit}{r_prior}{---}
\setres{cty_Nigeria}{DA_lit}{r_prior_se}{---}
\setres{cty_Nigeria}{DA_lit}{r_prob}{---}
\setres{cty_Nigeria}{DA_lit}{r_prob_se}{---}
\setres{cty_Nigeria}{DA_lit}{spread_penalty}{4.616}
\setres{cty_Nigeria}{DA_lit}{spread_penalty_se}{2.164}
\setres{cty_Nigeria}{DA_lit}{volume_ess}{---}
\setres{cty_Nigeria}{DA_transfer_others}{kappa}{-7.243}
\setres{cty_Nigeria}{DA_transfer_others}{kappa_prior_mean}{-7.243}
\setres{cty_Nigeria}{DA_transfer_others}{kappa_prior_mean_se}{4.723}
\setres{cty_Nigeria}{DA_transfer_others}{kappa_se}{4.723}
\setres{cty_Nigeria}{DA_transfer_others}{mse}{43.91}
\setres{cty_Nigeria}{DA_transfer_others}{mse_prior_mean}{43.91}
\setres{cty_Nigeria}{DA_transfer_others}{nrmse_pct}{36.0}
\setres{cty_Nigeria}{DA_transfer_others}{pred_variance}{0.00}
\setres{cty_Nigeria}{DA_transfer_others}{r_ff}{---}
\setres{cty_Nigeria}{DA_transfer_others}{r_ff_se}{---}
\setres{cty_Nigeria}{DA_transfer_others}{r_prior}{---}
\setres{cty_Nigeria}{DA_transfer_others}{r_prior_se}{---}
\setres{cty_Nigeria}{DA_transfer_others}{r_prob}{---}
\setres{cty_Nigeria}{DA_transfer_others}{r_prob_se}{---}
\setres{cty_Nigeria}{DA_transfer_others}{spread_penalty}{0.000}
\setres{cty_Nigeria}{DA_transfer_others}{spread_penalty_se}{0.000}
\setres{cty_Nigeria}{DA_transfer_others}{theta_star}{1.57, 1.21}
\setres{cty_Nigeria}{DA_transfer_others}{transfer}{others}
\setres{cty_Nigeria}{DA_transfer_others}{volume_ess}{---}
\setres{cty_Nigeria}{DA_uniform}{kappa}{-41.616}
\setres{cty_Nigeria}{DA_uniform}{kappa_prior_mean}{-28.349}
\setres{cty_Nigeria}{DA_uniform}{kappa_prior_mean_se}{12.569}
\setres{cty_Nigeria}{DA_uniform}{kappa_se}{18.739}
\setres{cty_Nigeria}{DA_uniform}{mse}{79.27}
\setres{cty_Nigeria}{DA_uniform}{mse_prior_mean}{65.62}
\setres{cty_Nigeria}{DA_uniform}{nrmse_pct}{43.5}
\setres{cty_Nigeria}{DA_uniform}{pred_variance}{13.65}
\setres{cty_Nigeria}{DA_uniform}{r_ff}{---}
\setres{cty_Nigeria}{DA_uniform}{r_ff_se}{---}
\setres{cty_Nigeria}{DA_uniform}{r_prior}{---}
\setres{cty_Nigeria}{DA_uniform}{r_prior_se}{---}
\setres{cty_Nigeria}{DA_uniform}{r_prob}{---}
\setres{cty_Nigeria}{DA_uniform}{r_prob_se}{---}
\setres{cty_Nigeria}{DA_uniform}{spread_penalty}{13.268}
\setres{cty_Nigeria}{DA_uniform}{spread_penalty_se}{6.219}
\setres{cty_Nigeria}{DA_uniform}{volume_ess}{---}
\setres{cty_Nigeria}{EU_deterministic}{kappa}{-8.104}
\setres{cty_Nigeria}{EU_deterministic}{kappa_se}{5.421}
\setres{cty_Nigeria}{EU_deterministic}{mse}{44.80}
\setres{cty_Nigeria}{EU_deterministic}{nrmse_pct}{36.3}
\setres{cty_Nigeria}{EU_deterministic}{r_ff}{---}
\setres{cty_Nigeria}{EU_deterministic}{r_ff_se}{---}
\setres{cty_Nigeria}{EU_deterministic}{r_prior}{---}
\setres{cty_Nigeria}{EU_deterministic}{r_prior_se}{---}
\setres{cty_Nigeria}{EU_deterministic}{r_prob}{---}
\setres{cty_Nigeria}{EU_deterministic}{r_prob_se}{---}
\setres{cty_Nigeria}{EU_deterministic}{theta_star}{1.22}
\setres{cty_Nigeria}{EU_deterministic}{volume_ess}{---}
\setres{cty_Nigeria}{EU_lit}{kappa}{-15.268}
\setres{cty_Nigeria}{EU_lit}{kappa_prior_mean}{-14.097}
\setres{cty_Nigeria}{EU_lit}{kappa_prior_mean_se}{7.593}
\setres{cty_Nigeria}{EU_lit}{kappa_se}{8.135}
\setres{cty_Nigeria}{EU_lit}{mse}{52.17}
\setres{cty_Nigeria}{EU_lit}{mse_prior_mean}{50.96}
\setres{cty_Nigeria}{EU_lit}{nrmse_pct}{36.0}
\setres{cty_Nigeria}{EU_lit}{pred_variance}{1.20}
\setres{cty_Nigeria}{EU_lit}{r_ff}{---}
\setres{cty_Nigeria}{EU_lit}{r_ff_se}{---}
\setres{cty_Nigeria}{EU_lit}{r_prior}{---}
\setres{cty_Nigeria}{EU_lit}{r_prior_se}{---}
\setres{cty_Nigeria}{EU_lit}{r_prob}{---}
\setres{cty_Nigeria}{EU_lit}{r_prob_se}{---}
\setres{cty_Nigeria}{EU_lit}{spread_penalty}{1.171}
\setres{cty_Nigeria}{EU_lit}{spread_penalty_se}{0.549}
\setres{cty_Nigeria}{EU_lit}{volume_ess}{---}
\setres{cty_Nigeria}{EU_transfer_others}{kappa}{-9.358}
\setres{cty_Nigeria}{EU_transfer_others}{kappa_prior_mean}{-9.358}
\setres{cty_Nigeria}{EU_transfer_others}{kappa_prior_mean_se}{5.937}
\setres{cty_Nigeria}{EU_transfer_others}{kappa_se}{5.937}
\setres{cty_Nigeria}{EU_transfer_others}{mse}{46.09}
\setres{cty_Nigeria}{EU_transfer_others}{mse_prior_mean}{46.09}
\setres{cty_Nigeria}{EU_transfer_others}{nrmse_pct}{35.4}
\setres{cty_Nigeria}{EU_transfer_others}{pred_variance}{0.00}
\setres{cty_Nigeria}{EU_transfer_others}{r_ff}{---}
\setres{cty_Nigeria}{EU_transfer_others}{r_ff_se}{---}
\setres{cty_Nigeria}{EU_transfer_others}{r_prior}{---}
\setres{cty_Nigeria}{EU_transfer_others}{r_prior_se}{---}
\setres{cty_Nigeria}{EU_transfer_others}{r_prob}{---}
\setres{cty_Nigeria}{EU_transfer_others}{r_prob_se}{---}
\setres{cty_Nigeria}{EU_transfer_others}{spread_penalty}{0.000}
\setres{cty_Nigeria}{EU_transfer_others}{spread_penalty_se}{0.000}
\setres{cty_Nigeria}{EU_transfer_others}{theta_star}{0.93}
\setres{cty_Nigeria}{EU_transfer_others}{transfer}{others}
\setres{cty_Nigeria}{EU_transfer_others}{volume_ess}{---}
\setres{cty_Nigeria}{EU_uniform}{kappa}{-30.723}
\setres{cty_Nigeria}{EU_uniform}{kappa_prior_mean}{-22.755}
\setres{cty_Nigeria}{EU_uniform}{kappa_prior_mean_se}{10.644}
\setres{cty_Nigeria}{EU_uniform}{kappa_se}{14.345}
\setres{cty_Nigeria}{EU_uniform}{mse}{68.06}
\setres{cty_Nigeria}{EU_uniform}{mse_prior_mean}{59.87}
\setres{cty_Nigeria}{EU_uniform}{nrmse_pct}{40.8}
\setres{cty_Nigeria}{EU_uniform}{pred_variance}{8.20}
\setres{cty_Nigeria}{EU_uniform}{r_ff}{---}
\setres{cty_Nigeria}{EU_uniform}{r_ff_se}{---}
\setres{cty_Nigeria}{EU_uniform}{r_prior}{---}
\setres{cty_Nigeria}{EU_uniform}{r_prior_se}{---}
\setres{cty_Nigeria}{EU_uniform}{r_prob}{---}
\setres{cty_Nigeria}{EU_uniform}{r_prob_se}{---}
\setres{cty_Nigeria}{EU_uniform}{spread_penalty}{7.969}
\setres{cty_Nigeria}{EU_uniform}{spread_penalty_se}{3.735}
\setres{cty_Nigeria}{EU_uniform}{volume_ess}{---}
\setres{cty_Nigeria}{EV_deterministic}{kappa}{-8.694}
\setres{cty_Nigeria}{EV_deterministic}{kappa_se}{5.509}
\setres{cty_Nigeria}{EV_deterministic}{mse}{45.41}
\setres{cty_Nigeria}{EV_deterministic}{nrmse_pct}{35.3}
\setres{cty_Nigeria}{EV_deterministic}{r_ff}{---}
\setres{cty_Nigeria}{EV_deterministic}{r_ff_se}{---}
\setres{cty_Nigeria}{EV_deterministic}{r_prior}{---}
\setres{cty_Nigeria}{EV_deterministic}{r_prior_se}{---}
\setres{cty_Nigeria}{EV_deterministic}{r_prob}{---}
\setres{cty_Nigeria}{EV_deterministic}{r_prob_se}{---}
\setres{cty_Nigeria}{EV_deterministic}{volume_ess}{---}
\setres{cty_Nigeria}{reference}{country_name}{Nigeria}
\setres{cty_Nigeria}{reference}{lookup_nrmse_pct}{34.7}
\setres{cty_Nigeria}{reference}{midpoint_mse}{36.46}
\setres{cty_Nigeria}{reference}{midpoint_nrmse_pct}{35.0}
\setres{cty_Nigeria}{reference}{missing_lotteries}{none}
\setres{cty_Nigeria}{reference}{n_lotteries}{14}
\setres{cty_Nigeria}{reference}{n_obs}{2{,}828}
\setres{cty_Nigeria}{reference}{n_subjects}{202}
\setres{cty_Nigeria}{reference}{oracle_mse}{35.43}
\setres{cty_Nigeria}{reference}{sample}{LV2019\_Nigeria}
\setres{cty_Peru}{CPT_deterministic}{kappa}{0.948}
\setres{cty_Peru}{CPT_deterministic}{kappa_se}{0.032}
\setres{cty_Peru}{CPT_deterministic}{mse}{23.47}
\setres{cty_Peru}{CPT_deterministic}{nrmse_pct}{27.4}
\setres{cty_Peru}{CPT_deterministic}{r_ff}{---}
\setres{cty_Peru}{CPT_deterministic}{r_ff_se}{---}
\setres{cty_Peru}{CPT_deterministic}{r_prior}{---}
\setres{cty_Peru}{CPT_deterministic}{r_prior_se}{---}
\setres{cty_Peru}{CPT_deterministic}{r_prob}{---}
\setres{cty_Peru}{CPT_deterministic}{r_prob_se}{---}
\setres{cty_Peru}{CPT_deterministic}{theta_star}{1.00, 0.50, 1.10}
\setres{cty_Peru}{CPT_deterministic}{volume_ess}{---}
\setres{cty_Peru}{CPT_lit}{kappa}{-0.701}
\setres{cty_Peru}{CPT_lit}{kappa_prior_mean}{-0.155}
\setres{cty_Peru}{CPT_lit}{kappa_prior_mean_se}{0.341}
\setres{cty_Peru}{CPT_lit}{kappa_se}{0.408}
\setres{cty_Peru}{CPT_lit}{mse}{32.24}
\setres{cty_Peru}{CPT_lit}{mse_prior_mean}{29.34}
\setres{cty_Peru}{CPT_lit}{nrmse_pct}{29.8}
\setres{cty_Peru}{CPT_lit}{pred_variance}{2.90}
\setres{cty_Peru}{CPT_lit}{r_ff}{---}
\setres{cty_Peru}{CPT_lit}{r_ff_se}{---}
\setres{cty_Peru}{CPT_lit}{r_prior}{---}
\setres{cty_Peru}{CPT_lit}{r_prior_se}{---}
\setres{cty_Peru}{CPT_lit}{r_prob}{---}
\setres{cty_Peru}{CPT_lit}{r_prob_se}{---}
\setres{cty_Peru}{CPT_lit}{spread_penalty}{0.545}
\setres{cty_Peru}{CPT_lit}{spread_penalty_se}{0.100}
\setres{cty_Peru}{CPT_lit}{volume_ess}{---}
\setres{cty_Peru}{CPT_transfer_others}{kappa}{0.806}
\setres{cty_Peru}{CPT_transfer_others}{kappa_prior_mean}{0.806}
\setres{cty_Peru}{CPT_transfer_others}{kappa_prior_mean_se}{0.130}
\setres{cty_Peru}{CPT_transfer_others}{kappa_se}{0.130}
\setres{cty_Peru}{CPT_transfer_others}{mse}{24.23}
\setres{cty_Peru}{CPT_transfer_others}{mse_prior_mean}{24.23}
\setres{cty_Peru}{CPT_transfer_others}{nrmse_pct}{27.8}
\setres{cty_Peru}{CPT_transfer_others}{pred_variance}{0.00}
\setres{cty_Peru}{CPT_transfer_others}{r_ff}{---}
\setres{cty_Peru}{CPT_transfer_others}{r_ff_se}{---}
\setres{cty_Peru}{CPT_transfer_others}{r_prior}{---}
\setres{cty_Peru}{CPT_transfer_others}{r_prior_se}{---}
\setres{cty_Peru}{CPT_transfer_others}{r_prob}{---}
\setres{cty_Peru}{CPT_transfer_others}{r_prob_se}{---}
\setres{cty_Peru}{CPT_transfer_others}{spread_penalty}{0.000}
\setres{cty_Peru}{CPT_transfer_others}{spread_penalty_se}{0.000}
\setres{cty_Peru}{CPT_transfer_others}{theta_star}{1.00, 0.60, 0.90}
\setres{cty_Peru}{CPT_transfer_others}{transfer}{others}
\setres{cty_Peru}{CPT_transfer_others}{volume_ess}{---}
\setres{cty_Peru}{CPT_uniform}{kappa}{-4.678}
\setres{cty_Peru}{CPT_uniform}{kappa_prior_mean}{-1.411}
\setres{cty_Peru}{CPT_uniform}{kappa_prior_mean_se}{0.461}
\setres{cty_Peru}{CPT_uniform}{kappa_se}{0.956}
\setres{cty_Peru}{CPT_uniform}{mse}{53.38}
\setres{cty_Peru}{CPT_uniform}{mse_prior_mean}{36.02}
\setres{cty_Peru}{CPT_uniform}{nrmse_pct}{33.6}
\setres{cty_Peru}{CPT_uniform}{pred_variance}{17.37}
\setres{cty_Peru}{CPT_uniform}{r_ff}{---}
\setres{cty_Peru}{CPT_uniform}{r_ff_se}{---}
\setres{cty_Peru}{CPT_uniform}{r_prior}{---}
\setres{cty_Peru}{CPT_uniform}{r_prior_se}{---}
\setres{cty_Peru}{CPT_uniform}{r_prob}{---}
\setres{cty_Peru}{CPT_uniform}{r_prob_se}{---}
\setres{cty_Peru}{CPT_uniform}{spread_penalty}{3.266}
\setres{cty_Peru}{CPT_uniform}{spread_penalty_se}{0.600}
\setres{cty_Peru}{CPT_uniform}{volume_ess}{---}
\setres{cty_Peru}{DA_deterministic}{kappa}{0.526}
\setres{cty_Peru}{DA_deterministic}{kappa_se}{0.175}
\setres{cty_Peru}{DA_deterministic}{mse}{25.71}
\setres{cty_Peru}{DA_deterministic}{nrmse_pct}{28.6}
\setres{cty_Peru}{DA_deterministic}{r_ff}{---}
\setres{cty_Peru}{DA_deterministic}{r_ff_se}{---}
\setres{cty_Peru}{DA_deterministic}{r_prior}{---}
\setres{cty_Peru}{DA_deterministic}{r_prior_se}{---}
\setres{cty_Peru}{DA_deterministic}{r_prob}{---}
\setres{cty_Peru}{DA_deterministic}{r_prob_se}{---}
\setres{cty_Peru}{DA_deterministic}{theta_star}{2.41, 2.14}
\setres{cty_Peru}{DA_deterministic}{volume_ess}{---}
\setres{cty_Peru}{DA_lit}{kappa}{-3.375}
\setres{cty_Peru}{DA_lit}{kappa_prior_mean}{-2.482}
\setres{cty_Peru}{DA_lit}{kappa_prior_mean_se}{0.762}
\setres{cty_Peru}{DA_lit}{kappa_se}{0.897}
\setres{cty_Peru}{DA_lit}{mse}{46.46}
\setres{cty_Peru}{DA_lit}{mse_prior_mean}{41.71}
\setres{cty_Peru}{DA_lit}{nrmse_pct}{35.8}
\setres{cty_Peru}{DA_lit}{pred_variance}{4.75}
\setres{cty_Peru}{DA_lit}{r_ff}{---}
\setres{cty_Peru}{DA_lit}{r_ff_se}{---}
\setres{cty_Peru}{DA_lit}{r_prior}{---}
\setres{cty_Peru}{DA_lit}{r_prior_se}{---}
\setres{cty_Peru}{DA_lit}{r_prob}{---}
\setres{cty_Peru}{DA_lit}{r_prob_se}{---}
\setres{cty_Peru}{DA_lit}{spread_penalty}{0.893}
\setres{cty_Peru}{DA_lit}{spread_penalty_se}{0.164}
\setres{cty_Peru}{DA_lit}{volume_ess}{---}
\setres{cty_Peru}{DA_transfer_others}{kappa}{0.370}
\setres{cty_Peru}{DA_transfer_others}{kappa_prior_mean}{0.370}
\setres{cty_Peru}{DA_transfer_others}{kappa_prior_mean_se}{0.236}
\setres{cty_Peru}{DA_transfer_others}{kappa_se}{0.236}
\setres{cty_Peru}{DA_transfer_others}{mse}{26.54}
\setres{cty_Peru}{DA_transfer_others}{mse_prior_mean}{26.54}
\setres{cty_Peru}{DA_transfer_others}{nrmse_pct}{28.8}
\setres{cty_Peru}{DA_transfer_others}{pred_variance}{0.00}
\setres{cty_Peru}{DA_transfer_others}{r_ff}{---}
\setres{cty_Peru}{DA_transfer_others}{r_ff_se}{---}
\setres{cty_Peru}{DA_transfer_others}{r_prior}{---}
\setres{cty_Peru}{DA_transfer_others}{r_prior_se}{---}
\setres{cty_Peru}{DA_transfer_others}{r_prob}{---}
\setres{cty_Peru}{DA_transfer_others}{r_prob_se}{---}
\setres{cty_Peru}{DA_transfer_others}{spread_penalty}{0.000}
\setres{cty_Peru}{DA_transfer_others}{spread_penalty_se}{0.000}
\setres{cty_Peru}{DA_transfer_others}{theta_star}{1.68, 1.44}
\setres{cty_Peru}{DA_transfer_others}{transfer}{others}
\setres{cty_Peru}{DA_transfer_others}{volume_ess}{---}
\setres{cty_Peru}{DA_uniform}{kappa}{-6.791}
\setres{cty_Peru}{DA_uniform}{kappa_prior_mean}{-4.224}
\setres{cty_Peru}{DA_uniform}{kappa_prior_mean_se}{0.985}
\setres{cty_Peru}{DA_uniform}{kappa_se}{1.396}
\setres{cty_Peru}{DA_uniform}{mse}{64.62}
\setres{cty_Peru}{DA_uniform}{mse_prior_mean}{50.97}
\setres{cty_Peru}{DA_uniform}{nrmse_pct}{39.7}
\setres{cty_Peru}{DA_uniform}{pred_variance}{13.65}
\setres{cty_Peru}{DA_uniform}{r_ff}{---}
\setres{cty_Peru}{DA_uniform}{r_ff_se}{---}
\setres{cty_Peru}{DA_uniform}{r_prior}{---}
\setres{cty_Peru}{DA_uniform}{r_prior_se}{---}
\setres{cty_Peru}{DA_uniform}{r_prob}{---}
\setres{cty_Peru}{DA_uniform}{r_prob_se}{---}
\setres{cty_Peru}{DA_uniform}{spread_penalty}{2.567}
\setres{cty_Peru}{DA_uniform}{spread_penalty_se}{0.472}
\setres{cty_Peru}{DA_uniform}{volume_ess}{---}
\setres{cty_Peru}{EU_deterministic}{kappa}{0.404}
\setres{cty_Peru}{EU_deterministic}{kappa_se}{0.264}
\setres{cty_Peru}{EU_deterministic}{mse}{26.37}
\setres{cty_Peru}{EU_deterministic}{nrmse_pct}{28.8}
\setres{cty_Peru}{EU_deterministic}{r_ff}{---}
\setres{cty_Peru}{EU_deterministic}{r_ff_se}{---}
\setres{cty_Peru}{EU_deterministic}{r_prior}{---}
\setres{cty_Peru}{EU_deterministic}{r_prior_se}{---}
\setres{cty_Peru}{EU_deterministic}{r_prob}{---}
\setres{cty_Peru}{EU_deterministic}{r_prob_se}{---}
\setres{cty_Peru}{EU_deterministic}{theta_star}{1.17}
\setres{cty_Peru}{EU_deterministic}{volume_ess}{---}
\setres{cty_Peru}{EU_lit}{kappa}{-0.938}
\setres{cty_Peru}{EU_lit}{kappa_prior_mean}{-0.712}
\setres{cty_Peru}{EU_lit}{kappa_prior_mean_se}{0.480}
\setres{cty_Peru}{EU_lit}{kappa_se}{0.513}
\setres{cty_Peru}{EU_lit}{mse}{33.50}
\setres{cty_Peru}{EU_lit}{mse_prior_mean}{32.30}
\setres{cty_Peru}{EU_lit}{nrmse_pct}{30.8}
\setres{cty_Peru}{EU_lit}{pred_variance}{1.20}
\setres{cty_Peru}{EU_lit}{r_ff}{---}
\setres{cty_Peru}{EU_lit}{r_ff_se}{---}
\setres{cty_Peru}{EU_lit}{r_prior}{---}
\setres{cty_Peru}{EU_lit}{r_prior_se}{---}
\setres{cty_Peru}{EU_lit}{r_prob}{---}
\setres{cty_Peru}{EU_lit}{r_prob_se}{---}
\setres{cty_Peru}{EU_lit}{spread_penalty}{0.226}
\setres{cty_Peru}{EU_lit}{spread_penalty_se}{0.042}
\setres{cty_Peru}{EU_lit}{volume_ess}{---}
\setres{cty_Peru}{EU_transfer_others}{kappa}{0.255}
\setres{cty_Peru}{EU_transfer_others}{kappa_prior_mean}{0.255}
\setres{cty_Peru}{EU_transfer_others}{kappa_prior_mean_se}{0.287}
\setres{cty_Peru}{EU_transfer_others}{kappa_se}{0.287}
\setres{cty_Peru}{EU_transfer_others}{mse}{27.16}
\setres{cty_Peru}{EU_transfer_others}{mse_prior_mean}{27.16}
\setres{cty_Peru}{EU_transfer_others}{nrmse_pct}{28.9}
\setres{cty_Peru}{EU_transfer_others}{pred_variance}{0.00}
\setres{cty_Peru}{EU_transfer_others}{r_ff}{---}
\setres{cty_Peru}{EU_transfer_others}{r_ff_se}{---}
\setres{cty_Peru}{EU_transfer_others}{r_prior}{---}
\setres{cty_Peru}{EU_transfer_others}{r_prior_se}{---}
\setres{cty_Peru}{EU_transfer_others}{r_prob}{---}
\setres{cty_Peru}{EU_transfer_others}{r_prob_se}{---}
\setres{cty_Peru}{EU_transfer_others}{spread_penalty}{0.000}
\setres{cty_Peru}{EU_transfer_others}{spread_penalty_se}{0.000}
\setres{cty_Peru}{EU_transfer_others}{theta_star}{0.94}
\setres{cty_Peru}{EU_transfer_others}{transfer}{others}
\setres{cty_Peru}{EU_transfer_others}{volume_ess}{---}
\setres{cty_Peru}{EU_uniform}{kappa}{-4.285}
\setres{cty_Peru}{EU_uniform}{kappa_prior_mean}{-2.743}
\setres{cty_Peru}{EU_uniform}{kappa_prior_mean_se}{0.788}
\setres{cty_Peru}{EU_uniform}{kappa_se}{1.027}
\setres{cty_Peru}{EU_uniform}{mse}{51.29}
\setres{cty_Peru}{EU_uniform}{mse_prior_mean}{43.10}
\setres{cty_Peru}{EU_uniform}{nrmse_pct}{36.2}
\setres{cty_Peru}{EU_uniform}{pred_variance}{8.20}
\setres{cty_Peru}{EU_uniform}{r_ff}{---}
\setres{cty_Peru}{EU_uniform}{r_ff_se}{---}
\setres{cty_Peru}{EU_uniform}{r_prior}{---}
\setres{cty_Peru}{EU_uniform}{r_prior_se}{---}
\setres{cty_Peru}{EU_uniform}{r_prob}{---}
\setres{cty_Peru}{EU_uniform}{r_prob_se}{---}
\setres{cty_Peru}{EU_uniform}{spread_penalty}{1.542}
\setres{cty_Peru}{EU_uniform}{spread_penalty_se}{0.283}
\setres{cty_Peru}{EU_uniform}{volume_ess}{---}
\setres{cty_Peru}{EV_deterministic}{kappa}{0.355}
\setres{cty_Peru}{EV_deterministic}{kappa_se}{0.281}
\setres{cty_Peru}{EV_deterministic}{mse}{26.62}
\setres{cty_Peru}{EV_deterministic}{nrmse_pct}{29.2}
\setres{cty_Peru}{EV_deterministic}{r_ff}{---}
\setres{cty_Peru}{EV_deterministic}{r_ff_se}{---}
\setres{cty_Peru}{EV_deterministic}{r_prior}{---}
\setres{cty_Peru}{EV_deterministic}{r_prior_se}{---}
\setres{cty_Peru}{EV_deterministic}{r_prob}{---}
\setres{cty_Peru}{EV_deterministic}{r_prob_se}{---}
\setres{cty_Peru}{EV_deterministic}{volume_ess}{---}
\setres{cty_Peru}{reference}{country_name}{Peru}
\setres{cty_Peru}{reference}{lookup_nrmse_pct}{27.3}
\setres{cty_Peru}{reference}{midpoint_mse}{28.51}
\setres{cty_Peru}{reference}{midpoint_nrmse_pct}{28.6}
\setres{cty_Peru}{reference}{missing_lotteries}{none}
\setres{cty_Peru}{reference}{n_lotteries}{14}
\setres{cty_Peru}{reference}{n_obs}{1{,}330}
\setres{cty_Peru}{reference}{n_subjects}{95}
\setres{cty_Peru}{reference}{oracle_mse}{23.20}
\setres{cty_Peru}{reference}{sample}{LV2019\_Peru}
\setres{cty_Poland}{CPT_deterministic}{kappa}{0.984}
\setres{cty_Poland}{CPT_deterministic}{kappa_se}{0.013}
\setres{cty_Poland}{CPT_deterministic}{mse}{10.19}
\setres{cty_Poland}{CPT_deterministic}{nrmse_pct}{17.3}
\setres{cty_Poland}{CPT_deterministic}{r_ff}{---}
\setres{cty_Poland}{CPT_deterministic}{r_ff_se}{---}
\setres{cty_Poland}{CPT_deterministic}{r_prior}{---}
\setres{cty_Poland}{CPT_deterministic}{r_prior_se}{---}
\setres{cty_Poland}{CPT_deterministic}{r_prob}{---}
\setres{cty_Poland}{CPT_deterministic}{r_prob_se}{---}
\setres{cty_Poland}{CPT_deterministic}{theta_star}{1.10, 0.70, 0.70}
\setres{cty_Poland}{CPT_deterministic}{volume_ess}{---}
\setres{cty_Poland}{CPT_lit}{kappa}{0.592}
\setres{cty_Poland}{CPT_lit}{kappa_prior_mean}{0.881}
\setres{cty_Poland}{CPT_lit}{kappa_prior_mean_se}{0.042}
\setres{cty_Poland}{CPT_lit}{kappa_se}{0.061}
\setres{cty_Poland}{CPT_lit}{mse}{14.12}
\setres{cty_Poland}{CPT_lit}{mse_prior_mean}{11.22}
\setres{cty_Poland}{CPT_lit}{nrmse_pct}{19.8}
\setres{cty_Poland}{CPT_lit}{pred_variance}{2.90}
\setres{cty_Poland}{CPT_lit}{r_ff}{---}
\setres{cty_Poland}{CPT_lit}{r_ff_se}{---}
\setres{cty_Poland}{CPT_lit}{r_prior}{---}
\setres{cty_Poland}{CPT_lit}{r_prior_se}{---}
\setres{cty_Poland}{CPT_lit}{r_prob}{---}
\setres{cty_Poland}{CPT_lit}{r_prob_se}{---}
\setres{cty_Poland}{CPT_lit}{spread_penalty}{0.289}
\setres{cty_Poland}{CPT_lit}{spread_penalty_se}{0.025}
\setres{cty_Poland}{CPT_lit}{volume_ess}{---}
\setres{cty_Poland}{CPT_transfer_others}{kappa}{0.962}
\setres{cty_Poland}{CPT_transfer_others}{kappa_prior_mean}{0.962}
\setres{cty_Poland}{CPT_transfer_others}{kappa_prior_mean_se}{0.020}
\setres{cty_Poland}{CPT_transfer_others}{kappa_se}{0.020}
\setres{cty_Poland}{CPT_transfer_others}{mse}{10.41}
\setres{cty_Poland}{CPT_transfer_others}{mse_prior_mean}{10.41}
\setres{cty_Poland}{CPT_transfer_others}{nrmse_pct}{16.7}
\setres{cty_Poland}{CPT_transfer_others}{pred_variance}{0.00}
\setres{cty_Poland}{CPT_transfer_others}{r_ff}{---}
\setres{cty_Poland}{CPT_transfer_others}{r_ff_se}{---}
\setres{cty_Poland}{CPT_transfer_others}{r_prior}{---}
\setres{cty_Poland}{CPT_transfer_others}{r_prior_se}{---}
\setres{cty_Poland}{CPT_transfer_others}{r_prob}{---}
\setres{cty_Poland}{CPT_transfer_others}{r_prob_se}{---}
\setres{cty_Poland}{CPT_transfer_others}{spread_penalty}{0.000}
\setres{cty_Poland}{CPT_transfer_others}{spread_penalty_se}{0.000}
\setres{cty_Poland}{CPT_transfer_others}{theta_star}{1.00, 0.60, 0.90}
\setres{cty_Poland}{CPT_transfer_others}{transfer}{others}
\setres{cty_Poland}{CPT_transfer_others}{volume_ess}{---}
\setres{cty_Poland}{CPT_uniform}{kappa}{-1.404}
\setres{cty_Poland}{CPT_uniform}{kappa_prior_mean}{0.325}
\setres{cty_Poland}{CPT_uniform}{kappa_prior_mean_se}{0.106}
\setres{cty_Poland}{CPT_uniform}{kappa_se}{0.219}
\setres{cty_Poland}{CPT_uniform}{mse}{34.18}
\setres{cty_Poland}{CPT_uniform}{mse_prior_mean}{16.81}
\setres{cty_Poland}{CPT_uniform}{nrmse_pct}{22.4}
\setres{cty_Poland}{CPT_uniform}{pred_variance}{17.37}
\setres{cty_Poland}{CPT_uniform}{r_ff}{---}
\setres{cty_Poland}{CPT_uniform}{r_ff_se}{---}
\setres{cty_Poland}{CPT_uniform}{r_prior}{---}
\setres{cty_Poland}{CPT_uniform}{r_prior_se}{---}
\setres{cty_Poland}{CPT_uniform}{r_prob}{---}
\setres{cty_Poland}{CPT_uniform}{r_prob_se}{---}
\setres{cty_Poland}{CPT_uniform}{spread_penalty}{1.729}
\setres{cty_Poland}{CPT_uniform}{spread_penalty_se}{0.150}
\setres{cty_Poland}{CPT_uniform}{volume_ess}{---}
\setres{cty_Poland}{DA_deterministic}{kappa}{0.917}
\setres{cty_Poland}{DA_deterministic}{kappa_se}{0.028}
\setres{cty_Poland}{DA_deterministic}{mse}{10.86}
\setres{cty_Poland}{DA_deterministic}{nrmse_pct}{19.0}
\setres{cty_Poland}{DA_deterministic}{r_ff}{---}
\setres{cty_Poland}{DA_deterministic}{r_ff_se}{---}
\setres{cty_Poland}{DA_deterministic}{r_prior}{---}
\setres{cty_Poland}{DA_deterministic}{r_prior_se}{---}
\setres{cty_Poland}{DA_deterministic}{r_prob}{---}
\setres{cty_Poland}{DA_deterministic}{r_prob_se}{---}
\setres{cty_Poland}{DA_deterministic}{theta_star}{1.56, 1.52}
\setres{cty_Poland}{DA_deterministic}{volume_ess}{---}
\setres{cty_Poland}{DA_lit}{kappa}{-0.256}
\setres{cty_Poland}{DA_lit}{kappa_prior_mean}{0.217}
\setres{cty_Poland}{DA_lit}{kappa_prior_mean_se}{0.146}
\setres{cty_Poland}{DA_lit}{kappa_se}{0.179}
\setres{cty_Poland}{DA_lit}{mse}{22.65}
\setres{cty_Poland}{DA_lit}{mse_prior_mean}{17.89}
\setres{cty_Poland}{DA_lit}{nrmse_pct}{22.9}
\setres{cty_Poland}{DA_lit}{pred_variance}{4.75}
\setres{cty_Poland}{DA_lit}{r_ff}{---}
\setres{cty_Poland}{DA_lit}{r_ff_se}{---}
\setres{cty_Poland}{DA_lit}{r_prior}{---}
\setres{cty_Poland}{DA_lit}{r_prior_se}{---}
\setres{cty_Poland}{DA_lit}{r_prob}{---}
\setres{cty_Poland}{DA_lit}{r_prob_se}{---}
\setres{cty_Poland}{DA_lit}{spread_penalty}{0.473}
\setres{cty_Poland}{DA_lit}{spread_penalty_se}{0.041}
\setres{cty_Poland}{DA_lit}{volume_ess}{---}
\setres{cty_Poland}{DA_transfer_others}{kappa}{0.906}
\setres{cty_Poland}{DA_transfer_others}{kappa_prior_mean}{0.906}
\setres{cty_Poland}{DA_transfer_others}{kappa_prior_mean_se}{0.037}
\setres{cty_Poland}{DA_transfer_others}{kappa_se}{0.037}
\setres{cty_Poland}{DA_transfer_others}{mse}{10.97}
\setres{cty_Poland}{DA_transfer_others}{mse_prior_mean}{10.97}
\setres{cty_Poland}{DA_transfer_others}{nrmse_pct}{18.7}
\setres{cty_Poland}{DA_transfer_others}{pred_variance}{0.00}
\setres{cty_Poland}{DA_transfer_others}{r_ff}{---}
\setres{cty_Poland}{DA_transfer_others}{r_ff_se}{---}
\setres{cty_Poland}{DA_transfer_others}{r_prior}{---}
\setres{cty_Poland}{DA_transfer_others}{r_prior_se}{---}
\setres{cty_Poland}{DA_transfer_others}{r_prob}{---}
\setres{cty_Poland}{DA_transfer_others}{r_prob_se}{---}
\setres{cty_Poland}{DA_transfer_others}{spread_penalty}{0.000}
\setres{cty_Poland}{DA_transfer_others}{spread_penalty_se}{0.000}
\setres{cty_Poland}{DA_transfer_others}{theta_star}{1.70, 1.44}
\setres{cty_Poland}{DA_transfer_others}{transfer}{others}
\setres{cty_Poland}{DA_transfer_others}{volume_ess}{---}
\setres{cty_Poland}{DA_uniform}{kappa}{-1.840}
\setres{cty_Poland}{DA_uniform}{kappa_prior_mean}{-0.481}
\setres{cty_Poland}{DA_uniform}{kappa_prior_mean_se}{0.220}
\setres{cty_Poland}{DA_uniform}{kappa_se}{0.315}
\setres{cty_Poland}{DA_uniform}{mse}{38.56}
\setres{cty_Poland}{DA_uniform}{mse_prior_mean}{24.91}
\setres{cty_Poland}{DA_uniform}{nrmse_pct}{27.7}
\setres{cty_Poland}{DA_uniform}{pred_variance}{13.66}
\setres{cty_Poland}{DA_uniform}{r_ff}{---}
\setres{cty_Poland}{DA_uniform}{r_ff_se}{---}
\setres{cty_Poland}{DA_uniform}{r_prior}{---}
\setres{cty_Poland}{DA_uniform}{r_prior_se}{---}
\setres{cty_Poland}{DA_uniform}{r_prob}{---}
\setres{cty_Poland}{DA_uniform}{r_prob_se}{---}
\setres{cty_Poland}{DA_uniform}{spread_penalty}{1.359}
\setres{cty_Poland}{DA_uniform}{spread_penalty_se}{0.118}
\setres{cty_Poland}{DA_uniform}{volume_ess}{---}
\setres{cty_Poland}{EU_deterministic}{kappa}{0.838}
\setres{cty_Poland}{EU_deterministic}{kappa_se}{0.058}
\setres{cty_Poland}{EU_deterministic}{mse}{11.66}
\setres{cty_Poland}{EU_deterministic}{nrmse_pct}{18.2}
\setres{cty_Poland}{EU_deterministic}{r_ff}{---}
\setres{cty_Poland}{EU_deterministic}{r_ff_se}{---}
\setres{cty_Poland}{EU_deterministic}{r_prior}{---}
\setres{cty_Poland}{EU_deterministic}{r_prior_se}{---}
\setres{cty_Poland}{EU_deterministic}{r_prob}{---}
\setres{cty_Poland}{EU_deterministic}{r_prob_se}{---}
\setres{cty_Poland}{EU_deterministic}{theta_star}{0.85}
\setres{cty_Poland}{EU_deterministic}{volume_ess}{---}
\setres{cty_Poland}{EU_lit}{kappa}{0.622}
\setres{cty_Poland}{EU_lit}{kappa_prior_mean}{0.742}
\setres{cty_Poland}{EU_lit}{kappa_prior_mean_se}{0.069}
\setres{cty_Poland}{EU_lit}{kappa_se}{0.079}
\setres{cty_Poland}{EU_lit}{mse}{13.82}
\setres{cty_Poland}{EU_lit}{mse_prior_mean}{12.62}
\setres{cty_Poland}{EU_lit}{nrmse_pct}{20.2}
\setres{cty_Poland}{EU_lit}{pred_variance}{1.21}
\setres{cty_Poland}{EU_lit}{r_ff}{---}
\setres{cty_Poland}{EU_lit}{r_ff_se}{---}
\setres{cty_Poland}{EU_lit}{r_prior}{---}
\setres{cty_Poland}{EU_lit}{r_prior_se}{---}
\setres{cty_Poland}{EU_lit}{r_prob}{---}
\setres{cty_Poland}{EU_lit}{r_prob_se}{---}
\setres{cty_Poland}{EU_lit}{spread_penalty}{0.120}
\setres{cty_Poland}{EU_lit}{spread_penalty_se}{0.010}
\setres{cty_Poland}{EU_lit}{volume_ess}{---}
\setres{cty_Poland}{EU_transfer_others}{kappa}{0.816}
\setres{cty_Poland}{EU_transfer_others}{kappa_prior_mean}{0.816}
\setres{cty_Poland}{EU_transfer_others}{kappa_prior_mean_se}{0.065}
\setres{cty_Poland}{EU_transfer_others}{kappa_se}{0.065}
\setres{cty_Poland}{EU_transfer_others}{mse}{11.88}
\setres{cty_Poland}{EU_transfer_others}{mse_prior_mean}{11.88}
\setres{cty_Poland}{EU_transfer_others}{nrmse_pct}{18.6}
\setres{cty_Poland}{EU_transfer_others}{pred_variance}{0.00}
\setres{cty_Poland}{EU_transfer_others}{r_ff}{---}
\setres{cty_Poland}{EU_transfer_others}{r_ff_se}{---}
\setres{cty_Poland}{EU_transfer_others}{r_prior}{---}
\setres{cty_Poland}{EU_transfer_others}{r_prior_se}{---}
\setres{cty_Poland}{EU_transfer_others}{r_prob}{---}
\setres{cty_Poland}{EU_transfer_others}{r_prob_se}{---}
\setres{cty_Poland}{EU_transfer_others}{spread_penalty}{0.000}
\setres{cty_Poland}{EU_transfer_others}{spread_penalty_se}{0.000}
\setres{cty_Poland}{EU_transfer_others}{theta_star}{0.95}
\setres{cty_Poland}{EU_transfer_others}{transfer}{others}
\setres{cty_Poland}{EU_transfer_others}{volume_ess}{---}
\setres{cty_Poland}{EU_uniform}{kappa}{-0.737}
\setres{cty_Poland}{EU_uniform}{kappa_prior_mean}{0.079}
\setres{cty_Poland}{EU_uniform}{kappa_prior_mean_se}{0.158}
\setres{cty_Poland}{EU_uniform}{kappa_se}{0.216}
\setres{cty_Poland}{EU_uniform}{mse}{27.48}
\setres{cty_Poland}{EU_uniform}{mse_prior_mean}{19.28}
\setres{cty_Poland}{EU_uniform}{nrmse_pct}{24.3}
\setres{cty_Poland}{EU_uniform}{pred_variance}{8.20}
\setres{cty_Poland}{EU_uniform}{r_ff}{---}
\setres{cty_Poland}{EU_uniform}{r_ff_se}{---}
\setres{cty_Poland}{EU_uniform}{r_prior}{---}
\setres{cty_Poland}{EU_uniform}{r_prior_se}{---}
\setres{cty_Poland}{EU_uniform}{r_prob}{---}
\setres{cty_Poland}{EU_uniform}{r_prob_se}{---}
\setres{cty_Poland}{EU_uniform}{spread_penalty}{0.816}
\setres{cty_Poland}{EU_uniform}{spread_penalty_se}{0.071}
\setres{cty_Poland}{EU_uniform}{volume_ess}{---}
\setres{cty_Poland}{EV_deterministic}{kappa}{0.781}
\setres{cty_Poland}{EV_deterministic}{kappa_se}{0.075}
\setres{cty_Poland}{EV_deterministic}{mse}{12.22}
\setres{cty_Poland}{EV_deterministic}{nrmse_pct}{18.9}
\setres{cty_Poland}{EV_deterministic}{r_ff}{---}
\setres{cty_Poland}{EV_deterministic}{r_ff_se}{---}
\setres{cty_Poland}{EV_deterministic}{r_prior}{---}
\setres{cty_Poland}{EV_deterministic}{r_prior_se}{---}
\setres{cty_Poland}{EV_deterministic}{r_prob}{---}
\setres{cty_Poland}{EV_deterministic}{r_prob_se}{---}
\setres{cty_Poland}{EV_deterministic}{volume_ess}{---}
\setres{cty_Poland}{reference}{country_name}{Poland}
\setres{cty_Poland}{reference}{lookup_nrmse_pct}{16.2}
\setres{cty_Poland}{reference}{midpoint_mse}{20.08}
\setres{cty_Poland}{reference}{midpoint_nrmse_pct}{23.1}
\setres{cty_Poland}{reference}{missing_lotteries}{none}
\setres{cty_Poland}{reference}{n_lotteries}{14}
\setres{cty_Poland}{reference}{n_obs}{1{,}245}
\setres{cty_Poland}{reference}{n_subjects}{89}
\setres{cty_Poland}{reference}{oracle_mse}{10.03}
\setres{cty_Poland}{reference}{sample}{LV2019\_Poland}
\setres{cty_Russia}{CPT_deterministic}{kappa}{0.967}
\setres{cty_Russia}{CPT_deterministic}{kappa_se}{0.011}
\setres{cty_Russia}{CPT_deterministic}{mse}{9.29}
\setres{cty_Russia}{CPT_deterministic}{nrmse_pct}{17.0}
\setres{cty_Russia}{CPT_deterministic}{r_ff}{---}
\setres{cty_Russia}{CPT_deterministic}{r_ff_se}{---}
\setres{cty_Russia}{CPT_deterministic}{r_prior}{---}
\setres{cty_Russia}{CPT_deterministic}{r_prior_se}{---}
\setres{cty_Russia}{CPT_deterministic}{r_prob}{---}
\setres{cty_Russia}{CPT_deterministic}{r_prob_se}{---}
\setres{cty_Russia}{CPT_deterministic}{theta_star}{1.30, 0.70, 0.60}
\setres{cty_Russia}{CPT_deterministic}{volume_ess}{---}
\setres{cty_Russia}{CPT_lit}{kappa}{0.526}
\setres{cty_Russia}{CPT_lit}{kappa_prior_mean}{0.809}
\setres{cty_Russia}{CPT_lit}{kappa_prior_mean_se}{0.058}
\setres{cty_Russia}{CPT_lit}{kappa_se}{0.076}
\setres{cty_Russia}{CPT_lit}{mse}{13.81}
\setres{cty_Russia}{CPT_lit}{mse_prior_mean}{10.91}
\setres{cty_Russia}{CPT_lit}{nrmse_pct}{19.1}
\setres{cty_Russia}{CPT_lit}{pred_variance}{2.90}
\setres{cty_Russia}{CPT_lit}{r_ff}{---}
\setres{cty_Russia}{CPT_lit}{r_ff_se}{---}
\setres{cty_Russia}{CPT_lit}{r_prior}{---}
\setres{cty_Russia}{CPT_lit}{r_prior_se}{---}
\setres{cty_Russia}{CPT_lit}{r_prob}{---}
\setres{cty_Russia}{CPT_lit}{r_prob_se}{---}
\setres{cty_Russia}{CPT_lit}{spread_penalty}{0.283}
\setres{cty_Russia}{CPT_lit}{spread_penalty_se}{0.024}
\setres{cty_Russia}{CPT_lit}{volume_ess}{---}
\setres{cty_Russia}{CPT_transfer_others}{kappa}{0.964}
\setres{cty_Russia}{CPT_transfer_others}{kappa_prior_mean}{0.964}
\setres{cty_Russia}{CPT_transfer_others}{kappa_prior_mean_se}{0.015}
\setres{cty_Russia}{CPT_transfer_others}{kappa_se}{0.015}
\setres{cty_Russia}{CPT_transfer_others}{mse}{9.32}
\setres{cty_Russia}{CPT_transfer_others}{mse_prior_mean}{9.32}
\setres{cty_Russia}{CPT_transfer_others}{nrmse_pct}{16.6}
\setres{cty_Russia}{CPT_transfer_others}{pred_variance}{0.00}
\setres{cty_Russia}{CPT_transfer_others}{r_ff}{---}
\setres{cty_Russia}{CPT_transfer_others}{r_ff_se}{---}
\setres{cty_Russia}{CPT_transfer_others}{r_prior}{---}
\setres{cty_Russia}{CPT_transfer_others}{r_prior_se}{---}
\setres{cty_Russia}{CPT_transfer_others}{r_prob}{---}
\setres{cty_Russia}{CPT_transfer_others}{r_prob_se}{---}
\setres{cty_Russia}{CPT_transfer_others}{spread_penalty}{0.000}
\setres{cty_Russia}{CPT_transfer_others}{spread_penalty_se}{0.000}
\setres{cty_Russia}{CPT_transfer_others}{theta_star}{1.00, 0.60, 0.90}
\setres{cty_Russia}{CPT_transfer_others}{transfer}{others}
\setres{cty_Russia}{CPT_transfer_others}{volume_ess}{---}
\setres{cty_Russia}{CPT_uniform}{kappa}{-1.522}
\setres{cty_Russia}{CPT_uniform}{kappa_prior_mean}{0.171}
\setres{cty_Russia}{CPT_uniform}{kappa_prior_mean_se}{0.132}
\setres{cty_Russia}{CPT_uniform}{kappa_se}{0.242}
\setres{cty_Russia}{CPT_uniform}{mse}{34.82}
\setres{cty_Russia}{CPT_uniform}{mse_prior_mean}{17.45}
\setres{cty_Russia}{CPT_uniform}{nrmse_pct}{22.0}
\setres{cty_Russia}{CPT_uniform}{pred_variance}{17.36}
\setres{cty_Russia}{CPT_uniform}{r_ff}{---}
\setres{cty_Russia}{CPT_uniform}{r_ff_se}{---}
\setres{cty_Russia}{CPT_uniform}{r_prior}{---}
\setres{cty_Russia}{CPT_uniform}{r_prior_se}{---}
\setres{cty_Russia}{CPT_uniform}{r_prob}{---}
\setres{cty_Russia}{CPT_uniform}{r_prob_se}{---}
\setres{cty_Russia}{CPT_uniform}{spread_penalty}{1.693}
\setres{cty_Russia}{CPT_uniform}{spread_penalty_se}{0.141}
\setres{cty_Russia}{CPT_uniform}{volume_ess}{---}
\setres{cty_Russia}{DA_deterministic}{kappa}{0.946}
\setres{cty_Russia}{DA_deterministic}{kappa_se}{0.021}
\setres{cty_Russia}{DA_deterministic}{mse}{9.51}
\setres{cty_Russia}{DA_deterministic}{nrmse_pct}{17.1}
\setres{cty_Russia}{DA_deterministic}{r_ff}{---}
\setres{cty_Russia}{DA_deterministic}{r_ff_se}{---}
\setres{cty_Russia}{DA_deterministic}{r_prior}{---}
\setres{cty_Russia}{DA_deterministic}{r_prior_se}{---}
\setres{cty_Russia}{DA_deterministic}{r_prob}{---}
\setres{cty_Russia}{DA_deterministic}{r_prob_se}{---}
\setres{cty_Russia}{DA_deterministic}{theta_star}{1.74, 1.80}
\setres{cty_Russia}{DA_deterministic}{volume_ess}{---}
\setres{cty_Russia}{DA_lit}{kappa}{-0.385}
\setres{cty_Russia}{DA_lit}{kappa_prior_mean}{0.078}
\setres{cty_Russia}{DA_lit}{kappa_prior_mean_se}{0.178}
\setres{cty_Russia}{DA_lit}{kappa_se}{0.209}
\setres{cty_Russia}{DA_lit}{mse}{23.16}
\setres{cty_Russia}{DA_lit}{mse_prior_mean}{18.41}
\setres{cty_Russia}{DA_lit}{nrmse_pct}{23.8}
\setres{cty_Russia}{DA_lit}{pred_variance}{4.75}
\setres{cty_Russia}{DA_lit}{r_ff}{---}
\setres{cty_Russia}{DA_lit}{r_ff_se}{---}
\setres{cty_Russia}{DA_lit}{r_prior}{---}
\setres{cty_Russia}{DA_lit}{r_prior_se}{---}
\setres{cty_Russia}{DA_lit}{r_prob}{---}
\setres{cty_Russia}{DA_lit}{r_prob_se}{---}
\setres{cty_Russia}{DA_lit}{spread_penalty}{0.463}
\setres{cty_Russia}{DA_lit}{spread_penalty_se}{0.039}
\setres{cty_Russia}{DA_lit}{volume_ess}{---}
\setres{cty_Russia}{DA_transfer_others}{kappa}{0.953}
\setres{cty_Russia}{DA_transfer_others}{kappa_prior_mean}{0.953}
\setres{cty_Russia}{DA_transfer_others}{kappa_prior_mean_se}{0.024}
\setres{cty_Russia}{DA_transfer_others}{kappa_se}{0.024}
\setres{cty_Russia}{DA_transfer_others}{mse}{9.44}
\setres{cty_Russia}{DA_transfer_others}{mse_prior_mean}{9.44}
\setres{cty_Russia}{DA_transfer_others}{nrmse_pct}{17.1}
\setres{cty_Russia}{DA_transfer_others}{pred_variance}{0.00}
\setres{cty_Russia}{DA_transfer_others}{r_ff}{---}
\setres{cty_Russia}{DA_transfer_others}{r_ff_se}{---}
\setres{cty_Russia}{DA_transfer_others}{r_prior}{---}
\setres{cty_Russia}{DA_transfer_others}{r_prior_se}{---}
\setres{cty_Russia}{DA_transfer_others}{r_prob}{---}
\setres{cty_Russia}{DA_transfer_others}{r_prob_se}{---}
\setres{cty_Russia}{DA_transfer_others}{spread_penalty}{0.000}
\setres{cty_Russia}{DA_transfer_others}{spread_penalty_se}{0.000}
\setres{cty_Russia}{DA_transfer_others}{theta_star}{1.69, 1.44}
\setres{cty_Russia}{DA_transfer_others}{transfer}{others}
\setres{cty_Russia}{DA_transfer_others}{volume_ess}{---}
\setres{cty_Russia}{DA_uniform}{kappa}{-2.000}
\setres{cty_Russia}{DA_uniform}{kappa_prior_mean}{-0.669}
\setres{cty_Russia}{DA_uniform}{kappa_prior_mean_se}{0.265}
\setres{cty_Russia}{DA_uniform}{kappa_se}{0.355}
\setres{cty_Russia}{DA_uniform}{mse}{39.72}
\setres{cty_Russia}{DA_uniform}{mse_prior_mean}{26.07}
\setres{cty_Russia}{DA_uniform}{nrmse_pct}{27.5}
\setres{cty_Russia}{DA_uniform}{pred_variance}{13.65}
\setres{cty_Russia}{DA_uniform}{r_ff}{---}
\setres{cty_Russia}{DA_uniform}{r_ff_se}{---}
\setres{cty_Russia}{DA_uniform}{r_prior}{---}
\setres{cty_Russia}{DA_uniform}{r_prior_se}{---}
\setres{cty_Russia}{DA_uniform}{r_prob}{---}
\setres{cty_Russia}{DA_uniform}{r_prob_se}{---}
\setres{cty_Russia}{DA_uniform}{spread_penalty}{1.331}
\setres{cty_Russia}{DA_uniform}{spread_penalty_se}{0.111}
\setres{cty_Russia}{DA_uniform}{volume_ess}{---}
\setres{cty_Russia}{EU_deterministic}{kappa}{0.840}
\setres{cty_Russia}{EU_deterministic}{kappa_se}{0.048}
\setres{cty_Russia}{EU_deterministic}{mse}{10.60}
\setres{cty_Russia}{EU_deterministic}{nrmse_pct}{18.5}
\setres{cty_Russia}{EU_deterministic}{r_ff}{---}
\setres{cty_Russia}{EU_deterministic}{r_ff_se}{---}
\setres{cty_Russia}{EU_deterministic}{r_prior}{---}
\setres{cty_Russia}{EU_deterministic}{r_prior_se}{---}
\setres{cty_Russia}{EU_deterministic}{r_prob}{---}
\setres{cty_Russia}{EU_deterministic}{r_prob_se}{---}
\setres{cty_Russia}{EU_deterministic}{theta_star}{0.90}
\setres{cty_Russia}{EU_deterministic}{volume_ess}{---}
\setres{cty_Russia}{EU_lit}{kappa}{0.563}
\setres{cty_Russia}{EU_lit}{kappa_prior_mean}{0.680}
\setres{cty_Russia}{EU_lit}{kappa_prior_mean_se}{0.081}
\setres{cty_Russia}{EU_lit}{kappa_se}{0.090}
\setres{cty_Russia}{EU_lit}{mse}{13.44}
\setres{cty_Russia}{EU_lit}{mse_prior_mean}{12.23}
\setres{cty_Russia}{EU_lit}{nrmse_pct}{19.7}
\setres{cty_Russia}{EU_lit}{pred_variance}{1.20}
\setres{cty_Russia}{EU_lit}{r_ff}{---}
\setres{cty_Russia}{EU_lit}{r_ff_se}{---}
\setres{cty_Russia}{EU_lit}{r_prior}{---}
\setres{cty_Russia}{EU_lit}{r_prior_se}{---}
\setres{cty_Russia}{EU_lit}{r_prob}{---}
\setres{cty_Russia}{EU_lit}{r_prob_se}{---}
\setres{cty_Russia}{EU_lit}{spread_penalty}{0.117}
\setres{cty_Russia}{EU_lit}{spread_penalty_se}{0.010}
\setres{cty_Russia}{EU_lit}{volume_ess}{---}
\setres{cty_Russia}{EU_transfer_others}{kappa}{0.844}
\setres{cty_Russia}{EU_transfer_others}{kappa_prior_mean}{0.844}
\setres{cty_Russia}{EU_transfer_others}{kappa_prior_mean_se}{0.051}
\setres{cty_Russia}{EU_transfer_others}{kappa_se}{0.051}
\setres{cty_Russia}{EU_transfer_others}{mse}{10.55}
\setres{cty_Russia}{EU_transfer_others}{mse_prior_mean}{10.55}
\setres{cty_Russia}{EU_transfer_others}{nrmse_pct}{18.8}
\setres{cty_Russia}{EU_transfer_others}{pred_variance}{0.00}
\setres{cty_Russia}{EU_transfer_others}{r_ff}{---}
\setres{cty_Russia}{EU_transfer_others}{r_ff_se}{---}
\setres{cty_Russia}{EU_transfer_others}{r_prior}{---}
\setres{cty_Russia}{EU_transfer_others}{r_prior_se}{---}
\setres{cty_Russia}{EU_transfer_others}{r_prob}{---}
\setres{cty_Russia}{EU_transfer_others}{r_prob_se}{---}
\setres{cty_Russia}{EU_transfer_others}{spread_penalty}{0.000}
\setres{cty_Russia}{EU_transfer_others}{spread_penalty_se}{0.000}
\setres{cty_Russia}{EU_transfer_others}{theta_star}{0.95}
\setres{cty_Russia}{EU_transfer_others}{transfer}{others}
\setres{cty_Russia}{EU_transfer_others}{volume_ess}{---}
\setres{cty_Russia}{EU_uniform}{kappa}{-0.880}
\setres{cty_Russia}{EU_uniform}{kappa_prior_mean}{-0.081}
\setres{cty_Russia}{EU_uniform}{kappa_prior_mean_se}{0.193}
\setres{cty_Russia}{EU_uniform}{kappa_se}{0.247}
\setres{cty_Russia}{EU_uniform}{mse}{28.23}
\setres{cty_Russia}{EU_uniform}{mse_prior_mean}{20.04}
\setres{cty_Russia}{EU_uniform}{nrmse_pct}{24.5}
\setres{cty_Russia}{EU_uniform}{pred_variance}{8.20}
\setres{cty_Russia}{EU_uniform}{r_ff}{---}
\setres{cty_Russia}{EU_uniform}{r_ff_se}{---}
\setres{cty_Russia}{EU_uniform}{r_prior}{---}
\setres{cty_Russia}{EU_uniform}{r_prior_se}{---}
\setres{cty_Russia}{EU_uniform}{r_prob}{---}
\setres{cty_Russia}{EU_uniform}{r_prob_se}{---}
\setres{cty_Russia}{EU_uniform}{spread_penalty}{0.799}
\setres{cty_Russia}{EU_uniform}{spread_penalty_se}{0.067}
\setres{cty_Russia}{EU_uniform}{volume_ess}{---}
\setres{cty_Russia}{EV_deterministic}{kappa}{0.824}
\setres{cty_Russia}{EV_deterministic}{kappa_se}{0.058}
\setres{cty_Russia}{EV_deterministic}{mse}{10.76}
\setres{cty_Russia}{EV_deterministic}{nrmse_pct}{18.9}
\setres{cty_Russia}{EV_deterministic}{r_ff}{---}
\setres{cty_Russia}{EV_deterministic}{r_ff_se}{---}
\setres{cty_Russia}{EV_deterministic}{r_prior}{---}
\setres{cty_Russia}{EV_deterministic}{r_prior_se}{---}
\setres{cty_Russia}{EV_deterministic}{r_prob}{---}
\setres{cty_Russia}{EV_deterministic}{r_prob_se}{---}
\setres{cty_Russia}{EV_deterministic}{volume_ess}{---}
\setres{cty_Russia}{reference}{country_name}{Russia}
\setres{cty_Russia}{reference}{lookup_nrmse_pct}{16.4}
\setres{cty_Russia}{reference}{midpoint_mse}{19.21}
\setres{cty_Russia}{reference}{midpoint_nrmse_pct}{22.2}
\setres{cty_Russia}{reference}{missing_lotteries}{none}
\setres{cty_Russia}{reference}{n_lotteries}{14}
\setres{cty_Russia}{reference}{n_obs}{979}
\setres{cty_Russia}{reference}{n_subjects}{70}
\setres{cty_Russia}{reference}{oracle_mse}{8.95}
\setres{cty_Russia}{reference}{sample}{LV2019\_Russia}
\setres{cty_Saudi_Arabia}{CPT_deterministic}{kappa}{0.962}
\setres{cty_Saudi_Arabia}{CPT_deterministic}{kappa_se}{0.012}
\setres{cty_Saudi_Arabia}{CPT_deterministic}{mse}{10.30}
\setres{cty_Saudi_Arabia}{CPT_deterministic}{nrmse_pct}{18.2}
\setres{cty_Saudi_Arabia}{CPT_deterministic}{r_ff}{---}
\setres{cty_Saudi_Arabia}{CPT_deterministic}{r_ff_se}{---}
\setres{cty_Saudi_Arabia}{CPT_deterministic}{r_prior}{---}
\setres{cty_Saudi_Arabia}{CPT_deterministic}{r_prior_se}{---}
\setres{cty_Saudi_Arabia}{CPT_deterministic}{r_prob}{---}
\setres{cty_Saudi_Arabia}{CPT_deterministic}{r_prob_se}{---}
\setres{cty_Saudi_Arabia}{CPT_deterministic}{theta_star}{0.90, 0.60, 1.40}
\setres{cty_Saudi_Arabia}{CPT_deterministic}{volume_ess}{---}
\setres{cty_Saudi_Arabia}{CPT_lit}{kappa}{0.014}
\setres{cty_Saudi_Arabia}{CPT_lit}{kappa_prior_mean}{0.301}
\setres{cty_Saudi_Arabia}{CPT_lit}{kappa_prior_mean_se}{0.134}
\setres{cty_Saudi_Arabia}{CPT_lit}{kappa_se}{0.145}
\setres{cty_Saudi_Arabia}{CPT_lit}{mse}{19.90}
\setres{cty_Saudi_Arabia}{CPT_lit}{mse_prior_mean}{17.00}
\setres{cty_Saudi_Arabia}{CPT_lit}{nrmse_pct}{23.3}
\setres{cty_Saudi_Arabia}{CPT_lit}{pred_variance}{2.90}
\setres{cty_Saudi_Arabia}{CPT_lit}{r_ff}{---}
\setres{cty_Saudi_Arabia}{CPT_lit}{r_ff_se}{---}
\setres{cty_Saudi_Arabia}{CPT_lit}{r_prior}{---}
\setres{cty_Saudi_Arabia}{CPT_lit}{r_prior_se}{---}
\setres{cty_Saudi_Arabia}{CPT_lit}{r_prob}{---}
\setres{cty_Saudi_Arabia}{CPT_lit}{r_prob_se}{---}
\setres{cty_Saudi_Arabia}{CPT_lit}{spread_penalty}{0.286}
\setres{cty_Saudi_Arabia}{CPT_lit}{spread_penalty_se}{0.029}
\setres{cty_Saudi_Arabia}{CPT_lit}{volume_ess}{---}
\setres{cty_Saudi_Arabia}{CPT_transfer_others}{kappa}{0.814}
\setres{cty_Saudi_Arabia}{CPT_transfer_others}{kappa_prior_mean}{0.814}
\setres{cty_Saudi_Arabia}{CPT_transfer_others}{kappa_prior_mean_se}{0.061}
\setres{cty_Saudi_Arabia}{CPT_transfer_others}{kappa_se}{0.061}
\setres{cty_Saudi_Arabia}{CPT_transfer_others}{mse}{11.80}
\setres{cty_Saudi_Arabia}{CPT_transfer_others}{mse_prior_mean}{11.80}
\setres{cty_Saudi_Arabia}{CPT_transfer_others}{nrmse_pct}{19.2}
\setres{cty_Saudi_Arabia}{CPT_transfer_others}{pred_variance}{0.00}
\setres{cty_Saudi_Arabia}{CPT_transfer_others}{r_ff}{---}
\setres{cty_Saudi_Arabia}{CPT_transfer_others}{r_ff_se}{---}
\setres{cty_Saudi_Arabia}{CPT_transfer_others}{r_prior}{---}
\setres{cty_Saudi_Arabia}{CPT_transfer_others}{r_prior_se}{---}
\setres{cty_Saudi_Arabia}{CPT_transfer_others}{r_prob}{---}
\setres{cty_Saudi_Arabia}{CPT_transfer_others}{r_prob_se}{---}
\setres{cty_Saudi_Arabia}{CPT_transfer_others}{spread_penalty}{0.000}
\setres{cty_Saudi_Arabia}{CPT_transfer_others}{spread_penalty_se}{0.000}
\setres{cty_Saudi_Arabia}{CPT_transfer_others}{theta_star}{1.00, 0.60, 0.90}
\setres{cty_Saudi_Arabia}{CPT_transfer_others}{transfer}{others}
\setres{cty_Saudi_Arabia}{CPT_transfer_others}{volume_ess}{---}
\setres{cty_Saudi_Arabia}{CPT_uniform}{kappa}{-2.354}
\setres{cty_Saudi_Arabia}{CPT_uniform}{kappa_prior_mean}{-0.640}
\setres{cty_Saudi_Arabia}{CPT_uniform}{kappa_prior_mean_se}{0.179}
\setres{cty_Saudi_Arabia}{CPT_uniform}{kappa_se}{0.298}
\setres{cty_Saudi_Arabia}{CPT_uniform}{mse}{43.90}
\setres{cty_Saudi_Arabia}{CPT_uniform}{mse_prior_mean}{26.54}
\setres{cty_Saudi_Arabia}{CPT_uniform}{nrmse_pct}{27.4}
\setres{cty_Saudi_Arabia}{CPT_uniform}{pred_variance}{17.37}
\setres{cty_Saudi_Arabia}{CPT_uniform}{r_ff}{---}
\setres{cty_Saudi_Arabia}{CPT_uniform}{r_ff_se}{---}
\setres{cty_Saudi_Arabia}{CPT_uniform}{r_prior}{---}
\setres{cty_Saudi_Arabia}{CPT_uniform}{r_prior_se}{---}
\setres{cty_Saudi_Arabia}{CPT_uniform}{r_prob}{---}
\setres{cty_Saudi_Arabia}{CPT_uniform}{r_prob_se}{---}
\setres{cty_Saudi_Arabia}{CPT_uniform}{spread_penalty}{1.714}
\setres{cty_Saudi_Arabia}{CPT_uniform}{spread_penalty_se}{0.172}
\setres{cty_Saudi_Arabia}{CPT_uniform}{volume_ess}{---}
\setres{cty_Saudi_Arabia}{DA_deterministic}{kappa}{0.883}
\setres{cty_Saudi_Arabia}{DA_deterministic}{kappa_se}{0.042}
\setres{cty_Saudi_Arabia}{DA_deterministic}{mse}{11.10}
\setres{cty_Saudi_Arabia}{DA_deterministic}{nrmse_pct}{18.3}
\setres{cty_Saudi_Arabia}{DA_deterministic}{r_ff}{---}
\setres{cty_Saudi_Arabia}{DA_deterministic}{r_ff_se}{---}
\setres{cty_Saudi_Arabia}{DA_deterministic}{r_prior}{---}
\setres{cty_Saudi_Arabia}{DA_deterministic}{r_prior_se}{---}
\setres{cty_Saudi_Arabia}{DA_deterministic}{r_prob}{---}
\setres{cty_Saudi_Arabia}{DA_deterministic}{r_prob_se}{---}
\setres{cty_Saudi_Arabia}{DA_deterministic}{theta_star}{1.47, 0.31}
\setres{cty_Saudi_Arabia}{DA_deterministic}{volume_ess}{---}
\setres{cty_Saudi_Arabia}{DA_lit}{kappa}{-1.435}
\setres{cty_Saudi_Arabia}{DA_lit}{kappa_prior_mean}{-0.967}
\setres{cty_Saudi_Arabia}{DA_lit}{kappa_prior_mean_se}{0.266}
\setres{cty_Saudi_Arabia}{DA_lit}{kappa_se}{0.295}
\setres{cty_Saudi_Arabia}{DA_lit}{mse}{34.59}
\setres{cty_Saudi_Arabia}{DA_lit}{mse_prior_mean}{29.85}
\setres{cty_Saudi_Arabia}{DA_lit}{nrmse_pct}{31.9}
\setres{cty_Saudi_Arabia}{DA_lit}{pred_variance}{4.75}
\setres{cty_Saudi_Arabia}{DA_lit}{r_ff}{---}
\setres{cty_Saudi_Arabia}{DA_lit}{r_ff_se}{---}
\setres{cty_Saudi_Arabia}{DA_lit}{r_prior}{---}
\setres{cty_Saudi_Arabia}{DA_lit}{r_prior_se}{---}
\setres{cty_Saudi_Arabia}{DA_lit}{r_prob}{---}
\setres{cty_Saudi_Arabia}{DA_lit}{r_prob_se}{---}
\setres{cty_Saudi_Arabia}{DA_lit}{spread_penalty}{0.469}
\setres{cty_Saudi_Arabia}{DA_lit}{spread_penalty_se}{0.047}
\setres{cty_Saudi_Arabia}{DA_lit}{volume_ess}{---}
\setres{cty_Saudi_Arabia}{DA_transfer_others}{kappa}{0.708}
\setres{cty_Saudi_Arabia}{DA_transfer_others}{kappa_prior_mean}{0.708}
\setres{cty_Saudi_Arabia}{DA_transfer_others}{kappa_prior_mean_se}{0.087}
\setres{cty_Saudi_Arabia}{DA_transfer_others}{kappa_se}{0.087}
\setres{cty_Saudi_Arabia}{DA_transfer_others}{mse}{12.87}
\setres{cty_Saudi_Arabia}{DA_transfer_others}{mse_prior_mean}{12.87}
\setres{cty_Saudi_Arabia}{DA_transfer_others}{nrmse_pct}{18.5}
\setres{cty_Saudi_Arabia}{DA_transfer_others}{pred_variance}{0.00}
\setres{cty_Saudi_Arabia}{DA_transfer_others}{r_ff}{---}
\setres{cty_Saudi_Arabia}{DA_transfer_others}{r_ff_se}{---}
\setres{cty_Saudi_Arabia}{DA_transfer_others}{r_prior}{---}
\setres{cty_Saudi_Arabia}{DA_transfer_others}{r_prior_se}{---}
\setres{cty_Saudi_Arabia}{DA_transfer_others}{r_prob}{---}
\setres{cty_Saudi_Arabia}{DA_transfer_others}{r_prob_se}{---}
\setres{cty_Saudi_Arabia}{DA_transfer_others}{spread_penalty}{0.000}
\setres{cty_Saudi_Arabia}{DA_transfer_others}{spread_penalty_se}{0.000}
\setres{cty_Saudi_Arabia}{DA_transfer_others}{theta_star}{1.70, 1.48}
\setres{cty_Saudi_Arabia}{DA_transfer_others}{transfer}{others}
\setres{cty_Saudi_Arabia}{DA_transfer_others}{volume_ess}{---}
\setres{cty_Saudi_Arabia}{DA_uniform}{kappa}{-3.441}
\setres{cty_Saudi_Arabia}{DA_uniform}{kappa_prior_mean}{-2.095}
\setres{cty_Saudi_Arabia}{DA_uniform}{kappa_prior_mean_se}{0.344}
\setres{cty_Saudi_Arabia}{DA_uniform}{kappa_se}{0.444}
\setres{cty_Saudi_Arabia}{DA_uniform}{mse}{54.92}
\setres{cty_Saudi_Arabia}{DA_uniform}{mse_prior_mean}{41.28}
\setres{cty_Saudi_Arabia}{DA_uniform}{nrmse_pct}{34.4}
\setres{cty_Saudi_Arabia}{DA_uniform}{pred_variance}{13.65}
\setres{cty_Saudi_Arabia}{DA_uniform}{r_ff}{---}
\setres{cty_Saudi_Arabia}{DA_uniform}{r_ff_se}{---}
\setres{cty_Saudi_Arabia}{DA_uniform}{r_prior}{---}
\setres{cty_Saudi_Arabia}{DA_uniform}{r_prior_se}{---}
\setres{cty_Saudi_Arabia}{DA_uniform}{r_prob}{---}
\setres{cty_Saudi_Arabia}{DA_uniform}{r_prob_se}{---}
\setres{cty_Saudi_Arabia}{DA_uniform}{spread_penalty}{1.347}
\setres{cty_Saudi_Arabia}{DA_uniform}{spread_penalty_se}{0.135}
\setres{cty_Saudi_Arabia}{DA_uniform}{volume_ess}{---}
\setres{cty_Saudi_Arabia}{EU_deterministic}{kappa}{0.879}
\setres{cty_Saudi_Arabia}{EU_deterministic}{kappa_se}{0.048}
\setres{cty_Saudi_Arabia}{EU_deterministic}{mse}{11.14}
\setres{cty_Saudi_Arabia}{EU_deterministic}{nrmse_pct}{18.5}
\setres{cty_Saudi_Arabia}{EU_deterministic}{r_ff}{---}
\setres{cty_Saudi_Arabia}{EU_deterministic}{r_ff_se}{---}
\setres{cty_Saudi_Arabia}{EU_deterministic}{r_prior}{---}
\setres{cty_Saudi_Arabia}{EU_deterministic}{r_prior_se}{---}
\setres{cty_Saudi_Arabia}{EU_deterministic}{r_prob}{---}
\setres{cty_Saudi_Arabia}{EU_deterministic}{r_prob_se}{---}
\setres{cty_Saudi_Arabia}{EU_deterministic}{theta_star}{1.22}
\setres{cty_Saudi_Arabia}{EU_deterministic}{volume_ess}{---}
\setres{cty_Saudi_Arabia}{EU_lit}{kappa}{0.018}
\setres{cty_Saudi_Arabia}{EU_lit}{kappa_prior_mean}{0.137}
\setres{cty_Saudi_Arabia}{EU_lit}{kappa_prior_mean_se}{0.166}
\setres{cty_Saudi_Arabia}{EU_lit}{kappa_se}{0.172}
\setres{cty_Saudi_Arabia}{EU_lit}{mse}{19.86}
\setres{cty_Saudi_Arabia}{EU_lit}{mse_prior_mean}{18.66}
\setres{cty_Saudi_Arabia}{EU_lit}{nrmse_pct}{23.6}
\setres{cty_Saudi_Arabia}{EU_lit}{pred_variance}{1.20}
\setres{cty_Saudi_Arabia}{EU_lit}{r_ff}{---}
\setres{cty_Saudi_Arabia}{EU_lit}{r_ff_se}{---}
\setres{cty_Saudi_Arabia}{EU_lit}{r_prior}{---}
\setres{cty_Saudi_Arabia}{EU_lit}{r_prior_se}{---}
\setres{cty_Saudi_Arabia}{EU_lit}{r_prob}{---}
\setres{cty_Saudi_Arabia}{EU_lit}{r_prob_se}{---}
\setres{cty_Saudi_Arabia}{EU_lit}{spread_penalty}{0.119}
\setres{cty_Saudi_Arabia}{EU_lit}{spread_penalty_se}{0.012}
\setres{cty_Saudi_Arabia}{EU_lit}{volume_ess}{---}
\setres{cty_Saudi_Arabia}{EU_transfer_others}{kappa}{0.743}
\setres{cty_Saudi_Arabia}{EU_transfer_others}{kappa_prior_mean}{0.743}
\setres{cty_Saudi_Arabia}{EU_transfer_others}{kappa_prior_mean_se}{0.085}
\setres{cty_Saudi_Arabia}{EU_transfer_others}{kappa_se}{0.085}
\setres{cty_Saudi_Arabia}{EU_transfer_others}{mse}{12.52}
\setres{cty_Saudi_Arabia}{EU_transfer_others}{mse_prior_mean}{12.52}
\setres{cty_Saudi_Arabia}{EU_transfer_others}{nrmse_pct}{18.5}
\setres{cty_Saudi_Arabia}{EU_transfer_others}{pred_variance}{0.00}
\setres{cty_Saudi_Arabia}{EU_transfer_others}{r_ff}{---}
\setres{cty_Saudi_Arabia}{EU_transfer_others}{r_ff_se}{---}
\setres{cty_Saudi_Arabia}{EU_transfer_others}{r_prior}{---}
\setres{cty_Saudi_Arabia}{EU_transfer_others}{r_prior_se}{---}
\setres{cty_Saudi_Arabia}{EU_transfer_others}{r_prob}{---}
\setres{cty_Saudi_Arabia}{EU_transfer_others}{r_prob_se}{---}
\setres{cty_Saudi_Arabia}{EU_transfer_others}{spread_penalty}{0.000}
\setres{cty_Saudi_Arabia}{EU_transfer_others}{spread_penalty_se}{0.000}
\setres{cty_Saudi_Arabia}{EU_transfer_others}{theta_star}{0.94}
\setres{cty_Saudi_Arabia}{EU_transfer_others}{transfer}{others}
\setres{cty_Saudi_Arabia}{EU_transfer_others}{volume_ess}{---}
\setres{cty_Saudi_Arabia}{EU_uniform}{kappa}{-1.966}
\setres{cty_Saudi_Arabia}{EU_uniform}{kappa_prior_mean}{-1.157}
\setres{cty_Saudi_Arabia}{EU_uniform}{kappa_prior_mean_se}{0.275}
\setres{cty_Saudi_Arabia}{EU_uniform}{kappa_se}{0.330}
\setres{cty_Saudi_Arabia}{EU_uniform}{mse}{39.97}
\setres{cty_Saudi_Arabia}{EU_uniform}{mse_prior_mean}{31.78}
\setres{cty_Saudi_Arabia}{EU_uniform}{nrmse_pct}{31.8}
\setres{cty_Saudi_Arabia}{EU_uniform}{pred_variance}{8.20}
\setres{cty_Saudi_Arabia}{EU_uniform}{r_ff}{---}
\setres{cty_Saudi_Arabia}{EU_uniform}{r_ff_se}{---}
\setres{cty_Saudi_Arabia}{EU_uniform}{r_prior}{---}
\setres{cty_Saudi_Arabia}{EU_uniform}{r_prior_se}{---}
\setres{cty_Saudi_Arabia}{EU_uniform}{r_prob}{---}
\setres{cty_Saudi_Arabia}{EU_uniform}{r_prob_se}{---}
\setres{cty_Saudi_Arabia}{EU_uniform}{spread_penalty}{0.809}
\setres{cty_Saudi_Arabia}{EU_uniform}{spread_penalty_se}{0.081}
\setres{cty_Saudi_Arabia}{EU_uniform}{volume_ess}{---}
\setres{cty_Saudi_Arabia}{EV_deterministic}{kappa}{0.808}
\setres{cty_Saudi_Arabia}{EV_deterministic}{kappa_se}{0.073}
\setres{cty_Saudi_Arabia}{EV_deterministic}{mse}{11.86}
\setres{cty_Saudi_Arabia}{EV_deterministic}{nrmse_pct}{17.8}
\setres{cty_Saudi_Arabia}{EV_deterministic}{r_ff}{---}
\setres{cty_Saudi_Arabia}{EV_deterministic}{r_ff_se}{---}
\setres{cty_Saudi_Arabia}{EV_deterministic}{r_prior}{---}
\setres{cty_Saudi_Arabia}{EV_deterministic}{r_prior_se}{---}
\setres{cty_Saudi_Arabia}{EV_deterministic}{r_prob}{---}
\setres{cty_Saudi_Arabia}{EV_deterministic}{r_prob_se}{---}
\setres{cty_Saudi_Arabia}{EV_deterministic}{volume_ess}{---}
\setres{cty_Saudi_Arabia}{reference}{country_name}{Saudi Arabia}
\setres{cty_Saudi_Arabia}{reference}{lookup_nrmse_pct}{18.0}
\setres{cty_Saudi_Arabia}{reference}{midpoint_mse}{20.05}
\setres{cty_Saudi_Arabia}{reference}{midpoint_nrmse_pct}{23.0}
\setres{cty_Saudi_Arabia}{reference}{missing_lotteries}{none}
\setres{cty_Saudi_Arabia}{reference}{n_lotteries}{14}
\setres{cty_Saudi_Arabia}{reference}{n_obs}{910}
\setres{cty_Saudi_Arabia}{reference}{n_subjects}{65}
\setres{cty_Saudi_Arabia}{reference}{oracle_mse}{9.91}
\setres{cty_Saudi_Arabia}{reference}{sample}{LV2019\_Saudi\_Arabia}
\setres{cty_South_Africa}{CPT_deterministic}{kappa}{0.974}
\setres{cty_South_Africa}{CPT_deterministic}{kappa_se}{0.018}
\setres{cty_South_Africa}{CPT_deterministic}{mse}{13.45}
\setres{cty_South_Africa}{CPT_deterministic}{nrmse_pct}{21.3}
\setres{cty_South_Africa}{CPT_deterministic}{r_ff}{---}
\setres{cty_South_Africa}{CPT_deterministic}{r_ff_se}{---}
\setres{cty_South_Africa}{CPT_deterministic}{r_prior}{---}
\setres{cty_South_Africa}{CPT_deterministic}{r_prior_se}{---}
\setres{cty_South_Africa}{CPT_deterministic}{r_prob}{---}
\setres{cty_South_Africa}{CPT_deterministic}{r_prob_se}{---}
\setres{cty_South_Africa}{CPT_deterministic}{theta_star}{1.10, 0.70, 0.70}
\setres{cty_South_Africa}{CPT_deterministic}{volume_ess}{---}
\setres{cty_South_Africa}{CPT_lit}{kappa}{0.613}
\setres{cty_South_Africa}{CPT_lit}{kappa_prior_mean}{0.901}
\setres{cty_South_Africa}{CPT_lit}{kappa_prior_mean_se}{0.046}
\setres{cty_South_Africa}{CPT_lit}{kappa_se}{0.062}
\setres{cty_South_Africa}{CPT_lit}{mse}{17.09}
\setres{cty_South_Africa}{CPT_lit}{mse_prior_mean}{14.19}
\setres{cty_South_Africa}{CPT_lit}{nrmse_pct}{22.7}
\setres{cty_South_Africa}{CPT_lit}{pred_variance}{2.90}
\setres{cty_South_Africa}{CPT_lit}{r_ff}{---}
\setres{cty_South_Africa}{CPT_lit}{r_ff_se}{---}
\setres{cty_South_Africa}{CPT_lit}{r_prior}{---}
\setres{cty_South_Africa}{CPT_lit}{r_prior_se}{---}
\setres{cty_South_Africa}{CPT_lit}{r_prob}{---}
\setres{cty_South_Africa}{CPT_lit}{r_prob_se}{---}
\setres{cty_South_Africa}{CPT_lit}{spread_penalty}{0.287}
\setres{cty_South_Africa}{CPT_lit}{spread_penalty_se}{0.025}
\setres{cty_South_Africa}{CPT_lit}{volume_ess}{---}
\setres{cty_South_Africa}{CPT_transfer_others}{kappa}{0.949}
\setres{cty_South_Africa}{CPT_transfer_others}{kappa_prior_mean}{0.949}
\setres{cty_South_Africa}{CPT_transfer_others}{kappa_prior_mean_se}{0.027}
\setres{cty_South_Africa}{CPT_transfer_others}{kappa_se}{0.027}
\setres{cty_South_Africa}{CPT_transfer_others}{mse}{13.70}
\setres{cty_South_Africa}{CPT_transfer_others}{mse_prior_mean}{13.70}
\setres{cty_South_Africa}{CPT_transfer_others}{nrmse_pct}{21.2}
\setres{cty_South_Africa}{CPT_transfer_others}{pred_variance}{0.00}
\setres{cty_South_Africa}{CPT_transfer_others}{r_ff}{---}
\setres{cty_South_Africa}{CPT_transfer_others}{r_ff_se}{---}
\setres{cty_South_Africa}{CPT_transfer_others}{r_prior}{---}
\setres{cty_South_Africa}{CPT_transfer_others}{r_prior_se}{---}
\setres{cty_South_Africa}{CPT_transfer_others}{r_prob}{---}
\setres{cty_South_Africa}{CPT_transfer_others}{r_prob_se}{---}
\setres{cty_South_Africa}{CPT_transfer_others}{spread_penalty}{0.000}
\setres{cty_South_Africa}{CPT_transfer_others}{spread_penalty_se}{0.000}
\setres{cty_South_Africa}{CPT_transfer_others}{theta_star}{1.10, 0.60, 0.80}
\setres{cty_South_Africa}{CPT_transfer_others}{transfer}{others}
\setres{cty_South_Africa}{CPT_transfer_others}{volume_ess}{---}
\setres{cty_South_Africa}{CPT_uniform}{kappa}{-1.349}
\setres{cty_South_Africa}{CPT_uniform}{kappa_prior_mean}{0.372}
\setres{cty_South_Africa}{CPT_uniform}{kappa_prior_mean_se}{0.132}
\setres{cty_South_Africa}{CPT_uniform}{kappa_se}{0.234}
\setres{cty_South_Africa}{CPT_uniform}{mse}{36.89}
\setres{cty_South_Africa}{CPT_uniform}{mse_prior_mean}{19.53}
\setres{cty_South_Africa}{CPT_uniform}{nrmse_pct}{25.4}
\setres{cty_South_Africa}{CPT_uniform}{pred_variance}{17.37}
\setres{cty_South_Africa}{CPT_uniform}{r_ff}{---}
\setres{cty_South_Africa}{CPT_uniform}{r_ff_se}{---}
\setres{cty_South_Africa}{CPT_uniform}{r_prior}{---}
\setres{cty_South_Africa}{CPT_uniform}{r_prior_se}{---}
\setres{cty_South_Africa}{CPT_uniform}{r_prob}{---}
\setres{cty_South_Africa}{CPT_uniform}{r_prob_se}{---}
\setres{cty_South_Africa}{CPT_uniform}{spread_penalty}{1.720}
\setres{cty_South_Africa}{CPT_uniform}{spread_penalty_se}{0.152}
\setres{cty_South_Africa}{CPT_uniform}{volume_ess}{---}
\setres{cty_South_Africa}{DA_deterministic}{kappa}{0.903}
\setres{cty_South_Africa}{DA_deterministic}{kappa_se}{0.035}
\setres{cty_South_Africa}{DA_deterministic}{mse}{14.16}
\setres{cty_South_Africa}{DA_deterministic}{nrmse_pct}{21.5}
\setres{cty_South_Africa}{DA_deterministic}{r_ff}{---}
\setres{cty_South_Africa}{DA_deterministic}{r_ff_se}{---}
\setres{cty_South_Africa}{DA_deterministic}{r_prior}{---}
\setres{cty_South_Africa}{DA_deterministic}{r_prior_se}{---}
\setres{cty_South_Africa}{DA_deterministic}{r_prob}{---}
\setres{cty_South_Africa}{DA_deterministic}{r_prob_se}{---}
\setres{cty_South_Africa}{DA_deterministic}{theta_star}{1.45, 1.28}
\setres{cty_South_Africa}{DA_deterministic}{volume_ess}{---}
\setres{cty_South_Africa}{DA_lit}{kappa}{-0.210}
\setres{cty_South_Africa}{DA_lit}{kappa_prior_mean}{0.260}
\setres{cty_South_Africa}{DA_lit}{kappa_prior_mean_se}{0.165}
\setres{cty_South_Africa}{DA_lit}{kappa_se}{0.197}
\setres{cty_South_Africa}{DA_lit}{mse}{25.40}
\setres{cty_South_Africa}{DA_lit}{mse_prior_mean}{20.65}
\setres{cty_South_Africa}{DA_lit}{nrmse_pct}{25.2}
\setres{cty_South_Africa}{DA_lit}{pred_variance}{4.75}
\setres{cty_South_Africa}{DA_lit}{r_ff}{---}
\setres{cty_South_Africa}{DA_lit}{r_ff_se}{---}
\setres{cty_South_Africa}{DA_lit}{r_prior}{---}
\setres{cty_South_Africa}{DA_lit}{r_prior_se}{---}
\setres{cty_South_Africa}{DA_lit}{r_prob}{---}
\setres{cty_South_Africa}{DA_lit}{r_prob_se}{---}
\setres{cty_South_Africa}{DA_lit}{spread_penalty}{0.470}
\setres{cty_South_Africa}{DA_lit}{spread_penalty_se}{0.041}
\setres{cty_South_Africa}{DA_lit}{volume_ess}{---}
\setres{cty_South_Africa}{DA_transfer_others}{kappa}{0.887}
\setres{cty_South_Africa}{DA_transfer_others}{kappa_prior_mean}{0.887}
\setres{cty_South_Africa}{DA_transfer_others}{kappa_prior_mean_se}{0.046}
\setres{cty_South_Africa}{DA_transfer_others}{kappa_se}{0.046}
\setres{cty_South_Africa}{DA_transfer_others}{mse}{14.32}
\setres{cty_South_Africa}{DA_transfer_others}{mse_prior_mean}{14.32}
\setres{cty_South_Africa}{DA_transfer_others}{nrmse_pct}{21.6}
\setres{cty_South_Africa}{DA_transfer_others}{pred_variance}{0.00}
\setres{cty_South_Africa}{DA_transfer_others}{r_ff}{---}
\setres{cty_South_Africa}{DA_transfer_others}{r_ff_se}{---}
\setres{cty_South_Africa}{DA_transfer_others}{r_prior}{---}
\setres{cty_South_Africa}{DA_transfer_others}{r_prior_se}{---}
\setres{cty_South_Africa}{DA_transfer_others}{r_prob}{---}
\setres{cty_South_Africa}{DA_transfer_others}{r_prob_se}{---}
\setres{cty_South_Africa}{DA_transfer_others}{spread_penalty}{0.000}
\setres{cty_South_Africa}{DA_transfer_others}{spread_penalty_se}{0.000}
\setres{cty_South_Africa}{DA_transfer_others}{theta_star}{1.70, 1.45}
\setres{cty_South_Africa}{DA_transfer_others}{transfer}{others}
\setres{cty_South_Africa}{DA_transfer_others}{volume_ess}{---}
\setres{cty_South_Africa}{DA_uniform}{kappa}{-1.779}
\setres{cty_South_Africa}{DA_uniform}{kappa_prior_mean}{-0.427}
\setres{cty_South_Africa}{DA_uniform}{kappa_prior_mean_se}{0.261}
\setres{cty_South_Africa}{DA_uniform}{kappa_se}{0.351}
\setres{cty_South_Africa}{DA_uniform}{mse}{41.23}
\setres{cty_South_Africa}{DA_uniform}{mse_prior_mean}{27.58}
\setres{cty_South_Africa}{DA_uniform}{nrmse_pct}{28.7}
\setres{cty_South_Africa}{DA_uniform}{pred_variance}{13.65}
\setres{cty_South_Africa}{DA_uniform}{r_ff}{---}
\setres{cty_South_Africa}{DA_uniform}{r_ff_se}{---}
\setres{cty_South_Africa}{DA_uniform}{r_prior}{---}
\setres{cty_South_Africa}{DA_uniform}{r_prior_se}{---}
\setres{cty_South_Africa}{DA_uniform}{r_prob}{---}
\setres{cty_South_Africa}{DA_uniform}{r_prob_se}{---}
\setres{cty_South_Africa}{DA_uniform}{spread_penalty}{1.352}
\setres{cty_South_Africa}{DA_uniform}{spread_penalty_se}{0.119}
\setres{cty_South_Africa}{DA_uniform}{volume_ess}{---}
\setres{cty_South_Africa}{EU_deterministic}{kappa}{0.840}
\setres{cty_South_Africa}{EU_deterministic}{kappa_se}{0.051}
\setres{cty_South_Africa}{EU_deterministic}{mse}{14.80}
\setres{cty_South_Africa}{EU_deterministic}{nrmse_pct}{22.4}
\setres{cty_South_Africa}{EU_deterministic}{r_ff}{---}
\setres{cty_South_Africa}{EU_deterministic}{r_ff_se}{---}
\setres{cty_South_Africa}{EU_deterministic}{r_prior}{---}
\setres{cty_South_Africa}{EU_deterministic}{r_prior_se}{---}
\setres{cty_South_Africa}{EU_deterministic}{r_prob}{---}
\setres{cty_South_Africa}{EU_deterministic}{r_prob_se}{---}
\setres{cty_South_Africa}{EU_deterministic}{theta_star}{0.84}
\setres{cty_South_Africa}{EU_deterministic}{volume_ess}{---}
\setres{cty_South_Africa}{EU_lit}{kappa}{0.646}
\setres{cty_South_Africa}{EU_lit}{kappa_prior_mean}{0.766}
\setres{cty_South_Africa}{EU_lit}{kappa_prior_mean_se}{0.065}
\setres{cty_South_Africa}{EU_lit}{kappa_se}{0.075}
\setres{cty_South_Africa}{EU_lit}{mse}{16.75}
\setres{cty_South_Africa}{EU_lit}{mse_prior_mean}{15.55}
\setres{cty_South_Africa}{EU_lit}{nrmse_pct}{23.6}
\setres{cty_South_Africa}{EU_lit}{pred_variance}{1.20}
\setres{cty_South_Africa}{EU_lit}{r_ff}{---}
\setres{cty_South_Africa}{EU_lit}{r_ff_se}{---}
\setres{cty_South_Africa}{EU_lit}{r_prior}{---}
\setres{cty_South_Africa}{EU_lit}{r_prior_se}{---}
\setres{cty_South_Africa}{EU_lit}{r_prob}{---}
\setres{cty_South_Africa}{EU_lit}{r_prob_se}{---}
\setres{cty_South_Africa}{EU_lit}{spread_penalty}{0.119}
\setres{cty_South_Africa}{EU_lit}{spread_penalty_se}{0.011}
\setres{cty_South_Africa}{EU_lit}{volume_ess}{---}
\setres{cty_South_Africa}{EU_transfer_others}{kappa}{0.813}
\setres{cty_South_Africa}{EU_transfer_others}{kappa_prior_mean}{0.813}
\setres{cty_South_Africa}{EU_transfer_others}{kappa_prior_mean_se}{0.067}
\setres{cty_South_Africa}{EU_transfer_others}{kappa_se}{0.067}
\setres{cty_South_Africa}{EU_transfer_others}{mse}{15.07}
\setres{cty_South_Africa}{EU_transfer_others}{mse_prior_mean}{15.07}
\setres{cty_South_Africa}{EU_transfer_others}{nrmse_pct}{22.1}
\setres{cty_South_Africa}{EU_transfer_others}{pred_variance}{0.00}
\setres{cty_South_Africa}{EU_transfer_others}{r_ff}{---}
\setres{cty_South_Africa}{EU_transfer_others}{r_ff_se}{---}
\setres{cty_South_Africa}{EU_transfer_others}{r_prior}{---}
\setres{cty_South_Africa}{EU_transfer_others}{r_prior_se}{---}
\setres{cty_South_Africa}{EU_transfer_others}{r_prob}{---}
\setres{cty_South_Africa}{EU_transfer_others}{r_prob_se}{---}
\setres{cty_South_Africa}{EU_transfer_others}{spread_penalty}{0.000}
\setres{cty_South_Africa}{EU_transfer_others}{spread_penalty_se}{0.000}
\setres{cty_South_Africa}{EU_transfer_others}{theta_star}{0.95}
\setres{cty_South_Africa}{EU_transfer_others}{transfer}{others}
\setres{cty_South_Africa}{EU_transfer_others}{volume_ess}{---}
\setres{cty_South_Africa}{EU_uniform}{kappa}{-0.683}
\setres{cty_South_Africa}{EU_uniform}{kappa_prior_mean}{0.129}
\setres{cty_South_Africa}{EU_uniform}{kappa_prior_mean_se}{0.180}
\setres{cty_South_Africa}{EU_uniform}{kappa_se}{0.234}
\setres{cty_South_Africa}{EU_uniform}{mse}{30.17}
\setres{cty_South_Africa}{EU_uniform}{mse_prior_mean}{21.98}
\setres{cty_South_Africa}{EU_uniform}{nrmse_pct}{25.7}
\setres{cty_South_Africa}{EU_uniform}{pred_variance}{8.20}
\setres{cty_South_Africa}{EU_uniform}{r_ff}{---}
\setres{cty_South_Africa}{EU_uniform}{r_ff_se}{---}
\setres{cty_South_Africa}{EU_uniform}{r_prior}{---}
\setres{cty_South_Africa}{EU_uniform}{r_prior_se}{---}
\setres{cty_South_Africa}{EU_uniform}{r_prob}{---}
\setres{cty_South_Africa}{EU_uniform}{r_prob_se}{---}
\setres{cty_South_Africa}{EU_uniform}{spread_penalty}{0.812}
\setres{cty_South_Africa}{EU_uniform}{spread_penalty_se}{0.072}
\setres{cty_South_Africa}{EU_uniform}{volume_ess}{---}
\setres{cty_South_Africa}{EV_deterministic}{kappa}{0.775}
\setres{cty_South_Africa}{EV_deterministic}{kappa_se}{0.079}
\setres{cty_South_Africa}{EV_deterministic}{mse}{15.46}
\setres{cty_South_Africa}{EV_deterministic}{nrmse_pct}{22.2}
\setres{cty_South_Africa}{EV_deterministic}{r_ff}{---}
\setres{cty_South_Africa}{EV_deterministic}{r_ff_se}{---}
\setres{cty_South_Africa}{EV_deterministic}{r_prior}{---}
\setres{cty_South_Africa}{EV_deterministic}{r_prior_se}{---}
\setres{cty_South_Africa}{EV_deterministic}{r_prob}{---}
\setres{cty_South_Africa}{EV_deterministic}{r_prob_se}{---}
\setres{cty_South_Africa}{EV_deterministic}{volume_ess}{---}
\setres{cty_South_Africa}{reference}{country_name}{South Africa}
\setres{cty_South_Africa}{reference}{lookup_nrmse_pct}{20.1}
\setres{cty_South_Africa}{reference}{midpoint_mse}{23.28}
\setres{cty_South_Africa}{reference}{midpoint_nrmse_pct}{26.6}
\setres{cty_South_Africa}{reference}{missing_lotteries}{none}
\setres{cty_South_Africa}{reference}{n_lotteries}{14}
\setres{cty_South_Africa}{reference}{n_obs}{994}
\setres{cty_South_Africa}{reference}{n_subjects}{71}
\setres{cty_South_Africa}{reference}{oracle_mse}{13.18}
\setres{cty_South_Africa}{reference}{sample}{LV2019\_South\_Africa}
\setres{cty_Spain}{CPT_deterministic}{kappa}{0.983}
\setres{cty_Spain}{CPT_deterministic}{kappa_se}{0.010}
\setres{cty_Spain}{CPT_deterministic}{mse}{7.92}
\setres{cty_Spain}{CPT_deterministic}{nrmse_pct}{15.5}
\setres{cty_Spain}{CPT_deterministic}{r_ff}{---}
\setres{cty_Spain}{CPT_deterministic}{r_ff_se}{---}
\setres{cty_Spain}{CPT_deterministic}{r_prior}{---}
\setres{cty_Spain}{CPT_deterministic}{r_prior_se}{---}
\setres{cty_Spain}{CPT_deterministic}{r_prob}{---}
\setres{cty_Spain}{CPT_deterministic}{r_prob_se}{---}
\setres{cty_Spain}{CPT_deterministic}{theta_star}{1.00, 0.80, 0.80}
\setres{cty_Spain}{CPT_deterministic}{volume_ess}{---}
\setres{cty_Spain}{CPT_lit}{kappa}{0.656}
\setres{cty_Spain}{CPT_lit}{kappa_prior_mean}{0.887}
\setres{cty_Spain}{CPT_lit}{kappa_prior_mean_se}{0.029}
\setres{cty_Spain}{CPT_lit}{kappa_se}{0.033}
\setres{cty_Spain}{CPT_lit}{mse}{12.02}
\setres{cty_Spain}{CPT_lit}{mse_prior_mean}{9.13}
\setres{cty_Spain}{CPT_lit}{nrmse_pct}{16.2}
\setres{cty_Spain}{CPT_lit}{pred_variance}{2.90}
\setres{cty_Spain}{CPT_lit}{r_ff}{---}
\setres{cty_Spain}{CPT_lit}{r_ff_se}{---}
\setres{cty_Spain}{CPT_lit}{r_prior}{---}
\setres{cty_Spain}{CPT_lit}{r_prior_se}{---}
\setres{cty_Spain}{CPT_lit}{r_prob}{---}
\setres{cty_Spain}{CPT_lit}{r_prob_se}{---}
\setres{cty_Spain}{CPT_lit}{spread_penalty}{0.231}
\setres{cty_Spain}{CPT_lit}{spread_penalty_se}{0.014}
\setres{cty_Spain}{CPT_lit}{volume_ess}{---}
\setres{cty_Spain}{CPT_transfer_others}{kappa}{0.932}
\setres{cty_Spain}{CPT_transfer_others}{kappa_prior_mean}{0.932}
\setres{cty_Spain}{CPT_transfer_others}{kappa_prior_mean_se}{0.011}
\setres{cty_Spain}{CPT_transfer_others}{kappa_se}{0.011}
\setres{cty_Spain}{CPT_transfer_others}{mse}{8.56}
\setres{cty_Spain}{CPT_transfer_others}{mse_prior_mean}{8.56}
\setres{cty_Spain}{CPT_transfer_others}{nrmse_pct}{14.7}
\setres{cty_Spain}{CPT_transfer_others}{pred_variance}{0.00}
\setres{cty_Spain}{CPT_transfer_others}{r_ff}{---}
\setres{cty_Spain}{CPT_transfer_others}{r_ff_se}{---}
\setres{cty_Spain}{CPT_transfer_others}{r_prior}{---}
\setres{cty_Spain}{CPT_transfer_others}{r_prior_se}{---}
\setres{cty_Spain}{CPT_transfer_others}{r_prob}{---}
\setres{cty_Spain}{CPT_transfer_others}{r_prob_se}{---}
\setres{cty_Spain}{CPT_transfer_others}{spread_penalty}{0.000}
\setres{cty_Spain}{CPT_transfer_others}{spread_penalty_se}{0.000}
\setres{cty_Spain}{CPT_transfer_others}{theta_star}{1.10, 0.60, 0.80}
\setres{cty_Spain}{CPT_transfer_others}{transfer}{others}
\setres{cty_Spain}{CPT_transfer_others}{volume_ess}{---}
\setres{cty_Spain}{CPT_uniform}{kappa}{-1.044}
\setres{cty_Spain}{CPT_uniform}{kappa_prior_mean}{0.342}
\setres{cty_Spain}{CPT_uniform}{kappa_prior_mean_se}{0.075}
\setres{cty_Spain}{CPT_uniform}{kappa_se}{0.116}
\setres{cty_Spain}{CPT_uniform}{mse}{33.32}
\setres{cty_Spain}{CPT_uniform}{mse_prior_mean}{15.96}
\setres{cty_Spain}{CPT_uniform}{nrmse_pct}{22.4}
\setres{cty_Spain}{CPT_uniform}{pred_variance}{17.37}
\setres{cty_Spain}{CPT_uniform}{r_ff}{---}
\setres{cty_Spain}{CPT_uniform}{r_ff_se}{---}
\setres{cty_Spain}{CPT_uniform}{r_prior}{---}
\setres{cty_Spain}{CPT_uniform}{r_prior_se}{---}
\setres{cty_Spain}{CPT_uniform}{r_prob}{---}
\setres{cty_Spain}{CPT_uniform}{r_prob_se}{---}
\setres{cty_Spain}{CPT_uniform}{spread_penalty}{1.386}
\setres{cty_Spain}{CPT_uniform}{spread_penalty_se}{0.083}
\setres{cty_Spain}{CPT_uniform}{volume_ess}{---}
\setres{cty_Spain}{DA_deterministic}{kappa}{0.957}
\setres{cty_Spain}{DA_deterministic}{kappa_se}{0.013}
\setres{cty_Spain}{DA_deterministic}{mse}{8.25}
\setres{cty_Spain}{DA_deterministic}{nrmse_pct}{16.3}
\setres{cty_Spain}{DA_deterministic}{r_ff}{---}
\setres{cty_Spain}{DA_deterministic}{r_ff_se}{---}
\setres{cty_Spain}{DA_deterministic}{r_prior}{---}
\setres{cty_Spain}{DA_deterministic}{r_prior_se}{---}
\setres{cty_Spain}{DA_deterministic}{r_prob}{---}
\setres{cty_Spain}{DA_deterministic}{r_prob_se}{---}
\setres{cty_Spain}{DA_deterministic}{theta_star}{1.26, 0.74}
\setres{cty_Spain}{DA_deterministic}{volume_ess}{---}
\setres{cty_Spain}{DA_lit}{kappa}{-0.031}
\setres{cty_Spain}{DA_lit}{kappa_prior_mean}{0.348}
\setres{cty_Spain}{DA_lit}{kappa_prior_mean_se}{0.089}
\setres{cty_Spain}{DA_lit}{kappa_se}{0.101}
\setres{cty_Spain}{DA_lit}{mse}{20.63}
\setres{cty_Spain}{DA_lit}{mse_prior_mean}{15.88}
\setres{cty_Spain}{DA_lit}{nrmse_pct}{21.1}
\setres{cty_Spain}{DA_lit}{pred_variance}{4.75}
\setres{cty_Spain}{DA_lit}{r_ff}{---}
\setres{cty_Spain}{DA_lit}{r_ff_se}{---}
\setres{cty_Spain}{DA_lit}{r_prior}{---}
\setres{cty_Spain}{DA_lit}{r_prior_se}{---}
\setres{cty_Spain}{DA_lit}{r_prob}{---}
\setres{cty_Spain}{DA_lit}{r_prob_se}{---}
\setres{cty_Spain}{DA_lit}{spread_penalty}{0.379}
\setres{cty_Spain}{DA_lit}{spread_penalty_se}{0.023}
\setres{cty_Spain}{DA_lit}{volume_ess}{---}
\setres{cty_Spain}{DA_transfer_others}{kappa}{0.936}
\setres{cty_Spain}{DA_transfer_others}{kappa_prior_mean}{0.936}
\setres{cty_Spain}{DA_transfer_others}{kappa_prior_mean_se}{0.017}
\setres{cty_Spain}{DA_transfer_others}{kappa_se}{0.017}
\setres{cty_Spain}{DA_transfer_others}{mse}{8.52}
\setres{cty_Spain}{DA_transfer_others}{mse_prior_mean}{8.52}
\setres{cty_Spain}{DA_transfer_others}{nrmse_pct}{15.8}
\setres{cty_Spain}{DA_transfer_others}{pred_variance}{0.00}
\setres{cty_Spain}{DA_transfer_others}{r_ff}{---}
\setres{cty_Spain}{DA_transfer_others}{r_ff_se}{---}
\setres{cty_Spain}{DA_transfer_others}{r_prior}{---}
\setres{cty_Spain}{DA_transfer_others}{r_prior_se}{---}
\setres{cty_Spain}{DA_transfer_others}{r_prob}{---}
\setres{cty_Spain}{DA_transfer_others}{r_prob_se}{---}
\setres{cty_Spain}{DA_transfer_others}{spread_penalty}{0.000}
\setres{cty_Spain}{DA_transfer_others}{spread_penalty_se}{0.000}
\setres{cty_Spain}{DA_transfer_others}{theta_star}{1.71, 1.48}
\setres{cty_Spain}{DA_transfer_others}{transfer}{others}
\setres{cty_Spain}{DA_transfer_others}{volume_ess}{---}
\setres{cty_Spain}{DA_uniform}{kappa}{-1.379}
\setres{cty_Spain}{DA_uniform}{kappa_prior_mean}{-0.290}
\setres{cty_Spain}{DA_uniform}{kappa_prior_mean_se}{0.139}
\setres{cty_Spain}{DA_uniform}{kappa_se}{0.176}
\setres{cty_Spain}{DA_uniform}{mse}{37.52}
\setres{cty_Spain}{DA_uniform}{mse_prior_mean}{23.87}
\setres{cty_Spain}{DA_uniform}{nrmse_pct}{28.5}
\setres{cty_Spain}{DA_uniform}{pred_variance}{13.65}
\setres{cty_Spain}{DA_uniform}{r_ff}{---}
\setres{cty_Spain}{DA_uniform}{r_ff_se}{---}
\setres{cty_Spain}{DA_uniform}{r_prior}{---}
\setres{cty_Spain}{DA_uniform}{r_prior_se}{---}
\setres{cty_Spain}{DA_uniform}{r_prob}{---}
\setres{cty_Spain}{DA_uniform}{r_prob_se}{---}
\setres{cty_Spain}{DA_uniform}{spread_penalty}{1.089}
\setres{cty_Spain}{DA_uniform}{spread_penalty_se}{0.066}
\setres{cty_Spain}{DA_uniform}{volume_ess}{---}
\setres{cty_Spain}{EU_deterministic}{kappa}{0.930}
\setres{cty_Spain}{EU_deterministic}{kappa_se}{0.024}
\setres{cty_Spain}{EU_deterministic}{mse}{8.59}
\setres{cty_Spain}{EU_deterministic}{nrmse_pct}{16.6}
\setres{cty_Spain}{EU_deterministic}{r_ff}{---}
\setres{cty_Spain}{EU_deterministic}{r_ff_se}{---}
\setres{cty_Spain}{EU_deterministic}{r_prior}{---}
\setres{cty_Spain}{EU_deterministic}{r_prior_se}{---}
\setres{cty_Spain}{EU_deterministic}{r_prob}{---}
\setres{cty_Spain}{EU_deterministic}{r_prob_se}{---}
\setres{cty_Spain}{EU_deterministic}{theta_star}{0.87}
\setres{cty_Spain}{EU_deterministic}{volume_ess}{---}
\setres{cty_Spain}{EU_lit}{kappa}{0.733}
\setres{cty_Spain}{EU_lit}{kappa_prior_mean}{0.829}
\setres{cty_Spain}{EU_lit}{kappa_prior_mean_se}{0.034}
\setres{cty_Spain}{EU_lit}{kappa_se}{0.038}
\setres{cty_Spain}{EU_lit}{mse}{11.06}
\setres{cty_Spain}{EU_lit}{mse_prior_mean}{9.85}
\setres{cty_Spain}{EU_lit}{nrmse_pct}{17.6}
\setres{cty_Spain}{EU_lit}{pred_variance}{1.20}
\setres{cty_Spain}{EU_lit}{r_ff}{---}
\setres{cty_Spain}{EU_lit}{r_ff_se}{---}
\setres{cty_Spain}{EU_lit}{r_prior}{---}
\setres{cty_Spain}{EU_lit}{r_prior_se}{---}
\setres{cty_Spain}{EU_lit}{r_prob}{---}
\setres{cty_Spain}{EU_lit}{r_prob_se}{---}
\setres{cty_Spain}{EU_lit}{spread_penalty}{0.096}
\setres{cty_Spain}{EU_lit}{spread_penalty_se}{0.006}
\setres{cty_Spain}{EU_lit}{volume_ess}{---}
\setres{cty_Spain}{EU_transfer_others}{kappa}{0.916}
\setres{cty_Spain}{EU_transfer_others}{kappa_prior_mean}{0.916}
\setres{cty_Spain}{EU_transfer_others}{kappa_prior_mean_se}{0.034}
\setres{cty_Spain}{EU_transfer_others}{kappa_se}{0.034}
\setres{cty_Spain}{EU_transfer_others}{mse}{8.77}
\setres{cty_Spain}{EU_transfer_others}{mse_prior_mean}{8.77}
\setres{cty_Spain}{EU_transfer_others}{nrmse_pct}{16.3}
\setres{cty_Spain}{EU_transfer_others}{pred_variance}{0.00}
\setres{cty_Spain}{EU_transfer_others}{r_ff}{---}
\setres{cty_Spain}{EU_transfer_others}{r_ff_se}{---}
\setres{cty_Spain}{EU_transfer_others}{r_prior}{---}
\setres{cty_Spain}{EU_transfer_others}{r_prior_se}{---}
\setres{cty_Spain}{EU_transfer_others}{r_prob}{---}
\setres{cty_Spain}{EU_transfer_others}{r_prob_se}{---}
\setres{cty_Spain}{EU_transfer_others}{spread_penalty}{0.000}
\setres{cty_Spain}{EU_transfer_others}{spread_penalty_se}{0.000}
\setres{cty_Spain}{EU_transfer_others}{theta_star}{0.95}
\setres{cty_Spain}{EU_transfer_others}{transfer}{others}
\setres{cty_Spain}{EU_transfer_others}{volume_ess}{---}
\setres{cty_Spain}{EU_uniform}{kappa}{-0.429}
\setres{cty_Spain}{EU_uniform}{kappa_prior_mean}{0.225}
\setres{cty_Spain}{EU_uniform}{kappa_prior_mean_se}{0.096}
\setres{cty_Spain}{EU_uniform}{kappa_se}{0.118}
\setres{cty_Spain}{EU_uniform}{mse}{25.62}
\setres{cty_Spain}{EU_uniform}{mse_prior_mean}{17.42}
\setres{cty_Spain}{EU_uniform}{nrmse_pct}{22.6}
\setres{cty_Spain}{EU_uniform}{pred_variance}{8.20}
\setres{cty_Spain}{EU_uniform}{r_ff}{---}
\setres{cty_Spain}{EU_uniform}{r_ff_se}{---}
\setres{cty_Spain}{EU_uniform}{r_prior}{---}
\setres{cty_Spain}{EU_uniform}{r_prior_se}{---}
\setres{cty_Spain}{EU_uniform}{r_prob}{---}
\setres{cty_Spain}{EU_uniform}{r_prob_se}{---}
\setres{cty_Spain}{EU_uniform}{spread_penalty}{0.654}
\setres{cty_Spain}{EU_uniform}{spread_penalty_se}{0.039}
\setres{cty_Spain}{EU_uniform}{volume_ess}{---}
\setres{cty_Spain}{EV_deterministic}{kappa}{0.890}
\setres{cty_Spain}{EV_deterministic}{kappa_se}{0.042}
\setres{cty_Spain}{EV_deterministic}{mse}{9.09}
\setres{cty_Spain}{EV_deterministic}{nrmse_pct}{16.3}
\setres{cty_Spain}{EV_deterministic}{r_ff}{---}
\setres{cty_Spain}{EV_deterministic}{r_ff_se}{---}
\setres{cty_Spain}{EV_deterministic}{r_prior}{---}
\setres{cty_Spain}{EV_deterministic}{r_prior_se}{---}
\setres{cty_Spain}{EV_deterministic}{r_prob}{---}
\setres{cty_Spain}{EV_deterministic}{r_prob_se}{---}
\setres{cty_Spain}{EV_deterministic}{volume_ess}{---}
\setres{cty_Spain}{reference}{country_name}{Spain}
\setres{cty_Spain}{reference}{lookup_nrmse_pct}{14.4}
\setres{cty_Spain}{reference}{midpoint_mse}{20.24}
\setres{cty_Spain}{reference}{midpoint_nrmse_pct}{23.9}
\setres{cty_Spain}{reference}{missing_lotteries}{none}
\setres{cty_Spain}{reference}{n_lotteries}{14}
\setres{cty_Spain}{reference}{n_obs}{1{,}120}
\setres{cty_Spain}{reference}{n_subjects}{80}
\setres{cty_Spain}{reference}{oracle_mse}{7.71}
\setres{cty_Spain}{reference}{sample}{LV2019\_Spain}
\setres{cty_Thailand}{CPT_deterministic}{kappa}{0.974}
\setres{cty_Thailand}{CPT_deterministic}{kappa_se}{0.011}
\setres{cty_Thailand}{CPT_deterministic}{mse}{15.55}
\setres{cty_Thailand}{CPT_deterministic}{nrmse_pct}{21.1}
\setres{cty_Thailand}{CPT_deterministic}{r_ff}{---}
\setres{cty_Thailand}{CPT_deterministic}{r_ff_se}{---}
\setres{cty_Thailand}{CPT_deterministic}{r_prior}{---}
\setres{cty_Thailand}{CPT_deterministic}{r_prior_se}{---}
\setres{cty_Thailand}{CPT_deterministic}{r_prob}{---}
\setres{cty_Thailand}{CPT_deterministic}{r_prob_se}{---}
\setres{cty_Thailand}{CPT_deterministic}{theta_star}{0.80, 0.70, 1.10}
\setres{cty_Thailand}{CPT_deterministic}{volume_ess}{---}
\setres{cty_Thailand}{CPT_lit}{kappa}{0.642}
\setres{cty_Thailand}{CPT_lit}{kappa_prior_mean}{0.886}
\setres{cty_Thailand}{CPT_lit}{kappa_prior_mean_se}{0.061}
\setres{cty_Thailand}{CPT_lit}{kappa_se}{0.070}
\setres{cty_Thailand}{CPT_lit}{mse}{19.50}
\setres{cty_Thailand}{CPT_lit}{mse_prior_mean}{16.60}
\setres{cty_Thailand}{CPT_lit}{nrmse_pct}{22.4}
\setres{cty_Thailand}{CPT_lit}{pred_variance}{2.90}
\setres{cty_Thailand}{CPT_lit}{r_ff}{---}
\setres{cty_Thailand}{CPT_lit}{r_ff_se}{---}
\setres{cty_Thailand}{CPT_lit}{r_prior}{---}
\setres{cty_Thailand}{CPT_lit}{r_prior_se}{---}
\setres{cty_Thailand}{CPT_lit}{r_prob}{---}
\setres{cty_Thailand}{CPT_lit}{r_prob_se}{---}
\setres{cty_Thailand}{CPT_lit}{spread_penalty}{0.243}
\setres{cty_Thailand}{CPT_lit}{spread_penalty_se}{0.022}
\setres{cty_Thailand}{CPT_lit}{volume_ess}{---}
\setres{cty_Thailand}{CPT_transfer_others}{kappa}{0.943}
\setres{cty_Thailand}{CPT_transfer_others}{kappa_prior_mean}{0.943}
\setres{cty_Thailand}{CPT_transfer_others}{kappa_prior_mean_se}{0.019}
\setres{cty_Thailand}{CPT_transfer_others}{kappa_se}{0.019}
\setres{cty_Thailand}{CPT_transfer_others}{mse}{15.91}
\setres{cty_Thailand}{CPT_transfer_others}{mse_prior_mean}{15.91}
\setres{cty_Thailand}{CPT_transfer_others}{nrmse_pct}{21.4}
\setres{cty_Thailand}{CPT_transfer_others}{pred_variance}{0.00}
\setres{cty_Thailand}{CPT_transfer_others}{r_ff}{---}
\setres{cty_Thailand}{CPT_transfer_others}{r_ff_se}{---}
\setres{cty_Thailand}{CPT_transfer_others}{r_prior}{---}
\setres{cty_Thailand}{CPT_transfer_others}{r_prior_se}{---}
\setres{cty_Thailand}{CPT_transfer_others}{r_prob}{---}
\setres{cty_Thailand}{CPT_transfer_others}{r_prob_se}{---}
\setres{cty_Thailand}{CPT_transfer_others}{spread_penalty}{0.000}
\setres{cty_Thailand}{CPT_transfer_others}{spread_penalty_se}{0.000}
\setres{cty_Thailand}{CPT_transfer_others}{theta_star}{1.10, 0.60, 0.80}
\setres{cty_Thailand}{CPT_transfer_others}{transfer}{others}
\setres{cty_Thailand}{CPT_transfer_others}{volume_ess}{---}
\setres{cty_Thailand}{CPT_uniform}{kappa}{-1.153}
\setres{cty_Thailand}{CPT_uniform}{kappa_prior_mean}{0.305}
\setres{cty_Thailand}{CPT_uniform}{kappa_prior_mean_se}{0.135}
\setres{cty_Thailand}{CPT_uniform}{kappa_se}{0.217}
\setres{cty_Thailand}{CPT_uniform}{mse}{40.88}
\setres{cty_Thailand}{CPT_uniform}{mse_prior_mean}{23.51}
\setres{cty_Thailand}{CPT_uniform}{nrmse_pct}{26.8}
\setres{cty_Thailand}{CPT_uniform}{pred_variance}{17.37}
\setres{cty_Thailand}{CPT_uniform}{r_ff}{---}
\setres{cty_Thailand}{CPT_uniform}{r_ff_se}{---}
\setres{cty_Thailand}{CPT_uniform}{r_prior}{---}
\setres{cty_Thailand}{CPT_uniform}{r_prior_se}{---}
\setres{cty_Thailand}{CPT_uniform}{r_prob}{---}
\setres{cty_Thailand}{CPT_uniform}{r_prob_se}{---}
\setres{cty_Thailand}{CPT_uniform}{spread_penalty}{1.458}
\setres{cty_Thailand}{CPT_uniform}{spread_penalty_se}{0.131}
\setres{cty_Thailand}{CPT_uniform}{volume_ess}{---}
\setres{cty_Thailand}{DA_deterministic}{kappa}{0.941}
\setres{cty_Thailand}{DA_deterministic}{kappa_se}{0.023}
\setres{cty_Thailand}{DA_deterministic}{mse}{15.94}
\setres{cty_Thailand}{DA_deterministic}{nrmse_pct}{21.5}
\setres{cty_Thailand}{DA_deterministic}{r_ff}{---}
\setres{cty_Thailand}{DA_deterministic}{r_ff_se}{---}
\setres{cty_Thailand}{DA_deterministic}{r_prior}{---}
\setres{cty_Thailand}{DA_deterministic}{r_prior_se}{---}
\setres{cty_Thailand}{DA_deterministic}{r_prob}{---}
\setres{cty_Thailand}{DA_deterministic}{r_prob_se}{---}
\setres{cty_Thailand}{DA_deterministic}{theta_star}{1.03, 0.24}
\setres{cty_Thailand}{DA_deterministic}{volume_ess}{---}
\setres{cty_Thailand}{DA_lit}{kappa}{-0.134}
\setres{cty_Thailand}{DA_lit}{kappa_prior_mean}{0.265}
\setres{cty_Thailand}{DA_lit}{kappa_prior_mean_se}{0.174}
\setres{cty_Thailand}{DA_lit}{kappa_se}{0.196}
\setres{cty_Thailand}{DA_lit}{mse}{28.74}
\setres{cty_Thailand}{DA_lit}{mse_prior_mean}{23.99}
\setres{cty_Thailand}{DA_lit}{nrmse_pct}{26.7}
\setres{cty_Thailand}{DA_lit}{pred_variance}{4.75}
\setres{cty_Thailand}{DA_lit}{r_ff}{---}
\setres{cty_Thailand}{DA_lit}{r_ff_se}{---}
\setres{cty_Thailand}{DA_lit}{r_prior}{---}
\setres{cty_Thailand}{DA_lit}{r_prior_se}{---}
\setres{cty_Thailand}{DA_lit}{r_prob}{---}
\setres{cty_Thailand}{DA_lit}{r_prob_se}{---}
\setres{cty_Thailand}{DA_lit}{spread_penalty}{0.399}
\setres{cty_Thailand}{DA_lit}{spread_penalty_se}{0.036}
\setres{cty_Thailand}{DA_lit}{volume_ess}{---}
\setres{cty_Thailand}{DA_transfer_others}{kappa}{0.905}
\setres{cty_Thailand}{DA_transfer_others}{kappa_prior_mean}{0.905}
\setres{cty_Thailand}{DA_transfer_others}{kappa_prior_mean_se}{0.024}
\setres{cty_Thailand}{DA_transfer_others}{kappa_se}{0.024}
\setres{cty_Thailand}{DA_transfer_others}{mse}{16.37}
\setres{cty_Thailand}{DA_transfer_others}{mse_prior_mean}{16.37}
\setres{cty_Thailand}{DA_transfer_others}{nrmse_pct}{21.4}
\setres{cty_Thailand}{DA_transfer_others}{pred_variance}{0.00}
\setres{cty_Thailand}{DA_transfer_others}{r_ff}{---}
\setres{cty_Thailand}{DA_transfer_others}{r_ff_se}{---}
\setres{cty_Thailand}{DA_transfer_others}{r_prior}{---}
\setres{cty_Thailand}{DA_transfer_others}{r_prior_se}{---}
\setres{cty_Thailand}{DA_transfer_others}{r_prob}{---}
\setres{cty_Thailand}{DA_transfer_others}{r_prob_se}{---}
\setres{cty_Thailand}{DA_transfer_others}{spread_penalty}{0.000}
\setres{cty_Thailand}{DA_transfer_others}{spread_penalty_se}{0.000}
\setres{cty_Thailand}{DA_transfer_others}{theta_star}{1.72, 1.50}
\setres{cty_Thailand}{DA_transfer_others}{transfer}{others}
\setres{cty_Thailand}{DA_transfer_others}{volume_ess}{---}
\setres{cty_Thailand}{DA_uniform}{kappa}{-1.597}
\setres{cty_Thailand}{DA_uniform}{kappa_prior_mean}{-0.451}
\setres{cty_Thailand}{DA_uniform}{kappa_prior_mean_se}{0.258}
\setres{cty_Thailand}{DA_uniform}{kappa_se}{0.327}
\setres{cty_Thailand}{DA_uniform}{mse}{46.17}
\setres{cty_Thailand}{DA_uniform}{mse_prior_mean}{32.52}
\setres{cty_Thailand}{DA_uniform}{nrmse_pct}{31.4}
\setres{cty_Thailand}{DA_uniform}{pred_variance}{13.65}
\setres{cty_Thailand}{DA_uniform}{r_ff}{---}
\setres{cty_Thailand}{DA_uniform}{r_ff_se}{---}
\setres{cty_Thailand}{DA_uniform}{r_prior}{---}
\setres{cty_Thailand}{DA_uniform}{r_prior_se}{---}
\setres{cty_Thailand}{DA_uniform}{r_prob}{---}
\setres{cty_Thailand}{DA_uniform}{r_prob_se}{---}
\setres{cty_Thailand}{DA_uniform}{spread_penalty}{1.146}
\setres{cty_Thailand}{DA_uniform}{spread_penalty_se}{0.103}
\setres{cty_Thailand}{DA_uniform}{volume_ess}{---}
\setres{cty_Thailand}{EU_deterministic}{kappa}{0.942}
\setres{cty_Thailand}{EU_deterministic}{kappa_se}{0.028}
\setres{cty_Thailand}{EU_deterministic}{mse}{15.93}
\setres{cty_Thailand}{EU_deterministic}{nrmse_pct}{21.7}
\setres{cty_Thailand}{EU_deterministic}{r_ff}{---}
\setres{cty_Thailand}{EU_deterministic}{r_ff_se}{---}
\setres{cty_Thailand}{EU_deterministic}{r_prior}{---}
\setres{cty_Thailand}{EU_deterministic}{r_prior_se}{---}
\setres{cty_Thailand}{EU_deterministic}{r_prob}{---}
\setres{cty_Thailand}{EU_deterministic}{r_prob_se}{---}
\setres{cty_Thailand}{EU_deterministic}{theta_star}{0.88}
\setres{cty_Thailand}{EU_deterministic}{volume_ess}{---}
\setres{cty_Thailand}{EU_lit}{kappa}{0.718}
\setres{cty_Thailand}{EU_lit}{kappa_prior_mean}{0.820}
\setres{cty_Thailand}{EU_lit}{kappa_prior_mean_se}{0.072}
\setres{cty_Thailand}{EU_lit}{kappa_se}{0.078}
\setres{cty_Thailand}{EU_lit}{mse}{18.59}
\setres{cty_Thailand}{EU_lit}{mse_prior_mean}{17.38}
\setres{cty_Thailand}{EU_lit}{nrmse_pct}{23.1}
\setres{cty_Thailand}{EU_lit}{pred_variance}{1.20}
\setres{cty_Thailand}{EU_lit}{r_ff}{---}
\setres{cty_Thailand}{EU_lit}{r_ff_se}{---}
\setres{cty_Thailand}{EU_lit}{r_prior}{---}
\setres{cty_Thailand}{EU_lit}{r_prior_se}{---}
\setres{cty_Thailand}{EU_lit}{r_prob}{---}
\setres{cty_Thailand}{EU_lit}{r_prob_se}{---}
\setres{cty_Thailand}{EU_lit}{spread_penalty}{0.101}
\setres{cty_Thailand}{EU_lit}{spread_penalty_se}{0.009}
\setres{cty_Thailand}{EU_lit}{volume_ess}{---}
\setres{cty_Thailand}{EU_transfer_others}{kappa}{0.941}
\setres{cty_Thailand}{EU_transfer_others}{kappa_prior_mean}{0.941}
\setres{cty_Thailand}{EU_transfer_others}{kappa_prior_mean_se}{0.035}
\setres{cty_Thailand}{EU_transfer_others}{kappa_se}{0.035}
\setres{cty_Thailand}{EU_transfer_others}{mse}{15.94}
\setres{cty_Thailand}{EU_transfer_others}{mse_prior_mean}{15.94}
\setres{cty_Thailand}{EU_transfer_others}{nrmse_pct}{21.8}
\setres{cty_Thailand}{EU_transfer_others}{pred_variance}{0.00}
\setres{cty_Thailand}{EU_transfer_others}{r_ff}{---}
\setres{cty_Thailand}{EU_transfer_others}{r_ff_se}{---}
\setres{cty_Thailand}{EU_transfer_others}{r_prior}{---}
\setres{cty_Thailand}{EU_transfer_others}{r_prior_se}{---}
\setres{cty_Thailand}{EU_transfer_others}{r_prob}{---}
\setres{cty_Thailand}{EU_transfer_others}{r_prob_se}{---}
\setres{cty_Thailand}{EU_transfer_others}{spread_penalty}{0.000}
\setres{cty_Thailand}{EU_transfer_others}{spread_penalty_se}{0.000}
\setres{cty_Thailand}{EU_transfer_others}{theta_star}{0.95}
\setres{cty_Thailand}{EU_transfer_others}{transfer}{others}
\setres{cty_Thailand}{EU_transfer_others}{volume_ess}{---}
\setres{cty_Thailand}{EU_uniform}{kappa}{-0.545}
\setres{cty_Thailand}{EU_uniform}{kappa_prior_mean}{0.143}
\setres{cty_Thailand}{EU_uniform}{kappa_prior_mean_se}{0.186}
\setres{cty_Thailand}{EU_uniform}{kappa_se}{0.227}
\setres{cty_Thailand}{EU_uniform}{mse}{33.64}
\setres{cty_Thailand}{EU_uniform}{mse_prior_mean}{25.45}
\setres{cty_Thailand}{EU_uniform}{nrmse_pct}{27.3}
\setres{cty_Thailand}{EU_uniform}{pred_variance}{8.20}
\setres{cty_Thailand}{EU_uniform}{r_ff}{---}
\setres{cty_Thailand}{EU_uniform}{r_ff_se}{---}
\setres{cty_Thailand}{EU_uniform}{r_prior}{---}
\setres{cty_Thailand}{EU_uniform}{r_prior_se}{---}
\setres{cty_Thailand}{EU_uniform}{r_prob}{---}
\setres{cty_Thailand}{EU_uniform}{r_prob_se}{---}
\setres{cty_Thailand}{EU_uniform}{spread_penalty}{0.688}
\setres{cty_Thailand}{EU_uniform}{spread_penalty_se}{0.062}
\setres{cty_Thailand}{EU_uniform}{volume_ess}{---}
\setres{cty_Thailand}{EV_deterministic}{kappa}{0.919}
\setres{cty_Thailand}{EV_deterministic}{kappa_se}{0.046}
\setres{cty_Thailand}{EV_deterministic}{mse}{16.20}
\setres{cty_Thailand}{EV_deterministic}{nrmse_pct}{22.0}
\setres{cty_Thailand}{EV_deterministic}{r_ff}{---}
\setres{cty_Thailand}{EV_deterministic}{r_ff_se}{---}
\setres{cty_Thailand}{EV_deterministic}{r_prior}{---}
\setres{cty_Thailand}{EV_deterministic}{r_prior_se}{---}
\setres{cty_Thailand}{EV_deterministic}{r_prob}{---}
\setres{cty_Thailand}{EV_deterministic}{r_prob_se}{---}
\setres{cty_Thailand}{EV_deterministic}{volume_ess}{---}
\setres{cty_Thailand}{reference}{country_name}{Thailand}
\setres{cty_Thailand}{reference}{lookup_nrmse_pct}{21.0}
\setres{cty_Thailand}{reference}{midpoint_mse}{27.15}
\setres{cty_Thailand}{reference}{midpoint_nrmse_pct}{28.0}
\setres{cty_Thailand}{reference}{missing_lotteries}{none}
\setres{cty_Thailand}{reference}{n_lotteries}{14}
\setres{cty_Thailand}{reference}{n_obs}{1{,}106}
\setres{cty_Thailand}{reference}{n_subjects}{79}
\setres{cty_Thailand}{reference}{oracle_mse}{15.24}
\setres{cty_Thailand}{reference}{sample}{LV2019\_Thailand}
\setres{cty_Tunisia}{CPT_deterministic}{kappa}{0.969}
\setres{cty_Tunisia}{CPT_deterministic}{kappa_se}{0.031}
\setres{cty_Tunisia}{CPT_deterministic}{mse}{19.42}
\setres{cty_Tunisia}{CPT_deterministic}{nrmse_pct}{24.6}
\setres{cty_Tunisia}{CPT_deterministic}{r_ff}{---}
\setres{cty_Tunisia}{CPT_deterministic}{r_ff_se}{---}
\setres{cty_Tunisia}{CPT_deterministic}{r_prior}{---}
\setres{cty_Tunisia}{CPT_deterministic}{r_prior_se}{---}
\setres{cty_Tunisia}{CPT_deterministic}{r_prob}{---}
\setres{cty_Tunisia}{CPT_deterministic}{r_prob_se}{---}
\setres{cty_Tunisia}{CPT_deterministic}{theta_star}{1.50, 0.60, 0.50}
\setres{cty_Tunisia}{CPT_deterministic}{volume_ess}{---}
\setres{cty_Tunisia}{CPT_lit}{kappa}{-0.107}
\setres{cty_Tunisia}{CPT_lit}{kappa_prior_mean}{0.408}
\setres{cty_Tunisia}{CPT_lit}{kappa_prior_mean_se}{0.226}
\setres{cty_Tunisia}{CPT_lit}{kappa_se}{0.302}
\setres{cty_Tunisia}{CPT_lit}{mse}{25.48}
\setres{cty_Tunisia}{CPT_lit}{mse_prior_mean}{22.58}
\setres{cty_Tunisia}{CPT_lit}{nrmse_pct}{26.8}
\setres{cty_Tunisia}{CPT_lit}{pred_variance}{2.90}
\setres{cty_Tunisia}{CPT_lit}{r_ff}{---}
\setres{cty_Tunisia}{CPT_lit}{r_ff_se}{---}
\setres{cty_Tunisia}{CPT_lit}{r_prior}{---}
\setres{cty_Tunisia}{CPT_lit}{r_prior_se}{---}
\setres{cty_Tunisia}{CPT_lit}{r_prob}{---}
\setres{cty_Tunisia}{CPT_lit}{r_prob_se}{---}
\setres{cty_Tunisia}{CPT_lit}{spread_penalty}{0.515}
\setres{cty_Tunisia}{CPT_lit}{spread_penalty_se}{0.087}
\setres{cty_Tunisia}{CPT_lit}{volume_ess}{---}
\setres{cty_Tunisia}{CPT_transfer_others}{kappa}{0.966}
\setres{cty_Tunisia}{CPT_transfer_others}{kappa_prior_mean}{0.966}
\setres{cty_Tunisia}{CPT_transfer_others}{kappa_prior_mean_se}{0.060}
\setres{cty_Tunisia}{CPT_transfer_others}{kappa_se}{0.060}
\setres{cty_Tunisia}{CPT_transfer_others}{mse}{19.44}
\setres{cty_Tunisia}{CPT_transfer_others}{mse_prior_mean}{19.44}
\setres{cty_Tunisia}{CPT_transfer_others}{nrmse_pct}{24.9}
\setres{cty_Tunisia}{CPT_transfer_others}{pred_variance}{0.00}
\setres{cty_Tunisia}{CPT_transfer_others}{r_ff}{---}
\setres{cty_Tunisia}{CPT_transfer_others}{r_ff_se}{---}
\setres{cty_Tunisia}{CPT_transfer_others}{r_prior}{---}
\setres{cty_Tunisia}{CPT_transfer_others}{r_prior_se}{---}
\setres{cty_Tunisia}{CPT_transfer_others}{r_prob}{---}
\setres{cty_Tunisia}{CPT_transfer_others}{r_prob_se}{---}
\setres{cty_Tunisia}{CPT_transfer_others}{spread_penalty}{0.000}
\setres{cty_Tunisia}{CPT_transfer_others}{spread_penalty_se}{0.000}
\setres{cty_Tunisia}{CPT_transfer_others}{theta_star}{1.00, 0.60, 0.90}
\setres{cty_Tunisia}{CPT_transfer_others}{transfer}{others}
\setres{cty_Tunisia}{CPT_transfer_others}{volume_ess}{---}
\setres{cty_Tunisia}{CPT_uniform}{kappa}{-3.614}
\setres{cty_Tunisia}{CPT_uniform}{kappa_prior_mean}{-0.526}
\setres{cty_Tunisia}{CPT_uniform}{kappa_prior_mean_se}{0.368}
\setres{cty_Tunisia}{CPT_uniform}{kappa_se}{0.819}
\setres{cty_Tunisia}{CPT_uniform}{mse}{45.20}
\setres{cty_Tunisia}{CPT_uniform}{mse_prior_mean}{27.84}
\setres{cty_Tunisia}{CPT_uniform}{nrmse_pct}{28.2}
\setres{cty_Tunisia}{CPT_uniform}{pred_variance}{17.37}
\setres{cty_Tunisia}{CPT_uniform}{r_ff}{---}
\setres{cty_Tunisia}{CPT_uniform}{r_ff_se}{---}
\setres{cty_Tunisia}{CPT_uniform}{r_prior}{---}
\setres{cty_Tunisia}{CPT_uniform}{r_prior_se}{---}
\setres{cty_Tunisia}{CPT_uniform}{r_prob}{---}
\setres{cty_Tunisia}{CPT_uniform}{r_prob_se}{---}
\setres{cty_Tunisia}{CPT_uniform}{spread_penalty}{3.087}
\setres{cty_Tunisia}{CPT_uniform}{spread_penalty_se}{0.521}
\setres{cty_Tunisia}{CPT_uniform}{volume_ess}{---}
\setres{cty_Tunisia}{DA_deterministic}{kappa}{0.801}
\setres{cty_Tunisia}{DA_deterministic}{kappa_se}{0.122}
\setres{cty_Tunisia}{DA_deterministic}{mse}{20.37}
\setres{cty_Tunisia}{DA_deterministic}{nrmse_pct}{24.9}
\setres{cty_Tunisia}{DA_deterministic}{r_ff}{---}
\setres{cty_Tunisia}{DA_deterministic}{r_ff_se}{---}
\setres{cty_Tunisia}{DA_deterministic}{r_prior}{---}
\setres{cty_Tunisia}{DA_deterministic}{r_prior_se}{---}
\setres{cty_Tunisia}{DA_deterministic}{r_prob}{---}
\setres{cty_Tunisia}{DA_deterministic}{r_prob_se}{---}
\setres{cty_Tunisia}{DA_deterministic}{theta_star}{2.25, 3.00}
\setres{cty_Tunisia}{DA_deterministic}{volume_ess}{---}
\setres{cty_Tunisia}{DA_lit}{kappa}{-2.063}
\setres{cty_Tunisia}{DA_lit}{kappa_prior_mean}{-1.219}
\setres{cty_Tunisia}{DA_lit}{kappa_prior_mean_se}{0.592}
\setres{cty_Tunisia}{DA_lit}{kappa_se}{0.720}
\setres{cty_Tunisia}{DA_lit}{mse}{36.48}
\setres{cty_Tunisia}{DA_lit}{mse_prior_mean}{31.73}
\setres{cty_Tunisia}{DA_lit}{nrmse_pct}{32.4}
\setres{cty_Tunisia}{DA_lit}{pred_variance}{4.75}
\setres{cty_Tunisia}{DA_lit}{r_ff}{---}
\setres{cty_Tunisia}{DA_lit}{r_ff_se}{---}
\setres{cty_Tunisia}{DA_lit}{r_prior}{---}
\setres{cty_Tunisia}{DA_lit}{r_prior_se}{---}
\setres{cty_Tunisia}{DA_lit}{r_prob}{---}
\setres{cty_Tunisia}{DA_lit}{r_prob_se}{---}
\setres{cty_Tunisia}{DA_lit}{spread_penalty}{0.844}
\setres{cty_Tunisia}{DA_lit}{spread_penalty_se}{0.142}
\setres{cty_Tunisia}{DA_lit}{volume_ess}{---}
\setres{cty_Tunisia}{DA_transfer_others}{kappa}{0.750}
\setres{cty_Tunisia}{DA_transfer_others}{kappa_prior_mean}{0.750}
\setres{cty_Tunisia}{DA_transfer_others}{kappa_prior_mean_se}{0.158}
\setres{cty_Tunisia}{DA_transfer_others}{kappa_se}{0.158}
\setres{cty_Tunisia}{DA_transfer_others}{mse}{20.66}
\setres{cty_Tunisia}{DA_transfer_others}{mse_prior_mean}{20.66}
\setres{cty_Tunisia}{DA_transfer_others}{nrmse_pct}{24.9}
\setres{cty_Tunisia}{DA_transfer_others}{pred_variance}{0.00}
\setres{cty_Tunisia}{DA_transfer_others}{r_ff}{---}
\setres{cty_Tunisia}{DA_transfer_others}{r_ff_se}{---}
\setres{cty_Tunisia}{DA_transfer_others}{r_prior}{---}
\setres{cty_Tunisia}{DA_transfer_others}{r_prior_se}{---}
\setres{cty_Tunisia}{DA_transfer_others}{r_prob}{---}
\setres{cty_Tunisia}{DA_transfer_others}{r_prob_se}{---}
\setres{cty_Tunisia}{DA_transfer_others}{spread_penalty}{0.000}
\setres{cty_Tunisia}{DA_transfer_others}{spread_penalty_se}{0.000}
\setres{cty_Tunisia}{DA_transfer_others}{theta_star}{1.67, 1.39}
\setres{cty_Tunisia}{DA_transfer_others}{transfer}{others}
\setres{cty_Tunisia}{DA_transfer_others}{volume_ess}{---}
\setres{cty_Tunisia}{DA_uniform}{kappa}{-4.891}
\setres{cty_Tunisia}{DA_uniform}{kappa_prior_mean}{-2.464}
\setres{cty_Tunisia}{DA_uniform}{kappa_prior_mean_se}{0.802}
\setres{cty_Tunisia}{DA_uniform}{kappa_se}{1.169}
\setres{cty_Tunisia}{DA_uniform}{mse}{52.38}
\setres{cty_Tunisia}{DA_uniform}{mse_prior_mean}{38.74}
\setres{cty_Tunisia}{DA_uniform}{nrmse_pct}{34.1}
\setres{cty_Tunisia}{DA_uniform}{pred_variance}{13.65}
\setres{cty_Tunisia}{DA_uniform}{r_ff}{---}
\setres{cty_Tunisia}{DA_uniform}{r_ff_se}{---}
\setres{cty_Tunisia}{DA_uniform}{r_prior}{---}
\setres{cty_Tunisia}{DA_uniform}{r_prior_se}{---}
\setres{cty_Tunisia}{DA_uniform}{r_prob}{---}
\setres{cty_Tunisia}{DA_uniform}{r_prob_se}{---}
\setres{cty_Tunisia}{DA_uniform}{spread_penalty}{2.426}
\setres{cty_Tunisia}{DA_uniform}{spread_penalty_se}{0.409}
\setres{cty_Tunisia}{DA_uniform}{volume_ess}{---}
\setres{cty_Tunisia}{EU_deterministic}{kappa}{0.363}
\setres{cty_Tunisia}{EU_deterministic}{kappa_se}{0.258}
\setres{cty_Tunisia}{EU_deterministic}{mse}{22.84}
\setres{cty_Tunisia}{EU_deterministic}{nrmse_pct}{26.4}
\setres{cty_Tunisia}{EU_deterministic}{r_ff}{---}
\setres{cty_Tunisia}{EU_deterministic}{r_ff_se}{---}
\setres{cty_Tunisia}{EU_deterministic}{r_prior}{---}
\setres{cty_Tunisia}{EU_deterministic}{r_prior_se}{---}
\setres{cty_Tunisia}{EU_deterministic}{r_prob}{---}
\setres{cty_Tunisia}{EU_deterministic}{r_prob_se}{---}
\setres{cty_Tunisia}{EU_deterministic}{theta_star}{0.97}
\setres{cty_Tunisia}{EU_deterministic}{volume_ess}{---}
\setres{cty_Tunisia}{EU_lit}{kappa}{-0.286}
\setres{cty_Tunisia}{EU_lit}{kappa_prior_mean}{-0.072}
\setres{cty_Tunisia}{EU_lit}{kappa_prior_mean_se}{0.358}
\setres{cty_Tunisia}{EU_lit}{kappa_se}{0.392}
\setres{cty_Tunisia}{EU_lit}{mse}{26.49}
\setres{cty_Tunisia}{EU_lit}{mse_prior_mean}{25.28}
\setres{cty_Tunisia}{EU_lit}{nrmse_pct}{27.6}
\setres{cty_Tunisia}{EU_lit}{pred_variance}{1.20}
\setres{cty_Tunisia}{EU_lit}{r_ff}{---}
\setres{cty_Tunisia}{EU_lit}{r_ff_se}{---}
\setres{cty_Tunisia}{EU_lit}{r_prior}{---}
\setres{cty_Tunisia}{EU_lit}{r_prior_se}{---}
\setres{cty_Tunisia}{EU_lit}{r_prob}{---}
\setres{cty_Tunisia}{EU_lit}{r_prob_se}{---}
\setres{cty_Tunisia}{EU_lit}{spread_penalty}{0.214}
\setres{cty_Tunisia}{EU_lit}{spread_penalty_se}{0.036}
\setres{cty_Tunisia}{EU_lit}{volume_ess}{---}
\setres{cty_Tunisia}{EU_transfer_others}{kappa}{0.393}
\setres{cty_Tunisia}{EU_transfer_others}{kappa_prior_mean}{0.393}
\setres{cty_Tunisia}{EU_transfer_others}{kappa_prior_mean_se}{0.253}
\setres{cty_Tunisia}{EU_transfer_others}{kappa_se}{0.253}
\setres{cty_Tunisia}{EU_transfer_others}{mse}{22.67}
\setres{cty_Tunisia}{EU_transfer_others}{mse_prior_mean}{22.67}
\setres{cty_Tunisia}{EU_transfer_others}{nrmse_pct}{26.2}
\setres{cty_Tunisia}{EU_transfer_others}{pred_variance}{0.00}
\setres{cty_Tunisia}{EU_transfer_others}{r_ff}{---}
\setres{cty_Tunisia}{EU_transfer_others}{r_ff_se}{---}
\setres{cty_Tunisia}{EU_transfer_others}{r_prior}{---}
\setres{cty_Tunisia}{EU_transfer_others}{r_prior_se}{---}
\setres{cty_Tunisia}{EU_transfer_others}{r_prob}{---}
\setres{cty_Tunisia}{EU_transfer_others}{r_prob_se}{---}
\setres{cty_Tunisia}{EU_transfer_others}{spread_penalty}{0.000}
\setres{cty_Tunisia}{EU_transfer_others}{spread_penalty_se}{0.000}
\setres{cty_Tunisia}{EU_transfer_others}{theta_star}{0.94}
\setres{cty_Tunisia}{EU_transfer_others}{transfer}{others}
\setres{cty_Tunisia}{EU_transfer_others}{volume_ess}{---}
\setres{cty_Tunisia}{EU_uniform}{kappa}{-2.947}
\setres{cty_Tunisia}{EU_uniform}{kappa_prior_mean}{-1.490}
\setres{cty_Tunisia}{EU_uniform}{kappa_prior_mean_se}{0.627}
\setres{cty_Tunisia}{EU_uniform}{kappa_se}{0.850}
\setres{cty_Tunisia}{EU_uniform}{mse}{41.45}
\setres{cty_Tunisia}{EU_uniform}{mse_prior_mean}{33.25}
\setres{cty_Tunisia}{EU_uniform}{nrmse_pct}{33.1}
\setres{cty_Tunisia}{EU_uniform}{pred_variance}{8.20}
\setres{cty_Tunisia}{EU_uniform}{r_ff}{---}
\setres{cty_Tunisia}{EU_uniform}{r_ff_se}{---}
\setres{cty_Tunisia}{EU_uniform}{r_prior}{---}
\setres{cty_Tunisia}{EU_uniform}{r_prior_se}{---}
\setres{cty_Tunisia}{EU_uniform}{r_prob}{---}
\setres{cty_Tunisia}{EU_uniform}{r_prob_se}{---}
\setres{cty_Tunisia}{EU_uniform}{spread_penalty}{1.457}
\setres{cty_Tunisia}{EU_uniform}{spread_penalty_se}{0.246}
\setres{cty_Tunisia}{EU_uniform}{volume_ess}{---}
\setres{cty_Tunisia}{EV_deterministic}{kappa}{0.392}
\setres{cty_Tunisia}{EV_deterministic}{kappa_se}{0.256}
\setres{cty_Tunisia}{EV_deterministic}{mse}{22.67}
\setres{cty_Tunisia}{EV_deterministic}{nrmse_pct}{26.4}
\setres{cty_Tunisia}{EV_deterministic}{r_ff}{---}
\setres{cty_Tunisia}{EV_deterministic}{r_ff_se}{---}
\setres{cty_Tunisia}{EV_deterministic}{r_prior}{---}
\setres{cty_Tunisia}{EV_deterministic}{r_prior_se}{---}
\setres{cty_Tunisia}{EV_deterministic}{r_prob}{---}
\setres{cty_Tunisia}{EV_deterministic}{r_prob_se}{---}
\setres{cty_Tunisia}{EV_deterministic}{volume_ess}{---}
\setres{cty_Tunisia}{reference}{country_name}{Tunisia}
\setres{cty_Tunisia}{reference}{lookup_nrmse_pct}{24.5}
\setres{cty_Tunisia}{reference}{midpoint_mse}{24.88}
\setres{cty_Tunisia}{reference}{midpoint_nrmse_pct}{27.3}
\setres{cty_Tunisia}{reference}{missing_lotteries}{none}
\setres{cty_Tunisia}{reference}{n_lotteries}{14}
\setres{cty_Tunisia}{reference}{n_obs}{1{,}036}
\setres{cty_Tunisia}{reference}{n_subjects}{74}
\setres{cty_Tunisia}{reference}{oracle_mse}{19.25}
\setres{cty_Tunisia}{reference}{sample}{LV2019\_Tunisia}
\setres{cty_UK}{CPT_deterministic}{kappa}{0.955}
\setres{cty_UK}{CPT_deterministic}{kappa_se}{0.010}
\setres{cty_UK}{CPT_deterministic}{mse}{10.51}
\setres{cty_UK}{CPT_deterministic}{nrmse_pct}{17.4}
\setres{cty_UK}{CPT_deterministic}{r_ff}{---}
\setres{cty_UK}{CPT_deterministic}{r_ff_se}{---}
\setres{cty_UK}{CPT_deterministic}{r_prior}{---}
\setres{cty_UK}{CPT_deterministic}{r_prior_se}{---}
\setres{cty_UK}{CPT_deterministic}{r_prob}{---}
\setres{cty_UK}{CPT_deterministic}{r_prob_se}{---}
\setres{cty_UK}{CPT_deterministic}{theta_star}{0.80, 0.70, 1.40}
\setres{cty_UK}{CPT_deterministic}{volume_ess}{---}
\setres{cty_UK}{CPT_lit}{kappa}{0.367}
\setres{cty_UK}{CPT_lit}{kappa_prior_mean}{0.612}
\setres{cty_UK}{CPT_lit}{kappa_prior_mean_se}{0.074}
\setres{cty_UK}{CPT_lit}{kappa_se}{0.078}
\setres{cty_UK}{CPT_lit}{mse}{17.48}
\setres{cty_UK}{CPT_lit}{mse_prior_mean}{14.58}
\setres{cty_UK}{CPT_lit}{nrmse_pct}{22.1}
\setres{cty_UK}{CPT_lit}{pred_variance}{2.90}
\setres{cty_UK}{CPT_lit}{r_ff}{---}
\setres{cty_UK}{CPT_lit}{r_ff_se}{---}
\setres{cty_UK}{CPT_lit}{r_prior}{---}
\setres{cty_UK}{CPT_lit}{r_prior_se}{---}
\setres{cty_UK}{CPT_lit}{r_prob}{---}
\setres{cty_UK}{CPT_lit}{r_prob_se}{---}
\setres{cty_UK}{CPT_lit}{spread_penalty}{0.244}
\setres{cty_UK}{CPT_lit}{spread_penalty_se}{0.019}
\setres{cty_UK}{CPT_lit}{volume_ess}{---}
\setres{cty_UK}{CPT_transfer_others}{kappa}{0.885}
\setres{cty_UK}{CPT_transfer_others}{kappa_prior_mean}{0.885}
\setres{cty_UK}{CPT_transfer_others}{kappa_prior_mean_se}{0.030}
\setres{cty_UK}{CPT_transfer_others}{kappa_se}{0.030}
\setres{cty_UK}{CPT_transfer_others}{mse}{11.34}
\setres{cty_UK}{CPT_transfer_others}{mse_prior_mean}{11.34}
\setres{cty_UK}{CPT_transfer_others}{nrmse_pct}{18.2}
\setres{cty_UK}{CPT_transfer_others}{pred_variance}{0.00}
\setres{cty_UK}{CPT_transfer_others}{r_ff}{---}
\setres{cty_UK}{CPT_transfer_others}{r_ff_se}{---}
\setres{cty_UK}{CPT_transfer_others}{r_prior}{---}
\setres{cty_UK}{CPT_transfer_others}{r_prior_se}{---}
\setres{cty_UK}{CPT_transfer_others}{r_prob}{---}
\setres{cty_UK}{CPT_transfer_others}{r_prob_se}{---}
\setres{cty_UK}{CPT_transfer_others}{spread_penalty}{0.000}
\setres{cty_UK}{CPT_transfer_others}{spread_penalty_se}{0.000}
\setres{cty_UK}{CPT_transfer_others}{theta_star}{1.10, 0.60, 0.80}
\setres{cty_UK}{CPT_transfer_others}{transfer}{others}
\setres{cty_UK}{CPT_transfer_others}{volume_ess}{---}
\setres{cty_UK}{CPT_uniform}{kappa}{-1.614}
\setres{cty_UK}{CPT_uniform}{kappa_prior_mean}{-0.151}
\setres{cty_UK}{CPT_uniform}{kappa_prior_mean_se}{0.116}
\setres{cty_UK}{CPT_uniform}{kappa_se}{0.181}
\setres{cty_UK}{CPT_uniform}{mse}{40.98}
\setres{cty_UK}{CPT_uniform}{mse_prior_mean}{23.62}
\setres{cty_UK}{CPT_uniform}{nrmse_pct}{26.0}
\setres{cty_UK}{CPT_uniform}{pred_variance}{17.37}
\setres{cty_UK}{CPT_uniform}{r_ff}{---}
\setres{cty_UK}{CPT_uniform}{r_ff_se}{---}
\setres{cty_UK}{CPT_uniform}{r_prior}{---}
\setres{cty_UK}{CPT_uniform}{r_prior_se}{---}
\setres{cty_UK}{CPT_uniform}{r_prob}{---}
\setres{cty_UK}{CPT_uniform}{r_prob_se}{---}
\setres{cty_UK}{CPT_uniform}{spread_penalty}{1.464}
\setres{cty_UK}{CPT_uniform}{spread_penalty_se}{0.112}
\setres{cty_UK}{CPT_uniform}{volume_ess}{---}
\setres{cty_UK}{DA_deterministic}{kappa}{0.927}
\setres{cty_UK}{DA_deterministic}{kappa_se}{0.019}
\setres{cty_UK}{DA_deterministic}{mse}{10.83}
\setres{cty_UK}{DA_deterministic}{nrmse_pct}{17.5}
\setres{cty_UK}{DA_deterministic}{r_ff}{---}
\setres{cty_UK}{DA_deterministic}{r_ff_se}{---}
\setres{cty_UK}{DA_deterministic}{r_prior}{---}
\setres{cty_UK}{DA_deterministic}{r_prior_se}{---}
\setres{cty_UK}{DA_deterministic}{r_prob}{---}
\setres{cty_UK}{DA_deterministic}{r_prob_se}{---}
\setres{cty_UK}{DA_deterministic}{theta_star}{1.06, -0.02}
\setres{cty_UK}{DA_deterministic}{volume_ess}{---}
\setres{cty_UK}{DA_lit}{kappa}{-0.704}
\setres{cty_UK}{DA_lit}{kappa_prior_mean}{-0.304}
\setres{cty_UK}{DA_lit}{kappa_prior_mean_se}{0.154}
\setres{cty_UK}{DA_lit}{kappa_se}{0.170}
\setres{cty_UK}{DA_lit}{mse}{30.19}
\setres{cty_UK}{DA_lit}{mse_prior_mean}{25.44}
\setres{cty_UK}{DA_lit}{nrmse_pct}{27.5}
\setres{cty_UK}{DA_lit}{pred_variance}{4.75}
\setres{cty_UK}{DA_lit}{r_ff}{---}
\setres{cty_UK}{DA_lit}{r_ff_se}{---}
\setres{cty_UK}{DA_lit}{r_prior}{---}
\setres{cty_UK}{DA_lit}{r_prior_se}{---}
\setres{cty_UK}{DA_lit}{r_prob}{---}
\setres{cty_UK}{DA_lit}{r_prob_se}{---}
\setres{cty_UK}{DA_lit}{spread_penalty}{0.400}
\setres{cty_UK}{DA_lit}{spread_penalty_se}{0.031}
\setres{cty_UK}{DA_lit}{volume_ess}{---}
\setres{cty_UK}{DA_transfer_others}{kappa}{0.831}
\setres{cty_UK}{DA_transfer_others}{kappa_prior_mean}{0.831}
\setres{cty_UK}{DA_transfer_others}{kappa_prior_mean_se}{0.033}
\setres{cty_UK}{DA_transfer_others}{kappa_se}{0.033}
\setres{cty_UK}{DA_transfer_others}{mse}{11.97}
\setres{cty_UK}{DA_transfer_others}{mse_prior_mean}{11.97}
\setres{cty_UK}{DA_transfer_others}{nrmse_pct}{17.7}
\setres{cty_UK}{DA_transfer_others}{pred_variance}{0.00}
\setres{cty_UK}{DA_transfer_others}{r_ff}{---}
\setres{cty_UK}{DA_transfer_others}{r_ff_se}{---}
\setres{cty_UK}{DA_transfer_others}{r_prior}{---}
\setres{cty_UK}{DA_transfer_others}{r_prior_se}{---}
\setres{cty_UK}{DA_transfer_others}{r_prob}{---}
\setres{cty_UK}{DA_transfer_others}{r_prob_se}{---}
\setres{cty_UK}{DA_transfer_others}{spread_penalty}{0.000}
\setres{cty_UK}{DA_transfer_others}{spread_penalty_se}{0.000}
\setres{cty_UK}{DA_transfer_others}{theta_star}{1.72, 1.52}
\setres{cty_UK}{DA_transfer_others}{transfer}{others}
\setres{cty_UK}{DA_transfer_others}{volume_ess}{---}
\setres{cty_UK}{DA_uniform}{kappa}{-2.356}
\setres{cty_UK}{DA_uniform}{kappa_prior_mean}{-1.206}
\setres{cty_UK}{DA_uniform}{kappa_prior_mean_se}{0.209}
\setres{cty_UK}{DA_uniform}{kappa_se}{0.266}
\setres{cty_UK}{DA_uniform}{mse}{49.79}
\setres{cty_UK}{DA_uniform}{mse_prior_mean}{36.14}
\setres{cty_UK}{DA_uniform}{nrmse_pct}{33.1}
\setres{cty_UK}{DA_uniform}{pred_variance}{13.65}
\setres{cty_UK}{DA_uniform}{r_ff}{---}
\setres{cty_UK}{DA_uniform}{r_ff_se}{---}
\setres{cty_UK}{DA_uniform}{r_prior}{---}
\setres{cty_UK}{DA_uniform}{r_prior_se}{---}
\setres{cty_UK}{DA_uniform}{r_prob}{---}
\setres{cty_UK}{DA_uniform}{r_prob_se}{---}
\setres{cty_UK}{DA_uniform}{spread_penalty}{1.150}
\setres{cty_UK}{DA_uniform}{spread_penalty_se}{0.088}
\setres{cty_UK}{DA_uniform}{volume_ess}{---}
\setres{cty_UK}{EU_deterministic}{kappa}{0.933}
\setres{cty_UK}{EU_deterministic}{kappa_se}{0.019}
\setres{cty_UK}{EU_deterministic}{mse}{10.76}
\setres{cty_UK}{EU_deterministic}{nrmse_pct}{17.5}
\setres{cty_UK}{EU_deterministic}{r_ff}{---}
\setres{cty_UK}{EU_deterministic}{r_ff_se}{---}
\setres{cty_UK}{EU_deterministic}{r_prior}{---}
\setres{cty_UK}{EU_deterministic}{r_prior_se}{---}
\setres{cty_UK}{EU_deterministic}{r_prob}{---}
\setres{cty_UK}{EU_deterministic}{r_prob_se}{---}
\setres{cty_UK}{EU_deterministic}{theta_star}{1.07}
\setres{cty_UK}{EU_deterministic}{volume_ess}{---}
\setres{cty_UK}{EU_lit}{kappa}{0.425}
\setres{cty_UK}{EU_lit}{kappa_prior_mean}{0.526}
\setres{cty_UK}{EU_lit}{kappa_prior_mean_se}{0.085}
\setres{cty_UK}{EU_lit}{kappa_se}{0.089}
\setres{cty_UK}{EU_lit}{mse}{16.79}
\setres{cty_UK}{EU_lit}{mse_prior_mean}{15.59}
\setres{cty_UK}{EU_lit}{nrmse_pct}{21.7}
\setres{cty_UK}{EU_lit}{pred_variance}{1.20}
\setres{cty_UK}{EU_lit}{r_ff}{---}
\setres{cty_UK}{EU_lit}{r_ff_se}{---}
\setres{cty_UK}{EU_lit}{r_prior}{---}
\setres{cty_UK}{EU_lit}{r_prior_se}{---}
\setres{cty_UK}{EU_lit}{r_prob}{---}
\setres{cty_UK}{EU_lit}{r_prob_se}{---}
\setres{cty_UK}{EU_lit}{spread_penalty}{0.102}
\setres{cty_UK}{EU_lit}{spread_penalty_se}{0.008}
\setres{cty_UK}{EU_lit}{volume_ess}{---}
\setres{cty_UK}{EU_transfer_others}{kappa}{0.908}
\setres{cty_UK}{EU_transfer_others}{kappa_prior_mean}{0.908}
\setres{cty_UK}{EU_transfer_others}{kappa_prior_mean_se}{0.029}
\setres{cty_UK}{EU_transfer_others}{kappa_se}{0.029}
\setres{cty_UK}{EU_transfer_others}{mse}{11.06}
\setres{cty_UK}{EU_transfer_others}{mse_prior_mean}{11.06}
\setres{cty_UK}{EU_transfer_others}{nrmse_pct}{17.7}
\setres{cty_UK}{EU_transfer_others}{pred_variance}{0.00}
\setres{cty_UK}{EU_transfer_others}{r_ff}{---}
\setres{cty_UK}{EU_transfer_others}{r_ff_se}{---}
\setres{cty_UK}{EU_transfer_others}{r_prior}{---}
\setres{cty_UK}{EU_transfer_others}{r_prior_se}{---}
\setres{cty_UK}{EU_transfer_others}{r_prob}{---}
\setres{cty_UK}{EU_transfer_others}{r_prob_se}{---}
\setres{cty_UK}{EU_transfer_others}{spread_penalty}{0.000}
\setres{cty_UK}{EU_transfer_others}{spread_penalty_se}{0.000}
\setres{cty_UK}{EU_transfer_others}{theta_star}{0.94}
\setres{cty_UK}{EU_transfer_others}{transfer}{others}
\setres{cty_UK}{EU_transfer_others}{volume_ess}{---}
\setres{cty_UK}{EU_uniform}{kappa}{-1.143}
\setres{cty_UK}{EU_uniform}{kappa_prior_mean}{-0.452}
\setres{cty_UK}{EU_uniform}{kappa_prior_mean_se}{0.163}
\setres{cty_UK}{EU_uniform}{kappa_se}{0.194}
\setres{cty_UK}{EU_uniform}{mse}{35.39}
\setres{cty_UK}{EU_uniform}{mse_prior_mean}{27.19}
\setres{cty_UK}{EU_uniform}{nrmse_pct}{28.1}
\setres{cty_UK}{EU_uniform}{pred_variance}{8.20}
\setres{cty_UK}{EU_uniform}{r_ff}{---}
\setres{cty_UK}{EU_uniform}{r_ff_se}{---}
\setres{cty_UK}{EU_uniform}{r_prior}{---}
\setres{cty_UK}{EU_uniform}{r_prior_se}{---}
\setres{cty_UK}{EU_uniform}{r_prob}{---}
\setres{cty_UK}{EU_uniform}{r_prob_se}{---}
\setres{cty_UK}{EU_uniform}{spread_penalty}{0.691}
\setres{cty_UK}{EU_uniform}{spread_penalty_se}{0.053}
\setres{cty_UK}{EU_uniform}{volume_ess}{---}
\setres{cty_UK}{EV_deterministic}{kappa}{0.933}
\setres{cty_UK}{EV_deterministic}{kappa_se}{0.022}
\setres{cty_UK}{EV_deterministic}{mse}{10.76}
\setres{cty_UK}{EV_deterministic}{nrmse_pct}{17.6}
\setres{cty_UK}{EV_deterministic}{r_ff}{---}
\setres{cty_UK}{EV_deterministic}{r_ff_se}{---}
\setres{cty_UK}{EV_deterministic}{r_prior}{---}
\setres{cty_UK}{EV_deterministic}{r_prior_se}{---}
\setres{cty_UK}{EV_deterministic}{r_prob}{---}
\setres{cty_UK}{EV_deterministic}{r_prob_se}{---}
\setres{cty_UK}{EV_deterministic}{volume_ess}{---}
\setres{cty_UK}{reference}{country_name}{UK}
\setres{cty_UK}{reference}{lookup_nrmse_pct}{16.7}
\setres{cty_UK}{reference}{midpoint_mse}{21.83}
\setres{cty_UK}{reference}{midpoint_nrmse_pct}{24.9}
\setres{cty_UK}{reference}{missing_lotteries}{none}
\setres{cty_UK}{reference}{n_lotteries}{14}
\setres{cty_UK}{reference}{n_obs}{1{,}120}
\setres{cty_UK}{reference}{n_subjects}{80}
\setres{cty_UK}{reference}{oracle_mse}{9.97}
\setres{cty_UK}{reference}{sample}{LV2019\_UK}
\setres{cty_USA}{CPT_deterministic}{kappa}{0.981}
\setres{cty_USA}{CPT_deterministic}{kappa_se}{0.010}
\setres{cty_USA}{CPT_deterministic}{mse}{8.67}
\setres{cty_USA}{CPT_deterministic}{nrmse_pct}{15.8}
\setres{cty_USA}{CPT_deterministic}{r_ff}{---}
\setres{cty_USA}{CPT_deterministic}{r_ff_se}{---}
\setres{cty_USA}{CPT_deterministic}{r_prior}{---}
\setres{cty_USA}{CPT_deterministic}{r_prior_se}{---}
\setres{cty_USA}{CPT_deterministic}{r_prob}{---}
\setres{cty_USA}{CPT_deterministic}{r_prob_se}{---}
\setres{cty_USA}{CPT_deterministic}{theta_star}{1.00, 0.80, 0.70}
\setres{cty_USA}{CPT_deterministic}{volume_ess}{---}
\setres{cty_USA}{CPT_lit}{kappa}{0.732}
\setres{cty_USA}{CPT_lit}{kappa_prior_mean}{0.957}
\setres{cty_USA}{CPT_lit}{kappa_prior_mean_se}{0.017}
\setres{cty_USA}{CPT_lit}{kappa_se}{0.022}
\setres{cty_USA}{CPT_lit}{mse}{11.89}
\setres{cty_USA}{CPT_lit}{mse_prior_mean}{8.99}
\setres{cty_USA}{CPT_lit}{nrmse_pct}{16.1}
\setres{cty_USA}{CPT_lit}{pred_variance}{2.90}
\setres{cty_USA}{CPT_lit}{r_ff}{---}
\setres{cty_USA}{CPT_lit}{r_ff_se}{---}
\setres{cty_USA}{CPT_lit}{r_prior}{---}
\setres{cty_USA}{CPT_lit}{r_prior_se}{---}
\setres{cty_USA}{CPT_lit}{r_prob}{---}
\setres{cty_USA}{CPT_lit}{r_prob_se}{---}
\setres{cty_USA}{CPT_lit}{spread_penalty}{0.225}
\setres{cty_USA}{CPT_lit}{spread_penalty_se}{0.012}
\setres{cty_USA}{CPT_lit}{volume_ess}{---}
\setres{cty_USA}{CPT_transfer_others}{kappa}{0.901}
\setres{cty_USA}{CPT_transfer_others}{kappa_prior_mean}{0.901}
\setres{cty_USA}{CPT_transfer_others}{kappa_prior_mean_se}{0.020}
\setres{cty_USA}{CPT_transfer_others}{kappa_se}{0.020}
\setres{cty_USA}{CPT_transfer_others}{mse}{9.71}
\setres{cty_USA}{CPT_transfer_others}{mse_prior_mean}{9.71}
\setres{cty_USA}{CPT_transfer_others}{nrmse_pct}{15.3}
\setres{cty_USA}{CPT_transfer_others}{pred_variance}{0.00}
\setres{cty_USA}{CPT_transfer_others}{r_ff}{---}
\setres{cty_USA}{CPT_transfer_others}{r_ff_se}{---}
\setres{cty_USA}{CPT_transfer_others}{r_prior}{---}
\setres{cty_USA}{CPT_transfer_others}{r_prior_se}{---}
\setres{cty_USA}{CPT_transfer_others}{r_prob}{---}
\setres{cty_USA}{CPT_transfer_others}{r_prob_se}{---}
\setres{cty_USA}{CPT_transfer_others}{spread_penalty}{0.000}
\setres{cty_USA}{CPT_transfer_others}{spread_penalty_se}{0.000}
\setres{cty_USA}{CPT_transfer_others}{theta_star}{1.10, 0.60, 0.80}
\setres{cty_USA}{CPT_transfer_others}{transfer}{others}
\setres{cty_USA}{CPT_transfer_others}{volume_ess}{---}
\setres{cty_USA}{CPT_uniform}{kappa}{-0.815}
\setres{cty_USA}{CPT_uniform}{kappa_prior_mean}{0.531}
\setres{cty_USA}{CPT_uniform}{kappa_prior_mean_se}{0.064}
\setres{cty_USA}{CPT_uniform}{kappa_se}{0.105}
\setres{cty_USA}{CPT_uniform}{mse}{31.85}
\setres{cty_USA}{CPT_uniform}{mse_prior_mean}{14.49}
\setres{cty_USA}{CPT_uniform}{nrmse_pct}{21.5}
\setres{cty_USA}{CPT_uniform}{pred_variance}{17.37}
\setres{cty_USA}{CPT_uniform}{r_ff}{---}
\setres{cty_USA}{CPT_uniform}{r_ff_se}{---}
\setres{cty_USA}{CPT_uniform}{r_prior}{---}
\setres{cty_USA}{CPT_uniform}{r_prior_se}{---}
\setres{cty_USA}{CPT_uniform}{r_prob}{---}
\setres{cty_USA}{CPT_uniform}{r_prob_se}{---}
\setres{cty_USA}{CPT_uniform}{spread_penalty}{1.346}
\setres{cty_USA}{CPT_uniform}{spread_penalty_se}{0.072}
\setres{cty_USA}{CPT_uniform}{volume_ess}{---}
\setres{cty_USA}{DA_deterministic}{kappa}{0.955}
\setres{cty_USA}{DA_deterministic}{kappa_se}{0.013}
\setres{cty_USA}{DA_deterministic}{mse}{9.01}
\setres{cty_USA}{DA_deterministic}{nrmse_pct}{17.0}
\setres{cty_USA}{DA_deterministic}{r_ff}{---}
\setres{cty_USA}{DA_deterministic}{r_ff_se}{---}
\setres{cty_USA}{DA_deterministic}{r_prior}{---}
\setres{cty_USA}{DA_deterministic}{r_prior_se}{---}
\setres{cty_USA}{DA_deterministic}{r_prob}{---}
\setres{cty_USA}{DA_deterministic}{r_prob_se}{---}
\setres{cty_USA}{DA_deterministic}{theta_star}{1.19, 0.82}
\setres{cty_USA}{DA_deterministic}{volume_ess}{---}
\setres{cty_USA}{DA_lit}{kappa}{0.196}
\setres{cty_USA}{DA_lit}{kappa_prior_mean}{0.564}
\setres{cty_USA}{DA_lit}{kappa_prior_mean_se}{0.072}
\setres{cty_USA}{DA_lit}{kappa_se}{0.085}
\setres{cty_USA}{DA_lit}{mse}{18.80}
\setres{cty_USA}{DA_lit}{mse_prior_mean}{14.06}
\setres{cty_USA}{DA_lit}{nrmse_pct}{19.8}
\setres{cty_USA}{DA_lit}{pred_variance}{4.75}
\setres{cty_USA}{DA_lit}{r_ff}{---}
\setres{cty_USA}{DA_lit}{r_ff_se}{---}
\setres{cty_USA}{DA_lit}{r_prior}{---}
\setres{cty_USA}{DA_lit}{r_prior_se}{---}
\setres{cty_USA}{DA_lit}{r_prob}{---}
\setres{cty_USA}{DA_lit}{r_prob_se}{---}
\setres{cty_USA}{DA_lit}{spread_penalty}{0.368}
\setres{cty_USA}{DA_lit}{spread_penalty_se}{0.020}
\setres{cty_USA}{DA_lit}{volume_ess}{---}
\setres{cty_USA}{DA_transfer_others}{kappa}{0.886}
\setres{cty_USA}{DA_transfer_others}{kappa_prior_mean}{0.886}
\setres{cty_USA}{DA_transfer_others}{kappa_prior_mean_se}{0.030}
\setres{cty_USA}{DA_transfer_others}{kappa_se}{0.030}
\setres{cty_USA}{DA_transfer_others}{mse}{9.91}
\setres{cty_USA}{DA_transfer_others}{mse_prior_mean}{9.91}
\setres{cty_USA}{DA_transfer_others}{nrmse_pct}{15.8}
\setres{cty_USA}{DA_transfer_others}{pred_variance}{0.00}
\setres{cty_USA}{DA_transfer_others}{r_ff}{---}
\setres{cty_USA}{DA_transfer_others}{r_ff_se}{---}
\setres{cty_USA}{DA_transfer_others}{r_prior}{---}
\setres{cty_USA}{DA_transfer_others}{r_prior_se}{---}
\setres{cty_USA}{DA_transfer_others}{r_prob}{---}
\setres{cty_USA}{DA_transfer_others}{r_prob_se}{---}
\setres{cty_USA}{DA_transfer_others}{spread_penalty}{0.000}
\setres{cty_USA}{DA_transfer_others}{spread_penalty_se}{0.000}
\setres{cty_USA}{DA_transfer_others}{theta_star}{1.72, 1.48}
\setres{cty_USA}{DA_transfer_others}{transfer}{others}
\setres{cty_USA}{DA_transfer_others}{volume_ess}{---}
\setres{cty_USA}{DA_uniform}{kappa}{-1.021}
\setres{cty_USA}{DA_uniform}{kappa_prior_mean}{0.036}
\setres{cty_USA}{DA_uniform}{kappa_prior_mean_se}{0.119}
\setres{cty_USA}{DA_uniform}{kappa_se}{0.158}
\setres{cty_USA}{DA_uniform}{mse}{34.51}
\setres{cty_USA}{DA_uniform}{mse_prior_mean}{20.87}
\setres{cty_USA}{DA_uniform}{nrmse_pct}{25.7}
\setres{cty_USA}{DA_uniform}{pred_variance}{13.65}
\setres{cty_USA}{DA_uniform}{r_ff}{---}
\setres{cty_USA}{DA_uniform}{r_ff_se}{---}
\setres{cty_USA}{DA_uniform}{r_prior}{---}
\setres{cty_USA}{DA_uniform}{r_prior_se}{---}
\setres{cty_USA}{DA_uniform}{r_prob}{---}
\setres{cty_USA}{DA_uniform}{r_prob_se}{---}
\setres{cty_USA}{DA_uniform}{spread_penalty}{1.058}
\setres{cty_USA}{DA_uniform}{spread_penalty_se}{0.056}
\setres{cty_USA}{DA_uniform}{volume_ess}{---}
\setres{cty_USA}{EU_deterministic}{kappa}{0.921}
\setres{cty_USA}{EU_deterministic}{kappa_se}{0.019}
\setres{cty_USA}{EU_deterministic}{mse}{9.45}
\setres{cty_USA}{EU_deterministic}{nrmse_pct}{16.5}
\setres{cty_USA}{EU_deterministic}{r_ff}{---}
\setres{cty_USA}{EU_deterministic}{r_ff_se}{---}
\setres{cty_USA}{EU_deterministic}{r_prior}{---}
\setres{cty_USA}{EU_deterministic}{r_prior_se}{---}
\setres{cty_USA}{EU_deterministic}{r_prob}{---}
\setres{cty_USA}{EU_deterministic}{r_prob_se}{---}
\setres{cty_USA}{EU_deterministic}{theta_star}{0.79}
\setres{cty_USA}{EU_deterministic}{volume_ess}{---}
\setres{cty_USA}{EU_lit}{kappa}{0.806}
\setres{cty_USA}{EU_lit}{kappa_prior_mean}{0.899}
\setres{cty_USA}{EU_lit}{kappa_prior_mean_se}{0.022}
\setres{cty_USA}{EU_lit}{kappa_se}{0.027}
\setres{cty_USA}{EU_lit}{mse}{10.94}
\setres{cty_USA}{EU_lit}{mse_prior_mean}{9.73}
\setres{cty_USA}{EU_lit}{nrmse_pct}{16.9}
\setres{cty_USA}{EU_lit}{pred_variance}{1.20}
\setres{cty_USA}{EU_lit}{r_ff}{---}
\setres{cty_USA}{EU_lit}{r_ff_se}{---}
\setres{cty_USA}{EU_lit}{r_prior}{---}
\setres{cty_USA}{EU_lit}{r_prior_se}{---}
\setres{cty_USA}{EU_lit}{r_prob}{---}
\setres{cty_USA}{EU_lit}{r_prob_se}{---}
\setres{cty_USA}{EU_lit}{spread_penalty}{0.093}
\setres{cty_USA}{EU_lit}{spread_penalty_se}{0.005}
\setres{cty_USA}{EU_lit}{volume_ess}{---}
\setres{cty_USA}{EU_transfer_others}{kappa}{0.856}
\setres{cty_USA}{EU_transfer_others}{kappa_prior_mean}{0.856}
\setres{cty_USA}{EU_transfer_others}{kappa_prior_mean_se}{0.039}
\setres{cty_USA}{EU_transfer_others}{kappa_se}{0.039}
\setres{cty_USA}{EU_transfer_others}{mse}{10.30}
\setres{cty_USA}{EU_transfer_others}{mse_prior_mean}{10.30}
\setres{cty_USA}{EU_transfer_others}{nrmse_pct}{15.1}
\setres{cty_USA}{EU_transfer_others}{pred_variance}{0.00}
\setres{cty_USA}{EU_transfer_others}{r_ff}{---}
\setres{cty_USA}{EU_transfer_others}{r_ff_se}{---}
\setres{cty_USA}{EU_transfer_others}{r_prior}{---}
\setres{cty_USA}{EU_transfer_others}{r_prior_se}{---}
\setres{cty_USA}{EU_transfer_others}{r_prob}{---}
\setres{cty_USA}{EU_transfer_others}{r_prob_se}{---}
\setres{cty_USA}{EU_transfer_others}{spread_penalty}{0.000}
\setres{cty_USA}{EU_transfer_others}{spread_penalty_se}{0.000}
\setres{cty_USA}{EU_transfer_others}{theta_star}{0.95}
\setres{cty_USA}{EU_transfer_others}{transfer}{others}
\setres{cty_USA}{EU_transfer_others}{volume_ess}{---}
\setres{cty_USA}{EU_uniform}{kappa}{-0.170}
\setres{cty_USA}{EU_uniform}{kappa_prior_mean}{0.465}
\setres{cty_USA}{EU_uniform}{kappa_prior_mean_se}{0.080}
\setres{cty_USA}{EU_uniform}{kappa_se}{0.103}
\setres{cty_USA}{EU_uniform}{mse}{23.53}
\setres{cty_USA}{EU_uniform}{mse_prior_mean}{15.33}
\setres{cty_USA}{EU_uniform}{nrmse_pct}{21.8}
\setres{cty_USA}{EU_uniform}{pred_variance}{8.20}
\setres{cty_USA}{EU_uniform}{r_ff}{---}
\setres{cty_USA}{EU_uniform}{r_ff_se}{---}
\setres{cty_USA}{EU_uniform}{r_prior}{---}
\setres{cty_USA}{EU_uniform}{r_prior_se}{---}
\setres{cty_USA}{EU_uniform}{r_prob}{---}
\setres{cty_USA}{EU_uniform}{r_prob_se}{---}
\setres{cty_USA}{EU_uniform}{spread_penalty}{0.635}
\setres{cty_USA}{EU_uniform}{spread_penalty_se}{0.034}
\setres{cty_USA}{EU_uniform}{volume_ess}{---}
\setres{cty_USA}{EV_deterministic}{kappa}{0.814}
\setres{cty_USA}{EV_deterministic}{kappa_se}{0.044}
\setres{cty_USA}{EV_deterministic}{mse}{10.83}
\setres{cty_USA}{EV_deterministic}{nrmse_pct}{15.4}
\setres{cty_USA}{EV_deterministic}{r_ff}{---}
\setres{cty_USA}{EV_deterministic}{r_ff_se}{---}
\setres{cty_USA}{EV_deterministic}{r_prior}{---}
\setres{cty_USA}{EV_deterministic}{r_prior_se}{---}
\setres{cty_USA}{EV_deterministic}{r_prob}{---}
\setres{cty_USA}{EV_deterministic}{r_prob_se}{---}
\setres{cty_USA}{EV_deterministic}{volume_ess}{---}
\setres{cty_USA}{reference}{country_name}{USA}
\setres{cty_USA}{reference}{lookup_nrmse_pct}{14.5}
\setres{cty_USA}{reference}{midpoint_mse}{21.34}
\setres{cty_USA}{reference}{midpoint_nrmse_pct}{23.7}
\setres{cty_USA}{reference}{missing_lotteries}{none}
\setres{cty_USA}{reference}{n_lotteries}{14}
\setres{cty_USA}{reference}{n_obs}{1{,}358}
\setres{cty_USA}{reference}{n_subjects}{97}
\setres{cty_USA}{reference}{oracle_mse}{8.43}
\setres{cty_USA}{reference}{sample}{LV2019\_USA}
\setres{cty_Vietnam}{CPT_deterministic}{kappa}{0.954}
\setres{cty_Vietnam}{CPT_deterministic}{kappa_se}{0.019}
\setres{cty_Vietnam}{CPT_deterministic}{mse}{12.54}
\setres{cty_Vietnam}{CPT_deterministic}{nrmse_pct}{19.4}
\setres{cty_Vietnam}{CPT_deterministic}{r_ff}{---}
\setres{cty_Vietnam}{CPT_deterministic}{r_ff_se}{---}
\setres{cty_Vietnam}{CPT_deterministic}{r_prior}{---}
\setres{cty_Vietnam}{CPT_deterministic}{r_prior_se}{---}
\setres{cty_Vietnam}{CPT_deterministic}{r_prob}{---}
\setres{cty_Vietnam}{CPT_deterministic}{r_prob_se}{---}
\setres{cty_Vietnam}{CPT_deterministic}{theta_star}{1.30, 0.70, 0.70}
\setres{cty_Vietnam}{CPT_deterministic}{volume_ess}{---}
\setres{cty_Vietnam}{CPT_lit}{kappa}{0.184}
\setres{cty_Vietnam}{CPT_lit}{kappa_prior_mean}{0.543}
\setres{cty_Vietnam}{CPT_lit}{kappa_prior_mean_se}{0.109}
\setres{cty_Vietnam}{CPT_lit}{kappa_se}{0.131}
\setres{cty_Vietnam}{CPT_lit}{mse}{18.77}
\setres{cty_Vietnam}{CPT_lit}{mse_prior_mean}{15.87}
\setres{cty_Vietnam}{CPT_lit}{nrmse_pct}{22.7}
\setres{cty_Vietnam}{CPT_lit}{pred_variance}{2.90}
\setres{cty_Vietnam}{CPT_lit}{r_ff}{---}
\setres{cty_Vietnam}{CPT_lit}{r_ff_se}{---}
\setres{cty_Vietnam}{CPT_lit}{r_prior}{---}
\setres{cty_Vietnam}{CPT_lit}{r_prior_se}{---}
\setres{cty_Vietnam}{CPT_lit}{r_prob}{---}
\setres{cty_Vietnam}{CPT_lit}{r_prob_se}{---}
\setres{cty_Vietnam}{CPT_lit}{spread_penalty}{0.358}
\setres{cty_Vietnam}{CPT_lit}{spread_penalty_se}{0.037}
\setres{cty_Vietnam}{CPT_lit}{volume_ess}{---}
\setres{cty_Vietnam}{CPT_transfer_others}{kappa}{0.965}
\setres{cty_Vietnam}{CPT_transfer_others}{kappa_prior_mean}{0.965}
\setres{cty_Vietnam}{CPT_transfer_others}{kappa_prior_mean_se}{0.029}
\setres{cty_Vietnam}{CPT_transfer_others}{kappa_se}{0.029}
\setres{cty_Vietnam}{CPT_transfer_others}{mse}{12.46}
\setres{cty_Vietnam}{CPT_transfer_others}{mse_prior_mean}{12.46}
\setres{cty_Vietnam}{CPT_transfer_others}{nrmse_pct}{19.6}
\setres{cty_Vietnam}{CPT_transfer_others}{pred_variance}{0.00}
\setres{cty_Vietnam}{CPT_transfer_others}{r_ff}{---}
\setres{cty_Vietnam}{CPT_transfer_others}{r_ff_se}{---}
\setres{cty_Vietnam}{CPT_transfer_others}{r_prior}{---}
\setres{cty_Vietnam}{CPT_transfer_others}{r_prior_se}{---}
\setres{cty_Vietnam}{CPT_transfer_others}{r_prob}{---}
\setres{cty_Vietnam}{CPT_transfer_others}{r_prob_se}{---}
\setres{cty_Vietnam}{CPT_transfer_others}{spread_penalty}{0.000}
\setres{cty_Vietnam}{CPT_transfer_others}{spread_penalty_se}{0.000}
\setres{cty_Vietnam}{CPT_transfer_others}{theta_star}{1.00, 0.60, 0.90}
\setres{cty_Vietnam}{CPT_transfer_others}{transfer}{others}
\setres{cty_Vietnam}{CPT_transfer_others}{volume_ess}{---}
\setres{cty_Vietnam}{CPT_uniform}{kappa}{-2.499}
\setres{cty_Vietnam}{CPT_uniform}{kappa_prior_mean}{-0.352}
\setres{cty_Vietnam}{CPT_uniform}{kappa_prior_mean_se}{0.178}
\setres{cty_Vietnam}{CPT_uniform}{kappa_se}{0.343}
\setres{cty_Vietnam}{CPT_uniform}{mse}{40.48}
\setres{cty_Vietnam}{CPT_uniform}{mse_prior_mean}{23.11}
\setres{cty_Vietnam}{CPT_uniform}{nrmse_pct}{25.0}
\setres{cty_Vietnam}{CPT_uniform}{pred_variance}{17.37}
\setres{cty_Vietnam}{CPT_uniform}{r_ff}{---}
\setres{cty_Vietnam}{CPT_uniform}{r_ff_se}{---}
\setres{cty_Vietnam}{CPT_uniform}{r_prior}{---}
\setres{cty_Vietnam}{CPT_uniform}{r_prior_se}{---}
\setres{cty_Vietnam}{CPT_uniform}{r_prob}{---}
\setres{cty_Vietnam}{CPT_uniform}{r_prob_se}{---}
\setres{cty_Vietnam}{CPT_uniform}{spread_penalty}{2.147}
\setres{cty_Vietnam}{CPT_uniform}{spread_penalty_se}{0.221}
\setres{cty_Vietnam}{CPT_uniform}{volume_ess}{---}
\setres{cty_Vietnam}{DA_deterministic}{kappa}{0.867}
\setres{cty_Vietnam}{DA_deterministic}{kappa_se}{0.045}
\setres{cty_Vietnam}{DA_deterministic}{mse}{13.25}
\setres{cty_Vietnam}{DA_deterministic}{nrmse_pct}{19.7}
\setres{cty_Vietnam}{DA_deterministic}{r_ff}{---}
\setres{cty_Vietnam}{DA_deterministic}{r_ff_se}{---}
\setres{cty_Vietnam}{DA_deterministic}{r_prior}{---}
\setres{cty_Vietnam}{DA_deterministic}{r_prior_se}{---}
\setres{cty_Vietnam}{DA_deterministic}{r_prob}{---}
\setres{cty_Vietnam}{DA_deterministic}{r_prob_se}{---}
\setres{cty_Vietnam}{DA_deterministic}{theta_star}{1.95, 1.74}
\setres{cty_Vietnam}{DA_deterministic}{volume_ess}{---}
\setres{cty_Vietnam}{DA_lit}{kappa}{-1.272}
\setres{cty_Vietnam}{DA_lit}{kappa_prior_mean}{-0.685}
\setres{cty_Vietnam}{DA_lit}{kappa_prior_mean_se}{0.266}
\setres{cty_Vietnam}{DA_lit}{kappa_se}{0.312}
\setres{cty_Vietnam}{DA_lit}{mse}{30.55}
\setres{cty_Vietnam}{DA_lit}{mse_prior_mean}{25.80}
\setres{cty_Vietnam}{DA_lit}{nrmse_pct}{29.0}
\setres{cty_Vietnam}{DA_lit}{pred_variance}{4.75}
\setres{cty_Vietnam}{DA_lit}{r_ff}{---}
\setres{cty_Vietnam}{DA_lit}{r_ff_se}{---}
\setres{cty_Vietnam}{DA_lit}{r_prior}{---}
\setres{cty_Vietnam}{DA_lit}{r_prior_se}{---}
\setres{cty_Vietnam}{DA_lit}{r_prob}{---}
\setres{cty_Vietnam}{DA_lit}{r_prob_se}{---}
\setres{cty_Vietnam}{DA_lit}{spread_penalty}{0.587}
\setres{cty_Vietnam}{DA_lit}{spread_penalty_se}{0.060}
\setres{cty_Vietnam}{DA_lit}{volume_ess}{---}
\setres{cty_Vietnam}{DA_transfer_others}{kappa}{0.871}
\setres{cty_Vietnam}{DA_transfer_others}{kappa_prior_mean}{0.871}
\setres{cty_Vietnam}{DA_transfer_others}{kappa_prior_mean_se}{0.055}
\setres{cty_Vietnam}{DA_transfer_others}{kappa_se}{0.055}
\setres{cty_Vietnam}{DA_transfer_others}{mse}{13.22}
\setres{cty_Vietnam}{DA_transfer_others}{mse_prior_mean}{13.22}
\setres{cty_Vietnam}{DA_transfer_others}{nrmse_pct}{20.1}
\setres{cty_Vietnam}{DA_transfer_others}{pred_variance}{0.00}
\setres{cty_Vietnam}{DA_transfer_others}{r_ff}{---}
\setres{cty_Vietnam}{DA_transfer_others}{r_ff_se}{---}
\setres{cty_Vietnam}{DA_transfer_others}{r_prior}{---}
\setres{cty_Vietnam}{DA_transfer_others}{r_prior_se}{---}
\setres{cty_Vietnam}{DA_transfer_others}{r_prob}{---}
\setres{cty_Vietnam}{DA_transfer_others}{r_prob_se}{---}
\setres{cty_Vietnam}{DA_transfer_others}{spread_penalty}{0.000}
\setres{cty_Vietnam}{DA_transfer_others}{spread_penalty_se}{0.000}
\setres{cty_Vietnam}{DA_transfer_others}{theta_star}{1.69, 1.44}
\setres{cty_Vietnam}{DA_transfer_others}{transfer}{others}
\setres{cty_Vietnam}{DA_transfer_others}{volume_ess}{---}
\setres{cty_Vietnam}{DA_uniform}{kappa}{-3.466}
\setres{cty_Vietnam}{DA_uniform}{kappa_prior_mean}{-1.778}
\setres{cty_Vietnam}{DA_uniform}{kappa_prior_mean_se}{0.363}
\setres{cty_Vietnam}{DA_uniform}{kappa_se}{0.502}
\setres{cty_Vietnam}{DA_uniform}{mse}{48.29}
\setres{cty_Vietnam}{DA_uniform}{mse_prior_mean}{34.65}
\setres{cty_Vietnam}{DA_uniform}{nrmse_pct}{32.1}
\setres{cty_Vietnam}{DA_uniform}{pred_variance}{13.65}
\setres{cty_Vietnam}{DA_uniform}{r_ff}{---}
\setres{cty_Vietnam}{DA_uniform}{r_ff_se}{---}
\setres{cty_Vietnam}{DA_uniform}{r_prior}{---}
\setres{cty_Vietnam}{DA_uniform}{r_prior_se}{---}
\setres{cty_Vietnam}{DA_uniform}{r_prob}{---}
\setres{cty_Vietnam}{DA_uniform}{r_prob_se}{---}
\setres{cty_Vietnam}{DA_uniform}{spread_penalty}{1.687}
\setres{cty_Vietnam}{DA_uniform}{spread_penalty_se}{0.174}
\setres{cty_Vietnam}{DA_uniform}{volume_ess}{---}
\setres{cty_Vietnam}{EU_deterministic}{kappa}{0.762}
\setres{cty_Vietnam}{EU_deterministic}{kappa_se}{0.077}
\setres{cty_Vietnam}{EU_deterministic}{mse}{14.10}
\setres{cty_Vietnam}{EU_deterministic}{nrmse_pct}{21.9}
\setres{cty_Vietnam}{EU_deterministic}{r_ff}{---}
\setres{cty_Vietnam}{EU_deterministic}{r_ff_se}{---}
\setres{cty_Vietnam}{EU_deterministic}{r_prior}{---}
\setres{cty_Vietnam}{EU_deterministic}{r_prior_se}{---}
\setres{cty_Vietnam}{EU_deterministic}{r_prob}{---}
\setres{cty_Vietnam}{EU_deterministic}{r_prob_se}{---}
\setres{cty_Vietnam}{EU_deterministic}{theta_star}{1.03}
\setres{cty_Vietnam}{EU_deterministic}{volume_ess}{---}
\setres{cty_Vietnam}{EU_lit}{kappa}{0.160}
\setres{cty_Vietnam}{EU_lit}{kappa_prior_mean}{0.309}
\setres{cty_Vietnam}{EU_lit}{kappa_prior_mean_se}{0.149}
\setres{cty_Vietnam}{EU_lit}{kappa_se}{0.161}
\setres{cty_Vietnam}{EU_lit}{mse}{18.97}
\setres{cty_Vietnam}{EU_lit}{mse_prior_mean}{17.76}
\setres{cty_Vietnam}{EU_lit}{nrmse_pct}{24.1}
\setres{cty_Vietnam}{EU_lit}{pred_variance}{1.20}
\setres{cty_Vietnam}{EU_lit}{r_ff}{---}
\setres{cty_Vietnam}{EU_lit}{r_ff_se}{---}
\setres{cty_Vietnam}{EU_lit}{r_prior}{---}
\setres{cty_Vietnam}{EU_lit}{r_prior_se}{---}
\setres{cty_Vietnam}{EU_lit}{r_prob}{---}
\setres{cty_Vietnam}{EU_lit}{r_prob_se}{---}
\setres{cty_Vietnam}{EU_lit}{spread_penalty}{0.149}
\setres{cty_Vietnam}{EU_lit}{spread_penalty_se}{0.015}
\setres{cty_Vietnam}{EU_lit}{volume_ess}{---}
\setres{cty_Vietnam}{EU_transfer_others}{kappa}{0.753}
\setres{cty_Vietnam}{EU_transfer_others}{kappa_prior_mean}{0.753}
\setres{cty_Vietnam}{EU_transfer_others}{kappa_prior_mean_se}{0.079}
\setres{cty_Vietnam}{EU_transfer_others}{kappa_se}{0.079}
\setres{cty_Vietnam}{EU_transfer_others}{mse}{14.18}
\setres{cty_Vietnam}{EU_transfer_others}{mse_prior_mean}{14.18}
\setres{cty_Vietnam}{EU_transfer_others}{nrmse_pct}{21.4}
\setres{cty_Vietnam}{EU_transfer_others}{pred_variance}{0.00}
\setres{cty_Vietnam}{EU_transfer_others}{r_ff}{---}
\setres{cty_Vietnam}{EU_transfer_others}{r_ff_se}{---}
\setres{cty_Vietnam}{EU_transfer_others}{r_prior}{---}
\setres{cty_Vietnam}{EU_transfer_others}{r_prior_se}{---}
\setres{cty_Vietnam}{EU_transfer_others}{r_prob}{---}
\setres{cty_Vietnam}{EU_transfer_others}{r_prob_se}{---}
\setres{cty_Vietnam}{EU_transfer_others}{spread_penalty}{0.000}
\setres{cty_Vietnam}{EU_transfer_others}{spread_penalty_se}{0.000}
\setres{cty_Vietnam}{EU_transfer_others}{theta_star}{0.94}
\setres{cty_Vietnam}{EU_transfer_others}{transfer}{others}
\setres{cty_Vietnam}{EU_transfer_others}{volume_ess}{---}
\setres{cty_Vietnam}{EU_uniform}{kappa}{-1.915}
\setres{cty_Vietnam}{EU_uniform}{kappa_prior_mean}{-0.901}
\setres{cty_Vietnam}{EU_uniform}{kappa_prior_mean_se}{0.280}
\setres{cty_Vietnam}{EU_uniform}{kappa_se}{0.362}
\setres{cty_Vietnam}{EU_uniform}{mse}{35.75}
\setres{cty_Vietnam}{EU_uniform}{mse_prior_mean}{27.55}
\setres{cty_Vietnam}{EU_uniform}{nrmse_pct}{29.1}
\setres{cty_Vietnam}{EU_uniform}{pred_variance}{8.20}
\setres{cty_Vietnam}{EU_uniform}{r_ff}{---}
\setres{cty_Vietnam}{EU_uniform}{r_ff_se}{---}
\setres{cty_Vietnam}{EU_uniform}{r_prior}{---}
\setres{cty_Vietnam}{EU_uniform}{r_prior_se}{---}
\setres{cty_Vietnam}{EU_uniform}{r_prob}{---}
\setres{cty_Vietnam}{EU_uniform}{r_prob_se}{---}
\setres{cty_Vietnam}{EU_uniform}{spread_penalty}{1.013}
\setres{cty_Vietnam}{EU_uniform}{spread_penalty_se}{0.104}
\setres{cty_Vietnam}{EU_uniform}{volume_ess}{---}
\setres{cty_Vietnam}{EV_deterministic}{kappa}{0.771}
\setres{cty_Vietnam}{EV_deterministic}{kappa_se}{0.076}
\setres{cty_Vietnam}{EV_deterministic}{mse}{14.02}
\setres{cty_Vietnam}{EV_deterministic}{nrmse_pct}{21.7}
\setres{cty_Vietnam}{EV_deterministic}{r_ff}{---}
\setres{cty_Vietnam}{EV_deterministic}{r_ff_se}{---}
\setres{cty_Vietnam}{EV_deterministic}{r_prior}{---}
\setres{cty_Vietnam}{EV_deterministic}{r_prior_se}{---}
\setres{cty_Vietnam}{EV_deterministic}{r_prob}{---}
\setres{cty_Vietnam}{EV_deterministic}{r_prob_se}{---}
\setres{cty_Vietnam}{EV_deterministic}{volume_ess}{---}
\setres{cty_Vietnam}{reference}{country_name}{Vietnam}
\setres{cty_Vietnam}{reference}{lookup_nrmse_pct}{19.5}
\setres{cty_Vietnam}{reference}{midpoint_mse}{20.26}
\setres{cty_Vietnam}{reference}{midpoint_nrmse_pct}{24.3}
\setres{cty_Vietnam}{reference}{missing_lotteries}{none}
\setres{cty_Vietnam}{reference}{n_lotteries}{14}
\setres{cty_Vietnam}{reference}{n_obs}{1{,}218}
\setres{cty_Vietnam}{reference}{n_subjects}{87}
\setres{cty_Vietnam}{reference}{oracle_mse}{12.18}
\setres{cty_Vietnam}{reference}{sample}{LV2019\_Vietnam}
\setres{gains}{CPT}{max_over_const}{0.88}
\setres{kern}{diagonal__CPT_lit}{r_prob}{0.955}
\setres{kern}{diagonal__CPT_uniform}{r_prob}{0.575}
\setres{kern}{diagonal__DA_lit}{r_prob}{0.953}
\setres{kern}{diagonal__DA_uniform}{r_prob}{0.899}
\setres{kern}{diagonal__EU_lit}{r_prob}{0.868}
\setres{kern}{diagonal__EU_uniform}{r_prob}{0.700}
\setres{kern}{linear__CPT_lit}{r_prob}{0.950}
\setres{kern}{linear__CPT_uniform}{r_prob}{0.530}
\setres{kern}{linear__DA_lit}{r_prob}{0.946}
\setres{kern}{linear__DA_uniform}{r_prob}{0.883}
\setres{kern}{linear__EU_lit}{r_prob}{0.842}
\setres{kern}{linear__EU_uniform}{r_prob}{0.642}
\setres{kern}{matern52__CPT_lit}{r_prob}{0.951}
\setres{kern}{matern52__CPT_uniform}{r_prob}{0.538}
\setres{kern}{matern52__DA_lit}{r_prob}{0.949}
\setres{kern}{matern52__DA_uniform}{r_prob}{0.889}
\setres{kern}{matern52__EU_lit}{r_prob}{0.851}
\setres{kern}{matern52__EU_uniform}{r_prob}{0.661}
\setres{kern}{rbf__CPT_lit}{r_prob}{0.951}
\setres{kern}{rbf__CPT_uniform}{r_prob}{0.533}
\setres{kern}{rbf__DA_lit}{r_prob}{0.948}
\setres{kern}{rbf__DA_uniform}{r_prob}{0.888}
\setres{kern}{rbf__EU_lit}{r_prob}{0.848}
\setres{kern}{rbf__EU_uniform}{r_prob}{0.655}
\setres{b3}{summary}{kappa_cpt_min}{0.471}
\setres{b3}{summary}{kappa_da_max}{0.291}
\setres{b3}{summary}{margin_cpt_over_eu_main}{0.043}
\setres{b3}{summary}{margin_da_over_eu_main}{0.050}
\setres{b3}{summary}{margin_over_eu}{0.042}
\setres{b3}{summary}{max_dr_cpt}{0.036}
\setres{b3}{summary}{max_dr_da}{0.029}
\setres{b3}{summary}{max_r_eu}{0.878}
\setres{b3}{summary}{min_r_cpt}{0.920}
\setres{b3}{summary}{min_r_cpt_da}{0.920}
\setres{b3}{summary}{min_r_da}{0.927}
\setres{b3}{summary}{n_cells}{23}
\setres{b3}{summary}{n_cpt}{13}
\setres{b3}{summary}{n_cpt_above_da_main}{6}
\setres{b3}{summary}{n_da}{9}
\setres{b3}{summary}{n_da_below_cpt_main}{4}
\setres{b3}{summary}{n_kappa_order_ok}{21}
\setres{b3}{summary}{one_minus_rff_cpt}{0.860}
\setres{b3}{summary}{one_minus_rff_da}{0.719}
\setres{b3}{CPT_alpha_wider}{kappa}{0.554}
\setres{b3}{CPT_alpha_wider}{r_prob}{0.944}
\setres{b3}{CPT_alpha_wider}{kappa_se}{0.057}
\setres{b3}{CPT_alpha_tighter}{kappa}{0.686}
\setres{b3}{CPT_alpha_tighter}{r_prob}{0.967}
\setres{b3}{CPT_alpha_tighter}{kappa_se}{0.046}
\setres{b3}{CPT_alpha_mean_down}{kappa}{0.512}
\setres{b3}{CPT_alpha_mean_down}{r_prob}{0.942}
\setres{b3}{CPT_alpha_mean_down}{kappa_se}{0.072}
\setres{b3}{CPT_alpha_mean_up}{kappa}{0.672}
\setres{b3}{CPT_alpha_mean_up}{r_prob}{0.964}
\setres{b3}{CPT_alpha_mean_up}{kappa_se}{0.043}
\setres{b3}{DA_alpha_wider}{kappa}{-0.099}
\setres{b3}{DA_alpha_wider}{r_prob}{0.946}
\setres{b3}{DA_alpha_wider}{kappa_se}{0.146}
\setres{b3}{DA_alpha_tighter}{kappa}{0.101}
\setres{b3}{DA_alpha_tighter}{r_prob}{0.966}
\setres{b3}{DA_alpha_tighter}{kappa_se}{0.130}
\setres{b3}{DA_alpha_mean_down}{kappa}{-0.509}
\setres{b3}{DA_alpha_mean_down}{r_prob}{0.945}
\setres{b3}{DA_alpha_mean_down}{kappa_se}{0.199}
\setres{b3}{DA_alpha_mean_up}{kappa}{0.291}
\setres{b3}{DA_alpha_mean_up}{r_prob}{0.968}
\setres{b3}{DA_alpha_mean_up}{kappa_se}{0.100}
\setres{b3}{CPT_gamma_wider}{kappa}{0.617}
\setres{b3}{CPT_gamma_wider}{r_prob}{0.952}
\setres{b3}{CPT_gamma_wider}{kappa_se}{0.050}
\setres{b3}{CPT_gamma_tighter}{kappa}{0.651}
\setres{b3}{CPT_gamma_tighter}{r_prob}{0.965}
\setres{b3}{CPT_gamma_tighter}{kappa_se}{0.050}
\setres{b3}{CPT_gamma_mean_down}{kappa}{0.684}
\setres{b3}{CPT_gamma_mean_down}{r_prob}{0.926}
\setres{b3}{CPT_gamma_mean_down}{kappa_se}{0.039}
\setres{b3}{CPT_gamma_mean_up}{kappa}{0.581}
\setres{b3}{CPT_gamma_mean_up}{r_prob}{0.977}
\setres{b3}{CPT_gamma_mean_up}{kappa_se}{0.060}
\setres{b3}{CPT_delta_wider}{kappa}{0.471}
\setres{b3}{CPT_delta_wider}{r_prob}{0.920}
\setres{b3}{CPT_delta_wider}{kappa_se}{0.062}
\setres{b3}{CPT_delta_tighter}{kappa}{0.696}
\setres{b3}{CPT_delta_tighter}{r_prob}{0.978}
\setres{b3}{CPT_delta_tighter}{kappa_se}{0.047}
\setres{b3}{CPT_delta_mean_down}{kappa}{0.614}
\setres{b3}{CPT_delta_mean_down}{r_prob}{0.957}
\setres{b3}{CPT_delta_mean_down}{kappa_se}{0.059}
\setres{b3}{CPT_delta_mean_up}{kappa}{0.646}
\setres{b3}{CPT_delta_mean_up}{r_prob}{0.955}
\setres{b3}{CPT_delta_mean_up}{kappa_se}{0.044}
\setres{b3}{DA_eta_wider}{kappa}{-0.190}
\setres{b3}{DA_eta_wider}{r_prob}{0.927}
\setres{b3}{DA_eta_wider}{kappa_se}{0.146}
\setres{b3}{DA_eta_tighter}{kappa}{0.059}
\setres{b3}{DA_eta_tighter}{r_prob}{0.979}
\setres{b3}{DA_eta_tighter}{kappa_se}{0.138}
\setres{b3}{DA_eta_mean_down}{kappa}{0.077}
\setres{b3}{DA_eta_mean_down}{r_prob}{0.948}
\setres{b3}{DA_eta_mean_down}{kappa_se}{0.123}
\setres{b3}{DA_eta_mean_up}{kappa}{-0.057}
\setres{b3}{DA_eta_mean_up}{r_prob}{0.962}
\setres{b3}{DA_eta_mean_up}{kappa_se}{0.150}
\setres{b3}{CPT_all_TN_family}{kappa}{0.654}
\setres{b3}{CPT_all_TN_family}{r_prob}{0.954}
\setres{b3}{CPT_all_TN_family}{kappa_se}{0.049}
\setres{b3}{DA_all_TN_family}{kappa}{0.023}
\setres{b3}{DA_all_TN_family}{r_prob}{0.956}
\setres{b3}{DA_all_TN_family}{kappa_se}{0.137}
\setres{b3}{EU_all_TN_family}{kappa}{0.530}
\setres{b3}{EU_all_TN_family}{r_prob}{0.878}
\setres{b3}{EU_all_TN_family}{kappa_se}{0.074}
\setres{pdAvg}{CPT_deterministic}{kappa}{0.976}
\setres{pdAvg}{CPT_deterministic}{r_prob}{1.000}
\setres{pdAvg}{CPT_deterministic}{kappa_gains_only}{0.486}
\setres{pdAvg}{CPT_deterministic}{r_prob_gains_only}{0.500}
\setres{pdAvg}{CPT_lit}{kappa}{0.174}
\setres{pdAvg}{CPT_lit}{r_prob}{0.937}
\setres{pdAvg}{CPT_lit}{kappa_prior_mean}{0.515}
\setres{pdAvg}{CPT_lit}{kappa_gains_only}{0.320}
\setres{pdAvg}{CPT_lit}{r_prob_gains_only}{0.478}
\setres{pdAvg}{CPT_uniform}{kappa}{-1.277}
\setres{pdAvg}{CPT_uniform}{r_prob}{0.595}
\setres{pdAvg}{CPT_uniform}{kappa_prior_mean}{0.117}
\setres{pdAvg}{CPT_uniform}{kappa_gains_only}{-0.248}
\setres{pdAvg}{CPT_uniform}{r_prob_gains_only}{0.291}
\setres{pdAvg}{DA_deterministic}{kappa}{0.749}
\setres{pdAvg}{DA_deterministic}{r_prob}{1.000}
\setres{pdAvg}{DA_deterministic}{kappa_gains_only}{0.313}
\setres{pdAvg}{DA_deterministic}{r_prob_gains_only}{0.500}
\setres{pdAvg}{DA_lit}{kappa}{-1.012}
\setres{pdAvg}{DA_lit}{r_prob}{0.937}
\setres{pdAvg}{DA_lit}{kappa_prior_mean}{-0.610}
\setres{pdAvg}{DA_lit}{kappa_gains_only}{0.010}
\setres{pdAvg}{DA_lit}{r_prob_gains_only}{0.478}
\setres{pdAvg}{DA_uniform}{kappa}{-1.237}
\setres{pdAvg}{DA_uniform}{r_prob}{0.876}
\setres{pdAvg}{DA_uniform}{kappa_prior_mean}{-0.138}
\setres{pdAvg}{DA_uniform}{kappa_gains_only}{-0.413}
\setres{pdAvg}{DA_uniform}{r_prob_gains_only}{0.452}
\setres{pdAvg}{EU_deterministic}{kappa}{0.603}
\setres{pdAvg}{EU_deterministic}{r_prob}{1.000}
\setres{pdAvg}{EU_deterministic}{kappa_gains_only}{0.308}
\setres{pdAvg}{EU_deterministic}{r_prob_gains_only}{0.500}
\setres{pdAvg}{EU_lit}{kappa}{0.006}
\setres{pdAvg}{EU_lit}{r_prob}{0.811}
\setres{pdAvg}{EU_lit}{kappa_prior_mean}{0.213}
\setres{pdAvg}{EU_lit}{kappa_gains_only}{0.261}
\setres{pdAvg}{EU_lit}{r_prob_gains_only}{0.438}
\setres{pdAvg}{EU_uniform}{kappa}{-1.174}
\setres{pdAvg}{EU_uniform}{r_prob}{0.710}
\setres{pdAvg}{EU_uniform}{kappa_prior_mean}{-0.640}
\setres{pdAvg}{EU_uniform}{kappa_gains_only}{-0.105}
\setres{pdAvg}{EU_uniform}{r_prob_gains_only}{0.360}
\setres{pdAvg}{EV_deterministic}{kappa}{0.439}
\setres{pdAvg}{EV_deterministic}{r_prob}{1.000}
\setres{pdAvg}{EV_deterministic}{kappa_gains_only}{0.247}
\setres{pdAvg}{EV_deterministic}{r_prob_gains_only}{0.500}
\setres{pdAvgAsis}{CPT_deterministic}{kappa}{0.976}
\setres{pdAvgAsis}{CPT_deterministic}{r_prob}{1.000}
\setres{pdAvgAsis}{CPT_deterministic}{kappa_gains_only}{0.486}
\setres{pdAvgAsis}{CPT_deterministic}{r_prob_gains_only}{0.500}
\setres{pdAvgAsis}{CPT_lit}{kappa}{0.191}
\setres{pdAvgAsis}{CPT_lit}{r_prob}{0.956}
\setres{pdAvgAsis}{CPT_lit}{kappa_prior_mean}{0.417}
\setres{pdAvgAsis}{CPT_lit}{kappa_gains_only}{0.320}
\setres{pdAvgAsis}{CPT_lit}{r_prob_gains_only}{0.478}
\setres{pdAvgAsis}{CPT_uniform}{kappa}{-1.277}
\setres{pdAvgAsis}{CPT_uniform}{r_prob}{0.595}
\setres{pdAvgAsis}{CPT_uniform}{kappa_prior_mean}{0.117}
\setres{pdAvgAsis}{CPT_uniform}{kappa_gains_only}{-0.248}
\setres{pdAvgAsis}{CPT_uniform}{r_prob_gains_only}{0.291}
\setres{pdAvgAsis}{DA_deterministic}{kappa}{0.749}
\setres{pdAvgAsis}{DA_deterministic}{r_prob}{1.000}
\setres{pdAvgAsis}{DA_deterministic}{kappa_gains_only}{0.313}
\setres{pdAvgAsis}{DA_deterministic}{r_prob_gains_only}{0.500}
\setres{pdAvgAsis}{DA_lit}{kappa}{-0.438}
\setres{pdAvgAsis}{DA_lit}{r_prob}{0.948}
\setres{pdAvgAsis}{DA_lit}{kappa_prior_mean}{-0.066}
\setres{pdAvgAsis}{DA_lit}{kappa_gains_only}{0.010}
\setres{pdAvgAsis}{DA_lit}{r_prob_gains_only}{0.478}
\setres{pdAvgAsis}{DA_uniform}{kappa}{-1.237}
\setres{pdAvgAsis}{DA_uniform}{r_prob}{0.876}
\setres{pdAvgAsis}{DA_uniform}{kappa_prior_mean}{-0.138}
\setres{pdAvgAsis}{DA_uniform}{kappa_gains_only}{-0.413}
\setres{pdAvgAsis}{DA_uniform}{r_prob_gains_only}{0.452}
\setres{pdAvgAsis}{EU_deterministic}{kappa}{0.603}
\setres{pdAvgAsis}{EU_deterministic}{r_prob}{1.000}
\setres{pdAvgAsis}{EU_deterministic}{kappa_gains_only}{0.308}
\setres{pdAvgAsis}{EU_deterministic}{r_prob_gains_only}{0.500}
\setres{pdAvgAsis}{EU_lit}{kappa}{0.017}
\setres{pdAvgAsis}{EU_lit}{r_prob}{0.872}
\setres{pdAvgAsis}{EU_lit}{kappa_prior_mean}{0.112}
\setres{pdAvgAsis}{EU_lit}{kappa_gains_only}{0.261}
\setres{pdAvgAsis}{EU_lit}{r_prob_gains_only}{0.438}
\setres{pdAvgAsis}{EU_uniform}{kappa}{-1.174}
\setres{pdAvgAsis}{EU_uniform}{r_prob}{0.710}
\setres{pdAvgAsis}{EU_uniform}{kappa_prior_mean}{-0.640}
\setres{pdAvgAsis}{EU_uniform}{kappa_gains_only}{-0.105}
\setres{pdAvgAsis}{EU_uniform}{r_prob_gains_only}{0.360}
\setres{pdAvgAsis}{EV_deterministic}{kappa}{0.439}
\setres{pdAvgAsis}{EV_deterministic}{r_prob}{1.000}
\setres{pdAvgAsis}{EV_deterministic}{kappa_gains_only}{0.247}
\setres{pdAvgAsis}{EV_deterministic}{r_prob_gains_only}{0.500}
\setres{pdAvgAdapt}{CPT_deterministic}{kappa}{0.976}
\setres{pdAvgAdapt}{CPT_deterministic}{r_prob}{1.000}
\setres{pdAvgAdapt}{CPT_deterministic}{kappa_gains_only}{0.486}
\setres{pdAvgAdapt}{CPT_deterministic}{r_prob_gains_only}{0.500}
\setres{pdAvgAdapt}{CPT_lit}{kappa}{0.174}
\setres{pdAvgAdapt}{CPT_lit}{r_prob}{0.937}
\setres{pdAvgAdapt}{CPT_lit}{kappa_prior_mean}{0.515}
\setres{pdAvgAdapt}{CPT_lit}{kappa_gains_only}{0.320}
\setres{pdAvgAdapt}{CPT_lit}{r_prob_gains_only}{0.478}
\setres{pdAvgAdapt}{CPT_uniform}{kappa}{-0.766}
\setres{pdAvgAdapt}{CPT_uniform}{r_prob}{0.621}
\setres{pdAvgAdapt}{CPT_uniform}{kappa_prior_mean}{0.774}
\setres{pdAvgAdapt}{CPT_uniform}{kappa_gains_only}{-0.248}
\setres{pdAvgAdapt}{CPT_uniform}{r_prob_gains_only}{0.291}
\setres{pdAvgAdapt}{DA_deterministic}{kappa}{0.749}
\setres{pdAvgAdapt}{DA_deterministic}{r_prob}{1.000}
\setres{pdAvgAdapt}{DA_deterministic}{kappa_gains_only}{0.313}
\setres{pdAvgAdapt}{DA_deterministic}{r_prob_gains_only}{0.500}
\setres{pdAvgAdapt}{DA_lit}{kappa}{-1.012}
\setres{pdAvgAdapt}{DA_lit}{r_prob}{0.937}
\setres{pdAvgAdapt}{DA_lit}{kappa_prior_mean}{-0.610}
\setres{pdAvgAdapt}{DA_lit}{kappa_gains_only}{0.010}
\setres{pdAvgAdapt}{DA_lit}{r_prob_gains_only}{0.478}
\setres{pdAvgAdapt}{DA_uniform}{kappa}{-0.878}
\setres{pdAvgAdapt}{DA_uniform}{r_prob}{0.885}
\setres{pdAvgAdapt}{DA_uniform}{kappa_prior_mean}{0.322}
\setres{pdAvgAdapt}{DA_uniform}{kappa_gains_only}{-0.413}
\setres{pdAvgAdapt}{DA_uniform}{r_prob_gains_only}{0.452}
\setres{pdAvgAdapt}{EU_deterministic}{kappa}{0.603}
\setres{pdAvgAdapt}{EU_deterministic}{r_prob}{1.000}
\setres{pdAvgAdapt}{EU_deterministic}{kappa_gains_only}{0.308}
\setres{pdAvgAdapt}{EU_deterministic}{r_prob_gains_only}{0.500}
\setres{pdAvgAdapt}{EU_lit}{kappa}{0.006}
\setres{pdAvgAdapt}{EU_lit}{r_prob}{0.811}
\setres{pdAvgAdapt}{EU_lit}{kappa_prior_mean}{0.213}
\setres{pdAvgAdapt}{EU_lit}{kappa_gains_only}{0.261}
\setres{pdAvgAdapt}{EU_lit}{r_prob_gains_only}{0.438}
\setres{pdAvgAdapt}{EU_uniform}{kappa}{-0.520}
\setres{pdAvgAdapt}{EU_uniform}{r_prob}{0.688}
\setres{pdAvgAdapt}{EU_uniform}{kappa_prior_mean}{0.210}
\setres{pdAvgAdapt}{EU_uniform}{kappa_gains_only}{-0.105}
\setres{pdAvgAdapt}{EU_uniform}{r_prob_gains_only}{0.360}
\setres{pdAvgAdapt}{EV_deterministic}{kappa}{0.439}
\setres{pdAvgAdapt}{EV_deterministic}{r_prob}{1.000}
\setres{pdAvgAdapt}{EV_deterministic}{kappa_gains_only}{0.247}
\setres{pdAvgAdapt}{EV_deterministic}{r_prob_gains_only}{0.500}
\setres{pdAvgShared}{CPT_deterministic}{kappa}{0.976}
\setres{pdAvgShared}{CPT_deterministic}{r_prob}{1.000}
\setres{pdAvgShared}{CPT_deterministic}{kappa_gains_only}{0.486}
\setres{pdAvgShared}{CPT_deterministic}{r_prob_gains_only}{0.500}
\setres{pdAvgShared}{CPT_lit}{kappa}{-0.005}
\setres{pdAvgShared}{CPT_lit}{r_prob}{0.913}
\setres{pdAvgShared}{CPT_lit}{kappa_prior_mean}{0.449}
\setres{pdAvgShared}{CPT_lit}{kappa_gains_only}{0.222}
\setres{pdAvgShared}{CPT_lit}{r_prob_gains_only}{0.456}
\setres{pdAvgShared}{CPT_uniform}{kappa}{-0.914}
\setres{pdAvgShared}{CPT_uniform}{r_prob}{0.631}
\setres{pdAvgShared}{CPT_uniform}{kappa_prior_mean}{0.791}
\setres{pdAvgShared}{CPT_uniform}{kappa_gains_only}{-0.396}
\setres{pdAvgShared}{CPT_uniform}{r_prob_gains_only}{0.301}
\setres{pdAvgShared}{DA_deterministic}{kappa}{0.749}
\setres{pdAvgShared}{DA_deterministic}{r_prob}{1.000}
\setres{pdAvgShared}{DA_deterministic}{kappa_gains_only}{0.313}
\setres{pdAvgShared}{DA_deterministic}{r_prob_gains_only}{0.500}
\setres{pdAvgShared}{DA_lit}{kappa}{-1.276}
\setres{pdAvgShared}{DA_lit}{r_prob}{0.931}
\setres{pdAvgShared}{DA_lit}{kappa_prior_mean}{-0.869}
\setres{pdAvgShared}{DA_lit}{kappa_gains_only}{-0.255}
\setres{pdAvgShared}{DA_lit}{r_prob_gains_only}{0.472}
\setres{pdAvgShared}{DA_uniform}{kappa}{-0.686}
\setres{pdAvgShared}{DA_uniform}{r_prob}{0.879}
\setres{pdAvgShared}{DA_uniform}{kappa_prior_mean}{0.659}
\setres{pdAvgShared}{DA_uniform}{kappa_gains_only}{-0.221}
\setres{pdAvgShared}{DA_uniform}{r_prob_gains_only}{0.446}
\setres{pdAvgShared}{EU_deterministic}{kappa}{0.603}
\setres{pdAvgShared}{EU_deterministic}{r_prob}{1.000}
\setres{pdAvgShared}{EU_deterministic}{kappa_gains_only}{0.308}
\setres{pdAvgShared}{EU_deterministic}{r_prob_gains_only}{0.500}
\setres{pdAvgShared}{EU_lit}{kappa}{-0.168}
\setres{pdAvgShared}{EU_lit}{r_prob}{0.750}
\setres{pdAvgShared}{EU_lit}{kappa_prior_mean}{0.148}
\setres{pdAvgShared}{EU_lit}{kappa_gains_only}{0.166}
\setres{pdAvgShared}{EU_lit}{r_prob_gains_only}{0.379}
\setres{pdAvgShared}{EU_uniform}{kappa}{-0.499}
\setres{pdAvgShared}{EU_uniform}{r_prob}{0.666}
\setres{pdAvgShared}{EU_uniform}{kappa_prior_mean}{0.408}
\setres{pdAvgShared}{EU_uniform}{kappa_gains_only}{-0.084}
\setres{pdAvgShared}{EU_uniform}{r_prob_gains_only}{0.339}
\setres{pdAvgShared}{EV_deterministic}{kappa}{0.439}
\setres{pdAvgShared}{EV_deterministic}{r_prob}{1.000}
\setres{pdAvgShared}{EV_deterministic}{kappa_gains_only}{0.247}
\setres{pdAvgShared}{EV_deterministic}{r_prob_gains_only}{0.500}
\setres{pdDiff}{CPT_deterministic__minus__DA_deterministic}{diff}{0.227}
\setres{pdDiff}{CPT_deterministic__minus__DA_deterministic}{diff_se}{0.038}
\setres{pdDiff}{CPT_deterministic__minus__DA_deterministic}{diff_over_se}{6.0}
\setres{pdDiff}{CPT_deterministic__minus__DA_deterministic}{corr}{0.05}
\setres{pdDiff}{CPT_deterministic__minus__EU_deterministic}{diff}{0.373}
\setres{pdDiff}{CPT_deterministic__minus__EU_deterministic}{diff_se}{0.067}
\setres{pdDiff}{CPT_deterministic__minus__EU_deterministic}{diff_over_se}{5.6}
\setres{pdDiff}{CPT_deterministic__minus__EU_deterministic}{corr}{0.02}
\setres{pdDiff}{CPT_deterministic__minus__EV_deterministic}{diff}{0.537}
\setres{pdDiff}{CPT_deterministic__minus__EV_deterministic}{diff_se}{0.069}
\setres{pdDiff}{CPT_deterministic__minus__EV_deterministic}{diff_over_se}{7.8}
\setres{pdDiff}{CPT_deterministic__minus__EV_deterministic}{corr}{0.07}
\setres{pdDiff}{CPT_lit__minus__DA_lit}{diff}{1.186}
\setres{pdDiff}{CPT_lit__minus__DA_lit}{diff_se}{0.098}
\setres{pdDiff}{CPT_lit__minus__DA_lit}{diff_over_se}{12.1}
\setres{pdDiff}{CPT_lit__minus__DA_lit}{corr}{0.97}
\setres{pdDiff}{CPT_lit__minus__EU_lit}{diff}{0.168}
\setres{pdDiff}{CPT_lit__minus__EU_lit}{diff_se}{0.027}
\setres{pdDiff}{CPT_lit__minus__EU_lit}{diff_over_se}{6.2}
\setres{pdDiff}{CPT_lit__minus__EU_lit}{corr}{1.00}
\setres{pdDiff}{EV_deterministic__minus__CPT_lit}{diff}{0.265}
\setres{pdDiff}{EV_deterministic__minus__CPT_lit}{diff_se}{0.019}
\setres{pdDiff}{EV_deterministic__minus__CPT_lit}{diff_over_se}{13.8}
\setres{pdDiff}{EV_deterministic__minus__CPT_lit}{corr}{0.96}
\setres{pdDiffAsis}{CPT_deterministic__minus__DA_deterministic}{diff}{0.227}
\setres{pdDiffAsis}{CPT_deterministic__minus__DA_deterministic}{diff_se}{0.038}
\setres{pdDiffAsis}{CPT_deterministic__minus__DA_deterministic}{diff_over_se}{6.0}
\setres{pdDiffAsis}{CPT_deterministic__minus__DA_deterministic}{corr}{0.05}
\setres{pdDiffAsis}{CPT_deterministic__minus__EU_deterministic}{diff}{0.373}
\setres{pdDiffAsis}{CPT_deterministic__minus__EU_deterministic}{diff_se}{0.067}
\setres{pdDiffAsis}{CPT_deterministic__minus__EU_deterministic}{diff_over_se}{5.6}
\setres{pdDiffAsis}{CPT_deterministic__minus__EU_deterministic}{corr}{0.02}
\setres{pdDiffAsis}{CPT_deterministic__minus__EV_deterministic}{diff}{0.537}
\setres{pdDiffAsis}{CPT_deterministic__minus__EV_deterministic}{diff_se}{0.069}
\setres{pdDiffAsis}{CPT_deterministic__minus__EV_deterministic}{diff_over_se}{7.8}
\setres{pdDiffAsis}{CPT_deterministic__minus__EV_deterministic}{corr}{0.07}
\setres{pdDiffAsis}{CPT_lit__minus__DA_lit}{diff}{0.629}
\setres{pdDiffAsis}{CPT_lit__minus__DA_lit}{diff_se}{0.069}
\setres{pdDiffAsis}{CPT_lit__minus__DA_lit}{diff_over_se}{9.1}
\setres{pdDiffAsis}{CPT_lit__minus__DA_lit}{corr}{0.94}
\setres{pdDiffAsis}{CPT_lit__minus__EU_lit}{diff}{0.174}
\setres{pdDiffAsis}{CPT_lit__minus__EU_lit}{diff_se}{0.027}
\setres{pdDiffAsis}{CPT_lit__minus__EU_lit}{diff_over_se}{6.5}
\setres{pdDiffAsis}{CPT_lit__minus__EU_lit}{corr}{0.99}
\setres{pdDiffAsis}{EV_deterministic__minus__CPT_lit}{diff}{0.247}
\setres{pdDiffAsis}{EV_deterministic__minus__CPT_lit}{diff_se}{0.019}
\setres{pdDiffAsis}{EV_deterministic__minus__CPT_lit}{diff_over_se}{13.1}
\setres{pdDiffAsis}{EV_deterministic__minus__CPT_lit}{corr}{0.96}
\setres{pdDiffAdapt}{CPT_deterministic__minus__DA_deterministic}{diff}{0.227}
\setres{pdDiffAdapt}{CPT_deterministic__minus__DA_deterministic}{diff_se}{0.038}
\setres{pdDiffAdapt}{CPT_deterministic__minus__DA_deterministic}{diff_over_se}{6.0}
\setres{pdDiffAdapt}{CPT_deterministic__minus__DA_deterministic}{corr}{0.05}
\setres{pdDiffAdapt}{CPT_deterministic__minus__EU_deterministic}{diff}{0.373}
\setres{pdDiffAdapt}{CPT_deterministic__minus__EU_deterministic}{diff_se}{0.067}
\setres{pdDiffAdapt}{CPT_deterministic__minus__EU_deterministic}{diff_over_se}{5.6}
\setres{pdDiffAdapt}{CPT_deterministic__minus__EU_deterministic}{corr}{0.02}
\setres{pdDiffAdapt}{CPT_deterministic__minus__EV_deterministic}{diff}{0.537}
\setres{pdDiffAdapt}{CPT_deterministic__minus__EV_deterministic}{diff_se}{0.069}
\setres{pdDiffAdapt}{CPT_deterministic__minus__EV_deterministic}{diff_over_se}{7.8}
\setres{pdDiffAdapt}{CPT_deterministic__minus__EV_deterministic}{corr}{0.07}
\setres{pdDiffAdapt}{CPT_lit__minus__DA_lit}{diff}{1.186}
\setres{pdDiffAdapt}{CPT_lit__minus__DA_lit}{diff_se}{0.098}
\setres{pdDiffAdapt}{CPT_lit__minus__DA_lit}{diff_over_se}{12.1}
\setres{pdDiffAdapt}{CPT_lit__minus__DA_lit}{corr}{0.97}
\setres{pdDiffAdapt}{CPT_lit__minus__EU_lit}{diff}{0.168}
\setres{pdDiffAdapt}{CPT_lit__minus__EU_lit}{diff_se}{0.027}
\setres{pdDiffAdapt}{CPT_lit__minus__EU_lit}{diff_over_se}{6.2}
\setres{pdDiffAdapt}{CPT_lit__minus__EU_lit}{corr}{1.00}
\setres{pdDiffAdapt}{EV_deterministic__minus__CPT_lit}{diff}{0.265}
\setres{pdDiffAdapt}{EV_deterministic__minus__CPT_lit}{diff_se}{0.019}
\setres{pdDiffAdapt}{EV_deterministic__minus__CPT_lit}{diff_over_se}{13.8}
\setres{pdDiffAdapt}{EV_deterministic__minus__CPT_lit}{corr}{0.96}
\setres{pdDiffShared}{CPT_deterministic__minus__DA_deterministic}{diff}{0.227}
\setres{pdDiffShared}{CPT_deterministic__minus__DA_deterministic}{diff_se}{0.038}
\setres{pdDiffShared}{CPT_deterministic__minus__DA_deterministic}{diff_over_se}{6.0}
\setres{pdDiffShared}{CPT_deterministic__minus__DA_deterministic}{corr}{0.05}
\setres{pdDiffShared}{CPT_deterministic__minus__EU_deterministic}{diff}{0.373}
\setres{pdDiffShared}{CPT_deterministic__minus__EU_deterministic}{diff_se}{0.067}
\setres{pdDiffShared}{CPT_deterministic__minus__EU_deterministic}{diff_over_se}{5.6}
\setres{pdDiffShared}{CPT_deterministic__minus__EU_deterministic}{corr}{0.02}
\setres{pdDiffShared}{CPT_deterministic__minus__EV_deterministic}{diff}{0.537}
\setres{pdDiffShared}{CPT_deterministic__minus__EV_deterministic}{diff_se}{0.069}
\setres{pdDiffShared}{CPT_deterministic__minus__EV_deterministic}{diff_over_se}{7.8}
\setres{pdDiffShared}{CPT_deterministic__minus__EV_deterministic}{corr}{0.07}
\setres{pdDiffShared}{CPT_lit__minus__DA_lit}{diff}{1.272}
\setres{pdDiffShared}{CPT_lit__minus__DA_lit}{diff_se}{0.114}
\setres{pdDiffShared}{CPT_lit__minus__DA_lit}{diff_over_se}{11.2}
\setres{pdDiffShared}{CPT_lit__minus__DA_lit}{corr}{0.95}
\setres{pdDiffShared}{CPT_lit__minus__EU_lit}{diff}{0.164}
\setres{pdDiffShared}{CPT_lit__minus__EU_lit}{diff_se}{0.027}
\setres{pdDiffShared}{CPT_lit__minus__EU_lit}{diff_over_se}{6.1}
\setres{pdDiffShared}{CPT_lit__minus__EU_lit}{corr}{1.00}
\setres{pdDiffShared}{EV_deterministic__minus__CPT_lit}{diff}{0.443}
\setres{pdDiffShared}{EV_deterministic__minus__CPT_lit}{diff_se}{0.025}
\setres{pdDiffShared}{EV_deterministic__minus__CPT_lit}{diff_over_se}{18.0}
\setres{pdDiffShared}{EV_deterministic__minus__CPT_lit}{corr}{0.96}
\setres{share}{CPT_uniform}{r_prior_gains}{0.5364}
\setres{share}{CPT_uniform}{r_prob_gains}{0.626}
\setres{share}{CPT_uniform}{r_prior_losses}{0.6030}
\setres{share}{CPT_uniform}{r_prob_losses}{0.680}
\setres{share}{CPT_uniform}{r_prob_avg}{0.653}
\setres{share}{CPT_lit}{r_prior_gains}{0.8966}
\setres{share}{CPT_lit}{r_prob_gains}{0.917}
\setres{share}{CPT_lit}{r_prior_losses}{0.9012}
\setres{share}{CPT_lit}{r_prob_losses}{0.920}
\setres{share}{CPT_lit}{r_prob_avg}{0.919}
\setres{share}{DA_uniform}{r_prior_gains}{0.8493}
\setres{share}{DA_uniform}{r_prob_gains}{0.901}
\setres{share}{DA_uniform}{r_prior_losses}{0.8339}
\setres{share}{DA_uniform}{r_prob_losses}{0.889}
\setres{share}{DA_uniform}{r_prob_avg}{0.895}
\setres{share}{DA_lit}{r_prior_gains}{0.9234}
\setres{share}{DA_lit}{r_prob_gains}{0.949}
\setres{share}{DA_lit}{r_prior_losses}{0.8981}
\setres{share}{DA_lit}{r_prob_losses}{0.932}
\setres{share}{DA_lit}{r_prob_avg}{0.941}
\setres{share}{EU_uniform}{r_prior_gains}{0.5146}
\setres{share}{EU_uniform}{r_prob_gains}{0.716}
\setres{share}{EU_uniform}{r_prior_losses}{0.4460}
\setres{share}{EU_uniform}{r_prob_losses}{0.689}
\setres{share}{EU_uniform}{r_prob_avg}{0.702}
\setres{share}{EU_lit}{r_prior_gains}{0.6364}
\setres{share}{EU_lit}{r_prob_gains}{0.787}
\setres{share}{EU_lit}{r_prior_losses}{0.5835}
\setres{share}{EU_lit}{r_prob_losses}{0.766}
\setres{share}{EU_lit}{r_prob_avg}{0.777}

\doublespacing

\singlespacing
\ifdefined\bwfigures
  \providecommand{\figfile}[1]{figures/#1_bw.png}
  \providecommand{\figfileapp}[1]{figures_appendix/#1_bw.png}
  \colorlet{figA}{black}
  \colorlet{figB}{black}
  \colorlet{figC}{black!55}
\else
  \providecommand{\figfile}[1]{figures/#1.png}
  \providecommand{\figfileapp}[1]{figures_appendix/#1.png}
  \definecolor{figA}{HTML}{D55E00}
  \definecolor{figB}{HTML}{0072B2}
  \colorlet{figC}{black!65}
\fi

\section{Introduction}
Economics has mostly judged its theories by how well they can be fitted to a given dataset.\footnote{The appendices referred to in the text are not part of this version; they are available from the author on request.} A flexible theory, however, can be made to fit many patterns of behavior, so a good fit can arise because the theory rules out little or because its predictions are precise and right. More recently, prediction competitions and machine-learning benchmarks have moved the focus to held-out behavior prediction \citep{erev2017,peysakhovich_naecker2017,fudenberg2019}, which separates fit from overfitting. However, it does not allow to distinguish a precise theory from a permissive one since the rule that is being evaluated is still selected on the data. 
Two recent measures have made both sides of the question measurable for a functional form: \citet{completeness2022} define the \emph{completeness} of a model as the
share of the achievable improvement in out-of-sample prediction that the
model delivers, and \citet{restrictiveness2026} define its
\emph{restrictiveness} as how poorly it approximates synthetic behavior drawn
from a class of admissible alternatives. Both score the best-fitting member of a given family of models, that is, all the prediction rules one functional form can produce.
 A theory, however, is rarely held as a family. For example, prospect theory is usually assumed to have a 
probability-weighting parameter well below one and a value function that is close to linear. The parameter values a literature considers plausible are part of the theory's content and they are what anyone who knows the literature would
use to predict a new sample before seeing it.

Thus, this paper proposes measures to assess the empirical content of a theory as it is held. A theory
in a given domain is operationalized as a combination of three components: a \emph{support restriction},
qualitative conditions such as monotonicity or first-order stochastic
dominance; a \emph{parameterization}, the prediction rules its functional
form can express; and a \emph{parameter distribution}, which says which
parameter values the theory deems plausible. Both
completeness and restrictiveness then take probabilistic versions, computed by drawing a rule from the
theory instead of fitting one. \emph{Probabilistic completeness} is the
expected out-of-sample error of a drawn rule, normalized between a
behavior-agnostic baseline and the error of the best achievable predictor.
\emph{Probabilistic restrictiveness} is the expected distance of conceivable
behavior from a drawn rule, separated into what the form
can reach at all, and how the distribution concentrates the theory's
predictions on the reachable set. The fitted theory is the special case of a
point mass on the same plane and the distance to the prior theory is a measure of the share of its completeness and restrictiveness attained through fitting on the dataset

The parameter prior of a theory can be interpreted in two ways: First, it can be interpreted as uncertainty about the one parameter value that describes a
representative agent. Each draw from the theory is then a candidate for that value and is scored by a loss, so completeness measures how close the draws come, on average, to the best predictor, and restrictiveness how far conceivable behavior lies from them. Alternatively, the distribution can be interpreted as representing a population of agents who differ in their parameters. The theory then predicts a distribution of outcomes in each situation, which can be evaluated by a proper scoring rule. Thus, completeness measures how close the predicted distribution comes to the observed one, and restrictiveness how far it moves from a reference under which every admissible outcome is equally likely. 

Further, we introduce \emph{scope} as the set of situations to which a theory is meant to apply. Prospect theory, for example, claims to apply to situations of choices over gains and over losses, with one value function reflected at zero. The proposed framework allows to compare theories with possibly different scope across domains. A theory's \emph{predictive content} aggregates its completeness and its restrictiveness under a relevance weighting across domains, with each domain counted only if the theory claims it. Theories of different scope then compare on a Pareto frontier, and breadth is only rewarded if a claimed domain adds to restrictiveness wherever the theory rules anything out there, and to completeness only where the theory predicts better than the baseline.

We provide estimators for probabilistic completeness and restrictiveness for any loss and discrepancy,
with closed forms for squared error and for the continuous ranked probability
score as well as a worked toy example.
Finally, we illustrate the usefulness of this framework by applying it to Cumulative Prospect Theory, Disappointment
Aversion and Expected Utility on the certainty equivalents of
\citet{https://doi.org/10.3982/ECTA7139} and evaluate each theory with a literature prior calibrated on the
meta-analysis of \citet{imai_nunnari_wu_vieider_2025}. We find that the share of a theory's completeness that comes from what it commits to rather than from fitting differs by family. A rule
drawn from the literature prior retains \res{gains}{CPT}{share_lit} percent of fitted CPT's
completeness, \res{gains}{EU}{share_lit} percent of fitted EU's and
\res{gains}{DA}{share_lit} percent of fitted DA's. And the flexibility--fit trade-off of
\citet{restrictiveness2026}, which holds for the functional forms, disappears if a theory is evaluated jointly with its literature prior, making CPT the sole theory on the Pareto frontier, level with DA in restrictiveness and far ahead in completeness.

The paper proceeds as follows. Section~\ref{sec:related} relates the
framework to the literature. Section~\ref{sec:framework} defines the
measures, Section~\ref{sec:comparing_theories} the comparison across
situations, domains and scopes, and Section~\ref{sec:estimation} their
estimation. Section~\ref{sec:toy} illustrates them on a toy example.
Section~\ref{sec:ex1} applies them to the gain lotteries of Bruhin et al.\
(2010) under both interpretations, and
Section~\ref{sec:scope} extends the scope to losses.
Section~\ref{sec:conclusion} concludes.

\section{Related Literature}
\label{sec:related}

Our paper builds most directly on a recent program that aims to make the
predictive content of economic theories measurable.
\citet{fudenberg2019} benchmark models of initial play in games
against machine-learning predictors trained on the same data and,
\citet{completeness2022} generalize the exercise.
As a complementary measure \citet{restrictiveness2026} introduce the \emph{restrictiveness} of a functional form. \citet{fudenberg2026_structural} extend the measure to semi- and
nonparametric and structural settings, constructing reference
distributions from Gaussian-process priors under shape restrictions. Finally,
\citet{andrews2025transfer} measure how well an estimated model carries
to new domains. 

Relative to this program we make three
contributions. First, we evaluate a theory through the full predictive
distribution it induces over prediction rules rather than at its single
best-fitting element. A theory in our framework is defined by a support restriction (coherence
conditions such as first-order stochastic dominance together with
outcome-range bounds), a parameterization, and a prior over parameters. The prior is treated as part of the theory's ex-ante claim. Because the
theory is evaluated at these ex-ante commitments, no parameter is fit to
the evaluation data. We thus measure properties of a theory before the data were drawn, scoring generalization capabilities 
rather than accommodation. Second, we decompose restrictiveness into two channels: (1) what the theory can reach (its support
restriction and functional form, evaluated against a reference distribution on
the admissible outcome ranges that is otherwise unrestricted, 
structurally comparable to the measure of \citet{restrictiveness2026} and (2) 
how the theory distributes its predictive mass over what it can reach
(a Kullback--Leibler divergence from the uniform distribution on the
reachable set). Because the reference distribution is not confined to coherent
behavior, qualitative conditions such as stochastic dominance, can be measured as part of a theory's content. A theory consisting of
these conditions alone thus becomes a special case of the framework. Third, we add a scope layer, aggregating a
theory's claimed domains under domain weights into a predictive-content
pair and comparing theories on a Pareto frontier, so that cross-domain
applicability becomes a measured component of predictive content.

 Our framework is related to, but distinct from Bayesian model comparison. Marginal likelihood 
scores a model by the probability it assigned to the data before seeing them,
and Bayes factors compare models on this basis
\citep{Jeffreys1961, KassRaftery1995}. Because a diffuse prior spreads mass
over parameter values that fit poorly, the marginal likelihood contains an
automatic penalty for flexibility, the ``Occam factor'' of
\citet{JefferysBerger1992} and \citet{MacKay1992}.
This is the standard
argument that Bayesian evidence trades fit against parsimony without an
explicit complexity term, and \citet{Box1980} uses the same prior-predictive
distribution for model criticism. Our measures share the premise that a
theory is to be judged at its prior rather than at its best fit, but differ
in three respects. First, the log marginal likelihood is the \emph{logarithm
of the prior expectation of the likelihood}, whereas probabilistic
completeness is built on the \emph{prior expectation of a loss},
 which under Gaussian error is the prior expectation of the log-likelihood. By Jensen's inequality
the former is governed by the best-fitting region of the prior and the latter by its entire support, so our measure penalises dispersion more heavily. Second, the marginal likelihood
collapses fit and restrictiveness into a single number and requires a full
probability model for the outcome. We keep the two dimensions separate,
require only a loss and a discrepancy, and measure restrictiveness against a
theory-independent reference distribution that involves no data at all,
whereas the Occam factor is a data-dependent ratio of posterior to prior
volume. Third, probabilistic completeness is normalised between a
behaviour-agnostic baseline and the best achievable predictor, so that it is
comparable across domains and datasets. 

Other earlier attempts to quantify the content of theories mainly focused on volume based measures for restrictiveness. \citet{selten1991} introduced a volume-counting measure that
operationalizes predictive success as the hit rate minus the size of the
predicted region. The measure can be characterized axiomatically for
theories that delineate a permitted subset of outcomes, and it remains
useful in such settings (e.g., \citealp{beatty_crawford2011}); but it is
uninformative for theories that take the form of parametric functional
forms, whose predictions trace a lower-dimensional set of measure zero:
the volume term then assigns every such theory maximal restrictiveness,
regardless of how much behavior it can in fact approximate. This
shortcoming, as \citet{restrictiveness2026} observe, is what makes
distance-based measures appealing. Restrictiveness prices a theory by
how far admissible behaviors lie from it, which discriminates among
theories of identical (zero) volume.

Our approach also relates to a literature concerned with selecting and
testing a model on the same sample. Using the data twice, i.e. once to
choose the element of a family and once to score it, inflates apparent
performance. Previous literature referred to this as ``double dipping''
 in neuroscience \citep{kriegeskorte2009}, ''data snooping'' in empirical finance and econometrics
\citep{lo_mackinlay1990,white2000} and  overfitting that motivates
machine-learning practice. 

The standard remedy is to evaluate on data not used for selection: cross-validation in statistics since \citet{stone1974}, and the
train-test paradigm at the core of machine learning, which entered
applied economics as a general-purpose tool with
\citet{mullainathan_spiess2017} and has been turned on behavioral
theories directly. \citet{peysakhovich_naecker2017} use machine-learning
fits as an upper bound on the explainable variation in choices under
risk and ambiguity, and the choice-prediction competitions synthesized
by \citet{erev2017} pit structured models against black-box algorithms
on held-out problems. The completeness benchmark of
\citet{completeness2022} already adopts this discipline. Our framework takes this logic one
step further. Held-out evaluation makes the accuracy estimate honest,
but the object being scored is still the data-selected element of the
family. Because we score the theory at its ex-ante parameter
distribution, no selection step occurs at all, and the completeness and
restrictiveness of a theory, as we define them, are properties the
theory possessed before any data were drawn.

Our framework is also a response to concerns about cumulative theory
appraisal that the replication and generalizability debates have put forward. \citet{mischel2008} diagnosed the ``toothbrush problem'', i.e. that no researcher is willing to use someone else's theory. This proliferation persists partly because the
field lacks a common ledger on which competing theories can be scored.
Large-scale replication projects supplied one kind of ledger for
empirical findings \citep{osc2015,camerer2016} but \citet{yarkoni2022}
argues that the deeper problem is generalizability. In particular, verbal theories
routinely claim far more than the sampled settings support. Our measures
speak to both concerns, probabilistic completeness and restrictiveness make it possible to compare competing theories on one scale, and the scope layer forces the claimed range of settings to be stated ex ante, which is precisely what the
generalizability critique demands of theory.

Finally, the paper gives quantitative expression to older methodological
positions. \citet{popper1959} identified empirical content with
prohibition but obtained only a partial order over theories;
\citet{samuelson1947} demanded operationally meaningful theorems; and
\citet{friedman1953} located a theory's worth in its predictions. We make the prohibition view measurable so that ``rules out more'' becomes
a number, and we measure the predictive criterion out of sample. That
models are deliberate simplifications underlies our insistence that scope has to be claimed explicitly, echoing the notion that scope is traded off rather than maximized as advocated by
\citet{keynes1891}. Further, our Pareto frontier can be seen as
the quantitative descendant of the generality--precision--realism
trade-off of \citet{levins1966}, formalized by
\citet{matthewson_weisberg2009}.

\section{Framework} \label{sec:framework}

An analyst wants to compare a set of theories $\mathcal{M}$ on how much each of
them rules out and how well each of them predicts. The first step is to fix what
the theories speak about. They predict a common outcome $Y \in \mathcal{Y}$, and
each theory $M$ conditions its prediction on a set of features $\mathcal{A}_M$
of the situation in which the outcome arises. A \emph{situation} is described by
the features that at least one theory conditions on,
$\mathcal{A} = \bigcup_{M \in \mathcal{M}} \mathcal{A}_M$, that is, by a vector
$x \in \mathbb{R}^k$ with $k = |\mathcal{A}|$. A theory with
$\mathcal{A}_M \subsetneq \mathcal{A}$ ignores the remaining features, and
variables that no theory in $\mathcal{M}$ conditions on are not part of $x$.

\begin{figure}[htbp]\centering
\begin{tikzpicture}[x=1cm,y=1cm,
    every node/.style={font=\fontsize{8.5}{10.5}\selectfont},
    lab/.style={font=\fontsize{8}{9.5}\selectfont,inner sep=1.5pt},
    leg/.style={font=\fontsize{8}{10}\selectfont,anchor=west,inner sep=0pt},
    legh/.style={font=\fontsize{8}{10}\selectfont\bfseries,anchor=west,inner sep=0pt},
    refdot/.style={fill=black!30}]

  \draw[line width=0.6pt,rounded corners=3pt,fill=black!2] (0,0) rectangle (12.6,6.9);
  \node[anchor=north west,inner sep=4pt] at (0.05,6.87) {domain $d$: all admissible rules $\mathcal F_d$};
  \foreach \px/\py in {0.5/5.6,0.9/0.7,1.4/4.9,1.8/0.45,2.6/6.1,3.3/0.9,4.4/6.3,4.9/0.5,5.9/6.2,6.4/1.0,
     7.6/6.4,8.1/0.6,9.1/5.9,10.4/6.3,11.9/5.6,12.1/1.5,0.45/2.6,0.4/4.0,10.7/0.7,12.15/3.7,2.4/1.3,
     11.2/4.6,1.0/1.8,11.4/2.4,9.4/1.1,11.0/3.3,3.9/5.6,7.3/0.45}{\fill[refdot] (\px,\py) circle (1pt);}

  \draw[fill=black!11,draw=black!55,line width=0.4pt]
    plot[smooth cycle,tension=0.75] coordinates {(1.5,1.7) (1.8,3.6) (3.2,5.0) (5.6,5.5) (7.6,5.0) (7.5,3.7) (5.5,2.6) (3.2,1.45)};
  \node[lab,anchor=west] at (2.15,4.25) {$\mathcal S_{M,d}$};
  \draw[line width=0.5pt,black!45,smooth,samples=40,domain=0:1,variable=\t]
    plot ({2.05+5.2*\t},{2.05+2.6*pow(\t,0.62)});
  \foreach \i in {0,...,39}{
    \pgfmathsetmacro{\ta}{\i/40}
    \pgfmathsetmacro{\tb}{(\i+1)/40}
    \pgfmathsetmacro{\dens}{exp(-((0.5*(\ta+\tb)-0.5)/0.12)^2/2)}
    \pgfmathsetmacro{\wid}{0.6+3.4*\dens}
    \pgfmathtruncatemacro{\gv}{30+68*\dens}
    \draw[black!\gv,line width=\wid pt,line cap=round]
      ({2.05+5.2*\ta},{2.05+2.6*pow(\ta,0.62)}) -- ({2.05+5.2*\tb},{2.05+2.6*pow(\tb,0.62)});}
  \node[lab,anchor=north] at (6.75,4.3) {$\mathcal R_{M,d}$};
  \draw[black!55,line width=0.3pt] (3.9,4.85)--(4.55,3.85);
  \node[lab,anchor=south] at (3.85,4.85) {$\Pi_{M,d}$};

  \draw[black,line width=0.45pt,densely dashed] (9.9,6.05)--(7.25,4.65);
  \fill[black!65] (9.9,6.05) circle (1.5pt);
  \node[lab,anchor=west] at (10.0,6.05) {$f'\sim\Lambda_d$};
  \node[lab,anchor=south,rotate=27.8] at (8.6,5.4) {$r^{f\!f}$};
  \fill[black] (12.48,0.0) rectangle (12.6,0.12);
  \node[lab,anchor=south east] at (12.45,0.1) {farthest rule $g^{far}$};

  \draw[black,line width=0.3pt,densely dotted] (5.15,2.95)--(4.75,3.75);      
  \draw[black,line width=0.3pt,densely dotted] (5.15,2.95)--(9.3,1.75);       
  \fill[white,draw=black,line width=0.55pt] (5.15,2.95) circle (2pt);
  \node[lab,anchor=north east] at (5.1,2.9) {$f^*$};
  \node[lab,anchor=west] at (5.3,3.3) {$\kappa^{prob}$};
  \fill[black] (9.3,1.75) circle (1.6pt);
  \node[lab,anchor=west] at (9.4,1.72) {$f_{\mathrm{base}}$};
  \fill[white,draw=black!50,line width=0.5pt] (8.3,3.35) circle (2pt);
  \node[lab,anchor=west,text=black!60] at (8.42,3.35) {$f^*$ in another environment};

  \begin{scope}[shift={(0,-0.55)}]
    \node[legh] at (0,0) {Domain $d=(\mathcal X_d,\mathcal Y_d)$};
    \draw[line width=0.6pt,rounded corners=1pt,fill=black!2] (0,-0.55) rectangle (0.32,-0.33);
    \node[leg] at (0.5,-0.44) {admissible rules $\mathcal F_d$};
    \fill[refdot] (0.16,-0.88) circle (1pt);
    \node[leg] at (0.5,-0.88) {draws from the reference distribution $\Lambda_d$};
    \fill[black] (0.10,-1.38) rectangle (0.22,-1.26);
    \node[leg] at (0.5,-1.32) {farthest rule $g^{far}$, for which $r^{f\!f}=1$};

    \node[legh] at (0,-1.95) {Environment $e=(\mathcal X_d,\mathcal Y_d,P_d)$};
    \fill[white,draw=black,line width=0.55pt] (0.16,-2.39) circle (2pt);
    \node[leg] at (0.5,-2.39) {best predictor $f^*$ under the process $P_d$};
    \fill[black] (0.16,-2.83) circle (1.6pt);
    \node[leg] at (0.5,-2.83) {baseline rule $f_{\mathrm{base}}$};

    \node[legh] at (6.5,0) {Theory $M$};
    \draw[fill=black!11,draw=black!55,line width=0.4pt] (6.5,-0.55) rectangle (6.82,-0.33);
    \node[leg] at (7.0,-0.44) {support restriction $\mathcal S_{M,d}$};
    \draw[line width=0.5pt,black!45] (6.5,-0.88)--(6.82,-0.88);
    \node[leg] at (7.0,-0.88) {reachable set $\mathcal R_{M,d}$};
    \draw[line width=3pt,black!90,line cap=round] (6.56,-1.32)--(6.76,-1.32);
    \node[leg] at (7.0,-1.32) {predictive distribution $\Pi_{M,d}$};

    \node[legh] at (6.5,-1.95) {Measures};
    \draw[line width=0.45pt,densely dashed] (6.5,-2.39)--(6.82,-2.39);
    \node[leg] at (7.0,-2.39) {$r^{f\!f}$: reference draws to $\mathcal R_{M,d}$, relative to $g^{far}$};
    \draw[line width=3pt,black!90,line cap=round] (6.56,-2.83)--(6.76,-2.83);
    \node[leg] at (7.0,-2.83) {$r^{prior}$: concentration of $\Pi_{M,d}$ on $\mathcal R_{M,d}$};
    \draw[line width=0.3pt,densely dotted] (6.5,-3.27)--(6.82,-3.27);
    \node[leg] at (7.0,-3.27) {$\kappa^{prob}$: $\Pi_{M,d}$ to $f^*$, relative to $f_{\mathrm{base}}$};
  \end{scope}
\end{tikzpicture}
\caption{\label{fig:framework} The components of the framework. A domain fixes the set of admissible
prediction rules, the reference distribution on it and the farthest rule. A
theory restricts the admissible rules to its support, its functional form
reaches a surface within the support, and its parameter distribution spreads
predictive mass over that surface. An environment adds a data-generating
process, which determines the best predictor $f^*$; a second environment on the
same domain differs only in where $f^*$ lies. Restrictiveness compares the
theory with the reference draws and needs no data; completeness compares it
with $f^*$.}
\end{figure}
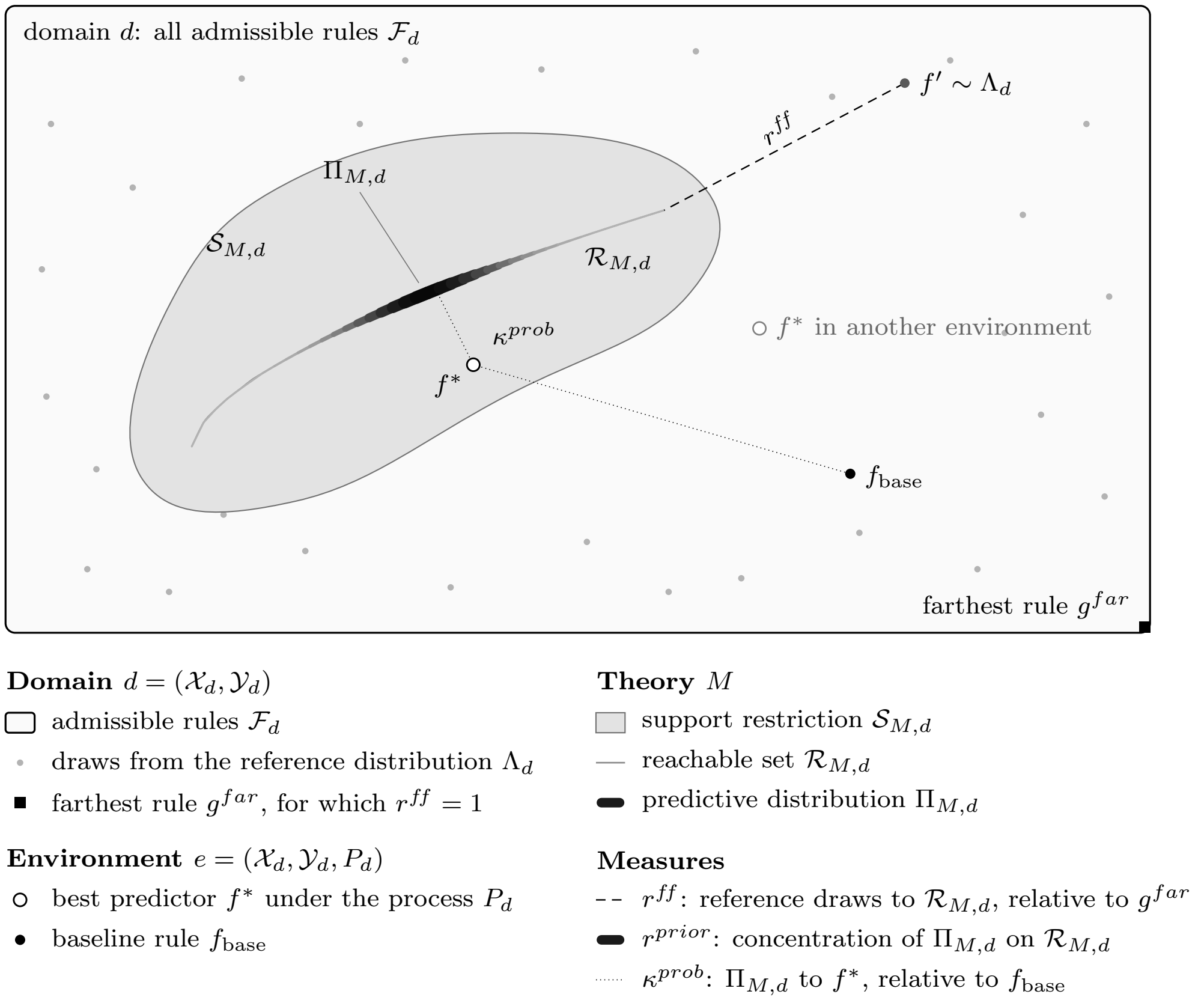

Given this description, a comparison rests on a specification of the situations
in which the theories are to be evaluated and of the outcomes that can be
recorded in each of them. We call such a specification a \emph{domain}. Domains
are the primitives of our formal framework. Figure~\ref{fig:framework} shows the
objects that the rest of this section introduces.

Formally, domains are indexed by $d \in \mathcal{D}$. A \emph{domain} is a pair
$d = (\mathcal{X}_d, \mathcal{Y}_d)$: a finite set of covariate values
$\mathcal{X}_d \subset \mathbb{R}^k$ at which predictions are compared, and an
outcome correspondence
$\mathcal{Y}_d : \mathcal{X}_d \rightrightarrows \mathcal{Y}$ specifying the
outcomes $\mathcal{Y}_d(x) \subseteq \mathcal{Y}$ admissible at each covariate
value. For $\mathcal{Y}_d(x)$ we assume that it is nonempty, compact and
convex.\footnote{Compactness is necessary to ensure that the reference
distribution and the normalization of restrictiveness in Section~\ref{sec:measuring_restrictiveness} exist.
Convexity is required such that conditional means are guaranteed to constitute
admissible predictions.} In most applications $\mathcal{Y}_d(x)$ is determined
by the instrument that records the outcome and satisfies these conditions.
Where outcomes have no natural bounds, as with earnings or expenditures, the analyst declares a range, and the reference distribution and the normalization are taken on it.

\paragraph{Example 1.} (Certainty equivalents for binary lotteries.) Suppose
the analyst compares Cumulative Prospect Theory (CPT) and Expected Utility (EU)
as theories of how people value binary lotteries. Both condition on the two
prizes of a lottery, $\underline z<\bar z$, and on the probability $p$ of the
larger prize, so
$\mathcal A_{CPT}=\mathcal A_{EU}=\mathcal A=\{\underline z,\bar z,p\}$, $k=3$,
and a situation is a lottery $x=(\underline z,\bar z,p)$. The outcome $Y$ is the certainty
equivalent a respondent reports for a lottery. It is elicited from a choice
list of sure amounts spanning the prize range, so the admissible outcomes at
lottery $x$ are $\mathcal Y_d(x)=[\underline z,\bar z]$. No response can record
a certainty equivalent outside this interval. The admissible set differs from
lottery to lottery, and it is nonempty, compact and convex. \\

The analyst then needs to decide which situations to include in her comparison, that is, the values of $x$ at which the
theories are to be compared, and how to group these situations into domains. A
theory can claim or disclaim a domain only as a whole, and a domain is also the
unit of calibration for a theory's parameters. The grouping therefore has to be
at least as fine as the boundaries of the theories' claims but it can be finer.

\paragraph{Example 1 (continued).} (Situations and domains.) The analyst takes
the lotteries of an experiment as the situations. If some of them are over
gains and others over losses, she can treat them as one domain or as two. With
two domains a theory may be calibrated separately for gains and for losses, and
a theory that speaks only to gains can claim the gains domain alone; with one
domain every theory is held to a single calibration across all lotteries.

We define a \emph{prediction rule} in domain $d$
is a function $f : \mathcal{X}_d \to \mathcal{Y}$ with
$f(x) \in \mathcal{Y}_d(x)$ for all $x$, and $\mathcal{F}_d$ denotes the set of
all such rules. An \emph{environment} is a domain together with a \emph{data-generating
process} (DGP) on it, $e = (\mathcal{X}_d, \mathcal{Y}_d, P)$, where $P$ is a
joint distribution over pairs $(X, Y)$ supported on
$\{(x, y) : x \in \mathcal{X}_d,\ y \in \mathcal{Y}_d(x)\}$.  For example an environment can be an experiment, a survey or a market.
Two environments belong to the same domain if and only if they have the same
prediction rules, $\mathcal{F}_d = \mathcal{F}_{d'}$. This means a rule stated for one can be scored in the other without translation.

What the analyst observes are \emph{samples}: for each environment in the data,
$n_d$ pairs $(X_i, Y_i)$ drawn from its DGP. Throughout, we consider one
environment per domain and index its DGP by the domain, writing $P_d$. A sample can serve a comparison only if it records every component of $x$.

\paragraph{Example 1 (continued).} (Prediction rules and environments.) A
prediction rule $f$ maps each lottery into its own prize range,
$f(x)\in[\underline z,\bar z]$; with $n$ lotteries, $\mathcal F_d$ is the
product of the $n$ prize ranges, a box in $\mathbb R^n$. An experiment that uses
these lotteries and choice lists is an environment on this domain. Its subject
pool, incentives and procedures determine the data-generating process $P_d$,
which captures the distribution of lotteries presented and the relationship
between lotteries and reported certainty equivalents. The same lotteries and
lists with other participants are a second environment on the same domain. The
same lotteries with certainty equivalents---for whatever reason--- restricted to values between $[\frac{\underline z}{2},2 \times \bar z]$ would constitute a separate domain, because other outcomes can be recorded.

\subsection{Theories as Predictive Distributions}
 \label{sec:theories_pred_dist}

We think of a theory $M$ as a set of assumptions that restricts the
relationship between covariates and outcomes. It specifies the set of
prediction rules in $\mathcal{F}_d$ that are consistent with it, and how
plausible each of them is. We therefore represent a theory by the probability
distribution over $\mathcal{F}_d$ that it induces.
This representation consists of three components for each domain $d$:

\begin{enumerate}
\item \textbf{Support restriction.} A subset
$\mathcal{S}_{M,d} \subseteq \mathcal{F}_d$, describing the prediction rules that satisfy
the restrictions the theory imposes irrespective of its parameters.
\item \textbf{Parameterization.} A parameter space
$\Theta_M \subseteq \mathbb{R}^{m}$ and a mapping
$\theta \mapsto f_\theta \in \mathcal{F}_d$ that assigns a prediction rule to
every parameter value. A parameter is a finite-dimensional vector that pins
down one prediction rule while the mapping constitutes the theory's functional form.
\item \textbf{Parameter distribution.} A probability distribution $\pi_{M,d}$
on $\{\theta \in \Theta_M : f_\theta \in \mathcal{S}_{M,d}\}$, capturing the
relative plausibility of different parameter values under the theory.
\end{enumerate}
The first two components jointly determine the theory's \emph{reachable set},
\[
\mathcal{R}_{M,d} = \{f_\theta : \theta \in \Theta_M\} \cap \mathcal{S}_{M,d},
\]
the set of prediction rules consistent with the theory as a whole. 

The third component, the parameter distribution, then distributes the theory's predictive mass over the reachable set. Together, the three elements induce a distribution $\Pi_{M,d}$ over
prediction rules in domain $d$ via the mapping $\theta \mapsto f_\theta$,
$f_\theta \sim \Pi_{M,d}$. This formulation accommodates both parametric and
nonparametric theories. In particular, a purely qualitative theory can be represented by letting
$\Theta_M = \mathcal{S}_{M,d}$ and $f_\theta = \theta$, so that the parameter
is the prediction rule itself.
We call a theory together with one parameter distribution a \emph{rung}. The rungs of a family
are ordered by how much of the evaluation data they use: none for the uniform distribution on the
admissible parameter values, other samples for a distribution calibrated on the literature, and
the sample's own training data for the fitted theory, the point mass at the best-fitting value
that the published measures score.

\paragraph{Example 1 (continued).} (Cumulative Prospect Theory)
Cumulative Prospect Theory (CPT) restricts the way lotteries over gains are
mapped into predicted certainty equivalents. The \emph{support restriction}
$\mathcal{S}_{M,d}$ consists of all admissible prediction rules that respect first-order stochastic dominance: $f(x) \geq f(x')$ whenever lottery $x$ dominates lottery $x'$. The \emph{parameterization}
maps $\theta = (\alpha, \gamma, \delta) \in \Theta_M = (0,1] \times (0,1]
\times (0,\bar\delta]$ to prediction rules $f_\theta(x) = v^{-1}\!\left(w(p)\,v(\bar{z}) + [1-w(p)]\,v(\underline{z})\right)$,
where $v(z) = z^\alpha$ and $w(p) = \delta p^\gamma / (\delta p^\gamma + (1-p)^\gamma)$; $\alpha$
determines the curvature of the value function over gains, and $\gamma$ and
$\delta$ the curvature and elevation of the probability weighting
function. The reachable set
$\mathcal{R}_{M,d}$ is the intersection of the form's image with
$\mathcal{S}_{M,d}$; for binary lotteries over gains, every rule generated by
the form already respects dominance, so the support
restriction does not bind and $\mathcal{R}_{M,d}$ coincides with the image of
the parameterization. The \emph{parameter distribution} $\pi_{M,d}$ may be
uniform over $\Theta_M$, reflecting agnosticism about parameter values, or
concentrated on empirically estimated ranges from prior literature.

\subsection{Scope}
\label{sec:framework_scope}

A theory need not apply to all domains the analyst is interested in. The
\emph{scope} of theory $M$ is the set of domains
\[
\mathcal{D}_M \subseteq \mathcal{D}
\]
in which its three components are defined, that is, in which it generates
predictions. A theory claims or disclaims a domain as a whole. Thus, if the
analyst wants to evaluate a theory that speaks to only some of the covariate
values of a domain, the domain needs to be split accordingly, for all theories
under comparison.

\paragraph{Example 1 (continued).} (Decisions Under Risk) CPT is designed to
model decisions under risk and uncertainty, so $\mathcal{D}_M$ includes domains
in which individuals evaluate risky prospects, such as financial or insurance
decisions. It excludes, for example, domains in which outcomes are
deterministic, such as choices between certain consumption bundles. Suppose we
want to compare CPT with a theory that speaks only to lotteries over gains. A
domain that contains lotteries over gains and over losses then has to be split
into a gains domain and a losses domain. 

\subsection{Levels and Interpretations}
\label{sec:levels_readings}
\begin{figure}[htbp]\centering
\tikzset{lr/.style={x=5.0cm,y=2.0cm,line cap=round,line join=round}}
\setlength{\tabcolsep}{4pt}
\begin{tabular}{cc}
\begin{tikzpicture}[lr]
  \node[font=\small\bfseries,anchor=south] at (0.5,1.42) {Representative-agent interpretation};
  \node[font=\small\bfseries,rotate=90,anchor=south] at (-0.16,0.5) {Population level};
  \draw[gray!90!black,thin] (0.5,0) -- (0.5,0.95);
  \node[font=\scriptsize,gray!90!black,align=center] at (0.5,1.09) {centre};
  \draw[black!40,thin] (0.62,0) -- (0.62,1.05);
  \fill[figA] (0.62,0) circle (1.4pt);
  \node[font=\scriptsize,anchor=south] at (0.62,1.07) {$f^*(x)$};
  \draw[figB!50,thin,dotted] (0.76,0) -- (0.76,0.92);
  \node[font=\scriptsize,anchor=south] at (0.76,0.94) {$\hat f(x)$};
  \draw[figB,thick,dashed] plot[domain=0.3:1.0,samples=100] (\x,{0.9*exp(-0.5*((\x-0.76)/0.12)^2)});
  \draw[black,thin] (0,0) -- (1,0);
  \node[font=\scriptsize,below,align=center,text width=5.2cm] at (0.5,-0.32) {(a) $\pi$ is uncertainty about the rule of a representative agent};
\end{tikzpicture}
&
\begin{tikzpicture}[lr]
  \node[font=\small\bfseries,anchor=south] at (0.5,1.42) {Heterogeneous-agent interpretation};
  \fill[gray!35] plot[domain=0.0:1.0,samples=100] (\x,{0.9*exp(-0.5*((\x-0.62)/0.16)^2)}) -- (1,0) -- (0,0) -- cycle;
  \node[font=\scriptsize,gray!90!black,align=center] (lab) at (0.22,1.08) {outcomes\\across agents};
  \draw[gray!90!black,thin] (lab.south) -- (0.46,0.42);
  \draw[gray!90!black,thin,dotted] (0,0.36) -- (1,0.36);
  \node[font=\scriptsize,gray!90!black,anchor=south west,align=left] at (0.0,0.35) {uniform};
  \draw[figB,thick] plot[domain=0.42:0.82,samples=80] (\x,{exp(-0.5*((\x-0.62)/0.06)^2)});
  \draw[figB,thick,dashed] plot[domain=0.0:1.0,samples=100] (\x,{0.9*exp(-0.5*((\x-0.62)/0.16)^2)});
  \draw[black,thin] (0,0) -- (1,0);
  \node[font=\scriptsize,below,align=center,text width=5.2cm] at (0.5,-0.32) {(b) $\pi$ represents the spread of rules across agents};
\end{tikzpicture}
\\[6pt]
\begin{tikzpicture}[lr]
  \node[font=\small\bfseries,rotate=90,anchor=south] at (-0.16,0.5) {Individual level};
  \fill[gray!35] plot[domain=0.0:1.0,samples=100] (\x,{0.9*exp(-0.5*((\x-0.62)/0.16)^2)}) -- (1,0) -- (0,0) -- cycle;
  \draw[gray!90!black,thin] (0.62,0) -- (0.62,0.95);
  \node[font=\scriptsize,gray!90!black,align=center] at (0.40,1.12) {population\\mean};
  \draw[figB,thick,dashed] plot[domain=0.0:1.0,samples=100] (\x,{0.9*exp(-0.5*((\x-0.62)/0.16)^2)});
  \draw[figC,thick] plot[domain=0.58:1.0,samples=100] (\x,{exp(-0.5*((\x-0.80)/0.065)^2)});
  \node[font=\scriptsize,align=center,anchor=south] at (0.84,1.04) {updated\\theory};
  \draw[black!40,thin,dotted] (0.80,0) -- (0.80,1.0);
  \draw[figC,thick] (0.80,-0.07) -- (0.80,0.07);
  \node[font=\scriptsize,below] at (0.78,-0.08) {$\hat y_{ij}$};
  \draw[black!40,thin] (0.85,0) -- (0.85,0.75);
  \fill[figA] (0.85,0) circle (1.4pt);
  \node[font=\scriptsize,below] at (0.88,-0.08) {$y_{ij}$};
  \draw[black,thin] (0,0) -- (1,0);
  \node[font=\scriptsize,below,align=center,text width=5.2cm] at (0.5,-0.32) {(c) $\pi$ is the prior about agent $i$'s rule};
\end{tikzpicture}
&
\begin{tikzpicture}[lr]
  \fill[gray!35] plot[domain=0.0:1.0,samples=100] (\x,{0.9*exp(-0.5*((\x-0.62)/0.16)^2)}) -- (1,0) -- (0,0) -- cycle;
  \node[font=\scriptsize,gray!90!black,align=center] at (0.17,0.92) {population\\distribution};
  \draw[gray!90!black,thin] (0.40,0.80) -- (0.50,0.6);
  \draw[figB,thick,dashed] plot[domain=0.0:1.0,samples=100] (\x,{0.9*exp(-0.5*((\x-0.62)/0.16)^2)});
  \draw[figC,thick] plot[domain=0.5:1.0,samples=100] (\x,{0.85*exp(-0.5*((\x-0.80)/0.09)^2)});
  \node[font=\scriptsize,align=center,anchor=south] at (0.84,1.0) {posterior\\predictive $Q_{ij}$};
  \fill[figA] (0.85,0) circle (1.4pt);
  \node[font=\scriptsize,below] at (0.88,-0.08) {$y_{ij}$};
  \draw[black,thin] (0,0) -- (1,0);
  \node[font=\scriptsize,below,align=center,text width=5.2cm] at (0.5,-0.32) {(d) $\pi$  represents the population the agent is drawn from};
\end{tikzpicture}
\end{tabular}
\caption{\label{fig:two_readings} Levels and interpretations of a theory's parameter distribution at one
covariate value $x$. The axis is the admissible range of the outcome. (a) Under the representative-agent interpretation a single
rule $f^*$ holds, the mean of the outcomes across agents, and every draw from a theory is evaluated
by its distance from $f^*(x)$. The baseline is the centre of the reference, and the best theory places all its mass at $f^*(x)$. (b) Under the heterogeneous-agent interpretation the outcomes across
agents form a distribution and a theory's predictive distribution is a claim about its shape. The
baseline is the reference distribution itself (dotted). The wider distribution scores better,
because it matches the grey distribution, while the narrower prior is penalized for not accounting
for the spread. (c) On the individual level, under the representative-agent interpretation the posterior mean $\hat y_{ij}$ is scored at
the agent's outcome $y_{ij}$ against the population mean. (d) Under the heterogeneous-agent interpretation
the posterior predictive distribution $Q_{ij}$, which carries the response noise, is scored at
$y_{ij}$ against the population distribution.}
\end{figure}
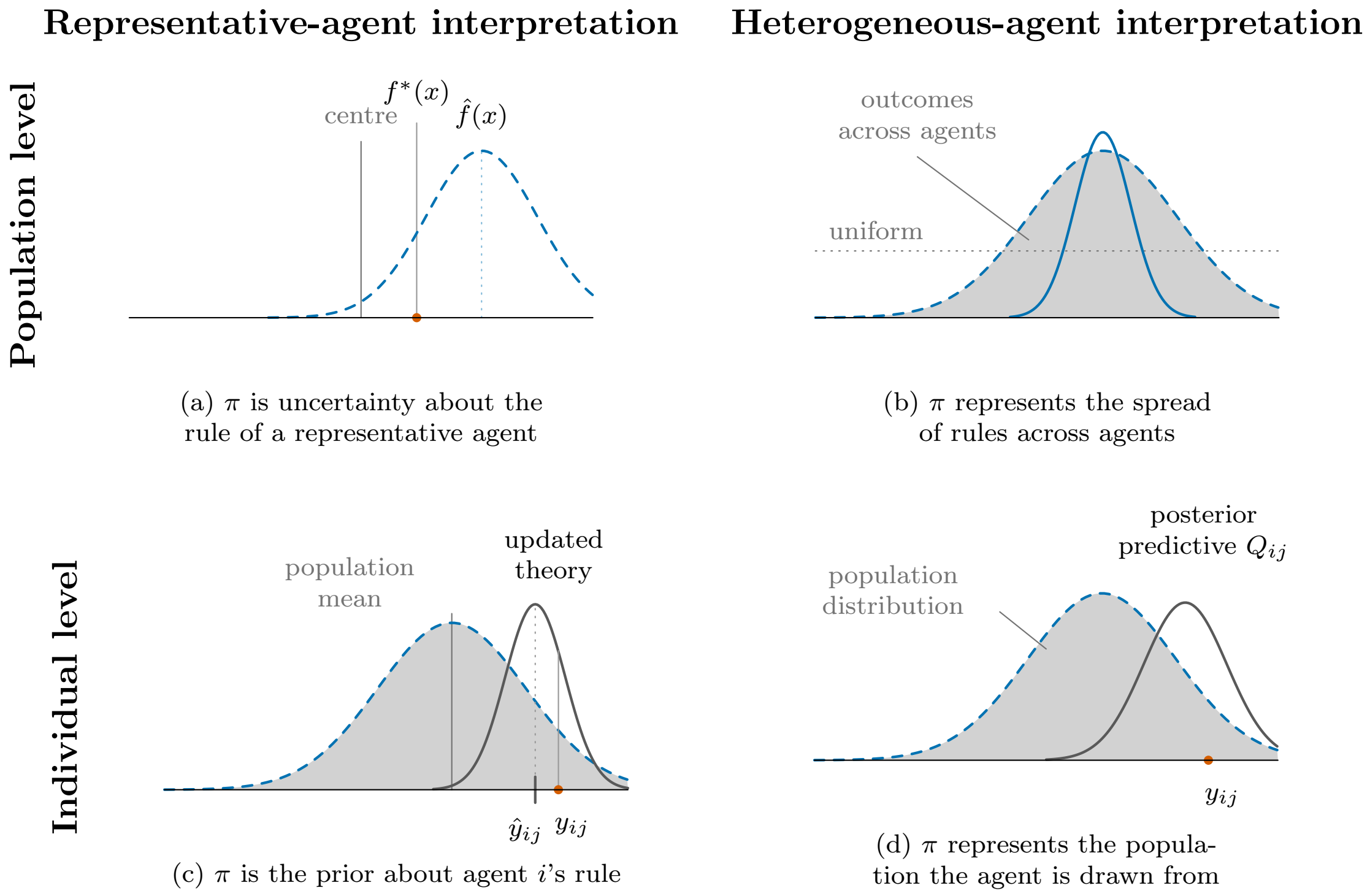

How a theory should be evaluated depends on how it is constructed or interpreted by the analyst. A theory represented by
$\Pi_{M,d}$ can be evaluated at the \emph{population level}, based on the statements it makes about a
group of agents, or at the \emph{individual level}, based on the statements it makes about a given
agent. Independently of the level, its parameter distribution can be interpreted in two ways: as encoding
uncertainty about the correct parameter of a representative agent, which we call the
\emph{representative-agent interpretation}, or as describing how the parameter is distributed across
agents who differ in it, which we call the \emph{heterogeneous-agent interpretation}.
Figure~\ref{fig:two_readings} illustrates the four combinations. 

\paragraph{The population level.} Panels (a) and (b) evaluate a theory based on its statements about a group of agents. Under the representative-agent interpretation (a) a single rule $f^*$ holds for every agent and each draw from the theory is a candidate for the value of the representative agent. Its predictive capabilities are assessed by its distance from $f^*(x)$. Thus, a theory that centers most of its mass on the true value scores better than one with the
same centre that is more widely dispersed. Under the heterogeneous-agent interpretation (b) the outcomes across agents form a distribution, and the theory's predictive distribution is a claim about the shape of the distribution of the agent's parameters. A more widespread theory that closely traces the distribution of values within the
population scores better than a narrower one that puts too much mass on a single value, so the ranking of (a) can reverse. A baseline represents a ``theory'' someone would hold who has no knowledge about the population at all. In terms of the representative agent interpretation this corresponds to the centre of the outcome range in (a). In (b) it is the uniform distribution on the admissible outcomes at the covariate value, the claim that agents' outcomes are spread evenly across the full possible range.

\paragraph{The individual level.} Panels (c) and (d) illustrate the evaluation on the individual agent level. Where an agent is observed
at several covariate values or where agents can be separated into meaningful sub-groups, a parameter distribution that is updated based on previous observations from agent $i$ or similar agents,  can serve as a prior about that agent $i$'s own rule, obtained via Bayesian updating.  The prediction for agent $i$'s outcome at a specific covariate value is then the posterior mean under the representative-agent interpretation (c), a point scored against the population mean, and the posterior
predictive distribution under the heterogeneous-agent interpretation (d). The baseline at the individual level encodes the ``theory'' someone who has observed the outcomes of all agents but has no specific information about agent $i$. In the representative-agent interpretation this is the population mean, in the heterogeneous-agent interpretation it is the population's distribution of outcomes.

\paragraph{Example 1 (continued).} (Levels and Interpretations) At the population level CPT is judged based on its statements about the subjects of an experiment as a group. At the individual level the distribution is the prior about one
subject's triple, updated on that subject's certainty equivalents at other lotteries; the prediction
for a remaining lottery is a value under the first interpretation and a distribution under the second. 

\subsection{Measuring Completeness Under the Representative-Agent Interpretation}
\label{sec:measuring_completeness}
We first evaluate the completeness of a theory under the representative-agent interpretation, and refer to it as \emph{probabilistic completeness}. Intuitively, a theory is more complete if it generates predictions that closely match observed outcomes in the domains where it applies. We extend the completeness measure of \citet{completeness2022} by evaluating predictive performance with respect to the full predictive distribution induced by the theory, rather than its best-fitting element. To distinguish the two, we refer to the measure of \citet{completeness2022} as \emph{fitted completeness}, $\kdet$.

Let $l : \mathcal Y\times\mathcal Y\to\mathbb R_+$ denote a loss function.
 
\begin{definition}[Probabilistic Predictive Error]\label{def:eprob}
For a prediction rule $f \in \mathcal{F}_d$, define its predictive error in
domain $d$ as
\[
e_{P_d}(f) = \mathbb{E}_{P_d}\bigl[l(f(X),Y)\bigr].
\]

The probabilistic predictive error of theory $M$ in domain $d$ is
\[
e^{prob}_{P_d}(M) = \mathbb{E}_{f \sim \Pi_{M,d}} \left[ e_{P_d}(f) \right].
\]
\end{definition}
 
\begin{definition}[Probabilistic Completeness]\label{def:kprob}
Let $f^* \in \arg\min_{f \in \mathcal{F}_d} e_{P_d}(f)$ denote the best possible predictor, and let $f_{\text{base}} \in  \mathcal{F}_d$ denote a baseline prediction rule that uses no theory-specific structure. The \emph{probabilistic completeness} of theory $M$ in domain $d$ is
\[
\kappa^{prob}(M,d)
=
\frac{
e_{P_d}(f_{\text{base}}) - e^{prob}_{P_d}(M)
}{
e_{P_d}(f_{\text{base}}) - e_{P_d}(f^*)
}.
\]
\end{definition}
 
The baseline $f_{\text{base}}$ serves as a reference point representing predictive performance achievable without the theory under evaluation. Its precise construction is discussed in Section~\ref{sec:estimation}.

\paragraph{Fitted Completeness as a Special Case.}
The two measures take different objects as input. Fitted completeness is a property of the reachable set. It scores the best-fitting rule, $f_{\theta^*}$ with $\theta^* \in \arg\min_{\theta} e_{P_d}(f_\theta)$, and does not depend on the parameter distribution, so all theories that share a functional form and support restriction have the same fitted completeness. Probabilistic completeness is a property of the distribution $\Pi_{M,d}$. The two coincide for one particular distribution, the point mass at $f_{\theta^*}$. In this case $e^{prob}_{P_d}(M) = e_{P_d}(f_{\theta^*})$, and probabilistic completeness equals fitted completeness. We call this distribution the \emph{fitted theory}. It is a theory in our sense, but one that can only be written down with knowledge of the data. Fitted completeness therefore corresponds to the probabilistic completeness of the fitted theory built on the same functional form.

\begin{proposition}[Bounds on completeness]\label{cor:bounds}
$\kprob(M,d)\le\kdet(M,d)$, with equality if $\PiM$ is a point mass at a
best-fitting rule. If the loss $l(\cdot,y)$ is convex, then also
$\kprob(M,d)\le\kappa(\fbar_M,d)$, where $\fbar_M(x)=\E_{f\sim\PiM}[f(x)]$ is
the theory's mean prediction.
\end{proposition}

\begin{proof}
$e^{prob}_{P_d}(M)$ is an average of $e_{P_d}(f_\theta)$ over the reachable set
and cannot fall below its minimum. $\bar f_M$ is a prediction rule because the
admissible outcome sets are convex. For fixed $(x,y)$, convexity of
$l(\cdot,y)$ and Jensen's inequality give
\[
l\bigl(\bar f_M(x),y\bigr)=l\bigl(\E_{f\sim\Pi_{M,d}}[f(x)],\,y\bigr)\;\le\;
\E_{f\sim\Pi_{M,d}}\bigl[l(f(x),y)\bigr].
\]
Taking expectations over $(X,Y)\sim P_d$ on both sides and exchanging the two integrals
by Fubini's theorem, which applies because the loss is non-negative and $\Pi_{M,d}$ is fixed
before the data, so that $f$ and $(X,Y)$ carry the product measure, yields
$e_{P_d}(\bar f_M)\le e^{prob}_{P_d}(M)$.
\end{proof}

 Intuitively, probabilistic and fitted completeness diverge when the theory's predictive distribution places substantial mass on prediction rules far from the best-fitting element. This occurs, for example, when a theory's functional form admits both well-fitting and poorly-fitting parameterizations, or when prior information is diffuse. In such cases, the probabilistic measure penalizes the theory for the full range of predictions that are consistent with it, whereas fitted completeness scores it based on the best achievable fit.

\paragraph{Example 1 (continued).} (Decisions Under Risk) Fitted completeness evaluates CPT by how well the theory can approximate the observed relationship between lotteries and certainty equivalents when parameters are chosen to minimize predictive error. In contrast, in our framework, predictive performance is evaluated by a weighted average over possible parameter realizations. A theory whose parameterization admits many poorly-fitting configurations will therefore receive a lower probabilistic completeness score than its fitted counterpart.

\subsection{Measuring Restrictiveness Under the Representative-Agent Interpretation}
\label{sec:measuring_restrictiveness}
We now evaluate the \emph{probabilistic restrictiveness} of a theory.
Intuitively, a theory is more restrictive if its structural assumptions
confine predictions to a narrow region of the space of possible relationships
between covariates and outcomes. Here we extend the functional form
restrictiveness measure of \citet{restrictiveness2026} by incorporating the
theory's parameter distribution into the assessment. To distinguish the two, we refer to the measure of \citet{restrictiveness2026} as \emph{functional-form restrictiveness}.

As in \citet{restrictiveness2026}, we consider a collection of
\emph{pseudo-true prediction rules}, that is, rules $f \in \mathcal{F}_d$.
Unlike in their approach, where the reference set is restricted by
background constraints that the theories under study satisfy, our reference
space is the set $\mathcal{F}_d$ of all admissible prediction rules. The only
constraints on it are those the analyst imposes by defining the domain. This
allows structural assumptions such as monotonicity, equilibrium conditions, or
other behavioral restrictions to be treated as part of the theory rather than
as constraints on the reference space.

Let $\Lambda_d$ denote a reference distribution over $\mathcal{F}_d$, and let
$\delta_d : \mathcal{F}_d \times \mathcal{F}_d \to \mathbb{R}_+$ measure the
discrepancy between two prediction rules. The framework accommodates any
discrepancy satisfying the conditions of \citet{restrictiveness2026}:
$\delta_d(f,f') = 0$ if and only if $f = f'$, with $\delta_d$ bounded and
continuous. Both objects belong to the domain: neither depends on the theory,
so that restrictiveness is comparable across theories, nor on a data-generating
process. Their construction is discussed in Section~\ref{sec:estimation}.

\paragraph{Sources of restrictiveness.}

A theory restricts the space of possible relationships through two channels,
onto which the three components of Section~\ref{sec:theories_pred_dist} map.
The first channel is \emph{reachability}: the support restriction
$\mathcal{S}_{M,d}$ and the functional form $\theta \mapsto f_\theta$ jointly
determine which prediction rules the theory can generate at all, confining
them to the reachable set $\mathcal{R}_{M,d}$. The reachable set corresponds to a generically
$m$-dimensional surface within the $n$-dimensional prediction space, where
$m = \dim(\Theta_M)$ and $n$ is the number of covariate values in $\mathcal X_d$. The second
channel is \emph{concentration}: the parameter distribution $\pi_{M,d}$
determines how the theory's predictive mass is spread over the reachable set,
and may concentrate it on a small region, leaving most of
$\mathcal{R}_{M,d}$ effectively unused. Restrictions stated in function space contribute to reachability, as do the
  bounds of the parameter space $\Theta_M$, which belong to the
  parameterization. Commitments stated as a distribution on the parameter space
  (informative priors, including priors whose support is a strict subset of
  $\Theta_M$) contribute to concentration The two channels are measured against
different references. Reachability is evaluated against the space of alternatives
$\Lambda_d$, and concentration against the uniform distribution on
$\mathcal{R}_{M,d}$. In the limit in which the parameter distribution
collapses to a point, the theory makes a single deterministic prediction and
is maximally restrictive regardless of its functional form.

\paragraph{Functional form restrictiveness (reachability).}

The joint contribution of support restrictions and functional form is captured
by an adaptation of the restrictiveness measure of \citet{restrictiveness2026},
which we refer to as ``functional form restrictiveness'' here. For a given pseudo-true prediction rule $f$, let
  $\theta^*(f)\in\arg\min_{\theta\in\Theta_M:\,f_\theta\in\mathcal S_{M,d}}
  \delta_d(f_\theta,f)$ denote a parameter value that minimizes the discrepancy
  between the theory's prediction rule and $f$.
Functional form restrictiveness is the theory's expected discrepancy from a reference rule, as a share of the largest expected discrepancy that any theory could have:
\[
\rff(M,d)\;=\;
\frac{\E_{f\sim\LM}\bigl[\delta_d(f_{\theta^*(f)},f)\bigr]}
     {\displaystyle\max_{g\in\mathcal F_d}\E_{f\sim\LM}\bigl[\delta_d(g,f)\bigr]}\,.
\]

The denominator is the measure's own maximum. Enlarging a reachable set can
only lower the expected discrepancy, so among all theories it is largest for a
theory that consists of a single prediction rule, namely the admissible rule
that is on average farthest from a reference rule,
\[
g^{far} \in \arg\max_{g\in\mathcal{F}_d} \E_{f\sim\LM}\bigl[\delta_d(g,f)\bigr],
\]
which exists because $\mathcal{F}_d$ is compact and $\delta_d$ continuous. As a
result, $\rff=1$ for the theory that consists of this rule alone,
$\mathcal{S}_{M,d}=\{g^{far}\}$. It corresponds to the tightest support restriction the domain
admits, placed as far from the reference distribution as possible. Under the
squared discrepancy the farthest rule and the denominator
have closed forms (Section~\ref{sec:estimation}).

\begin{proposition}[Range and nesting]\label{prop:range}\label{prop:nest}
(i) $0\le\rff(M,d)\le1$ for every theory, with $\rff=1$ for a theory that
consists of a single farthest rule. (ii) If
$\mathcal R_{M,d}\subseteq\mathcal R_{M',d}$, then $\rff(M,d)\ge\rff(M',d)$.
\end{proposition}
\begin{proof}
(i) For any $g_0\in\RM$,
$\E\min_{g\in\RM}\delta_d(g,f)\le\E\,\delta_d(g_0,f)\le\max_{g\in\mathcal F_d}\E\,\delta_d(g,f)$.
(ii) For every reference rule the minimum of $\delta_d(\cdot,f)$ over the larger
set is weakly smaller, and the denominator does not depend on the theory.
\end{proof}
\noindent Within a nested family the functional-form channel can therefore
never rank the more general theory as the more restrictive one, any such
reversal in $\rprob$ comes from the parameter distribution.

Because the optimization is constrained to the reachable set $\RM$, both the
support restrictions and the functional form's geometry contribute to $\rff$;
and because discrepancy is measured from where the reference puts its mass,
so does the \emph{location} of the reachable set: a theory that is far from where the reference distribution
puts its mass is, in this sense, more
restrictive. Restrictiveness is always a property of a theory relative to a
reference space.
This is the notion we want, for two reasons. First, restrictiveness is the
counterpart of completeness. Completeness is the loss a theory incurs on actual
behavior, relative to the baseline; restrictiveness is the loss it would incur
if conceivable behavior were the truth. A count of the rules a theory excludes
would pair with no loss, and it could not separate theories that exclude all
but one rule, since every single rule has reference measure zero. Second, the
location of a prediction is part of its content. The center of the reference is
the prediction of an analyst who knows the admissible range and nothing about
behavior, so a single rule at the center claims nothing beyond the range, while
a rule at the edge of the range claims that behavior is extreme and is wrong
about most of what is conceivable. Under the squared discrepancy the expected
discrepancy of a single rule is the dispersion of the reference plus the rule's
squared distance from the center (Section~\ref{sec:est_restrictiveness}), so
single rules span the interval from one quarter, at the center, to one, at the
farthest rule. A theory earns completeness by predicting better than the
baseline and restrictiveness by predicting differently from it. The reference
is the domain's, not the theory's: a theory that confines outcomes to part of
the admissible range is scored against the range the domain admits, so two
theories of one domain are measured on one scale.
This normalization differs from the fixed-baseline construction of
\citet{restrictiveness2026}, which places a single analyst-chosen rule
$f_{\text{base}}$ in the denominator. Appendix~\ref{app:relation_fudenberg} gives the reasons for normalizing by the farthest rule.

\paragraph{Prior restrictiveness (concentration).}

The second component measures how much the parameter distribution concentrates
predictions within the reachable set. As noted above, the reachable set of an $m$-parameter theory is generically an $m$-dimensional surface within the $n$-dimensional prediction space, where $n$ is the number of covariate values in $\mathcal X_d$. A distribution $\pi_{M,d}$ over parameters induces a distribution
$\Pi_{M,d}$ over prediction rules on this surface via the change-of-variables
formula: the density of $\Pi_{M,d}$ at a prediction rule $f_\theta$ with respect to surface area on $\mathcal R_{M,d}$ is 
\[
p_\Pi(f_\theta) = \frac{\pi_{M,d}(\theta)}{|J(\theta)|},
\]
where $J(\theta)$ is the $n \times m$ Jacobian matrix with entries
$J_{ji}(\theta) = \partial f_\theta(x_j)/\partial\theta_i$, and
$|J(\theta)| = \sqrt{\det(J(\theta)^\top J(\theta))}$ is the volume element
that accounts for how the parameterization stretches or compresses regions of
parameter space when mapped into function space.\footnote{The
change-of-variables formula requires the parameterization to be injective and
the Jacobian to have full column rank $\pi_{M,d}$-almost everywhere. If two
parameter values map to the same prediction rule, the density involves a sum
over preimages. If the rank drops, $|J(\theta)| = 0$ and the density is
undefined. Both conditions hold generically for the theories we consider.}

The natural reference is the uniform distribution $\Pi_{\text{uniform}}$ on
$\mathcal{R}_{M,d}$, with density
$p_{\Pi_{\text{uniform}}} = 1/\mathrm{Vol}(\mathcal{R}_{M,d})$, where the
volume is taken with respect to the surface measure on the reachable set and
is finite under the bounded parameter domains we impose. It represents
maximal agnosticism about which reachable prediction the theory favors. Prior
restrictiveness is defined as
\[
r^{prior}(M,d) = 1 - \exp\!\left(-D_{KL}(\Pi_{M,d} \,\|\, \Pi_{\text{uniform}})\right),
\]
where
\[
D_{KL}(\Pi_{M,d} \,\|\, \Pi_{\text{uniform}})
= \mathbb{E}_{\theta \sim \pi_{M,d}}\!\left[
\log \frac{p_\Pi(f_\theta)}{p_{\Pi_{\text{uniform}}}(f_\theta)}
\right],
\]
with the convention $r^{prior} = 1$ when the divergence is infinite, as for a
degenerate parameter distribution. The transformation $1 - \exp(-D_{KL})$ maps
the divergence from $[0,\infty]$ to $[0,1]$. $r^{prior} = 0$ when $\pi_{M,d}$
induces a uniform distribution over the reachable set (no restrictions beyond
the functional form), and $r^{prior} \to 1$ as the predictive distribution
concentrates on a vanishing fraction of it.

\begin{definition}[Probabilistic Restrictiveness]
\label{def:rprob}
The probabilistic restrictiveness of theory $M$ in domain $d$ is
\[
r^{prob}(M,d) = 1 - \big(1 - r^{ff}(M,d)\big)\big(1 - r^{prior}(M,d)\big).
\]
\end{definition}

\begin{proposition}[Range of $\rprob$]\label{prop:rprob}
$\rff(M,d)\le\rprob(M,d)\le1$. The lower bound is attained if and only if
$\PiM$ is uniform on $\RM$, the upper bound if and only if $\rff=1$ or $\PiM$ is
degenerate.
\end{proposition}
\begin{proof}
$\rprob-\rff=(1-\rff)\,\rprior$ and $\rprior=1-\exp(-D_{KL})\in[0,1]$, with
$D_{KL}=0$ if and only if $\PiM=\Pi_{\mathrm{uniform}}$.
\end{proof}

The two components combine multiplicatively through a pass-through
interpretation. The term $(1 - r^{ff})$ captures the permissiveness of the
functional form: how closely the model class can approximate arbitrary
prediction rules relative to the most restrictive rule the domain admits. A theory is
permissive only to the extent that its functional form can reach a wide range
of predictions and its parameter distribution spreads mass broadly
over that range.

Because $\mathcal R_{M,d}$, $\Pi_{M,d}$ and $\Pi_{\text{uniform}}$ are objects
in prediction space, the measure is invariant to reparametrisation. Replacing $\theta$ by $\varphi=g(\theta)$ for a smooth
one-to-one $g$, with the parameter distribution transformed accordingly,
only relabels the points of the reachable set. 
\paragraph{Aggregating the two channels.}
The product form of Definition~\ref{def:rprob} is one convention for combining the channels.
Any aggregator that is
increasing in $\rff$ and in $\rprior$ yields the same ranking of two theories
whenever one of them is weakly more restrictive in both channels. We therefore report both channels throughout, and call a restrictiveness
ranking \emph{channel-wise} when it holds in each.\label{rem:agg} 
\paragraph{Example 1 (continued).} (Restrictiveness of CPT) CPT restricts the
mapping from lotteries to certainty equivalents through both channels. On the
reachability side, its functional form confines predictions to a
three-dimensional reachable set indexed by $\theta = (\alpha, \gamma,
\delta)$. 

\subsection{Measuring Completeness Under the Heterogeneous-Agent Interpretation}
\label{sec:population_reading}

The loss described in the previous sections compares $f(x)$ with $y$, and, under a convex loss
this charges a theory for its spread. Under the heterogeneous-agent interpretation of Section~\ref{sec:theories_pred_dist} the parameter
distribution itself becomes a claim of the theory that parameters differ 
across agents sharing the functional form.\footnote{The form may itself be selected by a parameter, $\Theta_M=\bigsqcup_k{k}\times\Theta_k$ with $f_{(k,\theta_k)}=f^k_{\theta_k}$, in which case the parameter distribution carries the shares of agents following each form and $\Pi_{M,d}$ is a mixture across forms. Every definition of this subsection and the next applies to such a theory; its reachable set is the union of the forms' reachable sets, and the restrictiveness of the form channel is weakly lower than that of each form alone by Proposition~\ref{prop:rdist_split}(ii).} A theory that spreads mass widely
in a domain where agents in fact differ widely describes the domain correctly
and should not be penalized for it; a theory that concentrates mass where
agents differ makes a false claim and should be. Scoring such a claim requires a
loss defined on distributions. This subsection and the next restate
completeness and restrictiveness for this interpretation. They use the same objects as before, the
domain, the reference distribution $\LM$ and the theory's distribution $\PiM$,
and replace the loss and the discrepancy by a scoring rule and the distance it
generates. To distinguish the resulting measures from those introduced in the previous section, we refer to them as \emph{distributional completeness} and \emph{distributional restrictiveness}.

\paragraph{Scoring rules.} Let $\PiM(x)$ denote the theory's \emph{predictive
distribution} at $x$, the distribution of $f(x)$ when $f\sim\PiM$. A
\emph{scoring rule} $S(Q,y)$ assigns a loss to a predictive distribution $Q$ on
$\mathcal Y_d(x)$ when the outcome $y$ is observed. It is \emph{strictly
proper} if, for outcomes drawn from $P$, the expected score $\E_P[S(Q,Y)]$ is
minimized at $Q=P$ and at no other distribution \citep{gneiting2007}. Every
strictly proper score generates a divergence,
\[
d_S(Q,P)=\E_P\bigl[S(Q,Y)\bigr]-\E_P\bigl[S(P,Y)\bigr]\;\ge\;0 ,
\]
the expected excess score of predicting $Q$ when the outcomes follow $P$, which
is zero only at $Q=P$. For two families of predictive distributions on
$\mathcal X_d$ we write
\[
D_d(P,Q)=\sum_{j=1}^{n}\nu_j\,d_S\bigl(P(x_j),Q(x_j)\bigr),
\]
with weights $\nu_j\ge0$, $\sum_j\nu_j=1$, that say how much each covariate
value counts, as in the discrepancy $\delta_d$.

The definitions below apply to any strictly proper score.\footnote{Examples are the
continuous ranked probability score, $\rho(y,y')=|y-y'|$, for scalar outcomes;
the energy score, $\rho(y,y')=\lVert y-y'\rVert$, for vector outcomes; and the
Brier score, $\rho(y,y')=\mathbf 1\{y\ne y'\}$, for categorical outcomes. The
logarithmic score is strictly proper but not a kernel score: its divergence,
the Kullback--Leibler divergence, is unbounded on $\mathcal F_d$ and assigns an
infinite score to a point prediction at any value but the outcome, so the
definitions below apply to it while the range and bound results do not.} For
these scores $d_S$ is a squared metric, it is bounded because the admissible
sets are compact, and the score of a point mass at $f(x)$ is $\rho(f(x),y)$,
so point predictions and distributions are scored on one scale: a theory whose
parameter distribution is degenerate is scored by the loss $\rho$ as in
Section~\ref{sec:measuring_completeness}.

\paragraph{Completeness.}

\begin{definition}[Distributional Predictive Error and Completeness]\label{def:kdist}
The distributional predictive error of theory $M$ in domain $d$ is
\[
e^{dist}_{P_d}(M)=\E_{P_d}\bigl[S\bigl(\PiM(X),Y\bigr)\bigr].
\]
Let $P^*$ denote the family of conditional distributions of $Y$ given $X=x$,
which minimizes the expected score, and let $\LM$ denote the reference
distribution of Section~\ref{sec:measuring_restrictiveness}, evaluated
covariate value by covariate value. The \emph{distributional completeness} of
theory $M$ in domain $d$ is
\[
\kappa^{dist}(M,d)=1-\frac{e^{dist}_{P_d}(M)-e^{dist}_{P_d}(P^*)}
{e^{dist}_{P_d}(\LM)-e^{dist}_{P_d}(P^*)} .
\]
\end{definition}

The score is applied covariate value by covariate value.\footnote{Where every agent is observed
at every covariate value, a kernel score for vectors could instead be applied to an agent's whole
vector of outcomes, which would also score the dependence across covariate values. It requires
complete observation and a common scale for the outcomes at different covariate values, and it
is dominated by the marginal distributions. The measures of this
subsection and the next therefore concern the distribution of outcomes at each covariate value;
how the outcomes of one agent at different covariate values relate is not scored.} The best achievable predictor is the conditional distribution of the outcome,
the distributional counterpart of the conditional mean $f^*$. The baseline is
the reference distribution itself which corresponds to a distribution where any admissible outcome is equally likely and thus the prediction someone would make who  has no prior knowledge or theory about the data, just as
its center $f_{base}$ is the point prediction that uses none. The two interpretations
thus share one reference object, and completeness is anchored at its
distribution under this interpretation and at its center under the other.

Because the score is proper, the excess score of any family of predictive
distributions $Q$ over the truth is a distance,
$e^{dist}_{P_d}(Q)-e^{dist}_{P_d}(P^*)=D_d(Q,P^*)$ with
$\nu_j=P_d(X=x_j)$, so that
\[
\kappa^{dist}(M,d)=1-\frac{D_d(\PiM,P^*)}{D_d(\LM,P^*)} :
\]
the distance of the theory's predictive distributions from the true ones,
relative to the distance of the reference from them. This is the counterpart
of the pairing of loss and discrepancy under squared error
(Section~\ref{sec:estimation}), where completeness is a ratio of squared
distances between rules. Under the representative-agent interpretation the fitted theory is a
point mass at the best-fitting rule. Its counterpart here is the parameter
distribution whose predictive distributions come closest to the truth,
$\pi^{*}\in\arg\min_{\pi\in\Delta(\Theta_M)}e^{dist}_{P_d}(\Pi_\pi)$, where
$\Pi_\pi$ denotes the distribution over rules induced by $\pi$: the population
of agents sharing the form that best describes the outcomes. It need not be a
point mass, and its completeness bounds $\kappa^{dist}$ for every parameter
distribution on $\Theta_M$.

\paragraph{Example 1 (continued).} (Heterogeneous respondents) Under the
heterogeneous-agent interpretation, $\pi_{M,d}$ describes how $(\alpha,\gamma,\delta)$ are
distributed across the respondents of an experiment, and $\PiM(x)$ is the
distribution of certainty equivalents the theory predicts across respondents
at lottery $x$. The baseline predictive distribution at $x$ is the uniform
distribution on the prize range $[\underline z,\bar z]$, the best achievable
one the actual distribution of certainty equivalents across respondents. Distributional
completeness asks whether the spread of certainty equivalents the theory
predicts at each lottery matches the spread across respondents. A prior that
reproduces the dispersion of respondents is credited for it here and charged
for it under the representative-agent interpretation.

\subsection{Measuring Restrictiveness Under the Heterogeneous-Agent Interpretation}
\label{sec:population_restrictiveness}
Under the heterogeneous-agent interpretation a theory is a statement about how outcomes are distributed across the agents of a population. We measure its restrictiveness as the distance between the population the theory describes and the reference, under which every admissible outcome is equally likely across agents, relative to the distance between the reference and the most extreme claim, that every agent has the admissible outcome farthest from the reference (under the CRPS this is an endpoint of the admissible range).
Because the score is proper, this distance is the excess score the theory would incur if the population were in fact the reference, so restrictiveness and completeness share one scale.  The same distance is defined for the population itself as well. The distance of the distribution of outcomes across agents from the reference is the population's own restrictiveness, $r^*$. As a result a theory can be more or less restrictive than the truth, and it loses completeness when it over- or under-commits (Proposition~\ref{prop:rdist_bound}).
For a
kernel score $D_d(\,\cdot\,,\LM)$ is convex, so its maximum over families of
distributions supported on the admissible sets is attained at a point mass at
every covariate value, the admissible rule farthest from the reference in
$D_d$. We denote it $g^{far}$; under the continuous ranked probability score
and the squared discrepancy it is the farthest rule of
Section~\ref{sec:measuring_restrictiveness}.

\begin{definition}[Distributional Restrictiveness]\label{def:rdist}

\[
r^{dist}(M,d)=\frac{D_d(\PiM,\LM)}{D_d(g^{far},\LM)} .
\]
\end{definition}

\noindent The measure is $0$ for a theory whose predictive distributions are
those of the reference and $1$ for the most extreme point prediction. We define  the restrictiveness of the population as
\[r^{*}=D_d(P^*,\LM)/D_d(g^{far},\LM)
\]
with $P^*$ referring to the population.

\begin{proposition}[Range and the population's value]\label{prop:rdist_bound}
(i) $0\le r^{dist}(M,d)\le1$, with $r^{dist}=0$ if and only if the theory's
predictive distributions coincide with the reference at every covariate value
that carries weight, and $r^{dist}=1$ for the theory that consists of
$g^{far}$ alone. (ii) With $\nu_j=P_d(X=x_j)$,
\[
\kappa^{dist}(M,d)\;\le\;1-\Bigl(\sqrt{r^{dist}(M,d)/r^{*}}-1\Bigr)^{2}.
\]
\end{proposition}
\begin{proof}
(i) follows from $d_S\ge0$, with equality only at equal distributions, and
from the choice of $g^{far}$ as the maximizer of the numerator. (ii) For a
kernel score $\sqrt{D_d}$ is a norm distance between families of
distributions. The triangle inequality applied to the theory, the reference and
the truth gives
$\sqrt{D_d(\PiM,P^*)}\ge\bigl|\sqrt{D_d(\PiM,\LM)}-\sqrt{D_d(P^*,\LM)}\bigr|$;
dividing by $\sqrt{D_d(P^*,\LM)}$ and using the distance form of
$\kappa^{dist}$ yields the bound.
\end{proof}

\noindent A theory can be fully complete only if it is exactly as restrictive
as the population it describes. One that rules out more than the population
does, $r^{dist}>r^{*}$, is over-committed and loses completeness in
proportion; one that rules out less is under-committed and loses it as well.
Under the representative-agent interpretation no such benchmark exists, since the best
predictor is a single rule and every single rule is maximally restrictive.

\paragraph{Form and distribution.} Like probabilistic restrictiveness, $r^{dist}$ can be split
into a part due to the functional form, how close a population of agents who share the form can
come to the reference, and a part due to the parameter distribution. Appendix~\ref{app:dist_channels}
gives the construction and its properties. We report $r^{dist}$ as a single measure.

\paragraph{Example 1 (continued).} (Restrictiveness of a heterogeneous
population) The reference predictive distribution at lottery $x$ is the
uniform distribution on the prize range, and the farthest rule places a point
mass at the prize farther from its midpoint. Distributional restrictiveness
asks how far the spread of certainty equivalents CPT predicts across
respondents is from ``any amount in the prize range, equally likely''.

\section{Comparing Theories Across Domains}
 \label{sec:comparing_theories}

 We now combine scope, completeness and restrictiveness to compare theories across domains. Section~\ref{sec:within_domain} first makes
explicit how the measures on a domain aggregate the situations it contains and the following Sections explain how weighting setups can be used to compare across domains. The definitions are stated for the representative-agent interpretation; under the
heterogeneous-agent interpretation they apply with $\kappa^{dist}$ and $r^{dist}$ in place of
$\kappa^{prob}$ and $r^{prob}$.
\subsection{Situations Within a Domain}
\label{sec:within_domain}

 For completeness and distributional
restrictiveness, a statement about a domain corresponds to a weighted sum across the situations contained in the domain. The weights are implicitly determined by the environment, through the
distribution over situations that the data-generating process induces, and by the room for
improvement, the difference between the baseline and the best predictor at a given covariate
value.

\begin{proposition}[Aggregation across covariate values]\label{prop:aggregation}
Fix a domain $d$ with $\nu_j=P_d(X=x_j)$ and a theory whose predictive distributions are given.
For a covariate value $x_j$ let $\kappa^{prob}(M,x_j)$, $\kappa^{dist}(M,x_j)$ and
$r^{dist}(M,x_j)$ denote the measures on the singleton domain $\{x_j\}$, with the same predictive
distribution, baseline, reference and best predictor. Suppose the baseline and the best predictor
are determined covariate value by covariate value. Then each measure on $d$ is a weighted average
of its singleton values,
\begin{gather*}
\kappa^{prob}(M,d)=\sum_j w_j\,\kappa^{prob}(M,x_j),\qquad
\kappa^{dist}(M,d)=\sum_j w^{dist}_j\,\kappa^{dist}(M,x_j),\\
r^{dist}(M,d)=\sum_j v_j\,r^{dist}(M,x_j),
\end{gather*}
with weights proportional to $\nu_j$ times, respectively, the room for improvement
$\xi_j=\mathbb{E}_{P_d}[l(f_{\text{base}}(x_j),Y)-l(f^*(x_j),Y)\mid X=x_j]$, the room
$\xi^{dist}_j=\mathbb{E}_{P_d}[S(\LM(x_j),Y)-S(P^*(x_j),Y)\mid X=x_j]$, and the normaliser
$n_j=D(g^{far}(x_j),\LM(x_j))$, each assumed positive. None of the weights depends on $M$.
\end{proposition}
\begin{proof}
Each measure is a ratio whose numerator and denominator are $\nu$-weighted sums over covariate
values: $e_{P_d}(f)=\sum_j\nu_j\,\mathbb{E}_{P_d}[l(f(x_j),Y)\mid X=x_j]$, where the expectation
over $f\sim\Pi_{M,d}$ commutes with the sum; $e^{dist}_{P_d}$ likewise with the score; and
$D_d(Q,Q')=\sum_j\nu_j\,D(Q(x_j),Q'(x_j))$. The $j$th term of each numerator equals the $j$th
term of the denominator times the singleton value of the measure, by the definition of the
measure on $\{x_j\}$. A ratio of sums is the denominator-weighted average of the termwise ratios.
\end{proof}

A covariate value thus counts in proportion to how often it occurs and to how much a perfect
predictor could improve on the baseline there, or, for distributional restrictiveness, to the
distance the farthest rule can reach. Neither factor depends on the theory, so the weights are
common to every theory under comparison. This weighting is not a choice: it follows from defining completeness as a ratio of
totals, for any additive loss. A covariate value at which the baseline is already as
good as the best predictor receives no weight, and when the loss is measured in the units of the
outcome, situations with larger stakes carry more. The proposition does not apply to restrictiveness under the representative-agent interpretation. The closest rule of a theory has to
serve every situation of a domain at once, so $r^{ff}$ and $r^{prob}$ take the distance of a
pseudo-true rule to the reachable set in the product space over all covariate values of the
domain, and a form that can reach any single value has no restrictiveness at a single covariate
value. What these measures record is how a theory ties its predictions across covariate values
together; Appendix~\ref{app:grouping} bounds how the value changes when domains are merged.

\subsection{Dominance and Predictive Content}
The scope $\mathcal{D}_M$ of a theory is the set of domains it claims
(Section~\ref{sec:framework_scope}). Without loss of generality we assume that each domain carries one environment, so that
$\kappa^{prob}(M,d)$ is the completeness of $M$ under $P_d$. In cases where a domain has  several environments everything that follows applies to pairs of domain and
environment. Each environment then has its own completeness and its own weights over situations,
while the restrictiveness of the theory's channels is common to all
environments on a domain as it is data independent. Comparing two
theories domain by domain is meaningful because everything the measures depend
on besides the theory (e.g. the covariate values, the admissible outcomes, the
reference distribution, the discrepancy and the baseline) belongs to the
domain and is common to both. We begin with the benchmark case in which two
theories $M_1$ and $M_2$ share the same scope,
$\mathcal{D}_{M_1} = \mathcal{D}_{M_2}$. Differences between them then arise
solely from their completeness and restrictiveness within domains.
We say that $M_1$ \emph{strongly dominates} $M_2$ if
$\kappa^{prob}(M_1,d) \ge \kappa^{prob}(M_2,d)$ and $r^{prob}(M_1,d) \ge r^{prob}(M_2,d)$
for all $d \in \mathcal{D}_{M_1}$, with at least one strict inequality for some $d$.
The theory must thus deliver weakly higher predictive performance and weakly stronger structural restrictions in every domain, with strict improvement in at least one.

Whether $M_1$ is more restrictive than $M_2$ in a domain can depend on how the
two channels are aggregated (Section~\ref{sec:measuring_restrictiveness}). If $r^{ff}(M_1,d)\ge r^{ff}(M_2,d)$
and $r^{prior}(M_1,d)\ge r^{prior}(M_2,d)$ in every domain, it does not, and we
call the dominance \emph{channel-wise}. The same qualification applies to the
weighted criteria below.

Whenever one theory is more complete or restrictive only in some domains, strong dominance does not apply. To compare theories that strong dominance cannot rank, the analyst must take a stand on the relative importance of domains.  We capture this
with a relevance weighting, a probability measure $\mu$ over $\mathcal{D}$. This weighting may reflect the frequency with which the situations of a domain arise, their empirical importance, or their similarity to settings of interest. 
Aggregating a theory's within-domain completeness and restrictiveness against $\mu$ summarizes its performance in a way
that respects the analyst's view of which domains matter, and resolves comparisons
that the domain-by-domain criterion leaves open.
$M_1$ \emph{dominates} $M_2$ under $\mu$ if both inequalities hold for the expectations
$\mathbb{E}_{d\sim\mu}[\kappa^{prob}(\cdot,d)]$ and $\mathbb{E}_{d\sim\mu}[r^{prob}(\cdot,d)]$, with at least one
strict inequality.

The relevance weighting and the weighting within domains of Section~\ref{sec:within_domain}
combine into one weighting over situations (Corollary~\ref{cor:partition} in Appendix~\ref{app:grouping}).
The choice of how to partition into domains thus corresponds to a choice about the weighting. Within a cell the weighting is determined through the definition of the measure. Across domains, the analyst decides the weighting. The finest partition, i.e. viewing each situation as its own domain thus makes the full weighting dependent on the analyst. Similarly, choosing $\mu$ such that each cell's weight is proportional to its total room, $\mu_k\propto\sum_{i\in d_k}\nu_i\xi_i$, corresponds to viewing all situations as a single domain, whatever the partition.
Under the representative-agent interpretation the partition is instead a
parameter of the restrictiveness measure itself, since $r^{ff}$ credits the ties a theory imposes
across the situations of a domain. We provide a bound on the change of  $r^{ff}$ due to choices of domain partition in Appendix~\ref{app:grouping}.

More commonly, theories apply to different but overlapping sets of domains,
i.e. $\mathcal{D}_{M_1}\neq\mathcal{D}_{M_2}$ with $\mathcal{D}_{M_1} \cap \mathcal{D}_{M_2} \neq \emptyset$. A theory should receive no credit in a
domain it does not cover, so we gate each domain's contribution by the indicator
$\mathbf{1}\{d\in\mathcal{D}_M\}$. A theory with broad scope is then rewarded for its
breadth, and a theory that is silent in a domain neither gains nor loses there. The relevance weighting $\mu$ continues to operate alongside the indicator, and it
can now also reflect uncertainty about applicability. The scope $\mathcal{D}_M$ is a
hard set, so the indicator assigns a domain either full weight or none. In practice,
though, a theory often has core domains, where the analyst requires it to apply
fully, alongside domains that sit closer to the boundary of its claims. By placing
less mass on these borderline domains, the analyst can make the aggregate less
sensitive to them, without having to make a sharp in-or-out judgment for each one.

To compare theories in such cases, we define a theory's \emph{predictive content} to summarize its  performance as a single pair that aggregates across domains while keeping completeness and restrictiveness separate.

\begin{definition}[Predictive Content]\label{def:predictive_content}
Let $\mu$ be a probability measure over $\mathcal{D}$. The \emph{predictive content}
of a theory $M$ under $\mu$ is the pair
\[
C(M;\mu)=\big(C_\kappa(M;\mu),\,C_r(M;\mu)\big),
\]
where
\[
C_\kappa(M;\mu)=\mathbb{E}_{d\sim\mu}\!\big[\mathbf{1}\{d\in\mathcal{D}_M\}\,\kappa^{prob}(M,d)\big],
\qquad
C_r(M;\mu)=\mathbb{E}_{d\sim\mu}\!\big[\mathbf{1}\{d\in\mathcal{D}_M\}\,r^{prob}(M,d)\big]
\]
are the \emph{weighted completeness} and \emph{weighted restrictiveness} of $M$.
\end{definition}

When scopes differ, the plain weighting across all domains is no longer well-defined across both
theories, since each is silent on part of the domain space. Predictive content
resolves this by gating each domain with the indicator, so the comparison extends
directly.
Theory $M_1$ \emph{dominates} $M_2$ under $\mu$ if $C(M_1;\mu)\ge C(M_2;\mu)$,
componentwise with at least one strict inequality; with identical scopes this is the weighted
criterion above.

Broader scope is not automatically advantageous under this definition. Claiming
an additional domain raises a theory's weighted restrictiveness whenever the
theory rules anything out there, since $r^{prob}\ge0$. It raises weighted
completeness only if the theory's predictions in that domain improve upon the
baseline ($\kappa^{prob}>0$). 
If the predictions are worse than the baseline ($\kappa^{prob}<0$), the
additional domain \emph{reduces} weighted completeness, and the two versions of
the theory are not ranked. On the
completeness coordinate a theory is then better served by remaining silent in a
domain than by making predictions that underperform the baseline.

\section{Estimation}
\label{sec:estimation}

This section describes how we estimate the quantities defined in
Section~\ref{sec:framework} from finite data. We first describe the objects that are common to all estimators: the cross validation fold split, the parameter draws, the reference distribution and the standard error bootstrap. We then describe
the estimation of completeness and of restrictiveness under the representative-agent interpretation, for any loss and
discrepancy, and its special case of squared error. Then we proceed to describing estimation under
the heterogeneous-agent interpretation of Sections~\ref{sec:population_reading}
and~\ref{sec:population_restrictiveness}, for any kernel score, and its special case of the
continuous ranked probability score. The section closes with the estimation of predictive content,
which is the same under both interpretations.
\subsection{Preliminaries}
\label{sec:est_preliminaries}
\paragraph{Observations and folds.}
Throughout, $i$ indexes the sampling units, the agents, and $j$ the covariate values
$x_1,\dots,x_n$ of the domain. An observation is the outcome $y_{ij}$ of agent $i$ at covariate
value $x_j$, and $J_i$ denotes the set of covariate values at which agent $i$ is observed. The
$K$ folds partition the agents, so that all observations of an agent fall into the same fold, and
$N_k$ is the number of observations in fold $k$.

\paragraph{The lookup table.}
The completeness measures require an estimate of the irreducible error. Following
\citet{completeness2022}, we estimate it by a nonparametric lookup table that stores, for each
unique value of $x$, a summary of the training observations at that value: in each fold $k$ the
table is trained on the $K-1$ training folds and evaluated on the held-out fold. Under the representative-agent interpretation the table stores, for each value of $x$, the prediction that minimizes the average loss among the training observations at $x$, which estimates $f^*(x)$.   Under the heterogeneous-agent interpretation it stores the empirical distribution of the training observations at $x$, which estimates the conditional distribution $P^*(x)$. In experimental settings where the number of observations per unique covariate vector is large, this estimator closely approximates the best achievable error: the irreducible error under the representative-agent interpretation, and the expected score of the true conditional distribution under the heterogeneous-agent interpretation.

\paragraph{Parameter distribution.}
To operationalize the framework, we consider two specifications. The first is a \emph{uniform prior} over $\Theta_M$, which treats all parameter values consistent with the theory as equally plausible ex ante.  The second is an \emph{informative prior} that concentrates mass on parameter ranges estimated in prior empirical work. Both priors are defined on the reachable set: when the support restrictions
exclude part of the parameterization's image, draws violating them are
discarded via rejection sampling.
\paragraph{Monte Carlo integration.}
The probabilistic predictive error $e^{prob}_{P_d}(M) = \mathbb{E}_{f \sim \Pi_{M,d}}[e_{P_d}(f)]$ requires integration over the theory's predictive distribution $\Pi_{M,d}$. For most theories of interest, this expectation cannot be evaluated in closed form. We approximate it by drawing $B = 1{,}000$ parameter vectors $\theta^{(1)}, \ldots, \theta^{(B)} \sim \pi_{M,d}$ and replacing the expectation with the sample average over $\{f_{\theta^{(b)}}\}_{b=1}^{B}$. Because $\pi_{M,d}$ is a property of the theory rather than a data-driven estimate, the parameter draws are fixed throughout all subsequent computations. The one exception is
the fitted distribution $\pi^*$ of Section~\ref{sec:population_reading}, which is estimated from
the data; it is fitted within the folds and its draws change with the training fold.

\paragraph{Reference distribution.}
\label{sec:estimating_ff_ref}

The reference distribution $\Lambda_d$
determines the pseudo-true prediction rules against which the theory's
restrictions are evaluated. It is a distribution over $\mathcal F_d$, the
admissible prediction rules on the $n$ covariate values of $\mathcal X_d$, so
every draw satisfies $f(x_j)\in\mathcal Y_d(x_j)$ by construction.

We construct $\LM$ from a family of distributions indexed by a
covariance kernel $K$ over the covariate points. Our primary specification draws $f(x_j)$ independently across covariate values
and uniformly on $\mathcal{Y}_d(x_j)$. It is the maximum-entropy
distribution on the admissible set $\mathcal{F}_d=\prod_j\mathcal{Y}_d(x_j)$, in
the same spirit as the uniform reference $\Pi_{\text{uniform}}$ used for prior
restrictiveness.  We prefer it because it imposes no smoothness on the
reference functions.  

The reference distribution serves three purposes: (1) Its draws are the pseudo-true prediction rules needed to calculate functional from restrictiveness, (2) 
under the heterogeneous-agent interpretation its restriction to a single covariate value is the baseline of distributional completeness, and as a whole it is the reference population of distributional restrictiveness.
(Sections~\ref{sec:population_reading} and~\ref{sec:population_restrictiveness}). 

\paragraph{Standard errors.}
Standard errors for the completeness estimates are obtained by block
bootstrap, clustering on subjects, with $1{,}000$ bootstrap draws. Parameter
draws $\theta^{(1)}, \ldots, \theta^{(B)}$ and reference function draws
$f^{(1)}, \ldots, f^{(S)}$ are held fixed across bootstrap replicates. Restrictiveness,
by contrast, is a property of the theory and the reference distribution and
does not depend on observed outcomes. The only estimation noise that exists for restrictiveness is Monte
Carlo noise, which report alongside the restrictiveness estimates.

\subsection{Estimation Under the Representative-Agent Interpretation}
\label{sec:est_uncertainty}
\paragraph{Baseline.}

The baseline prediction rule $f_{base}$ represents the
predictive performance achievable without theory-specific structure. It enters only 
the completeness measure. Functional-form restrictiveness is normalized by
the largest expected discrepancy any admissible rule can have (Section \ref{sec:measuring_restrictiveness})
which involves no choice by the analyst. The two are nevertheless linked through
the reference distribution. When $f_{base}$ is the center of $\LM$, as in our
application, completeness is anchored at the \emph{best} uninformed rule and
restrictiveness at the \emph{worst}.
 When the admissible values are the same at every covariate a natural default is the unconditional mean,
$f_{\rm base}(x) = \bar{Y}$ for all $x \in \mathcal{X}_d$, where
$\bar{Y} = \mathbb{E}_{P_d}[Y]$, which ignores all covariate information and
requires no structural assumptions. When outcomes do not share a common
scale the unconditional mean mixes incommensurable
quantities and can fall outside an individual $x$'s support. In such cases
the analyst may replace $\bar Y$ with a domain-specific baseline that
respects the local scale of each $x$, provided it remains behavior-agnostic.

\subsubsection{Completeness}
For the fitted completeness of \citet{completeness2022}, the model's
best-fitting parameter $\theta^*$ must be estimated from training data and
evaluated out of sample, requiring cross-validation for both the model error
and the irreducible error. In our framework, the theory's
predictions $f_{\theta^{(b)}}$ are fully determined by the parameter draws
from $\pi_{M,d}$ and require no training data, and the baseline $f_{\rm mid}$
is computed from lottery characteristics alone. 

The baseline error and the
probabilistic predictive error are evaluated on the same held-out folds and
averaged. Because neither uses the training data, this coincides with their
full-sample average up to differences in fold size.
\[
\hat e^{prob}(M)=\frac1B\sum_{b=1}^{B}\hat e\bigl(f_{\theta^{(b)}}\bigr),\qquad
\hat e(f)=\frac1K\sum_{k=1}^{K}\frac1{N_k}\sum_{i\in k}\sum_{j\in J_i}l\bigl(f(x_j),y_{ij}\bigr).
\]
From the same draws we compute the theory's mean prediction,
$\hat{\fbar}(x)=B^{-1}\sum_b f_{\theta^{(b)}}(x)$, its error
$\hat e(\hat{\fbar})$ and completeness $\hat\kappa(\fbar)$.

\subsubsection{Restrictiveness}
\label{sec:est_restrictiveness}

The restrictiveness estimator has two components: functional form
restrictiveness $r^{ff}$ and
prior restrictiveness $r^{prior}$. We describe
each in turn.

\paragraph{Discrepancy function.}
We take $\delta(f,f') = n^{-1}\sum_{j=1}^{n}(f(x_j) - f'(x_j))^2$, the
mean-squared discrepancy between two prediction rules evaluated at the
observed covariate values.
The same equal weights, $\nu_j=1/n$, are used for the distance $D_d$ of the heterogeneous-agent interpretation;
the bound of Proposition~\ref{prop:rdist_bound} is stated for $\nu_j=P_d(X=x_j)$, and in a balanced
design, in which every covariate value is presented equally often, the two coincide.

\paragraph{Functional form restrictiveness.}
\label{sec:estimating_ff}
Following \citet{restrictiveness2026}, we draw $S = 1{,}000$ reference
functions $f^{(1)}, \ldots, f^{(S)} \sim \Lambda_{M,d}$.  For each reference function
$f^{(s)}$ we solve $\theta^{*(s)}\in\arg\min_{\theta\in\Theta_M:\,f_\theta\in
\mathcal S_{M,d}}\delta_d(f_\theta,f^{(s)})$. The estimator is the ratio of two
sample means,
\[
\hat r^{f\!f}(M,d)=
\frac{\frac1S\sum_s\delta_d\bigl(f_{\theta^{*(s)}},f^{(s)}\bigr)}
     {\frac1S\sum_s\delta_d\bigl(\hat g^{far},f^{(s)}\bigr)},
\qquad
\hat g^{far}\in\arg\max_{g\in\mathcal F_d}\frac1S\sum_s\delta_d\bigl(g,f^{(s)}\bigr).
\]
The denominator does not depend on the theory and is computed once per domain.

\begin{figure}[htbp]\centering
\begin{tikzpicture}[x=1cm,y=1cm,
    every node/.style={font=\fontsize{8.5}{10.5}\selectfont},
    lab/.style={font=\fontsize{8}{9.5}\selectfont,inner sep=1.5pt},
    ttl/.style={font=\fontsize{8.5}{10.5}\selectfont,anchor=south west,inner sep=0pt},
    ax/.style={line width=0.5pt},
    lead/.style={black!55,line width=0.3pt},
    nextstep/.style={-stealth,line width=0.5pt,black!70}]

  \def\showuniform{0}   
  \def\su{3.8}     
  \def\sd{0.5}     

  \begin{scope}[shift={(0.35,0)}]
    \node[ttl] at (-0.35,3.25) {\textbf{(a)} Parameter space $\Theta_M$};
    \fill[black!14] plot[domain=0:0.7,samples=71,variable=\t]
      ({\su*\t},{\sd*5.0309*exp(-((\t-0.51)/0.08)^2/2)}) -- ({\su*0.7},0) -- (0,0) -- cycle;
    \fill[black!45] plot[domain=0.5:0.6,samples=11,variable=\t]
      ({\su*\t},{\sd*5.0309*exp(-((\t-0.51)/0.08)^2/2)}) -- ({\su*0.6},0) -- ({\su*0.5},0) -- cycle;
    \draw[line width=0.6pt] plot[domain=0:0.7,samples=71,variable=\t]
      ({\su*\t},{\sd*5.0309*exp(-((\t-0.51)/0.08)^2/2)});
    \draw[ax] (0,0)--({\su*0.7+0.12},0);
    \draw[ax] (0,0)--(0,-0.06) ({\su*0.7},0)--({\su*0.7},-0.06);
    \node[lab,anchor=north] at (0,-0.07) {$0$};
    \node[lab,anchor=north] at ({\su*0.7},-0.07) {$0.7$};
    \node[lab,anchor=north] at ({\su*0.35},-0.07) {$\theta$};
    \node[lab,anchor=west] at (0.0,2.45) {$\pi_{M,d}(\theta)$};
    \draw[lead] (0.95,2.3)--(1.62,1.55);
    \node[lab,anchor=west] at (2.28,2.75) {$\pi\,d\theta$};
    \draw[lead] (2.36,2.62)--(2.1,1.7);
    \foreach \q in {0.385,0.41,0.43,0.445,0.46,0.47,0.48,0.49,0.497,0.505,0.512,0.52,0.527,0.535,0.545,0.555,0.565,0.58,0.595,0.615,0.64}{
      \draw[black,line width=0.45pt] ({\su*\q},-0.62)--({\su*\q},-0.80);}
    \node[lab,anchor=north west] at (-0.05,-0.84) {$\theta^{(b)}\sim\pi_{M,d}$\; ($B_H$ draws)};
    \ifnum\showuniform=1
    \foreach \q in {0.018,0.05,0.087,0.12,0.151,0.19,0.226,0.262,0.291,0.312,0.348,0.385,0.41,0.447,0.482,0.52,0.551,0.588,0.62,0.654,0.688}{
      \draw[black!50,line width=0.45pt] ({\su*\q},-1.42)--({\su*\q},-1.60);}
    \node[lab,anchor=north west] at (-0.05,-1.64) {$\theta^{(b)}_{\mathrm{unif}}$ uniform\; ($B_U$ draws)};
    \fi
  \end{scope}

  \begin{scope}[shift={(4.55,0)},x=4cm,y=4cm]
    \node[ttl] at (-0.09,0.8125) {\textbf{(b)} Reachable set $\mathcal R_{M,d}$};
    \draw[-stealth,line width=0.4pt,black!60] (0.56,0.03)--(0.72,0.03);
    \draw[-stealth,line width=0.4pt,black!60] (0.56,0.03)--(0.56,0.19);
    \node[lab,anchor=west,text=black!70] at (0.72,0.03) {$f(x_0)$};
    \node[lab,anchor=south,text=black!70] at (0.56,0.19) {$f(x_1)$};
    \draw[black!60,line width=0.9pt] plot[domain=0:0.7,samples=50,variable=\t] ({\t*\t},{\t});
    \draw[black,line width=2.2pt] plot[domain=0.5:0.6,samples=8,variable=\t] ({\t*\t},{\t});
    \foreach \q in {0,0.1,0.2,0.3,0.4,0.5,0.6,0.7}{
      \pgfmathsetmacro{\nn}{sqrt(1+4*\q*\q)}
      \pgfmathsetmacro{\nx}{0.024/\nn}
      \pgfmathsetmacro{\ny}{-0.024*2*\q/\nn}
      \draw[black,line width=0.5pt] ({\q*\q-\nx},{\q-\ny})--({\q*\q+\nx},{\q+\ny});}
    \node[lab,anchor=west] at (0.04,0.005) {$\theta=0$};
    \node[lab,anchor=west] at (0.525,0.7) {$\theta=0.7$};
    \node[lab,anchor=west] at (0.345,0.545) {$|J(\theta)|\,d\theta$};
    \draw[-stealth,line width=0.6pt] (0.04,0.2)--(0.152,0.48);
    \node[lab,anchor=east] at (0.085,0.45) {$J(\theta)$};
    \node[lab,anchor=north west,align=left] at (0.27,0.4) {equal steps\\ in $\theta$};
    \draw[lead] (0.27,0.34)--(0.115,0.3);
  \end{scope}

  \begin{scope}[shift={(9.0,0)}]
    \node[ttl] at (-0.2,3.25) {\textbf{(c)} \ifnum\showuniform=1 Densities\else Density\fi\ on $\mathcal R_{M,d}$};
    \fill[black!14] plot[domain=0:0.7,samples=71,variable=\t]
      ({\su*0.5*(\t*sqrt(1+4*\t*\t)+0.5*ln(2*\t+sqrt(1+4*\t*\t)))},
       {\sd*5.0309*exp(-((\t-0.51)/0.08)^2/2)/sqrt(1+4*\t*\t)}) -- ({\su*0.88666},0) -- (0,0) -- cycle;
    \fill[black!45] plot[domain=0.5:0.6,samples=11,variable=\t]
      ({\su*0.5*(\t*sqrt(1+4*\t*\t)+0.5*ln(2*\t+sqrt(1+4*\t*\t)))},
       {\sd*5.0309*exp(-((\t-0.51)/0.08)^2/2)/sqrt(1+4*\t*\t)}) -- ({\su*0.72261},0) -- ({\su*0.57390},0) -- cycle;
    \draw[line width=0.6pt] plot[domain=0:0.7,samples=71,variable=\t]
      ({\su*0.5*(\t*sqrt(1+4*\t*\t)+0.5*ln(2*\t+sqrt(1+4*\t*\t)))},
       {\sd*5.0309*exp(-((\t-0.51)/0.08)^2/2)/sqrt(1+4*\t*\t)});
    \ifnum\showuniform=1
    \draw[line width=0.45pt,densely dashed] (0,{\sd/0.88666})--({\su*0.88666},{\sd/0.88666});
    \node[lab,anchor=south west,align=left] at (0.0,0.78) {uniform\\ density $1/L$};
    \draw[lead] (0.28,0.8)--(0.28,{\sd/0.88666+0.02});
    \node[lab,anchor=west] at (-0.05,-0.71) {$D_{KL}=\log L-H$};
    \node[lab,anchor=west] at (-0.05,-1.51) {$r^{prior}=1-\exp(-D_{KL})$};
    \fi
    \draw[ax] (0,0)--({\su*0.88666+0.12},0);
    \draw[ax] (0,0)--(0,-0.06) ({\su*0.88666},0)--({\su*0.88666},-0.06);
    \node[lab,anchor=north] at (0,-0.07) {$0$};
    \node[lab,anchor=north] at ({\su*0.88666},-0.07) {$L$};
    \node[lab,anchor=north] at ({\su*0.44},-0.07) {position on $\mathcal R_{M,d}$};
    \ifnum\showuniform=0
    \draw[lead,stealth-stealth] (0,-0.71)--({\su*0.88666},-0.71);
    \node[lab,fill=white,inner sep=2pt] at ({\su*0.44},-0.71) {length $L$};
    \fi
    \node[lab,anchor=west] at (0.0,2.45) {$p_\Pi=\pi_{M,d}/|J|$};
    \draw[lead] (1.2,2.3)--(1.85,1.25);
    \node[lab,anchor=west] at (2.45,2.1) {same mass};
    \draw[lead] (2.75,1.97)--(2.5,1.2);
  \end{scope}

  \draw[nextstep] (3.22,1.35)--(4.2,1.35);
  \node[lab,anchor=south] at (3.71,1.4) {$\theta\mapsto f_\theta$};
  \draw[nextstep] (7.85,1.35)--(8.75,1.35);
  \node[lab,anchor=south,align=center] at (8.3,1.4) {along\\ $\mathcal R_{M,d}$};

  \draw[nextstep] (3.2,-0.71)--(3.75,-0.71);
  \node[lab,anchor=west] at (3.85,-0.71) {$\hat H$ = mean of $-\log\pi_{M,d}+\log|J|$};
  \ifnum\showuniform=1
  \draw[nextstep] (3.2,-1.51)--(3.75,-1.51);
  \node[lab,anchor=west] at (3.85,-1.51) {$\hat L$ = $\mathrm{Vol}(\Theta_M)\times{}$mean of $|J|$};
  \fi
\end{tikzpicture}
\caption{\label{fig:priorest} Estimating prior restrictiveness (schematic representation). (a) The parameter
distribution and the draws from it that enter $\hat H$. (b) The
parameterization maps equal steps in $\theta$ into pieces of the reachable set
of unequal length. (c) The dark strip in (a) is spread over a piece of
length $|J(\theta)|\,d\theta$, so the density of $\Pi_{M,d}$ per unit of length
is $\pi_{M,d}/|J|$. The two dark strips have the same area. Prior
restrictiveness compares this density with the uniform density $1/L$.}

\end{figure}
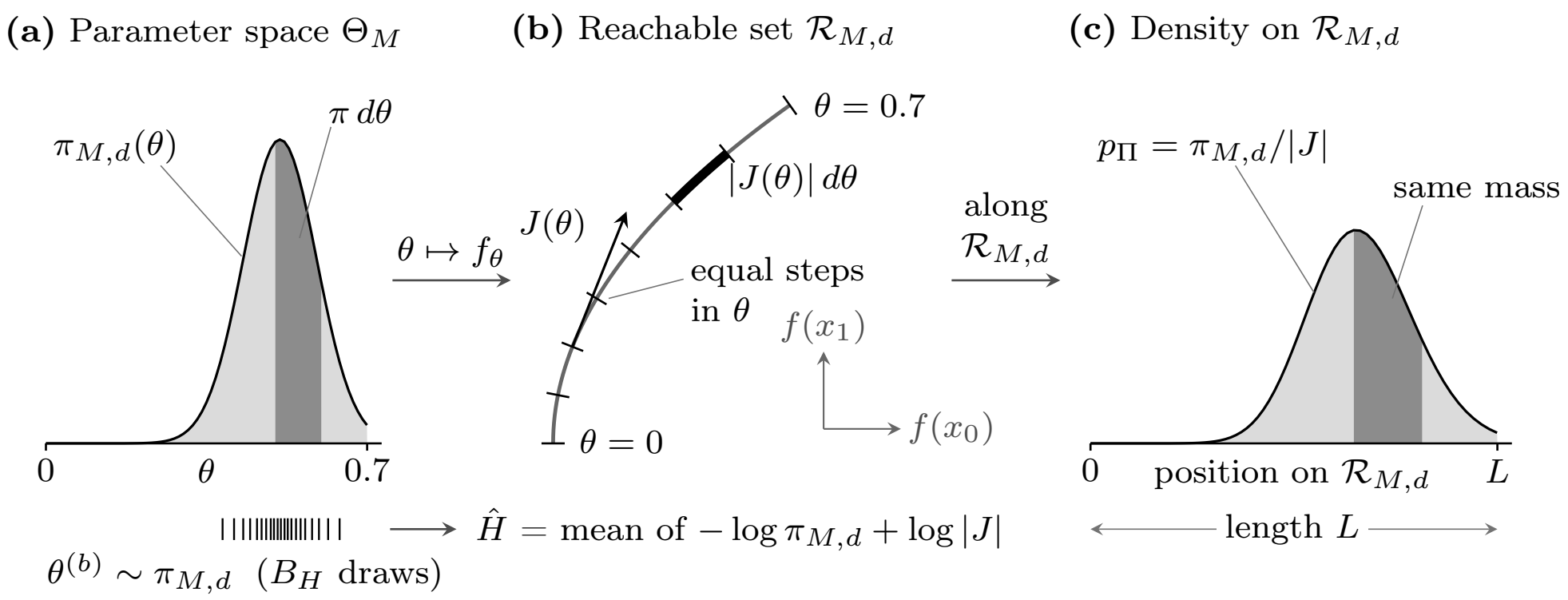

\paragraph{Prior restrictiveness.}
Prior restrictiveness requires the KL divergence between $\Pi_{M,d}$ and the
uniform distribution on the reachable set $\mathcal{R}_{M,d}$ (Section~\ref{sec:measuring_restrictiveness}).
Both are distributions on $\mathcal{R}_{M,d}$, an $m$-dimensional surface in
the space of prediction rules. Figure~\ref{fig:priorest} shows the objects involved.

 Write $f_\theta=(f_\theta(x_1),\dots,f_\theta(x_n))$
for the predictions of the theory at the $n$ covariate values, and let
$J(\theta)$ denote the $n\times m$ Jacobian of the parameterization
$\theta\mapsto f_\theta$. Its entry $(j,k)$ is
$\partial f_\theta(x_j)/\partial\theta_k$, so its $k$-th column records how the
$n$ predictions move when the $k$-th parameter changes. The volume element
\[
|J(\theta)|=\sqrt{\det\bigl(J(\theta)^\top J(\theta)\bigr)}
\]
is the factor that governs how the parameter volume $d\theta$ around $\theta$ is mapped to a piece of
$\mathcal{R}_{M,d}$ of $m$-dimensional volume $|J(\theta)|\,d\theta$. In the case of a single function parameter, $|J(\theta)|$ is the length of the vector
$df_\theta/d\theta$, or equivalently, the speed at which $f_\theta$ moves along the reachable
curve (panel~(b)). More generally, with $m=n$ it is the  $|\det J(\theta)|$ of a change
of variables.\footnote{Nothing here requires a finite set of covariate values.
With a continuum, $J(\theta)^\top J(\theta)$ is replaced by the $m\times m$
matrix $\int\partial_\theta f_\theta(x)\,\partial_\theta f_\theta(x)^\top
d\nu_d(x)$. Constant factors, such as equal weights $1/n$, multiply the volume
element and the volume $L$ alike and cancel in the divergence.}

The volume of the reachable set is
$L=\int_{\Theta_M}|J(\theta)|\,d\theta$, and the uniform distribution on
$\mathcal{R}_{M,d}$ has density $1/L$ with respect to the surface measure. The
parameter distribution places mass $\pi_{M,d}(\theta)\,d\theta$ on a piece of
volume $|J(\theta)|\,d\theta$, so the density of $\Pi_{M,d}$ is
$p_\Pi(f_\theta)=\pi_{M,d}(\theta)/|J(\theta)|$ (panels~(a) and~(c)).
The divergence therefore separates into two terms,
\begin{align*}
D_{KL}(\Pi_{M,d}\,\|\,\Pi_{\text{uniform}})
&=\E_{\theta\sim\pi_{M,d}}\!\left[\log\frac{\pi_{M,d}(\theta)/|J(\theta)|}{1/L}\right]
=\log L-H,\\
H&=\E_{\theta\sim\pi_{M,d}}\bigl[-\log\pi_{M,d}(\theta)+\log|J(\theta)|\bigr],
\end{align*}
where $H$ is the entropy of $\Pi_{M,d}$ with respect to the surface measure.

\emph{Estimation.} We estimate the two terms from two separate Monte Carlo
samples. For the entropy we draw $\theta^{(1)},\dots,\theta^{(B_H)}$ from
$\pi_{M,d}$, with $B_H=50{,}000$, and average,
\[
\hat H=\frac1{B_H}\sum_{b=1}^{B_H}
\Bigl[-\log\pi_{M,d}\bigl(\theta^{(b)}\bigr)+\log\bigl|J\bigl(\theta^{(b)}\bigr)\bigr|\Bigr].
\]
For the volume we draw
$\theta^{(1)}_{\text{unif}},\dots,\theta^{(B_U)}_{\text{unif}}$ uniformly from
$\Theta_M$, with $B_U=10^6$, and use that $L$ equals $\mathrm{Vol}(\Theta_M)$
times the mean of the volume element under the uniform distribution on
$\Theta_M$,
\[
\hat L=\mathrm{Vol}(\Theta_M)\cdot\frac1{B_U}\sum_{b=1}^{B_U}
\bigl|J\bigl(\theta^{(b)}_{\text{unif}}\bigr)\bigr|.
\]
The estimate of the divergence is $\hat D_{KL}=\log\hat L-\hat H$.\footnote{If
the support restriction excluded part of $\Theta_M$, both samples would be
restricted to the parameters whose rules satisfy it, and $\mathrm{Vol}(\Theta_M)$
replaced by the volume of that set. In our application the restriction does not
bind.}

Appendix~\ref{app:est_volume} describes the computation of the volume element and a diagnostic for the
volume estimate.

 The
estimators are $\hat r^{prior}(M,d)=1-\exp(-\hat D_{KL})$ and
$\hat r^{prob}(M,d)=1-(1-\hat r^{f\!f}(M,d))(1-\hat r^{prior}(M,d))$; their Monte Carlo
standard errors are described at the end of this subsection.

\paragraph{Standard errors.}
$\hat\kappa^{prob}(M,d)$ is bootstrapped as described in Section~\ref{sec:est_preliminaries}.
$\hat r^{prob}(M,d)$ carries no sampling uncertainty from the data, and
its only estimation error is Monte Carlo noise from the finite numbers of
reference draws (in $\hat r^{f\!f}$) and of prior and volume draws (in
$\hat r^{prior}$), which we quantify by the delta method
(Appendix~\ref{app:est_mc_se}).

\subsection{The special case of squared error}
\label{sec:est_squared_error}
The previous subsection holds for any loss and any discrepancy under the representative-agent interpretation. A common choice
is squared error for both, in which case the loss from predicting $y'$ when the
outcome is $y$ is $l(y',y)=\lVert y'-y\rVert^2$ and the discrepancy between two
prediction rules is
\[
\delta_d(f,g)=\sum_{j=1}^{n}\nu_j\,\bigl\lVert f(x_j)-g(x_j)\bigr\rVert^2,
\]
where the weights $\nu_j\ge0$, $\sum_j\nu_j=1$, say how much each covariate
value counts when two rules are compared.\footnote{The normalization of the
weights is immaterial for $\rff$, which is a ratio of discrepancies. The example
of Section~\ref{sec:toy} uses $\nu_j=1$, that is, $\delta_d(f,g)=\lVert f-g\rVert^2$.} Three
properties are specific to this case.

\paragraph{Pairing of loss and discrepancy.} The best predictor is the
conditional mean, $f^*(x)=\E_{P_d}[Y\mid X=x]$, so the lookup table of
Section~\ref{sec:est_preliminaries} is a table of cell means, and the expected loss of any rule exceeds
that of the best predictor by the rule's squared distance from it.
Completeness thereby becomes, like restrictiveness, a ratio of squared distances
between prediction rules,
\[
\kprob(M,d)=1-\frac{\E_{f\sim\PiM}\bigl[\delta_{P_d}(f,f^*)\bigr]}
                   {\delta_{P_d}(f_{base},f^*)},
\qquad
\delta_{P_d}(f,g)=\E_X\lVert f(X)-g(X)\rVert^2 ,
\]
the expected distance of the theory's rules from the best predictor, relative
to the distance of the baseline from it (see Appenidx \ref{app:squared_error_distances}for the full derivation). The distance
$\delta_{P_d}$ weights each covariate value by its probability under $P_d$. It
is the discrepancy $\delta_d$ of restrictiveness when $\nu_j=P_d(X=x_j)$ and loss and discrepancy are
then ``paired'' in the sense of \citet{restrictiveness2026} (Online Appendix~E).

\paragraph{Restrictiveness: the normalizer in closed form.} Let
$c=\E_{\LM}[f]$ denote the center of the reference distribution and
$V=\E_{\LM}[\delta_d(c,f)]$ its dispersion. For every prediction rule $g$,
$\E_{f\sim\LM}[\delta_d(g,f)]=V+\delta_d(g,c)$ (Lemma~\ref{lem:identity},
Appendix~\ref{app:squared_error_single}). The center is therefore the best single rule against the
reference distribution, the farthest rule is the admissible rule farthest from
the center, $g^{far}(x_j)\in\arg\max_{y\in\mathcal Y_d(x_j)}\lVert
y-c(x_j)\rVert$, and the denominator of $\rff$ is $V+\rho^2$ with
$\rho^2=\delta_d(g^{far},c)$. When the admissible outcomes are intervals and the
reference distribution is uniform, $\rho^2=3V$, so the center rule, the best
uninformed guess, scores $V/(V+\rho^2)=\tfrac14$ whatever the widths of the
intervals and the weights (Appendix~\ref{app:squared_error_intervals}. This anchors the scale of $\rff$, resulting in $0$
for a theory that can mimic every admissible rule, $\tfrac14$ for the best
uninformed point prediction, $1$ for the most extreme one. Appendix~\ref{app:squared_error_intervals} also
gives the estimator in closed form and a
factorization of $\rff$ into the location and the size of the reachable set.

\subsection{Estimation Under the Heterogeneous-Agent Interpretation}
\label{sec:est_population}
\paragraph{Baseline and best achievable predictor.}
The baseline is the reference distribution $\LM$ itself, evaluated covariate value by covariate
value (Section~\ref{sec:population_reading}); $f_{base}$ is its center. The best achievable
prediction at $x$ is the conditional distribution of $Y$ given $X=x$, and the lookup table of
Section~\ref{sec:est_preliminaries} stores, for each unique value of $x$, the empirical distribution
of the training observations.

\paragraph{The distance between two distributions.}
Both measures of this interpretation are built from one object, a distance between two distributions.
Consider two distributions $P$
and $Q$ on the admissible outcomes at one covariate value $x_j$, and let $Y,Y'$ denote two
independent draws from $P$ and $\tilde Y,\tilde Y'$ two independent draws from $Q$. For a kernel score with
kernel $\rho$, their divergence is
\[
d_S(P,Q)=\E\,\rho(Y,\tilde Y)-\tfrac12\,\E\,\rho(Y,Y')-\tfrac12\,\E\,\rho(\tilde Y,\tilde Y') ,
\]
the average distance between a draw from one distribution and a draw from the other, net of half
the dispersion of each. It is zero when the two distributions coincide and positive otherwise.
Completeness applies this distance to the theory and the observed outcomes, restrictiveness to
the theory and the reference; Appendix~\ref{app:est_kernel} gives the sample versions of the three terms.

\subsubsection{Completeness}
 The $B$ parameter draws $\theta^{(1)},\dots,\theta^{(B)}$ from the theory's parameter distribution $\pi_{M,d}$ (Section~\ref{sec:est_preliminaries}) give, at each covariate value $x$, $B$ predicted outcomes $f_{\theta^{(1)}}(x),\dots,f_{\theta^{(B)}}(x)$. Under the heterogeneous-agent interpretation
these are a simulated population: the outcomes of $B$ agents drawn from the theory. The score
compares this simulated population with the observed outcomes, separately at each covariate
value $x_j$ and for each agent $i$ observed there. For one observation the second
distribution is the single outcome $y_{ij}$, whose dispersion is zero, so the third term of
$\hat d_S$ vanishes and the distance reduces to the score $\widehat S(x_j,y_{ij})$, whose sample
form and interpretation are given in Appendix~\ref{app:est_kernel}.
The remaining steps are those of the representative-agent interpretation with the score in place of the
loss. Averaging over the held-out observations of each fold, and over folds, gives
\[
\hat e^{dist}(M)=\frac1K\sum_{k=1}^{K}\frac1{N_k}\sum_{i\in k}\sum_{j\in J_i}\widehat S(x_j,y_{ij}) .
\]
The baseline error $\hat e^{dist}(\LM)$ is the same expression with the reference draws
$f^{(1)}(x_j),\dots,f^{(S)}(x_j)$ of Section~\ref{sec:est_preliminaries} in place of the theory's
draws, and $S$ in place of $B$, or its closed form where one exists
(Section~\ref{sec:est_crps}). The best achievable error $\hat e^{dist}(\hat P^{*})$ is that of
the lookup table of Section~\ref{sec:est_preliminaries}, with the training observations at $x_j$
in the role of the draws. The estimate is
\[
\hat\kappa^{dist}(M,d)=1-\frac{\hat e^{dist}(M)-\hat e^{dist}(\hat P^{*})}
{\hat e^{dist}(\LM)-\hat e^{dist}(\hat P^{*})} ,
\]
and folds, bootstrap and the treatment of the draws are those of
Section~\ref{sec:est_preliminaries}.

The fitted distribution $\pi^*$ of Section~\ref{sec:population_reading} is
estimated on a finite grid of parameter values. A kernel score is convex in the
predictive distribution and the predictive distribution is linear in the grid
weights, so the held-out score is a convex function of the weights on the
simplex, which we minimize by exponentiated gradient descent. The solution is
sparse, with mass on a few parameter values, like the nonparametric
maximum-likelihood mixture of \citet{kiefer1956}. It is cross-validated like the
fitted value, fitted on the training folds and scored on the held-out fold.

\subsubsection{Restrictiveness}
\paragraph{Distributional restrictiveness.}
Distributional restrictiveness is built from the distance $D_d$ between two families of
predictive distributions (Section~\ref{sec:population_restrictiveness}), the weighted sum of the
divergences over the covariate values,
$D_d(P,Q)=\sum_{j=1}^{n}\nu_j\,d_S\bigl(P(x_j),Q(x_j)\bigr)$, with $\nu_j=1/n$.
For the theory's distance from the reference, the first distribution at $x_j$ is the theory's
predictive distribution, with draws $f_{\theta^{(b)}}(x_j)$, and the second is the reference
distribution at $x_j$, with the reference draws $f^{(s)}(x_j)$ of
Section~\ref{sec:est_preliminaries} (Appendix~\ref{app:est_kernel} gives the sample expression).
The denominator $\hat D_d(g^{far},\LM)$ is the same expression with the single value
$g^{far}(x_j)$, the farthest rule of Section~\ref{sec:population_restrictiveness} evaluated at
$x_j$, in place of the theory's draws, so that the theory's dispersion term vanishes, and
\[
\hat r^{dist}(M,d)=\frac{\hat D_d(\PiM,\LM)}{\hat D_d(g^{far},\LM)} .
\]
The population's own value $\hat r^{*}$ replaces the theory's draws by the observed outcomes
$y_{ij}$ at $x_j$. Where the divergence from the reference has a closed form, as under the score
we use (Section~\ref{sec:est_crps}), it replaces the averages over the reference draws.

\paragraph{Completeness and restrictiveness together.}
Completeness and restrictiveness are related under this interpretation since both are distances among the
theory, the reference and the truth. In Appendix~\ref{app:dist_geometry} we show how they relate and how this can be used to split a theory's incompleteness into a part due to the direction
in which it departs from the reference and a part due to how far it departs.

\paragraph{Standard errors.}
$\hat\kappa^{dist}(M,d)$ is bootstrapped like $\hat\kappa^{prob}(M,d)$, with the parameter draws
held fixed. $\hat r^{dist}(M,d)$ does not depend on the outcomes, so its only
estimation error is Monte Carlo noise from the $B$ parameter draws, from the $S$ reference draws
where no closed form is used. For comparison, we also report the
population's own value $r^{*}$, which does depend on the data. Its standard error is computed by the same block bootstrap over subjects as for completeness.

\subsection{The special case of the CRPS}
\label{sec:est_crps}
Section~\ref{sec:population_reading} leaves the scoring rule open. For a
scalar outcome we use the continuous ranked probability score
(CRPS; \citet{matheson1976}), the kernel score with $\rho(y,y')=|y-y'|$. With
$F_x$ the distribution function of $\PiM(x)$,
\[
\mathrm{CRPS}\bigl(\PiM(x),y\bigr)
=\int\bigl(F_x(z)-\mathbf 1\{y\le z\}\bigr)^2dz
=\E_{f\sim\PiM}\bigl|f(x)-y\bigr|-\tfrac12\,\E_{f,f'\sim\PiM}\bigl|f(x)-f'(x)\bigr| .
\]
The first term is the expected absolute error of a draw. The second returns
half of the theory's own dispersion, so that spread is charged only where the
outcomes do not fill it. For a point prediction the dispersion term vanishes
and the score is the absolute error. The divergence the CRPS generates is the
integrated squared difference of two distribution functions, so the distance
of Section~\ref{sec:population_reading} is
\[
D_d(P,Q)=\sum_{j=1}^{n}\nu_j\int\bigl(F^{P}_{x_j}(z)-F^{Q}_{x_j}(z)\bigr)^2dz ,
\]
and $\sqrt{D_d}$ is the $L^2$ distance between families of distribution
functions. 

\paragraph{Computation.} Under the CRPS none of these quantities needs draws from the reference.
When the admissible outcomes at $x$ are an interval, $\mathcal Y_d(x)=[a,b]$, and the reference is
uniform on it, the score of the reference at a single value $y\in[a,b]$ has an explicit expression,
\[
\mathrm{CRPS}\bigl(\LM(x),y\bigr)=\frac{(y-a)^2+(b-y)^2}{2(b-a)}-\frac{b-a}{6} ,
\]
the expected absolute distance of a uniform draw from $y$ minus half the expected distance between
two uniform draws. This expression is used twice. For completeness, $y$ is an observed outcome
$y_{ij}$, and the average over the observations is the baseline error. For restrictiveness, the
theory's draws take the place of $y$
\subsection{Estimating Predictive Content}
In practice, the analyst observes a finite collection of domains
$d_1, \ldots, d_D$. For each domain in which theory $M$ applies, we compute
domain-specific estimates $\hat{\kappa}^{prob}(M,d)$ and
$\hat{r}^{prob}(M,d)$ using the procedures above; under the heterogeneous-agent interpretation the same construction applies to
the pair $(\hat{\kappa}^{dist}(M,d),\hat{r}^{dist}(M,d))$. The predictive content
estimator replaces the population expectation with a weighted sample analog:
\[
\hat{C}_\kappa(M;\mu) = \sum_{d=1}^{D} \mu_d \,
\mathbf{1}\{d \in \mathcal{D}_M\} \, \hat{\kappa}^{prob}(M,d),
\qquad
\hat{C}_r(M;\mu) = \sum_{d=1}^{D} \mu_d \,
\mathbf{1}\{d \in \mathcal{D}_M\} \, \hat{r}^{prob}(M,d).
\]

Appendix~\ref{app:est_content} gives the standard errors of the two coordinates, the dominance test in
the sample, and the case of domains unlike any that were observed.

\section{Toy Examples}
\label{sec:toy}

\subsection{Setup}

\begin{figure}[t]
\centering

\begin{subfigure}{0.45\textwidth}\centering
\begin{tikzpicture}[unitsq,scale=1.0]
  \fill[gray!18] (0,0)--(0.7,0.7)--(0,0.7)--cycle;
  \draw[frame] (0,0) rectangle (1,1);
  \draw[black,thin] (0,0)--(0.7,0.7)--(0,0.7);
  \draw[thick,gray!55,densely dashed,smooth,samples=40,domain=0.7:1,variable=\t] plot ({\t*\t},{\t});
  \draw[curveM,figB,smooth,samples=50,domain=0:0.7,variable=\t] plot ({\t*\t},{\t});
  \node[font=\footnotesize] at (0.14,0.28) {$f_\theta$};
  \foreach \i in {0,...,33}{
    \pgfmathsetmacro{\ta}{0.7*\i/34}
    \pgfmathsetmacro{\tb}{0.7*(\i+1)/34}
    \pgfmathsetmacro{\dens}{exp(-((\ta-0.51)/0.08)^2/2)}
    \pgfmathsetmacro{\wid}{0.6+3.4*\dens}
    \pgfmathtruncatemacro{\gv}{18+80*\dens}
    \draw[figB!\gv,line width=\wid pt,line cap=round] ({\ta*\ta},{\ta}) -- ({\tb*\tb},{\tb});
  }
    \fill[black] (0.385,0.385) circle (1.6pt);

  \node[anchor=west,font=\footnotesize,inner sep=2pt] at (0.385,0.385) {$f_{\text{base}}$};
  \fill[figA] (0.15,0.62) circle (1.6pt);
  \node[anchor=west,font=\footnotesize,inner sep=2pt] at (0.15,0.62) {$f^*$};
  \node[anchor=north,font=\footnotesize] at (0,0) {$0$};
  \node[anchor=north,font=\footnotesize] at (1,0) {$1$};
  \node[anchor=east,font=\footnotesize] at (0,1) {$1$};
  \node[anchor=north,font=\footnotesize] at (0.5,-0.04) {$f(x_0)$};
  \node[anchor=south,rotate=90,font=\footnotesize] at (-0.05,0.5) {$f(x_1)$};
\end{tikzpicture}
\caption{Probabilistic}
\end{subfigure} \hfil
\begin{subfigure}{0.45\textwidth}\centering
\begin{tikzpicture}[unitsq,scale=1.0]
  \fill[gray!18] (0,0)--(0.7,0.7)--(0,0.7)--cycle;
  \draw[frame] (0,0) rectangle (1,1);
  \draw[black,thin] (0,0)--(0.7,0.7)--(0,0.7);
  \draw[thick,gray!55,densely dashed,smooth,samples=40,domain=0.7:1,variable=\t] plot ({\t*\t},{\t});
  \draw[curveM,figB,smooth,samples=50,domain=0:0.7,variable=\t] plot ({\t*\t},{\t});
  \node[font=\footnotesize] at (0.14,0.28) {$f_\theta$};
  \fill[black] (0.385,0.385) circle (1.6pt);
  \node[anchor=west,font=\footnotesize,inner sep=2pt] at (0.385,0.385) {$f_{\text{base}}$};
  \fill[figA] (0.15,0.62) circle (1.6pt);
  \node[anchor=west,font=\footnotesize,inner sep=2pt] at (0.15,0.62) {$f^*$};
  \fill[figB] (0.259,0.509) circle (1.6pt);
  \node[anchor=west,font=\footnotesize,inner sep=1pt] at (0.259,0.509) {$f_{\theta^*}$};
  \node[anchor=north,font=\footnotesize] at (0,0) {$0$};
  \node[anchor=north,font=\footnotesize] at (1,0) {$1$};
  \node[anchor=east,font=\footnotesize] at (0,1) {$1$};
  \node[anchor=north,font=\footnotesize] at (0.5,-0.04) {$f(x_0)$};
  \node[anchor=south,rotate=90,font=\footnotesize] at (-0.05,0.5) {$f(x_1)$};
\end{tikzpicture}
\caption{Fitted (degenerate prior)}
\end{subfigure} 
\caption{Completeness on the toy theory. $f^*=(0.15,0.62)$ marks the best possible
prediction. The baseline is $f_{\text{base}}=(0.385,0.385)$. (a) The probabilistic measure averages over the prior
$\pi_{M,d}$, a normal density centered near the best fit and truncated to the
supported range (shaded along the curve, darker where denser); the spread lengthens
the average gap between the theory's predictions and the best possible prediction.  (b) Fitted completeness is
the degenerate prior that places all mass on the closest rule on the curve,
$f_{\theta^*}$.}
\label{fig:completeness_toy}
\end{figure}
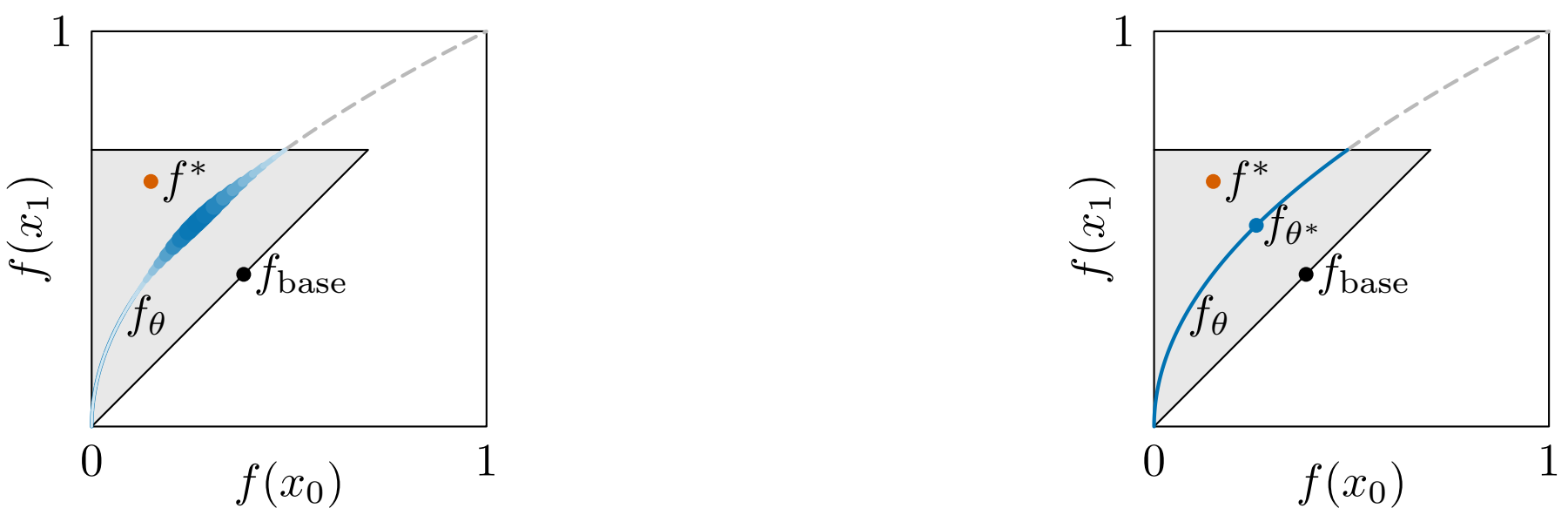

Consider a simple prediction problem with a binary covariate $x\in\{x_0,x_1\}$ and
an outcome $y\in[0,1]$, as illustrated in Figure \ref{fig:completeness_toy}. A prediction rule is a pair
$f=(f(x_0),f(x_1))=(y_0,y_1)\in[0,1]^2$, a point in the unit square. The domain
  is $d=(\mathcal X_d,\mathcal Y_d)$ with $\mathcal X_d=\{x_0,x_1\}$ and
  $\mathcal Y_d(x)=[0,1]$ at both covariate values, so that the set of admissible
  rules is $\mathcal F_d=[0,1]^2$. Let us assume
the underlying ground truth is $f^*(x_0)=0.15$ and $f^*(x_1)=0.62$. Concretely, the
outcome is generated as $Y=f^*(x)+\varepsilon$ with $\varepsilon$ mean-zero noise,
and the covariate is balanced, $\Pr(x_0)=\Pr(x_1)=\tfrac12$. The data are a large
sample of such pairs, enough that the conditional means are pinned down. The best
predictor is the conditional mean $f^*=(0.15,0.62)$, and the best covariate-free
predictor is the unconditional mean $f_{\text{base}}=(0.385,0.385)$, where
$0.385=\tfrac12(0.15)+\tfrac12(0.62)$.

Now suppose we have a theory with the three components described earlier.
First, a support restriction consisting of two intrinsic constraints on
prediction rules: monotonicity, $y_1 \ge y_0$, and a cap, $y_1 \le 0.7$,
yielding $\mathcal{S}_{M,d} = \{(y_0, y_1) : y_1 \ge y_0,\ y_1 \le 0.7\}$
(the shaded grey area in Figure~\ref{fig:completeness_toy}). Second, a
parameterization, the curve $f_\theta = (\theta^{\rho}, \theta)$ with
$\rho = 2$ and $\theta \in [0,1]$, the domain implied by the outcome range
(drawn as the curve in Figure~\ref{fig:completeness_toy}). 
The reachable set
$\mathcal{R}_{M,d}$ is the segment of this curve satisfying the support
restriction, which binds through the cap: $\theta \in [0, 0.7]$. Third, a
prior $\pi_{M,d}$ over the parameter, a normal density centered at
$\theta = 0.51$ with standard deviation $0.08$, truncated to $[0, 0.7]$.  The prior leaves the
reachable set unchanged and only reweights it, in this example placing most mass near the
best-fitting rule.

We evaluate the theory under both interpretations of its parameter distribution
(Section~\ref{sec:theories_pred_dist}): first under the representative-agent interpretation, in which the
prior expresses uncertainty about a single value of $\theta$, then under the heterogeneous-agent
interpretation, in which it describes how $\theta$ differs across agents.
\subsection{Completeness Under the Representative-Agent Interpretation}
Under the representative-agent interpretation a draw from the prior is scored as a prediction of the
rule, and the measure is probabilistic completeness
(Section~\ref{sec:measuring_completeness}). Probabilistic completeness is the ratio of the error reduction achieved by the
theory to the maximum achievable error reduction. In this example, under squared error loss with balanced covariates, the irreducible noise cancels in both numerator and denominator, so the ratio reduces to a
comparison of squared distances to the truth $f^*$ in prediction space. The
baseline error corresponds to $\|f_{\text{base}} - f^*\|^2 = (0.385-0.15)^2 + (0.385-0.62)^2 \approx 0.110$,
and the theory's probabilistic error corresponds to the prior-averaged squared
distance $\mathbb{E}_{\theta \sim \pi_{M,d}}\|f_\theta - f^*\|^2$, which we compute
numerically. Details are given in Appendix~\ref{app:toy_computation}.
Probabilistic completeness is $\kappa^{prob} \approx 0.66$.

Fitted completeness is the special case of a degenerate prior on the closest
rule on the curve, at $\theta^*\approx 0.51$. That rule's squared distance to $f^*$
is about $0.024$, so $\kappa^{fit} \approx  \frac{0.110 -0.024}{0.110-0}\approx 0.78$. Probabilistic completeness
coincides with it exactly when $\pi_{M,d}$ collapses to a point mass at the closest
rule, and stays close whenever the prior concentrates tightly around it. Here the
prior is concentrated but spread, so the two diverge modestly. 

\subsection{Restrictiveness Under the Representative-Agent Interpretation}

The reference space is the set of all
  admissible rules, $\mathcal F_d=[0,1]^2$, and the reference distribution
  $\Lambda_d$ is uniform on it. Functional-form restrictiveness is the expected
  squared distance from a reference rule $f'$ to the theory, as a share of the
  largest value this expectation can take for any theory. That maximum is
  attained by the four corners of the square, the rules farthest from its center
  $c=(0.5,0.5)$, and equals $V+\rho^2=\tfrac16+\tfrac12=\tfrac23$, where
  $V=\E\|f'-c\|^2=E(y_0-\tfrac12)^2+\E(y_1-\tfrac12)^2=2\cdot\mathrm{Var}(U[0,1])=2\cdot\tfrac1{12}=\tfrac16$ (the expected squared distance from the center of the reference space) and $\rho^2=\|(1,1)-c\|^2=(\tfrac12)^2+(\tfrac12)^2=\tfrac12$ (the squared distance of the center to the corner point). The center itself, the best
  uninformed guess, scores $\tfrac16\big/\tfrac23=\tfrac14$.

We isolate the three channels influencing probabilistic restrictiveness by layering them one at a time. 

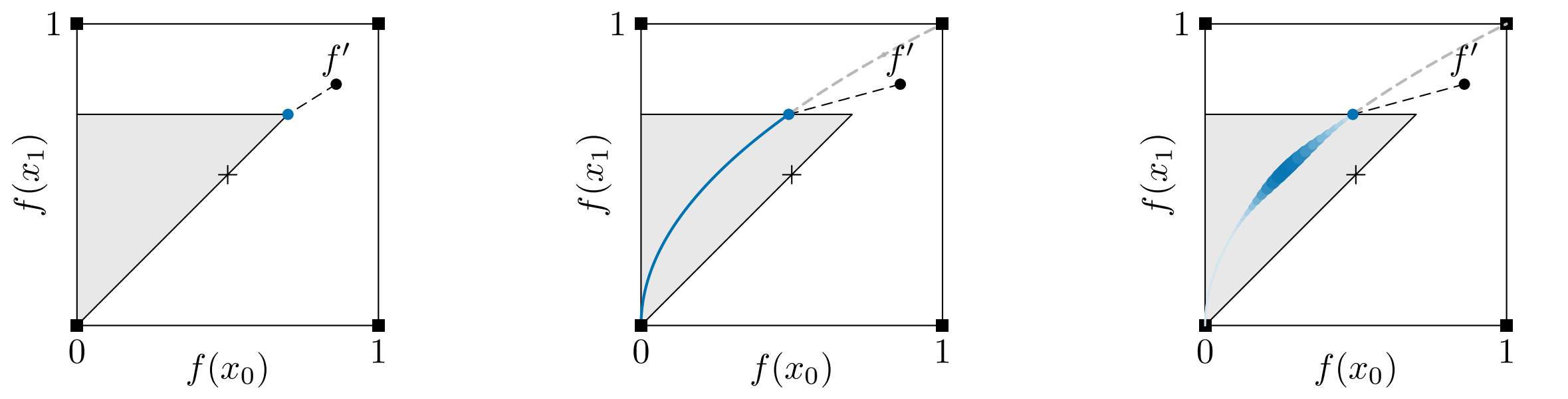
\begin{figure}[t]
\centering
\begin{subfigure}{0.32\textwidth}\centering
\begin{tikzpicture}[unitsq]

  \fill[gray!18] (0,0)--(0.7,0.7)--(0,0.7)--cycle;
  \draw[frame] (0,0) rectangle (1,1);
  \draw[black,thin] (0,0)--(0.7,0.7)--(0,0.7);
  \draw[distseg] (0.86,0.80) -- (0.70,0.70);
  \draw (0.47,0.5)--(0.53,0.5) (0.5,0.47)--(0.5,0.53);
\foreach \cx/\cy in {0/0,1/0,0/1,1/1}
{\fill (\cx-0.02,\cy-0.02) rectangle (\cx+0.02,\cy+0.02);}
  \fill[black] (0.86,0.80) circle (1.6pt);
  \fill[figB] (0.70,0.70) circle (1.6pt);
  \node[anchor=south,font=\footnotesize,inner sep=2pt] at (0.86,0.80) {$f'$};
  \node[anchor=north,font=\footnotesize] at (0,0) {$0$};
  \node[anchor=north,font=\footnotesize] at (1,0) {$1$};
  \node[anchor=east,font=\footnotesize] at (0,1) {$1$};
  \node[anchor=north,font=\footnotesize] at (0.5,-0.04) {$f(x_0)$};
  \node[anchor=south,rotate=90,font=\footnotesize] at (-0.05,0.5) {$f(x_1)$};
\end{tikzpicture}
\caption{Support restriction}
\end{subfigure}\hfill
\begin{subfigure}{0.32\textwidth}\centering
\begin{tikzpicture}[unitsq]
  \fill[gray!18] (0,0)--(0.7,0.7)--(0,0.7)--cycle;
  \draw[frame] (0,0) rectangle (1,1);
  \draw[black,thin] (0,0)--(0.7,0.7)--(0,0.7);
  \draw[thick,gray!55,densely dashed,smooth,samples=40,domain=0.7:1,variable=\t] plot ({\t*\t},{\t});
  \draw[curveM,figB,smooth,samples=50,domain=0:0.7,variable=\t] plot ({\t*\t},{\t});
  \fill[gray!60] (0.806,0.898) circle (0.7pt);
  \draw[distseg] (0.86,0.80) -- (0.49,0.70);
  \draw (0.47,0.5)--(0.53,0.5) (0.5,0.47)--(0.5,0.53);
\foreach \cx/\cy in {0/0,1/0,0/1,1/1}
{\fill (\cx-0.02,\cy-0.02) rectangle (\cx+0.02,\cy+0.02);}
  \fill[figB] (0.49,0.70) circle (1.6pt);
  \fill[black] (0.86,0.80) circle (1.6pt);
  \node[anchor=south,font=\footnotesize,inner sep=2pt] at (0.86,0.80) {$f'$};
  \node[anchor=north,font=\footnotesize] at (0,0) {$0$};
  \node[anchor=north,font=\footnotesize] at (1,0) {$1$};
  \node[anchor=east,font=\footnotesize] at (0,1) {$1$};
  \node[anchor=north,font=\footnotesize] at (0.5,-0.04) {$f(x_0)$};
  \node[anchor=south,rotate=90,font=\footnotesize] at (-0.05,0.5) {$f(x_1)$};
\end{tikzpicture}
\caption{Functional form}
\end{subfigure}\hfill
\begin{subfigure}{0.32\textwidth}\centering
\begin{tikzpicture}[unitsq]
  \fill[gray!18] (0,0)--(0.7,0.7)--(0,0.7)--cycle;
  \draw (0.47,0.5)--(0.53,0.5) (0.5,0.47)--(0.5,0.53);
\foreach \cx/\cy in {0/0,1/0,0/1,1/1}
{\fill (\cx-0.02,\cy-0.02) rectangle (\cx+0.02,\cy+0.02);}
  \draw[frame] (0,0) rectangle (1,1);
  \draw[black,thin] (0,0)--(0.7,0.7)--(0,0.7);
  \draw[thick,gray!55,densely dashed,smooth,samples=40,domain=0.7:1,variable=\t] plot ({\t*\t},{\t});
  \foreach \i in {0,...,33}{
    \pgfmathsetmacro{\ta}{0.7*\i/34}
    \pgfmathsetmacro{\tb}{0.7*(\i+1)/34}
    \pgfmathsetmacro{\dens}{exp(-((\ta-0.51)/0.08)^2/2)}
    \pgfmathsetmacro{\wid}{0.6+3.4*\dens}
    \pgfmathtruncatemacro{\gv}{18+80*\dens}
    \draw[figB!\gv,line width=\wid pt,line cap=round] ({\ta*\ta},{\ta}) -- ({\tb*\tb},{\tb});
  }
  \draw[distseg] (0.86,0.80) -- (0.49,0.70);
  \fill[figB] (0.49,0.70) circle (1.6pt);
  \fill[black] (0.86,0.80) circle (1.6pt);
  \node[anchor=south,font=\footnotesize,inner sep=2pt] at (0.86,0.80) {$f'$};
  \node[anchor=north,font=\footnotesize] at (0,0) {$0$};
  \node[anchor=north,font=\footnotesize] at (1,0) {$1$};
  \node[anchor=east,font=\footnotesize] at (0,1) {$1$};
  \node[anchor=north,font=\footnotesize] at (0.5,-0.04) {$f(x_0)$};
  \node[anchor=south,rotate=90,font=\footnotesize] at (-0.05,0.5) {$f(x_1)$};
\end{tikzpicture}
\caption{Parameter distribution}
\end{subfigure}

\vspace{0.6em}
\centering

\caption{Layering the three components of the toy theory. A single reference
rule $f'$, drawn from the uniform reference distribution $\Lambda_{d}$ on
the unit square, illustrates the ingredient of functional-form
restrictiveness: 
(a)~The support restriction confines predictions to the monotone triangle
capped at $y_1 \le 0.7$ (shaded); the nearest reachable point to $f'$ is the
corner where the diagonal meets the cap. (b)~Adding the functional form
confines predictions to the arc of the curve $(\theta^2, \theta)$ inside the
support; the segment of the curve with $\theta > 0.7$ (dashed grey) is
excluded, so the nearest reachable point is the arc's endpoint and the
distance is longer.  (c)~The prior concentrates mass along the in-support arc (darker where
denser) while leaving the reachable set unchanged; $r^{prior}$ measures this concentration relative to the uniform distribution on the arc.}
\label{fig:toy_restrictiveness}
\end{figure}

Consider first the support restriction alone, a purely qualitative theory,
represented by letting the parameter index admissible rules directly
($\Theta_M = \mathcal{S}_{M,d}$, $f_\theta = \theta$), so that the reachable set
is the triangle itself (Figure~\ref{fig:toy_restrictiveness}, panel~a). With a
uniform distribution over the triangle, $r^{prior} = 0$ and probabilistic
restrictiveness coincides with functional-form restrictiveness. Under the squared discrepancy it is the
  average squared distance of a reference rule to its nearest point in the
  triangle, $0.053$, divided by $\tfrac23$: $\rprob=\rff\approx0.08$
(calculation details in Appendix~\ref{app:toy_computation}).

We now add the functional form. It confines the theory further to the part of the
curve $f_\theta=(\theta^{2},\theta)$ lying within the support (Figure \ref{fig:toy_restrictiveness}, panel~b). No prior
is specified yet, so probabilistic restrictiveness again equals functional-form
restrictiveness. This is now the average squared distance of a reference rule to the nearest point
on the curve. The curve is one-dimensional, so reference rules lie farther away on
average than before, the average
  squared distance rises to $0.125$, so
  $\rprob=\rff\approx0.125\big/\tfrac23\approx0.19$.

Finally we add the parameter distribution, the truncated normal $\pi_{M,d}$
introduced above (Figure~\ref{fig:toy_restrictiveness}, panel~c, shown as shading
on the in-support arc). A parameter distribution that is uniform on the
  curve, by arc length, gives $\rprior=0$ and $\rprob=\rff$. One that is uniform
  in $\theta$ is not quite uniform on the curve, which stretches as $\theta$
  grows: it gives $\rprior\approx0.015$. Concentrating the prior near the best-fitting
rule raises $r^{prior}$ while leaving $r^{ff}$ unchanged, since the geometry of the
curve has not moved.

The divergence is measured between the theory's distribution over predictions
and the uniform distribution on the reachable arc. A parameter
density that is flat in $\theta$ is not uniform on the arc where the curve
stretches; the prior is therefore reweighted by the Jacobian $|J(\theta)|$
before the divergence is taken. This gives $r^{prior} \approx 0.49$, so that $r^{prob} = 1 - (1 - r^{ff})(1 - r^{prior}) \approx 0.58$.  The three layers take the theory from
  $0.08$ to $0.19$ to $0.58$. The derivation is given in Appendix~\ref{app:toy_computation}.

\subsection{Completeness Under the Heterogeneous-Agent Interpretation}
\label{sec:toy_distributional}

Under the heterogeneous-agent interpretation, the parameter distribution of the toy theory says that each agent has
a parameter $\theta$ drawn from $\pi_{M,d}$, so that the population of
outcomes lies on the curve. The truth is
then a population too. Let the agents' outcomes at the two covariate values
be drawn independently around $f^*$, with means $0.15$ and $0.62$ and
standard deviations $0.08$ and $0.10$ (Beta distributions;
Figure~\ref{fig:toy_distributional}a). For the probabilistic measures this changes
nothing: the conditional means are still $f^*$, and the dispersion of the
agents around them is irreducible error. For the distributional measures
(Sections~\ref{sec:population_reading} and~\ref{sec:population_restrictiveness}), computed here with
the CRPS, the dispersion is part of what a theory is evaluated on.

\input{figures/fig_toy_distributional}
Figures~\ref{fig:toy_distributional}b and c show, at $x_0$ and $x_1$, the distribution functions of the theory's predictive distribution, the push-forward of the prior through $f_\theta$ (dashed), next to those of the population (solid) and of the reference (the diagonal).
Distributional completeness compares the squared area between the theory's
and the population's curves with that between the diagonal and the
population's: $\kappa^{dist}\approx0.62$ for the prior of
the setup. The uniform prior on $[0,0.7]$ scores $0.41$,
and the fitted theory, the point at the closest rule $\theta^*=0.51$, scores $0.34$: under
the representative-agent interpretation it was the most complete ($\kappa^{fit}\approx0.78$),
under the heterogeneous-agent interpretation it is the least, because a point cannot cover agents who
differ. A prior that spreads along the curve covers them at $x_1$ but not at
$x_0$, where the curve's values $\theta^2$ sit to the right of most agents.

\subsection{Restrictiveness Under the Heterogeneous-Agent Interpretation}

Distributional restrictiveness is the squared
area between a theory's curves and the diagonal, relative to the farthest
rule's. The population itself is at $r^{*}\approx0.33$, the value a theory
would have if it reproduced the agents exactly; the prior of
the setup is at $0.22$, the uniform prior at $0.25$ and
the fitted theory at $0.34$. The two priors are less restrictive than
the population, because their curves sit nearer the diagonal than the
agents' do, and the bound of Proposition~\ref{prop:rdist_bound} is loose for
them ($0.97$ and $0.98$). The fitted theory is as restrictive as the population, so
the bound allows it to be complete, and its completeness of $0.34$ shows
that the bound is a ceiling.


\section{Application to Certainty Equivalents for Binary Lotteries}
\label{sec:ex1}
\label{sec:application_games}
We illustrate the framework by evaluating theories of risk preferences using
certainty equivalent data from \citet{https://doi.org/10.3982/ECTA7139}. This
application revisits the setting studied by \citet{completeness2022},
\citet{restrictiveness2026} and \citet{andrews2025transfer}, and extends it to the framework
developed in this work. We begin with a single domain, the Zurich 2003
experiment, so that the scope of all four theories is $\mathcal D_M=\{d\}$ and
predictive content reduces to the domain-level completeness and restrictiveness
pair $(\kappa^{prob}(M,d),\, r^{prob}(M,d))$. We report the theories under the
representative-agent interpretation (Section~\ref{sec:application_results}), under the
heterogeneous-agent interpretation (Section~\ref{sec:learn-crps}), and at the level of the individual
respondent (Appendix~\ref{app:individual}).
Table~\ref{tab:evaluation-map} lists what each interpretation scores and where the results
are reported; in brief, the results are these.

\label{sec:results_together}

CPT is the most complete theory under both interpretations in the population setting and in the individual setting (Appendix~\ref{app:individual}), with the rung fitted based on the respective loss function reaching the highest completeness level. Under the heterogeneous-agent interpretation the two held priors of CPT reach the same completeness from
different positions. The uniform prior is about as dispersed as the respondents but less well
directed; the types prior, the literature's claim about heterogeneity, is better directed but
commits more than the respondents do (Figure~\ref{fig:crps-plane}). The fitted distribution is
what matching both takes: it is as dispersed as the respondents and aligned with them.

DA and EU are separated from CPT more sharply at each step. Fitted, they describe the average
respondent moderately well. Their distributions describe the population poorly, since their forms
cannot produce its dispersion. And for the individual respondent they do worse than a predictor
that knows nothing about the person.

\begin{table}[htbp]
\centering
\caption{Map of the evaluation: what each interpretation scores, and where the results are reported.}
\label{tab:evaluation-map}
\small\setlength{\tabcolsep}{4pt}
\begin{tabular}{p{2.9cm} p{3.0cm} p{1.5cm} p{2.2cm} p{2.3cm} p{2.3cm}}
\toprule
 & The theory's claim & Loss & Baseline & Best predictor & Reported in \\
\midrule
Representative-agent interpretation & one rule for the average respondent, with uncertainty about its parameter & squared error & midpoint of the prize range & lookup table of the lottery means & Tables~\ref{tab:lotteries_completeness} and~\ref{tab:lotteries_restrictiveness}, Figure~\ref{fig:lotteries_pareto} \\
\addlinespace
Heterogeneous-agent interpretation & a distribution of rules across the respondents & CRPS & uniform distribution on the prize range & empirical distribution of the certainty equivalents & Table~\ref{tab:lotteries_crps_fit}, Figures~\ref{fig:crps-fit} and~\ref{fig:crps-plane} \\
\bottomrule
\end{tabular}
\end{table}

\subsection{Setting}

The data consist of reported certainty equivalents for 25 binary lotteries over
gains, elicited from 179 subjects in the Zurich 2003 experiment of
\citet{https://doi.org/10.3982/ECTA7139}; the lotteries and the data are documented in that paper
and its replication package. 
Each lottery $x=(\bar z,\underline z,p)$ is
characterized by two prizes $\bar z>\underline z\ge0$ and the probability $p$ of
the larger prize. The outcome $Y$ is the certainty equivalent a subject reports,
which the choice list confines to the prize range,
$\mathcal Y_d(x)=[\underline z,\bar z]$. A prediction rule assigns to each
lottery a certainty equivalent within its prize range, as in the running
example.

\begin{tcolorbox}[
  breakable, enhanced, colback=gray!4, colframe=black!60,
  title=\textbf{Example: Lottery 8 from \citet{https://doi.org/10.3982/ECTA7139}},
  fonttitle=\bfseries, boxrule=0.5pt, arc=2pt,
  left=8pt, right=8pt, top=6pt, bottom=6pt,
]
\textbf{Lottery.} Each subject in the experiment is asked to value a
\emph{binary lottery}: a gamble that pays the larger prize $\bar z = 20$
Swiss francs with probability $p = 0.95$, or the smaller prize
$\underline z = 0$ otherwise. The lottery's expected monetary payout is
$p\,\bar z + (1-p)\,\underline z = 19.00$ CHF. The baseline prediction is the prize-range midpoint,
$f_{\rm mid}(x) = (\underline z + \bar z)/2 = 10.00$ CHF, which presumes no
model of behavior. The reported \emph{certainty equivalent} (CE) of lottery
$x$ for subject $i$ is the amount of certain money the subject says they
would accept in lieu of playing the lottery. The 179 subjects report a
sample mean of $15.16$ and standard deviation $4.26$ (range $[1, 20]$). The
lookup-table estimate of $f^*(x)$ is $15.16$ (below the expected value,
reflecting risk aversion, and above the behavior-agnostic midpoint).
\end{tcolorbox}

\subsection{Theories}
\label{sec:ex1_theories}
We evaluate four theories of decision-making under risk: Cumulative Prospect
Theory (CPT), Disappointment Aversion (DA), Expected Utility (EU) and Expected
Value (EV).

\subsubsection{Functional Forms}
 All theories share the value function $v(z)=z^\alpha$.
\paragraph{Cumulative Prospect Theory.}
$f_\theta(x)=v^{-1}\bigl(w(p)\,v(\bar z)+[1-w(p)]\,v(\underline z)\bigr)$ with
$w(p)=\delta p^\gamma/(\delta p^\gamma+(1-p)^\gamma)$ and
$\theta=(\alpha,\gamma,\delta)\in[0.01,1]^2\times[0.1,3]$. $\alpha$ governs
diminishing marginal utility, $\gamma$ the curvature of probability weighting,
and $\delta$ its elevation.

\paragraph{Disappointment Aversion.} Following Gul (1991) in the specification of
Routledge and Zin (2010),
$f_\lambda(x)=v^{-1}\bigl(\tilde w(p)\,v(\bar z)+[1-\tilde w(p)]\,v(\underline z)\bigr)$
with $\tilde w(p)=p/(1+(1-p)\eta)$ and
$\lambda=(\alpha,\eta)\in[0.01,1]\times[-1,3]$. The weight $\tilde w$ belongs to
the better prize, $\eta>0$ underweights it (disappointment aversion), and
$\eta=0$ gives expected utility.

\paragraph{Expected Utility} is the common special case $\gamma=\delta=1$ of CPT and
$\eta=0$ of DA, $f_\alpha(x)=v^{-1}\bigl(p\,v(\bar z)+(1-p)\,v(\underline z)\bigr)$
with $\alpha\in[0.01,1]$.

\paragraph{Expected Value} is the member of EU  with $\alpha=1$ and has no free parameter, $f(x)=p\bar z+(1-p)\underline z$.

The given parameter boxes are those reported in the literature. $\alpha\le1$ encodes diminishing
sensitivity which corresponds to the box of \citet{restrictiveness2026}. Each form is defined on
a larger domain, with the curvature free up to $20$. The reachable set of every rung is the image of
that domain and defines the support of the
uniform prior and the truncation of the literature priors. The lower bounds on $\alpha$, $\gamma$ and $\delta$
keep the certainty-equivalent map well-conditioned near the edge of the
parameter space, the upper bounds on $\delta$ and $\eta$ make the domain
bounded. Appendix~\ref{app:parameter_box} reports the sensitivity of all measures to
each of these edges. For the fitted version of a theory, $\theta^*$ minimizes
the mean squared error on the training folds over the form's domain.

\subsubsection{Support restrictions}
\label{sec:ex1_support_restrictions}
For CPT, DA and EU the support restriction $\SM$ requires that prediction rules
respect first-order stochastic dominance (FOSD) (Hadar and Russell, 1969; Hanoch
and Levy, 1969). This means, $f(A)\ge f(B)$ whenever lottery $A$ dominates lottery $B$. That
certainty equivalents lie within the prize range,
$\underline z\le f(x)\le\bar z$, is something we consider part of the domain rather than
of any theory.  On the forms' domains, dominance is implied by the functional
forms of CPT, DA and EU. The restriction does not bind, and the reachable set
of each family is the image of its domain.

Dominance is not imposed on the reference distribution (Section~\ref{sec:ex1_reference_distribution}), so what it
excludes counts as content of the theories. To measure what this assertion alone is worth, we
add a fifth, purely qualitative theory, which we refer to as \emph{coherence only}. Its reachable
set is the set of all admissible rules that respect dominance,
$\mathcal R_{coh}=\SM$ (in the notation of Section \ref{sec:theories_pred_dist}, $\Theta=\SM$ and
$f_\theta=\theta$), and its parameter distribution is uniform on that set. It
nests CPT, DA, EU and EV, so by Proposition~\ref{prop:nest} it bounds their
functional-form restrictiveness from below.

 \subsubsection{Parameter Priors}
 \label{sec:ex1_parameter_distributions}
\paragraph{Uniform prior.} For each theory the \emph{uniform prior} places equal weight on the parameter box of
the functional forms above. The uniform prior serves as a diagnostic benchmark, reflecting the predictive content of the functional form, support restrictions and the set of admissible parameters alone. Comparing it to the literature prior thus gives an estimate of what the literature's parameter estimates add. $\alpha,\gamma\sim\mathrm{Uniform}[0.01,1]$ and
$\delta\sim\mathrm{Uniform}[0.1,3]$ for CPT;
$\alpha\sim\mathrm{Uniform}[0.01,1]$ and $\eta\sim\mathrm{Uniform}[-1,3]$ for DA;
$\alpha\sim\mathrm{Uniform}[0.01,1]$ for EU. It is held under both interpretations: agnosticism about the
parameter of a representative agent, or agents spread evenly over the box.

\paragraph{Literature prior (representative-agent interpretation).} The \emph{literature priors} are based on
estimates reported in the existing literature. They express uncertainty about the common parameter of a
sample, which is why it only makes sense to evaluate them under the representative-agent interpretation.

For CPT we center each parameter on
gains-specific estimates from the meta-analysis of
\citet{imai_nunnari_wu_vieider_2025}\footnote{The meta-analysis pools 812 estimates from 166 papers, among them \citet{https://doi.org/10.3982/ECTA7139}. The
literature prior is therefore not entirely independent of the data on which we evaluate it, although the
weight of a single study in the meta-analytic means is small.} and choose a concentration that
reflects how far the estimates differ across the studies and populations the
meta-analysis collects. The prior thus reflects the uncertainty an analyst has about the
common parameter of a new sample:
\[
\alpha \sim \text{Beta}(7, 3), \qquad
\gamma \sim \text{Beta}(4, 1.5), \qquad
\delta \sim \text{LogNormal}(0.05,\, 0.25^2).
\]
The informative prior for DA is calibrated to estimates from
\citet{choi2007}, \citet{harrison2023}, and \citet{abdellaoui2007},
weighting gains-specific evidence more heavily. We adopt
\[
\alpha \sim \text{Beta}(3.5, 2.5), \qquad
\eta \sim \text{TruncNormal}(0.30,\, 0.45^2;\, {-}1, 3).
\]

EU is the special case of CPT with $\gamma = \delta = 1$, so its
single free parameter $\alpha$ shares the same interpretation and
supporting evidence as CPT's utility-curvature parameter. For
structural consistency across the theory ladder, we adopt the
identical marginal:
\[
  \alpha \sim \text{Beta}(7,\, 3),
\]

Each informative marginal is truncated to the parameter box and
renormalized; the truncated mass is below $2\times10^{-5}$ for every marginal
of the main specification, so the box cuts nothing off the priors, and what
it contributes to their restrictiveness is the share of the reachable set's
volume it keeps (Section~\ref{sec:results_restrictiveness},
Appendix~\ref{app:parameter_box}).
The calibration of these priors and the sensitivity of completeness
and restrictiveness to alternative specifications are detailed in
Appendix~\ref{app:priors}.

\paragraph{Types prior (heterogeneous-agent interpretation).} Under the heterogeneous-agent interpretation a prior is a
claim about how parameters differ across the respondents of a sample. To build a literature informed prior, we thus need data on how individuals' preferences are distributed across the outcome range. 
\citet{https://doi.org/10.3982/ECTA7139} classify the respondents of each of their samples into
two types, about one fifth who weight probabilities almost linearly (corresponding to expected utility) and four fifths who distort
them, and report each type's share and parameters. We take the types of their other two samples, Zurich 2006 and Beijing 2005, as point masses with their shares, and weight the two samples equally. On the gains domain the published types are, for Zurich 2006, shares $0.22$ and $0.78$ at $(\alpha,\gamma,\delta)=(0.99,0.95,0.91)$ and $(0.90,0.43,0.86)$, and for Beijing 2005 shares $0.20$ and $0.80$ at $(1.08,0.91,0.89)$ and $(0.38,0.24,1.34)$. We refer to this object as the \emph{types prior}.
DA and EU have no published types. For them we estimate two types per sample on the same two samples with the finite mixture of \citet{https://doi.org/10.3982/ECTA7139} on the parameter grid and build the prior in the same way.

\subsection{Evaluation}
\label{sec:ex1_reference_distribution}
\label{sec:ex1_evaluation}
We evaluate every theory under the two interpretations of
Section~\ref{sec:levels_readings}: against the average respondent and against the distribution of
respondents. Rungs chosen with the data are fitted on the
training folds and scored on the held-out fold.

\paragraph{Representative-agent interpretation.}
Following \citet{completeness2022}, loss is squared error, and the discrepancy
between two prediction rules is the mean squared distance across lotteries,
that is, the squared discrepancy of Section~\ref{sec:est_restrictiveness} with equal weights.

The baseline is the midpoint of each lottery's prize range,
$f_{base}(x)=(\underline z+\bar z)/2$, corresponding to the best possible prediction of someone who knows the
admissible range of each certainty equivalent but nothing about behavior. \citet{restrictiveness2026} use the expected value
as their preferred baseline while noting (p.~11) that a less informed benchmark
is an even more theory-agnostic alternative. We interpret the expected value itself as a theory of choice under risk, and include it in our evaluation. The best achievable predictor $f^*$ is estimated by the lookup table of Section~\ref{sec:est_preliminaries}, the mean certainty equivalent of each lottery. The fitted rung under this interpretation is the
\emph{fitted theory}, the point mass at the best-fitting parameter value
(Section~\ref{sec:measuring_completeness}), the object evaluated by \citet{completeness2022,
restrictiveness2026}.

\paragraph{Heterogeneous-agent interpretation.}
Distributional completeness and restrictiveness are computed with the CRPS
(Section~\ref{sec:est_crps}). The predictive distribution of a rung at a lottery is its parameter
distribution pushed through the functional form, with no noise added, so that every spread is a
spread of parameter values (Appendix~\ref{app:dist_noise} repeats the evaluation with the response noise of the individual level added to every rung). The baseline is the reference distribution at the lottery, the uniform
distribution on the prize range, whose center is the midpoint. The best achievable predictor is
the empirical distribution of the certainty equivalents at the lottery. The fitted rung under this
interpretation is the \emph{fitted distribution}, the parameter distribution whose predictive distributions
come closest to the outcomes (Section~\ref{sec:population_reading}); the fitted theory is scored
under this interpretation as well, as the claim that every respondent follows one rule.

\paragraph{Reference distribution.}
Functional-form restrictiveness compares each theory's reachable set with a
reference distribution $\Lambda_d$ over all admissible certainty-equivalent
mappings. Following Section~\ref{sec:estimating_ff_ref}, the reference value of each of the 25
lotteries is drawn independently and uniformly on that lottery's own prize range
$[\underline z,\bar z]$, so that the center of $\Lambda_d$ is the midpoint rule,
the baseline of the representative-agent interpretation. $\rff$ is normalized by the expected discrepancy of
the farthest admissible rule (Section~\ref{sec:measuring_restrictiveness}), which for this reference is any rule
that predicts one of the two prizes for every lottery, such as ``the certainty
equivalent equals the better prize''. The prior channel  compares each theory's predictive distribution with the uniform
distribution on the reachable set of the form's domain, the same reference
for every rung of a family (Section~\ref{sec:est_restrictiveness}).

Appendix~\ref{app:kernels} reports
smooth Gaussian-process references as a robustness check, and Appendix~\ref{app:relation_fudenberg}
connects our numbers to the ones published by \citet{completeness2022} and \citet{restrictiveness2026}.

 \textbf{Implementation.} Completeness uses $K=10$ folds that are split by
subject and $B=1{,}000$ parameter draws; its standard errors come from
$1{,}000$ bootstrap replicates that resample subjects, with the parameter draws
held fixed and, for the fitted theories, the fit repeated in every replicate.
Functional-form restrictiveness uses $S=1{,}000$ reference functions. Prior
restrictiveness uses $B_H=50{,}000$ draws from the parameter distribution for
the entropy and $B_U=10^6$ uniform draws for the volume. Under the heterogeneous-agent interpretation
the folds and draws are the same, and the fitted distribution lives on a grid of parameter values
($20^3$ cells for CPT, $40^2$ for DA, $200$ for EU). \subsection{Results Under the Representative-Agent Interpretation}
\label{sec:application_results}

\subsubsection{Completeness}
\label{sec:results_completeness}

Table~\ref{tab:lotteries_completeness} reports completeness for CPT, DA, EU
and EV evaluated against the per-lottery midpoint baseline. Held-out errors come from 10-fold subject-holdout cross-validation. Standard errors cluster on the 179 subjects via a block bootstrap with 1{,}000 replicates. For fitted rows the bootstrap re-fits $\theta^*$ per replicate. For probabilistic rows the prior draws $\theta^{(1)}, \dots, \theta^{(B)}$ are held fixed across replicates and only the resampled outcomes vary. Each of CPT, DA and EU appears on three rungs (Section~\ref{sec:theories_pred_dist}): a fitted
point estimate ($\theta^*$, fitted on the form's domain; corresponding to what is evaluated in \citet{completeness2022}), and two probabilistic evaluations in which the
parameter is uncertain and a rule is drawn from the prior, once the uniform prior
on the parameter box and once the literature-calibrated prior. EV has no free parameters and appears in a
single row.
Under the fitted rung the parameter distribution collapses to a point mass at $\theta^*$,
so $\kappa^{fit} = \kappa^{prob}$ by construction.

\paragraph{Fitted Theories}
The fitted rows report each theory at its MLE parameter $\theta^*$,
refit on each training fold. CPT is highly complete, attaining $\kappa =
\res{gains}{CPT_deterministic}{kappa}$ and closing almost all of the distance between the midpoint baseline
and the irreducible-noise floor. Substituting EV for the midpoint as the baseline
reproduces the completeness values reported by 
\citet{completeness2022}(see Appendix~\ref{app:relation_fudenberg} for a more detailed comparison on the Zurich 2003 examples).
The other three fitted theories are less complete: DA
attains $\kappa = \res{gains}{DA_deterministic}{kappa}$, EU  $\kappa = \res{gains}{EU_deterministic}{kappa}$, and EV
$\kappa = \res{gains}{EV_deterministic}{kappa}$.

\paragraph{Uniform Priors}
Under a uniform prior on each theory's parameter
box, all three families score worse than the baseline. CPT
$\kprob=\res{gains}{CPT_uniform}{kappa}$, DA $\res{gains}{DA_uniform}{kappa}$,
EU $\res{gains}{EU_uniform}{kappa}$. A parameter vector drawn at random from a
theory's admissible space thus forecasts certainty equivalents that are predicting worse than
discarding the theory and predicting the na\"ive midpoint. EU sits highest
because its one-dimensional parameter space leaves little room for a
uniform prior to spread predictions, while, in this setup, CPT and DA average over larger boxes. Proposition~\ref{cor:bounds} locates the gap: the mean prediction of the uniform prior is
$\res{gains}{CPT_uniform}{kappa_prior_mean}$ complete for CPT, $\res{gains}{DA_uniform}{kappa_prior_mean}$ for DA and
$\res{gains}{EU_uniform}{kappa_prior_mean}$ for EU, so the gap to the fitted value reflects the dispersion of a family's predictions, not their location.
Literature-calibrated priors recover two thirds of fitted-theory performance
for CPT, most of it for EU and none of it for DA. For CPT the
literature prior yields $\kprob=\res{gains}{CPT_lit}{kappa}$, \res{gains}{CPT}{share_lit}\% of
fitted CPT's $\res{gains}{CPT_deterministic}{kappa}$. EU attains
$\res{gains}{EU_lit}{kappa}$, \res{gains}{EU}{share_lit}\% of its fitted value
$\res{gains}{EU_deterministic}{kappa}$. For DA the literature prior yields
$\kprob=\res{gains}{DA_lit}{kappa}$ (SE $\res{gains}{DA_lit}{kappa_se}$),
\res{gains}{DA}{share_lit}\% of its fitted value at the baseline.

\paragraph{The Role of Learning from Data}
The gap
$\Delta_{\text{learning}} \equiv \kappa^{fit} - \kappa^{prob}$
quantifies how much of a theory's completeness is based on the fitting sample, as opposed to structural.
For CPT under the literature prior,
$\Delta_{learning}=\res{gains}{CPT_deterministic}{kappa}-\res{gains}{CPT_lit}{kappa}=\res{gains}{CPT}{gap_lit}$.
A third of CPT's completeness is thus attributable to fitting rather than to
structure. Under a uniform prior the gap widens to
$\res{gains}{CPT}{gap_uniform}$. DA and EU show the same pattern with
$\Delta_{learning}=\res{gains}{DA}{gap_lit}$ and $\res{gains}{EU}{gap_lit}$ under the literature prior.On this design most of CPT's and EU's predictive power is structural, while DA's is almost entirely learned from the data, its literature prior sits at the baseline.

\begin{table}[htbp]
\centering
\caption{Completeness: Fitted vs.\ Probabilistic Evaluation}
\label{tab:lotteries_completeness}
\resizebox{\ifdim\width>\linewidth\linewidth\else\width\fi}{!}{
\begin{threeparttable}
\begin{tabular}{llcc}
\toprule
Model & Method & $\kappa^{prob}$ & MSE \\
\midrule
Coherence & Probabilistic: Uniform on $\mathcal{F}_{\mathrm{coh}}$ & -1.434 & 236.96 \\
  &  & (0.109) &  \\
\addlinespace
CPT & Fitted ($\theta^*$) & 0.972 & 63.70 \\
  &  & (0.008) &  \\
 & Probabilistic: Uniform $\pi_M$ & -0.495 & 169.35 \\
  &  & (0.129) &  \\
 & Probabilistic: Lit.\ $\pi_M$ & 0.641 & 87.56 \\
  &  & (0.050) &  \\
\addlinespace
DA & Fitted ($\theta^*$) & 0.626 & 88.64 \\
  &  & (0.062) &  \\
 & Probabilistic: Uniform $\pi_M$ & -0.827 & 193.20 \\
  &  & (0.212) &  \\
 & Probabilistic: Lit.\ $\pi_M$ & 0.020 & 132.27 \\
  &  & (0.136) &  \\
\addlinespace
EU & Fitted ($\alpha^*$) & 0.616 & 89.31 \\
  &  & (0.061) &  \\
 & Probabilistic: Uniform $\pi_M$ & -0.209 & 148.75 \\
  &  & (0.154) &  \\
 & Probabilistic: Lit.\ $\pi_M$ & 0.521 & 96.17 \\
  &  & (0.074) &  \\
\addlinespace
EV & Fitted (no free parameters) & 0.493 & 98.19 \\
  &  & (0.064) &  \\
\addlinespace
\multicolumn{3}{l}{\textit{Reference predictors}} \\
  & Midpoint baseline: $(\bar z + \underline z)/2$ & --- & 133.69 \\
  & Lookup table (irreducible-noise floor) & --- & 61.70 \\
\bottomrule
\end{tabular}
\begin{tablenotes}[flushleft]
\small
\item \textit{Notes:} Completeness evaluated against the per-lottery midpoint baseline. Probabilistic rows score a draw from the parameter distribution (the average loss over the distribution). Standard errors (in parentheses) cluster on the 179 subjects via a block bootstrap with 1{,}000 replicates. For the fitted rows the fitted $\theta^*$ is refit per replicate (each replicate re-runs the full 10-fold CV on the resampled subjects). Data: Bruhin et al.\ (2010), $n = 4{,}459$ observations.
\end{tablenotes}
\end{threeparttable}%
}
\end{table}

\subsubsection{Restrictiveness}
\label{sec:results_restrictiveness}
\begin{table}[htbp]
\centering
\caption{Restrictiveness under probabilistic evaluation.}
\label{tab:lotteries_restrictiveness}
\resizebox{\ifdim\width>\linewidth\linewidth\else\width\fi}{!}{
\begin{threeparttable}
\begin{tabular}{llccc}
\toprule
Model & Method & $r^{\mathrm{ff}}$ & $r^{\mathrm{prior}}$ & $r^{\mathrm{prob}}$ \\
\midrule
Coherence & Probabilistic: Uniform on $\mathcal{F}_{\mathrm{coh}}$ & 0.046 & 0.000 & 0.046 \\
  &                              &  &  & (0.001) \\
\addlinespace
CPT & Fitted ($\theta^*$) & 0.140 & 1.000 & 1.000 \\
 & Probabilistic: Uniform $\pi_M$ & 0.140 & 0.515 & 0.583 \\
  &                              &  &  & (0.003) \\
 & Probabilistic: Lit.\ $\pi_M$ & 0.140 & 0.949 & 0.956 \\
  &                              &  &  & (0.000) \\
\addlinespace
DA & Fitted ($\theta^*$) & 0.281 & 1.000 & 1.000 \\
 & Probabilistic: Uniform $\pi_M$ & 0.281 & 0.867 & 0.904 \\
  &                              &  &  & (0.001) \\
 & Probabilistic: Lit.\ $\pi_M$ & 0.281 & 0.938 & 0.956 \\
  &                              &  &  & (0.000) \\
\addlinespace
EU & Fitted ($\alpha^*$) & 0.336 & 1.000 & 1.000 \\
 & Probabilistic: Uniform $\pi_M$ & 0.336 & 0.578 & 0.720 \\
  &                              &  &  & (0.002) \\
 & Probabilistic: Lit.\ $\pi_M$ & 0.336 & 0.814 & 0.877 \\
  &                              &  &  & (0.001) \\
\addlinespace
EV & Fitted (no free parameters) & 0.484 & 1.000 & 1.000 \\
\bottomrule
\end{tabular}
\begin{tablenotes}[flushleft]
\small
\item \textit{Notes:} $r^{\mathrm{ff}}$ is the reach of the functional form on its own parameter domain, the same for every rung of a family; $r^{\mathrm{prior}}$ is the divergence of the rung's parameter distribution from the uniform distribution on that reachable set, so the parameter box of the literature, on which the uniform and the literature prior live, is priced there. A fitted rung is a point prediction, so $r^{\mathrm{prior}} = r^{\mathrm{prob}} = 1$. EV is a rule without parameters. MC standard errors (in parentheses under $r^{\mathrm{prob}}$) combine the simulation error of $r^{\mathrm{ff}}$ (delta method, $N_\Lambda = 1{,}000$ reference draws) and of $r^{\mathrm{prior}}$ (50{,}000 prior draws; 1{,}000{,}000 uniform draws for the manifold volume).
\end{tablenotes}
\end{threeparttable}%
}
\end{table}

Table~\ref{tab:lotteries_restrictiveness} reports, for each theory, the
functional-form restrictiveness $r^{\rm ff}$, the prior restrictiveness $r^{\rm prior}$ and the joint probabilistic
restrictiveness $r^{\rm prob}$ under the uniform and literature priors. The
fitted rungs and EV appear with $\rprior=\rprob=1$. In our setup, a fitted theory, i.e. a point-mass parameter
distribution is maximally restrictive whatever the form.

Appendix~\ref{app:squared_error_intervals} works through one reference draw.

\paragraph{Functional Form}
The ordering of $r^{ff}$ follows the nesting of the reachable sets
(Proposition~\ref{prop:nest}). The dominance-respecting set contains the
reachable sets of CPT and DA, each of which contains EU's, which contains the
single expected-value rule. Correspondingly
$\rff=\res{gains}{COH_uniform}{r_ff}$ for coherence only,
$\res{gains}{CPT_uniform}{r_ff}$ for CPT and $\res{gains}{DA_uniform}{r_ff}$ for
DA, $\res{gains}{EU_uniform}{r_ff}$ for EU and
$\res{gains}{EV_deterministic}{r_ff}$ for EV. Thus, dominance alone excludes only very little. Although none of the $S=1{,}000$ reference draws respects dominance, every reference rule is nonetheless close to a dominance-respecting rule.
The reaches are those of the forms on their domains; held to the literature's
box.

While the ordering coincides with the number of free parameters, the number of parameters does not determine the level of $r^{ff}$.
The expected value and the midpoint rule both have no parameters, but nonetheless differ by a factor of \res{anat}{steps}{EV_over_midpoint} (\res{gains}{EV_deterministic}{r_ff} against
0.250).\footnote{\citet{restrictiveness2026} (Section~6.4) make this point with
sub-specifications of CPT and DA, whose restrictiveness ranges from $0.28$ to
$0.92$ and differs widely among specifications with the same number of
parameters.}

\paragraph{Adding Priors}
The parameter distribution adds restrictiveness on
top of the functional-form floor. A parameter distribution that induces the
uniform distribution on the \emph{reachable set} would have $\rprior=0$. The uniform prior on the \emph{parameter box} does not have $\rprior=0$, for two
reasons: (1) the box keeps only part of the reachable set's volume,
\res{boxfit}{CPT_gains__gain}{box_volume_share_pct} percent for CPT,
\res{boxfit}{DA_gains__gain}{box_volume_share_pct} for DA and
\res{boxfit}{EU_gains__gain}{box_volume_share_pct} for EU
(Appendix~\ref{app:parameter_box}), and (2) the parameterization stretches it unevenly. We find
$\rprior=\res{gains}{CPT_uniform}{r_prior}$ for CPT,
$\res{gains}{EU_uniform}{r_prior}$ for EU and
$\res{gains}{DA_uniform}{r_prior}$ for DA.
Under the aggregation convention proposed in Section \ref{sec:est_restrictiveness}, calibration raises restrictiveness substantially. It lifts CPT from
$\rprob=\res{gains}{CPT_uniform}{r_prob}$ to $\res{gains}{CPT_lit}{r_prob}$, EU
from $\res{gains}{EU_uniform}{r_prob}$ to $\res{gains}{EU_lit}{r_prob}$, and DA
from $\res{gains}{DA_uniform}{r_prob}$ to $\res{gains}{DA_lit}{r_prob}$.
Measured against the functional-form floor, the literature prior adds
$\res{gains}{CPT}{prior_contrib_lit}$ for CPT,
$\res{gains}{DA}{prior_contrib_lit}$ for DA and
$\res{gains}{EU}{prior_contrib_lit}$ for EU. Thus, under the literature prior, most of the remainder $r^{ff}$ is removed by the prior.

The ranking of families therefore depends on the channel. On functional form
alone EU is the most restrictive of the three and CPT the least; on the
parameter distribution EU-Lit is the least restrictive of the literature rungs,
and among the uniform rungs CPT's is. Rung for rung, the only comparisons of EU
with CPT or DA that are channel-wise are EU's two rungs above CPT's uniform
rung; the others depend on how the channels are combined. Under the product
rule, with literature priors DA ($\res{gains}{DA_lit}{r_prob}$) and CPT
($\res{gains}{CPT_lit}{r_prob}$) are level, and both are above EU
($\res{gains}{EU_lit}{r_prob}$). This ranking does not hinge on the convention:
if the channels are weighted, $1-(1-\rff)^a(1-\rprior)^b$, EU overtakes CPT only
when the functional form receives more than
\res{anat}{CPT_lit__EU_lit}{weight_factor} times the weight of the parameter
distribution (i.e. $a \geq$\res{anat}{CPT_lit__EU_lit}{weight_factor}$b$), and DA only beyond \res{anat}{DA_lit__EU_lit}{weight_factor}
times. Under uniform priors EU is ahead of CPT on both channels, so its
ranking holds under any weighting. Of the \res{anat}{summary}{n_pairs} pairwise
rankings of the seven probabilistic rungs in Table~\ref{tab:lotteries_restrictiveness},
\res{anat}{summary}{n_channelwise} are channel-wise. 

\subsubsection{Joint Comparison}
\label{sec:results_joint}

\begin{figure}[t]
  \centering
  \includegraphics[width=\linewidth]{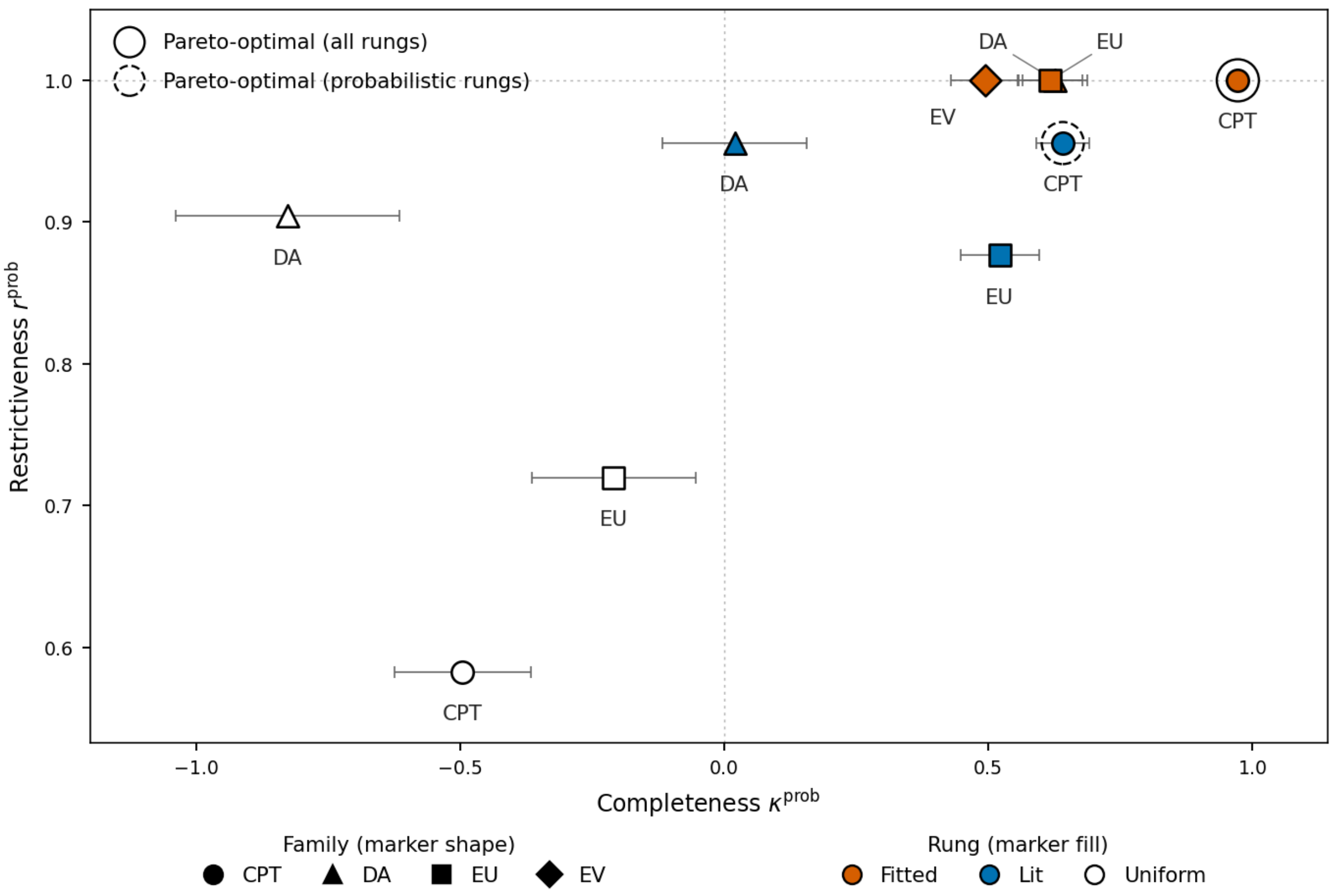}
\caption{Completeness $\kprob$ versus restrictiveness $\rprob$
for the theory-rungs on the Zurich 2003 gain lotteries of Bruhin et al.\ (2010). Marker shape encodes the
theory family, marker fill the rung. Solid ring: Pareto-optimal among all rungs;
dashed rings: Pareto-optimal among the probabilistic rungs. Horizontal bars are
subject-block bootstrap standard errors of $\kprob$; Monte Carlo standard errors
of $\rprob$ are smaller than the markers. The coherence-only rung, at
$(\res{gains}{COH_uniform}{kappa},\res{gains}{COH_uniform}{r_prob})$, lies off
the completeness axis and is not drawn. $\rprob$ combines the two channels  by the product rule of Definition~\ref{def:rprob}.}
  \label{fig:lotteries_pareto}
\end{figure}

Figure~\ref{fig:lotteries_pareto} plots the ten theory-rungs in the
$(\kappa^{\mathrm{prob}}, r^{\mathrm{prob}})$ plane.  

Over all rungs the Pareto frontier is the single point CPT-Fitted at
$(\kprob,\rprob)=(\res{gains}{CPT_deterministic}{kappa},1.00)$. All fitted rungs
sit on the upper edge $\rprob=1$ by convention, and CPT has the highest fitted
completeness. This says only that CPT fits best, since a fitted rung's
$\rprob$ carries no information about its functional form; the reach of the
form is the $\rff$ column of Table~\ref{tab:lotteries_restrictiveness}. The
substantive comparison is among the probabilistic rungs. There the frontier
is the single point CPT-Lit at
$(\res{gains}{CPT_lit}{kappa},\res{gains}{CPT_lit}{r_prob})$. DA-Lit, at
$(\res{gains}{DA_lit}{kappa},\res{gains}{DA_lit}{r_prob})$, is level with it
in restrictiveness and sits at the baseline in completeness.

Every uniform rung is dominated by the literature rung of its own family, and
the coherence-only theory, at
$(\res{gains}{COH_uniform}{kappa},\res{gains}{COH_uniform}{r_prob})$, is
dominated by every other rung. Both statements are channel-wise
(Section~\ref{sec:comparing_theories}). EU-Lit at
$(\res{gains}{EU_lit}{kappa},\res{gains}{EU_lit}{r_prob})$ is dominated by
CPT-Lit on both axes of the figure, but not channel-wise. EU is ahead on
functional form ($\rff=\res{gains}{EU_uniform}{r_ff}$ against
$\res{gains}{CPT_uniform}{r_ff}$). Thus it depends on the way $r^{ff}$ and $r^{prior}$ are aggregated whether EU is dominated by CPT or not. Under
uniform priors EU's rung is ahead of CPT's on both channels.

\subsection{Results Under the Heterogeneous-Agent Interpretation}
\label{sec:learn-crps}
\label{sec:results_dist}

\subsubsection{Completeness} 

\begin{figure}[t]
  \centering
  \includegraphics[width=\linewidth]{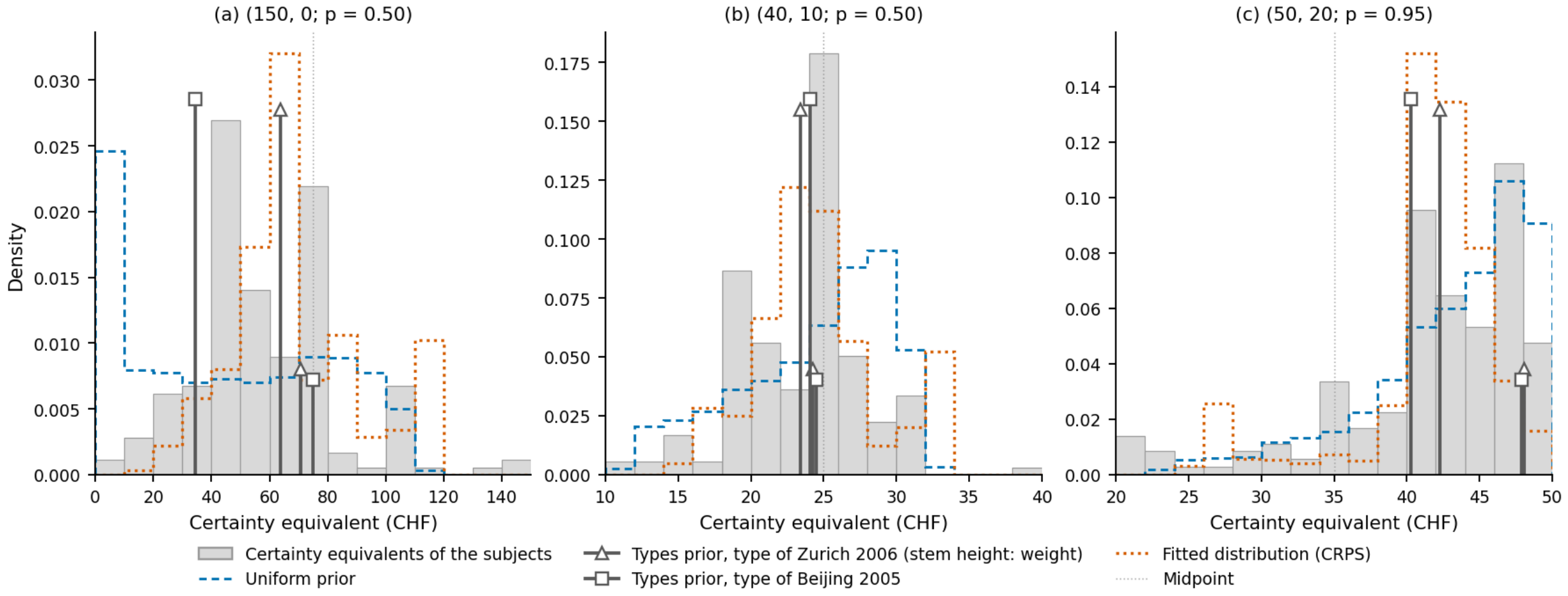}
  \caption{Three lotteries of Zurich 2003 (larger prize, smaller prize; probability of the larger prize), CPT. Bars: the certainty equivalents the subjects reported. Lines: the predictive distributions of the rungs, each parameter distribution pushed through the functional form.}
  \label{fig:crps-fit}
\end{figure}

Table~\ref{tab:lotteries_crps_fit} reports the overall completeness scores. The uniform
prior of CPT which encodes the belief that agents' risk profiles are distributed equally across the permitted range, attains $\kappa^{dist}=\res{crpsfit}{CPT_uniform}{kappa_crps}$.
The types prior, based on the types of \citet{https://doi.org/10.3982/ECTA7139}, reaches the same level of completeness ($\res{crpsfit}{CPT_types}{kappa_crps}$).
The two types of Zurich 2006 alone reach $\res{crpsfit}{CPT_types_zh06}{kappa_crps}$ and those
of Beijing $\res{crpsfit}{CPT_types_bj05}{kappa_crps}$, so pooling the two populations is what
makes the prior cover the subjects. Figure~\ref{fig:crps-fit} shows the
certainty equivalents the subjects reported against the predictive distributions of the CPT rungs for three example lotteries from the data. It also shows why the uniform prior and the types prior reach similar levels of completeness. Where the smaller prize is not zero (Figure \ref{fig:crps-fit}b, c), the uniform prior outperforms the types prior. The four types sit within a few francs of each other, at certainty equivalents the respondents choose often, but the respondents spread across the range, and a proper score charges a forecast for a lack of mass. The uniform prior's mode lies above the respondents' in both panels, because much of the box elevates the weight of the larger prize's probability, but its mean stays close to theirs and its spread covers them.
Where the smaller prize is zero (Figure \ref{fig:crps-fit}a), the uniform prior is drawn towards zero by the mass it puts on low curvature, and the types prior wins because the two samples complement each other. The Zurich types sit in the upper half of the respondents and the heavier Beijing type in the lower half, so each sample's types alone miss half of the respondents while the pooled prior spans them. Across the design the two priors win on complementary lotteries. The types priors of DA and EU,  reach
$\res{crpsfit}{DA_types}{kappa_crps}$ and $\res{crpsfit}{EU_types}{kappa_crps}$. For DA and EU the uniform prior does no better than the baseline, a certainty equivalent drawn uniformly from the prize range ($\res{crpsfit}{DA_uniform}{kappa_crps}$ and $\res{crpsfit}{EU_uniform}{kappa_crps}$). Since the uniform prior carries no information about the population beyond the parameter box, the gap comes from the functional forms and their boxes: much of CPT's completeness is there before anything about the distribution of its parameters is known, and DA's and EU's push-forward densities are too narrow for the observed degree of dispersion. The fitted distribution is the most complete rung under the
CRPS,
$\res{crpsfit}{CPT_crpsfit}{kappa_crps}$ for CPT,
$\res{crpsfit}{DA_crpsfit}{kappa_crps}$ for DA and
$\res{crpsfit}{EU_crpsfit}{kappa_crps}$ for EU.

\begin{table}[htbp]
\centering
\caption{The heterogeneous-agent interpretation on the gain lotteries of Zurich 2003: distributional completeness $\kappa^{dist}$ and distributional restrictiveness $r^{dist}$ of every rung. The last row gives the restrictiveness of the subjects' own distributions, $r^{*}$. Standard errors in parentheses.}
\label{tab:lotteries_crps_fit}
\resizebox{\ifdim\width>\linewidth\linewidth\else\width\fi}{!}{
\begin{threeparttable}
\begin{tabular}{ll cc}
\toprule
Model & Rung & $\kappa^{dist}$ & $r^{dist}$ \\
\midrule
CPT & Uniform prior & +0.700 (0.018) & 0.181 \\
 & Types prior & +0.696 (0.022) & 0.264 \\
 & Fitted distribution & +0.890 (0.015) & 0.166 \\
 & Fitted theory & +0.287 (0.043) & 0.359 \\
\addlinespace
DA & Uniform prior & -0.147 (0.088) & 0.457 \\
 & Types prior & +0.298 (0.069) & 0.449 \\
 & Fitted distribution & +0.532 (0.046) & 0.287 \\
 & Fitted theory & -0.143 (0.092) & 0.557 \\
\addlinespace
EU & Uniform prior & -0.012 (0.085) & 0.504 \\
 & Types prior & -0.064 (0.091) & 0.524 \\
 & Fitted distribution & +0.189 (0.076) & 0.475 \\
 & Fitted theory & -0.153 (0.096) & 0.571 \\
\addlinespace
EV & No free parameter & -0.183 (0.099) & 0.554 \\
\addlinespace
\midrule
Subjects & Their own distributions & & 0.185 (0.009) \\
\bottomrule
\end{tabular}
\begin{tablenotes}[flushleft]
\small
\item \textit{Notes:} Ten folds over subjects. The predictive distribution of a rung is its parameter distribution pushed through the functional form. Standard errors of $\kappa^{dist}$ and $r^{*}$ are from a block bootstrap.
\end{tablenotes}
\end{threeparttable}%
}
\end{table}

\begin{figure}[t]
  \centering
  \includegraphics[width=0.7\linewidth]{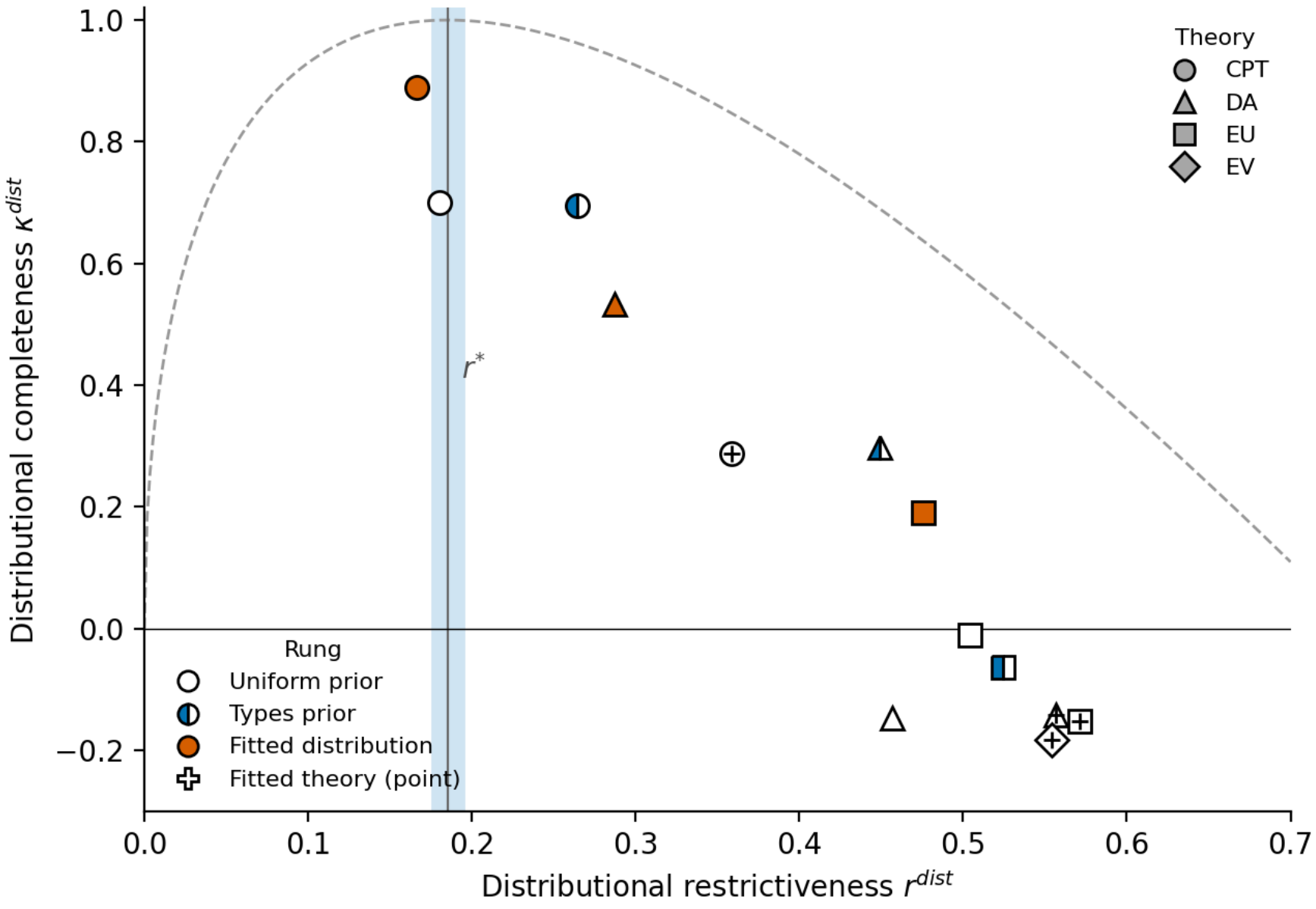}
  \caption{Distributional completeness $\kappa^{dist}$ against distributional restrictiveness $r^{dist}$ for the rungs of Table~\ref{tab:lotteries_crps_fit} on the Zurich 2003 gain lotteries. Marker shape encodes the theory family, marker fill the rung; a half-filled marker is the types prior and a cross marks the fitted theories. The vertical line is the restrictiveness of the subjects' own distributions, $r^{*}$, with a band of one standard error. The dashed curve is the ceiling $1-(\sqrt{r^{dist}/r^{*}}-1)^2$ of Proposition~\ref{prop:rdist_bound}: a rung can be fully complete only at $r^{*}$.}
  \label{fig:crps-plane}
\end{figure}

\paragraph{Restrictiveness} The subjects' own distributions have
$r^{*}=\res{crpsfit}{subjects}{r_dist}$ (standard error $\res{crpsfit}{subjects}{r_dist_se}$).
This means the population of Zurich 2003 is about a fifth as far from the reference as the farthest rule.
CPT's fitted distribution, at $\res{crpsfit}{CPT_crpsfit}{r_dist}$, is close to that value by construction. The fitted distribution is the mixture over the grid cells whose push-forward best matches the subjects' distributions lottery by lottery, so it
lands at the subjects' dispersion whenever the form can reach it. Thus, $r^{dist}$ being close to $r^*$ for CPT, shows that the functional form of CPT can accommodate this population. The more informative comparisons are the held priors, which are not
fitted, and DA and EU, whose fitted distributions stay far from $r^{*}$ because their forms
cannot produce the dispersion (Appendix~\ref{app:dist_channels}).
Figure~\ref{fig:crps-plane} plots every rung against the
ceiling that Proposition~\ref{prop:rdist_bound} puts on every theory, given how much more or less
restrictive it is than the population. The types prior sits between, at $\res{crpsfit}{CPT_types}{r_dist}$. \res{crpsfit}{CPT_types}{n_atoms} point masses cannot be as dispersed as the subjects, but they
commit less than a single point. Every rung of DA and of EU, the types priors included, is more restrictive than the population, from
$\res{crpsfit}{DA_crpsfit}{r_dist}$ for DA's fitted distribution to
$\res{crpsfit}{EU_deterministic}{r_dist}$ for EU's fitted theory.

\paragraph{Response noise.} The predictive distributions of this section contain no response
noise: every spread is a spread of parameter values, whereas the individual level adds a Gaussian
response model around the rule (Appendix~\ref{app:individual_estimation}). A family whose
push-forward is narrower than the subjects is therefore charged here for a dispersion that the
individual level grants it. Appendix~\ref{app:dist_noise} adds the response model of the
individual level to every rung of Table~\ref{tab:lotteries_crps_fit}, under the calibration the
individual level uses and at the larger level of the relative residual. With the individual
level's noise every rung gains completeness, most where its push-forward is narrowest: the fitted
theory of CPT rises from $\kappa^{dist}=\res{crpsN}{CPT_deterministic}{kappa_crps_none}$ to
$\res{crpsN}{CPT_deterministic}{kappa_crps_range}$ and EU's from
$\res{crpsN}{EU_deterministic}{kappa_crps_none}$ to
$\res{crpsN}{EU_deterministic}{kappa_crps_range}$, while CPT's uniform prior, whose push-forward
already spans the subjects, moves from $\res{crpsN}{CPT_uniform}{kappa_crps_none}$ to
$\res{crpsN}{CPT_uniform}{kappa_crps_range}$. CPT stays ahead of DA and EU on every rung, and
every rung of DA and EU stays more restrictive than the population. Two features of the exercise
argue for the noise-free table as the main specification. The noise level is each family's own
residual around its best-fitting rules, so it is largest for the family that fits worst (EU's
$\res{crpsN}{EU}{c_range_pct}$\% of the prize range against CPT's
$\res{crpsN}{CPT}{c_range_pct}$\%), and granting it rewards misfit. And a score applied lottery by
lottery cannot tell noise from heterogeneity, since both widen the marginal distribution at a
lottery: at the residual level a single CPT rule plus noise reaches
$\res{crpsN}{CPT_deterministic}{kappa_crps_rms}$ against
$\res{crpsN}{CPT_crpsfit}{kappa_crps_rms}$ for the fitted distribution, and sits at the
population's restrictiveness. The response model is part of the evaluation at the individual
level because the update needs a likelihood; at the population level the theories are scored on
the dispersion they produce themselves.

\section{Scope}
\label{sec:scope}

Section~\ref{sec:ex1} evaluated the theories on one set of lotteries
and one pool of subjects.  In this section, we evaluate them on more than just a single domain.

\subsection{Same Subjects, New Domain: Gains and Losses}

The domain of Section~\ref{sec:ex1}
contains only gains lotteries. However,  the subjects participating in the experiment of \citet{https://doi.org/10.3982/ECTA7139} were also evaluated on 25 loss lotteries. In this subsection, we extend the same theories to losses and evaluate them using the identical subject pool of Section \ref{sec:ex1}.

We compare two natural ways to adapt the gains-only theories to the loss domain: (1) extending the theory to be more general by adapting the parameter box and distributions to be able to accommodate both scales and trade off the fit on gains against the fit on losses and (2) combining the gains theory with a theory that is specific to the loss domain. In both cases, however, gain lotteries and the loss lotteries are treated as two separate
domains. Each has its own baseline (the midpoint $(z_1+z_2)/2$ of every
lottery's prize range), its own best predictor (the lookup table of its
lotteries), its own reference distribution (uniform on every prize range as in
Section~\ref{sec:ex1_reference_distribution}), and its own reachable set. The support restriction is dominance in both. We report the fitted theories and the literature priors; the uniform priors are reported in Appendix~\ref{app:losses-plane}. We aggregate the predictive content across the domains using equal weighting, i.e. the lotteries of the gains domain contribute as much as the lotteries from the loss domain to the overall score. Both interpretations of Section~\ref{sec:ex1} apply here as well, we first report the results for the representative-agent interpretation, followed by the results for the heterogeneous-agent interpretation (Section~\ref{sec:res-scope-dist}).

\subsubsection{Unified Theories}
\label{sec:onetheory}
In terms of functional forms, CPT, EU and EV
extend naturally to losses as long as every lottery is written as
$x=(z_1,z_2,p)$, with $z_1$ the prize that is larger in absolute value, $z_2$
the other prize and $p$ the probability of $z_1$. The value function is
reflected at zero, $v(z)=\operatorname{sign}(z)\,|z|^{\alpha}$. The formulas
of Section~\ref{sec:ex1} then hold for a lottery of either sign. On losses $\alpha<1$ implies a convex
value function, the diminishing sensitivity documented by \citet{tversky_kahneman_1992}, while  $\alpha>1$ corresponds to a concave one.

For DA the weight $\tilde w$ belongs to the better prize, which on gains is
$z_1$ and on losses is $z_2$, the smaller loss, with probability $1-p$. On
losses the certainty equivalent is therefore
\[
f_\lambda(x)=v^{-1}\bigl(\tilde w(1-p)\,v(z_2)+[1-\tilde w(1-p)]\,v(z_1)\bigr),
\]
and thus DA on losses is not the mirror image of DA on gains.

\paragraph{Parameter box.} 
While the parameter ranges for $\gamma$,
$\delta$ and $\eta$ are the same as in Section~\ref{sec:ex1}, we extend the range for $\alpha$ to $[0.01,2]$ ($\alpha\in[0.01,1]$ in Section~\ref{sec:ex1}). \citet{completeness2022} and \citet{andrews2025transfer} impose no upper bound and estimate a loss exponent of $0.975$ on all 50 lotteries at once, and in the meta-analysis of \citet{imai_nunnari_wu_vieider_2025} a quarter of the loss-domain estimates imply a concave value function with $\alpha \geq 1$ ($24$ percent, against $14$ percent of the
gains-domain estimates). Thus, on data including losses, setting the upper limit for $\alpha$ at one would likely be too restrictive and does not reflect the current literature consensus.

\paragraph{Parameter distributions.} The literature prior is calibrated on the
estimates across gains and losses.
Appendix~\ref{app:priors_losses} documents the calibration.

\subsubsection{Domain Specific Theories}
\label{sec:signed}\label{sec:calibrations}
Instead of specifying a theory with a unified functional form, parameter box and parameter distribution, another way to specify a theory for both domains, corresponds to creating a theory that is composed of a gains specific theory and a loss specific theory, combined by an indicator function.

Here, the functional forms and the two domains are those of
Section~\ref{sec:onetheory}; the parameter box, the parameter distribution and
the fit are specific to the sign.

\paragraph{Parameter box.} On gains the box of
Section~\ref{sec:ex1}, $\alpha\le1$. On losses the lifted box,
$\alpha\in[0.01,2]$, for the reasons given in Section~\ref{sec:onetheory}.

\paragraph{Parameter distributions.} On gains those of
Section~\ref{sec:ex1_parameter_distributions}. On losses the literature priors follow the recipe of Section~\ref{sec:ex1_parameter_distributions}
with the estimates for losses.
Details on calibration can be found in Appendix~\ref{app:priors_losses}.

\subsubsection{Results}

\paragraph{Gains vs. Losses}
\label{sec:res-losses}
Fitted, the loss domain behaves like the gains domain. The fitted loss specific theories are at least as complete on losses as the gains-specific theory versions on gains. CPT attains $\kappa=\res{pdLadapt}{CPT_deterministic}{kappa}$ ($\res{pdG}{CPT_deterministic}{kappa}$ on gains), DA
$\res{pdLadapt}{DA_deterministic}{kappa}$ ($\res{pdG}{DA_deterministic}{kappa}$ on gains) and EU
$\res{pdLadapt}{EU_deterministic}{kappa}$ (on gains
$\res{pdG}{EU_deterministic}{kappa}$). Literature priors lose all of their completeness on losses: every literature rung falls below the baseline there, whether adapted to losses or not. 
CPT-Lit (domain specific) attains $\kprob=\res{pdLadapt}{CPT_lit}{kappa}$ on losses against
$\res{pdG}{CPT_lit}{kappa}$ on gains, EU-Lit $\res{pdLadapt}{EU_lit}{kappa}$
against $\res{pdG}{EU_lit}{kappa}$ and DA-Lit $\res{pdLadapt}{DA_lit}{kappa}$
against $\res{pdG}{DA_lit}{kappa}$. The gains priors applied as they are would score higher, $\res{pdLasis}{CPT_lit}{kappa}$, $\res{pdLasis}{EU_lit}{kappa}$ and $\res{pdLasis}{DA_lit}{kappa}$, but still below the baseline. The main reason is that the loss lotteries would require a curvature of
$\res{boxfit}{CPT_losses__domain}{alpha_star}$ (the fitted value of CPT,
$\res{boxfit}{EU_losses__domain}{alpha_star}$ for EU), at the edge of the losses prior. Following \citet{imai_nunnari_wu_vieider_2025} it places a quarter of its mass above 1 and only about two
percent above $1.4$. The main driver of this effect value are the two largest lotteries, indicating that 
respondents are more risk averse at larger stakes than a power value function, with its
constant relative risk aversion, can express.
Value functions with rising relative risk aversion or a stake-dependent weighting function
accommodate it \citep{fehrduda2010stakes,bouchouicha2017stakes}. A prior conditioned on the elicitation method
would sit closer.

Regarding restrictiveness, CPT's functional form rules out about as much on
losses as on gains ($\rff=\res{pdLadapt}{CPT_uniform}{r_ff}$ against
$\res{pdG}{CPT_uniform}{r_ff}$), EU's more
($\res{pdLadapt}{EU_uniform}{r_ff}$ against $\res{pdG}{EU_uniform}{r_ff}$) and
DA's less ($\res{pdLadapt}{DA_uniform}{r_ff}$ against
$\res{pdG}{DA_uniform}{r_ff}$). The literature rungs rule out less on losses
than on gains: $\rprob=\res{pdLadapt}{CPT_lit}{r_prob}$ against
$\res{pdG}{CPT_lit}{r_prob}$ for CPT-Lit, $\res{pdLadapt}{EU_lit}{r_prob}$
against $\res{pdG}{EU_lit}{r_prob}$ for EU-Lit and
$\res{pdLadapt}{DA_lit}{r_prob}$ against $\res{pdG}{DA_lit}{r_prob}$ for
DA-Lit. On losses as on gains the frontier of the probabilistic rungs is CPT-Lit alone.

\begin{figure}[t]
  \centering
  \includegraphics[width=\linewidth]{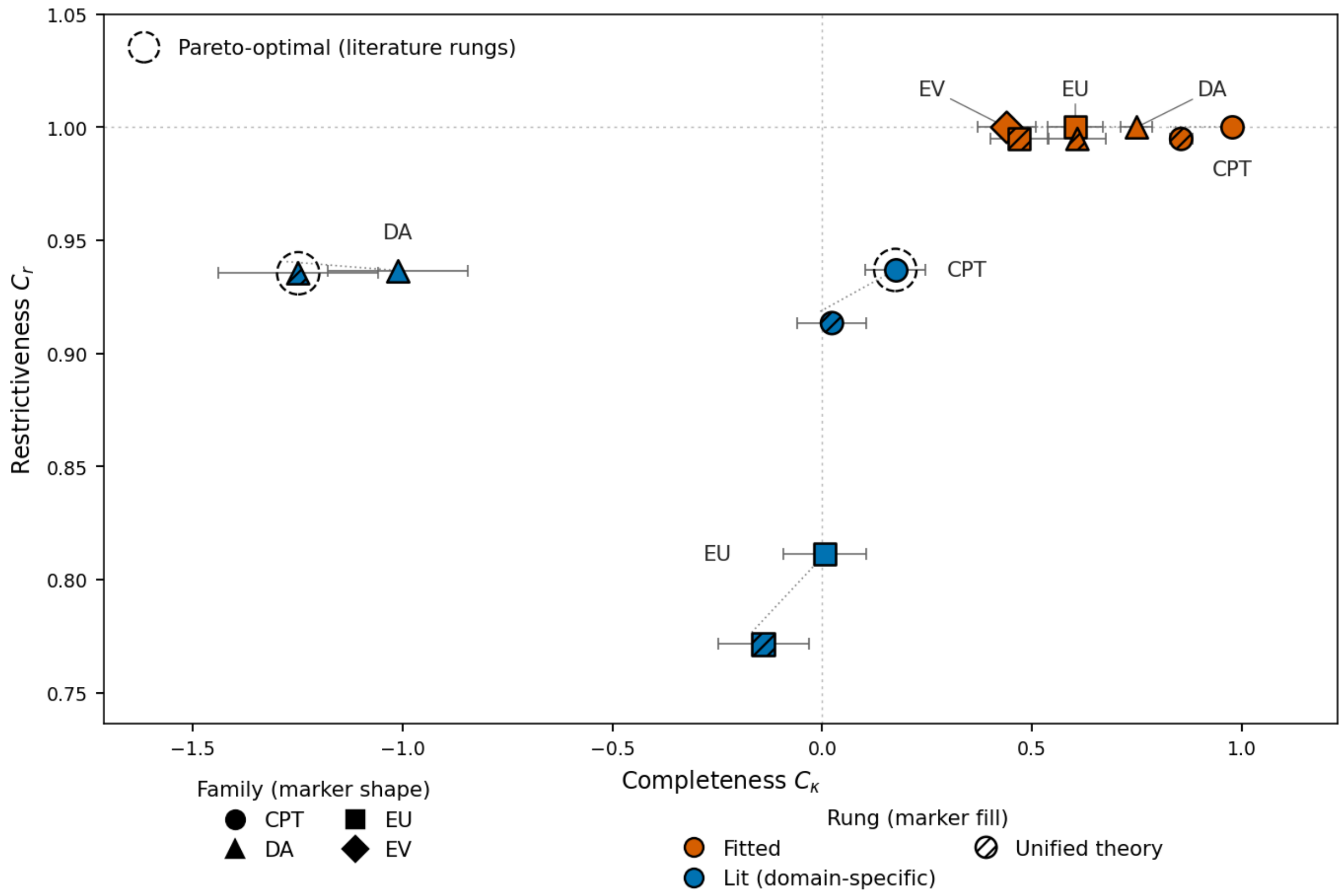}
  \caption{Predictive content $(C_\kappa,C_r)$ of every rung on the gain and the loss lotteries of Zurich 2003 (Bruhin et al.\ 2010), two domains with equal weights. Marker shape encodes the theory family, marker fill the rung. Black and grey markers are the domain-specific theories, the theory of Section~\ref{sec:ex1} on gains and the theory adapted to losses (Panel~A of Tables~\ref{tab:lotteries_completeness_panel} and~\ref{tab:lotteries_restrictiveness_panel}). Hatched markers are the unified theory (Panel~B: the lifted box on either sign, the prior calibrated on both, one fitted value), each joined to its counterpart by a dotted line. Dashed rings: Pareto-optimal among the literature rungs. Horizontal bars are standard errors of $C_\kappa$ from the joint bootstrap that resamples subjects once for both domains; Monte Carlo standard errors of $C_r$ are smaller than the markers. $C_r$ averages $\rprob$, which combines the two channels by the product rule of Definition~\ref{def:rprob}; fitted rungs and expected value have $\rprob=1$. The uniform rungs are shown in Appendix~\ref{app:losses-plane}.}
  \label{fig:predictive-content}
\end{figure}

\paragraph{Predictive Content}
\label{sec:res-content}
Figure~\ref{fig:predictive-content} plots every rung at its predictive
content $(C_\kappa,C_r)$ under the representative agent interpretation on the scope of both domains, the averages of its
completeness and its restrictiveness over the gain and the loss lotteries.

Among the literature rungs of the domain-specific theories CPT-Lit, at $(\res{pdAvgAdapt}{CPT_lit}{kappa},\res{pdAvgAdapt}{CPT_lit}{r_prob})$, is the Pareto frontier on its own. As in the single domain case, DA-Lit, at $(\res{pdAvgAdapt}{DA_lit}{kappa},\res{pdAvgAdapt}{DA_lit}{r_prob})$, is level with CPT-Lit in restrictiveness but predicts far less well. EU-Lit at
$(\res{pdAvgAdapt}{EU_lit}{kappa},\res{pdAvgAdapt}{EU_lit}{r_prob})$ is
dominated by CPT-Lit, by a margin in completeness of six standard errors ($\res{pdDiffAdapt}{CPT_lit__minus__EU_lit}{diff}$, SE
$\res{pdDiffAdapt}{CPT_lit__minus__EU_lit}{diff_se}$).
The parameter-free expected-value rule, at $(\res{pdAvgAdapt}{EV_deterministic}{kappa},1.00)$, dominates CPT-Lit. It is more complete by $\res{pdDiffAdapt}{EV_deterministic__minus__CPT_lit}{diff}$ (SE $\res{pdDiffAdapt}{EV_deterministic__minus__CPT_lit}{diff_se}$): fifteen hundredths behind CPT-Lit on gains and seven tenths ahead on losses, where a draw from the adapted prior falls below the baseline.

Claiming the loss lotteries pays for the fitted theories and for expected value, which predict better than the baseline on losses, but not for CPT-Lit. Had it claimed the gains domain only, its predictive content would be $(\res{pdAvgAdapt}{CPT_lit}{kappa_gains_only},\res{pdAvgAdapt}{CPT_lit}{r_prob_gains_only})$ instead of $(\res{pdAvgAdapt}{CPT_lit}{kappa},\res{pdAvgAdapt}{CPT_lit}{r_prob})$, so that the broader scope is more restrictive but less complete, and the two scopes are not ranked (Section~\ref{sec:comparing_theories}). Broader scope adds to both coordinates only where a theory predicts better than the baseline on the added domain.

\paragraph{Domain-Specific vs. Unified}
\label{sec:res-unified}\label{sec:panel-results}
Holding one theory for both domains costs predictive content. The unified fit
is less complete than the domain-specific fits, by fifteen hundredths for CPT ($C_\kappa=\res{share}{CPT}{kappa_avg}$ against
$\res{pdAvgAdapt}{CPT_deterministic}{kappa}$), seventeen for DA
($\res{share}{DA}{kappa_avg}$ against
$\res{pdAvgAdapt}{DA_deterministic}{kappa}$) and sixteen for EU
($\res{share}{EU}{kappa_avg}$ against
$\res{pdAvgAdapt}{EU_deterministic}{kappa}$). Under literature priors the
unified CPT-Lit, at
$(\res{pdAvgShared}{CPT_lit}{kappa},\res{share}{CPT_lit}{r_prob_avg})$, and the
unified EU-Lit, at
$(\res{pdAvgShared}{EU_lit}{kappa},\res{share}{EU_lit}{r_prob_avg})$, are
dominated by their domain-specific counterparts, since the pooled prior is
wider than the gains prior it replaces. The unified DA-Lit, at
$(\res{pdAvgShared}{DA_lit}{kappa},\res{share}{DA_lit}{r_prob_avg})$, is less
complete than its counterpart and rules out marginally more, because one
parameter value for both domains narrows what the functional form can reach;
it is the second ringed marker among the literature rungs of the figure.
Appendix~\ref{app:losses-plane} reports the comparison in each domain.

\begin{figure}[t]
  \centering
  \includegraphics[width=\linewidth]{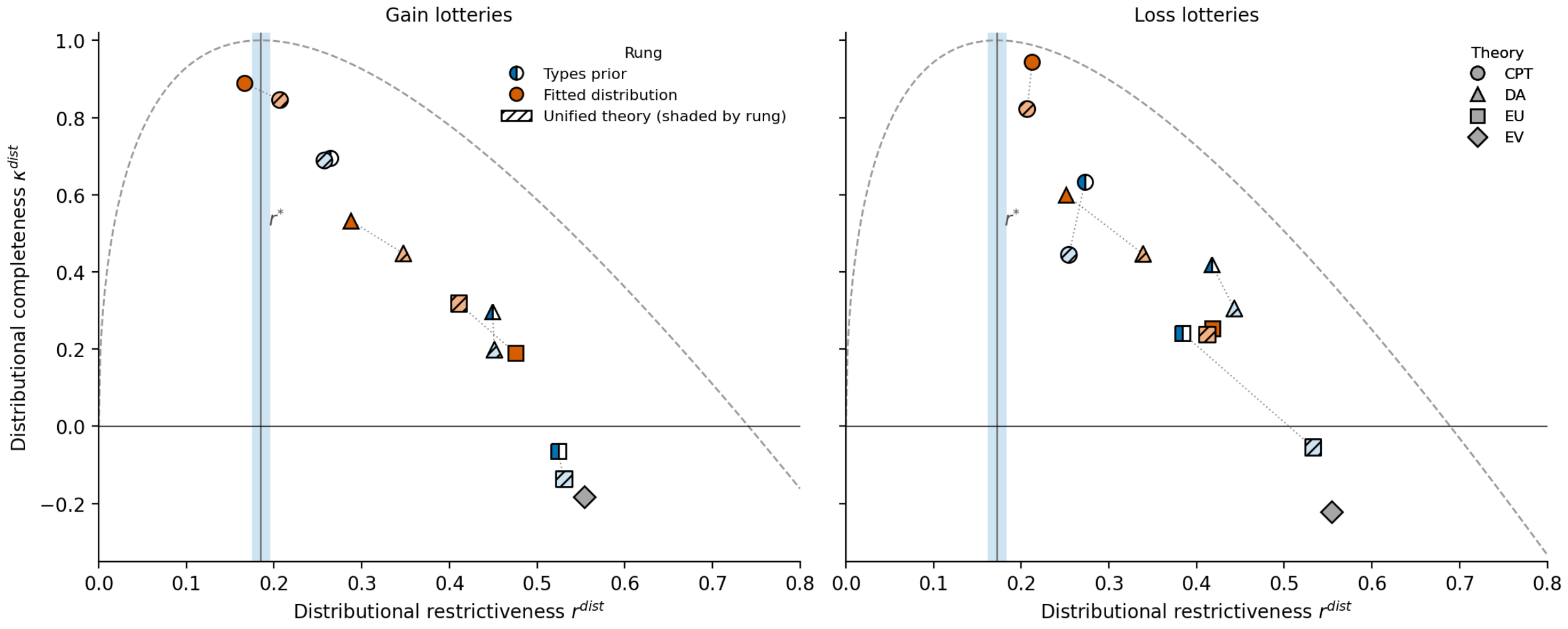}
  \caption{Distributional completeness $\kappa^{dist}$ against distributional restrictiveness $r^{dist}$ on the gain (left) and the loss (right) lotteries of Zurich 2003: the types prior and the fitted distribution of each theory and expected value (every rung is in Table~\ref{tab:lotteries_crps_panel}). Marker shape encodes the theory family and marker fill the rung, as in Figure~\ref{fig:crps-plane}. Hatched markers are the unified theory, shaded by rung and joined to their domain-specific counterparts by dotted lines; where the two coincide, as for CPT's uniform prior on losses, only the hatched marker shows. The vertical line is the subjects' own $r^{*}$ in the domain with a band of one standard error, the dashed curve the ceiling of Proposition~\ref{prop:rdist_bound}.}
  \label{fig:crps-scope-plane}
\end{figure}

\paragraph{The Heterogeneous-Agent Interpretation}
\label{sec:res-scope-dist}
Figure~\ref{fig:crps-scope-plane} repeats the comparison under the heterogeneous-agent interpretation of
Section~\ref{sec:learn-crps}, in which a rung's parameter distribution is interpreted as the population of
subjects and scored by the CRPS, for the types prior and the fitted distribution of each theory
and for expected value. Completeness and restrictiveness of every rung of
Table~\ref{tab:lotteries_crps_fit} in each domain, the uniform prior and the fitted theory
included, are reported in Table~\ref{tab:lotteries_crps_panel} in
Appendix~\ref{app:dist_channels}. The types prior on losses is the two-type mixture refitted on
the loss lotteries of Zurich 2006 and Beijing 2005, and for the unified theory on both signs of the
two samples with one parameter value per type. The unified fitted distribution is one distribution
fitted to both signs at once.

Fitted, the loss domain again behaves like the gains domain. CPT's fitted distribution reaches
$\kappa^{dist}=\res{csL}{CPT_crpsfit}{kappa_crps}$ on losses against
$\res{csG}{CPT_crpsfit}{kappa_crps}$ on gains, DA's $\res{csL}{DA_crpsfit}{kappa_crps}$ against
$\res{csG}{DA_crpsfit}{kappa_crps}$ and EU's $\res{csL}{EU_crpsfit}{kappa_crps}$ against
$\res{csG}{EU_crpsfit}{kappa_crps}$. Unlike the literature rungs under the representative-agent
interpretation, the literature priors perform similarly well on the loss domain. CPT's types prior attains
$\res{csL}{CPT_types}{kappa_crps}$ on losses compared to $\res{csG}{CPT_types}{kappa_crps}$ on gains.

The populations' own dispersion is similar on gains and on losses,  and $\res{csSG}{subjects}{r_dist}$ and$r^{*}=\res{csSL}{subjects}{r_dist}$, respectively.

One theory for both signs costs completeness under this interpretation too. CPT's unified fitted
distribution is less complete than the domain-specific ones on average
($\res{cuA}{CPT_crpsfit}{kappa_crps}$ against $\res{csA}{CPT_crpsfit}{kappa_crps}$), mostly on
losses ($\res{cuL}{CPT_crpsfit}{kappa_crps}$ against $\res{csL}{CPT_crpsfit}{kappa_crps}$), and so
is DA's ($\res{cuA}{DA_crpsfit}{kappa_crps}$ against $\res{csA}{DA_crpsfit}{kappa_crps}$). EU's
unified fitted distribution is the more complete ($\res{cuA}{EU_crpsfit}{kappa_crps}$ against
$\res{csA}{EU_crpsfit}{kappa_crps}$), because on gains the lifted box lets it place subjects at a
curvature above one, which the box of Section~\ref{sec:ex1} excludes
($\res{cuG}{EU_crpsfit}{kappa_crps}$ against $\res{csG}{EU_crpsfit}{kappa_crps}$ on gains): for EU
the box weighs more than the tie. CPT's types prior refitted on both signs loses most of its
completeness on losses ($\res{cuL}{CPT_types}{kappa_crps}$ against
$\res{csL}{CPT_types}{kappa_crps}$).
Restrictiveness behaves differently under this interpretation. The unified theory makes a claim across
lotteries and one parameter value serves every lottery of both signs. Thus, what a respondent does at a
loss lottery is pinned down by what they do at its mirror on the gains domain. Distributional restrictiveness cannot see such a claim. It compares the forecast with the reference one lottery at a time, so its value
on a domain is a weighted average of the lotteries' values (Proposition~\ref{prop:aggregation}),
and two theories with the same forecast at every lottery have the same $r^{dist}$ however their
forecasts are linked across lotteries. The unified theory therefore changes $r^{dist}$ only by
forecasting different populations at each lottery. CPT's unified fitted distribution sits at
$\res{cuG}{CPT_crpsfit}{r_dist}$ on gains and $\res{cuL}{CPT_crpsfit}{r_dist}$ on losses, against
$\res{csG}{CPT_crpsfit}{r_dist}$ and $\res{csL}{CPT_crpsfit}{r_dist}$ for the domain-specific fits,
close to the subjects on both signs. Under this interpretation the two theories are therefore compared on
completeness alone: for CPT and DA the domain-specific theories are the more complete, for EU, whose gains box is too narrow, the unified one.

\section{Conclusion}
\label{sec:conclusion}

A theory of behavior is usually accompanied by a belief about possible parameter values, and how stringent this belief it is implicitly part of the theory. We have proposed a way to measure this explicitly. A theory in our
framework is a functional form together with a distribution over its parameters, and its empirical content has two channels: what the form can
reach at all, and how the distribution spreads the theory's predictions over
what it can reach. Completeness and restrictiveness, defined for a fitted
functional form by \citet{completeness2022} and \citet{restrictiveness2026},
then have versions for the theory as it is held before the data are seen, computed by drawing a rule from the theory instead of fitting one. The same theory can be interpreted as a claim about a representative agent or about a population of heterogeneous agents, and it can be scored for what it says about one agent once some of that agent's choices are known. Finally, a
theory claims a scope, and the framework makes that claim measurable. A domain adds to
restrictiveness wherever the theory rules anything out, and to completeness
only where the theory beats the baseline.

On the lotteries of Bruhin et al.\ (2010) we find that dominance, the support restriction the three theories share, excludes very little of conceivable behavior. Which theory is better depends on which object is compared. On functional
form the most flexible family is the most complete and the least restrictive;
held with literature priors, CPT is the frontier on its own, at
$(\res{gains}{CPT_lit}{kappa},\res{gains}{CPT_lit}{r_prob})$. Interpreted as a
population and for the individual respondent the separation sharpens, since
the forms of DA and EU cannot produce the dispersion the subjects show, and across the sign of the prizes the literature priors do not carry.

What the framework adds to the measurement of theories is the object that it scores. Evaluating a theory in its fitted form answers how well a family of functions can be fitted to explain a datatset, while a theory in conjunction with support restrictions and a literature-informed parameter prior, evaluates what the current consensus of the field commits to. The gap between them is an indicator about the generalizability of a theory.  Measuring the
convergence of a literature in these terms, net of what the accumulation of
studies alone would produce and net of the analytic choices its studies
share, is the subject of ongoing work.

\setlength\bibsep{0pt}
\bibliographystyle{apalike}
\bibliography{bibliography}

\begin{thebibliography}{}

\bibitem[Abdellaoui and Bleichrodt, 2007]{abdellaoui2007}
Abdellaoui, M. and Bleichrodt, H. (2007).
\newblock Eliciting {G}ul's theory of disappointment aversion by the tradeoff method.
\newblock {\em Journal of Economic Psychology}, 28(6):631--645.

\bibitem[Andrews et~al., 2025]{andrews2025transfer}
Andrews, I., Fudenberg, D., Lei, L., Liang, A., and Wu, C. (2025).
\newblock The transfer performance of economic models.
\newblock arXiv:2202.04796.

\bibitem[Beatty and Crawford, 2011]{beatty_crawford2011}
Beatty, T. K.~M. and Crawford, I.~A. (2011).
\newblock How demanding is the revealed preference approach to demand?
\newblock {\em American Economic Review}, 101(6):2782--2795.

\bibitem[Bouchouicha and Vieider, 2017]{bouchouicha2017stakes}
Bouchouicha, R. and Vieider, F.~M. (2017).
\newblock Accommodating stake effects under prospect theory.
\newblock {\em Journal of Risk and Uncertainty}, 55(1):1--28.

\bibitem[Box, 1980]{Box1980}
Box, G. E.~P. (1980).
\newblock Sampling and {B}ayes' inference in scientific modelling and robustness.
\newblock {\em Journal of the Royal Statistical Society, Series A}, 143(4):383--430.

\bibitem[Bruhin et~al., 2010]{https://doi.org/10.3982/ECTA7139}
Bruhin, A., Fehr-Duda, H., and Epper, T. (2010).
\newblock Risk and rationality: Uncovering heterogeneity in probability distortion.
\newblock {\em Econometrica}, 78(4):1375--1412.

\bibitem[Camerer et~al., 2016]{camerer2016}
Camerer, C.~F., Dreber, A., Forsell, E., Ho, T.-H., Huber, J., Johannesson, M., Kirchler, M., Almenberg, J., Altmejd, A., Chan, T., Heikensten, E., Holzmeister, F., Imai, T., Isaksson, S., Nave, G., Pfeiffer, T., Razen, M., and Wu, H. (2016).
\newblock Evaluating replicability of laboratory experiments in economics.
\newblock {\em Science}, 351(6280):1433--1436.

\bibitem[Choi et~al., 2007]{choi2007}
Choi, S., Fisman, R., Gale, D., and Kariv, S. (2007).
\newblock Consistency and heterogeneity of individual behavior under uncertainty.
\newblock {\em American Economic Review}, 97(5):1921--1938.

\bibitem[Erev et~al., 2017]{erev2017}
Erev, I., Ert, E., Plonsky, O., Cohen, D., and Cohen, O. (2017).
\newblock From anomalies to forecasts: Toward a descriptive model of decisions under risk, under ambiguity, and from experience.
\newblock {\em Psychological Review}, 124(4):369--409.

\bibitem[Fehr-Duda et~al., 2010]{fehrduda2010stakes}
Fehr-Duda, H., Bruhin, A., Epper, T., and Schubert, R. (2010).
\newblock Rationality on the rise: Why relative risk aversion increases with stake size.
\newblock {\em Journal of Risk and Uncertainty}, 40(2):147--180.

\bibitem[Friedman, 1953]{friedman1953}
Friedman, M. (1953).
\newblock The methodology of positive economics.
\newblock In {\em Essays in Positive Economics}, pages 3--43. University of Chicago Press, Chicago.

\bibitem[Fudenberg et~al., 2026a]{restrictiveness2026}
Fudenberg, D., Gao, W., and Liang, A. (2026a).
\newblock How flexible is that functional form? quantifying the restrictiveness of theories.
\newblock {\em The Review of Economics and Statistics}, 108(1):194--209.

\bibitem[Fudenberg et~al., 2026b]{fudenberg2026_structural}
Fudenberg, D., Gao, W.~Y., and You, Z. (2026b).
\newblock Model restrictiveness in functional and structural settings.
\newblock arXiv:2602.07688.

\bibitem[Fudenberg et~al., 2022]{completeness2022}
Fudenberg, D., Kleinberg, J., Liang, A., and Mullainathan, S. (2022).
\newblock Measuring the completeness of economic models.
\newblock {\em Journal of Political Economy}, 130(4):956--990.

\bibitem[Fudenberg and Liang, 2019]{fudenberg2019}
Fudenberg, D. and Liang, A. (2019).
\newblock Predicting and understanding initial play.
\newblock {\em American Economic Review}, 109(12):4112–41.

\bibitem[Gneiting and Raftery, 2007]{gneiting2007}
Gneiting, T. and Raftery, A.~E. (2007).
\newblock Strictly proper scoring rules, prediction, and estimation.
\newblock {\em Journal of the American Statistical Association}, 102(477):359--378.

\bibitem[Harrison and Swarthout, 2023]{harrison2023}
Harrison, G.~W. and Swarthout, J.~T. (2023).
\newblock Cumulative prospect theory in the laboratory: A reconsideration.
\newblock In {\em Models of Risk Preferences: Descriptive and Normative Challenges}, volume~22 of {\em Research in Experimental Economics}, pages 37--129. Emerald Publishing Limited.

\bibitem[Imai et~al., 2025]{imai_nunnari_wu_vieider_2025}
Imai, T., Nunnari, S., Wu, J., and Vieider, F.~M. (2025).
\newblock Meta-analysis of prospect theory parameters.
\newblock CESifo Working Paper 12334, CESifo.

\bibitem[{J.~N. Keynes}, 1891]{keynes1891}
{J.~N. Keynes} (1891).
\newblock {\em The Scope and Method of Political Economy}.
\newblock Macmillan, London.

\bibitem[Jefferys and Berger, 1992]{JefferysBerger1992}
Jefferys, W.~H. and Berger, J.~O. (1992).
\newblock Ockham's razor and {B}ayesian analysis.
\newblock {\em American Scientist}, 80(1):64--72.

\bibitem[Jeffreys, 1961]{Jeffreys1961}
Jeffreys, H. (1961).
\newblock {\em Theory of Probability}.
\newblock Oxford University Press, Oxford, 3 edition.

\bibitem[Kass and Raftery, 1995]{KassRaftery1995}
Kass, R.~E. and Raftery, A.~E. (1995).
\newblock Bayes factors.
\newblock {\em Journal of the American Statistical Association}, 90(430):773--795.

\bibitem[Kiefer and Wolfowitz, 1956]{kiefer1956}
Kiefer, J. and Wolfowitz, J. (1956).
\newblock Consistency of the maximum likelihood estimator in the presence of infinitely many incidental parameters.
\newblock {\em Annals of Mathematical Statistics}, 27(4):887--906.

\bibitem[Kriegeskorte et~al., 2009]{kriegeskorte2009}
Kriegeskorte, N., Simmons, W.~K., Bellgowan, P. S.~F., and Baker, C.~I. (2009).
\newblock Circular analysis in systems neuroscience: The dangers of double dipping.
\newblock {\em Nature Neuroscience}, 12(5):535--540.

\bibitem[Levins, 1966]{levins1966}
Levins, R. (1966).
\newblock The strategy of model building in population biology.
\newblock {\em American Scientist}, 54(4):421--431.

\bibitem[Lo and MacKinlay, 1990]{lo_mackinlay1990}
Lo, A.~W. and MacKinlay, A.~C. (1990).
\newblock Data-snooping biases in tests of financial asset pricing models.
\newblock {\em Review of Financial Studies}, 3(3):431--467.

\bibitem[MacKay, 1992]{MacKay1992}
MacKay, D. J.~C. (1992).
\newblock Bayesian interpolation.
\newblock {\em Neural Computation}, 4(3):415--447.

\bibitem[Matheson and Winkler, 1976]{matheson1976}
Matheson, J.~E. and Winkler, R.~L. (1976).
\newblock Scoring rules for continuous probability distributions.
\newblock {\em Management Science}, 22(10):1087--1096.

\bibitem[Matthewson and Weisberg, 2009]{matthewson_weisberg2009}
Matthewson, J. and Weisberg, M. (2009).
\newblock The structure of tradeoffs in model building.
\newblock {\em Synthese}, 170(1):169--190.

\bibitem[Mischel, 2008]{mischel2008}
Mischel, W. (2008).
\newblock The toothbrush problem.
\newblock {\em APS Observer}, 21(11).

\bibitem[Mullainathan and Spiess, 2017]{mullainathan_spiess2017}
Mullainathan, S. and Spiess, J. (2017).
\newblock Machine learning: An applied econometric approach.
\newblock {\em Journal of Economic Perspectives}, 31(2):87--106.

\bibitem[{Open Science Collaboration}, 2015]{osc2015}
{Open Science Collaboration} (2015).
\newblock Estimating the reproducibility of psychological science.
\newblock {\em Science}, 349(6251):aac4716.

\bibitem[Peysakhovich and Naecker, 2017]{peysakhovich_naecker2017}
Peysakhovich, A. and Naecker, J. (2017).
\newblock Using methods from machine learning to evaluate behavioral models of choice under risk and ambiguity.
\newblock {\em Journal of Economic Behavior \& Organization}, 133:373--384.

\bibitem[Popper, 1959]{popper1959}
Popper, K.~R. (1959).
\newblock {\em The Logic of Scientific Discovery}.
\newblock Hutchinson, London.
\newblock First published in German as \emph{Logik der Forschung}, Vienna: Julius Springer, 1934.

\bibitem[Samuelson, 1947]{samuelson1947}
Samuelson, P.~A. (1947).
\newblock {\em Foundations of Economic Analysis}.
\newblock Harvard University Press, Cambridge, MA.

\bibitem[Selten, 1991]{selten1991}
Selten, R. (1991).
\newblock Properties of a measure of predictive success.
\newblock {\em Mathematical Social Sciences}, 21(2):153--167.

\bibitem[Stone, 1974]{stone1974}
Stone, M. (1974).
\newblock Cross-validatory choice and assessment of statistical predictions.
\newblock {\em Journal of the Royal Statistical Society, Series B}, 36(2):111--147.

\bibitem[Tversky and Kahneman, 1992]{tversky_kahneman_1992}
Tversky, A. and Kahneman, D. (1992).
\newblock Advances in prospect theory: Cumulative representation of uncertainty.
\newblock {\em Journal of Risk and Uncertainty}, 5(4):297--323.

\bibitem[White, 2000]{white2000}
White, H. (2000).
\newblock A reality check for data snooping.
\newblock {\em Econometrica}, 68(5):1097--1126.

\bibitem[Yarkoni, 2022]{yarkoni2022}
Yarkoni, T. (2022).
\newblock The generalizability crisis.
\newblock {\em Behavioral and Brain Sciences}, 45:e1.

\end{thebibliography}
\clearpage

\onehalfspacing


\end{document}